\documentclass[multi]{cambridge6Atight}

\newcommand{\bookversion}{2}

\newif\ifstandalone
\standalonefalse

\newif\ifpictures
\picturestrue

\usepackage[numbers]{natbib}

\usepackage{url,hyperref}
\usepackage{pdfpages}
\usepackage[utf8]{inputenc}
\usepackage[T1]{fontenc}
  \usepackage{rotating}
  \usepackage{floatpag}
  \rotfloatpagestyle{empty}

\usepackage[Lenny]{fncychap}

\usepackage{suffix}
\renewcommand\chapterauthor[1]{\authortoc{#1}\printchapterauthor{#1}}
\renewcommand\chapterauthor[1]{\printchapterauthor{#1}}
\WithSuffix\newcommand\chapterauthor*[1]{\printchapterauthor{#1}}

\makeatletter
\newcommand{\printchapterauthor}[1]{%
  {\parindent0pt\vspace*{-25pt}%
  \linespread{1.1}\large\scshape#1%
  \par\nobreak\vspace*{35pt}}
  \@afterheading%
}

\makeatother

\usepackage{amsthm}
\usepackage{amsmath}
\usepackage{amssymb}

\newtheorem{theorem}{Theorem}[chapter]
\newtheorem{proposition}{Proposition}[chapter]
\newtheorem{lemma}{Lemma}[chapter]
\newtheorem{corollary}{Corollary}[chapter]

\newtheorem{definition}{Definition}[chapter]
\newtheorem{fact}{Fact}[chapter]
\newtheorem{convention}{Convention}[chapter]
\newtheorem{example}{Example}[chapter]
\newtheorem{remark}{Remark}[chapter]
\newtheorem{problem}{Problem}[chapter]
\newtheorem{property}{Property}[chapter]
\newtheorem{claim}{Claim}[chapter]

\usepackage{mathptmx}
\usepackage{courier}

\usepackage{type1cm}         

\usepackage{graphicx}       
\usepackage{xcolor}      

\usepackage[bottom]{footmisc}

\usepackage[obeyFinal,colorinlistoftodos]{todonotes}

\usepackage{multirow,array,booktabs}
\usepackage{pdflscape}
\usepackage{afterpage}

\usepackage[small,full]{complexity}

\usepackage[norelsize,ruled,algochapter,vlined]{algorithm2e}

\usepackage{dsfont}

\usepackage{cleveref}

\Crefname{algocf}{Algorithm}{Algorithms}
\Crefname{algocf}{Algorithm}{Algorithms}
\Crefname{section}{Sect.}{Sects.}
\Crefname{subsection}{Sect.}{Sects.}
\Crefname{subsubsection}{Sect.}{Sects.}
\Crefname{section}{Section}{Sections}
\Crefname{subsection}{Section}{Sections}
\Crefname{subsubsection}{Section}{Sections}
\Crefname{chapter}{Chap.}{Chaps.}
\Crefname{chapter}{Chapter}{Chapters}
\Crefname{page}{p.}{pp.}
\Crefname{page}{Page}{Pages}
\Crefname{figure}{Fig.}{Figs.}
\Crefname{figure}{Figure}{Figures}
\Crefname{table}{Table}{Tables}
\Crefname{table}{Table}{Tables}
\Crefname{equation}{}{}
\Crefname{equation}{Equation}{Equations}
\Crefname{theorem}{Thm.}{Thms.}
\Crefname{theorem}{Theorem}{Theorems}
\Crefname{property}{Property}{Properties}
\Crefname{property}{Property}{Properties}
\Crefname{problem}{Pb.}{Pbs.}
\Crefname{problem}{Problem}{Problems}
\Crefname{lemma}{Lem.}{Lems.}
\Crefname{lemma}{Lemma}{Lemmata}
\Crefname{corollary}{Cor.}{Cors.}
\Crefname{corollary}{Corollary}{Corollaries}
\Crefname{claim}{Claim}{Claims}
\Crefname{claim}{Claim}{Claims}
\Crefname{corollary}{Corollary}{Corollaries}
\Crefname{proposition}{Prop.}{Props.}
\Crefname{proposition}{Proposition}{Propositions}
\Crefname{definition}{Def.}{Defs.}
\Crefname{definition}{Definition}{Definitions}
\Crefname{claim}{Claim}{Claims}
\Crefname{claim}{Claim}{Claims}
\Crefname{fact}{Fact}{Facts}
\Crefname{fact}{Fact}{Facts}
\Crefname{example}{Ex.}{Exs.}
\Crefname{example}{Example}{Examples}
\Crefname{remark}{Rmk.}{Rmks.}
\Crefname{remark}{Remark}{Remarks}
\Crefname{algorithm}{Alg.}{Algs.}
\Crefname{algorithm}{Algorithm}{Algorithms}

\usepackage{wasysym} 
\usepackage{bookmark}
\usepackage{makeidx}         
\usepackage{multicol}        

\usepackage{tikz}

\usepackage{subcaption}
\usepackage{mathtools}

\usepackage[notion,quotation,paper]{knowledge}

\usepackage{comment} 
\newcommand{\commentAlt}[1]{\ignorespaces} 
\newcommand{\commentLongAlt}[1]{\ignorespaces} 

\newcommand{\Eve}{\textrm{Eve}}
\newcommand{\Adam}{\textrm{Adam}}

\newcommand{\set}[1]{\left\{ #1 \right\}}
\newcommand{\bigO}{O}

\newcommand{\N}{\mathbb{N}}
\newcommand{\Z}{\mathbb{Z}}

\providecommand{\R}{\mathbb{R}}
\newcommand{\Rinfty}{\R \cup \set{\pm \infty}}
\providecommand{\Q}{\mathbb{Q}}
\newcommand{\Qinfty}{\Q \cup \set{\pm \infty}}

\newcommand{\argmax}{\textrm{argmax}}
\newcommand{\argmin}{\textrm{argmin}}

\newcommand{\Op}{\mathbb{O}}

\newcommand{\Prob}{\mathbb{P}} 
\newcommand{\Expectation}{\mathbb{E}} 
\newcommand{\dist}{\mathcal{D}} 
\newcommand{\Dist}{\dist} 
\newcommand{\supp}{\textrm{supp}} 

\newcommand{\Count}{\texttt{Count}}
\newcommand{\Incorrect}{\texttt{Incorrect}}

\newcommand{\game}{\mathcal{G}} 
\renewcommand{\Game}{\game} 
\newcommand{\arena}{\mathcal{A}} 
\newcommand{\Arena}{\arena} 

\newcommand{\col}{\mathfrak{c}} 

\newcommand{\mEve}{\mathrm{Eve}}
\newcommand{\mAdam}{\mathrm{Adam}}
\newcommand{\mRandom}{\mathrm{Random}}
\newcommand{\mMin}{\mathrm{Min}}
\newcommand{\mMax}{\mathrm{Max}}

\newcommand{\vertices}{V} 
\newcommand{\VE}{V_\mEve} 
\newcommand{\VA}{V_\mAdam} 
\newcommand{\VR}{V_\mRandom} 
\newcommand{\VMax}{V_\mMax} 
\newcommand{\VMin}{V_\mMin} 

\newcommand{\ing}{\textrm{In}}
\newcommand{\Ing}{\ing}
\newcommand{\out}{\textrm{Out}}
\newcommand{\Out}{\out}

\newcommand{\Degree}{\texttt{Degree}}
\newcommand{\dest}{\Delta} 

\newcommand{\WE}{W_\mEve} 
\newcommand{\WA}{W_\mAdam} 
\newcommand{\WMin}{W_\mMin} 

\newcommand{\Paths}{\textrm{Paths}} 
\newcommand{\play}{\pi} 
\newcommand{\first}{\textrm{first}} 
\newcommand{\last}{\textrm{last}} 

\newcommand{\mem}{\mathcal{M}} 
\newcommand{\Mem}{\mem} 

\newcommand{\Pre}{\textrm{Pre}} 
\newcommand{\PreE}{\textrm{Pre}_\mEve} 
\newcommand{\AttrE}{\textrm{Attr}_\mEve} 
\newcommand{\AttrA}{\textrm{Attr}_\mAdam} 
\newcommand{\AttrMax}{\textrm{Attr}_\mMax} 
\newcommand{\AttrMin}{\textrm{Attr}_\mMin} 
\newcommand{\rank}{\textrm{rank}}

\newcommand{\Win}{\textrm{Win}} 
\newcommand{\Lose}{\textrm{Lose}} 

\newcommand{\Value}{\textrm{val}}

\newcommand{\ValueMin}{\textrm{val}_\mMin} 
\newcommand{\ValueMax}{\textrm{val}_\mMax}
\newcommand{\val}{\Value} 

\newcommand{\Automaton}{\mathbf{A}} 

\newcommand{\Safe}{\mathtt{Safe}}
\newcommand{\Reach}{\mathtt{Reach}} 
\newcommand{\Buchi}{\mathtt{Buchi}} 
\newcommand{\CoBuchi}{\mathtt{CoBuchi}} 
\newcommand{\Parity}{\mathtt{Parity}} 
\newcommand{\Muller}{\mathtt{Muller}} 
\newcommand{\Rabin}{\mathtt{Rabin}} 
\newcommand{\Streett}{\mathtt{Streett}} 
\newcommand{\MeanPayoff}{\mathtt{MeanPayoff}} 
\newcommand{\DiscountedPayoff}{\mathtt{DiscountedPayoff}}
\newcommand{\Energy}{\mathtt{Energy}}
\newcommand{\TotalPayoff}{\mathtt{TotalPayoff}}
\newcommand{\ShortestPath}{\mathtt{ShortestPath}}
\newcommand{\Sup}{\mathtt{Sup}}
\newcommand{\Inf}{\mathtt{Inf}}
\newcommand{\LimSup}{\mathtt{LimSup}}
\newcommand{\LimInf}{\mathtt{LimInf}}

\newcommand{\decpb}[3][]{\begin{problem}[#1]\hfill\\[-1.7em]\begin{description}     
    \item[\textsc{input:}] {#2}
    \item[\textsc{output:}] {#3}
    \end{description}
  \end{problem}}

\newcommand{\slopefrac}[2]{\leavevmode\kern.1em
  \raise .5ex\hbox{\the\scriptfont0 #1}\kern-.1em
  /\kern-.15em\lower .25ex\hbox{\the\scriptfont0 #2}}

\newcommand{\kEXP}[1][\mathit{k}]{{\ComplexityFont{#1}}\ComplexityFont{EXP}}
\newcommand{\LOGSPACE}{\ComplexityFont{LOGSPACE}}
\newcommand{\logspace}{\LOGSPACE}
\newcommand{\QBF}{\ensuremath{\mathsf{QBF}}}

\usepackage{xspace}
\usetikzlibrary{arrows}
\usetikzlibrary{automata}
\usetikzlibrary{shapes}
\usetikzlibrary{calc}
\usetikzlibrary{patterns}
\usetikzlibrary{shapes.geometric}
\usetikzlibrary{positioning}
\tikzstyle{every node}=[font=\small]
\tikzstyle{eve}=[circle,minimum size=.3cm,draw=gray!90,inner sep=1pt,fill=gray!20,very thick]
\tikzstyle{adam}=[rounded corners=.5,regular polygon,regular polygon
  sides=4,minimum size=.4cm,draw=gray!90,inner sep=1pt,fill=gray!20,very thick]
\tikzstyle{every edge}=[draw,>=stealth',shorten >=1pt]
\tikzstyle{win}=[fill=black!50,draw=white!70!black]
\tikzstyle{lose}=[fill=white,draw=red!70!black]

\tikzstyle{state}=[draw,circle,minimum size=5mm]
\tikzstyle{accepting}=[double]

\colorlet{grey-10}{black!10!white}
\colorlet{grey-20}{black!20!white}
\colorlet{grey-30}{black!30!white}
\colorlet{grey-40}{black!40!white}
\colorlet{grey-50}{black!50!white}
\colorlet{grey-60}{black!60!white}

\colorlet{lgrey-back}{black!10!white}
\colorlet{lgrey-border}{black!40!white}
\colorlet{dgrey-back}{black!30!white}
\colorlet{dgrey-border}{black!80!white}

\colorlet{state-back}{lgrey-back}
\colorlet{state-border}{lgrey-border}

\tikzstyle{grey}=[fill=lgrey-back,draw=lgrey-border]
\tikzstyle{lgrey}=[grey]
\tikzstyle{dgrey}=[fill=dgrey-back,draw=dgrey-border]
\tikzstyle{white}=[fill=white,draw=black]
\tikzstyle{black}=[fill=black,draw=black]

\tikzstyle{@state}=[fill=state-back,draw=state-border,inner sep=0pt,line width=.8pt]
\tikzstyle{square16}=[@state,rectangle,minimum height=16mm,minimum width=16mm]
\tikzstyle{square10}=[@state,rectangle,minimum height=10mm,minimum width=10mm]
\tikzstyle{square5}=[@state,rectangle,minimum height=5mm,minimum width=5mm]
\tikzstyle{square4}=[@state,rectangle,minimum height=4mm,minimum width=4mm]
\tikzstyle{circle6}=[@state,circle,minimum size=6mm]
\tikzstyle{circle4}=[@state,circle,minimum size=4.3mm]
\tikzstyle{diamond7}=[@state,diamond,minimum height=7.5mm,minimum width=7.5mm]
\tikzstyle{diamond5}=[@state,diamond,minimum height=5mm,minimum width=5mm]
\tikzstyle{triangle7}=[@state,isosceles triangle,isosceles triangle apex angle=60,minimum height=7.5mm,minimum width=7.5mm]
\tikzstyle{triangle5}=[@state,isosceles triangle,isosceles triangle apex angle=60,minimum height=5mm,minimum width=5mm]
\tikzstyle{state}=[s-eve]
\tikzstyle{s-eve}=[circle6]
\tikzstyle{s-adam}=[square5]
\tikzstyle{s-random}=[triangle7]
\tikzstyle{s-eve-small}=[circle4]
\tikzstyle{s-adam-small}=[square4]
\tikzstyle{s-random-small}=[triangle5]
\tikzset{node distance=2.5cm}
\tikzset{every node/.style={anchor=base}}

\tikzset{>=latex,bend angle=20}
\tikzstyle{line}=[line width=.6pt]
\tikzstyle{arrow}=[->,line]
\tikzstyle{dblarrow}=[<->,line]
\tikzstyle{invarrow}=[<-,line]
\tikzstyle{initial}=[invarrow]
\tikzset{selfloop/.style={arrow,out={#1-30},in={#1+30},looseness=6}}

\tikzstyle{fillarea}=[line width=.6pt,line join=round]
\newlength{\hatchspread}
\newlength{\hatchthickness}
\tikzset{hatchspread/.code={\setlength{\hatchspread}{#1}},
  hatchthickness/.code={\setlength{\hatchthickness}{#1}}}
\tikzset{hatchspread=3pt,hatchthickness=0.4pt}
\pgfdeclarepatternformonly[\hatchspread]
   {custom north west lines}
   {\pgfqpoint{\dimexpr-2\hatchthickness}{\dimexpr-2\hatchthickness}}
   {\pgfqpoint{\dimexpr\hatchspread+2\hatchthickness}{2\dimexpr\hatchspread+2\hatchthickness}}
   {\pgfqpoint{\dimexpr\hatchspread}{2\dimexpr\hatchspread}}
   {
    \pgfsetlinewidth{\hatchthickness}
    \pgfpathmoveto{\pgfqpoint{\dimexpr-\hatchthickness}{2\dimexpr\hatchspread+\hatchthickness}}
    \pgfpathlineto{\pgfqpoint{\dimexpr\hatchspread+\hatchthickness}{\dimexpr-\hatchthickness}}
    \pgfusepath{stroke}
   }
\tikzstyle{NWlinesarea}=[line width=.6pt,pattern=custom north west lines, hatchspread=4pt]
\tikzstyle{hatcharea}=[NWlinesarea]

\newcommand{\niceloop}[1]{ 
\draw[-,shorten >=0,solid](#1.center) -- (#1.north) to[out=90,in=180] ($(#1.north east)+(0,0.5cm)$); 
\draw[solid] ($(#1.north east)+(0,0.5cm)$) to[out=0,in=0,min distance=0.7cm]  (#1);
}

\newcommand{\sco}{} 
\newcommand{\ma}[1][xshift=0]{ 
\renewcommand{\sco}{#1}
\maA
}
\newcommand{\name}{}
\newcommand{\maA}[1]{
\renewcommand{\name}{#1}
\maB
}

\newcommand{\maB}[3][-NoValue-]{
\expandafter\maC\expandafter{\sco}{\name}{#1}{#2}{#3}
}
\newcommand{\maC}[5]{
\begin{scope}[#1,solid,-,shorten >=0,shorten <=0]
\expandafter\draw (0,0) node[rectangle, minimum height=#4 cm,minimum width=#5 cm,draw] (\name) {};

\begin{scope}[shift={($(0,0)!0.5!(-#5,-#4)$)}]

\foreach\i in {0,...,#5}{
\draw (\i,0) -- (\i,#4);
}
\foreach \j in {0, ..., #4} {
\draw (0,\j) -- (#5,\j);
}
\foreach\i in {1,...,#5}{
\foreach\j in {1,...,#4}{
\draw ($(0,1+#4)+(\i,-\j)-(0.5,0.5)$) node[rectangle, minimum height=1 cm,minimum width=1 cm,draw] (\name -\j -\i) {};
}}

\ifstrequal{#3}{-NoValue-}{
\node (name #2) at ($(-0.5 , 0)!.5!(-0.5 , #4)$) {#2};
}{
\node (name #2) at ($(-0.5 , 0)!.5!(-0.5 , #4)$) {#3};
}

\end{scope}
\end{scope}

}

\begin{document}

\title[From Logic and Automata to Algorithms]{Games on Graphs}
\author{Nathana{\"e}l Fijalkow}

\frontmatter
\maketitle

\ifpictures
\includepdf{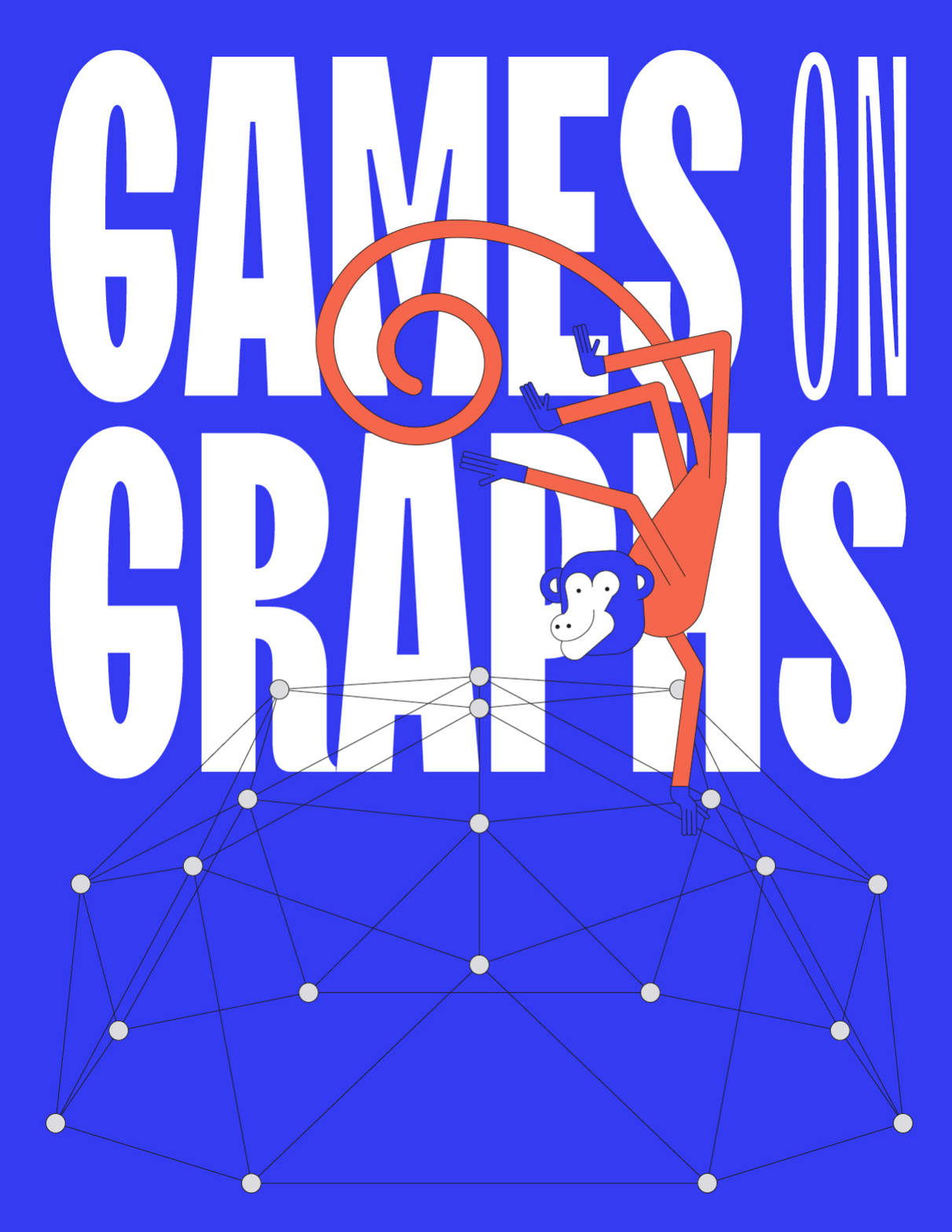}
\fi

\chapter*{Contributors (alphabetical order)}

\vskip-4em
\begin{itemize}
\setlength{\itemsep}{.15em}
\item[] \textbf{C. Aiswarya}	
Chennai Mathematical Institute, India \& CNRS IRL ReLaX

\item[]
\textbf{Guy Avni}	
University of Haifa, Department of Computer Science, Israel

\item[]
\textbf{Nathalie Bertrand}	
Univ of Rennes, Inria, CNRS, IRISA, France

\item[]
\textbf{Patricia Bouyer}	
University Paris-Saclay, CNRS, ENS Paris-Saclay, LMF, France

\item[]
\textbf{Romain Brenguier}	
SonarSource, Switzerland

\item[]
\textbf{Arnaud Carayol}	
Gustave Eiffel University, CNRS, LIGM, France

\item[]
\textbf{Antonio Casares}	
University of Warsaw, MIMUW, Poland

\item[]
\textbf{John Fearnley}
University of Liverpool, United Kingdom

\item[]
\textbf{Nathanaël Fijalkow}			
University of Bordeaux, CNRS, LaBRI, France
	
\item[]
\textbf{Paul Gastin}	
University Paris-Saclay, CNRS, ENS Paris-Saclay, LMF, France \& CNRS IRL ReLaX

\item[]
\textbf{Hugo Gimbert}			
University of Bordeaux, CNRS, LaBRI, France

\item[]
\textbf{Thomas A. Henzinger}		
Institute of Science and Technology Austria, Austria

\item[]
\textbf{Florian Horn} 		
Paris Cité University, CNRS, IRIF, France

\item[]
\textbf{Rasmus Ibsen-Jensen}			
University of Liverpool, United Kingdom

\item[]
\textbf{Nicolas Markey}	
Univ of Rennes, Inria, CNRS, IRISA, France

\item[]
\textbf{Benjamin Monmege}		
Aix-Marseille University, CNRS, LIS, France

\item[]
\textbf{Petr Novotný}
Masaryk University, Faculty of Informatics, Czech Republic

\item[]
\textbf{Pierre Ohlmann}	
Aix-Marseille University, CNRS, LIS, France

\item[]
\textbf{Mickael Randour}
F.R.S.-FNRS \& UMONS -- Université de Mons, Belgium

\item[]
\textbf{Ocan Sankur}		
Univ of Rennes, Inria, CNRS, IRISA, France

\item[]
\textbf{Sylvain Schmitz}
Paris Cité University, CNRS, IRIF, France

\item[]
\textbf{Olivier Serre}
Paris Cité University, CNRS, IRIF, France

\item[]
\textbf{Mateusz Skomra}
University of Toulouse, CNRS, LAAS, France

\item[]
\textbf{Nathalie Sznajder}	
Sorbonne University, CNRS, LIP6, France
			
\item[]
\textbf{Pierre Vandenhove}
UMONS -- Université de Mons, Belgium
\end{itemize}

\tableofcontents

\mainmatter


\chapter*{Preface}

\section*{What is this book about?}
The objective of this book is to give a comprehensive presentation of the research field concerned with infinite duration games on graphs. 
Historically, these game models appeared in the study of automata and logic, and they later became important for program verification and synthesis. 
They have many more applications, in particular some of the models investigated in this book were introduced and studied in neighbouring research fields such as optimisation, reinforcement learning, model theory, and set theory.

The primary objective in this book is algorithmic: constructing efficient algorithms for analysing different types of games.
Yet the goal is not to describe their implementation in full detail but rather to explain their theoretical foundations.
In this endeavour we often need to set the stage by proving properties of the corresponding games and most prominently of their winning strategies. So the language of this book is mathematics.

This book owes a lot to two reference textbooks:
\begin{itemize}
  \item Automata, Logics, and Infinite Games: A Guide to Current Research, edited by Erich Gr{\"{a}}del, Wolfgang Thomas, and Thomas Wilke~\cite{Gradel.Thomas.ea:2002}, and 
  \item Lectures in Game Theory for Computer Scientists, edited by Krzysztof R. Apt and Erich Gr{\"{a}}del~\cite{Apt.Gradel:2011}.
\end{itemize} 

This book does not claim to give a full account of all existing results or models in the literature, which is close to impossible for two reasons: the wealth of existing results and the constant flow of new ones.

\section*{What is this book not about?}
The games we play in this book are by far not the only ones, even among those played on graphs.
A first defining feature of the models we study here is that we use graphs to represent games, whereas in other fields games have been used as a tool for studying graphs. A prominent example is pursuit-evasion games, also called cops and robber games or graph searching games~\cite{Fomin.Thilikos:2008}, which have been used for characterising different measures on graphs~\cite{Thomas:1993}.
Technically, these games are special cases of models studied in this book, but the questions and techniques are very different from what we will do here. 
Another important field within combinatorics is called ``positional games''~\cite{Hefetz.Krivelevich.ea:2014}. Despite also using the word ``positional'', positional games have little connection with the models studied in this book.
A second defining feature of our games is that most of them have infinite duration, meaning they are played over infinitely many steps. This makes them technically very different from many models studied in other branches of algorithmic game theory: for instance a game tree is typically used for representing games of finite duration.

\section*{How to read}
All the material presented in this book is accessible to an advanced master student or a PhD student with a background in computer science or mathematics. The goal is at the same time to present all the basic and fundamental results commonly assumed by the research community working on games on graphs, and most of the latest prominent advances.
We assume familiarity with complexity theory and the notions of graphs and automata but as much as possible do not rely on advanced results in these fields.

The book is divided in five parts each including two to four chapters. At the end of each chapter is a section dedicated to bibliographic references. \Cref{1-chap:introduction} serves as an introduction. After that and to some extent each part is independent. As much as possible we avoid back references but some chapters naturally build on the previous ones in which case we clearly indicate this.

\section*{How to cite}

To cite the whole book, here is a bib item.

\begin{verbatim}
@book{gamesbook,
 title     = {Games on Graphs: 
    From Logic and Automata to Algorithms},
 author    = {Nathanaël Fijalkow and 
    C. Aiswarya and
    Guy Avni and 
    Nathalie Bertrand and 
    Patricia Bouyer and 
    Romain Brenguier and 
    Arnaud Carayol and 
    Antonio Casares and 
    John Fearnley and 
    Paul Gastin and 
    Hugo Gimbert and 
    Thomas A. Henzinger and 
    Florian Horn and 
    Rasmus Ibsen-Jensen and 
    Nicolas Markey and 
    Benjamin Monmege and 
    Petr Novotný and 
    Pierre Ohlmann and 
    Mickael Randour and 
    Ocan Sankur and 
    Sylvain Schmitz and 
    Olivier Serre and 
    Mateusz Skomra and
    Nathalie Sznajder and
    Pierre Vandenhove},
 editor    = {Nathanaël Fijalkow},
 publisher = {Online},
 year 	    = {2025},
}
\end{verbatim}
If you wish to only cite one chapter, here is an example.
\begin{verbatim}
@InCollection{timedgameschapter,
 title     = {Timed Games},
 author    = {Nicolas Markey, Ocan Sankur},
 booktitle = {Games on Graphs: 
    From Logic and Automata to Algorithms},
 editor    = {Nathanaël Fijalkow},
 year      = {2025}
}
\end{verbatim}

\section*{Acknowledgements}

The following people have contributed to early versions of the book in different ways, we thank them warmly for their comments, suggestions, discussions, bug fixes and reports:
Antoine Amarilli,
Hugo Francon,
Pierre Gaillard,
Kristoffer Arnsfelt Hansen,
Johannes Alexander Lehmann,
Th{\'e}o Matricon,
R{\'e}mi Morvan,
Karan Muvvala,
Damian Niwi{\'n}ski,
Wolfgang Thomas,
Georg Zetzsche.

Please send any comments (including typos) you may have to 
\begin{center}
\url{nathanael.fijalkow@gmail.com},
\end{center}
or directly to the relevant authors.

\section*{Versions}

A full version of the book is available on ArXiv. The current document is version \bookversion.
A printed published version will be available in 2025, published by Cambridge University Press.

\section*{Illustrations}

The illustrations (cover and each chapter) were realised by Podpunkt: 
\begin{center}
\url{http://podpunkt.pl/}
\end{center}


\ifpictures
\includepdf{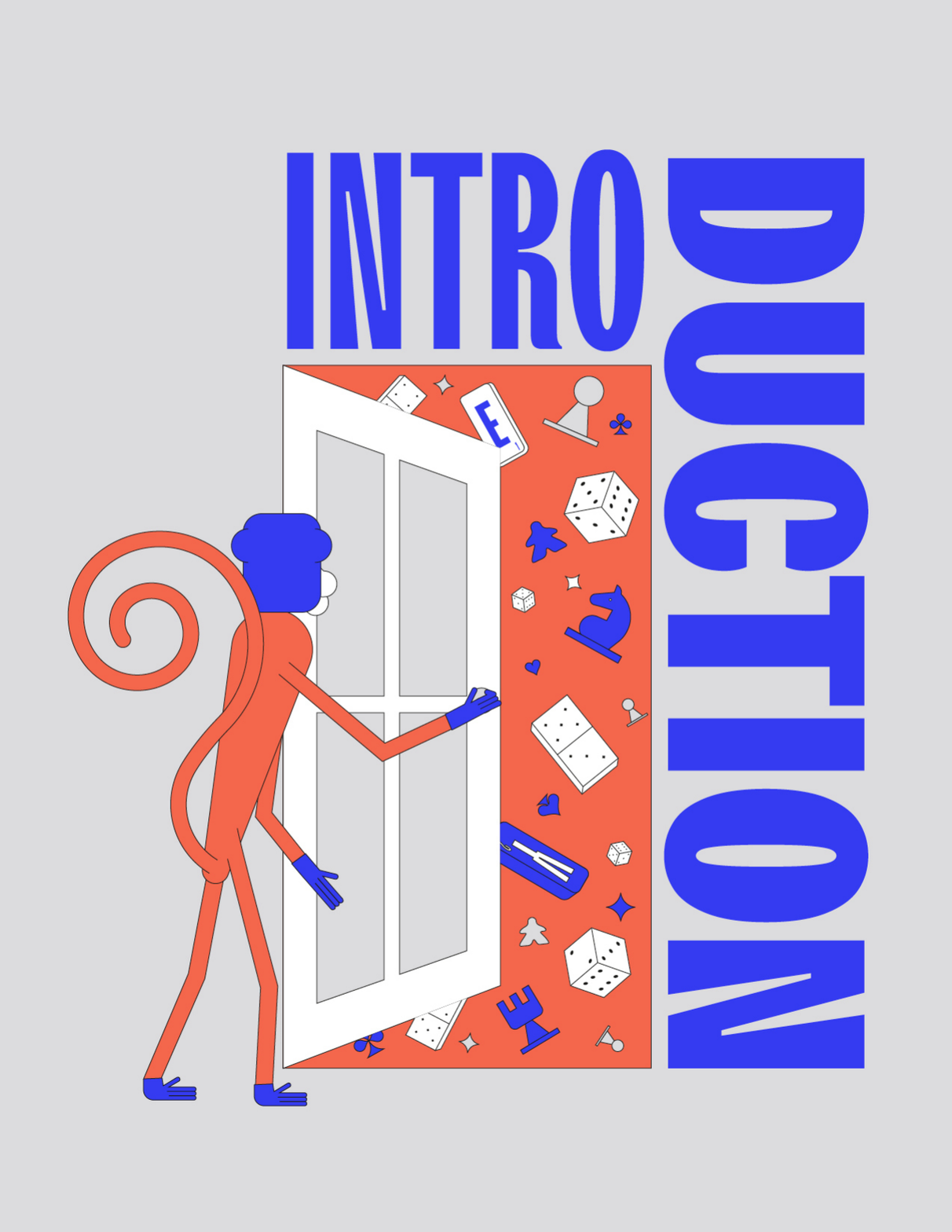}
\fi
\author[Nathana{\"e}l Fijalkow]{Nathana{\"e}l Fijalkow}
\copyrightline{Copyright by Nathana{\"e}l Fijalkow 2025, to be published by Cambridge University Press in the volume \textit{Games on Graphs} edited by Nathana\"el Fijalkow}

\chapter{Introduction}
\chapterauthor{Nathana{\"e}l Fijalkow}
\label{1-chap:introduction}

\newcommand{\GenBuchi}{\mathtt{GenBuchi}}

\Cref{1-sec:arenas} presents the definitions of arenas and strategies, which are common to almost all chapters in this book.
We start with reachability games in~\Cref{1-sec:attractors} and construct a linear time algorithm for solving them: this is the most fundamental result in this book.
Building on this first model,~\Cref{1-sec:qualitative_games} defines the class of qualitative games, which are games where the objective is boolean (win or lose), and define the main qualitative objectives.
We introduce in \Cref{1-sec:subgames} the notions of subgames and traps, and fixed-point algorithms, which are central to the study of games.
We recall the main two methods for proving the existence and computing fixed points in~\Cref{1-sec:fixed_points}.
We then move on to studying B{\"u}chi games in~\Cref{1-sec:buchi}, and construct a quadratic time algorithm for solving them.
Understanding generalized B{\"u}chi games requires the notions of automata and reductions, introduced in~\Cref{1-sec:automata},
which are then complemented by memory in~\Cref{1-sec:memory}.

We then move on to quantitative games in~\Cref{1-sec:quantitative_games}.
The family of value iteration algorithms is defined in general terms in~\Cref{1-sec:value_iteration}.
The other important family of algorithms is called strategy improvement algorithms, discussed in~\Cref{1-sec:strategy_improvement}.
To conclude this chapter, we specify the computational models that we use in~\Cref{1-sec:computation} and briefly discuss linear programming.

\subsection*{Usual notations}
We write $[i,j]$ for the interval $\set{i,i+1,\dots,j-1,j}$, and use parentheses to exclude extremal values,
so $[i,j)$ is $\set{i,i+1,\dots,j-1}$.
An alphabet $\Sigma$ is a finite set. 
We let $\Sigma^*$ denote the set of finite sequences of $\Sigma$ (also called finite words),
$\Sigma^+$ the subset of non-empty sequences, and $\Sigma^\omega$ the set of infinite sequences of $\Sigma$ (also called infinite words).
For a (finite or infinite) sequence $u = u_0 u_1 \cdots$, we let $u_i$ denote the $i$\textsuperscript{th} element of $u$
and $u_{< i}$ the prefix of $u$ of length $i$, \textit{i.e.} the finite sequence $u_0 u_1 \cdots u_{i-1}$.
Similarly $u_{\le i} = u_0 u_1 \cdots u_i$.
The length of $u$ is $|u|$.


\section{Arenas and strategies}
\label{1-sec:arenas}
In this section we define
\begin{itemize}
	\item two-player,
	\item turn-based,
	\item deterministic,
	\item perfect information
\end{itemize}
arenas, and the corresponding strategies.

\subsection*{Players}
The term two-player means that there are two players.
Many, many different names have been used: Eve and Adam, Player $0$ and Player $1$, 
Player I and Player II as in descriptive complexity,
{\'E}lo{\"i}se and Ab{\'e}lard, Circle and Square, corresponding to the graphical representation, 
Even and Odd, mostly for parity objectives, Player and Opponent, Pathfinder and Verifier in the context of automata,
Max and Min, which makes sense for quantitative objectives, and this is only a very partial list of names they have been given.
In the names Eve and Adam, the first letters refer to $\exists$ and $\forall$ suggesting a duality between them.
We will use the pronouns she / her for Eve and he / him for Adam, so we can speak of her or his strategy.
In this book, we will use Eve and Adam in qualitative contexts (win / lose), and Max and Min in quantitative contexts.

We speak of one-player games when there is only one player.
In the context of stochastic games, we refer to random as a third player, and more precisely as half a player.
Hence a two-and-a-half-player game is a stochastic game with two players,
and a one-and-a-half-player game is a stochastic game with one player.
The situation where there are more than two players is called multiplayer games.

\subsection*{Graphs}
We consider directed graphs:

\begin{definition}[Graphs]
\label{1-def:graphs}
A graph is a tuple $G = (V,E, \ing, \out)$ where $V$ is a set of vertices, $E$ is a set of edges, and $\ing,\out : E \to V$ define the incoming and outgoing vertices of edges.
\end{definition}

We say that $e$ is an outgoing edge of~$\ing(e)$ and an incoming edge to~$\out(e)$.
To introduce an edge, it is convenient to write $e : v \to v'$ to express that $v = \ing(e)$ and $v' = \out(e)$.
A path $\pi$ is a (non-empty) finite or infinite sequence:
\[
\pi = v_0 \rightarrow v_1 \rightarrow v_2 \rightarrow \cdots
\] 
We let $\first(\pi)$ denote the first vertex occurring in $\pi$ and $\last(\pi)$ the last one if $\pi$ is finite.
We say that $\pi$ starts from $\first(\pi)$ and if $\pi$ is finite that $\pi$ ends in $\last(\pi)$.
We write $\pi_{\le i}$ for the finite path $v_0 \rightarrow v_1 \rightarrow \cdots \rightarrow v_i$.
We sometimes talk of a path and let the context determine whether it is finite or infinite.
Note that a single vertex can be seen as a path.

We let $\Paths(G)$ denote the set of finite paths in the graph $G$, sometimes omitting $G$ when clear from the context.
To restrict to paths starting from $v$ we write $\Paths(G,v)$.
The set of infinite paths is $\Paths_\omega(G)$, and $\Paths_\omega(G,v)$ for those starting from~$v$.
We naturally see sets of infinite paths as subsets of $E^\omega$.

We use the standard terminology of graphs: for instance a vertex $v'$ is a successor of $v$ if there exists $e : v \rightarrow v' \in E$, 
and then $v$ is a predecessor of $v'$, a vertex $v'$ is reachable from $v$ if there exists a path starting from $v$ and ending in $v'$,
the out-degree of a vertex is its number of outgoing edges,
the in-degree is its number of incoming edges,
a simple path is a path with no repetitions of vertices,
a cycle is a path whose first and last vertex coincide,
it is a simple cycle if it does not strictly contain another cycle,
a self loop is an edge from a vertex to itself,
and a sink is a vertex with only self loops as outgoing edges.

\begin{remark}[Edges as subsets of pairs of vertices]
\label{1-rmk:edges_subsets_pairs}
Our definition with $\ing,\out : E \to V$ includes graphs where there can be multiple edges between a single pair of vertices.
Although this can be useful in some cases, we will also sometimes simply define the set of edges as $E \subseteq V \times V$,
which implies that an edge is fully determined by incoming and outgoing vertices.
\end{remark}

\begin{convention}
We assume that all vertices have an outgoing edge.
\end{convention}
This is for technical convenience, as it implies that we do not need to explain what happens when a path cannot be prolonged.

\subsection*{Arenas}
The arena is the place where the game is played, they have also been called game structures or game graphs.
In the turn-based setting we define here, the set of vertices is divided into vertices controlled by each player.
Since we are for now interested in \textit{two-player} games, 
we have $V = \VE \uplus \VA$, where $\VE$ is the set of vertices controlled by Eve and $\VA$ the set of vertices controlled by Adam.

\begin{definition}[Arenas]
\label{1-def:arenas}
An arena is a tuple $\arena = (G,\VE,\VA)$ where $G$ is a graph over the set of vertices $V$ and $V = \VE \uplus \VA$.
\end{definition}

We represent vertices in $\VE$ by circles, and vertices in $\VA$ by squares, and also say that $v \in \VE$
belongs to Eve, or that Eve owns or controls $v$, and similarly for Adam.
An arena is given by a graph and the sets $\VE$ and $\VA$.
In the context of games, vertices are often referred to as positions.

\vskip1em
The adjective \textit{finite} means that the graph is finite, \textit{i.e.} there are finitely many vertices and edges.
Unless otherwise stated we assume that graphs are finite.
We equivalently say that the arena or the game is finite.
\Cref{part:infinite} will study games over infinite graphs.

We oppose \textit{deterministic} to \textit{stochastic}: in the first definition we are giving here, 
there is no stochastic aspect in the game.
An important assumption, called \textit{perfect information}, says that the players see everything about how the game
is played out, in particular they see the other player's moves.

\begin{figure}
\centering
  \begin{tikzpicture}[scale=1.3]
    \node[s-eve] (init) at (0,0) {$v$};
    \node[s-eve] (v1) at (2,0) {$v_1$};
    \node[s-eve] (v2) at (4,0) {$v_2$};    
    \node[s-adam] (v3) at (6,0) {$v_3$};
    \node[s-adam] (v4) at (0,-1.5) {$v_4$};
    \node[s-eve] (v5) at (2,-1.5) {$v_5$};
    \node[s-adam] (v6) at (4,-1.5) {$v_6$};    
    \node[s-eve] (v7) at (6,-1.5) {$v_7$};
    \path[arrow]
      (init) edge (v1)
      (init) edge[bend left] (v4)
      (v4) edge[bend left] (init)
      (v4) edge node[above] {} (v5)
      (v5) edge[bend left] (v6)
      (v6) edge (v7)
      (v6) edge[bend left] (v5)
      (v5) edge[selfloop=90]  (v5)
      (v7) edge[selfloop]  (v7)
      (v1) edge (v2)
      (v1) edge (v6)
      (v2) edge (v6)
      (v2) edge[bend left] (v3)
      (v3) edge[bend left] (v2)
      (v3) edge (v7);
  \end{tikzpicture}
\caption{An example of an arena. In a qualitative context, circles are controlled by Eve and squares by Adam.}
\label{1-fig:arena_example}
\commentAlt{Figure~\ref{1-fig:arena_example}: A directed graph with 8 nodes, labeled v to v7. Nodes v, v1, v2, v5, and v7 are circles. Nodes v3, v4, and v6 are squares. See long description.} 
\commentLongAlt{Figure~\ref{1-fig:arena_example}: A directed graph illustrating transitions between states. Node 'v' (circle) points to 'v1' (circle), and 'v4' (square) points to 'v'. 'v' and 'v4' have a bidirectional edge. 'v1' points to 'v2' (circle) and 'v6' (square). 'v2' points to 'v3' (square) and 'v6'. 'v3' points to 'v2' and 'v7' (circle). 'v4' points to 'v5' (circle). 'v5' has a self-loop and points to 'v6'. 'v6' points to 'v5' and 'v7'. 'v7' has a self-loop.}
\end{figure}

\subsection*{Playing}
The interaction between the two players consists in moving a token on the vertices of the arena.
The token is initially on some vertex.
When the token is in some vertex~$v$, the player who controls the vertex chooses an edge $e : v \to v'$
and pushes the token along this edge to the next vertex $v'$.
The outcome of this interaction is the sequence of vertices and edges traversed by the token: it is a path. 
In the context of games a path is also called a play and as for paths usually written~$\play$.
We note that plays can be finite (but non-empty) or infinite.

\subsection*{Strategies}
The most important notion in this book is that of \textit{strategies} (sometimes called policies).
A strategy for a player is a full description of their moves in all situations.
Formally, a strategy is a function mapping finite plays to edges: 
\[
\sigma : \Paths \to E.
\]
Similarly, a strategy from $v$ is $\sigma : \Paths(v) \to E$.

\begin{remark}
In some game models there is a set of actions (which may depend on the set of vertices), and the strategy picks an action.
This is strictly equivalent, but it introduces another parameter: the number of actions, which is not the same as the number of edges.
Stochastic games are traditionally defined using actions, we will follow this convention in this book.
\end{remark}

\begin{convention}
We use $\sigma$ for strategies of Eve, and $\tau$ for strategies of Adam, 
so when considering a strategy $\sigma$ it is implicitly for Eve,
and similarly $\tau$ is implicitly a strategy for Adam.
\end{convention}

We say that a play $\play = v_0 \rightarrow v_1 \rightarrow \cdots$ is consistent with a strategy $\sigma$ if
for all $i$ such that $v_i \in \VE$ we have $\sigma(\play_{\le i}) = v_i \rightarrow v_{i+1}$.
The definition is easily adapted for strategies of Adam.

Once an initial vertex $v$ and two strategies $\sigma$ and $\tau$ have been fixed, 
there exists a unique infinite play starting from $v$ and consistent with both strategies, it is written~$\pi^{v}_{\sigma,\tau}$.
Note that the existence follows from our convention that all vertices have an outgoing edge.

\subsection*{Positional strategies}
A strategy can be a very complicated object, in particular it is infinite: in order to choose the next move the strategy considers everything played so far, meaning that the strategy depends upon the whole play.
For understanding a certain class of games a great insight is often to prove that simple strategies are (in some sense) enough.
A prominent class of simple strategies is \textit{positional} strategies: they carry no memory about the play constructed so far and in choosing an edge only look at the current vertex. The word memoryless is sometimes used in lieu of positional.
Formally, a positional strategy for Eve is a function 
\[
\widehat{\sigma} : \VE \to E.
\]
A positional strategy induces a strategy by $\sigma(\pi) = \widehat{\sigma}(\last(\pi))$.
In practice, we identify $\sigma$ and $\widehat{\sigma}$.


\section{Reachability games}
\label{1-sec:attractors}
\begin{figure}
\centering
  \begin{tikzpicture}[scale=1.3]
    \node[s-eve] (v0) at (0,0) {$v_0$};
    \node[s-adam] (v1) at (2,0) {$v_1$};
    \node[s-adam] (v2) at (4,0) {$v_2$};    
    \node[s-eve] (v3) at (6,0) {$v_3$};
    \node[s-adam] (v4) at (0,-1.5) {$v_4$};
    \node[s-adam] (v5) at (2,-1.5) {$v_5$};
    \node[s-eve] (v6) at (4,-1.5) {$v_6$};    
    \node[s-eve, win] (v7) at (6,-1.5) {$v_7$};
    
    \draw[dotted,rounded corners=4mm,fill=grey-10]
    ($(v0)+(-5mm,0)$) |- ($(v5)+(5mm,-5mm)$) -- ($(v1)+(5mm,5mm)$) -| ($(v0)+(-5mm,0)$);
    \draw[dotted,rounded corners=5mm,fill=grey-20]
    ($(v2)+(-5mm,0)$) |- ($(v7)+(13mm,-5mm)$) -- ($(v3)+(13mm,5mm)$) -| ($(v2)+(-5mm,0)$);

    \node[s-eve] (v0) at (0,0) {$v_0$};
    \node[s-adam] (v1) at (2,0) {$v_1$};
    \node[s-adam] (v2) at (4,0) {$v_2$};    
    \node[s-eve] (v3) at (6,0) {$v_3$};
    \node[s-adam] (v4) at (0,-1.5) {$v_4$};
    \node[s-adam] (v5) at (2,-1.5) {$v_5$};
    \node[s-eve] (v6) at (4,-1.5) {$v_6$};    
    \node[s-eve, win] (v7) at (6,-1.5) {$v_7$};

    \path[arrow]
      (v0) edge[bend right] (v1)
      (v0) edge[bend left] (v4)
      (v1) edge[bend right] (v0)
      (v1) edge (v2)
      (v1) edge (v4)
      (v2) edge[bend left] (v3)
      (v3) edge[bend left] (v2)
      (v2) edge (v7)
      (v3) edge (v7)
      (v4) edge[bend left] (v0)
      (v4) edge (v5)
      (v5) edge[selfloop=90]  (v5)
      (v5) edge[bend left] (v6)
      (v6) edge (v2)
      (v6) edge[bend left] (v5)
      (v7) edge[selfloop] (v7);
  \end{tikzpicture}
\caption{A reachability game. Eve tries to reach the vertex $v_7$, and Adam fights to prevent this from happening. The two dotted areas represent the winning regions of each player.}
\label{1-fig:reachability_game_example}
\commentAlt{Figure~\ref{1-fig:reachability_game_example}: A directed graph with 8 nodes, v0 to v7, split into two shaded regions. See long description.}
\commentLongAlt{Figure~\ref{1-fig:reachability_game_example}: A directed graph with two main regions. The left region (lighter shade, dotted border) contains nodes v0 (circle), v1 (square), v4 (square), and v5 (square). v0 and v1 have bidirectional edges. v0 and v4 have bidirectional edges. v1 points to v4. v4 points to v5. v5 has a self-loop. The right region (darker shade, solid border) contains nodes v2 (square), v3 (circle), v6 (circle), and v7 (darker circle). v2 and v3 have bidirectional edges. v3 points to v7. v6 points to v2. v2 points to v7. v7 has a self-loop. A single edge connects the two regions: v1 points to v2. There are also bidirectional edges between v5 and v6.}
\end{figure}

In a reachability game, the target is a set of vertices: the goal of Eve is to ensure that the play visits some vertex from the target set.
We refer to~\Cref{1-fig:reachability_game_example} for an example of a reachability game.
Formally, we let $\Win \subseteq V$ and define the reachability condition
\[
\Reach(\Win) = \set{\pi = v_0 \rightarrow v_1 \rightarrow v_2 \dots \in \Paths_\omega : \exists i \in \N, v_i \in \Win}.
\]
A game equipped with a reachability condition is called a reachability game.
We will see in~\Cref{1-sec:qualitative_games} that a more natural way of talking of reachability games is by defining the class of reachability objectives.

Since we consider zero-sum games, a play which is not winning is losing: this means that if Eve has a reachability condition, then Adam has a safety condition. Formally:
\[
\Safe(\Win) = \set{\pi = v_0 \rightarrow v_1 \rightarrow v_2 \dots \in \Paths_\omega : \forall i \in \N, v_i \notin \Win}.
\]
Reachability and safety conditions are dual: $\Paths_\omega \setminus \Reach(\Win) = \Safe(\Win)$.



\begin{theorem}[Positional determinacy and complexity of reachability games]
\label{1-thm:reachability}
Reachability objectives are uniformly positionally determined: for all reachability games $\game$, there is a positional strategy $\sigma$ for Eve winning from all vertices of $\WE(\game)$ and a positional strategy $\tau$ for Adam winning from all vertices of $\WA(\game)$.
There exists an algorithm for computing the winning regions of reachability games in linear time and space,
which also outputs uniform positional winning strategies for both players.
More precisely the time and space complexity are both $O(m)$.
\end{theorem}

As we will discuss in the proof, the positional determinacy result holds for infinite arenas at the price of additional technical details.

The complexity results are stated in the unit cost RAM model with machine word size $w = \log(m)$ with $m$ the number of edges.
We refer to \Cref{1-sec:computation} for more details about the model, which is in subtle ways different from the Turing model.
The complexity would be slightly different in the Turing model: an additional $\log(m)$ factor would be incurred for manipulating numbers of order $m$, which the unit cost RAM model allows us to conveniently hide.

In the literature the complexity $O(n + m)$ is often reported for solving reachability games.
Since we make the assumption that every vertex has an outgoing edge this implies that $n \le m$, so $O(n + m) = O(m)$.

\vskip1em
Let us introduce some notations.
For a subset $X \subseteq V$, we let $\PreE(X) \subseteq V$ the set of vertices from which Eve can ensure 
that the next vertex is in~$X$:
\[
\begin{array}{lll}
\PreE(X) & = & \set{u \in \VE : \exists u \xrightarrow{} v \in E, v \in X} \\
        & \cup & \set{u \in \VA : \forall u \xrightarrow{} v \in E,\ v \in X}.
\end{array}
\]
We define $\AttrE^0(\Win) = \emptyset$ and
\[
\AttrE^{k+1}(\Win) = \Win\ \cup\ \PreE(\AttrE^{k}(\Win)).
\]
This constructs a sequence $(\AttrE^{k}(\Win))_{k \in \N}$ of non-decreasing subsets of $V$.
If the game is finite and $n$ is the number of vertices, 
the sequence stabilises after at most $n$ steps, \textit{i.e.} $\AttrE^{n+1}(\Win) = \AttrE^{n}(\Win)$, and we write $\AttrE(\Win)$ for the limit, called the attractor. The following lemma shows how the attractor yields a solution to reachability games and directly implies \Cref{1-thm:reachability}.

\begin{lemma}[Characterisation of the winning region of reachability games using attractors]
\label{1-lem:reachability}
Let $\game$ a reachability game.
Then $\WE(\game) = \AttrE(\Win)$, and:
\begin{itemize}
	\item there exists a positional strategy $\sigma$ called the attractor strategy defined on $\AttrE(\Win)$
	which ensures to reach $\Win$ from any vertex in $\AttrE(\Win)$, 
	with the property that for any $k \in \N$ all plays consistent with $\sigma$ from $\AttrE^{k}(\Win)$ reach $\Win$ within $k-1$ steps 
	and remain in $\AttrE(\Win)$ until doing so;
	\item there exists a positional strategy $\tau$ called the counter-attractor strategy defined on $V \setminus \AttrE(\Win)$ which ensures never to reach $\Win$ from any vertex in $V \setminus \AttrE(\Win)$,
	with the property that all plays consistent with $\tau$ remain in $V \setminus \AttrE(\Win)$.
\end{itemize}
\end{lemma}

The following definition is very important: for $v \in V$, the rank of $v$ is the smallest $k \in \N$ such that $v \in \AttrE^{k}(\Win)$.

\begin{proof}
We first show that $\AttrE(\Win) \subseteq \WE(\game)$. 
We use the rank to define a positional strategy $\sigma$.
Let $u \in \VE$ of rank $k+1$, then $u \in \PreE(\AttrE^{k}(\Win))$, 
so there exists $u \xrightarrow{} v \in E$ such that $v \in \AttrE^{k}(\Win)$, 
define $\sigma(u) = u \xrightarrow{} v$.
If $u \in \VE$ has rank $1$, meaning $u \in \Win$, the game is already won.

We argue that $\sigma$ ensures $\Reach(\Win)$.
By construction in any play consistent with $\sigma$ at each step either we are in $\Win$ or the rank decreases by at least one.
Thus any play consistent with $\sigma$ from $\AttrE(\Win)$ reaches $\Win$.

\vskip1em
We now show that $\WE(\game) \subseteq \AttrE(\Win)$.
For this we actually show 
\[
V \setminus \AttrE(\Win) \subseteq \WA(\game).
\]
Indeed, $\WE(\game) \cap \WA(\game) = \emptyset$, because Eve and Adam cannot have a winning strategy from the same vertex.
So $\WA(\game) \subseteq V \setminus \WE(\game)$.

We define a positional strategy $\tau$ from $V \setminus \AttrE(\Win)$.
Let $u \in \VA$ in $V \setminus \AttrE(\Win)$, there exists $u \xrightarrow{} v \in E$ such that $v \in V \setminus \AttrE(\Win)$, 
define $\tau(u) = u \xrightarrow{} v$.
Similarly, if $u \in \VE$ in $V \setminus \AttrE(\Win)$, then for all $u \xrightarrow{} v \in E$, we have $v \in V \setminus \AttrE(\Win)$.
It follows that any play consistent with $\tau$ remains in $V \setminus \AttrE(\Win)$ hence never reaches $\Win$,
in other words $\tau$ ensures $\Safe(\Win)$ from $V \setminus \AttrE(\Win)$.
\end{proof}

It is instructive (especially in light of further developments) to understand the attractor computation as a fixed-point algorithm.
Let us define an operator on subsets of vertices:
\[
X \mapsto \Win \cup \PreE(X).
\]
We note that this operator is monotonic when equipping the powerset of vertices with the inclusion preorder:
if $X \subseteq X'$ then $\PreE(X) \subseteq \PreE(X')$.
Hence Kleene fixed-point theorem~\footnote{See~\Cref{1-sec:fixed_points} for fixed-point theorems and algorithms, and in particular~\Cref{1-thm:kleene}.} applies: this operator has a least fixed point, this is $\AttrE(\Win)$.

\begin{remark}[Extension of the argument to infinite reachability games]
As explained, if the game has $n$ vertices the sequence stabilises after at most $n$ steps.
Let us drop the finiteness assumption: if the game is infinite but has finite out-degree, meaning that for any vertex there is a finite number of outgoing edges, then the operator above preserves suprema so thanks to Kleene fixed-point theorem we have $\AttrE(\Win) = \bigcup_{k \in \N} \AttrE^k(\Win)$.
In full generality the operator does not preserve suprema and the use of ordinals is necessary:
we define the sequence $(\AttrE^{\alpha}(\Win))$ indexed by ordinals up to the cardinal of $\Game$,
the case of a limit ordinal $\alpha$ being $\AttrE^{\alpha}(\Win) = \bigcup_{\beta < \alpha} \AttrE^{\beta}(\Win)$.
We then show that $\AttrE(\Win)$ is the union of all elements in this sequence.
We do not elaborate further this most general case but note that the overhead is mostly technical, the proof above of \Cref{1-lem:reachability} can be adapted with little changes using a transfinite induction.
\end{remark}

\subsection*{The linear time attractor computation algorithm}
\Cref{1-algo:reachability} is an efficient implementation of the attractor computation, and more precisely it computes the ranks of all vertices: it returns a function $\mu : V \to \N \cup \set{\infty}$ such that $\mu(u)$ is the rank of $u$, as stated in the following theorem.

\begin{theorem}[Computing ranks for reachability games]
\label{1-thm:reachability_ranks}
There exists an algorithm for computing the ranks of all vertices in reachability games in linear time and space.
More precisely the time and space complexity are both $O(m)$.
\end{theorem}


The correctness of the algorithm hinges on the following invariant:
for $i \ge 1$, before the $i$\textsuperscript{th} iteration in the \texttt{Main} function,
\begin{itemize}
	\item $\mu$ has correctly computed the ranks of vertices strictly less than $i$,
	\item $\Incorrect$ is the set of vertices of rank $i-1$,
	\item for each $v \in \VA$, $\Count(v)$ is the number of outgoing edges of $v$ to vertices of ranks larger than or equal to $i$.
\end{itemize}
The function \texttt{Init} ensures these properties for $i = 1$.
To see that the invariant is preserved, note that each vertex $v$ is updated at most once, and therefore each edge $u \xrightarrow{} v$
is considered at most once, so $\Count$ is correctly updated.
To get the overall $O(m)$ complexity, we note that each vertex $v$ is updated at most once over the course of the algorithm,
implying that each edge is considered at most once.


\begin{remark}[RAM versus Turing models of computation]
\label{1-rmk:RAM}
We note that in the complexity analysis the cost of manipulating (and in particular incrementing) the counters for the number of edges is constant, which holds in the unit cost RAM model of computation.
As discussed above, the same algorithm analysed in the Turing model of computation would have an additional $O(\log(m))$ multiplicative factor in the time complexity to take this into account.
\end{remark}

\begin{algorithm}
 \KwData{A reachability game}
 \SetKwFunction{FInit}{Init}
 \SetKwFunction{FTreat}{Treat}
 \SetKwFunction{FUpdate}{Update}
 \SetKwFunction{FMain}{Main}
 \SetKwProg{Fn}{Function}{:}{}
 \DontPrintSemicolon

\Fn{\FInit{}}{
	\For{$u \in V$}{
		$\mu(u) \leftarrow \infty$

		\If{$u \in \VA$}{
			$\Count(u) \leftarrow \Degree(u)$
		}
	}	

	\For{$u \in \Win$}{
   		Add $u$ to $\Incorrect$
	}

%
%
}

\vskip1em
\Fn{\FUpdate{$v$}}{	
	\For{$u \xrightarrow{} v \in E$}{
		\If{$u \in \VA$}{
	        $\Count(u) \leftarrow \Count(u) - 1$
        
	        \If{$\Count(u) = 0$}{
    	    	Add $u$ to $\Incorrect'$
	        }	
		}

		\If{$u \in \VE$}{
			Add $u$ to $\Incorrect'$	
		}
	}
}

\vskip1em
\Fn{\FMain{}}{
	\FInit()    

	\For{$i = 1,2,\dots$}{
		$\Incorrect' \leftarrow \emptyset$

		\For{$v \in \Incorrect$}{

			$\mu(v) \leftarrow i$    

			\FUpdate($v$)    
		}
		\If{$\Incorrect' = \emptyset$}{

			\Return{$\mu$}
		}
		\Else{

			$\Incorrect \leftarrow \Incorrect'$
		}
	}
}
\caption{The linear time algorithm for reachability games.}
\label{1-algo:reachability}
\end{algorithm}

\section{Qualitative games}
\label{1-sec:qualitative_games}
Now that we have covered reachability games, we give more general definitions and introduce qualitative games, where a play is either won or lost. 
Later in this chapter we will consider quantitative games.
A qualitative condition $W$ separates winning from losing plays, in other words a play which belongs to $W$ is winning and otherwise it is losing. We also say that the play satisfies~$W$.

\begin{definition}[Qualitative conditions]
\label{1-def:qualitative_condition}
A qualitative condition is $W \subseteq \Paths_\omega$.
\end{definition}
As an example, we define the reachability condition as
\[
\Reach(\Win) = \set{\pi = v_0 \rightarrow v_1 \rightarrow v_2 \dots \in \Paths_\omega : \exists i \in \N, v_i \in \Win}.
\]
where $\Win \subseteq V$.
Often we define $W$ as a subset of $E^\omega$ since $\Paths_\omega$ can be seen as a subset of $E^\omega$.

\begin{definition}[Qualitative games]
\label{1-def:qualitative_games}
A qualitative game $\game$ is a tuple $(\arena,W)$ where $\arena$ is an arena and $W$ a qualitative condition.
\end{definition}

Now that we have the definitions of a qualitative game we can ask the main question: 
given a qualitative game $\game$ and a vertex $v$, who wins $\game$ from $v$?
A strategy $\sigma$ is called winning from $v$ if every play starting from $v$ consistent with $\sigma$ is winning, 
\textit{i.e.} satisfies $W$.
Another common terminology is that $\sigma$ ensures $W$.
In that case we say that Eve has a winning strategy from $v$ in $\game$, or equivalently that Eve wins $\game$ from $v$.
This vocabulary also applies to Adam: for instance 
a strategy $\tau$ is called winning from $v$ if every play starting from $v$ consistent with $\tau$ is losing, 
\textit{i.e.} does not satisfy $W$.

We let $\WE(\game)$ denote the set of vertices $v$ such that Eve wins $\game$ from $v$,
it is called winning region, or sometimes winning set. 
A vertex in $\WE(\game)$ is said winning for Eve.
The analogous notation for Adam is $\WA(\game)$.
We say that a strategy is optimal if it is winning from all vertices in $\WE(\game)$.

\subsection*{Computational problems}
We identify three computational problems for qualitative games.
The first is that of solving a game, which is the simplest one and since it induces a decision problem, allows us 
to make complexity theoretic statements.

\decpb[Solving the game]{A qualitative game $\game$ and a vertex $v$}{Does Eve win $\game$ from $v$?}

The second problem extends the previous one: most algorithms solve games for all vertices at once instead of only for the given initial vertex.
This is called computing the winning regions.

\decpb[Computing the winning regions]{A qualitative game $\game$}{Compute $\WE(\game)$ and $\WA(\game)$}

The third problem is constructing a winning strategy.

\decpb[Constructing a winning strategy]{A qualitative game $\game$ and a vertex $v$}{Construct a winning strategy for Eve from $v$}

Note that we did not specify how the games, winning regions, and winning strategies are represented: this will depend on the types of games we consider.

\subsection*{Objectives}
To reason about classes of games with the same conditions, we introduce the notions of objectives and colouring functions.
An objective and a colouring function together induce a condition.
The main point is that \textit{objectives are independent of the arenas}, so we can speak of the class of conditions induced by a given objective,
and by extension a class of games induced by a given objective, for instance reachability games.

We fix a set $C$ of colours. 
A qualitative objective is $\Omega \subseteq C^\omega$.
The link between an arena and an objective is given by a colouring function $\col : E \to C$ labelling edges of the graph by colours.
We extend $\col$ component wise to induce $\col : \Paths_\omega \to C^\omega$ mapping plays to sequences of colours:
\[
\col(v_0 \rightarrow v_1 \rightarrow \dots) = \col(v_0 \rightarrow v_1)\ \col(v_1 \rightarrow v_2) \dots
\]
A qualitative objective $\Omega$ and a colouring function $\col$ induce a qualitative condition $\Omega(\col)$ defined by:
\[
\Omega(\col) = \set{\play \in \Paths_\omega : \col(\play) \in \Omega}.
\] 
When $\col$ is clear from the context we sometimes say that a play $\play$ satisfies $\Omega$ but the intended meaning is that 
$\play$ satisfies $\Omega(\col)$, equivalently that $\col(\play) \in \Omega$.

We extend the notation $v \to v'$ in the presence of a colouring function: 
$e : v \xrightarrow{c} v'$ further indicates that $\col(e) = c$.
As we will see, a convenient abuse of notations consists in identifying a colour $c$ and the set of edges of that colour $\col^{-1}(c)$.

Let us return to the case of reachability and safety.
The reachability objective is defined over the set of colours $\set{0,1}$:
\[
\Reach = \set{\rho \in \set{0,1}^\omega : \exists i \in \N, \rho_i = 1},
\]
and the safety objective over the same set of colours:
\[
\Safe = \set{\rho \in \set{0,1}^\omega : \forall i \in \N, \rho_i = 0}.
\]
Reachability and safety objectives are dual: $\set{0,1}^\omega \setminus \Reach = \Safe$.

Now, to define the reachability condition $\Reach(\Win)$ with $\Win \subseteq V$, we define the colouring function
$\col(v \rightarrow v') = 1$ if $v \in \Win$ and $0$ otherwise.
Note that here it would have been more natural to define a colouring function on vertices, but colouring edges is more general than colouring vertices and will be useful for other objectives. Let us discuss this subtlety now.

As a last example, let us introduce the most important qualitative objective in this book: parity objectives.
The set of colours is $[1,d]$, which we call priorities in this context. 
We see $d$ as a parameter, the total number of priorities.
\[
\Parity = \set{\rho \in [1,d]^\omega \mid \text{ the largest priority appearing infinitely often in } \rho \text{ is even}}.
\]
We note that the parity objectives can be defined using either ``largest'' or ``smallest'', sometimes called ``max-parity'' and ``min-parity''.
This is strictly equivalent, as one can go from one convention to the other one by applying a transformation $p \mapsto d - p$ (assuming $d$ even).
Depending on the context, one variant can be technically more convenient than the other in terms of notations.

\subsection*{Labelling edges or vertices}

In our definition the colouring function labels edges: $\col : C \to E$.
It is sometimes convenient to label vertices instead: $\col : C \to V$, and in some cases it may feel more natural: for instance for reachability conditions as discussed above, the target is naturally a set of vertices rather than a set of edges.
But labelling edges is often technically easier, and more succinct, so it is the preferred convention in this book.
In addition to labelling vertices or edges, there is another aspect: allowing partial colouring functions, \textit{i.e.} allowing some vertices to remain uncoloured, or not.
To keep things well defined, if we consider partial colouring functions we make the assumption that every path contains infinitely many coloured vertices.
There are two consequences to these conventions: the complexity of algorithms, and the memory requirements,
both (often mildly) affected by the choice.

Let us state simple and general reduction results between the four conventions.
Clearly,
\[
\text{vertex labelling} \subseteq \text{partial vertex labelling} \subseteq \text{edge labelling} \subseteq \text{partial edge labelling}.
\]
Indeed, to go from (partial) vertex labelling to edge labelling it is enough to define
$\col'(e) = \col(\ing(e))$: the colour of an edge is the colour of its incoming vertex.
A path induces the same sequence of colours for both $\col$ and $\col'$, implying a strong equivalence between the two games.

Conversely, we can reduce from edge labelling to partial vertex labelling:

\begin{lemma}[From labelling edges to labelling vertices]
\label{1-lem:reduction_label_vertices_edges}
Let $G$ a graph with $n$ vertices and $m$ edges, $\col : E \to C$ an edge colouring function, we can construct a graph $G'$ with $n+m$ vertices and $2m$ edges and a \emph{partial} vertex colouring function $\col' : V' \to C$ such that there is a one to one correspondence between infinite paths in $G$ coloured by $\col$ and infinite paths in $G'$ coloured by $\col'$.
\end{lemma}

As a consequence, an algorithm using edge colouring functions of complexity $T(n,m)$ induces an algorithm for partial vertex colouring functions of complexity $T(n+m, 2m)$.

\begin{proof}
We add edges as intermediate vertices. 
Both the sets of vertices and of edges of $G'$ are (disjoint copies of) $V \cup E$.
For each edge $e : v \to v'$ of $G$, we add two edges in $G'$: one from $v$ to $e$, and one from $e$ to $v'$.
The colouring function $\col'$ is defined only on $E$ by $\col'(e) = \col(e)$.
The construction is illustrated in~\Cref{1-fig:reduction_edge_colouring}.
Infinite paths in $G$ and in $G'$ are in one to one correspondence, and the sequences of colours are the same, since colours appear exactly every second position in every infinite path of $G'$.
\end{proof}

\begin{figure}[!ht]
\centering
  \begin{tikzpicture}[scale=1.3]
    \node[s-eve] (v0) at (0,0)   {$v$};
    \node[s-eve] (v1) at (1.5,-0.05) {$v'$};
	\node        (t)  at (2.75,0)   {\Large becomes}; 
	
    \node[s-eve] (v0b) at (4,0) {$\begin{array}{c} v \\ \varepsilon \end{array}$};
    \node[s-eve] (e) at (5.5,0) {$\begin{array}{c} e \\ c \end{array}$};
    \node[s-eve] (v1b) at (7,0) {$\begin{array}{c} v' \\ \varepsilon \end{array}$};
    \path[arrow]
      (v0) edge node[above] {\LARGE $c$} (v1)
      (v0b) edge (e)
      (e) edge (v1b);
  \end{tikzpicture}
\caption{Reduction from edge colouring to vertex colouring.}
\label{1-fig:reduction_edge_colouring}
\commentAlt{Figure~\ref{1-fig:reduction_edge_colouring}: A diagram showing a transformation from a direct edge to a sequence of edges. See long description.}
\commentLongAlt{Figure~\ref{1-fig:reduction_edge_colouring}: The left side of the diagram shows a directed edge from node 'v' to node 'v'' with a label 'C' above the arrow. The right side, separated by the word "becomes", shows a sequence of three nodes connected by two directed edges. The first node is labeled 'v' above and 'epsilon' below. The first arrow points to a middle node labeled 'e' above and 'c' below. The second arrow points from the middle node to a final node labeled 'v'' above and 'epsilon' below.}
\end{figure}

There are no easy reductions from partial vertex labelling to vertex labelling, and the same for edge labelling, but in most cases an ad-hoc simple trick proves the two conventions to be (essentially) equivalent. For instance, if the objective contains a neutral letter, then the construction above can be replicated:
for an objective $\Omega \subseteq C^\omega$, we say that $\varepsilon \in C$ is a neutral letter if 
for all $\rho \in C^\omega$, let $\rho_{\varepsilon}$ the sequence obtained from $\rho$ by removing all $\varepsilon$,
then $\rho \in \Omega$ if and only if $\rho_{\varepsilon}$ is finite or in $\Omega$.


\subsection*{Determinacy and (bi-)positional determinacy}

\begin{fact}[Winning regions are disjoint]
\label{1-fact:winning_regions_disjoint}
For all qualitative games $\game$ we have $\WE(\game) \cap \WA(\game) = \emptyset$.
\end{fact}
\begin{proof}
Assume for the sake of contradiction that both players have a winning strategy from $v$, say $\sigma$ and $\tau$, then $\pi^{v}_{\sigma,\tau}$
would both satisfy $W$ and not satisfy $W$, a contradiction.
\end{proof}

It is however not clear that for every vertex $v$, \textit{some} player has a winning strategy from $v$.
One might imagine that if Eve picks a strategy, then Adam can produce a counter strategy beating Eve's strategy, 
and vice versa, if Adam picks a strategy, then Eve can come up with a strategy winning against Adam's strategy.

\begin{definition}
We say that a qualitative game $\game$ is determined in $v$ if either Eve or Adam has a winning strategy for $v$,
and it is determined whenever this holds for all vertices, equivalently: $\WE(\game) \cup \WA(\game) = V$.
A qualitative objective $\Omega$ is determined if all games with objective $\Omega$ are determined.
\end{definition}
Being determined can be understood as follows: the outcome can be determined before playing since one of them can ensure to win whatever is the strategy of the opponent.

\begin{theorem}[Borel determinacy]
\label{1-thm:borel_determinacy}
Borel objectives are determined.
\end{theorem}

The definition of Borel sets goes beyond the scope of this book. 
Suffice to say that all objectives studied in this book are (very simple) examples of Borel sets,
implying that our qualitative games are all determined 
(as long as we consider perfect information and turn-based games, the situation will change with more general models of games).

\vskip1em
\begin{definition}
We say that a qualitative objective $\Omega$ is \textit{positional} if 
for every game $\game$ with objective $\Omega$ and every vertex $v$,
if Eve has a winning strategy from $v$, then she has a positional winning strategy from~$v$.
\end{definition}

We say that $\Omega$ is \textit{positionally determined} if it is additionally determined, meaning that 
for every game $\game$ with objective $\Omega$ and every vertex $v$,
either Eve has a positional winning strategy from $v$, or Adam has a winning strategy from $v$.
In this (very typical) case we speak of a positional determinacy result.

\begin{definition}
We say that a qualitative objective $\Omega$ is \textit{bi-positional} if 
for every game $\game$ with objective $\Omega$ and every vertex $v$:
\begin{itemize}
	\item if Eve has a winning strategy from $v$, then she has a positional winning strategy from~$v$,
	\item if Adam has a winning strategy from $v$, then he has a positional winning strategy from~$v$.
\end{itemize}
\end{definition}
The definition simplifies for \textit{bi-positionally determined}: for every game $\game$ with objective $\Omega$ and every vertex $v$,
either of the two players has a positional winning strategy from $v$.
In this case we speak of a bi-positional determinacy result.

\begin{remark}
Another popular terminology replaces ``positionally determined'' and ``bi-positionally determined'' by ``half-positionally determined'' and ``positionally determined''. The best argument for the terminology we adopt in this book is that the main notion is positionality, which applies to one objective.
\end{remark}

\begin{remark}
Positional determinacy may hold only for some class of arenas (such as finite arenas or arenas with finite out-degree), in which case this is made explicit: positionally determined over finite arenas.
\end{remark}

\subsection*{Reasoning about positional strategies}

For reasoning about positional strategies it is useful to define the following notion.
Let $\Game$ be a game and $\sigma$ a positional strategy, we define $\Game[\sigma]$ the graph induced by $\sigma$ on $\Game$.
The set of vertices is $V$ and the set of edges is
\[
E[\sigma] = \set{e : v \to v' \in E : v \in \VA \text{ or } \left( v \in \VE \text{ and } \sigma(v) = e \right)}.
\]

\begin{fact}[Game induced by a positional strategy]
\label{1-fact:game_induced_positional_strategy}
Let $\Game$ be a game with condition $W$, $\sigma$ a positional strategy, and $v$ a vertex.
Then the strategy $\sigma$ is winning from $v$ if and only if all infinite paths in $\Game[\sigma]$ from $v$ satisfy $W$.
\end{fact}

As we discussed earlier, the task of solving a game does not include constructing winning strategies.
We present a general binary search technique for doing so for positional objectives.

\begin{lemma}[Binary search for constructing positional strategies]
\label{1-lem:constructing_winning_strategy}
Let $\Omega$ be a positional qualitative objective.
If there exists an algorithm $A$ for solving games with objective $\Omega$,
then there exists an algorithm for constructing positional winning strategies for Eve for games in this class 
using $O(n \cdot \log(\frac{m}{n}))$ calls to the algorithm~$A$.
\end{lemma}

\begin{proof}
Let $\Omega$ be a positional objective, $\Game$ a game with objective $\Omega$, and $v_0$ and initial vertex.
We first determine whether $v_0 \in \WE(\Game)$, which requires one call to a solving algorithm.
Now for each vertex $v \in \VE$ we need to define $\sigma(v)$.
We let $d(v)$ denote the out-degree of $v$.
We choose a subset of $\lfloor \frac{d(v)}{2} \rfloor$ outgoing edges of $v$, construct the game where we remove these edges,
and solve it using $v_0$ as initial vertex.
If Eve wins this game from $v_0$, then there is a positional winning strategy that picks one of the remaining outgoing edges of~$v$,
otherwise the same holds for the removed edges.
This binary search algorithm requires $O(\log(d(v)))$ calls to a solving algorithm for finding a winning positional move from $v$.
Doing so for all vertices requires 
\[
O\left( \sum_{v \in V} \log d(v) \right) = O\left(n \cdot \log \left( \frac{m}{n} \right) \right)
\]
calls to a solving algorithm.
\end{proof}

\subsection*{Uniformity}

When defining positionality, we need to further append one adjective: a qualitative objective $\Omega$ is uniformly positional if for every game $\game$ with objective $\Omega$, there exists a single positional winning strategy from $\WE(\game)$.
We define similarly being uniformly positionally determined, and bi-positional.

Being uniformly positional is a stronger property than being positional, but in most cases an objective satisfies either both or none. 

\subsection*{Prefix-independent qualitative objectives}

To get familiar with the notions introduced above, we prove a few useful properties.

A qualitative objective $\Omega$ is:
\begin{itemize}
	\item \textit{closed under adding prefixes} if for every finite sequence $\rho$ and for every infinite sequence $\rho'$,
if $\rho' \in \Omega$ then $\rho \rho' \in \Omega$;
	\item \textit{closed under removing prefixes} if for every finite sequence $\rho$ and for every infinite sequence $\rho'$,
if $\rho \rho' \in \Omega$ then $\rho' \in \Omega$;
	\item \textit{prefix-independent} if it is closed under both adding and removing prefixes;
in other words whether a sequence satisfies $\Omega$ does not depend upon finite prefixes.
\end{itemize}

Let $\play$ be a finite play consistent with $\sigma$, we write $\sigma_{\mid \play}$ for the strategy defined by
\[
\sigma_{\mid \play}(\play') = \sigma(\play \play').
\]

\begin{fact}[Winning for conditions closed under removing prefixes]
\label{1-fact:winning_prefix_independent_qualitative}
Let $\Game$ be a qualitative game with objective $\Omega$ closed under removing prefixes, $\sigma$ a winning strategy from $v$,
and $\play$ a finite play consistent with $\sigma$ starting from $v$.
Then $\sigma_{\mid \play}$ is winning from $v' = \last(\play)$.
\end{fact}
\begin{proof}
Let $\play'$ be an infinite play consistent with $\sigma_{\mid \play}$ from $v'$,
then $\play \play'$ is an infinite play consistent with $\sigma$ starting from $v$, 
implying that it is winning, and since $\Omega$ is closed under removing prefixes
the play $\play'$ is winning. Thus $\sigma_{\mid \play}$ is winning from~$v'$.
\end{proof}

\begin{corollary}[Reachable vertices of a winning strategy for objectives closed under removing prefixes]
\label{1-cor:reachable_vertices_prefix_independent}
Let $\Game$ be a qualitative game with objective $\Omega$ closed under removing prefixes and $\sigma$ a winning strategy from $v$.
Then all vertices reachable from $v$ by a play consistent with $\sigma$ are winning.
\end{corollary}
In other words, when playing a winning strategy the play does not leave the winning region.

\begin{lemma}[From positional to uniformly positional prefix-independent objectives]
\label{1-lem:from_positional_to_uniformly_positional}
Let $\Omega$ be a prefix-independent qualitative objective.
\begin{itemize}
	\item If $\Omega$ is positional, then it is uniformly positional.
	\item If $\Omega$ is bi-positional, then it is uniformly bi-positional.
\end{itemize}
\end{lemma}

\begin{proof}
Let us consider a game $\game$ with a qualitative prefix-independent objective $\Omega$.
For each $v \in \WE(\game)$, let $\sigma_v$ be a positional winning strategy from $v$.
Thanks to~\Cref{1-fact:winning_prefix_independent_qualitative} (here we use the closure under removing prefixes for $\Omega$),
the strategy $\sigma_v$ is winning from all vertices reachable by a play consistent with $\sigma_v$ starting from~$v$.
Without loss of generality let us assume that $\sigma_v$ is only defined on these vertices.

If $\game$ has a finite state space, we fix $\le$ a total order on the set of vertices\footnote{The argument we give in this proof extends to infinite games whose set of vertices can be well ordered, which always exists assuming the axiom of choice. A well-order is a total order such that every non-empty subset has a least element, which is exactly the property we need in this proof.}.
We let $\sigma$ be the positional strategy defined by $\sigma(u)$ is $\sigma_v(u)$ where $v$ is the least vertex (with respect to $\le$) such that $\sigma_v$ is defined on $u$. We say that $\sigma$ uses $\sigma_v$ at $u$.

We argue that $\sigma$ is winning from $\WE(\game)$. 
Consider a play consistent with $\sigma$ starting from some vertex in $\WE(\game)$ and look at the sequence of strategies it uses.
This sequence is non-increasing (with respect to $\le$), hence it is eventually constant (this argument extends naturally to well-orders).
In other words the play is eventually consistent with some strategy $\sigma_v$, implying that this suffix satisfies $\Omega$.
Since $\Omega$ is closed under adding prefixes this means that the play itself satisfies $\Omega$, so $\sigma$ is indeed winning.
\end{proof}

\section{Subgames and traps}
\label{1-sec:subgames}
Before we can continue our journey and study B{\"u}chi games, we need to introduce a technical but very useful notion: traps, which enable decomposing games, for instance for inductive reasoning. They will be central in many algorithms.

Let us consider a game $\Game$ and a set $X$ of vertices, the goal is to define the subgame induced by $X$.
Without any assumption it may violate our requirement that every vertex has an outgoing edge.
So let us assume that for every $v \in X$ there exists $v \to v' \in E$ such that $v' \in X$.
Now we can define the game $\Game[X]$ by restricting $\Game$ to the vertices in $X$
and say that $\Game[X]$ is the subgame of $\Game$ induced by $X$.
Formally, the arena is $\Arena[X]$ with $X$ the set of vertices and $E[X]$ is the set of edges 
such that both incoming and outgoing vertices are in $X$.
The assumption on $X$ ensures that every vertex of $\Game[X]$ has an outgoing edge.
Conditions defined on $\Game$ naturally induce conditions on $\Game[X]$.
However, the two games $\Game$ and $\Game[X]$ can be very different, and it is not clear whether a strategy ensuring a condition in $\Game[X]$ still ensures that condition in $\Game$. This is why we introduce traps.

\subsection*{Traps}
We say that $X$ is a trap for Adam if
\begin{itemize}
	\item for every $u \in X \cap \VE$, there exists $u \to v$ with $v \in X$, and 
	\item for every $u \in X \cap \VA$, for all $u \to v$, we have $v \in X$.
\end{itemize}
Intuitively, a trap for Adam is a subset of vertices which Eve can decide to stay in while Adam cannot force to leave.
The same notion can be defined for Eve.
Traps satisfy the property above so if $X$ is a trap then the subgame $\Game[X]$ described above is well defined, meaning 
that every vertex has an outgoing edge.
But they satisfy much stronger properties:

\begin{fact}
\label{1-fact:traps_winning}
Let $\Game$ be a game and $X$ a trap for Adam.
If $\sigma$ is a winning strategy for Eve in the subgame $\Game[X]$, then it induces a winning strategy in $\Game$ from the same vertices.
\end{fact}

\begin{proof}
Any play consistent with $\sigma$ in $\Game$ stays forever in $X$ because $X$ is a trap for Adam, hence is winning.
\end{proof}

The notion of traps is very useful when decomposing games.
We present some simple facts about traps, here stated for Adam but easily transposed for Eve.

\begin{fact}[Traps]
\label{1-fact:traps}
Let $\Game$ a game.
\begin{itemize}
	\item Let $P,Q$ two traps for Adam in $\Game$, then $P$ is a trap for Adam in $\Game[Q]$ (but $P \cap Q$ may not be a trap in $\Game$).
	\item Let $P$ a trap for Eve in $\Game$ and $Q$ a trap for Adam in the game $\Game$, 
	then $P \cap Q$ is a trap for Eve in $\Game[Q]$.
	\item Let $P$ a trap for Adam in $\Game$ and $Q$ a trap for Adam in $\Game[P]$,
	then $Q$ is a trap for Adam in $\Game$.
\end{itemize}
\end{fact}

\subsection*{Attractors induce traps}
Additionally to solving reachability games, the notion of attractors induces a common way of constructing traps and subgames.

\begin{fact}
\label{1-fact:attractors_induce_traps}
Let $\Game$ a game and $F$ a subset of vertices.
Then $V \setminus \AttrE(F)$ is a trap for Eve, and symmetrically $V \setminus \AttrA(F)$ is a trap for Adam.
\end{fact}

We introduce a useful notation: 
$\Game \setminus \AttrE(F)$ is the subgame $\Game[V \setminus \AttrE(F)]$, which is well defined thanks to the above.

\section{Fixed-point algorithms}
\label{1-sec:fixed_points}
Fixed-point algorithms are ubiquitous in the study of games, in particular they will be useful in our study of B{\"u}chi games in the next section.

Let $X$ be a set and $\Op : X \to X$ a function that we call an operator, we say that $x \in X$ is a fixed point of $\Op$ if $\Op(x) = x$.
Fixed points will appear extensively in this book. 
We describe here two general approaches for computing them: Kleene and Banach fixed-point theorems.

%

\subsection*{Kleene fixed-point theorem}
Let us consider a lattice $(X,\le)$: the binary relation $\le$ is a partial order, and every pair of elements has a least upper bound and a greatest lower bound. It is a complete lattice if every set has a least upper bound and a greatest lower bound.
A lattice is finite if the set $X$ is finite.
Note that finite lattices are always complete.

We write $\bot$ for the least element in $X$ and $\top$ for the greatest element.
An operator $\Op : X \to X$ is monotonic if for all $x,y \in X$ such that $x \le y$ we have $\Op(x) \le \Op(y)$,
and preserves suprema if $\Op(\sup_n x_n) = \sup_n \Op(x_n)$ for all increasing sequences $(x_n)_{n \in \N}$.
The twin notion 
preserves infima is defined accordingly.

We say that $x$ is a pre-fixed point if $\Op(x) \le x$ and a post-fixed point if $\Op(x) \ge x$.

\begin{theorem}[Kleene fixed-point theorem]
\label{1-thm:kleene}
Let $(X,\le)$ be a "complete lattice" and $\Op : X \to X$ a monotonic operator, then $\Op$ has a least fixed point
which is also the least pre-fixed point.

Furthermore:
\begin{itemize}
	\item if $X$ is finite the sequence defined by $u_0 = \bot$ and $u_{k+1} = \Op(u_k)$ is stationary and its limit is the least fixed point of $\Op$;
	\item if $\Op$ preserves suprema then the least fixed point of $\Op$ is $\sup \set{ \Op^k(\bot) :  k \in \N}$.
\end{itemize}
\end{theorem}

Under the same assumptions $\Op$ has a greatest fixed point which is the greatest post-fixed point and can be computed in similar ways,
replacing ``preserves suprema'' by ``preserves infima''.

The typical example of a "complete lattice" and one that we will use often in this book is the powerset of a set equipped with the inclusion between subsets. The least element is the empty set, the greatest element the full set, and least and greatest upper bounds are given by union and intersection.
An example of an infinite "complete lattice" is $\Rinfty$ equipped with the natural order.

%
%
%
%
%
%
%

\subsection*{Banach fixed-point theorem}

Let us consider a set $X$ equipped with some norm $\|\cdot\|$.
It is called a complete space if all Cauchy sequences converge.
The typical example of a complete space is $\R^d$ for some $d \in \N$ equipped with the infinity norm $\|x\| = \max_{i \in [1,d]} |x_i|$.

An operator $\Op : X \to X$ is contracting if there exists $\lambda < 1$ such that for all $x,y \in X$ we have
$\|\Op(x) - \Op(y)\| \le \lambda \cdot \|x - y\|$.

\begin{theorem}[Banach fixed-point theorem]
\label{1-thm:banach}
Let $(X,\|\cdot\|)$ be a complete space and $\Op : X \to X$ a contracting operator, then $\Op$ has a unique fixed point $x_*$.
For any $x_0 \in X$, the sequence $(\Op^k(x_0))_{k \in \N}$ converges towards $x_*$ and the rate of convergence is given by
\[
\|\Op^k(x_0) - x_*\| \le \frac{\lambda^k}{1 - \lambda} \cdot \|\Op(x_0) - x_0\|.
\]
\end{theorem}


\section{B{\"u}chi games}
\label{1-sec:buchi}
Let us study a second class of qualitative games: B{\"u}chi games.
In a B{\"u}chi game, the goal of Eve is to ensure that the play visits some vertex from a target set infinitely many times.
Formally, we define the objective $\Buchi$ over the set of colours $\set{1,2}$ as:
\[
\Buchi = \set{\rho \in \set{1,2}^\omega : \forall i \in \N, \exists j \ge i, \rho_j = 2}.
\]
Given a colouring function $\col : E \to \set{1,2}$, we interpret $\Buchi(\col)$ as follows: it is satisfied by paths visiting edges from $\col^{-1}(2)$ infinitely many times. 
It is customary to write $\Win = \col^{-1}(2)$ and talk about $\Buchi(\Win)$. In some cases, $\Win$ is a subset of vertices, which is a special case of our definitions: formally, this is equivalent to $\Buchi(\col)$ for $\col(v \rightarrow v') = 2$ if $v \in \Win$, and $1$ otherwise.
A game equipped with a B{\"u}chi condition is called a B{\"u}chi game.

Since we consider zero-sum games, a play which is not winning is losing: this means that if Eve has a B{\"u}chi objective, then Adam has a CoB{\"u}chi objective. Formally:
\[
\CoBuchi = \set{\rho \in \set{1,2}^\omega : \exists j \in \N, \forall i \ge j, \rho_i = 1}.
\]
B{\"u}chi and CoB{\"u}chi conditions are dual: $\set{1,2}^\omega \setminus \Buchi = \CoBuchi$.

\begin{theorem}[Positional determinacy and complexity of B{\"u}chi games]
\label{2-thm:Buchi}
B{\"u}chi objectives are uniformly positionally determined: for all B{\"u}chi games $\game$, there is a positional strategy $\sigma$ for Eve winning from all vertices of $\WE(\game)$ and a positional strategy $\tau$ for Adam winning from all vertices of $\WA(\game)$.
There exists an algorithm for computing the winning regions of B{\"u}chi games in quadratic time, more precisely $O(n \cdot m)$,
and linear space, more precisely $O(m)$.
\end{theorem}

As for reachability games, the uniform positional determinacy result holds for infinite games, at the price of additional technicalities in the proof. We present two different yet very similar algorithms.

\subsection*{A first algorithm}
The following lemma implies \Cref{2-thm:Buchi}.

\begin{lemma}[Fixed-point characterisation of the winning region for B{\"u}chi games]
\label{2-lem:Buchi}
Let $\Game$ be a B{\"u}chi game.
\begin{itemize}
	\item If $\AttrE(\Win) = V$, then $\WE(\Game) = V$.
	\item If $\AttrE(\Win) \neq V$, 
	let $\Game' = \Game \setminus \AttrA( V \setminus \AttrE(\Win) )$,
	then $\WE(\Game) = \WE(\Game')$.	
\end{itemize}
\end{lemma}

\begin{proof}
We prove the first item. 
Let $\sigma$ be an attractor strategy ensuring to reach $\Win$ from $\AttrE(\Win) = V$.
We argue that $\sigma$ ensures $\Buchi(\Win)$.
Indeed a play consistent with $\sigma$ can be divided into infinitely many finite plays,
each of them consistent with $\sigma$ until reaching $\Win$, and starting from scratch from the next vertex onwards.
Thus $\sigma$ is winning from $V$.

We now look at the second item.
We first prove that $\AttrA(V \setminus \AttrE(\Win)) \subseteq \WA(\Game)$.
Let $\tau_a$ denote an attractor strategy ensuring to reach $V \setminus \AttrE(\Win)$ from $\AttrA(V \setminus \AttrE(\Win))$,
and $\tau_c$ a counter-attractor strategy ensuring to never reach $\Win$ from $V \setminus \AttrE(\Win)$.
We construct the strategy $\tau$ as the disjoint union of $\tau_a$ and~$\tau_c$:
\[
\tau(v) = 
\begin{cases}
\tau_a(v) & \text{ if } v \in \AttrA(V \setminus \AttrE(\Win)) \setminus (V \setminus \AttrE(\Win)), \\
\tau_c(v) & \text{ if } v \in V \setminus \AttrE(\Win).
\end{cases}
\]
Note that $\tau$ is positional.
Any play consistent with $\tau$ is first consistent with $\tau_a$ until reaching $V \setminus \AttrE(\Win)$ and 
then is consistent with $\tau_c$ and stays there forever.
In this second phase it does not visit $\Win$, 
implying that the play visits $\Win$ finitely many times, so it is winning for Adam.
Thus we have proved that $\AttrA(V \setminus \AttrE(\Win)) \subseteq \WA(\Game)$,
implying $\WE(\Game) \subseteq V \setminus \AttrA(V \setminus \AttrE(\Win))$.

Now, consider a winning strategy from $\WE(\Game)$ in $\Game$, thanks to~\Cref{1-cor:reachable_vertices_prefix_independent} it remains in the winning region hence in $V \setminus \AttrA(V \setminus \AttrE(\Win))$. 
Therefore it induces a winning strategy in $\Game'$. This implies that $\WE(\Game) \subseteq \WE(\Game')$.

We now show the converse inclusion: that $\WE(\Game') \subseteq \WE(\Game)$.
Consider a winning strategy from $\WE(\Game')$ in $\Game'$, it induces a winning strategy in $\Game$
because $V \setminus \AttrA(V \setminus \AttrE(\Win))$ is a trap for Adam (\Cref{1-fact:traps_winning}).
\end{proof}

The algorithm is presented in pseudocode in \Cref{2-algo:Buchi_first}.
For the complexity analysis, the algorithm performs at most $n$ recursive calls
and each of them involves two attractor computations, implying the time complexity $O(n \cdot m)$.

\begin{algorithm}
 \KwData{A B{\"u}chi game.}
 \SetKwFunction{FSolve}{Solve}
 \SetKwProg{Fn}{Function}{:}{}
 \DontPrintSemicolon

\Fn{\FSolve{$\Game$}}{
	$X \leftarrow \AttrE(\Win)$

	\If{$X = V$}{
		\Return{$V$}
	}
	\Else{
		Let $\Game' = \Game \setminus \AttrA(V \setminus X)$
		
		\Return{$\FSolve{$\Game'$}$}
	}
}
\caption{The first quadratic time algorithm for solving B{\"uchi} games.}
\label{2-algo:Buchi_first}
\end{algorithm}

\vskip1em
Let us see how uniform positional determinacy follows from~\Cref{2-lem:Buchi}.
The shortest proof is by induction on the number of vertices, and remark that in both cases in~\Cref{2-lem:Buchi} this number decreases.

A more instructive proof proceeds by unfolding the fixed-point computation.
Let $\Game_0 = \Game$ the original game, applying~\Cref{2-algo:Buchi_first} yields a sequence of subgames
$\Game_1,\Game_2,\dots, \Game_p$. 
Let us write $V_k$ for the set of vertices of $\Game_k$. 
Thanks to~\Cref{2-lem:Buchi}, we have $\WE(\Game) = V_p$. 
The proof of~\Cref{2-lem:Buchi} constructs a positional winning strategy in $\Game_p$ from $\WE(\Game)$.
Applying iteratively the third item of~\Cref{1-fact:traps} and~\Cref{1-fact:attractors_induce_traps} shows that 
$\Game_p$ is a trap for Adam in $\Game$.
Thanks to~\Cref{1-fact:traps_winning} $\sigma$ induces a positional winning strategy in $\Game$ from $\WE(\Game)$.

The case of Adam is a bit more complicated. Recall that thanks to~\Cref{2-lem:Buchi}, we have $\WA(\Game) = V \setminus V_p$.
For $v \in V \setminus V_p$, the rank of $v$ is the smallest $k \in \N$ such that $v \in V \setminus V_k$.
Equivalently, $v \in V_{k-1} \setminus V_k$.
By definition, $V_{k+1} = V_k \setminus \AttrA^{\Game_k}(V_k \setminus \AttrE^{\Game_k}(\Win))$.

For each $k$, let $\tau_{a,k}$ denote an attractor strategy ensuring to reach $V_k \setminus \AttrE^{\Game_k}(\Win)$ from $\AttrA^{\Game_k}(V_k \setminus \AttrE^{\Game_k}(\Win))$,
and $\tau_{c,k}$ a counter-attractor strategy ensuring to never reach $\Win$ from $V_k \setminus \AttrE^{\Game_k}(\Win)$.
We construct the strategy $\tau$ in $\Game$ as the disjoint union of all $\tau_{a,k}$ and $\tau_{c,k}$:
\[
\tau(v) = 
\begin{cases}
\tau_{a,k}(v) & \text{if } \rank(v) = k \text{ and } v \notin V_k \setminus \AttrE^{\Game_k}(\Win),\\
\tau_{c,k}(v) & \text{if } \rank(v) = k \text{ and } v \in V_k \setminus \AttrE^{\Game_k}(\Win).
\end{cases}
\]
and argue that it ensures $\CoBuchi(\Win)$. 
Note that $\tau$ is the disjoint union of positional strategies, it is positional.
Consider a play consistent with $\tau$ starting from a vertex of rank $k$.
In the first phase, the play is consistent with $\tau_{a,k}$. If we were playing in $\Game_k$, this would go on until reaching $V \setminus \AttrE^{\Game_k}(\Win)$. But since we are here playing in $\Game$, there is another possibility: that Eve chooses an edge leading outside of $\Game_k$.
In that case necessarily the next vertex is in $\Game_{k-1}$, so it has smaller rank.
In the second phase, which starts upon reaching $V_k \setminus \AttrE^{\Game_k}(\Win)$, the play is consistent with $\tau_{c,k}$, and again two things can happen.
Either the play remains in $\Game_k$, so it is consistent with $\tau_{c,k}$ forever, in which case it never sees $\Win$ and therefore satisfies $\CoBuchi(\Win)$, or it exits $\Game_k$, necessarily to reach $\Game_{k-1}$, meaning that it reaches a vertex of smaller rank.
Along any play consistent with $\tau$ the rank never increases, implying that it is eventually consistent with some $\tau_{c,k}$ hence satisfies $\CoBuchi(\Win)$.

\subsection*{A second algorithm}
The following lemma induces a different algorithm with the same complexity.
We define the operator $\PreE^{\Win}$ on subsets of vertices: for $Y \subseteq V$,
\[
\begin{array}{lll}
\PreE^{\Win}(Y) & = & \set{v \in V_E : \exists e : v \xrightarrow{} v', e \in \Win \text{ and } v' \in Y} \\
& \cup & \set{v \in V_A : \forall e : v \xrightarrow{} v' \in E, e \in \Win \text{ and } v' \in Y}.
\end{array}
\]

\begin{lemma}[Second fixed-point characterisation of the winning region for Buchi games]
\label{2-lem:Buchi_second}
Let $\game$ a B{\"u}chi game.
Then $\WE(\game)$ is the greatest fixed point of the monotonic operator 
\[
Y \mapsto \AttrE \left( \PreE^{\Win}(Y) \right).
\]
\end{lemma}

\begin{proof}
Thanks to \Cref{1-thm:kleene} the greatest fixed point is also the greatest post-fixed point, so we need to show two properties:
\begin{itemize}
	\item $\WE(\Game)$ is a post-fixed point, meaning $\WE(\Game) \subseteq \AttrE \left( \PreE^{\Win}(\WE(\Game)) \right)$.
	\item For all post-fixed points $Y$, we have $Y \subseteq \WE(\Game)$.
\end{itemize}

We first show that $\WE(\Game) \subseteq \AttrE \left( \PreE^{\Win}(\WE(\Game)) \right)$.
We actually show that $V \setminus \AttrE \left( \PreE^{\Win}(\WE(\Game)) \right) \subseteq \WA(\Game)$, implying the inclusion by complementing.

Let $\tau$ be a counter-attractor strategy ensuring never to reach $\PreE^{\Win}(\WE(\Game))$ from $V \setminus \AttrE \left( \PreE^{\Win}(\WE(\Game)) \right)$. 
We additionally require that $\tau$ chooses edges outside of $\Win$ whenever possible:
for each $v \in \VA \cap (V \setminus \AttrE \left( \PreE^{\Win}(\WE(\Game)) \right))$,
if there exists $e : v \xrightarrow{} v'$ with $e \notin \Win$ and $v' \notin \AttrE \left( \PreE^{\Win}(\WE(\Game)) \right)$, define $\tau(v) = e$.

Let $\tau'$ a winning strategy from $\WA(\Game)$.
We play the following strategy: play $\tau'$ from $\WA(\Game)$ and $\tau$ otherwise.
Let us consider a play consistent with this strategy from $V \setminus \AttrE \left( \PreE^{\Win}(\WE(\Game)) \right)$,
and assume that it reaches an edge $v \xrightarrow{\Win} v'$.
If $v \in \VE$, since $v \notin \PreE^{\Win}(\WE(\Game))$ this implies that $v' \in \WA(\Game)$.
If $v \in \VA$, the additional property of $\tau'$ ensures that $v' \in \WA(\Game)$.
Hence after reaching $\Win$ we switch to a winning strategy, so the strategy is winning from $V \setminus \AttrE \left( \PreE^{\Win}(\WE(\Game)) \right)$.

\vskip1em
Let $Y$ a post-fixed point, meaning $Y \subseteq \AttrE \left( \PreE^{\Win}(Y) \right)$.
We show that $\WE(\game) \subseteq Y$.
Let $\sigma_a$ be an attractor strategy ensuring to reach $\PreE^{\Win}(Y)$ from $Y$.
We also define a strategy $\sigma_p$:
for $v \in \VE$, if $v \in \PreE^{\Win}(Y)$ there exists $e : v \xrightarrow{} v' \in E$ such that $e \in \Win$ and $v' \in Y$, let us define $\sigma_p(v) = e$.
We define the strategy $\sigma$ as follows:
\[
\sigma(v) = 
\begin{cases}
\sigma_a(v) & \text{if } v \in \AttrE(\PreE^{\Win}(Y)) \setminus \PreE^{\Win}(Y), \\
\sigma_p(v)	& \text{if } v \in \PreE^{\Win}(Y).
\end{cases}
\]
We argue that $\sigma$ ensures $\Buchi(\Win)$ from $Y$. 
Indeed a play consistent with $\sigma$ can be divided into infinitely many finite plays,
each of them consistent with $\sigma_a$ until reaching $\PreE^{\Win}(Y)$,
then one step consistent with $\sigma_p$ reaching $\Win$, before starting from scratch in~$Y$.
\end{proof}

\Cref{2-lem:Buchi} directly transfers to \Cref{2-algo:Buchi_second}.
We could also obtain uniform positional determinacy from~\Cref{2-lem:Buchi_second}, using a similar unfolding as for \Cref{2-lem:Buchi}.

\begin{algorithm}
 \KwData{A B{\"u}chi game.}

$Y \leftarrow V$ 
    
\Repeat{$Y = \AttrE \left( \PreE^{\Win}(Y) \right)$}{
	$Y \leftarrow \AttrE \left( \PreE^{\Win}(Y) \right)$ 
}

\Return{$Y$}
\caption{The second quadratic time algorithm for solving B{\"uchi} games.}
\label{2-algo:Buchi_second}
\end{algorithm}

\begin{remark}[Comparison of the two algorithms]
\label{2-rmk:comparison_algorithms}
Both algorithms have the same complexity but they are not equivalent: the number of recursive calls of the first algorithm
may be strictly smaller than the number of iterations of the repeat loop in the second algorithm.
Both can be extended into (different) algorithms for parity games and beyond; in this chapter we will work with the first algorithm.
\end{remark}


\section{Automata}
\label{1-sec:automata}
The study of games is deeply intertwined with automata over infinite words and trees.
We will not elaborate much on that aspect in this book, but in a few places we will use automata.
We define here (non-deterministic) automata over infinite words, and refer to~\cite{Thomas:1997}
for a survey on automata theory over infinite objects (words and trees) and logic,
and to~\cite{Pin:2021} for the most recent and complete textbook on automata theory.

\begin{definition}[Automata]
\label{1-def:automata}
Let $\Sigma$ be an alphabet.
An automaton over the alphabet $\Sigma$ is a tuple $\Automaton = (Q,q_0,\Delta,A)$ where:
\begin{itemize}
	\item $Q$ is a finite set of states,
	
	\item $q_0 \in Q$ is the initial state,

	\item $\Delta \subseteq Q \times \Sigma \times Q$ is the transition relation,
	
	\item $A \subseteq \Delta^\omega$ is the acceptance condition.
\end{itemize}
\end{definition}
The size of $\Automaton$ is its number of states, written $|\Automaton|$.
We assume that automata are complete: from any state $q$ and letter $a$,
there exists a transition $(q,a,q') \in \Delta$. 
This mirrors the convention for games that every vertex has an outgoing edge.

We use transition based acceptance conditions instead of state based acceptance conditions, for the same reasons as we use edge labelling colouring functions rather than vertex labelling.
This more succinct definition of automata naturally composes with games as we will see.

\vskip1em
For a (finite or infinite) word $w = w_0 w_1 \dots$, a run $\rho = (q_0,w_0,q_1)(q_1,w_1,q_2) \dots$ over $w$ is a sequence of consecutive transitions starting from the initial state $q_0$.
An infinite run is accepting if it belongs to~$A$, in which case we also say that it satisfies~$A$.
A word~$w$ is accepted if there exists an accepting run over~$w$. 
We let $L(\Automaton)$ denote the set of accepted words and call it the language defined by $\Automaton$,
or sometimes recognised by~$\Automaton$.

\vskip1em
An automaton is deterministic if for all states $q \in Q$ and letter $a \in \Sigma$, there exists at most one transition $(q,a,q') \in \Delta$.
In that case the transition relation becomes a transition function $\delta : Q \times \Sigma \to Q$.
The key property of deterministic automata is that for every word there exists exactly one run over it.

\vskip1em
We use the same approach as for games for defining classes of automata with the same conditions:
an objective $\Omega \subseteq C^\omega$ and a colouring function $\col : \Delta \to C$ 
induce an acceptance condition $\Omega[\col] \subseteq \Delta^\omega$.
For deterministic automata the colouring function becomes $\col : Q \times \Sigma \to C$.
As for games the objective qualifies the automaton, so we speak of a parity automaton if it uses a parity objective.

\begin{theorem}[Omega-regular languages]
\label{1-thm:omega_regular_languages}
Non-deterministic B{\"u}chi, CoB{\"u}chi, parity, and deterministic parity automata define the same class of languages called $\omega$-regular languages.
\end{theorem}

\subsection*{Generalized B{\"u}chi games}

Automata can be used to construct reductions between games.
To illustrate this, we study generalized B{\"u}chi games by reducing them to B{\"u}chi games.
We will define automata reductions in more generality after this example.

The generalized B{\"u}chi objective with parameter $k$ uses as set of colours $\set{1,2}^k$ and is defined by
\[
\GenBuchi = \set{\rho \in (\set{1,2}^k)^\omega : \forall p \in \set{1,\dots,k}, \forall i \in \N, \exists j \ge i, \rho_{j,p} = 2}.
\]
In words: $\GenBuchi$ is a conjunction of $k$ independent B{\"u}chi objectives.
Given a colouring function $\col : E \to \set{1,2}^k$, we define $F_p = \set{e \in E : \col(e)_p = 2}$, so that $\GenBuchi(\col)$ is satisfied by paths which see infinitely many times each $F_1,\dots,F_k$.

A simple way to solve generalized B{\"u}chi games is to construct a deterministic B{\"u}chi automaton recognising $\GenBuchi$.
Intuitively, the automaton reads the colours: in state $i$ it waits for the next occurrence of $F_i$, and moves to $i+1$ when this occurs (or $1$ if $i = k$).
The automaton accepts it is cycles over all $k$ components infinitely many times.
Formally, the set of states is $\set{1,\dots,k}$, the initial state is $1$, the input alphabet is the set of colours $\set{1,2}^k$, the transition function is 
\[
\delta(i,c) = 
\begin{cases}
i+1 & \text{ if } c_i = 2, \\
i & \text{ otherwise,}
\end{cases}
\]
and the acceptance condition is $\Buchi(\Win)$ with $\Win = \set{(k,c,1)}$, meaning the transition cycling back from component $k$ to the first component.

\begin{fact}
The deterministic B{\"u}chi automaton defined above recognises $\GenBuchi$.
\end{fact}

The corollary of this construction is that one can simply reduce generalized B{\"u}chi games to B{\"u}chi games, by composing a generalized B{\"u}chi game with the above automaton, yielding a B{\"u}chi game.
The algorithmic consequence is the following: the B{\"u}chi game has $n \cdot k$ vertices and $m \cdot k$ edges,
so we can solve generalized B{\"u}chi game in time $O(n \cdot m \cdot k^2)$ and space $O(m \cdot k)$.
(We note that a better complexity can be obtained using different techniques.)


\subsection*{Reductions between objectives using automata}
Let $\Omega$ a qualitative objective over the set of colours $C$, and $\Omega'$ a second qualitative objective.
We say that $\Omega$ reduces to $\Omega'$ if there exists a \textit{deterministic} automaton $\Automaton$ over the alphabet $C$ with acceptance objective $\Omega'$ defining $\Omega$, \textit{i.e.} such that $L(\Automaton) = \Omega$.

This implies that we can transform a game $\Game$ with objective $\Omega$ into an equivalent one $\Game \times \Automaton$ with objective $\Omega'$ by composing $\Game$ with $\Automaton$: 
the automaton reads the sequence of colours from $C$ induced by the play and 
produces a new sequence of colours which is accepted if it satisfies $\Omega'$.

Formally, let $\arena$ be an arena and 
\[
\Automaton = (Q,q_0,\delta : Q \times C \to Q,\Omega'(\col_\Automaton))
\]
a deterministic automaton with $\col_\Automaton : Q \times C \to C'$.
We construct the arena $\arena \times \Automaton$ as follows.
We first define the graph $G \times Q$ whose set of vertices is $V \times Q$ and set of edges is defined as follows:
for every edge $e : v \xrightarrow{c} v' \in E$ and state $q \in Q$ there is an edge $e[q] = (v,q) \xrightarrow{\col_\Automaton(q,c)} (v',\delta(q,c))$:
the second component computes the run of $\Automaton$ on the sequence of colours induced by the play.
The arena is $\arena \times \Automaton = (G \times Q, \VE \times Q, \VA \times Q)$.
The game is $\game \times \Automaton = (\arena \times \Automaton, \Omega'(\col'))$. 

The following lemma states two consequences to the fact that $\Omega$ reduces to $\Omega'$.

\begin{lemma}[Automata reductions]
\label{1-lem:automata_reduction}
If $\Omega$ reduces to $\Omega'$ through the automaton $\Automaton$, then 
Eve has a winning strategy in $\Game$ from $v_0$ if and only if she has a winning strategy in $\Game \times \Automaton$ from $(v_0,q_0)$.

Consequently, the following properties hold:
\begin{itemize}
  \item Assume that there exists an algorithm for solving games with objective $\Omega'$ with complexity $T(n,m)$. 
  Then there exists an algorithm for solving games with objectives $\Omega$ of complexity $T(n \cdot |\Automaton|, m \cdot |\Automaton|)$.

  \item If $\Omega'$ is determined with finite memory strategies of size $m$, then $\Omega$ is determined with finite memory strategies of size $m \cdot |\Automaton|$.
\end{itemize}
\end{lemma}
Since the next type of reduction extends this one and the two proofs are very similar we will prove this lemma as a corollary of the next one.

Reductions between objectives using automata are very general: 
they operate at the level of objectives and therefore completely ignore the arena.


\section{Memory}
\label{1-sec:memory}
%
%

Positional strategies are sometimes not enough. 
A more powerful class of strategies is \textit{finite memory} strategies.
Intuitively, a finite memory strategy uses a finite state machine called a memory structure 
to store the relevant pieces of information about the play constructed so far.

To define finite memory strategies formally, we fix a graph $G$.
A memory structure is $\mem = (M, m_0, \delta)$: the set $M$ is a set of (memory) states, 
$m_0 \in M$ is the initial state and $\delta : M \times E \to M$ is the update function.
The update function is extended to $\delta^* : E^* \to M$ by 
$\delta^*(\varepsilon) = m_0$ and $\delta^* (\rho e) = \delta(\delta^*(\rho), e)$.
The size of a memory structure is its number of states.
Note that a memory structure is a deterministic automaton over the alphabet $E$ but without specifying the acceptance condition.

We define a strategy using $\mem$ as a function 
\[
\widehat{\sigma} : \VE \times M \to E.
\]
It induces a strategy $\sigma$ via $\sigma(\pi) = \widehat{\sigma}(\last(\pi), \delta^*(\pi))$.
A common abuse of notations is to write $\sigma$ for $\widehat{\sigma}$.

We note that positional strategies correspond to strategies using the trivial memory structure consisting of only one state.

\vskip1em
We give in \Cref{1-fig:memory_required} an example of a game where Eve has a winning strategy using two memory states
but no positional winning strategy. The condition is $\Buchi(\Win_1) \wedge \Buchi(\Win_2)$, meaning that a play is winning if both $\Win_1$ and $\Win_2$ are visited infinitely many times. A positional strategy would either always choose the edge labelled $\Win_1$ or the one labelled $\Win_2$, hence does not satisfy the condition. Some memory is required to switch between the two.

Formally, let $e_1$ be the edge labelled by $\Win_1$ and $e_2$ be the one labelled by $\Win_2$.
We let $\mem = (\set{m_1,m_2},m_1,\delta)$ defined by 
$\delta(m_1,e_1) = m_2$ and $\delta(m_2,e_2) = m_1$.
In words, $m_1$ is the memory state saying that the last chosen edge was $e_1$ and $m_2$ correspondingly for $e_2$. 
We switch from $m_1$ to $m_2$ when traversing $e_1$ and conversely with $e_2$.
Then we define $\sigma(v_0,m_1) = e_1$ and $\sigma(v_0,m_2) = e_2$ inducing the strategy $\widehat{\sigma}$ using~$\Mem$.

\begin{figure}
\centering
  \begin{tikzpicture}[scale=1.3]
    \node[s-eve] (v0) at (0,0) {$v_0$};
    \path[arrow]
      (v0) edge[selfloop=180] node[left] {$\Win_1$} (v0)
      (v0) edge[selfloop=0] node[right] {$\Win_2$} (v0);
  \end{tikzpicture}
\caption{An example of a game where Eve has a winning strategy for $\Buchi(\Win_1) \wedge \Buchi(\Win_2)$ using two memory states
but no positional winning strategy.}
\label{1-fig:memory_required}
\commentAlt{Figure~\ref{1-fig:memory_required}: A single node labeled 'v0' with two self-loops, one labeled "Win1" on the left and another labeled "Win2" on the right.}
\end{figure}

Let us note that to transform this example into one where colours label vertices, we would need three states instead of one.
Memory requirements for edge-labelled games and vertex-labelled games are not equivalent.

\subsection*{Finite-memory determinacy}
We extend to finite memory the definitions of qualitative objectives being positional, positionally determined, and bi-positionally determined.
Let us fix a qualitative objective $\Omega$.
\begin{itemize}
	\item We say that $\Omega$ has finite memory if for every game $\game$ and every vertex $v$, 
if Eve has a winning strategy from $v$, then she has a winning strategy from $v$ using finite memory.
	\item We say that $\Omega$ is determined with finite memory if for every game $\game$ and every vertex $v$, 
either Eve has a finite memory winning strategy from $v$, or Adam has a winning strategy from $v$.
	\item We say that $\Omega$ is determined with finite memory for both players if for every game $\game$ and every vertex $v$, 
either of the two players has a finite memory winning strategy from~$v$.
\end{itemize}
There are several variants of this definition covering cases where the memory is constant or bounded, and uniform.

\subsection*{Reductions between conditions using memory structures}
Reductions using automata act at the level of objectives, while reductions using memory structures act between conditions: the main difference being that 
the memory structure reads the sequences of edges and produces a sequence of memory states. In that sense the latter are more general: the edges contain more information than the sequence of colours (which is what the automaton reads), and this information is dependent on the graph.

Formally, let $\arena$ an arena and $\mem$ a memory structure.
We construct the arena $\arena \times \mem$ as follows.
We first define the graph $G \times M$ whose set of vertices is $V \times M$ and set of edges $E_M$ is defined as follows:
for every edge $e : v \to v' \in E$ and state $m \in M$ there is an edge $e[m] = (v,m) \xrightarrow{} (v',\delta(m,e))$.
The arena is $\arena \times \mem = (G \times M, \VE \times M, \VA \times M)$.

\begin{fact}[Strategies with memory]
\label{1-fact:strategies_memory}
There is a one to one correspondence between plays in $\arena$ from $v_0$ and in $\arena \times \mem$ from $(v_0,m_0)$.
\end{fact}

To the play 
\[
\pi = e_0 e_1 \dots \text{ where } e_i = v_i \to v_{i+1}
\] 
we associate the play 
\[
\pi' = e_0[m_0]\ e_1[m_1] \dots,
\]
where for all $i$ we define inductively $m_{i+1} = \delta(m_i, e_i[m_i])$.

Let $W$ be a condition on $\arena$ and $W'$ a condition on $\arena \times \Mem$.
We say that $W$ reduces to $W'$ (through $\Mem$) if for all plays $\play$ in $\arena$,
we have 
\[
\play \in W \Longleftrightarrow \play' \in W'.
\]

Let $\Mem$ and $\Mem'$ be two memory structures over the same graph, 
we let $\Mem \times \Mem'$ denote the memory structure obtained by direct product.

\begin{lemma}[Memory structure reductions]
\label{1-lem:memory_structure_reduction}
If $W$ reduces to $W'$ through the memory structure $\mem$, then
Eve has a winning strategy in $\Game = (\Arena,W)$ from $v_0$ if and only if 
she has a winning strategy in $\Game \times \Mem = (\Arena \times \Mem, W')$ from $(v_0,m_0)$. 

More specifically, if Eve has a winning strategy in $\Game \times \Mem$ from $(v,m_0)$ using $\Mem'$ as memory structure, 
then she has a winning strategy in $\Game$ from $v$ using $\Mem \times \Mem'$ as memory structure.
In particular if the strategy in $\Game \times \Mem$ is positional, then the strategy in $\Game$ uses $\Mem$ as memory structure.
\end{lemma}
\begin{proof}
A winning strategy in $\Game$ directly induces a winning strategy in $\Game \times \Mem$ simply by ignoring the additional information
and thanks to the equivalence above because $W$ reduces to $W'$.
For the converse implication, let $\sigma$ be a winning strategy in $\Game \times \Mem$ using $\Mem'$ as memory structure.
Recall that $\sigma$ is defined through the function $\sigma : (\VE \times M) \times M' \to E_M$.
Let $p : E_M \to E$ mapping the edge $e_m$ to $e$.
We construct a strategy $\sigma'$ in $\Game$ using $\Mem \times \Mem'$ as memory structure by
\[
\sigma'(v, (m,m')) = p(\sigma((v,m), m')).
\]
The correspondence between plays in $\arena$ and $\arena \times \mem$ maps plays consistent with $\sigma$ to plays consistent with $\sigma'$,
which together with the fact that $W$ reduces to $W'$ implies that $\sigma'$ is a winning strategy in $\Game$ from $v$.
\end{proof}

To obtain \Cref{1-lem:automata_reduction} as a corollary of \Cref{1-lem:memory_structure_reduction}
we observe that a reduction between objectives using an automaton induces a reduction between the induced conditions using a memory structure.
Formally, let us assume that $\Omega$ reduces to $\Omega'$, 
and let $\Automaton = (Q, q_0, \delta, \Omega'(\col_\Automaton))$ such that $L(\Automaton) = \Omega$.
Let $\Game = (\Arena, \Omega(\col))$ be a game.

We first define the memory structure $\Mem = (Q, q_0, \delta')$. 
The transition function is $\delta' : Q \times E \to Q$, it is defined
by $\delta'(q,e) = \delta(q, \col(e))$.
We consider the arena $\Arena \times \Mem$, and define the colouring function $\col'(e_q) = \col_\Automaton(q, \col(e))$.

We note that $\Omega(\col)$ reduces to $\Omega'(\col')$: for all plays $\play$ in $\Arena$, we have 
$\play \in \Omega(\col)$ if and only if $\play' \in \Omega'(\col')$: this is a reformulation of the fact that~$L(\Automaton) = \Omega$.

We construct the game $\Game' = (\Arena \times \Mem, \Omega'(\col'))$.
Thanks to \Cref{1-lem:memory_structure_reduction} the two games have the same winner and a strategy in the latter induces a strategy in the former
by composing with the memory structure $\Mem$, implying \Cref{1-lem:automata_reduction}.

\section{Quantitative games}
\label{1-sec:quantitative_games}
In a quantitative context the two players are called Min and Max, leaving no ambiguity for their goals.
We adapt the notations to $V = \VMax \uplus \VMin$ if the players are Max and Min.
A strategy $\sigma$ is implicitly for Max, and $\tau$ is for strategies of Min.
Circles are controlled by Max and squares by Min.

\begin{itemize}
	\item A quantitative condition is $f : \Paths_\omega \to \Rinfty$: it assigns a real value (or plus or minus infinity) to a play, which can be thought of as a payoff or a score.
More generally, we sometimes consider quantitative conditions of the form $f : \Paths_\omega \to Y$ with $Y$ a totally ordered set.

	\item A "quantitative objective" $\Phi$ is a function $\Phi : C^\omega \to \Rinfty$. A colouring function $\col$ and a quantitative objective $\Phi$ induce a quantitative condition $\Phi(\col)$. A common abuse of notations is to write $\Phi(\pi)$ instead of $\Phi(\col)(\pi)$, making $\col$ implicit.
	
	\item A quantitative game $\game$ is a tuple $(\arena,f)$ where $\arena$ is an arena and $f$ a quantitative condition.	
\end{itemize}

Since we consider zero-sum games we can easily reverse the point of view by considering the quantitative game $(\arena,-f)$.
Indeed for the latter minimising $f$ is equivalent to maximising $-f$.
The term ``zero-sum'' comes from this fact: the total outcome for the two players is $f + (-f)$, meaning zero.

Quantitative games extend qualitative games: a qualitative condition can be thought of as a quantitative condition by considering its indicator function: $1$ for winning sequences and $0$ for losing sequences, in which case Eve becomes Max, and Adam becomes Min. Conversely, from a quantitative condition $f$ and a threshold $x \in \R$, we can define a qualitative condition: the set of plays having value at least $x$ under $f$:
\[
f_{\ge x} = \set{\play \in \Paths_\omega \mid f(\play) \ge x}.
\]

\subsection*{Five classical quantitative objectives}
\label{1-subsec:classical_quantitative}
We define five quantitative objectives, which are variations of each other: shortest path, mean payoff, total payoff, energy, and discounted payoff.
For all these objectives, the set of colours is (essentially) $C = \Z$, the set of integers.
A colour is called a weight, interpreted as a cost.
The cost of a finite path is the sum over the weights, which Max wants to maximise and Min to minimise.
The problem is that mathematically, the cost of infinite paths may not be defined: the five definitions below can be seen as five different ways to tackle this mathematical obstacle.

\subsubsection*{Shortest path}
The following definition resolves the problem above by adding a reachability objective, ensuring that paths of interest are finite.
The shortest path quantitative objective is defined over the set of colours $C = \Z \cup \set{\Win}$ by
\[
\ShortestPath(\rho) =
  \begin{cases}
    \sum_{i = 0}^{k-1} \rho_i & \text{for $k$ the first index such that } \rho_k = \Win, \\
    \infty & \text{if } \rho_k \neq \Win \text{ for all } k.
  \end{cases}
\]
Note that we are looking for a path of minimal cost, hence not necessarily the shortest in number of edges.

\subsubsection*{Mean payoff}
The mean-payoff quantitative objective represents the long-term average weight.
It comes in two different flavours: using the supremum limit
\[
\MeanPayoff^+(\rho) = \limsup_k \frac{1}{k} \sum_{i = 0}^{k-1} \rho_i.
\]
or the infimum limit
\[
\MeanPayoff^-(\rho) = \liminf_k \frac{1}{k} \sum_{i = 0}^{k-1} \rho_i.
\]
As we shall see, in most settings the two objectives will be equivalent.
For this reason, we often use $\MeanPayoff$ to denote $\MeanPayoff^-$.

\Cref{1-fig:mp_game_example} presents an example of a mean-payoff game. 
The weight of an edge is given by its label.
In this example the threshold is $0$, \textit{i.e.} the induced qualitative condition
is $\MeanPayoff_{\ge 0} = \set{\rho \in C^\omega : \MeanPayoff(\rho) \ge 0}$.
\begin{figure}
\centering
  \begin{tikzpicture}[scale=1.3]
    \node[s-eve] (v0) at (0,0) {$v_0$};
    \node[s-adam] (v1) at (2,0) {$v_1$};
    \node[s-eve] (v2) at (4,0) {$v_2$};    
    \node[s-eve] (v3) at (6,0) {$v_3$};
    \node[s-adam] (v4) at (0,-1.5) {$v_4$};
    \node[s-eve] (v5) at (2,-1.5) {$v_5$};
    \node[s-adam] (v6) at (4,-1.5) {$v_6$};    
    \node[s-eve] (v7) at (6,-1.5) {$v_7$};

    \draw[dotted,rounded corners=4mm,fill=grey-10]
    ($(v0)+(-5mm,1mm)$) |- ($(v5)+(5mm,-5mm)$) -- ($(v1)+(5mm,6mm)$) -| ($(v0)+(-5mm,1mm)$);
    \draw[dotted,rounded corners=5mm,fill=grey-20]
    ($(v2)+(-5mm,1mm)$) |- ($(v7)+(11mm,-5mm)$) -- ($(v3)+(11mm,6mm)$) -| ($(v2)+(-5mm,1mm)$);

    \node[s-eve] (v0) at (0,0) {$v_0$};
    \node[s-adam] (v1) at (2,0) {$v_1$};
    \node[s-eve] (v2) at (4,0) {$v_2$};    
    \node[s-eve] (v3) at (6,0) {$v_3$};
    \node[s-adam] (v4) at (0,-1.5) {$v_4$};
    \node[s-eve] (v5) at (2,-1.5) {$v_5$};
    \node[s-adam] (v6) at (4,-1.5) {$v_6$};    
    \node[s-eve] (v7) at (6,-1.5) {$v_7$};

    \path[arrow]
      (v0) edge[bend left] node[above] {$-1$} (v1)
      (v0) edge[bend left] node[right] {$-2$} (v4)
      (v1) edge[bend left] node[below] {$1$} (v0)
      (v1) edge[bend left] node[right] {$4$} (v5)
      (v2) edge[bend left] node[above] {$-3$} (v3)
      (v2) edge node[left] {$4$} (v6)
      (v3) edge[bend left] node[below] {$2$} (v2)
      (v3) edge node[right] {$1$} (v7)
      (v4) edge[bend left] node[left] {$1$} (v0)
      (v4) edge node[above] {$1$} (v5)
      (v5) edge[bend left] node[left] {$-2$} (v1)
      (v5) edge[bend left] node[above] {$-4$} (v6)
      (v6) edge[bend left] node[below] {$3$} (v5)
      (v6) edge node[above] {$5$} (v7)
      (v7) edge[selfloop] node[right] {$-6$} (v7);
  \end{tikzpicture}
\caption{An example of a mean-payoff game with threshold $0$.}
\label{1-fig:mp_game_example}
\commentAlt{Figure~\ref{1-fig:mp_game_example}: A directed graph with 8 nodes, v0 to v7, split into two shaded regions, with labeled edges. See long description.}
\commentLongAlt{Figure~\ref{1-fig:mp_game_example}: A directed graph divided into two shaded regions. The left region (lighter shade, dotted border) contains nodes v0 (circle), v1 (square), v4 (square), and v5 (circle). Edges with labels are as follows: v0 to v1 (label -1), v1 to v0 (label 1), v0 to v4 (label -2), v4 to v0 (label 1), v1 to v5 (label -2), v5 to v1 (label 4), v4 to v5 (label 1). The right region (darker shade, solid border) contains nodes v2 (circle), v3 (circle), v6 (square), and v7 (circle). Edges with labels are: v2 to v3 (label -3), v3 to v2 (label 2), v2 to v6 (label 4), v3 to v7 (label 1), v7 has a self-loop (label -6). There are also edges connecting the two regions: v5 to v6 (label -4), and v6 to v5 (label 3), and v6 to v7 (label 5).}
\end{figure}

\subsubsection*{Total payoff}
We define the total payoff quantitative objective over the set of colours $C = \Z$ by
\[
\TotalPayoff(\rho) = \limsup_{n} \sum_{i=0}^{n-1} \rho_i
\]

Total payoff is closely related to mean payoff, in the sense that it refines it. 
Indeed, the total payoff of a play is finite if and only if the mean payoff of
this play is null.
Hence, if a vertex in a game has value $0$ for mean payoff, computing its value for total payoff gives an additional insight:
it quantifies how the partial sums are fluctuating around the mean payoff.
For instance, this allows one to distinguish situation
$1,-1,1,-1,1,\ldots$ where the total payoff is $1$ from a close
situation $-1,1,-1,1,\ldots$ with total payoff $0$. 

\subsubsection*{Energy}
\label{1-subsubsec:energy}
The energy quantitative objective is defined over the set of colours $C = \Z$:
\[
\Energy(\rho) = \inf \set{\ell \in \N : \forall k \in \N,\ \ell + \sum_{i=0}^{k-1} \rho_i \ge 0}.
\]
The interpretation is the following: weights are energy consumptions (negative values) and recharges (positive values), and $\Energy(\rho)$ is the smallest initial budget $\ell$ such that all partial sums remain non-negative.
In a counter-intuitive way, the objective of Min is to collect higher (positive) weights.

We can observe that 
$\Energy(\rho) = \sup \set{\sum_{i=0}^{k-1} (-\rho_i) : k \in \N}$,
which motivates defining:
\[
\Energy^+(\rho) = \sup \set{\sum_{i=0}^{k-1} \rho_i : k \in \N}.
\]
The two objectives are clearly equivalent since one can reduce from one to the other by switching the signs of every weight.
The benefit of this definition is that it is intuitively simpler: Max aims at positive weights.

\subsubsection*{Discounted payoff}
The discounted-payoff quantitative objective is parameterised by a discount factor $\lambda \in (0,1)$.
It is defined by:
\[
\DiscountedPayoff_\lambda(\rho) = \lim_k \sum_{i = 0}^{k-1} \lambda^i \rho_i.
\]
Expanding the definition: $\DiscountedPayoff_\lambda(\rho) = \rho_0 + \lambda \rho_1 + \lambda^2 \rho_2 + \dots$.
The influence of a weight decays exponentially over time: the weight $\rho_i$ is multiplied by $\lambda^i$ which goes to $0$ as $i$ goes to infinity.
The discount factor ensures that the limit exists for sequences with bounded weights,
which holds for all plays since a (finite) game contains finitely many different weights.

\subsection*{Values in quantitative games}
Let $\game$ be a quantitative game and $v$ a vertex.
Given $x \in \R$ called a threshold, we say that a strategy $\sigma$ ensures $x$ from $v$ 
if every play $\pi$ starting from $v$ consistent with $\sigma$ has value at least $x$ under $f$,
\textit{i.e.} $f(\play) \ge x$.
In that case we say that Max has a strategy ensuring $x$ in $\game$ from $v$.
Analogously, we say that a strategy $\tau$ ensures $x$ from $v$ 
if every play $\play$ starting from $v$ consistent with $\tau$ has value at most $x$ under $f$,
\textit{i.e.} $f(\play) \le x$.

We let $\ValueMax^{\game}(v)$ denote the quantity
\[
\sup_{\sigma} \inf_{\tau} f(\play_{\sigma,\tau}^{v}),
\]
where $\sigma$ ranges over all strategies of Max and $\tau$ over all strategies of Min.
We also write $\ValueMax^{\sigma}(v) = \inf_{\tau} f(\play_{\sigma,\tau}^{v})$
so that $\ValueMax^{\game}(v) = \sup_{\sigma} \ValueMax^\sigma(v)$.
This is called the value of Max in the game $\game$ from $v$,
and represents the outcome that she can get arbitrarily close to against any strategy of Min.
Note that $\ValueMax^{\game}(v)$ is either a real number, $\infty$, or $-\infty$.

A strategy $\sigma$ such that $\ValueMax^\sigma(v) = \ValueMax^{\game}(v)$ is called optimal from $v$,
and it is simply optimal if the equality holds for all vertices.
Equivalently, $\sigma$ is "optimal from $v$" if for every play $\play$ consistent with $\sigma$ starting from $v$ 
we have $f(\play) \ge \ValueMax^{\game}(v)$.

There may not exist optimal strategies which is why we introduce the following notion.
Let us fix $\varepsilon > 0$. 
\begin{itemize}
	\item If $\ValueMax^{\game}(v) < \infty$, we say that a strategy $\sigma$ is $\varepsilon$-optimal if $\ValueMax^\sigma(v) \ge \ValueMax^{\game}(v) - \varepsilon$.
	\item Similarly, if $\ValueMax^{\game}(v) = \infty$, we say that a strategy $\sigma$ is $\varepsilon$-optimal if $\ValueMax^\sigma(v) > \frac{1}{\varepsilon}$.
\end{itemize}
Note that in both cases, there exist $\varepsilon$-optimal strategies for any $\varepsilon > 0$.

Symmetrically, we let $\ValueMin^{\game}(v)$ denote 
\[
\inf_{\tau} \sup_{\sigma} f(\play_{\sigma,\tau}^{v}).
\]
\begin{fact}[Comparison of values for Min and Max]
\label{1-fact:comparaison_values_eve_adam}
For all quantitative games $\game$ and vertex $v$ we have $\ValueMax^{\game}(v) \le \ValueMin^{\game}(v)$.
\end{fact}
\begin{proof}
For any function $F : X \times Y \to \Rinfty$, we have 
\[
\sup_{x \in X} \inf_{y \in Y} F(x,y) \le \inf_{y \in Y} \sup_{x \in X} F(x,y).
\]
\end{proof}

If this inequality is an equality, meaning 
\[
\ValueMax^{\game}(v) = \ValueMin^{\game}(v),
\]
we say that the game $\game$ is \textit{determined} in $v$,
and let $\Value^{\game}(v)$ denote the value in the game $\game$ from $v$.
If this holds for all vertices we say that $\game$ is determined.
A quantitative objective $\Phi$ is determined if all games with objective $\Phi$ are determined.
Similarly as for the qualitative case, being determined can be understood as follows: the outcome can be determined before playing, and in that case the outcome is the value.

We say that a quantitative objective $\Phi : C^\omega \to \Rinfty$ is Borel if for all $x \in \R$,
the qualitative objective $\Phi_{\ge x} \subseteq C^\omega$ is a Borel set.

\begin{corollary}[Borel determinacy for quantitative objectives]
\label{1-cor:borel_determinacy}
Quantitative Borel objectives are determined.
\end{corollary}
\begin{proof}
Let $f = \Phi(\col)$. If $\ValueMax^{\game}(v) = \infty$ then thanks to the inequality above $\ValueMin^{\game}(v) = \infty$ and the equality holds.
Assume $\ValueMax^{\game}(v) = -\infty$ and let $r$ be a real number.
(The argument is actually the same as for the finite case but for the sake of clarity we treat them independently.)
We consider $f_{\ge r}$.
By definition, this a qualitative Borel condition, so \Cref{1-thm:borel_determinacy} implies that it is determined.
Since Max cannot have a winning strategy for $f_{\ge r}$, as this would contradict the definition of $\ValueMax^{\game}(v)$,
this implies that Min has a winning strategy for $f_{\ge r}$, 
meaning a strategy $\tau$ such that every play $\play$ starting from $v$ consistent with $\tau$ satisfy $f(\play) < r$.
In other words, $\ValueMin^{\tau}(v) = \sup_{\sigma} f(\play_{\sigma,\tau}^{v}) \le r$, which implies that $\ValueMin^{\game}(v) \le r$.
Since this is true for any real number $r$, this implies $\ValueMin^{\game}(v) = -\infty$.

Let us now assume that $x = \ValueMax^{\game}(v)$ is finite and let $\varepsilon > 0$.
We consider $f_{\ge x + \varepsilon}$.
By definition, this a qualitative Borel condition, so \Cref{1-thm:borel_determinacy} implies that it is determined.
Since Max cannot have a winning strategy for $f_{\ge x + \varepsilon}$, as this would contradict the definition of $\ValueMax^{\game}(v)$,
this implies that Min has a winning strategy for $f_{\ge x + \varepsilon}$, 
meaning a strategy $\tau$ such that every play $\play$ starting from $v$ consistent with $\tau$ satisfy $f(\play) < x + \varepsilon$.
In other words, $\ValueMin^{\tau}(v) = \sup_{\sigma} f(\play_{\sigma,\tau}^{v}) \le x + \varepsilon$, which implies that $\ValueMin^{\game}(v) \le x + \varepsilon$.
Since this is true for any $\varepsilon > 0$, this implies $\ValueMin^{\game}(v) \le \ValueMax^{\game}(v)$.
As we have seen the converse inequality holds, implying the equality.
\end{proof}

Note that this determinacy result does not imply the existence of optimal strategies.

\subsection*{Computational problems for quantitative games}
As for qualitative games, we identify different computational problems.
The first is solving the game.

\decpb[Solving the game]{A "quantitative game" $\game$, a vertex $v$, and a threshold $x \in \Qinfty$}{Does Max have a strategy ensuring $x$ from $v$?}

A very close problem is the value problem.

\decpb[Solving the value problem]{A "quantitative game" $\game$, a vertex $v$, and a threshold $x \in \Qinfty$}{Is it true that $\Value^{\game}(v) \ge x$?}

The two problems of "solving a game" and the "value problem" are not quite equivalent: 
they become equivalent if we assume the existence of optimal strategies.

The "value problem" is directly related to computing the value.

\decpb[Computing the value]{A "quantitative game" $\game$ and a vertex $v$}{Compute $\Value^{\game}(v)$}

What "computing the value" means may become unclear if the value is not a rational number, making its representation complicated.
Especially in this case, it may be enough to approximate the value, which is indeed what the "value problem" gives us: 
by repeatedly applying an algorithm solving the value problem one can approximate the value to any given precision,
using a binary search.

\begin{lemma}[Binary search for computing the value]
\label{1-lem:binary_search_computing_value}
If there exists an algorithm $A$ for solving the value problem of a class of games, 
then there exists an algorithm for approximating the value of games in this class within precision $\varepsilon$ 
using $\log(\frac{1}{\varepsilon})$ calls to the algorithm $A$.
\end{lemma}

The following problem is global, in the same way as "computing the winning regions".

\decpb[Computing the value function]{A "quantitative game" $\game$}{Compute the value function $\Value^{\game} : V \to \Rinfty$}

Finally, we are sometimes interested in constructing optimal strategies provided they exist.

\decpb[Constructing an optimal strategy]{A "quantitative game" $\game$ and a vertex $v$}{Compute an optimal strategy from $v$}

A close variant is to construct $\varepsilon$-optimal strategies, usually with $\varepsilon$ given as input.

\subsection*{Positional determinacy}

\begin{definition}
We say that a quantitative objective $\Phi$ is \textit{positional} if 
for every game $\game$ with objective $\Phi$, every vertex $v$ and every threshold $x$,
if Max has a strategy ensuring $x$ from $v$, then she has a positional strategy ensuring $x$ from~$v$.
\end{definition}

We say that $\Phi$ is \textit{positionally determined} if it is additionally determined.
In case the existence of an optimal strategy is guaranteed, this is equivalent to the existence of a positional optimal strategy.

\begin{definition}
We say that a qualitative objective $\Phi$ is \textit{bi-positional} if 
for every game $\game$ with objective $\Phi$, every vertex $v$ and every threshold $x$:
\begin{itemize}
	\item if Max has a strategy ensuring $x$ from $v$, then she has a positional strategy ensuring $x$ from~$v$,
	\item if Min has a strategy ensuring $x$ from $v$, then he has a positional strategy ensuring $x$ from~$v$.
\end{itemize}
\end{definition}
The definition simplifies assuming the existence of optimal strategies: for every game $\game$ with objective $\Phi$ and every vertex $v$,
both players have a positional optimal strategy from~$v$.

\subsection*{Prefix-independent quantitative objectives}
A quantitative objective $\Phi$ is:
\begin{itemize}
	\item monotonic under adding prefixes if for every finite sequence $\rho$ and for every infinite sequence $\rho'$
	we have $\Phi(\rho') \le \Phi(\rho \rho')$;
	\item monotonic under removing prefixes if for every finite sequence $\rho$ and for every infinite sequence $\rho'$
	we have $\Phi(\rho') \ge \Phi(\rho \rho')$;
	\item prefix-independent if it is monotonic under both adding and removing prefixes.
\end{itemize}

\begin{fact}
\label{1-fact:winning_prefix_independent_quantitative}
Let $\Game$ be a quantitative game with objective $\Phi$ monotonic under removing prefixes,
$\sigma$ a strategy ensuring $x$ from $v$, and $\play$ a finite play consistent with $\sigma$ starting from $v$.
Then $\sigma_{\mid \play}$ ensures $x$ from $v' = \last(\play)$.
\end{fact}

\begin{proof}
Let $\play'$ be an infinite play consistent with $\sigma_{\mid \play}$ from $v'$,
then $\play \play'$ is an infinite play consistent with $\sigma$ starting from $v$, 
implying that $\Phi(\col)(\play \play') \ge x$, and since $\Phi$ is monotonic under removing prefixes
this implies that $\Phi(\col)(\play') \ge x$. Thus $\sigma_{\mid \play}$ ensures $x$ from $v'$.
\end{proof}

\begin{corollary}[Comparison of values along a play]
\label{1-cor:comparison_values_along_play}
Let $\Game$ be a quantitative game with objective $\Phi$ monotonic under removing prefixes and $\sigma$ an optimal strategy from $v$.
Then for all vertices $v'$ reachable from $v$ by a play consistent with $\sigma$ we have $\val^{\game}(v) \le \val^{\game}(v')$.
\end{corollary}
In other words, when playing an optimal strategy the value is non-decreasing along the play.


\section{Value iteration algorithms}
\label{1-sec:value_iteration}
In this section and the next we discuss two families of fixed-point algorithms for solving games.
The goal is to highlight the main ingredients for constructing algorithms in these two families.
If the descriptions below are too abstract it may be useful to see concrete instantiations, referenced below.

\subsection*{Quantitative games}

Let us consider a quantitative game $\Game$ with condition $f = \Phi[\col]$.
Assuming that $\Game$ is determined it admits a value function
\[
\Value^{\game} : V \to \Rinfty,
\]
which is defined as 
\[
\Value^{\game} = \sup_{\sigma}\ \inf_{\tau}\ f(\pi_{\sigma,\tau}^v) = \inf_{\tau}\ \sup_{\sigma}\ f(\pi_{\sigma,\tau}^v),
\]
where $\sigma$ ranges over strategies of Max, $\tau$ over strategies of Min, 
and $\pi_{\sigma,\tau}^v$ is the play consistent with $\sigma$ and $\tau$ from $v$.
In particular we write $\Value^{\sigma}(v)$ for $\inf_{\tau}\ f(\pi_{\sigma,\tau}^v)$
and $\Value^{\tau}(v)$ for $\sup_{\sigma}\ f(\pi_{\sigma,\tau}^v)$.

Let us write $Y = \Rinfty$, and note that $Y$ is a total order (and a fortiori a lattice) for the usual order over the reals.

\subsection*{Qualitative games}

For a qualitative game $\Game$ there is no notion of a value function so the first step in constructing a value iteration
algorithm is to define a meaningful notion of value function.
Let us assume that the condition is $\Omega[\col]$ over the set of colours $C$.
The first ingredient is a lattice $(Y,\le)$ together with a function $f : C^\omega \to Y$ for evaluating plays.
The value function is $\Value^{\game} : V \to Y$ defined as
\[
\Value^{\game}(v) = \sup_{\sigma}\ \inf_{\tau}\ f(\pi_{\sigma,\tau}^v),
\]
where $\sigma$ ranges over strategies of Max, $\tau$ over strategies of Min, 
and $\pi_{\sigma,\tau}^v$ is the play consistent with $\sigma$ and $\tau$ from $v$.
As above we write $\Value^{\sigma}(v)$ for $\inf_{\tau}\ f(\pi_{\sigma,\tau}^v)$
and $\Value^{\tau}(v)$ for $\sup_{\sigma}\ f(\pi_{\sigma,\tau}^v)$.

We let $\top$ denote the largest element in $Y$, and $\bot$ for the smallest.
The following principle implies that computing the value function in particular yields the winning regions,
relating the condition $\Omega[\col]$ to the function $f$:

\begin{property}[Characterisation of the winning regions]
\label{1-property:characterisation_winning_regions}
For all games $\Game$, for all vertices $v$, 
Eve wins from $v$ for the qualitative objective $\Omega[\col]$ if and only if $\Value^{\game}(v) \neq \top$.
\end{property}

This implies that the goal is now to compute or approximate the value function $\Value^{\game} : V \to Y$. Indeed choosing $Y = \Rinfty$ covers the quantitative case.



%
%

\subsection*{Fixed point}

Let us consider a game $\Game$. 
We let $F_V$ denote the set of functions $V \to Y$, it is a lattice when equipped with the component wise (partial) order induced by $Y$:
we say that $\mu \le \mu'$ if for all vertices $v$ we have $\mu(v) \le \mu'(v)$.
The main ingredient is an operator $\Op^{\Game} : F_V \to F_V$. The intent is to obtain the function $\val^{\game}$ as a fixed point of the operator $\Op^{\Game}$, using the two different approaches for fixed points we introduced above.
Whenever $\Game$ is clear from the context, we simply write $\Op$ instead of $\Op^{\Game}$.

\subsection*{Fixed point through monotonicity}
The first family of algorithms is based on Kleene fixed-point theorem as stated in \Cref{1-thm:kleene}.

\begin{property}[Fixed point through monotonicity]
\label{1-property:fixed_point_monotonicity}
For all games $\Game$, the operator $\Op^{\Game}$ is monotonic, and $\val^{\game}$ is the least fixed point of $\Op^{\Game}$.
\end{property}

\begin{remark}[Greatest versus least fixed point]
\label{1-rmk:greatest_least_fixed_point}
Whenever technically convenient or more intuitive, we will obtain the value function as the greatest fixed point of $\Op^{\Game}$.
For value iteration algorithms, only small adjustments are necessary to swap between least and greatest fixed point. 
This will be different for strategy improvement algorithms, in the next section.
\end{remark}

\Cref{1-thm:kleene} further states that $\Value^{\game}$ is the limit of the sequence $(\Op^k(\bot))_{k \in \N}$.
The pseudocode is given in~\Cref{1-algo:generic_value_iteration_Kleene_naive}.

\begin{algorithm}[ht]
 \DontPrintSemicolon

\For{$v \in V$}{
$\mu(v) \leftarrow \bot$
}
     
\Repeat{$\mu = \Op^{\Game}(\mu)$}{
	$\mu \leftarrow \Op^{\Game}(\mu)$
}

\Return{$\mu$}
\caption{A generic value iteration algorithm based on fixed point through monotonicity -- naive version.}
\label{1-algo:generic_value_iteration_Kleene_naive}
\end{algorithm}

\begin{theorem}[Generic value iteration algorithm through monotonicity]
\label{1-thm:complexity_value_iteration_Kleene}
Assume \Cref{1-property:fixed_point_monotonicity} (fixed point through monotonicity) and that $Y$ is finite.
Then the generic value iteration algorithm outputs $\Value^\Game$ within at most $n \cdot |Y|$ iterations.
\end{theorem}

A concrete instantiation of this theorem is for energy games, see~\Cref{5-subsec:value_iteration_energy}.

\begin{proof}
If $Y$ is a finite lattice then so is $F_V$.
For each $v$, the sequence $(\Op^k(\mu)(v))_{k \in \N}$ is non-decreasing, so it can be strictly decreased at most $|Y|$ times. At each iteration the value of at least one vertex is strictly decreased. Hence there are at most $n \cdot |Y|$ iterations.
\end{proof}

If $Y$ is not finite then the sequence $(\Op^k(\mu))_{k \in \N}$ converges towards $\Value^{\game}$ but further analysis is required to evaluate the convergence speed.

\subsection*{Fixed point through contraction}
The second family of value iteration algorithms is based on Banach fixed-point theorem as stated in \Cref{1-thm:banach},
let us fix as a goal to approximate $\Value^{\game}$.
We equip $F_V$ with a norm $\|\cdot\|$.

\begin{property}[Fixed point through contraction]
\label{1-property:fixed_point_contraction}
The operator $\Op^{\Game}$ is contracting in the complete space $(F_V,\|\cdot\|)$, 
and $\Value^{\game}$ is the unique fixed point of $\Op^{\Game}$.
\end{property}

The pseudocode of the algorithm is given in \Cref{1-algo:generic_value_iteration_Banach_naive}.

\begin{algorithm}[ht]
 \KwData{Desired precision $\varepsilon$.}
 \DontPrintSemicolon

Choose $\mu \in F_V$ 
     
\Repeat{$\|\Op^{\Game}(\mu) - \mu\| \le \varepsilon \cdot (1 - \lambda)$}{
	$\mu \leftarrow \Op^{\Game}(\mu)$
}

\Return{$\mu$}
\caption{A generic value iteration algorithm based on fixed point through contraction -- naive version.}
\label{1-algo:generic_value_iteration_Banach_naive}
\end{algorithm}

\begin{theorem}[Generic value iteration algorithm through contraction]
\label{1-thm:complexity_value_iteration_Banach}
Assume \Cref{1-property:fixed_point_contraction} (fixed point through contraction). 
Then the generic value iteration algorithm computes an $\varepsilon$-approximation of $\Value^\Game$ within at most 
$O \left( \frac{\log(\varepsilon)}{\log(\lambda)} \right)$ iterations,
where $\lambda \in (0,1)$ is the contraction factor of $\Op^{\Game}$.
\end{theorem}

A concrete instantiation of this theorem is for stochastic reachability games, see~\Cref{7-sec:value_iteration}.

\begin{proof}
Thanks to \Cref{1-thm:banach}, for any $\mu$ we have
\[
\|\Op^k(\mu) - \Value^{\game}\| \le \frac{\lambda^k}{1 - \lambda} \cdot \|\Op^{\Game}(\mu) - \mu\|.
\]
Let us write $\mu_k = \Op^k(\mu_0)$, and set the number of iterations to be 
$k = \frac{\log \left(\frac{\varepsilon \cdot (1 - \lambda)^2}{\|\Op^{\Game}(\mu_0) - \mu_0\|} \right)}{\log(\lambda)}$.
It has been chosen so that
\[
\frac{\lambda^k}{1 - \lambda} \cdot \|\Op^{\Game}(\mu_0) - \mu_0\| \le \varepsilon \cdot (1 - \lambda).
\]
Thanks to the inequality above for $\mu = \mu_0$, this implies that $\| \Op^{\Game}(\mu_k) - \mu_k \| \le \varepsilon \cdot (1 - \lambda)$.
Again thanks to the inequality above for $\mu = \mu_k$ this implies that
$\|\mu_k - \Value^{\game}\| \le \varepsilon$.
Thus, after $k$ iterations, the algorithm outputs an $\epsilon$-approximation of $\Value^{\game}$.
\end{proof}

\subsection*{Wrap up}
Let us wrap up this section: to construct a value iteration algorithm, one needs:
\begin{itemize}
	\item for qualitative games, a notion of values induced by a function $f : C^\omega \to Y$ satisfying \Cref{1-property:characterisation_winning_regions};
	\item in all cases, either a monotonic operator satisfying \Cref{1-property:fixed_point_monotonicity} or a contracting operator satisfying \Cref{1-property:fixed_point_contraction}.
\end{itemize}

\subsection*{Local operator}

In the general case above where nothing is known about the operator $\Op^{\Game}$, we cannot hope for a finer analysis.
However in most cases the operator is defined in the following way, called ``local''.
Fix a game $\Game$. Let us consider a function $\delta : Y \times C \to Y$, which can be thought of as the transition function of a deterministic automaton whose set of states is $Y$ and which reads colours. 
It induces an operator $\Op^{\Game} : F_V \to F_V$ defined by:
\[
\Op^{\Game}(\mu)(u) = 
\begin{cases}
\max \set{\delta( \mu(v), c) : u \xrightarrow{c} v \in E} & \text{ if } u \in \VMax, \\
\min \set{\delta( \mu(v), c) : u \xrightarrow{c} v \in E} & \text{ if } u \in \VMin.
\end{cases}
\]
A first interesting fact about this family of operators is that if $\delta$ is monotonic, meaning
for all $y,y' \in Y$, $c \in C$, if $y \le y'$ then $\delta(y,c) \le \delta(y',c)$, then $\Op^{\Game}$ is monotonic.
Let us formulate this assumption explicitly:

\begin{property}[Monotonicity of the $\delta$ function]
\label{1-property:fixed_point_monotonicity_strong}
The function $\delta$ is monotonic.
\end{property}

Let us examine the property that $\val^{\game}$ is a fixed point of $\Op^{\Game}$. 
We give sufficient conditions in the following lemma:

\begin{lemma}
\label{1-lem:sufficient_condition_fixed_point}
Let $\Game$ a game. Assume that:
\begin{itemize}
	\item For all $\rho \in C^\omega$ and $c \in C$, we have $f(c \cdot \rho) = \delta(f(\rho), c)$.
	\item The function $\delta$ is monotonic and continuous.
\end{itemize}
Then $\val^{\Game}$ is a fixed point of $\Op^{\Game}$.
\end{lemma}

\begin{proof}
We show that $\val^{\Game} \ge \Op^{\Game}(\val^{\Game})$.
Let $u \in \VMax$, we consider $\sigma,\tau$ two strategies from $v$.
We write $\sigma(u) = u \xrightarrow{c} v$. 
We let $\sigma',\tau'$ denote the strategies induced by $\sigma,\tau$ after playing $u \xrightarrow{c} v$,
$\pi^{\sigma,\tau}_u$ the play consistent with $\sigma$ and $\tau$ from $u$,
and $\pi^{\sigma',\tau'}_v$ the play consistent with $\sigma'$ and $\tau'$ from~$v$.
The first property implies that $f(\pi^{\sigma,\tau}_u) = \delta(f(\pi^{\sigma',\tau'}_v), c)$. 
Carefully using infimum and supremum as well as monotonicity and continuity of $\delta$, we obtain the series of inequalities:
\[
\begin{array}{llll}
f(\pi^{\sigma,\tau}_u) & = & \delta(f(\pi^{\sigma',\tau'}_v), c) \\
f(\pi^{\sigma,\tau}_u) & \ge & \inf_{\tau'} \delta(f(\pi^{\sigma',\tau'}_v), c) = \delta(\inf_{\tau'} f(\pi^{\sigma',\tau'}_v), c) \\
\inf_{\tau} f(\pi^{\sigma,\tau}_u) & \ge & \delta(\inf_{\tau'} f(\pi^{\sigma',\tau'}_v), c) \\
\sup_{\sigma} \inf_{\tau} f(\pi^{\sigma,\tau}_u) & \ge & \delta( \inf_{\tau'} f(\pi^{\sigma',\tau'}_v), c) \\
\sup_{\sigma} \inf_{\tau} f(\pi^{\sigma,\tau}_u) & \ge & \sup_{\sigma'} \delta(\inf_{\tau'} f(\pi^{\sigma',\tau'}_v), c) = \delta(\sup_{\sigma'} \inf_{\tau'} f(\pi^{\sigma',\tau'}_v), c) \\
\val^{\Game}(u) & \ge & \max \set{\delta(\val^{\Game}(v) : u \xrightarrow{c} v \in E}.
\end{array}
\]
The reasoning is similar for $u \in \VMin$, and the converse inequality is also proved along the same lines.
\end{proof}


Saying that $\val^{\game}$ is a fixed point of $\Op^{\Game}$ defined this way is tightly related to the fact the $\Game$ is "positionally determined", but not equivalent.
Indeed, the fact that $\val^{\game} = \Op^{\Game}(\val^{\game})$ means that from a vertex $u$, the value $\val^{\game}(u)$ can be computed locally, \textit{i.e.} by considering the maximum or the minimum over all outgoing edges.
It is very tempting to define the argmax and argmin (positional) strategies as follows
\[
\begin{array}{l}
u \in \VMax:\ \sigma(u) \in \argmax \set{\delta(\val^{\Game}(v),c) : u \xrightarrow{c} v \in E}, \\
u \in \VMin:\ \tau(u) \in \argmin \set{\delta(\val^{\Game}(v),c) : u \xrightarrow{c} v \in E}.
\end{array}
\]
and to claim that they are optimal. This is unfortunately often not the case, and one needs to be more careful to define optimal positional strategies. We refer to~\Cref{7-fig:counter_example_optimality} for a counter-example for the case of (stochastic) reachability games.

\vskip1em
Let us discuss~\Cref{1-property:fixed_point_monotonicity}, which states that $\val^{\game}$ is the least fixed point of $\Op^{\Game}$,
and specifically how to prove that such a property holds.
Under the assumptions of~\Cref{1-lem:sufficient_condition_fixed_point} we already know that $\val^{\Game}$ is a fixed point of $\Op^{\Game}$,
which implies that $\val^{\game}$ is larger than the least fixed point.
Let us mention another approach for proving this inequality: we define a sequence of objectives $(f_k)_{k \in \N}$
and show the following two properties.
\begin{itemize}
	\item Writing $\val^{\Game,f_k}$ for the values of the game equipped with objective $f_k$, we have $\val^{\Game,f_k} = \Op^k(\mu_0)$,
	meaning that the $k$\textsuperscript{th} iteration of $\Op$ computes the values for $f_k$.
	\item We have $f_0 \le f_1 \le \dots \le f_k \le f$.
\end{itemize}
The second property directly implies that $\val^{\Game,f_0} \le \val^{\Game,f_1} \le \dots \le \val^{\Game,f_k} \le \val^{\Game}$.
Since the least fixed point of $\Op$ is the limit of $\Op^k(\mu_0)$, by the inequality above it is smaller than $\val^{\Game}$.
This approach is not easier to use than~\Cref{1-lem:sufficient_condition_fixed_point}, but it may be useful as it explains what are the iterations of $\Op^{\Game}$. We refer to~\Cref{5-lem:least_fixed_point_energy} for an example.

Now let us discuss how to prove that $\val^{\Game}$ is the least fixed point.
For this, we consider a fixed point $\mu$ of $\Op^{\Game}$, and argue that $\val^{\Game} \le \mu$.
To this end, we extract from $\mu$ a strategy $\tau$ for Min, and show that $\val^{\tau} \le \mu$; 
since we know that $\val^{\Game} \le \val^{\tau}$, this implies $\val^{\Game} \le \mu$.
Again we refer to~\Cref{5-lem:least_fixed_point_energy} for an example.

\subsection*{Refined complexity analysis}

The definition of local operators allows us to refine the complexity analysis for the generic value iteration algorithm.
Let us write $\Delta$ for the complexity of computing $\delta(y,c)$ for given $y \in Y$ and $c \in C$, and of determining whether $y \le y'$ for $y,y' \in Y$. Note that $\Delta$ is typically very small and dominated by other factors.

In a naive implementation, at each iteration we go through each vertex $u$ and update it, which requires considering all outgoing edges $u \xrightarrow{c} v$, computing $\delta(\mu(v), c)$ and extracting the maximum or minimum.
Hence the cost of a single iteration is $O(m \cdot \Delta)$, and since the total number of iterations is bounded by $n \cdot |Y|$ we obtain an overall complexity $O(nm \cdot \Delta \cdot |Y|)$.
Let us see how this can be improved, removing the linear factor $n$.

We say that $u \xrightarrow{c} v \in E$ is incorrect if $\mu(u) < \delta( \mu(v), c)$,
and that $u$ is incorrect if either $u \in \VMin$ and all outgoing edges are incorrect, 
or $u \in \VMax$ and there exists an incorrect outgoing edge.
The algorithmic improvement of the upcoming algorithm over the naive implementation above is that instead of applying the operator $\Op^{\Game}$ to all vertices at each iteration we can keep track of incorrect vertices and only update them. This computes exactly the same sequence of functions, but at a lesser computational cost.
For $u \in \VMax$, it is easy to determine whether it is incorrect, as it relies on the existence of a single outgoing edge.
This is more complicated for $u \in \VMin$: the idea here is not to keep track of all incorrect edges, but rather to count them.


\begin{theorem}[Complexity analysis of the refined generic value iteration algorithm based on fixed point through monotonicity]
\label{1-thm:complexity_value_iteration_refined}
Assume \Cref{1-property:fixed_point_monotonicity,1-property:fixed_point_monotonicity_strong} (fixed point through monotonicity and monotonicity of $\delta$), and that $Y$ is finite.
Then the generic value iteration algorithm outputs $\Value^\Game$ in time $O(m \cdot \Delta \cdot |Y|)$.
\end{theorem}

The data structure consists of the following objects:
\begin{itemize}
	\item an element of $Y$ for each vertex, representing the current function $\mu : V \to Y$;
	\item a set $\Incorrect$ of vertices (the order in which vertices are stored and retrieved from the set does not matter);
	\item a table $\Count$ storing for each vertex of Min a number of edges.
\end{itemize}

The invariant of the algorithm satisfied before each iteration of the repeat loop is the following:
\begin{itemize}
	\item for $u \in \VMin$, the value of $\Count(u)$ is the number of incorrect outgoing edges of $u$;
	\item $\Incorrect$ is the set of incorrect vertices.
\end{itemize}
The invariant is satisfied initially thanks to the function $\texttt{Init}$.
Let us assume that we choose and remove $u$ from $\Incorrect$.
Since we modify only $\mu(u)$ the only potentially incorrect vertices are in $\Incorrect$ (minus $u$) and the incoming edges of $u$;
for the latter each of them is checked and added to $\Incorrect'$ when required.
By monotonicity, incorrect vertices remain incorrect so all vertices in $\Incorrect$ (minus $u$) are still incorrect.
Hence the invariant is satisfied.

The invariant implies that the algorithm indeed implements~\Cref{1-algo:generic_value_iteration_Kleene} hence returns the least fixed point, 
but it also has implications on the complexity.
Indeed one iteration of the repeat loop over some vertex $u$ involves 
\[
O\left( (|\Ing^{-1}(u)| + |\Out^{-1}(u)|) \cdot \Delta \right)
\]
operations,
the first term corresponds to updating $\mu(u)$ and $\Incorrect$,
which requires for each outgoing edge of $u$ to compute $\delta$,
and the second term corresponds to considering all incoming edges of $u$ and treating the incorrect ones.
Each vertex is updated at most $|Y|$ times, which yields an overall complexity bounded by
\[
O\left( 
\left(\sum_{u \in V} (|\Ing^{-1}(u)| + |\Out^{-1}(u)|) \cdot \Delta\right) \cdot |Y|
\right) 
= O(m \cdot \Delta \cdot |Y|).
\]
A remark here: this analysis does not improve the bound on the number of iterations, it only amortises the cost of updates.

The pseudocode for the case of fixed point through monotonicity is given in \Cref{1-algo:generic_value_iteration_Kleene}.
It can be easily adapted to the case of fixed point through contraction by adding a stopping criterion when $\|\Op^{\Game}(\mu) - \mu\|$ is small enough.


\begin{algorithm}
 \SetKwFunction{FInit}{Init}
 \SetKwFunction{FTreat}{Treat}
 \SetKwFunction{FUpdate}{Update}
 \SetKwFunction{FMain}{Main}
 \SetKwProg{Fn}{Function}{:}{}
 \DontPrintSemicolon

\Fn{\FInit{}}{

	\For{$u \in V$}{
		$\mu(u) \leftarrow \bot$
	}

	\For{$u \in \VMin$}{
        \For{$u \xrightarrow{c} v \in E$}{
        	\If{incorrect: $\mu(u) < \delta(\mu(v), c)$}{
		        $\Count(u) \leftarrow \Count(u) + 1$
        	}
        }
        
        \If{$\Count(u) = \Degree(u)$}{
        	Add $u$ to $\Incorrect$
        }
	}    
	
	\For{$u \in \VMax$}{
        \For{$u \xrightarrow{c} v \in E$}{
        	\If{incorrect: $\mu(u) < \delta(\mu(v), c)$}{
        		Add $u$ to $\Incorrect$
        	}
        }
    }
}

\vskip1em
\Fn{\FTreat{$u$}}{
	\If{$u \in \VMax$}{
		$\mu(u) \leftarrow \max \set{\delta( \mu(v), c) : u \xrightarrow{c} v \in E}$
	}

	\If{$u \in \VMin$}{
		$\mu(u) \leftarrow \min \set{\delta( \mu(v), c) : u \xrightarrow{c} v \in E}$
	}
}

\vskip1em
\Fn{\FUpdate{$u$}}{	
	\If{$u \in \VMin$}{
        $\Count(u) \leftarrow 0$
	}
	\For{$v \xrightarrow{c} u \in E$}{
		\If{$v \xrightarrow{c} u$ is incorrect}{	
			\If{$v \in \VMin$}{
	    	    $\Count(v) \leftarrow \Count(v) + 1$
        
	    	    \If{$\Count(v) = \Degree(v)$}{
    		    	Add $v$ to $\Incorrect'$
		        }	
			}

			\If{$v \in \VMax$}{
				Add $v$ to $\Incorrect'$	
			}
		}

	}
}

\vskip1em
\Fn{\FMain{}}{
	\FInit()    

	\For{$i = 0,1,2,\dots$}{
		$\Incorrect' \leftarrow \emptyset$

		\For{$u \in \Incorrect$}{

			\FTreat($u$)    

			\FUpdate($u$)    
		}
		\If{$\Incorrect' = \emptyset$}{

			\Return{$\mu$}
		}
		\Else{

			$\Incorrect \leftarrow \Incorrect'$
		}
	}
}
\caption{A generic value iteration algorithm based on fixed point through monotonicity -- refined version.}
\label{1-algo:generic_value_iteration_Kleene}
\end{algorithm}

%
%
%
%
%
%
%
%
%
%

\begin{theorem}[Complexity analysis of the refined generic value iteration algorithm based on fixed point through contraction]
\label{1-thm:complexity_value_iteration_refined_contraction}
Assume \Cref{1-property:fixed_point_contraction,1-property:fixed_point_monotonicity_strong} (fixed point through contraction, monotonicity of $\delta$). 
Then the generic value iteration algorithm computes an $\varepsilon$-approximation of $\Value^\Game$ within at most 
$O \left( \frac{\log(\varepsilon)}{\log(\lambda)} \right)$ iterations, where $\lambda \in (0,1)$ is the contraction factor of $\Op^{\Game}$.
The computational cost of a single iteration is $O(m \cdot \Delta)$, so the running time of the whole algorithm is $O(n m \cdot \Delta \cdot |Y|)$.
\end{theorem}



\section{Strategy improvement algorithms}
\label{1-sec:strategy_improvement}
Value iteration algorithms manipulate value functions and never construct any strategy, at least explicitly.
This is a key difference with strategy improvement algorithms (also called policy iteration algorithms) whose fundamental idea is to maintain and improve a strategy.
Let us consider a ``local'' operator $\Op^{\Game}$, induced by a function $\delta : Y \times C \to Y$:
\[
\Op^{\Game}(\mu)(u) = 
\begin{cases}
\max \set{\delta( \mu(v), c) : u \xrightarrow{c} v \in E} & \text{ if } u \in \VMax, \\
\min \set{\delta( \mu(v), c) : u \xrightarrow{c} v \in E} & \text{ if } u \in \VMin.
\end{cases}
\]
Let us recall from the previous section that we consider two scenarios:
\begin{itemize}
	\item in the quantitative case, we start from a quantitative condition $f : C^\omega \to \Rinfty$,
	\item in the qualitative case, we assume the existence of a function $f : C^\omega \to Y$,
	and thanks to \Cref{1-property:characterisation_winning_regions} computing the value function $\val^{\Game}$ yields the winning regions.
\end{itemize}
For value iteration algorithms it was enough for $Y$ to be a lattice. 
Here we need a stronger assumption: $Y$ is a total order.

Let us consider a game $\game$ and set as a goal to construct an optimal strategy for Max.
The key idea behind strategy improvement is to use $\val^{\sigma}$ to improve the strategy $\sigma$ 
by \emph{switching edges}, which is an operation that creates a new strategy.
This involves defining the notion of \emph{improving edges}:
let us consider a vertex $u \in \VMax$, we say that $e : u \xrightarrow{c} v$ is an improving edge if
\[
\delta(\val^{\sigma}(v),c) > \val^{\sigma}(u).
\]
Intuitively: according to $\val^{\sigma}$, playing $e$ is better\footnote{We need $Y$ to be totally ordered here.} than playing $\sigma(u)$.


Given a strategy $\sigma$ and a set of improving edges $S$ (for each $u \in \VMax$, $S$ contains at most one outgoing edge of $u$), we write $\sigma[S]$ for the strategy 
\[
\sigma[S](u) = 
\begin{cases}
e & \text{ if there exists } e : u \xrightarrow{c} v \in S,\\
\sigma(u) & \text{ otherwise}.
\end{cases}
\]

The difficulty is that an edge being improving does not mean that it is a better move than the current one in any context,
but only according to the value function $\val^{\sigma}$, so it is not clear that $\sigma[S]$ is better than $\sigma$.
Strategy improvement algorithms depend on the following two principles:
\begin{itemize}
	\item \textbf{Progress}: updating a strategy using improving edges is a strict improvement,
	\item \textbf{Optimality}: a strategy which does not have any improving edges is optimal.
\end{itemize}

The pseudocode of the algorithm is given in~\Cref{1-algo:strategy_improvement}.
The algorithm is non-deterministic, in the sense that both the initial strategy and at each iteration, the choice of improving edge can be chosen arbitrarily. 
A typical choice, called the ``greedy all-switches'' rule, choosing for each $u \in \VMax$ a maximal improving edge, meaning 
\[
\argmax \set{\delta(\val^{\sigma}(v),c) : u \xrightarrow{c} v \in E}.
\]

\begin{algorithm}
 \DontPrintSemicolon
 
 Choose an initial strategy $\sigma_0$ for Max
 
 \For{$i = 0,1,2,\dots$}{

 	Compute $\val^{\sigma_i}$ and the set of improving edges

	\If{$\sigma_i$ does not have improving edges}{
		\Return{$\sigma_i$}
	}

%

	Choose a non-empty set $S_i$ of improving edges 
	
	$\sigma_{i+1} \leftarrow \sigma_i[S_i]$
 } 
 \caption{The generic strategy improvement algorithm.}
\label{1-algo:strategy_improvement}
\end{algorithm}

Let us write $\sigma \le \sigma'$ if for all vertices $v$ we have $\val^{\sigma}(v) \le \val^{\sigma'}(v)$,
and $\sigma < \sigma'$ if additionally $\neg (\sigma' \le \sigma)$.
Unfortunately, there are no generic correctness proofs for the progress property for this algorithm (we refer to the reference section for further discussion). However, the optimality property can be proved at this level of generality, for the two approaches for fixed-point computations: this is the object of the next two subsections.

We refer to~\Cref{5-subsec:strategy_improvement_energy} for a concrete instantiation for energy games using monotonic fixed-point computations, and to~\Cref{5-subsec:strategy_improvement_discounted} for discounted-payoff games using contracting fixed-point computations.

\subsection*{Fixed point through monotonicity}

Let us start with an important remark: \Cref{1-property:fixed_point_monotonicity} (fixed point through monotonicity) assumes that the value function is computed as the least fixed point of the monotonic operator $\Op^{\Game}$. It is very important in the developments below that this is a least fixed point, and not a greatest fixed point: the proof cannot be easily adapted to this latter case.

\begin{theorem}[Optimality property for strategy improvement algorithms based on fixed point through monotonicity]
\label{1-thm:optimality_Kleene}
Assume \Cref{1-property:fixed_point_monotonicity,1-property:fixed_point_monotonicity_strong} (fixed point through monotonicity, monotonicity of $\delta$).
If $\sigma$ is a strategy that has no "improving edges", then $\sigma$ is optimal.
\end{theorem}

\begin{proof}
We prove the contrapositive: assume that $\sigma$ is not optimal, we show that it must have some improving edge.
The fact that $\sigma$ is not optimal means that $\val^{\sigma} < \val^{\Game}$.
Since $\val^{\Game}$ is the least fixed point of $\Op^{\Game}$, it is also its least pre-fixed point.
Therefore $\val^{\sigma}$ is not a pre-fixed point: $\neg (\val^{\sigma} \ge \Op^{\Game}(\val^{\sigma}))$.
Hence there exists $u \in V$ such that $\val^{\sigma}(u) < \Op^{\Game}(\val^{\sigma})(u)$.

We rule out the case that $u \in \VMin$: since $\val^{\sigma}$ is a fixed point of $\Op^{\Game[\sigma]}$, this implies that for $u \in \VMin$ we have $\val^{\sigma}(u) = \min \set{ \delta(\val^{\sigma}(v), c) : u \xrightarrow{c} v \in E}$, equal to $\Op^{\Game}(\val^{\sigma})(u)$.
Therefore $u \in \VMax$, implying that there exists $u \xrightarrow{c} v$ such that $\val^{\sigma}(u) < \delta(\val^{\sigma}(v), c)$.
This is the definition of $u \xrightarrow{c} v$ being an improving edge.
\end{proof}

\subsection*{Fixed point through contraction}

\begin{theorem}[Optimality property for strategy improvement algorithms based on fixed point through contraction]
\label{1-thm:optimality_Banach}
Assume \Cref{1-property:fixed_point_contraction} (fixed point through contraction).
Let $\sigma$ be a strategy that has no "improving edges", then $\sigma$ is optimal.
\end{theorem}

\begin{proof}
Let $\sigma$ be a strategy that has no "improving edges".
We claim that $\val^{\sigma}$ is a fixed point of $\Op^{\Game}$,
which thanks to \Cref{1-property:fixed_point_contraction} implies that $\val^{\sigma} = \val^{\Game}$, meaning that $\sigma$ is optimal.

Thanks to \Cref{1-property:fixed_point_contraction} $\val^{\sigma}$ is the unique fixed point of $\Op^{\Game[\sigma]}$,
so for $u \in \VMin$ we have 
$\val^{\sigma}(u) = \min \set{\delta( \val^{\sigma}(v), c) : u \xrightarrow{c} v \in E}$.

The fact that $\sigma$ has no improving edges reads:
for all $u \in \VMax$, for all $u \xrightarrow{c'} v' \in E$, 
$\delta( \val^{\sigma}(v'), c') \le \delta( \val^{\sigma}(v), c )$ where $\sigma(u) = u \xrightarrow{c} v$.
Since $\val^{\sigma}(u) = \delta( \val^{\sigma}(v), c)$, this implies that 
$\val^{\sigma}(u) = \max \set{\delta( \val^{\sigma}(v'), c) : u \xrightarrow{c} v' \in E}$.

The two equalities above witness that $\val^{\sigma}$ is the unique fixed point of $\Op^{\Game}$.
\end{proof}

%


\section{Computational models}
\label{1-sec:computation}
\subsection*{The Random Access Machine model of computation}
For complexity statements we consider the classical Turing model of computation.
However for algorithmic results the Turing model is a bit painful and unnatural hence it is customary to use the Random Access Machine (RAM) model instead.
Intuitively this corresponds to using a standard imperative programming language on a usual computer which can create, access, and update variables.
There are variants of the RAM model; to be specific the one we use and describe here is called `word RAM'.
The main reason to use the RAM model is to make our life easier by hiding some small computational costs which are inessential for our purposes.

\vskip1em
The memory is arranged in machine words whose size is a parameter $w$ to be fixed depending on the problem.
A machine word is a register which stores some information as a binary word of length $w$.
The first key assumption of the RAM model is that \textit{memory can be accessed in constant time}.
In other words, machine words are registers with a unique address and can be accessed either directly or indirectly.
A concrete implication is that checking whether an element belongs to a set is a single operation (if each element of the set can be stored in a single machine word).

\vskip1em
We consider an (often implicit) set of basic operations operating on a constant number of machine words; 
addition, multiplication, subtraction, division, and comparison of integers are typical examples.
The second key assumption is the `unit cost model', it says that the time complexity (also called cost) of basic operations is constant.
This convention implies that we can manipulate counters for small numbers with no additional complexity, this will be useful in many situations.

We note that this is unrealistic as it means that for instance we can compute the number $2^{2^n}$ by repeatedly squaring $2$: the complexity is $O(n)$ but this number uses $O(2^n)$ bits hence cannot be generated in polynomial time using a Turing machine.
We will not make use of such weaknesses in our algorithms.

\vskip1em
The size of an input is the number of machine words required to store it.
The most common choice for the machine word size is $w = \log(s)$ where $s$ is the size of the input as we want to at least be able to store an integer $x$ of order $s$ in one machine word.
However in situations in which there are numerical inputs, it is reasonable to assume that each input number fits into one machine word,
leading to a potentially larger $w$.
Note that an algorithm in the word RAM model with machine word size $w$ is allowed to use numbers that are larger than $2^w$, but such numbers should be split among several machine words.

\vskip1em
The time complexity is the number of steps performed by the machine, as a function of the input size, 
and the space complexity is the maximal number of machine words used throughout the computation.

\subsection*{Games representations in the RAM model}
The important parameters for algorithms on games are $n$ the number of vertices and $m$ the number of edges.
Note that our assumption that every vertex has an outgoing edge implies that $n \le m$.

The machine word size is always at least $w = \log(m)$, so that both a vertex and an edge can be stored in one machine word.
A graph is given by the list of vertices and the list of edges, which implies that its size is $O(n + m) = O(m)$.
An algorithm can go through all vertices, all edges, or all "successors" or "predecessors" of a given vertex, 
with no additional cost for the space complexity.

An arena additionally specifies for each vertex which player controls the vertex, which is a boolean value also stored in one machine word.
The representation of conditions and colouring functions is different for each and is discussed when introducing them.

\subsection*{Representations for quantitative objectives}
Let us consider a quantitative game $\Game$. In the five quantitative objectives defined in~\Cref{1-subsec:classical_quantitative},
the colouring function assigns a weight to every edge. Let $W$ denote the largest weight appearing in $\Game$ in absolute value.

Choosing the machine word size $w = \log(m) + \log(W)$ implies that either an edge 
or a vertex together with its weight can be stored in one machine word and that we can perform arithmetic operations on weights.
For all payoff games except discounted games, this means that the input size is $O(m)$.

For discounted-payoff games we additionally need to represent the discount factor $\lambda$, 
which we assume is a rational number $\lambda = \frac{a}{b}$.
Since we want to perform arithmetic operations on $\lambda$ it is convenient to store it on one machine word,
hence the choice for the machine word size $w = \log(m) + \log(W) + \log(b)$.

\subsection*{Polynomial versus strongly polynomial-time algorithms}
Let us consider a computational problem in which the input consists of a sequence of $N$ integers plus a number $n$ of other input bits.
We write $L$ for the total number of bits needed to encode the input integer numbers. 
We say that an algorithm runs in strongly polynomial time if: 
\begin{itemize}
	\item the number of arithmetic operations is bounded by a polynomial in the number of integers $N$ in the input instance; and
	\item the space used by the algorithm is bounded by a polynomial in the size $L + n$ of the input.
\end{itemize}
An equivalent definition using the unit cost word RAM model is that the algorithm uses machine word size $w = L + \log(n)$
and runs in polynomial time.

\subsection*{Linear programming}
We give here only the very essential definitions and results related to linear programming,
and refer to~\cite{Bertsimas.Tsitsiklis:1997} for a reference book on the topic.

A linear program (in canonical form) uses a set of real variables organised in a vector $x$ and is defined by

\begin{equation*}
\begin{array}{lr@{}}
\text{maximise }   & c^T x \\
\text{subject to } & A x \le b,
\end{array}
\end{equation*}

where $c$ and $b$ are rational vectors (with $c^T$ the transpose of $c$) and $A$ is a rational matrix. 
More explicitly: 

\begin{equation*}
\begin{array}{lr@{}c@{}r@{}l}
\text{maximize }   & c_1 x_1   & {}+{\dots}+{} & c_n x_n   &           \\
\text{subject to } & a_{11}x_1 & {}+{\dots}+{} & a_{1n}x_n & {}\le b_1 \\
                   &           &     \vdots    &           &           \\
                   & a_{m1}x_1 & {}+{\dots}+{} & a_{mn}x_n & {}\le b_m. 
\end{array}
\end{equation*}

Solving a linear program is finding an optimal assignment $x^*$ of the variables.

\begin{theorem}[Linear programming]
\label{1-thm:linear_programming}
There exists a polynomial-time algorithm for solving linear programs.
\end{theorem}


Note that we define here linear programs with a maximising objective, but the same problem with minimising $c^T x$ can be easily shown to be equivalent.

The statement above says that there exists a weakly polynomial-time algorithm; whether there exists a strongly polynomial algorithm for linear programming is a long standing open question.


\section*{Bibliographic references}
\label{1-sec:references}
The study of games, usually called game theory, has a very long history rooted in mathematics, logic, and economics, among other fields.
Foundational ideas and notions emerged from set theory with for instance backward induction by Zermelo~\cite{Zermelo:1913}, 
and topology with determinacy results by Martin~\cite{Martin:1975} (stated as \Cref{1-thm:borel_determinacy} in this chapter),
and Banach-Mazur and Gale-Stewart games~\cite{Gale.Stewart:1953}.

The topic of this book is a small part of game theory: we focus on infinite duration games played on graphs.
In this chapter we defined deterministic games, meaning games with no source of randomness, which will be the focus of \Cref{part:classic}.
\Cref{part:stochastic} introduces stochastic games, which were initially studied in mathematics.
We refer to~\Cref{7-sec:references} for more bibliographic references on stochastic games,
and focus in this chapter on references for deterministic games.

The model presented in this chapter emerged from the study of automata theory and logic, where it is used as a tool for various purposes.
Let us first discuss the role of games in two contexts: 
for solving the synthesis problem of reactive systems and for automata and logic over infinite trees.

\vskip1em
The synthesis problem for non-terminating reactive systems was formulated in general terms by Church~\cite{Church:1957} and is therefore also called Church's problem:
from a specification of a step-by step transformation of an input stream given in some logical formalism, 
construct a system satisfying the specification.
The first published paper solving Church's problem for monadic second-order logic was written by B{\"u}chi and Landweber~\cite{Buchi.Landweber:1969}, following a paper by Landweber~\cite{Landweber:1967} (then B{\"u}chi's PhD student) focusing on solving games.
However, the idea of casting the synthesis problem as a game between a system and its environment is due to McNaughton:
in the technical report~\cite{McNaughton:1965} McNaughton attempted to give a solution to the synthesis problem using games, initiating many of the most important ideas for analysing games. 
Unfortunately the proof contained an error which Landweber detected and communicated to McNaughton,
who then decided to let Landweber publish his complete solution.
One of the most difficult step in the solution of Church's problem for monadic second-order logic by B{\"u}chi and Landweber~\cite{Buchi.Landweber:1969} is the determinisation procedure from B{\"u}chi to Muller automata due to McNaughton~\cite{McNaughton:1966}.
We refer to Thomas' survey~\cite{Thomas:2009} for more details on some historical and technical aspects of the early papers on Church's synthesis problem.

\vskip1em
Games emerged in another aspect of automata theory: for understanding the difficult result of Rabin~\cite{Rabin:1969} saying that automata over infinite trees can be effectively complemented. 
This is the key step for proving Rabin's seminal result that the monadic second-order theory of the infinite binary tree is decidable.
The celebrated paper of Gurevich and Harrington~\cite{Gurevich.Harrington:1982} revisits Rabin's result by reducing the complementation question to a determinacy result for games. Interestingly, they credit McNaughton for `airing the idea' of using games in this context and then for exploiting it to Landweber~\cite{Landweber:1967}, B{\"u}chi and Landweber~\cite{Buchi.Landweber:1969}, and B{\"u}chi~\cite{Buchi:1977}.

\vskip1em
Both lines of work have been highly influential in automata theory and logic;
we refer to the reference section in~\Cref{3-chap:regular} for more bibliographic references on this connection.
They bind automata theory and logic to the study of games on graphs and provide motivations and questions many of which are still open today.

\vskip1em
Beyond these two examples there are many applications of games in theoretical computer science and logic in particular.
The following quote is due to Hodges~\cite{Hodges:1993}:
\begin{quotation}
An extraordinary number of basic ideas in model theory can be expressed in terms of games.
\end{quotation}
Let us mention model checking games, which are used for checking whether a model satisfies a formula.
They often form both a theoretical tool for understanding the model checking problem and proving its properties, as well as an algorithmic backend for effectively deciding properties of a logical formalism (we refer to~\cite{Gradel:2002} for a survey on model checking games).
Another important construction of a game for understanding logical properties is the Ehrenfeucht-Fra{\"i}ss{\'e} games~\cite{Ehrenfeucht:1961} whose goal is to determine whether two models are equivalent against a logical formalism.

\vskip1em
The interest in reachability objectives goes beyond automata theory and logic.
The attractor computation presented in~\Cref{1-sec:attractors} is inspired by the backward induction principle due to Zermelo~\cite{Zermelo:1913}, 
which was used to show that well founded games (\textit{i.e.} where all plays are finite) are determined.
The word `attractor' (together with `traps' and `subgames') first appeared in Zielonka's work 
on Muller games~\cite{Zielonka:1998}, but without the algorithmic point of view.
A naive implementation of the attractor would have a quadratic time complexity.
It is difficult to give credit for the linear time algorithm since the problem being very natural it has appeared in several contexts,
for instance in database theory as an inference algorithm by Beeri and Bernstein~\cite{Beeri.Bernstein:1979}
or in the framework of computing least fixed points over transition systems by Arnold and Crubill{\'e}~\cite{Arnold.Crubille:1988}.

\vskip1em
Value iteration and strategy improvement algorithms are the most common families of algorithms. The latter are also often called policy improvement or policy iteration. There are many frameworks presenting them in generic terms as we did in this chapter, see for instance~\cite{Colcombet.Fijalkow.ea:2022} for value iteration and~\cite{Ohlmann:2021} for strategy improvement algorithms.
The PhD thesis~\cite{Ohlmann:2021} shows a generic argument for the progress property to hold in the setting of monotonic universal graphs, but it considers the restricted version where only one edge can be switched at a time. 



\part{Classic}
\label{part:classic}

\ifpictures
\includepdf{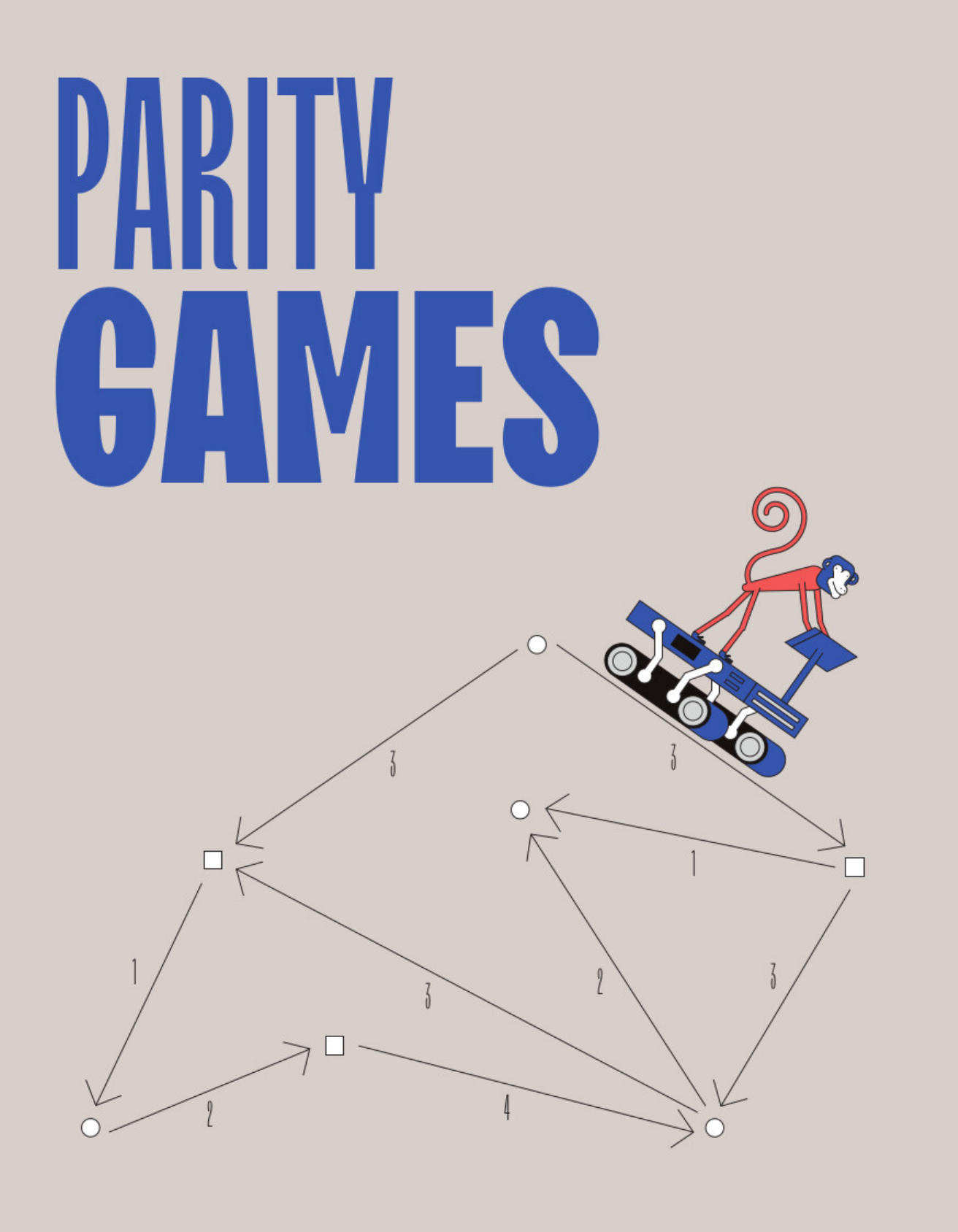}
\fi
\author[John Fearnley, Nathana{\"e}l Fijalkow]{John Fearnley, Nathana{\"e}l Fijalkow}
\copyrightline{Copyright by John Fearnley and Nathana{\"e}l Fijalkow 2025, to be published by Cambridge University Press in the volume \textit{Games on Graphs} edited by Nathana\"el Fijalkow}

\chapter{Parity Games}
\chapterauthor{John Fearnley, Nathana{\"e}l Fijalkow}
\label{2-chap:parity}

\renewcommand{\H}{\mathcal{H}} 

\providecommand{\F}{} 
\renewcommand{\F}{\mathcal{F}} 

\newcommand{\sinit}{\sigma_{\textnormal{init}}}

\newcommand{\siblank}{\mathtt{-}}

We will construct several algorithms for solving parity games.
\begin{itemize}
	\item We start in \Cref{2-sec:zielonka} by constructing an exponential-time algorithm called the McNaughton Zielonka algorithm. It belongs to the family of \emph{attractor decomposition algorithms}, which decompose a game through a sequence of attractor computations. Along the way, we obtain a simple proof for the bi-positional determinacy of parity objectives. 
	
	\item We explain in~\Cref{2-sec:zielonka_quasipolynomial} how to improve the previous algorithm to obtain a quasi-polynomial runtime.

	\item We continue in \Cref{2-sec:strategy_improvement} with an exponential-time \emph{strategy improvement algorithm}. The algorithm constructs a sequence of improving strategies until reaching an optimal strategy.
	
	\item We introduce in \Cref{2-sec:separation} the framework of \emph{separating automata} and give a quasi-polynomial-time algorithm as an instantiation of it. Separating automata formalise a family of algorithms for reducing parity games to safety games. The algorithm has mildly smaller runtime than the algorithm from the previous section.
	
	\item We proceed in \Cref{2-sec:value_iteration} with the construction of a quasi-polynomial time and quasi-linear space \emph{value iteration algorithms}, which finds an optimal strategy through the computation of a value function. This is based on the notion of universal trees.
\end{itemize}

As a conclusion \Cref{2-sec:relationships} discusses the relationships between the different algorithms: in what sense are separating automata and value iteration algorithms equivalent through the notion of universal trees, and how does this family compare to the other two families of algorithms described above.


\section{An exponential-time attractor decomposition algorithm}
\label{2-sec:zielonka}
Recall that the parity objective extends B{\"u}chi and coB{\"u}chi objectives:
\[
\Parity = \set{\rho \in [1,d]^\omega \mid \text{ the largest priority appearing infinitely often in } \rho \text{ is even}}.
\]

\begin{theorem}[Bi-positional determinacy and complexity of parity games]
\label{2-thm:parity}
Parity objectives are uniformly bi-positionally determined.
There exists an algorithm for computing the winning regions of parity games in exponential time,
and more precisely of complexity $O(m n^{d-1})$.
The space complexity of $O(nd)$.

Furthermore, solving parity games is in $\NP \cap \coNP$.
\end{theorem}

To prove \Cref{2-thm:parity} we first construct a recursive algorithm for computing the winning regions of parity games.
The algorithm is often called Zielonka's algorithm, or more accurately McNaughton Zielonka's algorithm.
We refer to the reference section~\Cref{2-sec:references} for a discussion on this nomenclature.
The $\NP \cap \coNP$ complexity bounds will be discussed at the end of this section.

The following lemma induces (half of) the recursive algorithm.
Identifying a colour and its set of vertices we write $d$ for the set of vertices of priority $d$.

\begin{lemma}[Fixed-point characterisation of the winning regions for parity games]
\label{2-lem:zielonka_even}
Let $\Game$ be a parity game with priorities in $[1,d]$, and $d$ even.
Let $\Game' = \Game \setminus \AttrE(d)$.
\begin{itemize}
	\item If $\WA(\Game') = \emptyset$, then $\WE(\Game) = V$.
	\item If $\WA(\Game') \neq \emptyset$, 
	let $\Game'' = \Game \setminus \AttrA( \WA(\Game') )$,
	then $\WE(\Game) = \WE(\Game'')$.	
\end{itemize}
\end{lemma}

Note that $\Game'$ has priorities in $[1,d-1]$ and that if $\WA(\Game') \neq \emptyset$, then $\Game''$ has less vertices than $\Game$.



\begin{proof}
We prove the first item. 
Let $\sigma_d$ be an attractor strategy ensuring to reach $d$ from $\AttrE(d)$.
Consider a winning strategy for Eve from $V \setminus \AttrE(d)$ in $\Game'$, it induces a strategy $\sigma'$ in $\Game$.
We construct a strategy $\sigma$ in $\Game$ as the disjoint union of $\sigma_d$ on $\AttrE(d)$ and of $\sigma'$ on $V \setminus \AttrE(d)$.
Any play consistent with $\sigma$ either enters $\AttrE(d)$ infinitely many times, 
or eventually remains in $V \setminus \AttrE(d)$ and is eventually consistent with $\sigma'$.
In the first case it sees infinitely many times $d$, which is even and maximal, hence satisfies $\Parity$, 
and in the other case since $\sigma'$ is winning the play satisfies $\Parity$.
Thus $\sigma$ is winning from $V$.

We now look at the second item.
Let $\tau_a$ denote an attractor strategy ensuring to reach $\WA(\Game')$ from $\AttrA(\WA(\Game'))$.
We consider a winning strategy for Adam from $\WA(\Game')$ in $\Game'$, it induces a strategy $\tau'$ in $\Game$.
Thanks to \Cref{1-fact:traps_winning} $\tau'$ is a winning strategy in $\Game$.
Consider now a winning strategy in the game $\Game''$ from $\WA(\Game'')$, it induces a strategy $\tau''$ in $\Game$.
The set $V \setminus \AttrA( \WA(\Game') )$ is not a trap for Eve, so we cannot conclude that $\tau''$ is a winning strategy in $\Game$,
and it indeed may not be.
We construct a strategy $\tau$ in $\Game$ as the (disjoint) union of the strategy $\tau_a$ on $\AttrA(\WA(\Game')) \setminus \WA(\Game')$,
the strategy $\tau'$ on $\WA(\Game')$ and the strategy $\tau''$ on $\WA(\Game'')$.
We argue that $\tau$ is winning from $\AttrA( \WA(\Game') ) \cup \WA(\Game'')$ in $\Game$.
Indeed, any play consistent with this strategy in $\Game$ either stays forever in $\WA(\Game'')$ hence is consistent with $\tau''$
or enters $\AttrA( \WA(\Game') )$, hence is eventually consistent with $\tau'$.
In both cases this implies that the play is winning.
Thus we have proved that $\AttrA( \WA(\Game') ) \cup \WA(\Game'') \subseteq \WA(\Game)$.

We now show that $\WE(\Game'') \subseteq \WE(\Game)$, which implies the converse inclusion.
Consider a winning strategy from $\WE(\Game'')$ in $\Game''$, it induces a strategy $\sigma$ in $\Game$.
By prefix-independence, any play consistent with $\sigma$ stays forever in $\WE(\Game'')$, implying that $\sigma$ is winning from $\WE(\Game'')$ in $\Game$.
\end{proof}

To get the full algorithm we need the analogous lemma for the case where the maximal priority is odd.
We do not prove the following lemma as it is the exact dual of the previous lemma, and the proof is the same swapping the two players.

\begin{lemma}[Dual fixed-point characterisation of the winning regions for parity games]
\label{2-lem:zielonka_odd}
Let $\Game$ be a parity game with priorities in $[1,d]$, and $d$ odd.
Let $\Game' = \Game \setminus \AttrA(d)$.
\begin{itemize}
	\item If $\WE(\Game') = \emptyset$, then $\WA(\Game) = V$.
	\item If $\WE(\Game') \neq \emptyset$, let $\Game'' = \Game \setminus \AttrE( \WE(\Game') )$,
	then $\WA(\Game) = \WA(\Game'')$.	
\end{itemize}
\end{lemma}

The algorithm is presented in pseudocode in \Cref{2-algo:zielonka}.

The proofs of \Cref{2-lem:zielonka_even} and \Cref{2-lem:zielonka_odd} also imply that parity games are bi-positionally determined.
Indeed, winning strategies are defined as disjoint unions of strategies constructed inductively.

\vskip1em
We now perform a complexity analysis.
Let us write $f(n,d)$ for the number of recursive calls performed by the algorithm on parity games with $n$ vertices and priorities in $[1,d]$.
We have $f(n,1) = f(0,d) = 0$, with the general induction:
\[
f(n,d) \le f(n,d-1) + f(n-1,d) + 2.
\]
The term $f(n,d-1)$ corresponds to the recursive call to $\Game'$ and the term $f(n-1,d)$ to the recursive call to $\Game''$.
We obtain $f(n,d) \le n \cdot f(n,d-1) + 2n$,
so $f(n,d) \le 2n (1 + n + \dots + n^{d-2}) = O(n^{d-1})$.
In each recursive call we perform two attractor computations so the number of operations
in one recursive call is $O(m)$.
Thus the overall time complexity is $O(m n^{d-1})$.

\vskip1em
We finish the proof of \Cref{2-thm:parity} by sketching the argument that solving parity games is in $\NP \cap \coNP$.
The first observation is that computing the winning regions of the one player variants of parity games can be done in polynomial time
through a simple graph analysis that we do not detail here.
The $\NP$ and $\coNP$ algorithms are the following: guess a winning positional strategy,
and check whether it is winning by computing the winning regions of the one player game induced by the strategy.
Guessing a strategy for Eve is a witness that the answer is yes so it yields an $\NP$ algorithm,
and guessing a strategy for Adam yields a $\coNP$ algorithm.

\begin{algorithm}
 \KwData{A parity game $\Game$ with priorities in $[1,d]$}
 \SetKwFunction{FSolveEven}{SolveEven}
 \SetKwFunction{FSolveOdd}{SolveOdd}
 \SetKwProg{Fn}{Function}{:}{}
 \DontPrintSemicolon

\Fn{\FSolveEven{$\Game$}}{
	$\Game' \leftarrow \Game \setminus \AttrE^{\Game}(d)$
	
	$X \leftarrow$ \FSolveOdd{$\Game'$} \tcp{$\Game'$ has one less priority}

	\If{$X = \emptyset$}{\Return{$V$}}
	\Else{
		$\Game'' \leftarrow \Game \setminus \AttrA^{\Game}(X)$
		
		\Return{\FSolveEven{$\Game''$}} \tcp{$\Game''$ has less vertices}
	}
}
\vskip1em
\Fn{\FSolveOdd{$\Game$}}{
	\If{$d = 1$}{\Return{$V$}}

	$\Game' \leftarrow \Game \setminus \AttrA^{\Game}(d)$
	
	$X \leftarrow$ \FSolveEven{$\Game'$} \tcp{$\Game'$ has one less priority}

	\If{$X = \emptyset$}{\Return{$V$}}
	\Else{
		$\Game'' \leftarrow \Game \setminus \AttrE^{\Game}(X)$
		
		\Return{\FSolveOdd{$\Game''$}} \tcp{$\Game''$ has less vertices}
	}
}
\vskip1em
\If{$d$ is even}{
	\FSolveEven{$\Game$}
}
\Else{
	\FSolveOdd{$\Game$}
}
\caption{A recursive algorithm for computing the winning regions of parity games.}
\label{2-algo:zielonka}
\end{algorithm}


\section{A quasi-polynomial time attractor decomposition algorithm}
\label{2-sec:zielonka_quasipolynomial}
\begin{theorem}[Quasi-polynomial McNaughton Zielonka algorithm]
\label{2-thm:quasipolynomial_mcnaughton_zielonka_algorithm}
There exists a quasi-polynomial-time algorithm for solving parity games, and more specifically of complexity $n^{O(\log n)}$.
\end{theorem}

\begin{algorithm}
 \KwData{A parity game $\Game$ with priorities in $[1,d]$}
 \SetKwFunction{FSolveEven}{SolveEven}
 \SetKwFunction{FSolveOdd}{SolveOdd}
 \SetKwProg{Fn}{Function}{:}{}
 \DontPrintSemicolon

\Fn{\FSolveEven{$\Game$}}{

	$i \leftarrow -1$ 
	
	$V_0 \leftarrow V$

	\Repeat{$\WA(\Game_i) = \emptyset$}{
		$i \leftarrow i + 1$

		Let $\Game_i$ the subgame of $\Game$ induced by $V_i \setminus \AttrE^{\Game_i}(d)$

		$\WE(\Game_i) \leftarrow$ \FSolveOdd{$\Game_i$}

		$V_{i+1} \leftarrow V_i \setminus \AttrA^{\Game}( \WA(\Game_i) )$ 
		}
		
	\Return{$V_i$}
}
\vskip1em
\Fn{\FSolveOdd{$\Game$}}{
	\If{$d = 1$}{
		\Return{$V$}	
	}
	
	$i \leftarrow -1$ 
	
	$V_0 \leftarrow V$

	\Repeat{$\WE(\Game_i) = \emptyset$}{
		$i \leftarrow i + 1$

		Let $\Game_i$ the subgame of $\Game$ induced by $V_i \setminus \AttrA^{\Game_i}(d)$

		$\WA(\Game_i) \leftarrow$ \FSolveEven{$\Game_i$}

		$V_{i+1} \leftarrow V_i \setminus \AttrE^{\Game}( \WE(\Game_i) )$ 
		}
		
	\Return{$V_i$}
}
\vskip1em
\If{$d$ is even}{
	\Return{\FSolveEven{$\Game$}}
}
\Else{
	\Return{\FSolveOdd{$\Game$}}
}
\caption{The recursive algorithm for computing the winning region of parity games.}
\label{2-algo:zielonka_even}
\end{algorithm}

We revisit the exponential recursive algorithm presented in \Cref{2-sec:zielonka}.
We refer to \Cref{2-algo:zielonka_even} for an equivalent presentation of this algorithm, 
where we make explicit all recursive calls involving the maximal priority $d$.
The benefit of doing this is to make the following observation:
during the $i$\textsuperscript{th} recursive call for $d$, the algorithm removes from the game $\Game$ the subset 
$X_i = \AttrA^{\Game}( \WA(\Game_i) )$. 
Note that $X_i$ is a trap for Eve in $\Game$ and a subset of the winning region of Adam in $\Game$:
we say that $X_i$ is a dominion for Eve.
More generally, given a game $\Game$, a set $X$ of vertices is a dominion for Eve if
it is a trap for Adam and a subset of the winning region of Eve.

Let $i_{\infty}$ be the final value of $i$, the sets $X_0,\dots,X_{i_{\infty}-1}$ are disjoint.
The \textit{key idea} is that this implies that at most one of them can have size more than half the total size.

To take advantage of this, we change the specification of the algorithm: 
the new algorithm takes as input a parity game and two parameters $s_{\mEve}$ and $s_{\mAdam}$.
As before, they are two mutually recursive procedures, $\textsl{SolveE}$ and $\textsl{SolveA}$.
At an intuitive level, the objective of $\textsl{SolveE}(\Game,s_{\mEve},s_{\mAdam})$ 
is to return a (non-empty whenever possible) "dominion" for Eve of size at most $s_{\mEve}$.

We spell out the pseudocode of $\textsl{SolveE}$ in \Cref{2-algo:quasipoly_zielonka_even}, leaving out the perfectly symmetric $\textsl{SolveA}$.
The base cases are when there is only one priority, in which case Eve wins everywhere if the priority is even, and Adam wins everywhere if the priority is odd.

\begin{algorithm}
 \KwData{A parity game $\Game$ with priorities in $[1,d]$, and $d$ even, two parameters $s_{\mEve}$ and $s_{\mAdam}$}
 \SetKwFunction{FTreat}{Treat}
 \SetKwProg{Fn}{Function}{:}{}

	Let $i = 0$ and $V_0 = V$
	
	Let $\H_0$ the subgame of $\Game$ induced by $V_0$

	Let $\Game_0$ the subgame of $\H_0$ induced by $V_0 \setminus \AttrE^{\H_0}(d)$

	\vskip1em
	\Fn{\FTreat{$X_i$}}{
		Let $V_{i+1} = V_i \setminus \AttrA^{\H_i}(X_i)$
	
		Let $\H_{i+1}$ the subgame of $\H_i$ induced by $V_{i+1}$

		Let $\Game_{i+1}$ the subgame of $\H_{i+1}$ induced by $V_{i+1} \setminus \AttrE^{\H_{i+1}}(d)$
				
		$i = i + 1$
}

	\vskip1em
	\tcp{recursive calls for small dominions}
	\While{$X_i = \textsl{SolveA}(\Game_i, s_{\mEve}, \lfloor s_{\mAdam} / 2 \rfloor) \neq \emptyset$}{
		\FTreat($X_i$)
	}
	
	\tcp{one recursive call for a large dominion}

	\If{$X_i = \textsl{SolveA}(\Game_i, s_{\mEve}, s_{\mAdam}) \neq \emptyset$}{
		\FTreat($X_i$) 
	}
	
	\tcp{recursive calls for small dominions}
	\While{$X_i = \textsl{SolveA}(\Game_i, s_{\mEve}, \lfloor s_{\mAdam} / 2 \rfloor) \neq \emptyset$}{
		\FTreat($X_i$) 
	}

	\Return{$V_i$}	
\caption{A recursive quasi-polynomial algorithm for computing the winning regions of parity games -- the procedure $\textsl{SolveE}$.}
\label{2-algo:quasipoly_zielonka_even}
\end{algorithm}

We need three simple facts about "traps".

\begin{fact}[Facts about traps]
\label{2-fact:traps}\hfill
\begin{itemize}
	\item Let $S$ be a trap for Eve in the game $\Game$ and $X$ a set of vertices, 
	then $S \setminus \AttrE(X)$ is a trap for Eve in the subgame of $\Game$ induced by $V \setminus \AttrE(X)$.
	\item Let $S$ be a trap for Eve in the game $\Game$ and $X$ a set of vertices such that $S \cap X = \emptyset$, 
	then $S \subseteq V \setminus \AttrA(X)$ and $S$ is a trap for Eve in the subgame of $\Game$ induced by $V \setminus \AttrA(X)$.
	\item Let $S$ be a trap for Eve in the game $\Game$ and $Z$ a trap for Eve in the subgame of $\Game$ induced by $S$,
	then $Z$ is a trap for Eve in $\Game$.
\end{itemize}
\end{fact}

The following lemma implies the correctness of the algorithm.

\begin{lemma}[Correctness of the quasi-polynomial McNaughton Zielonka algorithm]
\label{2-lem:correctness_quasipoly_zielonka}\hfill
\begin{itemize}
	\item For all dominions $S$ for Eve, if $|S| \le s_{\mEve}$, then $S \subseteq \textsl{SolveE}(\Game,s_{\mEve},s_{\mAdam})$.
	\item For all dominions $S$ for Adam, if $|S| \le s_{\mAdam}$, then $S \cap \textsl{SolveE}(\Game,s_{\mEve},s_{\mAdam}) = \emptyset$.
\end{itemize}
\end{lemma}

Indeed, $\WE(\Game)$ and $\WA(\Game)$ are dominions for Eve and Adam in $\Game$, 
so \Cref{2-lem:correctness_quasipoly_zielonka} implies that $\textsl{SolveE}(\Game,n,n) = \WE(\Game)$.

\begin{proof}
The proof is by induction on the number of priorities: indeed all recursive calls to $\textsl{SolveA}$ are for games with one less priority.
It follows that by induction hypothesis the following two properties hold, for all $i$.
\begin{itemize}
	\item For all dominions $S$ in $\Game_i$ for Adam, if $|S| \le s_{\mAdam}$, 
	then \[ S \subseteq \textsl{SolveA}(\Game_i,s_{\mEve},s_{\mAdam}). \]
	\item For all dominions $S$ in $\Game_i$ for Eve, if $|S| \le s_{\mEve}$, 
	then \[ S \cap \textsl{SolveA}(\Game_i,s_{\mEve},s_{\mAdam}) = \emptyset. \]
\end{itemize}
Since there will be an induction inside this induction, we refer to the induction above as the external induction,
and the second one as the internal induction.

We write $i_{\infty}$ for the final value of $i$, \textit{i.e.} such that 
$\textsl{SolveE}(\Game,s_{\mEve},s_{\mAdam}) = V_{i_{\infty}}$.
Note that the first item reads $S \subseteq V_{i_{\infty}}$.

Let $S$ be a dominion for Eve in $\Game$ such that $|S| \le s_{\mEve}$.
We show by internal induction on $i$ that 
$S \subseteq V_i$.

For $i = 0$ this is by definition.
We now assume that $S \subseteq V_i$.
Recall that $\Game_i$ is the subgame of $\H_i$ induced by $V_i \setminus \AttrE^{\H_i}(d)$.
It follows from the first item of \Cref{2-fact:traps} that $S \setminus \AttrE^{\H_i}(d)$
is a dominion for Eve in $\Game_i$.
Since $S \setminus \AttrE^{\H_i}(d)$ has size at most $s_{\mEve}$, 
the second item of the external induction hypothesis implies that 
$S \setminus \AttrE^{\Game_i}(d)$ has an empty intersection with $X_i = \textsl{SolveA}(\Game_i,s_{\mEve},s_{\mAdam})$,
implying that $S \cap X_i = \emptyset$.
It follows from the second item of \Cref{2-fact:traps} that $S \subseteq V_i \setminus \AttrA^{\H_i}(X_i) = V_{i+1}$.
This finishes the internal induction, and implies the first item.

\vskip1em
We now prove the second item.

Let $S$ be a (non-empty) dominion for Adam in $\Game$ such that $|S| \le s_{\mAdam}$.
We write $S_i = S \cap V_i$.
Note that the second item reads $S_{i_{\infty}} = \emptyset$.

We first show by internal induction on $i$ that 
$S_i$ is a dominion for Adam in $\H_i$.
For $i = 0$ this is by definition.
We now assume that $S_i$ is a dominion for Adam in $\H_i$.
Recall that $\H_{i+1}$ is the subgame of $\H_i$ induced by $V_{i+1} = V_i \setminus \AttrA^{\H_i}(X_i)$.
It follows from the first item of \Cref{2-fact:traps} applied to $S_i$ (swapping the roles of Eve and Adam) 
that $S_i \setminus \AttrA^{\H_i}(X_i) = S_{i+1}$ is a dominion for Adam in $\H_{i+1}$.
This finishes the internal induction.

We showed that $S_i$ is a dominion for Adam in $\H_i$ for each $i$.
To apply the external inductive hypothesis, we need to exhibit dominions for Adam in $\Game_i$ for each $i$.
We consider $\H'_i$ the subgame of $\H_i$ induced by $S_i$.
Let $Z_i = \WA^{\H'_i}(\Parity \cup \Reach(d))$: 
in words, $Z_i$ is the subset of vertices of $S_i$ from where Adam can ensure that the parity objective is not satisfied
and never to see priority $d$.
We prove two properties:
\begin{itemize}
	\item If $Z_i = \emptyset$, then $S_i = \emptyset$.

The fact that $Z_i$ is empty implies that Eve has a strategy $\sigma$ in $\Game_i$ which from $S_i$ ensures $\Parity$ or to see priority $d$.
Since $S_i$ is a dominion for Adam in $\H_i$, there exists a strategy $\tau$ for Adam which from $S_i$ ensures the complement of $\Parity$
and to stay in $S_i$. 
Playing $\sigma$ against $\tau$ yields a contradiction, since $\sigma$ ensures $\Parity$ if the play remains in $S_i$.

	\item $Z_i$ is a dominion for Adam in $\Game_i$.

That $Z_i$ is a subset of the winning region of Adam in $\Game_i$ is clear.
To see that $Z_i$ is a trap for Eve in $\Game_i$, we first note that since $S_i$ is a trap for Eve in $\H_i$
and $Z_i$ is a trap for Eve in $\H'_i$, the subgame of $\H_i$ induced by $S_i$, 
then $Z_i$ is a trap in $\H_i$ by the third item of \Cref{2-fact:traps}.
Now, since $Z_i$ has an empty intersection with $d$, by the second item of \Cref{2-fact:traps} 
this implies that $Z_i$ is a trap for Eve in the subgame of $\H_i$ induced by $V_i \setminus \AttrE^{\H_i}(d)$,
which is exactly $\Game_i$.
\end{itemize}

We are now fully equipped to prove that $S_{i_{\infty}} = \emptyset$.
Let $i_{\ell}$ be the value of $i$ for which the algorithm performs a recursive call for a large dominion.
Since the sequence $(S_i)_i$ is non-increasing, if $S_i$ is empty for some $i$, then $S_{i_{\infty}}$ as well.

The first while loop was exited for $i = i_{\ell}$, implying that 
\[
\textsl{SolveA}(\Game_{i_{\ell}},s_{\mEve},\lfloor s_{\mAdam} / 2 \rfloor) = \emptyset.
\]
We apply the external inductive hypothesis to the dominion $Z_{i_{\ell}}$ for Adam in $\Game_{i_{\ell}}$
for the parameter $\lfloor s_{\mAdam} / 2 \rfloor$: if $|Z_{i_{\ell}}| \le \lfloor s_{\mAdam} / 2 \rfloor$, then
$Z_{i_{\ell}} \subseteq \textsl{SolveA}(\Game_{i_{\ell}},s_{\mEve},\lfloor s_{\mAdam} / 2 \rfloor)$,
implying that $Z_{i_{\ell}}$ is empty, which by the first property above implies that $S_{i_{\ell}}$ is empty,
thus $S_{i_{\infty}}$ as well, proving the second item of \Cref{2-lem:correctness_quasipoly_zielonka}.
Excluding this case, we then analyse the situation where 
$|Z_{i_{\ell}}| > \lfloor s_{\mAdam} / 2 \rfloor$. 

We apply again the external inductive hypothesis to $Z_{i_\ell}$, but this time 
for the parameter $s_{\mAdam}$: 
if $|Z_{i_\ell}| \le s_{\mAdam}$, then $Z_{i_\ell} \subseteq X_{i_\ell}$.
Since $Z_{i_\ell} \subseteq S_{i_\ell} \subseteq S$ and $|S| \le s_{\mAdam}$, the premise is satisfied,
so $Z_{i_\ell} \subseteq X_{i_\ell}$.
Since $Z_{i_\ell}$ is non-empty, so is $X_{i_\ell}$: the search for a large dominion was successful, and in particular
$i$ is incremented at this stage, implying that $i_\ell < i_\infty$.

Consider $S_{i_\ell + 1} = S_{i_\ell} \setminus \AttrA^{\H_{i_\ell}}(X_{i_\ell})$.
In particular $S_{i_\ell + 1} \subseteq S_{i_\ell} \setminus X_{i_\ell} \subseteq S_{i_\ell} \setminus Z_{i_\ell}$, so
\[
|S_{i_\ell + 1}| \le |S_{i_\ell}| - |Z_{i_\ell}| \le \lfloor s_{\mAdam} / 2 \rfloor.
\]

The second while loop was exited for $i = i_{\infty}$, implying that 
\[
\textsl{SolveA}(\Game_{i_{\infty}},s_{\mEve},\lfloor s_{\mAdam} / 2 \rfloor) = \emptyset.
\]		
We apply the external inductive hypothesis to the dominion $Z_{i_{\infty}}$ for Adam in $\Game_{i_{\infty}}$
for the parameter $\lfloor s_{\mAdam} / 2 \rfloor$: if $|Z_{i_{\infty}}| \le \lfloor s_{\mAdam} / 2 \rfloor$, then
$Z_{i_{\infty}} \subseteq \textsl{SolveA}(\Game_{i_{\infty}},s_{\mEve},\lfloor s_{\mAdam} / 2 \rfloor)$.
The premise holds because $|S_{i_\ell + 1}| \le \lfloor s_{\mAdam} / 2 \rfloor$, the sequence $(S_i)_i$ is non-increasing, 
$i_\ell < i_\infty$, and $Z_{i_{\infty}} \subseteq S_{i_\infty}$.
Hence $Z_{i_\infty}$ is empty. Thanks to the first property above for $Z_i$, this implies that $S_{i_\infty}$ is empty.
This finishes the proof of the second item of \Cref{2-lem:correctness_quasipoly_zielonka}.
\end{proof}

We obtain an algorithm for computing the winning regions of parity games: we call $\textsl{SolveE}(\Game,n,n)$,
where $\Game$ is a parity game with $n$ vertices.

We now perform a complexity analysis.
Let us write $f(m,n,d,s_{\mEve},s_{\mAdam})$ for the complexity of the algorithm over parity games with $m$ edges, $n$ vertices, $d$ priorities,
and parameters $s_{\mEve}$ and $s_{\mAdam}$.
We have $f(m,n,1,s_{\mEve},s_{\mAdam}) = O(n)$ and $f(m,0,d,s_{\mEve},s_{\mAdam}) = f(m,n,d,0,s_{\mAdam}) = f(m,n,d,s_{\mEve},0) = O(1)$, 
with the general induction 
\[
\begin{array}{lll}
f(m,n,d,s_{\mEve},s_{\mAdam}) & \le & n \cdot f(m,n,d-1,s_{\mEve},\lfloor s_{\mAdam} / 2 \rfloor) \\
							  & & +\ f(m,n,d-1,s_{\mEve},s_{\mAdam}) \\
							  & & +\ O(nm).
\end{array}
\]
The term $n \cdot f(m,n,d-1,s_{\mEve},\lfloor s_{\mAdam} / 2 \rfloor)$ corresponds to the recursive calls for small dominions, 
the term $f(m,n,d-1,s_{\mEve},s_{\mAdam})$ to the recursive call for a large dominion,
and $O(nm)$ for the computation of the attractors.
We obtain
\[
f(m,n,d,n,n) \le m \cdot n^{1 + 2 \lfloor \log n \rfloor} \cdot \binom{d + 2 \lfloor \log n \rfloor}{2 \lfloor \log n \rfloor}.
\]
which implies a quasi-polynomial upper bound of $n^{O(\log n)}$.


\section{An exponential-time strategy improvement algorithm}
\label{2-sec:strategy_improvement}
In a nutshell, a strategy improvement algorithm constructs a sequence of strategies, the next one being an improvement over the current one,
until reaching an optimal strategy. One can identify from the literature at least four strategy improvement algorithms for solving parity games, all running in exponential time. Two of them are better explained as consequences of their counterparts for energy games and discounted-payoff games: we first reduce parity games to the corresponding games, and apply the strategy improvement algorithm in that setting. We refer the reader to \Cref{5-chap:payoffs} for them. The other two are specific to parity games, although they are closely related to the first two algorithms.

\begin{theorem}[Strategy improvement]
\label{2-thm:strategy_improvement}
There exists a strategy improvement algorithm for solving parity games in exponential time.
\end{theorem}

\paragraph{\bf Adding the option of stopping the game.}
Let $\game$ a parity game with $n$ vertices and priorities in $[1,d]$.
Let us give Eve an extra move $\siblank$ that indicates that the game should stop and that she can play from any vertex of hers.
So a strategy for Eve is now a function $\sigma : \VE \rightarrow E \cup \set{\siblank}$ 
where $\sigma(v) = \siblank$ indicates that Eve has chosen to stop the game, and $\sigma(v) \ne \siblank$ should be interpreted as normal.
Adam is not allowed to stop the game, so strategies for Adam remain unchanged.
We say that a play ending with $\siblank$ is stopped.

\paragraph{\bf Evaluating a strategy.}
The first question is: given a strategy $\sigma$, how to evaluate it?
Let us define $Y = \N^d \cup \set{\top}$.
We first explain how to evaluate plays, using a function
\[
f : [1,d]^\omega \cup [1,d]^* \to Y:
\]
\begin{itemize}
	\item Stopped plays: if $\rho \in [1,d]^*$, then $f(\rho)$ is the tuple $(x_1, x_2, \dots, x_d)$ where $x_p$ is the number of times that priority $p$ appears in $\rho$.
	\item Infinite plays: if $\rho \in [1,d]^\omega$, then $f(\rho)$ is $\top$.
\end{itemize}
The evaluation for finite plays can be computed using an automaton as follows: we define $\delta : Y \times [1,d] \to Y$ by 
\[
\delta((x_1, x_2, \dots, x_d),p) = (x_1, x_2, \dots, x_p + 1, x_{p+1}, \dots, x_d)
\]
and $\delta(\top,p) = \top$.
Extending to $\delta : Y \times [1,d]^* \to Y$, we obtain the following fact.

\begin{fact}
\label{2-fact:delta}
For $\rho \in [1,d]^*$ we have $f(\rho) = \delta((0,\dots,0), \rho)$.
\end{fact}


We now define a total order $\le$ on~$Y$.
First, $\top$ is the greatest element.
For $t,t' \in \N^d$, we say that $t < t'$ if and only if $p$ is the largest index such that $t_p \ne t'_p$ and 
\begin{itemize}
	\item either $p$ is even and $t_p < t'_p$,
	\item or $p$ is odd and $t_p > t'_p$.
\end{itemize}
Then $\le$ is the reflexive closure of~$<$: we have $t \le t'$ if $t < t'$ or $t = t'$.
These conditions specify which priorities Eve likes to see along a play.
Given a choice, Eve would prefer to see even priorities and to avoid odd priorities, but these preferences are hierarchical: 
assuming that $d$ is even, Eve would like to see as many edges of priority $d$ as possible. 
If two plays see $d$ the same number of times, Eve prefers the play that visits the fewest vertices of
priority $d-1$, and if two plays see $d$ and $d-1$ the same number of times,
then Eve would like to maximise the number of times that $d-2$ is visited, and so on recursively.

We can now define the value of a strategy $\sigma$: 
\[
\val^{\sigma}(v) = \inf_\tau f(\play^{\sigma,\tau}_v),
\]
where $\tau$ ranges over (general) strategies for Adam.

Since $Y$ is totally ordered, we can naturally cast the computation of the value function as a shortest path problem:
$\val^\sigma(v)$ is the value of the shortest path\footnote{We use the usual albeit sometimes confusing terminology from graph algorithms: ``shortest path'' means of smallest value with respect to $Y$, not in number of edges.} (with respect to $Y$) in $\Game[\sigma]$.
Thus, any algorithm for the shortest path problem can be applied, such as the Bellman-Ford algorithm.
In particular computing $\val^\sigma$ can be done in cubic time, and even more efficiently through a refined analysis which we do not perform here.

Recall that we say that a parity graph (without stopping option) satisfies parity from $v$ if all infinite paths from $v$ satisfy parity.
Then a strategy $\sigma$ is winning from $v$ if and only if the parity graph $\Game[\sigma]$ satisfies parity from $v$.
We extend this definition to parity graphs with the stopping option: a strategy $\sigma$ satisfies parity if all infinite plays consistent with $\sigma$ satisfy parity. Crucially, this does not require anything of stopped plays.
We say that a cycle is even if the maximal priority along the cycle is even, and it is odd otherwise. 
The characterisation of satisfying parity using cycles extends to this setting:
\begin{fact}[Strategy satisfying parity in the presence of stopped plays]
\label{2-fact:satisfying}
A strategy $\sigma$ satisfies parity if and only if all cycles in $\Game[\sigma]$ are even.
\end{fact}
The algorithm will only manipulate strategies satisfying parity.

\paragraph{\bf Improving a strategy.}
We reach the last item in the construction of the algorithm: the notion of improving edges.
Let $\sigma$ a strategy. 
Let us consider a vertex $u \in \VE$, we say that $e : u \xrightarrow{p} v$ is an improving edge if
\[
\delta(\val^{\sigma}(v),p) > \val^{\sigma}(u).
\]
Intuitively: according to $\val^{\sigma}$, playing $e$ is better than playing $\sigma(u)$.
Given a strategy $\sigma$ and a set of improving edges $S$, we write $\sigma[S]$ for the strategy 
\[
\sigma[S](u) = 
\begin{cases}
e & \text{ if there exists } e = u \xrightarrow{p} v \in S,\\
\sigma(u) & \text{ otherwise}.
\end{cases}
\]

\paragraph{\bf The algorithm.}
The algorithm starts with a specified initial strategy, which is the strategy
$\sigma_0$ where $\sigma_0(v) = \siblank$ for all vertices $v \in \VE$. 
It may not hold that $\sigma_0$ satisfies parity since $\game$ may contain odd cycles fully controlled by Adam.
This can be checked in linear time and the attractor to the corresponding vertices removed from the game.
After this preprocessing $\sigma_0$ indeed satisfies parity.

The pseudocode of the algorithm is given in \Cref{2-algo:strategy_improvement}.

\begin{algorithm}
 \KwData{A parity game $\game$}
 \SetKwBlock{Repeat}{repeat}{}
 \DontPrintSemicolon
 
 \For{$v\in \VE$}{
   $\sigma_0(v) \leftarrow \siblank$
 }

 $C \leftarrow \set{v \in V : v \text{ contained in an odd cycle in } \game[\sigma_0]}$

 $\Game \leftarrow \Game \setminus \AttrA(C)$
 
 \Repeat{
	\For{$i = 0,1,2,\dots$}{

		Compute $\val^{\sigma_i}$ and the set of improving edges

		\If{$\sigma_i$ does not have improving edges}{
			\Return{$\set{v \in V : \val^{\sigma_i}(v) = \top}$}
		}

	Choose a non-empty set $S_i$ of improving edges 
	
	$\sigma_{i+1} \leftarrow \sigma_i[S_i]$
	}
 } 

 \caption{The strategy improvement algorithm for parity games.}
\label{2-algo:strategy_improvement}
\end{algorithm}

\paragraph{\bf Proof of correctness.}
As per usual for strategy improvement algorithms, the correctness depend on the following two principles:
\begin{itemize}
	\item \textbf{Progress}: updating a strategy using improving edges is a strict improvement,
	\item \textbf{Optimality}: a strategy which does not have any improving edges is optimal.
\end{itemize}
The difficulty is that an edge being improving does not mean that it is a better move than the current one in any context,
but only according to the value function $\val^{\sigma}$, so it is not clear that $\sigma[S]$ is better than $\sigma$.
Let us write $\sigma \le \sigma'$ if for all vertices~$v$ we have $\val^{\sigma}(v) \le \val^{\sigma'}(v)$,
and $\sigma < \sigma'$ if additionally $\neg (\sigma' \le \sigma)$.


We start by stating a very simple property of $\delta$.

\begin{fact}[Key property of $\delta$]
\label{2-fact:properties_delta}
Let $t \in Y$ and $p_1,\dots,p_k \in [1,d]$ with $\delta(t,p_1 \dots p_k) \neq \top$.
The following holds:
\begin{itemize}
	\item $t \neq \delta(t,p_1 \dots p_k)$.
	\item $t < \delta(t,p_1 \dots p_k)$ if and only if $\max \set{p_1,\dots,p_k}$ is even.
	\item $t > \delta(t,p_1 \dots p_k)$ if and only if $\max \set{p_1,\dots,p_k}$ is odd.	
\end{itemize}
\end{fact}

The following lemma states the two important properties of $(Y,\le)$ and $\delta$.

\begin{lemma}
\label{2-lem:key_property}
Let $G$ a parity graph (with no stopping option).
\begin{itemize}
	\item If there exists $\mu : V \to Y$ such that for all vertices $u$ we have $\mu(u) \neq \top$
	and for all edges $u \xrightarrow{p} v \in E$ we have $\mu(u) \le \delta(\mu(v),p)$,
	then $G$ satisfies parity.
	\item If there exists $\mu : V \to Y$ such that for all vertices $v$ we have $\mu(u) \neq \top$
	and for all edges $u \xrightarrow{p} v \in E$ we have $\mu(u) \ge \delta(\mu(v),p)$,
	then $G$ satisfies the complement of parity.
\end{itemize}
\end{lemma}
\begin{proof}
We prove the first property, the second is proved in exactly the same way.
Thanks to the characterisation using cycles it is enough to show that all cycles in $G$ are even.
Let us consider a cycle in $G$:
\[
\pi = v_0 \xrightarrow{p_0} v_1 \xrightarrow{p_1} v_2 \cdots v_{k-1} \xrightarrow{p_{k-1}} v_0.
\]
For all $i \in [0,k-1]$ we have $\mu(v_i) \le \delta(\mu(v_{i+1 \mod k}),p_i)$.
By monotonicity of $\delta$ this implies $\mu(v_0) \le \delta(\mu(v_0),p_{k-1} \cdots p_0)$.
Thanks to \Cref{2-fact:properties_delta} this implies that the maximum priority in $\set{p_0,\dots,p_{k-1}}$ is even.
\end{proof}

Let us fix a strategy $\sigma$. 
We let $F^\sigma_V$ denote the set of functions $\mu : V \to Y$ such that 
$\mu(v) = (0,\dots,0)$ if $\sigma(v) = \siblank$, it is a lattice when equipped with the component wise (partial) order induced by $Y$:
we say that $\mu \le \mu'$ if for all vertices $v$ we have $\mu(v) \le \mu'(v)$.
We then define an operator $\Op^{\sigma} : F^\sigma_V \to F^\sigma_V$ by
\[
\Op^{\sigma}(\mu)(u) = 
\begin{cases}
\min \set{\delta( \mu(v), p) : u \xrightarrow{p} v \in E}	& \text{if } u \in \VA, \\
\delta( \mu(v), p) 											& \text{if } u \in \VE \text{ and } \sigma(u) = u \xrightarrow{p} v, \\
(0,\dots,0) 												& \text{if } u \in \VE \text{ and } \sigma(u) = \siblank.
\end{cases}
\]

\begin{lemma}
\label{2-lem:fixed_point}
For all $\sigma$ satisfying parity, $\val^{\sigma}$ is a fixed point of $\Op^{\sigma}$.
Further, $\val^{\sigma}(u) = \top$ if and only if all paths consistent with $\sigma$ from $u$ are infinite,
and if $\val^{\sigma}(u) < \top$, then there exists $\pi$ stopped play such that $\val^{\sigma}(u) = f(\pi)$.
\end{lemma}

We now rely on \Cref{2-lem:key_property,2-lem:fixed_point} to prove the two principles: progress and optimality.

\begin{lemma}[Progress]
\label{2-lem:progress}
Let $\sigma$ a strategy satisfying parity and $S$ a set of improving edges.
We let $\sigma'$ denote $\sigma[S]$. Then $\sigma'$ satisfies parity and $\sigma < \sigma'$.
\end{lemma}

\begin{proof}
We first argue that $\sigma'$ satisfies parity.
To this end let us consider the parity graph $\Game[\sigma']$.
We claim that for all edges $e = u \xrightarrow{p} v$ in $\Game[\sigma']$, we have $\val^\sigma(u) \le \delta(\val^\sigma(v),p)$:
\begin{itemize}
	\item Either $e$ is an edge in $\Game[\sigma]$ and this is because $\val^\sigma$ is a fixed point of $\Op^{\sigma}$,
	\item Or $e$ was an improving edge, in which case $\val^\sigma(u) < \delta(\val^\sigma(v),p)$.
\end{itemize}
Since $\sigma$ satisfies parity, for all vertices $u$ such that $\val^\sigma(u) = \top$ all paths starting from $u$ satisfy parity.
Let us write $G$ for the graph obtained from $\Game[\sigma']$ by removing all such vertices.
The first item of \Cref{2-lem:key_property} implies that $G$ satisfies parity, hence $\Game[\sigma']$ satisfies parity.

\vskip1em
At this point we know that $\sigma'$ satisfies parity, let us show that $\val^{\sigma} \le \val^{\sigma'}$.
Let $u_0 \in V$, let $\pi'$ consistent with $\sigma'$ from $u_0$ such that $\val^{\sigma'}(u_0) = f(\pi')$, we show that $\val^{\sigma}(u_0) \le f(\pi')$.

If $f(\pi') = \top$ this is clear, so let us assume that $f(\pi') < \top$, meaning $\pi'$ is finite.
We reason by induction on the number of improving edges in $\pi'$.
If there are none, then $\pi'$ is consistent with $\sigma$ and we are done. 
Let us consider the induction step: we write $\pi' = \pi'_1 \cdot u_1 \xrightarrow{p_1} v_1 \cdot \pi'_2$ with $u_1 \xrightarrow{p_1} v_1$ the first improving edge in $\pi'$.
By definition of an improving edge we have $\val^{\sigma}(u_1) < \delta(\val^{\sigma}(v_1),p_1)$. 
Note that $\val^{\sigma'}(v_1) = f(\pi'_2)$, 
by induction hypothesis this implies that $\val^{\sigma}(v_1) \le f(\pi'_2)$.
Since $\val^{\sigma'}(u_1) = \delta(\val^{\sigma'}(v_1),p_1)$, we obtain
$\val^{\sigma}(u_1) \le \val^{\sigma'}(u_1)$.
Since $\pi'_1$ does not contain any improving edges, it is consistent with $\sigma$, hence an easy induction implies that 
$\val^{\sigma}(u_0) \le \val^{\sigma'}(u_0)$.

\vskip1em
We now show that $\val^{\sigma} < \val^{\sigma'}$. 
Let us consider $e = u \xrightarrow{p} v$ an improving edge. 
Using $\val^{\sigma}(v) \le \val^{\sigma'}(v)$ and the monotonicity of $\delta$ we obtain that
$\delta(\val^\sigma(v),p) \le \delta(\val^{\sigma'}(v),p)$.
Since $\val^{\sigma'}$ is a fixed point of $\Op^{\sigma'}$ we have $\val^{\sigma'}(u) = \delta(\val^{\sigma'}(v),p)$
and since $e$ is improving we have $\val^\sigma(u) < \delta(\val^\sigma(v),p)$. 
This implies that $\val^\sigma(u) < \val^{\sigma'}(u)$.
\end{proof}

\begin{lemma}[Optimality]
\label{2-lem:optimality}
Let $\sigma$ a strategy satisfying parity that has no improving edges, then 
$\sigma$ is winning from all vertices of $\WE(\Game)$.
\end{lemma}

\begin{proof}
The fact that $\sigma$ satisfies parity means that it is a winning strategy
from all vertices $v$ such that $\val^\sigma(v) = \top$.
We now prove that Adam has a winning strategy from all vertices $v$ such that $\val^{\sigma}(v) \neq \top$.
We construct a strategy of Adam by
\[
\forall u \in \VA,\ \tau(u) \in \argmin \set{ \delta(\val^{\sigma}(v),p) : u \xrightarrow{p} v \in E}.
\]
We argue that $\tau$ ensures the complement of parity from all vertices $v$ such that $\val^{\sigma}(v) \neq \top$.
Let us consider $G$ the subgraph of $\Game[\tau]$ restricted to such vertices.
We argue that for all edges $u \xrightarrow{p} v$ in $G$, we have $\val^{\sigma}(u) \ge \delta(\val^{\sigma}(v),p)$.
Once this is proved we conclude using the second item of \Cref{2-lem:key_property} implying that $G$ satisfies the complement of parity.

The first case is when $u \in \VE$. 
Let $\sigma(u) = u \xrightarrow{p'} v'$.
Consider an edge $e = u \xrightarrow{p} v$, since it is not improving we have 
$\delta(\val^{\sigma}(v),p) \le \delta(\val^{\sigma}(v'),p')$.
Since $\val^\sigma$ is a fixed point of $\Op^{\sigma}$ we have $\val^\sigma(u) = \delta(\val^{\sigma}(v'),p')$,
implying the desired inequality.

The second case is when $u \in \VA$, it holds by definition of $\tau$ and because $\val^\sigma$ is a fixed point of $\Op^{\sigma}$.
\end{proof}

\paragraph{\bf Complexity analysis.}
One aspect of the algorithm we do not elaborate on here is the choice of improving edges at each iteration.
Many possible rules for choosing this set have been studied, as for instance the \emph{greedy all-switches} rule, choosing maximal improving edges.
This impacts the number of iterations of the algorithm, meaning the length of the sequence $\sigma_0,\sigma_1,\dots$. It is at most exponential since it is bounded by the number of strategies (which is bounded aggressively by $m^n$).
There are lower bounds showing that the sequence can be of exponential length, which apply to different rules for choosing sets of improving edges. Hence the overall complexity is exponential 

\paragraph{\bf The other strategy improvement algorithm for parity games.}
The main drawback of the algorithm we constructed is that it requires the initial strategy to be the one stopping at every vertex.
This is in contrast with many practical scenarios, where the fact that the algorithm can be initialised with any strategy (hence a potentially good one) is a key factor for performances.
As discussed at the beginning of this section, there are several other strategy improvement algorithms. All but one are obtained by reducing parity games to another class of games and applying a strategy improvement algorithm there. The remaining one is defined directly on parity games. The valuation of a play is a triple:
\begin{itemize}
	\item the maximum priority $p$ seen infinitely many times,
	\item the set of higher priorities seen before the first occurrence of $p$,
	\item the length of the prefix before the first occurrence of $p$.
\end{itemize}
The key benefit of this algorithm is that it can be initialised with any strategy.
We do not present here a correctness proof. 
Let us note that the algorithm is actually closely related to the one we built here, in the sense that after adding a stopping option the two algorithms coincide. 
Both algorithms have exponential complexity, with exponential lower bounds on the number of iterations.


\section{A quasi-polynomial time separating automata algorithm}
\label{2-sec:separation}
\subsection*{The separation framework}
We describe a general approach for reducing parity games to safety games.
\Cref{1-sec:automata} constructs reductions between objectives using automata: 
in the case at hand parity reduces to safety if there exists a deterministic automaton $\Automaton$ over the alphabet $[1,d]$
with acceptance objective $\Safe$ and defining $\Parity([1,d])$,
meaning $L(\Automaton) = \Parity([1,d])$.
With such an automaton in hand~\Cref{1-lem:automata_reduction} implies the following reduction:
from a parity game $\game$ construct the safety game $\game \times \Automaton$ satisfying that
Eve has a winning strategy in $\Game$ from $v_0$ if and only if she has a winning strategy in $\Game \times \Automaton$ from $(v_0,q_0)$.

\vskip1em
Unfortunately, it can be shown (using a topological argument) that there is no such automaton.
The separation framework defines a (weaker) sufficient condition for the reduction above to be correct.
Instead of insisting that $L(\Automaton) = \Parity([1,d])$,
it is enough to have $\WE(\arena,\Parity[\col]) = \WE(\arena,L(\Automaton)[\col])$.
To ensure this equality we will only require that $L(\Automaton)$ \textit{separates} winning plays from losing plays.
The two \textit{key} ideas are first to take advantage of the positionality of parity objectives 
by restricting further winning plays to \textit{positional} winning plays,
and second to require \[
\WE(\arena,\Parity[\col]) = \WE(\arena,L(\Automaton)[\col])
\]
not for all parity games, but only for parity games with $n$ vertices and priorities in $[1,d]$.

\vskip1em
A deterministic safety automaton over the alphabet $[1,d]$ is given by 
\[
\Automaton = (Q,q_0,\delta : Q \times [1,d] \to Q,\Safe[\col_\Automaton]),
\]
where $\col_\Automaton : Q \times [1,d] \to \set{\Win,\Lose}$.
A word $w \in [1,d]^\omega$ is accepted if the run $\rho$ over $w$ only contains transitions in $\Win$.
We use the following simplifying convention for safety automata: we distinguish a special rejecting state $\bot$
and assume that all transitions are accepting except the ones leading to $\bot$. 
Said differently, a word $w$ is accepted if and only if it the run over $w$ does not contain the state $\bot$.
Hence we do not need to specify $\col_\Automaton$:
in the next two sections by an automaton we mean a deterministic safety automaton
given by $\Automaton = (Q,q_0,\delta : Q \times [1,d] \to Q)$.

\vskip1em
Let us now give a sufficient condition for the automaton $\Automaton$ 
to imply the correctness of the reduction from the parity game $\Game$ to the safety game $\Game \times \Automaton$.
The condition that we define now is that the automaton $\Automaton$ is $(n,d)$-separating; it relies on the notion of parity graphs.
Recall a parity graph satisfies parity from $v$ if all infinite paths from $v$ satisfy parity,
and that a strategy $\sigma$ is winning from $v$ if and only if the parity graph $\Game[\sigma]$ satisfies parity from $v$.

We say that an automaton reads, accepts, or rejects a path $\pi$ in a parity graph, 
which is an abuse because what the automaton reads is the induced sequence of colours $\col(\pi)$.

\begin{definition}[Separating automata]
\label{2-def:separating_automata}
An automaton $\Automaton$ is $(n,d)$-\textit{separating} if the two following properties hold.
\begin{itemize}
	\item For all parity graphs $G$ with $n$ vertices and priorities in $[1,d]$ satisfying parity, 
	all paths from $v$ are accepted by $\Automaton$.
	
	\item All words accepted by $\Automaton$ satisfy parity.
\end{itemize}
\end{definition}
We define the objective $\Parity_{\mid n}$ over the set of colours $[1,d]$ as:
\[
\Parity_{\mid n} = 
\set{ \play : 
\begin{array}{l}
\play \text{ is a path starting in some parity graph} \\
\text{with } n \text{ vertices and priorities in } [1,d] \text{ satisfying } \Parity
\end{array}}.
\]
The definition of $(n,d)$-separating automata is illustrated in \Cref{2-fig:separation} and can be summarised
as $\Parity_{\mid n} \subseteq L(\Automaton) \subseteq \Parity$.

\begin{figure}[!ht]
\centering
  \begin{tikzpicture}
    \begin{scope}
      \draw[clip,use as bounding box] (0,0) ellipse (2cm and 1.2cm);
      \draw[dgrey] (0,0) ellipse (2cm and 1.2cm);
      \draw[white, very thick] (-2.5,0) ellipse (2.9cm and 2.5cm);
      \draw[lgrey] (-2.3,0) ellipse (2.1cm and 1.7cm);
      \draw[dgrey] (-2.5,0) ellipse (1.5cm and 1.3cm);
      \draw[hatcharea,hatchspread=6pt] (-2.3,0) ellipse (2.1cm and 1.7cm);
      \draw (0,0) ellipse (2cm and 1.2cm);
    \end{scope}
    \path[arrow]
      (2.2,1.6) node[] {$[1,d]^\omega \setminus$ \textsf{Parity}} edge (1.5,.3)
      (-.5,1.6) node[] {\textsf{Parity}} edge (-.1,.5)
      (-3,1.2) node[] {\textsf{Parity}$_{|n}$} edge (-1.4,.2)
      (-1.5,-1.5) node[below,anchor=base] {$L(\Automaton)$} edge (-.6,-.6);
  \end{tikzpicture}
\caption{The separation problem.}
\label{2-fig:separation}
\commentAlt{Figure~\ref{2-fig:separation}: A Venn diagram illustrating set relationships with labeled regions. See long description.}
\commentLongAlt{Figure~\ref{2-fig:separation}: An oval shape is divided into several regions. The leftmost region, labeled 'Parity_n', is dark grey with diagonal lines. Adjacent to it, moving right, is a region labeled 'L(A)', which is a lighter grey with vertical lines. Next is an unlabeled white crescent-shaped region labeled 'Parity' above it. The largest region on the right is a medium grey and labeled '[1,d] to the omega minus Parity'.}
\end{figure}

The following lemma shows the definition of separating automata in action.

\begin{lemma}[Game equivalence using separating automata]
\label{2-lem:separating_automata}
Let $\Automaton$ an $(n,d)$-separating automaton.
Then for all parity games $\game = (\arena,\Parity[\col])$ with $n$ vertices and priorities in $[1,d]$, 
we have
\[
\WE(\game) = \WE(\arena,L(\Automaton)[\col]).
\]
\end{lemma}

\begin{proof}
The inclusion $\WE(\game) \subseteq \WE(\arena,L(\Automaton)[\col])$ follows from positional determinacy and the inclusion 
$\Parity_{\mid n} \subseteq L(\Automaton)$.
Let $v \in \WE(\game)$.
Consider $\sigma$ a positional strategy ensuring parity from $v$ and construct the parity graph $\game[\sigma]$, it satisfies parity from $v$.
Hence the strategy $\sigma$ also ensures $L(\Automaton)[\col]$ from $v$, so $v \in \WE(\arena,L(\Automaton)[\col])$.

Conversely, $L(\Automaton) \subseteq \Parity$ implies $L(\Automaton)[\col] \subseteq \Parity[\col]$, which in turn implies 
\[
\WE(\arena,L(\Automaton)[\col]) \subseteq \WE(\game).
\]
\end{proof}

The last step is to explain how to solve a game with objective $L(\Automaton)$, as already discussed in~\Cref{1-sec:automata}.
Let $\Game = (\arena, L(\Automaton)[\col])$.
We construct a safety game by making the synchronised product of the arena with the automaton:
\[
\Game \times \Automaton = (\arena \times \Automaton, \Safe[\col']),
\]
where the safety condition ensures that the play is accepted by $\Automaton$.
Formally, we construct the arena $\arena \times \Automaton$ as follows.
We first define the graph $G \times Q$ whose set of vertices is $V \times Q$ and set of edges is defined as follows:
for every edge $u \xrightarrow{p} v \in E$ and state $q \in Q$ there is an edge from $(u,q)$ to $(v,\delta(q,p))$.
The arena is $\arena \times \Automaton = (G \times Q, \VE \times Q, \VA \times Q)$.
Using the convention for safety automata that the rejecting transitions are precisely those leading to the rejecting state $\bot$,
the colouring function is defined by $\col'(v,q) = \Win$ if $q \neq \bot$, and $\Lose$ otherwise.

\begin{fact}[Reduction to safety games using separating automata]
\label{2-fact:reduction}
Eve has a winning strategy in $\game$ from $v_0$ if and only if
she has a winning strategy in $\Game \times \Automaton$ from $(v_0,q_0)$.
\end{fact}

\begin{theorem}[Algorithm using separating automata]
\label{2-thm:algorithm_separating_automata}
Let $\Automaton$ an $(n,d)$-separating automaton.
There exists an algorithm for solving parity games of complexity $O(m \cdot |\Automaton|)$.
\end{theorem}
\begin{proof}
Let $\Game$ a parity game with $n$ vertices and priorities in $[1,d]$.
Thanks to~\Cref{2-lem:separating_automata} and \Cref{2-fact:reduction}, solving $\Game$
is equivalent to solving the safety game $\Game \times \Automaton$.
Thanks to~\Cref{1-thm:reachability} solving safety games can be done in time linear in the number of edges.
This yields an algorithm for solving parity games whose running time is $O(m \cdot |\Automaton|)$.
\end{proof}

In the remainder of this section we give a construction for a quasi-polynomial $(n,d)$-separating automaton.

\subsection*{The original separating automaton}
\begin{theorem}[The original separating automaton]
\label{2-thm:original_separating_automaton}
There exists an $(n,d)$-separating automaton of size $n^{O(\log d)}$,
inducing an algorithm for solving parity games of complexity $n^{O(\log d)}$.
\end{theorem}

\paragraph{\bf $i$-sequences.}
The key definition used by the original separating automaton is an
\emph{$i$-sequence}. Let $\pi = p_1, p_2, \dots, p_t$ a finite sequence of
priorities. An $i$-sequence is a set of \emph{indices} that splits $\pi$ into
sub-sequences. An $i$-sequence consists of exactly $2^i$ indices $1 \le j_1 <
j_2 < \dots < j_{2^i} \le t$, where each $j_k$ is an integer that refers to the
priority $p_{j_k}$ from the sequence $\pi$. An $i$-sequence is required to
satisfy the following properties. 
\begin{itemize} \item \textbf{Evenness.} Each
index (except possibly the last index) refers to an even priority, meaning that
$p_{j_k}$ is an even priority for all $k < 2^i$.

\item \textbf{Inner domination.} The subsequence of priorities between any two
indices $j_k$ and $j_{k+1}$ is dominated by either $p_{j_k}$ or $p_{j_{k+1}}$.
Formally, this means that whenever $j_k < l < j_{k+1}$, we have that $p_l \le
p_{j_k}$ or we have that $p_l \le p_{j_{k+1}}$.

\item \textbf{Outer domination.} The final subsequence between $p_{j_{2^i}}$ and
$p_t$ is dominated by $p_{j_{2^i}}$, meaning that for all $l > j_{2^i}$ we have
$p_l \le p_{j_{2^i}}$.
\end{itemize}

\begin{figure}[!ht]
    \begin{center}
    \begin{tikzpicture}

    \tikzstyle{seqc}=[draw, circle]

    \node [seqc] (1) {\Large $2$};
    \node [right of=1, node distance=1cm] (2) {\Large $1$};
    \node [seqc, right of=2, node distance=1cm] (3) {\Large $4$};
    \node [right of=3, node distance=1cm] (4) {\Large $3$};
    \node [right of=4, node distance=1cm] (5) {\Large $1$};
    \node [seqc, right of=5, node distance=1cm] (6) {\Large $2$};
    \node [seqc, right of=6, node distance=1cm] (7) {\Large $8$};
    \node [right of=7, node distance=1cm] (8) {\Large $7$};
    \node [right of=8, node distance=1cm] (9) {\Large $1$};

    \path[->,thick,bend left=45]
        (1) edge (3)
        (3) edge (6)
        (6) edge (7)
        ;
    \end{tikzpicture}
    \end{center}
    \caption{A $2$-sequence.}
\label{2-fig:isequence}
\commentAlt{Figure~\ref{2-fig:isequence}: A linear sequence of four nodes (2, 4, 2, 8) with numbers below them and two curved arrows indicating jumps. See long description.}
\commentLongAlt{Figure~\ref{2-fig:isequence}: Four nodes, each containing a number, are arranged horizontally: '2', '4', '2', '8'. Below these nodes, there are numbers '1', '3', '1', '7', '1'. A curved arrow points from the first '2' node to the '4' node, with '1' below the arrow. Another curved arrow points from the '4' node to the second '2' node, with '3' below the arrow. A straight arrow points from the second '2' node to the '8' node, with '1' below the arrow. The numbers '7' and '1' are to the right of the '8' node.}
\end{figure}

\Cref{2-fig:isequence} gives an example of a $2$-sequence. The circled
priorities are the indices used in the sequence. Note that there are exactly
$2^2 = 4$ indices used, and that every circled priority is even. Inner
domination is satisfied because every priority that is between two circled
priorities is lower than one of the two end points, and outer domination is
satisfied because the final circled priority $8$ is larger than all the
priorities that come after it.

\paragraph{\bf The relationship to parity games.}
The relationship between $i$-sequences and parity games is explained by the
following lemma.

\begin{lemma}[Completeness for the separating automaton]
\label{2-lem:isequencewin}
Suppose that Adam and Eve play positional strategies in the parity game,
and let $\pi$ the resulting play.
\begin{itemize}
\item If Eve wins the parity game, then there exists prefixes of $\pi$ that
contain arbitrarily long $i$-sequences.
\item If Adam wins the parity game, then no prefix of $\pi$ will contain a
$\lceil \log n \rceil$-sequence.
\end{itemize}
\end{lemma}
\begin{proof}
If Eve wins the parity game then the largest
priority occurring in $\pi$ infinitely often is even. Let $p$ this priority.
To construct a prefix containing an $i$-sequence, we find the first index $j$
after which no priority $q > p$ is seen. We take as our indices $j_1 < j_2 <
\dots < j_{2^i}$ the first $2^i$ occurrences of priority $p$ after index $j$.
Evenness is trivially satisfied, and both inner and outer domination are
satisfied because no priority larger than $p$ is seen after index $j$.

We prove the second claim by contradiction. Suppose that Adam wins the game, but
that there is a prefix of $\pi$ that contains a $\lceil \log n \rceil$-sequence.
Since the sequence indexes $2^{\lceil \log n \rceil} \ge n$ vertices of the
game, it must index the same vertex $v$ twice. Thus our $i$-sequence must
contain a cycle passing through $v$. Note that inner domination ensures that the
largest priority on the cycle that passes through $v$ is even. However, no even
cycle can be formed when Adam wins the game by playing a positional winning
strategy, and so we have arrived at our contradiction.
\end{proof}

To summarise, if Eve wins the game, then she has a strategy that ensures that
arbitrarily long $i$-sequences occur, while if Adam wins the game then he has a
strategy that ensures that no $\lceil \log n \rceil$-sequence occurs. So to
solve the parity game, it is sufficient to determine whether Eve can force a
$\lceil \log n \rceil$-sequence to occur.

\paragraph{\bf A data structure for recognising $i$-sequences.}
We will build a quasi-polynomial sized automaton that reads a sequence of
priorities, 
and determines whether that sequence of priorities contains a
$k$-sequence. 
The automaton is defined by a data structure that we call a \emph{record}, which
contains information about the $i$-sequences that have been seen so far in the
sequence.  

A record is a sequence $b_k, b_{k-1}, \dots, b_1, b_0$, where each
$b_i$ is either a priority, or the special symbol $\siblank$. The value of $b_i$
has the following meaning: 
\begin{itemize}
\item \textbf{Witnessing.} If $b_i \ne \siblank$, then we have seen an
$i$-sequence, and the final priority on that $i$-sequence is $b_i$. 
\item \textbf{Order.} If $b_i \ne \siblank$ and $b_j \ne \siblank$ and $j < i$,
then the first index of the $j$-sequence witnessed by $b_j$ occurs after the
last index of the $i$-sequence witnessed by $b_i$.
\end{itemize}
Note that, although each element $b_i$ records the existence of an
$i$-sequence, the record data structure \emph{does not} store the $2^i$ indices
of this $i$-sequence, it only stores the priority of the final index of that
sequence.

\begin{figure}[!ht]
    \begin{center}
    \begin{tikzpicture}
    \tikzstyle{seqc}=[draw, circle]
    \node [seqc] (1) {\Large $2$};
    \node [right of=1, node distance=0.9cm] (2) {\Large $1$};
    \node [seqc, right of=2, node distance=0.9cm] (3) {\Large $4$};
    \node [seqc, right of=3, node distance=0.9cm] (4) {\Large $2$};
    \node [seqc, right of=4, node distance=0.9cm] (5) {\Large $8$};
    \node [right of=5, node distance=0.9cm] (6) {\Large $7$};
    \node [seqc, right of=6, node distance=0.9cm] (7) {\Large $2$};
    \node [right of=7, node distance=0.9cm] (8) {\Large $1$};
    \node [seqc, right of=8, node distance=0.9cm] (9) {\Large $4$};
    \node [right of=9, node distance=0.9cm] (10) {\Large $1$};
    \node [seqc, grey, right of=10, node distance=0.9cm] (11) {\Large $2$};

    \path[->,thick,bend left=45]
        (1) edge (3)
        (3) edge (4)
        (4) edge (5)
        ;

    \path[->,thick,bend left=45]
        (7) edge (9)
        ;
    \end{tikzpicture}
    \end{center}
\caption{An example sequence that corresponds to the record $\siblank 8 4 2$.}
\label{2-fig:ds}
\commentAlt{Figure~\ref{2-fig:ds}: A linear sequence of nodes and numbers with two distinct colored sections, illustrating a transformation or comparison. See long description.}
\commentLongAlt{Figure~\ref{2-fig:ds}: The image displays two sequences of nodes and numbers. The left sequence shows four interconnected nodes: '2', '4', '2', '8'. Arrows connect them linearly. Below the first arrow is '1', below the second is '4', and below the third is '2'. The right sequence starts with a '2' node, followed by an arrow to a '4' node. Below the arrow is '1'. Following the '4' node, there's an isolated '1' and then a grey '2' node. A '7' separates the two sequences.}
\end{figure}

\Cref{2-fig:ds} shows an example sequence that is consistent with the
record that sets $b_3 = \siblank$, $b_2 = 8$, $b_1 = 4$, and $b_0 = 2$.
A $2$-sequence is represented by $b_2 = 8$, which is the last priority of
the $2$-sequence. Another $1$-sequence starts after the end of the
$2$-sequence, and it is represented by $b_1 = 4$. Likewise a 
$0$-sequence starts after the end of the $1$-sequence, and is represented by
$b_0 = 2$. There is no $3$-sequence in the example, and this is represented by
setting $b_3 = \siblank$.

\paragraph{\bf The update rule.}
Suppose that we have a record that represents the $i$-sequences in a
finite sequence of priorities, and that we then read the next priority $p$ in
that sequence. We need to update the record to take this priority into
account. We do this by applying the following \emph{update rule}.
The update rule consists of two steps, which occur one after the other.

\begin{itemize}
\item \textbf{Step 1.} In this step, we find the largest index $i$
such that $b_j$ is even for all $j \le i$. If $b_i = \siblank$, or $b_i < p$,
then we create a new record $b'_k$, $b'_{k-1}$, \dots, $b'_0$ by
setting:
\begin{equation*}
b'_j = \begin{cases}
b_j & \text{if $j > i$,} \\
p & \text{if $j = i$,} \\
\siblank & \text{if $j < i$.} 
\end{cases}
\end{equation*}
If there is no index $i$ that satisfies the conditions, then we do not modify
the record.

\item \textbf{Step 2.} In step 2, we take the output of step 1, and we find the
largest index $i$ such that $p > b_i$ and we create a new record $b'_k$,
$b'_{k-1}$, \dots, $b'_0$ by setting:
\begin{equation*}
b'_j = \begin{cases}
b_j & \text{if $j > i$,} \\
p & \text{if $j = i$,} \\
\siblank & \text{if $j < i$.} 
\end{cases}
\end{equation*}
Again, if there is no such index $i$, then the record is not modified.
\end{itemize}

\begin{figure}[!ht]
    \begin{center}
    \begin{tikzpicture}
    \tikzstyle{seqc}=[draw, circle]
    \node [seqc] (1) {\Large $2$};
    \node [right of=1, node distance=0.9cm] (2) {\Large $1$};
    \node [seqc, right of=2, node distance=0.9cm] (3) {\Large $4$};
    \node [seqc, right of=3, node distance=0.9cm] (4) {\Large $2$};
    \node [seqc, right of=4, node distance=0.9cm] (5) {\Large $8$};
    \node [right of=5, node distance=0.9cm] (6) {\Large $7$};
    \node [seqc, right of=6, node distance=0.9cm] (7) {\Large $2$};
    \node [right of=7, node distance=0.9cm] (8) {\Large $1$};
    \node [seqc, right of=8, node distance=0.9cm] (9) {\Large $4$};
    \node [right of=9, node distance=0.9cm] (10) {\Large $1$};
    \node [seqc, right of=10, node distance=0.9cm] (11) {\Large $2$};
    \node [seqc, right of=11, node distance=0.9cm] (12) {\Large $4$};

    \path[->,thick,bend left=45]
        (1) edge (3)
        (3) edge (4)
        (4) edge (5)
        (5) edge (7)
        (7) edge (9)
        (9) edge (11)
        (11) edge (12)
        ;

    \end{tikzpicture}
    \end{center}
    \caption{An example of a Step 1 update applied to the sequence and record from~\Cref{2-fig:ds}}
\label{2-fig:ds1}
\commentAlt{Figure~\ref{2-fig:ds1}: A linear sequence of interconnected nodes and numbers, depicting a flow or process. See long description.}
\commentLongAlt{Figure~\ref{2-fig:ds1}: The image displays a horizontal sequence of circles containing numbers, with arrows indicating transitions and numbers placed below the arrows. The sequence starts with a '2' node, an arrow to a '4' node (with '1' below), an arrow to a '2' node (with '2' below), an arrow to an '8' node (with '7' below). This is followed by a '2' node, an arrow to a '4' node (with '1' below), and finally an arrow to a '2' node (with '1' below) which then points to a '4' node (with no number below the final arrow). The overall pattern suggests a directed path through various numerical states or values.}
\end{figure}
Intuitively, Step 1 attempts to combine the $i$-sequences in the existing record
into a longer $i$-sequence. Suppose that we have read the sequence shown in~\Cref{2-fig:ds}, 
that we have compute the record $\siblank 8 4 2$, and that the next priority in the sequence is $4$.
\Cref{2-fig:ds1} shows the result of applying Step 1 to this situation.
Observe that $3$ is the largest index $i$ such that for all $j < i$ we have that
$b_j$ is even, so Step 1 will output the record $4 \siblank \siblank \siblank$.

So in this circumstance, Step 1 claims that we have now seen a $3$-sequence.
\Cref{2-fig:ds1} shows why this is correct: the $0$-sequence of $b_0$, the
$1$-sequence of $b_1$, and the $2$-sequence of $b_2$ can be merged together,
along with the new priority, to create a $3$-sequence. Observe that inner
domination in this new $3$-sequence is satisfied due to the outer domination
property for each of the $i$-sequences that it was constructed from.
For example, the $2$-sequence ends at priority $8$, and the $1$-sequence begins
at priority $2$, and we know that $8$ must dominate all priorities between the
$8$ and the $2$ because $8$ is required to dominate \emph{all} priorities that
follow it.

\begin{figure}[!ht]
    \begin{center}
    \begin{tikzpicture}
    \tikzstyle{seqc}=[draw, circle]
    \node [seqc] (1) {\Large $2$};
    \node [right of=1, node distance=0.9cm] (2) {\Large $1$};
    \node [seqc, right of=2, node distance=0.9cm] (3) {\Large $4$};
    \node [seqc, right of=3, node distance=0.9cm] (4) {\Large $2$};
    \node [right of=4, node distance=0.9cm] (5) {\Large $8$};
    \node [right of=5, node distance=0.9cm] (6) {\Large $7$};
    \node [right of=6, node distance=0.9cm] (7) {\Large $2$};
    \node [right of=7, node distance=0.9cm] (8) {\Large $1$};
    \node [right of=8, node distance=0.9cm] (9) {\Large $4$};
    \node [right of=9, node distance=0.9cm] (10) {\Large $1$};
    \node [right of=10, node distance=0.9cm] (11) {\Large $2$};
    \node [seqc, right of=11, node distance=0.9cm] (12) {\Large $9$};

    \path[->,thick,bend left=45]
        (1) edge (3)
        (3) edge (4);

    \path[->,thick,bend left=20]
        (4) edge (12)
        ;

    \end{tikzpicture}
    \end{center}
    \caption{An example of a Step 2 update applied to the sequence and record from~\Cref{2-fig:ds}}
\label{2-fig:ds2}
\commentAlt{Figure~\ref{2-fig:ds2}: A linear sequence of nodes and numbers, with one long curved arrow skipping several elements. See long description.}
\commentLongAlt{Figure~\ref{2-fig:ds2}: The image shows a series of nodes and numbers arranged horizontally. It begins with a '2' node, followed by an arrow to a '4' node (with '1' below the arrow). An arrow then connects the '4' node to a '2' node. After this '2' node, a long curved arrow originates and arcs over the numbers '8', '7', '2', '1', '4', '1', '2', before landing on a final '9' node. The numbers '8', '7', '2', '1', '4', '1', '2' are positioned linearly below the path of the long curved arrow.}
\end{figure}

Step 2 ensures that the outer domination property holds. 
In~\Cref{2-fig:ds2}, we show the result of applying Step 2 to the record
$\siblank 8 4 2$ that corresponds to the sequence shown in~\Cref{2-fig:ds},
when the next priority in the sequence is $9$. Observe that since $9 > 8$, the
outer domination property for the $2$-sequence ending at $8$ now fails to hold,
and likewise for the sequences ending at $4$ and $2$. Hence, Step 2 deletes the
$0$-sequence and $1$-sequence from the record, and updates the $2$-sequence to
end at $9$, thereby restoring outer domination. The resulting record is $\siblank
9 \siblank \siblank$.

\paragraph{\bf Correctness.} 
To compute a record for a particular sequence of priorities, we start with the
record $\siblank \siblank \dots \siblank$, and then process the sequence one
priority at a time, using the update rule that we have described. 

We must now argue that the record data structure and update rule is sufficient
to decide the winner of a parity game. The following lemma states that a record
will never falsely claim that an $i$-sequence has occurred.

\begin{lemma}[Correctness for the separating automaton]
\label{2-lem:correctness_separating_automata}
Let $b_k, b_{k-1}, \dots, b_0$ the record for a sequence of priorities $\pi$.
If $b_i \ne \siblank$, then $\pi$ contains an $i$-sequence.
\end{lemma}
\begin{proof}
This can be proved by induction over the components of the record. In fact we
will prove the slightly stronger order property that we mentioned earlier:
the $i$-sequence corresponding to
$b_i$ starts after the $j$ sequence corresponding to $b_j$ whenever $i < j$ and
$b_i \ne \siblank$ and $b_j \ne \siblank$.

The base case is trivially true, since the value of $b_0$ asserts the existence
of a 0-sequence, and any priority by itself is a $0$-sequence. So when Step 1 or
Step 2 updates $b_0$, the corresponding $0$-sequence is the new priority, and
this clearly starts after all other $i$-sequences in the record.

For the inductive step, we must prove that the two steps of the update rule are
correct. 
\begin{itemize}
\item For Step 1 updates, we can use the inductive hypothesis to argue that, for
each $j < i$, the $j$-sequence corresponding to $b_j$ exists, and that they
appear in order in the sequence, and that it ends before the sequence
corresponding to $b_{j-1}$ starts. Furthermore, the outer domination property
the $j$-sequence ensures that all priorities between the end of the
$j$-sequence and the start of the $(j-1)$-sequence are dominated by the last
priority in the $j$-sequence, which must be even according to the definition of
a Step 1 update.
Hence, we can combine all of the $j$-sequences with $j < i$ together, along with
the new priority, to create an $i$-sequence. This new $i$-sequence starts at
exactly the same point as the sequence corresponding to $b_{i-1}$, and so the
order property still holds.

\item For Step 2 updates, we only need to argue that the value of $b'_i$
correspond to an $i$ sequence. This can be constructed as we showed in~\Cref{2-fig:ds2}: 
take the $i$-sequence that corresponds to $b_i$, and
replace the final priority with the new priority. Observe that the final
priority of an $i$-sequence is permitted to be odd, and so this new sequence
satisfies all of the requirements of an $i$-sequence. Furthermore, the starting
point of this sequence has not changed, and so the order property is preserved.
\end{itemize}
\end{proof}

As a consequence of the lemma above, if Adam has a strategy to ensure that no
$k$-sequence occurs in the game, then Adam has a strategy to ensure that the
$b_k$ component of the record is never set so that $b_k \ne \siblank$.

It can be shown that the other direction is also true: if an $i$-sequence has
occurred, then there will be some index $j \ge i$ such that $b_j \ne \siblank$.
However, the proof is somewhat tedious, and this statement is actually stronger
than what we need. To argue that the record can determine the winner of a parity
game, the following weaker lemma suffices. 

\begin{lemma}[Weaker correctness for the separating automaton]
\label{2-lem:weaker_correctness_separating_automaton}
Let $\pi$ an infinite play that is winning for Eve. For all $k$, there exists
a prefix of $\pi$ such that $b_k \ne \siblank$.
\end{lemma}
\begin{proof}
Let $p$ the largest even priority that is seen infinitely often, and let $j$
be the first index after which no priority larger than $p$ is visited. We argue
that after index $j$ has been reached, the record will eventually set $b_i \ne
\siblank$ for all $i$.

To see why, observe that after index $j$ has been reached, Step 2 cannot replace
any component $b_j$ with $b_j = p$, since Step 2 can only overwrite the priority
in $b_j$ when the new priority $p'$ satisfies $p' > b_j$, but no priority $p' >
p$ is seen after index~$j$. 

On the other hand, Step 1 will always be triggered whenever we visit the
priority $p$. Step 1 will always set some component of the record to $p$, and as
we have observed this cannot be overwritten by Step 2. Moreover, since $p$ is
even, repeated application of Step 1 will
build a longer and longer $i$-sequences whose outer domination
priority is $p$. Thus, after we have made $2^k$ visits to $p$, we will have set
$b_k = p \ne \siblank$, if we have not done so already.
\end{proof}

Hence, if Eve wins the parity game, then she has a strategy to eventually ensure
that $b_k \ne \siblank$. Combining the two lemmas above, with~\Cref{2-lem:isequencewin} gives the following corollary.

\begin{corollary}[Correctness of the reduction for the separating automaton]
\label{2-cor:correctness_reduction_separating_automaton}
Suppose that we monitor the play of a parity game with a record $b_{\lceil \log
n \rceil}, \dots, b_0$. Eve has a strategy that ensures $b_{\lceil \log n
\rceil} \ne \siblank$ if and only if Eve wins the parity game.
\end{corollary}

\paragraph{\bf The size of the automaton.}
The record data structure and update rule can be encoded as a deterministic
finite automaton that reads the play. Each state of the automaton is associated
with some configuration of $b_{\lceil \log n \rceil}, \dots, b_0$, and the
transitions of the automaton are defined by the update rule. 

This automaton has quasi-polynomial size. The number of states used in the
automaton is the number of possible configurations of
$b_{\lceil \log n \rceil}, \dots, b_0$. Each $b_i$ can be one of the $d$
priorities in the game, or the symbol $\siblank$, and so there are $d + 1$
possible values that it can take. Moreover there are $\log n + 1$ components of
the record, so the total number of configurations is at most
$(d+1)^{\log n +1} = n^{O(\log d)}$.


\section{A quasi-polynomial time value iteration algorithm}
\label{2-sec:value_iteration}
\begin{theorem}[Value iteration]
\label{2-thm:value_iteration_quasipoly}
There exists a value iteration algorithm for solving parity games in time 
\[
O\left(m \cdot \log(n) \log(d) \cdot n^{2.45 + \log_2 \left( 1 + \frac{d/2-1}{\log_2(n)} \right)} \right),
\]
which is quasi-polynomial in general and polynomial if $d = O(\log_2(n))$.
The space complexity of the algorithm is quasi-linear, and more precisely $O(m + n \log_2(d))$.
\end{theorem}

The presentation follows the introduction to value iteration algorithms given in \Cref{1-sec:value_iteration}, although it does not technically rely on it.
Let $\game = (\arena,\Parity[\col])$ a parity game with $n$ vertices and priorities in $[1,d]$,
and without loss of generality $d$ is even.

The first step is to define a notion of value function $\val^\game : V \to Y$ with $(Y,\le)$ a lattice satisfying the characterisation principle:
for all vertices $u$ we have that Eve wins from $u$ if and only if $\val^\game(u) \neq \top$, where $\top$ is the largest element in $Y$.
The goal of the algorithm is to compute $\val^\game$, from which we then easily obtain the winning region thanks to the characterisation principle.


We let $F_V$ be the lattice of functions $V \to Y$ equipped with the component wise order induced by $Y$.
We are looking for a monotonic function $\delta : Y \times [1,d] \to Y$ inducing the operator $\Op : F_V \to F_V$ defined by:
\[
\Op(\mu)(u) = 
\begin{cases}
\min \set{\delta( \mu(v), p) : u \xrightarrow{p} v \in E} & \text{ if } u \in \VE, \\
\max \set{\delta( \mu(v), p) : u \xrightarrow{p} v \in E} & \text{ if } u \in \VA,
\end{cases}
\]
such that $\val^\game$ is the least fixed point of $\Op$.
The algorithm will then simply use Kleene's fixed-point theorem (\Cref{1-thm:kleene}) to compute $\val^\game$ by iterating the operator $\Op$.

Let us look at this question using the notion of progress measures, which are pre-fixed points of $\Op$,
meaning $\mu$ such that $\Op(\mu) \le \mu$. 
Since the least fixed point of $\Op$ is also its least pre-fixed point, an equivalent formulation of the characterisation principle above reads: for all vertices $u$ we have that Eve wins from $u$ if and only if there exists a progress measure $\mu$ such that $\mu(u) \neq \top$.

To summarise this discussion, we are looking for a lattice $(Y,\le)$ and a monotonic function $\delta : Y \times [1,d] \to Y$ 
such that for all parity games $\Game$ with $n$ vertices and priorities in $[1,d]$, 
for all vertices $u$ we have that Eve wins from $u$ if and only if there exists a progress measure $\mu$ such that $\mu(u) \neq \top$.
Our next step is to show how the notion of universal trees provides a class of solutions to this problem.

\subsection*{Universal trees}
The trees we consider have three properties: 
they are rooted, every leaf has the same depth, and the children of a node are totally ordered.
Formally, a tree of height $0$ is a leaf,
and a tree $t$ of height $h + 1$ is an ordered list $[t_1,\dots,t_k]$ of subtrees each of height~$h$.

We consider two parameters for trees: the height, and the size which is defined to be the number of leaves.

\begin{figure*}[!ht]
\centering
\begin{tikzpicture}
  \begin{scope}[xscale=1,yscale=1.6]
  \path (0,0) node[coordinate] (root) {};
  \foreach \x in {-2,...,2}
    {\draw (root) -- (\x,-1) node[coordinate] (n\x) {};}
  \foreach \s/\x/\n in {-2/-2/,
    -1/-1.4/,-1/-.6/,
    0/-.4/,0/-.2/,0/0/,0/.2/,0/.4/,
    1/.6/,1/1.4/,
    2/2/}
    {\draw (n\s) -- (\x,-2) node[below] {$\n$};}
  \end{scope}

  \begin{scope}[xscale=.6,yscale=1]
  \path (6,-1) node[coordinate] (root) {};
  \foreach \x in {-1,0,1}
    {\draw[red, very thick] (root) -- (6+\x,-2) node[coordinate] (m\x) {};}
  \foreach \s/\x in {-1/-1.4,
				    -1/-.6,
				    0/-.4,
				    0/.4,
				    1/1}
    {\draw[red, very thick] (m\s) -- (6+\x,-3);}

  \path (10,-1) node[coordinate] (root) {};
  \foreach \x in {-2,1}
    {\draw (root) -- (10+\x,-2) node[coordinate] (o\x) {};}
  \foreach \x in {-1,0,2}
    {\draw[red, very thick] (root) -- (10+\x,-2) node[coordinate] (o\x) {};}
  \foreach \s/\x in {-2/-2,
			    0/-.4,
			    0/-.2,
			    0/.2,
			    1/.6,
			    1/1.4}
    {\draw (o\s) -- (10+\x,-3);}
  \foreach \s/\x in {-1/-1.4,
    			-1/-.6,
			    0/0,
			    0/.4,
			    2/2}
    {\draw[red, very thick] (o\s) -- (10+\x,-3);}
  \end{scope}
\end{tikzpicture}
\caption{On the left, a tree of height $h = 2$, which is the smallest $(5,2)$-universal tree:
it has size $11$ (meaning it has $11$ leaves).
On the right, a tree of size $5$ and one possible embedding into the universal tree.}
\label{2-fig:example_universal}
\commentAlt{Figure~\ref{2-fig:example_universal}: Three tree-like diagrams, each with a central trunk branching upwards. See long description.}
\commentLongAlt{Figure~\ref{2-fig:example_universal}: The image shows three abstract tree structures. The leftmost structure is entirely composed of thin lines, with a broad base converging to a single point at the top, and a smaller, similar tree nested within its lower center. The middle structure is similar in form to the nested tree on the left but is drawn with thicker lines. The rightmost structure combines elements of the first two: it has the same broad-based, single-point top as the leftmost tree, but some of its branches and the inner nested tree are drawn with thicker lines.}
\end{figure*}

We say that a tree $t$ embeds into another tree $T$ if:
\begin{itemize}
	\item either both are leaves,
	\item or let $t = [t_1,\dots,t_k]$ and $T = [T_1,\dots,T_{k'}]$, 
	there exist $i_1 < \dots < i_k$ such that for all $j \in [1,k]$ we have that $t_j$ embeds into $T_{i_j}$.
\end{itemize}

\begin{definition}[Universal trees]
\label{2-def:universal_trees}
A tree is $(n,h)$-\textit{universal} if it embeds all trees of size $n$ and height $h$.
\end{definition}

We refer to \Cref{2-fig:example_universal} for an example of a $(5,2)$-universal tree.
A first example of an $(n,h)$-universal tree is the tree where each node has degree $n$:
formally we define it recursively by $T_{n,0}$ is a leaf, and $T_{n,h+1} = [\underbrace{T_{n,h},\dots,T_{n,h}}_{n \text{ copies}}]$.
It has size $n^h$.

\subsection*{A quasi-polynomial universal tree}
We present an inductive construction of a quasi-polynomial universal tree.

\begin{theorem}[Construction of a quasi-polynomial universal tree]
\label{2-thm:universal_tree}
There exists an $(n,h)$-universal tree with size $f(n,h)$ satisfying the following:
$$\begin{array}{lll}
f(n,h) & = & f(n,h-1) + f(\lfloor n/2 \rfloor,h) + f(n - 1 - \lfloor n/2 \rfloor,h), \\
f(n,0) & = & 1, \\
f(0,h) & = & 0.
\end{array}$$
An upper bound is given by
\[
f(n,h) \le n \cdot \binom{h - 1 + \lfloor \log_2(n) \rfloor}{\lfloor \log_2(n) \rfloor}.
\]
This is also bounded from above by $n^{2.45 + \log_2 \left(1 + \frac{h-1}{\log_2(n)} \right)}$, which is quasi-polynomial in $n$ and $h$ in general, and polynomial if $h = O(\log_2(n))$.
\end{theorem}


\begin{proof}
To construct the $(n,h)$-universal tree $T$, let:
\begin{itemize}
	\item $T_\text{left}$ be a $(\lfloor n/2 \rfloor,h)$-universal tree,
	\item $T_\text{middle}$ be a $(n,h-1)$-universal tree,
	\item $T_\text{right}$ be a $(n - 1 - \lfloor n/2 \rfloor,h)$-universal tree.
\end{itemize}
The intuitive construction of $T$ is as follows: 
we merge the roots of $T_\text{left}$ and $T_\text{right}$ and insert in-between them
a child of the root to which is attached $T_\text{middle}$.
Formally, let $T_\text{left} = [T^1_{\text{left}},\dots,T^k_{\text{left}}]$ and 
$T_\text{right} = [T^1_{\text{right}},\dots,T^{k'}_{\text{right}}]$,
we define $T$ as
\[
[T^1_{\text{left}},\dots,T^k_{\text{left}},\ T_\text{middle},\ T^1_{\text{right}},\dots,T^{k'}_{\text{right}}].
\]
The construction is illustrated in \Cref{2-fig:smallest_tree_construction}.

\begin{figure}[!ht]
\centering
\begin{tikzpicture}
  \begin{scope}[line width=.8pt]
  \foreach \sign/\dir in {-/left,/right}
    {\draw[fill=black!20!white] (0,0) -- (\sign 4,-5) -- (\sign 2,-5) -- (0,0);
     \path (\sign 2.75,-4.5) node {$T_{\textup{\dir}}$};}
  \draw[fill=blue!60!green!20!white] (0,-1) -- (-1.2,-5) -- (1.2,-5) -- (0,-1);
  \path (0,-4.5) node {$T_{\textup{middle}}$};
  \draw (0,0) -- (0,-1);
  \end{scope}
\end{tikzpicture}
\caption{The inductive construction.}
\label{2-fig:smallest_tree_construction}
\commentAlt{Figure~\ref{2-fig:smallest_tree_construction}: Three triangular regions labeled T_left, T_middle, and T_right, all converging to a single point at the top.}
\commentLongAlt{Figure~\ref{2-fig:smallest_tree_construction}: The image displays three distinct triangular regions. All three triangles share a common apex at the top of the image. The left triangle is shaded light grey and labeled "T_left". The middle triangle is shaded light blue and labeled "T_middle". The right triangle is also shaded light grey and labeled "T_right". The bases of the triangles are separated from each other.}
\end{figure}

\vskip1em
We argue that $T$ is $(n,h)$-universal.
Consider a tree $t = [t_1,\dots,t_k]$ with $n$ leaves.
The question is where to cut, \textit{i.e.} which subtree of $t$ gets mapped to $T_\text{middle}$.
Let $n(t_i)$ be the number of leaves in $t_i$. 
Since $t$ has $n$ leaves, we have $n(t_1) + \cdots + n(t_k) = n$.
There exists a unique $p \in [1,k]$ such that 
$n(t_1) + \cdots + n(t_{p-1}) \le \lfloor n/2 \rfloor
\text{ and } 
n(t_1) + \cdots + n(t_p) > \lfloor n/2 \rfloor$.
The choice of $p$ implies that $n(t_{p+1}) + \cdots + n(t_k) \le n - 1 - \lfloor n/2 \rfloor$.
To embed $t$ into $T$, we proceed as follows:
\begin{itemize}
	\item the tree $[t_1,\dots,t_{p-1}]$ has at most $\lfloor n/2 \rfloor$ leaves,
	so it embeds into $T_\text{left}$ by induction hypothesis;
	\item the tree $t_p$ has height $h-1$ and at most $n$ leaves, so in embeds into $T_\text{middle}$ by induction hypothesis;
	\item the tree $[t_{p+1},\dots,t_k]$ has at most $n - 1 - \lfloor n/2 \rfloor$ leaves,
	so it embeds into $T_\text{right}$ by induction hypothesis.
\end{itemize}
\end{proof}
\noindent The construction given in the proof yields the smallest $(5,2)$-universal tree illustrated in \Cref{2-fig:example_universal}.

\subsection*{Ordering the leaves}
Let us consider a tree $t$ of height $h$, and write $d = 2h$.
A leaf is given by a list of directions that we define now.
For technical convenience that will manifest itself later, the list of directions is indexed by odd numbers $p \in [1,d]$ downwards:
for example for $d = 10$ a leaf is $(D_9,D_7,D_5,D_3,D_1)$.

We write $Y_t$ for the set of leaves of $t$ and $\le$ for the lexicographic order on $Y_t$.
Note that its interpretation on the tree is: for two leaves $\ell,\ell'$, we have $\ell \le \ell'$ if and only if $\ell$ is to the left of $\ell'$.
The strict version of $\le$ is $<$.

We introduce a set of relations $\vartriangleleft_p$ over $Y_t$ for each $p \in [1,d]$.
For a leaf $\ell = (D_{d-1},\dots,D_1)$ we write $\ell_{\ge p}$ for the tuple $(D_{d-1},\dots,D_p)$,
which we call the $p$-truncated branch of $\ell$.
\begin{itemize}
	\item For $p$ odd, we say that $\ell \vartriangleleft_p \ell'$ 
	if $\ell_{\ge p}\ <\ \ell'_{\ge p}$.
	\item For $p$ even, we say that $\ell \vartriangleleft_p \ell'$ 
	if $\ell_{\ge p}\ \le\ \ell'_{\ge p}$.
\end{itemize}

To interpret $\vartriangleleft_p$ on the tree, we label the levels by priorities from bottom to top as in \Cref{2-fig:example_universal}.
Then $\ell \vartriangleleft_p \ell'$ if and only if the $p$-truncated branch of $\ell$ is to the left of the $p$-truncated branch of $\ell'$,
strictly if $p$ is odd, and non-strictly if $p$ is even.

\begin{figure*}[!ht]
\centering
\begin{tikzpicture}
  \begin{scope}[xscale=1.4,yscale=1.6]
  \path (0,0) node[coordinate] (root) {};
  \foreach \x in {-2,...,2}
    {\draw (root) -- (\x,-1) node[coordinate] (n\x) {};}

  \path (0,-0.5) node[coordinate] (h0) {};
  \draw[red, very thick] (h0) -- (0,-1) node[below] {};
  \path (0.5,-0.5) node[coordinate] (h1) {};
  \draw[red, very thick] (h1) -- (1,-1) node[below] {};

  \foreach \s/\x/\n in {
	-2/-2/,
    -1/-1.4/,
    -1/-.6/,
    0/-.4/,
    0/-.2/\ell_1,
    0/0/,
    0/.2/,
    0/.4/,
    1/.6/\ell_2,
    1/1.4/\ell_3,
    2/2/
    }
    {\draw (n\s) -- (\x,-2) node[below] {$\n$};}
  \foreach \s/\x/\n in {
    0/-.2/\ell_1,
    1/.6/\ell_2
    }
	{\draw[red, thick] (n\s) -- (\x,-2) node[below] {};}

  \foreach \y/\l in {0.2/4,.7/3,1.2/2,1.7/1}
	{
	\path (-2.5,-\y+0.2) node[coordinate] (l\l) {};
    \draw (2.5,-\y) node (r\l) {\begin{Huge}\l\end{Huge}};
	\draw[dash dot] (l\l) -- (r\l);
    }

  \end{scope}
\end{tikzpicture}
\caption{Illustration of the relations $\vartriangleleft_p$.}
\label{2-fig:example_relations}
\commentAlt{Figure~\ref{2-fig:example_relations}: A layered tree structure with horizontal dashed lines and labeled points at the bottom. See long description.}
\commentLongAlt{Figure~\ref{2-fig:example_relations}: The image depicts a hierarchical structure resembling a tree, with a single apex at the top. From this apex, multiple thin lines descend. Four horizontal dashed lines, labeled '4', '3', '2', and '1' from top to bottom, intersect these descending lines, indicating different levels. Below the lowest dashed line, some of the descending lines terminate, while others continue downwards, forming a smaller, nested tree-like structure. Within this nested structure, a few specific paths are highlighted with thicker lines. The points where the highlighted thicker lines terminate at the very bottom are labeled 'l1', 'l2', and 'l3'.}
\end{figure*}

We refer to \Cref{2-fig:example_relations} for some examples.
We see that $\ell_1 \vartriangleleft_3 \ell_2$.

\[
\ell_2 \vartriangleleft_1 \ell_3 \quad ; \quad  
\ell_2 \vartriangleleft_2 \ell_3 \quad ; \quad 
\ell_3 \vartriangleleft_2 \ell_2 \quad ; \quad 
\ell_1 \vartriangleleft_3 \ell_2 \quad ; \quad 
\ell_1 \vartriangleleft_2 \ell_2.
\]

\begin{lemma}[Properties of the tree orders]
\label{2-lem:properties_tree}
The relations $\vartriangleleft_p$ for $p \in [1,d]$ induced by a tree $t$ satisfy the following properties:
\begin{itemize}
	\item $\vartriangleleft_d$ is the full relation: for all $b,b'$ we have $b \vartriangleleft_d b'$;
	\item if $\ell \vartriangleleft_p \ell'$ and $\ell' \vartriangleleft_q \ell''$ then $\ell \vartriangleleft_{\max(p,q)} \ell''$;
	\item the relation $\vartriangleleft_p$ is non-reflexive if $p$ is odd;
	\item the relation $\vartriangleleft_1$ is total;
	\item for $p < d$ even we have $\ell \vartriangleleft_p \ell'$ if and only if $\neg (\ell' \vartriangleleft_{p+1} \ell)$.
\end{itemize}
\end{lemma}
%

The following observation rephrases the notion of embeddings between trees using the ordering on leaves.

\begin{fact}[Equivalence between embedding and orders]
\label{2-fact:embedding}
Let $t,T$ be two trees of height $h$ and $d = 2h$.
Then $t$ embeds into $T$ if and only if there exists a function $\mu : Y_t \to Y_T$
such that for all leaves $\ell,\ell'$, for all $p \in [1,d]$:
\[
\ell \vartriangleleft_p^t \ell' \implies \mu(\ell) \vartriangleleft_p^T \mu(\ell').
\]
\end{fact}

\subsection*{Progress measures}
We explain how a tree $t$ induces both a lattice $(Y_t,\le)$ and a monotonic function $\delta_t : Y_t \times [1,d] \to Y_t$.
The set $Y_t$ is the set of leaves of $t$ augmented with a new element $\top$, 
and $\le$ is the lexicographic order on leaves with $\top$ as greatest element.
For each $p \in [1,d]$ and $\ell \in Y_t$ we extend $\vartriangleleft_p$ with $\ell \vartriangleleft_p \top$.
We then define $\delta_t : Y_t \times [1,d] \to Y_t$ by
\[
\delta_t(\ell,p) = \min_{\le} \set{\ell' : \ell \vartriangleleft_p \ell'}.
\]
This in turn induces a monotonic operator $\Op_t : F_V \to F_V$ defined by:
\[
\Op_t(\mu)(u) = 
\begin{cases}
\min \set{\delta_t( \mu(v), p) : u \xrightarrow{p} v \in E} & \text{ if } u \in \VE, \\
\max \set{\delta_t( \mu(v), p) : u \xrightarrow{p} v \in E} & \text{ if } u \in \VA,
\end{cases}
\]

Let $\Game$ be a parity game, a progress measure is a function $\mu : V \to Y_t$ which is a pre-fixed point: $\Op_t(\mu) \le \mu$. 
Unfolding the definitions, this means that for all vertices $u$, we have
\[
\begin{array}{llll}
\exists u \xrightarrow{p} v \in E,\ & \delta_t( \mu(v), p) \le \mu(u) & \text{ if } u \in \VE, \\
\forall u \xrightarrow{p} v \in E,\ & \delta_t( \mu(v), p) \le \mu(u) & \text{ if } u \in \VA.
\end{array}
\]
The definition of $\delta_t$ further simplifies it to: for all vertices $u$, we have
\[
\begin{array}{llll}
\exists u \xrightarrow{p} v \in E,\ & \mu(v) \vartriangleleft_{p} \mu(u) & \text{ if } u \in \VE, \\
\forall u \xrightarrow{p} v \in E,\ & \mu(v) \vartriangleleft_{p} \mu(u) & \text{ if } u \in \VA.
\end{array}
\]

The following theorem is our first and main step towards proving the characterisation principle.

\begin{theorem}[Fundamental theorem for progress measures]
\label{2-thm:progress_measure}
Let $\Game$ be a parity game and $v$ a vertex.
Then Eve wins from $v$ if and only if there exists a tree $t$ and a progress measure $\mu : V \to Y_t$ such that $\mu(v) \neq \top$.
\end{theorem}

In order to prove \Cref{2-thm:progress_measure}, we first consider the case of parity graphs.
A progress measure in a parity graph is a function $\mu : V \to Y_t$ such that 
for all edges $u \xrightarrow{p} v \in E$ we have $\mu(v) \vartriangleleft_{p} \mu(u)$.

Recall that a graph satisfies parity from $v$ if all infinite paths from $v$ satisfy parity.
This is equivalent to asking whether all cycles reachable from $v$ are even, meaning the maximal priority appearing in the cycle is even.

\begin{lemma}[Fundamental lemma for progress measures over graphs]
\label{2-lem:progress_measure}
Let $G$ be a parity graph and $v$ a vertex.
Then $G$ satisfies parity from $v$ if and only if 
there exists a tree $t$ and a progress measure $\mu : V \to Y_t$ such that $\mu(v) \neq \top$.
\end{lemma}

\begin{proof}
Let us assume that there exists a tree $t$ and a progress measure $\mu : V \to Y_t$ such that $\mu(v) \neq \top$
and for all edges $u \xrightarrow{p} v \in E$ we have $\mu(v) \vartriangleleft_{p} \mu(u)$.
To show that $G$ satisfies parity from $v$ we show that any cycle reachable from $v$ is even.
Let us consider such a cycle:
\[
v_1 \xrightarrow{p_1} v_2 \xrightarrow{p_2} v_3 \cdots v_k \xrightarrow{p_k} v_1.
\]
Since the cycle is reachable from $v$ and $\mu(v) \neq \top$, this implies that $\mu(v_i) \neq \bot$ for $i \in [1,k]$.
Let us assume towards contradiction that its maximal priority is odd, and without loss of generality it is $p_1$.
Applying our hypothesis to each edge of the cycle we have
\[
\mu(v_1) \vartriangleleft_{p_k} \mu(v_k) \vartriangleleft_{p_{k-1}} \cdots 
\vartriangleleft_{p_2} \mu(v_k) \vartriangleleft_{p_1} \mu(v_1).
\]
The second item of \Cref{2-lem:properties_tree} implies that $\mu(v_1) \vartriangleleft_{p_1} \mu(v_1)$, 
which contradicts the third item since $\vartriangleleft_{p_1}$ is non-reflexive given that $p_1$ is odd.

\vskip1em
Let us now prove the converse implication.
We prove the following property by induction on the number of priorities:
for all graphs satisfying parity (without the usual assumption that every vertex has an outgoing edge),
there exists a tree $t$ and a progress measure $\mu : V \to Y_t$ such that $\mu(v) \neq \top$
for all vertices~$v \in V$.

Let $G$ a graph satisfying parity. Without loss of generality the largest priority $d$ in the graph is even.
Let us define $G'$ the graph obtained from $G$ by removing all edges with priority $d$.
We consider its decomposition in strongly connected components: let $G_1,\dots,G_k$ denote the strongly connected components of $G'$,
numbered so that for any edge $u \rightarrow v \in E(G')$, if $u \in V(G_i)$ then $v \in V(G_j)$ for $j \ge i$.
A stronger property holds for edges $u \xrightarrow{d-1} v \in E(G')$: if $u \in V(G_i)$ then $v \in V(G_j)$ for $j > i$.
Indeed, if $v \in V(G_i)$ then we could form a cycle in $G'$ whose largest priority is $d-1$, a contradiction.
Hence each $G_i$ has priorities in $[1,d-2]$, and being a subgraph of $G$ it satisfies parity.
By induction hypothesis, there exists a tree $t_i$ and a progress measure $\mu_i : V(G_i) \to Y_{t_i}$ such that $\mu_i(v) \neq \top$
for all vertices~$v \in V(G_i)$.

Let us define $t = [t_1,\dots,t_k]$, and the function $\mu : V(G) \to Y_t$ by 
$\mu(v) = \mu_i(v)$ if $v \in V(G_i)$. We claim that $\mu$ is a progress measure: let us consider an edge $u \xrightarrow{p} v \in E(G)$, then
\begin{itemize}
	\item if $p = d$, then $\mu(v) \vartriangleleft_{d} \mu(u)$ because $\vartriangleleft_{d}$ is the full relation;
	\item if $p = d-1$, then $\mu(v) \vartriangleleft_{d-1} \mu(u)$ because $u \in V(G_i)$ and $v \in V(G_j)$ for $j > i$;
	\item if $p < d-1$, then $\mu(v) \vartriangleleft_{p} \mu(u)$. Indeed $u \in V(G_i)$ and $v \in V(G_j)$ for $j \ge i$, so either $j = i$ and this follows from the fact that $\mu_i$ is a progress measure, or $j > i$ and we have $\mu(v) \vartriangleleft_{d-1} \mu(u)$ so a fortiori $\mu(v) \vartriangleleft_{p} \mu(u)$.
\end{itemize}

\end{proof}

We can now prove \Cref{2-thm:progress_measure}.

\begin{proof}
Assume that Eve wins from $v$ and let $\sigma$ be a positional strategy.
The parity graph $\Game[\sigma]$ satisfies parity from $v$, so thanks to \Cref{2-lem:progress_measure}
there exists a tree $t$ and a function $\mu : V \to Y_t$ such that $\mu(v) \neq \top$
and for all edges $u \xrightarrow{p} v \in E$ we have $\mu(v) \vartriangleleft_{p} \mu(u)$.
We remark that $\mu : V \to Y_t$ is actually a progress measure: the condition for $u \in \VE$ is ensured by the edge $\sigma(u)$,
and the condition for $v \in \VA$ by assumption on $\mu$.

\vskip1em
Conversely, assume that there exists a tree $t$ and a progress measure $\mu : V \to Y_t$.
It induces a positional strategy defined by $\sigma(u) = u \xrightarrow{p} v$ such that $\mu(v) \vartriangleleft_{p} \mu(u)$.
We argue that $\sigma$ is a winning strategy from any vertex $v$ such that $\mu(v) \neq \top$.
This is a consequence of \Cref{2-lem:progress_measure} for the parity graph $\Game[\sigma]$.
\end{proof}

\Cref{2-thm:progress_measure} is very close to the characterisation principle we are after,
the only difference being that the lattice $(Y_t,\le)$ depends on an existentially quantified tree $t$.
This is where we use universal trees:

\begin{corollary}[Fundamental corollary for progress measures]
\label{2-cor:progress_measure}
Let $\Game$ be a parity game with $n$ vertices and priorities in $[1,d]$, and $v$ a vertex.
Let $T$ be a $(n,d/2)$-universal tree.
Then Eve wins from $v$ if and only if there exists a progress measure $\mu : V \to Y_T$ such that $\mu(v) \neq \top$.
\end{corollary}

\begin{proof}
Assume that Eve wins from $v$, thanks to \Cref{2-thm:progress_measure} there exists a tree $t$ and a progress measure $\mu : V \to Y_t$ 
such that $\mu(v) \neq \top$.
Since $T$ is $(n,d/2)$-universal and~$t$ has at most $n$ leaves, $t$ embeds into $T$,
which thanks to \Cref{2-fact:embedding} implies that there exists $\mu' : Y_t \to Y_T$ respecting the relations $\vartriangleleft$.
We extend it to $\mu' : Y_t \to Y_T$ by $\mu'(\bot) = \bot$.
Then the composition $\mu' \circ \mu : V \to Y_T$ is a progress measure such that $(\mu' \circ \mu)(v) \neq \bot$. 

The converse implication is a direct consequence of \Cref{2-thm:progress_measure}.
\end{proof}

We have thus proved that the characterisation principle holds for any $(n,d/2)$-universal tree.

\subsection*{The algorithm}
Let us fix $T$ an $(n,d/2)$-universal tree.
It induces both a lattice $(Y_T,\le)$ and a monotonic function $\delta_T : Y_T \times [1,d] \to Y_T$,
which in turn induces a monotonic operator $\Op_T : F_V \to F_V$.
Since $T$ is fixed we do not specify the subscript $T$ for all these objects.


The last step is to construct an algorithm returning the minimal progress measure relying on Kleene's fixed-point theorem (stated as \Cref{1-thm:kleene}).
The generic algorithm is explained in \Cref{1-sec:value_iteration}, let us instantiate it here.


We say that an edge $u \xrightarrow{p} v$ is \textit{incorrect} if $\neg (\mu(v) \vartriangleleft_{p} \mu(u))$,
and a vertex $u$ is \textit{incorrect} if either $u \in \VE$ and all outgoing edges are incorrect or $u \in \VA$ and there exists an incorrect outgoing edge.

\begin{figure}[!ht]
\centering
\begin{tikzpicture}
  \node[s-adam] (u) at (0,-2) {$u$};
  \node[s-eve] (v) at (2,-1) {$v$};
  \node[s-eve] (v') at (2,-3) {$v'$};    

    \path[arrow]
      (u) edge node[above] {$1$} (v)
      (u) edge node[above] {$2$} (v');

  \begin{scope}[xscale=1.4,yscale=1.6]
  \path (5,0) node[coordinate] (root) {};
  \foreach \x in {-2,...,2}
    {\draw (root) -- (5+\x,-1) node[coordinate] (n\x) {};}
  \foreach \s/\x/\n in {
  	-2/-2/,
    -1/-1.4/,
    -1/-.6/,
    0/-.4/u,
    0/-.2/,
    0/0/v,
    0/.2/,
    0/.4/,
    1/.6/,
    1/1.4/v',
    2/2/
    }
    {\draw (n\s) -- (5+\x,-2) node[below] {$\n$};}

  \draw (n1) -- (5+.6,-2) node[red, below] {$u$};

  \node (old) at (5-.4,-2.25) {};    
  \node (new) at (5+.6,-2.25) {};    
  \path[arrow]
    (old) edge[bend right, red, thick] (new);

  \foreach \y/\l in {0.2/4,.7/3,1.2/2,1.7/1}
	{
	\path (2.5,-\y+0.2) node[coordinate] (l\l) {};
    \draw (7.5,-\y) node (r\l) {\begin{Huge}\l\end{Huge}};
	\draw[dash dot] (l\l) -- (r\l);
    }
  \end{scope}
\end{tikzpicture}
\caption{The operator $\Op$ in action: the updated value for $u$ is the minimal leaf $\ell$ (meaning the leftmost leaf) 
which satisfies $\mu(v) \vartriangleleft_1 \ell$ and $\mu(v') \vartriangleleft_2 \ell$.}
\label{2-fig:lifting}
\commentAlt{Figure~\ref{2-fig:lifting}: A diagram showing a simple graph and a hierarchical tree structure, with an arrow indicating a relationship. See long description.}
\commentLongAlt{Figure~\ref{2-fig:lifting}: The image consists of two parts. On the left, a square node labeled 'u' has two directed arrows originating from it. One arrow points to a circular node labeled 'v' with '1' above the arrow. The other arrow points to a circular node labeled 'v'' with '2' below the arrow. On the right, a large tree structure is depicted. It has a single apex at the top from which multiple lines branch downwards. Four horizontal dashed lines, labeled '4', '3', '2', and '1' from top to bottom, intersect the tree branches. At the bottom level, four labels are visible: 'u', 'v', 'u'', and 'v''. A curved arrow originates from the 'u'' label and points to the 'u' label, positioned just above the 'v' label.}
\end{figure}

The pseudocode for the algorithm is given in \Cref{2-algo:value_iteration}, 
where we let $\ell_{\min}$ denote the minimal leaf in $T$.

\begin{algorithm}[ht]
 \KwData{A parity game with $n$ vertices priorities in $[1,d]$ and a $(n,d/2)$-universal tree $T$.}
 \DontPrintSemicolon

\For{$v \in V$}{
$\mu(v) \leftarrow \ell_{\min}$
}
     
\Repeat{$\mu = \Op(\mu)$}{
$\mu \leftarrow \Op(\mu)$
}

\Return{$\mu$}
\caption{The value iteration algorithm.}
\label{2-algo:value_iteration}
\end{algorithm}

\begin{theorem}[Generic value iteration algorithm]
\label{2-thm:generic_value_iteration}
For all $(n, d/2)$-universal tree $T$, for all parity games $\game$ with $n$ vertices and priorities in $[1,d]$,
the value iteration algorithm over the tree $T$ returns the minimal progress measure $\mu$ for $\game$ over $T$.
\end{theorem}

Thanks to \Cref{2-cor:progress_measure}, the minimal progress measure yields a solution for parity games:
Eve wins from $v$ if and only if $\mu(v) \neq \top$.

\subsection*{Complexity analysis}
The number of times the operator $\Op$ is used is bounded by the number of leaves of $T$,
which we write $|T|$, implying that the total number of iterations is bounded by~$n \cdot |T|$.
To determine the overall complexity we need to discuss two aspects of the algorithm:
\begin{itemize}
	\item the data structure and in particular the choice of the vertex $u$ in the loop;
	\item the computation of $\Op$ and in particular the encoding of leaves of $T$.
\end{itemize}

Recall that a vertex $u \in \VE$ is incorrect if and only if all its outgoing edges are incorrect,
and a vertex $u \in \VA$ is incorrect if and only if it has a incorrect outgoing edge.
Hence checking whether a vertex $u$ is incorrect requires considering all of its outgoing edges $u \xrightarrow{p} v$
and checking whether $\mu(v) \vartriangleleft_{p} \mu(u)$.
Let us write $\Delta$ for the complexity of checking whether $\mu(v) \vartriangleleft_{p} \mu(u)$.
Hence checking whether $u$ is incorrect costs 
$O(|\Ing^{-1}(u)| \cdot \Delta)$, where $|\Ing^{-1}(u)|$ is the number of outgoing edges of $u$.
A naive implementation of \Cref{2-algo:value_iteration} would in each repeat loop go through every vertex $u$ 
to check whether it is incorrect.
This would incur the following cost for a single iteration 
\[
\sum_{u \in V} O(|\Ing^{-1}(u)| \cdot \Delta) = O(m \cdot \Delta).
\]
Thus the overall running time of the algorithm would be
\[
O((m \cdot \Delta) \cdot (n \cdot |T|)) = O(n \cdot m \cdot \Delta \cdot |T|).
\]
Typically $\Delta$ is small (we will see that for a well chosen universal tree $T$ it is polylogarithmic in $n$ and $d$),
and $T$ is the dominating factor (quasi-polynomial in $n$ and $d$ thanks to \Cref{2-thm:universal_tree}).

A less naive implementation maintains the list of incorrect vertices using a better data structure, saving a linear factor in the complexity.
We first explain this, and then discuss the cost $\Delta$ by choosing an appropriate encoding of the quasi-polynomial universal tree constructed in~\Cref{2-thm:universal_tree}.

\subsection*{Data structure}
We use a data structure similar to the attractor computation presented in \Cref{1-sec:attractors}, also presented in general terms in \Cref{1-sec:value_iteration}.
The pseudocode is given in \Cref{2-algo:value_iteration_data_structure}.

The data structure consists of the following objects:
\begin{itemize}
	\item a leaf of $T$ for each vertex, representing the current function $\mu : V \to Y$;
	\item a set $\Incorrect$ of vertices (the order in which vertices are stored and retrieved from the set does not matter);
	\item a table $\Count$ storing for each vertex of Eve a number of edges.
\end{itemize}
For our complexity analysis we use the unit cost RAM model, see \Cref{1-sec:computation} for details.
In the case at hand let us choose for the machine word size $w = \log_2(m) + \log_2(d)$, 
so that an edge together with its priority can be stored in one machine word.
The space complexity of this data structure depends on the encoding of $T$, which we will discuss later.

The invariant of the algorithm satisfied before each iteration of the repeat loop is the following:
\begin{itemize}
	\item for $u \in \VE$, the value of $\Count(u)$ is the number of incorrect edges of $u$;
	\item $\Incorrect$ is the set of incorrect vertices.
\end{itemize}
The invariant is satisfied initially thanks to the function $\texttt{Init}$.
Let us assume that we choose and remove $u$ from $\Incorrect$.
Since we modify only $\mu(u)$ the only potentially incorrect vertices are in $\Incorrect$ (minus $u$) and the incoming edges of $u$;
for the latter each of them is checked and added to $\Incorrect'$ when required.
By monotonicity, incorrect vertices remain incorrect so all vertices in $\Incorrect$ (minus $u$) are still incorrect.
Hence the invariant is satisfied.

The invariant implies that the algorithm indeed implements~\Cref{2-algo:value_iteration} hence returns the minimal progress measure, 
but it also has implications on the complexity.
Indeed one iteration of the repeat loop over some vertex $u$ involves 
\[
O\left( (|\Ing^{-1}(u)| + |\Out^{-1}(u)|) \cdot \Delta \right)
\]
operations,
the first term corresponds to updating $\mu(u)$ and $\Incorrect$,
which requires for each outgoing edge of $u$ to compute $\delta$,
and the second term corresponds to considering all incoming edges of $u$.
Since each vertex is updated at most $|T|$ times, the overall running time of the algorithm is bounded by
\[
\left( \sum_{u \in V} O(|\Ing^{-1}(u)| \cdot \Delta) \right) \cdot |T| = O(m \cdot \Delta \cdot |T|).
\]
Typically $\Delta$ is small (we will see that for a well chosen universal tree $T$ it is polylogarithmic in $n$ and $d$),
and $T$ is the dominating factor (quasi-polynomial in $n$ and $d$ thanks to \Cref{2-thm:universal_tree}).

\begin{algorithm}
 \KwData{A parity game with $n$ vertices priorities in $[1,d]$ and a $(n,d/2)$-universal tree $T$.}
 \SetKwFunction{FInit}{Init}
 \SetKwFunction{FTreat}{Treat}
 \SetKwFunction{FUpdate}{Update}
 \SetKwFunction{FMain}{Main}
 \SetKwProg{Fn}{Function}{:}{}
 \DontPrintSemicolon

\Fn{\FInit{}}{

	\For{$u \in V$}{
		$\mu(u) \leftarrow \ell_{\min}$
	}

	\For{$u \in \VE$}{
        \For{$u \xrightarrow{p} v \in E$}{
        	\If{incorrect: $\neg (\mu(v) \vartriangleleft_{p} \mu(u))$}{
		        $\Count(u) \leftarrow \Count(u) + 1$
        	}
        }
        
        \If{$\Count(u) = \Degree(u)$}{
        	Add $u$ to $\Incorrect$
        }
	}    
	
	\For{$u \in \VA$}{
        \For{$u \xrightarrow{p} v \in E$}{
        	\If{incorrect: $\neg (\mu(v) \vartriangleleft_{p} \mu(u))$}{
        		Add $v$ to $\Incorrect$
        	}
        }
    }
}

\vskip1em
\Fn{\FTreat{$u$}}{
	\If{$u \in \VE$}{
		$\mu(u) \leftarrow \min \set{\delta( \mu(v), p) : u \xrightarrow{p} v \in E}$
	}

	\If{$u \in \VA$}{
		$\mu(u) \leftarrow \max \set{\delta( \mu(v), p) : u \xrightarrow{p} v \in E}$
	}

}

\vskip1em
\Fn{\FUpdate{$u$}}{	
	\If{$u \in \VE$}{
        $\Count(u) \leftarrow 0$
	}
	\For{$v \xrightarrow{p} u \in E$}{
		\If{$v \xrightarrow{p} u$ is incorrect}{
			\If{$v \in \VE$}{
		        $\Count(v) \leftarrow \Count(v) + 1$
        
	    	    \If{$\Count(v) = \Degree(v)$}{
    	    		Add $v$ to $\Incorrect'$
		        }	
			}

			\If{$v \in \VA$}{
				Add $v$ to $\Incorrect'$	
			}
		}

	}
}

\vskip1em
\Fn{\FMain{}}{
	\FInit()    

	\For{$i = 0,1,2,\dots$}{
		$\Incorrect' \leftarrow \emptyset$

		\For{$u \in \Incorrect$}{

			\FTreat($u$)    

			\FUpdate($u$)    
		}
		\If{$\Incorrect' = \emptyset$}{

			\Return{$\mu$}
		}
		\Else{

			$\Incorrect \leftarrow \Incorrect'$
		}
	}
}
\caption{The value iteration algorithm with explicit data structure.}
\label{2-algo:value_iteration_data_structure}
\end{algorithm}

\subsection*{Encoding leaves}
Let us fix $T$ to be the quasi-polynomial universal tree constructed in \Cref{2-thm:universal_tree}.

In our definition of trees we say that a tree is an ordered list of subtrees $[t_1,\dots,t_k]$,
so we use $[1,k]$ with the natural order for ordering the subtrees.
Any other total order can be used to that effect, and a more appropriate order may lead to smaller encoding.
Indeed, using $[1,k]$ for ordering subtrees, if a tree has height $h$ and $n$ leaves then a leaf is a sequence of $h$ numbers in $[1,n]$,
so it uses $O(h \log_2(n))$ bits.

Let us consider an order well suited for encoding $T$.
We use $\set{0,1}^*$ the set of binary words and order them using the following three rules that apply for any $u,v \in \set{0,1}^*$:
\[
0u < \varepsilon < 1u \quad ; \quad (0u < 0v \Longleftrightarrow u < v) \quad ; \quad (1u < 1v \Longleftrightarrow u < v).
\]
For words of length at most $2$ the order is $00 < 0 < 01 < \varepsilon < 10 < 1 < 11$.

\begin{figure}[!ht]
\centering
\begin{tikzpicture}
  \begin{scope}[xscale=1.4,yscale=1.6]
  \path (0,0) node[coordinate] (root) {};
  \foreach \x/\n in {-2/00,
  				-1/0,
  				0/\varepsilon}
    {\draw (root) -- node[below] {$\n$} (\x,-1) node[coordinate] (n\x) {};}
  \foreach \x/\n in {1/10,
  				2/1}
    {\draw (root) -- node[below left] {$\n$} (\x,-1) node[coordinate] (n\x) {};}

  \foreach \s/\x/\n in {-2/-2/\varepsilon,
					-1/-1.4/0,
					-1/-.6/\varepsilon,
    				1/.6/0,
    				1/1.4/\varepsilon,
    				2/2/\varepsilon}
    {\draw (n\s) -- node[right] {$\n$} (\x,-2);}
  \foreach \s/\x/\n in {0/-.4/00,
				    0/-.2/0,
				    0/0/\varepsilon,
				    0/.2/10,
				    0/.4/1}
    {\draw (n\s) -- (\x,-2) node[below] {\begin{small}$\n$\end{small}};}
  \end{scope}
\end{tikzpicture}
\caption{The succinct encoding on the $(5,2)$-universal tree.}
\label{2-fig:tree_encoded}
\commentAlt{Figure~\ref{2-fig:tree_encoded}: A tree-like diagram with labeled branches, showing a hierarchical structure of values. See long description.}
\commentLongAlt{Figure~\ref{2-fig:tree_encoded}: The image displays a tree structure with a single root at the top. From this root, five main branches extend downwards, labeled from left to right: '00', '0', 'epsilon', '10', and '1'. Some of these branches further divide into sub-branches, each with its own label. The '00' branch leads to a 'epsilon' leaf. The '0' branch splits into two sub-branches labeled '0' and 'epsilon'. The 'epsilon' branch from the root leads to a cluster of five very close sub-branches, labeled from left to right: '000', 'epsilon', '101', 'epsilon'. The '10' branch splits into two sub-branches labeled '0' and 'epsilon'. The '1' branch leads to an 'epsilon' leaf. The overall structure forms a fan-like shape, with all branches eventually leading to leaves.}
\end{figure}

We can now revisit the construction of the universal tree by defining directly the set of leaves.
Recall that $T$ is obtained from $T_\text{left},T_\text{middle}$, and $T_\text{right}$. 
By induction hypothesis leaves in $T_\text{left}$ and $T_\text{right}$ are tuples of length $h-1$
and leaves in $T_\text{middle}$ tuples of length $h$.
The leaves of $T$ are:
\begin{itemize}
	\item leaves of $T_\text{left}$ where the first component is prefixed with a $0$;
	\item leaves of $T_\text{middle}$ augmented with a new component $\varepsilon$;
	\item leaves of $T_\text{right}$ where the first component is prefixed with a $1$.
\end{itemize}
We call this encoding the succinct encoding, it is illustrated in \Cref{2-fig:tree_encoded} for the $(5,2)$-universal tree.
The leftmost leaf is $(00,\varepsilon)$, and the middle leaf $(\varepsilon,\varepsilon)$.
In general, the inductive construction implies that every leaf is a tuple $(D_{d-1},\dots,D_1)$ 
such that the sum of the lengths of the directions $D_i$ is at most $\log_2(n)$.
Thus a leaf is encoded using $O(\log_2(h) \log_2(n))$ bits: for each of the $\log_2(n)$ bits we need $\log_2(h)$ bits to specify its component.

In terms of machine words of size $w = \log_2(n) + \log_2(d)$, this means that a leaf can be stored using $\log_2(d)$ machine words.
Hence the data structure uses $O(n \log_2(d))$ machine words, with together with the input size $O(m)$
means that the space complexity of the algorithm is $O(m + n \log_2(d))$.

\vskip1em
Using the succinct encoding and a tedious but simple case analysis we can compute $\delta(\ell,p)$ in time $O(\log_2(n) \log_2(d))$.
Putting everything together we obtain the overall complexity stated in \Cref{2-thm:value_iteration_quasipoly}.


\section{Comparing the three families of algorithms}
\label{2-sec:relationships}
	At the beginning of the chapter we described three families of algorithms: 
strategy improvement, attractor decomposition, and value iterations.

\vskip1em
Let us first clarify the relationship between the separation framework discussed in \Cref{2-sec:separation}
and the value iteration paradigm presented in \Cref{2-sec:value_iteration}.
Both are families of algorithms: 
\begin{itemize}
	\item An $(n,d)$-separating automaton $\Automaton$ induces an algorithm for solving parity games in time 
$O(m \cdot |\Automaton|)$ where $|\Automaton|$ is the size of $\Automaton$, meaning the number of states.
	\item An $(n,d/2)$-universal tree $T$ induces a value iteration algorithm for solving parity games in time 
proportional to $|T|$ where $|T|$ is the size of $T$, meaning the number of leaves (the exact complexity depends on the cost of computing $\delta$ in $T$, which is typically small).
\end{itemize}
These two families are in a strong sense equivalent:

\begin{theorem}[Equivalence between separating automata and universal trees]
\label{2-thm:equivalence_separating_automata_universal_trees}
\hfill
\begin{itemize}
	\item An $(n,d)$-separating automaton induces an $(n,d/2)$-universal tree of the same size;
	\item An $(n,d/2)$-universal tree induces an $(n,d)$-separating automaton of the same size.
\end{itemize}
\end{theorem}
We do not prove this theorem here but note that it can be stated more generally for any positionally determined objective,
replacing universal trees by the notion of universal graphs.
The main advantage of the value iteration presentation is the space complexity, which for a good choice of the universal tree
can be made very small (quasi-linear).

\vskip1em
In terms of complexity, the strategy improvement has exponential complexity, while both attractor decompositions and value iterations algorithms have quasi-polynomial complexity.
Let us make a finer comparison: the complexity of the attractor decomposition algorithm is a polynomial multiplied by the (non polynomial) term
\[
\binom{d - 1 + \lfloor \log(n) \rfloor}{\lfloor \log(n) \rfloor},
\]
while for the value iteration algorithm the complexity is a polynomial multiplied by the (also non polynomial) term
\[
\binom{d/2 - 1 + \lfloor \log(n) \rfloor}{\lfloor \log(n) \rfloor}.
\]
The key difference is that the former performs an induction using all priorities, while the latter considers only odd priorities hence the dependence in $d/2$.
Although our presentation of the attractor decomposition algorithm does not make it explicit, 
this class of algorithms is also related to the notion of universal trees; 
however an algorithm is induced not by one $(n,d/2)$-universal tree, but by two: one for each player,
which are then interleaved to organise the recursive calls of the algorithm.

\vskip1em
Since both value iteration and attractor decomposition algorithms are connected to the combinatorial notion of universal trees,
the next question is whether the construction given in \Cref{2-sec:value_iteration} is optimal.
The answer is unfortunately yes, there exists a lower bound on the size of universal trees which matches this construction up to polynomial factors.

\vskip1em
The last question we discuss here is whether there exists a quasi-polynomial strategy improvement algorithm.
In particular a natural attempt would be to use universal trees for this endeavour.
Unfortunately, the natural approach fails, as we explain now.
As we have seen, universal trees can be used to represent progress measures for Eve: given a graph, a vertex is assigned $\top$ if all paths satisfy the complement of parity, and some left of the tree otherwise. Hence they can be used for a strategy improvement algorithm iterating over strategies of Adam. Let us consider the exponential universal tree, we present in~\Cref{2-fig:counter_example_strategy_improvement} a game where the strategy improvement algorithm alternates between two strategies, hence does not terminate.
The minimal progress measures are illustrated on the right hand side, for both strategies. One can verify that in both cases, the missing edge is an improving edge.

\begin{figure}
\centering
  \begin{tikzpicture}[scale=.9]
  \begin{scope}[xscale=1.4,yscale=1.6]
    \node[s-adam] (v0) at (-2,-1) {$v_0$};
    \node[s-adam] (v1) at (0,-1) {$v_1$};
    \node[s-adam] (v2) at (2,-1) {$v_2$};
    \path[arrow]
      (v0) edge[bend right] node[below] {$2$} (v1)
      (v1) edge[bend right] node[above] {$1$} (v0)
      (v2) edge[bend left] node[below] {$4$} (v1)
      (v1) edge[bend left] node[above] {$3$} (v2);
  \end{scope}
  
  \begin{scope}[xscale=1.4,yscale=1.6]
  \path (5,0) node[coordinate] (root) {};
  \draw (root) -- (4,-1) node[coordinate] (n-1) {};
  \draw (root) -- (6,-1) node[coordinate] (n1) {};
  \foreach \s/\x/\n in {
    -1/-1.4/v_2,
    -1/-.6/,
    1/.6/v_0,
    1/1.4/v_1
    }
    {\draw (n\s) -- (5+\x,-2) node[below] {$\n$};}

  \foreach \y/\l in {0.2/4,.7/3,1.2/2,1.7/1}
	{
	\path (3.5,-\y+0.2) node[coordinate] (l\l) {};
    \draw (7,-\y) node (r\l) {\begin{Huge}\l\end{Huge}};
	\draw[dash dot] (l\l) -- (r\l);
    }
  \end{scope}
  \end{tikzpicture}
\caption{A game where the strategy improvement over the exponential universal tree does not terminate.}
\label{2-fig:counter_example_strategy_improvement}
\commentAlt{Figure~\ref{2-fig:counter_example_strategy_improvement}: A diagram showing a linear graph with bidirectional edges and a binary tree with labeled levels. See long description.}
\commentLongAlt{Figure~\ref{2-fig:counter_example_strategy_improvement}: The image consists of two separate diagrams. On the left, there is a linear sequence of three square nodes: v0, v1, and v2. There are bidirectional arrows between v0 and v1, with the arrow from v0 to v1 labeled '1' and the arrow from v1 to v0 labeled '2'. Similarly, there are bidirectional arrows between v1 and v2, with the arrow from v1 to v2 labeled '3' and the arrow from v2 to v1 labeled '4'. On the right, there is a binary tree structure. It has a single root at the top, branching down into two main branches, which then branch again, creating four terminal nodes at the bottom. Four horizontal dashed lines, labeled '4', '3', '2', and '1' from top to bottom, indicate different levels of the tree. The leftmost terminal node is labeled 'v2', and the second from the right is labeled 'v0', while the rightmost terminal node is labeled 'v1'.}
\end{figure}

\begin{figure}
\centering
  \begin{tikzpicture}[scale=.9]
  \begin{scope}[xscale=1.4,yscale=1.6]
    \node[s-adam] (v0) at (-2,-1) {$v_0$};
    \node[s-adam] (v1) at (0,-1) {$v_1$};
    \node[s-adam] (v2) at (2,-1) {$v_2$};
    \path[arrow]
      (v0) edge[bend right] node[below] {$2$} (v1)
      (v1) edge[bend right, dash dot] node[above] {$1$} (v0)
      (v2) edge[bend left] node[below] {$4$} (v1)
      (v1) edge[bend left] node[above] {$3$} (v2);
  \end{scope}
  
  \begin{scope}[xscale=1.4,yscale=1.6]
  \path (5,0) node[coordinate] (root) {};
  \draw (root) -- (4,-1) node[coordinate] (n-1) {};
  \draw (root) -- (6,-1) node[coordinate] (n1) {};
  \foreach \s/\x/\n in {
    -1/-1.4/v_2,
    -1/-.6/,
    1/.6/v_0;v_1,
    1/1.4/
    }
    {\draw (n\s) -- (5+\x,-2) node[below] {$\n$};}

  \foreach \y/\l in {0.2/4,.7/3,1.2/2,1.7/1}
	{
	\path (3.5,-\y+0.2) node[coordinate] (l\l) {};
    \draw (7,-\y) node (r\l) {\begin{Huge}\l\end{Huge}};
	\draw[dash dot] (l\l) -- (r\l);
    }
  \end{scope}
  \end{tikzpicture}
  \begin{tikzpicture}[scale=.9]
  \begin{scope}[xscale=1.4,yscale=1.6]
    \node[s-adam] (v0) at (-2,-1) {$v_0$};
    \node[s-adam] (v1) at (0,-1) {$v_1$};
    \node[s-adam] (v2) at (2,-1) {$v_2$};
    \path[arrow]
      (v0) edge[bend right] node[below] {$2$} (v1)
      (v1) edge[bend right] node[above] {$1$} (v0)
      (v2) edge[bend left] node[below] {$4$} (v1)
      (v1) edge[bend left, dash dot] node[above] {$3$} (v2);
  \end{scope}
  
  \begin{scope}[xscale=1.4,yscale=1.6]
  \path (5,0) node[coordinate] (root) {};
  \draw (root) -- (4,-1) node[coordinate] (n-1) {};
  \draw (root) -- (6,-1) node[coordinate] (n1) {};
  \foreach \s/\x/\n in {
    -1/-1.4/v_0;v_2,
    -1/-.6/v_1,
    1/.6/,
    1/1.4/
    }
    {\draw (n\s) -- (5+\x,-2) node[below] {$\n$};}

  \foreach \y/\l in {0.2/4,.7/3,1.2/2,1.7/1}
	{
	\path (3.5,-\y+0.2) node[coordinate] (l\l) {};
    \draw (7,-\y) node (r\l) {\begin{Huge}\l\end{Huge}};
	\draw[dash dot] (l\l) -- (r\l);
    }
  \end{scope}
  \end{tikzpicture}
\caption{The two strategies have opposing improving edges (dashed).}
\label{2-fig:counter_example_strategy_improvement_explanations}
\commentAlt{Figure~\ref{2-fig:counter_example_strategy_improvement_explanations}: Two pairs of diagrams, each showing a linear graph with labeled nodes and edges, and a corresponding binary tree with labeled levels. See long description.}
\commentLongAlt{Figure~\ref{2-fig:counter_example_strategy_improvement_explanations}: The image contains two identical sets of diagrams, stacked vertically. Each set consists of two parts. The left part is a linear graph with three square nodes: v0, v1, and v2. There are two arrows between v0 and v1: a solid arrow from v1 to v0 labeled '2', and a dashed arrow from v0 to v1 labeled '1'. There are also two arrows between v1 and v2: a solid arrow from v2 to v1 labeled '4', and a dashed arrow from v1 to v2 labeled '3'. The right part is a binary tree structure with a root at the top and four terminal leaves at the bottom. Four horizontal dashed lines, labeled '1' to '4' from bottom to top, indicate the levels. In the top set, the leftmost leaf of the tree is labeled 'v2', and the two rightmost leaves are collectively labeled 'v0, v1'. In the bottom set, the leftmost leaves are collectively labeled 'v0, v2', and the rightmost leaf is labeled 'v1'.}
\end{figure}


\section*{Bibliographic references}
\label{2-sec:references}
We refer to~\Cref{3-sec:references} for the role of parity objectives and how they emerged in automata theory as a subclass of Muller objectives.
Another related motivation comes from the works of Emerson, Jutla, and Sistla~\cite{Emerson.Jutla.ea:1993},
who showed that solving parity games is linear-time equivalent to the model-checking problem for modal $\mu$-calculus.
This logical formalism is an established tool in program verification, and a common denominator to a wide range of modal, temporal and fixpoint logics used in various fields.

\vskip1em
Let us discuss the progress obtained over the years for each of the three families of algorithms.

\vskip1em
\textit{Value iteration algorithms and separating automata}.
The heart of value iteration algorithms is the value function, which in the context of parity games and related developments for automata
have been studied under the name progress measures or signatures.
They appear naturally in the context of fixed-point computations so it is hard to determine who first introduced them.
Streett and Emerson~\cite{Streett.Emerson:1984} defined signatures for the study of the modal $\mu$-calculus,
and Stirling and Walker~\cite{Stirling.Walker:1989} later developed the notion.
Both the proofs of Emerson and Jutla~\cite{Emerson.Jutla:1991} and of Walukiewicz~\cite{Walukiewicz:1996} use signatures to show the positionality of parity games over infinite games.

Jurdzi{\'n}ski~\cite{Jurdzinski:2000} used this notion to give the first value iteration algorithm for parity games, 
with running time $O(m n^{d/2})$.
The algorithm is called `small progress measure lifting' and is an instance of the class of value iteration algorithms we construct 
in~\Cref{2-sec:value_iteration} by considering the universal tree of size $n^h$.
Bernet, Janin, and Walukiewicz~\cite{Bernet.Janin.ea:2002} investigated the notion of permissive strategies and obtained reductions from parity games to safety games. In essence, they constructed the separating automaton corresponding to the universal tree of size $n^h$, although the general framework of separating automata came later, introduced by Boja{\'n}czyk and Czerwi{\'n}ski~\cite{Bojanczyk.Czerwinski:2018}.

The new era for parity games started in 2017 when Calude, Jain, Khoussainov, Li, and Stephan~\cite{Calude.Jain.ea:2017} constructed a quasi-polynomial-time algorithm. 
Our presentation follows the technical developments of the subsequent paper by Fearnley, Jain, Schewe, Stephan, and Wojtczak~\cite{Fearnley.Jain.ea:2017} which recasts the algorithm as a value iteration algorithm.
Boja{\'n}czyk and Czerwi{\'n}ski~\cite{Bojanczyk.Czerwinski:2018} introduce the separation framework to better understand the original algorithm.
Soon after two other quasi-polynomial-time algorithms emerged.
Jurdzi{\'n}ski and Lazi{\'c}~\cite{Jurdzinski.Lazic:2017} showed that the small progress measure algorithm can be adapted to a `succinct progress measure' algorithm, matching (and slightly improving) the quasi-polynomial time complexity.
The presentation using universal tree that we follow in~\Cref{2-sec:value_iteration} and an almost matching lower bound on their sizes is due to Fijalkow~\cite{Fijalkow:2018}.
The connection between separating automata and universal trees was shown by Czerwi{\'n}ski, Daviaud, Fijalkow, Jurdzi{\'n}ski, Lazi{\'c}, and Parys~\cite{Czerwinski.Daviaud.ea:2019}. 

The third quasi-polynomial-time algorithm is due to Lehtinen~\cite{Lehtinen:2018}.
The original algorithm has a slightly worse complexity ($n^{O(\log(n))}$ instead of $n^{O(\log(d))}$),
but it was later improved to (essentially) match the complexity of the previous two algorithms~\cite{Parys:2020}.
Although not explicitly, the algorithm constructs an automaton with similar properties as a separating automaton,
but the automaton is non-deterministic.
Colcombet and Fijalkow~\cite{Colcombet.Fijalkow:2019} revisited the link between separating automata and universal trees
and proposed the notion of good for small games automata, capturing the automaton defined by Lehtinen's algorithm.
The equivalence result between separating automata, good for small games automata, and universal graphs, holds for any positionally determined objective, giving a strong theoretical foundation for the family of value iteration algorithms.

\vskip1em
\textit{Attractor decomposition algorithms}.
One of the many contributions of Zielonka's landmark paper~\cite{Zielonka:1998} was to follow McNaughton's approach for 
constructing a recursive algorithm for solving parity games, and show that it implies their bi-positionality.
We follow in~\Cref{2-sec:zielonka} Zielonka's presentation of the algorithm, which is sometimes called Zielonka algorithm but more accurately 
McNaughton Zielonka algorithm.

The McNaughton Zielonka's algorithm has complexity $O(m n^d)$.
Parys~\cite{Parys:2019} constructed the fourth quasi-polynomial-time algorithm as an improved take over McNaughton Zielonka's algorithm.
As for Lehtinen's algorithm, the original algorithm has a slightly worse complexity ($n^{O(\log(n))}$ instead of $n^{O(\log(d))}$).
A further improvement of the algorithm~\cite{Lehtinen.Parys.ea:2022} yields a time complexity $n^{O(\log(d))}$.
As discussed in~\Cref{2-sec:relationships} the complexity of this algorithm is quasi-polynomial and of the form $n^{O(\log(d))}$,
but a bit worse than the three previous algorithms since the algorithm is symmetric and has a recursion depth of $d$,
while the value iteration algorithms only consider odd priorities hence replace $d$ by $d/2$.

Jurdzi{\'n}ski, Morvan, and Thejaswini~\cite{Jurdzinski.Morvan.ea:2022} constructed a McNaughton Zielonka's algorithm which is generic in the following sense: it is parameterised by the choice of two universal trees, one for each player.
How the value iteration algorithm can simulate the attractor decomposition algorithm was explained by Ohlmann~\cite{Ohlmann:2021}.

\vskip1em
\textit{Strategy improvement algorithms}.
As we will see in~\Cref{5-chap:payoffs}, parity games can be reduced to mean-payoff games or energy games, as well as discounted-payoff games,
so any algorithm for solving them can be used for solving parity games.
In particular, the existing strategy improvement algorithms for these games can be run on parity games. 
V{\"o}ge and Jurdzi{\'n}ski~\cite{voge.Jurdzinski:2000} introduced the first discrete strategy improvement for parity games,
running in exponential time.
The algorithm we present is due to Bj{\"o}rklund, Sandberg, and Vorobyov~\cite{Bjorklund.Sandberg.ea:2003}, which was shown to run in sub-exponential time using a randomisation procedure.
Our proof of correctness is original.
The complexity was reduced to sub-exponential with deterministic algorithms by Jurdzi{\'n}ski, Paterson, and Zwick~\cite{Jurdzinski.Paterson.ea:2008}.
For some time there was some hope that the strategy improvement algorithm, for some well chosen policy on switching edges,
solves parity games in polynomial time.
Friedmann~\cite{Friedmann:2011} cast some serious doubts by constructing numerous exponential lower bounds applying to different variants of the algorithm.
Fearnley~\cite{Fearnley:2017} investigated efficient implementations of the algorithm, focusing on the cost of computing and updating the value function for a given strategy.

A natural question is whether there exists a quasi-polynomial strategy improvement algorithm.
Koh and Loho~\cite{Koh.Loho:2022} constructed a quasi-polynomial-time algorithm and presented it as a strategy improvement algorithm: the subtlety is that in their algorithm, an iteration considers a strategy $\sigma$ together with its valuation $\val^{\sigma}$ and builds a new strategy $\sigma'$ with the additional constraint that $\val^{\sigma} < \val^{\sigma'}$. In other words, it does not improve only based on the strategy $\sigma$. Hence in some sense it is closer to a value iteration algorithm.
As discussed in~\Cref{2-sec:relationships} the notion of universal trees cannot be used to achieve this: the example is due to Ohlmann~\cite{Ohlmann:2021}. Whether there exists a quasi-polynomial time strategy improvement algorithm for parity games remains to this day open.

\ifpictures
\includepdf{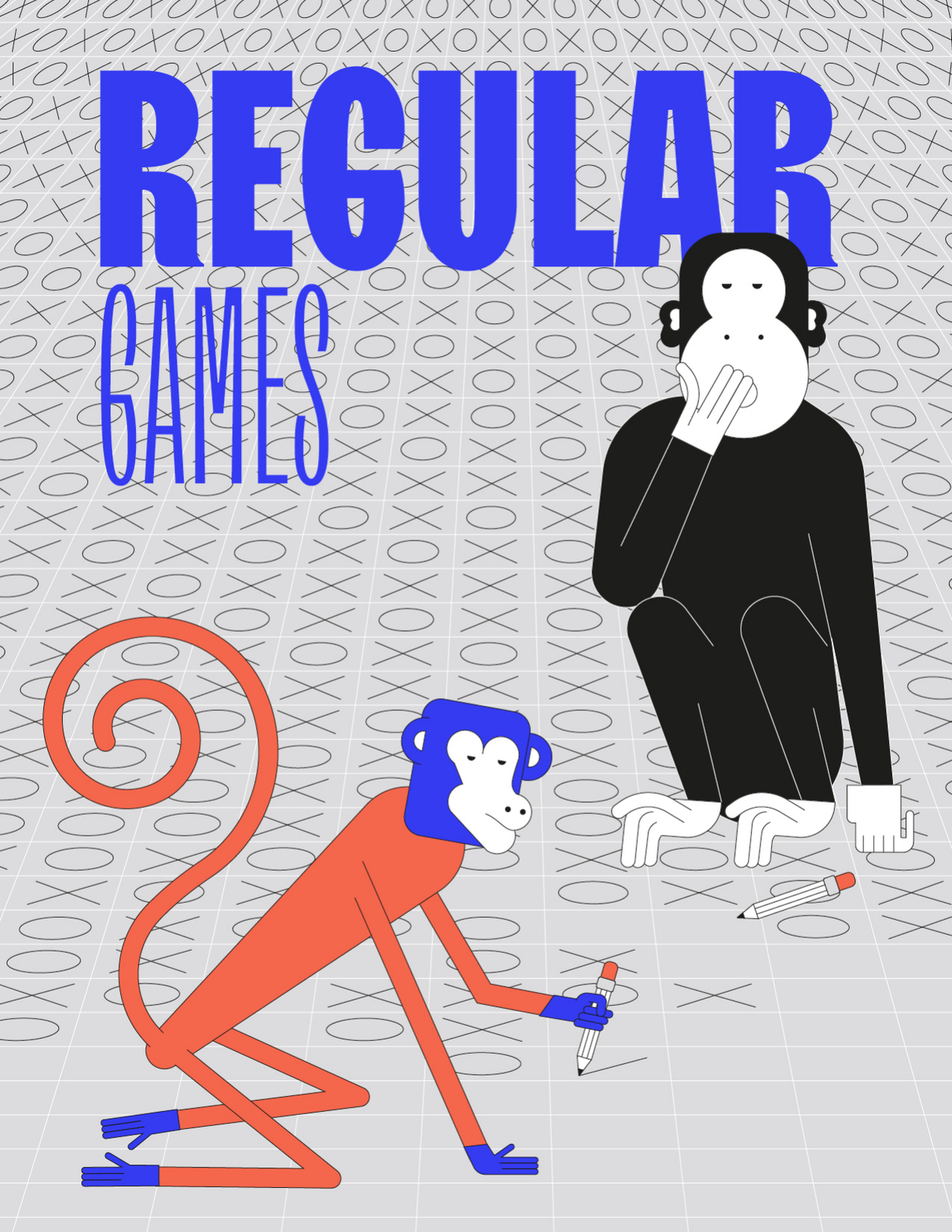}
\fi
\author[Nathana{\"e}l Fijalkow, Florian Horn]{Nathana{\"e}l Fijalkow, Florian Horn}
\copyrightline{Copyright by Nathana{\"e}l Fijalkow and Florian Horn 2025, to be published by Cambridge University Press in the volume \textit{Games on Graphs} edited by Nathana\"el Fijalkow}

\chapter{Regular Games}
\chapterauthor{Nathana{\"e}l Fijalkow, Florian Horn}
\label{3-chap:regular}

\newcommand{\QuantitativeReach}{\mathtt{QuantitativeReach}} 

\providecommand{\F}{\mathcal{F}}

\newcommand{\LAR}{\mathrm{LAR}}
\newcommand{\Zielonka}{\mathrm{Zielonka}}

\providecommand{\Count}{\texttt{Count}}
\renewcommand{\Count}{\texttt{Count}}

\providecommand{\ToUpdate}{\texttt{ToUpdate}}
\renewcommand{\ToUpdate}{\texttt{ToUpdate}}

\providecommand{\Degree}{\texttt{Degree}}
\renewcommand{\Degree}{\texttt{Degree}}

\newcommand{\depth}{\mathrm{depth}}
\newcommand{\support}{\mathrm{supp}}

This chapter considers the so-called regular games beyond parity games: Rabin / Streett games and Muller games.
In \Cref{3-sec:muller} we extend the recursive algorithm of parity games to Muller games, and discuss the computational complexities of solving Rabin, Streett, and Muller games.
Finally, \Cref{3-sec:zielonka} is devoted to the combinatorial notion of the Zielonka tree,
which beautifully explains the memory requirements for Muller games and gives additional insights into the structures of Rabin and parity objectives.

\begin{remark}[Finite versus infinite games]
\label{3-rmk:finite_infinite}
As in the rest of the book unless otherwise specified we consider finite games.
However all positionality and finite-memory determinacy results proved in this chapter hold for infinite games.
In all cases the proofs we give use the finiteness of the games. In many cases, the proofs can be extended to infinite games with a technical overhead involving in particular a transfinite induction. 
We refer to~\Cref{4-chap:memory} for refined results about memory determinacy.
\end{remark}


\section{Rabin, Streett, and Muller games}
\label{3-sec:muller}
The prefix-independent objectives we studied so far are B{\"u}chi, CoB{\"u}chi, and their joint extension the parity objectives.
The definition of the latter may seem a bit arbitrary; the study of Muller objectives will show how parity objectives naturally emerge as a well-behaved class of objectives.

\vskip1em
Let us start with a very general class of infinitary objectives, where infinitary means that the objective only considers the set of colours appearing infinitely many times.
For a sequence $\rho$, we let $\Inf(\rho)$ denote the set of colours appearing infinitely many times in $\rho$.
The Muller objective is over the set of colours $C = [1,d]$ and is parameterised by some $\F \subseteq 2^C$, \textit{i.e.} a family of subsets of $C$.
The objective is defined as follows:
\[
\Muller(\F) = \set{ \rho \in C^\omega : \Inf(\rho) \in \F }.
\]
Muller objectives include any objective specifying the set of colours appearing infinitely often.
There are different possible representations for a Muller objective, for instance using logical formulas or circuits.
We will here consider the most natural one which simply consists in listing the elements of $\F$.
Note that $\F$ can have size up to $2^{2^d}$, and each element of $\F$ (which is a subset of $C$)
requires up to $d$ bits to be identified, so the representation of $\F$ can be very large.

We note that the complement of a Muller objective is another Muller objective: 
$C^\omega \setminus \Muller(\F) = \Muller(2^C \setminus \F)$. 
In particular if Eve has a Muller objective then Adam also has a Muller objective.

\vskip1em
To define subclasses of Muller objectives we make assumptions on $\F \subseteq 2^C$.
We say that $\F$ is closed under union if whenever $X,Y \in \F$ then $X \cup Y \in \F$.
Let us define Streett objectives as the subclass of Muller objectives given by $\F$ closed under union.
The following purely combinatorial lemma gives a nice characterisation of these objectives.

\begin{lemma}[Characterisation of Streett among Muller objectives]
\label{3-lem:characterisation_Streett}
A collection $\F \subseteq 2^C$ is closed under union if and only if there exists a set of pairs $(R_i,G_i)_{i \in [1,d]}$
with $R_i,G_i \subseteq C$ such that $X \in \F$ is equivalent to for all $i \in [1,d]$, 
if $X \cap R_i \neq \emptyset$ then $X \cap G_i \neq \emptyset$.
\end{lemma}
We will see in \Cref{3-sec:zielonka} a natural and optimised way to construct these pairs using the Zielonka tree.
In the meantime let us give a direct proof of this result.

\begin{proof}
Let $\F$ closed under union.
We note that for any $S \notin \F$, either no subsets of $S$ are in $\F$ or there exists a maximal subset $S'$ of $S$ in $\F$:
indeed it is the union of all subsets of $S$ in $\F$.
It directly follows that for a subset $X$ we have the following equivalence:
$X \in \F$ if and only if for any $S \notin \F$, if $X \subseteq S$ then $X \subseteq S'$.
This is rewritten equivalently as: if $X \cap (C \setminus S') \neq \emptyset$ then $X \cap (C \setminus S) \neq \emptyset$.
Hence a suitable set of pairs satisfying the required property is 
$\set{(C \setminus S', C \setminus S) : S \notin \F}$.
\end{proof}
Thanks to this lemma we can give a direct definition of Streett objectives.
The set of colours is $C = [1,d]$, and we consider a family of subsets $G_1,R_1,\dots,G_d,R_d \subseteq C$.
\[
\Streett = \set{ \rho \in C^\omega : \forall i \in [1,d],\ R_i \cap \Inf(\rho) \neq \emptyset \implies G_i \cap \Inf(\rho) \neq \emptyset}.
\]
It is customary to call $R_i$ the $i$\textsuperscript{th} request and $G_i$ the corresponding response;
with this terminology the Streett objective requires that every request made infinitely many times must be responded to infinitely many times.

The Rabin objectives are the complement of the Streett objectives: 
\[
\Rabin = \set{ \rho \in C^\omega : \exists i \in [1,d],\ R_i \cap \Inf(\rho) \neq \emptyset \wedge G_i \cap \Inf(\rho) = \emptyset}.
\]

\subsection*{McNaughton algorithm: an exponential-time algorithm for Muller games}
\begin{theorem}[Finite-memory determinacy and complexity for Muller games]
\label{3-thm:muller}
Muller objectives are determined with finite memory strategies of size $d!$\footnote{See \Cref{3-rmk:finite_infinite} for the case of infinite games.}.
There exists an algorithm for computing the winning regions of Muller games in exponential time,
and more specifically of complexity $O(dm (dn)^{d-1})$, and in polynomial space, and more specifically $O(dm)$.
\end{theorem}
The complexity will be improved later in this chapter. The presentation of the recursive algorithm for computing the winning regions of Muller games follows the exact same lines as for parity games:
indeed, the Muller objective extends the parity objective, and specialising the algorithm for Muller games to parity games
yields the algorithm we presented above.

The following lemma induces the recursive algorithm for computing the winning regions of Muller games.

\begin{lemma}[Fixed-point characterisation of the winning regions for Muller games]
\label{3-lem:Muller_even}
Let $\Game$ be a Muller game such that $C \in \F$.
For each $c \in C$, let $\Game_c = \Game \setminus \AttrE(c)$.
\begin{itemize}
	\item If for all $c \in C$, we have $\WA(\Game_c) = \emptyset$, then $\WE(\Game) = V$.
	\item If there exists $c \in C$ such that $\WA(\Game_c) \neq \emptyset$,
	let $\Game' = \Game \setminus \AttrA( \WA(\Game_c) )$,
	then $\WE(\Game) = \WE(\Game')$.	
\end{itemize}
\end{lemma}

\begin{proof}
We prove the first item.

For each $c \in C$, let $\sigma_c$ be an attractor strategy ensuring to reach $c$ from $\AttrE(c)$,
and consider a winning strategy for Eve from $V \setminus \AttrE(c)$ in $\Game_c$, it induces a strategy $\sigma'_c$ in $\Game$.
We construct a strategy $\sigma$ in $\Game$ which will simulate the strategies above in turn; to do so it uses $C$ as top-level memory states.
(We note that the strategies $\sigma'_c$ may use memory as well, so $\sigma$ may actually use more memory than just $C$.)
The strategy $\sigma$ with memory $c$ simulates $\sigma_c$ from $\AttrE(c)$ and $\sigma'_c$ from $V \setminus \AttrE(c)$,
and if it ever reaches $c$ it updates its memory state to $c + 1$ and $1$ if $c = d$.
Any play consistent with $\sigma$ either updates its memory state infinitely many times, 
or eventually remains in $V \setminus \AttrE(c)$ and is eventually consistent with $\sigma'_c$.
In the first case it sees each colour infinitely many times, and since $C \in \F$ the play satisfies $\Muller(\F)$, 
and in the other case since $\sigma'_c$ is winning the play satisfies $\Muller(\F)$.
Thus $\sigma$ is winning from $V$.

We now look at the second item.

Let $\tau_a$ denote an attractor strategy from $\AttrA(\WA(\Game_c)) \setminus \WA(\Game_c)$.
Consider a winning strategy for Adam from $\WA(\Game_c)$ in $\Game_c$, it induces a strategy $\tau_c$ in $\Game$.
Thanks to~\Cref{1-fact:traps_winning}, this implies that $\tau_c$ is a winning strategy in $\Game$.
Consider now a winning strategy in the game $\Game'$ from $\WA(\Game')$, it induces a strategy $\tau'$ in $\Game$.
The set $V \setminus \AttrA( \WA(\Game_c) )$ may not be a trap for Eve, so we cannot conclude that $\tau'$ is a winning strategy in $\Game$,
and it indeed may not be.
We construct a strategy $\tau$ in $\Game$ as the (disjoint) union of the strategy $\tau_a$ on $\AttrA(\WA(\Game_c)) \setminus \WA(\Game_c)$,
the strategy $\tau_c$ on $\WA(\Game_c)$ and the strategy $\tau'$ on $\WA(\Game')$.
We argue that $\tau$ is winning from $\AttrA( \WA(\Game_c) ) \cup \WA(\Game')$ in $\Game$.
Indeed, any play consistent with this strategy in $\Game$ either stays forever in $\WA(\Game')$ hence is consistent with $\tau'$
or enters $\AttrA( \WA(\Game_c) )$, so it is eventually consistent with $\tau_c$.
In both cases this implies that the play is winning.
Thus we have proved that $\AttrA( \WA(\Game_c) ) \cup \WA(\Game') \subseteq \WA(\Game)$.

We now show that $\WE(\Game') \subseteq \WE(\Game)$, which implies the converse inclusion.
Consider a winning strategy from $\WE(\Game')$ in $\Game'$, it induces a strategy $\sigma$ in $\Game$.
Thanks to~\Cref{1-fact:traps_winning}, any play consistent with $\sigma$ stays forever in $\WE(\Game')$, 
implying that $\sigma$ is winning from $\WE(\Game')$ in $\Game$.
\end{proof}

To get the full algorithm we need the analogous lemma for the case where $C \notin \F$.
We do not prove it as it is the exact dual of the previous lemma, and the proof is the same swapping the two players.

\begin{lemma}[Dual fixed-point characterisation of the winning regions for Muller games]
\label{3-lem:Muller_odd}
Let $\Game$ be a Muller game such that $C \notin \F$.
For each $c \in C$, let $\Game_c = \Game \setminus \AttrA(c)$.
\begin{itemize}
	\item If for all $c \in C$, we have $\WE(\Game_c) = \emptyset$, then $\WA(\Game) = V$.
	\item If there exists $c \in C$ such that $\WE(\Game_c) \neq \emptyset$,
	let $\Game' = \Game \setminus \AttrE( \WE(\Game_c) )$,
	then $\WA(\Game) = \WA(\Game')$.	
\end{itemize}
\end{lemma}

The algorithm is presented in pseudocode in \Cref{3-algo:mcnaughton}.
We only give the case where $C \in \F$, the other case being symmetric.
The base case is when there is only one colour $c$, in which case Eve wins everywhere if $\F = \set{c}$
and Adam wins everywhere if $\F = \emptyset$.

We now perform a complexity analysis of the algorithm.
Let us write $f(n,d)$ for the number of recursive calls performed by the algorithm on Muller games with $n$ vertices and $d$ colours.
We have $f(n,1) = f(0,d) = 0$, with the general induction:
\[
f(n,d) \le d \cdot f(n,d-1) + f(n-1,d) + d + 1.
\]
The term $d \cdot f(n,d-1)$ corresponds to the recursive calls to $\Game_c$ for each $c \in C$ and the term $f(n-1,d)$ to the recursive call to $\Game'$.
We obtain $f(n,d) \le d n \cdot f(n,d-1) + (d+1)n$,
so $f(n,d) \le (d+1)n (1 + dn + (dn)^2 + \dots + (dn)^{d-2}) = O((dn)^{d-1})$.
In each recursive call we perform $d+1$ attractor computations so the number of operations in one recursive call is $O(dm)$.
Thus the overall time complexity is $O(dm (dn)^{d-1})$.


The proofs of \Cref{3-lem:Muller_even} and \Cref{3-lem:Muller_odd} also imply that Muller games are determined with finite memory of size $d!$.
We do not make it more precise here because an improved analysis of the memory requirements will be conducted in \Cref{3-sec:zielonka}
using a variant of this algorithm.

\begin{algorithm}
 \KwData{A Muller game $\Game$ over $C$}
 \SetKwFunction{FSolveIn}{SolveIn}
 \SetKwFunction{FSolveOut}{SolveOut}
 \SetKwProg{Fn}{Function}{:}{}

\Fn{\FSolveIn{$\Game$}}{ \tcp{Assumes $C \in \F$}
	\If{$C = \set{c}$}{
		\Return{$V$}
	}

	\For{$c \in C$}{
		
		$\Game_c \leftarrow \Game \setminus \AttrE^{\Game}(c)$

		\If{$C \setminus \set{c} \in \F$}{
			$\WE(\Game_c) \leftarrow \FSolveIn(\Game_c)$ \tcp{$\Game_c$ has one colour less}
		}
		\Else{
			$\WE(\Game_c) \leftarrow \FSolveOut(\Game_c)$ \tcp{$\Game_c$ has one colour less}			
		}
	}

	\If{$\forall c \in C, \WA(\Game_c) = \emptyset$}{
		\Return{$V$}
	}
	\Else{
		Let $c$ such that $\WA(\Game_c) \neq \emptyset$

		$\Game' \leftarrow \Game \setminus \AttrA^{\Game}( \WA(\Game_c) )$
			
		\Return{$\FSolveIn(\Game')$} \tcp{$\Game'$ has less vertices}
	}
}
\vskip1em
\Fn{\FSolveOut{$\Game$}}{
	\tcp{Symmetric to $\FSolveIn$, assumes $C \notin \F$}
}
\vskip1em
\If{$C \in \F$}{
	\FSolveIn{$\Game$}
}
\Else{
	\FSolveOut{$\Game$}
}
\caption{A recursive algorithm for computing the winning regions of Muller games.}
\label{3-algo:mcnaughton}
\end{algorithm}

\subsection*{Positional determinacy for Rabin games}

Before studying their complexity, we prove the following property of Rabin games.

\begin{theorem}[Positional determinacy for Rabin games]
	\label{3-thm:Rabin_positional_determinacy}
	Rabin games are uniformly "positionally determined".
\end{theorem}

We will actually give two different proofs of this result: the first, right below, is direct but only yields uniform positional determinacy over finite arenas (which suffices for the upcoming considerations about computational complexity).
A second proof in \Cref{3-thm:characterisation_Zielonka_tree} shows that Rabin games are uniformly positionally determined over arbitrary arenas through the more general study of the memory requirements of Muller conditions.


\begin{proof}
We proceed by induction over the following quantity: the total out-degree of vertices controlled by Eve, minus the number of vertices controlled by Eve.
Since we assume that every vertex has an outgoing edge, the base case is when each vertex of Eve has only one successor.
In that case Eve has only one strategy and it is positional, so the property holds.

In the inductive step, we consider a Rabin game $\game$ where Eve has a winning strategy~$\sigma$.
Let $v \in \VE$ with at least two successors.
We partition the outgoing edges of $v$ in two non-empty subsets which we call $E^v_1$ and $E^v_2$.
Let us define two games $\game_1$ and $\game_2$: the game $\game_1$ is obtained from $\game$ by removing the edges from $E^v_2$, and symmetrically for~$\game_2$.

We claim that Eve has a winning strategy either in $\game_1$ or in $\game_2$.
Let us assume towards a contradiction that this is not the case: then there exist $\tau_1$ and $\tau_2$ two strategies for Adam which are winning in $\game_1$ and $\game_2$ respectively.
We construct a strategy $\tau$ for Adam in $\game$ as follows: it has two ``modes'', $1$ and $2$.
Intuitively, modes serve as memory, allowing Adam to switch between $\tau_1$ and $\tau_2$.
The initial mode is $1$, and the strategy simulates $\tau_1$ from the mode $1$ and $\tau_2$ from the mode $2$. Whenever $v$ is visited, the mode is adjusted: if the outgoing edge is in $E^v_1$ then the new mode is $1$, otherwise it is $2$.
To be more specific: when simulating $\tau_1$, we play ignoring the parts of the play using mode $2$, so removing them yields a play consistent with $\tau_1$. The same goes for $\tau_2$.

Consider a play $\play$ consistent with $\sigma$ and $\tau$.
Since $\sigma$ is winning, the play $\play$ is winning. It can be decomposed following which mode the play is in:
\[
\begin{array}{ccccccccc}
\text{mode } 1 & \overbrace{v_0 \cdots v}^{\play_1^0} & &
\overbrace{v \cdots v}^{\play_1^1} & & \ \cdots \\
\text{mode } 2 && \underbrace{v \cdots v}_{\play_2^0}
& & \underbrace{v \cdots v}_{\play_2^1} & \ \cdots
\end{array}
\]
where $\play_1 = \play_1^0 \play_1^1 \cdots$ is consistent with $\tau_1$ and $\play_2 = \play_2^0 \play_2^1 \cdots$ is consistent with $\tau_2$.
Since $\tau_1$ and $\tau_2$ are winning strategies for Adam, $\play_1$ and $\play_2$ do not satisfy $\Rabin$.

\vskip1em
There are two cases: the decomposition is either finite or infinite.
If it is finite we get a contradiction: since $\play$ is winning and $\Rabin$ is prefix-independent, any suffix of $\play$ is winning as well, contradicting that it is consistent with either $\tau_1$ or $\tau_2$.

In the second case, to get a contradiction, we observe the following property of the Rabin objective:
for all finite sequences $(\rho_1^\ell)_{\ell \in \N}$ and $(\rho_2^\ell)_{\ell \in \N}$,
\begin{align} \label{3-eq:submixing_early}
\begin{array}{lccccccccccl}
\text{if} & \rho_1 & = & \rho_1^0 & & \rho_1^1 & & \cdots & \rho_1^\ell & & \cdots & \notin \Rabin \\
\text{and} & \rho_2 & = & & \rho_2^0 & & \rho_2^1 & \cdots & & \rho_2^\ell & \cdots & \notin \Rabin, \\
\text{then} & \rho & = & \rho_1^0 & \rho_2^0 & \rho_1^1 & \rho_2^1 & \cdots & \rho_1^\ell & \rho_2^\ell & \cdots & \notin \Rabin.
\end{array}
\end{align}
Indeed, since neither $\rho_1$ nor $\rho_2$ satisfies $\Rabin$, for all $i \in [1,d]$, if $R_i \in \Inf(\rho_1)$ (resp.\ $R_i \in \Inf(\rho_2)$), then $G_i \in \Inf(\rho_1)$ (resp.\ $G_i \in \Inf(\rho_2)$).
Since $\Inf(\rho) = \Inf(\rho_1) \cup \Inf(\rho_2)$, this implies that $\rho$ does not satisfy $\Rabin$.

This directly yields a contradiction: neither $\play_1$ nor $\play_2$ satisfies $\Rabin$, yet $\play$ does.
\end{proof}

The proof technique used above will be generalised in \Cref{4-chap:memory}, \Cref{4-thm:submixing_positional}, to show positional determinacy of prefix-independent quantitative objectives and a generalisation of property~\eqref{3-eq:submixing_early}.

%

\subsection*{The complexity of solving Rabin games}

\begin{theorem}[Complexity of solving Rabin games]
\label{3-thm:Rabin_complexity}
Solving Rabin games is $\NP$-complete.
\end{theorem}

\begin{proof}
The proof that solving Rabin games is in $\NP$ follows the same lines as for solving parity games: the algorithm guesses a positional strategy and checks whether it is indeed winning. This requires proving that solving Rabin games where Adam controls all vertices can be done in polynomial time, which is indeed true and easy to see so we will not elaborate further on this.

To prove the $\NP$-hardness we reduce the satisfiability problem for boolean formulas in conjunctive normal form ($\SAT$) to solving Rabin games. 

Let $\Phi$ be a formula in conjunctive normal form with $n$ variables $x_1 \ldots x_n$ and $m$ clauses $C_1 \dots C_m$, where each $C_j$ is of the form $\ell_{j_1} \vee \ell_{j_2} \vee \ell_{j_3}$:
\[
\Phi = \bigwedge_{j=1}^m \ell_{j_1} \vee \ell_{j_2} \vee \ell_{j_3}.
\]
A literal $\ell$ is either a variable $x$ or its negation $\bar{x}$, and we write $\bar{\ell}$ for the negation of a literal.
The question whether $\Phi$ is satisfiable reads: does there exist a valuation $\mathbf{v} : \set{x_1,\dots,x_n} \to \set{0,1}$
satisfying $\Phi$.

We construct a Rabin game $\game$ with $m+1$ vertices (one per clause, all controlled by Eve, plus a unique vertex controlled by Adam), 
$4m$ edges ($4$ per clause), and $2n$ Rabin pairs (one per literal).
We will show that the formula $\Phi$ is satisfiable if and only if Eve has a winning strategy in the Rabin game $\game$.

We first describe the Rabin condition. 
There is a Rabin pair $(R_\ell,G_\ell)$ for each literal~$\ell$, so the Rabin condition requires that there exists a literal $\ell$ such that $R_\ell$ is visited infinitely many times and $G_\ell$ is not.
Let us now describe the arena. 
A play consists in an infinite sequence of rounds, where in each round first Adam chooses a clause and second Eve chooses a literal in this clause. 
When Eve chooses a literal $\ell$ she visits $R_\ell$ and $G_{\bar{\ell}}$.
This completes the description of the Rabin game $\game$, it is illustrated in \Cref{3-fig:hardness_Rabin}.
Let us now prove that this yields a reduction from $\SAT$ to solving Rabin games.

\vskip1em
Let us first assume that $\Phi$ is satisfiable, and let $\mathbf{v}$ be a satisfying assignment: there is a literal $\ell$ in each clause satisfied by $\mathbf{v}$. 
Let $\sigma$ be the positional strategy choosing such a literal in each clause. 
We argue that in any play consistent with $\sigma$ there is at least one literal $\ell$ that Eve chooses infinitely many times without ever choosing $\bar{\ell}$: this implies that $R_\ell$ is visited infinitely often and $G_\ell$ is not.
Indeed, some clause is chosen infinitely many times, so the corresponding literal chosen by Eve is also chosen infinitely many times.
Since all the literals chosen by Eve satisfy the same assignment $\mathbf{v}$ she does not choose both a literal and its negation, so she never chooses $\bar{\ell}$. 
It follows that $\sigma$ is a winning strategy for Eve.

Conversely, let us assume that Eve has a winning strategy.
Thanks to \Cref{3-thm:Rabin_positional_determinacy} she has a positional winning strategy $\sigma$. 
We argue that $\sigma$ cannot choose some literal $\ell$ in some clause $C$ and the literal $\bar{\ell}$ in another clause $C'$.
If this would be the case, consider the strategy of Adam alternating between the two clauses $C$ and $C'$ and play it against $\sigma$:
both $\ell$ and $\bar{\ell}$ are chosen infinitely many times, and no other literals.
Hence $R_\ell, G_\ell, R_{\bar{\ell}}$, and $G_{\bar{\ell}}$ are all visited infinitely many times, 
implying that this play does not satisfy $\Rabin$, contradicting that $\sigma$ is winning.

There exists a valuation $\mathbf{v}$ which satisfies each literal chosen by Eve, implying that it satisfies $\Phi$ which is then satisfiable.
\end{proof}

\begin{figure}
\centering
  \begin{tikzpicture}[scale=1.3]
    \node[s-adam] (v0) at (2,2.5) {};
    \node[s-eve] (v1) at (0,3.6) {$\ x \vee y \vee z\ $};
    \node[s-eve] (v2) at (4.3,2.7) {$\ x \vee \bar{y} \vee \bar{z}\ $};
    \node[s-eve] (v3) at (1.7,0) {$\ \bar{x} \vee y \vee \bar{z}\ $};

    \path[arrow]
      (v0) edge[bend right] (v1)
      (v0) edge[bend right] (v2)
      (v0) edge[bend right] (v3)

      (v1) edge[bend left = 45] node[above, pos = 0.35] {$R_x$} node[right, pos = 0.55] {$G_{\bar{x}}$} (v0)
      (v1) edge[bend right] node[above, pos = 0.35] {$R_y$} node[above, pos = 0.65] {$G_{\bar{y}}$} (v0)
      (v1) edge[bend right = 45] node[below, pos = 0.25] {$R_z$} node[below, pos = 0.55] {$G_{\bar{z}}$} (v0)

      (v2) edge[bend left = 45] node[below, pos = 0.25] {$R_x$} node[below, pos = 0.55] {$G_{\bar{x}}$}(v0)
      (v2) edge[bend right] node[below, pos = 0.35] {$R_{\bar{y}}$} node[below, pos = 0.65] {$G_{y}$} (v0)
      (v2) edge[bend right = 45] node[above, pos = 0.35] {$R_{\bar{z}}$} node[above, pos = 0.65] {$G_{z}$} (v0)

      (v3) edge[bend left = 45] node[left, pos = 0.35] {$R_{\bar{x}}$} node[left, pos = 0.65] {$G_{x}$} (v0)
      (v3) edge[bend right] node[left, pos = 0.35] {$R_y$} node[left, pos = 0.65] {$G_{\bar{y}}$} (v0)
      (v3) edge[bend right = 45] node[right, pos = 0.35] {$R_{\bar{z}}$} node[right, pos = 0.65] {$G_{z}$} (v0);
  \end{tikzpicture}
\caption{The Rabin game for $\Phi = (x \vee y \vee z) \bigwedge (x \vee \bar{y} \vee \bar{z}) \bigwedge (\bar{x} \vee y \vee \bar{z})$.}
\label{3-fig:hardness_Rabin}
\commentAlt{Figure~\ref{3-fig:hardness_Rabin}: A directed graph with three circular nodes and one central square node, showing complex interactions. See long description.}
\commentLongAlt{Figure~\ref{3-fig:hardness_Rabin}: The image displays a directed graph with four nodes. Three circular nodes are positioned at the corners of an implied triangle, and a single square node is at the center. Each circular node is labeled with a logical expression: top-left is '(x OR y OR z)', bottom-left is '(x_bar OR y OR z_bar)', and right is '(x OR y_bar OR z_bar)'. Numerous labeled arrows connect the circular nodes to the central square node and vice-versa. For example, there's a bidirectional path between the top-left circle and the square node, with arrows labeled R_x, G_x_bar, R_y, G_y_bar, R_z, and G_z. Similar sets of labeled arrows connect the other two circular nodes to the central square node, indicating complex transitions or relationships based on logical variables x, y, and z, and their negations.}
\end{figure}

\subsection*{The complexity of solving Muller games}    

\begin{theorem}[Complexity of solving Muller games]
\label{3-thm:complexity_Muller}
Solving Muller games is $\PSPACE$-complete.
\end{theorem}

As for the previous reduction, in the Muller game constructed in the reduction below we label edges rather than vertices,
and some edges have more than one colour.
As for Rabin games this can be reduced to the original definition of colouring functions (labelling vertices with exactly one colour each) with a polynomial increase in size.

\begin{proof}
The $\PSPACE$ algorithm was constructed in \Cref{3-thm:muller}.

To prove the $\PSPACE$-hardness we reduce the evaluation of quantified boolean formulas in disjunctive normal form ($\QBF$) to solving Muller games. 

Let $\Psi$ be a quantified boolean formula in disjunctive normal form with $n$ variables $x_1 \ldots x_n$ and $m$ clauses $C_1 \dots C_m$, where each $C_j$ is of the form $\ell_{j_1} \wedge \ell_{j_2} \wedge \ell_{j_3}$:
\[
\Psi = \exists x_1,\forall x_2,\ldots,\exists x_n,\ \Phi(x_1,\dots,x_n) \text{ and } 
\Phi(x_1,\dots,x_n) = \bigvee_{j=1}^m \ell_{j_1} \wedge \ell_{j_2} \wedge \ell_{j_3}.
\]

We construct a Muller game $\Game$ with $m+1$ vertices (one per clause, all controlled by Adam, plus a unique vertex controlled by Eve), $4m$ edges ($4$ per clause), and $2n$ colours (one per literal).
We will show that the formula $\Psi$ evaluates to true if and only if Eve has a winning strategy in the Muller game $\Game$.

We first describe the Muller condition. 
The set of colours is the set of literals.
We let $x$ denote the lowest quantified variable such that $x$ or $\bar{x}$ is visited infinitely many times. 
The Muller condition requires that:
\begin{itemize}
	\item either $x$ is existential and only one of $x$ and $\bar{x}$ is visited infinitely many times,
	\item or $x$ is universal and both $x$ and $\bar{x}$ are visited infinitely many times,
\end{itemize}
and for all variables $y$ quantified after $x$, both $y$ and $\bar{y}$ are visited infinitely many times.
Formally, let $S_{> i} = \set{x_q, \bar{x_q} : q > p}$ and:
\[
\F = \set{ S_{> p},\ \set{x_p} \cup S_{> p},\ \set{\bar{x_p}} \cup S_{> p} : x_p \text{ existential}} \cup \set{ S_{\ge p} : x_p \text{ universal}}.
\]
Note that $\F$ contains $O(n)$ elements.

Let us now describe the arena. A play consists in an infinite sequence of rounds, where in each round first Eve chooses
a clause and second Adam chooses a literal $\ell$ in this clause corresponding to some variable $x_p$, 
and visits the colour $\ell$ as well as each colour in $S_{> p}$.

The reduction is illustrated in~\Cref{3-fig:hardness_Muller}.
Note that the edges from the vertex controlled by Eve to the other ones do not have a colour,
which does not fit our definitions.
For this reason we introduce a new colour $c$ and colour all these edges by $c$.
We define a new Muller objective by adding $c$ to each set in $\F$: 
since every play in the game visit $c$ infinitely many times, the two games are equivalent.
We note that this construction works for this particular game but not in general.

\vskip1em
For a valuation $\mathbf{v} : \set{x_1,\dots,x_n} \to \set{0,1}$ and $p \in [1,n]$,
we write $\Psi_{\mathbf{v},p}$ for the formula obtained from $\Psi$ by fixing the variables $x_1,\dots,x_{p-1}$
to $\mathbf{v}(x_1),\dots,\mathbf{v}(x_{p-1})$ and quantifying only over the remaining variables.
Let us say that a valuation $\mathbf{v}$ is \textit{positive} if for every $p \in [1,n]$, the formula $\Psi_{\mathbf{v},p}$ evaluates to true,
and similarly a valuation is \textit{negative} if for every $p \in [1,n]$, the formula $\Psi_{\mathbf{v},p}$ evaluates to false.

\vskip1em
Let us first assume that $\Psi$ evaluates to true. 
We construct a winning strategy $\sigma$ for Eve.
It uses \textit{positive} valuations over the variables $x_1,\ldots,x_n$ as memory states. 
Note that the fact that $\Psi$ evaluates to true implies that there exists a positive valuation. 
Let us choose an arbitrary positive valuation as initial valuation.
We first explain what the strategy $\sigma$ does and then how to update its memory.

Assume that the current valuation is $\mathbf{v}$, since it is positive there exists a clause satisfying $\mathbf{v}$, the strategy $\sigma$ chooses such a clause. 
Therefore, any literal that Adam chooses is necessarily true under $\mathbf{v}$.

The memory is updated as follows: 
assume that the current valuation is $\mathbf{v}$ and that Adam chose a literal corresponding to the variable $x_p$. 
If $x_p$ is existential the valuation is unchanged.
If $x_p$ is universal, we construct a new positive valuation as follows. 
We swap the value of $x_p$ in $\mathbf{v}$ and write $\mathbf{v}[x_p]$ for this new valuation. 
Since $\mathbf{v}$ is positive and $x_p$ is universally quantified, the formula $\Psi_{\mathbf{v}[x_p],p+1}$ evaluates to true,
so there exists a positive valuation $\mathbf{v}_{p+1} : \set{x_{p+1},\dots,x_n} \to \set{0,1}$ for this formula.
The new valuation is defined as follows:
\[
\mathbf{v}'(x_q) = 
\begin{cases}
\mathbf{v}(x_q) & \text{ if } q < p, \\[.5em]
\overline{\mathbf{v}(x_q)} & \text{ if } q = p, \\
\mathbf{v}_{p+1}(x_q) & \text{ if } q > p,
\end{cases}
\]
it is positive by construction.

Let $\play$ be a play consistent with $\sigma$ and $x_p$ be the lowest quantified variable chosen infinitely many times by Adam. 
First, all colours in $S_{> p}$ are visited infinitely many times (when visiting $x$ or $\bar{x}$).
Let us look at the sequence $(\mathbf{v}_i(x_p))_{i \in \N}$ where $\mathbf{v}_i$ is the valuation in the $i$\textsuperscript{th} round.
If $x_p$ is existential, the sequence is ultimately constant as it can only change when a lower quantified variable is visited.
If $x_p$ is universal, the value changes each time the variable $x_p$ is chosen.
Since any literal that Adam chooses is necessarily true under the current valuation, 
this implies that in both cases $\play$ satisfies $\Muller(\F)$.

\vskip1em
For the converse implication we show that if $\Psi$ evaluates to false, then there exists a winning strategy $\tau$ for Adam.
The construction is similar but using \textit{negative} valuations.
The memory states are negative valuations. The initial valuation is any negative valuation.
If the current valuation is $\mathbf{v}$ and Eve chose the clause $C$, since the valuation is negative $\mathbf{v}$ does not satisfy $C$,
the strategy $\tau$ chooses a literal in $C$ witnessing this failure. 
The memory is updated as follows: assume that the current valuation is $\mathbf{v}$ and that the strategy $\tau$ chose a literal corresponding to the variable $x_p$. 
If $x_p$ is universal the valuation is unchanged.
If $x_p$ is existential, we proceed as above to construct another negative valuation where the value of $x_p$ is swapped.

Let $\play$ be a play consistent with $\tau$ and $x$ be the lowest quantified variable chosen infinitely many times by Adam. 
As before, we look at the sequence $(\mathbf{v}_i(x))_{i \in \N}$ where $\mathbf{v}_i$ is the valuation in the $i$\textsuperscript{th} round.
If $x$ is existential, the value changes each time the variable $x$ is chosen.
If $x$ is universal, the sequence is ultimately constant.
Since any literal that Adam chooses is necessarily false under the current valuation, 
this implies that in both cases $\play$ does not satisfy $\Muller(\F)$.
\end{proof}

\begin{figure}
\centering
  \begin{tikzpicture}[scale=1.3]
\node[s-eve] (v0) at (2,2.5) {};
\node[s-adam] (v1) at (0,3.6) {$\ x \vee y \vee z\ $};
\node[s-adam] (v2) at (4.3,2.7) {$\ x \vee \bar{y} \vee \bar{z}\ $};
\node[s-adam] (v3) at (1.7,0) {$\ \bar{x} \vee y \vee \bar{z}\ $};

\path[arrow]
  (v0) edge[bend right] (v1)
  (v0) edge[bend right] (v2)
  (v0) edge[bend right] (v3)

  (v1) edge[bend left = 45] node[above, pos = 0.35] {$x$} node[above right, pos = 0.55] {$S_{> x}$} (v0)
  (v1) edge[bend right] node[above, pos = 0.35] {$y$} node[above, pos = 0.65] {$S_{> y}$} (v0)
  (v1) edge[bend right = 45] node[below, pos = 0.25] {$z$} node[below, pos = 0.55] {$S_{> z}$} (v0)

  (v2) edge[bend left = 45] node[below, pos = 0.25] {$x$} node[below, pos = 0.55] {$S_{> x}$}(v0)
  (v2) edge[bend right] node[below, pos = 0.35] {$\bar{y}$} node[below, pos = 0.65] {$S_{> y}$} (v0)
  (v2) edge[bend right = 45] node[above, pos = 0.35] {$\bar{z}$} node[above left, pos = 0.65] {$S_{> z}$} (v0)

  (v3) edge[bend left = 45] node[left, pos = 0.35] {$\bar{x}$} node[left, pos = 0.65] {$S_{> x}$} (v0)
  (v3) edge[bend right] node[left, pos = 0.35] {$y$} node[left, pos = 0.65] {$S_{> y}$} (v0)
  (v3) edge[bend right = 45] node[right, pos = 0.35] {$\bar{z}$} node[right, pos = 0.65] {$S_{> z}$} (v0);
  \end{tikzpicture}
\caption{The Muller game for $\Psi = \exists x, \forall y, \exists z, (x \wedge y \wedge z) \bigvee (x \wedge \bar{y} \wedge \bar{z}) \bigvee (\bar{x} \wedge y \wedge \bar{z})$. For a variable $v$ we write $S_{> v}$ for the set of literals corresponding to variables quantified after $v$, so for instance $S_{> x} = \set{y,\bar{y},z,\bar{z}}$.}
\label{3-fig:hardness_Muller}
\commentAlt{Figure~\ref{3-fig:hardness_Muller}: A directed graph with three rectangular nodes and one central circular node, showing complex interactions. See long description.}
\commentLongAlt{Figure~\ref{3-fig:hardness_Muller}: The image displays a directed graph with four nodes. Three rectangular nodes are positioned at the corners of an implied triangle, and a single circular node is at the center. Each rectangular node is labeled with a logical expression: top-left is '(x OR y OR z)', bottom-left is '(x_bar OR y OR z_bar)', and right is '(x OR y_bar OR z_bar)'. Numerous labeled arrows connect the rectangular nodes to the central circular node and vice-versa. For example, there's a bidirectional path between the top-left rectangle and the circular node, with arrows labeled x, S_x, y, S_y, z, and S_z. Similar sets of labeled arrows connect the other two rectangular nodes to the central circular node, indicating complex transitions or relationships based on logical variables x, y, and z, and their negations.}
\end{figure}


\section{Zielonka tree}
\label{3-sec:zielonka}
The Zielonka tree is a combinatorial structure associated with a Muller objective which very neatly exposes its properties.
As a warm-up we first present its predecessor the LAR construction, and then show the properties of Zielonka trees.
As we will see, one of the key features of the Zielonka tree of a Muller objective $\Muller(\F)$ is to characterise its exact memory requirements.

\subsection*{The latest appearance record}
Muller objectives can be reduced to parity objectives, see~\Cref{1-sec:automata} for an introduction to reductions between objectives.

\begin{theorem}[Latest Appearance Record (LAR) construction]
\label{3-thm:LAR}
Let $C = [1,d]$ be a set of colours and $\Muller(\F)$ a Muller objective.
There exists a deterministic parity automaton $\LAR_\F$ over the alphabet $C$ defining $\Muller(\F)$.
It has $d!$ states and has priorities in $[1,2d]$.
\end{theorem}

LAR stands for Latest Appearance Record.
In the literature the number of states is often $d \cdot d!$ instead of $d!$,
the multiplicative factor $d$ is saved since we consider transition-based acceptance conditions for automata.

\begin{proof}
We define the automaton $\LAR_\F$.
The set of states is the set of lists of all colours of $C$ without repetitions.
We represent a list by $(c_1,\dots,c_d)$.
The initial state is irrelevant because $\Muller(\F)$ is prefix-independent.
The transition function is defined as follows: 
$\delta(\ell, c)$ is $\ell'$ obtained from $\ell$ by pushing $c$ to the first position (hence shifting to the right the elements to the left of $c$). 
This is best understood on an example: 
\[
\delta( (4, 1, 2, 3), 2) = (2, 4, 1, 3).
\]
Let $j$ be the position of $c$ in $\ell$, the priority of this transition is defined by:
\[
\col((\ell,c,\ell')) =
\begin{cases}
2 j      & \text{ if } \ell([1,j]) \in \F, \\
2 j - 1  & \text{ otherwise.}
\end{cases}
\]

We now show that the automaton $\LAR_\F$ defines $\Muller(\F)$.
Let $\rho = c_0 c_1 \dots$ be an infinite word over the alphabet $C$. 
Let us consider the run of $\LAR_\F$ over~$\rho$:
\[
(\ell_0,c_0,\ell_1) (\ell_1,c_1,\ell_2) \dots
\]
Let us write $j_i$ for the position of $c_i$ in $\ell_i$.
We consider $\Inf(\rho)$ the set of colours appearing infinitely many times and write $j$ for its cardinal.
From some point onwards the lists $\ell_i$ are of the form 
\[
(\underbrace{c_1,\dots,c_j}_{\Inf(\rho)} ,\ \underbrace{c_{j + 1},\dots,c_d}_{C \setminus \Inf(\rho)}).
\]
From this point on $j_i$ is smaller than or equal to $j$, and it reaches $j$ infinitely many times.
It follows that the largest priority appearing infinitely many times in the run is $2 j$ if $\Inf(\rho) \in \F$
and $2 j - 1$ if $\Inf(\rho) \notin \F$.
Thus $\rho$ is accepted by $\LAR_\F$ if and only if $\Inf(\rho) \in \F$, as desired.
\end{proof}

\subsection*{The Zielonka tree}
\Cref{3-thm:LAR} implies a reduction from Muller games to parity games as explained in \Cref{1-sec:automata}.
This yields a small improvement from the complexity results we already obtained for Muller games in \Cref{3-thm:muller},
but not for the memory requirements.
One weakness of the LAR construction is that its size depends only on the number of colours, and not on the properties of $\F$.
The Zielonka tree is an improved take on the "LAR".

\begin{definition}[Zielonka tree]
\label{definition:zielonka_tree}
Let $\Muller(\F)$ be a Muller objective over the set of colours $C$.
The Zielonka tree $T_\F$ of $\Muller(\F)$ is a rooted tree with nodes labelled by subsets of colours, 
it is constructed inductively as follows:
\begin{itemize}
	\item the root is labelled $C$,
	\item the children of a node labelled $S$ are the maximal subsets $S_1, \dots, S_k$ of $S$ such that 
	$S_i \in \Muller(\F) \Longleftrightarrow S \notin \Muller(\F)$.
\end{itemize}
\end{definition}


\Cref{3-fig:Zielonka_tree_example} represents the Zielonka tree for $\Muller(\F)$ with 
\[
\F = \set{\set{2}, \set{3}, \set{4}, \set{1,2}, \set{1,3}, \set{1,3,4}, \set{2,3,4}, \set{1,2,3,4}}.
\]
We note that there are two nodes labelled $\set{1}$; in general there may be several nodes with the same label.
Also, not all branches have the same length.

\begin{figure}
\centering
  \begin{tikzpicture}[scale=1.1]
\node[s-eve, dashed] (1234) at (5,3.5) {$\ \set{1,2,3,4}\ $};

\node[s-adam, dashed] (123) at (2,2) {$\ \set{1,2,3}\ $};
\node[s-eve] (12g) at (1,1) {$\ \set{1,2}\ $};
\node[s-adam] (1g) at (1,0) {$\ \set{1}\ $};
\node[s-eve, dashed] (13) at (3,1) {$\ \set{1,3}\ $};
\node[s-adam, dashed] (1m) at (3,0) {$\ \set{1}\ $};

\node[s-adam] (124) at (5,2) {$\ \set{1,2,4}\ $};
\node[s-eve] (12d) at (4,1) {$\ \set{1,2}\ $};
\node[s-adam] (1d) at (4,0) {$\ \set{1}\ $};
\node[s-eve] (4g) at (6,1) {$\ \set{4}\ $};

\node[s-adam] (34) at (8,2) {$\ \set{3,4}\ $};
\node[s-eve] (3) at (7,1) {$\ \set{3}\ $};
\node[s-eve] (4d) at (9,1) {$\ \set{4}\ $};

\node (l1) at (10,0) {$\ 1\ $};
\node (l2) at (10,1) {$\ 2\ $};
\node (l3) at (10,2) {$\ 3\ $};
\node (l4) at (10,3.5) {$\ 4\ $};

\path
  (1234) edge (123)
  (1234) edge (124)
  (1234) edge (34)
  (123) edge (12g)
  (12g) edge (1g)
  (123) edge (13)
  (13) edge (1m)
  (124) edge (12d)
  (124) edge (4g)
  (12d) edge (1d)
  (34) edge (3)
  (34) edge (4d);  
  \end{tikzpicture}
\caption{The Zielonka tree for $\Muller(\F)$. By convention nodes labelled by a set in $\F$ are represented by a circle
and the others by a square.
The numbers on the right hand side and the dashed nodes (describing a branch) are both used in the proof of \Cref{3-thm:reduction_parity_Zielonka_tree}.}
\label{3-fig:Zielonka_tree_example}
\commentAlt{Figure~\ref{3-fig:Zielonka_tree_example}: A tree structure with numerical sets in nodes and a dashed circle indicating a new node, with levels labeled 1 to 4. See long description.}
\commentLongAlt{Figure~\ref{3-fig:Zielonka_tree_example}: A tree diagram with a root node at the top containing the set '{1,2,3,4}', marked with a dashed outline and at level 4. This root branches into two children nodes at level 3. The left child is a square with a dashed outline, containing the set '{1,2,3}'. The right child is a square containing the set '{3,4}'. The left child at level 3 further branches into two children at level 2: a circle containing '{1,2}' and a dashed circle containing '{1,3}'. The right child at level 3 branches into two children at level 2: a square containing '{1,2,4}' and a square containing '{3,4}'. The node '{1,2,4}' branches into two children at level 2: a circle containing '{1,2}' and a circle containing '{4}'. The node '{3,4}' branches into two children at level 2: a circle containing '{3}' and a circle containing '{4}'. The nodes at level 2 further branch to level 1. The circle '{1,2}' (leftmost at level 2) branches to a square containing '{1}'. The dashed circle '{1,3}' branches to a dashed square containing '{1}'. The circle '{1,2}' (middle at level 2) branches to a square containing '{1}'.}
\end{figure}

The first use of the Zielonka tree is to induce an improved reduction from Muller to parity objectives.
A branch in a tree is a path from the root to a leaf.

\begin{theorem}[Reduction from Muller to parity games using the Zielonka tree automaton]
\label{3-thm:reduction_parity_Zielonka_tree}
Let $C = [1,d]$ be a set of colours and $\Muller(\F)$ a Muller objective.
There exists a deterministic parity automaton $\Zielonka_\F$ over the alphabet $C$ defining $\Muller(\F)$.
Its number of states is the number of branches of $T_\F$ and its parity condition uses at most $d$ priorities (and exactly as many priorities as the height of the Zielonka tree).
\end{theorem}

Here again we take advantage of the fact that the acceptance conditions on automata are transition based;
using stated based transitions we would have added a multiplicative factor~$d$.

\begin{proof}
Without loss of generality $C \in \F$: if this is not the case we consider the complement $\Muller(2^C \setminus \F)$.
We number the levels of $T_\F$ from the leaves to the root such that nodes labelled by sets in $\F$ are even
and the other ones odd (this will be used for defining the parity condition).
See \Cref{3-fig:Zielonka_tree_example} for a possible numeration of the levels (on the right hand side), the other options being shifts of this numeration by an even number.

The set of states of $\Zielonka_\F$ is the set of branches of $T_\F$.
We represent a branch by $(S_1,\dots,S_k)$
where $S_1$ is the set labelling the root and $S_k$ the set labelling a leaf. Note that $k \le d$.
For the sake of simplicity we identify nodes with their labels, which is an abuse since two different nodes may have the same label
but will be convenient and harmless in our reasoning.

The initial state is irrelevant because $\Muller(\F)$ is prefix-independent.
We define the support $\support(b,c)$ of a branch $b$ and a colour $c$ to be the lowest node of $b$ which contains~$c$.
The transition function is defined as follows: 
$\delta(b,c)$ is the next branch (in the lexicographic order from left to right and in a cyclic way) which coincides with $b$ up to $\support(b,c)$.
The priority of this transition is given by the level on which $\support(b,c)$ sits.

This is best understood on an example: on \Cref{3-fig:Zielonka_tree_example} consider the branch $b$ represented by dashed nodes, reading the colour $2$, because $\support(b,2) = \set{1,2,3}$, we consider branches starting with $(\set{1,2,3,4}, \set{1,2,3})$.

The next branch after $b$ is $(\set{1,2,3,4}, \set{1,2,3},\set{1,2},\set{1})$ (because we cycle: the node after $\set{1,3}$ is $\set{1,2}$).
The priority of this transition is $3$ corresponding to the level where $\set{1,2,3}$ sits.

We now show that the automaton $\Zielonka_\F$ defines $\Muller(\F)$.
Let $\rho = c_0 c_1 \dots$ be an infinite word over the alphabet $C$.
Let us consider the run of $\Zielonka_\F$ over~$\rho$:
\[
(b_0,c_0,b_1) (b_1,c_1,b_2) \dots
\]
We consider $\Inf(\rho)$ the set of colours appearing infinitely many times.
Let us look at the largest prefix $(S_1,\dots,S_p)$ of a branch which is eventually common to all the branches $b_i$.
We make two claims:
\begin{itemize}
	\item $\Inf(\rho)$ is included in $S_p$;
	\item $\Inf(\rho)$ is not included in any child of $S_p$.
\end{itemize}
For the first claim, let $c \in \Inf(\rho)$, since eventually the branch $b_i$ starts with $(S_1,\dots,S_p)$,
the support of $b_i$ and $c$ is lower than or equal to $S_p$, meaning that $c \in S_p$.

For the second claim, we first note that by maximality of $(S_1,\dots,S_p)$ the support of $b_i$ and $c_i$ is infinitely many times $S_p$.
Indeed from some point onwards it is lower than or equal to $S_p$, and if it would be eventually strictly lower then the corresponding child of $S_p$ would be common to all branches $b_i$ from there on.
This implies that all children of $S_p$ appear infinitely many times in the branches $b_i$: each time the support of $b_i$ and $c_i$ is $S_p$, the branch switches to the next child of $S_p$.
Now since each child $S_{p+1}$ of $S_p$ is left infinitely many times this implies that there exists $c \in \Inf(\rho)$ with $c \notin S_{p+1}$.
Hence $\Inf(\rho)$ is not included in $S_{p+1}$.

By definition of the Zielonka tree, this implies that $\Inf(\rho) \in \F$ if and only if $S_p \in \F$,
thus $\rho$ is accepted by $\Zielonka_\F$ if and only if $\Inf(\rho) \in \F$, as desired.
\end{proof}

Since \Cref{3-thm:reduction_parity_Zielonka_tree} is a reduction from Muller to parity objectives,
it implies a reduction from Muller games to parity games as explained in \Cref{1-sec:automata},
improving over \Cref{3-thm:LAR}. 
Since solving parity games is in $\NP \cap \coNP$,
if we represent the Muller condition by a Zielonka tree then the automaton constructed in \Cref{3-thm:reduction_parity_Zielonka_tree}
is of polynomial size, implying the following result.

\begin{theorem}[Complexity of solving Muller games represented by the Zielonka tree]
\label{3-thm:complexity_Muller_games_representation_Zielonka_tree}
Solving Muller games where the condition is represented by a Zielonka tree is in $\NP \cap \coNP$.
\end{theorem}

As observed above different nodes of the Zielonka tree may be labelled by the same set of colours.
Hence it is tempting to represent a Muller condition not with its Zielonka tree but rather with the Zielonka DAG (Directed Acyclic Graph)
where nodes labelled by the same set of colours are identified.
However with this representation solving Muller games is again $\PSPACE$-complete:

\begin{theorem}[Complexity of solving Muller games represented by the Zielonka DAG]
\label{3-thm:Muller_games_DAG}
Solving Muller games where the condition is represented by a Zielonka DAG is $\PSPACE$-complete.
\end{theorem}

The algorithm presented in \Cref{3-thm:muller} runs in polynomial space for this representation.
To obtain the $\PSPACE$-hardness we observe that in the reduction from QBF constructed in \Cref{3-thm:complexity_Muller},
the Muller objective is of polynomial size when represented by a Zielonka DAG (but of exponential size when represented by a Zielonka tree).

\subsection*{The exact memory requirements}
The second and most interesting use of the Zielonka tree is for characterising the memory requirements for Eve.

Note that a node in the Zielonka tree $T_\F$ represents another Muller objective, over the set of colours labelling this node.
For instance in \Cref{3-fig:Zielonka_tree_example} the node labelled $\set{1,2,3}$ corresponds to $\Muller(\F')$ with
$\F' = \set{\set{2}, \set{3}, \set{1,2}, \set{1,3}}$.

\begin{definition}[Memory requirements for Muller objectives]
\label{3-def:memory_requirements_Muller_objectives}
Let $\Muller(\F)$ be a Muller objective over the set of colours $C$.
We define $m_\F$ by induction:
\begin{itemize}
	\item if the tree consists of a single leaf, then $m_\F = 1$;
	\item otherwise, let $\F_1,\dots,\F_k$ be the families induced by the children of the root,
	there are two cases:
	\begin{itemize}
		\item if $C \in \F$, then $m_\F$ is the \textit{sum} of $m_{\F_1},\dots,m_{\F_k}$;
		\item if $C \notin \F$, then $m_\F$ is the \textit{maximum} of $m_{\F_1},\dots,m_{\F_k}$.
	\end{itemize}
\end{itemize}
\end{definition}

For the Muller objective represented in \Cref{3-fig:Zielonka_tree_example}, we have $m_\F = 3$.

\begin{theorem}[Memory requirements for Muller games]
\label{3-thm:characterisation_Zielonka_tree}
Muller objectives $\Muller(\F)$ are determined with finite memory strategies of size $m_\F$ for Eve.
This bound is tight: there exists a game with objective $\Muller(\F)$ where Eve wins using $m_\F$ memory states
but not with less. Importantly, the lower bounds apply to partial edge-labelled colouring functions.
\end{theorem}

%

We will not construct the lower bound, meaning the game where Eve needs $m_\F$ memory states to win.
However, we will now prove the upper bound.
To this end we revisit the recursive algorithm presented in \Cref{3-lem:Muller_even} and \Cref{3-lem:Muller_odd}.
This algorithm was removing colours one by one and relying on the recursive solutions.
We show that we can adapt the algorithm to follow instead the structure of the Zielonka tree: 
for solving a Muller game, it is enough to recursively solve the induced Muller games
corresponding to the children of the root of the Zielonka tree.
The following lemma is an improved variant of \Cref{3-lem:Muller_even}.
The corresponding pseudocode is given in~\Cref{3-algo:McNaughton_zielonka}.

\begin{lemma}[Fixed-point characterisation of the winning regions for Muller games using the Zielonka tree]
\label{3-lem:McNaughton_Zielonka_even}
Let $\Game$ be a Muller game with objective $\Muller(\F)$ such that $C \in \F$.
Let $C_1, \dots, C_k$ be the maximal subsets of $C$ such that $C_i \notin \F$.
We let $\F_1,\dots,\F_k$ be the corresponding induced families,
and define $\Game_i$ be the subgame of $\Game$ induced by $V \setminus \AttrE(C_i)$
with objective $\Muller(\F_i)$.
\begin{itemize}
	\item If for all $i \in [1,k]$, we have $\WA(\Game_i) = \emptyset$, then $\WE(\Game) = V$.
	\item If there exists $i \in [1,k]$ such that $\WA(\Game_i) \neq \emptyset$,
	let $\Game'$ be the subgame of $\Game$ induced by $V \setminus \AttrA( \WA(\Game_i) )$,
	then $\WE(\Game) = \WE(\Game')$.
\end{itemize}
\end{lemma}

We will prove the memory requirement at the same time inductively.
Note that by duality, the case where $C \notin \F$ corresponds to the memory requirement for Adam when $C \in \F$:
\[
m_{2^C \setminus \F} = \max_{i \in [1,k]} m_{2^{C_i} \setminus \F_i}.
\]

\begin{proof}
We prove the first item.

For each $i \in [1,k]$, let $\sigma_i$ be an attractor strategy ensuring to reach $C_i$ from $\AttrE(C_i)$,
and consider a winning strategy for Eve from $V \setminus \AttrE(C_i)$ in $\Game_i$, it induces a strategy $\sigma'_i$ in $\Game$.
We construct a strategy $\sigma$ in $\Game$ which will simulate the strategies above in turn; to do so it uses $[1,k]$ as top-level memory states.
(We will look at more closely at the memory structure at the end of the proof.)
The strategy $\sigma$ with memory $i$ simulates $\sigma_i$ from $\AttrE(C_i)$ and $\sigma'_i$ from $V \setminus \AttrE(C_i)$,
and if it ever reaches a vertex in $C_i$ it updates its memory state to $i + 1$ and $1$ if $i = k$.
Any play consistent with $\sigma$ either updates its memory state infinitely many times, 
or eventually remains in $V \setminus \AttrE(C_i)$ and is eventually consistent with $\sigma'_i$.
In the first case it sees a colour from each $C_i$ infinitely many times, so by definition of the $C_i$'s and since $C \in \F$ 
the play satisfies $\Muller(\F)$, 
and in the other case since $\sigma'_i$ is winning the play satisfies $\Muller(\F)$.
Thus $\sigma$ is winning from $V$.

Let us now discuss how many memory states are necessary to implement the strategy $\sigma$.
By induction hypothesis, each of the strategies $\sigma'_i$ uses $m_{\F_i}$ memory states.
Using a disjoint union of the memory structures we implement $\sigma$ using $\sum_{i \in [1,k]} m_{\F_i}$ memory states,
corresponding to the definition of $m_\F$.

\vskip1em
We now look at the second item.

Consider a winning strategy for Adam from $\WA(\Game_i)$ in $\Game_i$, it induces a strategy $\tau_i$ in $\Game$.
Thanks to \Cref{1-fact:traps_winning} $\tau_i$ is a winning strategy in $\Game$.
Let $\tau_a$ denote an attractor strategy from $\AttrA(\WA(\Game_i)) \setminus \WA(\Game_i)$.
Consider now a winning strategy in the game $\Game'$ from $\WA(\Game')$, it induces a strategy $\tau'$ in $\Game$.
The set $V \setminus \AttrA( \WA(\Game_i) )$ may not be a trap for Eve, so we cannot conclude that $\tau'$ is a winning strategy in $\Game$,
and it indeed may not be.
We construct a strategy $\tau$ in $\Game$ as the (disjoint) union of the strategy $\tau_a$ on $\AttrA(\WA(\Game_i)) \setminus \WA(\Game_i)$,
the strategy $\tau_i$ on $\WA(\Game_i)$ and the strategy $\tau'$ on $\WA(\Game')$.
We argue that $\tau$ is winning from $\AttrA( \WA(\Game_i) ) \cup \WA(\Game')$ in $\Game$.
Indeed, any play consistent with this strategy in $\Game$ either stays forever in $\WA(\Game')$ hence is consistent with $\tau'$
or enters $\AttrA( \WA(\Game_i) )$, so it is eventually consistent with $\tau_i$.
In both cases this implies that the play is winning.
Thus we have proved that $\AttrA( \WA(\Game_c) ) \cup \WA(\Game') \subseteq \WA(\Game)$.

We now show that $\WE(\Game') \subseteq \WE(\Game)$, which implies the converse inclusion.
Consider a winning strategy from $\WE(\Game')$ in $\Game'$, it induces a strategy $\sigma$ in $\Game$.
Thanks to \Cref{1-fact:traps_winning} $\sigma$ is winning from $\WE(\Game')$ in $\Game$.

Let us now discuss how many memory states are necessary to implement the strategy $\tau$.
By induction hypothesis, the strategy $\tau_i$ uses $m_{2^{C_i} \setminus \F_i}$ memory states
and the strategy $\tau'$ uses $\max_{j \neq i} m_{2^{C_j} \setminus \F_j}$ memory states.
Since $\tau$ is a disjoint union of strategies the memory can be reused so we can implement $\tau$ using $\max_{i \in [1,k]} m_{2^{C_i} \setminus \F_i}$ memory states, corresponding to the definition of $m_{2^C \setminus \F}$.
\end{proof}

The corresponding lemma when $C \notin \F$ is stated below, its proof is analogous to the previous one by swapping the two players.

\begin{lemma}[Dual fixed-point characterisation of the winning regions for Muller games using the Zielonka tree]
\label{3-lem:McNaughton_Zielonka_odd}
Let $\Game$ be a Muller game such that $C \notin \F$.
Let $C_1, \dots, C_k$ be the maximal subsets of $C$ such that $C_i \in \F$.
We let $\F_1,\dots,\F_k$ be the corresponding induced Muller objectives,
and define $\Game_i$ be the subgame of $\Game$ induced by $V \setminus \AttrA(C_i)$ with objective $\Muller(\F_i)$.
\begin{itemize}
	\item If for all $i \in [1,k]$, we have $\WE(\Game_i) = \emptyset$, then $\WA(\Game) = V$.
	\item If there exists $i \in [1,k]$ such that $\WE(\Game_i) \neq \emptyset$,
	let $\Game'$ be the subgame of $\Game$ induced by $V \setminus \AttrE( \WE(\Game_i) )$,
	then $\WA(\Game) = \WA(\Game')$.	
\end{itemize}
\end{lemma}

\begin{algorithm}
 \KwData{A Muller game $\Game$ over $C$}
 \SetKwFunction{FSolveIn}{SolveIn}
 \SetKwFunction{FSolveOut}{SolveOut}
 \SetKwProg{Fn}{Function}{:}{}

\Fn{\FSolveIn{$\Game$}}{ \tcp{Assumes $C \in \F$}
	\If{$C = \set{c}$}{
		\Return{$V$}
	}

	Let $C_1,\dots,C_k$ the labels of the children of the root of the Zielonka tree of $\Muller(\F)$
	
	\For{$i \in [1,k]$}{
		
		$\Game_i \leftarrow \Game \setminus \AttrE^{\Game}(C_i)$

		$\WE(\Game_i) \leftarrow \FSolveOut(\Game_i)$ \tcp{The Zielonka tree of $\Game_i$ is the $i$-th subtree}			
	}

	\If{$\forall i \in [1,k], \WA(\Game_i) = \emptyset$}{
		\Return{$V$}
	}
	\Else{
		Let $i$ such that $\WA(\Game_i) \neq \emptyset$

		$\Game' \leftarrow \Game \setminus \AttrA^{\Game}( \WA(\Game_i) )$
			
		\Return{$\FSolveIn(\Game')$} \tcp{$\Game'$ has less vertices}
	}
}
\vskip1em
\Fn{\FSolveOut{$\Game$}}{
	\tcp{Symmetric to $\FSolveIn$, assumes $C \notin \F$}
}
\vskip1em
\If{$C \in \F$}{
	\FSolveIn{$\Game$}
}
\Else{
	\FSolveOut{$\Game$}
}
\caption{A recursive algorithm for computing the winning regions of Muller games following the Zielonka tree.}
\label{3-algo:McNaughton_zielonka}
\end{algorithm}

\subsection*{Revisiting Streett, Rabin, and parity objectives}
Let us look at the Streett, Rabin, and parity objectives under the new light shed by \Cref{3-thm:characterisation_Zielonka_tree}.
It is instructive to look at the Zielonka tree of a Rabin objective, illustrated in \Cref{3-fig:Zielonka_tree_Rabin}.
It has a simple recursive structure: the Zielonka tree of the Rabin objective for $d$ pairs contains $d$ copies
of the Zielonka tree of the Rabin objective for $d-1$ pairs.
Naturally, this implies that $m_{\Rabin} = 1$, so \Cref{3-thm:characterisation_Zielonka_tree} implies the positional determinacy result
stated in \Cref{3-thm:Rabin_positional_determinacy}.

\begin{figure}
\centering
  \begin{tikzpicture}[scale=1.3]
\node[s-adam] (123) at (2.5,3) {$\ C\ $};
\node[s-eve] (neq1) at (1,2) {$\ C \setminus \set{G_1}\ $};
\node[s-eve] (neq2) at (2.5,2) {$\ C \setminus \set{G_2}\ $};
\node[s-eve] (neq3) at (4,2) {$\ C \setminus \set{G_3}\ $};
\node[s-adam] (23) at (1,1) {$\ C \setminus \set{G_1,R_1}\ $};
\node[s-adam] (13) at (2.5,1) {$\ C \setminus \set{G_2,R_2}\ $};
\node[s-adam] (12) at (4,1) {$\ C \setminus \set{G_3,R_3}\ $};

\node (1) at (1,0.3) {};
\node (2) at (2.5,0.3) {};
\node (3) at (4,0.3) {};

\path
  (123) edge (neq1)
  (123) edge (neq2)
  (123) edge (neq3)
  (neq1) edge (23)
  (neq2) edge (13)
  (neq3) edge (12)
  (23) edge[dashed] (1)
  (13) edge[dashed] (2)
  (12) edge[dashed] (3);
  \end{tikzpicture}
\caption{The (beginning of the) Zielonka tree for $\Rabin$ with three pairs: 
$C = \set{G_1,R_1,G_2,R_2,G_3,R_3}$.}
\label{3-fig:Zielonka_tree_Rabin}
\commentAlt{Figure~\ref{3-fig:Zielonka_tree_Rabin}: A tree diagram showing a hierarchy of set operations, with a root 'C' branching into three sub-branches. See long description.}
\commentLongAlt{Figure~\ref{3-fig:Zielonka_tree_Rabin}: The image depicts a tree structure with a single square root node labeled 'C'. This root node branches downwards into three circular nodes, each representing a set difference: 'C \ {G1}', 'C \ {G2}', and 'C \ {G3}'. Each of these circular nodes then branches further down into a square node, representing a further set difference: 'C \ {G1, R1}', 'C \ {G2, R2}', and 'C \ {G3, R3}' respectively. Below these bottom square nodes, dashed lines indicate that the tree continues, implying further levels or processes.}
\end{figure}

Recall that we defined Streett objectives using closure under union, and Rabin objectives as the complement of Streett objectives.

\begin{theorem}[Positionally determined Muller objectives]
\label{3-thm:characterisation_positionally_determined_Muller_objectives}
Let $\Muller(\F)$ be a Muller objective.
\begin{itemize}
	\item $\Muller(\F)$ is positionally determined if and only if it is a Rabin objective;
	\item $\Muller(\F)$ is bi-positionally determined if and only if it is a parity objective.
\end{itemize}
\end{theorem}
This theorem gives a characterisation of Rabin and parity objectives: they form the class of Muller objectives which are respectively positional and bi-positional.

\begin{proof}
Thanks to \Cref{3-thm:characterisation_Zielonka_tree} the objective $\Muller(\F)$ is positionally determined if and only if $m_\F = 1$, which is equivalent to saying that all nodes labelled $S \in \F$ in the Zielonka tree of $\F$ have at most one child. Indeed, for such nodes the number $m$ is obtained as the sum of the numbers for the children, so there can be at most one, and conversely if this is the case then $m_\F = 1$.
This characterisation of the Zielonka tree is equivalent to the complement of $\F$ being closed under union:
\begin{itemize}
	\item Assume that the complement of $\F$ is closed under union and let $S \in \F$ be a node in the Zielonka tree of $\F$.
	Let $S_1,\dots,S_k$ be the children of $S$, by definition they are the maximal subsets of $S$ such that $S_i \notin \F$.
	The union $\bigcup_i S_i$ is a subset of $S$ and by closure under union of the complement of $\F$ it is in the complement of $\F$, 
	implying by maximality that it is one of the children, so they are all equal and $k = 1$.
	\item Conversely, assume that all nodes labelled $S \in \F$ in the Zielonka tree of $\F$ have at most one child.
	Let $S_1,S_2 \notin \F$, towards contradiction assume that $S_1 \cup S_2 \in \F$.
	By definition of the Zielonka tree, if $S_1 \cup S_2$ is included into a node $S \notin \F$, 
	then $S_1 \cup S_2$ is included into one of its children. 
	Starting from the root and applying this we find a node $S \in \F$ such that $S_1 \cup S_2 \subseteq S$
	and $S_1 \cup S_2 \not\subseteq S'$ with $S'$ the only child of $S$ 
	(the case where $S$ does not have any children is easy and treated separately).
	By definition of the Zielonka tree, since $S_1,S_2 \notin \F$ and $S_1,S_2 \subseteq S$, then $S_1,S_2 \subseteq S'$, implying that 
	$S_1 \cup S_2 \subseteq S'$, a contradiction.
\end{itemize}
We have proved the first equivalence: $\Muller(\F)$ is positionally determined if and only if the complement of $\F$ is closed under union, which is the definition of Rabin objectives.

For the second equivalence, we already have that $\Muller(\F)$ is bi-positionally determined if and only if all nodes in the Zielonka tree of $\F$ have at most one child. The Zielonka tree is in this case a chain:
\[
S_1 \subseteq S_2 \subseteq S_3 \subseteq S_4 \subseteq \cdots \subseteq S_{2d-1} \subseteq S_{2d} \subseteq C,
\]
with $S_{2i} \in \F$ and $S_{2i-1} \notin \F$.
Then $X \in \F$ is equivalent to asking that the largest $i \in [1,d]$ such that if $X \cap S_i \neq \emptyset$ is even.
Assigning priority $i$ to $S_i$ we get that $X \in \Muller(\F)$ if and only if 
the largest priority appearing infinitely many times in $X$ is even: 
this is the definition of the parity objective over the set of priorities $[1,2d]$.
Conversely, we observe that the Zielonka tree of a parity objective is indeed a chain.
\end{proof}


\section*{Bibliographic references}
\label{3-sec:references}

The objectives studied in this chapter are called $\omega$-regular,
let us discuss their relevance in automata theory and logic.
An important application of automata theory is to make logic effective: by translating, sometimes called compiling,
a logical formula into an equivalent automaton, we can solve problems such as satisfiability or model-checking by reducing them
to analysing automata and in particular their underlying graph structures.
In this context, the reachability objective is used for automata over finite words: 
the classical definition is that a run is accepting if the last state is accepting.
Monadic second-order logic over finite words can be effectively translated into finite automata, 
marking the beginning of a close connection between logic and automata theory.

Considering logics over infinite structures led to the study of automata over infinite structures such as words and trees.
The first objective to be studied in this context was B{\"u}chi objective, introduced by B{\"u}chi~\cite{Buchi:1962}: 
a run is accepting if it visits infinitely many times an accepting state.
Unfortunately the class of languages of infinite words recognised by deterministic B{\"u}chi automata is not closed under projection (corresponding in logic to existential quantification), said differently non-deterministic B{\"u}chi automata are strictly more expressive than deterministic ones
hence not equivalent to monadic second-order logic over infinite words.
Muller~\cite{Muller:1963} introduced the Muller objectives and attempted to prove the closure under projection for deterministic Muller automata. Alas, the proof had a flaw.
The first correct proof of the result is due to McNaughton~\cite{McNaughton:1966}.

The correspondence between monadic second-order logic and Muller automata was extended from infinite words to infinite binary trees
by Rabin~\cite{Rabin:1969}, yielding the celebrated decidability of monadic second-order logic over infinite trees.
Rabin introduced and worked with Rabin objectives; his proof is arguably very complicated and a lot of subsequent works focused on finding the right notions and tools for better understanding his approach.
Streett~\cite{Streett:1981} suggested to use the complement of Rabin objectives, now called Streett objectives, for translating temporal logics
to Streett automata.
As discussed in~\Cref{1-sec:references}, a key step was made by applying determinacy results for games
to complementation results for automata.
The parity objectives appeared in this context as a (and in fact, the) subclass of Muller objectives which is bi-positionally determined.
They have been defined (with some variants) independently by several authors: Wagner~\cite{Wagner:1979},
Mostowski~\cite{Mostowski:1984} who called them `Rabin chain', 
Emerson and Jutla~\cite{Emerson.Jutla:1991} who first used the name parity, 
and McNaughton~\cite{McNaughton:1993}.
The idea can be traced back to the `difference hierarchy' by Hausdorff~\cite{Hausdorff:1914}.
The proof of the bi-positionality was obtained independently by Mostowski~\cite{Mostowski:1991}, 
Emerson and Jutla~\cite{Emerson.Jutla:1991}, and McNaughton~\cite{McNaughton:1993} (the latter proof is for finite games).
Later Walukiewicz~\cite{Walukiewicz:2002} gave another very elegant proof.

\vskip1em
McNaughton~\cite{McNaughton:1993} introduced the idea of solving Muller games by induction on the colours,
leading to McNaughton algorithm as presented in~\Cref{3-sec:muller}.
To some extent, the recursive algorithms for solving B{\"u}chi, CoB{\"u}chi, and parity games are all special cases of McNaughton algorithm.

\vskip1em
Taking a step back in time, McNaughton already proposed the \emph{Latest Appearance Record} (LAR) discussed in~\Cref{3-sec:zielonka} 
for solving Muller games in his flawed attempt to solve the synthesis problem~\cite{McNaughton:1965} (see~\Cref{1-sec:references}).
The LAR was later used by Gurevich and Harrington~\cite{Gurevich.Harrington:1982} as memory for winning strategies in Muller games.
Thomas~\cite{Thomas:1995} showed that the LAR can be used to reduce Muller games to parity games.
Zielonka~\cite{Zielonka:1998} greatly contributed to the study of Muller objectives and their subclasses
through his illuminating analysis of Zielonka trees.
The characterisation result showing how Zielonka tree captures the exact memory requirements of Muller objectives is due to 
Dziembowski, Jurdzi{\'n}ski, and Walukiewicz~\cite{Dziembowski.Jurdzinski.ea:1997}.
Casares, Colcombet and Lehtinen~\cite{Casares.Colcombet.ea:2022} showed that the memory requirements of Muller objectives also correspond to the size of minimal good for games Rabin automata.
In 2021, Casares, Colcombet and Fijalkow~\cite{Casares.Colcombet.ea:2021,Casares.Colcombet.ea:2024} proved that the Zielonka tree parity automaton has a minimal number of states and uses a minimal number of priorities among deterministic parity automata recognising a given Muller objective.
They introduced the \emph{Alternating Cycle Decomposition}, a generalisation of the Zielonka tree that represents the connected subsets of a Muller game (or a Muller automaton) over a tree structure.
The Alternating Cycle Decomposition provides an optimal reduction from Muller games to parity games, more precisely, it allows to compute the smallest parity game that admits a well-behaved morphism to a given Muller game.

\vskip1em
The $\NP$-completeness stated in~\Cref{3-thm:Rabin_complexity} for solving Rabin games is due to Emerson and Jutla~\cite{Emerson.Jutla:1988}.
The study of the complexity of solving Muller games is due to Dawar and Hunter~\cite{Hunter.Dawar:2005}.
The $\PSPACE$-completeness results stated in~\Cref{3-thm:complexity_Muller,3-thm:Muller_games_DAG} only concern two representations for Muller objectives. There are several others, which are not equally succinct.
For all representations but one the $\PSPACE$-completeness result holds; the only exception is the explicit representation
where the condition is specified by listing all sets of vertices in $\F$.
Surprisingly, solving Muller games with the explicit representation is in $\P$ as shown by Horn~\cite{Horn:2008}.\footnote{The original paper contains a bug, which is fixed in~\cite{Liang.Khoussainov.ea:2023}.}

\vskip1em
The complexity results stated for Muller games are not optimal, and the state of the art is indeed quite complex: we refer to~\cite{Liang.Khoussainov.ea:2024} for a discussion on this topic.
For Muller games, extending the quasi-polynomial-time algorithm for parity games yields a better algorithm than the one presented here in some regime, as explained in~\cite{Calude.Jain.ea:2017}. An almost matching lower bound (conditional to the Exponential-Time Hypothesis) was proved in the same paper.
However, dynamic programming approaches yield competitive algorithms, see~\cite{Liang.Khoussainov.ea:2024}.
For Rabin games, there are two algorithms whose running times are not comparable: the original Piterman and Pnueli algorithm~\cite{Piterman.Pnueli:2006} and the recently introduced algorithm by Majumdar, Saglam, and Thejaswini~\cite{Majumdar.Saglam.ea:2024}.
An almost matching lower bound (again conditional to the Exponential-Time Hypothesis) was proved in~\cite{Casares.Pilipczuk.ea:2024}.


\ifpictures
\includepdf{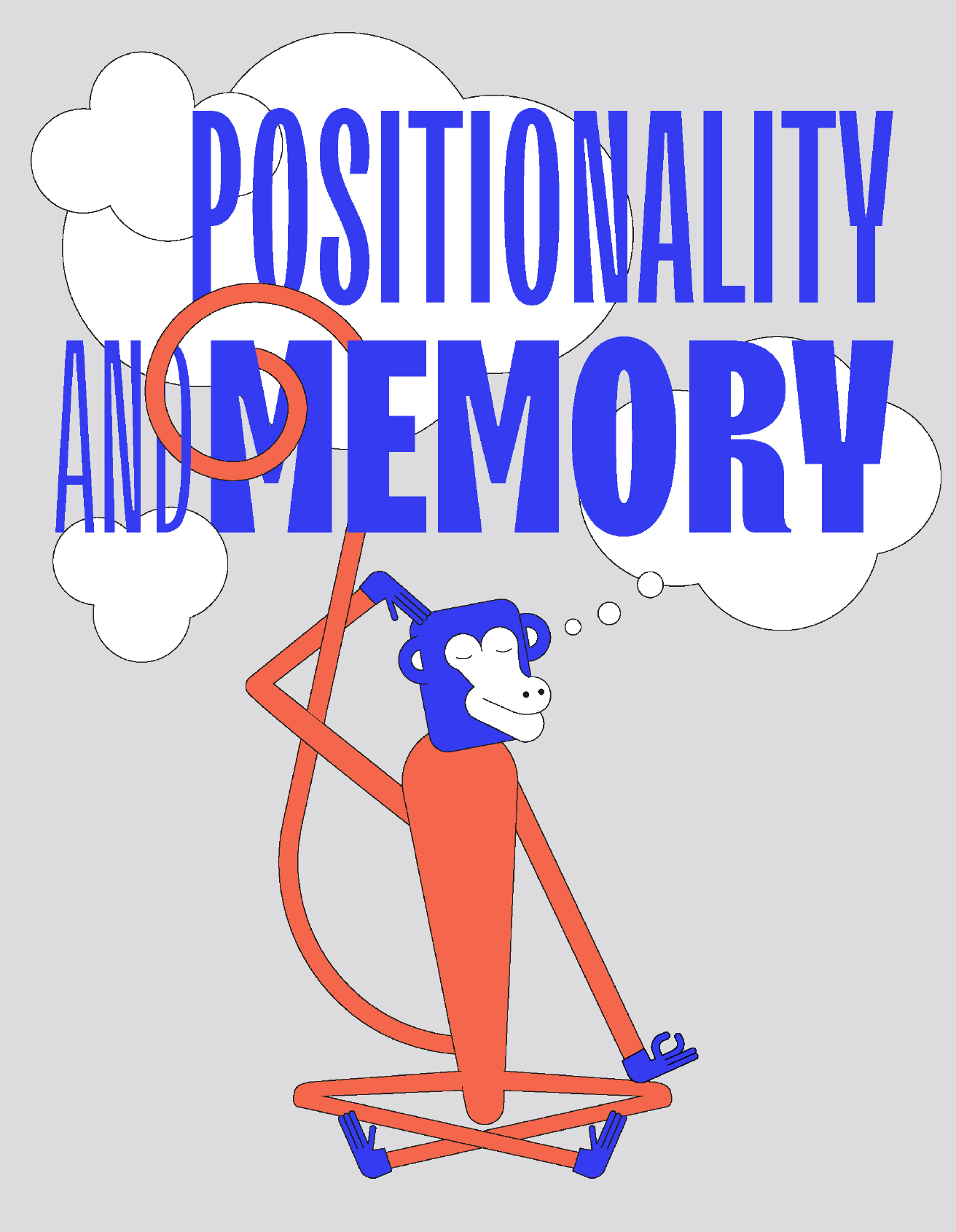}
\fi
\author[Antonio Casares, Pierre Ohlmann, Pierre Vandenhove]{Antonio Casares, Pierre Ohlmann, Pierre Vandenhove}
\copyrightline{Copyright by Antonio Casares, Pierre Ohlmann, and Pierre Vandenhove 2025, to be published by Cambridge University Press in the volume \textit{Games on Graphs} edited by Nathana\"el Fijalkow}

\chapter{Positionality and Memory}
\chapterauthor{Antonio Casares, Pierre Ohlmann, Pierre Vandenhove}
\label{4-chap:memory}

\usetikzlibrary{decorations.pathmorphing}
\tikzset{squish/.style={decorate,decoration={snake,amplitude=.4mm,segment length=4mm,post length=1mm,pre length=0mm}}}
\tikzset{squishloop/.style={decorate,decoration={snake,amplitude=.4mm,segment length=3mm,post length=1mm,pre length=0mm}}}

\newcommand{\re}[1]{\xrightarrow{#1}}

\newcommand{\pos}{\mathrm{pos}}

\newcommand{\PhiGeqx}{\Phi_{\geq x}}
\newcommand{\PhiGeqy}{\Phi_{\geq y}}
\newcommand{\PhiGneqx}{\Phi_{>x}}
\newcommand{\CEven}{C_{\mathrm{Even}}}
\newcommand{\COdd}{C_{\mathrm{Odd}}}
\newcommand{\smallerOdd}{\mathsf{smallerOdd}}

\newcommand{\rkUG}{\mathrm{rk}}

\newcommand{\residualO}[1]{#1^{-1}\Omega}
\newcommand{\automatonResO}{\mathbf{R}_\Omega}
\newcommand{\prefixClassifier}{\automatonResO}
\newcommand{\resSet}{\mathrm{Res}}
\newcommand{\resSetO}{\mathrm{Res}(\Omega)}

\newcommand{\Finite}{\mathrm{Finite}}
\newcommand{\Decreasing}{\mathrm{Decreasing}}
\newcommand{\MaxBounded}{\mathtt{MaxBounded}}
\newcommand{\First}{\mathtt{First}}
\newcommand{\TiltedEnergy}{\mathtt{TiltedEnergy}}
\newcommand{\Safety}{\mathtt{Safety}}

\newcommand{\nextmove}{\hat{\sigma}}
\newcommand{\genMemReq}{\mathrm{mem}_{\mathrm{gen}}}
\newcommand{\chromMemReq}{\mathrm{mem}_{\mathrm{chrm}}}
\newcommand{\arIndMemReq}{\mathrm{mem}_{\mathrm{arInd}}}
\newcommand{\genMemReqFin}{\mathrm{mem}^{\mathrm{fin}}_{\mathrm{gen}}}
\newcommand{\chromMemReqFin}{\mathrm{mem}^{\mathrm{fin}}_{\mathrm{chrm}}}
\newcommand{\arIndMemReqFin}{\mathrm{mem}^{\mathrm{fin}}_{\mathrm{arInd}}}

In this chapter, we study the subject of \emph{strategy complexity} for games, focusing on the notion of positionality.
We have already shown positionality of some objectives, such as $\Reach$ or $\Parity$.
The objective of this chapter is twofold.
First, we provide tools to show positionality of objectives, either over finite or over infinite arenas; in particular, we establish the positionality of many objectives from other chapters. 
Second, we aim to provide a comprehensive study of the subject of positionality in itself, establishing both necessary and sufficient conditions for an arbitrary objective to be {(bi-)}positional.

The chapter is organised as follows.

\begin{itemize}
	\item \Cref{4-sec:fundamental_positional} focuses on sufficient conditions for (bi-)positionality over finite arenas.
	One of the main results presented is a \emph{one-to-two-player lift}: in order to verify whether an objective is "bi-positional" over finite arenas, it suffices to check whether both players can play optimally in a positional way in finite one-player arenas.
	As a consequence of the results from this section, we establish the positionality over finite arenas of many of the classical objectives defined in \Cref{1-chap:introduction}, such as $\Parity$, $\Rabin$, $\MeanPayoff$, $\Energy$, and $\TotalPayoff$.
	
	\item In \Cref{4-sec:positional_infinite}, we provide a characterisation of "bi-positional" qualitative objectives over infinite arenas; in the case of "prefix-independent" objectives, these exactly correspond to "parity objectives".
	As a corollary, we obtain a one-to-two-player lift over infinite arenas. The techniques presented in this section focus on necessary conditions for positionality.
	
	\item In \Cref{4-sec:quantitative}, we discuss several natural definitions of positionality in the case of quantitative objectives. We show that they yield equivalent notions over finite arenas, and show how to reduce their study to the case of qualitative objectives.
	
	\item In \Cref{4-sec:positionality}, we focus on the study of positionality over infinite arenas, and present a characterisation of it via \emph{monotone universal graphs}.
	As a result, we establish the positionality over infinite arenas of $\MeanPayoff^-_{>0}$ and $\DiscountedPayoff_\lambda$.
	In \Cref{4-subsec:Kopcz_conj}, we discuss Kopczy\'nski's conjecture about the closure under union of prefix-independent positional objectives, and we present in \Cref{4-subsec:positional_omega-reg} results about "positionality" for the class of $\omega$-regular objectives.
	
	\item In \Cref{4-sec:memory_generalisations}, we introduce several models of memories for games and provide examples separating the different notions.
	We discuss how the results from the previous sections can be generalised to strategies with memory.	
\end{itemize}

The first four sections of this chapter can be read independently, and all give various kinds of tools to better approach positionality in different contexts.


\section{Fundamental positionality results over finite arenas}
\label{4-sec:fundamental_positional}
We aim here to understand some common underlying properties of objectives that make them "(bi-)positional" over finite "arenas".
Results from this section are meant to be widely applicable and easy to use.
We will focus on two results, which are sufficient to prove "(bi-)positionality" of most standard qualitative and "quantitative objectives".

The first result we present is a sufficient (but not necessary) property for "uniform positionality" over finite "arenas" of "prefix-independent" objectives.
The second one is a characterisation of "uniform bi-positionality" over finite "arenas", reducing the problem to the simpler study of "\emph{one-player} arenas".

\subsection{Positionality of submixing objectives}
We introduce one property of objectives which, along with "prefix-independence", entails "uniform positionality".

\begin{definition}[Submixing objective]
	An objective $\Phi\colon C^\omega \to \Rinfty$ is \emph{submixing} if for all sequences of non-empty finite words $\rho_1^0, \rho_1^1, \ldots \in C^+$ and $\rho_2^0, \rho_2^1, \ldots \in C^+$, we have
	\[
	\Phi(\rho_1^0\rho_2^0\rho_1^1\rho_2^1\dots)\leq \max \left\{\Phi(\rho_1^0\rho_1^1\dots), \;
	\Phi(\rho_2^0\rho_2^1\dots)\right\}.\]
\end{definition}

The "submixing" property means that two infinite words cannot be combined into a better infinite word (where ``better'' is with respect to the objective and to player Max) by intertwining them.
The "submixing" property can hold for an objective $\Phi$ but not its opposite $-\Phi$.
It can be rephrased as follows for "qualitative objectives" $\Omega\subseteq C^\omega$: if $\rho_1^0\rho_1^1\ldots \notin \Omega$ and $\rho_2^0\rho_2^1\ldots \notin \Omega$, then $\rho_1^0\rho_2^0\rho_1^1\rho_2^1\ldots \notin \Omega$.
Observe that it corresponds to the property of Rabin objectives highlighted in \eqref{3-eq:submixing_early} in \Cref{3-chap:regular}.

In the upcoming result, we also assume determinacy of the objective, which is a reasonable requirement thanks to \Cref{1-cor:borel_determinacy}.

\begin{theorem}["Prefix-independent" and "submixing" imply uniform "positionality"]
	\label{4-thm:submixing_positional}
	Every determined, "prefix-independent", and "submixing" objective is uniformly "positional" over finite "arenas".
\end{theorem}

\begin{proof}
	Let $\Phi\colon C^\omega \to \Rinfty$ be a determined, "prefix-independent", and "submixing" "quantitative objective".
	We show the result by induction over the following quantity of finite "arenas": the total out-degree of vertices controlled by Max, minus the number of vertices controlled by Max.
	For an "arena" $\arena = (G, \VMax, \VMin)$, this is the quantity
	$
		n(\arena) = \left(\sum_{v\in\VMax} d(v)\right) - |\VMax|,
	$
	where $d(v)$ is the out-degree of vertex $v$.
	Our base case occurs when $n(\arena)$ is $0$.
	Since we assume that every vertex has an outgoing edge, the base case corresponds to the situation in which each vertex of Max has exactly one outgoing edge.
	In this case, Max has only one strategy, which is positional and optimal.

	We move on to the inductive step.
	Let $\game = (\arena, \Phi(\col))$ be a game on a finite "arena"~$\arena$ such that $n(\arena) > 0$.
	Hence, there is an element $v \in \VMax$ with out-degree at least $2$; we pick such a $v$.
	We partition the outgoing edges of $v$ in two non-empty subsets which we call $E^v_1$ and~$E^v_2$.
	We define two games $\game_1$ and $\game_2$: the game $\game_1$ is obtained from $\game$ by removing the edges from $E^v_2$, and symmetrically for $\game_2$.
	For $i\in\set{1, 2}$, if $\arena_i$ is the "arena" on which $\game_i$ is played, observe that $n(\arena_i) < n(\arena)$.

	W.l.o.g., we assume that $\val^{\game_1}(v) \ge \val^{\game_2}(v)$.
	By induction hypothesis, we take $\sigma_1$ a positional optimal strategy of Max in $\game_1$.
	Observe that $\val^{\game_1, \sigma_1}(v) = \val^{\game_1}(v)$ (by optimality of~$\sigma_1$) and that $\val^{\game_1}(v) \le \val^{\game}(v)$ (as every strategy of Max in $\game_1$ can be seen as a strategy in $\game$, and Min has no additional power in $\game$).%
	\footnote{For convenience, we will sometimes consider the same positional strategy (here, $\sigma_1$) in different games over the same state space.
		We therefore extend notation $\val$ and write \textit{e.g.}\ $\val^{\game_1, \sigma_1}$ to avoid any ambiguity on the game being considered.}
	We show that we even have $\val^{\game_1}(v) = \val^{\game}(v)$; to do so, it is left to show that $\val^{\game}(v) \le \val^{\game_1}(v)$.
	This implies that the positional strategy $\sigma_1$, seen as a strategy on $\game$, is also optimal from $v$ in $\game$.

	For some $\epsilon > 0$, let $\tau_1$ and $\tau_2$ be strategies of Min ensuring a payoff of at most $\val^{\game_1}(v) + \epsilon$ from $v$ respectively in $\game_1$ and $\game_2$ ($\tau_1$ and $\tau_2$ exist as we assumed that $\Phi$ is determined, and that $\val^{\game_2}(v) \le \val^{\game_1}(v)$).
	We build a strategy $\tau$ of Min in $\game$ (defined only from $v$) based on $\tau_1$ and $\tau_2$.
	It has two modes, $1$ and $2$.
	The strategy simulates $\tau_1$ from the mode $1$ and $\tau_2$ from the mode $2$.
	Whenever $v$ is visited, the mode is updated by observing which outgoing edge is taken by Max: if the outgoing edge is in $E^v_1$, the new mode is $1$; otherwise, it is $2$.
	When simulating $\tau_1$, we ignore the parts of the play using mode $2$, so removing them yields a play consistent with $\tau_1$.
	The same goes for~$\tau_2$.

	Consider a play $\play$ from $v$ consistent with $\tau$.
	It can be decomposed following which mode the play is in.
	Play $\play$ looks like
	\[
	\begin{array}{ccccccccc}
		\text{mode } 1 & \overbrace{v \dots v}^{\play_1^0} & &
		\overbrace{v \dots v}^{\play_1^1} & & \ \cdots \\
		\text{mode } 2 && \underbrace{v \dots v}_{\play_2^0}
		& & \underbrace{v \dots v}_{\play_2^1} & \ \cdots
	\end{array}
	\]
	where $\play_1 = \play_1^0 \play_1^1 \dots$ is consistent with $\tau_1$ and $\play_2 = \play_2^0 \play_2^1 \dots$ is consistent with $\tau_2$ (in the graphical representation above, we assumed that Max starts with an edge in $E_1^v$, but this has no bearing on the proof).

	There are two cases: either exactly one of $\play_1$ and $\play_2$ is finite, or $\play_1$ and $\play_2$ are both infinite.
	The latter case happens if and only if $v$ is visited infinitely often along $\play$ and edges from both $E_1^v$ and $E_2^v$ are taken infinitely often.
	
	In the former case, if (w.l.o.g.) $\play_1$ is finite, then we have
	\begin{align*}
		\Phi(\play)
		&= \Phi(\play_2) &&\text{by "prefix-independence" of $\Phi$}\\
		&\le \val^{\game_1}(v) + \epsilon &&\text{as $\play_2$ is consistent with $\tau_2$}.
	\end{align*}
	If $\play_1$ and $\play_2$ are both infinite, then we have
	\begin{align*}
		\Phi(\play)
		&\le \max\set{\Phi(\play_1), \Phi(\play_2)} &&\text{as $\Phi$ is submixing}\\
		&\le \val^{\game_1}(v) + \epsilon &&\text{as $\play_1$ (resp.\ $\play_2$) is consistent with $\tau_1$ (resp.\ $\tau_2$)}.
	\end{align*}
	Therefore, $\val^{\game}(v) \le \val^{\game_1}(v) + \epsilon$ for all $\epsilon > 0$, so $\val^{\game}(v) \le \val^{\game_1}(v)$.
	This ends the proof that $\sigma_1$ is optimal from $v$ in $\game$.

	We now show that $\sigma_1$ is optimal from any other vertex as well.
	Let $v_0\in V$ be an initial vertex.
	We again have that $\val^{\game_1}(v_0) \le \val^\game(v_0)$ by construction; to get the equality, we show that $\val^\game(v_0) \le \val^{\game_1}(v_0)$.
	As $\val^{\game_1}(v_0) = \val^{\game_1, \sigma_1}(v_0) = \val^{\game, \sigma_1}(v_0)$, this will show that $\sigma_1$ is also optimal from $v_0$ in $\game$.

	Let $\sigma$ be any strategy of Max in $\game$.
	We modify $\sigma$ into another strategy $f(\sigma)$ in the following way: on every finite play $\play$, $f(\sigma)$ plays as $\sigma$ if $v$ is not (yet) visited along~$\play$, and plays as $\sigma_1$ if $v$ was already visited.
	We have that $\val^{\game, \sigma}(v_0) \le \val^{\game, f(\sigma)}(v_0)$.
	Indeed, any path consistent with $f(\sigma)$ that does not go through $v$ is also consistent with~$\sigma$, and switching to $\sigma_1$ when in $v$ guarantees a value at least as high (by optimality of $\sigma_1$ from $v$, and using "prefix-independence" to ignore the path from $v_0$ to $v$).
	Observe that $f(\sigma)$ can be seen as a strategy in $\game_1$, as no edge of $E_2^v$ is ever taken.
	Hence,
	\[
		\val^\game(v_0)
		= \sup_\sigma \val^{\game, \sigma}(v_0)
		\le \sup_{\sigma} \val^{\game, f(\sigma)}(v_0)
		= \sup_{\sigma} \val^{\game_1, f(\sigma)}(v_0)
		\le \val^{\game_1}(v_0).
	\]
	This ends the proof that the positional strategy $\sigma_1$ is optimal in $\game$.
\end{proof}

We now give a few examples of objectives to which this result provides an easy proof of positionality.

\begin{theorem} \label{4-thm:submixing_applications}
	The $\Buchi$, $\CoBuchi$, $\Parity$, $\Rabin$, $\MeanPayoff^+$, and $\Energy_{\ge +\infty}$ objectives are all "prefix-independent" and "submixing".
	In particular, thanks to \Cref{4-thm:submixing_positional}, they are "uniformly positional" over finite "arenas".
\end{theorem}

\begin{proof}[Partial proof]
	The fact that the Rabin objective is "prefix-independent" and submixing was already discussed as part of the proof of \Cref{3-thm:Rabin_positional_determinacy}.

	We now give arguments for the $\MeanPayoff^+$ objective.
	"Prefix-independence" of the $\MeanPayoff^+$ objective will be discussed again in \Cref{5-lem:prefix_independence_mean_payoff}; we focus here on showing that it is submixing.
	Let $\rho_1^0, \rho_1^1, \ldots$ and $\rho_2^0, \rho_2^1, \ldots$ be sequences of words in~$\Z^+$.
	Let $\rho_1 = \rho_1^0\rho_1^1\ldots$, $\rho_2 = \rho_2^0\rho_2^1\ldots$, and $\rho = \rho_1^0\rho_2^0\rho_1^1\rho_2^1\ldots$.
	We rename each colour sequentially in all three sequences as $\rho = c^0c^1\ldots$, $\rho_1 = c_1^0c_1^1\ldots$, and $\rho_2 = c_2^0c_2^1\ldots$, where all $c^i$, $c_1^i$, and $c_2^i$ are in $\Z$.
	For $k\in\N$ and $j\in\set{1, 2}$, let $n_j^k$ be the number of indices originating from $\rho_j$ among the first $k$ colours of $\rho$.
	Observe that $n_1^k + n_2^k = k$ for all $k\in\N$.
	For $k\in\N$, we have
	\begin{align*}
		\frac{1}{k}\sum_{i = 0}^{k - 1} c^i
		&= \frac{1}{k} \sum_{i = 0}^{n_1^k} c_{1}^i + \frac{1}{k} \sum_{i = 0}^{n_2^k} c_{2}^i \\
		&= \frac{n_1^k}{k} (\frac{1}{n_1^k}\sum_{i = 0}^{n_1^k} c_{1}^i) + \frac{n_2^k}{k} (\frac{1}{n_2^k} \sum_{i = 0}^{n_2^k} c_{2}^i) \\
		&\le \max\set{\frac{1}{n_1^k}\sum_{i = 0}^{n_1^k} c_{1}^i, \frac{1}{n_2^k} \sum_{i = 0}^{n_2^k} c_2^i},
	\end{align*}
	where the last inequality holds because the previous line is a convex combination of the two values.
	Taking the $\limsup$ as $k\to\infty$ of the first and the last expression, we obtain $\MeanPayoff^+(\rho) \le \max\set{\MeanPayoff^+(\rho_1), \MeanPayoff^+(\rho_2)}$.
\end{proof}

Note that $\Buchi$, $\CoBuchi$, $\Parity$, and $\Rabin$ objectives are even "positional" over \emph{infinite} "arenas"; this was shown using more specific techniques in \Cref{3-thm:characterisation_Zielonka_tree}.
However, positionality over infinite arenas does not hold for the $\MeanPayoff^+$ objective (see~\cite[Example~8.10.2]{Puterman:2005}).
In particular, there are "prefix-independent" "submixing" objectives that are not "positional" over \emph{infinite} "arenas".
There is therefore no hope of extending \Cref{4-thm:submixing_positional} to infinite "arenas" as is.

\begin{remark} \label{4-remark:mean_payoff_inf_not_submixing}
	Perhaps surprisingly, the dual "mean-payoff" objective $\MeanPayoff^-$ is not submixing (even if it is uniformly positional over finite "arenas", as we will see later in \Cref{4-thm:lift_applications}).
	To see it, take a sequence in $\set{0, 1}^\omega$ whose mean payoff oscillates between values close to $0$ and $1$: a way to build such a sequence is to have long stretches with only $0$'s or $1$'s, the lengths of which increase sufficiently fast.
	Its $\MeanPayoff^-$ is $0$.
	Now if you take the ``opposite'' sequence (inverting $1$'s and $0$'s), you get a sequence that also has a $\MeanPayoff^-$ of $0$, but that can be intertwined with the first into the sequence $(10)^\omega$, which has a better $\MeanPayoff^-$ of $\frac{1}{2}$.
	Hence, $\MeanPayoff^-$ is not submixing.
	In the above proof for $\MeanPayoff^+$, the argument that would fail for $\MeanPayoff^-$ is the application of $\liminf$ over $\max$ (unlike for $\limsup$, it does not hold that $\liminf_k \max\set{a_k, b_k} = \max\set{\liminf_k a_k, \liminf_k b_k}$ in general).
\end{remark}

\subsection{Reduction of bi-positionality to one-player arenas}
We show a second useful tool to establish the existence of positional optimal strategies.
This time, the result is about uniform \emph{bi-positionality} (\textit{i.e.}, positionality of both an objective~$\Phi$ and its opposite objective~$-\Phi$).
It reduces uniform bi-positionality over finite "arenas" to uniform bi-positionality over finite "\emph{one-player} arenas" of both players.
A \emph{one-player arena} is simply an "arena" in which the same player controls all vertices.
There are two kinds: an "arena" $(G, \VMax, \VMin)$ is a "one-player arena" of Max if $\VMin = \emptyset$, and is a "one-player arena" of Min if $\VMax = \emptyset$.

\begin{theorem}[One-to-two-player lift over finite arenas]
	\label{4-thm:1to2_lift}
	Every objective that is "uniformly bi-positional" over finite "one-player arenas" of both players is "uniformly bi-positional" over all finite "arenas".
\end{theorem}

We emphasise that the hypothesis of this result actually makes a requirement about both players: it requires that in all finite "one-player arenas" of both Eve/Max and Adam/Min, there exists a positional optimal strategy.
Having this hypothesis about both players is necessary to obtain a general result: there exist objectives for which Max (but not Min!) has positional optimal strategies in her finite one-player arenas, but for which memory is required in some finite two-player arenas (see, \textit{e.g.}, the example in~\cite[Proposition~2]{Kopczynski:2006}).
Note that unlike for \Cref{4-thm:submixing_positional}, the proof of \Cref{4-thm:1to2_lift} will not use the determinacy of the objective.

The point of \Cref{4-thm:1to2_lift} is that "one-player arenas" are usually easier to reason with: they are essentially graphs, in which no quantification over strategies of the opponent must be made.
A positional strategy from a given vertex in a finite "one-player arena" always induces an ultimately periodic play $\play_1\play_2^\omega$ (sometimes called a ``lasso'') in which~$\play_1$ is a path and $\play_2$ is a cycle, and both are \emph{simple} in the graph-theoretic sense (they do not go through the same vertex twice).
Proving that an objective is "bi-positional" over finite "one-player arenas" therefore reduces to showing that whenever there is an arbitrarily complex winning play for a player, there is also a winning play that is a ``simple lasso'' as described above.
Be careful that the hypothesis requires \emph{uniform} bi-positionality: uniformity requires no additional argument for "prefix-independent" objectives thanks to \Cref{1-lem:from_positional_to_uniformly_positional}, but needs to be shown in general.

The proof of \Cref{4-thm:1to2_lift} we will present shares many similarities with the one of \Cref{4-thm:submixing_positional}: it also proceeds by induction on the number of outgoing edges of "arenas", and the construction of a strategy of the opponent will be reminiscent of the previous proof.
However, it is more involved; one reason is that, in a given arena, we will need to show the existence of a positional optimal strategy for both players (not only for Max).
To this end, we first prove a sufficient condition for the optimality of two strategies.

Let $\game = (\arena, f)$ be a quantitative game and $v$ be a vertex of $\arena$.
We say that a strategy $\sigma$ of Max is a \emph{best response} to a strategy $\tau$ of Min from $v$ if $\val^\tau(v) = f(\play^v_{\sigma, \tau})$.
This means that $\sigma$ attains the supremum in the expression $\sup_{\sigma'} f(\play^v_{\sigma', \tau})$.
We can define symmetrically a best response of Min to a strategy $\sigma$ of Max.
A \emph{pair of best responses} from a vertex $v$ is a pair of strategies $(\sigma, \tau)$ such that $\sigma$ is a best response of Max to $\tau$ from $v$ and $\tau$ is a best response of Min to $\sigma$ from $v$.
We prove two properties of pairs of best responses: $(i)$~strategies that are part of such a pair are optimal, and $(ii)$~if strategies $\sigma$ of Max and $\tau$ of Min are each part of a pair of best responses, then $(\sigma, \tau)$ is also a pair of best responses.
This notion of \emph{best response} will be reused in a more general (non-zero-sum) setting in \Cref{15-chap:multiplayer}, but the following two lemmas need the zero-sum assumption.

\begin{lemma} \label{4-lem:best_response_pair_are_optimal}
	Let $(\sigma, \tau)$ be a pair of best responses from a vertex $v$.
	Then, $\sigma$ and $\tau$ are both optimal from $v$.
\end{lemma}
\begin{proof}
	We prove that $\sigma$ is optimal from $v$; the proof for $\tau$ is symmetric.
	Let $\sigma'$ be any strategy of Max.
	We show that $\val^{\sigma}(v) \ge \val^{\sigma'}(v)$.
	Indeed, we have
	\begin{align*}
		\val^\sigma(v)
		&= \inf_{\tau'} f(\play^v_{\sigma, \tau'}) \\
		&= f(\play^v_{\sigma, \tau}) &&\text{as $\tau$ is a best response to $\sigma$ from $v$} \\
		&\ge f(\play^v_{\sigma', \tau}) &&\text{as $\sigma$ is a best response to $\tau$ from $v$} \\
		&\ge \inf_{\tau'} f(\play^v_{\sigma', \tau'}) \\
		&= \val^{\sigma'}(v).
	\end{align*}
	This shows the optimality of $\sigma$ from $v$.
\end{proof}

\begin{lemma} \label{4-lem:best_responses_mix}
	Let $(\sigma, \tau')$ and $(\sigma', \tau)$ be two pairs of best responses from a vertex $v$.
	Then,~$(\sigma, \tau)$ is also a pair of best responses from $v$.
	In other words, the set of pairs of best responses is a Cartesian product.
\end{lemma}
\begin{proof}
	We prove that $\sigma$ is a best response to $\tau$ from $v$; the other direction is symmetric.
	Let $\sigma''$ be any strategy of Max.
	We show that $f(\play^v_{\sigma'', \tau}) \le f(\play^v_{\sigma, \tau})$.
	Indeed, we have
	\begin{align*}
		f(\play^v_{\sigma'', \tau})
		&\le f(\play^v_{\sigma', \tau}) &&\text{as $\sigma'$ is a best response to $\tau$}\\
		&\le f(\play^v_{\sigma', \tau'}) &&\text{as $\tau$ is a best response to $\sigma'$}\\
		&\le f(\play^v_{\sigma, \tau'}) &&\text{as $\sigma$ is a best response to $\tau'$}\\
		&\le f(\play^v_{\sigma, \tau}) &&\text{as $\tau'$ is a best response to $\sigma$}.
	\end{align*}
	This shows that $\sigma$ is a best response to $\tau$ from $v$.
\end{proof}

We can now prove \Cref{4-thm:1to2_lift}.
\begin{proof}[Proof of \Cref{4-thm:1to2_lift}]
	Let $\Phi\colon C^\omega \to \Rinfty$ be a "quantitative objective" that is "uniformly bi-positional" over finite "one-player arenas" of both players.
	Our proof shows that for every arena, there is a pair of positional strategies $(\sigma, \tau)$ that is a pair of best responses from every vertex.
	In particular, we obtain that $\sigma$ and $\tau$ are positional optimal strategies using \Cref{4-lem:best_response_pair_are_optimal}, which is what we want to show.

	Let $\game = (\arena, \Phi(\col))$ be a game played on the finite "arena" $\arena = (G, \VMax, \VMin)$, where $V = \VMax \cup \VMin$.
	We proceed by induction on the following quantity of "arenas":
	$
		n'(\arena) = \left(\sum_{v\in V} d(v)\right) - |V|,
	$
	where $d(v)$ is the out-degree of vertex $v$.
	As every vertex has at least one outgoing edge, the equality $n'(\arena) = 0$ means that every vertex has exactly one outgoing edge.
	In this case, both players have only one strategy (which is positional), so this pair of positional strategies is a pair of best responses.
	This constitutes the base case of the induction.

	We now assume that $n'(\arena) > 0$.
	We need to show the existence of a pair of positional strategies $(\sigma, \tau)$ that is a pair of best responses.
	We will focus on player Max and prove the following claim: if there is a vertex $v\in\VMax$ with at least two outgoing edges, then there exists a pair of best responses $(\sigma, \tau')$ such that $\sigma$ (but not necessarily~$\tau'$) is positional.
	A symmetric argument would show that if there is a vertex $v\in\VMin$ with at least two outgoing edges, then there is a pair of best responses $(\sigma', \tau)$ with $\tau$ positional.
	We show that this suffices to conclude.
	There are two cases to consider.
	\begin{itemize}
		\item If there exists a vertex with two outgoing edges both in $\VMax$ and in $\VMin$, then there is a pair of best responses $(\sigma, \tau')$ with $\sigma$ positional and there is a pair of best responses $(\sigma', \tau)$ with $\tau$ positional.
		We conclude using \Cref{4-lem:best_responses_mix} that $(\sigma, \tau)$ is a pair of best responses where both $\sigma$ and $\tau$ are positional.
		\item If $\VMin$ contains no such vertex, as $n'(\arena) > 0$, then $\VMax$ contains one.
		So there is a pair of best responses $(\sigma, \tau')$ such that $\sigma$ is positional.
		Observe that when~$\VMin$ contains no such vertex, Min has a single strategy which is positional, so~$\tau'$ is necessarily positional in the above pair.
		The case ``$\VMax$ contains no such vertex'' is symmetric.
	\end{itemize}
	The rest of the proof is devoted to the existence of a pair of best responses $(\sigma, \tau')$ such that $\sigma$ is positional when there is a vertex in $\VMax$ with at least two outgoing edges.

	We pick such a $v\in\VMax$ with at least two outgoing edges.
	We partition the outgoing edges of $v$ in two non-empty subsets which we call $E^v_1$ and~$E^v_2$.
	We define two games $\game_1$ and $\game_2$: the game $\game_1$ is obtained from $\game$ by removing the edges from $E^v_2$, and symmetrically for $\game_2$.
	For $i\in\set{1, 2}$, setting $\arena_i$ as the "arena" on which $\game_i$ is played, observe that $n'(\arena_i) < n'(\arena)$.

	Using the induction hypothesis, we obtain the existence of two pairs of positional best responses: $(\sigma_1, \tau_1)$ in $\game_1$ and $(\sigma_2, \tau_2)$ in $\game_2$.
	We will show that either $\sigma_1$ or $\sigma_2$, when viewed as a strategy on $\game$, is optimal in $\game$.
	To do so, we construct a "\emph{one-player} arena" $\widehat{\arena}$ of Max using $\arena$, $\tau_1$, and $\tau_2$ to which we will apply our hypothesis.

	We define $\widehat{\arena}$ as follows: we make two copies of $\arena$ (one labelled~$1$ and the other labelled~$2$), which we merge on vertex $v$.
	This means that every vertex $u$ of $\arena$ has two copies called $u_1$ and $u_2$, except for $v$ which has a single (unlabelled) copy.
	Edges in~$E^v_1$ are directed towards their original target in copy $1$, while edges in $E^v_2$ are directed towards copy $2$.
	We give the control of all vertices to Max, but we only keep a single outgoing edge of vertices previously controlled by Min: in copy $1$, we follow strategy~$\tau_1$, and in copy $2$, we follow strategy $\tau_2$.
	We illustrate this construction in \Cref{4-fig:lift_split}.

	\begin{figure}
		\centering
		\begin{tikzpicture}[scale=1.2,rotate=-90]
			\node (A) at (0,-1.2) {$\arena$};
			\node[s-eve] (v) at (0,0) {$v$};
			\node (v1) at (1.2,0.6) {};
			\node (v2) at (1,1.2) {};
			\node (v3) at (1,-1) {};

			\node[s-adam] (u) at (2.4,.4) {$u$};
			\node (u1) at ($(u)+(1,0.3)$) {};
			\node (u3) at ($(u)+(1.1,0)$) {};
			\node (u2) at ($(u)+(1,-0.3)$) {};

			\node[s-adam] (u') at (2.4,-.4) {$u'$};
			\node (u'1) at ($(u')+(-1,0.3)$) {};
			\node (u'2) at ($(u')+(-1,-0.3)$) {};

			\node[s-eve,draw=none,fill=none] (v0) at (4,0) {};

			\path let
			\p1 = ($(v)-(v0)$),
			\n1 = {veclen(\p1)}
			in
			(v0) -- (v)
			node[midway, sloped, draw, ellipse,
			minimum width=\n1*.9, minimum height=\n1*1.45] {};

			\path[arrow]
			(v) edge (v1)
			(v) edge (v2)
			(v) edge (v3)

			(u) edge node[right] {$\tau_1(u)$} (u1)
			(u) edge node[left] {$\tau_2(u)$} (u2)
			(u) edge (u3)

			(u') edge node[right] {$\tau_2(u')$} (u'1)
			(u') edge node[left] {$\tau_1(u')$} (u'2)
			;
		\end{tikzpicture}%
		\begin{tikzpicture}[scale=1.2,rotate=-90]
			\node (A) at (-.5,-1.2) {$\widehat{\arena}$};
			\node[s-eve] (v) at (0,0) {$v$};
			\node (v1) at (1.2,0.6) {};
			\node (v2) at (1,1.2) {};
			\node (v3) at (1,-1) {};

			\begin{scope}[rotate=-35]
				\node[s-eve] (u1) at (2.4,.4) {$u_1$};
				\node[s-eve] (u'1) at (2.4,-.4) {$u'_1$};
				\node (u11) at ($(u1)+(1,0.2)$) {};
				\node (u'21) at ($(u'1)+(-1,-0.3)$) {};
				\node[s-eve,draw=none,fill=none] (v01) at (4,-0) {};
			\end{scope}

			\begin{scope}[rotate=35]
				\node[s-eve] (u2) at (2.4,.4) {$u_2$};
				\node[s-eve] (u'2) at (2.4,-.4) {$u'_2$};
				\node (u22) at ($(u2)+(1,-0.4)$) {};
				\node (u'12) at ($(u'2)+(-1,0.3)$) {};
				\node[s-eve,draw=none,fill=none] (v02) at (4,0) {};
			\end{scope}

			\path let
			\p1 = ($(v)-(v01)$),
			\n1 = {veclen(\p1)}
			in
			(v01) -- (v)
			node[midway, sloped, draw, ellipse,
			minimum width=\n1*.65, minimum height=\n1*1.45] {};
			\path let
			\p1 = ($(v)-(v02)$),
			\n1 = {veclen(\p1)}
			in
			(v02) -- (v)
			node[midway, sloped, draw, ellipse,
			minimum width=\n1*.65, minimum height=\n1*1.45] {};

			\path[arrow]
			(v) edge (v1)
			(v) edge (v2)
			(v) edge (v3)

			(u1) edge node[above left=-3pt] {$\tau_1(u)$} (u11)
			(u2) edge node[above right=-3pt] {$\tau_2(u)$} (u22)

			(u'2) edge node[below left=-3pt] {$\tau_2(u')$} (u'12)
			(u'1) edge node[below right=-3pt] {$\tau_1(u')$} (u'21)
			;
		\end{tikzpicture}
		\caption{Construction of $\widehat{\arena}$ (right) from $\arena$ (left) in the proof of \Cref{4-thm:1to2_lift}.
		We assume that $E^v_1$ contains the leftward edge from $v$, and that $E^v_2$ contains the two rightward edges.
		We depict the transformation applied to two vertices of Min $u$ and $u'$: two copies of each vertex are created, only keeping the outgoing edge given respectively by $\tau_1$ and~$\tau_2$.}
		\label{4-fig:lift_split}
\commentAlt{Figure~\ref{4-fig:lift_split}: Two diagrams, labeled 'A' and 'A_tilde', illustrating hierarchical relationships and mappings between nodes. See long description.}
\commentLongAlt{Figure~\ref{4-fig:lift_split}: The image presents two conceptual diagrams, labeled 'A' on the left and 'A_tilde' on the right, both enclosed within large oval or amorphous shapes.

Diagram 'A' (left): Within a single large oval, there's a circular node 'v' at the top, branching downwards with three arrows. Below 'v', there are two square nodes, 'u'' and 'u', at roughly the same level. From 'u'', two arrows point upwards, labeled 'tau1(u')' and 'tau2(u')'. From 'u', two arrows point upwards and two arrows point downwards, labeled 'tau2(u)' and 'tau1(u)'.

Diagram 'A_tilde' (right): This diagram consists of two overlapping amorphous shapes. In the upper part of the left shape, a circular node 'v' branches downwards with three arrows. From the leftmost of these arrows, a new branch leads to a circular node 'u_1'' from which two arrows emanate, labeled 'tau1(u')' and 'tau2(u')'. Further down in the left shape, a circular node 'u_1' has an arrow pointing upwards labeled 'tau1(u)'. In the right amorphous shape, a circular node 'u_2'' has two arrows emanating, labeled 'tau1(u')' and 'tau2(u')'. Further down in the right shape, a circular node 'u_2' has two arrows emanating, labeled 'tau2(u)' and 'tau1(u)'. Node 'v' from the left part of the diagram is also present in the upper part of the right amorphous shape, suggesting a connection or shared element across the two overlapping regions.}
	\end{figure}

	Arena $\widehat{\arena}$ is a "one-player arena" of Max, and therefore (by hypothesis) Max has a positional optimal strategy $\widehat{\sigma}$ in $\widehat{\arena}$.
	We are first interested in the side ($1$ or $2$) that $\widehat{\sigma}$ picks in~$v$.
	We assume w.l.o.g.\ that $\widehat{\sigma}$ picks an edge towards copy $1$ (\textit{i.e.}, that $\widehat{\sigma}(v)\in E^v_1$).
	From this, we show that the positional strategy $\sigma_1$ is actually part of a pair of best responses in game~$\game$.

	We build a (non-necessarily positional) strategy $\tau$ of Min on $\game$ such that $(\sigma_1, \tau)$ is a pair of best responses from every vertex.
	The construction is very similar to the one in \Cref{4-thm:submixing_positional}, although we need to be more careful with the initialisation.
	Strategy $\tau$ has two modes, $1$ and $2$.
	The initial one is chosen as $1$ (this is not arbitrary: recall that side $1$ was preferred by Max in "arena" $\widehat{\arena}$ through strategy $\widehat{\sigma}$).
	Whenever $v$ is visited, the mode is updated: if the outgoing edge is in $E^v_1$, the new mode is $1$; otherwise, it is~$2$.
	The strategy uses edges prescribed by $\tau_1$ from the mode $1$ and by $\tau_2$ from the mode~$2$.
	Note that we rely on the positionality of $\tau_1$ and $\tau_2$ for this definition: a finite play may use edges both in $E^v_2$ and $E^v_1$, which do not exist in $\game_1$ and $\game_2$ respectively, but $\tau_1$ and $\tau_2$ ignore the past and only look at the current vertex.
	It is left to show that $(\sigma_1, \tau)$ is a pair of best responses in $\game$ from every vertex.
	Let $v_0$ be any vertex of $\arena$.

	We first show that $\tau$ is a best response to $\sigma_1$ from $v_0$.
	Let $\tau'$ be any strategy of Min in $\game$.
	We have that
	\begin{align*}
		\Phi(\play^{\game, v_0}_{\sigma_1, \tau'})
		&= \Phi(\play^{\game_1, v_0}_{\sigma_1, \tau'}) &\text{as $\sigma_1$ only chooses edges in $E^v_1$ from $v$}\\
		&\ge \Phi(\play^{\game_1, v_0}_{\sigma_1, \tau_1}) &\text{as $\tau_1$ is a best response to $\sigma_1$ in $\game_1$}\\
		&= \Phi(\play^{\game, v_0}_{\sigma_1, \tau}) &\text{as $\tau$ only uses mode $1$ against $\sigma_1$ in $\game$}.
	\end{align*}
	This shows that $\tau$ is a best response to $\sigma_1$ from $v_0$.

	We now show that $\sigma_1$ is a best response to $\tau$ in $\game$ from $v_0$.
	Let $\sigma'$ be any strategy of Max in $\game$.
	We denote $\widehat{\sigma'}$ for the strategy corresponding to $\sigma'$ in $\widehat{\game}$: it plays as $\sigma'$ would, ignoring in which copy of the "arena" it currently is.
	For convenience, we also denote $\widehat{\tau}$ the only ``empty'' strategy of Min in "arena" $\widehat{\arena}$ (recall that Min controls no vertex in~$\widehat{\arena}$).
	We have
	\begin{align*}
		\Phi(\play^{\game, v_0}_{\sigma', \tau})
		&= \Phi(\play^{\widehat{\game}, (v_0)_1}_{\widehat{\sigma'}, \widehat{\tau}}) &\text{by construction of $\widehat{\arena}$ and $\tau$}\\
		&\le \Phi(\play^{\widehat{\game}, (v_0)_1}_{\widehat{\sigma}, \widehat{\tau}}) &\text{by optimality of $\widehat{\sigma}$ in $\widehat{\game}$}\\
		&= \Phi(\play^{\game_1, v_0}_{\widehat{\sigma}, \tau_1}) &\text{as $\widehat{\sigma}$ always stays in copy $1$ from $(v_0)_1$ in $\widehat{\game}$}\\
		&\le \Phi(\play^{\game_1, v_0}_{\sigma_1, \tau_1}) &\text{as $\sigma_1$ is a best response to $\tau_1$ is $\game_1$}\\
		&= \Phi(\play^{\game, v_0}_{\sigma_1, \tau}) &\text{as $\tau$ only resorts to $\tau_1$ in $\game$ against $\sigma_1$}.
	\end{align*}
	This shows that $\sigma_1$ is a best response to $\tau$ in $\game$, ending the proof.
\end{proof}

\Cref{4-thm:1to2_lift} is a characterisation, as the other implication is immediate (the finite "one-player arenas" form a subset of the finite "arenas").
Therefore, uniform bi-positionality over finite "arenas" of all objectives satisfying this property can be established using this theorem.
It also implies that if an objective is \emph{not} "uniformly bi-positional", this is witnessed by some "one-player arena" for at least one of the two players.
We give some examples of objectives whose uniform bi-positionality had not been established yet and for which the use of \Cref{4-thm:1to2_lift} greatly simplifies the proof.

\begin{theorem} \label{4-thm:lift_applications}
	The $\Reach$, $\ShortestPath$, $\Energy$, $\MeanPayoff^-$, $\MeanPayoff^+$, and $\TotalPayoff$ objectives are "uniformly bi-positional" over finite "arenas".
\end{theorem}
\begin{proof}[Partial proof]
	We give arguments for the $\MeanPayoff^-$ objective.
	We show that in all finite "one-player arenas" of Max, Max has a positional optimal strategy.
	Symmetric arguments will provide the same result for "one-player arenas" of Min, and we can conclude using \Cref{4-thm:1to2_lift} that $\MeanPayoff^-$ is "uniformly bi-positional" over finite "arenas".

	Let $\arena$ be a finite "one-player arena" of Max (so Max controls all vertices of $\arena$).
	For each cycle of $\arena$, we define its \emph{mean payoff} as the average of its colours.
	Let $v_0$ be an initial vertex.
	We consider all cycles reachable from $v_0$.
	The main observation is that the maximal mean payoff among all these cycles is attained at a \emph{simple} cycle.
	Indeed, the mean payoff of an arbitrary cycle can always be expressed as a convex combination of mean payoffs of the simple cycles it contains.
	In particular, the mean payoff of an arbitrary cycle is less than or equal to the mean payoff of some simple cycle it contains.
	Moreover, in a finite arena, there are only finitely many simple cycles, so the maximum is attained.
	Based on this observation, here is therefore a positional optimal strategy from $v_0$: reach such a maximal simple cycle in a positional way (by "prefix-independence", the exact path does not matter) and then loop around it forever.
	No other play from $v_0$ can obtain a larger mean payoff (to see it, decompose any other play into a prefix and infinitely many cycles).

	Formally, we have now shown that from every vertex of $\arena$, there is a positional optimal strategy.
	We must still prove uniformity, \textit{i.e.}, the existence of a single positional strategy that is optimal from every vertex.
	In this particular case, as $\MeanPayoff^-$ is "prefix-independent", we can simply apply \Cref{1-lem:from_positional_to_uniformly_positional}.

	Observe that this proof also works with no change for $\MeanPayoff^+$.
\end{proof}

There is one previously defined objective that is "uniformly bi-positional" over finite arenas but for which we have not yet given a proof of this fact: the "discounted payoff".
Resorting to \Cref{4-thm:1to2_lift} to establish its bi-positionality is possible, but proving its positionality over one-player arenas is still arguably difficult.
Later in the book, two very different (and more direct) proofs will be given: the first uses \emph{universal graphs} (\Cref{4-subsec:finite-degree-arenas}); another, relying on Banach fixed-point theorem, will be given in \Cref{5-sec:discounted_payoff}.


\section{Bi-positionality over infinite arenas for qualitative objectives}
\label{4-sec:positional_infinite}
In this section, we prove that the only "prefix-independent" "qualitative objectives" that are "bi-positionally determined" over infinite arenas are, in some sense, "parity objectives".
"Bi-positionality" of the parity objective (over finite arenas) was derived in \Cref{4-thm:submixing_applications} (a proof for infinite arenas follows from \Cref{4-subsec:examples}).
We therefore focus on showing the necessity of the condition for bi-positionality.
We note that, by \Cref{3-thm:characterisation_positionally_determined_Muller_objectives}, we already know that "parity objectives" are the only "bi-positional" ones among "Muller objectives".

In \Cref{4-subsec:general-bipositional-infinite}, we generalise the characterisation of "bi-positionality" to objectives that are not necessarily "prefix-independent".
As a result, we obtain a one-to-two-player lift over infinite arenas (\Cref{4-cor:1-to-two-players_infinite_qualitative}), analogous to the one presented in the previous section.

\subsection{Prefix-independent qualitative objectives}

We say that a "qualitative objective" $\Omega\subseteq C^\omega$ (over a possibly infinite set of colours) is \emph{equivalent to a parity objective} if there exists $d\in \N$ and a mapping $\phi\colon C \to [1,d]$ such that, for all $\rho\in C^\omega$:
\[ \rho \in \Omega \; \iff \;\limsup \phi(\rho_i) \text{ is even}. \]

The main theorem of this section is the following.

\begin{theorem}\label{4-thm:char_positionality_infinite_qualitative_pref-indep}
	Every "prefix-independent" "qualitative objective" that is "bi-positional" over one-player infinite arenas of both players is "equivalent to a parity objective".
\end{theorem}

We note that "parity objectives" are uniformly "bi-positionally determined" over infinite arenas. Therefore, the previous theorem yields a characterisation of "bi-positionality" for "prefix-independent" objectives:

\begin{corollary}
	Let $\Omega$ be a "prefix-independent" "qualitative objective". The following are equivalent:
	\begin{itemize}
		\item $\Omega$ is "bi-positional" over infinite one-player arenas.
		\item $\Omega$ is "equivalent to a parity objective".
		\item $\Omega$ is "uniformly" "bi-positionally determined" over infinite arenas.
	\end{itemize}
\end{corollary}

The rest of the subsection is devoted to the proof of \Cref{4-thm:char_positionality_infinite_qualitative_pref-indep}.
We will use the following remark repeatedly.
\begin{remark}\label{4-rmk:swip_words_pref_indep}
	If $\Omega$ is "prefix-independent", then $(uv)^\omega\in \Omega$ if and only if $(vu)^\omega \in \Omega$, for all $u,v\in C^+$.
\end{remark}

\begin{lemma}\label{4-lem:mixing-even-words-accepting}
Let $\Omega\subseteq C^\omega$ be a "prefix-independent" "objective" that is "bi-positional" over infinite one-player arenas, and let $w_1,w_2,\ldots \in C^+$ be a sequence of words such that $w_i^\omega \in \Omega$ (resp.\ $w_i^\omega \notin \Omega$) 
for all $i$.
Then $w_1w_2w_3\ldots \in \Omega$ (resp.\ $w_1w_2w_3\ldots \notin \Omega$).
\end{lemma}
\begin{proof}
	We show the contrapositive. 
	Assume that $w_1w_2w_3\dots \notin \Omega$, and consider the one-player arena of Adam in which "Adam" controls one vertex $v$, from which there are self-loops labelled $w_i$ for each $i$.
	"Adam" can win by producing the word $w_1w_2\dots \notin \Omega$, so, by "bi-positionality", he has a "positional strategy" which always picks a same self-loop labelled $w_i$. Therefore, $w_i^\omega\notin \Omega$.
\end{proof}

\begin{lemma}\label{4-lem:strong-letters-bipositional}
	Let $\Omega\subseteq C^\omega$ be a "prefix-independent" "objective" that is "bi-positional" over infinite one-player arenas.
	Let $A, B\subseteq C$ such that for all $a\in A$ and $b\in B$, $(ab)^\omega\in \Omega$
	 (resp.\ $(ab)^\omega \notin \Omega$). 
	Then, $(AB^*)^\omega \subseteq \Omega$
	(resp.\ $(AB^*)^\omega \notin \Omega$).
\end{lemma}
\begin{proof}
	We show that for all $a\in A$, $(aB^*)^\omega \subseteq \Omega$, and conclude using \Cref{4-lem:mixing-even-words-accepting}.
	
	First, we prove that for all $u\in B^*$ there is $n\in \N$ such that $(a^nu)^\omega \in \Omega$ by induction on the length of $u$. If $|u|=1$, this follows by hypothesis for $n=1$.
	Let $u = u'b$, $b\in B$. By induction hypothesis, there is $n$ such that $(a^nu')^\omega\in \Omega$. Then, by \Cref{4-lem:mixing-even-words-accepting}, $((a^nu')(ba))^\omega \in \Omega$, and so $(a^{n+1}u)^\omega\in \Omega$.
	
	We now show that there is a fixed $k$ such that $(a^ku)^\omega \in \Omega$ for all $u\in B^*$. On the contrary, assume that for each $k$ there is $u_k\in B^*$ such that $(a^ku_k)^\omega \notin \Omega$. 
	Let $k_1,k_2,\ldots\in \N$ be a sequence such that $(a^{k_{i+1}}u_{k_i})^\omega \in \Omega$ (which exists by the previous claim), and consider the "one-player arena" of Eve in which she controls a vertex with self-loops labelled $a^ku_k$, for all $k$.
	She can win in this arena by producing the sequence $a^{k_1}(u_1a^{k_2})(u_2a^{k_2})\dots$, which belongs to $\Omega$ by \Cref{4-lem:mixing-even-words-accepting} and \Cref{4-rmk:swip_words_pref_indep}. However, she cannot win positionally, a contradiction.
	
	Finally, we prove that the minimal $k$ such that $(a^ku)^\omega \in \Omega$ for all $u\in B^*$ must be $k=1$. If $k>1$, there are words $u_1,u_2\in B^*$ such that $(a^{k-1}u_1)^\omega\notin \Omega$ and $(a^{k-1}u_2)^\omega\notin \Omega$. This implies that $a^{k-1}u_1u_2a^{k-1}\notin \Omega$, and so $(a^{k-2}(a^ku_1u_2))^\omega\notin \Omega$, a contradiction.
\end{proof}

We can now prove \Cref{4-thm:char_positionality_infinite_qualitative_pref-indep}.

\begin{proof}[Proof of \Cref{4-thm:char_positionality_infinite_qualitative_pref-indep}]
	Let 
	\[\CEven = \{c\in C \mid c^\omega \in \Omega\} \; \text{ and } \; \COdd = C \setminus \CEven.\]
	
	For $c\in \CEven$, we define
	\[ \smallerOdd(c) = \{ x\in \COdd \mid (cx)^\omega \in \Omega\}.\]
	We define the preorder over $\CEven$ given by $c\preceq c'$ if $\smallerOdd(c) \subseteq \smallerOdd(c')$.
	
	We claim that the preorder $\preceq$ is total over $\CEven$. Assume by contradiction that there are two incomparable elements $c,c'\in \CEven$. Then, there are colours $x,x'\in \COdd$ such that 
	\begin{align*}
		(cx)^\omega \in \Omega, & \quad (cx')^\omega \notin \Omega, \\
		(c'x)^\omega \notin \Omega, &\quad (c'x')^\omega \in \Omega.
	\end{align*}
	Consider the one-player arena of Eve in which she controls a vertex $v$ from which there are self-loops labelled $c'x$ and $cx'$. We show that she can "win" for objective $\Omega$ by producing the sequence $\rho = (c'xcx')^\omega = c'((xc)(x'c'))^\omega$.
	Indeed, by \Cref{4-lem:mixing-even-words-accepting}, $(xcx'c')^\omega \in \Omega$, so by "prefix-independence", $\rho\in \Omega$.
	However, she cannot "win positionally", a contradiction.
	
	Secondly, we prove that $\preceq$ admits finitely many equivalent classes. 
	On the contrary, assume that it contains either an infinite strictly increasing sequence or a strictly decreasing one. We assume w.l.o.g.\ that we are in the former case, so there is a sequence $c_1\prec c_2 \prec \dots$ of elements of $\CEven$. By definition, there are $x_1,x_2,\ldots\in \COdd$ such that $(c_ix_i)^\omega \in \Omega$ and $(c_ix_{i+1})^\omega \notin \Omega$, for all $i$.
	We consider the one-player arena of Eve in which she controls a vertex with self-loops labelled $c_ix_{i+1}$, for $i\geq 1$. She can win by producing the sequence $(c_1x_2)(c_2x_3)\ldots = c_1(x_2c_2)(x_3c_3)\ldots$, which belongs to $\Omega$ by \Cref{4-lem:mixing-even-words-accepting} and "prefix-independence".
	However, "Eve" cannot win this game positionally, a contradiction.
	
	We conclude that $\CEven$ admits a partition into finitely many subsets $C_1,\dots, C_d$ such that $C_1\prec C_2\prec \ldots \prec C_d$. 
	We define the sought mapping $\phi\colon C\to [1,2d]$ as follows:
	\begin{align*}
		&\text{For } c\in \CEven, &\;&\phi(c) = 2i & \text{ if } & c\in C_i,\\
		&\text{For } c\in \COdd, &\; &\phi(c) = 2i-1 & \text{ if } & c\in \smallerOdd(C_i)\setminus \smallerOdd(C_{i-1}) ,
	\end{align*}
	where we let $\smallerOdd(C_0) = \emptyset$.
	
	Finally, we prove that for $\rho\in C^\omega$, $\rho \in \Omega$ if and only if $i_{\max} = \limsup \phi(\rho_i)$ is even, as wanted.
	We prove this in the case where $i_{\max}$ is even; the odd case is analogous.
	
	We let $A= \{c\in C \mid \phi(c)= i_{\max}\}$ and $B = \{c\in C \mid \phi(c)< i_{\max} \}$. We note that $A\subseteq \CEven$, as $i_{\max}$ is assumed even.
	We claim that for all $a\in A$ and $b\in B$ we have that $(ab)^\omega \in \Omega$. 
	Indeed, if $b\in \CEven$, then $(ab)^\omega \in \Omega$ by \Cref{4-lem:mixing-even-words-accepting}. If $b\in \COdd$, then $b\in \smallerOdd(a)$, and $(ab)^\omega \in \Omega$ by definition.
	We deduce by \Cref{4-lem:strong-letters-bipositional} that $(AB^*)^\omega \subseteq \Omega$.
	Since $i_{\max}$ is the maximal index appearing infinitely often in $\rho$, $\rho$ contains a suffix in $(AB^*)^\omega$, so it belongs to $\Omega$.
	
	This concludes the proof that, for all $\rho\in C^\omega$:
	\[ \rho \in \Omega \; \iff \;\limsup \phi(\rho_i) \text{ is even}.\qedhere \]
\end{proof}

\begin{remark}
	\Cref{4-thm:char_positionality_infinite_qualitative_pref-indep} relies on the fact that games are edge-labelled; there are "objectives" that are "bi-positional" over state-labelled infinite arenas that are not "equivalent to a parity objective".
	One such example is the "Muller objective" ``see both colours $1$ and~$2$ infinitely often''.
	\end{remark}

\subsection{General qualitative objectives}\label{4-subsec:general-bipositional-infinite}

We now state a characterisation of "bi-positionality" over infinite arenas without assuming "prefix-independence", generalising the previous result. We only provide a partial proof of it.
The main ingredient in this characterisation consists in the \emph{residuals of an objective}, as introduced here.

\paragraph{Residuals.}
Let $\Omega\subseteq C^\omega$ be a qualitative objective. 
We define the \emph{residual of $\Omega$} with respect to a finite word $u\in C^*$ as:
\[ \residualO{u} = \set{w\in C^\omega \mid uw\in \Omega}.\]
We denote by $\resSet(\Omega)$ the set of residuals of $\Omega$.
We usually consider the order on the residuals given by the inclusion relation (which is a \emph{partial} order in general).

\begin{remark}
	An objective $\Omega$ is "prefix-independent" if and only if $\resSet(\Omega)$ is a singleton.
\end{remark}

\begin{remark}
	If $\Omega$ is $\omega$-regular, then $\resSet(\Omega)$ is finite.
	Unlike for regular languages of finite words, there are non-$\omega$-regular languages with finitely many residuals.
	For instance, there are "prefix-independent" objectives that are not $\omega$-regular, such as $\MeanPayoff^+$.
\end{remark}

We can associate to any objective $\Omega$ its \emph{automaton of residuals}, which is the "deterministic automaton" $\automatonResO = (Q, q_0, \Delta)$ given by:
\begin{itemize}
	\item $Q = \resSet(\Omega)$,
	\item $q_0 = \residualO{\epsilon}$,
	\item $\Delta = \set{(\residualO{u},a,\residualO{(ua)}) \mid u\in C^*,\, a\in C}$.
\end{itemize}
This is not formally an automaton as defined in \Cref{1-chap:introduction}: it is defined without an "acceptance condition" and it may have infinitely many states.

We say that $\Omega$ can be \emph{recognised by a parity automaton on top of its automaton of residuals} if there is some $d\in \N$ and some colouring $\col\colon \Delta \to [1,d]$ such that the automaton~$\automatonResO$ with the "parity condition" induced by $\col$ "recognises" $\Omega$.

We introduce now a property about the residuals of a language, called \emph{progress consistency}, that is necessary for "positionality".

\paragraph{Progress consistency.} We say that an objective $\Omega\subseteq C^\omega$ is \emph{progress consistent} if for all $u,w\in C^*$
\[ \residualO{u} \subsetneq \residualO{(uw)} \; \implies \; uw^\omega\in \Omega.\]

Intuitively, $\residualO{u} \subsetneq \residualO{(uw)}$ means that when producing $w$ after the word $u$, we ``make a progress towards $\Omega$'', in the sense that there are now strictly more winning continuations than after $u$.
The objective $\Omega$ is "progress consistent" if, in such a situation, repeating $w$ forever after $u$ yields a winning word.

\begin{example}
	Let $C = \{a,b\}$ and consider the objective
	\[ \Omega = \Reach(aa) = \set{w\in C^\omega \mid w \text{ contains the factor } aa}.\]
	
	The objective $\Omega$ has three residuals: $\residualO{\epsilon}\subsetneq \residualO{a} \subsetneq \residualO{(aa)}$.
	We claim that is not "progress consistent". Indeed, it suffices to take $u=\epsilon$ and $w = ba$. We have that $\residualO{\epsilon}\subsetneq \residualO{(ba)}$, but $(ba)^\omega\notin \Omega$.
	As we will see (\Cref{4-lem:necessity_progress_consistency}), "progress consistency" is a necessary condition for "positionality".
	Non-positionality of $\Omega$ can be shown by considering a game with a vertex controlled by Eve with self-loops labelled by $ba$ and by $ab$.
\end{example}

\paragraph{The characterisation.}
We can now state the main characterisation theorem.

\begin{theorem}\label{4-thm:char_positionality_infinite_qualitative_general}
	Let $\Omega\subseteq C^\omega$ be a "qualitative objective" that is "uniformly" "bi-positional" over infinite one-player arenas of both players. Then,
	\begin{itemize}
		\item $\resSet(\Omega)$ is finite and totally ordered by inclusion,
		\item both $\Omega$ and $C^\omega\setminus \Omega$ are "progress consistent", and
		\item the objective $\Omega$ can be "recognised" by a "parity automaton on top of" its "automaton of residuals" (in particular, $\Omega$ is "$\omega$-regular").
	\end{itemize}
	Furthermore, in this case, $\Omega$ is "uniformly" "bi-positionally determined" over all arenas.
\end{theorem}

We note that if $\Omega$ is "prefix-independent", then the first two conditions in the previous theorem are trivially satisfied. In this case, the third condition amounts to saying that $\Omega$ should be recognised by a parity automaton with a single state. This is equivalent to the fact that $\Omega$ is equivalent to a parity objective; we recover therefore \Cref{4-thm:char_positionality_infinite_qualitative_pref-indep}.

\begin{corollary}[One-to-two-player lift over infinite arenas]\label{4-cor:1-to-two-players_infinite_qualitative}
	Every "objective" that is "uniformly" "bi-positional" over infinite one-player arenas of both players is "uniformly" "positionally determined" over all infinite arenas.
\end{corollary}

We discuss the proof of necessity of the first two items of the characterisation.
These proofs exemplify some common techniques for proving non-positionality.
For a proof of necessity of the third item, we refer to~\cite{Bouyer.Randour.ea:2023}.
Sufficiency of the conditions can be established using monotone universal graphs, introduced in \Cref{4-sec:positionality}; we refer to~\cite{Casares.Ohlmann:2024} for a proof.

\paragraph{Necessity of well-order of residuals and progress consistency.}

\begin{lemma}
	If $\Omega\subseteq C^\omega$ is "uniformly" "positional" over infinite one-player arenas of Eve, then $\resSet(\Omega)$ is well-ordered by inclusion (that is, well-founded and totally ordered).
\end{lemma}
\begin{proof}
	We show the contrapositive.
	Assume first that the inclusion on $\resSet(\Omega)$ is not a total order. Then, there are two residuals $\residualO{u_1}$ and $\residualO{u_2}$ that are incomparable, that is, there are words $w_1, w_2\in C^\omega$ such that
	\begin{align*}
		u_1w_1\in \Omega, \quad&u_1w_2\notin \Omega, \\
		u_2w_1\notin \Omega, \quad&u_2w_2\in \Omega.
	\end{align*}
	
	It suffices to consider the (infinite sized) game in \Cref{4-fig:incomparable_residuals}. "Eve" can win from $v_1$ and~$v_2$; however, no "positional strategy" ensures that she will win from both these vertices. 
	\begin{figure}[!ht]
		\centering
		\begin{tikzpicture}			
			\node at (0,0.7) [s-eve] (1) {$v_1$};
			\node at (0,-0.7) [s-eve] (2) {$v_2$};
			\node at (2.1,0) [s-eve] (choice) {};
			\node at (4.8,0.65) [] (end1) {};
			\node at (4.8,-0.65) [] (end2) {};

			\path[->]
			(1) edge[squish] node[above, pos=0.45] {$u_1$} (choice)
			(2) edge[squish] node[above, pos=0.45] {$u_2$} (choice)
			(choice) edge[squish] node[above, pos=0.45] {$w_1$} (end1)
			(choice) edge[squish] node[above, pos=0.45] {$w_2$} (end2);
		\end{tikzpicture}
		\caption{Necessity of total order of residuals: game in which "Eve" can win, but not positionally when $u_iw_i\in \Omega$ and $u_iw_j\notin \Omega$ for $i\neq j$.
		Squiggly arrows are used as a shorthand for possibly multiple consecutive edges (they are labelled by an element of $C^*$).}
		\label{4-fig:incomparable_residuals}
\commentAlt{Figure~\ref{4-fig:incomparable_residuals}: A diagram showing two input nodes (v1, v2) with wavy arrows leading to a central node, and two output wavy arrows (w1, w2) from the central node.}
\commentLongAlt{Figure~\ref{4-fig:incomparable_residuals}: The image displays a central circular node with two inputs and two outputs. From the top-left, a circular node labeled 'v1' is connected to the central node by a wavy arrow labeled 'u1'. From the bottom-left, a circular node labeled 'v2' is connected to the central node by a wavy arrow labeled 'u2'. From the central node, two wavy arrows extend to the right: the top one is labeled 'w1' and the bottom one is labeled 'w2'. All arrows are directed towards or away from the central node, indicating flow or interaction.}
	\end{figure}
	
	Now, assume that $\resSetO$ is not well-founded, and let $\residualO{u_1}\supsetneq \residualO{u_2}\supsetneq \dots$ be an infinite strictly decreasing chain of "residuals". Then, there are words $w_1,w_2,\ldots \in C^\omega$ such that $u_iw_i\in \Omega$, but $u_iw_j\notin \Omega$ for all $j>i$.
	To conclude, consider the game from \Cref{4-fig:game_well_founded_residuals}. "Eve" can win from all vertices $v_i$, but no "positional strategy" guarantees a win from all of them.\qedhere
	
	\begin{figure}[!ht]
		\centering
	\begin{tikzpicture}
		\node at (0.3,1.5) [s-eve] (v1) {$v_1$};
		\node at (0.3,.5) [s-eve] (v2) {$v_2$};
		\node at (0.3,-0.5) [s-eve] (v3) {$v_3$};
		\node at (2.4,0) [s-eve] (1) {};
		
		\node at (4.6,1.7) [] (end1) {};
		\node at (4.9,0.5) [] (end2) {};
		\node at (4.8,-0.75) [] (end3) {};
		
		\node at (0.3,-1.5) [scale=1.4] (dots0) {$\vdots$};
		\node at (1.5,-1.3) [scale=1.4] (dots1) {$\vdots$};
		\node at (3.6,-1.3) [scale=1.4] (dots2) {$\vdots$};

		\path[->]
		(v1) edge[squish, out=00, in=110] node[above, pos=0.35] {$u_1$} (1)
		(v2) edge[squish, out=5, in=145] node[above, pos=0.45] {$u_2$} (1)
		(v3) edge[squish, out=-5, in=210] node[above, pos=0.46] {$u_3$} (1)
	
		(1) edge[squish] node[above, xshift=-2pt] {$w_1$} (end1)
		(1) edge[squish] node[above, pos=0.46,xshift=-1pt] {$w_2$} (end2)
		(1) edge[squish] node[above] {$w_3$} (end3);
	\end{tikzpicture}
	\caption{Necessity of well-foundedness of residuals: game in which "Eve" can win from every vertex~$v_i$, but no "positional strategy" ensures a win from all those vertices.}
	\label{4-fig:game_well_founded_residuals}
\commentAlt{Figure~\ref{4-fig:game_well_founded_residuals}: A diagram showing multiple input nodes with wavy arrows leading to a central node, and multiple wavy output arrows from the central node.}
\commentLongAlt{Figure~\ref{4-fig:game_well_founded_residuals}: The image displays a central circular node with multiple inputs and outputs. On the left, there are three visible circular input nodes, labeled 'v1', 'v2', and 'v3', followed by a vertical ellipsis, indicating more inputs. Wavy arrows connect these input nodes to the central node, labeled 'u1', 'u2', and 'u3' respectively. On the right, three wavy arrows emanate from the central node, labeled 'w1', 'w2', and 'w3' respectively, followed by a vertical ellipsis, indicating more outputs. All wavy arrows are directed towards or away from the central node, suggesting a fan-in/fan-out or multi-input/multi-output system.}
	\end{figure}
\end{proof}

\begin{corollary}
	If $\Omega\subseteq C^\omega$ is "uniformly" "bi-positional" over infinite "one-player arenas" (of both players), then $\resSet(\Omega)$ is finite and totally ordered by inclusion.
\end{corollary}
\begin{proof}
	By the previous lemma, both $\resSet(\Omega)$ and $\resSet(C^\omega \setminus\Omega) = \{C^\omega\setminus R \mid R\in \resSet(\Omega)\}$ are well-ordered, so $\resSet(\Omega)$ must be finite.
\end{proof}

\begin{lemma}\label{4-lem:necessity_progress_consistency}
	If $\Omega\subseteq C^\omega$ is "positional" over infinite one-player arenas of Eve, then it is "progress consistent".
\end{lemma}
\begin{proof}
	We prove the contrapositive. Assume that $\Omega$ is not "progress consistent". Then, there are words $u,w\in C^*$ such that $\residualO{u} \subsetneq \residualO{(uw)}$ but $uw^\omega\notin \Omega$. 
	The strict inclusion of the residuals implies that there is $w'\in C^\omega$ such that $uw'\notin \Omega$ and $uww'\in \Omega$. 
	Consider the game in \Cref{4-fig:game_progress_consistent}. "Eve" can win from $v_0$ by producing the output $uww'$, but she cannot win positionally.\qedhere
	
	\begin{figure}[!ht]
		\centering
		\begin{tikzpicture}
			\node at (0,0) [s-eve] (0) {$v_0$};
			\node at ($(0)+(2.1,0)$) [s-eve] (choice) {};
			\node at ($(0)+(4.2,0)$) [s-eve,draw=none,fill=none](end) {};
						
			\path[->]
			(0) edge[squish] node[above=1pt] {$u$} (choice)
			(choice) edge[squishloop, out=62, in=118, loop] node[above=2pt] {$w$} (choice)
			(choice) edge[squish] node[above=1pt] {$w'$} (end);
		\end{tikzpicture}
		\caption{Necessity of progress consistency: Game in which "Eve" can win from $v_0$, but not "positionally".}
		\label{4-fig:game_progress_consistent}
\commentAlt{Figure~\ref{4-fig:game_progress_consistent}: A diagram showing a linear flow from a node v0, through a central node with a self-loop, and out to the right. See long description.}
\commentLongAlt{Figure~\ref{4-fig:game_progress_consistent}: The image depicts a flow from left to right. A circular node labeled 'v0' is on the left, connected by a wavy arrow labeled 'u' to a central circular node. The central node has a wavy self-loop arrow above it, labeled 'w'. From the right side of the central node, a wavy arrow labeled 'w'' extends to the right, indicating an output or continuation of the flow.}
	\end{figure}
\end{proof}


\section{Positionality for quantitative objectives}
\label{4-sec:quantitative}
In the previous section, we studied positionality of qualitative objectives.
We can wonder whether neat characterisations such as the one presented in \Cref{4-thm:char_positionality_infinite_qualitative_pref-indep} can be obtained for quantitative objectives.
In this section, we will see in \Cref{4-prop:quantitative_characterisation_as_qualitative} that the study of positionality of quantitative objectives can be reduced to the study of qualitative ones.

However, in the case of quantitative objectives, various natural definitions of positionality are possible. 
We start by presenting the three main such definitions and studying their relations. We show that these three notions coincide over finite arenas and provide examples separating them over infinite arenas.
Each of them reduces to the study of qualitative objectives in a slightly different way.

\begin{definition}
	Let $\Phi\colon C^\omega \to \Rinfty$ be a "quantitative objective". We say that it is
	\begin{itemize}	
		\item \emph{positional for Max} if for every game $\game$ with "objective" $\Phi$, every "Max" "strategy"~$\sigma$, and every vertex $v$, "Max" has a "positional strategy" $\sigma'$ such that 
		\[ \val_{\mMax}^\sigma(v) \leq \val_{\mMax}^{\sigma'}(v), \]
		
		\item \emph{limit-optimal positional for Max} if for every game $\game$ with "objective" $\Phi$ and every vertex~$v$,
		\[\val_{\mMax}^\game(v) = \sup_{\sigma \text{ positional }} \val_{\mMax}^\sigma(v),\]
		
		\item \emph{positional and admitting optimal strategies for Max} if for every game $\game$ with "objective"~$\Phi$ and every vertex $v$, "Max" has a "positional" "strategy" $\sigma$ from $v$ such that 
		\[ \val_{\mMax}^\sigma(v) = \val_{\mMax}^\game(v) \; \text{ (\textit{i.e.}, $\sigma$ is optimal from $v$)}.\]	
	\end{itemize}	
	The "uniform" version of these three notions is defined in the expected way.
\end{definition}

\begin{remark}
	An objective can be "limit-optimal positional" but not "positional" if the following three conditions occur in some game:
	\begin{itemize}
		\item "Max" has an "optimal strategy" (strategy that reaches $\val_{\mMax}^\game(v)$),
		\item "Max" does not have a "positional" "optimal strategy", and
		\item "Max" has "positional" "$\epsilon$-optimal" strategies for all $\epsilon>0$.
	\end{itemize}
\end{remark}

\begin{proposition}\label{4-prop:notions_positionality_quantitative}
	If $\Phi$ is "positional", then $\Phi$ is "limit-optimal positional".
	Moreover, over finite arenas, if $\Phi$ is "limit-optimal positional", then $\Phi$ is "positional and admits optimal strategies".
\end{proposition}
\begin{proof}
	("Positionality" $\implies$~"limit-optimal positionality").
	Let $\epsilon>0$. We aim to prove that "Max" has a "positional strategy" ensuring a value at least $\val_{\mMax}^{\game}(v)-\epsilon$.
	By definition, "Max" has a (not necessarily "positional") "strategy" $\sigma$ ensuring a value greater than $\val_{\mMax}^{\game}(v)-\epsilon$. As $\Phi$ is assumed "positional", there exists a "positional" strategy $\sigma'$ such that 
	\[ \val_{\mMax}^{\game}(v)-\epsilon \leq \val_{\mMax}^\sigma(v) \leq \val_{\mMax}^{\sigma'}(v). \]
	
	(Over finite arenas, "limit-optimal positionality" $\implies$ "positionality and existence of optimal strategies"). 
	We note that over a finite arena, there are finitely many positional strategies. Therefore, if for every $\epsilon>0$ "Max" has a "positional strategy" ensuring a value greater than $\val_{\mMax}^{\game}(v)-\epsilon$, one of these strategies must ensure the "value" $\val_{\mMax}^{\game}$.
\end{proof}

\subsubsection*{Reduction to the qualitative case}
We now show that we can reduce the study of the positionality of quantitative objectives to the positionality of qualitative objectives. We introduce the following notation for this purpose. 

For a quantitative objective $\Phi\colon C^\omega \to \Rinfty$ and $x\in \Rinfty$, we let
\[\PhiGeqx = \{\rho\in C^\omega \mid \Phi(w)\geq x\} \quad \text{ and } \quad \PhiGneqx = \{\rho\in C^\omega \mid \Phi(w)> x\}.\]

We also define $\Phi_{<x} = C^\omega \setminus \PhiGeqx$ and $\Phi_{\leq x} = C^\omega \setminus \PhiGneqx$.

\begin{proposition}\label{4-prop:quantitative_characterisation_as_qualitative}
	Let $\Phi\colon C^\omega \to \Rinfty$ be a "quantitative objective".
	\begin{enumerate}		
		\item $\Phi$ is "positional" if and only if for all $x\in \Rinfty$, the qualitative objective $\PhiGeqx$ is "positional".
		
		\item $\Phi$ is "limit-optimal positional" if and only if for all $x\in \Rinfty$, the qualitative objective $\PhiGneqx$ is "positional".	
		
		\item $\Phi$ is "positional and admits optimal strategies" if and only if for all $x\in \Rinfty$, the qualitative objective $\PhiGeqx$ is "positional" and $\Phi(C^\omega)$ does not contain any infinite strictly increasing sequence.
	\end{enumerate}
	
	The equivalences remain valid over any subclass of arenas. 
	Moreover, if $\Phi$ is "uniformly" "positional" (resp. uniformly "limit-optimal positional"), so is $\PhiGeqx$ (resp.\ $\PhiGneqx$), for all $x$.
\end{proposition}
\begin{proof}
	In the following, $\game$ stands for a game labelled with colours in $C$, which can be viewed as a game using objective $\Phi$, $\PhiGeqx$, or $\PhiGneqx$.
	\begin{enumerate}
		\item Assume that $\Phi$ is "positional", and suppose that "Eve" has a winning "strategy" $\sigma$ for the "objective" $\PhiGeqx$ in $\game$ from a vertex $v$. By "positionality" of $\Phi$, there is "positional strategy" $\sigma'$ for "Max" such that $x\leq \val_{\mMax}^\sigma(v) \leq \val_{\mMax}^{\sigma'}(v)$. The "strategy" $\sigma'$ is therefore a "positional strategy" "winning" for $\PhiGeqx$ from $v$. The "uniform" case is analogous.
		
		Conversely, assume that $\PhiGeqx$ is "positional" for all $x\in \Rinfty$. Let $\sigma$ be a "strategy" for "Max" from $v$ and let $x=\val_{\mMax}^\sigma(v)$. Clearly, $\sigma$ is a winning strategy from $v$ for the objective $\PhiGeqx$. By "positionality" of $\PhiGeqx$, "Max" has a "positional strategy" $\sigma'$ such that $x= \val_{\mMax}^\sigma(v) \leq \val_{\mMax}^{\sigma'}(v)$, as wanted.	
		
		\item Let $\sigma$ be a "winning strategy" for $\PhiGneqx$ from a vertex $v$. By taking $\epsilon < \val_{\mMax}^{\sigma}(v) -x$ and applying "limit-optimal positionality", we conclude that "Max" has a "positional" strategy ensuring a value strictly greater than $x$, which is therefore winning for $\PhiGneqx$. The uniform case is analogous.
		
		Conversely, let $\epsilon>0$ and let $\sigma$ be an strategy ensuring a value strictly greater than $ x= \val_{\mMax}^{\sigma}(v) -\epsilon$.
		Then, by positionality of $\PhiGneqx$, "Max" has a "positional" strategy ensuring a value strictly greater than $x$.
		
		\item Assume that $\Phi$ is "positional and admits optimal strategies". The fact that $\PhiGeqx$ is "positional" for all $x$ is implied by the first item. Suppose by contradiction that there is a sequence of infinite words $w_1,w_2,\ldots\in C^\omega$ whose image by $\Phi$ is $x_1<x_2<\dots$. 
		Consider the infinite game in which "Max" controls a vertex $v$ from which she can choose to take a path labelled $w_i$, for every $i$. No strategy in this game ensures $\val_{\mMax}^{\game}(v)$, so $\Phi$ does not "admit optimal strategies".
		
		We show the converse. Consider a sequence of strategies ensuring values that tend to $x=\val_{\mMax}^{\game}(v)$. Since $\Phi(C^\omega)$ does not contain infinite strictly increasing sequences of values, such a sequence eventually stabilises at $x$, in particular, there is a "strategy" with value $x$. We conclude by "positionality" of $\PhiGeqx$.\qedhere
	\end{enumerate}
\end{proof}

\subsubsection*{Two-player bi-positionality for quantitative objectives}
We instantiate the previous results in the case of "bi-positionality".
We note that by \Cref{4-thm:char_positionality_infinite_qualitative_pref-indep,4-thm:char_positionality_infinite_qualitative_general}, "bi-positionality" of "qualitative objectives" is already characterised.
We obtain a weak version of the one-to-two-player lift of bi-positionality for quantitative objectives over infinite arenas, in which we can just deduce "non-uniform" bi-positionality over two-player arenas.

\begin{remark}
	Note that "bi-positionality" of $\Phi$ does not reduce to "bi-positionality" of $\Phi_{\geq x}$ for all $x$, but rather to "positionality" of both $\Phi_{\geq x}$ and $\Phi_{\leq x}$.
\end{remark} 

\begin{corollary}\label{4-cor:quantitative_bipositional}
	A "quantitative objective" $\Phi\colon C^\omega \to \Rinfty$ is "bi-positional" (over $X$ arenas) if and only if $\Phi(C^\omega)$ is finite and for all $x\in \Rinfty$ the qualitative objective $\PhiGeqx$ is "bi-positional" (over $X$ arenas).
\end{corollary}
\begin{proof}
	Necessity of these two conditions directly follows from \Cref{4-prop:quantitative_characterisation_as_qualitative}.
	To show sufficiency, we note that, in order to apply the previous proposition, we need to show that the objective $\Phi_{\leq x}$ is "positional", which does not follow directly from the fact that $\PhiGeqx$ is "bi-positional". However, we deduce it using the fact that $\Phi(C^\omega)$ must be finite: Assume that "Eve" "wins" a game $\game$ with objective $\Phi_{\leq x}$ from $v$. If $x\notin \Phi(C^\omega)$, then she can ensure the objective $\Phi_{< x}$, which is positional (by bi-positionality of $\Phi_{\geq x}$). If $x\in \Phi(C^\omega)$, then she wins for the objective $\Phi_{< x+\epsilon}$, for sufficiently small $\epsilon$, and we conclude as before.
\end{proof}

\begin{corollary}[Weak one-to-two-player lift for quantitative objectives]\label{4-cor:1-to-two-players_infinite_quantitative}
	Every "quantitative objective" that is "uniformly" "bi-positional" over infinite one-player arenas of both players is "bi-positional" over all infinite arenas.
\end{corollary}
\begin{proof}
	By \Cref{4-prop:quantitative_characterisation_as_qualitative} and \Cref{4-cor:quantitative_bipositional}, $\Phi(C^\omega)$ is finite and for all $x\in \Rinfty$, the "qualitative objective" $\PhiGeqx$ is "uniformly" "bi-positional" over one-player arenas. Therefore, by \Cref{4-cor:1-to-two-players_infinite_qualitative}, they are in fact "bi-positional" over all arenas.
	A further application of \Cref{4-cor:quantitative_bipositional} implies that $\Phi$ is "bi-positionally" over all infinite arenas.
\end{proof}

\subsubsection*{Separation over infinite arenas}
We provide two examples showing that the three notions under study are not equivalent over infinite arenas.

\begin{example}
	Let $C = \Z$ and consider the quantitative objective:
	\[\First(\rho_1\rho_2\rho_3\dots) = \rho_1. \]
	
	Clearly, $\First(C^\omega) = \Z$ contains infinite strictly increasing sequences.
	Therefore, by \Cref{4-prop:quantitative_characterisation_as_qualitative}, it does not "admit optimal strategies" (for any of the players).
	By the same proposition, we obtain that $\First$ is "bi-positional", as the qualitative objectives $\First_{\geq x} = \{\rho\in C^\omega \mid \rho_1 \geq x\}$ and $\First_{\leq x}$ are trivially "positional".
\end{example}

\begin{example}
	Consider the "limsup" objective over $C = \Z$:
	\[\LimSup(\rho) = \limsup\rho. \]
	
	We claim that $\LimSup$ is "limit-optimal bi-positional" but not "positional".
	
	For each $x\in \R$, $\LimSup_{>x}$ is a "B\"uchi" objective ("Eve" wins if numbers strictly larger than~$x$ are produced infinitely often), so it is "positional for her". Also, $\LimSup_{>\infty}$ is the trivial empty objective, and $\LimSup_{>-\infty}$ can be seen as a countable union of "B\"uchi objectives" ($\rho$ is winning if there is one negative number seen infinitely often, or if positive numbers are seen infinitely often). This latter objective will be shown to be "positional" in \Cref{4-subsec:examples}.
	Therefore, $\LimSup$ is "limit-optimal positional for Max".
	From the point of view of Min, $\LimSup_{<x}$ is a "coB\"uchi" objective for $x\in \R$, and $\LimSup_{<\infty} = \set{\rho\in C^\omega\mid \rho \text{ is bounded}}$ can be shown to be "positional" using \Cref{4-prop:countable_unions_sigma2}, so $\LimSup$ is also "limit-optimal positional for Min". We conclude that $\LimSup$ is "limit-optimal bi-positional".
	
	However, $\LimSup$ is not "positional", as $\LimSup_{\geq \infty}$ is not "positional". 	
\end{example}

Similarly, we can show using \Cref{4-cor:mean-payoff-positional} that the objective $\MeanPayoff^{-}$ over the set of colours $C=\Z$ is "limit-optimal positional"; however, it is not "positional"~\cite[Example~8.10.2]{Puterman:2005}.


\section{Positionality over infinite arenas}
\label{4-sec:positionality}
This section is concerned with understanding positional (qualitative or quantitative) objectives: those for which, on arbitrary arenas, if Eve wins, then she can do so with a positional strategy.
Just like for bi-positionality, it turns out that uniform positionality appears to be better behaved than the non-uniform counterpart.
In a nutshell, results about positionality are sparser and often more difficult than for bi-positionality.
No general simple characterisations akin to~\Cref{4-thm:1to2_lift,4-thm:char_positionality_infinite_qualitative_general} are available.

Most of the section is devoted to the relationship between positionality over arbitrary arenas and so-called \emph{monotone universal graphs}.
These are combinatorial objects which allow to reason about positional objectives.
\Cref{4-subsec:universal-graphs} introduces monotone universal graphs, then \Cref{4-subsec:characterisation-pos} states and proves a characterisation of positionality.
\Cref{4-subsec:examples} illustrates the technique with a number of examples, and then \Cref{4-subsec:finite-degree-arenas} shows how it adapts to finite-degree arenas.

An important conjecture about positionality, called Kopczy\'nski's conjecture, states that prefix-independent positional objectives are closed under countable unions; \Cref{4-subsec:Kopcz_conj} discusses results around this conjecture.
We end with a discussion on positionality of objectives which are $\omega$-regular in \Cref{4-subsec:positional_omega-reg}.

Throughout \Cref{4-sec:positionality}, it is more convenient, for technical reasons that will appear below, to take the point of view of player Min.
Therefore, by default, ``positionality'' means ``positionality for Min''.
We will establish general results about quantitative objectives, and freely allow ourselves to apply these to qualitative objectives $\Omega$, by identifying them with their indicator function.

\subsection{Universal graphs}\label{4-subsec:universal-graphs}

We now introduce the required vocabulary for talking about universal graphs.

\paragraph{Graphs and morphisms.}
We consider \emph{$C$-graphs}, which are (potentially infinite) directed graphs whose edges are labelled by colours in $C$, and which exclude sinks: all vertices have some outgoing edge.
Edges of $C$-graphs are denoted by $v \re c v'$.
We think of graphs as being arenas controlled by the opponent Max; therefore, given a quantitative objective $\Phi\colon C^\omega \to \Rinfty$, the \emph{value} $\Value(v)$ of a vertex $v$ in a graph is defined to be the supremum value of $\Phi(\pi)$, where $\pi$ ranges over infinite paths from $v$.

A \emph{morphism} from a $C$-graph $G$ to a $C$-graph $H$ is a map $h$ from vertices of $G$ to vertices of $H$ such that for all edges $v \re c v'$ in $G$, it holds that $h(v) \re c h(v')$ is an edge in $H$.
Note that $h$ needs not be injective; for example any graph admits a morphism towards the $1$-vertex graph with all possible self-loops.
We write $G \re h H$ when $h$ is a morphism from $G$ to $H$, and $G \to H$ when there is a morphism from $G$ to $H$.
Note that when $G \re h H$, paths in $G$ are mapped to paths in $H$ with the same labels, and in particular, the value of a vertex in $G$ is not larger than the value of its image in $H$.

\paragraph*{Universal graphs.}
We say that a morphism $G \re h H$ is \emph{$\Phi$-preserving}, where $\Phi\colon C^\omega \to \Rinfty$ is a quantitative objective, if for every vertex $v$ in $G$, the value of $h(v)$ in $H$ is in fact equal to the value of $v$ in $G$.
When $\Phi$ corresponds to a qualitative objective $\Omega$, note that a morphism is $\Phi$-preserving if and only if vertices satisfying $\Omega$ in $G$ are mapped to vertices satisfying $\Omega$ in $H$.

Given a quantitative objective $\Phi$ and a cardinal $\kappa$ (which is usually taken infinite in this section), we say that a graph $U$ is \emph{$(\kappa,\Phi)$-universal} if all graphs of cardinality $< \kappa$ admit a $\Phi$-preserving morphism towards $U$.

\begin{remark}
We often work with prefix-independent objectives $\Omega$.
In this case, graphs have no edges from vertices satisfying $\Omega$ to vertices which do not.
As a consequence, one may simply focus on vertices satisfying $\Omega$, and alter the definition as follows: a graph $U$ is $(\kappa,\Omega)$-universal if it satisfies $\Omega$, and all graphs of cardinality $<\kappa$ which satisfy $\Omega$ have a morphism towards $U$.
When considering prefix-independent objectives, for instance for $\CoBuchi$ and $\Buchi$ below, we use this definition.
\end{remark}

\begin{figure}[h]
    \begin{center}
        \includegraphics*[width=0.46\linewidth]{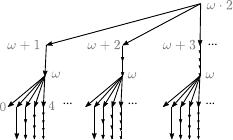}
        \caption{A well-founded tree of rank $\omega \cdot 2$. The ranks of a few vertices are displayed.}
        \label{4-fig:rank_of_a_tree}        
\commentAlt{Figure~\ref{4-fig:rank_of_a_tree}: A hierarchical diagram illustrating a process branching into parallel operations, each ending in a fan-out to multiple points. See long description.}
\commentLongAlt{Figure~\ref{4-fig:rank_of_a_tree}: The image depicts a hierarchical structure starting from a single node at the top right labeled 'omega * 2'. From this node, three arrows descend and fan out, connecting to three intermediate nodes labeled 'omega + 1', 'omega + 2', and 'omega + 3...', respectively. From each of these intermediate nodes, a single arrow points downwards to another node labeled 'omega'. Each of these 'omega' nodes then acts as a central point, from which multiple arrows branch downwards in a fan-out pattern. The leftmost 'omega' node's fan-out is partially labeled at its ends with '0', '4', and '...'. The other two 'omega' nodes also show fan-out patterns ending in '...' indicating continuation. The overall diagram suggests a process that diverges, undergoes parallel operations, and then each path further diverges into a multitude of final points.}
    \end{center}
\end{figure}

To prove universality results, we often use the notion of \emph{rank of a well-founded tree}, which is recalled now.
A tree $T$ is \emph{well-founded} if it does not admit any infinite branch.
The \emph{rank} of a vertex in a well-founded tree is defined to be zero for leaves, and one plus the supremum rank of its successors for non-leaves; the rank of a tree is the rank of its root.
Intuitively, the rank of a tree is an ordinal quantifying its depth.
We give a few examples in \Cref{4-fig:rank_of_a_tree}.

We now give some examples of universal graphs for standard objectives.

\begin{figure}[h]
    \begin{center}
        \includegraphics*[width=0.6\linewidth]{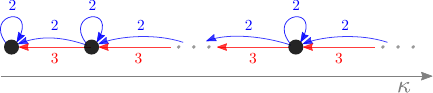}
        \caption{The graph $U$ for the coB\"uchi objective. Transitive edges are excluded for readability.}
        \label{4-fig:universal_graph_cobuchi}
        \commentAlt{Figure~\ref{4-fig:universal_graph_cobuchi}: A horizontal line with an arrow labeled 'k' on the right, and a repeating pattern of nodes and labeled self-loops and connections, extending indefinitely in both directions.}
    \end{center}
\end{figure}

\paragraph*{Example: CoBüchi.}
Let $C=\{2,3\}$ and $\Omega$ be the $\CoBuchi$ objective, meaning the set of words where $3$ has at most finitely many occurrences.
Fix a cardinal $\kappa$.
Consider the graph $U$ over $\kappa$ given by $u \re{3} u' \iff u > u'$ and $u \re{2} u' \iff u \geq u'$ (see \Cref{4-fig:universal_graph_cobuchi}).

Let us prove that $U$ is $(\kappa,\Omega)$-universal.
First, we show that $U$ satisfies the $\CoBuchi$ objective $\Omega$; this is clear since paths in $U$ are non-increasing and follow a strict decrease when a $3$ is seen.

Next, we should prove that any graph $G$ of cardinality $< \kappa$ which satisfies $\CoBuchi$ has a morphism towards $U$.
Take such a graph $G$ and fix a vertex $v$ in $G$.
Since $G$ satisfies $\CoBuchi$, paths from $v$ visit only finitely many $3$-edges; stated differently, the tree obtained by unfolding $G$ at $v$ and contracting $2$-edges is well-founded. 
We assign to $v$ the rank $h(v)$ of this tree.
Intuitively, $h(v)$ measures the maximal number of $3$-edges that appear on paths from $v$.
Since $G < \kappa$ we have $h(v)<\kappa$, so $h(v)$ is a vertex in $U$.

By definition, $h$ satisfies that whenever $v \re 3 v'$ is an edge in $G$, then $h(v) > h(v')$, and whenever $v \re 2 v'$ is an edge in $G$, then $h(v) \geq h(v')$.
Stated differently, $h$ defines a morphism from $G$ to $U$, as required.

\begin{figure}[h]
    \begin{center}
        \includegraphics*[width=0.45\linewidth]{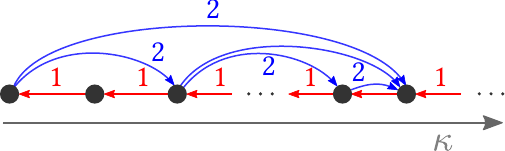}
        \caption{The graph $U$ for the B\"uchi objective. Most edges are excluded for readability.}
        \label{4-fig:universal_graph_buchi}
\commentAlt{Figure~\ref{4-fig:universal_graph_buchi}: A horizontal line with an arrow labeled 'k' on the right, and a repeating pattern of nodes with labeled arrows above and between them, extending indefinitely to the right.}
\commentLongAlt{Figure~\ref{4-fig:universal_graph_buchi}: The image displays a horizontal line with an arrow pointing right, labeled 'k'. Along this line, there is a sequence of filled circular nodes. From left to right, there are three visible nodes, followed by a dotted line indicating continuation, then two more visible nodes, and another dotted line indicating further continuation.

Arrows connect these nodes with labels:

A horizontal arrow points from the leftmost node to the second node, labeled '1'.
A horizontal arrow points from the third node to the second node, labeled '1'.
An arc-shaped arrow goes from the leftmost node to the third node, labeled '2'.
A horizontal arrow points from the third node to the node after the dotted line, labeled '1'.
An arc-shaped arrow goes from the third node to the last visible node on the right, labeled '2'.
A horizontal arrow points from the second-to-last visible node to the last visible node, labeled '1'.
An arc-shaped arrow from the second-to-last visible node to the very last visible node on the right, labeled '2'.
There is also a horizontal arrow from the last visible node pointing to the right, labeled '1', indicating continuation.}
    \end{center}
\end{figure}

\paragraph*{Example: Büchi.}
Let $C=\{1,2\}$ and $\Omega$ be the $\Buchi$ objective, meaning the set of words where $2$ has infinitely many occurrences.
Fix a cardinal $\kappa$.
Consider the graph $U$ over $\kappa$ given by $u \re{2} u'$ for every pair of vertices $u,u'$, and $u \re{1} u' \iff u > u'$ (see \Cref{4-fig:universal_graph_buchi}).

To prove that $U$ is $(\kappa,\Omega)$-universal, we start by showing that it satisfies the $\Buchi$ objective $\Omega$: all paths have infinitely many $2$'s.
Indeed, sequences of $1$'s force a strict decrease and are therefore finite.

Next, take a graph $G$ of cardinality $<\kappa$ which satisfies $\Omega$.
To any vertex $v$, we associate an ordinal $h(v)$ which counts the number of $1$'s before the next $2$; formally, $h(v)$ is the rank of the well-founded tree obtained from removing $2$-edges and then unfolding from $v$.
It satisfies $h(v) > h(v')$ whenever $v \re{1} v'$ is an edge, thus it is indeed a morphism from $G$ to~$U$.

\paragraph*{Monotone graphs.}
We consider \emph{ordered graphs}, which are simply graphs together with a linear order on their vertices.
We say that an ordered graph is \emph{monotone} if
\[
    u \geq u' \re c v \geq v' \implies u \re c v'.
\]
Note that the universal graphs for $\CoBuchi$ and $\Buchi$ above (\Cref{4-fig:universal_graph_cobuchi,4-fig:universal_graph_buchi}) are monotone.
Observe also that in a monotone graph, if $u \leq u'$, then the value of $u$ (with respect to some quantitative objective) is not greater than the value of $u'$.

\paragraph*{Neutral letters for quantitative objectives.}
Neutral letters are defined (and exemplified) in the preliminaries, for qualitative objectives.
For a quantitative objective $\Phi$, a letter $\varepsilon$ is neutral if for all $w \in C^\omega$,
\[
    \Phi(w) = \begin{cases}
        \Phi(w') &\text{ if the word $w'$ obtained from $w$ by removing $\varepsilon$'s is infinite} \\
        - \infty &\text{ otherwise}.
    \end{cases}
\]

\subsection{Characterisation of positionality}\label{4-subsec:characterisation-pos}

In~\Cref{4-fig:universal_graph_cobuchi,4-fig:universal_graph_buchi}, the proposed universal graphs provide a way to measure the quality of a vertex in a strategy: in \Cref{4-fig:universal_graph_cobuchi}, a vertex $v$ is of good quality if paths from $v$ visit few $3$-edges, and in \Cref{4-fig:universal_graph_buchi}, a vertex $v$ is of good quality if paths from $v$ visit a $2$-edge as quickly as possible.
Monotonicity of the universal graph formalises the intuition that this way of measuring the quality is sound.
To derive positionality by measuring strategies as above, we also require well-foundedness of the universal graph (any set of vertices admit a minimum), which formalises the requirement that for any vertex in the game (which corresponds to a set of vertices in the strategy), there is a best (\textit{i.e.} minimal) occurrence of this vertex in the strategy.

This intuition leads to the main result of this section.

\begin{theorem}\label{4-thm:characterisation_universal_graphs}
Let $\Phi\colon C^\omega \to \Rinfty$ be a quantitative objective.
If for all cardinals~$\kappa$, there exists a $(\kappa,\Phi)$-universal graph which is monotone and well-ordered, then $\Phi$ is uniformly positional (over arbitrary arenas).
The converse holds assuming $\Phi$ admits a neutral letter. 
\end{theorem}

\begin{remark}
In the statement above, $\kappa$ ranges over arbitrary cardinals.
Monotone universal graphs with respect to finite cardinals have also been studied, because they give rise to value iteration algorithms for solving the corresponding games.
For more details, see~\cite{Colcombet.Fijalkow.ea:2022}.
\end{remark}

\paragraph{Proof of the direct implication (from structure to positionality).}

Consider a game $\Game$, and let $\kappa$ be a cardinal strictly greater than $|\Paths(G)|$.
Our aim is to build a positional strategy which is uniformly optimal for Min in $\Game$.
Take a well-ordered monotone $(\kappa,\Phi)$-universal graph $U$.

To a Min strategy $\tau \colon \Paths(G) \to E$ from $v_0$ naturally corresponds a graph $G_\tau$ over $\Paths(G,v_0)$ with edges
\[
    \pi \re c \pi' \iff \pi' = \pi \re c v' \text{ and if $v \in \VE$ then } \sigma(\pi) = v \re c v'.
\]
For each strategy $\tau$, let $h_\tau$ be a $\Phi$-preserving morphism from $G_\tau$ to $U$.

Define $h$ to be the map assigning to each vertex $v$ in $G$ the minimal value of all possible~$h_\tau(\pi)$, where $\tau$ ranges over Min strategies and $\pi$ over paths ending in $v$.
Then $h(v)$ is a well-defined vertex of $U$ thanks to well-orderedness.
For each vertex $v$, we additionally fix a choice of a strategy $\tau(v)$ and a path $\pi(v)$ ending in $v$ where the minimum is met: $h_{\tau(v)}(\pi(v))=h(v)$.
Then we consider the positional strategy $\tau_\pos$ defined by assigning to $v \in \VMin$ the edge $\tau(v)(\pi(v))$.

We now show that $h$ defines a morphism from the graph $\Game[\tau_\pos]$ of $\tau_\pos$ to $U$.
Indeed, if $v \re c v'$ is an edge in $G$, then $\pi(v) \re c \pi(v)$ is an edge in $G_{\tau(v)}$, and therefore \[h_{\tau(v)}(\pi(v)) \re c h_{\tau(v)}(\pi(v) \re c v')\] is an edge in $U$.
Now the left term is $h(v)$ by definition, and the right term is $\geq h(v')$ in $U$ since $\pi(v)e$ is a path ending in $v'$.
We conclude that $h(v) \re c h(v')$ by monotonicity of $U$.

Since $h$ is a morphism, the value of $v$ in $\Game[\tau_\pos]$ is not larger than the value of $h(v)$ in $U$.
Now, for every $\varepsilon>0$, there is a strategy $\tau$ which is $\varepsilon$-optimal from $v$, which means that the value of the empty path $v$ in $G_\tau$ is at most the value of $v$ in $\Game$ plus $\epsilon$.
Since the morphism $h_\tau$ is $\Phi$-preserving, the same is true for the value of $h_\tau(v)$ in $U$ (here, $v$ stands for the empty path at $v$).
Now since $h(v)$ is smaller than $h(v)$ in $U$, it follows that the value of $v$ in $\Game[\tau_\pos]$ is not larger than the value of $v$ in $\Game$ plus $\varepsilon$.
We conclude by letting $\varepsilon$ go to zero.

\paragraph{Proof of the converse implication (from positionality to structure).}

We now prove the converse implication, so assume $\Phi$ is a uniformly positional quantitative objective which admits a neutral letter $\varepsilon$.
The crucial result is the following.

\begin{lemma}\label{4-lem:structuration}
Let $\Phi$ be a quantitative objective admitting a neutral letter $\varepsilon$ and let $H$ be a graph.
There exists a well-ordered monotone graph $H'$ with a $\Phi$-preserving morphism $H \to H'$.
\end{lemma}

It not hard to see that the lemma implies the theorem as follows.
Apply the lemma to $H$ being the disjoint union of all graphs of cardinality $<\kappa$, up to isomorphism.
Then, the obtained graph $H'$ is well-ordered, monotone, and $(\kappa,f)$-universal by composition of $\Phi$-preserving morphisms.

\begin{proof}[Proof of \Cref{4-lem:structuration}]
We build a game $\Game$ from the graph $H$ as follows.
The vertices of $\Game$ are those of $H$, which belong to Max, together with non-empty subsets of $H$, which are given to Min.
Edges within the copy of $H$ are just like in $H$, and we add edges of the form $v \re {\varepsilon} S$ and $S \re{\varepsilon} v$ whenever $v \in S$ (see \Cref{4-fig:choice_arena}).

\begin{figure}[ht]
    \begin{center}
        \includegraphics*[width=0.8 \linewidth]{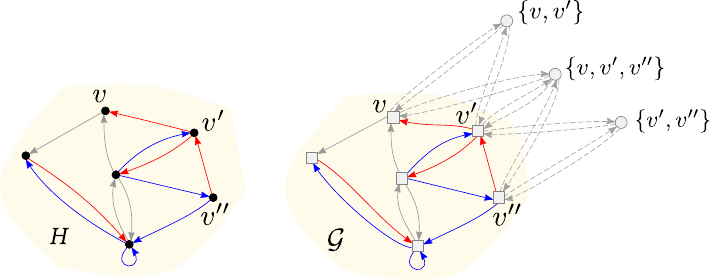}
        \caption{An example depicting the game in the proof of \Cref{4-lem:structuration}. Dashed edges are neutral. We represent only $3$ of the $2^6-1$ Min vertices.}
        \label{4-fig:choice_arena}
\commentAlt{Figure~\ref{4-fig:choice_arena}: Two related graphs, H and G, depicting transformations between nodes, with G having additional nodes representing sets. See long description.}
\commentLongAlt{Figure~\ref{4-fig:choice_arena}: The image displays two graphs, 'H' on the left and 'G' on the right, both contained within shaded, irregular shapes.

Graph H (left): This graph has five filled circular nodes. Three nodes are labeled 'v', 'v'', and 'v'''. There are multiple directed and bidirectional edges connecting these nodes, forming a complex network. One node has a self-loop.

Graph G (right): This graph appears to be an expansion or dual of graph H. It has square nodes corresponding to the circular nodes in H, also labeled 'v', 'v'', and 'v'''. It retains a similar core structure of directed and bidirectional edges and a self-loop, reflecting the connections in H. Additionally, G includes three new circular nodes, positioned above and to the right of the main graph. These new nodes represent sets of the original nodes: '{v, v''}', '{v, v'', v'''}', and '{v'', v'''}'. Directed arrows, some solid and some dashed, connect the square nodes within G to these new circular set nodes, indicating relationships between individual nodes and the sets they belong to.}
    \end{center}
\end{figure}

First, we claim that for any vertex $v$ of $H$, its value in $\Game$ is just as in $H$.
Indeed, the value is not smaller in $\Game$ since Max can just follow any path while remaining in the copy of $H$.
However, it is not larger since Min can choose to play back to vertex $v$, whenever the play reaches a subset $S$ by an edge $v \re \varepsilon S$; this has the neutral effect of adding two $\varepsilon$'s.

Now thanks to uniform positionality, this means that Min has a positional strategy $\tau$ ensuring the optimal value from each vertex.
Note that $\tau$ corresponds to a choice function over vertices of $H$: it assigns a chosen vertex to any non-empty set of vertices.
We let $H_1$ be the graph obtained from $H$ by adding every edge of the form $v \re \varepsilon v'$, whenever $v \in S$ and $\tau(S)=S \re{\varepsilon} v'$.

We claim that values in $H_1$ are just the same as in $H$.
Clearly, values are not smaller in $H_1$ since we just added edges, creating more paths.
Conversely, any path $\pi_1$ in $H_1$ can be converted to a path $\pi$ in $H$, where the label of $\pi$ is obtained from the label of $\pi_1$ by replacing some occurrences of $\varepsilon^2$ by $\varepsilon$; this does not affect the value of $\Phi$.
Note that in~$H_1$, the relation $\re \varepsilon$ has the property that for every non-empty subset $S$ of vertices, there is $v' \in S$ (namely, the one such that $\tau(S)=S \re \varepsilon v'$) such that for each $v \in S$, it holds that $v \re \varepsilon v'$; this property is already close to being a well-order, it simply lacks transitivity and anti-symmetry.

We let $H_2$ be obtained from $H_1$ by adding every edge of the form $v \re c v'$, whenever $v \re \varepsilon \dots \re \varepsilon \re c \re \varepsilon \dots \re \varepsilon v'$ occurs in $H_1$.
Again, it is easy to see that thanks to neutrality, values in $H_2$ are the same as in $H_1$.
Now over $H_2$, the relation $\re \varepsilon$ is transitive and moreover it satisfies the monotonicity requirement: $v \re \varepsilon \re c \re \varepsilon v'$ implies $v \re c v'$.

There remains to guarantee anti-symmetry, which poses no issue.
Indeed, equivalent vertices $v \re \varepsilon v' \re \varepsilon v$ in $H_2$ have the same incoming and outgoing edges; therefore, the quotient $H'=H_2 \ / \re{\varepsilon}$ is well-defined, well-ordered and monotone.
Now we have $H \re{} H_1 \re{} H_2 \re{} H'$ which concludes the proof of the lemma, and the theorem.
\end{proof}

\paragraph*{About neutral letters.}
The proof of the converse in \Cref{4-thm:characterisation_universal_graphs} used a neutral letter.
A quantitative objective $\Phi$ over $C$ admits a unique extension to an objective $\Phi^\varepsilon$ over $C\cup\{\varepsilon\}$ for which $\varepsilon$ is a neutral letter, obtained by setting $\Phi^\varepsilon(u')=\Phi(u)$ if $u'$ is obtained from an infinite word $u$ by adding (arbitrarily many) occurrences of $\varepsilon$, and $\Phi^\varepsilon(u \varepsilon^\omega)=-\infty$ for finite words $u$.

It follows from \Cref{4-thm:characterisation_universal_graphs} that existence of well-ordered monotone $(\kappa,\Phi)$-universal graphs is equivalent to uniform positionality of $\Phi^\varepsilon$.
However, it is not known whether uniform positionality of $\Phi$ entails uniform positionality of $\Phi^\varepsilon$.

\subsection{Examples}\label{4-subsec:examples}

We now give a few examples to illustrate how to use \Cref{4-thm:characterisation_universal_graphs} to establish positionality.

\paragraph*{Parity.}

We consider the parity objective over $C=\{1,2,\dots,d\}$, where $d$ is even, which generalises both the $\CoBuchi$ and $\Buchi$ examples from \Cref{4-subsec:universal-graphs}.
Fix a cardinal $\kappa$.
We write elements of $\kappa^{d/2}$ as tuples indexed by odd numbers $d-1,d-3,\dots,3,1$, and order them lexicographically.
Consider the graph $U$ over $\kappa^{d/2}$ given by edges
\[
    (u_{d-1},\dots,u_1) \re{p} (u'_{d-1},\dots,u'_1) \iff \begin{cases}
        (u_{d-1},\dots,u_{p+1}) \geq (u'_{d-1},\dots,u'_{p+1}) &\text{ if $p$ is even} \\
        (u_{d-1},\dots,u_{p}) > (u'_{d-1},\dots,u'_{p}) &\text{ if $p$ is odd}.
    \end{cases}
\]
We claim that $U$ is $(\kappa,\Parity)$-universal.

First, we should prove that $U$ satisfies $\Parity$.
Take an infinite path in $U$ and assume towards contradiction that the maximal priority seen infinitely often is an odd number $k$.
Then $(u_{d-1},\dots,u_k)$ only decreases along the path, with infinitely many strict decreases; this is not possible.

Second, we should prove that any graph of cardinality $< \kappa$ which satisfies $\Parity$ has a morphism into $U$, so take such a graph $G$.
To each vertex $v$ of $G$, we associate a tuple $h(v)=(h_{d-1}(v),\dots,h_3(v),h_1(v))$, where $h_k(v)$ is an ordinal counting the number of occurrences of the odd priority $k$ before a greater priority over paths from $v$ in $G$.
Formally, $h_k(v)$ is defined to be the rank of the well-founded tree obtained from $G$ by contracting edges with priority $<k$, cutting edges with priority $>k$ and unfolding at $v$.
It is a direct check that $h$ defines a morphism from $G$ to $U$.

Thanks to \Cref{4-thm:characterisation_universal_graphs}, this gives uniform positionality of $\Parity$.
In fact, this proof coincides with the original proof of Emerson and Jutla (see the bibliographic references at the end of the chapter), rephrased in the language of universal graphs.

\paragraph*{Energy.}
Recall the quantitative objective
\[
    \Energy^+(\rho)=\sup_k \sum_{i=0}^{k-1} \rho_i
\]
over $C=\Z$ (recall also that we take the point of view of Min).
Consider the graph $U$ over $\N \cup \{\infty\}$ with an edge $u \re{t} u'$ if and only if $u \geq u' + t$.
Note that vertex $\infty$ in $U$ has all possible outgoing edges (including self-loops), and no incoming edges besides self-loops.
In particular, it has value $\infty$.

We claim that in fact, for all $u \in \N \cup \{\infty\}$, vertex $u$ has value $u$ in $U$. This is already argued above for $u=\infty$, so let $u \in \N$.
Then paths from $u$ are of the form $u=u_0 \re{t_0} u_1 \re{t_1} \dots$, where the $u_i$'s belong to $\N$, and for each $i$, $u_i \geq u_{i+1} + t_i$.
This gives for any $k$ that $\sum_{i=0}^{k-1} t_i=u_0 - u_k \leq u_0=u$, therefore the value of $u$ is smaller than $u$.
Conversely, the path $u \re{u} 0 \re{0} 0 \re 0 \dots$ has value $u$ and therefore the value of $u$ is $u$.

Next, we prove that any graph has an $\Energy^+$-preserving morphism towards $U$, and therefore $U$ is $(\kappa,\Energy^+)$-universal for all cardinals $\kappa$.
Indeed, take a graph $G$, and map each vertex $v$ to its value.
Then consider an edge $e=v \re t v'$, and a path $\pi'$ from $v'$ witnessing the value $u'$ of $v'$.
Then $e \pi'$ is a path from $v$ with value $\geq u' + t $.
Therefore, the value $u$ of $v$ satisfies $u \geq u' + t$, so the map above is indeed a morphism; it is $\Energy^+$-preserving by definition.

\paragraph*{Finitely many letters occur infinitely often.}
In practice, for proving (or disproving) positionality of a given objective $\Omega$, it is often not too difficult to construct well-ordered monotone universal graphs by trial and error.
We give a detailed example on how such a method can help to \emph{disprove} the positionality of a given objective.

\begin{figure}[h]
    \begin{center}
        \includegraphics*[width=0.3 \linewidth]{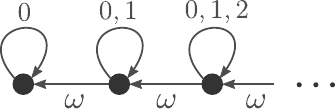}
        \caption{A first candidate for a universal graph.}
        \label{4-fig:first_candidate}
\commentAlt{Figure~\ref{4-fig:first_candidate}: A linear sequence of nodes with self-loops and connections, extending indefinitely to the right, illustrating a repetitive pattern of labeled transitions.}
\commentLongAlt{Figure~\ref{4-fig:first_candidate}: The image displays a horizontal line with three filled circular nodes visible, followed by a dotted line indicating continuation to the right. Each node has a self-loop with a label above it: the leftmost node's self-loop is labeled '0', the middle node's self-loop is labeled '0, 1', and the rightmost node's self-loop is labeled '0, 1, 2'. Directed arrows connect the nodes horizontally from right to left, and each of these arrows is labeled 'omega' below the line. This suggests a chain of states or processes with internal actions and a common transition mechanism between them.}
    \end{center}
\end{figure}

Consider the objective $\Omega=\{w \in \N^\omega \mid \Inf(w) \text{ is finite}\}$, where $\Inf(w) \subseteq \N$ is the set of letters occurring infinitely often in $w$.
We aim for a well-ordered monotone universal graph for $\Omega$ (for some large infinite cardinal).
First, notice that the graph $G_n$ with a single vertex $g_n$ and self-loops labelled by all integers $\leq n$ satisfies the objective.
Further, the graph $G$ comprised of the union of the $G_n$'s with all possible edges pointing from $g_i$ to $g_j$ when $i>j$ (see \Cref{4-fig:first_candidate}) also satisfies $\Omega$.

This gives us a first candidate for a universal graph: could it be that $G$ is universal?
Assume that it is the case.
Then in each (small) graph $H$ satisfying $\Omega$, there is a vertex $h_0$ which is mapped to a minimal position $g_n$ in $G$; observe that any path from $h_0$ should only be comprised of edges with label $\leq n$.
So is it the case that all graphs satisfying $\Omega$ admit a vertex from which all reachable edges are $\leq n$?

\begin{figure}[h]
    \begin{center}
        \includegraphics*[width=0.8 \linewidth]{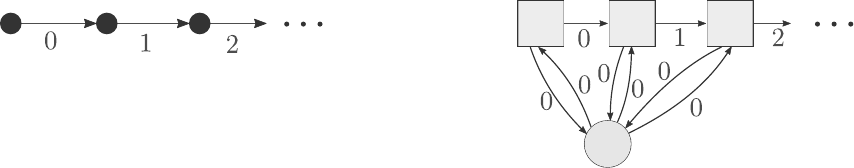}
        \caption{On the left the graph $P$, on the right a game disproving positionality of $\Omega$.}
        \label{4-fig:difficult_graph}
\commentAlt{Figure~\ref{4-fig:difficult_graph}: Two diagrams: a linear sequence of nodes and a diagram showing multiple rectangular nodes connected to a central circular node. See long description.}
\commentLongAlt{Figure~\ref{4-fig:difficult_graph}: The image presents two distinct diagrams. The left diagram shows a horizontal sequence of three filled circular nodes, with arrows pointing from left to right. The arrows are labeled '0', '1', and '2' respectively. A dotted line to the right indicates that the sequence continues indefinitely. The right diagram shows a horizontal sequence of three square nodes, also with arrows pointing from left to right, labeled '0', '1', and '2' respectively, and a dotted line indicating continuation. Below these square nodes, there is a central circular node. Multiple arrows, all labeled '0', connect each of the visible square nodes to this central circular node, with arrows going both from the square nodes to the circle and from the circle to the square nodes.}
    \end{center}
\end{figure}

No, for instance the graph $P$ comprised of only the path $p_0 \re 0 p_1 \re 1 p_2 \re 2 \dots$ satisfies $\Omega$ but does not admit such a vertex.
Therefore, $P \nrightarrow G$ and $G$ is not universal, so we should find a way of integrating $P$ to our construction.
A closer look at $P$ reveals that it seems difficult to endow it with a well-order so as to make it monotone: the only reasonable possible well-order over $P$ would be $p_0 < p_1 < \dots$, and making $P$ monotone would require adding edges $p_i \re i p_0$, and therefore including a bad path $p_0 \re 0 p_1 \re 1 p_0 \re 0 p_1 \re 1 p_2 \re 2 p_0 \re 0 \dots$.
By applying a (simplified variant of) the gadget in the proof of \Cref{4-lem:structuration}, one obtains the game from \Cref{4-fig:difficult_graph}, which disproves $\Omega$'s positionality.

\paragraph*{$\omega$-Büchi.}

We end with a more involved construction of a universal graph.
Consider the objective $\Omega=\{w \in \N^\omega \mid \Inf(w) \neq \emptyset\}$, which is a countable union of Büchi objectives.
Fix an ordinal $\alpha$, and consider elements of $\omega^\alpha$, which are finitely supported $\alpha$-sequences of natural numbers ordered with least-important coordinate first.
Let $U_\alpha$ be the graph over $\omega^\alpha$ given by edges
\[
    u \re{n} u' \iff u> u' \text{ or } \exists \lambda, [u_\lambda > n \text{ and } u_{\geq \lambda} \geq u'_{\geq \lambda}],
\]
where $s'=s_{\geq \lambda}$ is the sequence given by $s'_\beta=s_{\beta+\lambda}$.
It is not difficult to check that $U$ is monotone.
We claim that $U$ is $(\kappa,\Omega)$-universal.
\begin{claim}
The graph $U$ satisfies $\Omega$.
\end{claim}

To prove the claim, take an infinite path $\pi= u^0 \re {n_0} u^1 \re{n_1} \dots$ in $U$, and assume for contradiction that $w=n_0 n_1 \dots \notin \Omega$.
Let $\lambda_0$ be maximal such that $k_0=u^0_{\lambda_0}$ is non-zero.
Since $w \notin \Omega$, there is some suffix of the path which does not contain any edge $\leq k_0$.
Then observe that on this suffix, $u^i_{\geq \lambda_0}$ can not increase when $i$ grows, so it converges to some fixed value; let $i_0$ be large enough so that $u^{i_0}_{\geq \lambda_0}$ is constant.

Then iterate this reasoning by taking $\lambda_1$ to be the maximal coordinate $<\lambda_0$ such that $k_1=u^{i_0}(\lambda_1)$ is non-zero, restricting to a suffix which does not contain edges $\leq k_1$ and so on.
This way we build an infinite sequence $\lambda_0>\lambda_1>\dots$, obtaining a contradiction that proves the claim.

Towards proving that small graphs satisfying $\Omega$ have a morphism into $G_\alpha$ for some $\alpha$, define ranks of graphs as the smallest ordinal measure satisfying that
\begin{enumerate}
\item the rank of edgeless graphs is $0$;
\item if $G \to G'$ then $\rkUG(G) \leq \rkUG(G')$;
\item for any $n$, the graph $G'$ obtained from $G$ by adding all possible edges with weight $< n$ has rank $\rkUG(G') \leq \rkUG(G) +1$;
\item the graph $G_0+G_1+\dots$ obtained from an arbitrary ordinal sequence $G_0,G_1,\dots$ by adding to their disjoint union all edges from $G_\lambda$ to $G_\beta$ for $\lambda>\beta$, has rank $\rkUG(G_0+G_1+\dots) \leq \sum_{\lambda} G_\lambda$.
\end{enumerate}
We let $\rkUG(G)=\infty$ if it cannot be bounded by the above rules.
\begin{claim}
If $\rkUG(G)=\infty$ then $G$ does not satisfy $\Omega$.
\end{claim}
The claim is proved as follows.
Given a graph $G$ and a vertex $v$, let $G^v$ denote the restriction of $G$ to vertices reachable from $v$.
Let $G=G_0$ be a graph with $\rkUG(G)=\infty$.
First, we prove that for some $v$ we have $\rkUG(G^v)=\infty$.
Indeed, otherwise, well-order the vertices $v_0,v_1,\dots$ and note that $G \to G^{v_0}+G^{v_1}+\dots$; therefore, $\rkUG(G)$ should be defined thanks to rule 4, a contradiction.

Thus, pick a vertex $v_0$ such that $\rkUG(G^{v_0})=\infty$.
Then, let $G_1$ be obtained from $G^{v_0}$ be removing all $0$-edges, note that thanks to rule 3, it holds that $\rkUG(G_1)=\infty$.
Further, let $v_1$ be such that $\rkUG(G_1^{v_1})=\infty$.
Iterating this process, we construct by induction a path $v_0 \re{w_0} v_1 \re{w_1} \dots$ such that $w_i$ contains only edges $\geq i$.
Therefore, $G$ does not satisfy $\Omega$.

We continue with the following claim.
\begin{claim}
If $\rkUG(G)=\alpha$, then $G \to U_\alpha$.
\end{claim}
To prove the claim, we proceed by induction on the definition of the rank.
There are four cases.
\begin{enumerate}
\item If $G$ is edgeless, then $G$ maps into the edgeless graph with a single vertex $U_0$.
\item If $G \to G'$ and, by induction, $G' \to U_\alpha$, then $G \to U_\alpha$.
\item If $G$ is obtained from $G'$ by adding all edges of weight $< n$ and $G' \re {h'} U_\alpha$, then define $h\colon G \to \omega^{\alpha+1}$ by setting $h(v)=(h'(v),n)$, meaning that we append an extra element $u_{\alpha}=n$ to each sequence $u=h(v)=(u_\lambda)_{\lambda<\alpha}$.
It is a direct check that $h'$ defines a morphism $G \to U_{\alpha+1}$.
\item If $G=G_0+G_1+\ldots$ with $\alpha=\rkUG(G)=\rkUG(G_0)+\rkUG(G_1)+\ldots=\alpha_0+\alpha_1+\ldots$ and for all $\lambda$, $G_\lambda \re{h_\lambda} U_\lambda$.
Without loss of generality, we assume that the $h_\lambda(v)$'s are always non-zero.

Then we define $u=h(v)$, when $v \in G_\lambda$ and $u'=h_\lambda(v)$, by setting $u_\lambda=u'_\beta$ if $\lambda=\sum_{\lambda'<\lambda} \alpha_{\lambda'} + \beta$ and $u_\lambda=0$ otherwise.
If $v \re n v'$ is an edge in $G$, then either $v \in G_\lambda$ and $v' \in G_{\lambda'}$ with $\lambda>\lambda'$, in which case $h(v)>h(v')$, or $v,v' \in G_\lambda$ for some $\lambda$, in which case $h_{\lambda}(v) \re {n} h_\lambda(v')$ in $U_{\alpha_\lambda}$.
In both cases, we obtain that $h(v) \re n h(v')$ is an edge in $U_\alpha$, and thus $G \re{h} U_\alpha$.
\end{enumerate}

Finally, an easy induction yields $|\rkUG(G)| \leq |G|$.
Therefore, it follows from the two claims above that $U_{\kappa}$ is $(\kappa,\Omega)$-universal.

\subsection{Positionality over finite-degree arenas}\label{4-subsec:finite-degree-arenas}

One may wonder what happens if the well-foundedness hypothesis is dropped in \Cref{4-thm:characterisation_universal_graphs}.
It turns out that this provides a characterisation of objectives which are uniformly positional over arenas in which vertices owned by Min (resp.\ Eve) have finitely many outgoing edges.
We say that such arenas have \emph{finite Min-degree} (resp.\ \emph{Eve-degree}).

\begin{theorem}\label{4-thm:characterisation_universal_graphs_finite_degree}
    Let $\Phi\colon C^\omega \to \Rinfty$ be an objective.
    If for all $\kappa$, there exists a monotone $(\kappa,\Phi)$-universal graph, then $\Phi$ is uniformly positional over arenas with finite Min-degree.
    The converse holds assuming $\Phi$ admits a neutral letter. 
\end{theorem}

\begin{proof}[Proof sketch]
    We start just as in the proof of \Cref{4-thm:characterisation_universal_graphs}: take a game $\game$ and a monotone universal graph $U$ for some large enough cardinal so that $G_\tau$ has a $\Phi$-preserving morphism ${h_\tau}$ towards $U$ for any strategy $\tau$.

    Consider a Min vertex $v$.
    Since $v$ has only finitely-many outgoing edges, there exists an outgoing edge $\sigma(v)$ from $v$ such that in any strategy $\tau$ and for any path $\pi$ ending in $v$, there exists a strategy $\tau'$ and a path $\pi'$ ending in $v$ such that $\tau'(\pi')=\sigma(v)$ and $h_{\tau'}(\pi') \leq h_{\tau}(\pi)$.
    Then it is relatively straightforward to check that $\sigma$ is a positional optimal strategy.

    For the converse, we mimic the proof of \Cref{4-thm:characterisation_universal_graphs} except that in the game $\Game$, we keep only Min vertices corresponding to finite sets (we could also keep only the pairs).
    In particular, Min vertices have finite degree.
    Then the same proof provides a totally ordered (but not necessarily well-founded) universal graph.
\end{proof}

We give two natural examples of objectives which are positional on finite Min-degree arenas but not on arbitrary ones.

\paragraph*{Example: Discounted payoff.}
Let $\lambda \in (0,1)$ and consider the discounted-payoff objective over $C=[-N,N] \subseteq \R$ given by
\[
    \DiscountedPayoff_\lambda(\rho)=\lim_k \sum_{i=0}^{k-1} \rho_i \lambda^i \in [-S,S],
\]
where $S=N(1-\lambda)^{-1}$.
It is not positional over arbitrary arenas; to see this, simply consider the game with a single vertex belonging to Eve, and all loops with weights $>0$ (in fact, this game does not even admit optimal non-positional strategies).
However, when restricted to finite-degree arenas, positional optimal strategies exist.
The standard proof, which is presented in~\Cref{5-chap:payoffs}, uses Banach's fixed-point theorem (and applies to arenas with finite Eve-degree), but this can also be seen as a non-well-founded monotone universal graph.

Let $U$ be the graph over $[-S,S] \subseteq \R$ given by $u \re t u' \iff t \leq u - \lambda u'$.
Clearly, $U$ is monotone (but not well-founded).
It is an easy check that for any graph $G$, the map assigning its value to each vertex defines a (value-preserving) morphism $G \to U$, and therefore $U$ is $(\kappa,\DiscountedPayoff)$-universal for any $\kappa$.

\paragraph*{Example: $\omega$-coBüchi.}
Consider the objective $\Omega=\{w \in \N^\omega \mid \Inf(w)=\emptyset\}$, which is dual to the $\omega$-Büchi objective studied in \Cref{4-subsec:examples}.
To see that $\Omega$ is not positional on arbitrary arenas, simply consider the arena with a single Eve-vertex and all possible loops.
We now prove that it is positional on arenas with finite Eve-degree via \Cref{4-thm:characterisation_universal_graphs_finite_degree}.

Fix a cardinal $\kappa$ and consider the set $\kappa^\omega$ of maps $u\colon \omega \to \kappa$.
Given such a $u$ and $i \in \omega$, we let $u_{<i}$ (resp.\ $u_{\leq i}$) denote the restriction of $u$ to $[0,i)$ (resp.\ $[0,i]$).
We order $\kappa^\omega$ lexicographically:
\[
    u > u' \iff \exists i < \omega, [u_{<i} = u'_{<i} \text{ and } u(i) \geq u'(i)].
\]
We raise the reader's attention on the fact that the above is not the standard ordinal exponentiation because maps are not assumed to be finitely supported.
In fact, this order is not well-founded; for instance, $(1,1,1,\dots)>(0,1,1,\dots)>(0,0,1,\dots)>\dots$.

Let $U$ be the graph over $\kappa^\omega$ given by $u \re n u' \iff u_{\leq n} > u'_{\leq n}$; clearly, $U$ is monotone.
Observe that $U$ satisfies $\Omega$: if an infinite path $u^0 \re{n_0} u^1 \re{n_1} \dots$ does not satisfy $\Omega$, then for the smallest $n$ which repeats infinitely often, we have $u^0_{< n} \geq u^1_{< n} \geq \dots$ with infinitely many strict inequalities, which is a contradiction. 

There remains to prove that any graph $<\kappa$ satisfying $\Omega$ has a morphism towards $U$.
Let~$G$ be such a graph.
Then for each $i$, we let $h(v)(i)$ be the ordinal counting the number of occurrences of $i$-edges from $v$ (formally, it is the rank of the well-founded tree obtained by unfolding at $v$ and contracting $\neq i$-edges).
Then, it is a direct check that $h$ defines a morphism $G \to U$.

\subsection{Kopczy\'nski's conjecture}\label{4-subsec:Kopcz_conj}

Kopczy\'nski conjectured that prefix-independent positional objectives are closed under (finite or countable) unions, and gave an example showing that uncountable unions of prefix-independent positional objectives may fail to be positional.
Note that prefix-independence is required: for instance, the qualitative objectives $\{a^\omega\}$ and $\{b^\omega\}$ are positional, but their union is not.
Recall also that uniform positionality is granted for prefix-independent positional objectives by \Cref{1-lem:from_positional_to_uniformly_positional}.

We now discuss some results around this conjecture.

\paragraph*{The conjecture fails over finite arenas.}
We first present a proof that the conjecture fails over finite arenas.
This requires the notion of \emph{ordered group}, which we now recall.

An \emph{ordered group} is a group $G$ together with a linear order $\leq$ such that left-multiplication is order-preserving: for all group elements $h,g,g'$, it holds that $g \leq g' \implies hg \leq hg'$.
From an ordered group $(G,\leq)$, we define an objective $\Decreasing_{(G,\leq)}$ over $C=G$, comprised of all words admitting a decreasing sequence of prefixes; formally,
\[
    \Decreasing_{(G,\leq)} = \{w \in G^\omega \mid (\prod_{i=0}^{k-1} w_i)_{k \in \N} \text{ has an infinite decreasing subsequence}\}.
\]
For any ordered group $G$, the objective $\Decreasing_{(G,\leq)}$ is bi-positional over finite arenas, which may be shown using \Cref{4-thm:1to2_lift}.

The free group with two generators $F_2$ is defined to be the set of finite words over alphabet $\{a,a^{-1},b,b^{-1}\}$, up to addition or removal of infixes of the form $xx^{-1}$ or $x^{-1}x$ for $x \in \{a,b\}$.
The group multiplication is concatenation.
The following proposition uses the non-trivial fact that free groups admit orderings.

\begin{proposition}\label{4-prop:kopczynski_fails_finite_arenas}
Let $(F_2,\leq)$ be the ordered free group with two generators.
The objective
\[
    \Omega=\Decreasing_{(F_2,\leq)} \cup \Decreasing_{(F_2,\geq)}
\]
is not positional over finite arenas.
\end{proposition}

\begin{proof}
Observe that $\Omega$ is comprised of all words $w \in F_2^\omega$ such that prefixes of $w$ take infinitely many different values when multiplied in $F_2$.
Indeed, clearly words whose prefixes take finitely many values do not belong to $\Omega$; conversely, a word with infinitely many prefixes has either a strictly growing or a strictly decreasing subsequence by Ramsey's theorem.
Consider the arena from \Cref{4-fig:kozachinskiy}.

\begin{figure}[ht]
    \begin{center}
        \includegraphics*[width=0.43\linewidth]{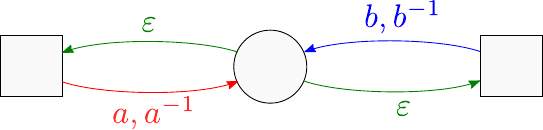}
        \caption{The arena for the proof of \Cref{4-prop:kopczynski_fails_finite_arenas}.}
        \label{4-fig:kozachinskiy}
\commentAlt{Figure~\ref{4-fig:kozachinskiy}: A central circular node connected to two square nodes on either side by bidirectional arrows with various labels.}
\commentLongAlt{Figure~\ref{4-fig:kozachinskiy}: The image shows a central circular node with two square nodes on either side. To the left, a bidirectional arrow connects the central circle to the left square. The top arrow is labeled 'epsilon', and the bottom arrow is labeled 'a, a^-1'. To the right, a bidirectional arrow connects the central circle to the right square. The top arrow is labeled 'b, b^-1', and the bottom arrow is labeled 'epsilon'.}
    \end{center}
\end{figure}

Clearly, Eve wins by alternating both sides, as prefixes of the resulting path admit no cancellations, and therefore take infinitely many values.
However, she cannot win positionally: if she commits to, say, playing to the left, then Adam can simply alternate between $a$ and $a^{-1}$.
\end{proof}

\paragraph*{The conjecture holds for countable unions of $\Sigma_2^0$ objectives.}

An objective $\Omega \subseteq C^\omega$ belongs to $\Sigma_2^0$ if and only if it can be written as $\Omega=\Finite(L)$ for some language $L \subseteq C^*$ of finite words, where
\[
    \Finite(L) = \{w \in C^\omega \mid \text{ finitely many prefixes of $w$ belong to $L$}\}.
\]
We have the following positive result.

\begin{proposition}\label{4-prop:countable_unions_sigma2}
Let $\Omega_0,\Omega_1,\ldots \subseteq C^\omega$ be countably many prefix-independent positional objectives in $\Sigma_2^0$ admitting neutral letters.
Then, their union is positional.
\end{proposition}

\begin{proof}
Let $\Omega$ be the union of $\Omega_0,\Omega_1,\dots$.
Fix a cardinal $\kappa$; we aim to build a monotone well-ordered $(\kappa,\Omega)$-universal graph.
For each $i$, let $L_i \subseteq C^*$ be such that $\Omega_i=\Finite(L_i)$, and let $U_i$ be a monotone well-ordered $(\kappa,\Omega_i)$-universal graph.
Then, let $\hat U$ be obtained from the disjoint copy of the $U_i$'s by adding all edges from $U_j$ to $U_i$ whenever $j>i$.
Clearly, $\hat U$ is monotone and well-founded.
Moreover, any infinite path in $\hat U$ ultimately remains in some~$U_i$, therefore $\hat U$ satisfies $\Omega$ by prefix-independence.

We define $U$ to be obtained by setting $\kappa$ disjoint copies of $\hat U$ next to one another, and adding all edges pointing from each copy to (strictly) smaller ones.
Again, $U$ is monotone and clearly satisfies $\Omega$.
There remains to prove that any graph $<\kappa$ satisfying $\Omega$ has a morphism towards $U$.
Take such a graph $G$, and let $G^v$ denote the restriction of $G$ to vertices reachable from some vertex $v$.

We first prove that for some vertex $v$ in $G$, there is a morphism $G^v \to U_i$ for some $i$.
Indeed, assume that this is not the case, and pick a vertex $v_0$.
Since $G^{v_0}$ does not have a morphism towards $U_0$, and $U_0$ is $(\kappa,\Omega_0)$-universal, it must be that $G^{v_0}$ does not satisfy $\Omega_0$: there is a path from $v_0$ with infinitely many prefixes in $L_0$.
Let $\pi_0\colon v_0 \re{w_0} v_1$ be a path from~$v_0$ such that $w_0 \in L_0$.

Likewise, $G^{v_1}$ does not satisfy $\Omega_1$, and thus there is a path from $v_1$ with label $w'_1 \notin \Omega_1$, which implies by prefix-independence that $w_0w'_1 \notin \Omega_1$, and therefore there is a finite path $\pi_1\colon v_1 \re{w_1} v_2$ such that $w_0w_1 \in L_1$.
We continue this construction in a round-robin fashion (for instance, $010120123\dots$) to construct a path from $v_0$ whose label $w$ has infinitely many prefixes in each $L_i$.
This contradicts the fact that $G$ satisfies~$\Omega$.

Therefore, there is some vertex $v_0$ such that $G^{v_0} \to U_i$ for some $i$, and $G^{v_0} \re{h_0} \hat U$.
Now, we map $G$ to the first copy of $\hat U$ in $U$, and conclude by transfinite induction\footnote{In~\cite{Ohlmann:2023}, the property satisfied by $\hat U$ is called \emph{almost universality}. One can always construct a universal graph from an almost universal graph by transfinite induction~\cite[Lemma 4.5]{Ohlmann:2023}.}.
\end{proof}

As a consequence, we may easily establish positionality of mean-payoff objectives over arbitrary arenas.

\begin{corollary}\label{4-cor:mean-payoff-positional}
    The qualitative objective
    \[
        \MeanPayoff^{+}_{<0} = \{\rho \in \Z^\omega \mid \limsup_k \dfrac{1}{k} \sum_{i=0}^{k-1} \rho_i <0\}
    \]
    is uniformly positional over arbitrary arenas.
\end{corollary}

We raise the reader's attention on the fact that the choice of $\limsup$ as well as the strict equality are required for the result to hold.

\begin{proof}
For $n>0$, consider the objective 
\[
    \TiltedEnergy_{n}=\{\rho \in \Z^\omega \mid \exists N, \forall k, \sum_{i=0}^{k-1} (\rho_i+1/n) < N\}.
\]
For each $n$, renaming letters using the map $w \mapsto nw -1$ embeds $\TiltedEnergy_n$ into $\Energy^+_{<\infty}$, and therefore these objectives are positional.
Now, it suffices to observe that these objectives belong to $\Sigma_2^0$, are prefix-independent, and that $\MeanPayoff^{+}_{<0}$ is their countable union.
\end{proof}

Another class of objectives where Kopczy\'nski's conjecture is known to hold is $\omega$-regular objectives; more discussion below.

\subsection{Positionality of $\omega$-regular objectives}\label{4-subsec:positional_omega-reg}

Recall that an objective is $\omega$-regular if it can be recognised by a deterministic parity automaton.
In this section, we consider parity automata with $\varepsilon$-transitions; these are denoted by $q \re{\varepsilon:x} q'$, where $x \in \N$ is some priority.

We say that an automaton with priorities in $\{0,\dots,d+1\}$, where $d$ is even, is \emph{$\varepsilon$-complete} if
\begin{itemize}
\item the relations $\re{\varepsilon:1},\re{\varepsilon:3},\dots,\re{\varepsilon:d+1}$ all define total preorders, each refining the previous one; and 
\item for each even $x \in \{0,2,\dots,d\}$, $q \re{\varepsilon:x}q'$ holds if and only if $q' \re{\varepsilon:x+1} q$ does not hold. Stated differently, the relation $\re{\varepsilon:x}$ is the strict preorder corresponding to the (non-strict) preorder $\re{\varepsilon:x+1}$.
\end{itemize}

\begin{figure}[h]
    \begin{center}
        \includegraphics*[width=\linewidth]{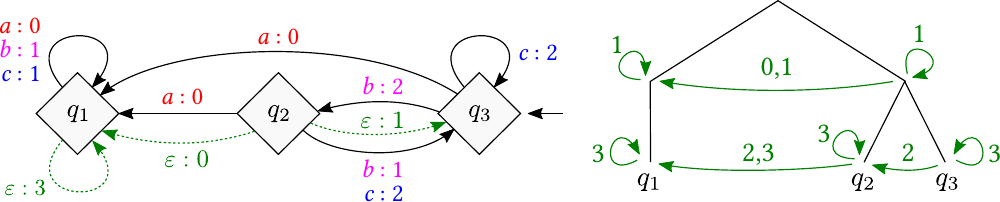}
        \caption{On the left, an automaton recognising the set of words with infinitely many $a$'s, or with no occurrence of $a$ and finitely many occurrences of the factor $bb$. All $\varepsilon$-transitions depicted on the right-hand side may be added to the automaton (like the three dashed ones) without augmenting the language.
        Therefore, the automaton is $\varepsilon$-completable.}
        \label{4-fig:epsilon-complete}
    \end{center}
\end{figure}

The decreasing sequence of total preorders is better visualised as a tree of height $d/2$ whose leaves correspond to states of the automaton; we refer to \Cref{4-fig:epsilon-complete} below for an example.
On an intuitive level, $q \re {\varepsilon:x} q'$ means that ``$q$ is much better than $q'$'', since one may, at any point, move from $q$ to $q'$ and be rewarded with an even priority $x$ on the way.
On the contrary, $q' \re{\varepsilon:x+1} q$ means that ``$q'$ is not much worse than $q$'', since one may at any point move from $q'$ to $q$ at the cost of reading an odd priority $x+1$.
In other words, in a "$\varepsilon$-complete automaton", one may say that $q$ and $q'$ are comparable for priority $x$.

An automaton is $\varepsilon$-completable if one may add $\varepsilon$-transitions so as to make it $\varepsilon$-complete without augmenting the language.

Positionality is well-understood for $\omega$-regular objectives thanks to the following characterisation.

\begin{theorem}\label{4-thm:positionality_omega_reg}
Let $\Omega$ be an $\omega$-regular objective.
The following are equivalent:
\begin{itemize}
\item $\Omega$ is positional over finite arenas of Eve;
\item $\Omega$ is recognised by a deterministic $\varepsilon$-completable automaton;
\item $\Omega$ is positional over arbitrary (including infinite) arenas.
\end{itemize}
\end{theorem}

For instance, the language defined by the automaton in \Cref{4-fig:epsilon-complete} is positional.
Note that \Cref{4-thm:positionality_omega_reg} implies finite-to-infinite and one-to-two-player lifts for $\omega$-regular objectives.
Moreover, positionality of an $\omega$-regular language given by a deterministic parity automaton is decidable in polynomial time.
Last, a stronger variant of Kopczy\'nski's conjecture in known to hold for $\omega$-regular objectives, as a consequence of \Cref{4-thm:positionality_omega_reg}.

\begin{proposition}
Let $\Omega_1$ and $\Omega_2$ be $\omega$-regular positional objectives, and assume that $\Omega_1$ is prefix-independent.
Then $\Omega_1 \cup \Omega_2$ is positional.
\end{proposition}

This proposition is incomparable with \Cref{4-prop:countable_unions_sigma2}, since $\omega$-regular objectives span a strict subclass of $\Delta_3^0$ which does not contain $\Sigma_2^0$.


\section{Generalisations to memory requirements}
\label{4-sec:memory_generalisations}
In this chapter, we have so far focused on general results for bi-positionality or positionality over finite or infinite arenas.
In this last section, we show that many of these results can be generalised to results about "finite memory" (cf.\ \Cref{1-sec:memory}).

We will mostly skip proofs, as the simpler arguments for the positional cases already highlight multiple of the technical challenges.
References to the complete proofs are found in the bibliographic references at the end of the chapter.

Before addressing these results, we spend some time discussing the definition of \emph{memory} and some natural variations of it in \Cref{4-subsec:memory_models}.

\subsection{Memory models} \label{4-subsec:memory_models}

Let $X$ be a set (which will usually be either the set of edges of a "game", or the set of colours labelling those edges). 
An \emph{$X$-memory structure} $\mem$ is a "deterministic" "automaton" over $X$ without an acceptance condition.
Formally, an "$X$-memory structure" is a tuple $\mem = (M, m_0, \delta)$, where $M$ is a set of memory states, 
$m_0 \in M$ is the initial state, and $\delta \colon M \times X \to M$ is the update function, which is extended to $\delta^* \colon X^* \to M$ by 
$\delta^*(\varepsilon) = m_0$ and $\delta^* (\rho x) = \delta(\delta^*(\rho), x)$.
The \emph{size} of a memory structure is its number of states.

\subsubsection{Different types of memory}

\paragraph{General memory.}
This first definition is the one given in~\Cref{1-chap:introduction}.
Let $\game$ be a "game".
A \emph{general memory structure} for $\game$ is an "$E$-memory structure" $\mem = (M, m_0, \delta)$, where $E$ is the set of edges of the arena of $\game$.
Let $\sigma\colon \Paths \to E$ be a "strategy" for "Eve"/"Max".
We say that $\sigma$ is \emph{based on $\mem$} if there is a function
\[ \nextmove\colon \VE \times M \to E \]
such that $\sigma(\pi) = \nextmove(\last(\pi), \delta^*(\pi))$.
In this case, we use expressions such as ``$\sigma$ uses $|M|$ memory states''.

Let $\Phi\colon C^\omega \to \Rinfty$ be an "objective". We define the \emph{general memory requirements} of $\Phi$, written $\genMemReq(\Phi)$, as the minimal $m\in \N\cup\{\infty\}$ such that
for every game $\game$ using $\Phi$ as an "objective", "Max" has an "optimal strategy" "using $m$ memory states".
We note that this induces a definition of "memory requirements" for "qualitative objectives".
Also, observe that we adopt a \emph{uniform} version of the memory: the strategy using $m$ memory states is optimal from any given vertex.
As in the case of "positionality", we may talk about the \emph{memory requirements of $\Phi$ over a subclass of arenas} (such as finite or without uncoloured edges).
The general memory requirements over finite arenas of $\Phi$ are denoted $\genMemReqFin(\Phi)$.

We note that the definition of "memory requirements" is an asymmetric one; we take the point of view of "Max"/"Eve" and focus on her optimal strategies.

\paragraph{Chromatic memory.}
A \emph{chromatic memory} is a memory structure that can only observe the information about the \emph{colours} produced in a game, and not about the particular \emph{edges} that are taken.

Formally, let $\game$ be a "game" and $\col\colon E\to C$ a colouring of its edges.
A \emph{chromatic memory structure} for $\game$ (with respect to $C$) is a "$C$-memory structure".
The update function $\delta\colon M\times C \to M$ of such a "$C$-memory structure" induces a function $\hat{\delta}\colon M\times E\to M$ by $\hat{\delta}(m, e) = \delta(m, \col(e))$.
Therefore, a "chromatic memory structure" $\mem$ for $\game$ induces a "general memory structure" $\hat{\mem}$.
We say that a "strategy" $\sigma$ in $\game$ is \emph{based on the "chromatic memory" $\mem$} if it is based on the "general memory" $\hat{\mem}$.

We define the \emph{chromatic memory requirements} of an objective $\Phi\colon C^\omega \to \Rinfty$, denoted $\chromMemReq(\Phi)$, as the minimal $m\in \N\cup\{\infty\}$ such that
for every game $\game$ using $\Phi$ as an "objective", "Max" has an "optimal strategy" based on a "chromatic memory" with $m$ states.
The \emph{chromatic memory requirements} over finite arenas of $\Phi$, defined analogously, are denoted~$\chromMemReqFin(\Phi)$.

\paragraph{Arena-independent memory.}
Let $\Phi\colon C^\omega \to \Rinfty$ be an "objective".
A "$C$-memory structure" $\mem$ is an \emph{arena-independent memory} for $\Phi$ if for all games $\game$ using $\Phi$ as objective, "Max" has an "optimal strategy" based on $\mem$.

We write $\arIndMemReq(\Phi)$ for the minimal size of an "arena-independent memory" for $\Phi$ (we let $\arIndMemReq(\Phi)=\infty$ if such an "arena-independent memory" does not exist).
Similarly, we let $\arIndMemReqFin(\Phi)$ be the minimal size of an "arena-independent memory" that "Max" can use to "implement" "optimal strategies" in finite arenas.
 
We say that \emph{$\mem$ suffices for $\Phi$ for a player over a class of arenas} if, for all arenas from this class, this player has a "strategy" based on $\mem$ that is "optimal" for $\Phi$ from every vertex.
This definition includes the "uniformity" of the strategies.

\subsubsection{Comparison between the models}
We have presented three models of memory: "general", "chromatic", and "arena-independent", by successively adding further restrictions. It is therefore clear that, for any objective~$\Phi$:
\[ \genMemReq(\Phi) \leq \chromMemReq(\Phi) \leq \arIndMemReq(\Phi). \]

We show that the second inequality is in fact an equality: we can assume that a single "chromatic memory structure" of minimal size suffices to play optimally in all games.
However, the first inequality might be strict and arbitrary large, already for "Muller objectives".

\begin{proposition}\label{4-prop:chromatic_equals_arInd}
	Let $\Phi\colon C^\omega \to \Rinfty$ be an "objective". 
	Then, \[\chromMemReq(\Phi) = \arIndMemReq(\Phi).\]
	Moreover, if $C$ is finite, $\chromMemReqFin(\Phi) = \arIndMemReqFin(\Phi)$.
\end{proposition}
\begin{proof}
	It is clear that $\chromMemReq(\Phi) \leq \arIndMemReq(\Phi)$ (resp.\ $\chromMemReqFin(\Phi) \leq \arIndMemReqFin(\Phi)$). We show that this inequality cannot be strict.
	Let $k = \chromMemReq(\Phi)$, and assume by contradiction that $k<\arIndMemReq(\Phi)$ (resp.\ $k<\arIndMemReqFin(\Phi)$).
	Let $\mathrm{Memories}_{\leq k}$ be the set of all "$C$-memory structures" with at most $k$ states. We remark that if $C$ is finite, so is the set~$\mathrm{Memories}_{\leq k}$.
	As $k<\arIndMemReq(\Phi)$, for each memory structure $\mem$ of size less than or equal to~$k$, there is a game $\game_\mem$ for which "Max" has no "optimal strategy" based on $\mem$.
	Let~$\game'$ be the disjoint union of all the games $\game_\mem$, for $\mem\in \mathrm{Memories}_{\leq k}$ (we note that, if $C$ is finite, this is a finite arena). 
	As $k = \chromMemReq(\Phi)$, there is a "chromatic memory" $\mem'$ with no more than $k$ states implementing an "optimal strategy" in $\game'$. 
	In particular, $\mem'$ implements an "optimal strategy" in $\game_{\mem'}$, a contradiction.
\end{proof}

\begin{proposition}\label{4-prop:chromatic_neq_general}
	There is a "Muller objective" $\Omega\subseteq C^\omega$ with $\genMemReq(\Omega) < \chromMemReq(\Omega)$.
\end{proposition}
\begin{proof}
	Let $C_n = \set{1,\dots,n}$ be a set of $n$ colours, and consider the family of subsets
	\[ \mathcal{F}_n = \set{A\subseteq C \mid |A| = 2 }. \]
	The "Muller objective" associated to $\mathcal{F}_n$ is
	\[ \Muller(\mathcal{F}_n) = \set{\rho\in C^\omega \mid \rho \text{ visits exactly two different colours infinitely often}}. \]
	Using the characterisation from \Cref{3-thm:characterisation_Zielonka_tree}, it is an easy exercise to check that the "general memory requirements" of $\Muller(\mathcal{F}_n)$ are exactly $2$, for all $n\geq 2$.
	
	We claim that the "chromatic memory requirements" of $\Muller(\mathcal{F}_n)$ are at least $n$. 
	We will show that this objective admits no "arena-independent memory" of size strictly less than~$n$, and conclude by \Cref{4-prop:chromatic_equals_arInd}.
	
	Assume by contradiction that $\mem = (M,m_0,\delta)$ is an "arena-independent memory" for $\Muller(\mathcal{F}_n)$ and $|M|<n$. For each $c\in C_n$, let $m_c$ be a state of $\mem$ in a cycle labelled~$c^+$. Since $|M|<n$, there must be two different colours $a,b\in C_n$ such that $m_a=m_b$. Let $k_1, k_2>0$ such that $\delta^*(m_a,a^{k_1}) = m_a$ and $\delta^*(m_a,b^{k_2}) = m_a$. Let $u\in C^*$ be such that $\delta^*(m_0,u) = m_a$.
	Consider the game $\game$ in \Cref{4-fig:memory_2_dif_colours}. It is clear that "Eve" "wins" $\game$, but not with a strategy based on $\mem$, as the memory structure would always be at the state $m_a$ whenever we are in vertex $v$, so it would always take the same edge from there.\qedhere
	\begin{figure}[!ht]
		\centering
		\begin{tikzpicture}			
			\draw ($(0,0)$) node[s-eve] (choice) {$v$};
			\draw ($(choice)-(2.1,0)$) node[s-eve] (0) {};
			
			\path[->]
			(0) edge[squish] node[above] {$u$} (choice)
			(choice) edge[squishloop, out=85, in=30,loop] node[anchor=south west, pos=0.5] {$a^{k_1}$} (choice)
			(choice) edge[squishloop, out=-85, in=-30,loop] node[anchor=north west, pos=0.5] {$b^{k_2}$} (choice);
		\end{tikzpicture}
		\caption{Game won by "Eve" by alternating the loops $a^{k_1}$ and $b^{k_2}$.}
		\label{4-fig:memory_2_dif_colours}
\commentAlt{Figure~\ref{4-fig:memory_2_dif_colours}: A diagram showing a circular node connected to a central node labeled 'v', which has two wavy self-loops with labels.}
\commentLongAlt{Figure~\ref{4-fig:memory_2_dif_colours}: The image depicts a circular node on the left, connected by a wavy arrow labeled 'u' to a central circular node labeled 'v'. The central node 'v' has two wavy self-loops: one above, labeled 'a^k1', and one below, labeled 'b^k2'. Both self-loops are directed back to the node 'v'.}
	\end{figure}
\end{proof}

In \Cref{3-thm:characterisation_Zielonka_tree} was presented a characterisation of the "general memory requirements" of Muller objectives over "partially-coloured arenas" (\textit{i.e.}, arenas with a \emph{partial edge labelling}; cf.\ Section~\ref{1-sec:qualitative_games}).
We show that this last hypothesis is necessary, as the memory requirements over partially or fully-coloured arenas might differ.

\begin{proposition}\label{4-prop:separation-memory-epsilon}
	There is a "Muller objective" $\Omega\subseteq C^\omega$ such that the "general memory requirements" of $\Omega$ over fully-coloured arenas are strictly smaller than its "general memory requirements" over partially-coloured arenas.
\end{proposition}
\begin{proof}
	Let $C_n = \set{1,\dots,n}$ be a set of $n$ colours, and consider the family of subsets
	\[ \mathcal{F}_n = \set{A\subseteq C \mid |A| >1 }. \]
	The "Muller objective" $\Muller(\mathcal{F}_n)$ consists therefore in the set of sequences that contain at least two different colours infinitely often.
	Its "Zielonka tree" is depicted in \Cref{4-fig:Zielonka_tree_atLeastTwo}. 
	The characterisation from \Cref{3-thm:characterisation_Zielonka_tree} implies that the general memory requirements of $\Muller(\mathcal{F}_n)$ over partially-coloured arenas are $n$.
	
	\begin{figure}
		\centering
		\begin{tikzpicture}[scale=1.1]
			\node[s-eve] (root) at (5,3.5) {$\ \set{1,2,\dots,n}\ $};
			
			\node[s-adam, minimum width=0.7cm, minimum height=0.7cm] (1) at (2,2) {$\ \set{1}\ $};
			
			\node[s-adam, minimum width=0.7cm, minimum height=0.7cm] (2) at (3.5,2) {$\ \set{2}\ $};
			
			\node[scale=2] (dots) at (5,2) {$\cdots$};
			
			\node[s-adam, minimum width=0.7cm, minimum height=0.7cm] (n) at (7,2) {$\ \set{n}\ $};
			
			\path
			(root) edge (1)
			(root) edge (2)
			(root) edge (n);
		\end{tikzpicture}
		\caption{The Zielonka tree for $\mathcal{F}_n = \set{A\subseteq C \mid |A| >1 }$.}
		\label{4-fig:Zielonka_tree_atLeastTwo}
\commentAlt{Figure~\ref{4-fig:Zielonka_tree_atLeastTwo}: A tree diagram with a circular root node containing a set, branching to multiple square leaf nodes, each containing a single element set.}
\commentLongAlt{Figure~\ref{4-fig:Zielonka_tree_atLeastTwo}: The image displays a tree structure with a single circular root node at the top. This root node contains the set '{1, 2, ..., n}'. From this root, branches extend downwards to multiple square leaf nodes. The visible leaf nodes from left to right are '{1}', '{2}', followed by an ellipsis '...', and finally '{n}'. This indicates that the root node is connected to 'n' individual leaf nodes, each representing a single element from the set in the root.}
	\end{figure}
	
	We claim that the general memory requirements of $\Muller(\mathcal{F}_n)$ over fully-coloured arenas are $2$.
	
	Let $\game$ be a fully-coloured game using the objective $\Muller(\mathcal{F}_n)$.
	Thanks to "prefix-independence", we can restrict our study to Eve's "winning region", so we assume without loss of generality that she wins from every vertex in $\game$.
	We associate to each vertex $v\in \VE$ a colour $c(v)\in C_n$ such that there is an outgoing transition from $v$ coloured with $c(v)$ (such a colour exists by the fully-coloured assumption). We obtain in this way a partition $\VE= V_1 \uplus \dots \uplus V_n$, with $v\in V_{c(v)}$.
	We fix a mapping $\sigma_0\colon\VE\rightarrow E$ picking one outgoing edge labelled with $c(v)$ from $v$.
	For each $x\in C_n$, we consider the reachability objective $\Omega_{\neg x} =$ ``reach a colour different from $x$''.
	Since Eve can ensure to reach two different colours in $\game$ from any vertex, she can in particular ensure the objective $\Omega_{\neg x}$.
	We fix a positional strategy $\sigma_x$ ensuring to see some colour $y\neq x$ and coinciding with $\sigma_0$ outside $V_x$. 
	
	We define a "general memory structure" for $\game$ with two states $m_0, m_1$ as follows: the state~$m_0$ is used to remember that we have to see the colour $x$ corresponding to the component $V_x$ that we are in. As soon as we arrive to a vertex controlled by Eve, we use the next transition to accomplish this and we change to state $m_1$ in $\mem$. The state $m_1$ serves to follow the positional strategy $\sigma_x$ reaching one colour different from $x$. We change to state~$m_0$ if we arrive to some state in $\VE$ not in $V_x$ (this ensures that we will see one colour different from $x$), or if Eve produces a colour different from $x$ staying in the component $V_x$. If she does not produce this colour and we do not go to a vertex in $\VE \setminus V_x$, that means that (since $\sigma_x$ ensures that we will see a colour different from $x$), Adam will take some transition coloured with some colour different from $x$.
\end{proof}

\subsection{One-to-two-player lift with memory}
Recall \Cref{4-thm:1to2_lift}: if an objective is uniformly bi-positional over finite \emph{one-player} arenas, then it is uniformly bi-positional over finite arenas.
This actually also holds if we replace ``positional strategies'' in this statement by ``strategies using memory $\mem$'' where $\mem$ is a fixed chromatic memory structure.

\begin{theorem}[One-to-two-player lift with memory] \label{4-thm:1to2_lift_memory}
	Let $\Phi$ be an objective and $\mem$ be a chromatic memory structure.
	If $\mem$ "suffices for $\Phi$ for both players over finite one-player arenas", then $\mem$ "suffices for $\Phi$ for both players over finite arenas".
\end{theorem}

It may seem surprising that we consider the same chromatic memory structure for both players; for a given objective, memory requirements of each player may vary wildly (see for instance Rabin and Streett games).
However, even if memory requirements are different in one-player games, this result still gives a useful information: if chromatic memory structure $\mem_1$ (resp.\ $\mem_2$) suffices to play optimally in the finite one-player arenas of Max (resp.\ Min), then the direct product $\mem_1 \times \mem_2$ also suffices for both players in their one-player arenas (any strategy based on $\mem_1$ or $\mem_2$ is also be based on $\mem_1 \times \mem_2$).
Hence, we can conclude that $\mem_1 \times \mem_2$ suffices to play optimally for both players in all finite arenas.
Under these specific hypotheses, it is unknown whether $\mem_1$ (resp.\ $\mem_2$) always suffices for Max (resp.\ Min) in all finite arenas.
In other words, it is unknown whether the product with the memory structure for the one-player arenas of the opponent is necessary in the two-player arenas of a player.

\begin{example} \label{4-ex:FMlift}
	Let $C = \set{\bot, \top} \times \Z$.
	We consider a quantitative objective $\Phi \colon C^\omega \to \Rinfty$ that is in some way a lexicographic combination of a reachability objective and a mean-payoff objective, with the priority given to the reachability objective along the first component:
	\[
		\Phi((a_1, z_1)(a_2, z_2)\ldots) =
		\begin{cases}
			-\infty &\text{if $a_i = \bot$ for all $i\ge 1$},\\
			\MeanPayoff^+(z_1z_2\ldots) &\text{otherwise}.
		\end{cases}
	\]
	We want to study the memory requirements of this objective using \Cref{4-thm:1to2_lift_memory}.
	We show that the memory structure $\mem$ with two states remembering whether $\top$ has already been seen suffices for both players over finite arenas.
	Memory structure $\mem$ is depicted in \Cref{4-fig:FMlift} along with two examples of one-player arenas (one of each player) in which $\mem$ suffices to play optimally, but positional strategies would not suffice.
	
	Over her finite one-player arenas, Max can always play optimally with~$\mem$.
	From any given state, if $\top$ was not seen yet, Max aims for the occurrence of $\top$ that then allows for the largest possible mean payoff, which can be obtained with a positional strategy.
	If $\top$ was already seen, Max simply aims for maximising the mean payoff.
	
	Over his finite one-player arenas, Min can also play optimally with $\mem$: if $\top$ was not seen yet, Min plays in a positional way for $\Safety(\top)$ if possible and, if not or if $\top$ was already seen, Min aims for minimising the mean payoff.
	
	By \Cref{4-thm:1to2_lift_memory}, memory structure $\mem$ suffices to play optimally over finite arenas for both players.
\end{example}
\begin{figure}
	\centering
	\begin{tikzpicture}
		\draw (0,0) node[diamond7] (0) {};
		\draw ($(0)+(2.2,0)$) node[diamond7] (1) {};
		\path[->]
		($(0.west)-(.5,0)$) edge[] (0)
		(0) edge[] node[above] {$\set{\top}\times\Z$} (1)
		(0) edge[out=60,in=120,loop] node[anchor=south] {$\set{\bot}\times\Z$} (0)
		(1) edge[out=-30,in=30,loop] node[anchor=west] {$C$} (1);
		
		\draw ($(1)+(3.1,0)$) node[s-eve] (2) {};
		\path[->]
		(2) edge[out=60,in=120,loop] node[anchor=south] {$(\bot,1)$} (2)
		(2) edge[out=-120,in=-60,loop] node[anchor=north] {$(\top,0)$} (2);
		
		\draw ($(2)+(2.2,0)$) node[s-adam] (3) {};
		\draw ($(3)+(2.2,0)$) node[s-adam] (4) {};
		\path[->]
		(3) edge[] node[above] {$(\top,0)$} (4)
		(4) edge[out=-120,in=-60,loop] node[anchor=north] {$(\bot,1)$} (4)
		(4) edge[out=60,in=120,loop] node[anchor=south] {$(\top,0)$} (4);
	\end{tikzpicture}
	\caption{Memory structure $\mem$ used in \Cref{4-ex:FMlift} (left).
		Arena in which Max needs~$\mem$ to play optimally for $\Phi$ (center).
		Arena in which Min needs~$\mem$ to play optimally in a uniform way for $\Phi$ (right).
		Note that positional strategies suffice to play optimally in a non-uniform way over finite one-player arenas of Min in general.}
	\label{4-fig:FMlift}
\commentAlt{Figure~\ref{4-fig:FMlift}: A set of four automata-like diagrams, each showing nodes with self-loops or transitions to other nodes, with various labels. See long description.}
\commentLongAlt{Figure~\ref{4-fig:FMlift}: The image contains four distinct diagrams arranged in two rows.

Top Left: A diamond-shaped node with an incoming arrow from the left, indicating a starting point. This node has a self-loop labeled '{$ \perp $} x Z'. A directed arrow connects this node to another diamond-shaped node on the right, labeled '{$ T $} x Z'. The second diamond node has a self-loop labeled 'c'.
Top Right: A circular node with a bidirectional self-loop. The top arrow is labeled '($\perp$,1)' and the bottom arrow is labeled '($\top$,0)'.}
\end{figure}

In \Cref{4-thm:1to2_lift_memory}, we considered the hypothesis of a fixed memory structure for both players, which is stronger than the existence of optimal finite-memory strategies for both players (for which the memory structure and its size may depend on the arena).
If we consider the weaker assumption with finite-memory strategies, such a general one-to-two-player lift does not hold~\cite[Section~3.4]{Bouyer.Roux.ea:2022}.

\subsection{From memory to automata}
We have discussed in \Cref{4-sec:positional_infinite} the central place of the parity objectives: they are bi-positional over arbitrary arenas (\Cref{2-thm:parity}), they are the only such "prefix-independent" qualitative objectives (\Cref{4-thm:char_positionality_infinite_qualitative_pref-indep}), and, more generally, any bi-positional objective can be recognised by a parity automaton built on top of its automaton of residuals (\Cref{4-thm:char_positionality_infinite_qualitative_general}).
In this section, we show that we can express a more general connection between the parity automata and, not only positionality, but memory requirements.

As a consequence of \Cref{1-lem:automata_reduction} (\Cref{1-chap:introduction}) and the "bi-positionality" of parity objectives, we obtain that a deterministic parity automaton recognising an objective $\Omega$ can be used (as a memory structure) to implement optimal strategies for both players in any game with objective $\Omega$.
In particular, any $\omega$-regular objective is finite-memory determined.

\begin{proposition}\label{4-prop:parity_is_memory}
	Let $\Automaton = (Q, q_0, \delta, A)$ be a deterministic parity automaton.
	Let $\mem_\Automaton = (Q, q_0, \delta)$ be the chromatic memory structure underlying $\Automaton$.
	Memory structure $\mem_\Automaton$ "suffices for objective $L(\Automaton)$ for both players over all arenas".
\end{proposition}

We claim that it is possible to go in the other direction.
For a given objective, from a chromatic memory structure $\mem$ that suffices for both players, it is possible to recover (with a small blow-up) a deterministic parity automaton recognising the objective.
This result makes use of the \emph{"automaton of residuals" $\prefixClassifier$} of an objective $\Omega$, defined in \Cref{4-subsec:general-bipositional-infinite}.

\begin{theorem} \label{4-thm:from_memory_to_parity_automaton}
	Let $\Omega$ be a qualitative objective and $\mem$ be a chromatic memory structure.
	If $\mem$ "suffices for $\Omega$ for both players over all arenas", then $(i)$ $\Omega$ has finitely many "residuals" and $(ii)$ $\Omega$ is recognised by a deterministic parity automaton on top of $\prefixClassifier \times \mem$.
\end{theorem}

A corollary of this result is a characterisation of the $\omega$-regular languages through a strategic perspective.

\begin{corollary}
	A qualitative objective $\Omega$ is $\omega$-regular if and only if some "chromatic memory structure" "suffices for $\Omega$ for both players over all arenas".
\end{corollary}
\begin{proof}
	Any $\omega$-regular objective is recognised by a deterministic parity automaton.
	Such an automaton, as discussed in \Cref{4-prop:parity_is_memory}, can be used as a memory structure to play optimally in all arenas for both players.
	
	Conversely, if $\mem$ "suffices for $\Omega$ for both players over all arenas", then $\Omega$ can be recognised by a deterministic parity automaton by \Cref{4-thm:from_memory_to_parity_automaton}.
	Hence, $\Omega$ is $\omega$-regular.
\end{proof}

\subsection{Universal graphs and memory}

In this section, we illustrate how monotone universal graphs (defined in \Cref{4-sec:positionality}) can be extended to prove upper bounds on the memory of given objectives.

We now consider \emph{partially ordered $C$-graphs}, which are $C$-graphs together with a partial order on their vertices.
Such a graph is \emph{monotone} if
\[
    u \geq u' \re c v \geq v' \implies u \re c v',
\]
just as in \Cref{4-sec:positionality}.
The \emph{width} of a partially ordered $C$-graph is the least upper bound on the size of its antichains.

Next, we should adapt the definition of \emph{universality}.
Given a quantitative objective $\Phi$, we say that a tree-to-graph morphism $h\colon T \to G$ \emph{preserves the value of the root} if the value of $h(t_0)$ in $G$, where $t_0$ is the root of $T$, is not greater than the value of $t_0$ in $T$.
Given a cardinal $\kappa$, we say that a graph $U$ is \emph{$(\kappa,\Phi)$-universal for trees} if all trees of cardinality $< \kappa$ admit a morphism towards $U$ which preserves the value of the root.

\begin{remark}
For a well-ordered (\textit{i.e.} well-founded and totally ordered) monotone graph, as in \Cref{4-thm:characterisation_universal_graphs}, and for an infinite cardinal $\kappa$, the notions of being $(\kappa,\Phi)$-universal and $(\kappa,\Phi)$-universal for trees coincide.
\end{remark}

\begin{remark}
We often work with prefix-independent qualitative objectives $\Omega$.
In this case, the definitions can be simplified as follows: a graph $U$ is $(\kappa,\Omega)$-universal for trees if it satisfies $\Omega$ and all trees of cardinality $< \kappa$ which satisfy $\Omega$ have a morphism towards $U$.
\end{remark}

Monotone universal graphs characterise a variant of memory called \emph{$\varepsilon$-memory}, for which the player is not allowed to update the memory state when an uncoloured edge $\varepsilon$ is read.
We may now state the characterisation.

\begin{theorem}\label{4-thm:universal_graphs_for_memory}
Let $\Phi \colon C^\omega \to \Rinfty$ be a quantitative objective and $m \in \N$.
If for all cardinals~$\kappa$, there exists a $(\kappa,\Phi)$-universal graph which is monotone, well-founded, and of width $\leq m$ (resp.\ which is monotone and of width $\leq m$), then $\Phi$ requires $\leq m$ states of $\varepsilon$-memory over arbitrary arenas (resp.\ over arenas with finite Min-degree).
The converse holds assuming $\Phi$ is qualitative and admits a neutral letter.
\end{theorem}

The proof is an extension of that of \Cref{4-thm:characterisation_universal_graphs} to the setting of memory.

\begin{example}\label{4-ex:example_memory_muller}
Consider the Muller objective
$
    \Omega = \{w \in \{a,b\}^\omega \mid \Inf(w) = \{a,b\}\}.
$
It is easy to see that $\Omega$ requires at least two states of memory, for instance over the one-vertex game controlled by Eve with the two possible self-loops.
Fix a cardinal $\kappa$, and consider the graph $U$ over $\{a,b\} \times \kappa$ given by edges 
\[
    (x,\lambda) \re y (x',\lambda') \iff [x=y=x' \text{ and } \lambda > \lambda'] \text{ or } x \neq y = x',
\]
and the order $(x,\lambda) \geq (x',\lambda') \iff [x = x' \text{ and } \lambda \geq \lambda']$ (see \Cref{4-fig:universal_graph_muller}).

\begin{figure}[h]
\begin{center}
\includegraphics*[width=0.75\linewidth]{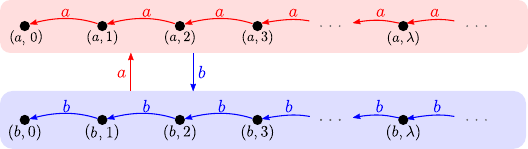}
\caption{The universal graph $U$ in \Cref{4-ex:example_memory_muller}.}
\label{4-fig:universal_graph_muller}
\end{center}
\end{figure}

Clearly $U$ is monotone and has width $2$.
One can check that $U$ satisfies $\Omega$, and a proof that $U$ is $(\kappa,\Omega)$-universal for trees follows the same lines as for the B\"uchi objective (see \Cref{4-subsec:universal-graphs}).
We conclude that $\Omega$ requires two states of memory (and even of $\varepsilon$-memory).
\end{example}

As an application of \Cref{4-thm:universal_graphs_for_memory}, we may characterise the memory requirements of \emph{closed} objectives.
An objective $\Omega \subseteq C^\omega$ is \emph{closed} if it can be written as 
\[
    \Omega=\Safety(S)=\{w \in C^\omega \mid w'\notin S \text{ for any prefix } w' \text{ of } w\},
\]
for some language $S \subseteq C^*$.

\begin{theorem}\label{4-thm:memory_of_closed_objectives}
Let $\Omega$ be a closed objective, and consider its set of residuals $\resSetO$ ordered by inclusion.
\begin{enumerate}
\item If $\resSetO$ has antichains of size $m$, then $\Omega$ requires $\geq m$ states of memory, already over finite-degree arenas.
\item If antichains of $\resSetO$ have size $\leq m$, then $\Omega$ requires $\leq m$ states of memory (even of $\varepsilon$-memory) on arenas of finite Eve-degree.
\item If, moreover, $\resSetO$ is well-founded, then the previous statement extends to arbitrary arenas.
\end{enumerate}
\end{theorem}

\begin{proof}
\begin{enumerate}
\item Take an antichain $\{u_1^{-1}\Omega,\dots,u_m^{-1}\Omega\}$, and for each $i \neq j$, fix an infinite word $w_{i,j} \in u_i^{-1} \Omega \setminus u_j^{-1}\Omega$.
Consider the game with $m+1$ Adam vertices $v_0,v_1,\ldots,v_m$ and one Eve vertex $e$, with all transitions $v_0 \re{u_i} e$, $e \re{\varepsilon} v_i$ and $v_i \re{w_{i,j}}$ (additional vertices within these disjoint finite and infinite paths are added adequately).
See \Cref{4-fig:safety-lower-bound}.

\begin{figure}[ht]
    \begin{center}
    \begin{tikzpicture}
		\draw (0,0) node[s-adam] (v0) {$v_0$};
		\draw ($(v0)+(2.2,0)$) node[s-eve] (e) {$e$};
        \draw ($(v0)!.5!(e)$) node[scale=.8,yshift=-4pt] (vdots0) {$\vdots$};

		\draw ($(e)+(2.2,1.5)$) node[s-adam] (v1) {$v_1$};
        \draw ($(e)!.5!(v1)$) node[scale=.8,yshift=-6pt,xshift=10pt] (vdots1) {$\vdots$};
		\draw ($(e)+(2.2,0)$) node[s-adam] (vi) {$v_i$};
		\draw ($(e)+(2.2,-1.5)$) node[s-adam] (vm) {$v_m$};
        \draw ($(e)!.5!(vm)$) node[scale=.8,yshift=-2pt,xshift=10pt] (vdotsm) {$\vdots$};

		\draw ($(vi)+(2.5,.8)$) node[rotate=17.7] (vi1) {\footnotesize$\cdots$};
		\draw ($(vi)+(2.5,0)$) node (vi2) {\footnotesize$\cdots$};
		\draw ($(vi)+(2.5,-.8)$) node[rotate=-17.7] (vi3) {\footnotesize$\cdots$};

        \draw ($(vi)!.5!(vi1)$) node[scale=.7,yshift=-7pt,xshift=30pt] (vdotsi1) {$\vdots$};
        \draw ($(vi)!.5!(vi3)$) node[scale=.7,yshift=-1pt,xshift=30pt] (vdotsi3) {$\vdots$};

		\draw ($(v1)+(1.25,.4)$) node[rotate=17.7] (v11) {\tiny$\cdots$};
		\draw ($(v1)+(1.25,0)$) node (v12) {\tiny$\cdots$};
		\draw ($(v1)+(1.25,-.4)$) node[rotate=-17.7] (v13) {\tiny$\cdots$};

		\draw ($(vm)+(1.25,.4)$) node[rotate=17.7] (vm1) {\tiny$\cdots$};
		\draw ($(vm)+(1.25,0)$) node (vm2) {\tiny$\cdots$};
		\draw ($(vm)+(1.25,-.4)$) node[rotate=-17.7] (vm3) {\tiny$\cdots$};

		\path[->]
		(v0) edge[squish,out=23,in=157] node[above=1pt] {$u_1$} (e)
		(v0) edge[squish,out=-23,in=-157] node[below=1pt] {$u_m$} (e)
		(e) edge[out=45,in=-170] node[above left=-2pt] {$\varepsilon$} (v1)
		(e) edge[] node[above] {$\varepsilon$} (vi)
		(e) edge[out=-45,in=170] node[below left=-2pt] {$\varepsilon$} (vm)
        (vi) edge[squish] node[above,xshift=8pt,yshift=2pt] {\small$w_{i,1}$} (vi1)
        (vi) edge[squish] node[above=-1.3pt,xshift=8pt] {\small$w_{i,j}$} (vi2)
        (vi) edge[squish] node[below,xshift=8pt,yshift=-2pt] {\small$w_{i,m}$} (vi3)
        (v1) edge[squish] (v11)
        (v1) edge[squish] (v12)
        (v1) edge[squish] (v13)
        (vm) edge[squish] (vm1)
        (vm) edge[squish] (vm2)
        (vm) edge[squish] (vm3);
    \end{tikzpicture}
    \end{center}
    \caption{The game for the proof of the first item.}
    \label{4-fig:safety-lower-bound}
\commentAlt{Figure~\ref{4-fig:safety-lower-bound}: A diagram showing a central circular node 'e' with multiple wavy inputs from a source node and multiple outputs branching out to various destinations. See long description.}
\commentLongAlt{Figure~\ref{4-fig:safety-lower-bound}: The image depicts a central circular node labeled 'e'. On the left, a square node 'v0' is shown, from which multiple wavy arrows point towards 'e'. These inputs are labeled 'u1' to 'um', with a dotted line in between to indicate more inputs. From the central node 'e', three main branches extend to the right, each terminating in a square node labeled 'v1', 'vi', and 'vm' respectively, with dotted lines indicating more intermediate branches. Each of these branches is connected to 'e' by a curved arrow labeled 'epsilon'. From each of these square nodes ('v1', 'vi', 'vm'), multiple wavy arrows emanate and extend to the right, followed by dotted lines, indicating multiple outputs. For the 'vi' node, these output arrows are specifically labeled 'wi,1', 'wi,j', and 'wi,m'. The overall diagram resembles a neuron or a processing unit with multiple inputs and outputs.}
\end{figure}

Then, Eve wins with $m$ states of memory by choosing the edge $e \re \varepsilon v_i$ whenever Adam picked $v_0 \re {u_i} e$.
However, if less than $m$ memory states are used by a strategy~$\sigma$, then by the pigeonhole principle, there are $i \neq j$ and $k$ such that $v_0 \re {u_i} e \re \varepsilon v_k$ and $v_1 \re {u_j} e \re \varepsilon v_k$ are consistent with $\sigma$, and thus, assuming without loss of generality that $k \neq i$, Adam wins by playing $v_0 \re {u_i} e \re \varepsilon v_k \re {w_{k,i}}$.

\item Consider the partially ordered graph $U$ over $\resSetO \setminus \{\emptyset\} \cup \{\top\}$ (ordered by inclusion and with maximal element $\top$) with all possible outgoing edges from $\top$, and edges $[u] \re c [v] \iff v^{-1} \Omega \subseteq (uc)^{-1} \Omega$.
Then $U$ is monotone, and has width $\leq m$.
It is an easy check that vertex $[u]$ satisfies the objective $u^{-1} \Omega$ in $U$; in particular, $[\epsilon]$ satisfies $\Omega$.

Therefore, thanks to \Cref{4-thm:universal_graphs_for_memory}, it suffices to prove that any tree $T$ has a morphism towards $U$ which preserves the value of the root.
Take a tree $T$ and let $t_0$ denote the root.
If $t_0$ does not satisfy $\Omega$, then preserving the value of the root is trivial, so simply map all $T$ to $\top$.

So assume that $t_0$ satisfies $\Omega$, and define let $h(t)=[u]$ where $u$ labels the unique path from $t_0$ to $t$.
Then $h$ is a morphism, and $h(t_0)=[\varepsilon]$ which satisfies $\Omega$, so $h$ preserves the value of the root.

\item Follows directly from the previous item, and \Cref{4-thm:universal_graphs_for_memory}.\qedhere
\end{enumerate}
\end{proof}


\section*{Bibliographic references}
\label{4-sec:references}
Arguably the first result related to positionality was Shapley's proof of existence of stationary strategies in stochastic discounted games~\cite{Shapley:1953} (we refer to \Cref{7-chap:stochastic} for more details on stochastic games).
The relevance of the question of strategy complexity was highlighted by Rabin in 1958, who showed that there are games in which one player can force a victory, but for which no computable winning strategy exists~\cite{Rabin:1958}. In 1969, Büchi and Landweber proved that $\omega$-regular winning objectives are finite-memory determined over finite games~\cite{Buchi.Landweber:1969}.
In 1979, Ehrenfeucht and Mycielski provided the first proof of bi-positionality of mean-payoff games~\cite{Ehrenfeucht.Mycielski:1979}. 

Interest in strategy complexity increased due to a quest for a simplified proof of Rabin's theorem~\cite{Rabin:1969}: In 1982, Gurevich and Harrington (and independently B\"uchi~\cite{Buchi:1977}) explicitly described "memory structures" for Muller games over infinite tree-shaped arenas; in 1991, Emerson and Jutla proved bi-positionality of parity games~\cite{Emerson.Jutla:1991} (the same result was obtained independently by Mostowski~\cite{Mostowski:1991}), and in 1994, Klarlund proved the "positionality" of "Rabin objectives"~\cite{Klarlund:1991}.
In 1998, Zielonka showed that the class of positional (resp.\ bi-positional) Muller objectives is exactly the class of Rabin objectives (resp.\ parity objectives), both over finite and infinite games~\cite{Zielonka:1998}.
Extending this result, Dziembowski, Jurdziński, and Walukiewicz characterised the precise "general" memory requirements of Muller objectives~\cite{Dziembowski.Jurdzinski.ea:1997}.

After the results of Zielonka, there was a growing interest in understanding positionality.
In 2005, Gimbert and Zielonka characterised objectives that are bi-positional over finite arenas by means of two properties called \emph{monotonicity} and \emph{selectivity}~\cite{Gimbert.Zielonka:2005}. As a consequence, they obtained the one-to-two-player lift stated in \Cref{4-thm:1to2_lift}.
The proof we presented here is closer to the one in~\cite{Gimbert.Zielonka:2009}; it is conceptually simpler (avoids going through monotonicity and selectivity) and, as is shown in that paper, can be generalised to a result about stochastic games as well.
In 2006, Colcombet and Niwiński proved that parity objectives are the only "prefix-independent" "bi-positional" objectives over infinite arenas~\cite{Colcombet.Niwinski:2006} (corresponding to \Cref{4-thm:char_positionality_infinite_qualitative_pref-indep}).
The generalisation of this result to non-prefix-independent objectives (\Cref{4-thm:char_positionality_infinite_qualitative_general}), was first stated by Casares and Ohlmann~\cite{Casares.Ohlmann:2024}; the necessity of recognisability by the automaton of residuals for positionality (and more generally, for finite-memory determinacy) was proven by Bouyer, Randour and Vandenhove~\cite{Bouyer.Randour.ea:2023}.

The first thorough study of "positionality" was carried out in Kopczyński's PhD thesis~\cite{Kopczynski:2008} (where it is called \emph{half-positionality}).
He presents several sufficient conditions for "positionality" of "prefix-independent" qualitative objectives, including the "submixing" property ("submixing" "qualitative objectives" are therein called \emph{concave}).
The result we showed in \Cref{4-thm:submixing_positional} is an extension of one such result to quantitative objectives.
The sufficient condition of \Cref{4-thm:submixing_positional} has been weakened in~\cite{Bianco.Faella.ea:2011} for "qualitative objectives", mostly relaxing "prefix-independence" to a property allowing finite prefixes to influence the outcome of the game, but in a totally ordered fashion.
\Cref{4-thm:submixing_positional} has been generalised to MDPs~\cite{Gimbert:2007} (also extending it for the first time to "quantitative objectives") and then to stochastic games~\cite{Gimbert.kelmendi:2023}: if an objective is prefix-independent and submixing, it is even "positional" (in \emph{pure} strategies) over finite \emph{stochastic} arenas.
The results from \Cref{4-sec:quantitative} on quantitative objectives are novel.

Universal graphs were first introduced by Colcombet and Fijalkow~\cite{Colcombet.Fijalkow:2019} as a generalisation of the notion of universal trees discussed in \Cref{2-sec:value_iteration}.
The characterisation of positionality based on monotone universal graphs (\Cref{4-thm:characterisation_universal_graphs}) is due to Ohlmann~\cite{Ohlmann:2023}.

A central question in Kopczyński's thesis~\cite{Kopczynski:2006} is whether "prefix-independent" "positional" objectives are closed under union, discussed in \Cref{4-subsec:Kopcz_conj}.
The counterexample for finite arenas we presented is described by Kozachinskiy~\cite{Kozachinskiy:2023}.
Ohlmann and Skrzypczak proved the conjecture for $\Sigma^0_2$ objectives~\cite{Ohlmann.Skrzypczak:2024}, and Casares and Ohlmann proved it for $\omega$-regular languages~\cite{Casares.Ohlmann:2024}.

We mention two other tools for positionality that were omitted from this chapter.
First, for "prefix-independent" objectives, the work of Aminof and Rubin~\cite{Aminof.Rubin:2017} gives a sufficient condition for "uniform positional determinacy" that focuses on the properties of winning cycles.
Second, positionality of the discounted-payoff objective and generalisations of it was discussed at length in~\cite{Kozachinskiy:2023}, which gives three different proofs of its positionality.

Kopczyński introduced the idea of chromatic and arena-independent memory~\cite{Kopczynski:2006}, and showed that the chromatic and arena-independent memory requirements coincide for all objectives (stated in \Cref{4-prop:chromatic_equals_arInd}). 
He left open the question on whether these moreover coincide with the general memory requirements. The separation between general and chromatic memory requirements (\Cref{4-prop:chromatic_neq_general}) is due to Casares~\cite{Casares:2022}, as well as the example from \Cref{4-prop:separation-memory-epsilon}.
The extension of the one-to-two-player lift to constant chromatic memory (\Cref{4-thm:1to2_lift_memory}) was shown in~\cite{Bouyer.Roux.ea:2022}, and it also admits an extension to stochastic games~\cite{Bouyer.Oualhadj.ea:2023}.
The link between chromatic memory structures and deterministic parity automata (\Cref{4-thm:from_memory_to_parity_automaton}) was shown in~\cite{Bouyer.Randour.ea:2023}.
The generalisation of monotone universal graphs to characterise the memory of objectives was presented in~\cite{Casares.Ohlmann:2023,Casares.Ohlmann:2025}.

\section*{Acknowledgments}
Antonio Casares was supported by the Polish National Science Centre (NCN) grant “Polynomial finite state computation” (2022/46/A/ST6/00072).

\ifpictures
\includepdf{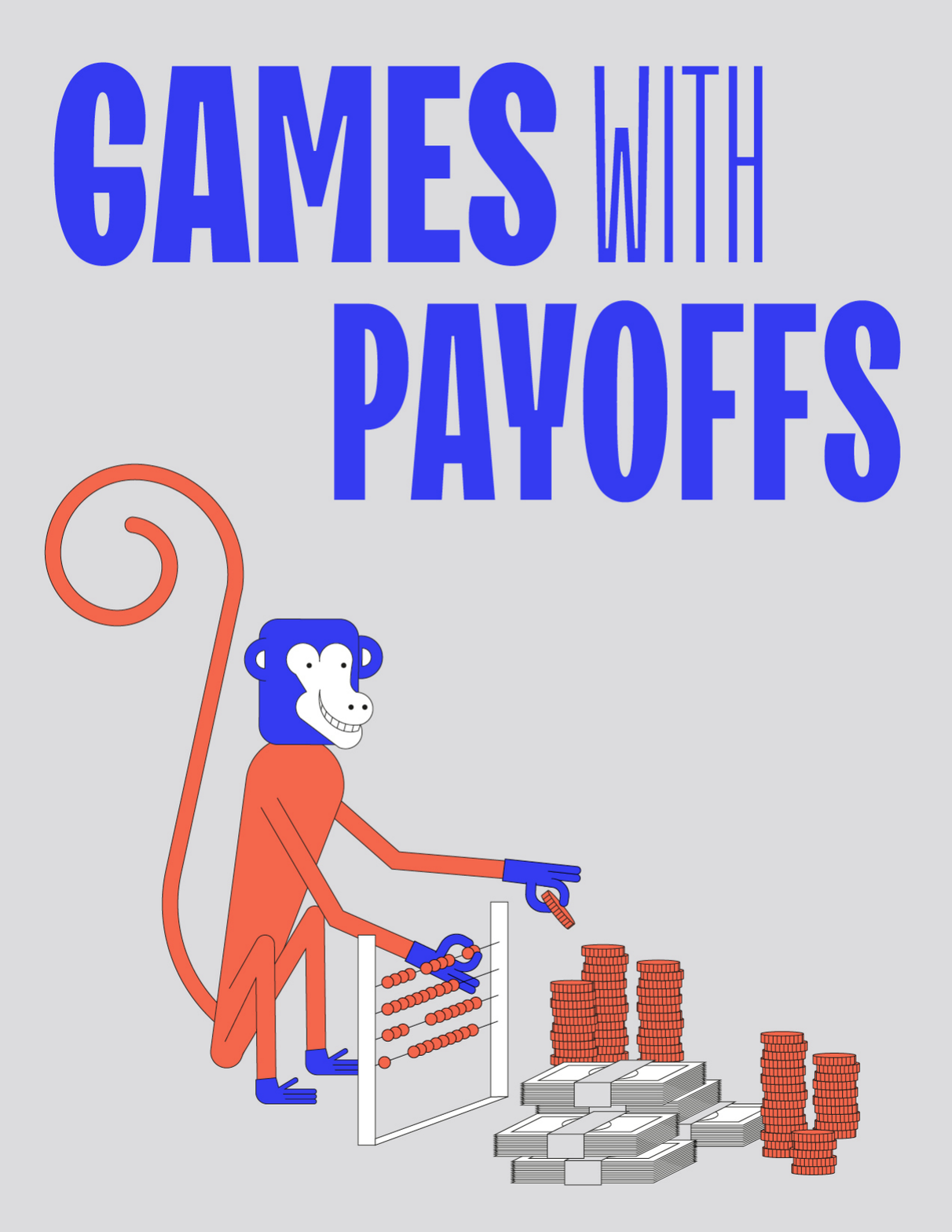}
\fi
\author[Nathana{\"e}l Fijalkow, Benjamin Monmege]{Nathana{\"e}l Fijalkow, Benjamin Monmege}
\copyrightline{Copyright by Nathana{\"e}l Fijalkow and Benjamin Monmege 2025, to be published by Cambridge University Press in the volume \textit{Games on Graphs} edited by Nathana\"el Fijalkow}

\chapter{Games with Payoffs}
\chapterauthor{Nathana{\"e}l Fijalkow, Benjamin Monmege}
\label{5-chap:payoffs}

\newcommand{\FC}{\mathrm{FC}\xspace} 
\newcommand{\Cycles}{\mathrm{Cycles}\xspace} 
\newcommand{\Mean}{\mathrm{Mean}\xspace} 
\newcommand{\FirstCycle}{\mathrm{FirstCycle}\xspace} 
\newcommand{\SuffixAllCycles}{\mathrm{SuffixAllCycles}\xspace} 
\newcommand{\FirstCycleReset}{\mathrm{FirstCycleReset}\xspace} 

\newcommand{\Payoff}{\mathtt{Payoff}} 

\providecommand{\siblank}{\mathtt{-}}
\renewcommand{\siblank}{\mathtt{-}}

\providecommand{\Lift}{\textrm{Lift}}
\renewcommand{\Lift}{\textrm{Lift}}

\newcommand{\Rbar}{\overline{\R}}
\newcommand{\downward}[1]{\mathop{\downarrow_{#1}}}
\newcommand{\gval}{\mathrm{gr}\text{-}\val}

This chapter considers quantitative objectives defined using payoffs. 
Adding quantities can serve two goals:
the first is for refining qualitative objectives by quantifying how well, how fast, or at what cost a qualitative objective is satisfied,
and the second is to define richer specifications and preferences over outcomes.
\begin{itemize}
	\item We start in~\Cref{5-sec:qualitative} by studying extensions of the classical qualitative objectives. Among two strategies in a reachability game that guarantee to reach a target in ten steps or in a billion steps, we would certainly prefer the first one from a pragmatic point of view.
%
	
	\item We study \emph{mean-payoff games} in~\Cref{5-sec:mean_payoff}, and \emph{energy games} along the way.
	We present three algorithms for solving them, the first two based on \emph{value iteration} and the third on \emph{strategy improvement}.
	Along the way we show that parity games reduce to mean-payoff games.

	\item We study \emph{discounted-payoff games} in~\Cref{5-sec:discounted_payoff}.
	We construct a strategy improvement algorithm for computing the value function.
	We also show that mean-payoff games reduce to discounted-payoff games, so the previous algorithm yields an algorithm for computing the value function of a mean-payoff game.

	\item We study \emph{shortest path games} in~\Cref{5-sec:shortest_path}.
	They extend reachability games by requiring that Min reaches her target with minimal cost, 
	which if the weights are all equal means \emph{as soon as possible}.
	
	\item We study \emph{total payoff games} in~\Cref{5-sec:total_payoff}.
\end{itemize}

\section*{Notations}
\label{5-sec:notations}
In this chapter all objectives we consider use the set of colours $C = \Z$ the set of integers 
(or $C = \Z \cup \set{\Win}$ for the shortest path objective), 
so a colour is called a weight interpreted as a payoff for Max.
We will study different quantitative objectives corresponding to different ways of aggregating the weights.
We let $\ValueMax^\game(v)$ and $\ValueMin^\game(v)$ be the values for Max and Min in the game $\game$: 
this is the best each player can unilaterally guarantee from the vertex $v$, no matter which strategy the other player uses.
All the quantitative objectives we will study in this chapter are Borel, hence determined thanks to \Cref{1-cor:borel_determinacy}: 
$\ValueMax^\game(v) = \ValueMin^\game(v)$ for all vertices $v$. 
We thus let $\Value^\game(v)$ denote this value (and $\Value(v)$ if $\Game$ is clear from the context).

For complexity statements we use the unit cost RAM model as defined in~\Cref{1-sec:computation}.
Let us make some preliminary remarks, and consider a game $\Game$ with an objective using the set of colours $C = \Z$.
We let $W$ denote the largest weight appearing in $\Game$ in absolute value.
Choosing the machine word size $w = \log(m) + \log(W)$ implies that an edge together with its weight can be stored in one machine word and that we can perform arithmetic operations on weights, hence for most objectives we use the machine word size $w = \log(m) + \log(W)$.
The only exception will be the discounted-payoff objectives, which additionally have a discount factor $\lambda \in (0,1)$
that needs to be given in the input.



\section{Refining qualitative objectives with quantities}
\label{5-sec:qualitative}
In this section we define quantitative objectives extending the qualitative objectives $\Safe$, $\Reach$, $\Buchi$, and $\CoBuchi$.
The four quantitative objectives we will define in this section return as outcome some weight in the sequence (for instance, the maximum weight).
This is in contrast with the $\MeanPayoff$ and $\DiscountedPayoff$ objectives that we will study later,
which perform \emph{arithmetic operations} on the sequence of weights.

A first way to compute a payoff from a sequence of weights $\rho \in \Z^\omega$ is to consider the maximum weight in the sequence:
\[
\Sup(\rho) = \sup_i \rho_i.
\]
This extends the qualitative objective $\Reach(\Win)$ in the following sense: 
the objective $\Reach(\Win)$ corresponds to the quantitative objective $\Sup$ using two weights: $0$ for $\Lose$ and $1$ for $\Win$.
The outcome of a sequence is $1$ if and only if the sequence contains $\Win$.
It refines $\Reach(\Win)$ by specifying (numerical) preferences.

The dual objective is to consider the smallest weight:
\[
\Inf(\rho) = \inf_i \rho_i.
\]
The qualitative objective $\Safe(\Win)$ corresponds to the quantitative objective $\Inf$
using two weights: $0$ for $\Win$ and $1$ for $\Lose$.
The outcome of a sequence is $0$ if and only if the sequence does not contain $\Lose$.

Similarly the following quantitative objectives refine $\Buchi$ and $\CoBuchi$:
\[
  \LimSup(\rho) = \limsup_i \rho_i,\qquad 
  \LimInf(\rho) = \liminf_i \rho_i.
\]

The analyses and algorithms for solving games with $\Reach$, $\Safe$, $\Buchi$, and $\CoBuchi$ objectives extend to these four quantitative objectives.

\begin{theorem}[Sup, Inf, LimSup, LimInf objectives]
\label{5-thm:sup-inf-limsup-liminf}
Games with objectives $\Sup$, $\Inf$, $\LimSup$, and $\LimInf$ are uniformly bi-positionally determined.
There exists an algorithm for computing the value function of those games in polynomial time and space.
More precisely, let $k$ be the number of different weights in the game,
the time complexity is $O(m)$ for objectives $\Sup$ and $\Inf$, and
$O(knm)$ for objectives $\LimSup$ and $\LimInf$,
and for all algorithms the space complexity is $O(m)$.
\end{theorem}

\begin{proof}
We sketch the algorithm for the objective $\Sup$, the other cases are similar.
Let $c_1 < \dots < c_k$ be the ordered enumeration of all weights in the game.
The set of vertices of value $c_k$ is $\AttrMax(c_k)$, which can be computed in linear time.
We then construct the subgame $\Game'$ of $\Game$ induced by $V \setminus \AttrMax(c_k)$,
and continue recursively: $\Game'$ has one less weight.

A naive complexity analysis yields a time complexity $O(km)$, but it is easily refined to $O(m)$ 
by revisiting the attractor computation and showing that each edge in the whole game is treated at most once
throughout the recursive attractor computations.
This complexity analysis does not extend to $\LimSup$ and $\LimInf$ objectives, where the complexity is multiplied by $k$.
\end{proof}

\section{Mean-payoff games}
\label{5-sec:mean_payoff}
A natural approach for aggregating an infinite sequence of weights is to consider the arithmetic mean.
Since the sequence $(\frac{1}{k} \sum_{i = 0}^{k-1} \rho_i)_{k \in \N}$ may not converge,
we can either consider the limit superior or the limit inferior, leading to the two definitions:
\[
\MeanPayoff^+(\rho) = \limsup_k \frac{1}{k} \sum_{i = 0}^{k-1} \rho_i \quad ; \quad 
\MeanPayoff^-(\rho) = \liminf_k \frac{1}{k} \sum_{i = 0}^{k-1} \rho_i.
\]
Note that $\MeanPayoff^+(-\rho) = - \MeanPayoff^-(\rho)$, where in $-\rho$ we take the opposite of each weight.
In other words, $\MeanPayoff^+$ and $\MeanPayoff^-$ are dual objectives.


\vskip1em
In this section we study mean-payoff games.
We will first prove that they are prefix-independent, then that they are bi-positionally determined,
and then construct algorithms for solving them and compute the value function.
The best known time complexity for both problems is pseudo-polynomial, meaning polynomial when the numerical inputs are given in unary.

\begin{lemma}[Prefix-independence] 
\label{5-lem:prefix_independence_mean_payoff}
$\MeanPayoff^+$ is prefix-independent.
\end{lemma}
\begin{proof}
We show that $\MeanPayoff^+(\rho_0 \rho_1 \dots) = \MeanPayoff^+(\rho_p \rho_{p+1} \dots)$
for a fixed $p \in \N$:
\begin{align*}
\limsup_k \frac{1}{k} \sum_{i = 0}^{k-1} \rho_i &= 
\limsup_k \left(
\underbrace{\frac{1}{k} \sum_{i = 0}^{p-1} \rho_i}_{\rightarrow 0 \text{ for } k \rightarrow \infty} + 
\underbrace{\frac{k-p}{k}}_{\rightarrow 1 \text{ for } k \rightarrow \infty} \cdot \frac{1}{k-p} \sum_{i = p}^{k-1} \rho_i
\right) \\
&=
\liminf_k \frac{1}{k-p} \sum_{i = 0}^{p-1} \rho_{p + i}.
\end{align*} 
\end{proof}
Note that by duality this implies that $\MeanPayoff^-$ is also prefix-independent.

In the setting we consider in this chapter, meaning two player zero-sum games of perfect information, 
the two objectives are equivalent, which is a corollary of their bi-positional determinacy~\Cref{4-thm:lift_applications}.

As an example, consider the mean-payoff game represented in~\Cref{5-fig:MP}. 
As we will show later, the positional strategies defined below are optimal strategies for Max and Min:
\[
\begin{array}{c}
\sigma^*(v_0) = v_0 \xrightarrow{4} v_1 \quad ; \quad \sigma^*(v_2) = v_2 \xrightarrow{4} v_3 \\
\tau^*(v_1) = v_1 \xrightarrow{2} v_2 \quad ; \quad \tau^*(v_3) = v_3 \xrightarrow{-1} v_1 \quad ; \quad \tau^*(v_4) = v_4 \xrightarrow{-2} v_0.
\end{array}
\] 
The play consistent with $\sigma^*$ and $\tau^*$ starting from $v_0$ is the lasso word 
\[
v_0 \xrightarrow{4} v_1\ \left( v_1 \xrightarrow{2} v_2 \xrightarrow{4} v_3 \xrightarrow{-1} v_1 \right)^\omega,
\]
which has as mean payoff the average weight of the cycle $v_1 \xrightarrow{2} v_2 \xrightarrow{4} v_3 \xrightarrow{-1} v_1$, \textit{i.e.}~$5/3$.
If we restrict ourselves to positional strategies for both players, we may easily convince ourselves that this is the best that both players may aim for. Indeed,
\begin{itemize}
\item the loop around $v_4$ has average weight $2$;
\item the simple cycle alternating between $v_0$ and $v_4$ has average
  weight $1.5$;
\item the simple cycle alternating between $v_0$ and $v_1$ has average
  weight $2$;
\item the loop around $v_2$ has average weight $1$, and
\item the simple cycle alternating between $v_0$, $v_1$, $v_2$ and
  $v_3$ has average weight $2$.
\end{itemize}
Therefore, we may check that no player can obtain a better mean payoff than $5/3$, from its own point of view: for instance, if Max switches her decision in $v_0$, hoping to get value $2$, she will get a lower value $1.5$ since Min still prefers cycle alternating between $v_0$ and $v_4$ to his self-loop around $v_4$. 
This is the case for all starting vertices, which proves that all vertices have the same value $5/3$. 

\begin{figure}[tbp]
  \centering
  \begin{tikzpicture}[node distance=2.5cm]
    \node[s-eve](0) {$v_0$};%
    \node[s-adam,below of=0](1) {$v_1$};%
    \node[s-eve,left of=1](2) {$v_2$};%
    \node[s-adam,left of=0](3) {$v_3$};%
    \node[s-adam,right of=0](4) {$v_4$};%

    \path[arrow] (0) edge[bend right=20] node[left] {4} (1)%
    (1) edge[bend right=20] node[right] {0} (0)%
    (1) edge node[below] {2} (2)%
    (2) edge[selfloop=180] node[left] {1} (2)%
    (2) edge node[left] {4} (3)%
    (3) edge node[below left] {$-1$} (1)%
    (3) edge node[above] {$-2$} (0)%
    (0) edge[bend left=20] node[above] {$5$} (4)%
    (4) edge[bend left=20] node[below] {$-2$} (0)
    (4) edge[selfloop=-90] node[below] {2} (4);%
  \end{tikzpicture}
\caption{A mean-payoff game.}
\label{5-fig:MP}
\commentAlt{Figure~\ref{5-fig:MP}: A directed graph with five nodes (v0 to v4) of mixed shapes (circles and squares) and labeled edges, showing various connections and self-loops. See long description.}
\commentLongAlt{Figure~\ref{5-fig:MP}: The image displays a directed graph with five nodes. Nodes v0 and v2 are circles, while v1, v3, and v4 are squares.

Node v0 (circle) has a self-loop labeled '0'.
Node v1 (square) has an incoming arrow from v0 labeled '4' and an outgoing arrow to v2 labeled '2'.
Node v2 (circle) has a self-loop labeled '1' and an incoming arrow from v1 labeled '2'.
Node v3 (square) has an outgoing arrow to v0 labeled '-2' and an outgoing arrow to v1 labeled '-1'. It also has an incoming arrow from v2 labeled '4'.
Node v4 (square) has a self-loop labeled '2'. It has an incoming arrow from v0 labeled '5' and an outgoing arrow to v0 labeled '-2'.}
\end{figure}

Let us draw some corollaries from bi-positional determinacy of mean-payoff games.

\begin{corollary}[Limit superior and limit inferior mean-payoff games]\hfill
\label{5-cor:rational-MP}
\begin{itemize}
	\item Limit superior and limit inferior mean-payoff games are equivalent:
	for every arena $\arena$ and colouring function $\col : E \to \Z$,
	let 
	\[
	\game_{+} = (\arena,\MeanPayoff^+(\col))\text{ and } \game_{-} = (\arena,\MeanPayoff^-(\col)),
	\]
	then $\val^{\game_+} = \val^{\game_-}$.
	This implies that a positional strategy is optimal in $\game_+$ if and only if it is optimal in $\game_-$.
	
	Since they are equivalent we speak of mean-payoff games without specifying whether the objective is $\MeanPayoff^+$ or $\MeanPayoff^-$,
	and write $\MeanPayoff$ instead.

	\item For a mean-payoff game with $n$ vertices and weights in $[-W,W]$, the mean-payoff values are rational numbers in $[-W,W]$ whose denominators are at most $n$.
\end{itemize}
\end{corollary}
\begin{proof}
Thanks to \Cref{4-thm:lift_applications}, there exist $\sigma_+$ and $\tau_+$ optimal positional strategies in $\game_+$
and $\sigma_-$ and $\tau_-$ optimal positional strategies in $\game_-$ (for the latter by duality).
By definition $\val^{\game_+}(v) = \MeanPayoff^+(\pi^v_{\sigma_+,\tau_+})$ and $\val^{\game_-}(v) = \MeanPayoff^-(\pi^v_{\sigma_-,\tau_-})$.
Since $\MeanPayoff^- \le \MeanPayoff^+$ we already have $\val^{\game_-} \le \val^{\game_+}$.

For two positional strategies $\sigma$ and $\tau$, the play $\pi^v_{\sigma,\tau}$ is a lasso, meaning of the form $\pi c^\omega$
with $\pi$ a simple path and $c$ a simple cycle, 
implying that $\MeanPayoff^+(\pi^v_{\sigma,\tau}) = \MeanPayoff^-(\pi^v_{\sigma,\tau})$,
let us write $\MeanPayoff(\pi^v_{\sigma,\tau})$ for this value.

We have:
\[
\MeanPayoff(\pi^v_{\sigma_+,\tau_+}) 
\le \MeanPayoff(\pi^v_{\sigma_+,\tau_-})
\le \MeanPayoff(\pi^v_{\sigma_-,\tau_-}),
\]
where the first inequality is by optimality of $\tau_-$ and the second inequality by optimality of $\sigma_-$.
Hence $\val^{\game_+} \le \val^{\game_-}$, and finally $\val^{\game_+} = \val^{\game_-}$.

\vskip1em
For the second item, recall that $\val^\game(v) = \MeanPayoff(\pi^v_{\sigma,\tau})$
with $\sigma$ and $\tau$ optimal positional strategies.
Let us write $\pi^v_{\sigma,\tau} = \pi c^\omega$ with $\pi$ a simple path and $c$ a simple cycle, 
then by prefix-independence $\MeanPayoff(\pi^v_{\sigma,\tau}) = \Mean(c)$,
thus $\val^\game(v)$ is the mean of at most $n$ weights from $\game$.
\end{proof}

\subsection*{Solving mean-payoff games in $\NP\cap\coNP$}
The bi-positional determinacy of mean-payoff games easily gives an upper bound on the complexity of \emph{solving} these games.

\begin{theorem}[Complexity of mean payoff]
\label{5-thm:MP-NPcoNP}
Solving mean-payoff games is in $\NP\cap\coNP$.
\end{theorem}
\begin{proof}
The first ingredient for this proof is a polynomial-time algorithm for solving the one player variants of mean-payoff games.
Indeed, they correspond to the minimum cycle mean problem in a weighted graph,
which can be solved in polynomial time by a dynamic programming algorithm.
The second ingredient is the bi-positional determinacy result proved in~\Cref{4-thm:lift_applications}.

Let us show the $\NP$ membership. 
Consider a mean-payoff game $\Game$, a vertex $v$ and a threshold~$x \in \Qinfty$.
Thanks to~\Cref{4-thm:lift_applications}, we know that there exists a positional optimal strategy for Max.
With a non-deterministic Turing machine, we may guess a positional strategy for Max, and check that it ensures $x$ in $\Game$ from $v$. 

Let us now show the $\coNP$ membership. 
By determinacy of mean-payoff games, whether Max \emph{cannot} ensure $x$ in $\Game$ from $v$ 
is equivalent to whether Min can ensure $x$ in $\Game$ from $v$.
Again thanks to~\Cref{4-thm:lift_applications}, we know that there exists a positional optimal strategy for Min.
With a non-deterministic Turing machine, we may guess a positional strategy for Min, and check that it ensures $x$ in $\Game$ from $v$. 
\end{proof}

We can turn the non-deterministic algorithm given in~\Cref{5-thm:MP-NPcoNP} into a deterministic algorithm with exponential complexity since there are exponentially many positional strategies.

\subsection*{Reducing parity games to mean-payoff games}

We now show that solving mean-payoff games is at least as hard as solving parity games.

\begin{theorem}[Reducing parity games to mean-payoff games]
\label{5-thm:parity2MP}
  Solving parity games reduce in polynomial time to solving mean-payoff games with threshold $0$.
\end{theorem}
\begin{proof}
Let $\game = (\arena, \Parity(\col))$ a parity game with $n$ vertices and priorities in $[1,d]$.
We construct a mean-payoff game $\game' = (\arena, \MeanPayoff[\col'])$ using the same arena and the colouring function:
\[
\col'(e) = (-n)^{\col(e)}.
\]
Note that $\col'(e)$ is of polynomial size since $\log(|\col'(e)|) = \col(e) \log(n) \leq d \log(n)$.

The key property relating $\col'$ and $\col$ is the following:
let $c$ a simple cycle
\[
v_0 \xrightarrow{p_0} p_1 \xrightarrow{p_2} v_2 \cdots v_{k-1} \xrightarrow{p_{k-1}} v_0,
\]
then the largest priority in $c$ is even 
if and only if the mean value with respect to $\col'$ is non-negative.
Indeed, if the largest priority in $c$ is $p$ even, then it contributes $n^p$ and all other values are greater than or equal to $-n^{p-1}$,
and since there are at most $n$ in total, the largest priority dominates the others.

\vskip1em
We claim that for all vertices $v$, $v \in \WE(\Game)$ if and only if $\val^{\Game'} \ge 0$.
Let $v \in \WE(\game)$, and $\sigma$ a positional strategy winning from $v$ in $\Game$, we show that $\sigma$ ensures mean payoff at least $0$ in $\Game'$ from $v$.
Indeed, in $\Game[\sigma,v]$ all cycles are even, which thanks to the above implies that all cycles in $\Game'[\sigma,v]$ are non-negative.
The converse implication is similar: a positional strategy $\sigma'$ in $\Game'$ ensuring mean payoff at least $0$ from $v$ has the property that all cycles in $\Game'[\sigma,v]$ are non-negative, hence that all cycles in $\Game[\sigma,v]$ are even, therefore that it is winning in $\Game$ from $v$.
\end{proof}

Note that we did not construct a reduction between objectives as defined in~\Cref{1-sec:automata}:
indeed it is not true that $\Parity$ reduces to $\MeanPayoff_{\ge 0}$, the reduction depends on the number $n$ of vertices.
As a corollary of~\Cref{5-thm:MP-NPcoNP}, this polynomial reduction gives an alternative proof of the fact
that solving parity games is in $\NP \cap \coNP$.

\subsection*{A first value iteration algorithm using finite horizon payoffs}

Let us give a first algorithm for solving mean-payoff games using the value iteration paradigm.
The idea is to consider the game where we only play for a fixed number of steps $k$, and the payoff is the sum of the weights.
We compute iteratively the optimal values for increasing values of $k$, and show that for $k$ large enough it allows us to obtain the values of mean payoff up by a simple rounding procedure.

Let $\game = (\arena,\MeanPayoff[\col])$ a mean-payoff game with $n$ vertices and weights in $[-W,W]$.
Recall that for a vertex $u$, the mean-payoff value is defined as 
\[
\val^{\Game}(u) = \sup_{\sigma}\ \inf_{\tau}\ \MeanPayoff[\col](\pi^{\sigma,\tau}_u).
\]
Our goal is to compute these values, or compare them to a threshold. 

Let us fix a number $k$ and define the following objective summing the first $k$ weights:
\[
\Payoff_k(\rho) = \sum_{i = 0}^{k-1} \rho_i.
\]
Let us define $\val^{\Game,k} : V \to \Z$ the value function for the $\Payoff_k$ objective:
\[
\val^{\Game,k}(u) = \sup_{\sigma}\ \inf_{\tau}\ \Payoff_k[\col](\pi^{\sigma,\tau}_u).
\]
We let $F_V$ be the set of functions $V \to \Z$, we define the operator $\Op^{\Game} : F_V \to F_V$ by:
\[
\Op^{\Game}(\mu)(u) = 
\begin{cases}
\max \set{\mu(v) + w : u \xrightarrow{w} v \in E} & \text{ if } u \in \VMax, \\
\min \set{\mu(v) + w : u \xrightarrow{w} v \in E} & \text{ if } u \in \VMin.
\end{cases}
\]
Let $\mu_0$ defined by $\mu_0(u) = 0$ for all $u$ and $\mu_{k+1} = \Op^{\Game}(\mu_k)$.

\begin{lemma}
The following holds for all $k$.
\begin{itemize}
	\item We have $\mu_k = \val^{\Game,k}$.
	\item For all $u \in V$ we have $k \cdot \val^{\Game}(u) - 2nW \le \val^{\Game,k} \le k \cdot \val^{\Game}(u) + 2nW$.
\end{itemize}
\end{lemma}

\begin{proof}
The proof of the first item is an easy induction on $k$.
We turn to the second item. Let us consider $\sigma$ a positional optimal strategy for Max. 
We claim that for all $k$, the strategy $\sigma$ ensures from $u$ that $\Payoff_k[\col]$ is at least $(k - n) \cdot \val^{\Game}(u) - nW$.
We first note that all cycles reachable from $u$ in $\Game[\sigma]$ have values (meaning, average value of the weights) at least $\val^{\Game}(u)$. 
Now, consider a path of length $k$, and iteratively remove cycles from it. What remains is a most $n$ edges, which contribute in the worst case to $-nW$. The remaining $k - n$ edges are included in some cycle, hence contribute $(k - n) \cdot \val^{\Game}(u)$.
Since $\val^{\Game}(u) \le W$, we obtain a lower bound of $k \cdot \val^{\Game}(u) - 2nW$.

The argument is symmetrical for Adam.
\end{proof}

A direct corollary is as follows.

\begin{corollary}
We have $\lim_k \frac{1}{k} \cdot \val^{\Game,k} = \val^{\Game}$.
\end{corollary}

However, we can make this effective using~\Cref{5-cor:rational-MP}, which places bounds on the mean-payoff values.
This yields two very simple algorithms, one for computing the mean-payoff values, and the other one for solving mean-payoff games.

\begin{theorem}
The following holds.
\begin{itemize}
	\item There exists an algorithm running in time $O(n^3 m W)$ for computing the mean-payoff values.
	\item There exists an algorithm running in time $O(n^2 m W)$ for solving the mean-payoff games, meaning determining whether $\val^{\Game}(u) \ge c$ for some threshold $c$.
\end{itemize}
\end{theorem}

\begin{proof}
Both arguments rely on~\Cref{5-cor:rational-MP}. 
\begin{itemize}
	\item Let us fix $k = 4n^3 W$. We can compute the values $\val^{\Game,k}$ in time $O(n^3 m W)$ as explained above.
	The value of $k$ was chosen in such a way that
	\[
	\val^{\Game,k}(u) - \frac{1}{2n(n-1)} < \val^{\Game}(u) < \val^{\Game,k}(u) + \frac{1}{2n(n-1)}.
	\]
	Indeed, since $\val^{\Game}(u)$ is a rational number whose denominator is at most $n$, the minimum distance between two possible values of $\val^{\Game}(u)$ is at most $\frac{1}{n(n-1)}$. Hence the exact value of $\val^{\Game}(u)$ is the unique rational number with a denominator at most $n$ that lies in the interval 
\[
\left[\ \val^{\Game,k}(u) - \frac{1}{2n(n-1)}\ ,\ \val^{\Game,k}(u) + \frac{1}{2n(n-1)}\ \right],
\]
which is easily computed.
	\item The distance between $c$ and the closest rational number with a denominator at most $n$ is $\frac{1}{n}$, so to determine whether $\val^{\Game}(u) \ge c$ it is enough to compute the values $\val^{\Game,k}$ for $k = 4n^2 W$.
\end{itemize}
\end{proof}

\subsection*{A second value iteration algorithm using energy}
\label{5-subsec:value_iteration_energy}

Let us give a second algorithm, based on energy games. The construction of the algorithm will follow closely the high-level presentation of value iteration algorithms given in \Cref{1-sec:value_iteration}.
Let $\game = (\arena,\MeanPayoff[\col])$ a mean-payoff game with $n$ vertices and weights in $[-W,W]$.

The key insight is to use the energy objective. 
Recall that the "energy" quantitative objective is defined over the set of colours $C = \Z$:
\[
\Energy(\rho) = \inf \set{\ell \in \N : \forall k \in \N,\ \ell + \sum_{i=0}^{k-1} \rho_i \ge 0}.
\]
The interpretation is the following: weights are energy consumptions (negative values) and recharges (positive values), and $\Energy(\rho)$ is the smallest initial budget $\ell$ such that Min can ensure that the energy level remains non-negative forever.
Let us be careful here: maximising the mean-payoff objective corresponds to minimising the energy objective. 
We formalise the relationship between mean payoff and energy in the following lemma.

\begin{lemma}[Relating mean payoff and energy objectives]
\label{5-lem:energy_mean_payoff}
Let $G$ be a mean-payoff graph.
Then $G$ satisfies $\MeanPayoff^{-} \ge 0$ if and only if it satisfies $\Energy < \infty$.
\end{lemma}

\begin{proof}
Let us say that a cycle in $G$ is non-negative if the sum of its weights is non-negative.
We consider the following properties:
\begin{enumerate}
	\item[(i)] $G$ satisfies $\MeanPayoff^{-} \ge 0$.
	\item[(ii)] All cycles in $G$ are non-negative.
	\item[(iii)] $G$ satisfies $\Energy < \infty$.
\end{enumerate}
We prove the implications $(i) \Rightarrow (ii)$, then $(ii) \Rightarrow (iii)$, and finally $(iii) \Rightarrow (i)$.

$(i) \Rightarrow (ii)$ is clear.

$(ii) \Rightarrow (iii)$. Let us consider an infinite path, and strike out all edges involved in a cycle in it. At most $n$ edges are not stricken out, incurring at most $-nW$ in energy drop. Since cycles are non-negative, the lowest level in a cycle is also lower bounded by $-nW$. Hence the energy level of the infinite path is at most $2nW$.
 
$(iii) \Rightarrow (i)$. Assume that $G$ satisfies $\Energy < \infty$, this implies that all partial sums are greater than or equal to a constant $\ell$.
This implies that the means of the partial sums are lower bounded by $\frac{\ell}{k}$, which converges to $0$ when $k$ goes to infinity.
Therefore $G$ satisfies $\MeanPayoff^{-} \ge~0$.
\end{proof}

It is tempting to claim that a stronger property hold, namely 
$\MeanPayoff^{-}(\rho) \ge 0$ if and only if $\Energy(\rho) < \infty$. 
This is not the case.

\begin{corollary}
\label{5-cor:energy_mean_payoff}
Let $\game$ a mean-payoff game. We define $\game'$ the energy game induced by $\game$. Then for all vertices $u$, we have $\val^{\Game}(u) \ge 0$ if and only if $\val^{\Game'}(u) < \infty$.
Consequently:
\begin{itemize}
	\item An algorithm computing the energy values induces an algorithm for solving mean-payoff games with the same complexity.
	\item An algorithm computing the energy values in time $T(n,m,W)$ induces an algorithm for computing the mean-payoff values in time $O(\log(n^2 W) \cdot T(n,m,W))$.
\end{itemize}
\end{corollary}

\begin{proof}
We consider $\sigma$ a positional optimal strategy for Max in $\Game$, we have $\val^{\Game} = \val^{\Game,\sigma}$. 
Let $u$ such that $\val^{\Game}(u) \ge 0$, we consider the subgraph of $\Game[\sigma,u]$ of vertices reachable from $u$.
Thanks to~\Cref{5-lem:energy_mean_payoff} applied to this graph, since $\val^{\Game,\sigma}(u) \ge 0$ we have $\val^{\Game',\sigma}(u) < \infty$. This implies that $\sigma$ (seen as a strategy for Min in $\Game'$) ensures $\Energy < \infty$, hence $\val^{\Game'}(u) < \infty$.

Conversely, we consider $\tau$ a positional optimal strategy for Min in $\Game'$, we have $\val^{\Game'} = \val^{\Game',\tau}$. 
Let $u$ such that $\val^{\Game'}(u) < \infty$, we consider the subgraph of $\Game[\tau,u]$ of vertices reachable from $u$.
Thanks to~\Cref{5-lem:energy_mean_payoff} applied to this graph, since $\val^{\Game',\tau}(u) < \infty$ we have $\val^{\Game,\tau}(u) \ge 0$. This implies that $\tau$ (seen as a strategy for Max in $\Game$) ensures $\MeanPayoff \ge 0$, hence $\val^{\Game}(u) \ge 0$.

\vskip1em
The first consequence is immediate. For the second, we use a simply binary search. Indeed, to determine whether $\val^{\Game}(u) \ge c$, we can shift all weights by $c$, thus reducing to the question whether $\val^{\Game}(u) \ge 0$, and compute the energy values. 
Thanks to~\Cref{5-cor:rational-MP} the mean-payoff values are rational numbers in $[-W,W]$ whose denominators are at most $n$, so this requires $O(\log(n^2W))$ calls.
\end{proof}

Therefore, we set as a goal to compute the energy values.

\begin{theorem}[Value iteration algorithm]
\label{5-thm:MP-value-iteration-Brim}
There exists a value iteration algorithm which computes the energy values in time $O(n m W)$.
\end{theorem}

We define $Y = \N \cup \set{\infty}$, equipped with the natural total order on integers.
We let $F_V$ be the lattice of functions $V \to Y$ equipped with the component wise order induced by $Y$.
We define a function $\delta : Y \times [-W,W] \to Y$ by $\delta(\ell,w) = \max(\ell - w, 0)$.
This induces an operator $\Op^{\Game} : F_V \to F_V$:
\[
\Op^{\Game}(\mu)(u) = 
\begin{cases}
\min \set{\delta(\mu(v),w) : u \xrightarrow{w} v \in E} & \text{ if } u \in \VMin, \\
\max \set{\delta(\mu(v),w) : u \xrightarrow{w} v \in E} & \text{ if } u \in \VMax.
\end{cases}
\]
We note that $\Op^{\Game}$ is a monotonic operator, therefore it has a least fixed point. 

\begin{lemma}[Least fixed point for energy games]
\label{5-lem:least_fixed_point_energy}
Let $\game$ an energy game with $n$ vertices and weights in $[-W,W]$.
Then the least fixed point of $\Op^{\Game}$ computes the energy values of $\Game$.
\end{lemma}

\begin{proof}
We first argue that the energy values, meaning the function $\val^{\Game} : V \to \Z \cup \set{\infty}$, form a fixed point of the operator $\Op^{\Game}$.
The fact that $\val^{\Game} = \Op^{\Game}(\val^{\Game})$ is a routine verification, which follows from two properties (see~\Cref{1-lem:sufficient_condition_fixed_point}).
\begin{itemize}
	\item For all $\rho$ sequences of weights, we have $\Energy(w \cdot \rho) = \delta(\Energy(\rho), w)$.
	\item The function $\delta$ is monotonic and continuous.
\end{itemize}
It already yields one inequality: $\val^{\Game}$ is larger than or equal to the least fixed point of $\Op^{\Game}$.
Let us give another proof of the same inequality. 
We recall that thanks to Kleene fixed-point theorem (\Cref{1-thm:kleene}), the least fixed point of $\Op^{\Game}$ is computed as follows:
\[
\forall u \in V,\ \mu_0(u) = 0 \quad ; \quad \mu_{b+1} = \Op^{\Game}(\mu_b).
\]
We have $\mu_0 \le \mu_1 \le \dots$, and since $F_V$ is a finite lattice, for some $k$ we have that $\mu_k$ is the least fixed point of $\Op^{\Game}$.
The crux here is to understand what are the values $\mu_b$ for $b = 0,1,\dots$. Let us define the truncated energy objective:
\[
\Energy_b(\rho) = \inf \set{\ell \in \N : \forall k \in [0,b],\ \ell + \sum_{i=0}^{k-1} \rho_i \ge 0}.
\]
The interpretation is the following: $\Energy_b(\rho)$ is the smallest initial budget $\ell$ such that Min can ensure that the energy level remain non-negative for the first $b$ steps.
A simple induction on $b$ shows that $\mu_b$ is the values for $\Energy_b$ in $\Game$.
Note that $\Energy_0 \le \Energy_1 \le \dots \le \Energy$, hence $\mu_b = \val^{\Energy_b} \le \val^{\Game}$, the desired inequality.

\vskip1em
Let us now prove the converse inequality: $\val^{\Game}$ is smaller than or equal to the least fixed point of $\Op^{\Game}$.
For this, we consider a fixed point $\mu$ of $\Op^{\Game}$, and argue that $\val^{\Game} \le \mu$.
To this end, we extract from $\mu$ a strategy $\tau$ for Min, and show that $\val^{\tau} \le \mu$; 
since we know that $\val^{\Game} \le \val^{\tau}$, this implies $\val^{\Game} \le \mu$.
We define $\tau$ as an argmin strategy:
\[
\begin{array}{l}
u \in \VMin:\ \tau(u) \in \argmin \set{\delta(\mu(v),w) : u \xrightarrow{w} v \in E}.
\end{array}
\]
Let us define the graph $G = \Game[\tau]$, by definition of $\tau$ for all edges $u \xrightarrow{w} v$ in $G$ we have
$\mu(u) \ge \delta(\mu(v),w)$.
We claim that this implies that for all paths $\rho$ from $u$ in $G$, we have $\Energy(\rho) \le \mu(u)$.
To this end, we show that for all $k \in \N$ we have $\mu(u) + \sum_{i = 0}^{k-1} \rho_i \ge 0$.
We proceed by induction on $k$. This is clear for $k = 0$, let us assume that it holds for $k$ and show that it also does for $k+1$.
Let us write $\rho_0 = u \xrightarrow{w} v$.
By the property above, we have $\mu(u) \ge \delta(\mu(v),w) = \max(\mu(v) - w, 0) \ge \mu(v) - w$.
Hence 
\[
\mu(u) + \sum_{i = 0}^k \rho_i = \mu(u) + w + \sum_{i = 1}^k \rho_i \ge \mu(v) + \sum_{i = 0}^{k-1} \rho_i \ge 0,
\]
where the last inequality is by induction hypothesis for the path $\rho_1 \dots \rho_{k-1}$ from $v$.
This concludes the induction.
\end{proof}

\Cref{5-lem:least_fixed_point_energy} does not immediately yield a value iteration algorithm because the lattice $Y$ is infinite.
However, let us note that by bi-positional determinacy, the value of a vertex is the value of a path consisting of a prefix of length at most $n$
and a simple cycle, which is $\infty$ if the cycle is negative and in $[0,nW]$ otherwise.
Hence we can equivalently define $Y = [0,nW] \cup \set{\infty}$ and $\delta(\ell,w) = \max(\ell - w, 0)$ if $\ell - w \le nW$, and $\infty$ otherwise. It is clear that the least fixed points for both operators coincide thanks to the remark above, and now that we have a finite lattice we obtain the value iteration algorithm presented in~\Cref{5-algo:value_iteration_energy}.

\begin{algorithm}
 \KwData{An energy game}

\For{$u \in V$}{
$\mu(u) \leftarrow 0$
}
     
\Repeat{$\mu = \Op^{\Game}(\mu)$}{
$\mu \leftarrow \Op^{\Game}(\mu)$
}

\Return{$\mu$}
\caption{The value iteration algorithm for energy games.}
\label{5-algo:value_iteration_energy}
\end{algorithm}

The value iteration algorithm for energy games is conceptually very simple, its pseudocode is given in~\Cref{5-algo:value_iteration_energy}.
Since for each vertex its value can only increase, and at each iteration at least one vertex increases, the total number of iterations before reaching the fixed point is at most $O(n^2 W)$.
However to obtain the announced complexity of $O(n m W)$, one needs to be a little bit more subtle: in the naive version the computational cost of an iteration of $\Op^{\Game}$ is $O(m)$, since we need to look at each vertex and each outgoing edge.
This can be improved using a more involved data structure keeping track of vertices to be updated. We detail this below.

\subsection*{Refined value iteration algorithm for energy values}

The pseudocode is given in \Cref{5-algo:refined_value_iteration_energy}.
For an edge $u \xrightarrow{w} v$ we say that it is incorrect if $\mu(u) < \delta( \mu(v), w)$.
A vertex $u \in \VMax$ is incorrect if it has an outgoing edge which is incorrect, and a vertex $u \in \VMin$ is incorrect if all of its outgoing edges are incorrect.
The key idea of the data structure we are building is not to keep track of all incorrect edges for vertices in $\VMin$, but rather to count them.

The data structure consists of the following objects:
\begin{itemize}
	\item a value of $Y$ for each vertex, representing the current function $\mu : V \to Y$;
	\item a set $\Incorrect$ of vertices (the order in which vertices are stored and retrieved from the set does not matter);
	\item a table $\Count$ storing for each vertex of Min a number of edges.
\end{itemize}
For our complexity analysis we use the unit cost RAM model, see \Cref{1-sec:computation} for details.
In the case at hand let us choose for the machine word size $w = \log_2(m) + \log_2(W)$, 
so that an edge together with its weight can be stored in one machine word.

The invariant of the algorithm satisfied before each iteration of the repeat loop is the following:
\begin{itemize}
	\item for $u \in \VMin$, the value of $\Count(u)$ is the number of incorrect edges of $u$;
	\item $\Incorrect$ is the set of incorrect vertices.
\end{itemize}
The invariant is satisfied initially thanks to the function $\texttt{Init}$.
Let us assume that we choose and remove $u$ from $\Incorrect$.
Since we modify only $\mu(u)$ the only potentially incorrect vertices are in $\Incorrect$ (minus $u$) and the incoming edges of $u$;
for the latter each of them is checked and added to $\Incorrect'$ when required.
By monotonicity, incorrect vertices remain incorrect so all vertices in $\Incorrect$ (minus $u$) are still incorrect.
Hence the invariant is satisfied.

The invariant implies that the algorithm indeed implements~\Cref{5-algo:value_iteration_energy} hence returns the minimal fixed point, 
but it also has implications on the complexity.
Indeed one iteration of the repeat loop over some vertex $u$ involves 
\[
O\left( (|\Ing^{-1}(u)| + |\Out^{-1}(u)|)\right)
\]
operations:
the first term corresponds to updating $\mu(u)$ and $\Incorrect$, which requires for each outgoing edge of $u$ to compute $\delta$,
and the second term corresponds to considering all incoming edges of $u$.
Since each vertex is updated at most $nW$ times, the overall running time for the algorithm is
\[
O\left( 
\left( \sum_{u \in V} (|\Ing^{-1}(u)| + |\Out^{-1}(u)|) \right) \cdot nW
\right) 
= O(m n W).
\]

\begin{algorithm}
 \SetKwFunction{FInit}{Init}
 \SetKwFunction{FTreat}{Treat}
 \SetKwFunction{FUpdate}{Update}
 \SetKwFunction{FMain}{Main}
 \SetKwProg{Fn}{Function}{:}{}
 \DontPrintSemicolon

\Fn{\FInit{}}{

	\For{$u \in V$}{
		$\mu(u) \leftarrow 0$
	}

	\For{$u \in \VMin$}{
        \For{$u \xrightarrow{w} v \in E$}{
        	\If{incorrect: $\mu(u) < \delta(\mu(v), w)$}{
		        $\Count(u) \leftarrow \Count(u) + 1$
        	}
        }
        
        \If{$\Count(u) = \Degree(u)$}{
        	Add $u$ to $\Incorrect$
        }
	}    
	
	\For{$u \in \VMax$}{
        \For{$u \xrightarrow{w} v \in E$}{
        	\If{incorrect: $\mu(u) < \delta(\mu(v), w)$}{
        		Add $u$ to $\Incorrect$
        	}
        }
    }
}

\vskip1em
\Fn{\FTreat{$u$}}{
	\If{$u \in \VMax$}{
		$\mu(u) \leftarrow \max \set{\delta( \mu(v), w) : u \xrightarrow{w} v \in E}$
	}

	\If{$u \in \VMin$}{
		$\mu(u) \leftarrow \min \set{\delta( \mu(v), w) : u \xrightarrow{w} v \in E}$
	}
}

\vskip1em
\Fn{\FUpdate{$u$}}{	
	\If{$u \in \VMin$}{
        $\Count(u) \leftarrow 0$
	}
	\For{$v \xrightarrow{w} u \in E$ which is incorrect}{
		\If{$v \in \VMin$}{
	        $\Count(v) \leftarrow \Count(v) + 1$
        
	        \If{$\Count(v) = \Degree(v)$}{
    	    	Add $v$ to $\Incorrect'$
	        }	
		}

		\If{$v \in \VMax$}{
			Add $v$ to $\Incorrect'$	
		}

	}
}

\vskip1em
\Fn{\FMain{}}{
	\FInit()    

	\For{$i = 0,1,2,\dots$}{
		$\Incorrect' \leftarrow \emptyset$

		\For{$u \in \Incorrect$}{

			\FTreat($u$)    

			\FUpdate($u$)    
		}
		\If{$\Incorrect' = \emptyset$}{

			\Return{$\mu$}
		}
		\Else{

			$\Incorrect \leftarrow \Incorrect'$
		}
	}
}
\caption{The refined value iteration algorithm for energy games.}
\label{5-algo:refined_value_iteration_energy}
\end{algorithm}

\subsection*{A strategy improvement algorithm for energy games}
\label{5-subsec:strategy_improvement_energy}

As explained above, solving energy games yields algorithms for solving mean-payoff games.
So, we continue our investigation of energy games, and now construct a strategy improvement algorithm for them.

\begin{theorem}[Strategy improvement algorithm for energy games]
\label{5-thm:strategy_improvement_energy}
There exists a strategy improvement algorithm for solving energy games in exponential time.
\end{theorem}

We rely on the high-level presentation of strategy improvement algorithms given in \Cref{1-sec:strategy_improvement}, although it is not necessary to have read that part.
We have already proved in~\Cref{5-lem:least_fixed_point_energy} that the energy values correspond to the least fixed point of the operator $\Op^{\Game}$ defined by:
\[
\Op^{\Game}(\mu)(u) = 
\begin{cases}
\min \set{\delta(\mu(v),w) : u \xrightarrow{w} v \in E} & \text{ if } u \in \VMin, \\
\max \set{\delta(\mu(v),w) : u \xrightarrow{w} v \in E} & \text{ if } u \in \VMax.
\end{cases}
\]
Recall that $Y = [0, nW] \cup \set{\infty}$, and $F_V$ is the lattice of functions $V \to Y$ equipped with the component wise order induced by $Y$.
The function $\delta : Y \times [-W,W] \to Y$ is defined by $\delta(\ell,w) = \max(\ell - w, 0)$ if $\ell - w \le nW$, and $\infty$ otherwise. 

\paragraph{\bf Improving a strategy.}
Let $\sigma$ a positional strategy for Max, and a vertex $u \in \VMax$, we say that $e : u \xrightarrow{w} v$ is an improving edge if
\[
\delta(\val^{\sigma}(v),w) > \val^{\sigma}(u).
\]
Intuitively: according to $\val^{\sigma}$, playing $e$ is better than playing $\sigma(u)$.

Given a strategy $\sigma$ and a set of improving edges $S$ (for each $u \in \VMax$, $S$ contains at most one outgoing edge of $u$), we write $\sigma[S]$ for the strategy 
\[
\sigma[S](u) = 
\begin{cases}
e & \text{ if there exists } e = u \xrightarrow{w} v \in S,\\
\sigma(v) & \text{ otherwise}.
\end{cases}
\]
The difficulty is that an edge being improving does not mean that it is a better move than the current one in any context,
but only according to the value function $\val^{\sigma}$, so it is not clear that $\sigma[S]$ is better than $\sigma$.
Strategy improvement algorithms depend on the following two principles:
\begin{itemize}
	\item \textbf{Progress}: updating a strategy using improving edges is a strict improvement,
	\item \textbf{Optimality}: a strategy which does not have any improving edges is optimal.
\end{itemize}

Let us write $\sigma \le \sigma'$ if for all vertices~$v$ we have $\val^{\sigma}(v) \le \val^{\sigma'}(v)$,
and $\sigma < \sigma'$ if additionally $\neg (\sigma' \le \sigma)$.

\paragraph{\bf The algorithm.}
The pseudocode of the algorithm is given in \Cref{5-algo:strategy_improvement_energy}.

\begin{algorithm}
 \KwData{An energy game $\game$}
 \DontPrintSemicolon
 
 Choose an initial strategy $\sigma_0$ for Max
 
 \For{$i = 0,1,2,\dots$}{

 	Compute $\val^{\sigma_i}$ and the set of improving edges

	\If{$\sigma_i$ does not have improving edges}{
		\Return{$\sigma_i$}
	}

	Choose a non-empty set $S_i$ of improving edges 
	
	$\sigma_{i+1} \leftarrow \sigma_i[S_i]$
 } 
 \caption{The strategy improvement algorithm for energy games.}
\label{5-algo:strategy_improvement_energy}
\end{algorithm}

\paragraph{\bf The potential reduction point of view.}
A first important insight into the algorithm is through so-called potential reductions.
From a game $\Game$ and a strategy $\sigma$ with value $\val^{\sigma}$, we define $\Game_{\sigma}$ as follows.
The two games are identical, except for the weights: 
if $u \xrightarrow{w} v$ in $\Game$, then $u \xrightarrow{w + \val^{\sigma}(v) - \val^{\sigma}(u)} v$ in $\Game_\sigma$.

\begin{fact}
\label{5-fact:potential_reduction}
We have $\val^{\Game} = \val^{\sigma} + \val^{\Game_{\sigma}}$.
\end{fact}

\begin{proof}
This follows from the observation that given any finite play 
\[
\pi = v_0 \xrightarrow{w_0} v_1 \cdots v_{k-1} \xrightarrow{w_{k-1}} v_k
\]
in $\Game$, writing the corresponding play $\pi' = v_0 \xrightarrow{w'_0} v_1 \cdots v_{k-1} \xrightarrow{w'_{k-1}} v_k$ in $\Game_{\sigma}$, we have a telescoping sum:
\[
\sum_{i = 0}^{k-1} w'_i = \val^{\sigma}(v_k) - \val^{\sigma}(v_0) + \sum_{i = 0}^{k-1} w_i.
\]
\end{proof}

The benefit of this point of view is to interpret the notion of improving edges in $\Game_\sigma$.

\begin{fact}
\label{5-fact:potential_reduction_improving_edges}
Let $e = u \xrightarrow{w} v$ an edge in $\Game$.
\begin{itemize}
	\item if $u \in \VMax$, then $e$ is an improving edge if and only if its weight in $\Game_\sigma$ is negative.
	\item if $u \in \VMax$ and $\sigma(u) = e$ then the weight of $e$ in $\Game_{\sigma}$ is zero.
	\item if $u \in \VMin$, then the weight of $e$ in $\Game_{\sigma}$ is non-negative.
\end{itemize}
\end{fact}
The second and third properties directly follow from the fact that $\val^{\sigma}$ is a fixed point of $\Op^{\Game[\sigma]}$.

\begin{figure}
\centering
  \begin{tikzpicture}[scale=.9]
    \node[s-eve] (v0) at (0,0) {$v_0$};
    \node[fill=red!30!white] at (0,-.8) {$7$};
    \node[s-adam] (v1) at (2,1) {$v_1$};
    \node[fill=red!30!white] at (2,0.2) {$0$};
    \node[s-adam] (v2) at (2,-1) {$v_2$};
    \node[fill=red!30!white] at (2,-1.8) {$4$};
    \node[s-adam] (v3) at (4,-1) {$v_3$};
    \node[fill=red!30!white] at (4,-1.8) {$0$};
    \node at (7,0) {The game $\Game$ and its values in red};
    \path[arrow]
      (v0) edge node[above] {$-2$} (v1)
	  (v1) edge[selfloop=0] node[right] {$5$} (v1)
      (v0) edge node[below] {$-3$} (v2)
      (v2) edge[bend left] node[above] {$-4$} (v3)
      (v3) edge[bend left] node[below] {$5$} (v2);
  \end{tikzpicture}
  \begin{tikzpicture}[scale=.9]
	\node at (0,2) {};
    \node[s-eve] (v0) at (0,0) {$v_0$};
    \node[fill=red!30!white] at (0,-.8) {$2$};
    \node[s-adam] (v1) at (2,1) {$v_1$};
    \node[fill=red!30!white] at (2,0.2) {$0$};
    \node[s-adam] (v2) at (2,-1) {$v_2$};
    \node[fill=red!30!white] at (2,-1.8) {$4$};
    \node[s-adam] (v3) at (4,-1) {$v_3$};
    \node[fill=red!30!white] at (4,-1.8) {$0$};
    \node at (6.8,0) {The game $\Game[\sigma]$ and its values in red};
    \path[arrow]
      (v0) edge node[above] {$-2$} (v1)
	  (v1) edge[selfloop=0] node[right] {$5$} (v1)
      (v2) edge[bend left] node[above] {$-4$} (v3)
      (v3) edge[bend left] node[below] {$5$} (v2);
  \end{tikzpicture}
  \begin{tikzpicture}[scale=.9]
	\node at (0,2) {};
    \node[s-eve] (v0) at (0,0) {$v_0$};
    \node[fill=red!30!white] at (0,-.8) {$5$};
    \node[s-adam] (v1) at (2,1) {$v_1$};
    \node[fill=red!30!white] at (2,0.2) {$0$};
    \node[s-adam] (v2) at (2,-1) {$v_2$};
    \node[fill=red!30!white] at (2,-1.8) {$0$};
    \node[s-adam] (v3) at (4,-1) {$v_3$};
    \node[fill=red!30!white] at (4,-1.8) {$0$};
    \node at (7,0) {The game $\Game_\sigma$ and its values in red};
    \path[arrow]
      (v0) edge node[above] {$0$} (v1)
	  (v1) edge[selfloop=0] node[right] {$5$} (v1)
      (v0) edge node[below] {$-5$} (v2)
      (v2) edge[bend left] node[above] {$0$} (v3)
      (v3) edge[bend left] node[below] {$1$} (v2);
  \end{tikzpicture}
\caption{An example of potential reduction.}
\label{5-fig:potential_reduction_example}
\end{figure}

\paragraph{\bf Proof of correctness.}
We now rely on \Cref{5-lem:least_fixed_point_energy} to prove the two principles: progress and optimality.

\begin{lemma}[Progress for the strategy improvement algorithm for energy games]
\label{5-lem:progress_energy}
Let $\sigma$ a strategy and $S$ a set of improving edges.
We let $\sigma'$ denote $\sigma[S]$.
Then $\sigma < \sigma'$.
\end{lemma}

\begin{proof}
We use the potential reduction point of view.
Let us apply~\Cref{5-fact:potential_reduction} to the game $\Game[\sigma']$: 
we have $\val^{\sigma'} = \val^{\sigma} + \val^{G'}$.
Thanks to~\Cref{5-fact:potential_reduction_improving_edges} in $G'$ all weights are non-positive, implying that $\val^{G'} \ge 0$,
and even $\val^{G'} > 0$ because the weights corresponding to improving edges are negative.
\end{proof}

\begin{lemma}[Optimality for the strategy improvement algorithm for energy games]
\label{5-lem:optimality_energy}
Let $\sigma$ be a strategy that has no improving edges, then $\sigma$ is optimal.
\end{lemma}

\begin{proof}
We prove the contrapositive: assume that $\sigma$ is not optimal, we show that it must have some improving edge.
The fact that $\sigma$ is not optimal means that $\val^{\sigma} < \val^{\Game}$.
Since $\val^{\Game}$ is the least fixed point of $\Op^{\Game}$, it is also its least pre-fixed point.
Therefore $\val^{\sigma}$ is not a pre-fixed point: $\neg (\val^{\sigma} \ge \Op^{\Game}(\val^{\sigma}))$.
Hence there exists $u \in V$ such that $\val^{\sigma}(u) < \Op^{\Game}(\val^{\sigma})(u)$.

We rule out the case that $u \in \VMin$: since $\val^{\sigma}$ is a fixed point of $\Op^{\Game[\sigma]}$, this implies that for $u \in \VMin$ we have $\val^{\sigma}(u) = \min \set{ \delta(\val^{\sigma}(v), w) : u \xrightarrow{w} v \in E}$, equal to $\Op^{\Game}(\val^{\sigma})(u)$.
Therefore $u \in \VMax$, implying that there exists $u \xrightarrow{w} v$ such that $\val^{\sigma}(u) < \delta(\val^{\sigma}(v), w)$.
This is the definition of $u \xrightarrow{w} v$ being an improving edge.
\end{proof}

\paragraph{\bf Complexity analysis.}
The computation of $\val^\sigma$ for a strategy $\sigma$ can be seen to be a shortest path problem. 
Thus, any algorithm for the shortest path problem can be applied, such as the Bellman-Ford algorithm.
In particular computing $\val^\sigma$ can be done in polynomial time, and even more efficiently through a refined analysis.


An aspect of the algorithm we did not develop is choosing the set of improving edges.
Many possible rules for choosing this set have been studied, as for instance the \emph{greedy all-switches} rule. 

The next question is the number of iterations, meaning the length of the sequence
$\sigma_0,\sigma_1,\dots$. It is at most exponential since it is bounded by the number of strategies (which is bounded aggressively by $m^n$).
There are lower bounds showing that the sequence can be of exponential length, which apply to different rules for choosing improving edges.
Hence the overall complexity is exponential; we do not elaborate further here. 
We refer to \Cref{5-sec:references} for bibliographic references and a discussion on the family of strategy improvement algorithms.

\section{Discounted-payoff games}
\label{5-sec:discounted_payoff}
From a practical point of view, the modelling of a real-world
situation via mean-payoff games requires that only the long-term
behaviour is important. Since mean payoff only depends on the limit of
the play, it cannot be used to model the beginning of the execution:
the mean-payoff objective is \emph{prefix-independent}. 
In economical studies, there is a tendency to make the prefixes count more, since
they represent short-term implications of the actions taken, even if
long-term behaviours also matter. The common payoff used to model this
preference to prefixes is the discounted payoff that associates to a play $\rho$ the value
\[
\DiscountedPayoff(\rho) = (1 - \lambda) \cdot \sum_{i=0}^{\infty} \lambda^i\, \rho_i,
\] 
where $\lambda$ is a parameter in the interval $(0,1)$, ensuring the convergence of the infinite series (since weights $\play_i$ are bounded). 
We assume that $\lambda$ is a rational number, it is part of the input of a discounted-payoff game.
The coefficient $1-\lambda$ before
the series is just to counterbalance the fact that if all weights in
the game are $1$, we would like the payoff to be 1 too, which then
holds since $\sum_{i=0}^\infty \lambda^i = \frac 1{1-\lambda}$. When
$\lambda$ tends to $0$, only the prefixes (and even the first weight)
matters. On the contrary, when $\lambda$ tends to $1$, discounted payoff looks more and more like mean payoff. 
To grasp an intuition why this holds, consider a play that results from
positional strategies in a mean-payoff game. The weights encountered
during the play then ultimately follow a periodic sequence
$w_0,w_1,\ldots,w_{r-1},w_0,w_1,\ldots,w_{r-1},w_0,\ldots$ with
average payoff $\frac 1 r\sum_{i=0}^{r-1}w_i$. Grouping the terms of
the series $(1-\lambda)\sum_{i=0}^{\infty} \lambda^i \, w_i$ by
batches of $r$ terms, we then obtain
\[(1-\lambda)\sum_{i=0}^{\infty} \lambda^{ri} \sum_{j=0}^{r-1}
  \lambda^jw_j=\frac{1-\lambda}{1-\lambda^r}\sum_{j=0}^{r-1}
  \lambda^jw_j= \frac
  1{1+\lambda+\cdots+\lambda^{r-1}}\sum_{j=0}^{r-1} \lambda^jw_j\]
that tends towards the average-payoff $\frac 1 r\sum_{j=0}^{r-1} w_j$
when $\lambda$ tends to $1$.

The game of~\Cref{5-fig:MP} can also be equipped with a discounted-payoff condition. 
If $\lambda$ is close to $1$, for instance $\lambda=0.9$, then optimal strategies are the same as for the
mean-payoff objective: 
\[
\begin{array}{c}
\sigma^*(v_0) = v_0 \xrightarrow{4} v_1 \quad ; \quad \sigma^*(v_2) = v_2 \xrightarrow{4} v_3 \\
\tau^*(v_1) = v_1 \xrightarrow{2} v_2 \quad ; \quad \tau^*(v_3) = v_3 \xrightarrow{-1} v_1 \quad ; \quad \tau^*(v_4) = v_4 \xrightarrow{-2} v_0.
\end{array}
\] 
The play consistent with $\sigma^*$ and $\tau^*$ starting from $v_0$ is the lasso word 
\[
v_0 \xrightarrow{4} v_1 \cdot \big( v_1 \xrightarrow{2} v_2 \xrightarrow{4} v_3 \xrightarrow{-1} v_1 \big)^\omega,
\]
which has as discounted payoff $(1-\lambda)\left(4+\frac{2\lambda+4\lambda^2-\lambda^3}{1-\lambda^3}\right)$, it is approximately $1.7$ when $\lambda=0.99$. 
Recall that the mean-payoff optimal value of vertex $v_0$ was $5/3\approx 1.67$. 
However, the situation completely changes when $\lambda$ decreases. 
When $\lambda=0.5$ for instance, Min changes his decision in vertex $v_1$ and his optimal move is $v_1 \xrightarrow{0} v_0$.
For a really low value of $\lambda$, for instance $\lambda=0.1$, the decisions again change drastically for both players: now the optimal (positional) strategies become 
\[
\begin{array}{c}
\sigma^*(v_0) = v_0 \xrightarrow{4} v_1 \quad ; \quad \sigma^*(v_2) = v_2 \xrightarrow{4} v_3 \\
\tau^*(v_1) = v_1 \xrightarrow{0} v_0 \quad ; \quad \tau^*(v_3) = v_3 \xrightarrow{-2} v_0 \quad ; \quad \tau^*(v_4) = v_4 \xrightarrow{-2} v_0.
\end{array}
\] 


\subsection*{Computing the values using a contracting fixed point}

Let us consider a discounted-payoff game $\Game$ with $n$ vertices and weights in $[-W,W]$.
We define $Y$ as $\Rinfty$, equipped with the natural total order on the reals.
We let $F_V$ be the lattice of functions $V \to Y$ equipped with the component wise order induced by~$Y$.
We equip $F_V$ with the infinity norm: $\|\mu\| = \max_{u \in V} |\mu(u)|$.

We define a function $\delta : Y \times [-W,W] \to Y$ by $\delta(x,w) = \lambda \cdot x + (1 - \lambda) \cdot w$. 
To understand the definition of $\delta$, the key observation is that the discounted payoff can be computed recursively:
\[
\begin{array}{lll}
\DiscountedPayoff(\rho) & = & (1 - \lambda) \cdot \sum_{i=0}^{\infty} \lambda^i\, \rho_i \\
 & = & (1 - \lambda) \cdot \rho_0 + \lambda \cdot \DiscountedPayoff(\rho_{\ge 1}).
\end{array}
\]
The function $\delta$ induces an operator $\Op^{\Game} : F_V \to F_V$:
\[
\Op^{\Game}(\mu)(u) = 
\begin{cases}
\max \set{\delta(\mu(v),w) : u \xrightarrow{w} v \in E} & \text{ if } u \in \VMax, \\
\min \set{\delta(\mu(v),w) : u \xrightarrow{w} v \in E} & \text{ if } u \in \VMin.
\end{cases}
\]
We note that $\Op^{\Game}$ is a monotonic operator. Even more interesting, it is contracting:

\begin{fact}
The operator $\Op^{\Game}$ is contracting with contraction factor $\lambda$, meaning that for all $\mu,\mu' \in F_V$:
\[
\|\Op^{\Game}(\mu) - \Op^{\Game}(\mu')\| \le \lambda \cdot \|\mu - \mu'\|.
\]
\end{fact}


By Banach fixed-point theorem, see~\Cref{1-thm:banach}, a direct consequence of this fact is that $\Op^{\Game}$ has a unique fixed point.
The following theorem is the cornerstone of the study of discounted-payoff games.

\begin{lemma}[Values as unique fixed point]
\label{5-thm:values_discounted_contracting_fixed_point}
The discounted-payoff values are the unique fixed point of the operator $\Op^{\Game}$.
Moreover, the unique fixed point induces a pair of optimal positional strategies:
\[
\begin{array}{l}
u \in \VMax:\ \sigma(u) \in \argmax \set{\delta(\val^{\Game}(v),w) : u \xrightarrow{w} v \in E} \\
u \in \VMin:\ \tau(u) \in \argmin \set{\delta(\val^{\Game}(v),w) : u \xrightarrow{w} v \in E}.
\end{array}
\]
\end{lemma}


\begin{proof}
Since we already know that $\Op^{\Game}$ has a unique fixed point, it is enough to show that $\val^{\Game}$ is a fixed point of $\Op^{\Game}$.
This follows from two properties, see~\Cref{1-lem:sufficient_condition_fixed_point}:
\begin{itemize}
	\item For all $\rho$ sequences of weights, we have 
	\[
	\DiscountedPayoff(w \cdot \rho) = \delta(\DiscountedPayoff(\rho), w).
	\]
	\item The function $\delta$ is monotonic and continuous.
\end{itemize}

We now move to the second point: the values imply optimal positional strategies.
Let us define $\sigma,\tau$ two positional strategies:
\[
\begin{array}{l}
u \in \VMax:\ \sigma(u) \in \argmax \set{\delta(\val^{\Game}(v),w) : u \xrightarrow{w} v \in E} \\
u \in \VMin:\ \tau(u) \in \argmin \set{\delta(\val^{\Game}(v),w) : u \xrightarrow{w} v \in E}.
\end{array}
\]
To show that $(\sigma,\tau)$ is a pair of optimal strategies, we claim that
$\val^{\sigma} = \val^{\tau} = \val^{\Game}$. 
Indeed, both $\val^{\sigma}$ and $\val^{\tau}$ are fixed points of $\Op^{\Game}$, and since there exists a unique fixed point the claim follows.
\end{proof}

Let us illustrate the operator $\Op^{\Game}$ on the discounted-payoff game of~\Cref{5-fig:MP}.
It is convenient to  the contracting operator is:
\[
\begin{array}{lll}
\Op^{\Game}(\mu)(v_0) & = & \max \big((1-\lambda) \cdot 4    + \lambda \cdot \mu(v_1),\ (1-\lambda) \cdot 5    + \lambda \cdot \mu(v_4) \big) \\
\Op^{\Game}(\mu)(v_1) & = & \min \big((1-\lambda) \cdot 0    + \lambda \cdot \mu(v_0),\ (1-\lambda) \cdot 2    + \lambda \cdot \mu(v_2)\big) \\
\Op^{\Game}(\mu)(v_2) & = & \max \big((1-\lambda) \cdot 1    + \lambda \cdot \mu(v_2),\ (1-\lambda) \cdot 4    + \lambda \cdot \mu(v_3)\big) \\
\Op^{\Game}(\mu)(v_3) & = & \min \big((1-\lambda) \cdot (-2) + \lambda \cdot \mu(v_0),\ (1-\lambda) \cdot (-1) + \lambda \cdot \mu(v_1)\big) \\
\Op^{\Game}(\mu)(v_4) & = & \min \big((1-\lambda) \cdot (-2) + \lambda \cdot \mu(v_0),\ (1-\lambda) \cdot 2    + \lambda \cdot \mu(v_4)\big)
\end{array}
\]

A careful analysis gives the fixed points for all values of $\lambda\in (0,1)$, which in turn allows us to find the associated
optimal positional strategies $\sigma^*$ and~$\tau^*$ on the various intervals of values for $\lambda$, summarised in the following table:
\[
  \begin{array}{|c|c|c|c|c|}\hline
    \lambda & (0, \lambda_1]
    & (\lambda_1,\lambda_2] & (\lambda_2,\lambda_3]
    & (\lambda_3,1) \\\hline
    \sigma^*(v_0) & v_4 & v_4 &  v_1   & v_1  \\\hline
    \tau^*(v_1) & v_0 &  v_0 &   v_0 &   v_2  \\\hline
    \sigma^*(v_2) & v_3  & v_3   & v_3 &   v_3 \\\hline
    \tau^*(v_3) &v_0 &  v_1  & v_1 &   v_1  \\\hline
    \tau^*(v_4) & v_0 &  v_0  & v_0 &  v_0\\\hline
  \end{array}\]
The frontiers are at $\lambda_1 = 1-\sqrt 2/2 \approx 0.293$, $\lambda_2 = 1/2$, and $\lambda_3 \approx 0.841$. 
For instance, on interval $(0,\lambda_1]$, Min gets discounted payoff
$\frac{5\lambda-2}{1+\lambda}$ when starting 
in vertex $v_3$, while switching his decision in interval
$(\lambda_1,\lambda_2]$ allows him to secure
$\frac{-2\lambda^3+6\lambda^2-1}{1+\lambda}$: this gives the
explanation for the value of $\lambda_1$ which allows one to equal the
two values. A similar reasoning provides the values of
$\lambda_2$ and $\lambda_3$.

\subsection*{Solving discounted-payoff games is in $\NP \cap \coNP$}

The first step in proving $\NP \cap \coNP$ upper bounds is to solve the one-player variants in polynomial time.

\begin{lemma}[One player discounted-payoff games]
\label{5-lem:one-player-DP}
  There exists a polynomial-time algorithm for computing the optimal values of one-player discounted-payoff games.
\end{lemma}

\begin{proof}
Let us consider a discounted-payoff game where only Max has moves.
We claim that $\val^{\sigma}$ (seen as a vector $(x_u)_{u \in V}$) is the unique solution of the following linear program
\begin{equation*}
\begin{array}{lcr}
\text{maximise }   & \sum_{u \in V} x_u \\
\text{subject to } & x_u \le (1 - \lambda) \cdot w + \lambda x_v & \text{ for } u \xrightarrow{w} v \in E,
\end{array}
\end{equation*}
Indeed, vectors satisfying the constraints are exactly pre-fixed points of $\Op^{\sigma}$, and since $\Op^{\sigma}$ is monotonic
the least fixed point of $\Op^{\sigma}$ coincide with its least pre-fixed point.
Linear programming can be solved in polynomial time, see~\Cref{1-thm:linear_programming}.

A dual argument solves the case where only Min has moves.
\end{proof}

As for mean-payoff (or parity) games, the existence of positional
optimal (or winning) strategies for both players, and the ability to
solve in polynomial time the one-player version of these games, allows
us to obtain easily an $\NP\cap\coNP$ complexity to solve
discounted-payoff games. The use of the above contracting operator even
ensures that the Turing machines guessing and checking the optimal
strategies may indeed be designed as unambiguous (instead of just
non-deterministic). Calling $\UP$ the class of problems that can be
solved by an unambiguous Turing machine running in polynomial time,
and $\coUP$ the class of problems whose complement are in $\UP$, we
then obtain the theorem:
\begin{theorem}[Complexity]\label{5-thm:disc-up}
  Solving discounted-payoff games is in $\UP\cap\coUP$.
\end{theorem}
\begin{proof}
  Using the previous result, we know that the value of a discounted-payoff game is the \emph{unique} solution of the fixed-point equation $\mu = \Op^{\Game}(\mu)$.   
  Therefore, guessing $\mu$ and checking it is indeed a fixed point of $\Op^{\Game}$ can be done by an unambiguous Turing machine. 
  To ensure that the machine runs in polynomial time, it only remains to show that the solution is of polynomial size.
  Let us fix $\sigma,\tau$ a pair of positional optimal strategies.
  Let us rewrite the equation in a matrix form:
  \begin{itemize}
  	\item We write $\vec x$ for a vector of values, $\vec x \in \R^V$.
  	\item We define the matrix $Q \in \{0,1\}^{V \times V}$: the entry $Q_{u,v}$ is $1$ if $u \in \VMin$ and $\sigma(u) = u \xrightarrow{w} v$ or $u \in \VMax$ and $\tau(u) = u \xrightarrow{w} v$, and $0$ otherwise.
	\item We define the vector $\vec c \in \Z^V$: the entry $c_u$ is the weight of the edge $u \xrightarrow{w} v$ chosen by the strategies $\sigma$ and $\tau$.
  \end{itemize}
  With these notations, the equation rewrites
  \[
  \vec x = (1- \lambda) \cdot \vec c + \lambda \cdot Q \cdot \vec x.
  \] 
  Letting $\lambda = a/b$ the rational discount factor, the above equation
  rewrites into 
  \begin{equation} 
  A \cdot \vec x = (b-a) \cdot \vec c
  \label{5-eq:1} 
  \end{equation} 
  with $A = b \cdot I - a \cdot Q$ ($I$ being the identity matrix). 
  Therefore, $A$ is a matrix that has at most two non-zero elements in each row: each of these non-zero elements can
  be written using at most $N=\max(\log_2 a,\log_2 b)$ bits (therefore
  polynomial in the representation of the game), and are therefore
  bounded in absolute value by $2^N$. By induction on the size of the
  matrix, we can then show that the determinant of $A$ is at most
  $4^{n \cdot N}$. 
  The solution to \Cref{5-eq:1}, using Cramer's formula, reads $x_v = \det (A_v) / \det (A)$ where $A_v$ is the matrix obtained from $A$ by replacing the $v$-th column with the vector $(b-a) \cdot \vec c$. 
  Therefore, all components of $\vec x$ can be written with only a polynomial number of bits with respect to the size of the weights in the
  game and $N$.

  The $\coUP$ membership follows, as in \Cref{5-thm:MP-NPcoNP}, from a dual reasoning for Min. 
\end{proof}

\subsection*{A value iteration algorithm for discounted-payoff games}

Let us recall that thanks to Banach fixed-point theorem, see~\Cref{1-thm:banach}, the fixed point is obtained as the limit of the following sequence:
\[
\forall u \in V,\ \mu_0(u) = 0 \quad ; \quad \mu_{k+1} = \Op^{\Game}(\mu_k).
\]
We have $\lim_k \mu_k = \val^{\Game}$.
To make sense of this sequence, let us define the following condition for each $k$:
\[
\DiscountedPayoff_k(\rho) = (1 - \lambda) \cdot \sum_{i = 0}^{k-1} \lambda^i \rho_i.
\]
We define $\val^{\Game,k} : V \to \Rinfty$ the value function by 
\[
\val^{\Game,k}(u) = \sup_{\sigma}\ \inf_{\tau}\ \DiscountedPayoff_k[\col](\pi^{\sigma,\tau}_u)
\]
\begin{fact}
For all $k$, we have $\val^{\Game,k} = \mu_k$.
\end{fact}

The remaining question is therefore to design a rounding procedure to compute the exact values.
The following lemma places bounds on the discounted-payoff values, in a similar manner as for~\Cref{5-cor:rational-MP} for mean payoff.

\begin{lemma}[Upper bound on values in discounted-payoff games]
\label{5-lem:rational-discounted}
Let us write $\lambda = \frac a b\in (0,1)$.
The discounted-payoff values are rational numbers in $[-W,W]$ whose denominators are at most $D = b^{n-1} \prod_{j=1}^{n}(b^j - a^j)$.
\end{lemma}

\begin{proof}
Let $\sigma,\tau$ a pair of optimal positional strategies. The corresponding play consists of a prefix of length at most $n$ and a simple cycle,
hence the sequence of weights is:
\[
w_0,w_1,\ldots,w_{k-1},(w_{k},\ldots,w_\ell)^\omega,
\]
with $k,\ell \leq n$.
  \begin{align*}
    \Value^{\Game}(v) &= (1-\lambda) \left[\sum_{i=0}^{k-1} \lambda^i w_i +
               \lambda^{k}\sum_{m=0}^\infty
               \lambda^{(\ell-k+1)m}\sum_{i=0}^{\ell-k}
               \lambda^iw_{k+i}\right]\\
             &= \frac{b-a}b \left[\sum_{i=0}^{k-1} \frac{b^{k-1-i}a^i}{b^{k-1}} \cdot w_i +
               \frac{\lambda^{k}}{1-\lambda^{\ell-k+1}}\sum_{i=0}^{\ell-k}
               \frac{b^{\ell-k-i}a^i}{b^{\ell-k}} \cdot w_{k+i} \right]\\
             &= \frac {N_1}{b^{k}} + 
               \frac{a^{k}b^{\ell-k+1}}{b^{k+1}(b^{\ell-k+1}-a^{\ell-k+1})}
               \frac{N_2}{b^{\ell-k}} \qquad \text{(with $N_1,N_2\in \Z$)}\\
             &= \frac{N_3}{b^{k}(b^{\ell-k+1}-a^{\ell-k+1})} \qquad \text{(with
               $N_3\in \Z$)}\\
             &= \frac{N}{b^{n-1}\prod_{j=1}^{n}(b^j-a^j)} \qquad \text{(with
               $N\in \Z$)} 
  \end{align*}
\end{proof}
It follows from~\Cref{5-lem:rational-discounted} that if we have an approximation $\eta$ of $\Value^{\Game}(v)$ such that 
$|\Value^{\Game}(v)-\eta|<\frac 1 {2D}$, we can recover the value as $\Value^{\Game}(v) = \frac{\lfloor D\eta+1/2\rfloor}D$. 

Let 
\[
K = \left\lceil \frac{1}{-\log_2\lambda} \left(\frac{n(n+3)}{2}\log_2b + \log_2 W+2\right) 
     \right\rceil
\]
\begin{lemma}[Number of steps of value iteration]
\label{5-lem:number-steps-VI-discounted}
For the value $K$ above, we have $\|F^K(\vec 0)-\Value\| < \frac 1 {2D}$.
\end{lemma}
\begin{proof}
  First, we bound $D$ by $b^{n+\frac{n(n+1)}2}$, so that
  $\frac{n(n+3)}{2}\log_2b\geq \log_2D$. Therefore
  $K\geq \frac{1}{-\log_2\lambda} (\log_2D + \log_2 W+\log_24) =
  \log_{1/\lambda}(4DW)$. This implies that
  $\lambda^KW\leq \frac{1}{4D}< \frac 1 {2D}$. 
  Since $\Op^{\Game}$ is contracting with contraction factor $\lambda$, we have
  $\|F^K(\vec 0) - \Value^{\Game} \|_\infty \leq \lambda^K \cdot \|\Value^{\Game}\|_\infty$. 
  Since $|\Value^{\Game}(v)| \le W$, we obtain the desired inequality.
\end{proof}


\begin{algorithm}
 \KwData{A discounted-payoff game $\game$ with discount factor
 $\lambda= a/b\in(0,1)$, and $F$ the contracting operator}

 \For{$u \in V$}{
	 $\mu(u) \leftarrow 0$
	 }

 $K \leftarrow \left\lceil \frac{1}{-\log_2\lambda} \left(\frac{n(n+3)}{2}\log_2b +
     \log_2 W+2\right) \right\rceil$ ;

 \For{$i = 1$ to $K$}
 {
   $\mu \leftarrow \Op^{\Game}(\mu)$
 }

 \Return{$\mu$}
\caption{The value iteration algorithm for discounted-payoff games.}
\label{5-algo:DP-value-iteration}
\end{algorithm}

\begin{theorem}[Value iteration algorithm for discounted-payoff games]
\label{5-thm:DP-value-iteration}
  There exists a value iteration algorithm computing in pseudo-polynomial time the discounted-payoff values. 
\end{theorem}

One can check that $K$ is polynomial in the size of the arena, but not in the discount factor $\lambda$.
Indeed, consider that $\lambda = 1 - \frac 1 b$, with $b \in \N\setminus \{0\}$. 
Then, we may store $\lambda$ with $\log_2 b$ bits, yet $\frac 1{-\log_2\lambda} \sim_{b\to \infty} b\ln 2$ is exponential in $\log_2b$.
As a consequence, the value iteration algorithm runs in pseudo-polynomial time. 


\subsection*{Strategy improvement algorithm for discounted-payoff games}
\label{5-subsec:strategy_improvement_discounted}

\begin{theorem}[Strategy improvement for discounted-payoff games]
\label{5-thm:DP-strategy-improvement-correctness}
There exists a strategy improvement algorithm computing the discounted-payoff values in exponential time. 
\end{theorem}

We construct a strategy improvement algorithm, following the presentation in~\Cref{1-sec:strategy_improvement}.
Let us consider a (positional) strategy $\sigma$ for Max.
We compute $\val^{\sigma}$ (for instance by solving a linear program as in~\Cref{5-lem:one-player-DP}).
Now there are two cases:
\begin{itemize}
	\item Either $\val^{\sigma} = \Op^{\Game}(\val^{\sigma})$, in which case we have found a fixed point of $\Op^{\Game}$. By uniqueness, this is the discounted-payoff values, and $\sigma$ is an optimal strategy for Max.
	\item Or $\val^{\sigma} \neq \Op^{\Game}(\val^{\sigma})$, in which case we need to improve the strategy $\sigma$.
\end{itemize}
Let us consider the second case. Let $u \in \VMax$, we say that $u \xrightarrow{w} v \in E$ is an improving edge if
$\val^{\sigma}(u) < \delta(\val^{\sigma}(v), w)$. Given a set of improving edges $S$ (for each $u \in \VMax$, $S$ contains at most one outgoing edge of $u$), we write $\sigma[S]$ for the strategy 
\[
\sigma[S](u) = 
\begin{cases}
e & \text{ if there exists } e = u \xrightarrow{w} v \in S,\\
\sigma(v) & \text{ otherwise}.
\end{cases}
\]
Strategy improvement algorithms depend on the following two principles:
\begin{itemize}
	\item \textbf{Progress}: updating a strategy using improving edges is a strict improvement,
	\item \textbf{Optimality}: a strategy which does not have any improving edges is optimal.
\end{itemize}

Let us write $\sigma \le \sigma'$ if for all vertices~$v$ we have $\val^{\sigma}(v) \le \val^{\sigma'}(v)$,
and $\sigma < \sigma'$ if additionally $\neg (\sigma' \le \sigma)$.

\paragraph{\bf The algorithm.}
The pseudocode of the algorithm is given in \Cref{5-algo:strategy_improvement_discounted}.

\begin{algorithm}
 \KwData{A discounted-payoff game $\game$}
 \DontPrintSemicolon
 
 Choose an initial strategy $\sigma_0$ for Max
 
 \For{$i = 0,1,2,\dots$}{

 	Compute $\val^{\sigma_i}$ and the set of improving edges

	\If{$\sigma_i$ does not have improving edges}{
		\Return{$\sigma_i$}
	}

	Choose a non-empty set $S_i$ of improving edges 
	
	$\sigma_{i+1} \leftarrow \sigma_i[S_i]$
 } 
 \caption{The strategy improvement algorithm for discounted-payoff games.}
\label{5-algo:strategy_improvement_discounted}
\end{algorithm}

\paragraph{An example}
Let us consider the discounted-payoff game of~\Cref{5-fig:MP} with $\lambda=0.5$,
and start from the strategy $\sigma(v_0) = v_0 \xrightarrow{5} v_4$ and $\sigma(v_2) = v_2 \xrightarrow{1} v_2$. 

We compute $\val^{\sigma}$ by solving the linear program:
\begin{equation*}
\begin{array}{lcll}
\text{maximise }   & & & x_0 + x_1 + x_2 + x_3 + x_4 \\
\text{subject to } 
	& x_0   & = & (1-\lambda) \cdot 5    + \lambda \cdot x_4 \\
    & x_1 & \leq & (1-\lambda) \cdot 0    + \lambda \cdot x_0 \\
    & x_1 & \leq & (1-\lambda) \cdot 2    + \lambda \cdot x_2 \\
    & x_2  & = &  (1-\lambda) \cdot 1    + \lambda \cdot x_2 \\
    & x_3 & \leq & (1-\lambda) \cdot (-2) + \lambda \cdot x_0 \\
    & x_3 & \leq & (1-\lambda) \cdot (-1) + \lambda \cdot x_1 \\
    & x_4 & \leq & (1-\lambda) \cdot (-2) + \lambda \cdot x_0 \\
    & x_4 & \leq & (1-\lambda) \cdot 2    + \lambda \cdot x_4
\end{array}
\end{equation*}
Feeding this linear program to a solver we obtain the solution 
\[
\vec x = \left( \frac 8 3, \frac 4 3,1, \frac 1 6, \frac 1 3 \right).
\]
The only improving edge is $v_2 \xrightarrow{4} v_3$, so we iterate with the strategy
$\sigma'$ defined by $\sigma'(v_0) = v_0 \xrightarrow{5} v_4$ and $\sigma'(v_2) = v_2 \xrightarrow{4} v_3$. 
The new vector of values found by solving the new linear program is 
\[
\vec{x'} = \left( \frac 8 3, \frac 4 3, \frac{25}{12}, \frac 1 6, \frac 1 3 \right),
\]
which is the (unique) fixed point of~$\Op^{\Game}$. 

\paragraph{\bf Proof of correctness.}
We now prove the two principles: progress and optimality.

\begin{lemma}[Progress for the strategy improvement algorithm for discounted-payoff games]
\label{5-lem:progress_discounted}
Let $\sigma$ a strategy and $S$ a set of improving edges.
We let $\sigma'$ denote $\sigma[S]$.
Then $\sigma < \sigma'$.
\end{lemma}

\begin{proof}
Let $\tau,\tau'$ optimal positional strategies in $\Game[\sigma]$ and $\Game[\sigma']$.
As in the proof of~\Cref{5-thm:disc-up}, letting $Q$, $\vec c$, $Q'$, and $\vec {c'}$ the respective matrices and cost vectors described by the pairs of strategies $(\sigma,\tau)$ and
  $(\sigma',\tau')$, we have
  \[
  \Value^\sigma = (1-\lambda) \cdot \vec c + \lambda \cdot Q \cdot \Value^\sigma 
  \qquad \text{ and } \qquad 
  \Value^{\sigma'} = (1-\lambda) \cdot \vec {c'} + \lambda \cdot Q' \cdot  \Value^{\sigma'} \,.
  \]
Therefore:
  \[
  \Value^{\sigma'} - \Value^\sigma = 
  \lambda \cdot Q' \cdot (\Value^{\sigma'} - \Value^{\sigma}) + 
  \underbrace{\lambda \cdot (Q'-Q) \cdot \Value^\sigma + (1-\lambda) \cdot (\vec{c'}-\vec c)}_{= \vec \delta} \,.
  \] 
So:
  \[
  (I - \lambda \cdot Q') \cdot (\Value^{\sigma'} - \Value^\sigma) = \vec \delta \,.
  \] 
Since $Q'$ is a positive matrix with coefficients in $\{0,1\}$, the series $\sum_i \lambda^i \cdot Q'^i$ converges, which shows that $I - \lambda \cdot Q'$ is invertible of inverse $\sum_{i=0}^\infty \lambda^i \cdot Q'^i$. 
In particular, the inverse $(I-\lambda Q')^{-1}$ has only non-negative coefficients, and its diagonal coefficients are positive. 
Therefore, to show that $\Value^{\sigma'} - \Value^\sigma = (I - \lambda \cdot Q')^{-1} \cdot \vec \delta$ is non-negative with at least one positive coefficient, it suffices to show that $\vec\delta$ is non-negative with at least one positive coefficient. 
Let $u \in V$:
\begin{itemize}
  \item If $u \in \VMax$, let $\sigma(u) = u \xrightarrow{w} v$ and $\sigma'(u) = u \xrightarrow{w'} v'$, we have
    \[
    \delta_u = \lambda \cdot (\Value^{\sigma}(v') - \Value^\sigma(v)) + (1 - \lambda)(w' - w).
    \]
    If $\sigma(u) = \sigma'(u)$, then $\delta_u = 0$. 
    Otherwise, $\delta_u > 0$ by definition of an improving edge.
  \item If $u \in \VMin$, let $\tau(u) = u \xrightarrow{w} v$ and $\tau'(u) = u \xrightarrow{w'} v'$, we have
    \[
    \delta_u = \lambda \cdot (\Value^{\sigma}(v') - \Value^\sigma(v)) + (1 - \lambda)(w' - w).
    \]
    By definition of $\tau$ we have $\val^{\sigma}(u) = \delta(\val^{\sigma}(v),w) \le \delta(\val^{\sigma}(v'),w')$.
    This implies $\lambda \cdot \val^{\sigma}(v) + (1 - \lambda) \cdot w \le \lambda \cdot \val^{\sigma}(v') + (1 - \lambda) \cdot w'$,
    so $\delta_u \ge 0$.   
\end{itemize}
\end{proof}

\begin{lemma}[Optimality for the strategy improvement algorithm for discounted-payoff games]
\label{5-lem:optimality_discounted}
Let $\sigma$ be a strategy that has no improving edges, then $\sigma$ is optimal.
\end{lemma}

\begin{proof}
Let $\sigma$ be a strategy that has no "improving edges".
We claim that $\val^{\sigma}$ is a fixed point of $\Op^{\Game}$,
which implies that $\val^{\sigma} = \val^{\Game}$, meaning that $\sigma$ is optimal.
Since $\val^{\sigma}$ is a fixed point of $\Op^{\Game[\sigma]}$, for $u \in \VMin$ we have 
$\val^{\sigma}(u) = \min \set{\delta( \val^{\sigma}(v), w) : u \xrightarrow{w} v \in E}$.
The fact that $\sigma$ has no improving edges reads:
for all $u \in \VMax$, for all $u \xrightarrow{w'} v' \in E$, 
$\delta( \val^{\sigma}(v'), w') \le \delta( \val^{\sigma}(v), w)$ where $\sigma(u) = u \xrightarrow{w} v$.
Since $\val^{\sigma}(u) = \delta( \val^{\sigma}(v), w)$, this implies that 
$\val^{\sigma}(u) = \max \set{\delta( \val^{\sigma}(v'), w) : v \xrightarrow{w} v' \in E}$.
The two equalities above witness that $\val^{\sigma}$ is the unique fixed point of $\Op^{\Game}$.
\end{proof}

The value iteration and strategy improvement algorithms are incomparable:
\begin{itemize}
	\item the value iteration algorithm has a runtime pseudo-polynomial, and more precisely polynomial with respect to the number of vertices and the binary encoding of the weights of the arena, but exponential with respect to the binary encoding of
$\lambda$,
	\item the strategy improvement algorithm has a runtime exponential with respect to the number of vertices, but polynomial with respect to the
binary encoding of $\lambda$ and the weights of the arena.
\end{itemize}

\subsection*{Reducing mean-payoff games to discounted-payoff games}
\label{5-sec:mean_payoff-values}


Recall that~\Cref{5-cor:rational-MP} states that the mean-payoff value $\Value^{\Game}(v)$ is a rational number with denominator at most $n$. 
The minimal distance between two such rational numbers is $\frac 1{n-1}-\frac 1{n} = \frac{1}{n(n-1)}$, 
implying that a $\frac{1}{2n(n-1)}$ approximation $\beta$ of $\Value^{\Game}(v)$ is enough
to apply a rounding procedure finding the only such rational in the interval $[\beta - \frac{1}{2n(n-1)}, \beta + \frac{1}{2n(n-1)}]$. 
By interpreting the mean-payoff game as a discounted-payoff game with a nicely chosen $\lambda$, we are able to find such a good approximation:

\begin{theorem}[Discounted-payoff approximation]
\label{5-thm:MP-Zwick-Paterson}
  Let $\game$ a mean-payoff game. 
  Let $\lambda\in(0,1)$, we define $\game_\lambda$ the discounted-payoff game obtained from $\game$.
  We let $\Value$ denote the mean-payoff values and $\Value_\lambda(v)$ the discounted-payoff values with $\lambda$ as discount factor.
  Then 
  \[
  \|\Value - \Value_\lambda\|_\infty \leq (1-\lambda) \cdot 2n \cdot W.
  \]
\end{theorem}

\begin{proof}
  Let $u \in V$. We prove the inequality
  $\Value_\lambda(u) - \Value^{\Game}(u) \geq -(1-\lambda) \cdot 2n \cdot W$ 
  by reasoning on Max's strategies: a similar reasoning on Min's strategies allows one to obtain the other inequality
  $\Value_\lambda(u) - \Value^{\Game}(u) \leq (1-\lambda) \cdot 2n \cdot W$.

  Let $\sigma,\tau$ a pair of optimal positional strategies for the mean-payoff game.
  The play starting from $u$ consistent with $\sigma$ and $\tau$ is a finite prefix and a simple cycle, 
  hence the sequence of weights is:
\[
\rho = w_0,w_1,\ldots,w_{k-1},(w_{k},\ldots,w_\ell)^\omega,
\]
with $k,\ell \leq n$. Note that $\Value^{\Game}(u) = \frac 1 {\ell-k+1} \sum_{i=k}^\ell w_i$. 
Playing $\sigma$ in the discounted-payoff game, we obtain that $\Value_\lambda(u) \geq \DiscountedPayoff(\rho)$. 

We now compute precisely $\DiscountedPayoff(\rho)$ as follows:
\[
      (1 - \lambda) \cdot \sum_{i=0}^{k-1} \lambda^i \cdot \underbrace{w_i}_{\geq -W}
      +
      (1 - \lambda) \cdot \lambda^k \cdot \sum_{m=0}^\infty \lambda^{(\ell-k+1)m} \sum_{i=0}^{\ell-k}\lambda^iw_{k+i}\\
\]
The first term is greater than or equal to $-(1-\lambda^k) \cdot W$,
and the second term is equal to 
\[
\frac{(1 - \lambda) \cdot \lambda^k}{1 - \lambda^{\ell-k+1}} \sum_{i=0}^{\ell-k}\lambda^i \cdot w_{k+i}.
\]
By shifting all weights by $W$, we can rewrite the sum as:
  \[
  \sum_{i=0}^{\ell-k}\lambda^i \cdot w_{k+i} = 
  \sum_{i=0}^{\ell-k}\lambda^i \cdot (w_{k+i}+W) - W \cdot \sum_{i=0}^{\ell-k} \lambda^i
  \]
  By using the fact that $w_{k+i}+W$ is non-negative and $\lambda^i\geq \lambda^{\ell-k}$, we obtain
  \begin{align*}
    \sum_{i=0}^{\ell-k}\lambda^i \cdot w_{k+i}
    &\geq \lambda^{\ell-k} \cdot \sum_{i=0}^{\ell-k}(w_{k+i}+W) -W \cdot \frac{1-\lambda^{\ell-k+1}}{1-\lambda} \\
    &= \lambda^{\ell-k}
      \sum_{i=0}^{\ell-k}w_{k+i} + (\ell-k+1) \cdot \lambda^{\ell-k} \cdot W
      -W \cdot \frac{1-\lambda^{\ell-k+1}}{1-\lambda}
  \end{align*}
  Therefore, since $\Value^{\Game}(u) = \frac 1 {\ell-k+1} \cdot \sum_{i=k}^\ell w_i$,
  we obtain
  \[\sum_{i=0}^{\ell-k}\lambda^i \cdot w_{k+i}
  \geq
    \lambda^{\ell-k} \cdot (\ell-k+1) \cdot (\Value^{\Game}(u) + W) - W \cdot \frac{1-\lambda^{\ell-k+1}}{1-\lambda} \,.\]
  Together with the first term, we have
  \[
  \DiscountedPayoff(\play)
    \geq -W 
    + \frac{(1-\lambda) \cdot (\ell-k+1)}{1-\lambda^{\ell-k+1}} \cdot \lambda^\ell \cdot (\Value^{\Game}(u) + W)
  \] 
  Since
  $\frac {1-\lambda^{\ell-k+1}}{1-\lambda} = \sum_{i=0}^{\ell-k}\lambda^i < \ell-k+1$ and $\Value^{\Game}(u) + W\geq 0$, 
  we have
  \[
  \DiscountedPayoff(\play) \geq -W + \lambda^\ell \cdot (\Value^{\Game}(u) + W)
  \] 
  Finally, since $\ell\leq n$ we have $\lambda^\ell \geq \lambda^n > 1 - n  \cdot (1-\lambda)$, using the fact that
  $\frac{1-\lambda^{n}}{1-\lambda} = \sum_{i=0}^{n-1}\lambda^i < n$. 
  Therefore, using again $\Value^{\Game}(u) \leq W$,
  \begin{align*}
    \DiscountedPayoff(\play)
    &\geq -W + (1 - n \cdot (1-\lambda)) \cdot (\Value^{\Game}(u) + W)\\
    &= - n \cdot (1-\lambda) \cdot (W + \Value^{\Game}(u)) + \Value^{\Game}(u)\\
    &\geq -2n \cdot (1-\lambda) \cdot W + \Value^{\Game}(u)
  \end{align*}
  We obtain
  \[
  \Value_\lambda(u)-\Value^{\Game}(u) \geq -2n(1-\lambda) \cdot W.
  \] 
\end{proof}

A direct corollary of \Cref{5-thm:MP-Zwick-Paterson} is an algorithm for computing the mean-payoff values:
picking $\lambda = 1-\frac 1{4n^2(n-1)W}$, we obtain a good enough approximation of the mean-payoff values by solving the associated discounted-payoff game. From a complexity point of view, this value iteration algorithm runs in polynomial time in the size of the arena, but exponential with respect to the representation of $\lambda$. 



The previous reductions implies an improved theoretical complexity for mean-payoff and parity games.

\begin{corollary}[Complexity]\label{5-col:UP}
  Solving mean-payoff games and parity games is in $\UP\cap\coUP$.
\end{corollary}
\begin{proof}
  The reduction from mean-payoff to discounted-payoff games allows to lift the $\UP\cap\coUP$ complexity of~\Cref{5-thm:disc-up}. 
  Moreover, the reduction of~\Cref{5-thm:parity2MP} implies the same complexity for parity games. 
\end{proof}

\section{Shortest path games}
\label{5-sec:shortest_path}
The quantitative objective $\Sup$ generalises the qualitative objective $\Reach$ by stating numerical preferences on the target. 
Another quantitative extension of the reachability objective is to quantify the cost of a path towards the target: we define the quantitative objective $\ShortestPath$ over the set of colours $C = \Z \cup \set{\Win}$ by
\[
\ShortestPath(\rho) =
  \begin{cases}
    \sum_{i = 0}^{k-1} \rho_i & \text{for $k$ the first index such that } \rho_k = \Win, \\
   \infty & \text{if } \rho_k \neq \Win \text{ for all } k.
  \end{cases}
\]
We interpret the weights as costs and Min is trying to reach the target with the smallest possible cost.
Note that we use the same abusive terminology as for the shortest path graph problem: the cost of a path is the sum of the weights along it
(until the first occurrence of $\Win$) and we are looking for a path of minimal cost, hence not necessarily the shortest in number of edges.

We fix a shortest path game $\Game$.
Without loss of generality we assume that $\Win$ appears only in a sink.
Recall that by definition:
\[
\val^{\Game}(u) = \sup_{\sigma} \inf_{\tau} \ShortestPath(\pi^v_{\sigma,\tau}).
\]
Hence for a vertex $u$ there are three possibilities: 
\begin{itemize}
	\item $\val^{\Game}(u) = \infty$, meaning that Min cannot ensure to reach $\Win$,
	\item $\val^{\Game}(u) \in \Z$, meaning that Min can ensure to reach $\Win$ with a finite cost (bounded from below),
	\item $\val^{\Game}(u) = -\infty$, meaning that Min can ensure to reach $\Win$ with arbitrarily negative cost.
\end{itemize}
Note that if all weights are one, then $\val^{\Game}$ is the rank defined in the attractor computation, see~\Cref{1-sec:attractors}.

Detecting whether $\val^{\Game}(u) = \infty$ is easy:

\begin{lemma}[Detection of $\infty$ values]
\label{5-lem:detecting_infinity}
Let $\Game$ a shortest path game and $u$ a vertex.
Then $\val^{\Game}(u) < \infty$ if and only if $u \in \WMin(\Reach(\Win))$.
Consequently, if $\val^{\Game}(u) < \infty$, then $\val^{\Game}(u) \le nW$.
\end{lemma}

\begin{proof}
The first equivalence is clear. For the second statement, consider a positional strategy ensuring $\Reach(\Win)$, it ensures to reach $\Win$ within at most $n$ steps, hence incurs a cost bounded by $n$ times the largest weight.
\end{proof}

Before moving to algorithms, let us illustrate two difficulties:
\begin{itemize}
	\item As illustrated in \Cref{5-fig:optimal_strategies_shortest_path_game}, Min does not have optimal strategies in general.
	\item As illustrated in \Cref{5-fig:memory_shortest_path_max}, even if Min has an optimal strategy, it may not be positional.
\end{itemize} 

\begin{figure}
\centering
  \begin{tikzpicture}[scale=1.3]
    \node[s-adam] (v0) at (0,0) {$v_0$};
    \node[s-adam] (v1) at (2,0) {$v_1$};
    \path[arrow]
      (v0) edge[selfloop=180] node[left] {$-1$} (v0)
      (v0) edge node[above] {$0$} (v1)
      (v1) edge[selfloop=0] node[right] {$\Win$}(v1);
  \end{tikzpicture}
\caption{An example of a shortest path game with negative weights where Min does not have an optimal strategy.
Indeed $\val^{\Game}(v_0) = -\infty$ since for any $k$, Min has a strategy ensuring that $\ShortestPath$ is $-k$
by using $k$ times the self loop $-1$ before $\Win$.
However, if Min never sees $\Win$ the outcome is $\infty$.}
\label{5-fig:optimal_strategies_shortest_path_game}
\commentAlt{Figure~\ref{5-fig:optimal_strategies_shortest_path_game}: A directed graph showing two square nodes, v0 and v1, with a transition between them and self-loops on each node.}
\commentLongAlt{Figure~\ref{5-fig:optimal_strategies_shortest_path_game}: The image displays a directed graph with two square nodes, 'v0' on the left and 'v1' on the right. Node 'v0' has a self-loop labeled '-1'. A directed arrow connects 'v0' to 'v1', with the arrow labeled '0'. Node 'v1' also has a self-loop, labeled 'Win'.}
\end{figure}

We discuss~\Cref{5-fig:memory_shortest_path_max}, which is a shortest path game where Min has an optimal strategy, but it cannot be made positional.
First, Min has two positional strategies: $\tau_1(v_0) = v_0 \xrightarrow{-1} v_1$ and $\tau_2(v_0) = v_0 \xrightarrow{0} v_2$. 
Strategy $\tau_1$ does not ensure to reach the target, since Max can enforce the cycle $v_0 \xrightarrow{-1} v_1 \xrightarrow{0} v_0$ forever and
obtain payoff $\infty$. 
Strategy $\tau_2$ guarantees a payoff of $0$. 
However, Min can be smarter by threatening Max.
If Min plays once $\tau_1$, and then switches to $\tau_2$, he guarantees a payoff of $-1$. 
Doing so twice, he guarantees a payoff of $-2$. 
This reasoning is valid for playing up to $50$ times $\tau_1$, showing to Min can ensure a payoff of $-50$. 
However it is not valid beyond: against the strategy that plays $51$ times $\tau_1$, Max's optimal decision is to stop the game with 
$v_1 \xrightarrow{-50} v_2$, ensuring a payoff of $-50$. Indeed and more generally (as we will see), Max has a positional optimal strategy, which is to choose $v_1 \xrightarrow{-50} v_2$ right from the beginning. 


\begin{figure}[tbp]
  \centering
    \begin{tikzpicture}
    \node[s-eve](2){$v_2$};%
    \node[s-adam,above left of=2,xshift=5mm](0){$v_0$};%
    \node[s-eve,above right of=2,xshift=-5mm](1){$v_1$};%

    \path[->] (0) edge[bend left] node[above]{$-1$} (1)%
    (1) edge[bend left] node[below]{0} (0)%
    (0) edge node[below left]{$0$} (2)%
    (1) edge node[below right]{$-50$} (2)%
    (2) edge[selfloop=-90] node[below]{$\Win$}(2);%
  \end{tikzpicture}
\caption{A shortest path game where Min needs memory to play optimally.}
\label{5-fig:memory_shortest_path_max}
\commentAlt{Figure~\ref{5-fig:memory_shortest_path_max}: A directed graph with three nodes (v0, v1, v2) of mixed shapes (square and circles) and labeled edges, including a self-loop.}
\commentLongAlt{Figure~\ref{5-fig:memory_shortest_path_max}: The image displays a directed graph with three nodes. Node 'v0' is a square, while 'v1' and 'v2' are circles.

A bidirectional arrow connects 'v0' and 'v1'. The arrow from 'v0' to 'v1' is labeled '-1', and the arrow from 'v1' to 'v0' is labeled '0'.
An arrow from 'v0' points to 'v2', labeled '0'.
An arrow from 'v1' points to 'v2', labeled '-50'.
Node 'v2' has a self-loop labeled 'Win'.}
\end{figure}

\subsection*{Detection of $-\infty$ values using mean-payoff games}

We call `detecting $-\infty$ values in shortest path games' the following decision problem: given a shortest path game $\Game$ and a vertex $u$, do we have $\val^{\Game}(u) = -\infty$?

\begin{theorem}[Detection of $-\infty$ values using mean-payoff games]
\label{5-thm:-infty-MP}
Detecting $-\infty$ values in shortest path games is polynomial time equivalent to solving mean-payoff games.
\end{theorem}

We construct two simple reductions.

\begin{lemma}
\label{5-lem:reduction_shortest_path_to_mean_payoff}
Let $\game$ a shortest path game, and $\game'$ the mean-payoff game obtained from~$\game$ by replacing the self loop with $\Win$ by a self loop with $0$.
Assume that for all vertices $u$ we have $\val^{\game}(u) \neq \infty$. 
Then for all vertices $u$, we have $\val^{\game}(u) = -\infty$ if and only if $\val^{\game'}(u) < 0$.
As a consequence, if $\val^{\game}(u) > -\infty$, then $\val^{\game}(u) \ge -nW$.
\end{lemma}

Note that as stated in \Cref{5-lem:detecting_infinity}, detecting vertices with value $\infty$ can be done in polynomial time,
hence it does not reduce the generality of the reduction.

\begin{proof}
Let us fix $\sigma$ an positional optimal strategy for Max in the mean-payoff game $\Game'$.
Since for all vertices $u$ we have $\val^{\game}(u) \neq \infty$, thanks to~\Cref{5-lem:detecting_infinity} there exists a strategy $\sigma_0$ which ensures $\ShortestPath \le nW$ from all vertices.

Assume that $\val^{\Game}(u) = -\infty$. 
Let $\tau_M$ a strategy ensuring $\ShortestPath < -nW$, and look at a play consistent with $\sigma$ and $\tau_M$.
It contains necessarily a negative cycle, implying that $\sigma$ does not ensure $\MeanPayoff \ge 0$.
Since $\sigma$ is optimal, this implies that $\val^{\Game'} < 0$.

For the converse implication, assume now that $\val^{\Game'}(u) < 0$.
This implies that all cycles consistent with $\sigma$ are negative. Consequently, for all paths consistent with $\sigma$, the sum of the weights diverges to $-\infty$.
Let us fix $M$, and consider the strategy $\sigma_M$ which plays like $\sigma$ until the sum of the weights reaches below $M - nW$, and then switches to $\sigma_0$. This strategy ensures $\ShortestPath \le M$. Thus $\val^{\Game}(u) = -\infty$.

We now explain how it follows that if $\val^{\game}(u) > -\infty$, then $\val^{\game}(u) \ge -nW$.
Since the positional optimal strategy $\sigma$ ensures $\MeanPayoff \ge 0$, it also ensures that the sum of the weights remains larger than $-nW$ at all times.
\end{proof}

\begin{lemma}
Let $\game$ a mean-payoff game.
We construct $\game'$ a shortest path game where after each edge, Min has a choice to stop the game with colour $\Win$.
Then for all vertices $u$ in the original game $\game$, we have $\Value^\game(u) < 0$ if and only if $\val^{\game'}(u) = -\infty$.
\end{lemma}

\begin{proof}
Note that Min can ensure to reach $\Win$ from anywhere, so Min has a strategy to reach $\Win$, implying that for all $u$ we have $\val^{\game'}(u) \neq \infty$. 

Let $\game''$ the mean-payoff game obtained from $\game'$ as in \Cref{5-lem:reduction_shortest_path_to_mean_payoff}, we have 
that $\val^{\game'}(u) = -\infty$ if and only if $\val^{\game''}(v) < 0$. 
To conclude, it only remains to show that $\val^{\game}(u) < 0$ if and only if $\val^{\game''}(u) < 0$. 
The game $\Game''$ is exactly as $\Game$, except that anytime Min can stop the game and settle for a mean payoff of $0$.
Hence when asking whether the mean-payoff value is $< 0$, this option is not relevant.
\end{proof}

\subsection*{A value iteration algorithm for shortest path games}

\begin{theorem}[Positional determinacy in shortest path games]
\label{5-thm:shortest_path_positional_determinacy_max}
Shortest path games are positionally determined over finite arenas.
\end{theorem}

This half-positional determinacy result will follow from the correctness of a value iteration algorithm, which works for shortest path games where no vertices has value $-\infty$. Thanks to the above, this can be computed and removed from the game. We say that such a game is normalised.
Note that it is enough to prove half-positional determinacy for normalised games, since on the vertices having value $-\infty$ the strategy of Max is irrelevant.

\begin{theorem}[Value iteration algorithm]
\label{5-thm:shortest_path_value_iteration}
There exists a value iteration algorithm for computing the value function of normalised shortest path games in pseudo-polynomial time and space.
\end{theorem}

Our first lemma shows the existence of optimal strategies.
\begin{lemma}[Optimal strategies]
\label{5-lem:optimal_strategies_shortest_path_games}
Let $\Game$ be a normalised shortest path game, then there exists an optimal strategy for Min.
\end{lemma}
\begin{proof}
Thanks to the assumption that the game is normalised and~\Cref{5-lem:reduction_shortest_path_to_mean_payoff}, the values are lower bounded by $-nW$, which implies that the infimum is indeed a minimum.
\end{proof}
\Cref{5-fig:optimal_strategies_shortest_path_game} shows that the assumption that the game is normalised in \Cref{5-lem:optimal_strategies_shortest_path_games} is necessary.

We follow the high-level presentation of value iteration algorithms given in \Cref{1-sec:value_iteration}.
Let us define $Y = \Z \cup \set{-\infty,\infty}$ equipped with the natural order and the function 
$\delta : Y \times (\Z \cup \set{\Win}) \to Y$ by
\[
\delta(x, w) = 
\begin{cases}
0 & \text{ if } w = \Win, \\
x + w & \text{ if } w \in \Z.
\end{cases}
\]
We let $F_V$ be the lattice of functions $\mu : V \to Y$ equipped with the component wise order induced by $Y$.
Note that $\delta$ is monotonic, it induces the monotonic operator $\Op^{\Game} : F_V \to F_V$ defined by:
\[
\Op^{\Game}(\mu)(u) = 
\begin{cases}
\max \set{\delta( \mu(v), w) : u \xrightarrow{w} v \in E} & \text{ if } u \in \VMax, \\
\min \set{\delta( \mu(v), w) : u \xrightarrow{w} v \in E} & \text{ if } u \in \VMin.
\end{cases}
\]
Thanks to \Cref{1-thm:kleene}, the operator $\Op^{\Game}$ has a greatest fixed point which is also the greatest post-fixed point of $\Op^{\Game}$.
Recall that a post-fixed point of $\Op^{\Game}$ is a function $\mu \in F_V$ such that $\Op^{\Game}(\mu) \ge \mu$, it is also called a progress measure.
Unfolding the definitions, for all vertices $u$, we have
\[
\begin{array}{llll}
\exists u \xrightarrow{w} v \in E,\ & \mu(u) \le \delta( \mu(v), w) & \text{ if } u \in \VMax, \\
\forall u \xrightarrow{w} v \in E,\ & \mu(u) \le \delta( \mu(v), w) & \text{ if } u \in \VMin.
\end{array}
\]
\begin{lemma}[Shortest path values as greatest fixed point]
\label{5-lem:SP-greatest-fixed point}
Let $\Game$ be a normalised shortest path game, then $\val^{\Game}$ is the greatest fixed point of $\Op^{\Game}$.
\end{lemma}

\begin{proof}
We show two properties:
\begin{itemize}
	\item $\val^{\Game}$ is a progress measure;
	\item $\val^{\Game}$ is larger than the greatest fixed point of $\Op^{\Game}$.
\end{itemize}
Since the greatest fixed point of $\Op^{\Game}$ is also the greatest progress measure, the first item implies that it is larger than $\val^{\Game}$,
and the second item states the converse inequality.

The first item is a routine verification, which follows from two properties (see~\Cref{1-lem:sufficient_condition_fixed_point}).
\begin{itemize}
	\item For all $\rho$ sequences of weights, we have 
	\[
	\ShortestPath(w \cdot \rho) = \delta(\ShortestPath(\rho), w).
	\]
	\item The function $\delta$ is monotonic and continuous.
\end{itemize}

We now show the second item: $\val^{\Game}$ is larger than or equal to the greatest fixed point of $\Op^{\Game}$.
For this, we consider a fixed point $\mu$ of $\Op^{\Game}$, and argue that $\val^{\Game} \ge \mu$.
To this end, we extract from $\mu$ a strategy $\sigma$ for Max, and show that $\val^{\sigma} \ge \mu$; 
since we know that $\val^{\Game} \ge \val^{\sigma}$, this implies $\val^{\Game} \ge \mu$.
We define $\sigma$ as an argmax strategy:
\[
\begin{array}{l}
u \in \VMax:\ \sigma(u) \in \argmax \set{\delta(\mu(v),w) : u \xrightarrow{w} v \in E}.
\end{array}
\]
Let us define the graph $G = \Game[\sigma]$, by definition of $\sigma$ for all edges $u \xrightarrow{w} v$ in $G$ we have
$\mu(u) \le \delta(\mu(v),w)$.
We claim that this implies that for all paths $\rho$ from $u$ in $G$, we have $\ShortestPath(\rho) \ge \mu(u)$.
If $\ShortestPath(\rho) = \infty$, this is clear, so let us consider the finite case, and proceed by induction on the length $k$ of the path before reaching $\Win$. 
This is clear for $k = 0$, since both values are $0$.
Let us assume that it holds for $k$ and show that it also does for $k+1$.
Let us write $\rho_0 = u \xrightarrow{w} v$.
By the property above, we have $\mu(u) \le \delta(\mu(v),w) = \mu(v) + w$.
Hence 
\[
\sum_{i = 0}^k \rho_i = w + \sum_{i = 1}^k \rho_i \ge w + \mu(v) \ge \mu(u),
\]
where the first inequality is by induction hypothesis for the path $\rho_1 \dots \rho_{k-1}$ from $v$.
This concludes the induction.
\end{proof}

\Cref{5-lem:SP-greatest-fixed point} does not immediately yield a value iteration algorithm because the lattice $Y$ is infinite.
However, let us note that by half-positional determinacy, the value of a vertex is either $\infty, -\infty$, or in $[-nW,nW]$.
Hence we can equivalently define $Y = [-nW,nW] \cup \set{-\infty,\infty}$ and $\delta : Y \times (\Z \cup \set{\Win}) \to Y$ by
\[
\delta(x, w) = 
\begin{cases}
0 & \text{ if } w = \Win, \\
x + w & \text{ if } w \in \Z \text{ and } x + w \in [-nW,nW], \\
\infty & \text{ if } w \in \Z \text{ and } x + w  > nW, \\
-\infty & \text{ if } w \in \Z \text{ and } x + w  < -nW.
\end{cases}
\]
It is clear that the greatest fixed points for both operators coincide thanks to the remark above, and now that we have a finite lattice we can use the generic monotonic fixed-point computation (see~\Cref{1-thm:kleene}) to compute $\val^{\Game}$, which yields a pseudo-polynomial time and space algorithm.

\subsection*{Shortest path games with non-negative weights}

\begin{theorem}[Polynomial-time algorithm for shortest path games with non-negative weights]
\label{5-thm:SP-poly-non-negative}
There exists an algorithm for computing the values in shortest path games with non-negative weights
running in time $\bigO(m+n\log(n))$. 
\end{theorem}

\begin{algorithm}
 \KwData{A shortest path game with non-negative weights.}
 \SetKwFunction{FInit}{Init}
 \SetKwFunction{FTreat}{Treat}
 \SetKwFunction{FUpdate}{Update}
 \SetKwFunction{FMain}{Main}
 \SetKwProg{Fn}{Function}{:}{}
 \DontPrintSemicolon
 
\Fn{\FInit{}}{
	\For{$u \in V$}{
		$\mu(u) \leftarrow \infty$
	}	

	\For{$u \in \VMin$}{		
		\If{$\exists u \xrightarrow{\Win} v \in E$}{
			$\mu(u) \leftarrow 0$					

	   		Add $u$ to $S$
		}
	}

	\For{$u \in \VMax$}{
		\If{$\forall u \xrightarrow{\Win} v \in E$}{
			$\mu(u) \leftarrow 0$					

    	    Add $u$ to $S$
		}
		\Else{	
			\For{$u \xrightarrow{w} v \in E$}{
				$\mu(u \xrightarrow{w} v) \leftarrow \infty$
    		}
    	}
	}	
}

\vskip1em
\Fn{\FMain{}}{
	\FInit()    

	\Repeat{$S \text{ is empty}$}{

		Extract $v \in \argmin \set{\mu(v) : v \in S}$

		\For{$u \xrightarrow{w} v \in E$}{
		
			\If{$u \in \VMin$ and $\mu(u) > \delta(\mu(v),w)$}{
				$\mu(u) \leftarrow \delta(\mu(v),w)$ 

				Update $\mu(u)$ in $S$	
			   
			}

			\If{$u \in \VMax$ and $\mu(u \xrightarrow{w} v) > \mu(v)$}{
				$\mu(u \xrightarrow{w} v) \leftarrow \mu(v)$
				
				$x \leftarrow \max \set{\mu(u \xrightarrow{w'} v') : u \xrightarrow{w'} v' \in E}$
				
				\If{$\mu(u) > x$}{
					$\mu(u) \leftarrow x$

					Update $\mu(u)$ in $S$	
				} 
			}
    	}
	}

	\Return{$\mu$}
}
\caption{A polynomial-time algorithm for shortest path games with non-negative weights.}
\label{5-algo:value_iteration_shortest_path_non_negative}
\end{algorithm}

In the case where all weights are non-negative, the value iteration algorithm can be made much more efficient,
and in particular in polynomial time. To understand this, let us start by noting that the one-player case where only Min has moves is the classical shortest path problem (towards $\Win$) for graphs. 
We extend Dijkstra's algorithm, see~\Cref{5-algo:value_iteration_shortest_path_non_negative} for the pseudocode.
%
%

Let us denote by $S_i, \mu_i(u)$, $\mu_i(u \xrightarrow{w} v)$ the values in iteration $i$.
We define the invariants satisfied by the algorithm.
\begin{enumerate}
	\item $\mu_i(u)$ is value of $u$ in the shortest path game~$\game_i$ obtained by replacing $u \xrightarrow{w} v$ with $u \xrightarrow{\infty} v$ if both $u$ and $v$ are still in $S_i$;
	\item $\min \set{ \mu_i(u) : u \in S_i} \geq \max \set{\Value^{\game}(u) : u \notin S_i}$.
\end{enumerate}
The second invariant generalises the greedy property of Dijkstra's algorithm.

Since the weight of every edge in $\game_i$ is non-increasing, the values $\Value^{\game_i}(u)$ are also non-increasing. 
The invariants show imply that $\mu_i(u) = \Value^{\game_i}(u)$ for all $u \in S_i$ and $\mu_i(u) = \Value_{\game}(u)$ for all $u \notin S_i$. 
We refer to~\cite{Khachiyan.Boros.ea:2008} for the detailed proofs of the invariants.
A careful analysis, using (minimum) Fibonacci heaps, as in Dijkstra's algorithm, allows one to obtain an overall complexity $\bigO(m+n\log(n))$.

\section{Total payoff games}
\label{5-sec:total_payoff}
Recall that the total payoff objective is defined by
\[
\TotalPayoff(\rho) = \limsup_k \sum_{i=0}^{k-1} \rho_i.
\]
Contrary to the shortest path objective, total payoff games do not include a reachability objective. 
In particular, all plays will be infinite and their payoff is the superior limit of the partial sums: 
we need this superior limit since partial sums might not have a limit (consider for instance the sequence of weights $1,-1,1,-1,1,\ldots$ whose partial sums alternate between $1$ and $0$).



It is easy to show that solving total payoff games is in $\NP \cap \coNP$: they are bi-positionally determined,
and the one-player games can be solved in polynomial time using shortest paths algorithms.
However, constructing a value iteration is not easy. A natural attempt is to define the following operator:
\[
\Op^{\Game}(\mu)(u) = 
\begin{cases}
\max \set{\mu(v) + w : u \xrightarrow{w} v \in E} & \text{ if } u \in \VMax, \\
\min \set{\mu(v) + w : u \xrightarrow{w} v \in E} & \text{ if } u \in \VMin.
\end{cases}
\]
One can show that the value function is indeed a fixed point of $\Op^{\Game}$.
But it is neither the greatest nor the least fixed point.
Consider for example the total payoff game represented in~\Cref{5-fig:totalpayoff}. 
The fixed points of $\Op^{\Game}$ are $(a,a+1,a+2)$ with $a \in \Rinfty$: 
in particular, the greatest fixed point is $(+\infty, +\infty, +\infty)$
and the least fixed point is $(-\infty, -\infty, -\infty)$. 
However, the value function is $(0,1,2)$.

We obtain a pseudo-polynomial-time algorithm by reduction to a shortest path game of pseudo-polynomial size.

\begin{theorem}[A reduction from total payoff games to shortest path games]
\label{5-thm:TP-optimal-strategies}
Let $\Game$ a total payoff game, we can construct a shortest path game $\Game'$
such that for all vertices $u$ from $\Game$, we have $\val^{\Game}(u) = \val^{\Game'}(u)$.
The game $\Game'$ has size pseudo-polynomial in the size of $\Game$.
\end{theorem}

\begin{proof}
Let us fix $K = n \cdot \big((2n-1) \cdot W + 1\big)$.
The game $\Game'$ consists in $K$ consecutive copies of $\Game$.
In each copy, after each move, Min can offer Max the following alternative:
either the game stops and reaches $\Win$, or we move to the next copy.
For the chosen $K$, one can show that the values of $\Game$ coincide with the values of the first copy of $\Game'$.


Instead of explicitly constructing $\Game'$, we can run the value iteration algorithm for shortest path games in each copy, improving the space complexity of the algorithm.
\end{proof}

\begin{figure}[tbp]
  \centering
  \begin{tikzpicture}
    \node[s-eve](0){$v_0$};%
    \node[s-adam,left of=0](1) {$v_1$};%
    \node[s-adam,right of=0] (2) {$v_2$};%
    
    \path[->] (0) edge[bend right] node[above]{$-1$} (1)%
    (1) edge[bend right] node[below]{$1$} (0)%
    (0) edge[bend left] node[above]{$-2$} (2)%
    (2) edge[bend left] node[below]{$2$} (0);%
  \end{tikzpicture}

  \caption{A total payoff game}
  \label{5-fig:totalpayoff}
\commentAlt{Figure~\ref{5-fig:totalpayoff}: A central circular node v0 connected to two square nodes, v1 and v2, on either side by bidirectional arrows with numerical labels.}
\commentLongAlt{Figure~\ref{5-fig:totalpayoff}: The image displays a central circular node labeled 'v0'. To its left is a square node labeled 'v1', and to its right is a square node labeled 'v2'. A bidirectional arrow connects 'v1' and 'v0'. The arrow from 'v1' to 'v0' is labeled '1', and the arrow from 'v0' to 'v1' is labeled '-1'. Similarly, a bidirectional arrow connects 'v0' and 'v2'. The arrow from 'v0' to 'v2' is labeled '-2', and the arrow from 'v2' to 'v0' is labeled '2'.}
\end{figure}

\section*{Bibliographic references}
\label{5-sec:references}
This chapter has been the occasion to start revealing a ladder of reductions going from parity games through mean-payoff games and to discounted-payoff games. In~\Cref{7-chap:stochastic}, the last reduction from discounted-payoff games to simple stochastic games will complete this chain of reductions.

\paragraph*{Mean-payoff games.}
Mean-payoff games have been first studied by Ehrenfeucht and Mycielski in~\cite{Ehrenfeucht.Mycielski:1979} where positional determinacy is
shown. The one-player variants had already been studied by Karp~\cite{Karp:1978}. It is much later that Zwick and Paterson~\cite{zwick.paterson:1996} first obtained the pseudo-polynomial value iteration algorithm to solve them, while introducing discounted-payoff games (that had been first studied in a probabilistic setting as will be studied in~\Cref{6-chap:mdp} and
\Cref{7-chap:stochastic}). The NP and coNP upper bound, together with strategy improvement algorithm, is due to Pure~\cite{Puri:1995}.


\paragraph*{Strategy improvement algorithms.}
There are several strategy improvement algorithms for computing the mean-payoff values.
Bj{\"o}rklund and Vorobyov~\cite{Bjorklund.Vorobyov:2007} constructed one such algorithm for computing the energy values, presented via the notion of \emph{longest shortest path problem}. 
A more direct approach was construct by Filar and Vrieze~\cite{Filar.Vrieze:1996}, involving a pair of values, called gain and bias.
Strategy improvement methods (for parity games or payoff games) are very closely related to the simplex method for solving linear programs, see for instance~\cite{Allamigeon.Benchimol.ea:2014}.
Lifshits and Pavlov~\cite{Lifshits.Pavlov:2007} develop yet another strategy improvement algorithm for mean-payoff games using potential reduction techniques, which were originally described by Gallai~\cite{Gallai:1958} in the context of networks-related problems, and also developed by Gurvich, Karzanov, and Khachiyan~\cite{Gurvich.Karzanov.ea:1988} in the context of mean-payoff games. This has been further explored, see for instance~\cite{Ohlmann:2022}.

\paragraph*{Energy games.}

\paragraph*{Value iteration algorithm.}
The value iteration algorithm for energy games has been developed by Brim, Chaloupka, Doyen, Gentilini, and Raskin~\cite{Brim.Chaloupka.ea:2011}. 
Comin and Rizzi~\cite{Comin.Rizzi:2017} have shown how to adapt the algorithm to compute optimal strategies as well with a running time $\bigO(n^2 m W)$. This removes the $\log(n)+\log(W)$ term due to binary search.

\paragraph*{Shortest path games.}
As we have seen, computing the values for shortest path games can be done in polynomial time when weights are non-negative, this has been proved by Khachiyan and coauthors~\cite{Khachiyan.Boros.ea:2008}.
No polynomial solution is known for the general case. 
The best known at the time this chapter is written is a polynomial time fragment consisting of \emph{divergent shortest path games}~\cite{Busatto-Gaston.Monmege.ea:2017} in which the arena does not contain any cycle with total weight $0$: this a priori weak property indeed partitions the strongly connected components of the arena into the ones where all cycles are positive, and the ones where all cycles are negative; in each of these components, it is shown why the value iteration algorithm converges in polynomial time. 
\Cref{5-thm:-infty-MP} shows the equivalence between mean-payoff games and detecting vertices of values $-\infty$ in shortest path games.
It is open whether solving shortest path games where no vertices have value $-\infty$ can be done in polynomial time.

\paragraph*{Total payoff games.}
The analysis of total payoff games and shortest path games is due to~\cite{Brihaye.Geeraerts.ea:2017}.



\part{Stochastic}
\label{part:stochastic}

\ifpictures
\includepdf{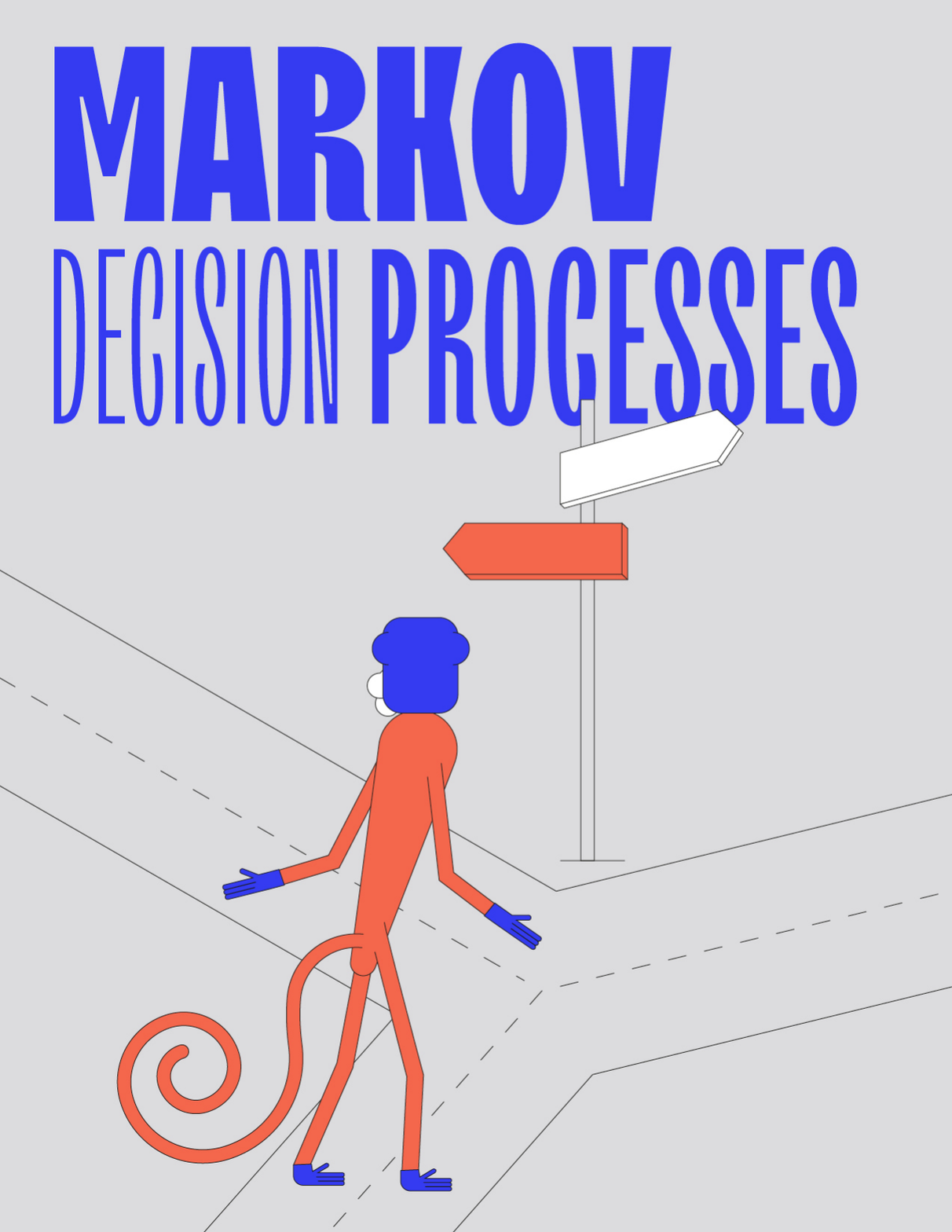}
\fi
\author[Petr Novotn{\'y}]{Petr Novotn{\'y}}
\copyrightline{Copyright by Petr Novotn{\'y} 2025, to be published by Cambridge University Press in the volume \textit{Games on Graphs} edited by Nathana\"el Fijalkow}

\chapter{Markov Decision Processes}
\chapterauthor{Petr Novotn{\'y}}
\label{6-chap:mdp}


\newcommand{\Max}{\text{Max} } 


\newcommand{\expv}{\mathbb{E}} 
\newcommand{\discProbDist}{f} 
\newcommand{\sampleSpace}{S} 
\newcommand{\sigmaAlg}{\mathcal{F}} 
\newcommand{\probm}{\mathbb{P}} 
\newcommand{\rvar}{X} 


\newcommand{\actions}{A} 
\newcommand{\coloring}{c} 
\newcommand{\probTranFunc}{\Delta} 
\newcommand{\edges}{E} 
\newcommand{\colors}{C} 
\newcommand{\mdp}{\mathcal{M}} 
\newcommand{\vinit}{v_0} 
\newcommand{\cylProb}{p} 
\newcommand{\emptyPlay}{\epsilon} 
\newcommand{\objective}{\Omega} 
\newcommand{\genColour}{\textsc{c}} 
\newcommand{\quantObj}{f} 
\newcommand{\quantObjExt}{\bar{\quantObj}} 
\newcommand{\indicator}[1]{\mathbf{1}_{#1}} 
\newcommand{\eps}{\varepsilon} 
\newcommand{\maxc}{\max_{\coloring}} 

\newcommand{\winPos}{W_{>0}}
\newcommand{\winAS}{W_{=1}}
\newcommand{\cylinder}{\mathit{Cyl}}

\newcommand{\PrePos}{\text{Pre}_{>0}}
\newcommand{\PreAS}{\text{Pre}_{=1}}

\newcommand{\PreOPPos}{\mathcal{P}_{>0}}
\newcommand{\OPAS}{\mathcal{P}_{=1}}

\newcommand{\safeOP}{\mathit{Safe_{=1}}}
\newcommand{\closed}{\mathit{Cl}}

\newcommand{\reachOP}{\mathcal{V}}
\newcommand{\discOP}{\mathbb{O}}
\newcommand{\valsigma}{\vec{x}^{\sigma}}

\newcommand{\lp}{\mathcal{L}}
\newcommand{\lpdisc}{\lp_{\mathit{disc}}}
\newcommand{\lpreach}{\lp_{\mathit{reach}}}
\newcommand{\lpmp}{\lp_{\mathit{mp}}}
\newcommand{\lpsol}[1]{\bar{\vec{#1}}}
\newcommand{\lpsolg}[1]{\bar{#1}}
\newcommand{\lpmpdual}{\lpmp^{\mathit{dual}}}

\newcommand{\actevent}[3]{\actions^{#1}_{#2,#3}} 

\newcommand{\MeanPayoffSup}{\MeanPayoff^{\;+}}
\newcommand{\MeanPayoffInf}{\MeanPayoff^{\;-}}

\newcommand{\mcprob}{P}
\newcommand{\invdist}{\vec{z}}

\newcommand{\hittime}{T}

\newcommand{\playPay}{\textsf{p-Payoff}}
\newcommand{\stepPay}{\textsf{s-Payoff}}
\newcommand{\Pay}{\textsf{Payoff}}

\newcommand{\mec}{M}
\newcommand{\OPS}{\mathcal{S}_{=1}}

\newcommand{\smallmp}{\mathit{mp}}
\newcommand{\vgood}{v_{\mathit{good}}}
\newcommand{\vbad}{v_{\mathit{bad}}}
\newcommand{\finact}{fin}
\newcommand{\mecs}{\mathit{MEC}}

\newcommand{\slice}[2]{#1_{#2-}}
\newcommand{\ReachOp}{\mathcal{R}}

\newcommand{\dPayoffStep}[1]{\DiscountedPayoff^{\;(#1)}}
\newcommand{\solvset}{S}

In this chapter we study Markov decision processes (MDPs), a standard model for decision making under uncertainty. MDPs are also called `$1\frac{1}{2}$-player games', since they can be viewed as a game in which only one player makes strategic choices, while the other player, which we call Nature, behaves according to some fixed probabilistic model. The chapter surveys the basic notions pertaining to MDPs, and algorithms for the following MDP-related problems:

\begin{itemize}
\item positive and almost-sure reachability and safety,
\item discounted payoff,
\item mean payoff in strongly connected MDPs,
\item decomposition of MDPs into maximal end-components (MECs),
\item reductions of general mean-payoff MDPs, B{\"u}chi MDPs, and parity MDPs to general reachability MDPs (via the MEC decomposition),
\item solving general reachability in MDPs.
\end{itemize}

\section*{Notations}
\label{6-sec:notations}
We write vectors in boldface: $ \vec{x}, \vec{y}, $ etc. For a vector $ \vec{x} $ indexed by a set $ I $ (\textit{i.e.} $ \vec{x}\in \mathbb{R}^I $) we denote by $ \vec{x}_i $ the value of the component whose index is  $i\in I  $. 

A (discrete) probability distribution over a finite or countably infinite set $A$ is a function $\discProbDist \colon A \rightarrow [0,1]$ such that $\sum_{a\in A}\discProbDist(a)=1$. The support of such a distribution $\discProbDist$ is the set of all $a\in A$ with $\discProbDist(a)>0$. A distribution $f$ is called Dirac if its support has size 1.
We denote by $\dist(A)$ the set of all probability distributions over $A$.


We also deal with probabilities over uncountable sets of events. This is accomplished via the standard notion of a \emph{probability space.}

\begin{definition}[Probability space]
\label{6-def:probspace}
A probability space is a triple
$(\sampleSpace,\sigmaAlg,\probm)$ where
\begin{itemize}
\item $\sampleSpace$ is a non-empty set of \emph{events} (so called
\emph{sample space}). 

\item $\sigmaAlg$ is a sigma-algebra over $\sampleSpace$, \emph{\textit{i.e.}} a collection of subsets of $\sampleSpace$ that contains the empty set
$\emptyset$ and that is closed under complementation and countable unions. The members of $\sigmaAlg$ are called $\sigmaAlg$-measurable 
sets.

\item $\probm$ is a probability measure on $\sigmaAlg$, \textit{i.e.} a function
$\probm\colon \sigmaAlg\rightarrow[0,1]$ such that:
\begin{enumerate}
\item $\probm(\emptyset)=0$;

\item for all $A\in \sigmaAlg$ it holds $\probm(\sampleSpace \setminus
A)=1-\probm(A)$; and

\item for all countable sequences of pairwise disjoint sets $A_1,A_2,\dots \in \sigmaAlg$ (\textit{i.e.}, $A_i \cap A_j = \emptyset$ for all $i\neq j$)
we have $\sum_{i=1}^{\infty}\probm(A_i)=\probm(\bigcup_{i=1}^{\infty} A_i)$.
\end{enumerate}
\end{itemize}
\end{definition}


A random variable in the probability space $(\sampleSpace,\sigmaAlg,\probm)$ is an $\sigmaAlg$-measurable function $\rvar\colon \Omega \rightarrow \R \cup
\{-\infty,\infty\}$, \textit{i.e.},
a function such that for every $a\in \R \cup \{ -\infty,\infty\}$ the set
$\{\omega\in \Omega\mid \rvar(\omega)\leq a\}$ belongs to $\mathcal{F}$. We denote by $\expv[\rvar]$ the expected value of a random variable $\rvar$~(see Chapter 5 in~\cite{Billingsley:1995} for a formal definition).


We first give a syntactic notion of an MDP which is an analogue of the notion of an "arena" for games.
There is a key technical albeit minor difference: we use a set of actions, meaning that a strategy picks actions rather than edges.

\begin{definition}[MDP]
\label{6-def:MDP}
A Markov decision process is a tuple $(\vertices,\edges,\probTranFunc,\coloring)$. The meaning of $\vertices$, $\edges$, and $\coloring$ is the same as for games, \textit{i.e.} $\vertices$ is a finite set of vertices, $\edges\subseteq \vertices\times\vertices$ is a set of edges and $\coloring\colon \edges \rightarrow \colors$ a mapping of edges to a set of colours. However, the meaning of $\probTranFunc$ is now different: $\probTranFunc$ is a partial probabilistic transition function of type $\probTranFunc\colon \vertices \times \actions \rightarrow \dist(\edges)$, such that the support of $\probTranFunc(v,a)$ only contains edges outgoing from $v$.
 We usually write $\probTranFunc(v'\mid v,a)$ as a shorthand for $\probTranFunc(v,a)((v,v'))$, \textit{i.e.} the probability of transitioning from $v$ to $v'$ under action $a$.
\end{definition}


We also stipulate that for each edge $(v_1,v_2)$ there exists an action $a\in \actions$ such that $\probTranFunc(v_2\mid v_1,a)>0$. Edges not satisfying this can be always removed without changing the semantics of the MDP, which is defined below. We denote by $ p_{\min} $ the smallest non-zero edge probability in a given MDP, \textit{i.e.} $ p_{\min} = \min\{x>0 \mid \exists u,v \in \vertices, a \in \actions \text{ s.t. } x = \probTranFunc(v\mid u,a)\}. $

We denote by $\edges_\genColour$ the set of edges coloured by $\genColour$. Also, for MDPs where $\colors$ is some set of numbers, we use $\maxc$ to denote the number $\max_{e\in 
	\edges}|\coloring(e)|$.
In the setting of MDPs it is technically convenient to encode regular objectives (Reachability, B{\"u}chi,$\dots$) by colours on \emph{vertices} as opposed to edges. Hence, when discussing these objectives, we assume that the colouring function $\coloring$ has the type $\vertices \rightarrow \colors$.



\paragraph{Plays and strategies in MDPs}


 The way in which a play is generated in an MDP is similar to games, but now encompasses a certain degree of randomness. There is a single player, say \Max, who controls all the vertices. \Max's interaction with the `world' described by an MDP is probabilistic. One reason is the stochasticity of the transition function, the other is the fact that in MDP settings, it is usually permitted for \Max to use randomised strategies. Formally, a randomised strategy is a function $\sigma : E^* \to \dist(A)$, which to each finite play assigns a probability distribution over actions. 
 We typically shorten $\sigma(\play)(a)$ to $\sigma(a\mid \play)$.
 
%
In this section, we will refer to randomised strategies simply as `strategies'. The strategies known from the game setting will be called  deterministic strategies. Formally, a deterministic strategy can be viewed as a special type of a randomised strategy which always selects a Dirac distribution over the edges. We shorten `positional randomised/deterministic' to MR and MD, respectively.

Now a play in an MDP is produced as follows: in each step, when the finite play produced so far (\textit{i.e.} the history of the game token's movement) is $\play$, \Max chooses an action $a$ randomly according to the distribution $\sigma(\play)$. Then, an edge outgoing from $\last(\play)$ is chosen randomly according to $\probTranFunc(\last(\play),a)$ and the token is pushed along the selected edge. As shown below, this intuitive process can be formalised by constructing a special probability space whose sample space consists of infinite plays in the MDP.


\paragraph{Formal semantics of MDPs}

Formally, to each MDP $\mdp$, each (\Max's) strategy $\sigma$ in $\mdp$, and 
each initial vertex $\vinit$ we assign a probability space 
$(\sampleSpace_{\mdp},\sigmaAlg_{\mdp},\probm^{\sigma}_{\mdp,\vinit})$. To 
explain the individual components, we need the notion of a cylinder set. A 
basic cylinder determined by a finite play $\play$ is the set of 
all infinite plays in $\mdp$ having $\play$ as a prefix. Now the above 
probability space consists of the following components:
\begin{itemize}
	\item $\sampleSpace_{\mdp}$ is the set of all infinite plays in $\mdp$;
	\item $\sigmaAlg_{\mdp}$ is the \emph{Borel} sigma-algebra over 
	$\Omega_{\mdp}$; this is the smallest sigma-algebra containing all the 
	basic cylinder sets determined by finite plays in $\mdp$. The sets in 
	$\sigmaAlg_{\mdp}$ are called events. Note that the smallest sigma-algebra of the desired property is guaranteed to exist, since an intersection of an arbitrary number of sigma-algebras is again a sigma algebra.
	\item $\probm^{\sigma}_{\mdp,\vinit}$ is the unique probability measure 
	arising from the \emph{cylinder construction} detailed below. We use 
	$\expv^{\sigma}_{\mdp,\vinit}$ to denote the expectation operator 
	associated to the measure $\probm^{\sigma}_{\mdp,\vinit}$.
\end{itemize}

Since the sample space $\sampleSpace_{\mdp}$ is uncountable, we construct the 
probability measure by first specifying a probability of certain simple sets of 
runs and then using an appropriate \emph{measure-extension} theorem to extend 
the probability measure, in a unique way, to all sets in $\sigmaAlg_{\mdp}$.
The standard cylinder construction  
proceeds as follows: for each finite play $\play$ we define the probability 
$\cylProb(\play)$ such that

\begin{itemize}
\item for an empty play $\emptyPlay$ we put $\cylProb(\emptyPlay)=1$;
\item for a non-empty play $\play=\play_0\cdots \play_{k}$ initiated in 
$\vinit$ we put 
\[\cylProb(\play) = \cylProb(\play_{< k})\cdot \Big(\sum_{a \in \actions} 
\sigma(a\mid \play_{< k})\cdot \probTranFunc(\last(\play)\mid 
\last(\play_{< k}),a) 
\Big), \]
where we use the convention that $\last(\play_{< 0})=\vinit$;
\item for all other $\play$ we have $\cylProb(\play)=0$.
\end{itemize}

Now using an appropriate measure-extension theorem (such as Hahn-Kol\-mo\-go\-rov theorem~\cite[Corollary 2.5.4 and Proposition  2.5.7]{Rosenthal:2006}, or Carath\'eodory theorem~\cite[Theorem 1.3.10]{Ash.Doleans-Dade:2000}) one can show that there is a unique probability measure 
$\probm^{\sigma}_{\mdp,\vinit} $ on $\sigmaAlg_{\mdp}$ such that for every cylinder set $\cylinder(\play)$ determined by some finite play $\play$ we have $\probm_{\vinit}^\sigma(\cylinder(\play))=\cylProb(\play)$. (Abusing the notation, we write $\probm^{\sigma}_{\mdp,\vinit}(\play)$ for the probability of this cylinder set). 
There are some intermediate steps to be performed before an extension theorem can be applied, and we omit these due to space constraints. Full details on the cylinder construction can be found, \textit{e.g.} in~\cite{Ash.Doleans-Dade:2000}.

While the construction of the probability measure 
$\probm^{\sigma}_{\mdp,\vinit}$ might seem a bit esoteric, in the context of 
MDP verification we do not usually need to be concerned with all the delicacies 
behind the associated probability space. The sets of plays that we work with 
typically arise from the basic cylinder sets by means of countable unions, 
intersections, and simple combinations thereof; such sets by definition belong 
to the 
sigma-algebra $\sigmaAlg_{\mdp}$, and their probabilities can be inferred using 
basic probabilistic reasoning. Nevertheless, one should keep in mind that all the 
probabilistic argumentation rests on solid formal grounds. 


In the standard MDP literature~\cite{Puterman:2005}, the plays are often defined as alternating sequence of vertices and actions. Here we stick to the edge-based definition inherited from deterministic games. Still, we would sometimes like to speak about quantities such as `probability that action $a$ is taken in step $i$'. To this end, we introduce, for each strategy $\sigma$, each action $a$,  and each $i\geq 0$,  a random variable $\actevent{\sigma}{a}{i}$ such that $\actevent{\sigma}{a}{i}(\play)=\sigma(\play_{< i})(a)$. It is easy to check that  $\expv^\sigma_v[\actevent{\sigma}{a}{i}]$ is the probability that action $a$ is played in step $i$ when using strategy $\sigma$.

\paragraph{Objectives in MDPs}

Similarly to plays, the notions of both qualitative and quantitative objectives 
are inherited from the non-stochastic world of games. However, since plays in 
MDPs are generated stochastically, even for a fixed strategy $\sigma$ there is 
typically no single infinite play that would constitute the outcome of 
$\sigma$. A concrete $\sigma$ might yield different outcomes, depending on the 
results of random events during the interaction with the MDP. Hence, we need a 
more general way of evaluating strategies in MDPs. 

In the game setting, a qualitative objective was given as a set $\objective
\subseteq \colors^{\omega}$. In the MDP setting, we require that such 
$\objective$ is measurable in the sense that the set $\coloring^{-1}(\objective) = \{\play \in \sampleSpace_{\mdp} \mid \coloring(\play) \in \objective \}$ belongs to $\sigmaAlg_{\mdp}$. We can then talk about a 
probability that the produced play satisfies $\objective$. For instance, for a 
colour $\genColour$ the objective $ \Reach(\genColour) $ is indeed measurable, since $ \coloring^{-1}(\objective) $ can be written as a countable union of all basic cylinders that are determined by finite plays ending in a vertex coloured by $ \genColour $. Indeed, all the qualitative objectives studied in previous chapters can be shown measurable in a similar way, and we encourage the reader to prove this as an exercise.
Hence, the expression $\probm^{\sigma}_{\mdp,\vinit}(\Reach(\genColour))$ denotes the probability that a vertex of colour $\genColour$ is reached when using strategy $\sigma$ from vertex $\vinit$. 
Naturally, \Max aims at maximising this probability. We refer to \Max as ``she'', and in subsequent chapters when studying two-player stochastic games the opponent Min will be ``he''.

The situation is more complex for quantitative objectives. As shown in the previous chapter, 
when working with quantitative objectives, the set of colours $\colors$ is typically the set of real numbers (or a subset thereof), and the quantitative objective is given by an `aggregating function' $\quantObj\colon \colors^\omega \rightarrow \R$, which can be extended into a function $\quantObjExt\colon \edges^\omega \rightarrow \R $ by putting $ \quantObjExt(\play) = \quantObj( \coloring(\play_0)\coloring(\play_1)\cdots) $.
In the MDP setting, we 
require that $\quantObjExt$ is $\sigmaAlg_{\mdp}$-measurable, which 
means that for each $x\in \R$ the set $\{\pi\in \edges^\omega\mid 
\quantObj(\coloring(\play_0)\coloring(\play_1)\cdots) \leq x\}$ belongs to 
$\sigmaAlg_{\mdp}$ (again this holds for all the objectives studied 
in the previous chapters). Then there are two ways in which we can define the expected payoff achieved by strategy $ \sigma $ from a vertex $ v $.
First, we can treat 
$\quantObjExt$ as a random variable 
in the probability space 
$(\sampleSpace_{\mdp},\sigmaAlg_{\mdp},\probm^{\sigma}_{\mdp,v})$. Then the play-based payoff of $\sigma$ from $ v $, which we denote by $ \playPay_\quantObj(v,\sigma) $, is the expected value of this random variable, \textit{i.e.} $ \playPay_\quantObj(v,\sigma) = \expv_{v}^\sigma [\quantObjExt] $. That is, we compute the expected payoff over all plays. This approach subsumes also qualitative objectives: For such an objective $\objective$ we can consider an indicator random 
variable $\indicator{\objective}$, such that $\indicator{\objective}(\play)=1$ 
of 
$\play\in\Omega$ and $\indicator{\objective}(\play)=0$ otherwise. Then 
$\probm^{\sigma}_{\mdp,v}(\objective) = 
\expv^{\sigma}_{\mdp,v}[\indicator{\objective}] = \playPay_{\indicator{\objective}}(v,\sigma)$.



The second approach to quantitative objectives in MDPs, common \textit{e.g.} in the operations research literature, is step-based: for each time step $i$ we compute the expected one-step reward (\textit{i.e.} colour) encountered in that step and then  aggregate  these one-step expectations. Formally, the step-based payoff of $ \sigma $ from $ v $ is $ \stepPay_f(v,\sigma) = \quantObj(\expv_{v}^\sigma[\coloring(\play_0)] \expv_{v}^\sigma[\coloring(\play_1)]\cdots] ) $, where for each $ i $ we treat the expression $ \coloring(\play_i) $ as a random variable returning the colour (\textit{i.e.} a number) which labels the $ i $-th edge of the randomly produced play (recall here that we index edges from~$ 0 $). 


Depending on the concrete quantitative objective and on the shape of $\sigma$, the path- and step-based payoffs from a given vertex might or might not be equal. Nevertheless, in this chapter we study only objectives for which these two semantics yield the \emph{same optimisation criteria:} no matter which of the two semantics we use, the optimal values will be the same and strategy that is optimal w.r.t. one of the semantics is also optimal for the other one. Hence, we will fix the play-based approach as the default one, writing just $ \Pay_f(v,\sigma)$ instead of $ \playPay_f(v,\sigma) $. We will prove the equivalence with step-based payoff where necessary. Also, we will drop the subscript $ f $ when the payoff function is known from the context.



\paragraph{Optimal strategies and decision problems}

Let us fix an MDP $\mdp$ and an objective given by a random variable 
$\quantObj$. The value of a vertex $v\in\vertices$ is the number 
$\Value(v)=\sup_{\sigma} \Pay_f(v,\sigma)$. We let $\Value(\mdp)$ denote the $|\vertices|$-dimensional vector whose component 
indexed by $v$ equals $\Value(v)$.

We say that a strategy $\sigma$ is $\eps$-optimal in $v$, for some $\eps\geq 0$, if $\Pay_f(v,\sigma) \geq \Value(v) - \eps$. A $0$-optimal strategy is simply called optimal. 

For qualitative objectives, there are additional modes of objective satisfaction. Given such an objective $\objective$, we say that a strategy $\sigma$ is almost-surely winning from $v$ if $\expv^{\sigma}_{\mdp,v}[\indicator{\objective}]=1$, \textit{i.e.} if the run produced by $\sigma$ falls into $\objective$ with probability $1$. We also say that $\sigma$ is positively winning from $ v $ if $\expv^{\sigma}_{\mdp,v}[\indicator{\objective}]>0$. For strategies that are winning in the non-stochastic game sense, \textit{i.e.} that \emph{cannot} produce a run not belonging to $\objective$, are usually called surely winning to distinguish them from the above concepts. We denote by $\winPos(\mdp,\objective)$ and $\winAS(\mdp,\objective)$ the sets of all vertices of $\mdp$ from which there exists a positively or almost-surely winning strategy for the objective $\objective$, respectively.


The problems pertaining to the existence of almost-surely or positively winning strategy are often called \emph{qualitative problems} in the MDP literature, while the notion \emph{quantitative problems} covers the general notion of optimising the expectation of some random variable. We do not use such a nomenclature here so as to avoid confusion with qualitative vs. quantitative objectives as defined in \Cref{1-chap:introduction}. Instead, we will refer directly to, \textit{e.g.} `almost-sure reachability' while using the term `optimal reachability' to refer to the expectation-maximisation problem.

%

%

\section{Positive and almost-sure reachability and safety in MDPs}
\label{6-sec:reachability}
We start our study of algorithmic problems for MDPs with the reachability objectives: we write $\Reach(\Win)$ for $\Win$ a set of colours. By a small abuse, we also write $\Win$ for the set of vertices of colours~$\Win$.



\paragraph{Positive reachability}  
Analogously to attractor computations in reachability games (cf. \Cref{1-sec:attractors}), we define a one-step \emph{positive probability} predecessor operator $\PrePos$ as follows: for $U \subseteq \vertices$ we put

\[
\PrePos(U) = \{v \in \vertices \mid \exists a \in \actions, \exists u \in U: \probTranFunc(u\mid v,a)>0 \}.
\]

\noindent
We define an operator $\PreOPPos$ on subsets of vertices: for $X\subseteq\vertices$ we have
$$\PreOPPos(X) = \Win \cup \PrePos(X).$$
We note that this operator is "monotonic" when equipping the powerset of vertices with the inclusion preorder:
if $X \subseteq X'$ then $\PreOPPos(X) \subseteq \PreOPPos(X')$.
Hence \Cref{1-thm:kleene} applies: this operator has a least fixed point computed by the following sequence: we let $X_0 = \emptyset$ and $X_i = \PreOPPos(X_{i-1})$. This constructs a sequence $(X_i)_{i \in \N}$ of non-decreasing subsets of $V$. 
Considering the graph induced by the MDP, it is see to see that $X_i$ is the set of vertices from which a vertex of $\Win$ is reachable via a finite play of length at most $i-1$.
Hence the sequence stabilises after at most $n-1$ steps.

We have the following simple characterisation of the positively winning set:

\begin{theorem}[Characterisation of the positively winning set]
\label{6-thm:positive-char}
Let $\mdp$ an MDP. Then the positively winning region $\winPos(\mdp,\Reach(\Win))$ is the least fixed point of the operator $\PreOPPos$.
Consequently, a vertex $v$ belongs to $\winPos(\mdp,\Reach(\Win))$ if and only if there exists a (possibly empty) finite play from $v$ to a vertex of colour $\Win$.
Moreover, there exists a uniform positional deterministic strategy that is positively winning from every vertex in $\winPos(\mdp,\Reach(\Win))$.
\end{theorem}
\begin{proof}
Let us write $X_0 = \emptyset$ and $X_{i+1} = \PreOPPos(X_i)$, thanks to \Cref{1-thm:kleene} this sequence of subsets of states converges to the least fixed point $X_\infty$ of $\PreOPPos$. Since it is non-decreasing, the sequence is stationary and is reached after at most $n$ iterations, meaning $X_n = X_\infty$. To simplify notations let us write $W_{> 0} = \winPos(\mdp,\Reach(\Win))$.
We show two properties:
\begin{itemize}
	\item For all $i$, we have $X_i \subseteq W_{> 0}$.
	\item $W_{> 0}$ is a pre-fixed point of $\PreOPPos$.
\end{itemize}
The first property implies that $X_\infty \subseteq W_{> 0}$, and the second the converse implication.

We prove the first property by induction on $i$. The case $i = 0$ is clear.
Let $v \in X_{i+1}$, either $v \in \Win$ and then it is in $W_{> 0}$, or $v$ in $\PrePos(X_i)$.
Choosing the action witnessing that inclusion yields a positive probability to land in $X_i$. By induction hypothesis, $X_i \subseteq W_{> 0}$. Hence we have constructed a strategy winning positively from $v$, implying that $v \in W_{> 0}$.

We now prove the second property: $W_{> 0} \subseteq \Win \cup \PrePos(W_{> 0})$.
Let $v \in W_{> 0}$, by definition either $v \in \Win$ or there exists an action which yields a positive probability of winning, implying that $v \in \PrePos(W_{> 0})$.

%
%
%
So far we have proved that $W_{> 0}$ is the least fixed point of $\PreOPPos$.
We conclude the proof by constructing a positional deterministic strategy $\sigma$ positively winning from all of $W_{> 0}$.
For a vertex $v \in W_{> 0}$, let $\rank(v)$ be the smallest $i$ such that $v \in X_i$. 
For each $v \in W_{> 0}$, either $v \in \Win$ or there exists an action $a$ and vertex $u$ such that $\probTranFunc(u \mid v,a)>0$ and $\rank(u) < \rank(v)$. Define $\sigma(v) = a$. A simple induction on the rank shows that $\sigma$ is positively winning from every vertex of $W_{> 0}$.
\end{proof}

\noindent
As for complexity, we can focus on the problem of determining whether a given 
vertex belongs to $\winPos(\mdp,\Reach(\Win))$. 

\begin{corollary}[Complexity of deciding positive reachability]
\label{6-cor:pos-complexity}
The problem of deciding whether a given vertex of a given MDP belongs to 
$\winPos(\mdp,\Reach(\Win))$ is $\NL$-complete. Moreover, the set $\winPos(\mdp,\Reach(\Win))$ can be computed in linear time.
\end{corollary}
\begin{proof}
\Cref{6-thm:positive-char} gives a blueprint for a logspace reduction from this problem to the \emph{s-t-connectivity} problem for directed graphs, and vice versa. The latter problem is well known to be $\NL$-complete~\cite{Savitch:1970}. Moreover, the set of states from which a target colour is reachable can be computed by a simple graph search (\textit{e.g.} by BFS), hence in linear time.
\end{proof}



\paragraph{Almost-sure safety} 
We define the \emph{almost-sure predecessor operator $\PreAS$}: for $X \subseteq \vertices$ we have 
$$\PreAS(X) = \{v \in \vertices \mid \exists a \in \actions, \forall t \in \vertices: \probTranFunc(t\mid v,a)>0 \Rightarrow t \in X \}.$$
It is clearly monotonic: if $X \subseteq X'$, then $\PreAS(X) \subseteq \PreAS(X')$.
One might be tempted to mimic the positive reachability case in order to solve the almost-sure reachability case: compute the least fixed point of the monotonic $X \mapsto \Win \cup \PreAS(X)$ and hope that it yields $\winAS(\mdp,\Reach(\Win))$. 
But this is not correct: consider an MDP with two vertices, $u,v$, the latter one being coloured by $\Win$. 
We have only one action $a$: in $v$, the action self loops on $v$, while in $u$ playing the action either moves us to $v$ or leaves us in $u$, both options having probability $\frac{1}{2}$. 
The probability that we \emph{never} reach $v$ from $u$ is equal to $\lim_{n\rightarrow \infty} \left(\frac{1}{2}\right)^n = 0$, and hence 
$\winAS(\mdp,\Reach(\Win))=\{u,v\}$. 
However, the least fixed of the operator above is $\set{v}$, excluding $u$. 
Note that there indeed exists an infinite play which \emph{can} be generated by the strategy and which never visits $v$, but the probability of generating such a play is $0$. 

Let us define the operator $\P_{=1}$ on subsets of vertices: for $X \subseteq \vertices$ we have
\[
\P_{=1}(X) = \Win \cap \PreAS(X).
\]
Clearly $\P_{=1}$ is monotonic.

\begin{lemma}[Characterisation of the almost-sure safety winning region]
\label{6-lem:safety-iteration}
Let $\mdp$ an MDP. Then the almost-sure safety winning region $\winAS(\mdp,\Safe(\Win))$ is the greatest fixed point of the operator $\P_{=1}$.
Moreover, there exists a positional deterministic strategy that is almost-surely winning from every vertex of $\winAS(\mdp,\Safe(\Win))$.
\end{lemma}

\begin{proof}
Let us write $X_0 = V$ and $X_{i+1} = \P_{=1}(X_i)$, thanks to \Cref{1-thm:kleene} this sequence of subsets of states converges to the greatest fixed point $X_\infty$ of $\P_{=1}$. Since it is non-increasing, the sequence is stationary and is reached after at most $n$ iterations, meaning $X_n = X_\infty$. To simplify notations let us write $W_{=1} = \winAS(\mdp,\Safe(\Win))$.
We show two properties:
\begin{itemize}
	\item For all $i$, we have $W_{=1} \subseteq X_i$.
	\item $W_{=1}$ is a post-fixed point of $\P_{=1}$.
\end{itemize}
The first property implies that $W_{=1} \subseteq X_\infty$, and the second the converse implication.

We prove the first property by induction on $i$. The case $i = 0$ is clear.
Let $v \in W_{=1}$. Clearly we must have that $v \in \Win$. Since $v \in W_{=1}$, there exists an action ensuring to stay in $W_{=1}$ with probability $1$.
By induction hypothesis $v \in X_i$, so this implies that $v \in \P_{=1}(X_i)$.

We now prove the second property: $W_{=1} \supseteq \Win \cap \PreAS(W_{=1})$.
Let $v \in \Win \cap \PreAS(W_{=1})$, there exists an action ensuring to win with probability $1$, implying that $v \in W_{=1}$.

So far we have proved that $W_{= 1}$ is the greatest fixed point of $\P_{=1}$.
We conclude the proof by constructing a positional deterministic strategy $\sigma$ almost-surely winning from all of $W_{= 1}$.
For a vertex $v \in W_{= 1}$, there exists an action $a$ such that if $\probTranFunc(u \mid v,a) > 0$ then $u \in W_{= 1}$. Define $\sigma(v) = a$. Clearly $\sigma$ is almost-surely winning from every vertex of $W_{= 1}$.
\end{proof}

\begin{algorithm}
	\KwData{An MDP $ \mdp $, a colour $\Win$}
	\SetKwFunction{KwOPAS}{$\OPAS$}
	\SetKwProg{Fn}{Function}{:}{}
	
		$X \leftarrow V$ \;
		\Repeat{$X' = X$}{
			$X' \leftarrow X$\;
			
			$X \leftarrow \Win \cap \PreAS(X)$
		}
		\Return{$X$} 
\caption{An algorithm computing $\winAS(\mdp,\Safe(\Win))$}
\label{6-algo:safety}
\end{algorithm}

\begin{corollary}[Complexity of the almost-sure safety winning set]
\label{6-cor:safety-main}
The problem of deciding whether a given vertex of a given MDP belongs to $\winAS(\mdp,\Safe(\Win))$ can be solved in strongly polynomial time, together with a positional deterministic strategy which is almost-surely winning for all vertices in $\winPos(\mdp,\Safe(\Win))$.
\end{corollary}

\Cref{6-algo:safety} computes the set $\winAS(\mdp,\Safe(\Win))$ in strongly polynomial time.

\begin{proof}
The algorithm makes at most linearly many iterations, each of which has at most linear complexity. Hence, the complexity is at most quadratic. The strong polynomiality is testified by the fact that the algorithm only tests whether a probability of a given transition is positive or not, and the exact values of positive probabilities are irrelevant. 
\end{proof}

%

\paragraph{Almost-sure reachability} 
With almost-sure safety solved, we can now solve almost-sure reachability. 
Consider the one-step \emph{almost-sure safety} operator $\safeOP$ acting on sets of vertices:
\begin{align*}
\safeOP(X)= X \cap \PreAS(X).
\end{align*}
This operator gives rise to the notion of a closed set, which is important for the study of safety objectives in MDPs.


\begin{definition}[Closed set in an MDP]
\label{6-def:closed_set_MDP}
A set $X$ of vertices is closed if $\safeOP(X)=X$. 
The sub-MDP of an MDP $ \mdp $ induced by the closed subset $X$ is the MDP $\mdp_X = (X,\edges_X,\probTranFunc_X,\coloring_X)$ defined as follows:
\begin{itemize}
	\item $\edges_X$ is obtained from $\edges$ by removing edges incident to a vertex from $\vertices\setminus X$;
	\item $\probTranFunc_X$ is obtained from $\probTranFunc\subseteq \vertices\times\actions\times\dist(\edges)$ by removing all triples $(v,a,f)$ where either $v\not \in X$ or where the support of $f$ is not contained in $X$;
	\item $\coloring_X$ is a restriction of $\coloring$ to $X$.
\end{itemize}
We denote by $\closed(\mdp)$ the set of all closed sets in $\mdp.$
\end{definition}

One can show that a set is closed if \Max has a strategy ensuring that she stays in the set forever.

\begin{algorithm}
	\KwData{An MDP $ \mdp $, a colour $ \Win $}
\SetKwFunction{FTreat}{Treat}
$W \leftarrow \vertices$
\Repeat{$W' \neq W$}{
	$W' \leftarrow W$
	$Z \leftarrow \winPos(\mdp_W,\Reach(\Win))$
	$W \leftarrow \winAS(\mdp_W,\Safe(W \setminus Z))$
}

\Return{$W$} \tcp*{A positive winning strategy in $\mdp_W$ is now almost-surely winning from $W$ in $\mdp$.}
\caption{An algorithm computing $\winAS(\mdp,\Reach(\Win))$}
\label{6-algo:reach-as}
\end{algorithm}

The pseudocode for solving almost sure reachability is given in~\Cref{6-algo:reach-as}. Note that in the first iteration, the algorithm computes the set $\winAS(\mdp,\Safe(Z))$ where $Z =  \winPos(\mdp,\Reach(\Win))$. 

We might be tempted to think that $\winAS(\mdp,\Safe(Z)) = \winAS(\mdp,\Reach(\Win))$, but this is not the case. 
To see this, consider an MDP with three states $ u,v,t $ and two actions $ a,b $ such that $ t $ is coloured by $ \Win $, both actions self loop in $ v $ and $ t $, and $ \probTranFunc(t \mid u,a) = \probTranFunc(v \mid u,a) = \frac{1}{2} $ while $ \probTranFunc(u \mid u,b) = 1 $. Then 
\[
\winAS(\mdp,\Reach(\Win)) = \{t\}, \text{ but } \winAS(\mdp,\Safe( \winPos(\mdp,\Reach(\Win)))) = \{u,t\}.
\]
However, iterating the computation works, as shown in the following theorem.

\begin{theorem}[Algorithm for the almost-sure reachability winning set]
\label{6-thm:as-char}
The problem of deciding whether a given vertex of a given MDP belongs to $\winAS(\mdp,\Reach(\Win))$ can be solved in strongly polynomial time, together with a positional deterministic strategy which is almost-surely winning for all vertices in $\winPos(\mdp,\Reach(\Win))$.
\end{theorem}

The pseudocode is given in~\Cref{6-algo:reach-as}.

\begin{proof}
Since the set $ W $ can only decrease in each iteration, the algorithm terminates.
We prove that upon termination, $W$ equals $\winAS(\mdp,\Reach(\Win))$.
	
We start with the $ \subseteq $ direction. We have $W \subseteq \winPos(\mdp_W,\Reach(\Win))$. By \Cref{6-thm:positive-char} there exists an MD strategy $\sigma$ in $\mdp_W$ which is positively winning from each vertex of $W$. We show that the same strategy is also almost-surely winning from each vertex of $W$ in $\mdp_W$ and thus also from each vertex of $W$ in $\mdp$, which also proves the second part of the theorem. 
Let $v$ be any vertex of $W$ and denote $|W|$ by $\ell$. Since $\sigma$ is positional, it guarantees that a vertex of $\Win$ is reached with a positive probability in at most $\ell$ steps (see also the construction of $ \sigma $ in the proof of \Cref{6-thm:positive-char}), and since it is also deterministic, it guarantees that the probability $p$ of reaching $\Win$ in at most $\ell$ steps is at least $p_{\min}^{\ell}$, where  $p_{\min}$ is the smallest non-zero edge probability in $\mdp_W$. Now imagine that $\ell$ steps have elapsed and we have not yet reached $\Win$. This happens with a probability at most $(1-p_{\min}^\ell)$. However, even after these $\ell$ steps we are still in $W$, since $ \sigma  $ is a strategy in $ \mdp_w $. Hence, the probability that we do not reach $\Win$ within the first $2\ell$ steps is bounded by $(1-p_{\min}^\ell)^{2}$. To realise why this is the case, note that any finite play $\play$ of length $2\ell$ can be split into two halves, $\play',\play''$ of length $\ell$, and then $\probm^{\sigma}_{v}(\cylinder(\play))=\probm^{\sigma}_{v}(\cylinder(\play'))\cdot\probm^{\sigma}_{\last(\play')}(\cylinder(\play''))$ (here we use the fact that $\sigma$ is positional). Using this and some arithmetic, one can show that, denoting $\mathit{Avoid}_i$ the set of all plays that avoid the vertices of $\Win$ in steps $\ell\cdot(i-1)$ to $\ell\cdot(i)-1$, it holds $$\probm^{\sigma}_{v}(\mathit{Avoid}_1\cap \mathit{Avoid}_2) \leq \probm^{\sigma}_{v}(\mathit{Avoid}_1)\cdot \max_{u\in W\setminus \Win}\probm^{\sigma}_{u}(\mathit{Avoid}_1)\leq (1-p_{\min}^\ell)^{2}.$$

\noindent
One can then continue by induction to show that $\probm^{\sigma}_{v}(\bigcap_{i=1}^j \mathit{Avoid}_i)\leq (1-p_{\min}^\ell)^{j},$ and hence
$$\probm^\sigma_{v}(\Reach(\Win))= 1-\probm^{\sigma}_{v}(\bigcap_{i=1}^\infty \mathit{Avoid}_i) \leq 1-\lim_{j\rightarrow \infty}(1-p_{\min}^\ell)^{j}= 1-0=1.$$

Now we prove the $ \supseteq $ direction. Denote $X=\winAS(\mdp,\Reach(\Win))$. We prove that $ W \supseteq X $ is an invariant of the iteration. Initially this is clear. Now assume that this holds before an iteration takes place. It is easy to check that $X$ is closed, so $\mdp_{X}$ is well-defined. We prove that $ X \subseteq \winAS(\mdp_W,\Safe(W\setminus Z)) $, where $ Z $ is defined during the iteration. A strategy in $\mdp$ that reaches $\Win$ with probability 1 must never visit a vertex from $\vertices\setminus X$ with a positive probability. Hence, each such strategy can be viewed also as a strategy in $\mdp_{X}$. It follows that  $X=\winAS(\mdp_X,\Reach(\Win)) = \winPos(\mdp_{X},\Reach(\Win)) \subseteq \winPos(\mdp_{W},\Reach(\Win)) = Z$, the middle inclusion following from induction hypothesis. Now by \Cref{6-lem:safety-iteration} and \Cref{6-cor:safety-main}, we have that the set $ \winAS(\mdp_W,\Safe(W\setminus Z)) $ is the largest closed set contained in $ Z $. But $ X $ is also closed, and as shown above, it is contained in $ Z $. Hence,  $ X \subseteq \winAS(\mdp_W,\Safe(W \setminus Z)) $.

The complexity follows form \Cref{6-cor:pos-complexity} and \Cref{6-cor:safety-main}; and also from the fact that the main loop must terminate in $ \leq |\vertices| $ steps. The strong polynomiality again follows from the algorithm being oblivious to precise probabilities.
\end{proof}



We also have a complementary hardness result. 

\begin{theorem}[Complexity of the almost-sure reachability winning set]
\label{6-thm:as-complexity}
The problem of determining whether a given vertex of a given MDP belongs to 
$\winAS(\mdp,\Reach(\Win))$ is $\P$-complete.
\end{theorem}
\begin{proof}
	We proceed by a reduction 
	from the \emph{circuit value problem (CVP)}.
	An instance of \emph{CVP} is a directed acyclic graph $\mathcal{C}$, 
	representing a boolean circuit: each internal node represents either an OR gate 
	or an AND gate, while each leaf node is labelled by \emph{true} or 
	\emph{false}. Each internal node is guaranteed to have exactly two children. 
	Each node of $\mathcal{C}$ evaluates to a unique truth value: the value of a 
	leaf is given by its label and the value of an internal node $v$ is given by 
	applying the logical operator corresponding to the node to the truth values of 
	the two children of $v$, the evaluation proceeding in a backward topological order. The task is to decide whether a given node $w$ of 
	$\mathcal{C}$ evaluates to \emph{true}. \emph{CVP} was shown to be 
	$\P$-hard (under logspace reductions) in~\cite{Ladner:1975}. 
	In~\cite{Chatterjee.Doyen.ea:2010}, the following logspace reduction 
	from CVP to almost-sure reachability in MDPs is presented: given a boolean circuit 
	$\mathcal{C}$, construct an MDP $\mdp_{\mathcal{C}}$ whose vertices correspond 
	to the gates 
	of $\mathcal{C}$. There are two actions, call them $\mathit{left}$ and $\mathit{right}$. In each vertex corresponding to an OR gate $g$, the 
	$\mathit{left}$ action transitions with probability 1 to the vertex 
	representing the left child of $g$, and similarly for the action 
	$\mathit{right}$ 
	and the right child. In a vertex corresponding to an AND gate $g$, both actions behave the same: the transition into each of the two children 
	of $g$ with probability $\frac{1}{2}$. Vertices corresponding to leafs have self loop as the only outgoing edges, and 
	moreover, they are coloured with the respective labels in $\mathcal{C}$. It is 
	easy to check that a gate of $\mathcal{C}$ evaluates to $\mathit{true}$ if and 
	only if the corresponding vertex belongs to 
	$\winAS(\mdp_{\mathcal{C}},\Reach(\mathit{true}))$.
\end{proof}

\paragraph{Positive safety} We conclude this section by a discussion of positive safety. 

\begin{theorem}[Algorithm for the positive safety winning set]
\label{6-thm:pos-safety-main}
Let $ \mdp_{\bar\Win} $ be an MDP obtained from $ \mdp $ by changing all $ \Win $-coloured vertices to sinks (\textit{i.e.} all actions in these vertices are self loops). Then 
\[
\winPos(\mdp,\Safe(\Win)) = \winPos(\mdp_{\bar\Win},\Reach(\winAS(\mdp_{\bar\Win},\Safe(\Win))) ).
\] 
In particular, the set $ \winPos(\mdp,\Safe(\Win)) $ can be computed in a strongly polynomial time and there exists a positional deterministic strategy, computable in strongly polynomial time, that is positively winning from every vertex of $\winPos(\mdp,\Safe(\Win))$.
\end{theorem}
\begin{proof}
Clearly $ \winPos(\mdp,\Safe(\Win)) = \winPos(\mdp_{\bar{\Win}},\Safe(\Win)) $ and $ \winAS(\mdp,\Safe(\Win)) = \winAS(\mdp_{\bar{\Win}},\Safe(\Win)) $; and moreover the corresponding winning strategies easily transfer between the two MDPs (for a safety objective, the behaviour after visiting a $ \Win $-coloured state is inconsequential).
Hence, putting $ Z = \winAS(\mdp_{\bar{\Win}},\Safe(\Win)) $, it is sufficient to show that  $ \winPos(\mdp_{\bar{\Win}},\Safe(\Win)) =\winPos(\mdp_{\bar\Win},\Reach(Z))  $  

The $ \supseteq $  inclusion is clear as well as the construction of the witnessing MD strategy (in the vertices of that are outside of $ Z $, we behave as the positively winning MD strategy for reaching $ Z $, while inside $ Z $ we behave as the almost-sure winning strategy for $ \Safe(\Win) $). 

For the other inclusion, let $ X = \vertices \setminus \winPos(\mdp_{\bar{\Win}},\Reach(Z)) $. We prove that $ X\subseteq \vertices \setminus \winPos(\mdp_{\bar{\Win}},\Safe(\Win)) $. Assign ranks to vertices inductively as follows: each vertex coloured by $ \Win $ gets rank $ 0 $. Now if ranks $ \leq i $ have been already assigned, then a vertex  $ v $ is assigned rank $i+1  $ if it  does  not already have a lower rank but for all actions $ a\in\actions $ there exists a vertex $ u $ of rank $ \leq i $ s.t. $ \probTranFunc(u \mid v,a) >0$. Then each vertex in $ X $ is assigned a finite rank: indeed, the set of vertices without a rank is closed and does not contain $ \Win $-coloured vertices, hence it is contained in $ Z $. 
Now fix any strategy $ \sigma $ starting in a vertex $v \in X $. By definition of $ X $, $ \sigma $ never reaches $ Z $ and hence never visits an unranked state. At the same time, whenever $ \sigma $ is in a ranked state, there is, by definition of ranks, a probability at least $p_{\min}  $ (the minimal edge probability in $ \mdp $) of transitioning to a lower-ranked state in the next step. Hence, in every moment, the probability of $ \sigma $ reaching a $ \Win $-coloured state within the next $ |\vertices| $ steps is at least $ p_{\min}^{|\vertices|} $. By a straightforward adaptation of the second part of the proof of \Cref{6-thm:as-char}, $ \sigma $ eventually visits $ \Win $ with probability 1. Since $ \sigma $ was arbitrary, this shows that $ v\not \in\winPos(\mdp_{\bar{\Win}},\Safe(\Win)).  $

The complexity follows from the results on positive reachability and almost-sure safety.
\end{proof}

%

\section{Discounted payoff in MDPs}
\label{6-sec:discounted}
In this section, we consider MDPs with edges coloured by rational numbers and with the objective $\DiscountedPayoff$. We start the chapter by proving that using the play-based semantics for the discounted-payoff objective yields no loss of generality. 

\begin{lemma}[Equivalence of the definitions for discounted payoffs]
\label{6-lem:disc-step-one}
In a discounted-payoff MDP, for each strategy $ \sigma $ and each vertex $ v $ we have 
\[
\playPay(v, \sigma) = \stepPay(v, \sigma).
\]
\end{lemma}
\begin{proof}
We have 
\begin{align*} \playPay(v,\sigma) &= \expv^\sigma_v[(1-\lambda)\lim_{k \rightarrow \infty} \sum_{i=0}^{k-1}\lambda^i \coloring(\play_i) ] = (1-\lambda) \lim_{k \rightarrow \infty} \expv^\sigma_v[\sum_{i=0}^{k-1}\lambda^i \coloring(\play_i) ] 
\\
&= (1-\lambda)\cdot\lim_{k \rightarrow \infty} \sum_{i=0}^{k-1}\lambda^i\expv^\sigma_v[ \coloring(\play_i) ] = \stepPay(v, \sigma).
\end{align*}
%
Here, the last equality on the first line follows from the dominated convergence theorem~\cite[Theorem 1.6.9]{Ash.Doleans-Dade:2000} and the following equality comes from the linearity of expectation. (To apply the domination convergence theorem, we use the fact that for each $k$ we have $\DiscountedPayoff^{k}(\play) \le \maxc$.)
\end{proof}

\subsection*{Optimal values and positional optimality}

 In this subsection we give a 
characterisation of the value vector $\Value(\mdp)$ and prove that there always exists a 
positional deterministic strategy that is optimal in every vertex. Our 
exposition follows (in a condensed form) the one in~\cite{Puterman:2005}, the techniques 
being somewhat similar to the ones in the previous chapter.

%

We define an operator $\discOP \colon \R^{\vertices}\rightarrow \R^{\vertices}$ as follows: each vector $\vec{x}$ is mapped to a vector 
$\vec{y} = \discOP(\vec{x})$ such that:
$$
\vec{y}_v = \max_{a \in \actions} \sum_{u\in \vertices} \probTranFunc(u \mid  v,a) 
\cdot\left((1-\lambda)\cdot\coloring(v,u) + \lambda\cdot \vec{x}_u \right).
$$

\begin{lemma}[Contraction mapping]
\label{6-lem:fixpoint}
The operator $\discOP$ is a contraction mapping. Hence, $\discOP$ has a unique 
fixed point $\vec{x}^*$ in $\R^{\vertices}$, and $\vec{x}^* = 
\lim_{k\rightarrow \infty} \discOP^k(\vec{x})$, for any 
$\vec{x}\in\R^{\vertices}$.
\end{lemma}
\begin{proof}
The proof proceeds by a computation analogous to the one in the first half of the proof of~\Cref{5-thm:values_discounted_contracting_fixed_point}; we just need to reason about actions rather than edges (and of course, use the formula defining $\discOP$ instead of the one for games). The second part follows from the Banach fixed-point theorem.
\end{proof}

We aim to prove that $\Value(\mdp)$ is the unique fixed point $\vec{x}^*$ of 
$\discOP$. We start with an auxiliary definition.

\begin{definition}[Safe actions]
\label{6-def:disc-safe-act}
Let $\vec{x}\in \R^{\vertices}$. We say that an action $a$ is $\vec{x}$-safe in 
a vertex $v$ if
\begin{equation}
\label{6-eq:disc-safe-act}
a= \underset{a' \in \actions}{\arg\max} \sum_{u\in \vertices} 
\probTranFunc(u\mid 
v,a') 
\cdot\left((1-\lambda)\cdot\coloring(v,u) + \lambda\cdot \vec{x}_u \right).
\end{equation}
\noindent
A strategy $\sigma$ is $\vec{x}$-safe, if for 
each play $ \play $ ending in a vertex $v$, all actions that are selected with a positive 
probability by $\sigma$ for $\play$ are $\vec{x}$-safe in $v$.
\end{definition}


Given a strategy $\sigma$ we define $\valsigma$ to be the vector such that $\vec{x}_{v}^\sigma = 
\playPay(v,\sigma)$. For positional strategies, $\valsigma$ can be computed efficiently as follows:
Each positional strategy $\sigma$ defines a \emph{linear} operator $\discOP^{\sigma}$ which maps each vector 
$\vec{x}\in \R^{\vertices}$ to a vector $\vec{y}=\discOP^{\sigma}(\vec{x})$ 
such that $$\vec{y}_v = \sum_{a\in \actions} \sigma(a\mid v) \cdot 
\left(\sum_{u \in \vertices} 
\probTranFunc(u \mid v,a) \cdot( (1-\lambda)\cdot \coloring(v,u) + 
\lambda \cdot \vec{x}_u )\right).$$  

\begin{lemma}[Operator using a fixed strategy]
\label{6-lem:disc-val-sigma}
Let $\sigma$ be a positional strategy using rational probabilities. Then the operator $\discOP^{\sigma}$ has a unique fixed point, which is equal to $\valsigma$ and which can be computed in polynomial time.
\end{lemma}
\begin{proof}
The operator $\discOP^{\sigma}$ can be seen as an instantiation of $\discOP$ in an MDP where there is no choice, since the action probabilities are chosen according to $\sigma$. \Cref{6-lem:fixpoint} shows that 
$\vec{x}^\sigma$ is a fixed point of $\discOP^\sigma$. Since $\discOP^{\sigma}$ is again a contraction, it has a unique fixed point; and since it is linear, the fixed point can be computed in polynomial time, \textit{e.g.} by Gaussian elimination (in its polynomial bit-complexity variant known as Bareiss algorithm~\cite{Bareiss:1968}).
\end{proof}

We now prove that there is a positional strategy ensuring outcome given by the fixed point of $\discOP$. Hence, the fixed point gives a lower bound on the values of vertices.

\begin{lemma}[Any fixed point induces an MD strategy]
\label{6-lem:disc-val-lower}
Let $\vec{x}^*$ be the unique fixed point of $\discOP$. 
Then there exists an MD strategy that is $\vec{x}^*$-safe. Moreover, for each $\vec{x}^*$-safe positional strategy it holds that  
$\playPay(v,\sigma) =\vec{x}_v^*$ for each $v\in \vertices$.
\end{lemma}
\begin{proof}
 Note that for each $\vec{x}\in \R^{\vertices}$ and each $v\in 
\vertices$ there always exists at least one action that is $\vec{x}$-safe in 
$v$. Hence, there is a positional deterministic strategy which 
in each $v$ chooses an arbitrary (but fixed) action that is $\vec{x}^*$-safe in 
$v$. 

Now let $ \sigma $ be an arbitrary $\vec{x}^*$-safe positional strategy.
By \Cref{6-lem:disc-val-sigma}, the vector $\valsigma$ is the unique fixed point of $\discOP^\sigma$.
 But since $\sigma$ 
is $\vec{x}^*$-safe, $\vec{x^*}$ is also a fixed point of $\discOP^\sigma$. 
Hence, $\vec{x}^* = \vec{x}^\sigma$.
\end{proof}

It remains to prove that $\vec{x}^*$ gives, for each vertex, an upper 
bound on the expected discounted payoff achievable by any strategy from that 
vertex. We introduce some additional notation to be used in the proof of the 
next lemma, as well as in some later results: namely, we denote by 
$\dPayoffStep{k}(\play)$ the discounted payoff accumulated during the first $k$ steps of $\play$, \textit{i.e.} the number 
$(1-\lambda)\sum_{i=0}^{k-1} \lambda^i
\, \coloring(\play_i)$. The following lemma can be proved by an easy induction.

\begin{lemma}[Properties of the sequence of iterates]
\label{6-lem:disc-iterates}
For each $k\geq 0$ and each vertex $v$ we have 
$$\sup_{\sigma}\expv^{\sigma}_{v}[\dPayoffStep{k}] = 
(\discOP^k(\vec{0}))_v$$ 
(here $\vec{0}$ is a $|\vertices|$-dimensional vector of zeroes).
\end{lemma}

The previous lemma is used to prove the required upper bound on $\Value(v)$.

\begin{lemma}[Upper bound on the optimal value]
\label{6-lem:disc-val-upper}
For each vertex $v$ it holds 
$\Value(v)\leq \vec{x}^*_v$, where $\vec{x}^*$ is the 
unique fixed point of $\discOP$.
\end{lemma}
\begin{proof}
%
We have $\DiscountedPayoff(\play) = \lim_{k\rightarrow 
\infty}\dPayoffStep{k}(\play)$ (for each $\play$) and hence, by 
dominated 
convergence theorem we have $\expv^\sigma_v[\DiscountedPayoff] = 
\lim_{k\rightarrow 
\infty}\expv^\sigma_v[\dPayoffStep{k}]$. 
Hence,
\begin{align}
\Value(v) &= \sup_{\sigma}\expv^\sigma_v[\DiscountedPayoff] \nonumber\\
&= \sup_{\sigma}\lim_{k\rightarrow \infty}\expv^\sigma_v[\dPayoffStep{k}] 
\label{6-eq:disc-limit-transition}
\end{align}

\noindent
It remains to prove that the RHS of~\Cref{6-eq:disc-limit-transition} is not 
greater than $\vec{x}^*= \lim_{k\rightarrow 
\infty}\discOP^k(\vec{0})=\lim_{k\rightarrow \infty} 
\sup_{\sigma}\expv^\sigma_v[\dPayoffStep{k}]$ (the last inequality follows 
by~\Cref{6-lem:disc-iterates}). It suffices to 
show that for each $\sigma'$ we have 
\[
\lim_{k\rightarrow 
\infty}\expv^{\sigma'}_v\dPayoffStep{k}] \leq \lim_{k\rightarrow 
\infty}\sup_{\sigma}\expv^\sigma_v[\dPayoffStep{k}].
\]
But this immediately 
follows from the fact that the inequality holds for each concrete $k$.
\end{proof}

The following theorem summarises the results.

\begin{theorem}[Characterisation of the optimal values]
\label{6-thm:disc-val-char-mem}
The vector of values $\Value(\mdp)$ in a discounted-payoff MDP $\mdp$ is the 
unique fixed point $\vec{x}^*$ of the operator $\discOP$. Moreover, there 
exists a 
positional deterministic strategy that is optimal in every vertex.
\end{theorem}
\begin{proof}
The characterisation of $\Value(\mdp)$ follows directly from~\Cref{6-lem:disc-val-lower,6-lem:disc-val-upper}. The MD 
optimality follows from~\Cref{6-lem:disc-val-lower}.
\end{proof}

In the rest of this section we discuss several algorithms for computing the 
values and optimal strategies in discounted-payoff MDPs.

\subsection*{Value iteration}

The value iteration algorithm works in the same way as in the case of 
discounted-payoff games: we simply iterate the operator $\discOP$. 
We know that $\Value(\mdp)=\lim_{k\rightarrow \infty}\discOP^k(\vec{0})$, and hence, iterating $\discOP$ yields an 
approximation of $\Value(\mdp)$. The iteration might not reach the fixed 
point (\textit{i.e.} $\Value(\mdp)$) in a finite number of steps, but we can provide 
simple bounds on the precision of the approximation.

\begin{lemma}[Properties of the iterates]
\label{6-lem:disc-val-it-convergence}
For each $k\in \N$, $\|\Value(\mdp)-\discOP^k(\vec{0}) \|_{\infty} \leq 
\lambda^k \cdot \maxc$.
\end{lemma}
\begin{proof}
This follows immediately from~\Cref{6-lem:disc-iterates} and from the fact that 
for each play $\play$, $\sum_{i=k}^{\infty}\lambda^i\cdot 
\coloring(\play_i)\leq \frac{1}{1-\lambda}\cdot\lambda^k \cdot \maxc$.
\end{proof}

Similar analysis can be applied to strategies induced by the approximate 
vectors.

\begin{lemma}[Relating iterates and optimal strategies]
\label{6-lem:disc-val-it-eps-strategies}
Let $\eps>0$ be arbitrary and let 
$$k(\eps)=\left\lceil\frac{\log_2\left(\frac{\eps(1-\lambda)}{4\maxc}\right)}{\log_2(\lambda)}
 \right\rceil .$$ Then, every 
$\discOP^{k(\eps)}(\vec{0})$-safe positional strategy is $\eps$-optimal in 
every vertex.
\end{lemma}
\begin{proof}
Let $\sigma$ be any $\discOP^{k(\eps)}(\vec{0})$-safe positional strategy and 
let $\discOP^{\sigma}$ be the corresponding operator. We have that
\begin{align}
\|\Value(\mdp) - \valsigma \|_{\infty} &= \|\Value(\mdp) 
-\discOP^{k(\eps)}(\vec{0}) +\discOP^{k(\eps)}(\vec{0}) - \valsigma 
\|_{\infty} \nonumber
\\
&\leq \|\Value(\mdp) -\discOP^{k(\eps)}(\vec{0}) 
\|_{\infty} + \| \discOP^{k(\eps)}(\vec{0}) - \valsigma
\|_{\infty}. \label{6-eq:disc-val-it-bound}
\end{align}
\noindent
The first term in~\Cref{6-eq:disc-val-it-bound} is $\leq \eps/2$ 
by the choice of $k(\eps)$ and~\Cref{6-lem:disc-val-it-convergence}. We prove 
that the second term 
in~\Cref{6-eq:disc-val-it-bound} is also bounded by $\eps/2$. Note that we 
have $\valsigma=\discOP^{\sigma}(\valsigma)$ (as was already proved 
in~~\Cref{6-lem:disc-val-lower}) and $\discOP(\discOP^{k(\eps)}(\vec{0})) = 
\discOP^\sigma(\discOP^{k(\eps)}(\vec{0}))$ (since $\sigma$ is 
$\discOP^{k(\eps)}(\vec{0})$-safe). Using this we get
\begin{align*}
\| \discOP^{k(\eps)}(\vec{0}) - \valsigma
\|_{\infty} & = \| \discOP^{k(\eps)}(\vec{0}) -\discOP^{k(\eps)+1}(\vec{0}) + 
\discOP^{k(\eps)+1}(\vec{0}) - \valsigma
\|_{\infty}\\
&\leq \| \discOP^{k(\eps)}(\vec{0}) -\discOP^{k(\eps)+1}(\vec{0}) \|_{\infty} + 
\|\discOP^{k(\eps)+1}(\vec{0}) - \valsigma
\|_{\infty}
\\
&=\| \discOP^{k(\eps)}(\vec{0}) -\discOP^{k(\eps)+1}(\vec{0}) \|_{\infty} + 
\|\discOP^\sigma(\discOP^{k(\eps)}(\vec{0})) - 
\discOP^\sigma(\valsigma)\|_{\infty}\\
&\leq \| \discOP^{k(\eps)}(\vec{0}) -\discOP^{k(\eps)+1}(\vec{0}) \|_{\infty} + 
\lambda\cdot\|(\discOP^{k(\eps)}(\vec{0})) - 
\valsigma\|_{\infty}
\end{align*}

\noindent
Re-arranging yields $\| \discOP^{k(\eps)}(\vec{0}) - \valsigma
\|_{\infty} \leq \frac{1}{1-\lambda}\cdot\| 
\discOP^{k(\eps)}(\vec{0}) - 
\discOP^{k(\eps)+1}
\|_{\infty} $.

It follows from~\Cref{6-lem:disc-val-it-convergence}  that 
$\|\discOP^{k(\eps)}(\vec{0}) -\discOP^{k(\eps)+1}(\vec{0}) \|_{\infty} 
\leq 2\cdot\lambda^{k(\eps)}\cdot \max|\coloring|\leq 
\frac{(1-\lambda)\eps}{2}$, the last 
inequality holding by the choice of $k(\eps)$. Plugging this into the 
formula above yields $\| \discOP^{k(\eps)}(\vec{0}) - \valsigma
\|_{\infty} \leq\frac{\eps}{2}$, as required. 
\end{proof}


Using a value-gap result 
(similar to the game case, but proved using a different technique), one can 
show that sufficiently precise iterates can be used to compute an \emph{optimal} strategy. 
This is summarised in the following lemma due to~\cite{Tseng:1990}.

\begin{lemma}[Sufficiently many iterates yield an optimal strategy]
\label{6-lem:disc-vi-optimal-strategy}
Let $d$ be the least common multiple of denominators of the following numbers: $\lambda$, all   
transition probabilities, and all edge colourings in $\mdp$. Next, let $\eps^* = 
\frac{1}{d^{(3|\vertices|+3)}\cdot |\vertices|^{\frac{\vertices}{2}}}$.
Then, any $\discOP^{k(\eps^*)}(\vec{0})$-safe positional deterministic strategy 
is 
optimal in every 
vertex.
\end{lemma}
\begin{proof}
 Let $\sigma^*$ be any MD optimal strategy (it is guaranteed 
to exist by~\Cref{6-thm:disc-val-char-mem}). By the same theorem, we have that 
$\Value(\mdp)=\discOP^{\sigma}(\Value(\mdp))$. By the definition of 
$\discOP^{\sigma}$, we can write the above equation as $\Value(\mdp)= 
(1-\lambda)\cdot \vec{c}+\lambda \cdot P\cdot \Value(\mdp)$, where $\vec{c}$ is 
a vector whose each 
component 
is a 
sum of several terms, each term being a product of an edge colour and of a 
transition probability; and $P$ is a matrix containing 
transition 
probabilities. Multiplying the equation by $d^3$ yields $d^3\Value(\mdp)= 
d^3(1-\lambda)\cdot \vec{c}+d^3\lambda \cdot P\cdot \Value(\mdp)$. Since this equation has a unique fixed point (due to 
$\discOP^\sigma$ being a contraction), the matrix $A = d^3(I - \lambda P)$ (where $ I $ is the unit matrix) is 
regular, and moreover, composed of integers (due to the choice of $ d $). By Cramer's rule, each entry of $\Value(\mdp)$ is equal to 
$\det(B)/\det(A)$, where $B$ is a matrix obtained by replacing some column of 
$A$ with $d^3(1-\lambda)\vec{c}$ (which is again an integer vector, due to the multiplication by $ d^3 $). Hence, each entry of $\Value(\mdp)$ is a rational number with denominator $\det(A)$. Hadamard's inequality~\cite{Garling:2007} implies $|\det(A)|\leq d^{3|\vertices|}{|\vertices|}^{\frac{|\vertices|}{2}}$.

Now let $\sigma$ be any $\discOP^{k(\eps^*)}(\vec{0})$-safe MD strategy. By~\Cref{6-lem:disc-val-it-eps-strategies}, $\sigma$ is $\eps^*$-optimal. We prove that all actions used by $\sigma$ are $\vec{x}^*$-safe, which means that $\sigma$ is optimal by~\Cref{6-lem:disc-val-lower}. Assume that in some vertex $v$ the strategy $\sigma$ uses action $a$ that is not $\vec{x}^*$-safe. Denote $\vec{y}=\discOP^\sigma(\vec{x}^*)$. We have $ |\vec{y}_v - \vec{x}^*_v| > 0 $, since otherwise $a$ would be $\vec{x}^*$-safe. But then we can obtain a lower bound on the difference by investigating the bitsize of the numbers involved:
\begin{align*}
|\vec{y}_v - \vec{x}^*_v| &= \left|\frac{d^3}{d^3}\vec{y}_v - \frac{d^3}{d^3}\vec{x}^*_v\right|
\\
&=\frac{1}{d^3}\left|\sum_{u \in \vertices} 
\underbrace{d\cdot\probTranFunc(u \mid v,a)}_{\text{integer}} \cdot( \underbrace{d^2(1-\lambda)\cdot \coloring(u,v)}_{\text{integer}} + 
\underbrace{d^2\cdot\lambda \cdot \vec{x}^*_u ) - d^3\vec{x}^*}_{\text{int. multiples of $1/\det(A)$}}\right| \\
&\geq \frac{1}{d^3\cdot \det(A)}\geq \frac{1}{d^{(3|\vertices|+3)}\cdot{|\vertices|}^{\frac{|\vertices|}{2}}}.
\end{align*}

\noindent
Now put $\vec{z}=\discOP^\sigma(\discOP^{k(\eps)}(\vec{0}))$. We have the following (using, on the first line, the fact that $|a+b| \geq |a|-|b|$):
\begin{align*}
	|\vec{z}_v - \vec{x}^*_v| &=
 |\vec{z}_v-\discOP^\sigma(\vec{x}^*)_v+\discOP^\sigma(\vec{x}^*)_v-\vec{x}^*_v | 
\geq |\discOP^\sigma(\vec{x}^*)_v-\vec{x}^*_v | - |\vec{z}_v-\discOP^\sigma(\vec{x}^*)_v |  \\
	&\geq \frac{1}{d^{(3|\vertices|+3)}\cdot |\vertices|^{\frac{|\vertices|}{2}}} - |\discOP^\sigma(\discOP^{k(\eps)}(\vec{0}))_v-\discOP^\sigma(\vec{x}^*)_v |\quad (\text{as shown above})\\
	&\geq \frac{1}{d^{(3|\vertices|+3)}\cdot |\vertices|^{\frac{|\vertices|}{2}}} - |\discOP^{\sigma}(\discOP^{k(\eps^*)}(\vec{0}) - \vec{x}^*)_v|  \quad (\text{since $\discOP^{\sigma}$ is linear})\\
	&\geq \eps^* - \lambda\cdot\|\discOP^{k(\eps^*)}(\vec{0}) - \vec{x}^* \|_{\infty}
	> \eps^* - \frac{\eps^*}{2}  \quad (\text{\Cref{6-lem:disc-val-it-convergence}})\\&=\frac{\eps^*}{2}.
\end{align*}

\noindent
In particular, it must hold that $\vec{z}_v< \vec{x}^*_v$. Otherwise we would have 
\[
\discOP^{k(\eps^*)+1}(\vec{0})_v \geq \discOP^{\sigma}(\discOP^{k(\eps^*)}(\vec{0}))_v \geq \vec{x^*}_v + \frac{\eps^*}{2},
\]
a contradiction with $\discOP^{k(\eps^*)+1}(\vec{0})$ being an $\frac{\eps^*}{4}$-ap\-prox\-imation of $\vec{x}^*$ (by \Cref{6-lem:disc-val-it-convergence} and the choice of $k(\eps^*)$).

At the same time, $|\discOP(\discOP^{k(\eps^*)}(\vec{0}))_v - \vec{x}^*|\leq \frac{\eps^*}{4}$, due to the choice of $k(\eps^*)$. Since $\vec{z}_v \leq \vec{x}_v^*$, we get $\vec{z}_v < \vec{x}_v^* - \frac{\eps}{2} \leq \discOP(\discOP^{k(\eps^*)}(\vec{0}))_v$, a contradiction with $\sigma$ being $\discOP^{k(\eps^*)}(\vec{0})$-safe. 
\end{proof}

\begin{corollary}[Complexity of discounted-payoff MDPs with a fixed discount factor]
\label{6-cor:VI-optimal-strategy-comp}
An optimal MD strategy in discounted-payoff MDPs with a fixed discount factor can be computed in polynomial time. 
\end{corollary}
\begin{proof}
The number $1/\eps^*$, where $\eps^*$ is from \Cref{6-lem:disc-vi-optimal-strategy}, has bitsize polynomial in the size of the MDP when the discount factor is fixed. Hence, the number $k(\eps^*)$ defined as in \Cref{6-lem:disc-val-it-eps-strategies} has a polynomial magnitude, so it suffices to perform only polynomially many iterations. Since each iteration requires polynomially many arithmetic operations that involve only summation and multiplication by a constant, the result follows.
\end{proof}

\subsection*{Strategy improvement, linear programming, and (strongly) polynomial time}

The strategy (or policy) improvement (also called strategy/policy iteration in the literature) for MDPs works similarly as for games, see \Cref{6-algo:disc-strategy-improvement}. In the algorithm, we use $\discOP_{a,v}(\vec{x})$ as a shortcut for $ \sum_{u \in \vertices}\probTranFunc(u\mid v,a)\left((1-\lambda)\cdot\coloring(v,u) + \lambda\cdot \vec{x}_u \right)$.
The computation of $\vec{x}^{\sigma_i}$ uses \Cref{6-lem:disc-val-sigma}.


\begin{algorithm}
	\KwData{A discounted-payoff MDP $ \mdp $}

	$i \leftarrow 0$\;
	$ \sigma_i \leftarrow \text{arbitrary MD strategy} $\;
	\Repeat{$ \sigma_{i} = \sigma_{i-1} $}{
	compute $ \vec{x}^{\sigma_i} = \left(\expv^{\sigma_i}_v[\DiscountedPayoff]\right)_{v\in \vertices} $\;
	\ForEach{$ v \in \vertices $}{
		$ \mathit{Improve}(v) \leftarrow \sigma_{i}(v) $\;
		\ForEach{$ a \in \actions $}{
			\lIf{$\discOP_{a,v}(\vec{x}^{\sigma_i}) >\discOP_{a,\mathit{Improve}(v)}(\vec{x}^{\sigma_i})$}{
				$\mathit{Improve}(v) \leftarrow a$
				}
			}
		$\sigma_{i+1}(v) \leftarrow \mathit{Improve}(v)$
		}
		$ i \leftarrow i+1 $
	}
\Return{$ \sigma_i $}
	
	
	\caption{An algorithm computing an optimal MD strategy in a discounted-payoff MDP}
	\label{6-algo:disc-strategy-improvement}	
\end{algorithm}

\begin{theorem}[Strategy improvement for discounted-payoff MDPs]
\label{6-thm:disc-strat-it}
The strategy improvement algorithm for discounted-payoff MDPs terminates in a finite (and at most exponential) number of steps and returns an optimal MD strategy.
\end{theorem}
\begin{proof}
First we need to show that whenever $\sigma_{i+1}\neq \sigma_i$, then  $\vec{x}^{\sigma_{i+1}} \geq \vec{x}^{\sigma_i}$ and $\vec{x}^{\sigma_{i+1}} \neq \vec{x}^{\sigma_i}$ (we write this by $\vec{x}^{\sigma_{i+1}} \succ\vec{x}^{\sigma_i}$). So fix some $ i $ s.t. an improvement is performed in the $ i $-th iteration of the repeat-loop. We have $\discOP_{\sigma_{i+1}}(\vec{x}^{\sigma_i})\succ\discOP_{\sigma_{i}}(\vec{x}^{\sigma_i})= \vec{x}^{\sigma_i}$, \textit{i.e.} $\discOP_{\sigma_{i+1}}(\vec{x}^{\sigma_i})-\vec{x}^{\sigma_i} \succ 0$. Let $P$, $\vec{c}$ be the matrix and vector such that the equation $\vec{x}=\discOP_{\sigma_{i+1}}(\vec{x})$ can be written as $\vec{x}= (1-\lambda)\cdot \vec{c}+\lambda \cdot P\cdot\vec{x}$. Since the equation $\vec{x}=\discOP_{\sigma_{i+1}}(\vec{x})$ has a unique fixed point (as $ \discOP_{\sigma_{i+1}} $ is a contraction), the matrix $ I-\lambda P $ is invertible. Then $\discOP_{\sigma_{i+1}}(\vec{x}^{\sigma_i})-\vec{x}^{\sigma_i} \succ 0$ can be written as  $(1-\lambda)\vec{c} + (\lambda P - I)\vec{x}^{\sigma_i} \succ 0 $, or equivalently $\vec{x}^{\sigma_i}\prec (1-\lambda)\vec{c}\cdot(I-\lambda P)^{-1}.$ But the RHS of this inequality is equal to the fixed point of $\discOP_{\sigma_{i+1}}$, \textit{i.e.} to $\vec{x}^{\sigma_{i+1}} .$

Now there are only finitely (exponentially) many MD strategies and since$\vec{x}^{\sigma_{i+1}} \succ\vec{x}^{\sigma_i}$, we have that no strategy is considered twice. Hence, the algorithm eventually terminates. Upon termination, there is no improving action, so the algorithm has found a strategy $\sigma$ whose value vector $\valsigma$ is the fixed point of $\discOP$. Such a strategy is optimal by \Cref{6-thm:disc-val-char-mem}. 
\end{proof}

While each of the steps (1.)--(4.) can be performed in polynomial time, the 
worst-case number of iterations is exponential~\cite{Hollanders.Delvenne.ea:2012}. However, the 
algorithm has nice properties in the case when the discount factor is fixed, as we'll see below. It is also intimately connected to the linear programming approach.

We can indeed aim to directly solve 
the equation $\vec{x} = \discOP(\vec{x})$, thus obtaining the fixed point of 
$\discOP$, by using a suitable LP. While the operator $\discOP$ is not 
in itself linear, solving the equation can be encoded into a linear  program. 
The main idea can be described as follows: given some numbers $y,z$, the 
solution $\bar{x}$ to the equation $x=\max\{y,z\}$ is exactly the smallest 
solution to the set of inequalities $x\geq y$, $x\geq z$. Hence, to solve the 
equation  $\vec{x} = \discOP(\vec{x})$, we set up the following linear program 
$\lpdisc$ with variables $x_v$, $v \in \vertices$.

\begin{equation*}
\begin{array}{llr}
\text{minimise }     & \sum_{v \in \vertices} x_v \\
\textrm{subject to } & x_v \geq \sum_{u \in \vertices} \probTranFunc(u \mid v,a) \cdot \left((1-\lambda) \cdot \coloring(v,u) + \lambda\cdot x_u \right) & \text{for all $v \in \vertices$ and $a \in \actions$.}
\end{array}
\label{6-eq:disc-lp}
\end{equation*}


\begin{lemma}[Properties of the linear program]
\label{6-lem:disc-lp}
The linear program $\lpdisc$ has a unique optimal solution 
$\bar{\vec{x}}$ that satisfies $\bar{\vec{x}} = \Value(\mdp)$.
\end{lemma}
\begin{proof}
Let $\vec{x}^* = \Value(\mdp)$ be the unique fixed point of $\discOP$. Clearly 
setting $x_v = \vec{x}^*_v$ yields a feasible solution of $\lpdisc$. 
We show 
that $\vec{x}^*$ is actually an optimal solution, by proving that for each 
feasible solution $\vec{x}$ it holds $\vec{x} \geq \vec{x}^*$. (This also 
shows 
the uniqueness, since it says that an optimal solution is the infimum of the 
set of 
all feasible solutions.) First, note that for any feasible solution $\vec{x}$ 
it holds $\discOP(\vec{x})\leq \vec{x}$, by the construction of $\lpdisc$. 
Next, if $\vec{x}$ is a feasible solution, then $\discOP(\vec{x})$ is also a 
feasible solution; otherwise, there would be some $v$ and $a\in \actions$ 
such that 
\begin{align*}\discOP(\vec{x})_v &< \sum_{u\in \vertices} \probTranFunc(u\mid 
v,a)\cdot\left((1-\lambda)\cdot \coloring(v,u) + \lambda\cdot 
\discOP(\vec{x})_u \right) \\ &\leq \sum_{u\in \vertices} \probTranFunc(u\mid 
v,a)\cdot\left((1-\lambda)\cdot \coloring(v,u) + \lambda\cdot \vec{x}_u 
\right) \leq \discOP(\vec{x})_v.
\end{align*}
Here, the first inequality on the second line follows from 
$\discOP(\vec{x})\leq \vec{x}$, while the second inequality follows from the 
definition of $\discOP$. But $\discOP(\vec{x})_v < \discOP(\vec{x})_v$ is an 
obvious contradiction. So $\discOP(\vec{x})$ is indeed a feasible solution and 
by applying the argument above, we get $\discOP^2(\vec{x}) \leq 
\discOP(\vec{x})$. By a simple induction, $\discOP^{i+1}(\vec{x})\leq 
\discOP^{i}(\vec{x})\leq \vec{x}$ for each $i\geq 0.$ Hence, also $\vec{x}^* = 
\lim_{i\rightarrow \infty} \discOP^i(\vec{x}) \leq \vec{x}$ (the first equality 
comes from \Cref{6-lem:fixpoint}).
\end{proof}

Linear programming can be solved in polynomial time (\Cref{1-thm:linear_programming}). Hence, we get the following.

\begin{theorem}[Properties of discounted-payoff MDPs]
\label{6-thm:disc-polytime-lp}
The following holds for discounted-payoff MDPs:
\begin{enumerate}
\item The value of each vertex as well as an MD optimal 
strategy can be computed in polynomial time. 
\item The problem whether the value of a given vertex $v$ is at least a given constant 
(say~1) is $\P$-complete (under logspace reductions). The hardness result holds 
even for a fixed discount factor.
\end{enumerate}
\end{theorem}
\begin{proof}(1.)
The first part comes directly from~\Cref{6-lem:disc-lp}. Once the optimal value 
vector $\Value(\mdp)$ is computed, we can choose any $\Value(\mdp)$-safe MD 
strategy as the optimal one 
(\Cref{6-lem:disc-val-lower}).

(2.) Let $\lambda$ be the fixed discount factor. We show the lower 
bound, by extending the reduction 
from the CVP problem used for almost-sure reachability. First, we 
transform the input circuit into an MDP in the same way as in the reachability 
case, and we let $v$ be the vertex corresponding to a gate we wish to evaluate. 
Assume, for a while, that each path from $v$ to a target state has the same 
length $\ell$. Then we simply assign reward 
$\frac{1}{(1-\lambda)\cdot\lambda^{\ell -1}}$ to each edge 
entering a target state, and $0$ to all other edges. It is easy to check that 
the value of $v$ in the resulting discounted-payoff MDP is equal to the value of $v$ 
in the reachability MDP. If the reachability MDP $\mdp$ does not have the 
`uniform 
path length' property, we modify it by producing $|\vertices|$ copies of 
itself so that each new vertex carries, apart from the original name, an index 
from $\{1,\dots,n\}$. The transition function in the new MDP mimics the 
original one, but from vertices with index $j<n$ we transition to the 
appropriate vertices of index $j+1$. The target vertices in the new MDP are 
those vertices of index $n$ that correspond to a target vertex of the 
original MDP (this does not break down the reduction, as target vertices in the original vertices can be assumed to have no outgoing edges other than self loops). This new MDP has the desired property and can be produced in a 
logarithmic space.
\end{proof}

The previous theorem shows that discounted-payoff MDPs can be solved in 
polynomial time even if the discount factor is not fixed (\textit{i.e.}, it is taken as 
a part of the input). This is an important difference from the two-player 
setting. However, the proof, resting on polynomial-time solvability of linear 
programming, leaves opened a question whether the discounted-payoff MDPs be solved in strongly polynomial time.  An answer was provided by Ye~\cite{Ye:2011}: already the classic simplex 
algorithm of Dantzig solves $\lpdisc$ in a strongly 
polynomial time in MDPs with a fixed discount factor. Formally, Ye proved that 
the number of iterations taken by the simplex method is bounded by 
$\frac{|\vertices|^2\cdot (|\actions|-1)}{1-\lambda}\cdot 
\log(\frac{|\vertices|^2}{1-\lambda})$, with each iteration requiring  
$\mathcal{O}(|\vertices|^2\cdot |\actions|)$ arithmetic operations. This has 
also an impact on the strategy improvement method: it can be shown that strategy 
improvement in discounted-payoff MDPs is really just a `re-implementation' of the 
simplex algorithm using a different syntax. Hence, the strongly polynomial 
complexity bound for a fixed discount factor holds there as well.

\begin{theorem}[Discounted-payoff MDPs with a fixed discount factor]
\label{6-thm:discounted_payoff_MDP_fixed_discount_factor}
For MDPs with a fixed discount factor, the value of each vertex as well as an 
optimal MD strategy can be computed in a strongly polynomial time.
\end{theorem}

\section{Mean payoff in MDPs: General properties and linear programming}
\label{6-sec:mean_payoff_properties}
We will use the $\liminf$ variant of mean payoff:

\[
\MeanPayoffInf(\play) = \liminf_n \frac{1}{n} \sum_{i = 0}^{n-1} c(\play_i)
\]

In particular, the play-based expected mean payoff achieved by $ \sigma $ from $ v $ is defined as $ \playPay(v, \sigma) = \expv_v^\sigma[\MeanPayoffInf] $, while for the step-based counterpart we have 
\[
\stepPay(v, \sigma) = \liminf_{n \rightarrow \infty} \frac{1}{n} \sum_{i=0}^{n-1} \expv_v^\sigma[\coloring(\play_i)].
\]

The use of $ \liminf $ is natural in the formal verification setting: taking the $\liminf$ rather than $ \limsup $ emphasises the worst-case behaviour along a play. However, all the results in this section hold also for $\limsup$-based mean payoff, though some proofs are more complex. See the bibliographic remarks for more details.
%

In general, $\stepPay(v,\sigma)$ can be different from $ \playPay(v, \sigma) $. However, we have the following simple consequence of the dominated convergence theorem:
\begin{lemma}
	\label{6-lem:limit-defined}
Let $U$ be the set of all plays $\play$  for which $\lim_{n\rightarrow \infty} \frac{1}{n} \sum_{i=0}^{n-1}[\coloring(\play_i)]$ is undefined. If $\vinit,\sigma$ are such that $\probm^\sigma_{\vinit}(U) = 0$, then $\stepPay(\vinit,\sigma) = \playPay(\vinit, \sigma) $.
\end{lemma}

In particular, for any finite-memory strategy $ \sigma $, the two values coincide, since applying such a strategy turns an MDP into a finite Markov chain where the existence of the limit can be inferred using decomposition into strongly connected components and applying the Ergodic theorem.
We will show that in our case of finite-state MDPs, the two approaches coincide not only at the levels of finite-memory strategies, but also as optimality criteria: that is, no matter which of the two semantics we use, the optimal values are the same and a strategy that is optimal w.r.t. one of the semantics is also optimal for the other one. We caution the reader that such a result does not automatically transfer to more complex settings, such as infinite-state MDPs or multi-objective optimisation.

To simplify the subsequent notation, we define the \emph{expected one-step} reward of a vertex-action pair $(v,a)$ to be the number $\sum_{w\in \vertices} \probTranFunc(w\mid v,a)\cdot \coloring(v,w)$. Overloading the notation, we denote this quantity by $\coloring(v,a)$.

In mean-payoff MDPs, a crucial role is played by the linear program $\lpmp$ defined in~\Cref{6-eq:mp-lin}.
The variables are $x_{(v,a)}$ for $v \in \vertices$  and $a \in \actions$.


\begin{equation}
\label{6-eq:mp-lin}
\begin{aligned}
\text{maximise }     & \sum_{v \in \vertices, a \in \actions} x_{(v,a)} \cdot \coloring(v,a) \\
\textrm{subject to } & \sum_{u \in \vertices, a \in \actions} x_{(u,a)} \cdot \probTranFunc(v\mid u,a) = \sum_{a \in \actions} x_{v,a} & \text{for all $v \in \vertices$} \\
                     & \sum_{v \in \vertices, a \in \actions} x_{v,a} = 1 & & \text{for all $v \in \vertices, a \in \actions$} \\
                     & x_{v,a} \geq 0 & \text{for all $v \in \vertices, a \in \actions$} 
\end{aligned}
\end{equation}

Let us name the equations:

\begin{equation}
\begin{array}{lr}
\sum_{u \in \vertices, a \in \actions} x_{(u,a)} \cdot \probTranFunc(v\mid u,a) = \sum_{a \in \actions} x_{v,a} & \text{for all $v \in \vertices$}
\end{array}
\label{6-eq:mdp-flow}
\end{equation}

\begin{equation}
\begin{array}{lr}
\sum_{v \in \vertices, a \in \actions} x_{v,a} = 1 & \text{for all $v \in \vertices, a \in \actions$}
\end{array}
\label{6-eq:mdp-freq-1}
\end{equation}

\begin{equation}
\begin{array}{lr}
x_{v,a} \geq 0 & \text{for all $v \in \vertices, a \in \actions$} 
\end{array}
\label{6-eq:mdp-freq-nonnegp}
\end{equation}


There is a correspondence between feasible solutions of $\lpmp$ and the strategies in the corresponding MDP. Intuitively, in a solution corresponding to a strategy $\sigma$, the value of a variable $x_{v,a}$ describes the expected frequency with which $\sigma$ visits $v$ and plays $a$ upon such a visit. These frequencies must obey the \emph{flow constraints}~\Cref{6-eq:mdp-flow} and must represent a probability distribution, which is ensured by~\Cref{6-eq:mdp-freq-1} and~\Cref{6-eq:mdp-freq-nonnegp}. The objective function then represents the expected mean payoff. Of high importance is also the linear program which is \emph{dual} to $\lpmp$. This dual program, denoted $\lpmpdual$, is defined in~\Cref{6-eq:mp-dual}. (For a refresher on linear programming and duality, see~\Cref{1-sec:computation}).

\begin{equation*}
\begin{array}{llr}
\text{minimise }     & g \\
\textrm{subject to } & g - \coloring(v,a) + \sum_{u \in \vertices} \probTranFunc(u \mid v,a) \cdot y_u \geq y_v & \text{for all $v \in \vertices, a \in \actions$}
\end{array}
\label{6-eq:mp-dual}
\end{equation*}


A feasible solution of $ \lpmp $ is a vector $\vec{x} \in \R^{\vertices\times \actions} $ s.t. setting $ x_{(v,a)}=\vec{x}_{(v,a)} $ for all $ (v,a) $ satisfies the constraints in $ \lpmp $. A feasible solution of $ \lpmpdual  $ is a tuple $ (g,\vec{y}) $, where $ g\in\R $ (using the same notation for the number and the variable should not cause  confusion here) and $ \vec{y}\in \R^{\vertices}$ s.t. setting the corresponding variables to numbers prescribed by $ g $ and $ \vec{y} $ satisfies the constraints.

The variable $g$ in $\lpmpdual$ is often called \emph{gain} while the vector of $y$-variables is called \emph{bias.} This is because it provides information on how much does the payoff (under some strategy) accumulated up to a certain step deviate from the estimate provided by the mean-payoff value of that strategy. This is illustrated in the following lemma, which forms the first step of our analysis. 

\begin{lemma}[Properties of feasible solutions]
\label{6-lem:dual-bound-step}
Let $({g}, \vec{y})$ be a feasible solution of $\lpmpdual$ and let $Y^{(i)}$, where $i\geq 0$, be a random variable such that $Y^{(i)}(\play)= \vec{y}_{\ing(\pi_i)}.$ Then for each strategy $\sigma$, each vertex $\vinit$, and each $n\geq 0$ it holds $\expv^\sigma_{\vinit}[\sum_{i=0}^{n-1}\coloring(\play_i)]\leq n\cdot {g}- \vec{y}_{\vinit} +\expv^\sigma_{\vinit} [Y^{(n)}]$.	
\end{lemma}
\begin{proof}
By induction on $n$. For $n=0$, both sides are equal to 0. Now assume that the inequality holds for some $n\geq 0$. By the induction hypothesis
\begin{align}
\expv^\sigma_{\vinit}[\sum_{i=0}^{n}\coloring(\play_i)] 
= \expv^\sigma_{\vinit}[\sum_{i=0}^{n-1}\coloring(\play_i)] + \expv^\sigma_{\vinit}[\coloring(\play_n)] \leq n{g} - \vec{y}_{\vinit} +\expv^\sigma_{\vinit} [Y^{(n)}] + \expv^\sigma_{\vinit}[\coloring(\play_n)] \label{6-eq:mpdual-1}
\end{align}

\begingroup
\allowdisplaybreaks
\noindent
We now obtain a bound for the third term on the RHS of~\Cref{6-eq:mpdual-1}. In the following, we denote by $\Pi_n$ the set of all plays of length $n$. Then we have
\begin{align*}
\expv^\sigma_v [Y^{(n)}]&= \sum_{v\in\vertices}  \vec{y}_v\cdot\probm^\sigma_{\vinit}(\ing(\pi_n)=v)
=\sum_{v\in\vertices}  \vec{y}_v\cdot \bigg(\sum_{\substack{\play'\in \Pi_n \\ \last(\play')=v}}\probm^\sigma_{\vinit}(\play') \bigg) \\
&=\sum_{v\in\vertices}  \vec{y}_v\cdot \bigg(\sum_{\substack{\play'\in \Pi_n \\ \last(\play')=v}}\probm^\sigma_{\vinit}(\play')\cdot\big(\underbrace{\sum_{a\in\actions}\sigma(a\mid \play')}_{=1} \big)\bigg) \\
&= \sum_{\substack{v\in\vertices \\ a \in \actions}}  \vec{y}_v\cdot \bigg(\sum_{\substack{\play'\in \Pi_n\\ \last(\play')=v}}\probm^\sigma_{\vinit}(\play')\cdot\sigma(a\mid \play')\bigg)\\
& {\leq}\sum_{\substack{v\in\vertices \\ a\in \actions}} \bigg( {g} -\coloring(v,a) + \sum_{u\in\vertices}\probTranFunc(u\mid v,a)\cdot  \vec{y}_u\bigg)\cdot\bigg(\sum_{\substack{\play'\in \Pi_n \\ \last(\play')=v}}\probm^\sigma_{\vinit}(\play')\cdot\sigma(a\mid \play')\bigg)\\
&= {g}\cdot\underbrace{\sum_{\substack{v\in\vertices \\ a\in \actions}}\sum_{\substack{\play'\in \Pi_n \\ \last(\play')=v}}\probm^\sigma_{\vinit}(\play')\cdot\sigma(a\mid \play')}_{=1}\\&\quad-\underbrace{\sum_{\substack{v\in\vertices \\ a\in \actions}}\sum_{\substack{\play'\in \Pi_n \\ \last(\play')=v}}\probm^\sigma_{\vinit}(\play')\cdot\sigma(a\mid \play')\cdot\coloring(v,a)}_{=\expv^\sigma_{\vinit}[\coloring(\play_n)]}\\
&\quad +\underbrace{\sum_{\substack{v,u\in \vertices \\ a \in \actions}}\sum_{\substack{\play'\in \Pi_n \\ \last(\play')=v}} \probm^\sigma_{\vinit}(\play')\cdot\sigma(a\mid \play')\cdot \probTranFunc(u\mid v,a)\cdot  \vec{y}_u}_{=\expv^\sigma_{\vinit}[Y^{(n+1)}]}.
\end{align*}
\endgroup
The first inequality follows from~\Cref{6-eq:mp-dual}.

Plugging this into~\Cref{6-eq:mpdual-1} yields the desired $\expv^\sigma_{\vinit}[\sum_{i=0}^{n}\coloring(\play_i)] \leq (n+1) {g}- \vec{y}_{\vinit} + \expv^\sigma_{\vinit}[Y^{(n+1)}]$. 
\end{proof}

\begin{corollary}[Feasible solutions imply upper bounds]
\label{6-cor:mp-value-bound}
Let $g$ be the objective value of some feasible solution of $\lpmpdual$. Then for every strategy $\sigma$ and every vertex $\vinit$ it holds $\playPay(\vinit,\sigma) \leq \stepPay(\vinit,\sigma) \leq g$.
\end{corollary}
\begin{proof}
Let $ ({g}, \vec{y})$ be any feasible solution of $\lpmpdual$.
By \Cref{6-lem:dual-bound-step} we have, for every $n\geq 0$, that $\expv^\sigma_{\vinit}[\frac{1}{n}\sum_{i=0}^{n-1}\coloring(\play_i)]\leq g - \frac{ \vec{y}_{\vinit}}{n}+ \frac{1}{n}\expv^\sigma_{\vinit} [Y^{(n)}]$. Since $ \vec{y}_{\vinit}$ is a constant and $|\expv^\sigma_{\vinit} [Y^{(n)}]|$ is bounded by the constant $ \max_{v\in \vertices}|\vec{y}_v|$ that is independent of $n$, the last two terms on the RHS vanish as $n$ goes to $\infty$. 
Hence, we also have that $\stepPay(\vinit,\sigma) = \liminf_{n\rightarrow \infty} \expv^\sigma_{\vinit}[\frac{1}{n}\sum_{i=0}^{n-1}\coloring(\play_i)] \leq g$. It remains to show that we have $\playPay(\vinit,\sigma) \leq \stepPay(\vinit,\sigma)$, but this immediately follows from the Fatou's lemma~\cite[Theorem 1.6.8]{Ash.Doleans-Dade:2000}.
\end{proof}

\begin{corollary}[Existence of optimal solutions for the linear programs]
\label{6-cor:lpmp-optimal-exists}
Both the linear programs $\lpmp$ and $\lpmpdual$ have a feasible solution. Hence, both have an optimal solution and the optimal values of the objective functions in these programs are equal.
\end{corollary}
\begin{proof}
One can easily construct a feasible solution for $\lpmpdual$ by setting all the $y$-variables to $0$ and $g$ to $\maxc$. By the duality theorem for linear programming, to show that also $\lpmp$ has feasible solution it suffices to show that the objective function of $\lpmpdual$ is bounded from below. But this follows from \Cref{6-cor:mp-value-bound}, since there is at least one strategy $\sigma$ giving us the lower bound (in particular, the objective function is bounded from below by $-\maxc$). The second part follows immediately by linear programming duality.
\end{proof}



\section{Mean payoff optimality in strongly connected MDPs}
\label{6-sec:mean_payoff_strongly_connected}
As shown in the previous section, the optimal solution of any of the programs $\lpmp$, $\lpmpdual$ gives us an upper bound on the optimal value. In this sub-section we show that in strongly connected MDPs: a) a value of every vertex is the same; b) from a solution of $\lpmp$ one can extract a positional deterministic strategy $\sigma$ whose expected mean payoff is well defined (\textit{i.e.}, the preconditions of \Cref{6-lem:limit-defined} are satisfied)) and equal to the objective value of the solution. Moreover, if the solution in question is optimal, then $ \sigma $ is optimal for both $\playPay$- and $\stepPay$-semantics.

\begin{definition}[Strongly connected MDPs]
\label{6-def:scc-mdp}
An MDP is \emph{strongly connected} if for each pair of vertices $u,v$ there exists a strategy which, when starting in $u$, reaches $v$ with a positive probability. 
\end{definition}

For the rest of this section we fix an optimal solution $\lpsol{x}_{v,a}$ of $\lpmp$. We denote by $\solvset$ the set of all vertices for which there exists action $a$ s.t. $\lpsol{x}_{v,a}>0.$ From the shape of $\lpmp$ it follows that $\solvset$ is non-empty and closed, and hence we can consider a sub-MDP $\mdp_{\solvset}$ induced by $\solvset$. In $\mdp_{\solvset}$ we then define a positional randomised strategy $\sigma$ by putting $$\sigma(a\mid v)=\frac{\lpsol{x}_{(v,a)}}{\sum_{b\in \actions}\lpsol{x}_{(v,b)}}.$$

Fixing a strategy $\sigma$ yields a \emph{Markov chain} $\mdp_\solvset^{\sigma}$. Markov chain can be viewed as an MDP with a single action (and hence, with no non-determinism). $\mdp_{\solvset}^\sigma$ in particular can be viewed an MDP with the same vertices, edges, and colouring as $\mdp_\solvset$, but with a single action (as non-determinism was already resolved by $\sigma$). The probability of transitioning from a vertex $u$ to a vertex $v$ in a Markov chain is denoted by $\mcprob_{u,v}$. In $\mdp_{\solvset}^{\sigma}$ we have $\mcprob_{u,v}=\sum_{a\in \actions} \probTranFunc(v\mid u,a)\cdot\sigma(a\mid u)$, the right-hand side being computed in the original MDP $\mdp$. Both $\mdp_\solvset$ and $\mdp_{\solvset}^{\sigma}$ have the same sets of plays and for each initial vertex, the probability measure induced by $\sigma$ in $\mdp$ equals the probability measure arising (under the unique policy) in $\mdp_{\solvset}^{\sigma}$. Hence, to prove anything about $\sigma$ it suffices to analyse $\mdp_{\solvset}^{\sigma}$.

\paragraph{A refresher on Markov chains.} We review some fundamental notions of Markov chain theory~\cite{Norris:1998}. A Markov chain that is strongly connected is called \emph{irreducible}. The one-step transition probabilities in a Markov chain can be arranged into a square matrix $\mcprob$, which has one row and one column for each vertex. The cell in the row corresponding to a vertex $u$ and in the column corresponding to a vertex $v$ bears the value $\mcprob_{u,v}$ defined above. An easy induction shows that the matrix $\mcprob^k$ contains $k$-step transition probabilities. That is, the probability of being in $v$ after $k$ steps from vertex $u$ is equal to the $(u,v)$-cell of $\mcprob^k$, which we denote by $\mcprob^{(k)}_{u,v}$.

A vertex $u$ of a Markov chain is \emph{recurrent} if, when starting from $u$, it is revisited infinitely often with probability $1$. On the other hand, if the probability that $u$ is re-visited only finitely often is one, then the vertex is \emph{transient}. It is known~\cite[Theorem 1.5.3]{Norris:1998} that each vertex of a finite Markov chain is either recurrent or transient, and that these two properties can be equivalently characterised as follows: vertex $u$ is recurrent if and only if  $\sum_{k=0}^{\infty}\mcprob^{(k)}_{u,u}=\infty$, otherwise it is transient.

An \emph{invariant distribution} in a Markov chain with a vertex set $\vertices$ is a $|\vertices|$-dimensional non-negative row vector $\invdist$ which adds up to $1$ and satisfies $ \invdist\cdot \mcprob = \invdist$.

The following lemma holds for arbitrary finite Markov chains.

\begin{lemma}[Invariant distributions witness recurrent states]
\label{6-lem:MC-inv-rec}
Let $\invdist$ be an invariant distribution and $v$ a vertex such that $\invdist_v > 0$. Then $v$ is recurrent.	
\end{lemma}
\begin{proof}
Let $n$ be the number of vertices in the chain and $p_{\min}$ the minimum non-zero entry of $\mcprob$.
Assume, for the sake of contradiction, that $v$ is transient. We show that in such a case, for each vertex $u$ it holds $\lim_{k\rightarrow\infty} \mcprob^{(k)}_{u,v} = 0$. For $u=v$ this is immediate, since the sum $\sum_{k=0}^{\infty}\mcprob^{(k)}_{v,v}$ converges for  transient $v$. Otherwise, let $f_{u,v,i}$ be the probability that a play starting in $u$ visits $v$ for the \emph{first time} in exactly $i$ steps. Then $\mcprob^{(k)}_{u,v}=\sum_{i=0}^k f_{u,v,i}\cdot \mcprob^{(k-i)}_{v,v}$. Now when starting in a vertex from which $v$ is reachable with a positive probability, at least one of the following events happens with probability $\geq p_{\min}^n$ in the first $n$ steps: either we reach a vertex from which $v$ is not reachable with positive probability, or we reach $v$. If neither of the events happens, we are, after $n$ steps, still in a vertex from which $v$ can be reached with a positive probability. In such a case, the argument can be inductively repeated (analogously to the proof of \Cref{6-thm:as-char}) to show that $f_{u,v,i}\leq (1-p_{\min}^n)^{\lfloor\frac{i}{n}\rfloor}\leq (1-p_{\min}^n)^{\frac{i-n}{n}}$.

Since $\sum_{k=0}^{\infty}\mcprob^{(k)}_{v,v}$ converges, for each $\eps>0$ there exists $j_\eps$ such that $\sum_{i=j_{\eps}}^{\infty}\mcprob^{(i)}_{v,v} < \frac{\eps}{2}$. Similarly, there exists $\ell_\eps$ such that $$\sum_{i=\ell_{\eps}}^{\infty}{(1-p_{\min}^n)^{\frac{i-n}{n}}} = \frac{(1-p_{\min}^n)^{\frac{\ell_\eps}{n}}}{\left(1-(1-p_{\min}^n)^{\frac{1}{n}}\right)\cdot(1-p_{\min}^n)}< \frac{\eps}{2},$$ and hence $\sum_{i=\ell_{\eps}}^{\infty} f_{u,v,i}< \frac{\eps}{2}.$

Now we put $m_{\eps}=\max\{j_\eps,\ell_\eps\}$. For any $k\geq 2m_{\eps}$ we have $\mcprob^{(k)}_{u,v}=\sum_{i=0}^k f_{u,v,i}\cdot \mcprob^{(k-i)}_{v,v} \leq \sum_{i=m_{\eps}}^{k}f_{u,v,i} + \sum_{i=0}^{m_{\eps}}\mcprob^{(k-i)}_{v,v}\leq\sum_{i=m_{\eps}}^{k}f_{u,v,i} + \sum_{i=m_{\eps}}^{k}\mcprob^{(i)}_{v,v}<\eps$ (note that all the series involved are non-negative). This proves that $\mcprob^{(k)}_{u,v}$ vanishes in the limit.

Finally, we derive the contradiction. Since $\invdist$ satisfies $\invdist\cdot \mcprob = \invdist$, we also have $\invdist\cdot \mcprob^k = \invdist$ for all $k$. Hence, the $v$-component of $\invdist\cdot \mcprob^k$ is equal to $\invdist_v>0$. But as shown above, the $v$-column of $\mcprob^k$ converges to the all-zero vector as $k\rightarrow \infty$, so also $(\invdist\cdot \mcprob^k)_v$ vanishes in the limit, a contradiction.
\end{proof}

\noindent
\paragraph{Towards the optimality of $ \sigma $.} We now turn back to the chain $\mdp_{\solvset}^{\sigma}$, where the positional strategy $ \sigma $ is obtained from the optimal solution of $ \lpmp $. In general, $ \mdp_{\solvset}^{\sigma} $ does not have to be irreducible. Hence, we use the following lemma and its corollary to extract an irreducible sub-chain, to which we can apply known results of Markov chain theory.

\begin{lemma}[Invariant distributions witness recurrent MDPs]
\label{6-lem:mc-rec}
Let $\bar{\invdist}$ be a vector such that for each $v\in \solvset$ it holds $\bar{\invdist}_v=\sum_{a\in \actions} \lpsol{x}_{v,a}$. Then $\bar{\invdist}$ is an invariant distribution of $\mdp_{\solvset}^{\sigma}$. Consequently, all vertices of $\mdp_{\solvset}^{\sigma}$ are recurrent.
\end{lemma}
\begin{proof}
The first part follows directly from the fact that $\lpsol{x}_{v,a}$ is a feasible solution of $\lpmp$. The second part follows from~\Cref{6-lem:MC-inv-rec} and from the fact that $\bar\invdist$ is positive (by the definition of $\solvset$).
\end{proof}

\begin{corollary}[Extraction of strongly connected components]
\label{6-cor:mp-scc-extraction}
The set $\solvset$ can be partitioned into subsets $\solvset_1,\solvset_2,\dots,\solvset_m$ such that each $\solvset_i$ induces a strongly connected sub-chain of $\mdp_{\solvset}^{\sigma}$.
\end{corollary}
\begin{proof}
Let $v\in\solvset$ be arbitrary and let $U_v\subseteq \solvset$ be the set of all vertices reachable with positive probability from $v$ in $\mdp_{\solvset}^{\sigma}$. Then $v$ is reachable (in $\mdp_{\solvset}^{\sigma}$) with positive probability from each $u\in U_v$: otherwise, there would be a positive probability of never revisiting $v$, a contradiction with each vertex being recurrent in $\mdp_{\solvset}^{\sigma}$ (\Cref{6-lem:mc-rec}). Hence, $U_v$ induces a strongly connected `sub-MDP' (or sub-chain) of $\mdp_{\solvset}^{\sigma}$. It is easy to show that if $U_v \neq U_w$ for some $v\neq w $, then the two sets must be disjoint.
\end{proof}

Hence, we can extract from $\solvset$ a set $Q$ inducing a strongly-connected sub-chain of $\mdp_{\solvset}^{\sigma}$, which we denote $\mdp^{\sigma}_{Q}$. The set $Q$ also induces a strongly connected sub-MDP of $\mdp$ denoted by $\mdp_Q$. The chain $\mdp^{\sigma}_{Q}$ arises by fixing, in $\mdp_Q$, a strategy formed by a restriction of $\sigma$ to $Q$. We use the following powerful theorem to analyse $\mdp^{\sigma}_{Q}$.

\begin{theorem}[Ergodic theorem]
\label{6-thm:ergodic} In a strongly connected Markov chain (with a finite set of vertices $\vertices$) there exists a unique invariant distribution $\invdist$. Moreover, for every vector $\vec{h}\in \R^{\vertices}$ the following equation holds with probability 1:
\[
\lim_{n \rightarrow \infty} \frac{1}{n} \sum_{i=0}^{n-1}\vec{h}_{\ing(\pi_i)} = \sum_{v\in\vertices} \invdist_v\cdot \vec{h}_v.
\]
(In particular, the limit is well-defined with probability 1).
\end{theorem}

We refer to Theorem~1.10.2 in~\cite{Norris:1998} for a proof of the Ergodic theorem.

We can use the Ergodic theorem to shows that the expected mean payoff achieved by $\sigma$ in $\mdp_{Q}$ matches the optimal value of $ \lpmp $, in a very strong sense: the probability of a play having a mean payoff equal to this optimal value is 1 under $ \sigma $.

\begin{theorem}
\label{6-cor:mp-scc-optimality}
Let $\sigma_Q$ be the restriction of $\sigma$ to $Q$. Then for every $v\in Q$ it holds that $\probm^{\sigma_Q}_{\mdp_Q,v}(\MeanPayoffInf = r^*)=1$, where $r^*$ is the is the optimal value of $\lpmp$. 
\end{theorem}
\begin{proof}
Let $ \vec{w}\in\R^{\vertices \times\actions} $ be a vector such that $\vec{w}_{(v,a)}=\lpsol{x}_{(v,a)}/\sum_{(q,a)\in Q\times\actions} \lpsol{x}_{(q,a)}$ for every $(v,a)\in Q\times \actions$, and $\vec{w}_{(v,a)}=0$ for all other $(v,a)$. We claim that $ \vec{w} $ is also an optimal solution of $\lpmp$. 

To prove feasibility, note that setting $\vec{w}_{(v,a)}=0$ for each $v\in \vertices\setminus Q$ does not break the constraints~\Cref{6-eq:mdp-flow}. This is because $Q$ induces a strongly connected sub-chain of $\mdp_{\solvset}^{\sigma}$, and hence there are no $v\in \vertices$, $u\in \vertices\setminus Q$ such that $\lpsol{x}_{(u,a)}\cdot \probTranFunc(v\mid u,a)>0$. Next,~\Cref{6-eq:mdp-flow} is invariant w.r.t. multiplication of variables by a constant, so normalising the remaining values preserves~\Cref{6-eq:mdp-flow} and ensures that~\Cref{6-eq:mdp-freq-1} holds. 

To prove optimality, assume that the objective value of $\vec{w}$ is smaller than $r^*$. Then we can mirror the construction from the previous paragraph and produce a feasible solution ${\hat{\vec{w}}_{(v,a)}}$ whose $(Q\times\actions)$-indexed components are zero and the rest are normalised components of $\lpsol{x}$. Then $r^*$ is a convex combination of the objective values of $\vec{w}$ and $\hat{\vec{w}}$, so $\hat{\vec{w}}$ must have a strictly larger value than $r^*$, a contradiction with the latter's optimality.

We now plug $ \vec{w} $ into the ergodic theorem as follows: As in~\Cref{6-lem:mc-rec}, it easy to prove that setting $\invdist_v=\sum_{a\in\actions}\vec{w}_{(v,a)}$ yields an invariant distribution. Now put $\vec{h}_v=\sum_{a\in\actions}\sigma(a\mid v)\cdot \coloring(v,a) (=  \sum_{w \in \vertices} \mcprob_{v,w}\cdot \coloring(v,w))$. 
From the Ergodic theorem we get that $\lim_{n \rightarrow \infty} \frac{1}{n} \sum_{i=0}^{n-1}\vec{h}_{\ing(\pi_i)}$ almost-surely exists and equals 
\begin{align}
\sum_{v\in Q} \invdist_v\cdot \vec{h}_v &= \sum_{v\in \vertices} \left(\big(\sum_{d\in\actions}\vec{w}_{(v,d)}\big)\cdot \big(\sum_{a\in\actions}\sigma(a\mid v)\cdot \coloring(v,a) \big)\right) \nonumber\\
&= \sum_{v\in Q} \left(\Bigg( \frac{\sum_{d\in\actions}\lpsol{x}_{(v,d)}}{\sum\limits_{\substack{q\in Q\\ b\in \actions}} \lpsol{x}_{(q,b)}} \Bigg)\cdot \Bigg( \frac{\sum_{a\in\actions}\lpsol{x}_{(v,a)}\cdot\coloring(v,a)}{\sum\limits_{d\in  \actions} \lpsol{x}_{(v,d)}} \Bigg) \right) \nonumber\\
&= \frac{1}{\sum\limits_{\substack{q\in Q\\ b\in \actions}} \lpsol{x}_{(q,b)}}\cdot\sum\limits_{\substack{v\in Q\\ a\in\actions}} \lpsol{x}_{(v,a)}\cdot \coloring(v,a) = \sum\limits_{\substack{v\in Q\\ a\in\actions}} \vec{w}_{(v,a)}\cdot\coloring(v,a) =r^*.\label{6-eq:ergodic-use}
\end{align}

It remains to take a step from the left-hand side of~\Cref{6-eq:ergodic-use} towards the mean payoff. To this end, we construct a new Markov chain $\mdp_Q'$ from $\mdp_Q$ by `splitting' every edge $(u,v)$ with a new dummy vertex $d_{u,v}$ (\textit{i.e.}, $d_{u,v}$ has one edge incoming from $u$ with probability $\mcprob_{u,v}$ and one edge outgoing to $v$ with probability $1$). In $\mdp_Q'$ we define a vector $\vec{h}'$ s.t. for each vertex $d_{u,v}$ the vector $ \vec{h}' $ has the $ d_{u,v} $-component equal to $\coloring(u,v)$, while the components corresponding to the original vertices are zero. It is easy to check that $\mdp_Q'$  is strongly connected and that it has an invariant distribution $\invdist'$ defined by $\invdist'_v=\invdist_v/2$ for $v$ in $Q$ and $\invdist'_{d_{u,v}}=\frac{\invdist_u\cdot\mcprob_{u,v}}{2}$ for $(u,v)$ an edge of $\mdp_Q$.
Also, by easy induction, for each play $\play$ of length $n$ in $\mdp_Q$ it holds $\frac{1}{n}\sum_{i=0}^{n-1}\coloring(\play_i) = \frac{1}{n}\sum_{i=0}^{2n-1}\vec{h}'_{\ing(\play_i')}$, where $\play'$ is the unique play in $\mdp_Q'$ obtained from $\play$ by splitting edges with appropriate dummy vertices. Hence, 
\begin{equation}
\label{6-eq:mc-opt-limit}
\lim_{n\rightarrow \infty}\frac{1}{n}\sum_{i=0}^{n-1}\coloring(\play_i) = 2\cdot \lim_{n\rightarrow \infty}\frac{1}{n}\sum_{i=0}^{n-1}\vec{h}'_{\ing(\play_i')},\end{equation} provided that both limits exist. By the ergodic theorem  applied to $\mdp_Q'$, we have that the RHS  limit in~\Cref{6-eq:mc-opt-limit} is defined with probability 1 and equal to
\begin{align*}
\underbrace{\sum_{v\in Q} \invdist'_v \cdot \vec{h}'_{v}}_{=0} + \sum_{u,v\in Q} \invdist'_{d_{u,v}}\cdot \vec{h}'_{d_{u,v}} = \frac{1}{2}\sum_{u\in Q}\invdist_u\cdot\left( \sum_{v\in Q}\mcprob_{u,v}\cdot \coloring(u,v)\right)\\ =\frac{1}{2}\sum_{u\in Q} \invdist_u\cdot \vec{h}_u=\frac{r^*}{2},
\end{align*}
\noindent
the last equality being shown above. Plugging this into~\Cref{6-eq:mc-opt-limit} yields that if a limit on the LHS (\textit{i.e.}, the mean payoff of a play) is well-defined with probability 1, then it is equal to $r^*$ also with probability 1. But if there was a set $L$ of positive probability in $\mdp_Q$ with $\lim_{n \rightarrow \infty}\frac{1}{n}\sum_{i=0}^{n-1}\coloring(\play_i)$ undefined for each $\play\in L$, by splitting the plays in $L$ we would obtain a positive-probability set of plays in $\mdp_Q'$ in which $\lim_{n \rightarrow \infty}\frac{1}{n}\sum_{i=0}^{n-1}\vec{h}'_{\ing(\play_i')}$ is also undefined, a contradiction with the Ergodic theorem. 
\end{proof}

So far, we have constructed an optimal strategy $\sigma_Q$ but only on the part $Q$ of the original MDP $\mdp$. To conclude the construction, we define a positional strategy $\sigma^*$ in $\mdp$ as follows: we fix a positional deterministic strategy $\sigma_{=1}$ that is winning, from each vertex of $\mdp,$ for the objective of almost-sure reaching of $Q$ (such a strategy exists since $\mdp$ is strongly connected, see also \Cref{6-thm:as-char}. Then we put $\sigma^*(v)=\sigma_{=1}(v)$ if $v\not\in Q$ and $\sigma^*(v)=\sigma_Q(v)$ otherwise. Hence, starting in any vertex, $\sigma^*$ eventually reaches $Q$ with probability 1 and then it starts behaving as $\sigma_Q$. The optimality of such a strategy follows from the prefix-independence of mean payoff, as argued in the next theorem.

\begin{theorem}[Prefix-independence of mean payoff]
\label{6-thm:mp-valcomp} For any sequence of numbers $c_0,c_1,\dots$ and any $k\in\N$ it holds $\liminf_{n\rightarrow \infty}\frac{1}{n}\sum_{i=0}^{n-1}c_i = \liminf_{m\rightarrow \infty}\frac{1}{m}\sum_{i=0}^{m-1}c_{k+i}$. As a consequence, 
for every vertex $v$ in $\mdp$ it holds $\probm^{\sigma_Q}_{\mdp_Q,v}(\MeanPayoffInf=r^*)=1,$ where $r^*$ is the optimal value of $\lpmp$. Hence, $\expv^{\sigma^*}_v[\MeanPayoffInf]= r^*$.
\end{theorem}
\begin{proof}
We have
\begin{align*}
\liminf_{n\rightarrow \infty}\frac{c_0 + \cdots c_{n-1}}{n} &= \liminf_{n\rightarrow \infty}\left(\underbrace{\frac{k}{n}}_{\mathrlap{\text{vanishes for } n\rightarrow \infty}}\cdot\frac{c_0 + \cdots + c_{k-1}}{k} + \underbrace{\frac{n-k}{n}}_{\mathrlap{\rightarrow 1 \text{ for } n\rightarrow \infty}}\cdot\frac{c_k+\cdot+c_{n-1}}{n-k} \right)\\
&=\liminf_{m\rightarrow\infty} \frac{c_k+\dots+c_{k+m-1}}{m}.
\end{align*} A similar argument holds for $\limsup.$

With probability 1, a play has an infinite suffix consisting of plays from $\mdp_Q^{\sigma}$, and thus also $\MeanPayoffInf$ and $\MeanPayoffSup$ determined by this suffix. By \Cref{6-cor:mp-scc-optimality}, these quantities are equal to $r^*$ with probability 1.
\end{proof}

\noindent
The following theorem summarises the computational aspects.

\begin{theorem}[Complexity of solving strongly connected mean-payoff MDPs]
\label{6-thm:mp-rand-opt-main}
In a strongly connected mean-payoff MDP, one can compute, in polynomial time, a positional randomised strategy which is optimal from every vertex, as well as the (single) optimal value of every vertex.
\end{theorem}
\begin{proof}
We obtain, in polynomial time, an optimal solution of $\lpmp$, with the optimal objective value being the optimal value of every vertex (\Cref{6-thm:mp-valcomp}). We then use this optimal solution $\lpsol{x}$ to construct the strategy $\sigma$ and the  Markov chain $\mdp_{\solvset}^{\sigma}$. From this chain we extract a strongly connected subset of vertices $Q$ (in polynomial time, by a simple graph reachability analysis). With the subset in hand, we can construct strategies $\sigma_Q$ and $\sigma_{=1}$, all polynomial-time computations (see \Cref{6-thm:as-char}). These two strategies are then combined to produce the optimal strategy $\sigma^*$.
\end{proof}

\subsection*{Deterministic optimality in strongly connected MDPs}

It remains to prove that we can actually compute a positional~\emph{deterministic} strategy that is optimal in every vertex. Looking back at the construction that resulted in \Cref{6-thm:mp-rand-opt-main}, we see that the optimal strategy $\sigma^*$ might be randomised because the computed optimal solution $\lpsol{x}$ of $\lpmp$ can contain two components $(v,a)$, $(v,b)$ with $a\neq b$ and both $\lpsol{x}_{(v,a)}$ and $\lpsol{x}_{(v,b)}$ being positive. To prove positional deterministic optimality, we will show that there is always an optimal solution which yields a deterministic strategy, and that such a solution can be obtained in polynomial time.

The previous section implicitly defined two mappings: First, a mapping $\Psi$, which maps every solution $ \vec{x} $ of $\lpmp$ to a positional strategy in some sub-MDP of $\mdp$, by putting $\Psi(\vec{x}) = \sigma$ where $\sigma(a\mid v) = \vec{x}_{(v,a)}/\sum_{b\in \actions}\vec{x}_{(v,b)}$. Second, mapping $\Xi$, which maps each positional strategy $\sigma$ that induces a strongly connected Markov chain to a solution $\Xi(\sigma)$ of $\lpmp$ such that $\Xi(\sigma)_{(v,a)}=\invdist_v\cdot \sigma(a\mid v)$, where $\invdist$ is the unique invariant distribution of the chain induced by $\sigma$.
\begin{lemma}[Correspondence between solutions and strategies]
\label{6-lem:sol-strat-correspondence}
Let $X$ be the set containing exactly those solutions $\vec{x}$ of $\lpmp$ for which the strategy  $\Psi(\vec{x})$ induces a strongly connected Markov chain. Then the mappings $\Psi$ and $\Xi$ are bijections between $X$ and the set of all positional strategies in some sub-MDP of $\mdp$ that induce a strongly connected Markov chain.
\end{lemma}
\begin{proof}
A straightforward computation shows that $\Xi\circ\Psi$ and $\Psi\circ\Xi$ are identity functions on the respective sets.
\end{proof}

\begin{definition}[Pure solutions]
\label{6-def:pure-lp}
A solution $\vec{x}$ of $\lpmp$	is \emph{pure} if for every vertex $v$ there is at most one action $a$ such that $\vec{x}_{(v,a)}>0$.
\end{definition}

\noindent
The following lemma follows from the way in which strategies $\sigma$ and $\sigma^*$ were constructed in the previous sub-section.

\begin{lemma}[Solutions of the linear programs]
\label{6-lem:pure-lpsol}
Let $\lpsol{x}$ be a pure optimal solution of $\lpmp$ and denote $ S = \{v \in \vertices\mid \exists a \text{ s.t. }\lpsol{x}_{(v,a)}>0\} $. Then the strategy $\sigma=\Psi(\lpsol{x})$ is an MD strategy in $\mdp_{\solvset}$. Hence, in such a case, the strategy $\sigma^*$ constructed from $ \sigma $ as in \Cref{6-thm:mp-rand-opt-main} is an optimal MD strategy in $\mdp$.
\end{lemma}

It remains to show how to find a pure optimal solution of $\lpmp$. To this end we exploit some fundamental properties of linear programs.

A linear program is in the \emph{standard} (or equational) form if its set of constraints can be expressed as $A\cdot \vec{x} = \vec{b}$, $\vec{x}\geq 0$, where $\vec{x}$ is a vector of variables, $\vec{b}$ is a non-negative vector, and $A$ is a matrix of an appropriate dimension. In this notation, all the vectors are column vectors, \textit{i.e.} $A$ has one column per each variable. Note that $\lpmp$ is a program in the standard form. A feasible solution $\vec{x}$ of such a program is \emph{basic} if the columns of $A$ corresponding to variables whose value is positive in $\vec{x}$ form a linearly independent set of vectors. Since the maximal number of linearly independent columns equals the maximal number of linearly independent rows (a number called a \emph{rank} of $A$), we know that each basic feasible solution has at most as many positive entries as there are rows of $A$. 

The next two lemmas prove some fundamental properties of basic feasible solutions.

\begin{lemma}[Properties of basic feasible solutions: uniqueness]
\label{6-lem:basic-cond-unique}
Assume that a linear program in a standard form has two basic feasible solutions $\vec{x},\vec{x}'$ such that both solutions have the same set of non-zero components, and the cardinality of this set equals the number of equality constraints in the program. Then $\vec{x}=\vec{x}'$.
\end{lemma}
\begin{proof}
Write $A\cdot \vec{x} = \vec{b}$ the equational constraints of the LP.
If $\vec{x}$ is a basic feasible solution, then it solves the equation $A_{N} \cdot \vec{x}_N = \vec{b}$, where $A_N$ ($  N$ stands for `non-zero') is obtained from $A$ by removing all columns corresponding to zero components of $\vec{x}$, and   $\vec{x}_N$ is obtained from $\vec{x}$ by removing all zero components. 

Since $\vec{x}$ has as many non-zero components as there are rows of $A$, it follows that $A_N$ is a square matrix. Since $\vec{x}$ is a basic solution, $A_N$ is regular (its columns are linearly independent) and $\vec{x}=A_{N}^{-1}\cdot \vec{b}$ is uniquely determined by $A_N$. Repeating the same argument for $\vec{x}'$ yields $\vec{x}'=A_{N}^{-1}\cdot \vec{b}= \vec{x}$.
\end{proof}

\begin{lemma}[Properties of basic feasible solutions]
\label{6-lem:basic-sol}
If a linear program in a standard form has an optimal solution, then it has also a basic optimal solution. Moreover, a basic optimal solution can be found in polynomial time.
\end{lemma}
\begin{proof}[Sketch]
The existence of a basic optimal solution is a well-known linear programming fact, \textit{e.g.} the standard simplex algorithm works by traversing the set of basic feasible solutions until it finds an optimal one~\cite{Matousek:2007}. For computing an optimal basic solution, we can use one of the polynomial-time interior-point methods for linear programming, such as the path-following method~\cite{Karmarkar:1984}. While these methods work by traversing the interior of the polyhedron of feasible solutions, they converge, in polynomial time, to a point that is closer to the optimal basic solution than to all the other basic solutions. By a process called \emph{purification,} such a point can be then converted to the closest basic solution, \textit{i.e.} to the optimal one~\cite{Gonzaga:1992}.
\end{proof}


\begin{theorem}[Complexity of computing optimal deterministic strategies in strongly connected mean-payoff MDPs]
\label{6-thm:lpmp-basic-dim}
One can find, in polynomial time, an optimal deterministic strategy in a given strongly connected  mean-payoff MDP.
\end{theorem}
\begin{proof}
First, we use~\Cref{6-lem:basic-sol} to find a basic optimal solution $\lpsol{x}$ of $\lpmp$.
We check if it is pure. If yes, we are done. Otherwise,
%
there is $v\in \vertices$ and two distinct actions $a,b$ such that $\lpsol{x}_{(v,a)}>0$ and $\lpsol{x}_{(v,b)}>0.$ Let $ S = \{v \in \vertices\mid \exists a \text{ s.t. }\lpsol{x}_{(v,a)}>0\} $. By~\Cref{6-cor:mp-scc-extraction}, we can partition $\solvset$ into several subsets, each of which induces a strongly connected sub-MDP of $\mdp$. Let $Q$ be a class of this partition containing $v$. We have that the optimal mean-payoff value of every vertex in $\mdp_Q$ is the same as in $\mdp$. This is because, 
as in the beginning of the proof of~\Cref{6-cor:mp-scc-optimality}, we can transform $\lpsol{x}$ into another optimal solution of the same value as $\lpsol{x}$ which has non-zero entries only for components indexed by $(q,a)$ with $q\in Q$. All these computations can be easily implemented in polynomial time. 

We argue that $Q$ is a strict subset of $\vertices$. Indeed, assume that $Q=\vertices$. Then $\lpsol{x}$ induces a randomised strategy $\sigma$ in $\mdp$. Moreover, since $\lpsol{x}$ is a basic solution, it has at most $|\vertices|+1$ positive entries, and since it is non-pure, it must have exactly $n+1$ positive entries, \textit{i.e.} \Cref{6-lem:basic-cond-unique} is applicable to $\lpsol{x}$, since $\lpmp$ has exactly $|\vertices|+1$ constraints. Now we define a new strategy $\sigma'$ in $\mdp$ by slightly changing the behaviour in $v$. To this end, choose some $\eps>0$ and put $\sigma'(a\mid v)=\sigma(a\mid v)-\eps$ and $\sigma'(b\mid v)=\sigma(b\mid v)+\eps$; we choose $\eps$ small enough so that both quantities are still non-zero. The chain $\mdp^{\sigma'}$ is still strongly connected. Now let $\vec{x}' = \Xi(\sigma')$. Then $\vec{x}'$ is a solution of $\lpmp$ which is still basic, with a set of non-zero components being the same as in $\lpsol{x}$. At the same time, $\vec{x}'\neq \lpsol{x}$, since $\sigma\neq {\sigma'}$ and $\Xi$ is a bijection (\Cref{6-lem:sol-strat-correspondence}). But this is a contradiction with \Cref{6-lem:basic-cond-unique}.

Hence, $\mdp_Q$ is a strict sub-MDP of $\mdp$ in which the value of every vertex is the same as in the original MDP. We can perform a recursive call of the aforementioned computation on $\mdp_Q$ (compute basic optimal solution of $\lpmp$, check purity, possibly extract and recurse on a sub-MDP).
The depth of recursion is bounded by $|\vertices|$, so the running time is polynomial. Since each sub-MDP obtained during the recursion is non-empty, and the size of the MDPs decreases, the recursion must eventually terminate with a basic optimal solution (in some sub-MDP $\mdp'$) that is pure. This yields a positional deterministic strategy in $\mdp'$ whose value is equal to the optimal value in $\mdp.$ Such a strategy can be extended to whole $\mdp$ by solving almost sure reachability to $ \mdp' $, as described in the previous sub-section.
%
%
%
%
\end{proof}

\section{End components}
\label{6-sec:end_components}
To solve mean-payoff optimisation in general MDPs, as well as general optimisation problems for $\omega$-regular objectives, we need to introduce a crucial notion of an \emph{end component}.

\begin{definition}
	\label{6-def:ec}
An \emph{end component (EC)} of an MDP is any set $\mec$ of vertices having the following two properties:
\begin{itemize}
 \item For each $u \in \mec$ there exists an action $ a $ that is \emph{$ \mec$-safe} in $ u $, \textit{i.e.} satisfies that for all vertices  $v$ with $ \probTranFunc(v \mid u,a) > 0 $ it holds $ v \in \mec $.
 \item For each pair of distinct vertices $ u,v \in \mec$ there is a path from $ u $ to $ v $ visiting only the states from $\mec$.
\end{itemize}
In other words, $ \mec $ is an EC of $ \mdp $ if and only if $ \mec $ is a closed set and the sub-MDP $ \mdp_\mec $ is strongly connected. 
\end{definition}


\noindent
From the player's point of view, the following property of ECs is important.

\begin{lemma}
		\label{6-lem:EC-sweep}
Let $\mec$ be an EC and $v \in \mec$. Then there is an MD strategy $\sigma$ which, when starting in a vertex inside $\mec$, never visits a vertex outside of $ \mec $ and at the same time ensures that with probability one, the vertex $v$ is visited infinitely often. Moreover, $\sigma$ can be computed in polynomial time.
\end{lemma}
\begin{proof}
%
%
From \Cref{6-thm:as-char} we know that we can compute, in polynomial time, an MD strategy $\sigma$ in the sub-MDP $ \mdp_\mec $ ensuring that $v$ is reached with probability 1 from any initial vertex in $ \mec $. Indeed, this is because $\mec = \winPos(\mdp_M,\Reach(v))$, due to the second condition in the definition of a MEC. Since $\mec$ is closed, this strategy never leaves $\mec$. Whenever, the strategy leaves $v$, it guarantees that we return to $v$ with probability 1. Hence, for each $k$, the probability of event $V_k$ --- visiting $v$ at least $k$ times --- is $1$. Since $V_{k+1}\subseteq V_k$, it follows that also the probability of $\bigcap_{i=1}^\infty V_i$ is equal to $1$, which is what we aimed to prove.
\end{proof}

The main reason for introducing ECs is that they are crucial for understanding the limiting behaviour of MDPs.

\begin{definition}
\label{6-def:inf}
We denote by $\Inf(\play)$ the set of vertices that appear infinitely often along a play $ \play $.
\end{definition}

\begin{lemma}
\label{6-lem:EC-inf}
For any $ \vinit $ and $ \sigma $ it holds $ \probm^\sigma_{\vinit} ( \{\play \mid \Inf(\play) \text{ is an EC }  \}) = 1 $. 
\end{lemma}
\begin{proof}
Assume the converse. Then in some MDP there is a set of vertices $ X $ which is not an EC but satisfies $ \probm^\sigma_{\vinit}(\Inf = X ) > 0 $. Since $ X $ is not an EC, there is a vertex $ v \in X $ in which any (even randomised) choice of action results in leaving $ X $ with probability at least $  p_{\min} > 0$ (recall that $ p_{\min} $ is the smallest non-zero edge probability in the MDP).

Let $ \mathit{Stay}_k $ be the set of plays in $\{\Inf = X \}$ which, from step $ k $ on, never visit a vertex outside of $ X $. Since $ \{\Inf = X \}  = \bigcup_{i=1}^{\infty} \mathit{Stay}_i$, by union bound we get $ \probm^\sigma_{\vinit}(\mathit{Stay}_{k_0})>0 $ for some  $ k_0\in \N $. Let $ \mathit{Vis}_j $ denote the set of all plays in $ \mathit{Stay}_{k_0} $ that visit $ v $ at least $ j $ times \emph{after} the step $ k_0 $. Since $  \mathit{Stay}_{k_0} \subseteq \{\Inf = X\}$, we have $  \mathit{Stay}_{k_0} \cap \mathit{Vis}_j = \mathit{Stay}_{k_0} $ for each $ j $. But an easy induction shows that  $ \probm_{\vinit}^\sigma (\mathit{Stay}_{k_0} \cap \mathit{Vis}_{j+1} ) \leq \probm_{\vinit}^\sigma(\{\Ing(\play_{k_0}) \in X \})\cdot p_{\min}^j$, since every visit to $ v $ brings a  risk at least $p_{\min}$ of falling out of $ X $. The latter number converges to zero, so $ \probm_{\vinit}^\sigma (\mathit{Stay}_{k_0}) = \lim_{j\rightarrow \infty}\probm_{\vinit}^\sigma (\mathit{Stay}_{k_0}) = \lim_{j\rightarrow \infty}\probm_{\vinit}^\sigma (\mathit{Stay}_{k_0} \cap\mathit{Vis}_{j+1}) =  0$, a contradiction.
\end{proof}

In general, there can be exponentially many end components in an MDP (\textit{e.g.} for a complete underlying graph and one action per edge, each subset of vertices is an EC). However, we can usually restrict to analysing \emph{maximal} ECs.

\begin{definition}
	\label{6-def:mec}
An end component $ \mec $ is a \emph{maximal end component (MEC)} if no other end-component $ \mec' $ is a superset of $ \mec $. We denote by $\mecs(\mdp)$ the set of all MECs of $\mdp.$
\end{definition} 

If two ECs have a non-empty intersection, then their union is again an EC. Hence, every EC is contained in exactly one MEC, and the total number of MECs is bounded by $ |\vertices| $, since two distinct MECs must be disjoint. Moreover, the decomposition of an MDP into MECs can be computed in polynomial time.

\begin{algorithm}
	\KwData{An MDP $ \mdp $}
	\SetKwFunction{FTreat}{Treat}
	\SetKwProg{Fn}{Function}{:}{}
	
	$\textit{List} \leftarrow \emptyset$ \tcp*{List of found MECs}
	
	$ G \leftarrow (\vertices,\edges) $ \tcp*{The underlying graph of $ \mdp $} 
	
	\While{$G$ is non-empty}{
	Decompose $ G $ into strongly connected components\;
	\ForEach{bottom SCC $ B $ of $ G $}{
		$ B $ is a MEC of $ \mdp $, add it to $ \textit{List} $}
	\( \mathcal{B} \gets \) the union of bottom SCCs of \( G \) \;
	$ R \leftarrow \vertices \setminus \winAS(\mdp,\Safe(\vertices\setminus \mathcal{B})) $\;
	remove vertices in $R$ from $G$ along with adjacent edges
	}

	\Return{$\textit{List}$}
	\caption{Algorithm for MEC decomposition of an MDP.}
	\label{6-algo:MEC-decomposition}
\end{algorithm}

\begin{theorem}
\label{6-thm:MEC-decomposition-complexity}
The set of all MECs in a given MDP can be computed in polynomial time.
\end{theorem}
\begin{proof}
There are several known algorithms, a simple one is pictured in \Cref{6-algo:MEC-decomposition}. Each iteration goes as follows: we first take the underlying directed graph of the MDP and find its strongly connected components using some of the well-known polynomial algorithms~\cite{Cormen.Leiserson.ea:2009}. We identify the bottom SCCs, \textit{i.e.} those from which there is no outgoing edge in the graph. It is easy to see that each such SCC must form a MEC of $\mdp$, and conversely, each MDP has at least one MEC that is also a bottom SCC of its underlying graph. For the union of such bottom SCCs (denoted \( \mathcal{B} \) in the pseudocode) we compute a \emph{random attractor}, \textit{i.e.} the set of vertices of $\mdp$ from which $\mathcal{B}$ cannot be (almost surely) avoided under any strategy. To this end, we compute, in polynomial time, the almost-surely winning set $ \winAS(\mdp,\Safe(\vertices \setminus \mathcal{B})) $ which is the largest (w.r.t. set inclusion) subset of $\vertices$ from which the player can ensure to stay in $\vertices \setminus \mathcal{B}$ forever (\textit{i.e.} the complement of the random attractor of $\mathcal{B}$). The computation can be done in polynomial time by \Cref{6-cor:safety-main}. No vertex of the random attractor of $\mathcal{B}$ can belong to a MEC: such a MEC \( M \) would be disjoint from those in $ \mathcal{B} $ (a non-disjoint union of two ECs is again an EC) but the player cannot force avoiding $ \mathcal{B} $ from \( M \), a contradiction with \( M \) being a closed set. Hence, all MECs of $\mdp$ which are not a bottom SCC of $G$ are subsets of $\winAS(\mdp,\Safe(\vertices \setminus \mathcal{B})) = \vertices \setminus R$, so we can remove all vertices in $R$ from the graph and continue to the next iteration (note that removing vertices in $R$ from $\mdp$ again yields a MDP, since  $\winAS(\mdp,\Safe(\vertices \setminus \mathcal{B}))$ is a closed set). The main loop performs at most $|\vertices|$ iterations, which yields the polynomial complexity.
\end{proof}

\section{Reductions to optimal reachability}
\label{6-sec:reductions}
The MEC decomposition can be used to reduce several optimisation problems (including general mean-payoff optimisation) to optimising reachability probability. Recall that in the optimal reachability problem, we are given an MDP $\mdp$ (with coloured vertices) and a colour $\Win \in\colors$. The task is to find a strategy $\sigma$ that maximises $ \probm^\sigma_{\vinit}(\Reach(\Win))$, the probability of reaching a vertex coloured by $\Win$. The main result on reachability MDPs, which we prove in \Cref{6-sec:optimal_reachability}, is as follows:

\begin{theorem}[Solving reachability MDPs]
\label{6-thm:quant-reachability-main}
In reachability MDPs, the value of each vertex is rational and computable in polynomial time. Moreover, we can compute, in polynomial time, a positional deterministic strategy that is optimal in every vertex.
\end{theorem}

\subsection*{From optimal B{\"u}chi to reachability}

In B{\"u}chi MDPs, the vertices are assigned colours from the set $\{1,2\}$ and our aim is to find a strategy maximising $ \probm^\sigma_{\vinit}(\Buchi)$, \textit{i.e.} maximising the probability that a vertex coloured by $2$ is visited infinitely often.
We say that a MEC $\mec$ of a B{\"u}chi MDP is \emph{good} if it contains a vertex coloured by 2.

\begin{theorem}[Solving B{\"u}chi MDPs]
\label{6-thm:quant-buchi}
In B{\"u}chi MDPs, the value of each vertex is rational and computable in polynomial time. Moreover, we can compute, in polynomial time, a positional deterministic strategy that is optimal in every vertex.
\end{theorem}
\begin{proof}
Let $\mdp_b$ be a B{\"u}chi MDP and let $\mdp_r$ be a reachability MDP obtained from $\mdp_b$ by repainting each vertex belonging to a good MEC with the colour $\Win$. Note that $\mdp_r$ can be computed in polynomial time by performing the MEC decomposition of $\mdp_b$ (\Cref{6-algo:MEC-decomposition}) and checking goodness of each MEC.

We prove that the value of each vertex in $\mdp_b$ is equal to the value of the corresponding vertex in $\mdp_r$.

First, fix any $\sigma$ and $\vinit$ (due to equality of underlying graphs, we can view these as a strategy/initial vertex both in $\mdp_b$ and $\mdp_r$). By \Cref{6-lem:EC-inf}, the probability of visiting infinitely often a vertex outside of a MEC is 0. Hence, the probability of visiting infinitely often a vertex coloured by 2 (in $\mdp_b$) is the same as the probability of visiting infinitely often a vertex coloured by 2 which belongs to (a necessarily good) MEC, which is in turn bounded from above by the probability that $\sigma$ visits (in $\mdp_r$) a vertex coloured by $\Win$.

Conversely, let $\sigma^*$ be the MD reachability-optimal strategy in $\mdp_r$ (which exists by~\Cref{6-thm:quant-reachability-main}). We construct a strategy $\sigma$ in $\mdp_b$ which achieves, in every vertex, the same B{\"u}chi-value as the reachability value achieved in that vertex by $\sigma^*$ in $\mdp_r$. Outside of any good MEC, $\sigma$ behaves exactly as $\sigma^*$. Inside a good MEC $\mec$, $\sigma$ behaves as the MD strategy from \Cref{6-lem:EC-sweep}, ensuring that some fixed vertex in $\mec$ of colour $2$ is almost-surely visited infinitely often. Since $\sigma$ is stitched together from MD strategies on non-overlapping domains, it is positional deterministic and it ensures that once a good MEC is reached, the B{\"u}chi condition is satisfied almost-surely.

The construction of $\sigma$ in the aforementioned paragraph is effective: given the optimal strategy $\sigma^*$ for reachability, $\sigma$ can be constructed in polynomial time.
\end{proof}

\subsection*{From optimal parity to optimal reachability}

In parity MDPs, the vertices are labelled by colours form the set $\{1,\dots,d\}$ (w.l.o.g. we stipulate that $d\leq |\vertices|$) and the goal is to find a strategy maximising $ \probm^\sigma_{\vinit}(\Parity),$ \textit{i.e.} maximising the probability that the largest priority appearing infinitely often along a play is even.

\begin{theorem}[Solving parity MDPs]
\label{6-thm:parity-main}
In Parity MDPs, the value of each vertex is rational and computable in polynomial time. Moreover, we can compute, in polynomial time, a positional deterministic strategy that is optimal in every vertex.
\end{theorem}
\begin{proof}
Let $\mdp_p$ be a parity MDP. We will proceed similarly to~\Cref{6-thm:quant-buchi}, constructing a reachability MDP $\mdp_r$ with the same underlying graph as $\mdp_p$.

To this end, let $\mdp_i$ be the largest sub-MDP of $\mdp_p$ containing only the vertices of priority $\leq i$. Formally, we set $\vertices_i = \winAS(\mdp_p,\Safe(\coloring^{-1}(\{i+1,\ldots,d\})) )$ and define $\mdp_i$ to be the sub-MDP induced by $\vertices_i$ (note that $\mdp_i$ might be empty). We say that a vertex of $\mdp_p$ is $i$-good if it is contained in some MEC $\mec$ of $\mdp_i$ such that the largest vertex priority inside $\mec$ is equal to $i$. We say that a vertex is even-good if it is $i$-good for some even $i$. We set up a reachability MDP $\mdp_r$ by taking $\mdp_p$ and re-colouring each its even-good vertex with colour $\Win$. To do this, we need to compute, for each even priority $i$, the MDP $\mdp_i$ and its MEC-decomposition. This can be done in polynomial time (\Cref{6-algo:MEC-decomposition}). 

We again prove that the value of every vertex in $\mdp_p$ is equal to the value of the corresponding vertex in $\mdp_r$.

Let $\sigma$ and $\vinit$ be arbitrary. By~\Cref{6-lem:EC-inf}, $\probm^\sigma_{\mdp_p,\vinit}(\Parity)$ is equal to the probability that $\Inf(\play)$  is an EC in which the largest priority is even. But each such EC is also an EC of some $\mdp_i$ with even $i$, and thus is also contained in a MEC of a $\mdp_i$ in which the largest priority is $ i $. Hence, $\probm^\sigma_{\mdp_p,\vinit}(\Parity)\leq \probm^\sigma_{\mdp_r,\vinit}(\Reach(\Win))$.

Conversely, let $\sigma^*$ be the MD reachability-optimal strategy in $\mdp_r$. We construct an MD strategy $\sigma$ in $\mdp_p$ as follows: in a vertex $v$ which is not even-good, $\sigma$ behaves as $\sigma^*$. For a vertex $v$ that is even-good, we identify the smallest even $i$ such that $v$ is $i$-good. 
This means that $v$ belongs to some MEC $\mec$ of $\mdp_i$ in which the largest priority is $i$. 
By \Cref{6-lem:EC-sweep}, we can compute, in polynomial time, an MD strategy $\sigma_M$ which ensures that the largest-priority vertex in $(\mdp_i)_\mec$ is visited infinitely often, and we set $\sigma(v)$ to $\sigma_M(v)$. Note that given $\sigma^*$, the strategy $\sigma$ can be constructed in polynomial time. It remains to show that $\probm^\sigma_{\mdp_p,\vinit}(\Parity)\geq \probm^{\sigma^*}_{\mdp_r,\vinit}(\Reach(\Win))$.

By the construction of $\sigma$, once we reach a vertex which is $i$-good for some even $i$, all the following vertices will be $j$-good for some even $j\leq i$. From this and from \Cref{6-lem:EC-inf} it follows that $\probm^{\sigma^*}_{\mdp_r,\vinit}(\Reach(\Win))$ is equal to the probability that $\sigma$ produces a play $\play$ with the following property: $\exists i \text{ even}$ such that all but finitely many vertices on $\play$ are $i$-good but are not $j$-good for any even $j<i$. This can be in turn rephrased as the probability that $\Inf(\play)$ is an EC whose all vertices are $i$-good for some even $i$ but none of them is $j$-good for an even $j<i$; we call such an EC \emph{$i$-definite}. But within such an EC, $\sigma$ forever behaves as $\sigma_M$ for some MEC $\mdp$ of $\mdp_i$ in which the maximal priority is $i$. Hence, once an $i$-definite EC is reached, the strategy almost-surely ensures that priority $i$ is visited infinitely often and ensures that no larger priority is ever visited. It follows that 
$\probm^{\sigma^*}_{\mdp_r,\vinit}(\Reach(\Win)) = \probm^{\sigma}_{\mdp_p,\vinit}(\inf(\play) \text{ is } i \text{-definite for even }i ) = \probm^{\sigma}_{\mdp_p,\vinit}(\Parity).$
\end{proof}

\subsection*{From general mean payoff to optimal reachability}

We already know how to solve strongly connected mean-payoff MDPs. We now combine this result with MEC decomposition to reduce the general (not strongly connected) mean-payoff optimisation to MDP reachability.

We start with a strengthening of \Cref{6-thm:mp-valcomp}.

\begin{lemma}
\label{6-lem:MEC-mp-strict-bound}
Let $\mdp$ be a strongly connected mean-payoff MDP and $r^*$ the value of each of its vertices. Then, for each $\sigma$ and $\vinit$ we have $\probm^\sigma_{\vinit}(\MeanPayoffInf > r^*) = 0 $.
\end{lemma}
\begin{proof}
Assume that the statement is not true. Then there exist $\sigma,\vinit$ as well as numbers $\epsilon,\delta>0 $ and $n_0 \in \N$ s.t. the probability of the following set of plays $X_{\epsilon,n_0}$ is at least $\delta$: a play $\play$ belongs to $X_{\epsilon,n_0}$ if for every $n\geq n_0$ it holds $\frac{1}{n}\sum_{i=0}^{n-1}\coloring(\play_i) \geq x^* + \eps$. We construct a new strategy $\sigma'$, which proceeds in a series of episodes. Every episode starts in $\vinit$, and for the first $n_0$ steps of the, episode $\sigma'$ mimics $\sigma$. After that, it checks, in every step $n$, whether the payoff accumulated since the start of the episode is at least $n\cdot(r^* + \eps)$. If this holds, we mimic $\sigma$ for one more step. If the inequality is violated, we immediately `restart', \textit{i.e.} return to $\vinit$ (can be performed with probability $1$ due to the MDP being strongly connected) and once in $\vinit$, start a new episode which mimics $\sigma$ from the beginning. By our assumption, the probability of not performing a reset in a given episode is at least $\delta>0$. Hence, with probability $1$ we witness only finitely many resets, after which we produce a play whose suffix has mean payoff at least $r^* + e$. By prefix-independence of mean payoff (\Cref{6-thm:mp-valcomp}), $\expv^{\sigma'}_{\vinit} [\MeanPayoffInf] \geq r^* + \eps,$ a contradiction.
\end{proof}

We will need to strengthen the previous lemma so that it applies not only to strongly connected MDPs, but also to MECs in some larger MDPs. The strengthening is performed in the following two lemmas. The first lemma says that once we exit a MEC, with some positive probability we will never return.

\begin{lemma}
\label{6-lem:MEC-noreturn}
Let $ \mec $ be a MEC of an MDP $ \mdp $ and let $ v\in \mec $, $ a\in \actions $ be such that $ a $ \emph{is not} $ \mec $-safe in $ v $. Then there exists $ t $ s.t. $ \probTranFunc(t\mid v,a)>0 $ and  $ t \not \in \winAS(\mdp,\Reach(\mec)) $.
\end{lemma}
\begin{proof}
Assume that $ a $ is not $ \mec $-safe in $ v $ and that all $ t $'s with $ \probTranFunc(t\mid v,a)>0 $ belong to $ \winAS(\mdp,\Reach(\mec)) $. Fix the MD strategy $  \sigma $ which is almost-surely winning for reaching $ \mec $ from each vertex of $ \winAS(\mdp,\Reach(\mec)) $ (\Cref{6-thm:as-char}). For each $ t $ s.t.  $ \probTranFunc(t\mid v,a)>0 $, let $ \mec_t $ denote the set of vertices which can be (with a positive probability) visited under $ \sigma $. Put $ \mec' = \mec \cup (\bigcup_{t\in \vertices,\probTranFunc(t\mid v,a)>0}\mec_t )$. Then $ \mec' $ is closed, since $ \mec $ is closed and since for every $ u $ in some $ \mec_t $ there exists an action (the one selected by $ \sigma $ for $ u $) under which we surely stay in $ \mec_t $. Moreover, the $ \mec'$-induced sub-MDP is strongly connected: each $ t $ with $ \probTranFunc(t\mid v,a)>0 $ is reachable from within $ \mec $ (through $ v $) and thus each vertex in some $\mec_t $ is reachable from $ \mec $. In turn, from each vertex in some $ \mec_t $ (where $ \probTranFunc(t\mid v,a)>0 $) we can reach $ \mec $ without leaving $ \mec_t $, due to the definition of $ \sigma $. Hence, $ \mec' $ is a MEC which strictly contains $ \mec $, a contradiction with the maximality of $ \mec. $
\end{proof} 

Given a play $\play$ and strategy $\sigma$, we define a \emph{slice} of $\sigma$ as a strategy $\slice{\sigma}{\play}$ such that for each $\play'$ starting in $\last(\play)$ it holds $\slice{\sigma}{\play}(\play') = \sigma(\play\play')$, while on other plays $\slice{\sigma}{\play}$ just mimics $\sigma$.

\begin{lemma}
		\label{6-lem:MEC-stable}
	Let $\mec$ be a MEC of $\mdp$ and $r^*$ the mean-payoff value of every vertex in the strongly connected sub-MDP induced by $\mec$. Then the set $E$ of all plays that have $\Inf(\play)\subseteq\mec$ and at the same time mean payoff greater than $r$ has probability zero under any strategy $\sigma$.
\end{lemma}
\begin{proof}
 Assume, for contradiction, that there is a strategy $\sigma$ and $\delta > 0$ such that the probability of $E$ under $ \sigma  $ is at least $\delta$. Note that we do not immediately have a contradiction with \Cref{6-lem:MEC-mp-strict-bound}, since $\sigma$ might leave $\mec$ (and then return back). 
 
 We say that a play $ \play $ \emph{cheats} in step $ i $ if it is inside $ \mec $ in $ i $-th step and outside of $ \mec $ in the next step (which can only be caused by an $ \mec $-unsafe action being played). From \Cref{6-lem:MEC-noreturn} we have that there is $ p>0 $ s.t. upon every exit from $ \mec $ we return with probability at most $ (1-p) $. Hence, the probability that a play cheats infinitely often is $ 0 $. It follows that there is $ k\in \N $ s.t. $ \probm_{\vinit}^\sigma(\play \text{ cheats after } \geq k \text{ steps}) \leq (\delta\cdot p_{\min})/4 $, where $ p_{\min} $ is the smallest non-zero edge probability in $ \mdp $. 
 
 Whenever we are in some $ v\in \mec $ and play an action that is not $ \mec $-safe in $ v $, this results into a cheat with  probability at least $ p_{\min} $. Thus, the total probability that this happens after at least $ k $ steps, \textit{i.e.} the quantity 
 \begin{equation}
 \label{6-eq:mec-cheat}
 q = \sum_{i \geq k}\;\sum_{v\in \mec}\;\sum_{a \text{ not $ \mec $-safe in v}}\expv^\sigma_{\vinit}[ \actevent{\sigma}{a}{i}\cdot\indicator{ \out(\play_i)= v} ] , 
 \end{equation}
 is bounded by $ \probm_{\vinit}^\sigma(\play \text{ cheats after more than } k \text{ steps})/p_{\min} \leq \delta/4$.
 
 Let's go back to $ E $ now. On each play in $ E $ there is a step $ i $ from which on the play stays in $ \mec $ forever: we say that the play is $ i $-definite and we denote by $E_k$ the set of all $ i $-definite plays in $ E $. By union bound, there is $ \ell \in \N, \ell \geq k $ s.t. $ \probm_{\vinit}^\sigma(E_\ell)  \geq \delta/2$. 
 
 We define a new strategy $ \sigma' $ as follows: on each play prefix, $ \sigma' $ by default mimics $ \sigma $, except for the case when at least $ \ell $ steps have elapsed, the current vertex $ v $ is in $ \mec $, and $ \sigma $ prescribes to play, with positive probability, an action which is not $ \mec $ safe in $ v $. In such a case, $ \sigma $ is overridden and we play any action that is $ \mec $-safe in $ v $ instead (after which we return to simulating $ \sigma $, until the override kicks in again). The probability that such an override happens is bounded by the quantity $ q $ from~\Cref{6-eq:mec-cheat}, and hence by $ \delta/4 $. Since  $ \probm_{\vinit}^\sigma(E_\ell)  \geq \delta/2$, at least half the measure of $ E_{\ell} $ stays untouched by the overrides; hence  $ \probm_{\vinit}^{\sigma'}(E_\ell)\geq \delta/4 $.
 
 We are ready to apply the final argument. There are only finitely many plays of length $ \ell $. Hence, by union bound, there is a play $ \play $ of length $ \ell $ such that $\probm_{\vinit}^{\sigma'}(E_\ell \cap \cylinder(\play))>0$. Consider the strategy  $\slice{\sigma'}{\play}$. 
 Starting in $ \last(\play) $, we have that $\slice{\sigma'}{\play}$ never leaves $ \mec $, due to the overrides in $ \sigma' $. Hence, $\slice{\sigma'}{\play}$ can be seen as a strategy in the strongly connected MDP $ \mdp_\mec $. Now consider the set $ E'=\{\play'\mid \play'\exists\play''\in E \text{ s.t. } \play''=\play\play'\} $. Then $ \probm_{\last(\play)}^{\slice{\sigma'}{\play}}(E') = \probm_{\vinit}^{\sigma'}(E_\ell \cap \cylinder(\play))>0 $; but due to the prefix-independence of mean payoff, all plays in $ E' $ have payoff $ > r^* $, a contradiction with \Cref{6-lem:MEC-mp-strict-bound}.
\end{proof}

\begin{theorem}
\label{6-thm:general-mp-main}
	In mean-payoff MDPs, the value of each vertex is rational and computable in polynomial time. Moreover, we can compute, in polynomial time, a positional deterministic strategy that is optimal in every vertex.
\end{theorem}
\begin{proof}
First, note that we can w.l.o.g. restrict to MDPs in which each edge is coloured by a number between $0$ and $ 1 $. To see this, let $\mdp$ be an MDP and $a,b$ any two numbers, with $a$ non-negative. We can construct an MDP $\mdp'$ by re-colouring each edge $(u,v)$ of $\mdp$ with colour $a\cdot \coloring(u,v)+b$, where $\coloring$ is the original colouring in $\mdp$. It is then easy to see that for each strategy $\sigma$ it holds $\expv_{\mdp,\vinit}^\sigma[\MeanPayoffInf]=(\expv_{\mdp',\vinit}^\sigma[\MeanPayoffInf]/a)-b$, so a strategy optimising the mean payoff in $\mdp'$ is also optimal in $\mdp$. Hence, we always can re-scale the colouring into the unit interval while preserving the optimisation criterion.

So now let $\mdp_\smallmp$ be a mean-payoff MDP with edge-colouring $\coloring$. We construct, in polynomial time, a new reachability MDP $\mdp_r$ as follows: first, we compute the MEC decomposition of $\mdp_\smallmp$ (\Cref{6-algo:MEC-decomposition}). Let $\mec_1,\dots,\mec_k$ be all the resulting MECs. For each MEC $\mec_i$ we compute the optimal mean-payoff value $r_i^*$ in the sub-MDP induced by $\mec_i$ (which is shared by all vertices of this sub-MDP, by \Cref{6-thm:mp-valcomp}), along with the corresponding positional deterministic optimal strategy. We already know how to do this in polynomial time (\Cref{6-thm:mp-rand-opt-main,6-thm:lpmp-basic-dim}). Now we add new vertices $\vgood$, $\vbad$, both with self loops, and edges incoming to these vertices from each vertex that belongs to some MEC of $\mdp_\smallmp$. The vertex $\vgood$ is the only vertex coloured by $\Win$ in $\mdp_r$. Finally, we add a new action $\finact$ which behaves as follows: For each vertex $v$ belonging to a MEC $\mec_i$ we set $\probTranFunc(\vgood\mid v,\finact) = r^*_i$ and $\probTranFunc(\vbad\mid v,\finact) = 1-r^*_i $. In a non-MEC vertex $ v $, we put $ \probTranFunc(v,\finact) = \probTranFunc(v,a) $ for some $ a\in \actions $, $ a\neq \finact $, so that no new behaviour is introduced in these vertices.

We show that for any original vertex (\textit{i.e.} all vertices but $\vgood,\vbad$) the optimal values in both MDPs are the same and the optimal strategies are easily transferable from one MDP to the other.

First, let $\sigma$ be an $\eps$-optimal strategy in $\mdp_\smallmp$. 
We have 
\[
\begin{array}{lll}
\expv^\sigma_{\vinit}[\MeanPayoffInf] 
& =    & \sum_{i=1}^k\expv^\sigma_{\vinit}[\MeanPayoffInf\cdot \indicator{\Inf\subseteq\mec_i}] \\
& \leq & \sum_{i=1}^k \expv^\sigma_{\vinit}[r_i^*\cdot \indicator{\Inf=\mec_i}] \\
& =    & \sum_{i=1}^k r_i^* \cdot \probm_{\vinit}^\sigma(\Inf=\mec_i)
\end{array}
\]
Here the first equation follows from \Cref{6-lem:EC-inf} and the subsequent inequality from \Cref{6-lem:MEC-stable}. Moreover, for each $i$ there is a number $n_0^i$ such that the probability of all plays that stay inside $\mec_i$ in all the steps from $n_0^i$ to infinity is at least $\probm_{\vinit}^\sigma(\Inf\subseteq\mec_i) - \frac{\eps}{k} $. Let $n_0 = \max_{1\leq i \leq k} n^i_0$.

We construct a reachability strategy $\sigma_r$ which mimics $\sigma$ for the first $n_0$ steps. After $n_0$ steps it performs a switch: if the current vertex is in some $\mec_i$ we immediately play the action $\finact$, otherwise we start to behave arbitrarily. 
We have 
\[
\begin{array}{lll}
\probm_{\vinit}^{\sigma_r}(\Reach(\Win)) 
& \geq & \sum_{i=1}^{k} r_i^* \cdot \probm_{\vinit}^{\sigma_r}(\last(\play_{\leq n_0}) \in \mec_i ) \\
& \geq & \sum_{i=1}^k r_i^* \cdot \probm_{\vinit}^\sigma(\Inf\subseteq\mec_i) - \eps \\
& \geq & \expv^\sigma_{\vinit}[\MeanPayoffInf] -\eps.
\end{array}
\]
This is the last equality shown in the previous paragraph. Since $\sigma$ is $\eps$-optimal for mean payoff, $\probm_{\vinit}^{\sigma_r}(\Reach(\Win))$ is at most $2\eps$ away from the mean-payoff value of $ v $. Since $\eps>0$ was chosen arbitrarily, we get that the reachability value in $\mdp_{r}$ is at least as large as the mean-payoff value in $\mdp_{\smallmp}$.

Conversely, let $\sigma^*$ be the optimal MD strategy in $\mdp_r$. We say that $\sigma^*$ ends in a vertex $v$ if $\sigma^*(v)=\finact$. We can assume that if $\sigma^*$ ends in some $v \in \mec_i$ then it ends in all vertices of $\mec_i$. This is because whenever $\sigma^*$ ends in some vertex $v \in \mec_i$, the reachability value of $v$ must be equal to $r^*_i$, otherwise playing $\finact$ would not be optimal here. But the optimal reachability value in every vertex of a given MEC is the same (due to \Cref{6-lem:EC-sweep}), so if playing $\finact$ is optimal in some vertex of $\mec_i$, it is optimal in all such vertices. Now we can define an MD strategy $\sigma_{\smallmp}$ in $\mdp_\smallmp$ to initially mimic $\sigma^*$, and upon encountering any MEC $\mec_i$ in which $\sigma^*$ ends, immediately switch to the MD strategy that is optimal in the mean-payoff sub-MDP $\mdp_i$. We have $\expv^{\sigma_{\smallmp}}_{\vinit}[\MeanPayoffInf]  =  \sum_{i=1}^{k} \probm^{\sigma^*}_{\vinit}(\text{end in }\mec_i)\cdot r^*_i = \probm^{\sigma^*}_{\vinit} (\Reach(\Win)). $ Since $\sigma^*$ as well as the optimal strategies in all $\mec_i$ can be computed in polynomial time (\Cref{6-thm:quant-reachability-main,6-thm:lpmp-basic-dim}), we get the result.
\end{proof}

\section{Optimal reachability}
\label{6-sec:optimal_reachability}
In this final section, we prove \Cref{6-thm:quant-reachability-main}. The proof bears many similarities to the methods for discounted-payoff MDPs, hence we only sketch the process and point out the key differences. Throughout the section we assume that \emph{targets are sinks}, \textit{i.e.} that a vertex coloured by $\Win$ has only a self loop as the single outgoing edge. Modifying an MDP to accommodate this does not influence reachability probabilities in any way.

Consider the reachability operator $\ReachOp\colon[0,1]^{\vertices}\rightarrow [0,1]^{\vertices}$ such that for $\vec{y} = \ReachOp(\vec{x})$ it holds

\[
\vec{y}_v = 
\begin{cases}
 \max_{a\in \actions} \sum_{u\in \vertices} \probTranFunc(u \mid v,a)\cdot x_u & \coloring(v) \neq \Win \\
 1 & \coloring(v) = \Win.
\end{cases} 
\]

\begin{lemma}
\label{6-lem:quant-reach-operator-fixed point}
Let $\vec{x}$ be any initial vector such that \( \vec{x} \) is component wise comparable with \( \ReachOp(\vec{x}) \). Then the limit $\lim_{k \rightarrow \infty}\ReachOp^k(\vec{x})$ exists. Moreover, if $\vec{x} \leq \ReachOp(\vec{x})$, then the limit is equal to the least fixed point of $\ReachOp$ that is greater than or equal to $\vec{x}$; if $\ReachOp(\vec{x})\leq \vec{x}$, then the limit is equal to the greatest fixed point of $\ReachOp$ that is less than or equal to $\vec{x}$.
\end{lemma}
\begin{proof}
The existence of the limit follows from the monotonicity of $\ReachOp$.
In addition, it can be easily checked that the set $[0,1]^{\vertices}$ is a directed complete partial order and that $\ReachOp$ is a Scott-continuous operator on this set. Hence, the result follows from the Kleene's theorem (see also Tarski-Kantorovich principle).
\end{proof}

We denote by $\Reach^k(\Win)$ the set of all plays that reach $\Win$ within the first $k$ steps. Clearly, for each $\sigma$ and $\vinit$ we have $\lim_{k \rightarrow \infty} \probm^\sigma_{\vinit}(\Reach^k(\Win)) = \probm^\sigma_{\vinit}(\Reach(\Win))$.

\begin{lemma}
\label{6-lem:quant-reach-step-operator}
For each $k\in \N$ and $v\in \vertices$, $\ReachOp^k(\vec{0})_v = \sup_{\sigma}\probm^\sigma_v(\Reach^k(\Win))$. In particular, the vector $\vec{x}^* = \lim_{k \rightarrow\infty} \ReachOp^k(\vec{0})$ is the least fixed point of $ \ReachOp $ and it is equal to the vector of reachability values. 
\end{lemma}
\begin{proof}
The first part can be proved by a straightforward induction, the second part follows by \Cref{6-lem:quant-reach-operator-fixed point} and a simple limiting argument.
\end{proof}

Similarly to \Cref{6-def:disc-safe-act} we say that an action $a$ is $\vec{x}$-safe in $v$ if it holds that $a= \underset{a' \in \actions}{\arg\max} \sum_{u\in \vertices} 
\probTranFunc(u\mid v,a') \cdot\vec{x}_u.$ Recall that a strategy $\sigma$ is $\vec{x}$-safe if all actions selected in a vertex with non-zero probability are $\vec{x}$-safe in that vertex. 

\begin{lemma}
\label{6-lem:quant-reach-value-distribution}
Let $ \vec{x}^* $ be as in \Cref{6-lem:quant-reach-step-operator}. 
Next, let $Z^{(n)}$ be a random variable which for a given time step $n$ looks at the current vertex $v$ after $n$ steps and returns the value $\vec{x}^*_v$. Then for every $\vec{x}^*$-safe strategy $\sigma$ it holds $\expv^\sigma_{\vinit}[Z^{(n)}] = \vec{x}^*_{\vinit}$. Moreover, it holds $\expv^\sigma_{\vinit}[Z^{(n)}\cdot \indicator{\coloring{(\Out{(\play_{n})})}=\Win}] = \probm^\sigma_{\vinit}(\Reach^n(\Win)).$
\end{lemma}
\begin{proof}
By an easy induction on $n$, using the fact that target states are  sinks.
\end{proof}

Now an analogue of \Cref{6-lem:disc-val-lower} does not hold for reachability: a strategy playing only $\vec{x}^*$-safe actions might not be optimal (indeed, it might not reach $\Win$ at all). Instead, we proceed as follows: Let $\mdp^*$ be an MDP in which we `disable', in each state $v$, all actions that are not $\vec{x}^*$-safe in $v$. This can be formally done by adding a new non-target sink vertex $ \mathit{sink} $, an edge from each original vertex to $ \mathit{sink} $, and stipulating that each action $a$ that is disabled in a vertex $ v $ chooses, when played in $ v $ in $ \mdp^*$, the edge leading to $ \mathit{sink} $ with probability 1. 

\begin{lemma}
\label{6-lem:quant-reach-pruning-unsafe}
The vectors of reachability values $ \Value(\mdp)$ and $\Value(\mdp^*) $ are equal.
In particular, $ \winPos(\mdp,\Reach(\Win)) = \winPos(\mdp^*,\Reach(\Win)).$
\end{lemma}
\begin{proof}
Let $\vec{x}^*$ again denote the vector of optimal values in $\mdp$. If all actions in $ \mdp $ are $\vec{x}^*$-safes, then the lemma clearly holds. Otherwise there is some $\delta\in(0,1)$ such that for each action $a$ that is not $\vec{x}^*$-safe in some vertex $ v $ it holds $\sum_{u\in \vertices} \probTranFunc(u\mid v,a) \cdot\vec{x}_u \leq \vec{x}^*_v - \delta$.

Let $\epsilon \in(0,\delta)$ be arbitrary and fix an $\epsilon$-optimal strategy $\sigma$ in $\mdp$. We will show that there is a $(2\eps/\delta)$-optimal  strategy $\sigma'$ which only uses $\vec{x}^*$-safe actions. Since $\eps$ can be chosen arbitrarily close to $0$, this shows that $\vec{x}^*$-safe strategies can get arbitrarily close to the value, hence $\Value(\mdp^*)=\Value(\mdp)$.

The strategy $\sigma'$ initially mimics $\sigma$ up to the first point in time when an action that is not $\vec{x}^*$-safe in the current vertex is to be selected. At this point $\sigma'$ switches to behave as any $\vec{x}^*$-safe strategy. To analyse the value achieved by $\sigma'$, we need to bound the probability of the event $\mathit{NonSafe}$ that the switch occurs. By the same reasoning as in \Cref{6-lem:quant-reach-value-distribution}, we can show that for all $n$ it holds $\probm^\sigma_{\vinit}(\Reach^n(\Win)) \leq \expv^\sigma_{\vinit}[Z^{(n)}] \leq  \vec{x}^*_{\vinit}- \delta\cdot\probm_{\vinit}^\sigma(\mathit{NonSafe^{(n)}})$, where $\mathit{NonSafe^{(n)}}$ is the probability that a switch occurs in the first $n$ steps. By taking $n$ to the limit we get $\probm^\sigma_{\vinit}(\Reach(\Win)) \leq \vec{x}^*_{\vinit}- \delta\cdot\probm_{\vinit}^\sigma(\mathit{NonSafe})$. At the same time $\vec{x}^*_{\vinit}-\eps\leq  \probm^\sigma_{\vinit}(\Reach(\Win))$. Combining these two inequalities yields $\probm_{\vinit}^\sigma(\mathit{NonSafe}) \leq  \frac{\eps}{\delta}.$ Now clearly $\probm_{\vinit}^{\sigma'}(\Reach(\Win)) \geq \vec{x}^*_{\vinit} - \eps - \probm_{\vinit}^{\sigma}(\mathit{NonSafe}) \geq \vec{x}^*_{\vinit} - \eps - \eps/\delta \geq \vec{x}^*_{\vinit} -2\eps/\delta$.
\end{proof}

\begin{lemma}
\label{6-lem:quant-reach-strat-construction}
Given the vector $\vec{x}^*$ of optimal reachability values, we can compute, in polynomial time, the optimal MD reachability strategy in $\mdp$.
\end{lemma}
\begin{proof}
Given $\vec{x}^*$, we construct the MDP $\mdp^*$ and compute the winning strategy $\sigma$ for positive reachability in $\mdp^*$. We already know that $\sigma$ can be taken positional and computed in polynomial time (\Cref{6-thm:positive-char}). We claim that $\sigma$ is an optimal reachability strategy in $\mdp$. By \Cref{6-lem:quant-reach-pruning-unsafe} it suffices to show that $\sigma$ is optimal in $\mdp^*$. Let $W$ be the winning region for positive reachability in $\mdp^*$. Since $\sigma$ is positional, with probability $1$ we reach either $\Win$ or a vertex of value $0$ (from which we cannot return to $W$ anymore); in other words, for almost all plays $\play$ we have that $\indicator{\Out(\play_n)\in W}$ eventually equals $\indicator{\coloring(\Out(\play_n)) = \Win}$. Hence, using \Cref{6-lem:quant-reach-value-distribution} we get $\vec{x}^*_{\vinit} = \lim_{n\rightarrow\infty}\expv^\sigma_{\vinit}[Z^{(n)}] = \expv^\sigma_{\vinit}[\lim_{n\rightarrow\infty} Z^{(n)}] = \expv^\sigma_{\vinit}[\lim_{n\rightarrow\infty} Z^{(n)}\cdot\indicator{\Out(\play_n)\in W}] = \expv^\sigma_{\vinit}[\lim_{n\rightarrow\infty} Z^{(n)}\cdot \indicator{\coloring(\Out(\play_{n})) = \Win } ] = \probm_{\vinit}^\sigma(\Reach(\Win))$. Here, the third equality holds since $ \vec{x}^*_v $ is zero for $ v\not\in W $, while the  swapping of expectations and limits can be performed due to the dominated convergence theorem.
\end{proof}

To finish the proof of \Cref{6-thm:quant-reachability-main}, it remains to prove that the vector of optimal values $\vec{x}^*$ can be computed in polynomial time. We again employ linear programming and define the linear program $\lpreach$ with variables $x_v$, $v\in \vertices$.

\begin{equation*}
\begin{array}{llr}
\text{minimise }     & \sum_{v \in \vertices} x_v \\
\textrm{subject to } & x_v = 1 & \text{if $\coloring({v}) = \Win$}\\
					 & x_v = 0 & \text{if $v \not \in \winPos(\mdp,\Reach(\Win))$} \\
					 & x_v \geq \sum_{u \in \vertices} \probTranFunc(u\mid v,a)\cdot x_u	& \text{for all other $v \in \vertices, a\in \actions$.}
\end{array}
\label{6-eq:reach-lp}
\end{equation*}


\begin{lemma}
\label{6-lem:quant-reach-lp}
The linear program $\lpreach$ in \Cref{6-eq:reach-lp} has a unique optimal solution 
 $\bar{\vec{x}}$ such that $\bar{\vec{x}} = \vec{x}^*$.
\end{lemma}
\begin{proof}
Clearly $\vec{x}^*$ is a feasible solution of $\lpreach$. Similarly to \Cref{6-lem:disc-lp} we prove that each feasible solution $\vec{x}$ of $\lpreach$ satisfies $\vec{x}\geq \vec{x}^*$. We can proceed analogously  to \Cref{6-lem:disc-lp}, just replacing the operator $\discOP$ with $\ReachOp$. The proof can be mimicked up to the point where we get that $\lim_{k\rightarrow \infty} \ReachOp^k (\vec{x}) \leq \vec{x}$ (the limit exists by \Cref{6-lem:quant-reach-operator-fixed point}). Since $\ReachOp(\vec{x})\leq \vec{x}$ for each feasible solution $\vec{x}$, from \Cref{6-lem:quant-reach-operator-fixed point} we get that the limit is a fixed point of $\ReachOp$, and in hence it is greater or equal to the least fixed point of $\ReachOp$, \textit{i.e.} $\vec{x}^*$ (\Cref{6-lem:quant-reach-step-operator}). Hence, also $\vec{x} \geq \lim_{k\rightarrow \infty} \ReachOp^k (\vec{x}_0)\geq \vec{x}^*$. 
\end{proof}

\noindent
\Cref{6-lem:quant-reach-strat-construction,6-lem:quant-reach-lp} give us \Cref{6-thm:quant-reachability-main}.

\section*{Bibliographic references}
\label{6-sec:references}
There is a broad field of study related to Markov decision processes, with a history going as far as  1950's~\cite{Bellman:1957}. It is beyond the scope of this chapter to provide a comprehensive overview of the related literature. Nonetheless, in this section we provide pointers to the most significant works connected to our techniques as well as to works that can serve as a starting point for a further study.

One of the most widely used references for MDP-related research is the textbook by Puterman~\cite{Puterman:2005}. The textbook views MDPs from an operations research point-of-view, focusing on finite-horizon, discounted, total reward, and average-reward (an alternative name for mean-payoff) objectives. Regular objectives fall outside of the book's focus, though reachability can be viewed as a special case of the ``positive bounded total reward'' objectives studied in the book. An in-depth study of the textbook will impart to its reader the knowledge of many useful techniques for MDP analysis, though a reader who is a newcomer to MDPs might feel somewhat intimidated by its sheer volume and generality. In this chapter, we follow Puterman's exposition mainly in the discounted payoff, albeit in a rather condensed form. 

For mean-payoff MDPs,~\cite{Puterman:2005} follows similar blueprint as in the discounted case: first characterising the optimal values via a suitable optimality equation and then deriving the value iteration, strategy improvement, and linear programming methods from this characterisation. We use the linear programming as our foundational stone, focusing on the relationship between strategies and feasible solutions of the program. We note that value and strategy iteration for mean-payoff MDPs come with super-polynomial lower bounds, see, \textit{e.g.}~\cite{Fearnley:2010}, or~\cite{Puterman:2005}, where it is shown that strategy improvement converges at least as fast as value iteration.

Also,~\cite{Puterman:2005} makes the initial analysis of mean-payoff MDPs in the context of \emph{unichain} MDPs, and then extends to arbitrary MDPs, with strongly connected MDPs treated as a special case of the latter. While unichain is an important theoretical concept, in the context of formal methods and automata it is preferable to work with strongly connected MDPs. We also note that all the results in the mean-payoff sections hold also for $\MeanPayoffSup$. Almost all of the proofs are the same, with an important exception of \Cref{6-cor:mp-value-bound}, where Fatou's lemma cannot be used to prove that $\playPay(\vinit,\sigma) \leq \stepPay(\vinit,\sigma)$. Instead, we could use \emph{martingale techniques} here. Martingales are an important concept in probability theory~\cite{Williams:1991}, with applications \textit{e.g.} in analysis of infinite-state MDPs and stochastic games~\cite{Brazdil.Brozek.ea:2011}. We can use martingales to strengthen \Cref{6-lem:dual-bound-step} by showing that the probability  of $\sum_{i=0}^{n-1}\coloring(\play_i) \geq \sqrt{n}\cdot \expv^\sigma_{\vinit}[\sum_{i=0}^{n-1}\coloring(\play_i)]$ converges (with an exponential rate of decay) to $0$ as $n\rightarrow \infty$, which allows us to prove the required bound for $\limsup$.

 The notion of a (M)EC as well as many techniques we use in the EC section are due to de Alfaro, whose thesis \cite{Alfaro:1997} details the evolution of the concept and its relation to similar notions. The algorithm for MEC decomposition is taken from~\cite{Chatterjee.Henzinger:2011}, where more advanced algorithms as well as use of MECs in parity MDPs are discussed.

For an overview of literature related to verification of temporal properties in MDPs, we refer the reader to the monograph~\cite{Baier.Katoen:2008}.

MDPs are also used as a prime model in reinforcement learning (RL), one of the classical yet rapidly evolving sub-fields of AI. For RL-centric view of MDPs, we point the reader towards the textbooks \cite{Sutton.Barto:2018}.

\ifpictures
\includepdf{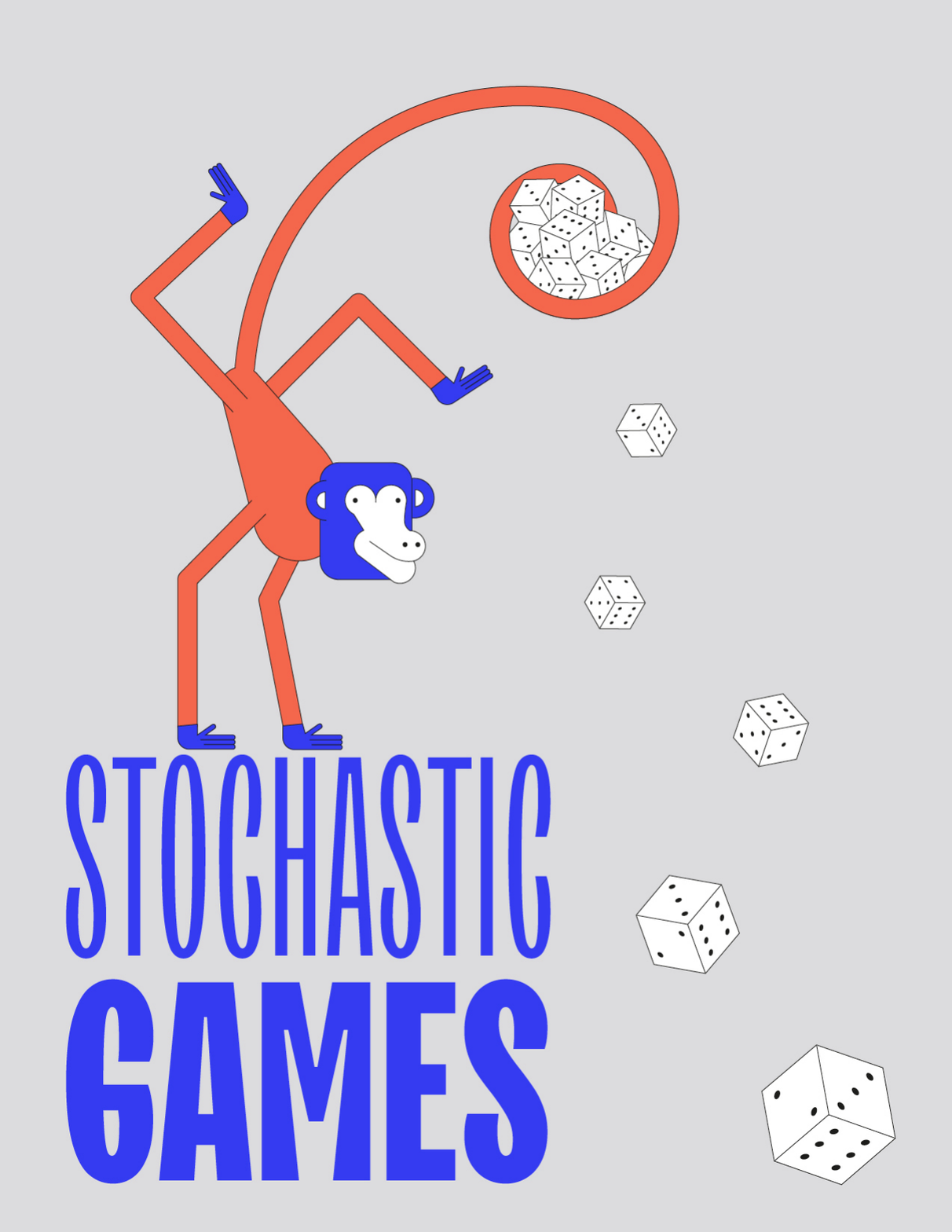}
\fi
\author[Nathalie Bertrand, Patricia Bouyer, Nathana{\"e}l Fijalkow, Mateusz Skomra]{Nathalie Bertrand, Patricia Bouyer, Nathana{\"e}l Fijalkow, Mateusz Skomra}
\copyrightline{Copyright by Nathalie Bertrand, Patricia Bouyer, Nathana{\"e}l Fijalkow, and Mateusz Skomra 2025, to be published by Cambridge University Press in the volume \textit{Games on Graphs} edited by Nathana\"el Fijalkow}

\chapter{Stochastic Games}
\chapterauthor{Nathalie Bertrand, Patricia Bouyer, Nathana{\"e}l Fijalkow, Mateusz Skomra}
\label{7-chap:stochastic}

\providecommand{\PrePos}{\text{Pre}_{>0}}
\providecommand{\PreAS}{\text{Pre}_{=1}}
\newcommand{\AttrPos}{\textrm{Attr}_{> 0}} 

\providecommand{\QuantitativeReach}{\mathtt{QuantReach}}
\providecommand{\QuantitativeReachShort}{\mathtt{QR}}

\newcommand{\adist}{\ensuremath{f}}

\newcommand{\vwin}{\Win}
\newcommand{\vlose}{\Lose}

\newcommand{\perm}{\pi}
\newcommand{\DetAtt}{\ensuremath{\textrm{DetAtt}}}

In this chapter, we introduce and review results on stochastic games
with two players. On the one hand, they extend two-player games with
random vertices; on the other hand, they extend Markov decision
processes with a second player. 

When equipped with a simple reachability objective, the objective of Max is to maximise the probability to reach the target. 
Most of the chapter focuses on stochastic games with reachability objectives. 
We show uniform positional determinacy in~\Cref{7-sec:determinacy}, using a fixed-point characterisation of the values.
\Cref{7-sec:normalisation} deals with a number of normalisation steps that are often useful when constructing algorithms for stochastic games.
This immediately yields a value iteration algorithm, discussed in~\Cref{7-sec:value_iteration}, and with a bit more work a strategy improvement algorithm, defined in~\Cref{7-sec:strategy_improvement}.
We construct two algorithms based on permutations of random vertices in~\Cref{7-sec:permutation_based_algorithms}, and conclude in~\Cref{7-sec:reductions} showing that many other stochastic games reduce to stochastic games with reachability objectives.

\section*{Notations}
\label{7-sec:notations}
There are two classical models for stochastic arenas, which we present now.
In this chapter, we will use the first one: using random vertices.
We present it first, and then quickly introduce the second one and explain why they are equivalent.

\subsection*{The model with random vertices}

Let us first define stochastic arenas.

\begin{definition}[Stochastic arenas]
\label{7-def:stochastic_arenas}
A stochastic arena is $\arena = (G,\VMax,\VMin,\VR,\delta)$ where
\begin{itemize}
	\item $G$ is a graph over the set of vertices $V$ and $V = \VMax \uplus \VMin \uplus \VR$ partitions the vertices into those controlled by Max, Min, and random vertices.
	\item $\delta : \VR \to \dist(E)$ is the probabilistic transition function. 
\end{itemize}
\end{definition}

\begin{figure}
\centering
\begin{tikzpicture}[scale=1.3]
\node[s-eve] (init) at (0,0) {$v_0$};
\node[s-random] (v1) at (2,0) {$v_1$};
\node[s-eve] (v2) at (4,0) {$v_2$};    
\node[s-random] (v3) at (6,0) {$v_3$};
\node[s-random] (v4) at (0,-1.5) {$v_4$};
\node[s-eve] (v5) at (2,-1.5) {$v_5$};
\node[s-adam] (v6) at (4,-1.5) {$v_6$};    
\node[s-eve] (v7) at (6,-1.5) {$v_7$};

\path[arrow] (init) edge (v1)
(init) edge[bend left] (v4)
(v4) edge[bend left] node[left] {$\frac 3 4$} (init)
(v4) edge node[above] {$\frac 1 4$} (v5)
(v5) edge[bend left] (v6)
(v6) edge (v7)
(v6) edge[bend left] (v5)
(v5) edge[selfloop=90]  (v5)
(v7) edge[selfloop]  (v7)
(v1) edge node[above] {$\frac 2 3$} (v2)
(v1) edge node[above right] {$\frac 1 3$} (v6)
(v2) edge (v6)
(v2) edge[bend left] (v3)
(v3) edge[bend left] node[below] {$\frac 1 2$} (v2)
(v3) edge node[right] {$\frac 1 2$} (v7)
;    

\end{tikzpicture}
\caption{Example of a stochastic arena: circle nodes are controlled by Max, square nodes by Min, and triangle nodes are random.}
\label{7-fig:ex-stoch-arena}
\commentAlt{Figure~\ref{7-fig:ex-stoch-arena}: A directed graph with 8 nodes of mixed shapes (circles, triangles, squares) and labeled edges indicating probabilities. See long description.}
\commentLongAlt{Figure~\ref{7-fig:ex-stoch-arena}: A directed graph with 8 nodes, labeled v0 to v7. Nodes v0, v2, v5, and v7 are circles. Nodes v1, v3, and v4 are triangles. Node v6 is a square. Edges are labeled with fractions.

A bidirectional arrow connects v0 (circle) and v4 (triangle). The arrow from v0 to v4 is labeled '3/4'. The arrow from v4 to v0 is not explicitly labeled on the image, but implies a corresponding inverse or alternative transition.
An arrow from v0 points to v1 (triangle).
An arrow from v1 (triangle) points to v2 (circle), labeled '2/3'.
An arrow from v1 (triangle) points to v6 (square), labeled '1/3'.
A bidirectional arrow connects v2 (circle) and v3 (triangle). The arrow from v2 to v3 is not explicitly labeled. The arrow from v3 to v2 is labeled '1/2'.
An arrow from v2 (circle) points to v6 (square).
An arrow from v3 (triangle) points to v7 (circle), labeled '1/2'.
An arrow from v4 (triangle) points to v5 (circle), labeled '1/4'.
Node v5 (circle) has a self-loop.
A bidirectional arrow connects v5 (circle) and v6 (square).
An arrow from v6 (square) points to v7 (circle).
Node v7 (circle) has a self-loop.}
\end{figure}

\begin{definition}[Stochastic games]
\label{7-def:stochastic_games}
Let $\arena$ a stochastic arena.
A qualitative stochastic game is $\game = (\arena,W)$ with $W \subseteq \Paths_\omega$ a qualitative condition,
and a quantitative stochastic game is $\game = (\arena,f)$ with $f : \Paths_\omega \to \Rinfty$ a quantitative condition.
\end{definition}

Most of the chapter will be devoted to stochastic reachability games, which are induced by $\Reach(\Win)$.
For simplicity we assume that $\Win$ is a sink (meaning a single vertex with a self-loop), the general case where $\Win$ is a subset of edges can be easily reduced to this case.
The following definitions are natural extensions of the ones given in~\Cref{1-chap:introduction} for the stochastic setting.

A \emph{strategy} for Max is a function $\sigma : \Paths \to \dist(E)$, and similarly for Min.
Note that a strategy is allowed to randomise over its actions. 
A pure strategy does not use randomisation: $\sigma : \Paths \to E$,
and a positional strategy does not use memory: $\sigma : \VMax \to \dist(E)$.

When a pair of strategies $(\sigma,\tau)$ and an initial vertex $u$ is fixed, we obtain a stochastic process: we write
$\Prob_{\sigma,\tau}^u$ for the probability measure on infinite plays. 
For instance, for $W$ a condition, we write $\Prob_{\sigma,\tau}^u(W)$ for the probability that the infinite path satisfies $W$ when Max plays $\sigma$, Min plays $\tau$, and we start from $u$. Note that we are implicitly assuming that $W$ is measurable.
For $f : \Paths_\omega \to \Rinfty$, we write $\Expectation_{\sigma,\tau}^u[f]$ for the expectation of $f$.

Let us consider a qualitative stochastic game $\Game = (\arena, W)$.
The \emph{"value" for Max} in $\game$ from $u$ is defined as 
$\ValueMax^\game(u) = \sup_{\sigma} \inf_{\tau} \Prob_{\sigma,\tau}^u(W)$, 
and symmetrically the value for Min is 
$\ValueMin^\game(u) = \inf_{\tau} \sup_{\sigma} \Prob_{\sigma,\tau}^u(W)$. 
Clearly enough, $\ValueMax^\game(u) \leq \ValueMin^\game(u)$.  

Let us now consider a quantitative stochastic game $\Game = (\arena, f)$.
The \emph{"value" for Max} in $\game$ from $u$ is defined as 
$\ValueMax^\game(u) = \sup_{\sigma} \inf_{\tau} \Expectation_{\sigma,\tau}^u[f]$, 
and symmetrically the value for Min is 
$\ValueMin^\game(u) = \inf_{\tau} \sup_{\sigma} \Expectation_{\sigma,\tau}^u[f]$. 
Clearly enough, $\ValueMax^\game(u) \leq \ValueMin^\game(u)$. 

We say that the game is \emph{determined} if $\ValueMax^\game(u) = \ValueMin^\game(u)$,
and in that case define the \emph{value} of $u$ in $\game$ as $\Value^\game(u)$. 
A strategy $\sigma$ is optimal from $u$ if $\val^{\sigma}(u) = \val^{\game}(u)$, and simply optimal if it is optimal from all vertices.

We say that the qualitative $\Omega$ is positional for Max
if for all stochastic games with objective $\Omega$, for all vertices $v$ and for all thresholds $x$,
if Max has a strategy ensuring $x$ from $v$, then he has a positional strategy ensuring $x$ from $v$.
We add the adjective purely to indicate that the strategy is pure, and uniformly to express that the same strategy can be used from all vertices.
We define as expected the notion for Min, and for bi-positional, as well as positionally determined and bi-positionally determined.
The same notions are naturally defined for qualitative and quantitative objectives.

\begin{remark}
For stochastic reachability games we also often assume that the probabilistic transition function is
\[
\delta : \VR \to \dist(V),
\]
and similarly strategies are functions $\sigma,\tau : \Paths \to \dist(V)$.
\end{remark}

Back to the example in~\Cref{7-fig:ex-stoch-arena}, 
let us consider the stochastic reachability game $\Game = (\arena, \Reach(\set{v_7}))$.
Assume that Max and Min play the following pure positional strategies:
$\sigma(v_0) = v_1$, $\sigma(v_2) = v_3$, $\sigma(v_5) = v_5$ and $\tau(v_6) = v_5$. 
Under such a strategy profile, starting in $v_0$, the probability to reach $v_7$ is 
$\Prob_{\sigma,\tau}^{v_0}(\Reach(\set{v_7})) = \frac 2 3$. 
One can show that both strategies are optimal, and $\ValueMax^\game(v_0) = \ValueMin^\game(v_0) = \frac 2 3$.

\subsection*{The model with explicit actions}

We now introduce a second model for stochastic arenas, using the notion of actions. 
This is the natural extension of the model used in~\Cref{6-chap:mdp} to two-player games.
We let $A$ be a (finite) set of actions, which is the set of choices the players can make at each step of the game.

\begin{definition}[Stochastic arenas and games -- explicit actions]
\label{7-def:stochastic_games_action}
A stochastic arena with explicit actions is $\arena = (G,\VMax,\VMin,\VR,\Delta)$ where 
\begin{itemize}
	\item $G$ is a graph over the set of vertices $V$ and $V = \VMax \uplus \VMin$,
	\item $\Delta : A \to \dist(E)$ maps actions to distributions of edges.
\end{itemize}
\end{definition}

All notions are adapted (or extended from~\Cref{6-chap:mdp}) in a natural way: for instance a strategy is a function
$\sigma : \Paths \to \dist(A)$.

The transformations between the two models are transparent. 
The key difference is that the model with random vertices allows us to naturally define a parameter: the number of random vertices, which will be important in~\Cref{7-sec:permutation_based_algorithms}.
A different parameter arises in the model with actions: the number of actions.
At the end of the day, the choice of models is a matter of taste and technical convenience.

\section{Fixed-point characterisation and positional determinacy}
\label{7-sec:determinacy}
In the same way as Martin's theorem implies determinacy for (essentially) all (non stochastic) games we consider in this book,
the following theorem by Maitra and Sudderth establishes determinacy for stochastic games:

\begin{theorem}[Determinacy of stochastic games with bounded quantitative objectives]
Stochastic games with bounded measurable quantitative objectives are determined.
\end{theorem}

Thanks to this theorem, the value is well defined in all games we consider.

\begin{theorem}[Pure bi-positional determinacy for stochastic reachability games]
\label{7-thm:determinacy}
Stochastic reachability games are uniformly purely bi-positionally determined.
\end{theorem}

Note that~\Cref{7-thm:determinacy} does not make any assumption on the game.

Before giving the proof of \Cref{7-thm:determinacy}, we establish two preliminary results. Let $\Game$ a stochastic reachability game.  
Let $Y$ the set of functions $\mu : V \to [0,1]$, it is a lattice when equipped with the component wise order.
We define the operator $\Op^{\Game} : Y \to Y$ by:
\[
\Op^{\Game}(\mu)(u) = 
\left\{\begin{array}{l@{~~}l}
    1 & \text{if}\ u = \Win, \\
    \max \set{\mu(v) : u \rightarrow v \in E} & \text{if}\ u \in \VMax, \\
    \min \set{\mu(v) : u \rightarrow v \in E} & \text{if}\ u \in \VMin, \\
    \sum_{v \in V} \delta(u)(v) \cdot \mu(v) & \text{if}\ u \in \VR.
\end{array}\right.
\]
Since $\Op^{\Game}$ is monotonic, it has a least fixed point, which is also the least pre-fixed point.

\begin{theorem}
\label{7-thm:least_fixed_point}
Let $\Game$ a stochastic reachability game. Then $\Game$ is determined and the least fixed point of $\Op^{\Game}$ computes the values of $\Game$. Furthermore, any uniform pure positional strategy $\tau$ for Min that satisfies
\[
u \in \VMin:\ \tau(u) \in \argmin \set{\val^{\Game}(v) : u \rightarrow v \in E}
\]
is optimal.
\end{theorem}

\begin{proof}
To begin, let us recall that thanks to Kleene fixed-point theorem (\Cref{1-thm:kleene}), the least fixed point of $\Op^{\Game}$ is computed as follows:
\[
\forall u \in V,\ \mu_0(u) = 0 \quad ; \quad \mu_{k+1} = \Op^{\Game}(\mu_k).
\]
We have $\mu_0 \le \mu_1 \le \dots$, and since $\Op^{\Game}$ preserves suprema, $\mu^* = \lim_k \mu_k$ is the least fixed point of $\Op^{\Game}$.
The crux here is to understand what are the values $\mu_k$ for $k = 0,1,\dots$. Let us define the truncated reachability objective:
\[
\Reach_{\le k}(\Win) = \set{\play : \exists i \le k, \play_i = \Win}.
\]
A simple induction on $k$ shows that $\mu_{k+1}$ is the values for $\Reach_{\le k}$ in $\Game$ and that both players have optimal deterministic strategies for $\Reach_{\le k}$. For any $\varepsilon > 0$, let $k$ be such that $\| \mu_{k+1} - \mu^*\|_{\infty} \le \varepsilon$ and let $\sigma$ be an optimal strategy of Max for $\Reach_{\le k}$. Then, for any $u$ and any strategy $\tau$ of Min we have $\Prob_{\sigma,\tau}^u(\Reach(\Win)) \ge \Prob_{\sigma,\tau}^u(\Reach_{\le k}(\Win)) \ge \mu^*(u) - \varepsilon$, so $\ValueMax^\game(u) \ge \mu^{*}(u)$.

To prove that $\ValueMin^\game(u) \le \mu^{*}(u)$, consider any pure positional strategy $\tau$ for Min that satisfies
\[
u \in \VMin:\ \tau(u) \in \argmin \set{\mu^*(v) : u \rightarrow v \in E}.
\]
Let $\sigma$ be any strategy of Max. We prove by induction that for all $k$, the following holds:
\[
\Expectation_{\sigma,\tau}^u[\mu^*(\pi_k)] \le \mu^*(u).
\]
In words, if we consider the distribution of vertices obtained after $k$ steps, then the expectation of $\mu^*$ on that distribution is smaller than or equal to $\mu^*(u)$.
This is clear for $k = 0$, with the inequality being an equality. 
For $k > 0$, we show that for any vertex~$u$ 
we have $\Expectation_{\sigma,\tau}^u[\mu^*(\pi_1)] \le \mu^*(u)$:
this is the one-step special case of the property above.
We distinguish four cases:
\begin{itemize}
	\item if $u = \Win$, this is clear;
	\item if $u \in \VMax$, then for all $u \rightarrow v \in E$ we have $\mu^*(v) \le \mu^*(u)$, implying the property;
	\item if $u \in \VMin$, then for $\tau(u) = u \rightarrow v$ we have $\mu^*(v) = \mu^*(u)$, implying the property;
	\item if $u \in \VR$, then $\sum_{v \in V} \delta(u)(v) \cdot \mu^*(v) = \mu^*(u)$, again implying the property.
\end{itemize}
To do the induction step for higher $k$, note that for every $v$ the conditional expected value $\Expectation_{\sigma,\tau}^u[\mu^*(\pi_{k+1}) \mid \pi_1 = v]$ is equal to $\Expectation_{\sigma',\tau}^{v}[\mu^*(\pi_{k})] \le \mu^*(v)$ for some $\sigma'$, which gives 
\[
\Expectation_{\sigma,\tau}^u[\mu^*(\pi_{k+1})] \le \sum_{v}\Prob^u_{\sigma,\tau}(\pi_1 = v)\mu^*(v) = \Expectation_{\sigma,\tau}^u[\mu^*(\pi_{1})] \le \mu^*(u)
\]
as claimed. In particular, for any $\sigma$ and any $k$ we get
\[
\Prob^u_{\sigma,\tau}(\Reach_{\le k}(\Win)) \le \Expectation_{\sigma,\tau}^u[\mu^*(\pi_k)] \le \mu^*(u),
\]
because $\mu^*(\Win) = 1$ and $\mu^* \ge 0$. Taking the limit when $k$ goes to infinity, this yields $\Prob^u_{\sigma,\tau}(\Reach(\Win)) \le \mu^*(u)$. Since $\sigma$ was arbitrary, we get $\ValueMin^\game(u) \le \mu^{*}(u)$, which implies that $\mu^*$ is the value of the game and that the strategy $\tau$ is optimal.
\end{proof}

Let us raise an important point here:
the set of values does not directly imply a pair of pure and positional optimal strategies.
It is very tempting to define them as follows
\[
\begin{array}{l}
u \in \VMax:\ \sigma(u) \in \argmax \set{\val^{\Game}(v) : u \rightarrow v \in E}, \\
u \in \VMin:\ \tau(u) \in \argmin \set{\val^{\Game}(v) : u \rightarrow v \in E}.
\end{array}
\]
and to claim that they are optimal. The optimality of $\tau$ follows from \Cref{7-thm:least_fixed_point}, but $\sigma$ may not be optimal, as shown in~\Cref{7-fig:counter_example_optimality}.
In that example, all vertices have value $1$, but the strategy $\sigma(v_0) = v_1$ is not optimal, although it can be obtained by the definition above. Some notion of progress is missing: the optimal strategy must make progress towards the target.
Note that this example is actually not a stochastic game: the optimality claim for $\sigma$ already fails for (non-stochastic) reachability games.

\begin{figure}
\centering
  \begin{tikzpicture}[scale=1.3]
    \node[s-eve] (v0) at (0,0) {$v_0$};
    \node[s-eve] (v1) at (-2,0) {$v_1$};
    \node[s-eve] (vwin) at (2,0) {$\vwin$};
    \path[arrow]
      (v0) edge[bend left] (v1)
      (v1) edge[bend left] (v0)
      (v0) edge (vwin)
      (vwin) edge[selfloop=0] (vwin);
  \end{tikzpicture}
\caption{An example of a game where some argmax strategy is not optimal.}
\label{7-fig:counter_example_optimality}
\commentAlt{Figure~\ref{thelabel}: A linear sequence of three circular nodes, with bidirectional arrows between the first two and a self-loop on the last.}
\commentLongAlt{Figure~\ref{thelabel}: The image displays three circular nodes arranged horizontally. From left to right, they are labeled 'v1', 'v0', and 'Win'. A bidirectional arrow connects 'v1' and 'v0'. A single directed arrow points from 'v0' to 'Win'. The 'Win' node has a self-loop.}
\end{figure}

To construct optimal pure and positional strategies for Max, we need to be more precise.
Let us fix $0 < \lambda < 1$ and consider the quantitative objective
\[
\QuantitativeReach(\rho) = 
  \begin{cases}
    0 & \text{if } \rho_i \neq \Win \text{ for all } i, \\
    \lambda^i & \text{for $i$ the first index such that } \rho_i = \Win.
  \end{cases}
\]
It refines the reachability objective by quantifying the number of steps required to reach $\Win$. A stochastic reachability game can be naturally interpreted as a limit of stochastic quantitative reachability games when $\lambda \to 1$. We now extend~\Cref{7-thm:least_fixed_point} to quantitative games.
Let $Y$ be the set of functions $\mu : V \to [0,1]$.
We define the operator $\Op^{\Game} : Y \to Y$ by:
\[
\Op^{\Game}(\mu)(u) = 
\left\{\begin{array}{l@{~~}l}
    1 & \text{if}\ u = \Win, \\
    \lambda \cdot \max \set{\mu(v) : u \rightarrow v \in E} & \text{if}\ u \in \VMax, \\
    \lambda \cdot \min \set{\mu(v) : u \rightarrow v \in E} & \text{if}\ u \in \VMin, \\
    \lambda \cdot \sum_{v \in V} \delta(u)(v) \cdot \mu(v) & \text{if}\ u \in \VR.
\end{array}\right.
\]
Since $\Op^{\Game}$ is monotonic, it has a least fixed point, which is also the least pre-fixed point.

\begin{theorem}
\label{7-thm:least_fixed_point_quantitative}
Let $\Game$ a stochastic quantitative reachability game. Then $\Game$ is uniformly purely bi-positionally determined, $\Op^{\Game}$ has a unique fixed point, and this point computes the values of $\Game$. Furthermore, any uniform pure positional strategies $\sigma,\tau$ that satisfy
\[
\begin{array}{l}
\forall u \in \VMax:\ \sigma(u) \in \argmax \set{\val^{\Game}(v) : u \rightarrow v \in E}, \\
\forall u \in \VMin:\ \tau(u)  \in \argmin \set{\val^{\Game}(v) : u \rightarrow v \in E}.
\end{array}
\]
are optimal.
\end{theorem}
\begin{proof}
Let us define the quantitative objective $\QuantitativeReach_{\le k}$:
\[
\QuantitativeReach_{\le k}(\rho) = 
  \begin{cases}
    0 & \text{if } \rho_i \neq \Win \text{ for all } i \le k, \\
    \lambda^i & \text{for $i \le k$ the first index such that } \rho_i = \Win.
  \end{cases}
\]
Let $\mu^*$ be any fixed point of $\Op^{\Game}$ and take any pure positional strategy $\sigma$ such that $\sigma(u) \in \argmax \set{\mu^*(v) : u \rightarrow v \in E}$ for all $u$. We prove by induction that for all $k \ge 1$, the following holds:
\[
\forall \tau,\ \Expectation_{\sigma,\tau}^u[\lambda^k \mu^*(\pi_k) + (1 - \lambda) \sum_{i = 1}^{k}\lambda^{i-1}\QuantitativeReach_{\le k-i}(\pi)] \ge \mu^*(u).
\]
To shorten the notation, we write $\QuantitativeReachShort$ instead of $\QuantitativeReach$. To prove the claim for $k = 1$ we distinguish four cases:
\begin{itemize}
	\item if $u = \Win$, then $\mu^*(u) = 1$;
	\item if $u \in \VMax$, then for $\sigma(u) = u \rightarrow v$ we have $\mu^*(v) = \lambda^{-1}\mu^*(u)$;
	\item if $u \in \VMin$, then for all $u \rightarrow v \in E$ we have $\mu^*(v) \ge \lambda^{-1}\mu^*(u)$;
	\item if $u \in \VR$, then $\sum_{v \in V} \delta(u)(v) \cdot \mu^*(v) = \lambda^{-1}\mu^*(u)$.
\end{itemize}
Therefore $\Expectation_{\sigma,\tau}^u[\lambda \mu(\pi_1) + (1 - \lambda) \QuantitativeReachShort_{\le 0}(\pi)] \ge \mu^*(u)$ for all $u$. To prove the induction step, note that the claimed inequality is satisfied as equality if $u = \Win$. Otherwise, for every $v$ there is $\tau'$ such that $\Expectation_{\sigma,\tau}^u[\mu(\pi_{k+1}) \mid \pi_1 = v] = \Expectation_{\sigma,\tau'}^v[\mu(\pi_{k})]$, $\Expectation_{\sigma,\tau}^u[\QuantitativeReachShort_{\le 0}(\pi) \mid \pi_1 = v] = 0$, and 
\[
\Expectation_{\sigma,\tau}^u[\QuantitativeReachShort_{\le j}(\pi) \mid \pi_1 = v] = \lambda \Expectation_{\sigma,\tau'}^v[\QuantitativeReachShort_{\le j-1}(\pi)]
\]
for all $j \ge 1$. Hence, by the induction assumption we get 
\[
\Expectation_{\sigma,\tau}^u [\lambda^{k+1} \mu(\pi_{k+1}) + (1 - \lambda) \sum_{i = 1}^{k+1}\lambda^{i-1}\QuantitativeReachShort_{\le k+1-i}(\pi) \mid \pi_1 = v] \ge \lambda \mu^*(v).
\]
Therefore, 
\[
\Expectation_{\sigma,\tau}^u [\lambda^{k+1} \mu(\pi_{k+1}) + (1 - \lambda) \sum_{i = 1}^{k+1}\lambda^{i-1}\QuantitativeReachShort_{\le k+1-i}(\pi)] \ge \lambda \Expectation_{\sigma,\tau}^u[\mu(\pi_1)]
\]
and the right-hand side is equal to $\Expectation_{\sigma,\tau}^u[\lambda \mu(\pi_1) + (1 - \lambda) \QuantitativeReachShort_{\le 0}(\pi)]$ because $\Expectation_{\sigma,\tau}^u[\QuantitativeReachShort_{\le 0}(\pi)] = 0$. To finish the proof, we want to show that the limit
\[
\lim_{k \to \infty} \Expectation_{\sigma,\tau}^u[\lambda^k \mu^*(\pi_k) + (1 - \lambda) \sum_{i = 1}^{k}\lambda^{i-1}\QuantitativeReachShort_{\le k-i}(\pi)]
\]
exists and is equal to $\Expectation_{\sigma,\tau}^u[\QuantitativeReachShort(\pi)]$. Since $0 \le \mu^*(u) \le 1$ for all $u$, we get the equality $\lim_{k \to \infty} \Expectation_{\sigma,\tau}^u[\lambda^k \mu^*(\pi_k)] = 0$. Since $\Expectation_{\sigma,\tau}^u[\QuantitativeReachShort_{\le k - i}(\pi)] \le \Expectation_{\sigma,\tau}^u[\QuantitativeReachShort(\pi)]$ for all $k,i$, we get 
\[
\Expectation_{\sigma,\tau}^u[(1 - \lambda) \sum_{i = 1}^{k}\lambda^{i-1}\QuantitativeReachShort_{\le k-i}(\pi)] \le (1 - \lambda^{k})\Expectation_{\sigma,\tau}^u[\QuantitativeReachShort(\pi)].
\]
This shows that 
\[
\limsup_{k \to \infty}\Expectation_{\sigma,\tau}^u[\lambda^k \mu^*(\pi_k) + (1 - \lambda) \sum_{i = 1}^{k}\lambda^{i-1}\QuantitativeReachShort_{\le k-i}(\pi)] \le \Expectation_{\sigma,\tau}^u[\QuantitativeReachShort(\pi)].
\]
Furthermore, the monotone convergence theorem gives the equality 
\[
\Expectation_{\sigma,\tau}^u[\QuantitativeReachShort(\pi)] = \lim_{k \to \infty} \Expectation_{\sigma,\tau}^u[\QuantitativeReachShort_{\le k}(\pi)].
\]
Fix $\varepsilon > 0$ and $k_0$ such that $\Expectation_{\sigma,\tau}^u[\QuantitativeReachShort(\pi)] - \varepsilon \le \Expectation_{\sigma,\tau}^u[\QuantitativeReachShort_{\le k_0}(\pi)]$. Then, for every $k > k_0$ we have the bound
\[
\Expectation_{\sigma,\tau}^u[\sum_{i = 1}^{k}\lambda^{i-1}\QuantitativeReachShort_{\le k-i}(\pi)] \ge \Expectation_{\sigma,\tau}^u[\sum_{i = 1}^{k - k_0}\lambda^{i-1}\QuantitativeReachShort_{\le k-i}(\pi)],
\]
so $\Expectation_{\sigma,\tau}^u[\sum_{i = 1}^{k}\lambda^{i-1}\QuantitativeReachShort_{\le k-i}(\pi)] \ge \frac{1 - \lambda^{k-k_0}}{1-\lambda}(\Expectation_{\sigma,\tau}^u[\QuantitativeReachShort(\pi)] - \varepsilon).$
By taking $k$ to infinity, we get $\Expectation_{\sigma,\tau}^u[\QuantitativeReachShort(\pi)] - \varepsilon \le \liminf_{k \to \infty}\Expectation_{\sigma,\tau}^u[\lambda^k \mu^*(\pi_k) + (1 - \lambda) \sum_{i = 1}^{k}\lambda^{i-1}\QuantitativeReachShort_{\le k-i}(\pi)]$. Since $\varepsilon$ was arbitrary, we get the existence of the limit. Therefore, we have shown that $\Expectation_{\sigma,\tau}^u[\QuantitativeReachShort(\pi)] \ge \mu^*(u)$ for every $\tau$. The proof for Min is analogous, proving that $\mu^*$ is the value of the game. Since the value is unique, $\Op^{\Game}$ has a unique fixed point. Moreover, our proof shows the optimality of the strategies $(\sigma,\tau)$ constructed as in the statement of the theorem.
\end{proof}

We can now use stochastic quantitative reachability games to prove the bi-positional determinacy of reachability games.

\begin{proof}[Proof of \Cref{7-thm:determinacy}]
Let $\Game$ be a stochastic reachability game. By \Cref{7-thm:least_fixed_point}, $\Game$ has a value and an optimal pure positional strategy for Min. We need to show that Max also has such a strategy. To do so, for $0 < \lambda < 1$, let $\Game_{\lambda}$ be the quantitative stochastic reachability game with parameter $\lambda$ played on the same arena as $\Game$. By \Cref{7-thm:least_fixed_point_quantitative}, $\Game_{\lambda}$ is bi-positionally determined. Furthermore, we have $0 \le \val^{\Game_\lambda}(u) \le 1$ for every $u$. Therefore, we can find an increasing sequence $\lambda_1 < \lambda_2 < \dots$ such that $\lim_{k \to \infty} \lambda_k = 1$, $\val^{\Game_{\lambda_k}}$ converges to some point, and every game $\Game_{\lambda_k}$ has the same optimal pure positional strategy $\sigma$ of Max. We will show that $\sigma$ is optimal in $\Game$. Let $\tau$ be any strategy of Min. Observe that for every $u$ and every $k$ we have
\[
\begin{array}{lll}
\val^{\Game_{\lambda_k}}(u) &\le& \Expectation_{\sigma,\tau}^u[\QuantitativeReach(\pi)] = \sum_{i = 0}^{\infty}\lambda^i\Prob_{\sigma,\tau}^u(\Reach_{=i}(\Win)) \\
& \le & \Prob_{\sigma,\tau}^u(\Reach(\Win)),
\end{array}
\]
where $\Reach_{=i}(\Win) = \{\pi \colon \pi_i = \Win, \pi_j \neq \Win \text{ for } j < i\}$. By taking $\tau$ to be the optimal pure positional strategy for Min in $\Game$, we get $\lim_{k \to \infty} \val^{\Game_{\lambda_k}}(u) \le \val^{\Game}(u)$. Furthermore, note that for every $\xi : V \to [0,1]$, we have $\|\Op^{\Game_{\lambda}}(\xi) - \Op^{\Game}(\xi)\|_{\infty} \le (1 - \lambda)$. Since $\val^{\Game_{\lambda_k}}$ is the fixed point of $\Op^{\Game_{\lambda}}$ by \Cref{7-thm:least_fixed_point_quantitative}, we get $\|\val^{\Game_{\lambda_k}} - \Op^{\Game}(\val^{\Game_{\lambda_k}})\|_{\infty} \le (1 - \lambda_k)$ for all $k$. Hence, the continuity of $\Op^{\Game}$ implies that $\val^{\Game_{\lambda_k}}$ converges to a fixed point of $\Op^{\Game}$. Therefore, by \Cref{7-thm:least_fixed_point}, we get $\lim_{k \to \infty} \val^{\Game_{\lambda_k}} = \val^{\Game}$, which implies that $\Prob_{\sigma,\tau}^u(\Reach(\Win)) \ge \val^{\Game}(u)$ for any $\tau$.
\end{proof}

\section{Normalisation: stopping, binary, simple}
\label{7-sec:normalisation}
\subsection{Normalised games}
\label{7-subsec:normalised_games}

We say that $\sigma$ is almost-surely winning from $u$ is for all strategies $\tau$ of Min,
we have $\Prob_{\sigma,\tau}^u(W) = 1$.
The almost-sure winning region is the set of vertices from where Max has an almost-surely winning strategy.
Similarly, $\sigma$ is positively winning from $u$ if for all strategies $\tau$ of Min,
we have $\Prob_{\sigma,\tau}^u(W) > 0$, and the positive winning region is the set of vertices from where Max has a positively winning strategy.
Let us write $W_{> 0}(\Game)$ for the positively winning region, and $W_{= 1}(\Game)$ for the almost surely winning region.

\begin{theorem}
\label{7-thm:positively_winning_reachability}
There exists an algorithm for computing the positively winning region of stochastic reachability games in time $O(m)$.
\end{theorem}

Analogously to attractor computations in reachability games (cf. \Cref{1-sec:attractors}), we define a one-step \emph{positive probability} predecessor operator $\PrePos$ as follows: for $X \subseteq V$ we put
\[
\begin{array}{lll}
\PrePos(X) & = & \set{u \in \VMax : \exists u \xrightarrow{} v \in E, v \in X} \\
        & \cup & \set{u \in \VMin : \forall u \xrightarrow{} v \in E,\ v \in X} \\
        & \cup & \set{u \in \VR : \exists u \xrightarrow{} v \in E,\ v \in X \text{ and } \delta(u)(u \xrightarrow{} v) > 0}.
\end{array}
\]
Let us define an operator on subsets of vertices:
\[
X \mapsto \Win \cup \PrePos(X).
\]
We note that this operator is "monotonic" when equipping the powerset of vertices with the inclusion preorder:
if $X \subseteq X'$ then $\PrePos(X) \subseteq \PrePos(X')$.
Hence \Cref{1-thm:kleene} applies: this operator has a least fixed point computed by the following sequence: 
we let $X_0 = \emptyset$ and $X_i = \Win \cup \PrePos(X_{i-1})$. 
This constructs a sequence $(X_i)_{i \in \N}$ of non-decreasing subsets of $V$. 
Hence the sequence stabilises after at most $n-1$ steps, let us write $\AttrPos(\Win)$ for the limit.
We have the following simple characterisation of the positively winning set:

\begin{lemma}[Characterisation of the positively winning set]
\label{7-lem:positive-char}
Let $\Game$ a stochastic reachability game. 
The positively winning region is the least fixed point of the operator $X \mapsto \Win \cup \PrePos(X)$.
\end{lemma}

\begin{proof}
We show by induction that for all $i$, we have $X_i \subseteq W_{> 0}(\Game)$. The case $i = 0$ is clear.
Let $v \in X_{i+1} \setminus X_i$, either $v \in \Win$ and then it is in $W_{> 0}(\Game)$, or $v$ in $\PrePos(X_i)$.
In each of the three cases ($v \in \VMax,\VMin,\VR$) we have a positive probability to land in $X_i$ at the next step. More precisely, if $v \in \VMin$, then the next visited vertex is in $X_i$, if $v \in \VR$ then the next visited vertex is in $X_i$ with positive probability, and if $v \in \VMax$, then Max can choose an edge that goes to $X_i$. Moreover, by induction hypothesis, $X_i \subseteq W_{> 0}(\Game)$. Hence, there exists a strategy of Max winning positively from $v$, implying that $v \in W_{> 0}(\Game)$. This implies that $\AttrPos(\Win) \subseteq W_{> 0}(\Game)$.

To prove the opposite inclusion, suppose that $v \in V \setminus \AttrPos(\Win)$. Since $\AttrPos(\Win)$ is a fixed point of the operator $X \mapsto \Win \cup \PrePos(X)$, we have the following cases. If $v \in \VMax \cup \VR$, then the next visited vertex is in $V \setminus \AttrPos(\Win)$ and if $v \in \VMin$, then Min can choose an edge that goes to $V \setminus \AttrPos(\Win)$. Hence, Min has a strategy that guarantees that $\AttrPos(\Win)$ is never reached from $V \setminus \AttrPos(\Win)$. Since $\AttrPos(\Win)$ contains $\Win$, we get that $v \notin W_{> 0}(\Game)$. Hence $W_{> 0}(\Game) \subseteq \AttrPos(\Win)$.
\end{proof}

Similarly as~\Cref{1-sec:attractors}, we can compute the least fixed point of the operator  $X \mapsto \Win \cup \PrePos(X)$ in time and space $O(m)$. To avoid repetitions, let us note that we can actually directly reduce to non-stochastic reachability games.
Let $\Game$ a stochastic reachability game, we construct a (non-stochastic) reachability game $\Game'$ as follows.
Max controls the vertices of Max and the random vertices, and Min the vertices of Min.
This means that for each random vertex, Max chooses an outgoing edge with positive probability.
We claim that Max has a positively winning strategy in $\Game$ from $u$ if and only if Max has a winning strategy in $\Game'$ from $u$.	
Indeed, the attractor computation in $\Game'$ is exactly the positive attractor computation in $\Game$.

\begin{theorem}
There exists an algorithm for computing the almost surely winning region of stochastic reachability games in time $O(n \cdot m)$.
\end{theorem}

In the proof, we use the following notation: $\AttrMin(F)$ is the attractor of Min of the subset $F \subseteq V$ in a non-stochastic reachability game in which Min controls both the vertices of Min and the random vertices.

\begin{lemma}[Fixed-point characterisation of the almost surely winning region for stochastic reachability games]
\label{7-lem:fixed_point_almost_sure}
Let $\Game$ be a stochastic reachability game.
\begin{itemize}
	\item If $\AttrPos(\Win) = V$, then $W_{=1}(\Game) = V$.
	\item If $\AttrPos(\Win) \neq V$ and
	we define $\Game' = \Game \setminus \AttrMin( V \setminus \AttrPos(\Win) )$,
	then $W_{=1}(\Game) = W_{=1}(\Game')$.	
\end{itemize}
\end{lemma}

\begin{proof}
We prove the first item. 
Let $\sigma$ be a positively winning pure and positional strategy from $\AttrPos(\Win) = V$.
We argue that $\sigma$ wins almost surely everywhere.
Note that there exist $N \in \N$ and $\eta > 0$ such that $\sigma$ ensures to reach $\Win$ within $N$ steps with probability at least $\eta$ from any starting vertex.
A play consistent with $\sigma$ can be divided into infinitely many finite plays of length $N$. Since each of them has probability at least $\eta$ to reach $\Win$, the infinite play has probability $1$ to reach $\Win$ by the Borel-Cantelli lemma.

We now prove the second item. To begin, let $V' = V \setminus \AttrMin(V \setminus \AttrPos(\Win))$. Note that if $v$ is a Min or a random vertex belonging to $V'$, then all the outgoing edges of $v$ go to $V'$. Likewise, if $v$ is a Max vertex belonging to $V'$, then it has at least one outgoing edge that goes to $V'$. Furthermore, $\Win \in V'$. Hence, $\Game'$ is a well defined stochastic reachability game. Moreover, an almost surely winning strategy $\sigma$ in $\Game'$ induces an almost surely winning strategy in $\Game$, so $W_{=1}(\Game') \subseteq W_{=1}(\Game)$.

To prove the opposite inclusion, we show that $\AttrMin(V \setminus \AttrPos(\Win)) \subseteq W_{< 1}(\Game)$,
the set of vertices where Min can ensure to reach $\Win$ with probability less than $1$. In the proof of \Cref{7-lem:positive-char} we have shown that Min has a strategy $\tau_c$ ensuring that $\Win$ is never reached from $V \setminus \AttrPos(\Win)$. Let $\tau_a$ denote an attractor strategy ensuring to reach $V \setminus \AttrPos(\Win)$ from $\AttrMin(V \setminus \AttrPos(\Win))$ in the non-stochastic reachability game in which Min controls both the vertices of Min and the random vertices. We construct the strategy $\tau$ as the disjoint union of $\tau_a$ and $\tau_c$. More precisely, for every $v \in \VMin \cap \AttrMin(V \setminus \AttrPos(\Win))$ we put
\[
\tau(v) = 
\begin{cases}
\tau_a(v) & \text{ if } v \in \AttrMin(V \setminus \AttrPos(\Win)) \setminus (V \setminus \AttrPos(\Win)), \\
\tau_c(v) & \text{ if } v \in V \setminus \AttrPos(\Win).
\end{cases}
\]
Any play consistent with $\tau$ and starting from $\AttrMin(V \setminus \AttrPos(\Win))$ reaches the set $V \setminus \AttrPos(\Win)$ with positive probability. If that happens, then the play becomes consistent with $\tau_c$ and it never reaches $\Win$. Thus, we have proved that $W_{=1}(\Game) \subseteq V'$. Moreover, an almost surely winning strategy of Max cannot use an edge that goes to $\AttrMin(V \setminus \AttrPos(\Win))$, and so it induces an almost surely winning strategy in $\Game'$. Hence $W_{=1}(\Game) \subseteq W_{=1}(\Game')$.
\end{proof}

The algorithm is presented in pseudocode in \Cref{7-algo:almost-sure}.
For the complexity analysis, the algorithm performs at most $n$ recursive calls
and each of them involves two positive attractor computations, implying the time complexity $O(n \cdot m)$.

\begin{algorithm}
 \KwData{A stochastic reachability game.}
 \SetKwFunction{FSolve}{Solve}
 \SetKwProg{Fn}{Function}{:}{}
 \DontPrintSemicolon

\Fn{\FSolve{$\Game$}}{
	$X \leftarrow \AttrPos(\Win)$

	\If{$X = V$}{
		\Return{$V$}
	}
	\Else{
		Let $\Game' = \Game \setminus \AttrMin(V \setminus X)$
		
		\Return{$\FSolve{$\Game'$}$}
	}
}
\caption{The quadratic time algorithm for the almost-sure winning region in stochastic reachability games.}
\label{7-algo:almost-sure}
\end{algorithm}

Another point of view on this algorithm is to directly reduce the problem to B{\"u}chi games.
Let $\Game$ a stochastic reachability game, we construct a (non-stochastic) B{\"u}chi game $\Game'$ as follows.
Max controls the vertices of Max, and Min the vertices of Min and the random vertices.
For each random vertex, Min can either choose an outgoing edge, or let Max choose one herself.
If he chooses, the edge has priority $2$, and if she does, the edge has priority $1$.
We add a self-loop over $\Win$ with priority $2$.
We claim that Max has an almost-surely winning strategy in $\Game$ from $u$ if and only if Max has a winning strategy in $\Game'$ from $u$.
Indeed, a closer look at the fixed-point computations reveals that the algorithm for B{\"u}chi games in $\Game'$ performs the exact same steps as~\Cref{7-algo:almost-sure}.

\begin{definition}[Normalised stochastic reachability games]
We say that a stochastic reachability game is normalised if there is a unique almost-surely winning vertex $\vwin$ and a unique vertex $\vlose$ which is not positively winning, and both are sinks.
\end{definition}

\begin{lemma}[Reduction to normalised games]
Let $\Game$ a stochastic reachability game, we can compute a normalised game $\Game'$ in time $O(n \cdot m)$ so that
for all vertices $u$ from $\Game$, we have $\val^{\Game}(u) = \val^{\Game'}(u)$ (unless it is $0$ or $1$, and then it is merged into $\vwin$ or $\vlose$).
\end{lemma}

We compute both the almost-surely and positively winning regions, and replace the set of almost-surely vertices by $\vwin$, and the complement of the set of positively winning vertices by $\vlose$.

\subsection{Simple stochastic games}
\label{7-subsec:ssg}

\begin{definition}[Binary and simple stochastic games]
\label{7-def:binary_stochastic_games}
We say that a stochastic reachability game is binary if every vertex except for $\vwin$ and $\vlose$ has out-degree two. A simple stochastic game is a stochastic reachability game such that:
\begin{itemize}
	\item it is binary,
	\item it is normalised,
	\item for every random vertex $u \in \VR$ we have $\delta(u) = \frac{1}{2} \cdot v + \frac{1}{2} \cdot v'$
	for some $v,v'$.
\end{itemize} 
\end{definition}

\begin{lemma}[Reduction from stochastic games to simple stochastic games]
\label{7-lem:reduction_ssg}
Let $\Game$ a stochastic reachability game, we can compute a simple stochastic game in polynomial time such that
for all vertices $u$ from $\Game$, we have $\val^{\Game}(u) = \val^{\Game'}(u)$.
The game $\Game'$ has $O(n \cdot (\log(n) + k))$ vertices, where $k$ is the number of bits required to represent probabilities in $\game$. \end{lemma}

\begin{proof}
Let $v \in \VR$ be a random vertex with $k$ outgoing edges, with probabilities $p_1, \dots, p_k$, leading to $v_1,\dots,v_k$. 
We first introduce intermediary vertices in order to build a binary tree, whose leaves are $v_1, \dots, v_k$, root is $v$, and probabilities are set at each level of the tree in order to recover $p_1, \dots, p_k$ on the respective branches. 
This introduces $O(\log(k))$ fresh vertices, and is illustrated on an example on~\Cref{7-fig:gen2binary}.

The same process can be applied to vertices in $\VMax$ and $\VMin$ (without probabilities) to reduce to out-degree $2$.

\begin{figure}[htbp]
\centering
\begin{tikzpicture}[scale=1.5]
\node[s-random-small] (init) at (-4,-.5) {$u$};
\node[s-random-small] (rv) at (-4,-1.5) {$v_2$};
\node[s-adam] (av) at (-5,-1.5) {$v_1$};    
\node[s-eve] (ev) at (-3,-1.5) {$v_3$};

\path[arrow] (init) edge[selfloop,out=90] node [right] {$\frac 1 5$} (init)
(init) edge node[left] {$\frac 1 {10}$} (av)
(init) edge node[left] {$\frac 3 {10}$} (rv)
(init) edge node[right] {$\frac 2 {5}$} (ev)
;

\node[s-random-small] (root) at (0,0) {$u$};
\node[s-random-small] (left) at (-.75,-1) {};
\node[s-random-small] (right) at (.75,-1) {};
\node[s-adam] (lleft) at (-1.5,-2) {$v_1$};
\node[s-random-small] (rleft) at (0,-2) {$v_2$};
\node[s-eve] (rright) at (.75,-2) {$v_3$};

\path[arrow] (root) edge node[left] {$\frac 2 5$}  (left)
(root) edge node[right] {$\frac 3 5$}  (right)
(left) edge node[left] {$\frac 1 4$} (lleft)
(left) edge node[right] {$\frac 3 4$} (rleft)
(right) edge node[right] {$\frac 2 3$} (rright)
(right) edge[out=70,in=0] node[right] {$\frac 1 3$} (root)
;
\end{tikzpicture}
\caption{From general random vertices to binary ones.}
\label{7-fig:gen2binary}
\commentAlt{Figure~\ref{7-fig:gen2binary}: Two tree-like diagrams side-by-side, each originating from a root node 'u' and branching into different shapes, with labeled edges representing fractions. See long description.}
\commentLongAlt{Figure~\ref{7-fig:gen2binary}: The image displays two distinct tree diagrams.

Left Diagram:
A triangular node labeled 'u' is at the top. It has a self-loop labeled '1/3'.
Three arrows emanate from 'u':

One points left to a square node 'v1', labeled '1/10'.
Another points downwards to a triangular node 'v2', labeled '3/10'.
A third points right to a circular node 'v3', labeled '2/5'.
Right Diagram:
A triangular node labeled 'u' is at the top.
Two arrows emanate from 'u':

One points left and downwards to an unlabeled triangular node, labeled '2/5'. This unlabeled node then branches further:
An arrow points left to a square node 'v1', labeled '1/4'.
An arrow points right to a triangular node 'v2', labeled '3/4'.
The other arrow from 'u' points right and downwards to another unlabeled triangular node, labeled '3/5'. This unlabeled node then branches:
An arrow points downwards to a circular node 'v3', labeled '2/3'.
Additionally, there is a curved arrow from the unlabeled triangular node on the right (which receives '3/5' from 'u') pointing back to the 'u' node, labeled '1/3'.}
\end{figure}

It remains to explain how to simulate a distribution $\delta(u) = \frac{p}{q} \cdot v + \frac{q-p}{q} \cdot v'$ using only probabilities $\frac 1 2$.
Let us denote the binary encodings (with most significant bit first) of $p$ and $q - p$ as $a_1 \cdots a_t$ resp. $b_1 \cdots b_t$.
We build the following gadget. 
The input vertex is $u$ and for every $i \in [2, t+1]$, it has two exit edges with accumulated probabilities $2^{-i}$. 
Now, if $a_i = 1$, one outgoing edge leads to $v$, and similarly if $b_i = 1$, then one outgoing edge leads to $v'$.
The remaining edges are redirected to $u$.

The transformation is illustrated in~\Cref{7-fig:simul}, with $p = 11$ and $q = 14$. The binary encodings are $1011$ for $p$ and $0011$ for $p-q = 3$.  
For simplicity some vertices are represented several times to avoid intricate transitions. 
One can check that this gadget indeed simulates probabilities $\frac p q$ to $v$ and $\frac {q-p} q$ to $v'$.

\begin{figure}[htbp]
\centering
\begin{tikzpicture}

\node[s-random-small] (rv) at (-3,-.5) {$u$};
\node[s-eve] (ev) at (-4,-1.5) {$v$};
\node[s-adam] (av) at (-2,-1.5) {$v'$};

\path[-latex'] 
(rv) edge node[left] {$\frac {11} {14}$} (ev)
(rv) edge node[right] {$\frac 3 {14}$} (av)
;

\node[s-random-small] (root) at (0,0) {$u$};
\node[s-random-small] (l02) at (2,0) {};
\node[s-random-small] (l03) at (4,0) {};
\node[s-random-small] (l04) at (6,0) {};
\node[s-random-small] (l11) at (0,-1) {};
\node[s-random-small] (l12) at (2,-1) {};
\node[s-random-small] (l13) at (4,-1) {};
\node[s-random-small] (l14) at (6,-1) {};

\node[s-eve] (l21) at (-.5,-2) {$v$};
\node (l21t) at (-.5,-2.8) {$\frac 1 4$};
\node[s-random-small] (l22) at (.5,-2) {$u$};
\node (l22t)  at (.5,-2.8) {$\frac 1 4$};
\node[s-random-small] (l23) at (1.5,-2) {$u$};
\node (l23t)  at (1.5,-2.8) {$\frac 1 8$};
\node[s-random-small] (l24) at (2.5,-2) {$u$};
\node (l24t)  at (2.5,-2.8) {$\frac 1 8$};
\node[s-eve] (l25) at (3.5,-2) {$v$};
\node (l25t)  at (3.5,-2.8) {$\frac 1 {16}$};
\node[s-adam] (l26) at (4.5,-2) {$v'$};
\node(l26t)  at (4.5,-2.8) {$\frac 1 {16}$};
\node[s-eve] (l27) at (5.5,-2) {$v$};
\node (l27t)  at (5.5,-2.8) {$\frac 1 {32}$};
\node[s-adam] (l28) at (6.5,-2) {$v'$};
\node (l28t)  at (6.5,-2.8) {$\frac 1 {32}$};

\path[-latex']
(root) edge (l11)
(root) edge (l02)
(l02) edge (l03)
(l02) edge (l12)
(l03) edge (l04)
(l03) edge (l13)
(l11) edge (l21)
(l11) edge (l22)
(l12) edge (l23)
(l12) edge (l24)
(l13) edge (l25)
(l13) edge (l26)
(l14) edge (l27)
(l14) edge (l28)
(l04) edge (l14)
(l04) edge[bend right] (root)
;

\end{tikzpicture}
\caption{From binary random vertices to binary uniform ones.}
\label{7-fig:simul}
\end{figure}
\end{proof}

\subsection{Stopping games}
The next lemmas will be useful in multiple proofs throughout the next sections. Note that if $(\sigma, \tau)$ is a pair of pure positional strategies of Max and Min, then the stochastic process induced on $V$ by fixing $(\sigma,\tau)$ is a Markov chain in which the state $\Win$ is absorbing. We denote by $C^{\sigma,\tau}$ the set of recurrent states in this Markov chain and we put $C^{\sigma} = \bigcup_{\tau} C^{\sigma,\tau}$.

\begin{lemma}\label{7-le:policy_from_inequality}
Let $\Game$ be a stochastic reachability game and $\mu : V \to \R$ be any function. Take a pair of pure positional strategies $(\sigma,\tau)$ such that 
\[
\begin{array}{l}
\forall u \in \VMax:\ \sigma(u) \in \argmax \set{\mu(v) : u \rightarrow v \in E},\\
\forall u \in \VMin:\ \tau(u) \in \argmin \set{\mu(v) : u \rightarrow v \in E}.
\end{array}
\]
Then, we have the following two properties.
\begin{enumerate}
\item If $\mu$ satisfies $\mu \le \Op^{\Game}(\mu)$, $\mu(\Win) = 1$, and $\mu(u) = 0$ for all $u \in C^{\sigma} \setminus \Win$, then $\val^{\Game} \ge \val^{\sigma} \ge \mu$.
\item If $\mu$ satisfies $\mu \ge  \Op^{\Game}(\mu)$ and $\mu \ge 0$, then $\val^{\Game} \le \val^{\tau} \le \mu$.
\end{enumerate}
\end{lemma}
Before giving a proof, we note that in this lemma the operator $\Op^{\Game}$ is defined not only on functions $\mu : \vertices \to [0,1]$, but more generally on functions $\mu : \vertices \to \R$.
\begin{proof}
We start by proving the first property. If $C^{\sigma} = \vertices$, then the claim is trivial. Otherwise, let $\tau'$ be any pure positional strategy of Min and $P$ be the transition matrix of the Markov chain obtained by fixing $(\sigma, \tau')$. By definition of $\mu$ and $\sigma$ we have $\mu \le \Op^{\Game}(\mu) \le P\mu$. Let $Q$ denote the matrix obtained from $P$ by removing the rows and columns indexed by $C^{\sigma,\tau'}$. Furthermore, let $z$ be the vector defined as $z_u = P_{u,\Win}$ for all $u \in \vertices \setminus C^{\sigma,\tau'}$. Then, the theory of absorbing Markov chains, see, \textit{e.g.}, \cite[Chapter~III]{Kemeny.Snell:1976}, implies that $Q^N \to 0$, $(I - Q)^{-1} = I + Q + Q^2 + \dots$, and $\Prob_{\sigma,\tau'}^u(\Reach(\Win)) = ((I - Q)^{-1}z)_u$ for every $u \in \vertices \setminus C^{\sigma,\tau'}$. Let $\rho : (\vertices \setminus C^{\sigma,\tau'}) \to \R$ denote the function obtained by restricting $\mu$ to the vertices in $\vertices \setminus C^{\sigma,\tau'}$. Then, $\mu \le P \mu$ implies that $\rho \le z + Q \rho$ (note that here we use the fact that $\mu(\Win) = 1$ and $\mu(u) = 0$ for $u \in C^{\sigma,\tau'}$). Hence, for every $N \ge 1$ we get $\rho \le (I + Q + \dots + Q^N)z + Q^{N+1} \rho \to (I - Q)^{-1}z$. In particular, $\Prob_{\sigma,\tau'}^u(\Reach(\Win)) \ge \mu(u)$ for all $u$. Since $\tau'$ was arbitrary, positional determinacy implies that $\val^{\Game} \ge \val^{\sigma} \ge \mu$.

The proof of the second property is analogous. Note that $\mu(\Win) \ge 1$ by definition. Let $\sigma'$ be any pure positional strategy of Max and $P$ be the transition matrix of the Markov chain obtained by fixing $(\sigma', \tau)$. If $C^{\sigma',\tau} = \vertices$, then $\Prob_{\sigma',\tau}^u(\Reach(\Win)) = 0$ for all $u \neq \Win$ so $\mu(u) \ge \Prob_{\sigma',\tau}^u(\Reach(\Win))$ for all $u$. Otherwise, let $Q,z,\rho$ be defined as in the previous proof. By definition of $\tau,\mu$ we get $\mu \ge P\mu$, which implies that $\rho \ge z + Q\rho$ (note that this implication only uses the fact that $\mu(\Win) \ge 1$ and $\mu \ge 0$). As previously, this inequality implies that $\mu(u) \ge \Prob_{\sigma',\tau}^u(\Reach(\Win))$ for all $u$, which gives the claim.
\end{proof}

The second lemma characterises absorption probabilities in Markov chains with rational transition probabilities. We refer to~\cite{Auger.Montjoye.ea:2021,Skomra:2021} for the proof.

\begin{lemma}\label{7-le:rational_absorption}
Suppose that $P$ is a square stochastic matrix with rational entries whose common denominator is $D$. Furthermore, let $K$ denote the number of rows of $P$ with at least two non-zero entries and suppose that $P_{vv} = 1$ for some index $v$. Consider a Markov chain with transition matrix given by $P$. Then, there exists a natural number $M \le D^K$ such that for every state $u$ the probability that the chain starting at $u$ reaches $v$ is a rational number of the form $a_{uv}/M$, where $a_{uv} \in \N$.
\end{lemma}

To apply the lemma above to stochastic reachability games, we use the following notation. We suppose that $D$ is the common denominator of all transition probabilities of $\Game$ and that $K$ is the number of \emph{significant} random vertices, \textit{i.e.}, random vertices $u$ such that $\delta(u)(v) > 0$ for at least two different $v$.

\begin{corollary}\label{7-cor:value_denominator}
The value of the game $\Game$ is a vector of rational numbers whose lowest common denominator is not greater than $D^K$.
\end{corollary}
\begin{proof}
Let $(\sigma,\tau)$ be a pair of optimal pure positional strategies in $\Game$. Then, $\val^{\Game}(u) = \Prob_{\sigma,\tau}^u(\Reach(\Win))$ for all $u$. Hence, the claim follows from \Cref{7-le:rational_absorption}.
\end{proof}

\begin{remark}
We note that, as discussed in \cite{Auger.Montjoye.ea:2021,Skomra:2021}, the bound $M \le D^K$ in \Cref{7-le:rational_absorption} and \Cref{7-cor:value_denominator} can be improved in the following way: if we denote by $d_u$ the lowest common denominator of the transition probabilities of a random vertex $u$, then $M \le \prod_u d_u \le D^K$. For the sake of simplicity, we use the bound $M \le D^K$ in our proofs, but this can be replaced by $M \le \prod_u d_u$.
\end{remark}

We now define the stopping stochastic reachability games. To do so, we suppose that the graph of the game is equipped with a special state $\Lose$ that is distinct from $\Win$ and that has only one outgoing edge, which is a loop going back to $\Lose$. In particular, if a game reaches $\Lose$, then it never leaves this state, and player Max cannot reach $\Win$ any more.

\begin{definition}[Stopping stochastic reachability games]
\label{7-def:stopping_stochastic_reachability_games}
A stochastic reachability game is stopping if for every vertex $u \in V$ and every pair of pure positional strategies $(\sigma,\tau)$ we have $\Prob_{\sigma,\tau}^u(\Reach(\{\Win,\Lose\})) =1 $.
\end{definition}

The following lemma shows that the condition of~\Cref{7-def:stopping_stochastic_reachability_games} can be replaced by a seemingly weaker one.

\begin{lemma}\label{7-le:stopping_game_weaker}
A stochastic reachability game is stopping if for every vertex $u \in V$ and every pair of pure positional strategies $(\sigma,\tau)$ we have $\Prob_{\sigma,\tau}^u(\Reach(\{\Win,\Lose\})) > 0$.
\end{lemma}
\begin{proof}
Suppose that a Markov chain defined by $(\sigma, \tau)$ can reach the set $\{\Win,\Lose\}$ from any starting state $u$. Since the states $\Win, \Lose$ are absorbing, this implies that they are the only recurrent states of the chain, so $\Prob_{\sigma,\tau}^u(\Reach(\{\Win,\Lose\})) =1$.
\end{proof}


\begin{theorem}[Fixed-point characterisation for stopping stochastic reachability games]
\label{7-thm:fixed_point_characterisation_stopping_ssg}
Let $\game$ be a stopping stochastic reachability game. Then, the operator $\Op^{\Game}$ has a unique fixed point $\mu$ such that $\mu(\Lose) = 0$, namely $\mu = \val^{\Game}$. Moreover, pure positional strategies $(\sigma,\tau)$ of Max and Min are optimal if and only if they satisfy
\[
\begin{array}{l}
\forall u \in \VMax:\ \sigma(u) \in \argmax \set{\val^{\Game}(v) : u \rightarrow v \in E}, \\
\forall u \in \VMin:\ \tau(u)  \in \argmin \set{\val^{\Game}(v) : u \rightarrow v \in E}.
\end{array}
\]
\end{theorem}
\begin{proof}
Let $\mu$ be any fixed point of $\Op^{\Game}$ such that $\mu(\Lose) = 0$ (by~\Cref{7-thm:least_fixed_point}, $\val^{\Game}$ is one such point). Let $\sigma,\tau$ be a pair of pure positional strategies such that 
\[
\sigma(u) \in \argmax \set{\mu(v) : u \rightarrow v \in E}
\]
for all $u \in \VMax$ and $\tau(u) \in \argmin \set{\mu(v) : u \rightarrow v \in E}$ for all $u \in \VMin$. Since the game is stopping, we have $C^{\sigma} = \{\Win, \Lose\}$. In particular, the pair $(\mu, \sigma)$ satisfies the conditions of~\Cref{7-le:policy_from_inequality} and therefore $\mu \le \val^{\Game}$. Likewise, the pair $(\mu, \tau)$ satisfies the conditions of~\Cref{7-le:policy_from_inequality}, so $\mu \ge \val^{\Game}$. Thus, $\mu = \val^{\Game}$ and~\Cref{7-le:policy_from_inequality} shows that $\sigma,\tau$ are optimal. Suppose now that $\tau$ is a pure positional strategy of Min such that $\tau(u) \notin \argmin \set{\val^{\Game}(v) : u \rightarrow v \in E}$ for some $u \in \VMin$. Then, $\val^{\Game} \neq \Op^{\Game[\tau]}(\val^{\Game})$. Hence, by \Cref{7-thm:least_fixed_point}, $\val^{\tau} \neq \val^{\Game}$ and $\tau$ is not optimal. The proof for player Max is analogous.
\end{proof}

\begin{lemma}[Reduction to stopping games]
\label{7-lem:reduction_stopping_games}
Let $\Game$ a stochastic reachability game. We can compute a stopping game $\Game'$ such that any pair of optimal pure stationary strategies in $\Game'$ is also optimal in $\Game$. Moreover, for all vertices $u$ in $\Game$, we have $\val^{\Game}(u) > \frac{1}{2}$ if and only if $\val^{\Game'}(u) > \frac{1}{2}$.
\end{lemma}
\begin{proof}
For every $0 < \varepsilon < 1$, we construct the game $\Game^{\varepsilon}$ in the following way. First, we add a state $\Lose$ to $\Game$. Then, for every edge $u \rightarrow v \in E$ such that $u \neq \Lose$ we add a new random vertex $w_{uv}$ to the graph. We remove the edge $u \rightarrow v$ and add the edges $u \rightarrow w_{uv}$, $w_{uv} \rightarrow v$, and $w_{uv} \to \Lose$, as in~\Cref{7-fig:reduction_to_stopping}. We also define the transition function $\delta^{\varepsilon}$ of the new game as follows. For every random state $u$ of the original game we put $\delta^{\varepsilon}(u)(w_{uv}) = \delta(u)(v)$ for all $v$, and for every new state $w_{uv}$ we put $\delta^{\epsilon}(w_{uv})(\Lose) = \varepsilon$ and $\delta^{\epsilon}(w_{uv})(v) = 1 - \varepsilon$. 

\begin{figure}[htbp]
\centering
\begin{tikzpicture}[scale=1.5]
\node[s-adam] (u) at (-5,1) {$u$};    
\node[s-eve] (v) at (-3.5,1) {$v$};
\node at (-2.5,1) {becomes};

\path[arrow] 
(u) edge (v);

\node[s-adam] (u2) at (-1.5,1) {$u$};    
\node[s-eve] (v2) at (.95,1) {$v$};
\node[s-random-small] (uv) at (-.4,1) {$w_{uv}$};
\node[s-eve] (lose) at (-.4,-.2) {$\vlose$};

\path[arrow] 
(u2) edge (uv)
(uv) edge node[above] {$1 - \varepsilon$} (v2)
(uv) edge node[right] {$\varepsilon$} (lose)
(lose) edge[selfloop=0] (lose);
;

\end{tikzpicture}
\caption{From reachability games to stopping games.}
\label{7-fig:reduction_to_stopping}
\commentAlt{Figure~\ref{7-fig:reduction_to_stopping}: A diagram showing a transformation from a direct edge between two nodes to an expanded structure involving an intermediate node and an alternative "Lose" state.}
\commentLongAlt{Figure~\ref{7-fig:reduction_to_stopping}: The image displays a transformation represented by two diagrams separated by the word "becomes".

The left diagram shows a simple directed edge from a square node labeled 'u' to a circular node labeled 'v'.

The right diagram shows the transformed structure. It starts with the same square node 'u', which now points to a triangular node labeled 'w_uv'. From 'w_uv', two paths diverge:

An arrow labeled '1 - epsilon' points to the original circular node 'v'.
An arrow labeled 'epsilon' points downwards to a new circular node labeled 'Lose'. The 'Lose' node has a self-loop.}
\end{figure}

We also consider the game $\Game^{0}$ obtained by putting $\varepsilon = 0$, \textit{i.e.}, a game that is identical to the original game $\Game$ except for the fact that we added an isolated vertex $\Lose$ to the graph and we put a dummy vertex on every edge of the graph. In particular, the game $\Game^{0}$ has the same optimal pure stationary strategies as $\Game$. Furthermore, every vertex of $\Game$ has the same value in $\Game$ and in $\Game^0$. Let $D$ denote the lowest common denominator of all transition probabilities in $\Game$. Furthermore, let $K$ be the number of random vertices in $\Game$. By \Cref{7-cor:value_denominator}, the values of the game $\Game^{0}$ are rational numbers with lowest common denominator not higher than $D^{K}$. Let $\varepsilon = \min\{\frac{1}{4},D^{-2K}\}$. We will show that $\Game' = \Game^{\varepsilon}$ satisfies the claim.

To begin, note that $\Game'$ is stopping. Indeed, if we fix any pair of pure positional strategies $(\sigma,\tau)$ in $\Game'$ and suppose that the resulting Markov starts at some state $u$, then this chain can reach the set $\{\Win, \Lose\}$, so $\Game'$ is stopping by~\Cref{7-le:stopping_game_weaker}. Furthermore, note that if $\xi$ is any vector with entries in $[0,1]$ and such that $\xi_{\Lose} = 0$, then $\Op^{\Game'}(\xi) \le \Op^{\Game^0}(\xi) \le \varepsilon + \Op^{\Game'}(\xi)$, where the notation $\varepsilon + \Op^{\Game'}(\xi)$ means that we add $\varepsilon$ to all coordinates of $\Op^{\Game'}(\xi)$. Let $\xi = \val^{\Game'}$ and suppose that $\sigma,\tau$ are optimal pure positional strategies of Max and Min in $\Game'$. We want to show that $(\sigma, \tau)$ are also optimal in $\Game^0$. By~\Cref{7-thm:fixed_point_characterisation_stopping_ssg}, we have $\sigma(u) \in \argmax \set{\xi(v) : u \rightarrow v \in E}$ for all $u \in \VMax$ and $\tau(u) \in \argmin \set{\xi(v) : u \rightarrow v \in E} = \argmin \set{\varepsilon + \xi(v) : u \rightarrow v \in E}$ for all $u \in \VMin$. Since $\xi \ge 0$,~\Cref{7-le:policy_from_inequality} implies that $\val^{\Game^0} \le \varepsilon + \xi$ and that the same is true if Min uses $\tau$ in $\Game^0$. Suppose that $u$ is any state in $\Game^0$ different than $\Win$ and that there exists a pure positional strategy $\tau'$ of Min such that $u$ is recurrent in the Markov chain induced by fixing $(\sigma,\tau')$ in $\Game^0$. In this situation, the chain starting at $u$ cannot reach $\Win$. Hence, if we consider the Markov chain induced by fixing $(\sigma,\tau')$ in $\Game'$, then this chain also cannot reach $\Win$ from $u$. By optimality of $\sigma$ this implies that $\xi_u = 0$. Since $\xi_{\Win} = 1$,~\Cref{7-le:policy_from_inequality} implies that $\val^{\Game'} \le \val^{\Game^0}$ and the same is true if Max uses $\sigma$ in $\Game^0$. Hence, we established the inequality $\val^{\Game'} \le \val^{\Game^0} \le \varepsilon + \val^{\Game'}$. Let $u$ be any state. Note that, by definition of $\varepsilon$, the interval $[\val^{\Game'}(u), \varepsilon + \val^{\Game'}(u)]$ can contain at most one rational number of denominator at most $D^{K}$. Moreover, $\val^{\Game^0}(u)$ is such a number. If we now consider  the one-player game obtained from $\Game^{0}$ by fixing $\tau$, then \Cref{7-cor:value_denominator} shows that the values of this game are also rational numbers of denominator at most $D^K$. Moreover, as shown above, these values are between $\val^{\Game^0}$ and $\varepsilon + \val^{\Game'}$. Hence, they must be equal to $\val^{\Game^0}$ and $\tau$ is optimal in $\Game^0$. By the same reasoning, $\sigma$ is optimal in $\Game^0$. To show the last claim, note that $\val^{\Game'}(u) > \frac{1}{2}$ implies $\val^{\Game^0}(u) > \frac{1}{2}$. Conversely, since $\val^{\Game^0}(u)$ has denominator at most $D^K$, if $\val^{\Game^0}(u) > \frac{1}{2}$, then $\val^{\Game^0}(u) > \varepsilon + \frac{1}{2}$, so $\val^{\Game'}(u) > \frac{1}{2}$.
\end{proof}



%
%

\section{A value iteration algorithm}
\label{7-sec:value_iteration}
We now present a value iteration algorithm for stochastic reachability games. This algorithm will rely on a appropriate rounding procedure. In this procedure, we want to approximate a given rational number from above by a number with bounded denominator. To make this more precise, given a positive natural number $M$, we denote by $\mathcal{Q}_M$ the set
\[
\mathcal{Q}_M = \set{\frac{p}{q} \colon p \in \{0,1, \dots, M\}, q \in \{1,2,\dots,M\}}.
\]

\begin{theorem}\label{7-thm:rounding}
Given a rational number $\alpha \in [0,1]$ and $M \ge 1$ we can find the smallest number $\beta \in \mathcal{Q}_M$ that satisfies $\alpha \le \beta$ in $O(\log M)$ arithmetic operations.
\end{theorem}
\begin{proof}
Let $\beta$ be the smallest number in $\mathcal{Q}_M$ that satisfies $\alpha \le \beta$. Note that if $\gamma$ is any number in $\mathcal{Q}_M$, then it is trivial to decide if $\beta \le \gamma$. Indeed, if $\alpha \le \gamma$, then $\beta \le \gamma$ and if $\alpha > \gamma$, then $\beta > \gamma$. Hence, the exact value of $\beta$ can be found in $O(\log M)$ operations using the rational search technique described in~\cite{Kwek.Mehlhorn:2003}.
\end{proof}

We denote by $\textsc{RoundUp}(\mu,M)$ the procedure that takes as an input $M \ge 1$ and a function $\mu : \vertices \to [0,1] \cap \Q$ and rounds up all values of $\mu$ using the procedure from \Cref{7-thm:rounding}. The pseudocode of the value iteration algorithm for stochastic reachability games is given in \Cref{7-algo:value_iteration}.

\begin{algorithm}[ht]
 \KwData{A stochastic reachability game.}
 \DontPrintSemicolon

Choose $\mu$ such that $\mu \le \val^{\game}$ and $\mu \le \Op^{\Game}(\mu)$ 
     
\Repeat{$\Op^{\Game}(\mu) = \mu$}{
	$\mu \leftarrow \textsc{RoundUp}(\Op^{\Game}(\mu),D^{K})$
}

\Return{$\mu$}
\caption{The value iteration algorithm.}
\label{7-algo:value_iteration}
\end{algorithm}

%
%
%

To analyse the algorithm, we use the following lemmas. To simplify the notation, we denote $\mathbb{B}(\mu) = \textsc{RoundUp}(\Op^{\Game}(\mu),D^{K})$ for every function $\mu \colon V \to [0,1] \cap \Q$.

\begin{lemma}\label{7-le:monotonic_rounding}
The operator $\mu \mapsto \mathbb{B}(\mu)$ is monotonic.
\end{lemma}
\begin{proof}
Suppose that $0 \le \alpha_1 \le \alpha_2 \le 1$ are two rational numbers. If $\beta_i \in \mathcal{Q}_M$ is the smallest number such that $\beta_i \ge \alpha_i$ for $i \in \{1,2\}$, then we have $\beta_2 \ge \alpha_2 \ge \alpha_1$, so $\beta_2 \ge \beta_1$ by definition. Hence, the operator $\mu \to \textsc{RoundUp}(\mu,D^K)$ is monotonic. Therefore, $\mathbb{B}(\mu)$ is monotonic as a composition of monotonic operators.
\end{proof}

\begin{lemma}\label{7-le:value_iter_nondecreasing}
If $\mu_0 \le \Op^{\Game}(\mu_0)$, then the sequence $\mu_{k+1} = \mathbb{B}(\mu_k)$ produced by the value iteration algorithm is nondecreasing, $\mu_k \le \mu_{k+1}$.
\end{lemma}
\begin{proof}
We have $\mu_0 \le \Op^{\Game}(\mu_0) \le \mathbb{B}(\mu_0) = \mu_1$ by the assumption. Hence, the monotonicity of $\mathbb{B}$ gives $\mu_k \le \mu_{k+1}$ for all $k$.
\end{proof}

\begin{theorem}[Starting condition for value iteration]
\label{7-thm:start_value_iter}
Suppose that $\mu_0 = 0$ or $\mu_0 = \val^{\sigma}$ for some pure positional strategy $\sigma$ of Max. Then, $\mu_0$ satisfies the initial condition of value iteration, \textit{i.e.}, $\mu_0 \le \val^{\Game}$ and $\mu_{0} \le \Op^{\Game}(\mu_0)$.
\end{theorem}
\begin{proof}
Both claims are trivial if $\mu_0 = 0$. Moreover, $\val^{\sigma} \le \val^{\Game}$ by positional determinacy. To prove the last claim, note that $\val^{\sigma} = \Op^{\Game[\sigma]}(\val^{\sigma})$ by~\Cref{7-thm:least_fixed_point} and $\Op^{\Game[\sigma]}(\val^{\sigma}) \le \Op^{\Game}(\val^{\sigma})$ by definition of $\Op^{\Game}$.
\end{proof}

\begin{theorem}[Correctness of value iteration]
\label{7-thm:value_iteration}
The value iteration algorithm halts in at most $nD^{2K}$ iterations and outputs a function $\mu$ such that $\mu = \val^{\Game}$.
\end{theorem}
\begin{proof}
Let $\mu_0$ be such that $\mu_0 \le \val^{\Game}$ and $\mu_0 \le \Op^{\Game}(\mu_0)$. By \Cref{7-le:value_iter_nondecreasing}, the sequence $\mu_{k+1} = \mathbb{B}(\mu_k)$ is nondecreasing. Furthermore, by \Cref{7-thm:least_fixed_point}, $\val^{\Game}$ is a fixed point of $\Op^{\Game}$ and by \Cref{7-cor:value_denominator} it is also a fixed point of $\textsc{RoundUp}(x,D^K)$, so it is a fixed point of $\mathbb{B}$. Hence, by the monotonicity of $\mathbb{B}$ we have $\mu_k \le \val^{\Game}$ for all $k$. In particular, if the algorithm halts, then it outputs a fixed point of $\Op^{\Game}$ that is not higher than $\val^{\Game}$. Therefore, by \Cref{7-thm:least_fixed_point}, the algorithm outputs $\val^{\Game}$ when it halts. To show that the algorithm halts, consider a sequence $\xi_k$ defined as $\xi_0 = \mu_0$ and $\xi_{k+1} = \Op^{\Game}(\xi_k)$, \textit{i.e.}, a sequence produced by value iteration without rounding. The monotonicity of $\Op^{\Game}$ shows that $\xi_k$ is nondecreasing, $\xi_k \le \xi_{k+1}$, and \Cref{7-thm:least_fixed_point} shows that $\xi_k \le \val^{\Game}$ for all $k$. In particular, the sequence $\mu_k$ converges to $\mu^{*} = \sup_k \mu_k$ (where we apply the supremum to all coordinates of the sequence), $\xi_k$ converges to $\xi^{*} = \sup_k \xi_k$, and we have $\mu^* \le \val^{\Game}$ and $\xi^* \le \val^{\Game}$. Furthermore, by induction we have $\xi_k \le \mu_k$ for all $k$. Indeed, $\xi_0 \le \mu_0$ by definition, and if $\xi_k \le \mu_k$, then $\xi_{k+1} = \Op^{\Game}(\xi_k) \le \Op^{\Game}(\mu_k) \le \mathbb{B}(\mu_k) = \mu_{k+1}$. Hence, we get $\xi^* \le \mu^* \le \val^{\Game}$. Moreover, by continuity of $\Op^{\Game}$, the sequence $\xi_k$ also converges to $\Op^{\Game}(\xi^*)$, which means that $\xi^*$ is a fixed point of $\Op^{\Game}$. Since $\xi^* \le \val^{\Game}$, \Cref{7-thm:least_fixed_point} shows that $\xi^* = \mu^* = \val^{\Game}$. To finish the proof, note that $\mu_k$ is a vector of rational numbers in $[0,1]$ with denominators bounded by $D^K$. Since there are only finitely many such numbers, the sequence $(\mu_k)$ stabilises after finitely many steps, \textit{i.e.}, there exists the smallest $k^*$ such that $\mu_{k} = \mu^* = \val^{\Game}$ for all $k\ge k^*$. Hence, the algorithm halts in $k^*$ iterations. To bound $k^*$ note that for $k < k^*$ we have $\mu_k \le \mu_{k+1}$ and $\mu_k \neq \mu_{k+1}$. Therefore, at least one coordinate of $\mu_{k+1}$ is strictly larger than in $\mu_k$. Since these coordinates are rational numbers of denominator at most $D^{K}$, they differ by at least $D^{-2K}$. Given that the coordinates of $\mu_k$ are in $[0,1]$, we conclude that $k^* \le nD^{2K}$.
\end{proof}

\section{A strategy improvement algorithm}
\label{7-sec:strategy_improvement}
Let us consider a stochastic reachability game $\game$ and set as a goal to construct an optimal strategy for Max.
The key idea behind strategy improvement is to use $\val^{\sigma}$ to improve the strategy $\sigma$ 
by \emph{switching edges}, which is an operation that creates a new strategy.
This involves defining the notion of \emph{improving edges}:
let us consider a vertex $u \in \VMax$, we say that $e : u \xrightarrow{} v$ is an improving edge if
\[
\val^{\sigma}(v) > \val^{\sigma}(u).
\]
Intuitively: according to $\val^{\sigma}$, playing $e$ is better than playing $\sigma(u)$.


Given a strategy $\sigma$ and a set of improving edges $S$ (for each $u \in \VMax$, $S$ contains at most one outgoing edge of $u$), we write $\sigma[S]$ for the strategy 
\[
\sigma[S](u) = 
\begin{cases}
e & \text{ if there exists } e = u \xrightarrow{} v \in S,\\
\sigma(v) & \text{ otherwise}.
\end{cases}
\]

The difficulty is that an edge being improving does not mean that it is a better move than the current one in any context,
but only according to the value function $\val^{\sigma}$, so it is not clear that $\sigma[S]$ is better than $\sigma$.
Strategy improvement algorithms depend on the following two principles:
\begin{itemize}
	\item \textbf{Progress}: updating a strategy using improving edges is a strict improvement,
	\item \textbf{Optimality}: a strategy which does not have any improving edges is optimal.
\end{itemize}

The pseudocode of the algorithm is given in~\Cref{7-algo:strategy_improvement}.
The algorithm is non-deterministic, in the sense that both the initial strategy and at each iteration, the choice of improving edge can be chosen arbitrarily. 
A typical choice, called the ``greedy all-switches'' rule, chooses for each $u \in \VMax$ a maximal improving edge, meaning 
\[
\argmax \set{\val^{\sigma}(v) : u \xrightarrow{} v \in E}.
\]

\begin{algorithm}
 \DontPrintSemicolon
 
 Choose an initial strategy $\sigma_0$ for Max
 
 \For{$i = 0,1,2,\dots$}{

 	Compute $\val^{\sigma_i}$ and the set of improving edges

	\If{$\sigma_i$ does not have improving edges}{
		\Return{$\sigma_i$}
	}

%

	Choose a non-empty set $S_i$ of improving edges 
	
	$\sigma_{i+1} \leftarrow \sigma_i[S_i]$
 } 
 \caption{The strategy improvement algorithm.}
\label{7-algo:strategy_improvement}
\end{algorithm}

Before proving the correctness of this algorithm, we show that computing $\val^{\sigma_i}$ at each iteration can be done in polynomial time. Note that the opposite problem (computing $\val^{\tau}$ given $\tau$) is the reachability problem in a MDP as discussed in \Cref{6-chap:mdp}. In particular, \Cref{6-thm:quant-reachability-main} gives a polynomial-time algorithm for this problem based on linear programming. Finding $\val^{\sigma}$ given $\sigma$ is similar.

\begin{lemma}
Let $\sigma$ be a pure positional strategy of player Max. Then, $\val^{\sigma}$ is the solution of the linear program
\[
\begin{array}{lll}
\text{maximise }     & \sum_{u \in \vertices} x_u \\
\textrm{subject to } & x_u = 1 &\text{if $u = \Win$,}\\
					 & x_{u} = x_{\sigma(u)}  &\text{if $u \in \VMax$,} \\
					 & x_u \le x_v &\text{if $u \in \VMin$, $u \to v \in E$,}\\
					 & x_u = \sum_{v} \delta(u)(v)x_v &\text{if $u \in \VR$,}\\
					 & x_u = 0 &\text{if $u \in \vertices \setminus W_{>0}(\Game^{[\sigma]})$.}\\
\end{array}
\]
In particular, $\val^{\sigma}$ can be found in polynomial time.
\end{lemma}
\begin{proof}
By applying \Cref{7-thm:least_fixed_point} to $\Op^{\Game[\sigma]}$ we get that $\val^{\sigma}$ is a feasible solution of the program. Moreover, note that every feasible solution $x$ satisfies $x \le \Op^{\Game[\sigma]}(x)$, $x_{\Win} = 1$, and $x_{u} = 0$ for all $u \in \vertices \setminus W_{>0}(\Game^{[\sigma]})$. Since $(C^{\sigma} \setminus \Win) \subseteq (\vertices \setminus W_{>0}(\Game^{[\sigma]}))$, \Cref{7-le:policy_from_inequality} shows that $x \le \val^{\sigma}$. Hence, $\val^{\sigma}$ can be found by computing $W_{>0}(\Game^{[\sigma]})$ (\Cref{7-thm:positively_winning_reachability}) and solving the linear program.
\end{proof}

Let us write $\sigma \le \sigma'$ if for all vertices $u$ we have $\val^{\sigma}(u) \le \val^{\sigma'}(u)$,
and $\sigma < \sigma'$ if additionally $\neg (\sigma' \le \sigma)$. We make the following observation.

\begin{lemma}\label{7-le:switches_to_operators}
Let $\sigma$ be any strategy of Max and $S$ a set of improving edges. We let $\sigma' = \sigma[S]$. Then, we have $\val^{\sigma}\le \Op^{\Game[\sigma']}(\val^{\sigma}) \le  \Op^{\Game}(\val^{\sigma})$. Furthermore, the inequality $\val^{\sigma}(u) \le \Op^{\Game}(\val^{\sigma})(u)$ is strict if and only if $u$ is a vertex of Max that has at least one improving edge. Likewise, the inequality $\val^{\sigma}(u) \le \Op^{\Game[\sigma']}(\val^{\sigma})(u)$ is strict if $u$ has an outgoing edge in $S$. Moreover, if $S$ is constructed by the greedy all-switches rule, then $\Op^{\Game[\sigma']}(\val^{\sigma}) = \Op^{\Game}(\val^{\sigma})$.
\end{lemma}
\begin{proof}
By \Cref{7-thm:least_fixed_point} we have $\val^{\sigma} = \Op^{\Game[\sigma]}(\val^{\sigma})$. Hence, the inequality $\val^{\sigma}(u) \le \Op^{\Game}(\val^{\sigma})(u)$ follows from the definition of $\Op^{\Game}$ and this inequality is strict if and only if $u$ is a vertex of Max that has at least one improving edge. Furthermore, the definition of $\sigma'$ implies that $\Op^{\Game[\sigma']}(\val^{\sigma})$ is sandwiched between $\val^{\sigma}$ and $\Op^{\Game}(\val^{\sigma})$ and that it satisfies $\val^{\sigma}(u) < \Op^{\Game[\sigma']}(\val^{\sigma})(u)$ for every $u$ that has an outgoing edge in $S$. The last claim follows from the definition of the greedy all-switches rule.
\end{proof}

\begin{theorem}[Progress property for the strategy improvement]
\label{7-thm:progress}
Let $\sigma$ a strategy and $S$ a set of improving edges.
We let $\sigma' = \sigma[S]$. Then $\sigma < \sigma'$.
\end{theorem}

\begin{proof}
By~\Cref{7-le:switches_to_operators}, we have $\val^{\sigma} \le \Op^{\Game[\sigma']}(\val^{\sigma})$ and this inequality is strict for at least one coordinate. Hence, by~\Cref{7-thm:least_fixed_point}, $\val^{\sigma'} \neq \val^{\sigma}$. It remains to prove that $\val^{\sigma'} \ge \val^{\sigma}$. To do so, we use~\Cref{7-le:policy_from_inequality}. Denote $\mu = \val^{\sigma}$, let $\tau$ be any pure positional strategy of Min and let $P$ be the transition matrix of the Markov chain induced by fixing $(\sigma',\tau)$. By definition, we have $\mu \le \Op^{\Game[\sigma']}(\mu) \le P \mu$. Suppose that $C \neq \{\Win\}$ is a recurrent class of this Markov chain. Then, this class has a stationary distribution (see, \textit{e.g.},~\cite[Chapter~V]{Kemeny.Snell:1976}). In other words, there exists a nonnegative vector $\pi \ge 0$ such that $\pi_u > 0$ for $u \in C$, $\pi_u = 0$ for $u \notin C$, and $\pi^{T}P = \pi$. Since $\pi^T \mu \le \pi^T P \mu = \pi^T \mu$ we get $\mu(u) = \Op^{\Game[\sigma']}(\mu)(u) = (P\mu)_u$ for all $u \in C$. In particular, the strategies $\sigma'$ and $\sigma$ are the same on all vertices in $C \cap \VMax$. Hence, $C$ is also a recurrent class in the Markov chain obtained by fixing $(\sigma, \tau)$ and so $\val^{\sigma}(u) = 0$ for every $u \in C$. Since $\tau$ was arbitrary, we get $\mu(u) = 0$ for all $u \in C^{\sigma'} \setminus \{\Win\}$. We also have $\mu(\Win) = 1$ and by~\Cref{7-le:policy_from_inequality} we conclude that $\mu \le \val^{\sigma'}$.
\end{proof}

\begin{theorem}[Optimality property for the strategy improvement]
\label{7-thm:optimality}
Let $\sigma$ be a strategy that has no "improving edges", then $\sigma$ is optimal.
\end{theorem}

\begin{proof}
Let $\sigma$ be a strategy that has no improving edges. Then, by \Cref{7-le:switches_to_operators} we have $\val^{\sigma} = \Op^{\Game}(\val^{\sigma})$, and so, by \Cref{7-thm:least_fixed_point}, $\val^{\sigma} \ge \val^{\Game}$, which gives $\val^{\sigma} = \val^{\Game}$.
\end{proof}

\begin{theorem}[Comparison of strategy improvement and value iteration algorithms]
\label{7-thm:comparison_SI_VI}
Let us write $\sigma_0 < \sigma_1 < \sigma_2 < \cdots$ for the sequence of positional strategies in an execution of the strategy improvement algorithm with the greedy all-switches rule over $\Game$ and $\mu_0 \le \mu_1 \le \mu_2 \le \cdots$ for the sequence of functions computed by the corresponding value iteration algorithm over $\Game$ initialised at $\val^{\sigma_0}$:
for all $k$, we have $\mu_0 = \val^{\sigma_0}$ and $\mu_{k+1} = \mathbb{B}(\mu_k)$. Then for all $k$, we have $\mu_k \le \val^{\sigma_k}$.
\end{theorem}

\begin{proof}
For every $k$ we have $\val^{\sigma_k} \le \val^{\sigma_{k+1}}$. Hence, by \Cref{7-thm:least_fixed_point} and the monotonicity of $\Op^{\Game[\sigma_{k+1}]}$ we get $\Op^{\Game[\sigma_{k+1}]}(\val^{\sigma_k}) \le \val^{\sigma_{k+1}}$. Furthermore, by~\Cref{7-le:switches_to_operators} we have $\Op^{\Game[\sigma_{k+1}]}(\val^{\sigma_k}) = \Op^{\Game}(\val^{\sigma_k})$. Thus $\Op^{\Game}(\val^{\sigma_k}) \le \val^{\sigma_{k+1}}$ for all $k$. Moreover, by \Cref{7-cor:value_denominator}, $\val^{\sigma_k}$ is a fixed point of $x \mapsto \textsc{RoundUp}(x,D^K)$ for all $k$. Therefore, $\mathbb{B}(\val^{\sigma_k}) \le \val^{\sigma_{k+1}}$ for all $k$. By induction and the monotonicity of $\mathbb{B}$, if $\mu_k \le \val^{\sigma_k}$, then $\mu_{k+1} = \mathbb{B}(\mu_k) \le \mathbb{B}(\val^{\sigma_k}) \le \val^{\sigma_{k+1}}$.
\end{proof}

By combining \Cref{7-thm:comparison_SI_VI,7-thm:value_iteration}, we get an $n \cdot D^{2K}$ bound on the number of iterations of the strategy improvement algorithm using the greedy all-switches rule. The following theorem improves this bound for \emph{all} switching rules.

\begin{theorem}
The strategy improvement algorithm stops in at most $|\VMax| \cdot D^K$ iterations.
\end{theorem}
\begin{proof}
Let $\sigma_0 < \sigma_1 < \sigma_2 < \cdots$ be the sequence of positional strategies in an execution of the strategy improvement algorithm. We have $\val^{\sigma_k} \le \val^{\sigma_{k+1}}$ for all $k$. As in the proof of~\Cref{7-thm:comparison_SI_VI}, by \Cref{7-thm:least_fixed_point} and the monotonicity of $\Op^{\Game[\sigma_{k+1}]}$ we get $\Op^{\Game[\sigma_{k+1}]}(\val^{\sigma_k}) \le \val^{\sigma_{k+1}}$. Furthermore,~\Cref{7-le:switches_to_operators} shows that $\val^{\sigma_k} \le \Op^{\Game[\sigma_{k+1}]}(\val^{\sigma_k})$ and that this inequality is strict for every vertex of Max that has an outgoing edge in $S$. Let $u$ be such a vertex. We claim that $ \Op^{\Game[\sigma_{k+1}]}(\val^{\sigma_k})(u) - \val^{\sigma_k}(u) \ge D^{-K}$. Indeed, by~\Cref{7-cor:value_denominator}, $\val^{\sigma}$ is a vector of rational numbers with common denominator at most $D^{K}$. Since $\Op^{\Game[\sigma_{k+1}]}(\val^{\sigma_k})(u) = \val^{\sigma_k}(v)$ for some vertex $v$, we get $ \Op^{\Game[\sigma_{k+1}]}(\val^{\sigma_k})(u) - \val^{\sigma_k}(u) \ge D^{-K}$. In particular, $\val^{\sigma_{k+1}}(u) \ge \val^{\sigma_k}(u) + D^{-K}$. Therefore, the strategy improvement algorithm increases the value of at least one vertex of Max by at least $D^{-K}$. Since the value is bounded by $1$, the algorithm must stop within $|\VMax| \cdot D^K$ iterations.
\end{proof}

\section{Algorithms based on permutations of random vertices}
\label{7-sec:permutation_based_algorithms}
We present two algorithms, both based on the same key idea: to order the random vertices and to restrict ourselves to strategies aiming at reaching the best random vertices (with respect to the order on random vertices).
For the remainder of this section we fix a stochastic normalised reachability game $\Game$.

\subsubsection{Permutation of random vertices}

Let us write $\VR = \set{v_1,\ldots,v_k}$. 
We represent total orders on $\VR$ using permutations $\perm: \VR \to \VR$. 
We write $\perm_i = \perm^{-1}(v_i)$ for the $i$-th element in the order defined by $\perm$.
Intuitively, $\perm$ represents a preference order for Max on random vertices.

A permutation $\perm$ induces what we call the $\perm$-regions:
\[
\left\{
\begin{array}{l}
W_\perm^{k+1}  = \{\vwin\} \\
W_\perm^i =
\AttrMax(\set{\perm_i,\ldots,\perm_k,\vwin}) \setminus \bigcup_{j > i} W_\perm^j \quad i \in [1,k] \\ 
W_\perm^0 = V \setminus \bigcup_{j > 0} W_\perm^j
\end{array}
\right.
\]
The idea is that once we fix the order on random vertices, the rest of the game is deterministic: Max and Min only aim at the best (with respect to the permutation $\perm$) random vertex they can get.
More precisely, $W_\perm^i$ is the set of vertices where Max can ensure to reach $\set{\perm_i,\ldots,\perm_k,\vwin}$ before any other random vertex (but no subset $\set{\perm_j,\ldots,\perm_k,\vwin}$).

\begin{fact}
We have $W_\perm^0 = \set{\Lose}$.
\end{fact}
Indeed, recall that $\vlose$ is the unique vertex with value $0$.
We write $W_\perm^{\ge i}$ for $\bigcup_{j \ge i} W_\perm^j$.
The $\perm$-regions induce the $\pi$-strategies $\sigma_\perm$ (for Max) and $\tau_\perm$ (for Min):
\begin{itemize}
\item on $W_\perm^i$, $\sigma_\perm$ is a pure and positional attractor strategy to $\{\perm_i,\ldots,\perm_k,\vwin\}$;
\item on $W_\perm^i$, $\tau_\perm$ is a pure and positional counter-attractor strategy ensuring never to reach $\{\perm_{i+1},\ldots,\perm_k,\vwin\}$.
\end{itemize}
We can then define for every $u \in \vertices$:
\[
\begin{array}{rcl}
\val_\perm(u) &=& \Prob_{\sigma_\perm,\tau_\perm}^u(\Reach(\Win)).
\end{array}
\]
It can be easily computed using systems of linear equations.

The key property of permutation-based algorithms is stated below:

\begin{theorem}[Existence of an optimal permutation]
\label{7-thm:optimal_permutation}
There exists a permutation $\perm$ such that $\sigma_\perm$ is optimal for Max and $\tau_\perm$ is optimal for Min. 
\end{theorem}
The remainder of this section is devoted to a proof of~\Cref{7-thm:optimal_permutation}.
We later explore how this yields algorithms.

\subsubsection{Live and self-consistent permutations}

\begin{definition}[Live and self-consistent permutations]
Let $\perm$ a permutation.
\begin{itemize}
	\item We say that $\perm$ is \emph{self-consistent} if
\[
\val_\perm(\perm_1) \le \val_\perm(\perm_2) \le \ldots \le \val_\perm(\perm_k).
\]
	\item We say that $\perm$ is \emph{live} if for every $i \in [1,k]$:
\[
\delta(\perm_i)\big(W_\perm^{\ge i+1}\big)>0.
\]
\end{itemize}
\end{definition}
Intuitively, if $\perm$ is self-consistent the order given by $\perm$ coincides with the preference of Max,
and if $\perm$ is live there is a positive probability to reach a preferable (for Max) $\perm$-region.
The following result directly implies~\Cref{7-thm:optimal_permutation}.

\begin{lemma}
\label{7-lem:optimal_permutation}
The following properties hold.
\begin{itemize}
	\item If a permutation $\perm$ is live and self-consistent, then the $\perm$-strategies are optimal.
	\item There exists a live and self-consistent permutation.
\end{itemize}
\end{lemma}

\subsubsection{Live and self-consistent permutations induce optimal strategies}
We prove the first item of~\Cref{7-lem:optimal_permutation}.

\begin{lemma}[Correctness of live and self-consistent permutations]
\label{7-lem:self-consistent_permutations}
If $\perm$ is self-consistent, then $\val_{\perm}$ is a fixed point of the operator $\Op$.
Consequently, $\val^{\Game} \le \val_{\perm}$.
\end{lemma}

\begin{proof}
We prove the following properties:
\begin{enumerate}
\item For $i \in [1,k]$ and $u \in W_\perm^i$ we have $\val_\perm(u) = \val_\perm(\perm_i)$.
\item For $u \in \VMax$ we have $\val_\perm(u) = \val_\perm(\sigma_\perm(u)) = \max \set{\val_\perm(v) : u \rightarrow v \in E}$.
\item For $u \in \VMin$ we have $\val_\perm(u) = \val_\perm(\tau_\perm(u)) = \min \set{\val_\perm(v) : u \rightarrow v \in E}$.
\end{enumerate}
For $u \in W_\perm^i$, up to the first visit to a random vertex, the strategy profile $(\sigma_\perm,\tau_\perm)$ generates a unique path. 
So we can speak of the first random vertex encountered from $u$ when applying $(\sigma_\perm,\tau_\perm)$. 
By definition of $\sigma_\perm$ (attractor to $\{\perm_i,\ldots,\perm_k,\vwin\}$) and $\tau_\perm$ (counter-attractor strategy avoiding $\{\perm_{i+1},\ldots,\perm_k,\vwin\}$), this random vertex can only be $\perm_i$. According values follow, proving the first item.

Assume $u \in \VMax \cap W_\perm^i$. 
By definition of $\sigma_\perm$ (being an attractor strategy), $\sigma_\perm(u) \in W_\perm^i \cup \{\perm_i,\ldots,\perm_k,\vwin\}$. 
Dually, since $u \notin W_\perm^{\ge i+1}$, we have $\sigma_\perm(u) \notin \{\perm_{i+1},\ldots,\perm_k,\vwin\}$. 
Hence, $\sigma_\perm(u) \in W_\perm^i \cup \{\perm_i\} = W_\perm^i$, and we get that $\val_\perm(u) = \val_\perm(\perm_i) = \val_\perm(\sigma_\perm(u))$.
Assume towards contradiction that there exists $u \rightarrow v \in E$ such that $\val_\perm(v) > \val_\perm(u)$. 
Since we know that $\val_\perm(u) = \val_\perm(\perm_i)$, by self-consistence this implies that $v \in W_\perm^j$ with $j > i$
(with $\val_\perm(v) = \val_\perm(\perm_j)$). 
This implies that $u \in \AttrMax(v) \subseteq \AttrMax(\{\perm_j,\ldots,\perm_k,\vwin\})$, which is not the case, since $u \notin W_\perm^{\ge i+1} \supseteq W_\perm^j$. Contradiction.

The reasoning is symmetric for $u \in \VMin$.
\end{proof}

The converse inequality $\val^{\Game} \ge \val_{\pi}$ is not true for general or self-consistent permutations, but it does hold when adding the liveness property.
A key technical property of the $\pi$-strategies $\sigma_\perm$ when $\pi$ is live is that it induces a stopping MDP:
Min cannot prevent the game converging to $\vlose$ or $\vwin$.

\begin{lemma}[Live permutations imply stopping MDPs]
\label{7-lem:live_stopping}
Let $\perm$ be a live permutation. Then, for every strategy $\tau$ of Min, for every $u$, we have $\Prob^u_{\sigma_\perm,\tau} (\Reach(\{\vlose,\vwin\})) = 1$.
\end{lemma}

\begin{proof}
Let $\alpha = \min_{i \in [1,k]} \delta(\perm_i) \big(W_\perm^{\ge i+1}\big)$. By definition of a live permutation we have $\alpha>0$.
We write $V_i$ for the random variable representing the $i$-th state of a run.
By definition of $\alpha$, for $i \in [1,k]$ and $\ell \ge 0$, we have
$\Prob^u_{\sigma_\perm,\tau}\Big(V_{\ell+1} \in W_\perm^{\ge i+1} \mid V_\ell = \perm_i\Big) \ge \alpha$
and $\Prob^u_{\sigma_\perm,\tau}\Big(\exists h < |W_\perm^i|,\ V_{\ell + h} \in \{\perm_i,\ldots,\perm_k,\vwin\} \mid 
V_\ell \in W_\perm^i\Big) = 1$, since $\sigma_\perm$ plays according to attractor strategies in the corresponding $\pi$-regions.
This implies that $\Prob^u_{\sigma_\perm,\tau}\Big(V_{\ell + n} = \vwin \mid V_\ell \ne \vlose\Big) \ge \alpha^k$, where $n$ is the number of vertices in the game. We can rewrite this as:
$\Prob^u_{\sigma_\perm,\tau}\Big(\forall \ell' \in [\ell, \ell + n],\ V_{\ell'} \ne \vwin \mid V_\ell \ne \vlose\Big) \le 1-\alpha^k$.
Iterating, we get that for every $i$,
\[
\Prob^u_{\sigma_\perm,\tau}\Big(\forall \ell \in [0,i \cdot n],\ V_{\ell} \notin \set{\vwin,\vlose} \mid V_0 \ne \vlose \Big) \le (1-\alpha^k)^i.
\]
Hence by taking the limit $\Prob^u_{\sigma_\perm,\tau}(\forall \ell \ge 0,\ V_\ell \notin \set{\vwin,\vlose}) = 0$, which equivalently reads $\Prob^u_{\sigma_\perm,\tau}(\exists \ell \ge 0,\ V_\ell \in \set{\vwin,\vlose}) = 1$.
\end{proof}

We can now show that if $\perm$ is a live and self-consistent permutation, then $\val_\perm = \val^{\Game}$.
By~\Cref{7-lem:self-consistent_permutations} $\val_\perm$ is a fixed point of the operator $\Op^{\Game}$, hence of $\Op^{\Game[\sigma_\perm]}$.
By~\Cref{7-lem:live_stopping}, the MDP obtained when restricting to $\sigma_\perm$ is stopping, hence (the special case for MDPs of)~\Cref{7-thm:fixed_point_characterisation_stopping_ssg} implies that $\Op^{\Game[\sigma_\perm]}$ has a unique fixed point, which is $\val^{\Game}$.
Hence $\val_\perm = \val^{\Game}$.

\subsubsection{Existence of a live and self-consistent permutation}
We now prove the second item of~\Cref{7-lem:optimal_permutation}.

\begin{lemma}[Live and value non-decreasing imply self-consistent]
\label{7-lem:live_non-dec_self-consistent}
Let $\perm$ be a live permutation such that $\val^{\Game}(\perm_1) \le \val^{\Game}(\perm_2) \le \dots \le \val^{\Game}(\perm_k)$. 
Then $\perm$ is self-consistent.
\end{lemma}

\begin{proof}
We show that for every vertex $u$ we have $\val^{\Game}(u) = \val_\perm(u)$, which implies the result.
We first note that for every $i \in [1,k]$, for every $u \in W_\perm^i$ we have $\val^{\Game}(u) = \val^{\Game}(\perm_i)$.

Let us define a $\sigma^*$ from $u$: it plays the attractor strategy to $\{\perm_i,\ldots,\perm_k,\vwin\}$ until $\vwin$ or a random vertex is reached, and in the latter case, switches to an optimal strategy. 
Clearly for every strategy $\tau$ for Min we have
$\Prob_{\sigma^*,\tau}^u(\Reach(\{\vwin\})) \ge \min_{i \in [i,k]} \val^{\Game}(\perm_j) = \val^{\Game}(\perm_i)$. 
Hence $\val^{\Game}(u) \ge \val^{\Game}(\perm_i)$.

Conversely, let us define a strategy $\tau^*$ from $u$: it plays the counter-attractor strategy to $\{\perm_{i+1},\ldots,\perm_k,\vwin\}$ until $\vlose$ or a random vertex is reached, and in the latter case, switches to an optimal strategy. 
It may never hit $\vlose$ or a random vertex, but this is good to Min. 
Clearly for every strategy $\sigma$ for Max we have
$\Prob_{\sigma,\tau^*}^u(\Reach(\{\vwin\})) \le \max_{j \in [1,i]} \val^{\Game}(\perm_j) = \val^{\Game}(\perm_i)$. 
Hence $\val^{\Game}(u) \le \val^{\Game}(\perm_i)$.
\end{proof}

It remains to show that there always exists a live permutation satisfying the hypothesis of~\Cref{7-lem:live_non-dec_self-consistent}.
To do so, we show the following structural property of the game, which will help building an appropriate live permutation.

\begin{lemma}
\label{7-lem:structure}
Let $X \subseteq V$ such that $\vwin \in X$, and $Y = V \setminus \AttrMax(X)$. 
Then either $Y = \{\vlose\}$, or there exists a random vertex $u \in Y$ such that 
$\val^{\Game}(u) = \max\{\val^{\Game}(v) : v \in Y\}$ and $\delta(u)\Big(\AttrMax(X)\Big) > 0$.
\end{lemma}

\begin{proof}
Let $m = \max\{\val^{\Game}(u) : u \in Y\}$, we write $Z = \set{u \in Y : \val^{\Game}(u) = m}$. 
We show that if there are no random vertices $u \in Z$ such that $\delta(u)\Big(\AttrMax(X)\Big)>0$, 
then $Z = \{\vlose\}$.
To do so, we show that if $u \in Z$, then $\val^{\Game}(u) = 0$. 
Let us assume towards a contradiction that $\val^{\Game}(u) > 0$.

Let $\tau$ be a counter-attractor (pure and positional) strategy for Min to not reach $\AttrMax(X)$ from $Y$.
The following properties are direct consequences of the definitions: for any $u \in Z$,
\begin{itemize}
	\item if $u \in \VMin$ and $\tau(u) = u \rightarrow v \in E$, then $v \in Z$,
	\item if $u \in \VMax$ and $u \rightarrow v \in E$, then $v \in Y$,
	\item if $u \in \VR$ and $\delta(u)(v) > 0$, then $v \in Z$.
\end{itemize}
%


We now define a strategy $\tau'$ which plays from $u$ as $\tau$ until $Z$ is left, and then switches to an optimal strategy for Min.
Let $\sigma$ a strategy for Max. 
Thanks to the properties above, a play consistent with $\sigma$ and $\tau'$ from $u$ either stays forever in $Z$, or 
reaches a vertex $v$ such that $\val^{\Game}(v) < \val^{\Game}(u) = m$. 
Let us write $\beta = \max \{\val^{\Game}(v) \mid v \in Y \setminus Z\} < m$.
We then have that $\Prob^u_{\sigma,\tau'}(\Reach(\{\vwin\})) \le \beta$, 
because the probability to reach $\vwin$ if staying forever in $Z$ is $0$.
Hence $\val^{\Game}(u) \le \beta < \val^{\Game}(u)$, a contradiction.
\end{proof}

We can now prove the existence of a live permutation such that $\val^{\Game}(\perm_1) \le \val^{\Game}(\perm_2) \le \dots \le \val^{\Game}(\perm_k)$. We define the permutation $\perm$ inductively, by repeatedly using~\Cref{7-lem:structure}.
For every $i \in [k,1]$ we define $\perm_i$ by applying~\Cref{7-lem:structure} to $X = \{\perm_{i+1},\ldots,\perm_k,\vwin\}$.
By construction,
\begin{itemize}
	\item $\val^{\Game}(\perm_i) = \max \{\val^{\Game}(v) \mid v \in V \setminus \AttrMax(\{\perm_{i+1},\ldots,\perm_k,\vwin\})\}$;
	\item $\delta(\perm_i)\Big(\AttrMax(\{\perm_{i+1},\ldots,\perm_k,\vwin\})\Big) > 0$.
\end{itemize}
It follows that $\perm$ is live, and the hypothesis of~\Cref{7-lem:live_non-dec_self-consistent} is satisfied. 
Hence $\perm$ is self-consistent. This concludes the proof of the second item of~\Cref{7-lem:optimal_permutation}.

\subsubsection{Strategy enumeration algorithm}
\label{7-subsec:last}

\Cref{7-thm:optimal_permutation} induces a simple algorithm for computing the values and optimal strategies for both players in a
stochastic reachability game: enumerate all permutations of random vertices, and for each of them, check whether it is live and
self-consistent; stop when one is found. Note that indeed given $\perm$, it is easy to compute the $\perm$-regions and strategies and check for liveness and self-consistency, in polynomial time.
However, as such, this requires to enumerate all permutations of random vertices, and there are $|\VR|!$ of them. Hence the overall complexity of finding the values and the optimal strategies is exponential.


\subsubsection{Strategy improvement algorithm}
\label{7-subsec:algo-strat-improv}

We will describe a strategy improvement algorithm, which may avoid enumerating all permutations. 
Note that there is nevertheless no guarantee that the overall complexity will be better than the strategy enumeration algorithm.

The algorithm consists in the following steps:
\begin{itemize}
	\item Initialisation step: Compute a live permutation $\perm$
	\item Improvement step: Given a live permutation $\perm$, compute a live and self-consistent permutation in $\game[\sigma_\perm]$.
\end{itemize}

We argue (not in full details) that the following properties are satisfied:
\begin{enumerate}
	\item We can compute an initial live permutation in polynomial time.
	\item For every live permutation $\perm$, we can compute in polynomial time a live and self-consistent permutation $\perm'$ in
$\game[\sigma_\perm]$.
	\item The above mentioned permutation $\perm'$ is live in $\game$ as well.
	\item The improvement step is an improvement:
	\begin{itemize}
		\item $\val^{\game[\sigma_\perm]} \le \val^{\game[\sigma_{\perm'}]}$, and
		\item If $\val^{\game[\sigma_\perm]} = \val^{\game[\sigma_{\perm'}]}$, then $\perm'$ is self-consistent in $\Game$.
	\end{itemize}
\end{enumerate}

The first property is based on the inductive construction following~\Cref{7-lem:structure}.

For the second property, we know as a consequence of~\Cref{7-thm:optimal_permutation} that there exists a live and self-consistent permutation $\perm'$ in $\game[\sigma_{\perm}]$.
Note that for this we need to argue that $\game[\sigma_{\perm}]$ is normalised. 

For the third property, we note that the $\perm'$-regions in $\game[\sigma_\perm]$ are included in the $\perm'$-regions in $\game$, which implies the result.

The last property is harder to argue; it expresses the fact that the
new permutation $\perm'$ improves over $\perm$.

This last property ensures both termination of the algorithm: indeed, it is ensured by the finiteness of the number of permutations, and by
the improvement characterisation above. Note that it may be the case that in the worst-case, all permutations will be enumerated. No lower nor upper bound is known.

\section{Reductions to simple stochastic games}
\label{7-sec:reductions}
The rest of this chapter focused on stochastic reachability games. As we will see in this section, 
they are powerful enough to encode other classes of games, such as non-stochastic discounted-payoff games, and stochastic parity, mean-payoff, and discounted-payoff games.


\subsection{From discounted-payoff games to stochastic reachability games}

\begin{theorem}[Reducing discounted-payoff games to stochastic reachability games]
\label{7-thm:discounted_games_to_stochastic_reachability_games}
Solving (non-stochastic) discounted-payoff games reduces in polynomial time to solving stochastic reachability games.
\end{theorem}

\begin{figure}[htbp]
\centering
\begin{tikzpicture}[scale=1.5]
\node[s-adam] (u) at (-5,1) {$u$};    
\node[s-eve] (v) at (-3.5,1) {$v$};
\node at (-2.5,1) {becomes};

\path[arrow] 
(u) edge node[above] {$w$} (v);

\node[s-adam] (u2) at (-1.5,1) {$u$};    
\node[s-eve] (v2) at (.95,1) {$v$};
\node[s-random-small] (uv) at (-.4,1) {$e_v$};
\node[s-eve] (lose) at (-.4,-.3) {$\vlose$};
\node[s-eve] (win) at (-.4,2.2) {$\vwin$};

\path[arrow] 
(u2) edge (uv)
(uv) edge node[above] {$\lambda$} (v2)
(uv) edge node[right] {$(1-\lambda) \cdot (1 - w)$} (lose)
(uv) edge node[right] {$(1-\lambda) \cdot w$} (win)
(lose) edge[selfloop=0] (lose);
;

\end{tikzpicture}
\caption{From discounted-payoff games to stochastic reachability games.}
\label{7-fig:reduction_discounted_to_stochastic}
\commentAlt{Figure~\ref{7-fig:reduction_discounted_to_stochastic}: A diagram showing a transformation from a direct edge between two nodes to an expanded structure involving an intermediate node with multiple weighted outputs. See long description.}
\commentLongAlt{Figure~\ref{7-fig:reduction_discounted_to_stochastic}: The image displays a transformation represented by two diagrams separated by the word "becomes".

The left diagram shows a simple directed edge from a square node labeled 'u' to a circular node labeled 'v', with the arrow labeled 'w'.

The right diagram shows the transformed structure. It starts with the same square node 'u', which now points to a triangular node labeled 'e_v'. From 'e_v', three paths diverge:

An arrow labeled 'lambda' points right to the original circular node 'v'.
An arrow labeled '(1 - lambda) * w' points upwards to a new circular node labeled 'Win'.
An arrow labeled '(1 - lambda) * (1 - w)' points downwards to a new circular node labeled 'Lose'. The 'Lose' node has a self-loop.}
\end{figure}

\begin{proof}
Let $\Game$ a discounted-payoff game.
Using a linear transformation we can assume that all weights are rational numbers in $[0,1]$.
We construct $\Game'$ a stochastic reachability game by adding to vertices $v_{\Win}$ and $v_{\Lose}$,
and each edge $e = u \xrightarrow{w} v$ in $\Game$ is redirected to a new random vertex $v_e$
with probabilistic distribution
\[
\delta(v_e) = \lambda \cdot v + (1 - \lambda) \cdot w \cdot \vwin + (1 - \lambda) \cdot (1 - w) \cdot \vlose.
\]
The reachability condition is $\Reach(\vwin)$.
(Note that the game $\Game'$ is stopping.)

We claim that for every $u \in V$, we have
\[
\val^{\Game}(u) = \val^{\Game'}(u).
\]
To this end, we rely on~\Cref{5-thm:values_discounted_contracting_fixed_point}, which states that the discounted-payoff values are the unique fixed point of the operator $\Op^{\Game} : F_V \to F_V$ defined by
\[
\Op^{\Game}(\mu)(u) = 
\begin{cases}
\max \set{\lambda \cdot \mu(v) + (1 - \lambda) \cdot w : u \xrightarrow{w} v \in E} & \text{ if } u \in \VMax, \\
\min \set{\lambda \cdot \mu(v) + (1 - \lambda) \cdot w : u \xrightarrow{w} v \in E} & \text{ if } u \in \VMin.
\end{cases}
\]
We show that $(\val^{\Game'}(u))_{u \in V}$ is a fixed point of the operator $\Op^{\Game}$.

Thanks to~\Cref{7-thm:least_fixed_point}, the values in $\Game'$ satisfy the following:
\[
\left\{\begin{array}{llll}
    \val^{\Game'}(\vwin) & = & 1, \\
    \val^{\Game'}(\vlose) & = & 0, \\
    \val^{\Game'}(u) & = & \max \set{\val^{\Game'}(v) : u \rightarrow v \in E} & \text{if}\ u \in \VMax, \\
    \val^{\Game'}(u) & = & \min \set{\val^{\Game'}(v) : u \rightarrow v \in E} & \text{if}\ u \in \VMin, \\
    \val^{\Game'}(v_e) & = & \lambda \cdot \val^{\Game'}(v) \\
                       & + & (1 - \lambda) \cdot w \cdot \val^{\Game'}(\vwin) \\
                       & + & (1 - \lambda) \cdot (1 - w) \cdot \val^{\Game'}(\vlose)
\end{array}\right.
\]
Eliminating the values $\val^{\Game'}(v_e)$ from these equations, we obtain
\[
\left\{\begin{array}{llll}
    \val^{\Game'}(u) & = & \max \set{\lambda \cdot \val^{\Game'}(v) + (1 - \lambda) \cdot w : u \rightarrow v \in E} & \text{if}\ u \in \VMax, \\
    \val^{\Game'}(u) & = & \min \set{\lambda \cdot \val^{\Game'}(v) + (1 - \lambda) \cdot w : u \rightarrow v \in E} & \text{if}\ u \in \VMin.
\end{array}\right.
\]
Hence $(\val^{\Game'}(u))_{u \in V}$ is a fixed point of the operator $\Op^{\Game}$.
Since $\Op^{\Game}$ has a unique fixed point $\val^{\Game}$, this implies the desired equality.
\end{proof}

We note that~\Cref{7-thm:discounted_games_to_stochastic_reachability_games} easily extends to discounted stochastic games: 
from a discounted stochastic game, one can build a stochastic reachability game with the same values.

\subsection{From stochastic mean-payoff games to stochastic discounted-payoff games}

Both bipositional determinacy for mean-payoff games and the reduction to discounted-payoff games (presented in~\Cref{5-thm:MP-Zwick-Paterson}) can be lifted to stochastic games.

\begin{theorem}[Stochastic mean-payoff games are bi-positionally determined]
\label{7-thm:stochastic_mean-payoff_games}
Stochastic mean-payoff games are uniformly purely bi-positionally determined.
\end{theorem}

\begin{theorem}[Reducing stochastic mean-payoff games to stochastic discounted-payoff games]
\label{7-thm:reduction_mean-payoff_discounted}
  Let $\game$ a stochastic mean-payoff game. 
  Let $\lambda \in (0,1)$, we define $\game_\lambda$ the stochastic discounted-payoff game obtained from $\game$.
  There exists a discount factor $\lambda$ computable in polynomial time such that any pair of optimal pure positional
strategies in the discounted-payoff game with discount factor $\lambda$ is also a pair of optimal strategies in the mean-payoff game.
\end{theorem}


\subsection{From stochastic parity to stochastic mean-payoff}

Both bi-positional determinacy for parity games and the reduction to mean-payoff games (presented in~\Cref{5-thm:parity2MP}) can be lifted to stochastic games.


\begin{theorem}[Stochastic parity games are bi-positionally determined]
\label{7-thm:stochastic_parity_games}
Stochastic parity games are uniformly purely bi-positionally determined.
\end{theorem}

\begin{theorem}[Reducing stochastic parity games to stochastic mean-payoff games]
\label{7-thm:stochastic_parity_games_to_stochastic_mean_payoff_games}
Solving stochastic parity games reduces in polynomial time to solving stochastic mean-payoff games.
\end{theorem}


The reduction is similar as the one presented in~\Cref{5-thm:parity2MP}, but requires additional technical care.
In particular, it relies on the fact that we can compute the almost-surely and positively winning regions.

\begin{theorem}
\label{7-thm:qualitative_winning_parity_games}
The following holds.
\begin{itemize}
	\item There exists a polynomial-time algorithm for computing the positively winning region of stochastic parity games.
	\item There exists a polynomial-time algorithm for computing the almost-surely winning region of stochastic parity games.
\end{itemize}
\end{theorem}

Let $\Game$ a stochastic parity game.
We let $p_{\min}$ the minimal non-zero probability that appears in $\Game$, and $n$ the number of vertices.

We construct $\Game'$ a stochastic mean-payoff game.
As a first step, we compute the almost-sure and positive winning regions, we replace the almost-sure winning region with a sink vertex $\vwin$ with a self-loop with weight $1$, and the complement of the positive winning region with a sink vertex $\vlose$ with a self-loop with weight $-1$.
For each edge $u \xrightarrow{p} v$ in $\Game$, we have an edge 
$u \xrightarrow{\left( -2n \cdot p_{\min}^{-n} \right)^k} v$ in $\Game'$.

We claim without proof that for all $u \in V$, we have
\[
\val^{\Game}(u) = \frac 1 2 \cdot \left( \val^{\Game'}(u) + 1 \right).
\]

\section*{Bibliographic references}
\label{7-sec:references}
The study of stochastic games started long before the algorithmic perspectives we develop here. 
This chapter took inspiration from the reference survey chapter by Ku{\v{c}}era~\cite{Kucera:2011}.
The introduction of stochastic games is due to Condon~\cite{Condon:1992}, where she proves (a weak form of) positional determinacy, and constructs strategy improvement algorithms.
Moving back in time, positional determinacy for stochastic discounted-payoff games was already proved (as a special case) by Shapley~\cite{Shapley:1953}, and for stochastic mean-payoff games by Liggett and Lippman~\cite{Liggett.Lippman:1969}.
The Borel determinacy result for stochastic games was proved by Maitra and Sudderth~\cite{Maitra.Sudderth:1998}.
The last section about reductions to stochastic reachability games is inspired by~\cite{Andersson.Miltersen:2009}, which clarifies a lot the relationships between the different classes.

As for parity games, mean-payoff games, and discounted-payoff games, the major open question is whether stochastic reachability games can be solved in polynomial time. Strongly polynomial bound for strategy improvement have been obtained when the discount factor is fixed~\cite{Hansen.Miltersen.ea:2013}.
The best complexity to date is sub-exponential time randomised algorithms. A very general point of view on these algorithms is given by reducing them to LP-type problems~\cite{Halman:2007}.

A more general point of view on strategy improvement algorithms was developed by Auger, de Montjoye, and Strozecki~\cite{Auger.Montjoye.ea:2021}.
Constructing strategy improvement algorithms for more general classes of stochastic games is non-trivial, see for instance the case of stochastic mean-payoff games~\cite{Akian.Cochet-Terrasson.ea:2013}.

The permutation-based algorithm is due to Gimbert and Horn~\cite{Gimbert.Horn:2008}. There is a wealth of algorithms we did not cover in this chapter, let us mention the accelerated value iteration for stochastic reachability games~\cite{Ibsen-Jensen.Miltersen:2012}, the value iteration algorithm for ergodic mean-payoff games~\cite{Allamigeon.Gaubert.ea:2025}, and the pumping algorithm for stochastic mean-payoff games~\cite{Boros.Elbassioni.ea:2019}.

Stochastic games have also been related to other models of computations, for instance constraint satisfaction problems~\cite{Bodirsky.Mamino:2016}, and complexity classes related to fixed-point computations~\cite{Etessami.Yannakakis:2010}.


\part{Information}
\label{part:information}

\ifpictures
\includepdf{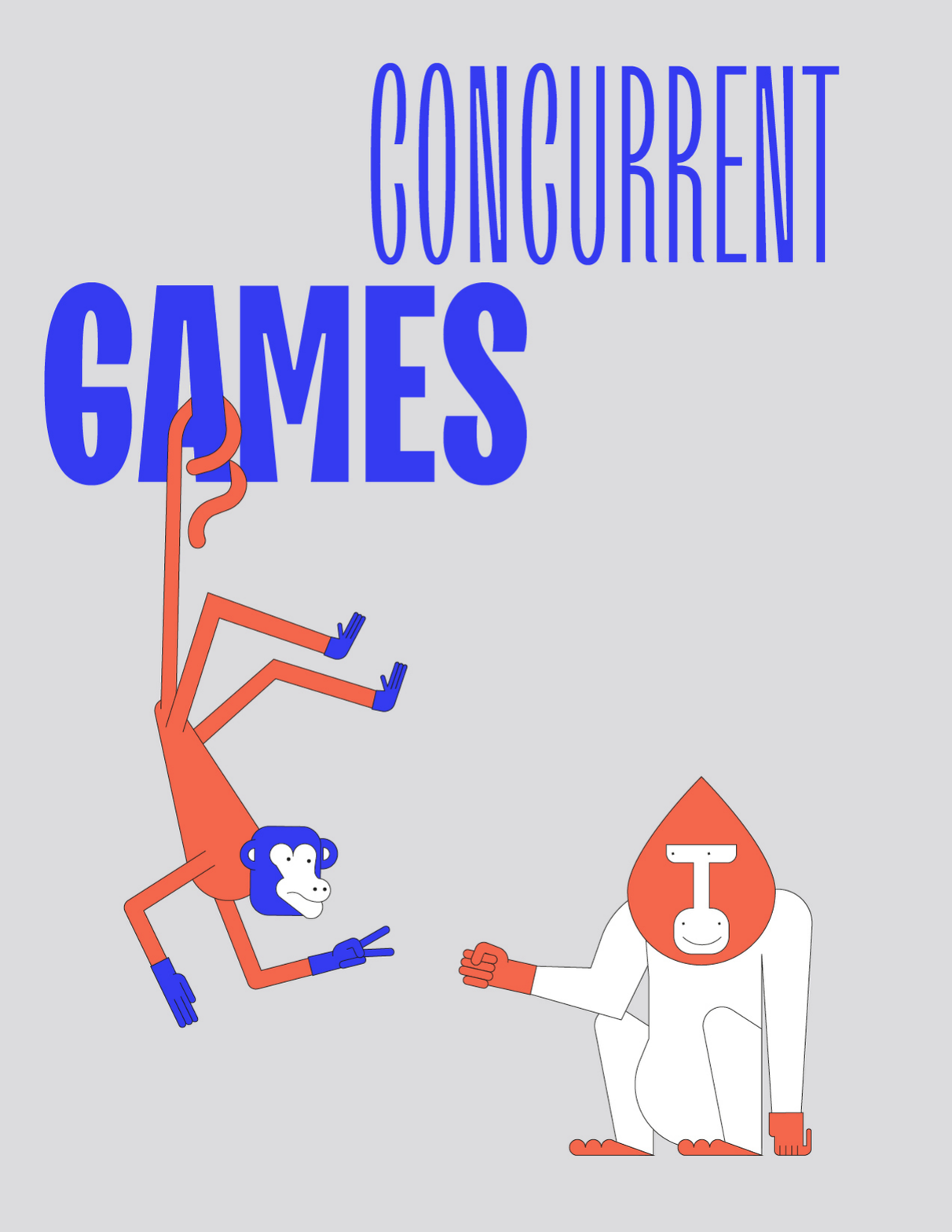}
\fi
\author[Rasmus Ibsen-Jensen]{Rasmus Ibsen-Jensen}
\copyrightline{Copyright by Rasmus Ibsen-Jensen 2025, to be published by Cambridge University Press in the volume \textit{Games on Graphs} edited by Nathana\"el Fijalkow}

\chapter{Concurrent Games}
\chapterauthor{Rasmus Ibsen-Jensen}
\label{8-chap:concurrent}

\newcommand{\ValueOp}{\text{valOp}}
\newcommand{\rk}{\text{rk}}
\newcommand{\crgLim}{\mathcal{A}_1}

This chapter considers concurrent games. The concurrent games we consider are extensions of the games considered in \Cref{3-chap:regular} and \Cref{5-chap:payoffs}, but where the choice of which edge to choose in a round is determined not by the choice of the owner of the vertex (indeed the vertices in concurrent games have no owners), but by the outcome of a matrix game corresponding to the vertex and played in that round. 
A matrix game is in turn a generalisation of rock-paper-scissors, where each player picks an action simultaneously and then their pair of actions determines the outcome.

We will consider concurrent discounted, reachability and mean-payoff games and the definitions of the different objectives is as in the introduction. 
The chapter is divided into four sections:
\begin{enumerate}
\item Matrix games
\item Concurrent discounted-payoff games
\item Concurrent reachability games
\item Concurrent mean-payoff games
\end{enumerate}
As we go through the sections in this chapter, the complexity of the strategies and the computational complexity of solving the games rises: indeed, since the games are generalisations of rock-paper-scissors, the strategies used requires randomness, but towards the end, no optimal or finite-memory $\epsilon$-optimal strategies exists in general and even the principle of sunken cost does not apply! 
Even with all this, the related questions about values are solvable in polynomial space and thus also in exponential time even in the last section.
The results we will focus on characterises the complexity of the both the strategies as well as the computational complexity.
In each section we first give some positive result and some number of negative results. Each negative result also applies to the classes of games considered in the latter sections and each positive result applies to the classes considered in earlier sections (however, the positive results of latter sections will have worse complexity than the positive results from earlier sections).
As mentioned, the strategies for this chapter require randomness and not too surprisingly, this implies that there is little difference between having stochastic or deterministic transition functions.

\section{Notations}
\label{8-sec:notations}
The definition of arena $\arena$ in this chapter is $\arena=(G,\dest)$, where $G=(V,E)$ is a graph and $\dest:V\times A\times A\rightarrow \Dist(E)$. In particular, we are not using the sets $\VA$ and $\VE$.
The games are played similarly to before and formally as follows: 
There is a token, initially on the initial vertex. 
Whenever the token is on some vertex $v$, 
Eve selects an action $r$ in $A$ and Adam selects an action $c$ in $A$. The edge $e=(v,c,w)$ is then drawn from the distribution $\dest(v,r,c)$ and the token is pushed from $v$ to $w$.
 In general, the game continues like that forever.

We will use the following simplifying assumptions in this chapter:
\begin{enumerate}
\item We will assume that all colours are in $\{0,1\}$, except for the section on Matrix games where we additionally also allow $-1$ (to be able to easily illustrate the game rock-paper-scissors). This simplifies some expressions, but generally, the dependency on the number of colours is not too bad comparatively.
\item To make illustrations easier, we assume that for any pair of edges $e,e'$ in $\dest(v,a,a')$ for any $v,a,a'$, we have that $c(e)=c(e')$, \textit{i.e.} the colour does not depend on which edge is picked from $\dest(v,a,a')$, but only $v,a,a'$. This assumption does not matter for the types of games considered.
\end{enumerate}

We will overload the notation slightly for notational convenience, in that for any $v,a,a'$, we will write $c(v,a,a')$ for $c(e)$ where $e\in \dest(v,a,a')$ (note that the second assumption ensures that this is well-defined, \textit{i.e.} there is only one such colour).

A vertex $v$ is absorbing if and only if each player has only 1 action in $v$ and $\Delta(v,1,1)=v$.

To describe the complexity of good stationary strategies in concurrent games, we will use the notion of patience. Given a probability distribution $d\in \Dist$ the distribution has patience $p$ if $p=\min_{i\in \supp(d)} d(i)$ (\textit{i.e.} the patience is the smallest, non-zero probability that an event may happen according to $d$).
In essence, if you have low enough patience you can typically guess the strategy and check whether it is a good strategy (when you fix a strategy, the game becomes a Markov decision process, which are relative easy to work with), the game can solved in $\NP\cup \coNP$. However, some times the patience is huge and writing down a good strategy, in binary, cannot be done in polynomial space (it is quite surprising in some sense that even with this property, finding the values in the games remain in $\PSPACE$).

We will illustrate a stochastic arena $\arena=(G,\dest)$ as follows:
For each non-absorbing vertex $v$, there is matrix.
 Entry $(i,j)$ of the matrix illustrating $v\in V$ describes what happens if, when the token is on $v$, Eve plays $i$ and Adam $j$. The entry contains a colour $c$, which is $c(v,i,j)$, and 
there is an arrow from entry $(i,j)$ of $v$ to $w$ if there is an edge   
$e=(v,c,w)$ in $\dest(v,i,j)$. 
 The arrow corresponding to $e$ is denoted with the probability $\dest(v,i,j)(e)$. 
Especially, to simplify the illustrations we will do as follows: If $|\supp(\dest(v,i,j))|=1$, we do not include the probability (which is 1). Also, in that case, let $e=(v,c,w)=\dest(v,i,j)$ 
and 
if $v=w$, we omit the arrow and if $w$ is absorbing we write $c^*$ in entry $i,j$ of $v$, where $c$ is the colour $c(w,1,1)$ (in this case, we omit the number $c(e)$ from the illustration, but in none of our illustrations does this number matter for what we try to illustrate).

\section{Matrix games}
\label{8-sec:matrix_games}
A matrix game is a game defined from a $(R\times C)$-matrix $M$  of numbers for some $R,C$.
The game is played as follows: Eve picks a row $r$ and Adam picks a column $c$ simulations like in rock-paper-scissors. Adam then pays Eve $M[r,c]$, \textit{i.e.} the content of the entry defined by being in row $r$ and column $c$.
A strategy in such a game for Eve (resp. Adam) consists of a distribution over the rows (resp. columns). 
There is an illustration of rock-paper-scissors as a matrix game in~\Cref{8-fig:rps}.

\begin{figure}
\center
\begin{tikzpicture}[node distance=3cm,-{stealth},shorten >=2pt]
\ma{main}[]{3}{3};

\node at (main-1-1.center) {0};
\node at (main-2-2.center) {0};
\node at (main-3-3.center) {0};
\node at (main-1-2.center) {-1};
\node at (main-2-3.center) {-1};
\node at (main-3-1.center) {-1};
\node at (main-1-3.center) {1};
\node at (main-2-1.center) {1};
\node at (main-3-2.center) {1};

\end{tikzpicture}
\caption{Rock-paper-scissors. The colour is 1 if Eve wins, 0 if they draw and -1 if Adam wins. Also, the actions are ordered as in the name of the game}\label{8-fig:rps}
\commentAlt{Figure~\ref{8-fig:rps}: A 3x3 grid (matrix) containing numerical values.}
\commentLongAlt{Figure~\ref{8-fig:rps}: A square grid divided into three rows and three columns, forming nine cells. Each cell contains an integer:

Top row, from left to right: 0, -1, 1
Middle row, from left to right: 1, 0, -1
Bottom row, from left to right: -1, 1, 0}
\end{figure}

The following theorem lists some known results for matrix games:
\begin{theorem}\label{8-lem:mat}
Each $(m\times n)$-matrix game $M$ is determined and there exists optimal strategies for each player. 
\begin{itemize}
\item The value and an optimal strategy for each player can be found in polynomial time and the problem is equivalent to linear programming.

\item Let $c>0$ be some constant. Consider the matrix $cM$ where each entry of $M$ has been multiplied by $c$. Then, the value of $cM$ is $cv$.
\item Let $c$ be some constant. Consider the matrix $M+c$ where each entry of $M$ is $c$ larger (additively). Then, the value of $M+c$ is $v+c$.
\item The value of matrix games are monotone in the entries.
\end{itemize}
\end{theorem}
We will omit the proof of the existence of values, optimal strategies and the first bullet.
The second bullet can be viewed as changing currency and clearly, this does not affect the optimal strategy.
The third bullet can be viewed as getting a reward before playing the game, and again, clearly this does not affect how to play it.
The last bullet can be seen from that each pair of strategies must give a higher reward if the entries of the matrix is higher.
This is especially true if you consider the optimal strategy for Eve in $M$ together with an arbitrary strategy for Adam, which then shows that the value is higher.


Given a matrix $M$, we will by $\Value[M]$ denote the value of the matrix game $M$. 

Perhaps interestingly, an illustration of a matrix $M$ can be viewed as a game arena $\arena$ (for concurrent games) with only one non-absorbing vertex. In each type of games considered in this section (apart from concurrent reachability games, where no game can be illustrated as a matrix with non-star entries different from 0), the value of the game with that arena matches $\Value[M]$ and the optimal strategies for each player is to play an optimal strategy from $M$ in each round. One can also consider a game arena $\arena^*$ with an illustration similar to $M$, but where there is a star in each entry (and $c(v,i,j)=0$ for the unique non-absorbing state $v$ and any pair of actions $i,j$).
Again, the value is $\Value[M]$ (except for the case of discounted objectives, where the value is $(1-\delta)\Value[M]$) and the optimal strategies for each player is again to play an optimal strategy from $M$. 

One could easily be lead to believe that in games (called repeated games with absorbing states) that can be illustrated as a single matrix $M$ with some entries stared and others not, the value would be similar to $\Value[M]$ and the optimal strategy would again be to play the optimal strategy from $M$. 
However, this is very much not true and indeed, many of the games in this chapter, illustrating how complex concurrent games can be, are repeated games with absorbing states! In particular repeated games with absorbing states may (1)~have irrational values and probabilities in optimal strategies (with any objective), (2)~have no optimal strategies (for reachability and mean-payoff objectives) and (3)~have no $\epsilon$-optimal finite-memory or $\epsilon$-optimal Markov strategies (for mean-payoff objectives)!

\section{Concurrent discounted-payoff games}
\label{8-sec:discounted}
In this section we focus on concurrent discounted-payoff games. 
The key property of these games is that to a high degree, only the relative early part of the play matters.
We will first argue that the value iteration algorithm works and especially converges to the value of the game and then that there are stationary optimal strategies in concurrent discounted-payoff games.
While the value iteration algorithm also works for the games considered in the latter sections, we will not explicitly show it there, since the proofs become much more complex. The argument here however will allow us to show quite a few more statements in essence as corollaries of the theorem that value iteration works.

The value iteration algorithm is based on the concept of finite-horizon (or time limited) games. It is also sometimes referred to as dynamic programming.
Specifically, apart from the usual definition of a game, there is an additional integer $T$, denoting how many rounds are remaining initially, and a vector $v$, assigning a reward to each node if the game ends in that node with 0 rounds remaining. After round $T$ the reward is 0. 
I.e. for $T=0$, the outcome reward from node $x$ is $v_x$ in the first round and 0 in each later round.
Let $\ValueOp^T(v)$ be the vector that assigns to each node its value in the game with time-limit $T$ with vector $v$.

In general this formulation leads to a simple dynamic algorithm that computes $\ValueOp^T(v)$ inductively in $T$. 
We have that $\ValueOp^0(v)=v$ and given $\ValueOp^{T-1}(v)$ it is easy to compute $\ValueOp^T(v)$ because, if Eve selects row $i$ and Adam column $j$ in node $x$ in the first round, the outcome is \[
\sum_{v\in V}\ValueOp^{T-1}(v)\dest(x,i,j)(v)
\]
and thus $(\ValueOp^T(v))_x$ is the value of the matrix $M^{T,x,v}$, where entry $i,j$ is \[
\sum_{v\in V}\ValueOp^{T-1}(v)\dest(x,i,j)(v)
\]

It is common to start with the all-0 vector for $v$ when using the value iteration algorithm.

The following lemma shows many interesting properties of concurrent discounted-payoff games.

\begin{lemma}[Concurrent discounted-payoff games]
\label{8-cor:long}
Concurrent discounted-payoff games have the following properties:
\begin{itemize}
\item The value iteration algorithm converges for any initial vector $v$.
\item The value iteration algorithm has an unique fixed point, independent of the initial vector $v$.
\item There are optimal stationary strategies in concurrent discounted-payoff games and the unique fixed point of the value iteration algorithm is the value (thus, the games are determined).
\item The value of a concurrent discounted game can approximated in PPAD.
\item There are $\epsilon$-optimal stationary strategies with patience below $\frac{m\log(\epsilon/2)}{\log(1-\gamma)\epsilon}$.
\end{itemize}
\end{lemma}
\begin{proof}
The first item comes from considering the vectors $v$ and $\ValueOp^1(v)$. We thus have that $\ValueOp^{T+1}(v)\in [\ValueOp^{T}(v)-(1-\gamma)^T,\ValueOp^{T}(v)+(1-\gamma)^T]$ for all $T$. The statement then comes from that $\sum_{i=1}^\infty (1-\gamma)^i$ is a converging sum.

The second item comes from considering two fixed points, $u,v$. I.e., $\ValueOp^1(v)=v$ and thus $\ValueOp^T(v)=v$ for all $T$. Similar for $u$.
But, $v=\ValueOp^T(v)\in [\ValueOp^T(u)-(1-\gamma)^T, \ValueOp^T(u)+(1-\gamma)^T]=[u-(1-\gamma)^T, u+(1-\gamma)^T]$. Since it is true for all $T$, we have that $u=v$.

\begin{claim}
Consider some $T$ and the strategy for Eve that plays the first $T$ steps following an optimal strategy in the finite-horizon game of length $T$ with vector $v$, followed by playing arbitrarily. Then, the outcome is above $\ValueOp^T(v)- (1-\gamma)^T\max_{i} v_i$.
\end{claim}
\begin{proof}
For any strategy for Adam, the expected reward for the first $T$ rounds is at least the expected reward in the finite-horizon game. In each remaining round, the reward is at least $0$ in the real game, but $v_i$ in round $T$ for some $i$ followed by 0's in the finite-horizon game.
Since the outcome is $\ValueOp^T(v)$ in the finite-horizon game, the real outcome is then as described.
\end{proof}
One can show a similar statement for Adam.
For any $\epsilon>0$ one can pick a big enough $T$ such that $(1-\gamma)^T\max_{i} v_i\leq \epsilon$.

Let $v^*$ be the unique fixed point of the value iteration algorithm. 
Thus, $v^*=\ValueOp^T(v^*)$ for all $T$. Pick some optimal strategies $\sigma_x,\tau_x$ in $M^{T,x,v^*}$ for each $x$. Let $\sigma^*,\tau^*$ be the strategies that play $\sigma_x,\tau_x$ whenever in node $x$ in each round.
The strategy $\sigma,\tau$ are optimal in $\ValueOp^T(v^*)$ for each $T$, because $v^*$ is a fixed point. 
But, for each $\epsilon>0$,  the strategy $\sigma$ ensures outcome at least $v-\epsilon$ and $\tau$ ensures outcome at most $v+\epsilon$ using the claim. Hence, the third item follows.

The fourth item follows from that the value iteration algorithm is a contraction.

For the fifth item, consider the strategy used in the claim. 
Let $T$ be $\log(\epsilon/2)/\log(1-\gamma)$, \textit{i.e.} $T$ is such that \[
\gamma \sum_{i=T}^{\infty}(1-\gamma)^i=\epsilon/2
\]
or in words, $T$ is such that the total outcome of each step after the $T$-th step is at most $\epsilon/2$.
Intuitively, if we modify the strategy very little, then the change is unlikely to come up in the first $T$ steps. More precisely, we will modify our strategy so that the probability that change will matter is less than $\epsilon/2$. That implies that the outcome differs by at most $\epsilon$ from the value.
We will use this intuition together with the argument for the third item to give a bound on the patience of $\epsilon$-optimal strategies. Fix some optimal stationary strategy $\sigma$ for Eve and an arbitrary stationary strategy $\tau$ for Adam. Let $\sigma'$ be a stationary strategy obtained from $\sigma$ rounded greedily  so that each probability is a rational number with denominator \[
q=mT/\epsilon=\frac{m\log(\epsilon/2)}{\log(1-\gamma)\epsilon}.
\] We will argue that $\sigma'$ is $\epsilon$-optimal.

The rounding proceeds inductively as follows for each node $x$:
The numbers $p_i$ are the original probability and the numbers $p_i'$ are the new probabilities.
For each $i$, the number $p_i'$ is defined as follows: If $\sum_{j=1}^{i-1}(p_i-p_i')>0$, then round up (\textit{i.e.} $p_i'$ is the smallest rational with denominator $q$ so that $p_i<p_i'$) and otherwise round down, except the last number $p_\ell'$, which is such that $\sum_{j=1}^{\ell}p_i'=1$.
Note that this ensures that $-1/q<\sum_{j=1}^{i-1}(p_i-p_i')<1/q$. It also ensures that $|p_i-p_i'|<1/q$ for all $i$ (including for $i=\ell$).

For all nodes $x$ and rounds $T'\leq T$ we will define some random variables.
Specifically, the random variables denote what happen in round $T'$ if in node $x$.
The random variable $X_{x,T'}$ (resp. $Y_{x,T'}$) denotes the action picked by Eve if Eve follows $\sigma$ (resp. $\sigma'$).
The random variable $Z_{x,T'}$ denotes the action picked by Adam.
For each action pair $(i,j)$  the random variable $W_{x,i,j,T'}$ denotes the node entered in round $T'+1$, if Eve picks $i$ and Adam $j$. 
(As a side note: Each of the random variables are distributed the same way independent of $T'$).
Each of these random variables are independent of each other, except that (as we will define later) the random variables $X_{x,T'}$ and $Y_{x,T'}$ for each $x,T'$ are very much not independent of each other.

We see that we can view the first $T$ steps of the play when Eve follows $\sigma$ by only considering the outcome of $X_{x,T'}$, $Z_{x,T'}$ and $W_{x,i,j,T'}$ for all $T'$ and $x$ (even stronger: We only need to consider one $x,i,j$ for each $T'$, because the token is on only one node at a time). Similarly, for $\sigma'$, but using $Y_{x,T'}$ instead of $X_{x,T'}$.
For this to work, note that each random variable should be independent, except that the random variables $X_{x,T'}$ and $Y_{x',T''}$ need not be independent of each other for any $x',T''$. This is precisely the property we had for them!
For each $x,T'$,  we will then use a coupling $C_{x,T'}=(X'_{x,T'},Y'_{x,T'})$, a distribution over $[m]^2$, such that $X'_{x,T}$ is distributed as $X_{x,T'}$ and $Y'_{x,T'}$ is distributed as $Y_{x,T'}$. We will use a classic result for distributions, called the Coupling Lemma.

To introduce the Coupling Lemma, first, we need the notion of total variation distance. Given two distributions, $\Delta$ and $\Delta'$ over a set $S$, the total variation distance $t$ between $\Delta$ and $\Delta'$ is \[
t(\Delta,\Delta')=\frac{1}{2}\sum_{x\in S} |\Delta(x)-\Delta'(x)|
\] 

\begin{lemma}
For any distributions $\Delta$ and $\Delta'$ over a set $S$, we have 
\begin{itemize}
\item for all couplings $(X,Y)$ of $\Delta$ and $\Delta'$, that 
\[
t(\Delta,\Delta)\leq \Pr[X\neq Y]
\]
\item that there is a coupling $(X',Y')$ of $\Delta$ and $\Delta'$ satisfying that \[
t(\Delta,\Delta)= \Pr[X'\neq Y']
\]
\end{itemize}
\end{lemma}

Because of our rounding, we have that $t(X'_{x,T'},Y'_{x,T'})<\frac{m}{2q}$. 
Using that with the coupling lemma (the second part to be precise), lets us find a coupling $C_{x,T'}=(X'_{x,T'},Y'_{x,T'})$ 
such that $\Pr[X'_{x,T'}\neq Y'_{x,T'}]<\frac{m}{2q}$.

Consider the plays $\play_1,\play_2$ for when Eve follows $\sigma$ or $\sigma'$ respectively.
We can view the first $T$ steps of these plays by considering $X'_{x,T'}$ instead of $X_{x,T'}$ and similar when Eve follows $\sigma'$.
We can therefore see that the first $T$ steps two plays are different with probability 
$=p<\frac{mT}{2q}=\epsilon/2$
 using union bounds.

We therefore see that the value for the path $\play_1$ cannot differ from the value of $\play_2$ with more than $p\gamma\sum_{i=1}^T(1-\gamma)^i=p$. I.e. in the worst case, if $\play_1$ and $\play_2$ differs, the reward is 1 in each step for $\play_1$ but 0 in each step for $\play_2$.
Also, the rewards in the steps after step $T$ can also differ by at most $1$ and by our choice of $T$, we have that outcome contributed from these remaining steps are worth less than $\epsilon/2$ as well.
Hence, we see that $\sigma'$ obtains the same as $\sigma$ except for $\epsilon$ against any strategy $\tau$ and is thus $\epsilon$-optimal.
\end{proof}

There is a classic problem in geometry called the sum-of-square-roots problem. The problem is defined as follows:
Let $a,b_1,b_2,\dots,b_n$ be natural numbers. Is $\sum_{i=1}^n\sqrt{b_i}>a$? 

The problem comes up for decision problems about distances in Euclidean space. It is not known to be in $\P$ or $\NP$ for that matter, but is in the fourth level of the countering hierarchy, slightly inside $\PSPACE$. The issue is in essence that it is not known how good an approximation of $\sqrt{b_i}$ is necessary to decide the strict inequality. 

We will use the sum-of-square-roots problem to give an informal hardness argument, in that finding the exact value of a concurrent game is in general harder than solving the sum-of-square-roots problem. 

Consider the following game $G$:
There are three vertices, $\{0,1,s\}$ where $0$ and $1$ are absorbing, with colour 0 and 1 respectively.
The vertex $s$ is such that (1)~$c(s,i,j)=0$, (2)~$\dest(s,i,i)=1$ (for $i\in \{1,2\}$), (3)~$\dest(s,2,1)=0$ and (4)~$\dest(s,1,i)$ is the uniform distribution over $s$ and $0$. The game is illustrated in \Cref{8-fig:sqroot}.

Consider an optimal stationary strategy in $G$ for Eve. Let $p$ be the probability with which she plays the first action. If Adam knows that Eve will follow this strategy, the game devolves into a MDP. We know from that for such there exists optimal positional strategies and thus Adam is either going to play the left or right column always. Clearly, $0<p<1$ because $p=0$ or 1 means that either playing the left or right column with probability 1 would ensure that no positive reward ever happens.

\begin{figure}

\center
\begin{tikzpicture}[node distance=3cm,-{stealth},shorten >=2pt]
\ma{s}[$s:$]{2}{2};

\node at (s-1-1.center) {$1^*$};
\node at (s-2-2.center) {$1^*$};
\node at (s-2-1.center) {$0^*$};
\draw (s-1-2.center)  to ($(s-1-2.center)+(0.5cm,0)$)  arc (-90:180:0.5cm) node[pos=.5, right] {$1/2$};
\node at ($(s-1-2.center)-(0.2cm,0)$) {0};

\ma[shift={($(s)+(3 cm,0.5)$)}]{s0}[]{1}{1};
\node at ($(s0-1-1.east)+(0.5cm,0)$) {$:0$};
\node at (s0-1-1.center) {$0^*$};

\draw (s-1-2.center) -- (s0) node[below,pos=0.5] {$1/2$};

\end{tikzpicture}
\caption{Concurrent discounted game with value $v_s=-2+\sqrt{4+2(1-\gamma)}$}
\label{8-fig:sqroot}
\commentAlt{Figure~\ref{8-fig:sqroot}: A 2x2 grid labeled 's', with arrows and a self-loop connecting one cell to itself and an external cell. See long description.}
\commentLongAlt{Figure~\ref{8-fig:sqroot}: The image displays a 2x2 grid labeled 's:' on its left. The cells contain values: top-left is '1*', top-right is '0', bottom-left is '0*', and bottom-right is '1*'. An arrow originates from the top-right cell (containing '0') and points to an external square cell on the right, which contains '0*' and is labeled ': 0'. A curved arrow from the top-right cell forms a self-loop, bending upwards and then back down to the same cell, labeled '1/2'. Another curved arrow also originates from the top-right cell and points to the external '0*' cell, labeled '1/2'.}
\end{figure}

Let $v_0=0,v_1=1,v_s$ be the values of the three vertices. If he plays the left column, the outcome is $p(1-\gamma)$.
If he plays the right column, the outcome is $p/2(1-\gamma)v_s+(1-p)(1-\gamma)$. Observe that the former is increasing in $p$ and the latter is decreasing (since clearly, $0<(1-\gamma)v_s<v_s$). Also, both are continues. Thus, the optimum is for $p(1-\gamma)$ to be equal to $p/2(1-\gamma)v_s+(1-p)(1-\gamma)$ and both equal to $v_s$.
We will first isolate $v_s$ in $v_s=p/2(1-\gamma)v_s+(1-p)(1-\gamma)$.
\begin{align*}
v_s&=p/2(1-\gamma)v_s+(1-p)(1-\gamma)\Rightarrow (1-p/2(1-\gamma))v_s=(1-p)(1-\gamma)\Rightarrow \\
v_s&=\frac{(1-p)(1-\gamma)}{1-p/2(1-\gamma)}\enspace .
\end{align*}

Note that $p,\gamma<1$ thus, $1-p/2(1-\gamma)\neq 0$.
We then have the equality 
\begin{align*}
\frac{(1-p)(1-\gamma)}{1-p/2(1-\gamma)}&=p(1-\gamma)\Rightarrow\\(1-p)(1-\gamma)&=p(1-\gamma)(1-p/2(1-\gamma))\Rightarrow\\
0&=\frac{1-\gamma}{2} p^2+2p-1\Rightarrow \\
p&=\frac{-2\pm\sqrt{4+2(1-\gamma)}}{1-\gamma} \enspace .
\end{align*}

We see that $\frac{-2-\sqrt{4+2(1-\gamma)}}{1-\gamma}<0$. Thus, $p=\frac{-2+\sqrt{4+2(1-\gamma)}}{1-\gamma}$.
Also, \[v_s=-2+\sqrt{4+2(1-\gamma)}\enspace .\] 
It is straightforward to modify the construction to get any square-root desired for a fixed $\gamma$.

By making such a construction for each number $\sqrt{b_i}$, we can make another vertex $s^*$ that has the value of $(1-\gamma)\frac{\sum_{i=1}^n\sqrt{b_i}}{n}$ with a single action for each player and that goes to a uniformly random vertex. 
Observe that decreasing each reward by $x$, reduces the value of each vertex by $x$. Reduce each reward by $\frac{an}{1-\gamma}$.
We can then decide the sum-of-square-roots problem by deciding whether the value of $s^*$ is strictly above $0$. 

We get the following lemma.

\begin{lemma}
The (exact) decision problem for the value is sum-of-square-root hard for concurrent discounted-payoff games.
\end{lemma}

We will use this game  $G$ as an example to illustrate how to make the $\dest$-function deterministic for concurrent games while having the same value and a similar optimal strategy.

\begin{figure}

\center
\begin{tikzpicture}[node distance=3cm,-{stealth},shorten >=2pt]
\ma{s}[$s:$]{3}{3};

\node at (s-1-1.center) {$1^*$};
\node at (s-2-1.center) {$1^*$};
\node at (s-3-2.center) {$1^*$};
\node at (s-3-3.center) {$1^*$};
\node at (s-3-1.center) {$0^*$};
\node at (s-1-2.center) {0};
\node at (s-2-3.center) {0};
\node at (s-2-2.center) {$0^*$};
\node at (s-1-3.center) {$0^*$};

\end{tikzpicture}
\caption{Alternate concurrent discounted game with value $v_s=-2+\sqrt{4+2(1-\gamma)}$}
\label{8-fig:sqroot2}
\commentAlt{Figure~\ref{8-fig:sqroot2}: A 3x3 grid (matrix) labeled 's:' containing numerical values, some with an asterisk.}
\commentLongAlt{Figure~\ref{8-fig:sqroot2}: A square grid divided into three rows and three columns, forming nine cells. The grid is labeled 's:' on its left side. Each cell contains a number, some followed by an asterisk:

Top row, from left to right: '1*', '0', '0*'
Middle row, from left to right: '1*', '0*', '0'
Bottom row, from left to right: '0*', '1*', '1*'}
\end{figure}

Consider the following game $G'$:
There are three vertices, $\{0,1,s\}$ where $0$ and $1$ are absorbing, with colour 0 and 1 respectively.
The vertex $s$ is such that (1)
$c(s,i,j)=0$ for all $i,j$, (2)~$\dest(s,i,j)=s$ for $i+1=j$ (\textit{i.e.} for $(i,j)\in \{(1,2),(2,3)\}$), (3)~$\dest(s,i,j)=0$ for $i+j=4$ (\textit{i.e.} the ``other'' diagonal, $(i,j)\in \{(3,1),(2,2),(1,3)\}$) and (4)~$\dest(s,i,j)=1$ otherwise (\textit{i.e.} for $(i,j)\in \{(1,1),(2,1),(3,2),(3,3)\}$).
 The game is illustrated in \Cref{8-fig:sqroot2}.
 
We will argue that the value of $G'$ is equal to that of $G$.
 We clearly have that the value of $s$ is in $(0,1)$.
Consider a stationary strategy $\sigma$ for Eve
such that $\sigma(1)\neq \sigma(2)$. Let $p_i=\sigma(i)$ for $i\in \{1,2,3\}$. 
Let $\sigma'$ be such that $\sigma'(3)=p_3$ and otherwise, $\sigma'(i)=\frac{p_1+p_2}{2}$ for $i\in\{1,2\}$. Let $p_i'=\sigma'(i)$ for $i\in \{1,2,3\}$.
\begin{claim}
The strategy $\sigma'$ is at least as good as $\sigma$.
\end{claim}
\begin{proof}
If Adam plays 1, then the expected outcome is 
$p_1+p_2=p'_1+p'_2$ no matter if Eve plays $\sigma$ or $\sigma'$. 
If he plays $i$ for $i\in\{2,3\}$, the expected outcome is $
\frac{p_{4-i}}{1-p_{i-1}}$ if Eve plays $\sigma$ and otherwise, if she plays $\sigma'$, the expected outcome is   
$\frac{p'_1}{1-p'_2}=\frac{p'_2}{1-p'_1}$. 
Note that $\frac{p'_1}{1-p'_2}>\min_{i\in\{2,3\}\frac{p_{4-i}}{1-p_{i-1}}}$ and thus, $\sigma'$ is at least as good a strategy as $\sigma$.
\end{proof}

A similar argument shows that for any strategy $\tau$ for Adam the similar strategy $\tau'$ where $\tau'(1)=\tau(1)$ and $\tau'(i)=\frac{\tau(2)+\tau(3)}{2}$ for $i\in\{2,3\}$ is at least as good as $\tau$.
Consider that the players follows such stationary strategies $\sigma'$ and $\tau'$.
Let $\sigma$ be 
\[\sigma(i)=\begin{cases} \sigma'(1)+\sigma'(2)&\text{if }i=1\\
\sigma'(3)&\text{if }i=2\enspace .\end{cases}\]
Similarly, let 
 $\tau$ be 
\[\tau(i)=\begin{cases} \tau'(1)&\text{if }i=1\\
\tau'(2)+\tau'(3)&\text{if }i=2\enspace .\end{cases}\]
But playing $\sigma$ and $\tau$ in $G$ gives the same outcome as playing $\sigma'$ and $\tau'$ in $G'$ as can be seen as follows: In either game, with probability
\[
\sigma'(1)\tau'(2)+\sigma'(2)\tau'(3)=\frac{\sigma(1)\tau(2)}{2}\] we play again with a reward of 0, with probability 
\[\sigma'(1)\tau'(3)+\sigma'(2)\tau'(2)+\sigma'(3)\tau'(1)=
\frac{\sigma(1)\tau(2)}{2}
+\sigma(2)\tau(1)
\]
we get absorbed in 0 after a reward of 0 and with probability \[
(\sigma'(1)+\sigma'(2))\tau'(1)+(\tau'(2)+\tau'(3))\sigma'(3)
\] we get absorbed in 1 after a reward of 0.
But this is in particular the case if the players play optimally and thus, the value is the same in the two games.

Before, in Corollary \Cref{8-cor:long}, we argued that the patience of $\epsilon$-optimal stationary strategies was $q=\frac{m\log(\epsilon/2)}{\log(1-\gamma)\epsilon}$.
Giving a similar exponential bound for the optimal stationary strategies is harder than solving the sum-of-square-roots problem, as we will argue next.
Assume that we had an exponential bound for optimal stationary strategies.

\begin{figure}

\center
\begin{tikzpicture}[node distance=3cm,-{stealth},shorten >=2pt]
\ma{s}[$s':$]{2}{2};

\node at (s-1-1.center) {$1^*$};
\node at (s-2-1.center) {$0^*$};
\node at (s-1-2.center) {$0^*$};

\ma[shift={($(s)+(3 cm,-0.5)$)}]{ss}[]{1}{1};
\node at ($(ss-1-1.east)+(0.5cm,0)$) {$:s^*$};
\node at ($(s-2-2.center)-(0.2cm,0)$) {0};

\draw [yscale=-1] (s-2-2.center)  to ($(s-2-2.center)+(0.5cm,0)$)  arc (-90:180:0.5cm);
\draw (s-2-2.center) -- (ss);

\end{tikzpicture}
\caption{Concurrent discounted game that implies that if there is an exponential lower bound on patience, then the sum-of-square-roots problem is in P}
\label{8-fig:exact-hard}
\commentAlt{Figure~\ref{8-fig:exact-hard}: A 2x2 grid labeled 's'' with arrows and a self-loop connecting one cell to itself and an external cell labeled 's^s'. See long description.}
\commentLongAlt{Figure~\ref{8-fig:exact-hard}: The image displays a 2x2 grid labeled 's':' on its left. The cells contain values: top-left is '1*', top-right is '0*', bottom-left is '0*', and bottom-right is '0'. An arrow originates from the bottom-right cell (containing '0') and points to an external square cell on the right, which is labeled ': s^s'. A curved arrow also originates from the bottom-right cell and forms a self-loop, bending downwards and then back up to the same cell.}
\end{figure}

Consider an arbitrary yes-instance of the sum-of-square-roots problem, giving a vertex $s^*$. Reduce each reward by $a$ and in the new game let $s^*_a$ be the vertex corresponding to $s^*$. 
We will now create a game that uses the previous game as a sub-game.
The game has 1 additional vertex $s'$, which is a 2x2-matrix, such that $c(s',i,j)=0$ and $\dest(s,1,1)=1$ and $\dest(s,2,2)=s^*$ and $\dest(s,i,j)=0$ for $i\neq j$.
There is an illustration in~\Cref{8-fig:exact-hard}, using the vertex $s^*$ as above. 
Using an argument like above, we see that the probability $p$ to play the top action in the vertex $s'$ is such that $p(1-\gamma)=(1-p)(1-\gamma)x$, where $x$ is the value of $s^*$. Thus, $x=\frac{p}{1-p}$. If $p$ only needs to be exponential small, then $x$ is exponentially small as well. This is true for any yes-instance of the sum-of-square-roots problem and thus, we only need polynomially many digits to decide the problem. We can find polynomially many digits of $\sqrt{b_i}$ for each $i$ in polynomial time. We get the following lemma.

\begin{lemma}
Giving an exponential lower-bound on patience for optimal stationary strategies in concurrent discounted-payoff games implies that the sum-of-square-roots problem is in $\P$
\end{lemma}

\section{Concurrent reachability games}
\label{8-sec:reachability}
In this section we consider concurrent reachability games. 
Intuitively, unlike concurrent discounted-payoff games, these games cares only about the final part of the play.
This, while perhaps not clear directly from the definitions, makes the games somewhat harder. For instance, the value iteration algorithm requires double-exponential time.
Note that like for concurrent discounted-payoff games, if we force Eve to follow some strategy, the game reduces to a MDP and there exists optimal positional strategies.

We will not prove the following known lemma. We will however, show the weaker statement that the decision problem for the value can be done in $\PSPACE$.

\begin{lemma}[Properties of concurrent reachability games]
\label{8-lem:reach_determined}
Concurrent reachability games are determined. Also, finding the value of a concurrent reachability game can be done in TFNP[NP].
\end{lemma}

We will argue that there might not be optimal strategies for Eve in concurrent reachability games. 
The game we will use will later be a member of a family of games that requires high patience to play well.

The snowball game (or purgatory 1) is defined as follows:
There are 3 vertices, the goal vertex $\Win$, $\bot$ (an absorbing vertex) and a vertex 1, which has a 2x2 matrix, such that $\dest(x,r,c)$ is a Dirac distribution over (1)
$\Win$ for $r=c$, (2) $1$ for $r<c$ and (3) $\bot$ for $r>c$.
When we illustrate the game, we write view the goal vertex $\Win$ as being an absorbing vertex with colour 1.
There is an illustration of the snowball game in \Cref{8-fig:snowball}. 

\begin{figure}

\center
\begin{tikzpicture}[node distance=3cm,-{stealth},shorten >=2pt]
\ma{s}[$1:$]{2}{2};

\node at (s-1-1.center) {$1^*$};
\node at (s-2-1.center) {$0^*$};
\node at (s-1-2.center) {$0$};
\node at (s-2-2.center) {$1^*$};

\end{tikzpicture}
\caption{The snowball game or purgatory 1, in which no optimal strategy exists for Eve}
\label{8-fig:snowball}
\commentAlt{Figure~\ref{8-fig:snowball}: A 2x2 grid (matrix) labeled '1:' containing numerical values, some with an asterisk.}
\commentLongAlt{Figure~\ref{8-fig:snowball}: A square grid divided into two rows and two columns, forming four cells. The grid is labeled '1:' on its left side. Each cell contains a number, some followed by an asterisk:

Top row, from left to right: '1*', '0'
Bottom row, from left to right: '0*', '1*'}
\end{figure}

For any $\epsilon>0$, consider the stationary strategy for Eve that plays the first action with probability $1-\epsilon$. 
If Adam plays the left column always, play will reach $\Win$ with pr. $1-\epsilon$. If Adam plays the right column always, play reaches $\Win$ with pr. 1. Hence, the strategy for Eve is $\epsilon$-optimal and the value of the vertex is 1.

On the other hand, if Adam plays the right column whenever Eve plays the first action with pr. 1, and otherwise plays the first action, 
then in the last round in the vertex there must be a positive pr. that play goes to vertex 0. Hence, Eve has no optimal strategy.

\begin{lemma}[No optimal strategies in concurrent reachability games]
\label{8-lem:no_opt_reach}
Eve need not have an optimal strategy in a concurrent reachability game.
\end{lemma}

The following lemma states some classical results for concurrent reachability games that we will not prove.
\begin{lemma}[Classical results for concurrent reachability games]
\label{8-lem:reach_class}
\begin{itemize}
\item For any $\epsilon>0$, there are always $\epsilon$-optimal stationary strategies for Eve and optimal stationary strategies for Adam. 
\item The value iteration algorithm converges to the optimal value and is defined exactly like for concurrent discounted-payoff games, except that $\gamma=0$ and the target vertex has value 1 in the first iteration.
\item The values $v$ are the least fixed point (\textit{i.e.}, every other fixed point $v'$ is such that for all $i$ $v_i\leq v'$) of the value iteration operator $\ValueOp$.
\end{itemize}
\end{lemma}
(Note that the games are not symmetric, in that Eve tries to reach a node and Adam tries to stay away from it, and in particular, even though Eve need not have an optimal strategy in a concurrent reachability game, Adam always has one.)

The decision problem for the existential first order theory over the reals is the following decision problem:
Given a function $F:\R^n\rightarrow \{\text{true},\text{false}\}$, is there an vector $v$ such that $F(v)$ is true?
The function $F$ must be an well-formed (\textit{i.e.} connected with logical `and', `or' and `not') quantifier-free formula over polynomial inequalities.
\textit{E.g.} $x^2y+z\geq 5\wedge \neg (xz\leq 3)$ would be such a function.

\begin{lemma}
The decision problem for the existential theory over the reals is in PSPACE
\end{lemma} 

We will now, given a number $c$, encode the problem whether the value in a concurrent reachability game is  $<c$, starting from some vertex $x$.
The idea is that we can describe a fixed point of the value iteration operator, \textit{i.e.}
we can describe that $v_i=\Value[M^i(v)]_i$. Since we know that the values are the least fixed point, we can then just add the condition that $v_x<c$. 
We can describe the value of a matrix game by guessing a strategy for each player, $\sigma$ for Eve and $\tau$ for Adam, and then checking that these strategies are optimal by showing that the outcome obtained by following Eve's strategy is equal to what is obtained by following Adam's.
\textit{I.e.} we check that for Eve's strategy, the least outcome obtained when Adam plays any given column is $v_i$ and similar for Adam.
We describe that as $v_i=\min_{a\in [m]} (\sigma M^i_{*,a}(v))$ and $v_i=\max_{a\in [m]} (\tau M^i_{a,*}(v))$, where $M^i_{*,j}(v)$ is the $j$-th column of $M^i(v)$ and $M^i_{j,*}(v)$ is the $j$-th row for all $j$. 
We can express that $x=\min(a_1,a_2,\dots,a_n)$ for any number $n$ and any polynomials $a_1,a_2,a_n$, in the first order theory of the reals by stating that \[
x\leq a_1\wedge x\leq a_2\wedge \dots \wedge x\leq a_n \wedge (x=a_1\vee x=a_2\vee \dots \vee x=a_n) \enspace .\]
Similarly, one can also describe that $x=\max(a_1,a_2,\dots,a_n)$ for any number $n$ and any polynomials $a_1,a_2,a_n$.

We can similarly make a statement that $v_x\leq c$. Using that PSPACE is equal to co-PSPACE, we also get that we can find if $v_x\geq c$ and $v_x>c$ and therefore also $v_x=c$ and $v_x\neq c$, all in PSPACE.

\begin{lemma}
Decision problems for the values in a concurrent reachability game is in PSPACE.
\end{lemma}

The set of vertices that have value 0 can be found in polynomial time. This is because, the set of vertices that have value 0 in the time limited game of length $n$ has also value 0 in all other time limited games. That this is so is easy to see by considering that Eve plays an $\epsilon$-optimal strategy for $v>\epsilon$, where $v$ is the value of the vertex with the lowest value. The game then devolves to a MDP for Adam, and the statement is true for such.

\begin{lemma}[Value 0 vertices in concurrent reachability games]
\label{8-lem:find_0_reach}
The set of vertices of value 0 in a concurrent reachability game can be found in polynomial time.
\end{lemma}

Next, we argue that we can find the set of vertices $S_1$ that have value 1 in polynomial time as well.
For notational convenience, for stationary strategies $\sigma,\tau$ we will, for a set $x$ and a set of vertices $S$ use \[
F^{\sigma,\tau}(x,S):=\sum_{r\in A_1}\sum_{c\in A_2}\sum_{x'\in S}\sigma(x)(r)\tau(x)(c)\delta(x,r,c)(x')\enspace ,\]
 \textit{i.e.} the probability when the players follows $\sigma,\tau$ to go from $x$ directly to a vertex in $S$.

For a set $S$, containing $\Win$ and a non-empty subset $S'\subseteq(S\setminus \{\Win\})$ and a vertex $x\in S'$, let the {\em subset property} be the following:
For each $\epsilon>0$, there is a strategy $\sigma$ for Eve, such that for any strategy $\tau$ for Adam, $F^{\sigma,\tau}(x,S\setminus S')\cdot \epsilon >F^{\sigma,\tau}(x,V\setminus S)$.

For a set $S$, containing $\Win$, let the {\em value-1-property} be that the subset property is satisfied for each $S'\subseteq(S\setminus \{\Win\})$ (with some $x\in S'$).
For a set $S$ satisfying the value-1-property, we will define some subsets.
Let $S^0$ be the set consisting of $\Win$.
Let $S^i$, for each $i\geq 1$ be the set of vertices such that for each $x\in S^i$,
the vertex $x$ satisfies the subset property for $S'$ and $S=\bigcup_{j=0}^{i-1}S^j$.
Let $\ell$ be the largest number such that $S^\ell$ is non-empty (note that $S^i$, for $i>\ell$ must then be empty).

We will be using the following lemma.

\begin{lemma}
The set of vertices $S_1$ of value 1 satisfies the value-1-property.
\end{lemma}
\begin{proof}
The proof is by contradiction. Thus, there is an $S$ (not containing $\Win$) such that $S_1$ and $S$ does not satisfy the subset property for any $x\in S$. \textit{I.e.} for some constant $\epsilon>0$,
$F^{\sigma,\tau}(x,S_1\setminus S)\epsilon \leq F^{\sigma,\tau}(x,V\setminus S_1)$ for any strategy $\sigma$ for Eve and some strategy $\tau_{\sigma}$ for Adam and for every $x\in S$.
But then, all vertices in $S$ have value $\leq (1-\epsilon)+\epsilon V_{\max}$ where $v_{\max}<1$ is the largest value of a vertex in $V\setminus S_1$.
This is because to get to $\Win$ from $S$, it must leave $S$ and in that step, the probability to go to a vertex in  $V\setminus S_1$ (from which one cannot obtain more than $v_{\max}$) is at least the constant $\epsilon$.  
\end{proof}

We will argue that this is a precise characterisation of $S_1$ next.

\begin{lemma}
\label{lem:sufficent_for_value1}
Consider a set $S'$ satisfying the value-1-property, then each vertex of $S'$ has value 1.
\end{lemma}
\begin{proof}
We will for all $i$ and any $\epsilon>0$ construct a strategy $\sigma_{i,\epsilon}$ for Eve that starting from a vertex in $\bigcup_{j=i}^n S^j$ will eventually get to $S^{i-1}$ with probability at least $1-\epsilon$ (especially, the strategy $\sigma_{1,\epsilon}$ is $\epsilon$-optimal). We will do it using backwards induction in $i$ and thus start from $i=\ell$.

Note that the base case for $i=\ell$ follows directly from the condition.
We now for any $\epsilon>0$ will find a strategy $\sigma_{i,\epsilon}$, given that we have a strategy $\sigma_{i+1,\epsilon'}$ for any $\epsilon'>0$.
The idea is that the subset property gives a strategy $\sigma$ such that for all $\tau$, $F^{\sigma,\tau}(x,S'\setminus S)\epsilon/2 >F^{\sigma,\tau}(x,V\setminus S')$, for $S=\bigcup_{j=i}^{n} S^j$.
Note that in expectation, following this strategy, we go to some other vertex in $S$ with at least some fixed probability $p$ (that could be quite close to 1).
Hence, in expectation, we need to be in such a vertex $1/(1-p)$ times before entering either $S'\setminus S$ or $V\setminus S'$.
We therefore follow the strategy $\sigma_{i+1,\epsilon'}$ in $\bigcup_{j=i+1}^n S^j$, where $\epsilon'=\frac{\epsilon}{2(1-p)}$.
The inductive construction then follows by applying union bound over the $1/(1-p)$ times we are in $S^i$.
\end{proof}
Note that the lemmas together show that $S_1$ is the largest set satisfying the value-1-property.

Our algorithm, $\crgLim$, for finding $S_1$ is then as follows:
Assign to each vertex $x$ a rank $\rk_x$, a value in $0,1,\dots,n,\bot$, starting with $\rk_x=0$.
Let $\overline{S}^i$ be the set of vertices of rank $i$.
Let $\overline{S}_1=V\setminus S^{\bot}$.
We increment (where a rank is less than another if it is a smaller number. Also, all other ranks are below $\bot$) the rank of a non-goal vertex $x\in \overline{S}^i$, whenever it does not satisfy the subset property for $S=\overline{S}_1$ and $S'=\bigcup_{j=i}^n \overline{S}^j$.
Note that no vertex can satisfy the subset property for rank 0, since $S'$ is all vertices.
Whenever a stable configuration is reached, output $\overline{S}_1$.

\begin{lemma}
The output of the $\crgLim$\ algorithm is correct.
\end{lemma}
\begin{proof}
The idea is that we want $S^i=\overline{S}^i$ at termination.
Note that the subset property is harder to satisfy for a vertex $x$ if we remove vertices from $S$ or from $(S\setminus S')$.
Thus, if at some time we have that $x$ does not satisfy the subset property for $S=\overline{S}_1$ and $S'=\bigcup_{j=i}^n \overline{S}^j$, then it does not do so for $S$ being any subset of $\overline{S}_1$ or $(S\setminus S')$ being any subset of $\bigcup_{j=0}^{i-1} \overline{S}^j$.
However, initially $S_1\supseteq\overline{S}_1$ (being all vertices) and $\bigcup_{j=i}^n S^j\supseteq \bigcup_{j=0}^{i-1} \overline{S}^j$ for all $i$. But we must have that $S_1\supseteq\overline{S}_1$ and $\bigcup_{j=i}^n S^j\supseteq \bigcup_{j=0}^{i-1} \overline{S}^j$ for all $i$, at all latter points as well, since in the last iteration it was satisfied, for all $i$ and all $x\in S^i$, we have that the subset property is satisfied for $S=\overline{S}_1$ and $S'=\bigcup_{j=i}^n \overline{S}^j$, because we have that $S\supseteq S_1$ and $(S\setminus S')=\bigcup_{j=1}^{i-1} \overline{S}^j\supseteq \bigcup_{j=1}^{i-1} S^j$.
On the other hand, eventually no vertex gets it rank incremented (since there are a finite number of ranks and vertices) and the algorithm terminates with a set $\overline{S}_1\supseteq S_1$ satisfying the value-1-property. Since $S_1$ is the largest such set, we have that $\overline{S}_1=S_1$.
\end{proof}

We will now consider the running time of the algorithm.
We will consider that we can decide whether a pair of sets $S$ and $S'$ satisfies the subset property for a vertex $x$ can be solved in $O(k)$ time.
Let $S_x$ be the set of vertices that can be visited in a play immediately after $x$. Observe that $|S_x|$ is a lower bound on the number of arrows from matrix $x$ in our illustration of the game.
Consider a vertex $x$ and a rank $i$, and we want to find an upper bound on the computation we do on $x$ while it has rank $i$.
Clearly, we only need to consider incrementing the rank of $x$ whenever a vertex in $S_x$ has changed its value and only if it is changed to either $i$ (from $i-1$) or to $\bot$ (from $n$), because otherwise, the sets $S$ and $S'$ have not changed. 
Thus, we only do at most $2|S_x|+1$ checks whether $x$ satisfies the subset property.
There are $n+1$ ranks, so in total for $x$, we use at most $(n+1)(2|S_x|+1)$ checks.
Hence, in total over all $x$, we do $O(n\sum_{x} |S_x|)$ checks.

\begin{lemma}
The runtime of the $\crgLim$ algorithm is $O(nk\sum_{x} |S_x|)$.
\end{lemma}

We will then finally consider how to check whether $x$ satisfies the subset property for a pair of sets $S$ and $S'$.
We will do so by constructing a strategy $\sigma=\sigma(\epsilon)$ for Eve satisfying the property for any fixed $\epsilon>0$.
We will construct the strategy  $\sigma$ from some sequence of pairs of sets (of rows and columns) $(R_1,C_1),(R_2,C_2),\dots,(R_{\ell},C_{\ell})$. We will let $C_i^*=\bigcup_{j=1}^i C_j$ and similar for $R^*_i$.
For convenience, we also define $C_0^*$ as the empty set of columns.
We will define $R_i$ from $C_{i-1}^*$ as each row $r\not\in R_{i-1}^*$ such that, for all $c\not\in C^*_{i-1}$, we have that $F^{r,c}(x,V\setminus S)=0$.
We will define $C_i\not\in C_{i=1}^*$ from $R_i$ as each column $c$ such that there is a $r\in R_i$ such that 
$F^{r,c}(x,S\setminus S')>0$.
The set $R^{*}_{\ell}$ is the first set such that $R^*_{\ell+1}$ is empty (clearly, by construction all sets $R_i,C_i$ for $i>\ell$ would also be empty).

\begin{lemma}
There is a strategy $\sigma(\epsilon)$ for all $\epsilon>0$ if and only if $C_{\ell}^*$ is the set of all columns.
\end{lemma}
\begin{proof}
We will first argue that if $C_{\ell}^{*}$ is not all columns $C$, then the strategy $\tau$ that plays uniformly over $C'=(C\setminus C_{\ell}^*)$ shows that no strategy $\sigma(\epsilon)$ exists for small enough $\epsilon>0$. This is because any row $r$ such that $F^{r,c}(x,S\setminus S')>0$ for some $c\in C'$ is also such that $F^{r,c'}(x,V\setminus S)>0$ for some column $c'\in C'$. This is because otherwise $r$ would be in $R^i$ for some $i$ and then $c$ would be in $C_{\ell}^*$. Hence, the probability $F^{r,c'}(x,V\setminus S)$ cannot be more than a constant factor smaller than $F^{r,c}(x,S\setminus S')$.

Otherwise, if $C_{\ell}^*=C$, then let $\sigma(\epsilon)$ be the strategy that picks an $i$ from the distribution $\dist$ and then plays an action in $R^i$ uniformly at random. The distribution $\dist$ is such that for all $j\in \{1,\dots,\ell-1\}$ we have $\Pr^{\dist}[i=j]\epsilon\delta_{\min}/m=\Pr^{\dist}[i>j]$. 

To argue that $\sigma=\sigma(\epsilon)$ satisfies the subset property for $x,S,S'$
consider each column $c$. We have that $c\in C^j$ for some $j$. 
Let $p$ be the pr. with which a row in $R^j$ is played by $\sigma$ and thus, $F^{\sigma,c}(x,S\setminus S')\geq p\delta_{\min}>0$.
Any row $r$ such that $F^{r,c}(x,V\setminus S)>0$ must be outside $R^*_j$ by construction (and if such a row exists $j<\ell$). 
We play such rows with pr. $\leq \Pr^{\dist}[i>j]$ and thus $\Pr^{\dist}[i>j]\geq F^{\sigma,c}(x,V\setminus S)$.
We have that $pm>\Pr^{\dist}[i=j]$ (strict because $j<\ell$) and thus, \[
F^{\sigma,c}(x,S\setminus S')\epsilon\geq p\delta_{\min}\epsilon>\delta_{\min}\epsilon/m\Pr^{\dist}\nolimits[i=j]\geq 
\Pr^{\dist}\nolimits[i>j]\geq F^{\sigma,c}(x,V\setminus S) \enspace .\]
This completes the proof of the lemma.
\end{proof}

Our algorithm for checking if a vertex $x$ satisfies the subset property for sets $S,S'$ is as follows:
We will construct the sequence of sets $(R_1,C_1),(R_2,C_2),\dots,(R_{\ell},C_{\ell})$.
To do so we will use a data structure.
The data structure has the following properties:
Initially, for each column $c$, we will make a list $L_c$ of the rows $r$ such that $F^{r,c}(x,V\setminus S)>0$.
We will also have a counter for each row $r$ that initially contains how many such columns there are.
Finally for each row $r$, there is a list $L_r$ of columns such that $F^{r,c}(x,S\setminus S')>0$.

The algorithm then uses the data structure as follows:
Let $i\leftarrow 1$.
Add all the rows with the counter at 0 to $R^i$. 
If $R^i$ is the empty set, return whether $C_{i-1}^*$ is all columns.
For each row $r$ in $R^i$ go through $c\in L_r$ and subtract 1 from the counter of each row in $L_c$. If a counter reach 0, add it to $R^{i+1}$.
Increment $i$.
Go to line 3.

The total running time the algorithm is $O(\sum_{r,c}|\supp(\dest(x,r,c))|)$.

\begin{lemma}
\label{8-lem:subset_property}
We can check whether a vertex $x$ satisfies the subset property for sets $S$ and $S'$ in time \[O(\sum_{r,c}|\supp(\dest(x,r,c))|).\]
\end{lemma}

We therefore get that 
\begin{lemma}
\label{8-lem:val1}
We can compute the set of vertices of value 1 in time 
\[O\left( n\sum_{x} |S_x|\sum_{r,c}|\supp(\dest(x,r,c))| \right).\]
\end{lemma}

Next, we will give a lower bound for patience: we prove that in some games, the patience for every $\epsilon$-optimal stationary strategy must be high. 
For a number $k$, let purgatory $k$ be the following game:
There are $2+k$ vertices, $\Win$ (which is vertex 0), one vertex $\bot$ which is absorbing and each other vertex $i\in \{1,\dots, k\}$ has a 2x2 matrix, such that $\dest(x,r,c)$ is a Dirac distribution over (1)
$i-1$ for $r=c$, (2) $k$ for $r<c$ and (3) $\bot$ for $r>c$.
There is an illustration of Purgatory $4$ in \Cref{8-fig:purgatory}. 

\begin{figure}
\center
\begin{tikzpicture}[node distance=3cm,-{stealth},shorten >=2pt]

\ma{s2}[$3:$]{2}{2};
\ma[shift={($(s2)+(0,3cm)$)}]{s1}[$4:$]{2}{2};
\ma[shift={($(s2)+(3cm,0)$)}]{s3}[$2:$]{2}{2};
\ma[shift={($(s3)+(3cm,0)$)}]{s4}[$1:$]{2}{2};

\draw (s2-1-2.center) to (s1);
\draw (s3-1-2.center) to[bend right] (s1);
\draw (s4-1-2.center) to[bend right] (s1);

\foreach \x/\y in {2/3,3/4} {
\draw (s\x-2-2.center) to (s\y);
\draw (s\x-1-1.center) to[out=70] (s\y);
}
\foreach \x in {1,2,3,4} \node at (s\x-2-1.center) {$0^*$};
\draw (s1-2-2.center) to (s2-1-2);
\draw (s1-1-1.center) to[out=200] (s2-1-1);

\niceloop{s1-1-2};

\node at (s4-1-1.center) {$1^*$};
\node at (s4-2-2.center) {$1^*$};

\end{tikzpicture}
\caption{Purgatory $4$. For clarity, the colours are omitted, except that $0^*$ corresponds to an edge to an absorbing vertex different from $\Win$ and $1^*$ corresponds to an edge to $\Win$}
\label{8-fig:purgatory}
\commentAlt{Figure~\ref{8-fig:purgatory}: A network of interconnected 2x2 grids, some with internal divisions or values, and labeled levels. See long description.}
\commentLongAlt{Figure~\ref{8-fig:purgatory}: The image displays four 2x2 grids, arranged in a descending and then ascending pattern, resembling a complex state machine. Each grid is labeled with a number followed by a colon, indicating its level or sequence.

Top grid: Labeled '4:', it has a top-right cell containing a vertical line, and a bottom-right cell containing a vertical line. The bottom-left cell contains '0*'. There's a self-loop on its top-right cell. Arrows point from this grid to the grid below it, and to the leftmost grid at level 3.
Leftmost middle grid: Labeled '3:', its bottom-left cell contains '0*'. Arrows connect it to the grid above and to the grid to its right.
Middle right grid: Labeled '2:', its bottom-left cell contains '0*'. An arrow connects it to the grid to its right.
Rightmost grid: Labeled '1:', its top-left cell contains '1*', bottom-left contains '0*', and bottom-right contains '1*'.
Multiple curved arrows indicate transitions between these grids, creating a complex flow. For instance, arrows connect grid 4 to grid 3, grid 3 to grid 2, and grid 2 to grid 1. There are also arrows going from grid 1 back to grid 4, and from grid 2 back to grid 4, and from grid 3 back to grid 4. Some cells within the grids also have internal arrows indicating further transitions.}
\end{figure}

It is easy to see that all vertices but $\bot$ are in $S_1$ using the value 1 property.

\begin{lemma}[Properties of the purgatory game]
\label{8-lem:purgatory}
For any $0<\epsilon<1/2$ and any $k\geq 1$, there is a unique strategy for Eve with least patience which is $\epsilon$-optimal in purgatory $k$. That strategy has patience $\epsilon^{-2^{k-1}}$
\end{lemma}
\begin{proof}
We will find the best strategy with patience $1/p$ for any number $0<p<1/2$.
It is clear that the best strategy with patience $1/p$ is to play the strategy that maximises the pr. of eventually reaching $i$ from vertex $j$ for all $j>i$ while having patience $1/p$.
Let $x_k=1-p$ and let $x_i=1-\sqrt{1-x_{i+1}}$. We will argue that $x_i$ is the probability of eventually reaching vertex $i-1$ from vertex $k$ for all $i\in \{1,\dots,k\}$ and that there is a unique best strategy with patience $1/p$.

The unique best strategy in vertex $k$ is to play the top action with pr. $1-p$ and the bottom action with pr. $p$. This ensures that the pr. $x_k$ of reaching $k-1$ from $k$ is $1-p$ if Adam plays the left column.

Consider now vertex $i\in \{1,\dots,k-1\}$.
For the purpose of finding good strategies in $i$, we can view vertex $i$, when Eve plays her best strategy in $j>i$ and Adam plays a best response, as a smaller reachability game with 3 vertices, \textit{i.e.} $i-1$ (as $\Win$), $\bot$ and $i$, where $\dest(i,r,c)$ is (1) a Dirac distribution over $i-1$ for $r=c$, (2) a Dirac distribution over $\bot$ for $r>c$ and (3) a distribution that goes to $\bot$ with pr. $1-x_{i-1}$ and to $i$ with the remaining pr. for $r<c$. See \Cref{8-fig:purgatory}.

Let $p_i$ be the probability with which a strategy $\sigma$ plays the top row in vertex $i$.
We can then consider the game as a MDP, since we have fixed a stationary strategy for one of the players.
It is clear that the pr. to reach $i-1$ from $i$ if Adam plays the right column is strictly increasing in $p_i$ and the pr. to reach $i-1$ from $i$ if Adam plays the left column is strictly decreasing in $p_i$. We will consider the strategy such that the pr. of reaching $i-1$ is equal no matter which column Adam plays (by the previous statement, this strategy must then be optimal).
Observe that the pr. of reaching $i-1$ if Adam always plays the left column is $p_i$ (which is then also the pr. to reach $i-1$ from $i$) and if he always plays the right column it is  $p_i x_i p_i+(1-p_i)$ (using that the pr. of reaching $i-1$ from $i$ is $p_i$).
Thus, we have that 
\[
p_i=p_i x_i p_i+(1-p_i)\Rightarrow 0=p_i^2/2 x_i-p_i+1/2 \Rightarrow p_i=\frac{1\pm \sqrt{1-x_i}}{x_i}\enspace .\] We see that $\frac{1+ \sqrt{1-x_i}}{x_i}>1$ and thus the solution is $p_i=\frac{1- \sqrt{1-x_i}}{x_i}$ or that $(x_{i-1}=)x_ip_i=1- \sqrt{1-x_i}$.

Consider now that the strategy is exactly $\epsilon$-optimal, implying that $x_1=1-\epsilon$. 
We will argue that $p=\epsilon^{2^{k-1}}$.
We will do so by arguing  using induction in $i$ that $x_i=1-\epsilon^{2^{k-1}}$, since $x_k=1-p$ (this also shows that the strategy is indeed using patience $1/p$ since the probabilities in the other vertices, which are $p_i=\frac{x_{i-1}}{x_i}$, are strictly above $1/p$).
We have already noted that  $x_1=1-\epsilon=1-\epsilon^{2^0}$.
We will next argue that $x_i=1-\epsilon^{2^{k-1}}$, for $i\geq 2$ using that $x_{i-1}=1-\epsilon^{2^{k-2}}$.
We have that \[
1-\epsilon^{2^{k-2}}=x_{i-1}=1-\sqrt{1-x_{i}}\Rightarrow \sqrt{1-x_{i}}= \epsilon^{2^{k-2}}\Rightarrow
1-\epsilon^{2^{k-1}}=x_i\enspace .
\]
This completes the proof of the lemma.
\end{proof}

Concurrent reachability games are not symmetric in the players. \textit{E.g.} Adam always have an optimal strategy but Eve might not. We will next argue that Adam still requires double exponential patience to play well.

Consider the following game called purgatory duel $k$ which can be viewed as a symmetric version of purgatory $k$.
There are $3+2k$ vertices, $\Win$ (which is vertex 0), one vertex $\bot$ which is absorbing (and is also vertex $0'$), and the start vertex $s$ and each other vertex $\{1,\dots, k,1',\dots,k'\}$ has a 2x2 matrix. Each vertex $x\in \{1,\dots, k\}$ is such that $\dest(x,r,c)$ is a Dirac distribution over (1) $x-1$ for $r=c$, (2) $s$ for $r<c$ and (3) $\bot$ for $r>c$.
 Each vertex $x'\in \{1',\dots, k'\}$ is such that $\dest(x',r,c)$ is a Dirac distribution over (1) $x-1'$ for $r=c$, (2) $s$ for $r<c$ and (3) $\Win$ for $r>c$. The start vertex is 1x1 matrix and is such that $\dest(s,r,c)$ is a uniform distribution over $k$ and $k'$.
There is a illustration of Purgatory Duel $2$ in~\Cref{8-fig:purgatoryduel}.

\begin{figure}
\center
\begin{tikzpicture}[node distance=3cm,-{stealth},shorten >=2pt]

\ma[shift={($(4.5cm,4.5cm)$)}]{s}[$s:$]{1}{1};
\ma[shift={($(3cm,0)$)}]{s2'}[$2':$]{2}{2};
\ma[shift={($(s2')+(3cm,0)$)}]{s1'}[$1':$]{2}{2};
\ma[shift={($(0,3cm)$)}]{s2}[$2:$]{2}{2};
\ma[shift={($(s2)+(0,3cm)$)}]{s1}[$1':$]{2}{2};

\foreach \x in {1,2,1',2'} \draw (s\x-1-2.center) to (s);
\foreach \x/\y in {1/0,2/0,1'/1,2'/1} \node at (s\x-2-1.center) {$\y ^*$};
\foreach \x/\y in {1/1,1'/0} \foreach \z in {1,2} \node at (s\x-\z-\z.center) {$\y ^*$};
\draw (s2-1-1.center) to (s1);
\draw (s2-2-2.center) to[out=30,in=-30] (s1);
\draw (s2'-2-2.center) to (s1');
\draw (s2'-1-1.center) to[out=60,in=120] (s1');
\draw (s.center) to[out=225,in=100] node[pos=0.6,right] {$1/2$} (s2');
\draw (s.center) to[out=225,in=-10] node[pos=0.6,above] {$1/2$} (s2) ;

\end{tikzpicture}
\caption{Purgatory duel $2$. For clarity, the colours are omitted, except that $0^*$ corresponds to an edge to an absorbing vertex different from $\Win$ and $1^*$ corresponds to an edge to $\Win$}
\label{8-fig:purgatoryduel}
\end{figure}

We will say that a strategy $\sigma$ for Eve mirrors a strategy $\tau$ for Adam, if $\sigma(i)=\tau(i')$ and $\sigma(i')=\tau(i)$ for all $i$.
Similarly, $\tau$ mirrors $\sigma$.

\begin{lemma}
The value of vertex $s$ is $1/2$. Also, for any $\epsilon>0$, any $(1/2-\epsilon)$-optimal strategy $\tau$ for Adam does not follow a Dirac distribution in $i'$ for any $i'\in\{1',\dots,k'\}$. Finally, every $\epsilon$-optimal strategy is a mirror of an $\epsilon$-optimal strategy
\end{lemma}
\begin{proof}
First, the value of vertex $s$ is at most $1/2$. This is because Adam can mirror any strategy $\sigma$ for Eve. This ensures that any play reaching $\Win$ is mirrored by an equally likely play reaching $\bot$. Thus, then the players follows these strategies, the pr. to reach $\Win$ is equal to the pr. to reach $\bot$ (there might also be some positive pr. to not reach neither, but Adam also wins those plays). 

Fix $\epsilon>0$ and consider a strategy $\tau$ for Eve that plays a Dirac distribution in $i'\in\{1',\dots,k'\}$. Then $\tau$ is not $(1/2-\epsilon)$-optimal. We can see that as follows: Let $\sigma$ be the strategy for Eve that plays $r=1$ when in $j'\in \{(i+1)',\dots,k'\}$ and the action which is not equal to $\tau(i')$ when in $i'$. This ensures that the play will always reach either $\Win$ or $s$ from $k'$. But then  Eve can play an $(\epsilon/2)$-optimal strategy for purgatory $k$ in $\{1,\dots,k\}$, ensuring that $\Win$ is reached with pr. at least $1-\epsilon/2$. But then $\tau$ is not $(1/2-\epsilon)$-optimal.

Consider that Adam is following a strategy $\tau$ that is not playing a Dirac distribution in $i'\in\{1',\dots,k'\}$ and Eve is playing an arbitrary strategy $\sigma$. Then, eventually play reaches $\Win$ or $\bot$ with pr. 1,  because in every $k+1$ steps, either $s$ is visited or either $\Win$ or $\bot$ is reached, and after $s$ has been visited, $k'$ is next half the time and from $k'$ $\bot$ is reached with positive pr.

Consider an $\epsilon$-optimal strategy $\tau$ for Adam, for $\epsilon<1/2$. 
Then, let Eve's mirror strategy to $\tau$ be $\sigma_{\tau}$. Now, either $\Win$ or $\bot$ is reached and because the strategies mirrors each other, the pr. to reach $\Win$ is equal to that of reaching $\bot$. Thus, we see that the value is at most $1/2$, implying that it is exactly $1/2$.

It also follows that the strategies that are $\epsilon$-optimal mirrors each other.
\end{proof}

We will now argue that Eve's (and thus Adam's) $\frac{1}{4}$-optimal strategies requires high patience.

To do so we will use the following lemma, showing that you can sometimes modify a concurrent game (or any of its special cases) and get a game with less value. While the proof is explicitly for concurrent reachability games, the proof is basically identical for concurrent discounted and mean-payoff games.
In a game $G$, for a vertex $v$ and a duration $T$, let $v^T_G$ be the value of the time-limited game with duration $T$.

\begin{lemma}
\label{8-lem:change_succ}
Consider a concurrent reachability game $G$ and a pair of vertices $u,v$, such that for all $T$, we have that $u^T_G\geq v^T_G$. Consider a vertex $w$ such that for a pair of actions, $(r,c)$ we have that $v\in \supp(\dest(w,r,c))$. Consider an alternate game $G'$ equal to $G$, except that some of the probability mass is moved from $v$ to $u$ when playing $(r,c)$ in $w$, \textit{i.e.} $0<\dest(G,w,r,c)(v)-\dest(G',w,r,c)(v)=\dest(G',w,r,c)(u)-\dest(G,w,r,c)(u)$.
Then for all vertices $z$ we have that $z^T_G\leq z^T_{G'}$
\end{lemma}
\begin{proof}
The proof is by induction in $T$.
The proof is trivial for $T=0$, because $z^T_G=0= z^T_{G'}$ for all non-goal vertices (and the goal vertex $\Win$ has value 1).
For $T\geq 1$, we have that $z^{T-1}_G\leq z^{T-1}_{G'}$.
But, matrix games are monotone in their entries, so it follows directly that for $z\neq w$ we have that $z^{T}_G\leq z^{T}_{G'}$.
Consider the matrix for $w^T_G$ compared to $w^T_{G'}$. All entries but the one for $(r,c)$ are smaller directly by induction. We also have that $v^T_{G}\leq u^T_{G}\leq u^T_{G'}$, the first inequality by definition and the second by induction. We thus see that all entries in $w^T_{G}$ are smaller than in $w^T_{G'}$.
The lemma follows.
\end{proof}

We are now ready to find the patience in concurrent reachability games.

\begin{lemma}
Any $(1/4)$-optimal strategy, for either player, in purgatory duel $k$ has patience at least $(3/4)^{-2^{k-1}}$ for each $k$
\end{lemma}
\begin{proof}
We will show that the lemma is true for Eve's strategies and that it is true for Adam's follows from \Cref{8-lem:change_succ}.
Consider an $(1/4)$-optimal strategy $\sigma$ for Eve. Fixing this strategy for Eve, we get a MDP for Adam. Clearly, in this MDP $G'$, we have that $0=\bot^T_{G'}\leq s^T_{G'}$ for all $T$. We can thus apply \Cref{8-lem:change_succ} to changing $\dest(1',1,1)$ from $\bot$ to $s$. In the resulting game $G''$, 
we still have that $0=\bot^T_{G''}\leq s^T_{G''}$ and thus, we can change $\dest(1',2,2)$ from $\bot$ to $s$.
Let the next game be $G^*$.
Thus, for any $i'\in \{1',\dots,k'\}$, the plays from $i'$ to $\bot$ in $G^*$ all goes through $s$. Note that Adam can ensure that the play reaches $s$ from $i'$ and thus, when he plays optimally, do so.
Thus, whenever $s$ is entered and Adam plays optimally, $k$ is enter eventually with pr. 1.
For the purpose of the value, we can thus disregard $s$ and vertices in $\{1',\dots,k'\}$ and just view each edge going to $s$ as going to $k$ instead.
But the resulting game is purgatory $k$ (in which Eve has fixed his strategy) and Eve is playing a strategy that gives value at least $1/4$, which requires  at least $(3/4)^{-2^{k-1}}$ patience, by \Cref{8-lem:purgatory}.
\end{proof}

\section{Concurrent mean-payoff games}
\label{8-sec:mean_payoff}
In this section we consider concurrent mean-payoff games. 
We will show that in general, any  $\epsilon$-optimal strategy in some concurrent mean-payoff games are quite complex. 
We will first, however, show that finding the value of a concurrent mean-payoff game can be done in polynomial space.

\begin{lemma}[Properties of concurrent mean-payoff games]
\label{8-lem:class_meanpayoff}
Concurrent mean-payoff games are determined and the value is the limit of the value of the corresponding time-limited game as well as the limit of the corresponding discounted game, for the discount factor going to 0 from above.
There is a polynomial-time algorithm similar to \Cref{8-lem:val1}, for finding the set of vertices where a finite memory strategy suffice to ensure $1-\epsilon$ (recall that all rewards are in $\{0,1\}$).
For any fixed number $n$, there is a polynomial-time algorithm for approximating the value in a concurrent mean-payoff game with $n$ vertices (\textit{i.e.} the running time is polynomial in the number of actions)
\end{lemma}
We will not show this lemma, but simply note that the $\epsilon$-optimal strategies known for general concurrent mean-payoff games  can be viewed as playing the corresponding discounted game with a variable discount factor that depends on how `nice' the rewards has been up to now. Basically, in each round you play the optimal strategy in the corresponding discounted game with a discount factor $\gamma$. Whenever 
 your rewards are close to or better than the value, you decrease $\gamma$ towards 0 and in each round your rewards are much worse than the value you let $\gamma$ increase, except not bigger than the initial $\gamma$ in the first round. Much of this section will argue that many natural candidates for simpler types of strategies does not work.

We will show that approximating the value, however, can, as mentioned, be done in polynomial space. The proof relies on Proposition~22 from~\cite{Hansen.Koucky.ea:2011}, stating the following:
\begin{proposition}
Let $\epsilon=2^{-j}$, where $j$ is some positive integer, and the probabilities be rational numbers where the numerator and denominator have bitsize at most $\tau$. Also, let $\lambda=\epsilon^{\tau m^{O(n^2)}}$. Consider some state $s$ and let the value of that state in the $\lambda$ discounted game be $v_{\lambda}$ and the value in mean-payoff game be $v$, then $|v-v_{\lambda}|<\epsilon$.
\end{proposition}

We will use that to again reduce to the existential theory over the reals. 
For a fixed discount factor $\gamma$, we can easily express the value of the corresponding discounted game, like we expressed the value of a concurrent reachability game.
We have that the value $v$ is then $v=\lim_{\gamma\rightarrow 0^+} f(\gamma)$, where $f$ is the found expression.
\textit{I.e.} for any $\epsilon$, there is a $\gamma'$ such that for all $\gamma<\gamma$, we have that $|f(\gamma)-v|\leq \epsilon$.
Also, that $v>c$ means that there is $\epsilon$, such that $v-\epsilon>c$.

The problem is thus to come up with a polynomial sized formula to express that $\lambda$ is $\epsilon^{\tau m^{O(n^2)}}=2^{-j \tau m^{O(n^2)}}$.

That can be done as follows, using $\ell=O(n^2)\cdot \log(m)+\log(j\tau)$ many variables, $v_0,v_1,\dots v_{\ell-1}$:
\[
v_0=1/2
\]
and for all $0<i< \ell$, we have that
\[
v_i=v_{i-1}\cdot v_{i-1}.
\]
Using induction, we see that $v_i=2^{-2^{i}}$, \textit{i.e.}, $v_1=1/2=2^{-2^0}$ and \[
v_i=v_{i-1}\cdot v_{i-1}=2^{-2^{i-1}}\cdot 2^{-2^{i-1}}=2^{-2^{i-1}-2^{i-1}}=2^{-2^{i}}\]
In particular, \[
v_{\ell-1}=2^{-2^{\ell}}=2^{-2^{O(n^2)\cdot \log(m)+\log(j\tau)}}=2^{-j\tau m^{O(n^2)}}
\] is the value we wanted for $\lambda$.
Thus, for a given number $v$, we can test if the value of a concurrent  $\lambda$-discounted game is above $v+\epsilon$, which, using the proposition above, implies that $v$ is below the value of the corresponding concurrent mean-payoff game. On the other hand, the proposition also implies that if the value of the concurrent  $\lambda$-discounted game is below $v-\epsilon$, then the value of the concurrent mean-payoff game is below $v$. Being able to answer these questions lets you easily approximate the value of a concurrent mean-payoff game using binary search. 

We get the following lemma.
\begin{lemma}
Approximating the value of a concurrent mean-payoff game can be in done in polynomial space
\end{lemma}

We will now consider a specific, well-studied example of a concurrent mean-payoff game, since it shows that many natural kinds of strategies do not suffice in general.
The game is called the big match and is defined as follows:
There are 3 vertices, $\{0,s,1\}$, where the vertices in $\{0,1\}$ are absorbing, and with value equal to their name.
The last vertex $s$ has a 2x2-matrix and for all $i,j$ for $i\neq j$, we have that 
$c(s,1,1)=1$, and for $i\neq 1\neq j$ we have that $c(s,1,1)=0$.
Also,  $\dest(s,1,i)=s$ for each $i$, $\dest(s,2,1)=0$ and $\dest(s,2,2)=1$. There is an illustration in \Cref{8-fig:bm}.
The value of the Big Match is $1/2$.

\begin{figure}

\center
\begin{tikzpicture}[node distance=3cm,-{stealth},shorten >=2pt]
\ma{s}[$s:$]{2}{2};

\node at (s-1-1.center) {$1$};
\node at (s-2-1.center) {$0^*$};
\node at (s-1-2.center) {$0$};
\node at (s-2-2.center) {$1^*$};

\end{tikzpicture}
\caption{The Big Match}
\label{8-fig:bm}
\commentAlt{Figure~\ref{8-fig:bm}: A 2x2 grid (matrix) labeled 's:' containing numerical values, some with an asterisk.}
\commentLongAlt{Figure~\ref{8-fig:bm}: A square grid divided into two rows and two columns, forming four cells. The grid is labeled 's:' on its left side. Each cell contains a number, some followed by an asterisk:

Top row, from left to right: '1', '0'
Bottom row, from left to right: '0*', '1*'}
\end{figure}

Consider a finite-memory strategy $\sigma$ for Eve. We will argue that $\sigma$ cannot guarantee $\epsilon$ (any strategy can guarantee $-1$, since the colours are between $0$ and $1$) for any $0<\epsilon$. Let $\tau$ be the stationary strategy for Adam that plays $1$ with pr. $\epsilon/2$.
Then playing $\sigma$ against $\tau$, we get an Markov chain, where the vertex space is pairs of memory states and game vertices. 
In Markov chains, eventually, with pr. 1, a set of vertices $S$ is reached such that the set of vertices visited infinitely often is $S$. Such a set is called ergodic.
The set $S$ can clearly only contain 1 game vertex, since whenever $s$ is left, it is never entered again.
Hence, if $S$ contains $s$, the pr. that play will ever reach $\{0,1\}$ is 0.
In the MC we get from the players playing $\sigma$ and $\tau$, let $T_{\epsilon/2}$ be such that with pr. $\epsilon/2$ some ergodic set has been reached. 
Let $\tau'$ be the strategy that plays $\tau$ for $T_{\epsilon/2}$ and afterwards plays $2$. 

When $\sigma$ is played against $\tau'$, either we reach $\{0,1\}$ and Adam plays 1 only finitely many times, while in $s$ (the latter because there are only finitely many numbers below $T_{\epsilon/2}$). Thus, for Eve to win a play, the play needs to reach vertex 1. There are two ways to do so, either Eve stops before $T_{\epsilon/2}$ or after. In the former case, the pr. to reach $1$ is only $\epsilon/2$ (because Adam needs to play $2$ at the time, which is only done with pr. $\epsilon/2$). The latter only happens with pr. $\epsilon/2$ by definition of $T_{\epsilon/2}$ (because, Adam could play $2$ for an arbitrary number of steps while following $\tau$ but $s$ would not be left anyway).

We get the following lemma.

\begin{lemma}[Properties of the Big Match]
\label{8-lem:no_finite_meanpayoff}
No finite memory strategy can guarantee more than $0$ in the Big Match.
\end{lemma}

The principle of sunken cost states that, when acting rationally, one should disregard cost already paid. We will next argue that this does not apply (naively) to the Big Match.
A strategy following the principle of sunken cost would not depend on past cost paid and thus, in each step $T$, there is a pr. $p_T$ of stopping for Eve.
Such strategies are called Markov strategies in the Big Match.
Fix some Markov strategy $\sigma$ for Eve. We will argue, like before, that $\sigma$ cannot guarantee more than $\epsilon$ for any $\epsilon>0$.
Note that Eve does not depend on the choices of Adam and thus, either she stops with pr. 1 or she does not.
In the former case, Adam just plays $1$ forever. When Eve stops, the vertex reached is thus $-1$.
Alternately, if Eve does not stop with pr. 1, there must be a time $T$, such that she only stops with pr. $\epsilon$ after $T$ (this is actually also the case even if she stops with pr. 1). 
Adam's strategy is then to play $1$ for $T$ steps and $2$ thereafter. Observe that the pr. to reach $1$ is thus at most $\epsilon$, in that it must be that Eve stops after $T$. If she does not stop (or stops in $0$), there will be only finitely many $1$'s.

We see the following:
\begin{lemma}[Properties of the Big Match]
\label{8-lem:no_markov_meanpayoff}
No Markov strategy can guarantee more than $0$ in the Big Match
\end{lemma}

\section*{Bibliographic references}
\label{8-sec:references}
We will now give the references for this chapter, split into a few paragraphs, each corresponding to a section in the chapter.

John von Neumann's work on matrix games~\cite{Neumann.Morgenstern:1944} (also called normal form games), showing that they have a value and there exists optimal stationary strategies, is typically considered the founding work in game theory. Besides that paper, Dantzig~\cite{Dantzig:1965} showed the equivalence to linear programming, and thus that they can be solved in polynomial time using \textit{e.g.} Khachiyan's~\cite{Khachiyan:1979} work on the ellipsoid method.
There are also some results on how complex the strategies for matrix games are: For any $\epsilon>0$, there exists an $\epsilon$-optimal strategy that plays uniformly over a multi-set of actions of size $\lceil (\ln n)/\epsilon^2\rceil$ as shown by Lipton and Young~\cite{Lipton.Young:1994} (this is a stronger requirement than patience).
Also, as shown by Feder, Nazerzadeh and Saberi~\cite{Feder.Nazerzadeh.ea:2007} there exists games such that any $\epsilon$-optimal strategy has support at least $\Omega(\frac{\log n}{\epsilon^2})$ (note that if, for some $x$, the support is $\Omega(x)$ then patience is also $\Omega(x)$).
Finally,  as shown by Hansen, Ibsen-Jensen, Podolskii and Tsigaridas~\cite{Hansen.Ibsen-Jensen.ea:2013}, there is an optimal strategy in any matrix game with patience less than $(n+2)^{\frac{n+2}{2}}/2^{n+1}$ and for each $k$ there exists games with $n=m=2^k$ such that any optimal strategy has patience at least $n^{n/2}/2^{n(1+o(1))}$ (there are also results for $m$ and $n$ not equal to $2^k$ for some $k$, but not quite as tight to the upper bound).

Shapley~\cite{Shapley:1953} first considered concurrent games and focused on the class of concurrent discounted-payoff games. For these, he showed that they have a value and that there are optimal stationary strategies, using in essence the proof we used for the first 3 items of \Cref{8-cor:long}.
The proof of the fourth item, \textit{i.e.} that the value can be approximated in PPAD, comes from the work of Etessami  and Yannakakis~\cite{Etessami.Yannakakis:2010}. The proof of the fifth item, an upper bound on the patience of $\epsilon$-optimal strategies appears in~\cite{Ibsen-Jensen:2012}.

Everett~\cite{Everett:1957} was the first to consider concurrent reachability games (formally, he considered a slight generalisation).
In that paper, he showed that the games have a value and $\epsilon$-optimal stationary strategies (\textit{i.e.} the first part of \Cref{8-lem:reach_determined} and \Cref{8-lem:reach_class}). He also used the snowball game to show that Eve does not always have an optimal strategy (\textit{i.e.} \Cref{8-lem:no_opt_reach}).
Finally, he introduced the notion of patience for strategies.
It was shown by Himmelberg, Parthasarathy, Raghavan and Vleck~\cite{Parthasarathy:1973} that Adam has an optimal strategy.
Frederiksen and Miltersen~\cite{Frederiksen.Miltersen:2013} showed that for each vertex $x$ and action $a$ (except for one action for each vertex), there is a number $c_{x,a}$ and an integer $d_{x,a}$, such that 
for any $\epsilon>0$, the strategy that plays each action $a'$ in vertex $x'$ with pr. $c_{x',a'} \epsilon^{d_{x',a'}}$ is $\epsilon$-optimal (the last action in each vertex is played with the remaining pr.).
They used that to show that approximating the value (since the value can be irrational, it seems reasonable to approximate) can be done in TFNP[NP], slightly inside PSPACE.
Finding the set of vertices of value~0, \textit{i.e.} \Cref{8-lem:find_0_reach}, is folklore.
Finding the set of vertices of value~1, on the other hand, \textit{i.e.} \Cref{8-lem:val1}, is by de Alfaro, Henzinger and Kupferman~\cite{Alfaro.Henzinger.ea:1998} (their proof is different).
Hansen, Kouck{\'y} and Miltersen~\cite{Hansen.Koucky.ea:2009} showed that purgatory $k$ requires patience $\epsilon^{2^{k-1}}$ for any $1>\epsilon>0$ for Eve. 
Later, Chatterjee, Hansen and Ibsen-Jensen~\cite{Chatterjee.Hansen.ea:2017} showed that purgatory duel $k$ requires patience $(3/4)^{2^{k-1}}$ for any $0\leq \epsilon<1/4$ for either player.

Gillette~\cite{Gillette:1957} was the first to consider concurrent mean-payoff games and introduced the Big Match game we use as an example. He showed that the Big Match does not have stationary strategies ensuring more than $0$. 
Later, Blackwell and Ferguson~\cite{Blackwell.Ferguson:1968} showed that the Big Match have a value and that value is $1/2$ by showing that some strategies that depends on the full history is $\epsilon$-optimal. They also showed that no optimal Markov strategy (\textit{i.e.} \Cref{8-lem:no_markov_meanpayoff}) can ensure more than $0$ in that game.
Next, Kohlberg~\cite{Kohlberg:1974} extended this to show that all repeated games with absorbing states have a value.
Finally, Mertens and Neyman~\cite{Mertens.Neyman:1981} showed that all concurrent mean-payoff games have a value (\textit{i.e.} the first part of \Cref{8-lem:class_meanpayoff}).
The strategies employed in all these papers kept track on the sum of over the rounds of the values of the vertex in that round minus the colour in that round (the strategy by Mertens and Neyman set the memory to 0 if it should have been negative though).
Finding the set of vertices where for every $\epsilon>0$, a finite memory strategy can ensure value $1-\epsilon$ was done by Chatterjee and Ibsen-Jensen~\cite{Chatterjee.Ibsen-Jensen:2015} (the middle part of \Cref{8-lem:class_meanpayoff}).
Furthermore, finding the values in a game with a fixed number of vertices in polynomial time was done by Hansen, Kouck{\'y}, Lauritzen, Miltersen and Tsigaridas~\cite{Hansen.Koucky.ea:2011}, informally speaking by doing binary search for the values.
That finite-memory strategies cannot ensure more than $0$ in the Big Match, \textit{i.e.} \Cref{8-lem:no_finite_meanpayoff} seems to be folklore.
Hansen, Ibsen-Jensen and Kouck{\'y}~\cite{Hansen.Ibsen-Jensen.ea:2016} considered extending Markov strategies with a finite amount of space and showed that if the memory is a deterministic function of the history, then no such strategy can ensure more than $-1$ in the Big Match. They also showed, for any fixed $\epsilon>0$, that in round $T$ one only needs $O(\log \log T)$ bits of memory to play $\epsilon$-optimal in any absorbing game.
Finally, Hansen, Ibsen-Jensen and Neyman~\cite{Hansen.Ibsen-Jensen.ea:2018} showed that Markov strategies extended with a single bit of space suffice to play the Big Match $\epsilon$-optimally, for any $\epsilon>0$ (naturally, the strategy used randomisation to update the memory state).

\ifpictures
\includepdf{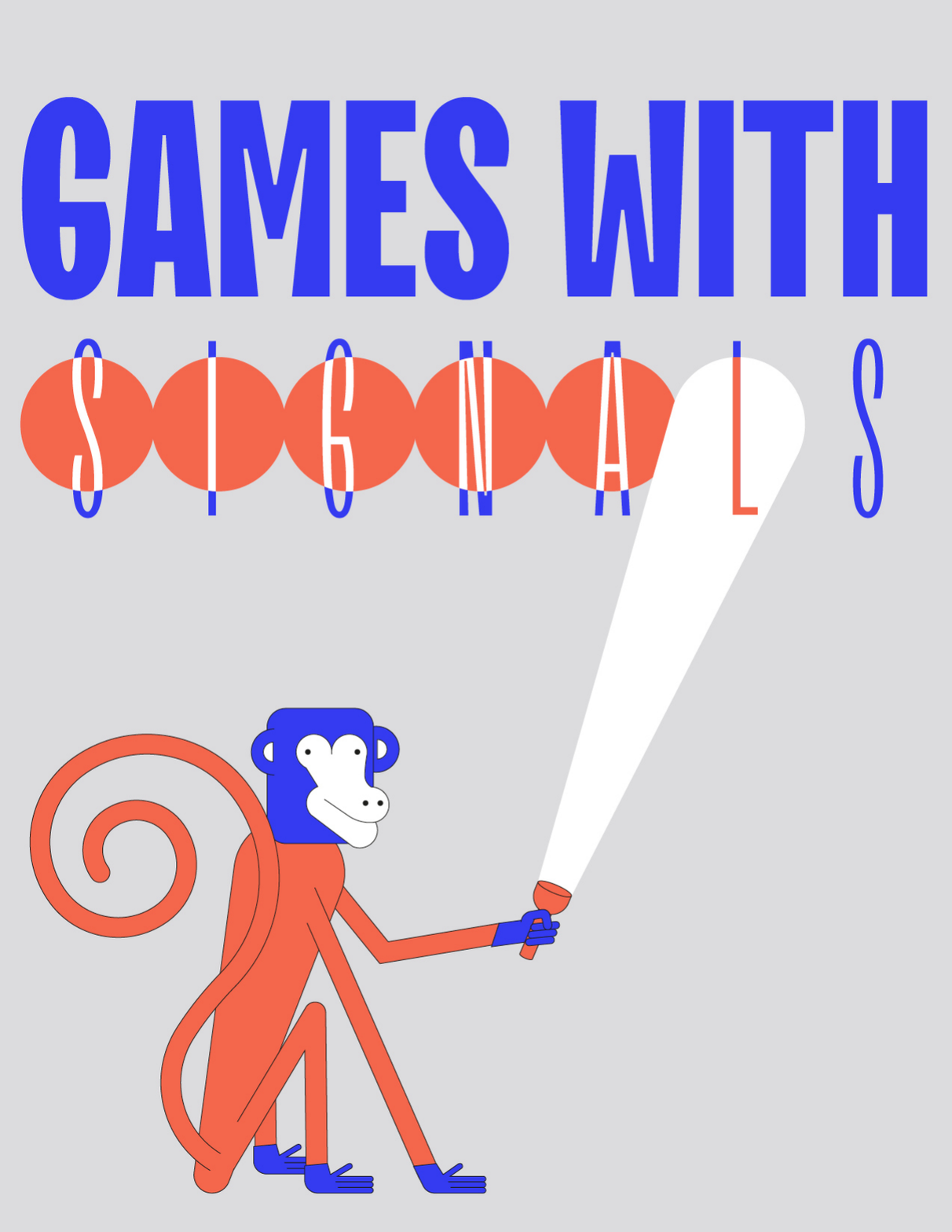}
\fi
\author[Hugo Gimbert]{Hugo Gimbert}
\copyrightline{Copyright by Hugo Gimbert 2025, to be published by Cambridge University Press in the volume \textit{Games on Graphs} edited by Nathana\"el Fijalkow}

\chapter{Games with Signals}
\chapterauthor{Hugo Gimbert}
\label{9-chap:signal}

\newcommand{\pay}{ {\tt pay}}
\newcommand{\probimp}[3]{\mathbb{P}^{#1}_{#2}\left({#3}\right)}
\newcommand{\rand}{{\tt rand}}
\newcommand{\Isafe}{{\tt ISafe}}
\newcommand{\LL}{\mathcal{L}}
\newcommand{\KK}{\mathcal{K}}
\newcommand{\LLE}{\LL_{\text{Eve},=1}}
\newcommand{\LLA}{\LL_{\text{Adam},>0}}
\newcommand{\can}{\textsf{max}}

\newcommand{\targets}{TT}
\newcommand{\bh}{\setminus}
\newcommand{\signauxdeux}{T}
\newcommand{\actionsun}{A}

\newcommand{\Strat}{\text{Strat}}

\newcommand{\Act}{\text{Act}}

\newcommand{\ini}{\delta_0}

\newcommand{\win}{{\tt Win}}
\newcommand{\winreach}{{\tt Reach}}
\newcommand{\winsafe}{{\tt Safety}}
\newcommand{\winbuchi}{{\tt Buchi}}
\newcommand{\wincobuchi}{{\tt CoBuchi}}

\newcommand{\states}{V}
\newcommand{\ar}{\mathcal{A}}
\newcommand{\action}{a}
\newcommand{\belun}{\mathcal{B}_{\text{Eve}}}
\newcommand{\beldeux}{\mathcal{B}_{\text{Adam}}}
\newcommand{\deuxbelun}{\mathcal{B}^{(2)}_{Eve}}
\newcommand{\tp}{\Delta}
\newcommand{\parties}[1]{\ensuremath{\mathcal{P}(#1)}}

\paragraph{Imperfect information.}
This chapter presents a few results about zero-sum games with imperfect information.
Those games are a generalization of concurrent games in order to take into account the possibility that players might be imperfectly informed about the current state of the game
and the actions taken by their opponent, or even their own action. We will also discuss situations where players may forget what they used to know.

Before providing formal definitions of games with imperfect information,
we give several examples.

\paragraph{Simple poker.}
Our first example is a finite duration game which is a simplified version of poker,
inspired by Borel and von Neumann simplified poker~\cite{Ferguson.Ferguson:2003}.
This game is played with  $4$ cards $\{\spadesuit,\heartsuit,\clubsuit,\diamondsuit\}$.

\begin{itemize}
\item The goal of Eve and Adam is to win
the content of a pot in which, initially, they both put $1$ euro.
\item Eve receives a private random card, unknown by Adam.
\item Eve decides whether to ``check'' or ``raise''.
If she ``checks'' then she wins the pot iff her card is $\spadesuit$.
\item If Eve raises then Adam has two options: ``fold''
or ``call''. If Adam folds then Eve receives the pot.
If Adam raises then both player add two euros in the pot
and Eve wins the pot iff her card is $\spadesuit$.
\end{itemize}

A natural strategy for Eve is to raise when she has a spade and otherwise
check. Playing so, she reveals her card to Adam,
and we will see that the optimal behaviour for her
consists in ``bluffing'' from time to time, \textit{i.e.} raise although her card is not a spade.

\paragraph{The distracted logician.}
Our second example is another finite duration game.
A logician is driving home. For that he should go through two crossings,
and turn left at the first one and right at the second one.
This logician is very much absorbed in his thoughts,
trying to prove that $P\neq NP$,
and is thus pretty distracted: upon taking a decision, he cannot  tell
whether he already saw a crossing or not.

This simple example is useful to discuss the observability of actions
and make a distinction between
mixed strategies and behavioral strategy.
 
\paragraph{Network controller.}
The following example is inspired from collision regulation
in Ethernet protocols: the controller of a network card
has to share an Ethernet layer with
another network card, controller by another controller,
possibly malicious.

When sending a data packet,
the controller selects a delay in microseconds between $1$ and $512$
and transmits this delay to the network card.
The other controller does the same.
The network cards try to send their data packet at the chosen dates.
Choosing the same date results in a data collision, and the process is repeated until
there is no collision, at that time the data can be sent.

The chosen delay has to be kept hidden from the opponent.
This way, it can be chosen randomly,
which ensures that the data will eventually be sent with probability $1$,
whatever does the opponent.
  
\paragraph{Guess my set.}
Our fourth example is an infinite duration game,
parametrized by some integer $n$.
The play is
divided into three phases.
\begin{itemize}
\item In the first phase, Eve secretly chooses a subset
$X \subsetneq \{1, \ldots,2n\}$ of size $n$
among the $\binom{2n}{n}$ possibilities.
\item In the second phase, Eve discloses to Adam
$\frac{1}{2}\binom{2n}{n}$ pairwise distinct sets of size
$n$ which are all different from $X$. 
\item In the third phase, Adam aims at guessing $X$ by trying up to
$\frac{1}{2} \binom{2n}{n}$ sets of size $n$. 
If Adam succeeds in guessing $X$,
the game restarts from the beginning. Otherwise, 
Eve wins.
\end{itemize}

Clearly Adam has a strategy to prevent forever
Eve to win: try up one by one all those sets
that were not disclosed by Eve.
This strategy uses a lot of memory:
Adam has to remember the whole sequence of $\frac{1}{2} \binom{2n}{n}$
 sets disclosed by Eve.
We will see that a variant of this game can be represented 
in a compact way, using a number of states polynomial in $n$.
As a consequence, playing optimally a game with imperfect-information and infinite duration
might require a memory of size doubly-exponential in the size of the game.

\section{Notations}
\label{9-sec:notations}
We consider \emph{stochastic games with signals}, that are a standard tool in game theory to model imperfect information in stochastic games~\cite{Rosenberg.Solan.ea:2006}.
When playing a stochastic game with signals, players cannot observe
the actual state of the game, nor the actions played by themselves or
their opponent: the only source of information of a player are private
signals they receive throughout the play.  Stochastic games with
signals subsume standard stochastic games~\cite{Shapley:1953}, repeated
games with incomplete information~\cite{Aumann:1964}, games with imperfect
monitoring~\cite{Rosenberg.Solan.ea:2006}, concurrent games~\cite{Alfaro.Henzinger:2000} and
deterministic games with imperfect information on one side~\cite{Reif:1984}.

Like in previous chapters, $V$, $C$ and $A$  denote respectively the sets of vertices, colours and actions.
\begin{definition}
An imperfect information arena $\arena$ is a tuple $(S,\Delta)$ where 
\begin{itemize}
	\item $S$ is the set of \emph{signals}
	\item $\Delta : V \times A \times A \to \dist(V \times S \times S \times C)$
	 maps the current vertex and a pair of actions to a probability distribution
	 over vertices, pairs of signals and colours.
\end{itemize}
\end{definition}

Initially, the game is in a state $v_0 \in V$ chosen according to a probability distribution
$\ini\in\dist(V)$ known by both players; the initial state is
$v_0$ with probability $\ini(v_0)$.  At each step $n \in \mathbb{N}$, both players
simultaneously choose some actions $a,b \in A$
 They respectively receive signals
$s,t \in S$ ,
 and the game moves to a
new state $v_{n+1}$.  This happens with probability
$\Delta(v_{n},a,b)(v_{n+1},c,d)$.
{This fixed probability is known by both players,
as well as the whole description of the game.}

A \emph{play} is a sequence $(v_0,a_0,b_0,s_0,t_0,c_0),(v_1,a_1,b_1,s_1,t_1,c_1),(v_2\ldots$
such that for every $n$, the probability $\Delta(v_{n},a_n,b_n)(v_{n+1},s_n,t_n,c_n)$
is positive.

A sequence of signals for a player
is \emph{realisable} for Eve if it appears in a play,
we denote $R_E \subseteq S^*$ the set of these sequence.
Similarly for Adam.

\paragraph{An example.}
The simplified poker can be 
modelled as a stochastic game with signals.
Actions of players are \emph{public signals}
sent to both players.
Also their the payoff of Eve is publicly announced,
when non-zero. 
Upon choosing whether to call or fold,
Adam cannot distinguish between states
$\spadesuit${\tt Raised} and $\blacksquare${\tt Raised},
in both cases he received the sequence of signals $\circ,{\tt raise}$.
A graphical representation is provided on~\Cref{11-fig:poker}.


\begin{figure}
   \centering
   \begin{tikzpicture}[scale=1.0]

\node (root) at (0,0) {{\tt Start}};
\node[below left =of root] (spade) {$\spadesuit${\tt Play}};
\node[below right =of root] (nospade) {$\blacksquare${\tt Play}};
\node[below left = 2cm and 0.5cm of spade] (spaderaise) {$\spadesuit${\tt Raised}};
\node[below right = 2cm and 0.5cm of nospade] (nospaderaise) {$\blacksquare${\tt Raised}};
\node(end) at (0,-5.5) {{\tt End}};

\path[->](root) edge node[near start,left,align=center]  
{$(\cdot,\cdot)\frac{1}{4}$
\\Eve receives $\spadesuit$
\\Adam receives $\circ$
} (spade);
\path[->](root) edge node[near start,right,align=center]  {$(\cdot,\cdot)\frac{3}{4}$\\Eve receives $ \blacksquare$\\Adam receives $\circ$} (nospade);
\path[->](spade) edge node[near start,left,align=center]  {$({\tt raise},\cdot)
$
} (spaderaise);
\path[->](nospade) edge node[near start,right,align=center]  {$({\tt raise},\cdot)
$
} (nospaderaise);
\path[->, bend left=20](spade) edge 
node[very near start,right] {$({\tt check},\cdot)$}
node[right] {{\bf +1}} (end);

\path[->, bend right=20](nospade) edge 
node[very near start, left]  {$({\tt check},\cdot)
$} 
node[left] {{\bf -1}} 
(end);
\path[->,bend left=-10](spaderaise) edge 
node[above]  {$(\cdot,{\tt call})~~{\bf +3}
$} 
(end);
\path[->,bend left=-50](spaderaise) edge 
node[below]  {$(\cdot,{\tt fold},)~~{\bf +1}
$} 
(end);

\path[->,bend left=10](nospaderaise) edge 
node[above]  {$(\cdot,{\tt call})~~{\bf -3} 
$} 
(end);
\path[->,bend left=50](nospaderaise) edge 
node[below]  { $(\cdot,{\tt fold})~~{\bf +1}
$} 
(end);

;

   \end{tikzpicture}
   \caption{The simplified poker game.}
   \label{11-fig:poker}
\commentAlt{Figure~\ref{11-fig:poker}: A decision tree-like diagram illustrating a game or process with starting points, actions, outcomes, and an end state. See long description.}
\commentLongAlt{Figure~\ref{11-fig:poker}: The image depicts a flow diagram starting from a "Start" node at the top.
From "Start", two branches diverge:

Left branch: Labeled '(., 1/4)' and "Eve receives [spade symbol], Adam receives circle". This leads to a node labeled "[spade symbol] Play".
Right branch: Labeled '(., 1/4)' and "Eve receives [square symbol], Adam receives circle". This leads to a node labeled "[square symbol] Play".
From "[spade symbol] Play" (left branch):

An arrow labeled '(raise, .)' points down to a node labeled "[spade symbol] Raised".
A curved arrow labeled '(check, .)' and with a '+1' value points to an "End" node at the bottom center.
From "[square symbol] Play" (right branch):

An arrow labeled '(raise, .)' points down to a node labeled "[square symbol] Raised".
A curved arrow labeled '(check, .)' and with a '-1' value points to the "End" node at the bottom center.
From "[spade symbol] Raised" (left side):

A curved arrow labeled '(., call)' and with a '+3' value points to the "End" node.
A curved arrow labeled '(., fold)' and with a '+1' value also points to the "End" node.
From "[square symbol] Raised" (right side):

A curved arrow labeled '(., call)' and with a '-3' value points to the "End" node.
A curved arrow labeled '(., fold)' and with a '+1' value also points to the "End" node.
The "End" node serves as a common termination point for all paths.}
 \end{figure}

The game is played with  $4$ cards $\{\spadesuit,\heartsuit,\clubsuit,\diamondsuit\}$.
We exploit the symmetry of payoffs with respect to $\{\heartsuit,\clubsuit,\diamondsuit\}$ and identify these three colours 
as a single one, denoted $\blacksquare$, received initially by Eve with probability $\frac{3}{4}$.
The set of vertices
is an initial vertex ${\tt Start}$,
a terminal vertex ${\tt End}$
plus the four states
\[
\{\spadesuit,\blacksquare\} \times 
 \{{\tt Play,Raised}\}\enspace.
 \]
The set of colours are possible payoffs $C=\{0,-1,+1,-3,+3\}$.

The set of actions $A$ 
is the union of 
actions of Eve 
$A_E=\{{\tt \cdot, check,raise}\}$
and actions of Adam
$A_A=\{{\tt \cdot, call, fold}\}$.

The set of signals is $\{\circ , \spadesuit, \blacksquare\}$ plus
$\{{\tt check},{\tt raise},{\tt call},{\tt fold}\}\times \{0,-1,+1,-3,+3\}$.

The rules of the game,
are defined by the set of \emph{legal} transitions.
Let $c \in \{\spadesuit,\blacksquare\}$.
The following transitions are legal.
\begin{align*}
&~\Delta({\tt Start},{\tt \cdot},{\tt \cdot})((c,{\tt Play}),c,\circ,0)=
\begin{cases}
\frac{1}{4}& \text{ if } c= \spadesuit\\
\frac{3}{4}& \text{ if } c= \blacksquare\enspace.
\end{cases}\\
&~\Delta((c,{\tt Play}),{\tt check},{\tt \cdot})({\tt End},{\tt check}_x,{\tt check}_x,x)=1
\text{ where } x=
\begin{cases}
+1 & \text{ if } c=\spadesuit\\
-1& \text{ if } c=\blacksquare.
\end{cases}
\\
&~\Delta((c,{\tt Play}),{\tt raise},{\tt \cdot})((c,{\tt Raised}),{\tt raise_0},{\tt raise_0},0)=1\\
&~\Delta((c,{\tt Raised}),{\tt \cdot},{\tt call})({\tt End},{\tt call}_x,{\tt call}_x,x)=1 
\text{ where } x=
\begin{cases}
+3 & \text{ if } c=\spadesuit\\
-3 & \text{ if } c=\blacksquare.
\end{cases}
\\
&~\Delta((c,{\tt Raised}),{\tt \cdot},{\tt fold})({\tt End},{\tt fold_1},{\tt fold_1},+1)=1\\
&~\text{state ${\tt End}$ is absorbing with payoff $0$.}
\end{align*}

To simplify the notations,
we assumed in the general case
that players share the same set of actions and signals.
As a consequence, other transitions than the legal ones
are possible. 
One can use a threat to guarantee that Eve plays
${\tt check}$ and ${\tt raise}$ after receiving her card,
by setting a heavy loss of $-10$ if she plays another action instead.
Same thing to enforce that Adam plays ${\tt call}$ or ${\tt fold}$
after receiving the signal ${\tt raise}$.
When targeting applications,
legal moves should be explicitly
specified, typically using an automaton
to compute the set of legal actions
depending on the sequence of signals.

\paragraph{Strategies: behavioral, mixed and general.}

Intuitively, players make their decisions based upon the sequence of
signals they receive, which is formalised with strategies. 
There are several natural classes of strategies to play games with signals, as discussed in~\cite{Cristau.David.ea:2010} and Section 4 in~\cite{Bertrand.Genest.ea:2017}.

A behavioural strategy of Eve associates
with every realisable sequence of signals a probability distribution
over actions:  
\[
\sigma: R_E \to \dist(A)\enspace.
\]
When Eve plays $\sigma$, after having received a sequence of signals
$s_0,\ldots,s_n$ she chooses action $a$ with probability
$\sigma(s_0,\ldots,s_n)(a)$. 
Strategies of Adam are the same, except they are defined on $R_A$.

Remark that in general a player may not observe which actions he actually played,
for example $S$ might be a singleton 
in which case the players only knows the number of steps so far.

A game has \emph{observable actions} if there exists a mapping
 $\Act:S \to A$ 
 such that
\[
\Delta(v,a,b)(w,s,t)>0 
\implies
(a=\Act(s) \land b=\Act(t))\enspace. 
\]

In~\cite[Lemma 4.6 and 4.7]{Bertrand.Genest.ea:2017} it was shown that without loss of generality,
one can consider games where actions are observable and players 
play behavioural strategies. The discussion is technical and beyond the scope of this book.

\section{Finite duration}
\label{9-sec:finite_duration}
We start with some results on the very interesting class of game
with finite duration.

A game has \emph{finite duration}
if there is a set of absorbing vertices $L$, called \emph{leaves},
such that every play eventually reaches $L$.
In other words, the directed graph $(V,E)$ induced by all pairs $(v,w)$
such that 
$\exists a,b\in A, s,t \in S, \Delta(v,a,b)(w,s,t) > 0$
is acyclic, except for self loops on leaves.

Moreover, $C$ is the set of real numbers,
colours are called \emph{payoffs}.
At the moment the play $\pi$ reaches a leaf $\ell\in L$
for the first time,
the game is essentially over:
Eve receives the sum of payoffs seen to far,
denoted ${\tt pay}(\pi)$ and all future payoffs are $0$.
Such plays are called \emph{terminal plays}.

Once a terminal play occurs, the game is over.
For this reason, in this section we restrict realisable sequences of signals
to the ones occurring in terminal plays and their prefixes.
This guarantees finiteness of $R_E$ and $R_A$ since
\[
R_E \cup R_A \subseteq S^{\leq n}\enspace.
\]

An initial distribution $\ini$ and two strategies $\sigma$ and $\tau$ of Eve and Adam naturally induce a probability distribution $\mathbb{P}_{\ini}^{\sigma,\tau}$
on the set of terminal plays starting in one of the vertices $v_0, \ini(v_0)>0$.
Players have opposite interests:
Eve seeks to maximize her expected payoff
\[
\mathbb{E}_{\ini}^{\sigma,\tau}= \sum_{\text{ terminal plays }\pi} 
\mathbb{P}_{\ini}^{\sigma,\tau}(\pi) \cdot {\pay}(\pi)\enspace,
\]
while Adam wants to minimise it.

\subsection{Existence and computability of the value}
\label{9-subsec:value}

Next theorem gathers several folklore results.

\begin{theorem}[Finite duration games]
\label{9-thm:finiteimperfecthaveval}
A game with finite duration and imperfect information has a value:
for every initial distribution $\ini$,
\[
\sup_\sigma \inf_\tau \mathbb{E}_{\ini}^{\sigma,\tau}
~=~
 \inf_\tau \sup_\sigma \mathbb{E}_{\ini}^{\sigma,\tau}\enspace.
\]
This value is denoted $\val(\ini)$
and is computable~\footnote{provided payoffs are presented in a way
compatible with linear solvers, typically 
rational values.}.
Both players have optimal strategies.
\end{theorem}

\paragraph{Reduction to normal form.}
The main ingredient for proving this theorem is a transformation
of the game into a matrix game called its \emph{normal form}.

The intuition is that a player,
instead of choosing progressively her actions
as she receives new signals,
may choose once for all at the beginning of the game
how to react to every possible sequence of signals
she might receive in the future.

Fix an initial distribution $\ini$.
In the normal form version the game,
Eve  picks 
a \emph{deterministic} strategy
$\sigma : R_E \to A$
while simultaneously
Adam picks
$\tau : R_A \to A$.
Then the game is over
and Eve receives payoff
$\mathbb{E}_{\ini}^{\sigma,\tau}$.
There are finitely many such deterministic strategies,
thus the normal form game is a \emph{matrix game}.
See~\Cref{8-sec:matrix_games} for more details about matrix games.

\paragraph{An example.}
In the simplified poker example,
the reduction is as follows.

We rely on the formal description of the game at the end of~\Cref{9-sec:notations}
and perform two simplifications.
First, we only consider strategies playing moves according to the rules,
other strategies are strategically useless.

Deterministic strategies of Eve are
mappings $\sigma : \{\spadesuit,\blacksquare\}
\to\{ {\tt check},{\tt raise}\}$.
Adam has only two deterministic strategies:
after the sequence $\circ {\tt Raised}$,
he should choose between
actions ${\tt call}$ and ${\tt fold}$.

The normal form is
\[
\begin{array}{c|c|c}
&  {\tt call} &{\tt fold}\\
\hline
\spadesuit\to {\tt check},  \blacksquare\to {\tt check}
& -0.5 & -0.5\\
\hline
\spadesuit\to {\tt raise},  \blacksquare\to {\tt check}
& 0 & -0.5\\
\hline
\spadesuit\to {\tt raise},  \blacksquare\to {\tt raise}
& -1.5 & +1\\
\hline
\spadesuit\to {\tt check},  \blacksquare\to {\tt raise}
& -2 & +1\\
\end{array}
\]
The first line corresponds to Eve never raising,
thus her odds are +1 euro at $25\%$ and -1 at $75\%$ thus an expected payoff of $-0.5$.
The third line corresponds to Eve always raising.
If Adam calls then her odds are +3 at $25\%$
and -3 at $75\%$, on average $-1.5$.
If Adam folds, she gets payoff +1.

Remark that the rows where Eve checks with $\spadesuit$
are dominated by the corresponding row where Eve does not.
Thus checking with $\spadesuit$ (slow playing) has no strategic interest,
and  by elimination of weakly dominated strategies,
the normal form game is equivalent to:
\[
\begin{array}{c|c|c}
&  {\tt call} &{\tt fold}\\
\hline
\spadesuit\to {\tt raise},  \blacksquare\to {\tt check}
& 0 & -0.5\\
\hline
\spadesuit\to {\tt raise},  \blacksquare\to {\tt raise}
& -1.5 & 1\\
\end{array}
\]
The value of this game is $-\frac{1}{4}$.
Eve has a unique optimal strategy which consists in playing the top row with probability
$\frac{5}{6}$.
In other words, she should bluff with probability $\frac{1}{6}$ when she receives $\blacksquare$.
Adam has a unique optimal strategy which consists in calling or folding
with equal probability $\frac{1}{2}$\enspace.

\paragraph{Proof of~\Cref{9-thm:finiteimperfecthaveval}.}
The example illustrates
the correspondence between behavioural strategies in the finite-duration game on one side
and mixed strategies in the normal form game on the other.
In the general case, the correspondence can be stated as follows.

\begin{lemma}
\label{9-lem:impinffinite}
Denote $\Strat$ the set of behavioural strategies,
$\Strat_d$ the subset of deterministic strategies
and $\dist(\Strat_d)$ the set of strategies in the normal form game.
\begin{enumerate}
\item There is a mapping 
$
\Phi : \Strat \to \dist(\Strat_d)
$ 
which preserves payoffs:
\[
\forall \sigma,\tau \in \Strat,
\mathbb{E}_{\ini}^{\sigma,\tau}
=
\sum_{\sigma',\tau' \in \Strat_d}\Phi(\sigma)(\sigma')\cdot\Phi(\tau)(\tau') 
\cdot\mathbb{E}_{\ini}^{\sigma',\tau'}\enspace.
\]
\item Since actions are observable,
there is a mapping 
$
\Phi' : \dist(\Strat_d) \to \Strat 
$ 
which preserves payoffs:
\[
\forall \Sigma,T \in \dist(\Strat),
\sum_{\sigma',\tau' \in \Strat_d}\Sigma(\sigma') T(\tau')
\mathbb{E}_{\ini}^{\sigma',\tau'}
=
\mathbb{E}_{\ini}^{\Phi'(\sigma),\Phi'(\tau)}
\enspace.
\]
\item $\Phi'\circ \Phi$ is the identity.
\end{enumerate}
\end{lemma}

We assumed earlier that each player can observe
its own actions. This hypothesis is necessary for ii) and iii)
to hold in general.

\begin{proof}
We start with i).
Intuitively,
all random choices of actions performed by
a behavioural strategy $\sigma$ of Eve can be done at the beginning of the play.
Playing $\sigma$ 
is equivalent to playing each deterministic strategy $\sigma'$ 
with probability
\[
\Phi(\sigma)(\sigma') = 
\Pi_{u \in R_E} \sigma(u)(\sigma'(u))\enspace.
\]

We prove ii).
Let $\Sigma\in\dist(\Strat)$. The definition of the behavioural strategy
$\sigma=\Phi'(\Sigma)$ is as follows.
Let $s_0\ldots s_k$ be a finite sequence of signals.
Since actions are observable, this defines unambiguously
the sequence of corresponding actions $a_0\ldots a_k$
where $a_i = \Act(s_i)$.
We set $\sigma(s_0\ldots s_k)(a)$ to be the probability that a 
deterministic strategy
chosen with $\Sigma$ chooses action $a$ after signals
$s_0\ldots s_k$, conditioned on the fact that it has already
chosen action $a_0\ldots a_k$:
\[
\sigma(s_0\ldots s_k)(a) 
=
\Sigma\left(\sigma'(s_0\ldots s_k)=a \mid \forall 0\leq i \leq k,
 \sigma'(s_0\ldots s_{i-1})=\Act(s_i)\right)\enspace,
\]
where the vertical pipe denotes a conditional probability.
\end{proof}

We proceed with the proof of~\Cref{9-thm:finiteimperfecthaveval}.
According to~\Cref{8-lem:mat},
the normal form has a value and optimal strategies
for each player. 
Denote $\val_N$ the value
and $\Sigma^\sharp$ and $T^\sharp$ the optimal strategies.
Let $\sigma^\sharp=\Phi'(\Sigma^\sharp)$.
Then $\sigma^\sharp$ ensures a payoff
of at least $\val_N$ in the imperfect information game,
because for every strategy $\tau$,
\[
\mathbb{E}_{\ini}^{\sigma^\sharp,\tau}
=
\mathbb{E}_{\ini}^{\Phi'(\Sigma^\sharp),\Phi'(\Phi(\tau))}
=
\sum_{\sigma',\tau' \in \Strat_d}\Sigma^\sharp(\sigma') \Phi(\tau)(\tau')
\mathbb{E}_{\ini}^{\sigma',\tau'}
\geq \val_N\enspace,
\]
where the first equalities are applications of~\Cref{9-lem:impinffinite}
and the inequality is by optimality of $\Sigma^\sharp$.
Symmetrically, 
$\tau^\sharp=\Phi'(T^\sharp)$ guarantees 
$\forall \sigma,\mathbb{E}_{\ini}^{\sigma,\tau^\sharp}\leq\val_N$. 
Thus the value of the game with finite duration
is $\val_N$ and $\sigma^\sharp$
and $\tau^\sharp$ are optimal.\qed

\subsection{The Koller-Meggido-von Stengel reduction to linear programming}
\label{9-subsec:reduction_linear_programming}

The reduction of a finite-duration game with imperfect information
to its normal form proves that the value exists and
is computable.
However the corresponding algorithm is computationally
very expensive, it requires solving
a linear program of size roughly doubly-exponential in the size 
of the original game, since the normal form is a matrix
index by $A^{R_E} \times A^{R_A}$ and the set of signal sequences
might contain all sequences of $S$ of length $\leq n$.

Koller, Meggido and von Stengel did provide a
more efficient direct reduction to linear programming.
Strategies of Eve in the normal form game live
in $\R^{A^{R_E}}$
while her strategies in the game with imperfect information
live in a space
with exponentially fewer dimensions, namely
$\R^{R_E\times A}$.
The direct reduction avoids this dimensional blowup.

\begin{theorem}
The value of a game with imperfect information
can be computed by a linear program with
$|R_E| + |R_A|$ variables.
\end{theorem}

As a consequence, in the particular case where the game graph is a tree
then $|R_E|\leq n$ and $|R_A|\leq n$
and the value can be computed in polynomial time, like stated in~\cite{Koller.Megiddo.ea:1994}.

\begin{proof}
The construction of the linear program relies on three key ideas.

First, representing a behavioral strategy $\sigma:R_E \to \dist(A)$
 of Eve as a \emph{plan} $\pi:R_E  \to [0,1]$
 recursively defined by $\pi(\epsilon) = 1$
 and for every $s_0\cdots s_n \in R_E, s\in S$,
 \begin{align*}
& \pi(s_0\cdots s_n\cdot s) = \pi(s_0\cdots s_n) \cdot
 \sigma(s_0\cdots s_n)(\Act(s))\enspace.
 \end{align*}
Remind that actions are observable and $\Act(s)$
denotes the action that Eve has just played
before receiving signal $s$.
In the linear program, plans are represented by variables 
$\left(p_r\right)_{r \in R_E}$. 
Valuations corresponding to plans can be characterised by 
the following equalities.
First, $p_\epsilon = 1$.
Second, for every 
$s_0\ldots s_{n-1}s,s_0\ldots s_{n-1}s' \in R_E$,
\begin{align*}
(\Act(s)=\Act(s')) \implies \left(p_{s_0\ldots s_{n-1}s}= p_{s_0\ldots s_{n-1}s'}\right)\enspace.
\end{align*}
We denote $p_{s_0\ldots s_{n-1}a}$ the common value of
all $p_{s_0\ldots s_{n-1}s}$ with $a=\Act(s)$.
The third equality is 
$p_{s_0\ldots s_{n-1}}=\sum_{a\in A} p_{s_0\ldots s_{n-1}a}$ \enspace.

 The second key idea is to introduce variables evaluating the contribution of a 
 (realisable) sequence
 of signals of Adam to the total expected payoff Eve.
 These contributions are represented by variables $(v_r)_{r \in R_A}$.
 
 The third key idea is to aggregate the product of transition
 probabilities along a play.
For every play $(v_0,a_0,b_0,s_0,t_0,c_0),\ldots,(v_k,a_k,b_k,s_k,t_k,c_k)$
 we denote $\mathbb{E}(\pi)$ the product of all transition
probabilities of $\pi$ and $r_{E}(\pi)$ the sequence of signals of Eve in this play:
\begin{align*}
&\mathbb{E}(\pi) = \ini(v_0)\cdot \Delta(v_0,a_0,b_0,s_0,t_0,c_0)
  \cdots \Delta(v_k,a_k,b_k,s_k,t_k,c_k)\\
 & r_{E}(\pi) = s_0,s_1,\ldots,s_k\enspace.
\end{align*}

We show that the following linear program with variables
  $(p_r)_{r \in R_E}$, $(v_r)_{r\in R_A}$
  has an optimal solution which equals to $\val(\ini)$.
 For every sequences of signals $r \in R_A$
 we denote $T_A(r)$ the (possibly empty)
 set of terminal plays whose sequence of signals for Adam is $r$.

\begin{align}
&\text{Maximise $v_{\epsilon}$ subject to}
\notag\\
\notag\\
\notag&\text{$\left(p_r\right)_{r \in R_E}$ is a plan of Eve}
\notag\\
\notag\\
\notag\forall r \in R_A,
\forall a \in A,&
\\
&
\label{9-eq:implp2}
v_{r} \leq \sum\limits_{\substack{rs \in R_A\\s \in S, \Act(s)=a}}
v_{rs}~+~\sum\limits_{\pi \in T(r)} \mathbb{E}(\pi) \cdot \pay(\pi) \cdot 
p_{r_E(\pi)}
\end{align}

For our purpose,
it is enough to establish 
that the optimal solution of the LP
is 
\[
\val(\ini) = \sup_\sigma \min\limits_{\tau\text{ deterministic}} \mathbb{E}_{\ini}^{\sigma,\tau}\enspace.
\]
The reason is that in a matrix game,
for every fixed strategy of Eve,
Adam can minimise the payoff by playing a single action
with probability $1$.
Thus, according to the reduction to normal form seen in the previous chapter,
for every strategy $\sigma$ of Eve,
there is a \emph{deterministic} strategy $\tau$ of Adam
which minimises $\mathbb{E}_{\ini}^{\sigma,\tau}$.

We show first that for every feasible solution 
$(p_r)_{r \in R_E}$, $(v_r)_{r\in R_A}$ of the linear program,
the strategy $\sigma$ corresponding to the plan $(p_r)_{r \in R_E}$
guarantees that for every \emph{deterministic} strategy $\tau$,
$\mathbb{E}_{\ini}^{\sigma,\tau} \geq v_\epsilon$.
Since $\tau$ is deterministic
then $\mathbb{E}_{\ini}^{\sigma,\tau}$
is the sum of all $\mathbb{E}(\pi) \cdot \pay(\pi) \cdot 
p_{r_E(\pi)}$ over plays $\pi$ played according to $\tau$
thus a trivial induction shows $\mathbb{E}_{\ini}^{\sigma,\tau}\geq v_\epsilon$.

We show now that to every strategy $\sigma$ of Eve,
and to every deterministic optimal answer $\tau$ of Adam, 
corresponds a feasible solution of the program
such that $v_\epsilon = \mathbb{E}_{\ini}^{\sigma,\tau}$.
Let  $(p_r)_{r \in R_E}$ the plan corresponding to $\sigma$.
For every $r\in R_A$ define $v_r$ be the expected payoff of Eve
in an auxiliary game where she plays $\sigma$
and Adam plays $\tau$ and the payoff of Eve is turned to $0$
whenever Adam signals sequence does not start with $r$.
We show that the linear constraint~\Cref{9-eq:implp2} holds for every $r\in R_A$
 and action $a$.
Since $\tau$ is deterministic then~\Cref{9-eq:implp2} is an equality
whenever $a=\tau(r)$.
And since $\tau$ is an optimal answer to $\sigma$,
it is locally optimal in the sense where playing an action
different from $\tau(r)$ after $r$ cannot be profitable to Adam,
hence~\Cref{9-eq:implp2} holds.
Finally, $(p_r)_{r \in R_E}$, $(v_r)_{r\in R_A}$ is a feasible solution.
\end{proof}

\paragraph{An example.}

The following linear program computes the value
of the simplified poker example.

\begin{center}
Maximise $v_{\epsilon}$ subject to
\end{center}
\begin{align*}
\forall r \in R_E,~& 0\leq p_r \leq 1\\
&p_{\spadesuit,{\tt check} } +  p_{\spadesuit,{\tt raise} } = 1\\
&p_{\blacksquare,{\tt check} } +  p_{\blacksquare,{\tt raise} } = 1\\
&v_\epsilon \leq v_\circ \leq v_{\circ, {\tt check}} + v_{\circ, {\tt raise}}\\
&v_{\circ, {\tt check}}\leq  \frac{1}{4} \cdot p_{\spadesuit,{\tt check} } \cdot (+1) 
+ \frac{3}{4} \cdot p_{\blacksquare,{\tt check} } \cdot (-1)\\
&v_{\circ, {\tt raise}} \leq \frac{1}{4} \cdot p_{\spadesuit,{\tt raise} } \cdot (+1) + \frac{3}{4} \cdot p_{\blacksquare,{\tt raise} } \cdot (+1)\\
&v_{\circ, {\tt raise}} \leq \frac{1}{4} \cdot p_{\spadesuit,{\tt raise} } \cdot (+3) + \frac{3}{4} \cdot p_{\blacksquare,{\tt raise} } \cdot (-3)
\end{align*}
Setting $x=p_{\spadesuit,{\tt check} }$
and $y=  p_{\blacksquare,{\tt check} }$,
the solution is
\begin{align*}
&
\frac{1}{4}\max_{(x,y)\in[0,1]^2}
\left(
{x - 3y}
+
\min\left(
{(1-x) +  3 (1-y)},
{3(1-x) - 9(1-y)}
 \right)\right)\\
 =&
 \frac{1}{4}\max_{(x,y)\in[0,1]^2}
\min\left(
4 - 6y,
-6 -2x + 6y   
\right)
=
 \frac{1}{4}\max_{y\in[0,1]}
\min\left(
4 - 6y,
-6 + 6y   
\right)
 \enspace,
\end{align*}
which is maximal when $y=\frac{5}{6}$
and the solution is $-\frac{1}{4}$.

\paragraph{Nose scratch variant.}
Assume now that Eve does not have the perfect poker face:
whenever she has $\spadesuit$ she scratches
her nose with probability $\frac{1}{2}$ whereas
in general it happens only with probability $\frac{1}{6}$.
Only Adam is aware of this sign,
which he receives
as a private signal $s$ (scratch) or $n$ (no scratch).

Compared to the perfect poker face situation,
 the situation is slightly better for Adam:
 the value drops from $-\frac{1}{4}$
to $(-\frac{1}{4} -\frac{1}{10})$.
The optimal bluff frequency of Eve decreases
 from $\frac{1}{6}$ to $\frac{1}{10}$.
Computation details follow.

\begin{center}
Maximise $v_{\epsilon}$ subject to
\end{center}
\begin{align*}
\forall u \in R_E,~& 0\leq p_u \leq 1\\
&p_{\spadesuit,{\tt c} } +  p_{\spadesuit,{\tt r} } = 1~~~~~p_{\blacksquare,{\tt c} } +  p_{\blacksquare,{\tt r} } = 1\\
&v_\epsilon \leq v_{\tt s} + v_{\tt n}~~~~~v_{\tt s} \leq v_{{\tt sc}} + v_{{\tt sr}}~~~~~
 v_{\tt n} \leq v_{{\tt nc}} + v_{{\tt nr}}\\
&v_{{\tt sc} } \leq 
 \frac{1}{4}\cdot\frac{1}{2} \cdot p_{\spadesuit,{\tt c} } \cdot (+1) 
+ \frac{3}{4}\cdot \frac{1}{6} \cdot p_{\blacksquare,{\tt c} } \cdot (-1)\\
&v_{{\tt nc} } \leq 
 \frac{1}{4}\cdot\frac{1}{2} \cdot p_{\spadesuit,{\tt c} } \cdot (+1) 
+ \frac{3}{4} \cdot\frac{5}{6} \cdot p_{\blacksquare,{\tt c} } \cdot (-1)\\
&v_{{\tt sr}} \leq \frac{1}{4}\cdot \frac{1}{2}\cdot p_{\spadesuit,{\tt r} } \cdot (+1) 
+ \frac{3}{4} \cdot\frac{1}{6}\cdot p_{\blacksquare,{\tt r} } \cdot (+1)\\
&v_{{\tt sr}} \leq \frac{1}{4}\cdot \frac{1}{2}\cdot p_{\spadesuit,{\tt r} } \cdot (+3) 
+ \frac{3}{4} \cdot\frac{1}{6}\cdot p_{\blacksquare,{\tt r} } \cdot (-3)\\
&v_{{\tt nr}} \leq \frac{1}{4}\cdot \frac{1}{2}\cdot p_{\spadesuit,{\tt r} } \cdot (+1) 
+ \frac{3}{4} \cdot\frac{5}{6}\cdot p_{\blacksquare,{\tt r} } \cdot (+1)\\
&v_{{\tt nr}} \leq \frac{1}{4}\cdot \frac{1}{2}\cdot p_{\spadesuit,{\tt r} } \cdot (+3) 
+ \frac{3}{4} \cdot\frac{5}{6}\cdot p_{\blacksquare,{\tt r} } \cdot (-3)
\end{align*}
Set $y=p_{\blacksquare,{\tt c} }$.
Some elementary simplifications lead to the equivalent program:
\begin{align*}
\max_{0\leq y \leq 1} \frac{1}{8} \left(\min\left( 8 -12y,-10 +8y
,  6  - 8y,-12  +12y \right) \right) 
\end{align*}
The optimum is reached when $8y-10=8-12y$, \textit{i.e.} when $p_{\blacksquare,{\tt c} }=\frac{9}{10}$ and is equal to $-\frac{7}{20}=-\frac{1}{4}-\frac{1}{10}$ .

\section{Infinite duration}
\label{9-sec:infinite_duration}
Games with infinite duration and imperfect information
are a natural model for applications such as synthesis 
of controllers of embedded systems.
This is illustrated by the example of the network controller.
Whereas in the previous section games of finite-duration
were equipped with real-valued payoffs, 
here we focus on B{\"u}chi conditions.

\subsection{Playing games with infinite duration and imperfect information}

Notations used for games of finite duration are kept.
On top of that we need to define how probabilities are measured
and the winning conditions.

\paragraph{Measuring probabilities.}
The choice of an initial distribution
$\ini\in\dist(V)$ 
and two strategies
$\sigma:  R_E \to \dist(A)$
and $\tau:  R_A \to \dist(A)$
for Eve and Adam
defines a Markov chain on the set of all finite plays.
This in turn defines a probability measure
$\mathbb{P}_{\ini}^{\sigma,\tau}$ on the Borel-measurable
subsets of $\Delta^\omega$.
The random variables $V_n,A_n,B_n,S_{n}$ and $T_{n}$ denote
respectively the $n$-th state, action of Eve, action of Adam, 
signal received by Eve and Adam,
and we denote $\pi_n$ the finite play 
$\pi_n = V_0,A_0,B_0,S_0,T_0,V_1,\ldots,S_{n},T_{n},V_{n+1}$.

The probability measure $\mathbb{P}_{\ini}^{\sigma,\tau}$ is the only
probability measure over $\Delta^\omega$ such that
for every $v\in V$, 
$\mathbb{P}^{\sigma,\tau}_{\ini}(V_0 = v) = \ini(v)$
and for every $n\in\N$,
\begin{multline*}
\mathbb{P}^{\sigma,\tau}_{\ini}(V_{n+1}, S_{n}, T_{n} \mid \pi_n) \\
= \sigma(S_0\cdots S_{n-1})(A_{n}) \cdot \tau(T_0\cdots T_{n-1})(B_n) \cdot \Delta(V_n,A_n,B_n)(V_{n+1},S_{n},T_{n})\enspace,
\end{multline*}

where we use standard notations for conditional probability measures.

\paragraph{Winning conditions.}

The set of colours is $C=\{0,1\}$.
The reachability, safety, B{\"u}chi and coB{\"u}chi condition
 condition are defined as follows:
 \begin{align*}
 &\winreach=\{\exists n\in\N, C_n  = 1\}\\
&\winsafe=\{\forall n\in\N, C_n = 0\}\\
&\winbuchi=\{\forall m \in \N, \exists n \geq m, C_n=1\}\\
&\wincobuchi = \{\exists m \in \N, \forall n \geq m, C_n = 0\}\enspace.
\end{align*}

When the winning condition is $\win$,
Eve and Adam use strategies
$\sigma$ and $\tau$ and the initial distribution is $\ini$,
then Eve wins the game with probability:
\[
\mathbb{P}^{\sigma,\tau}_{\ini}(\win)\enspace.
\]
Eve wants to maximise this probability, while Adam wants
to minimise it.  

\subsection{The value problem.}

The value problem is computationally intractable
for games with infinite duration and imperfect information.
This holds even for the very simple case
of blind one-player games with reachability conditions.
Those are games where the set of
signals is a singleton and actions of Adam have no influence
on the transition probabilities. These games can be seen
as probabilistic automata, hence the undecidability result of Paz applies.

\begin{theorem}[Undecidability for blind one-player games]
\label{9-thm:undecidability}
Whether Eve has a strategy to win with probability $\geq \frac{1}{2}$ is undecidable, even in blind one-player games.
\end{theorem}

Actually, the value might not even exist.

\begin{proposition}[The value may not be defined]
\label{9-prop:no_value_Buchi}
There is a game with infinite duration imperfect information and B{\"u}chi condition
in which 
\[
\sup_\sigma \inf_\tau \mathbb{P}^{\sigma,\tau}_{\ini}(\winbuchi)
=
\frac{1}{2}
<
1
=
\inf_\tau \sup_\sigma  \mathbb{P}^{\sigma,\tau}_{\ini}(\winbuchi)\enspace.
\]
\end{proposition}

The value however exists for games with reachability condition.
Although the value problem is not decidable, there are some other interesting decision problems to consider.

\subsection{Winning with probability $1$ or $>0$}

\paragraph{Winning almost-surely or positively.}
  A strategy $\sigma$ for Eve is \emph{almost-surely winning}
  from an initial distribution $\ini$ if
\begin{equation*}
\label{9-eq:as}
  \forall \tau,
  \mathbb{P}^{\sigma,\tau}_{\ini}(\win)=1\enspace.
\end{equation*}
When such an almost-surely strategy $\sigma$ exists, the initial distribution $\ini$
is said to be almost-surely winning (for Eve).

A less enjoyable situation for Eve is when she only has a
positively winning strategy.
  A strategy $\sigma$ for Eve is \emph{positively winning} from
  an initial distribution $\ini$ if
\begin{equation*}
\label{9-eq:pos}
  \forall \tau,
  \mathbb{P}^{\sigma,\tau}_{\ini}(\win)>0\enspace.
\end{equation*}
When such a strategy $\sigma$ exists, the initial distribution $\delta$
is said to be positively winning (for Eve).
Symmetrically, a
strategy $\tau$ for Adam is positively winning if it guarantees
$\forall \sigma, \mathbb{P}^{\sigma,\tau}_{\ini}(\win)<1$.

The worst situation for Eve is when her opponent has an
almost-surely winning strategy $\tau$, which thus ensures $\mathbb{P}^{\sigma,\tau}_{\ini}(\win)=0$
whatever strategy $\sigma$ is chosen by Eve.

\paragraph{Qualitative determinacy.}

\begin{theorem}[Qualitative determinacy]
\label{9-thm:qualitative_determinacy}
Stochastic games with signals and reachability, safety and B{\"u}chi
winning conditions are qualitatively determined:
either Eve wins almost-surely winning
or Adam wins positively.
Formally, in those games,
\[
\left(\forall \tau, \exists \sigma,\mathbb{P}^{\sigma,\tau}_{\ini}(\win)=1\right)
\implies
\left(\exists \sigma,\forall \tau ,\mathbb{P}^{\sigma,\tau}_{\ini}(\win)=1\right)\enspace.
\]
\end{theorem}

The proof of this result is given in the next section.

Since reachability and safety games are dual, a consequence of~\Cref{9-thm:qualitative_determinacy}, is that in a reachability game, every initial
distribution is either almost-surely winning for Eve,
almost-surely winning for Adam, or positively
winning for both players.
When a safety condition is satisfied almost-surely for a fixed profile of strategies,
it trivially implies that the safety condition is
satisfied by all consistent plays,
thus for safety games winning \emph{surely} is the same than winning almost-surely.

By contrast, co-B{\"u}chi games are \emph{not} qualitatively determined:
\begin{lemma}
There is a co-B{\"u}chi game in which neither Eve has an almost-surely winning strategy
nor Adam has a positively winning strategy.
\end{lemma}
\begin{proof}
In this game, Eve observes
everything, Adam is blind (he only observes his own actions),
and Eve's objective is to visit only finitely many times the ${\large \frownie}$-state. The initial state is $\large{\frownie}$. The set of actions is $\{a,b,c,d\}$.
All transitions are deterministic.


On one hand, no strategy $\Sigma$
is almost-surely winning for Eve
for her co-B{\"u}chi objective.
{
Since both players can observe their actions,
it is enough to prove that no behavioral
strategy
$\sigma\in C^*\to \Delta(I)$ of Eve is almost-surely winning.
Fix strategy $\sigma$ and assume towards contradiction that $\sigma$ is almost-surely winning. 
We define a strategy $\tau$
such that
$\probimp{\sigma,\tau}{\frownie}{ \winbuchi} > 0$.
Strategy $\tau$ starts by playing only $c$.
The probability to be in state $\frownie$ at step $n$ is
$x^{0}_n = \probimp{\sigma,c^\omega}{\frownie}{V_n=\frownie}$ and since $\sigma$ is almost-surely winning then $x^{0}_n \to_n 0$ thus there exists  $n_0$ such that 
$x^{0}_{n_0}\leq \frac{1}{2}$.
Then $\tau$ plays $d$ at step $n_0$.
Assuming the state was $2$ when $d$ was played, 
the probability to be in state $\frownie$ at step $n\geq n_0$ is
$x^{1}_n = \probimp{\sigma,c^{n_0}dc^\omega}{\frownie}{V_{n}=\frownie\mid V_{n_0}=\frownie}$
and since $\sigma$ is almost-surely winning there exists $n_1$ such that
$x^{1}_{n_1}\leq  \frac{1}{4}$.
Then $\tau$ plays $d$ at step $n_1$.
By induction we keep defining $\tau$ this way so that
$\tau=c^{n_0-1}d c^{n_1 - n_0 - 1}dc^{n_2 - n_1 - 1}d \cdots $.
and for every $k\in \N$,
\[
\probimp{\sigma,\tau}{\frownie}{
V_{n_{k+1}}=\frownie
\text{ and }
V_{n_{k+1}-1}=2
\mid 
V_{n_{k}}=\frownie
} \geq 1 - \frac{1}{2^{k+1}}.
\]
Thus finally
$\probimp{\sigma,\tau}{\frownie}{\winbuchi} \geq
\Pi_{k} (1 - \frac{1}{2^{k+1}})>0$ which contradicts the hypothesis.

}

On the other hand, Adam does not have a positively winning
strategy either.
{
Intuitively, Adam cannot win positively because as time passes, either the play reaches state $1$ or the chances that Adam plays action $d$ drop to $0$. When these chances are small, 
Eve can play action $c$ and she bets no more $d$ will be played and the play will stay safe in state $2$. If Eve loses her bet
then again she waits until the chances to see another $d$ are small and then plays action $c$. Eve may lose a couple of bets but almost-surely she eventually is right and the $\wincobuchi$ condition is almost-surely fulfilled.

Finally neither Eve wins almost-surely nor Adam wins positively.
}
\end{proof}

\paragraph{Decidability}

\begin{theorem}
\label{9-thm:main}
Deciding whether the initial distribution of a B{\"u}chi games,
is almost-surely winning for Eve is
2\EXP-complete.
For safety games, the same problem is \EXP-complete.
\end{theorem}

Concerning winning positively a {\em safety or co-B{\"u}chi game}, one
can use~\Cref{9-thm:qualitative_determinacy} and the determinacy property: Adam
has a positively winning strategy in the above game if and only if
Eve has no almost-surely winning strategy. Therefore, deciding
when Adam has a positively winning strategy can also be done, with
the same complexity.

\begin{theorem}
\label{9-thm:main2}
For reachability and B{\"u}chi games where either Eve is perfectly informed about the state
or Adam is
better informed than Eve, deciding whether the initial distribution is
almost-surely winning for Eve is \EXP-complete.
In safety games
Eve is perfectly
informed {about the state}, the decision problem is in \P.
\end{theorem}

\subsection{Qualitative determinacy: proof of~\Cref{9-thm:qualitative_determinacy}}

\paragraph{Beliefs.}
The
\emph{belief} of a player
is the set of possible states of the game, according
to the signals received by the player. 

\begin{definition}[Belief]
{Let $\arena$ be an arena with observable actions.}
  From an initial set of states $L\subseteq\states$, the belief of
  Eve after having received signal $s$ is:
  \begin{multline*}
\belun(L,s) =
 \{ v\in\states \mid \exists l\in L, t\in S \text{ such that } \tp(l,s,t)(v,\Act(s),\Act(t))>0\}\enspace.  
  \end{multline*}

Remark that in this definition we use the fact that actions of Eve are observable,
thus when he receives a signal $s\in C$ Eve can deduce he played action
$\action_1(c)\in I$.
The belief of Eve after having received a sequence of signals $s_1,\ldots,s_n$ is defined inductively by:
\[
\belun(L,s_1,s_2,\ldots,s_n)
 = \belun(\belun(L,s_1,\ldots,s_{n-1}),s_n).\enspace
\]
Beliefs of Adam are defined similarly.
{
Given an initial distribution $\delta$,
we denote
$\belun^n$ the random variable defined by}
\begin{align*}
&{\belun^{0} = \supp(\delta)}\\
&{\belun^{n+1} =\belun(\supp(\delta),C_1,\ldots,C_{n+1})
= \belun(\belun^n,C_{n+1})
\enspace.}
\end{align*}
\end{definition}

We will also rely on the notion of \emph{belief of belief}, called
here \emph{2-belief}, which, roughly speaking, represents for one
player the set of possible beliefs for his (or her) adversary,
as well as the possible current state.
\begin{definition}[2-Belief]
{Let $\ar$ be an arena with observable actions.}
  From an initial set ${\mathcal{L}} \subseteq \states
  \times \parties{\states}$ of pairs composed of a state and a belief
  for Adam, the 2-belief of Eve after having received signal $c$ is the subset of 
$\states
  \times \parties{\states}$ defined by:
\[
  \deuxbelun({\mathcal{L}},s) = \{ (v,\beldeux(L,t)) \mid
\exists  (\ell,L) \in {\mathcal{L}},  t\in S,
  \tp(v,s,t)(\ell,\Act(s),\Act(t))
  >0\} \enspace.
  \]

From an initial set ${\mathcal{L}} \subseteq \states
\times \parties{\states}$ of pairs composed of a state and a belief
for Adam, the 2-belief of Eve after having  received a sequence of
signals $s_1,\ldots,s_n$ is defined inductively by:
\[
\deuxbelun({\mathcal{L}},s_1,s_2,\ldots,s_n) =
\deuxbelun\left(\deuxbelun\left({\mathcal{L}},s_1,\ldots,s_{n-1}\right),s_n\right)\enspace.
\]
\end{definition}

There are natural definitions of $3$-beliefs (beliefs on beliefs on beliefs)
and even $k$-beliefs however for our purpose, $2$-beliefs are enough,
in the following sense:
in B{\"u}chi games the positively winning sets of Adam
can be characterised by fixed-point equations on sets of $2$-beliefs,
and some positively winning strategies of Adam with finite-memory
can be implemented using $2$-beliefs.

\paragraph{Supports positively winning supports.}

Note that whether an initial distribution $\ini$ is almost-surely or
positively winning depends only on its support, because
$\mathbb{P}^{\sigma,\tau}_{\ini}(\win)
=\sum_{v\in V}\ini(v)\cdot\mathbb{P}^{\sigma,\tau}_{\ini}(\win \mid V_0= v)$.
As a consequence, we will say that a support
$L\subseteq V$ is almost-surely or positively winning for a
player if there exists a distribution with support $L$ which has the
same property.

In the sequel, we will denote $\LLE$ the set of supports almost-surely winning for Eve
and  $\LLA$ those positively winning for Adam.

Then the qualitative determinacy theorem is a corollary of the following lemma.
\begin{lemma}
In every B{\"u}chi game, every non-empty support
which does not belong to $\LLA$ belongs to $\LLE$.
\end{lemma}

The proof of this lemma relies on the definition of a strategy
called the maximal strategy.
We prove that this strategy is almost-surely winning from any initial
distribution which is not positively winning for Adam. 

\begin{definition}[Maximal strategy]
\label{9-def:maximalstrategy}
For every non-empty support $L\subseteq V$ we define
 the set  of {$L$-safe} actions for Eve as
\[
\Isafe(L) = \left\{ a \in A \mid  \forall s \in S, (\Act(s)=a) \implies (\belun(L,s)\not\in\LLA
  )\right\}\enspace,
\]
in other words these are the actions which Eve can play without taking the risk
that her belief is positively winning for Adam.

The \emph{maximal strategy} is the strategy of Eve
which plays the uniform distribution
on $\Isafe(\belun)$
when it is not empty and plays the uniform distribution on $A$ otherwise.
It is denoted $\sigma_{\can}$.

\end{definition}

To play her maximal strategy at step $n$,
Eve only needs to keep track of her belief $\belun^n$,
thus $\sigma_{\can}$ can be implemented by Eve 
using a finite-memory device which keeps track of the current belief.
Such a strategy is said to be \emph{belief-based}.
We will use several times  the following technical lemma about belief-based strategies.

\begin{lemma}
\label{9-lem:borelcantelli}
{Fix a B{\"u}chi game.}
Let $\LL \subseteq \parties{V}$ 
and $\sigma$ a strategy for player $1$.
Assume that
$\sigma$ is a belief
strategy,
$\LL$ is downward-closed
(\textit{i.e.} $L\in\LL \land L' \subseteq L \implies L'\in \LL$)
and for every $L\in\LL\setminus \{\emptyset\}$ and every strategy $\tau$,
\begin{align}
& \probimp{\sigma,\tau}{\delta_L}{\Reach} > 0\enspace,\\
 \label{9-eq:belstab}
&\probimp{\sigma,\tau}{\delta_L}{\forall n\in\N,\belun^n\in \LL} = 1\enspace.
\end{align}
Then $\sigma$ is almost-surely winning for the B{\"u}chi game from any support 
$L\in \LL\setminus \{\emptyset\}$. 
\end{lemma}

\begin{proof}
Since $\LL$ is downward-closed then $\forall L\in \LL,\forall l\in L, \{l\}\in\LL$
thus~\Cref{9-eq:pos} implies 
\begin{equation}
\forall L\in \LL,\forall l\in L,
 \probimp{\sigma,\tau}{\delta_L}{\Reach\mid V_0=l} > 0\enspace.
 \label{9-eq:pos2}
\end{equation}

Once $\sigma$ is fixed then the game is a one-player game with state space $V\times 2^V$ and imperfect information and~\Cref{9-eq:pos2} implies that in this one-player game,
\begin{equation}
\label{9-eq:sigmamproperty}
\forall L\in \LL,\forall l\in L, \forall \tau,
 \probimp{\tau}{\delta_L}{\Reach \mid V_0=l} > \varepsilon\enspace,
\end{equation}
where $N=|K|\cdot 2^{|K|}$
and $\varepsilon = p_{\min}^{|K|\cdot 2^{|K|}}$
and $p_{\min}$ is the minimal non-zero transition probability.
Moreover~\Cref{9-eq:belstab} implies that
in this one-player game the second component of the state space is always in $\LL$, whatever strategy $\tau$ is played by player $2$.
Remind the definition
 \begin{align*}
 \winreach=\{\exists n\in\N, C_n  = 1\}\enspace.
\end{align*}
As a consequence, in this one-player game
for every $m\in\N$,
and every behavioral strategy $\tau$ and every 
$l\in V$,
\begin{equation}
\label{9-eq:sigmamproperty2}
\probimp{\tau}{\delta_L}{\exists m \leq n \leq m+ N, C_n = 1 \mid K_m = l} \geq \varepsilon,
\end{equation}
whenever $\probimp{\tau}{\delta_L}{V_m=l} > 0$.

We use the Borel-Cantelli Lemma to conclude the proof.
According to~\Cref{9-eq:sigmamproperty2},
for every $\tau$, $L\in\overline{\LL}$, 
$m\in \N$,
\begin{equation}
\probimp{\tau}{\delta_L}{\exists n, mN \leq n < (m+ 1)N, C_n=1 \mid V_{mN}} \geq \varepsilon,
\end{equation}
which implies for every behavioral strategy $\tau$ and $k,m\in\N$,
\[
\probimp{\tau}{\delta_L}{\forall n,  \left((m\cdot N) \leq n < ((m+k) \cdot N) \implies  C_n \neq 1 \right)}\leq  \left(1 - \varepsilon\right)^k\enspace.
\]
Since $\sum_k \left(1 - \varepsilon\right)^k$ is finite,
we can apply Borel-Cantelli Lemma for the events 
$(\{\forall n, m\cdot N \leq n < (m+k) \cdot N \implies  C_n\neq 1\})_k$
and we get
$
\probimp{\tau}{\delta_L}{\forall n, m\cdot N \leq n  \implies  C_n\neq 1}=0
$
thus
\begin{equation}
\label{9-eq:assss}
\probimp{\tau}{\delta_L}{\winbuchi}=1\enspace.
\end{equation}

As a consequence $\sigma$ is almost-surely winning for the 
B{\"u}chi game.
\end{proof}

An important feature of the maximal strategy is the following.
\begin{lemma}
In a B{\"u}chi game
with observable actions,
let $\delta\in\Delta(K)$ be an initial distribution which is not positively winning for Adam, \textit{i.e.} $\supp(\delta)\not\in\LLA$.
Then for every strategy $\tau$ of Adam
\begin{equation}
\label{9-eq:LLstable}
\mathbb{P}^{\sigma_\can,\tau}_{\delta}(\forall n \in \N, \belun^n \not\in\LLA )=1\enspace.
\end{equation}
\end{lemma}
\begin{proof}
We only provide a sketch of proof.
The proof is an induction based on the fact that for every non-empty subset $L\subseteq V$,
\[
(L\not \in \LLA) \implies (\Isafe(L)\neq \emptyset)\enspace.
\]
Assume a contrario that $\Isafe(L) = \emptyset$ for some $L\not \in \LLA$.
Then for every action $a\in A$ there exists a signal $s_a\in S$
such that $\belun(L,s_a)\neq \emptyset$ and $\belun(L,s_a) \in \LLA$.
Since $\belun(L,s_a)\neq \emptyset$, the definition of the belief operator implies:
\[
\exists v_a\in L, w_a\in V,  t_a \in T, \text{ such that }  \tp(w_a,s_a,t_a)(v_a,\Act(s),\Act(t_a)) > 0\enspace.
\]

But then Adam can win positively from $L$ with the following strategy.
At the first round, Adam plays randomly any action in $A$.
At the next round, Adam picks up randomly a belief in  $\LLA$ and 
plays forever the corresponding positively winning strategy.
Remark that this strategy of Adam is not described as a behavioural strategy
but rather as a finite-memory strategy. Since actions are observable,
such a finite-memory strategy can be turned into a behavioural one,
see~\cite[Lemma 4.6 and 4.7]{Bertrand.Genest.ea:2017}.

Why is Adam strategy positively winning from $L$?
Whatever action $a\in A$ is played by Eve,
with positive probability she will receive signal $s_a$,
because Adam might play the action $\Act(t_a)$.
Since $\belun(L,s_a) \in \LLA$ then there Adam might with positive probability
play a strategy positively winning when the initial belief
of Eve is $\belun(L,s_a)$. Thus whatever action Eve chooses,
she might lose with positive probability.

\end{proof}

\medskip 

The notion of maximal strategy being defined,
we can complete the proof of~\Cref{9-thm:qualitative_determinacy}.
For that, we show that
$\sigma_{\can}$
is almost-surely
winning from every support not in $\LLA$.

Reachability and safety conditions can be easily encoded as B{\"u}chi conditions,
thus it is enough to consider  B{\"u}chi games.

The first step is to prove that for every $L\in\LLE$,
for every strategy $\tau$ of Adam,
\begin{equation}
\label{9-eq:LLpasM}
\probimp{\sigma_{\can},\tau}{\delta_L}{\winsafe} < 1 \enspace.
\end{equation}
We prove~\Cref{9-eq:LLpasM} by contradiction.
Assume~\Cref{9-eq:LLpasM} does not hold for some $L\in \LLE$
and strategy $\tau$:
\begin{equation}
\label{9-eq:winsafe}
\probimp{\sigma_{\can},\tau}{\delta_L}{\winsafe } = 1\enspace.
\end{equation}
Under this assumption we use $\tau$ to build a strategy positively winning from $L$,
which will contradict the hypothesis 
$L\in\LLA$.
Although $\tau$ is surely winning from $L$ against the particular strategy $\sigma_{\can}$,
there is no reason for $\tau$ to be positively winning from $L$
against all other strategies of player $1$.

However we can rely on $\tau$
in order to define another strategy 
$\tau'$ for Adam positively winning from $L$.
The strategy $\tau'$ is a strategy
which gives positive probability to play $\tau$
all along the play,
as well as any strategy in the family
of strategies
$(\tau_{n,B})_{n\in\N,B\in \LLA}$ defined as follows.
For every $B\in\LLA$ we choose a strategy $\tau_B$ positively winning from $B$.
Then
$\tau_{n,B}$ is the strategy which plays 
the uniform distribution on $A$ for the first $n$ steps then forgets past signals and switches definitively to $\tau_B$. 

A possible way to implement the  strategy $\tau'$
is as follows.
At the beginning of the play
player $2$ tosses a fair coin. If the result is head then he plays $\tau$. Otherwise he keeps 
tossing coins and as long as the coin toss is head, player $2$ plays randomly an action in $J$ .
The day the coin toss is tail, he picks up randomly some $B\in\LLA$ and starts playing $\tau_B$.

Remark that this strategy of Adam is not described as a behavioural strategy
but, since actions are observable,
such a finite-memory strategy can be turned into a behavioural one,
see~\cite[Lemma 4.6 and 4.7]{Bertrand.Genest.ea:2017}.


Now that $\tau'$ is defined, we prove it is positively winning from $L$.
Let $E$ be the event 
`player $1$ plays only actions that are safe with respect to her belief', \emph{\textit{i.e.}}
\[
E = \{ \forall n\in \N, A_n \in \Isafe_\LL(\belun^n)\}\enspace.
\]
Then for every behavioral strategy $\sigma$:
\begin{itemize}
  \item Either $\probimp{\sigma,\tau'}{\delta_L}{E}=1$. In this case 
\[
\probimp{\sigma,\tau'}{\delta_L}{\winsafe}> 0\enspace,
\]
because for every {finite play $\play=v_0a_0b_0s_1t_1v_1\cdots v_n$,}
\[
\left(\probimp{\sigma,\tau'}{\delta_L}{\play} > 0\right)
\implies
\left(\probimp{\sigma_{\can},\tau'}{\delta_L}{\play} > 0\right)
\implies
\winsafe\enspace,
\]
where the first implication holds because, by definition of $\sigma_{\can}$ and $E$,
for every $s_1\cdots s_n\in CS^*, \supp(\sigma(s_1\cdots s_n))\subseteq \supp(\sigma_{\can}(s_1\cdots s_n))$
while the second implication is from~\Cref{9-eq:winsafe}.
Thus $\probimp{\sigma,\tau}{\delta_L}{\winsafe}= 1$ and we get
$\probimp{\sigma,\tau'}{\delta_L}{\winsafe} > 0$ by definition of
$\tau'$.
\item Or $\probimp{\sigma,\tau'}{\delta_L}{E}<1$.
Then by definition of $E$ there exists $n\in\N$
such that 
\[
\probimp{\sigma,\tau'}{\delta_L}{A_n  \not\in
\Isafe_\LL(\belun^n)}>0\enspace.
\]

By definition of $\Isafe_\LL$ it implies
$\probimp{\sigma,\tau'}{\delta_L}{\belun^{n+1}  \in \LL}>0$,
thus there exists $B\in \LL$ such that
$\probimp{\sigma,\tau'}{\delta_L}{\belun^{n+1} =B}>0$.
By definition of $\tau'$ we get
$\probimp{\sigma,\tau_{n+1,B}}{\delta_L}{\belun^{n+1} =B}>0$,
because whatever finite play $v_0,\ldots, v_{n+1}$ leads with positive probability to
the event $\{\belun^{n+1} =B\}$,
the same finite play can occur with 
$\tau_{n+1,B}$ since $\tau_{n+1,B}$ plays every possible action for the $n+1$ first steps.
Since $\tau_{n+1,B}$ coincides with $\tau_\rand$ for the first $n+1$ steps then
by definition of beliefs,
$\probimp{\sigma,\tau_{n+1,B}}{\delta_L}{\belun^{n+1} =B}>0$
and $B\subseteq \{ k\in K\mid \probimp{\sigma,\tau_{n+1,B}}{\delta_L}{K_{n+1}=k\mid \belun^{n+1} =B}>0\}$.
%
Using the definition of $\tau_B$ we get 
\[
\probimp{\sigma,\tau_{n+1,B}}{\delta_L}{\wincobuchi}>0.
\]
As a consequence by definition of $\tau'$
we get $\probimp{\sigma,\tau'}{\delta_L}{\wincobuchi} >0$.
\end{itemize}
In both cases, for every $\sigma$,
$\probimp{\sigma,\tau'}{\delta_L}{\wincobuchi } >0 $
thus $\tau'$ is positively winning from $L$.
This contradicts the
hypothesis $L\in \LLE$. As a consequence we get~\Cref{9-eq:LLpasM} by contradiction.

\medskip

Using~\Cref{9-eq:LLpasM}, we apply~\Cref{9-lem:borelcantelli} to the collection 
$\overline{\LLA}$ and the strategy $\sigma_{\can}$.
The collection $\overline{\LLA}$ is downward-closed because $\LLA$ is upward-closed: if a support is positively winning for Adam then any greater support is positively winning as well, using the same positively winning strategy.

Thus $\sigma_{\can}$ is almost-surely winning for the B{\"u}chi game from every support in $\overline{\LLA}$ \textit{i.e.} every support which is not positively winning for Adam, hence the game is qualitatively determined.

\subsection{Decidability: proof of~\Cref{9-thm:main} and~\Cref{9-thm:main2}}
\label{9-subsec:proof}

\subsubsection{A na{\"i}ve algorithm}
As a corollary of the proof of qualitative determinacy
(\Cref{9-thm:qualitative_determinacy}), we get a maximal strategy $\sigma_\can$
for player $1$ (see~\Cref{9-def:maximalstrategy}) to win
almost-surely B{\"u}chi games.
\begin{corollary}
\label{9-cor:asmem}
  If player $1$ has an almost-surely winning strategy in a B{\"u}chi
  game {with observable actions} then the maximal strategy $\sigma_{\can}$ is almost-surely
  winning.
\end{corollary}

A simple algorithm to decide for which player a game is winning can be derived from~\Cref{9-cor:asmem}: this simple algorithm enumerates all possible belief strategies
and test each one of them to see if it is almost-surely winning. The test reduces to checking positive winning in one-player co-B{\"u}chi games and can be done in exponential time.

As there is a doubly exponential number of {belief} strategies, this can be done in time doubly exponential. 
This algorithm also appears in \cite{Gripon.Serre:2009}.
This settles the upper bound for~\Cref{9-thm:main}. 

Matching complexity lower bounds are established in~\cite{Bertrand.Genest.ea:2017}, proving that this enumeration algorithm is optimal for worst case complexity.  While optimal in the worst case, this algorithm is {likely to be inefficient in practice}.  For instance, if player
  $1$ has no almost-surely winning strategy, then this algorithm will
  enumerate every single of the doubly exponential many {possible belief}
  strategies.  Instead, we provide fixed-point algorithms which do not
  enumerate every possible strategy in~\Cref{9-thm:qdec1} for
  reachability games and~\Cref{9-thm:qdec2} for B{\"u}chi games.
  Although they should perform better on games with particular
  structures, these fixed-point algorithms still have a worst-case
  2-\EXP complexity.

\subsubsection{A fixed-point algorithm for reachability games}

We turn now to the fixed-points algorithms which compute the set of supports that
are almost-surely or positively winning for various objectives.
\begin{theorem}[Deciding positive winning in reachability games]
\label{9-thm:qdec1} 
In a reachability game each initial distribution $\delta$ is either positively winning for player $1$ or surely winning for player $2$, and this depends only on $\supp(\delta)\subseteq \states$.
  The corresponding partition of $\parties{\states}$ is computable in
  time $\mathcal{O}\left(|G| \cdot 2^{|\states|}\right)$, where $|G| $ denotes
  the size of the description of the game,
  as the largest fixed point of a monotonic operator
$\Phi:\parties{\parties{V}}\to \parties{\parties{V}}$
computable in time linear in $|G| $.
\end{theorem}

We denote $\targets$ the set of vertices whose colour is $1$.

\begin{proof}

Let $\LL_\infty\subseteq \parties{\states\bh\targets}$
be the greatest fixed point of the monotonic operator
$\Phi:\parties{\parties{\states\bh\targets}}\to \parties{\parties{\states\bh\targets}}$ defined by:
\begin{equation}
\label{9-eq:defphi}
\Phi(\LL)=\{L\in \LL \mid
\exists j_L\in J, \forall d\in\signauxdeux, (\action_2(d)=j_L)\implies (\beldeux(L,d)\in \LL \cup \{\emptyset)\}\}\enspace,
\end{equation}
in other words $\Phi(\LL)$ is the set of supports
such that player $2$ has an action which
ensure his next belief will be in $\LL$,
whatever signal $d$ he might receive.
Let $\sigma_{\rand}$ be the strategy for player $1$ that plays randomly any action.

We are going to prove that:
\begin{enumerate}
\item every support in $\LL_\infty$ is surely winning for player $2$,
\item and $\sigma_{\rand}$ is positively winning from any support $L\subseteq\states$ which is not in $\LL_\infty$.
\end{enumerate}

We start with proving the first item.
To win surely from any support $L\in\LL_\infty$, player $2$ uses the following
belief strategy $\tau_B$: when the current belief of player $2$ is $L\in\LL_\infty$ then player $2$
plays an action $j_L$ defined as in~\Cref{9-eq:defphi}.
By definition of $\Phi$ and since $\LL_\infty$ is a fixed point of $\Phi$,
there always exists such an action.
When playing with the belief strategy $\tau_B$,
starting from a support in $\LL_\infty$,
the beliefs of player $2$ stay in $\LL_\infty$
and never intersect $\targets$ because $\LL_\infty\subseteq \parties{\states\bh\targets}$.
{According to~\Cref{9-eq:belstab} of beliefs (\Cref{9-lem:borelcantelli})},
this guarantees the play never visits $\targets$,
whatever strategy is used by player $1$.

We now prove the second item.
Let
$\LL_0=\parties{\states\bh\targets}\supseteq
\LL_1=\Phi(\LL_0)\supseteq \LL_2=\Phi(\LL_1)\ldots$ and $\LL_\infty$
be the limit of this sequence, the greatest fixed point of $\Phi$.
  We
prove that for any support $L\in\parties{\states}$, if
$L\not\in\LL_\infty$ then: 
\begin{equation}
\label{9-eq:postoprove} 
\text{$\sigma_{\rand}$ is positively winning for player $1$ from $L$}\enspace.  
\end{equation}
If $L\cap\targets
\not=\emptyset$,~\Cref{9-eq:postoprove} is obvious.  To deal with
the case where {$L\cap \targets =\emptyset$}, we define for every
$n\in\N$, $\KK_n = \parties{\states\bh\targets} \bh \LL_n$, and we
prove by induction on $n\in\N$ that for every $L\in\KK_n$, for every
initial distribution $\delta_L$ with support $L$, for every {behavioral} strategy $\tau$, 
\begin{equation}
\label{9-eq:topo} 
\probimp{\sigma_{\rand},\tau}{\delta_L}{\exists m, 2\leq m\leq n+1, V_m\in\targets }>0 \enspace.
\end{equation}
For $n=0$,~\Cref{9-eq:topo} is obvious because $\KK_0=\emptyset$.  Suppose
that for some $n\in\N$, \Cref{9-eq:topo} holds for every $L'\in\KK_n$,
and let $L\in\KK_{n+1}\bh \KK_n$.
Then by definition of $\KK_{n+1}$, 
\begin{equation}
\label{9-eq:LLLn}
L\in\LL_{n}\bh\Phi(\LL_n)\enspace.
\end{equation}
Let $\delta_L$ be an initial
distribution with support $L$ and $\tau$ any behavioral strategy for player $2$.
Let $J_0\subseteq J$ be the support of $\tau(\delta_L)$ and $j_L\in J_0$.  According
to~\Cref{9-eq:LLLn}, by definition of $\Phi$, there exists a signal
$d\in D$ such that $\action_2(d)=j_L$ and
 $\beldeux(L,d)\not \in \LL_n$ and $\beldeux(L,d)\neq \emptyset$.
{According to~\Cref{9-eq:belstab} of beliefs (\Cref{9-lem:borelcantelli}),} 
 $\forall k \in \beldeux(L,d),\probimp{\sigma_{\rand},\tau}{\delta_L}{V_2 =k\land D_1=d}  > 0$.
   If
$\beldeux(L,d)\cap\targets \not= \emptyset$ then according to
the definition of beliefs,
$\probimp{\sigma_{\rand},\tau}{\delta_L}{V_2\in\targets}>0$.  Otherwise
$\beldeux(L,d)\in\parties{\states\bh\targets}\bh\LL_n=\KK_n$ hence
distribution $\delta_{d}:k\to \probimp{\sigma_{\rand},\tau}{\delta_L}{V_2 =k\mid D_1=d}$
has its support in $\KK_n$. By inductive hypothesis, for every
behavioral strategy $\tau'$,
\[\probimp{\sigma_{\rand},\tau'}{\delta_{d}}{\exists m\in\N, 2\leq
  m\leq n+1, V_m\in\targets}>0\]
hence using the shifting lemma and the
definition of $\delta_{d}$,
\[
\probimp{\sigma_{\rand},\tau}{\delta}{\exists m\in\N,
  3\leq m\leq n+2, V_m\in\targets}>0\enspace,\]
which completes the proof of the inductive
step.

Hence~\Cref{9-eq:topo} holds for every behavioural strategy $\tau$. 
Thus~\Cref{9-eq:topo} holds as well for every general strategy $\tau$.

To compute
the partition of supports between those positively winning for player $1$
and those surely winning for player $2$,
it is enough to compute
the largest fixed point of $\Phi$.
Since $\Phi$ is monotonic, and each application of the operator
can be computed in time linear in the size of the game ($|G|$)
and the number of supports ($2^{|\states|}$)
the overall computation can be achieved in time $|G| \cdot 2^{|\states|}$.
To compute the strategy $\tau_B$, it is enough to compute
for each $L\in\LL_\infty$ one action $j_L$ such that
$(\action_2(d)=j_L)\implies (\beldeux(L,d)\in\LL_\infty)$.
\end{proof}

As a byproduct of the proof one obtains the following bounds on time
and probabilities before reaching a target state, when player $1$ uses
the uniform positional strategy $\sigma_{\rand}$.  From an initial
distribution positively winning for the reachability objective, for
every strategy $\tau$, 
\begin{equation}
\label{9-eq:bounds}
\probimp{\sigma_{\rand},\tau}{\delta}{\exists n\leq 2^{\mid \states
    \mid}, C_n = 1}\geq \left(
  \frac{1}{p_{\min}\mid\actionsun\mid}\right)^{2^{\lvert\states\lvert}}\enspace,
\end{equation}
where $p_{\min}$ is the smallest non-zero transition probability.

\subsubsection{A fixed-point algorithm for B{\"u}chi games}

To decide whether player $1$ wins almost-surely a B{\"u}chi game,
we provide an algorithm which runs in doubly-exponential time.
It uses the algorithm for reachability games as a sub-procedure.

\begin{theorem}[Deciding almost-sure winning in B{\"u}chi games]
  \label{9-thm:qdec2} In a B{\"u}chi game each initial distribution
  $\delta$ is either almost-surely winning for player $1$ or
  positively winning for player $2$, and this depends only on
  $\supp(\delta)\subseteq \states$.
The corresponding partition of $\parties{\states}$ is computable in
time $\mathcal{O}\left(2^{2^{|G|}}\right)$, where $|G|$ denotes the size of the description of the game,
as a projection of the greatest
fixed point $\LL_\infty$
of a monotonic operator
\[\Psi:
\parties{\parties{\states}\times\states}
\to
\parties{\parties{\states}\times\states}
\enspace.
\]
The operator $\Psi$ is computable using as a nested fixed point the operator $\Phi$ of~\Cref{9-thm:qdec1}.
 The almost-surely winning belief strategy of player $1$ and the positively winning $2$-belief strategy of player $2$  can be extracted 
from $\LL_\infty$.
\end{theorem}


\smallskip
We sketch the main ideas of the proof of~\Cref{9-thm:qdec2}.


First, suppose that from \emph{every} initial support, player $1$ can
win positively the  reachability game.
Then she can do so using a belief strategy and according to~\Cref{9-lem:borelcantelli},
this strategy guarantees almost-surely the B{\"u}chi condition.

In general though player $1$ is not in such an easy situation and
there exists a support $L$ which is \emph{not} positively winning
for her for the reachability objective.
Then by qualitative determinacy, player $2$ has a strategy to achieve surely her safety objective
from $L$, which is \emph{a fortiori}
surely winning for her co-B{\"u}chi objective as well.

We prove that in case player $2$ can \emph{force with positive
  probability the belief of player $1$} to be $L$ eventually from another
support $L'$, then player $2$
{ has a general strategy to win positively from $L'$}.
This is not completely obvious because in general player $2$ cannot
know exactly \emph{when} the belief of player $1$ is $L$ (he can only
compute the 2-Belief, letting him know all the possible beliefs player
1 can have).  However player $2$ can make blind guesses,
and be right with $>0$ probability.
For winning positively from $L'$, player $2$ plays
totally randomly until he guesses randomly that the belief of player
$1$ is $L$, at that moment he switches to a strategy surely winning
from $L$.  Such a strategy is far from being optimal, because player
$2$ plays randomly and in most cases he makes a wrong guess about the
belief of player $1$.  However 
there is a non zero probability for his guess to be right.

Hence, player $1$ should surely avoid her belief to be $L$
or $L'$ if she wants to win almost-surely.
However, doing so player $1$ may prevent the play from
reaching target states, which may create another positively winning
support for player $2$, and so on. This is the basis of our fixed-point algorithm.

Using these ideas, we prove that the set
$\LL_\infty\subseteq \parties{\states}$ of supports almost-surely
winning for player $1$ for the B{\"u}chi objective is the largest set of
initial supports from which:
\begin{multline}
\label{9-eq:dag}
\tag{$\dag$}
\textrm{player $1$ has a strategy
  which win positively the reachability game}\\
\textrm{and also ensures at the same time
  her belief to stay in } \LL_\infty .
\end{multline}

Property \Cref{9-eq:dag} can be reformulated as a reachability
condition in a new game whose states are states of the original game
augmented with beliefs of player $1$, kept hidden to player $2$.

The fixed-point characterisation suggests the following algorithm for
computing the set of supports positively winning for player $2$:
$\parties{\states}\bh\LL_\infty$ is the limit of the sequence
$\emptyset=\LL_0'\subsetneq \LL_0'\cup \LL_1''\subsetneq\LL_0'\cup
\LL_1'\subsetneq \LL_0'\cup \LL_1'\cup \LL_2''\subsetneq\ldots
\subsetneq \LL_0'\cup \cdots \cup \LL'_m
=\parties{\states}\bh\LL_\infty$, where
\begin{itemize}
\item from supports in $\LL''_{i+1}$ player $2$ can surely guarantee the safety objective,
under the hypothesis that player $1$ 
{guarantees for sure} her beliefs to stay outside $\LL'_i$,
\item from supports in $\LL'_{i+1}$ player $2$ can ensure with positive probability the belief of player $1$ to be in $\LL''_{i+1}$ eventually,
under the same hypothesis.
\end{itemize}

The overall strategy of player $2$ positively winning for the co-B{\"u}chi objective
consists in playing randomly for some time until he decides to pick
up randomly a belief $L$ of player $1$ in some $\LL''_i$,
bets that the current belief of player $1$ is $L$ and that player $1$
guarantees for sure
her future beliefs 
will stay outside $\LL'_i$.
He forgets
the signals he has received up to that moment and switches
definitively to a strategy which guarantees the first item.  With positive
probability, player $2$ 
guesses correctly the belief of player $1$ at the right moment, and
future beliefs of player $1$ will stay in $\LL'_i$, in which case the
co-B{\"u}chi condition holds and player $2$ wins.

In order to ensure the first item, player $2$ makes use of the hypothesis
about player $1$ beliefs staying outside $\LL_i'$. For that player $2$ needs to keep track of all the possible beliefs of player $1$, hence the doubly-exponential memory.
The reason is player $2$ can infer
from this data structure some information about the possible actions played by player $1$: in case
for every possible belief of player $1$ an action $i\in I$ creates a risk to reach $\LL'_i$
then player $2$ knows for sure this action is not played by player $1$.
This in turn helps player $2$ to know which are the possible states of the game.
Finally, when player $2$ estimates the state of the game using his $2$-beliefs,
this gives a potentially more accurate estimation of the possible states than simply computing his $1$-beliefs.

The positively winning $2$-belief strategy of player $2$ has a particular structure.
All memory updates are deterministic except for one: from
the initial memory state $\emptyset$,
whatever signal is received there is non-zero chance that the memory state stays $\emptyset$ but it may as well 
be updated to several other memory states.
\smallskip


\ifpictures
\includepdf{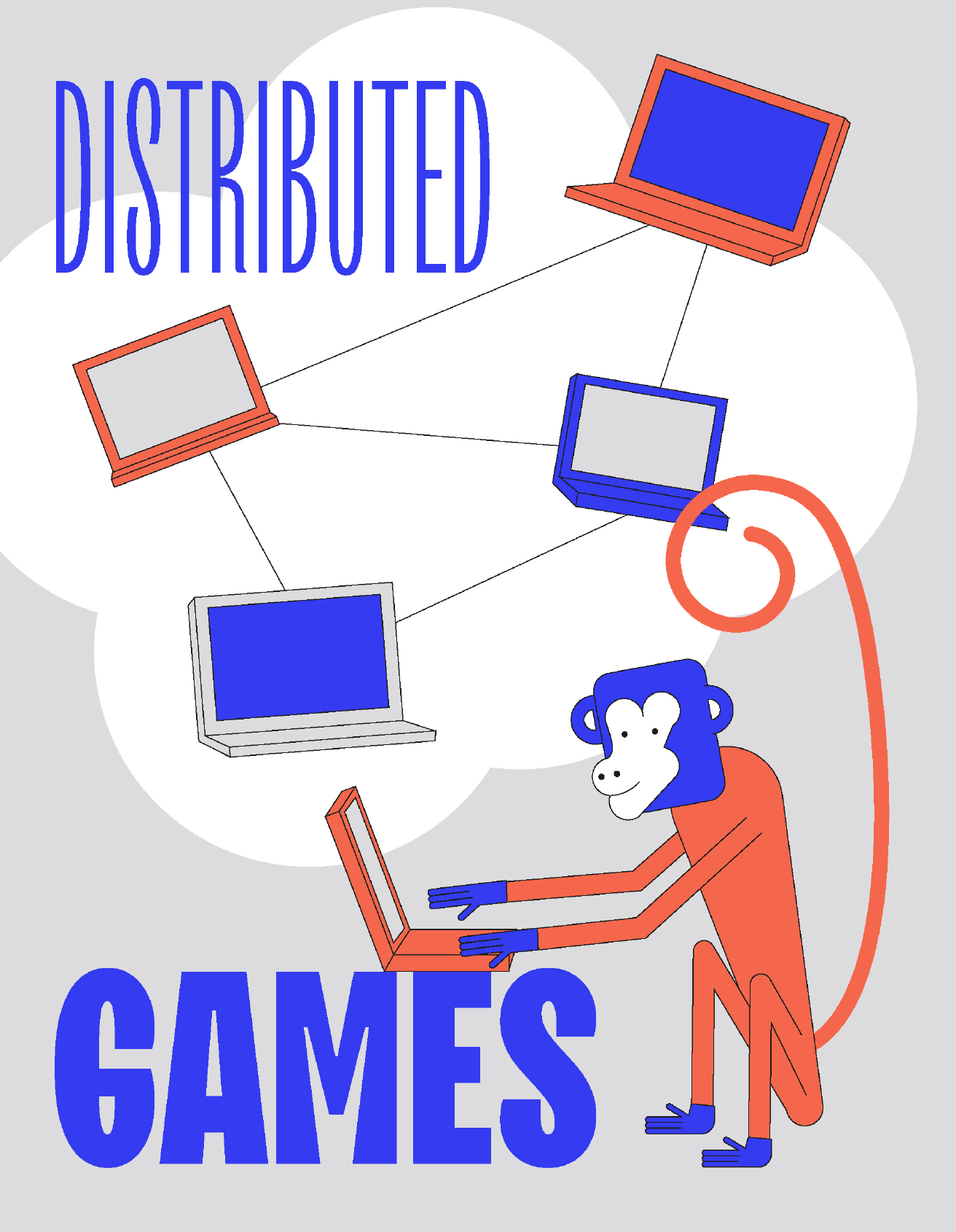}
\fi
\author[C.\ Aiswarya, Paul Gastin, Nathalie Sznajder]{C.\ Aiswarya, Paul Gastin, Nathalie Sznajder}
\copyrightline{Copyright by C.\ Aiswarya, Paul Gastin, and Nathalie Sznajder 2025, to be published by Cambridge University Press in the volume \textit{Games on Graphs} edited by Nathana\"el Fijalkow}

\chapter{Synchronous Distributed Games}
\chapterauthor{C.\ Aiswarya, Paul Gastin, Nathalie Sznajder}
\label{10-chap:distributed}

\graphicspath{{10_Distributed/fig/}}
%
%
%

\newcommand{\Red}[1]{\textcolor{red}{#1}}

\newcommand{\loc}{\mathbb{L}}
\newcommand{\players}{\mathbb{P}}
\newcommand{\archi}{\mathbb{A}}
\newcommand{\env}{\mathsf{env}}
\newcommand{\dom}{\mathsf{dom}}
\newcommand{\rdom}{\mathsf{r}}
\newcommand{\wdom}{\mathsf{w}}
\newcommand{\tower}{\mathsf{Tower}}
\newcommand{\move}{\to}
\newcommand{\da}{{\downarrow}}
\newcommand{\proj}[2]{{#1}_{\mid #2}}
\newcommand\sem[1]{[\![ #1 ]\!]}

\newcommand{\skp}{\mathsf{s}}
\newcommand{\skpout}{\$}
\newcommand{\nxt}{\mathsf{x}}
\newcommand{\ap}{\mathsf{AP}}
\newcommand{\True}{\top}
\newcommand{\False}{\bot}
\newcommand{\LTL}{\mathsf{LTL}}
\newcommand{\X}{\mathop{\mathsf{X}\vphantom{a}}\nolimits}
\newcommand{\Y}{\mathop{\mathsf{Y}\vphantom{a}}\nolimits}
\newcommand{\U}{\mathbin{\mathsf{U}}}
\newcommand{\SU}{\mathbin{\mathsf{SU}}}
\renewcommand{\S}{\mathbin{\mathsf{S}}}
\newcommand{\YS}{\mathbin{\mathsf{SS}}}
\newcommand{\Release}{\mathbin{\mathsf{R}}}
\providecommand{\F}{\mathop{\mathsf{F}\vphantom{a}}\nolimits}
\newcommand{\always}{\mathop{\mathsf{G}\vphantom{a}}\nolimits}
\newcommand{\Past}{\mathop{\mathsf{P}\vphantom{a}}\nolimits}
\newcommand{\History}{\mathop{\mathsf{H}\vphantom{a}}\nolimits}
\newcommand{\XZero}{\mathop{\mathsf{XZero}\vphantom{a}}\nolimits}
\newcommand{\leftend}{{\triangleright}} 
\newcommand{\blank}{\flat} 
\newcommand{\succTM}{\vdash}
\newcommand{\xor}{\otimes}
\newcommand{\limplies}{\rightarrow} 
\newcommand{\lequiv}{\leftrightarrow}

\newcommand{\denote}[1]{\sem{#1}}

\newcommand{\MM}{\mathfrak{M}}
\newcommand{\mm}{\mathfrak{m}}
\newcommand{\initmm}{\mm^{0}}
\newcommand{\initmmp}{\mm'^{0}}
\newcommand{\memory}{\Mem}

\newcommand{\Lang}{\mathcal{L}}
\newcommand{\Trans}{\mathsf{Trans}}
\newcommand{\Branch}{\mathsf{Branch}}
\newcommand{\clr}{\chi}
\newcommand{\qin}{q^\mathsf{in}}
\newcommand{\din}{d^\mathsf{in}}

\newcommand{\nta}{A}
\newcommand{\QA}{Q_{\nta}}
\newcommand{\IA}{I_{\nta}}
\newcommand{\clrA}{\clr_{\nta}}
\newcommand{\TransA}{\Trans_{\nta}}
\newcommand{\rhoA}{\rho_{\nta}}

\newcommand{\ata}{\mathcal{B}}
\newcommand{\QB}{Q_{\ata}}
\newcommand{\IB}{I_{\ata}}
\newcommand{\clrB}{\clr_{\ata}}
\newcommand{\TransB}{\Trans_{\ata}}
\newcommand{\rhoB}{\rho_{\ata}}

\newcommand{\size}{\mathsf{sz}}

\newcommand{\Ev}{\Eve\xspace}
\newcommand{\Ad}{\Adam\xspace}

\newcommand{\merge}[2]{{#1}\oplus{#2}} 
\newcommand{\archipr}{\archi_{p \succ r}}
\newcommand{\archiprf}{\archipr^f}
\newcommand{\fused}[1]{#1^f}
\newcommand{\fusedarchipr}{\archipr^f}
\newcommand{\fusedarchi}{\archi^f}

\newcommand{\factors}[2]{#2\textsf{-factors}(#1)}
\newcommand{\residue}[2]{#2\textsf{-residue}(#1)}
\newcommand{\sigmaR}{\tau}

Let us start with an introductory example game:
\paragraph{Odd Sum} Consider two players Alice and Bob sitting in a line.  They have two
dice with them, $A$ and $B$ respectively.  There is another player Eve, an antagonist,
with her dice $E$.  The game proceeds in rounds - in each round all players will place
(not throw) their dice.  The objective for the team of players Alice and Bob is to ensure
that the sum of the values of the dice is an odd number in each round.  The challenge is
that Alice can only observe the value of the die $E$, and Bob can only observe the value
of the die $A$ (and not $E$ in particular).  No communication is possible among the
players.  To make things worse (and practical) they observe the values of the respective
dice in the previous round.  Is there a winning strategy for the team of Alice and Bob?

The observability and control of the players are depicted in the figure below, which we
call the \textit{distributed architecture}.  Note in particular, that this figure is not
the game arena.  The actual game arena will consist of the Cartesian product of the values
of the three dice.  In this game, all players update their respective components in a
\textit{synchronous} manner.  In particular the vertices of the arena are not partitioned
among players, and it is not a turn-based game.  The players also have only partial
observation as depicted in the figure below.  We do not depict the antagonist in the
architecture.

\medskip
\hfil	\includegraphics[page=27]{10_gastex-pictures-pics.pdf}
\medskip

We can note that there is no winning strategy for Alice and Bob in the above game.  The
reason is that, whatever be the strategy of the players, the values of the dice $A$ and $B$
in the very first round is predetermined by their strategies, hence also their sum and 
parity.  Among the values of die $E$ that Eve may choose, half will result in an even 
sum.

Hence, for the sake of feasibility, the objectives must respect  \textit{delay}.  Consider the following
modified objective.  For each round $i$, $E(i) + A(i+1) + B(i+2)$ is odd.  Here $X(i)$ denotes
the value of die $X$ in round $i$.  Is there a winning strategy for the players Alice and
Bob with the modified objective?\footnote{ A winning strategy would be that player Bob
keeps $B$ always $2$.  Player Alice plays $1$ (resp.\ $2$) on $A$ in round $i+1$ if
$E(i)$ is even (resp.\ odd).}

Let us now consider a more complex objective where, in addition to the odd sum required in
the above paragraph, there is one more condition: if $E(i)$ is $x$, then in a future round
$j>i$, $B(j)$ must be $x$.  Do the players have a winning strategy for this
objective?\footnote{ One strategy would be that B cycles through all the six values of the
dice.  That is, $B(i) = 1+ (i \mod 6)$.  Keeping this strategy of Bob in mind, Alice will
play $1$ on $A$ in round $i+1$ if $E(i)+i+1$ is even, and $2$ otherwise.}

Consider yet another variant, where the environment has two dice $E$ and $F$.  As before,
Alice can observe only $E$.  Player Bob can observe both $A$ and $F$.  This arrangement of
players indicating their observation is depicted in the architecture below.  
The objective of the game is to maintain $E(i) + F(i) + B(i+2)$ odd for all $i$, with an
additional caveat that $A(i)$ should be at most 2.  Do the players have a winning
strategy?\footnote{One strategy can be as follows.  
  Alice plays 2 on $A(i+1)$ if $E(i)$ is even and 1 if $E(i)$ is odd. 
  Bob sets $B$ to $1$ (resp.\ $2$) if the sum of the last observed value of $A$ and
  the second last observed value of $F$ is even (resp.\ odd).}
If they do, how much memory is needed?\footnote{For the above strategy, 
  Alice does not require memory.  Bob requires two bits of memory, as he needs to remember
  the parity of the previous two values of $F$.}

\medskip
\hfil	\includegraphics[page=28]{10_gastex-pictures-pics.pdf}
\medskip

In this chapter we will be studying synchronous distributed games such as the Odd Sum game
described above.  Note that in this type of games we are not interested in a winning
strategy for Eve (or environment, the antagonist).

\bigskip

In a synchronous distributed game, the (global) arena is \emph{distributed} over several
locations/variables.  We let $\loc$ be the finite set of locations. 
In the above example, $\loc = \{F, E, A, B\}$. 
Each location $\ell\in\loc$ has a set $S_{\ell}$ of \emph{local} vertices/states/values.
In the example above, $S_{\ell} = \{1, 2, \dots 6\}$ for each $\ell \in \{F, E, A, B\}$. 
We assume that $|S_{\ell}|\geq2$ for each location $\ell\in\loc$.  The set of global
vertices/states of the induced \emph{global} arena is $S=\prod_{\ell\in\loc}S_{\ell}$.

In a synchronous distributed game, we also have a finite set $\players$ of players
(disjoint from $\loc$).  In the example above, $\players = \{\text{Alice, Bob}\}$.
Hence, a distributed game is a multiplayer game.  The players in $\players$ play as a team
against an environment $\env\notin\players$ (or Eve, in the game above) in order to
fulfil some winning condition.  Hence, the players in $\players$ are not antagonist, on
the contrary they cooperate in order to win the game.  Collectively, the set of players is
called \emph{the players} or \emph{the system}.

Each player in $\players$ has only a partial view of the global play, and their
strategy depends on this partial view.  Hence, a distributed game is a game with partial
information.  The specificity is the fact that the partial view of a player is determined
by the distributed nature of the game, described by the architecture.

The objective of the game can be any regular set of $\omega$-sequences of vertices in the
global arena $S=\prod_{\ell\in\loc}S_{\ell}$.

\section{Preliminaries}
\label{10-sec:prelims}

\subsection{Automata on infinite trees}
\label{10-sec:tree-automata}

In this section, we recall the definitions and results on tree automata that will be used 
for solving synchronous distributed games. The reader is referred to the chapter 
\emph{Automata on infinite trees} by Christof Löding \cite{Loding:2021}, which contains 
all the results needed here, and also  references to original works.

Let $D$ be a finite set of directions and $L$ be a finite set of labels.
A $D$-directed $L$-labelled (finite or infinite) tree, $(D,L)$-tree for short, is a
(partial) map $t\colon D^{*}\to L$ with a prefix closed domain: if $v\in\dom(t)$ and $u$
is a prefix of $v$ then $u\in\dom(t)$.  
A node of $t$ is an element $u\in\dom(t)$ and its label is $t(u)$.
The children of node $u\in\dom(t)$ are the nodes $ud\in\dom(t)$ with $d\in D$. 
We only consider nonempty trees, \textit{i.e.}, $\dom(t)\neq\emptyset$, and 
$\varepsilon\in\dom(t)$ is the \emph{root} of $t$.
We say that a $(D,L)$-tree $t$ is \emph{full} if $\dom(t)=D^{*}$.

\paragraph*{Tree Automata} 

A parity alternating tree automaton (ATA) over $D$-directed $L$-labelled trees is described
by $\ata=(Q,\Trans,I,\clr)$ where $Q$ is the finite set of states,
$I\subseteq Q$ is the set of initial states, 
$\clr\colon Q\to\mathbb{N}$ is the \emph{priority} function,
and the transition function $\Trans$ maps a pair $(q,\ell)\in Q\times L$ to a positive
boolean combination of conditions. A condition is a  pair in $D\times Q$.  
We assume that $\Trans(q,\ell)$ is written
in disjunctive normal form.  For instance, we may have
$\Trans(q,\ell)=((d_{1},q_{1})\wedge(d_{1},q_{2})\wedge(d_{2},q_{1})) \vee(d_{2},q_{3})
\vee((d_{1},q_{4})\wedge(d_{2},q_{5}))$.  Note that, in a disjunct of a transition,
multiple states may be imposed to the same direction.  For instance, in the first disjunct
above, both $q_{1}$ and $q_{2}$ are sent in direction $d_{1}$.

Let $\ata=(Q,\Trans,I,\clr)$ be an ATA over $(D,L)$-trees.  The size $|\ata|$ of the
ATA is dominated by the size $|\Trans|$ of its transition function.  Hence, we define
$|\ata|=|Q|+|D|+|L|+|\Trans|$ where $|\Trans|=\sum_{(q,\ell)\in Q\times L}1+|\Trans(q,\ell)|$ 
and $|\Trans(q,\ell)|$ is the number of conditions occurring in $\Trans(q,\ell)$.

An ATA on \emph{full} $(D,L)$-trees is called a \emph{non-deterministic} tree automaton
(NTA) if transition disjuncts send exactly one state in each direction.
In other words, each transition disjunct can be seen as a total function from $D$ to $Q$.
In the example above, only the last disjunct $(d_{1},q_{4})\wedge(d_{2},q_{5})$ satisfies
this condition.

\begin{figure}[tbp]
  \centering
  \includegraphics[page=24]{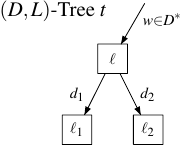}
  \hfil\hfil
  \includegraphics[page=25]{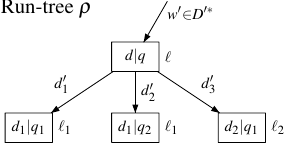}
  \caption{(left) Extract of a $(D,L)$-tree $t$. 
  \\
  (right) Extract of a run-tree $\rho$ of an ATA on the tree $t$.
  \\
  At this node, the run $\rho$ uses the disjunct
  $(d_{1},q_{1})\wedge(d_{1},q_{2})\wedge(d_{2},q_{1})$ of $\Trans(q,\ell)$.
  \\
  For clarity, on the right of each node of $\rho$ is the label of the corresponding 
  node of $t$.
  }
  \label{10-fig:tree-run-tree-ATA}
\commentAlt{Figure~\ref{10-fig:tree-run-tree-ATA}: Two tree diagrams, one labeled "(D,L)-Tree t" and the other "Run-tree rho", illustrating a transformation of tree structures with labeled nodes and edges. See long description.}
\commentLongAlt{Figure~\ref{10-fig:tree-run-tree-ATA}: The image displays two tree diagrams side-by-side, representing different types of trees.

Left Diagram: Labeled "(D,L)-Tree t".
It has a square root node labeled 'l'. An arrow points to this root node from above, labeled 'w in D*'.
From the root 'l', two branches descend:

A left branch with an arrow labeled 'd1' points to a square leaf node labeled 'l1'.
A right branch with an arrow labeled 'd2' points to a square leaf node labeled 'l2'.
Right Diagram: Labeled "Run-tree rho".
It has a square root node labeled 'd'/q'. An arrow points to this root node from above, labeled 'w' in D*'. Another label, 'epsilon', is to the right of this root node.
From the root 'd'/q', three branches descend:

A left branch with an arrow labeled 'd1'' points to a square leaf node labeled 'd1|q1'.
A middle branch with an arrow labeled 'd2'' points to a square leaf node labeled 'd1|q2'. Below this node, there is a label 'l1'.
A right branch with an arrow labeled 'd3'' points to a square leaf node labeled 'd2|q1'. Below this node, there is a label 'l2'.}
\end{figure}

A run of the ATA $\ata$ on a $(D,L)$-tree $t$ is another tree $\rho$ using 
a (possibly different) set $D'$ of directions to allow disjuncts in transitions to send 
several states to the same direction.  In our example $\Trans(q,\ell)$ above, the
first disjunct sends two states in direction $d_{1}$ and one state in direction $d_{2}$,
while the second disjunct does not send any state in direction $d_{1}$.
Formally, a run of $\ata$ on $t$ is a $(D',D\times Q)$-tree $\rho$ satisfying 
the two conditions below. For a node $w\in D'^{*}$ of $\rho$, we denote by $f_{\rho}(w)\in D^{*}$ 
the corresponding node of $t$ defined inductively by: $f_{\rho}(\varepsilon)=\varepsilon$ and 
$f_{\rho}(wd')=f_{\rho}(w)d$ if $\rho(wd')=(d,q)$ for some $q\in Q$.
\begin{itemize}
  \item the root label is $\rho(\varepsilon)=(d,q)$ where $q\in I$ is an initial
  state and $d\in D$. Here $d$ is immaterial. Only $q$, the state labelling the root of $\rho$, is meaningful.

  \item  for each node $w\in D'^{*}$ of $\rho$ labelled $(d,q)$, the conjunction of the 
  labels of the children of $w$ must be some disjunct of $\Trans(q,\ell)$ where 
  $\ell=t(f_{\rho}(w))$ is the label of the corresponding node of $t$.
\end{itemize}
An extract of a tree $t$ and a run-tree $\rho$ of an ATA is given in 
\Cref{10-fig:tree-run-tree-ATA}.

Since in an NTA, each disjunct in a transition sends exactly one state in each direction,
we may use the same set of directions $D'=D$ for the run-tree $\rho$ of an NTA over the
tree $t$.  In that case, the directions from the labels of $\rho$ are redundant and are
usually removed when dealing with NTAs only.  But the direction labels are needed for
ATAs.

A run $\rho$ is accepting if every infinite branch $B\in D'^{\omega}$ of $\rho$ satisfies
the parity condition: the maximal priority which occurs infinitely often in $\clr(B)$ is
even, where $\clr(B)$ is the infinite sequence of priorities obtained by applying $\clr$ to
the state labels of the nodes of $B$.
The language of an ATA $\ata$, denoted by $\Lang(\ata)$, is the set of trees on which
$\ata$ has an accepting run.  We say that a tree language $T$ is \textit{regular} if there
exists an ATA $\ata$ such that $T=\Lang(\ata)$.

\begin{theorem}[Simulation]\label{10-thm:ata2nta}
  For each alternating parity tree automaton $\ata$ with $n$ states and $c$ priorities,
  one can construct a language-equivalent non-deterministic parity tree automaton $\ata'$
  with $2^{\mathcal{O}(nc\log(nc))}$ states and $\mathcal{O}(nc)$ priorities.
  The automaton $\ata'$ can be constructed from $\ata$ in \textsc{ExpTime}.
%
\end{theorem}

The construction of the simulating NTA $\ata'$ from the ATA $\ata$ presented in
\cite{Loding:2021} relies on the determinization of a Büchi word automaton.

\begin{theorem}[\cite{Piterman:2007}]\label{10-thm:NBWtoDPW}
  Given a non-deterministic Büchi word automaton (NBW) $\mathcal{A}$ with $n$ states
  over an input alphabet $\Sigma$, we can construct in time $\mathcal{O}(n^{2n}|\Sigma|)$
  a deterministic parity word automaton (DPW) $\mathcal{A}'$ with $2n^{n}n!$ states and
  $2n$ priorities such that $\Lang(\mathcal{A}')=\Lang(\mathcal{A})$
  (resp.\ $\Lang(\mathcal{A}')=\Sigma^{\omega}\setminus\Lang(\mathcal{A})$).
\end{theorem}

We will need to check emptiness of a non-deterministic tree automaton.  Moreover, when a
regular tree language $T$ is nonempty, it contains simple trees, called \emph{regular}
trees since they can be generated by deterministic finite (word) automata.  Actually, a
full tree $t\colon D^{*}\to L$ is \emph{regular} if it can be defined using a finite
memory structure.\footnote{A memory structure is a tuple $\Mem=(\MM,\mu,\initmm)$ where
$\MM$ is the set of memory states, $\initmm\in\MM$ is the initial memory state, and
$\mu\colon\MM\times D\to\MM$ is the (complete) update function.  It is \emph{finite} if
$\MM$ is finite.  An output function $\tau\colon\MM\to L$ induces a regular tree $t\colon
D^{*}\to L$ defined by $t(u)=\tau(\mu(\initmm,u))$ for all $u\in D^{*}$, where the update 
function $\mu$ is extended to $\Mem\times D^{*}$ as usual.}

\begin{theorem}[Emptiness]\label{10-thm:nta-emptiness}
  Let $\ata$ be a non-deterministic parity tree automaton with $n$ states and $c$ 
  priorities.
  One can check whether $\Lang(\ata)$ is empty or not in time $\mathcal{O}(|\ata|^{c})$ 
  or in time $|\ata|^{\mathcal{O}(\log(c))}$.
  Moreover, when $\Lang(\ata)\neq\emptyset$, one can construct a regular tree 
  $t\in\Lang(\ata)$ using a finite memory structure having at most $n$ memory states.
\end{theorem}

To prove the theorem above, we use the parity \emph{emptiness game} $\game_{\ata}$
associated with a parity NTA $\ata$.  Then, $\Lang(\ata)\neq\emptyset$ if and only if some
initial state of $\ata$ is winning for \Ev in $\game_{\ata}$.  Moreover, from a uniform
positional strategy of \Ev, we construct easily the memory structure allowing to define a
regular tree accepted by $\ata$.

We cannot directly apply the emptiness game to parity \emph{alternating} tree automata,
but combining \Cref{10-thm:ata2nta,10-thm:nta-emptiness}, we can check emptiness of an ATA
in exponential time.  On the other hand, we can use a parity \emph{membership game}
$\game_{\ata,t}$ to check whether a tree $t$ is accepted by a parity ATA $\ata$.  When the
tree $t$ is regular, the membership game is \emph{finite}.  This allows to check 
membership of a regular tree by a parity alternating automaton by solving a parity game.

\begin{theorem}[Membership]\label{10-thm:ata-membership}
  Let $\ata$ be an alternating parity tree automaton with $n$ states and $c$ priorities. 
  Let $t$ be a regular tree defined with a memory structure $\Mem$ with $m$ states.
  One can check whether $t\in\Lang(\ata)$ in time $\mathcal{O}((m\times|\ata|)^{c})$.
\end{theorem}

Finally, an infinite branch of a $(D,L)$-tree $t\colon D^{*}\to L$ induces an infinite
word over $D\times L$.  More precisely, an infinite branch is an infinite word
$B=d_{0}d_{1}d_{2}\cdots\in D^\omega$ over the directions of the tree, and adding the
label of each node to the next direction on the branch results in an infinite word
$\overline{B}^{t}=(d_{0},\ell_{0})(d_{1},\ell_{1})(d_{2},\ell_{2})\cdots$ with
$\ell_{i}=t(d_{0}\cdots d_{i-1})$ for all $i\geq0$. 
If $W\subseteq(D\times L)^{\omega}$ is a word language, then we let $\Branch(W)$ be the
set of all full $(D,L)$-trees $t$ such that $\overline{B}^{t}\in W$ for all branches $B\in
D^{\omega}$ of $t$. We will use the following easy result.

\begin{theorem}[Branch]\label{10-thm:words2trees}
  If $W\subseteq(D\times L)^{\omega}$ is a regular word language, then $\Branch(W)$ is a
  regular tree language.
  Moreover, if $W$ is accepted by a \emph{deterministic} parity word automaton $\ata$ with
  $n$ states and $c$ priorities, then $\Branch(W)$ is accepted by a
  \emph{deterministic}\footnote{A tree automaton is deterministic if for all $(q,\ell)$,
  the transition function $\Trans(q,\ell)$ consists of a single disjunct $\bigwedge_{d\in
  D}(d,q_{d})$.} parity tree automaton $\ata'$ with $n$ states and $c$ priorities.
  The automaton $\ata'$ can be constructed from $\ata$ in linear time.
\end{theorem}

\subsection{Linear Temporal Logic}
\label{10-sec:ltl}

In the chapter, we will sometimes use the linear time temporal logic ($\LTL$) to express
winning conditions of games.  Usually, winning conditions describe regular sets of plays.
Since $\LTL$ is a convenient way to express some regular properties, we will use this
formalism in our examples.  Moreover, to establish the complexity lower bound for
decidable distributed games, we need to be precise on the description of the input: there,
winning conditions will be given by $\LTL$ formulas.  Also, the undecidability
results already hold when the winning conditions are given by $\LTL$ formulas, that only
allow to express a fragment of the regular properties.  We give here the syntax and
semantics of $\LTL$ with both future and past modalities.

Given a nonempty set $\ap$ of atomic propositions, the syntax of $\LTL$ formulas is given
below (with $p\in\ap$):
\begin{align*}
  \varphi ::= p \mid \neg \varphi \mid \varphi \vee \varphi \mid \varphi \U\varphi 
  \mid \varphi \S\varphi \mid \X \varphi \mid \Y\varphi
\end{align*}

Formulas are interpreted over infinite sequences of elements in $\Sigma=2^{\ap}$.  Given
$w=w_0w_1\dots\in\Sigma^\omega$, we let
$$
\begin{array}{ll}
  w,i \models p &\text{if $p\in w_i$}\\
  w,i \models \X\varphi &\text{if $w, i+1\models \varphi$}\\
  w,i \models \Y\varphi &\text{if $i>0$ and $w, i-1\models\varphi$}\\
  w,i \models \varphi_1 \U \varphi_2 &\text{if there exists $j\geq i$ 
  such that $w,j\models \varphi_2$ and $w,k \models\varphi_1$ for all $i\leq k <j$}\\
  w,i \models \varphi_1 \S \varphi_2 &\text{if there exists $j\leq i$ 
  such that $w,j\models \varphi_2$ and $w,k \models\varphi_1$ for all $j<k\leq i$.}
\end{array}
$$
Boolean operators are interpreted in the usual way. 
We say that $w\models\varphi$ if $w,0\models\varphi$ and we let
$\sem{\varphi}=\set{w\in\Sigma^{\omega}\mid w\models\varphi}$.

We use classic abbreviations: $\True := p\vee\neg p$, $\False :=\neg\True$, $\F\varphi :=
\True \U \varphi$, $\G\varphi := \neg\F\neg\varphi$.
We also use \emph{strict} versions of \emph{until} and \emph{since}:
$\varphi_1\SU\varphi_2 := \X(\varphi_1\U\varphi_2)$ and
$\varphi_1\YS\varphi_2 := \Y(\varphi_1\S\varphi_2)$.

In the following, formulas will be specifically interpreted over sequences of global
states $\pi=s^{0}s^{1}s^{2}\cdots\in S^\omega$, with $S=\prod_{\ell\in\loc}S_\ell$.
Unless stated otherwise, the set of atomic propositions we consider will then be
$\ap=\bigcup_{\ell\in\loc}\ap_\ell$, where $\ap_\ell=\{a_\ell\mid a\in S_\ell\}$, for all
$\ell\in\loc$.  In that case, we say that $\pi,i\models a_\ell$ if the $\ell$-component of
$s^{i}$ is $s^{i}_{\ell}=a$.

\subsection{Notation}
\label{10-sec:notation}

Let $I$ be an index set and $(S_{i})_{i\in I}$ be a tuple of sets, one for each index.
When $J\subseteq I$ is a subset of indices, we let $S_{J}=\prod_{i\in J}S_{i}$.
When $I$ is understood, we often simply write $S$ instead of $S_{I}$.
Also, if $s=(s_{i})_{i\in I}\in S$ is a full tuple, then we 
denote by $s_{J}=(s_{i})_{i\in J}\in S_{J}$ its projection on the indices in $J$.
This is extended to finite or infinite sequences $\pi=s^{0}s^{1}s^{2}\cdots$ in 
$S^\infty=S^{*}\cup S^{\omega}$ by $\proj{\pi}{J}=s^{0}_{J}s^{1}_{J}s^{2}_{J}\cdots$.


\section{Synchronous distributed games}
\label{10-sec:synchronous}

In a \emph{synchronous} game, the play evolves in \emph{synchronous rounds}. During a 
round, the value at each location is updated according to the choices of the environment 
and the players. Hence, a play is a finite or infinite 
sequence $\pi=s^{0}s^{1}s^{2}\cdots$ of global states. 

\begin{figure}[tbp]
  \centering
  \includegraphics[page=26]{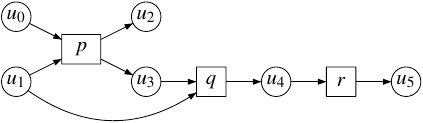}
  \caption{An architecture with a linear information preorder.}
  \label{10-fig:order3}
\commentAlt{Figure~\ref{10-fig:order3}: A directed graph with a central square node 'p' branching to multiple circular nodes, and a linear sequence of alternating square and circular nodes.}
\commentLongAlt{Figure~\ref{10-fig:order3}: The image displays a directed graph with a central square node labeled 'p'.

From 'p', three arrows emanate: one to 'u0' (circle), one to 'u2' (circle), and one to 'u3' (circle).
An arrow points from 'u1' (circle) to 'p'.
A curved arrow points from 'u1' (circle) to 'q' (square).
A linear sequence of nodes extends from 'u3': 'u3' (circle) points to 'q' (square), 'q' points to 'u4' (circle), 'u4' points to 'r' (square), and 'r' points to 'u5' (circle).}
\end{figure}

Each location in $\loc$ is controlled either by some player in $\players$ or by the
environment $\env$.  Moreover, each player observes only a subset of locations.  This is
formalised by a bipartite graph $(\loc\uplus\players,E)$ with
$E\subseteq(\loc\times\players)\cup(\players\times\loc)$, called an architecture.
See \Cref{10-fig:order3,10-fig:order1,10-fig:linear-order1,10-fig:information-fork} for
examples of architectures.
An architecture induces two maps $\rdom,\wdom\colon\players\to 2^{\loc}$ defining the
\emph{read domain} and the \emph{write domain} of each player, by setting
$\rdom(p)=E^{-1}(p)\subseteq\loc$ and $\wdom(p)=E(p)\subseteq\loc$.  The read and write
domains are extended to sets $P\subseteq\players$ of players by $\rdom(P)=\bigcup_{p\in
P}\rdom(p)$ and $\wdom(P)=\bigcup_{p\in P}\wdom(p)$.
An \emph{architecture} is a bipartite graph $\archi=(\loc\uplus\players,E)$
with the following restrictions
%
\begin{enumerate}
  \item the \emph{write domains} of the players are nonempty and pairwise disjoint: 
  $\wdom(p)\neq\emptyset$ and $\wdom(p)\cap\wdom(q)=\emptyset$ if $p\neq q$,

  \item each location is either written by some player or read by 
  some player (or both): $\loc=\wdom(\players)\cup\rdom(\players)$.
\end{enumerate}

A location is \emph{internal} if it is both written by some player and read by some
player.  The set of internal locations is therefore $\wdom(\players)\cap\rdom(\players)$.
The external locations can be seen as being either written by the environment
$\wdom(\env)=\loc\setminus\wdom(\players)$ or read by the environment
$\rdom(\env)=\loc\setminus\rdom(\players)$.  We also say that locations in $\wdom(\env)$
are the external inputs to the players (from the environment) whereas locations in
$\rdom(\env)$ are the external outputs from the players (to the environment).

A \emph{distributed arena}
is a tuple $\arena=(\archi,(S_{\ell})_{\ell\in\loc})$ where $\archi=(\loc\uplus\players,E)$ is
an architecture and $S_{\ell}$ with $|S_{\ell}|\geq2$ is the set of local states\footnote{A
location with a single state as no influence on the game and could be removed.} 
for location $\ell\in\loc$.  The distributed arena is \emph{finite} if the set
$S=\prod_{\ell\in\loc}S_{\ell}$ of global states is finite.
A \emph{play} of the distributed arena $\arena$ is a finite or infinite sequence
$\pi=s^{0}s^{1}s^{2}\cdots \in S^{\infty}$ of global states.\footnote{In some cases, 
  it could be convenient to restrict the possible moves that players may use in a play.
  This could be done by adding to the arena a tuple 
  $({\move}_{p})_{p\in\players}$ where ${\move}_{p}\subseteq S_{\rdom(p)}\times S_{\wdom(p)}$
  defines the possible moves of player $p\in\players$. However, this restriction could 
  be enforced by the winning condition, so we prefer to keep the arena simple.}

Each player only observes the locations in its read domain\footnote{Assume that some 
  player $p\in\players$ has an empty read domain: $\rdom(p)=\emptyset$. By convention, 
  we let $S_{\rdom(p)}=\{\mathsf{tick}\}$ (we can think of $\mathsf{tick}$ 
  as the empty tuple $()$ in the empty product $S_{\emptyset}$). 
  This convention corresponds to the fact that in a \emph{synchronous} distributed game, 
  each player $p$ is aware of the global discrete time and writes values to locations in 
  $\wdom(p)$ in each round, even if its read domain is empty.}.
Hence, a (local) strategy for player $p\in\players$ is a map
$\sigma_{p}\colon(S_{\rdom(p)})^{*}\to S_{\wdom(p)}$.
A sequence $\pi=s^{0}s^{1}s^{2}\cdots \in S^{\infty}$ is consistent with $\sigma_{p}$ if
player $p$ follows its local strategy with delay $1$. 
Formally, $\pi$ is consistent with $\sigma_{p}$ if
$s^{n}_{\wdom(p)}=\sigma_{p}\big(\proj{(s^{0}s^{1}\cdots s^{n-1})}{\rdom(p)}\big)$ for all
$n\geq0$.  Notice that, the values written by $p$ during the $n$-th round depend on the
sequence of values $p$ has observed on its read domain, up-to and including the values in
the $(n-1)$-th round since we assume that player $p$ reacts with a delay 1.
%
%
%
%

A \emph{distributed strategy} for the system is a tuple
$\sigma=(\sigma_{p})_{p\in\players}$ of local strategies for each player.  A sequence
$\pi=s^{0}s^{1}s^{2}\cdots \in S^{\infty}$ is consistent with
$\sigma=(\sigma_{p})_{p\in\players}$ if it is consistent with each $\sigma_{p}$ for
$p\in\players$.

There are no constraints on the values of the locations controlled by the environment.
Intuitively, the environment chooses values for the external input locations in
$\wdom(\env)$ and players react by writing to their respective locations according to
their local strategies.
Notice that, a sequence of states consistent with the strategy is fully determined by the
external inputs provided by the environment.  Formally, if we have two sequences of states
$\pi\neq\pi'$ consistent with $\sigma$ then
$\proj{\pi}{\wdom(\env)}\neq\proj{\pi'}{\wdom(\env)}$.

The winning condition of the distributed game is a subset $W\subseteq S^{\omega}$.
A play of $\arena$ is \emph{winning} if it belongs to $W$.
A distributed strategy $\sigma=(\sigma_{p})_{p\in\players}$ is \emph{winning} if all
plays $\pi\in S^{\omega}$ consistent with $\sigma$ are \emph{winning}.

As in classic two-player games, the winning set is usually regular, it may be defined by a
parity condition induced by a colouring function of the global states or by any 
equivalent formalism such as MSO logic. In \Cref{10-sec:complexity} we prove 
complexity lower bound for decidable architectures with winning conditions given in linear 
temporal logic ($\LTL$). For the complexity, the choice of the formalism to give the 
winning condition is indeed important. In \Cref{10-sec:undecidability-synchronous}, we prove 
some undecidability results also with winning conditions given in $\LTL$.

A \emph{synchronous distributed game} is a tuple $\game=(\arena,W)$
consisting of a finite distributed arena
$\arena=(\archi,(S_{\ell})_{\ell\in\loc})$ on
architecture $\archi=(\loc\uplus\players,E)$ and a regular winning condition
$W\subseteq S^{\omega}$.

\begin{problem}[Solving the game]~
  \begin{description}     \label{10-prob:existWS}
    \item[\textsc{input:}] A synchronous distributed game $\game$
    
    \item[\textsc{output:}] Existence of a winning distributed strategy
    $\sigma=(\sigma_{p})_{p\in\players}$ for the players.
  \end{description}
\end{problem}

\Cref{10-prob:existWS} is not decidable in general.  

\begin{theorem}\label{10-thm:undecidable-P1}
  \Cref{10-prob:existWS} is neither recursively enumerable nor co-recursively enumerable.
\end{theorem}

\Cref{10-thm:undecidable-P1} follows from \Cref{10-thm:undecidable-A1} given in
\Cref{10-sec:undecidability-synchronous}, which is a slightly stronger version where we
restrict to distributed games on the architecture $\archi_{1}$ of
\Cref{10-fig:information-fork} and to winning conditions given in linear temporal logic.

The undecidability argument requires two players in the architecture who are mutually
ignorant about each other inputs, and it exploits this ignorance.  However, if the
architecture does not exhibit this -- that is, for every pair of players, one of the
players can ``infer'' the inputs to the other player, \Cref{10-prob:existWS} becomes
decidable.  This notion of "information ordering" between the players is formalised in
\Cref{10-sec:information-preorder}.  This is a syntactic condition on the architecture.

\medskip

In general, a strategy for player $p$ is an \emph{infinite} object:
a map $\sigma_{p}\colon(S_{\rdom(p)})^{*}\to S_{\wdom(p)}$. Notice that $\sigma_{p}$ is 
an $S_{\rdom(p)}$-directed $S_{\wdom(p)}$-labelled tree. It is also important to study 
the existence (and to construct) simpler strategies, for instance using finite memory 
(regular trees).

A \emph{memory structure} for player $p$ is a tuple $\memory_{p}=(\MM_{p},\mu_{p},\initmm_{p})$ 
where $\MM_{p}$ is the set of memory states, $\initmm_{p}\in\MM_{p}$ is the 
initial memory state, and $\mu_{p}\colon\MM_{p}\times S_{\rdom(p)}\to\MM_{p}$ is the 
(complete) update function. The memory structure is \emph{finite} if $\MM_{p}$ is finite.

We define inductively the memory state $\memory_{p}(u)$ 
reached after reading a sequence $u\in(S_{\rdom(p)})^{*}$: 
$\memory_{p}(\varepsilon)=\initmm_{p}$ and $\memory_{p}(ua)=\mu_{p}(\memory_{p}(u),a)$ 
for $u\in(S_{\rdom(p)})^{*}$ and $a\in S_{\rdom(p)}$.

A strategy for player $p$ using memory structure $\memory_{p}$ is a map
$\sigma_{p}\colon\MM_{p}\to S_{\wdom(p)}$. It induces a strategy 
$\widetilde{\sigma_{p}}\colon(S_{\rdom(p)})^{*}\to S_{\wdom(p)}$ defined by
$\widetilde{\sigma_{p}}(u)=\sigma_{p}(\memory_{p}(u))$ for $u\in(S_{\rdom(p)})^{*}$.
We abuse notation and simply write $\sigma_{p}$ instead of $\widetilde{\sigma_{p}}$.

A distributed strategy $\sigma=(\sigma_{p})_{p\in\players}$ is \emph{with finite memory}
if each local strategy $\sigma_{p}$ is defined using a \emph{finite} memory structure
$\memory_{p}$. 

\begin{problem}[Solving the game with finite memory]~
  \begin{description}   \label{10-prob:existWSfiniteMemory}
    \item[\textsc{input:}] A synchronous distributed game $\game$
    
    \item[\textsc{output:}] Existence of a winning distributed strategy
    $\sigma=(\sigma_{p})_{p\in\players}$ \emph{with finite memory} for the players.
  \end{description}
\end{problem}

\Cref{10-prob:existWSfiniteMemory} is also undecidable in general.

\begin{theorem}\label{10-thm:undecidable-P2}
  \Cref{10-prob:existWSfiniteMemory} is recursively enumerable but not co-recursively enumerable.
\end{theorem}

We prove a slightly stronger version (\Cref{10-thm:undecidable-P2bis}) in
\Cref{10-sec:undecidability-synchronous} where we restrict to distributed games
on the architecture $\archi_{1}$ of \Cref{10-fig:information-fork} and to winning
conditions given in linear temporal logic.

There are synchronous distributed games having a winning distributed strategy but no
winning distributed strategies with finite memory (\Cref{10-cor:WSwithInfiniteMem}).  But
when the answer to \Cref{10-prob:existWSfiniteMemory} is positive, we also wish to
construct a winning distributed strategy with finite memory.

\begin{problem}[Constructing a winning distributed strategy with finite memory]~
	\begin{description}      \label{10-prob:constructWS}
		\item[\textsc{input:}] A synchronous distributed game $\game$
    
		\item[\textsc{output:}] Construct finite memory structures $(\Mem_{p})_{p\in\players}$
		and a winning distributed strategy $\sigma=(\sigma_{p})_{p\in\players}$ for the
		players using these memory structures.
	\end{description}
\end{problem}

A first step in solving a distributed game $\game=(\arena,W)$ with arena
$\arena=(\archi,(S_{\ell})_{\ell\in\loc})$ over architecture
$\archi=(\loc\uplus\players,E)$ is to ask what would happen if the game was \emph{not
distributed}.  Intuitively, this means that all players in $\players$ are fused together
in a single player which writes to all output locations in
$Z=\wdom(\players)=\bigcup_{q\in\players}\wdom(q)$ and has immediate access to all
\emph{external} input locations in $Y=\loc\setminus\wdom(\players)=\wdom(\env)$.  A
\emph{global} strategy for this (imaginary) \emph{global} player is a map $\tau\colon
S_{Y}^{*}\to S_{Z}$.  We prove below in \Cref{10-lem:combined-strategies} that from a
\emph{distributed} strategy $\sigma=(\sigma_{q})_{q\in\players}$ we can construct a
corresponding \emph{global} strategy $\tau=\sigma_{\players}$ with the same set of
consistent plays.  Hence, the \emph{distributed} strategy $\sigma$ is winning for $W$ if
and only if the induced \emph{global} strategy $\sigma_{\players}$ is winning for the same
winning condition $W$.

The first step in solving the distributed game $\game$ is to check
whether there exists a \emph{global} strategy $\tau\colon S_{Y}^{*}\to S_{Z}$ which is
winning for $W$.  This amounts to solving a \emph{classic} two-player game.  If the answer
is negative, then we know that there are no winning \emph{distributed} strategies for
$\game$.  But in the positive case, when there are winning \emph{global} strategies for
$W$, we still do not have the answer to \Cref{10-prob:existWS}.  This is because not all
global strategies $\tau$ are \emph{distributable}, \textit{i.e.}, are induced by a
\emph{distributed} strategy.

\begin{example}\label{10-ex:distributable}
  We give several examples to illustrate the problems that arise when trying to distribute
  a global strategy.  These examples are given using the architecture $\archi$ depicted
  on~\Cref{10-fig:order3} and the arena $\arena=(\archi,(S_{u})_{u\in\loc})$, with
  $S_{u}=\mathbb{B}=\{0,1\}$ for all $u\in\loc$.  We simply write $S_{i}$ instead of
  $S_{u_{i}}$ and for a state $s\in S_{\loc}$, we simply write $s_{i}=s_{u_{i}}$ the
  component on location $u_{i}$.
  The winning conditions are specified by means of $\LTL$ formulas, in which the atomic
  propositions will be $\ap=\bigcup_{i=1}^{5}\ap_{i}$ with $\ap_{i}=\{0_{i},1_{i}\}$.  A
  state $s\in S_{\loc}$ satisfies an atomic proposition $b_{i}$ if $s_{i}=b$.  
  
  As first example, consider the simple specification $\varphi_1=\G(1_{0}\lequiv\X 1_{4})$
  that requires that each value written on location $u_0$ by the environment is copied
  with delay 1 on location $u_4$ by the players.  It is easy to see that there are
  \emph{global} strategies satisfying this winning condition.
  However, a distributed strategy induces a delay (at least) 2 between locations $u_{0}$ and
  $u_{4}$.  Hence, it is easy to check that there are no \emph{winning distributed} strategy
  for the game $\game=(\arena,\varphi_1)$.  The \emph{global} strategies
  that are \emph{winning} for $\varphi_{1}$ are not \emph{distributable}, and the reason in 
  this example is that a global strategy may react with a shorter delay than a distributed 
  strategy.
  
  Another example of specification illustrates when different requirements are solvable
  globally but become conflicting when the strategies have to be local.  Take the
  specification $\varphi_2=\G(1_{0}\lequiv\X\X 1_{4}) \wedge 
  \G((1_{0}\S 1_{1})\lequiv\X 1_{3})$.
  Again, a global strategy that copies the value of $u_0$ on $u_4$ with delay 2, and always
  makes $u_3$ correspond to the truth value of $1_{0} \S 1_{1}$ with delay 1 is winning.  It
  is easy to see that we may satisfy the first requirement $\varphi'_{2}=\G(1_{0}\lequiv\X\X
  1_{4})$ with a \emph{distributed} strategy, and we may also satisfy the second requirement
  $\varphi''_{2}=\G((1_{0}\S 1_{1})\lequiv\X 1_{3})$ with a \emph{distributed} strategy.
  But we cannot satisfy both simultaneously with a \emph{distributed} strategy.
  Indeed, let $\sigma=(\sigma_{p},\sigma_{q},\sigma_{r})$ be a distributed strategy and
  consider two plays $\pi,\pi'$ consistent with $\sigma$ and such that
  $\proj{\pi}{u_{0}}\in 0\mathbb{B}^{\omega}$, 
  $\proj{\pi}{u_{1}}\in 00\mathbb{B}^{\omega}$, 
  $\proj{\pi'}{u_{0}}\in 1\mathbb{B}^{\omega}$, 
  $\proj{\pi'}{u_{1}}\in 00\mathbb{B}^{\omega}$.
  If $\pi$ satisfies $\varphi''_{2}$ then we have
  $\proj{\pi}{u_{3}}\in b00\mathbb{B}^{\omega}$ for some $b\in\mathbb{B}$.
  If $\pi'$ satisfies $\varphi''_{2}$ then we also have 
  $\proj{\pi}{u_{3}}\in b00\mathbb{B}^{\omega}$ for the same $b\in\mathbb{B}$.
  Now, since $\pi,\pi'$ are consistent with $\sigma_{q}$, we deduce that the first three 
  bits of $\proj{\pi}{u_{4}}$ and $\proj{\pi'}{u_{4}}$ are equal, say $b_{1}b_{2}b_{3}$.
  Condition $\varphi'_{2}$ requires that $b_{3}$ be the first value written on $u_{0}$, 
  hence we cannot have $\pi,\pi'$ both satisfying $\varphi_{2}$.
  Again, this example shows that the winning global strategies for $\varphi_{2}$ are not 
  distributable.
\end{example}

The following lemma shows how to combine strategies $(\sigma_{q})_{q\in P}$ of a subset
$P$ of players into an equivalent $P$-global strategy $\sigma_{P}$ for an imaginary single
player corresponding to all players in $P$ being fused together.  It is important to
notice that the output locations of the combined $P$-global strategy is the union
$Z=\wdom(P)=\bigcup_{q\in P}\wdom(q)$ of all output locations of players in $P$, whereas the
input locations of $\sigma_{P}$ are restricted to the \emph{external} inputs of $P$ only,
\textit{i.e.}, $Y=\rdom(P)\setminus\wdom(P)$ where $\rdom(P)=\bigcup_{q\in P}\rdom(q)$.
A $P$-global strategy is a map $\tau\colon S_{Y}^{*}\to S_{Z}$. Taking delay 1 into 
account, we say that a sequence $\pi=s^{0}s^{1}s^{2}\cdots \in S_{Y\cup Z}^\infty$ is 
consistent with $\tau$ if 
$\proj{s^{n}}{Z}=\tau(\proj{(s^{0}\cdots s^{n-1})}{Y})$ for all $n\geq0$.

\begin{lemma}\label{10-lem:combined-strategies}
  Let $P\subseteq\players$ be a set of players and let $(\sigma_{q})_{q\in P}$ be local 
  strategies for the players in $P$.
  Let $Z=\wdom(P)$ and $Y=\rdom(P)\setminus\wdom(P)$.
  
  We define two maps $\widehat{\sigma_{P}}\colon S_{Y}^{\infty}\to S_{Y\cup Z}^{\infty}$ and
  $\sigma_{P}\colon S_{Y}^{*}\to S_{Z}$ such that for all sequences
  $\pi=s^{0}s^{1}s^{2}\cdots \in S_{Y\cup Z}^\infty$,
  $\pi$ is consistent with $(\sigma_{q})_{q\in P}$
  if and only if 
  $\pi=\widehat{\sigma_{P}}(\proj{\pi}{Y})$ 
  if and only if
  $\pi$ is consistent with $\sigma_{P}$.
  
  The map $\sigma_{P}$ is called the composition of $(\sigma_{q})_{q\in P}$ and is denoted
  $\bigoplus_{q\in P}\sigma_{q}$.  Moreover, if the strategies $(\sigma_{q})_{q\in P}$
  have finite memory, then $\bigoplus_{q\in P}\sigma_{q}$ also has finite memory.
\end{lemma}

\begin{proof}
  We define $\widehat{\sigma_{P}}$ by induction on the length of the input sequence.  We
  set $\widehat{\sigma_{P}}(\varepsilon)=\varepsilon$, and, for all $u\in S_{Y}^{*}$ and
  $a\in S_{Y}$, we define $\widehat{\sigma_{P}}(u\cdot a)
  =\widehat{\sigma_{P}}(u)\cdot(a,b)$ where $b\in S_{Z}$ is defined by
  $b_{\wdom(q)}=\sigma_{q}(\proj{\widehat{\sigma_{P}}(u)}{\rdom(q)})$ for all $q\in P$.
  This is well-defined since $\rdom(q)\subseteq Y\cup Z$.
  We extend $\widehat{\sigma_{P}}$ to infinite sequences $v\in S_{Y}^{\omega}$ by
  continuity: $\widehat{\sigma_{P}}(v)$ is the unique $\omega$-sequence $w\in S_{Y\cup
  Z}^{\omega}$ such that $\widehat{\sigma_{P}}(u)$ is a prefix of $w$ for all finite 
  prefixes $u$ of $v$. 
  
  It is easy to check that a sequence $\pi=s^{0}s^{1}s^{2}\cdots \in S_{Y\cup Z}^\infty$
  is consistent with $(\sigma_{q})_{q\in P}$ if and only if
  $\pi=\widehat{\sigma_{P}}(\proj{\pi}{Y})$.
  
  Then, for $u\in S_{Y}^{*}$, we define $\sigma_{P}(u)=b\in S_{Z}$ by
  $b_{\wdom(q)}=\sigma_{q}(\proj{\widehat{\sigma_{P}}(u)}{\rdom(q)})$ for all $q\in P$.
  Notice that for all $u\in S_{Y}^{*}$ and $a\in S_{Y}$ we have
  $\widehat{\sigma_{P}}(u\cdot a)=\widehat{\sigma_{P}}(u)\cdot(a,\sigma_{P}(u))$.
  Hence, it is easy to check that a sequence $\pi=s^{0}s^{1}s^{2}\cdots\in S_{Y\cup Z}^\infty$
  is consistent with $\sigma_{P}$ if and only if $\pi=\widehat{\sigma_{P}}(\proj{\pi}{Y})$.
  
  It remains to prove that, if for each $q\in P$ the strategy $\sigma_{q}$ uses a
  finite memory $\memory_{q}=(\MM_{q},\mu_{q},\initmm_{q})$, then the strategy
  $\sigma_{P}$ can also be defined using a finite memory structure
  $\memory_{P}=(\MM_{P},\mu_{P},\initmm_{P})$.  Let $\MM_{P}=\prod_{q\in P}\MM_{q}$ and
  $\initmm_{P}=(\initmm_{q})_{q\in P}$.  Before defining the update function $\mu_{P}$, we
  define a strategy $\sigma'_{P}\colon\MM_{P}\to S_{Z}$ using the memory structure
  $\memory_{P}$.  For $\mm=(\mm_{q})_{q\in P}\in\MM_{P}$ we define $\sigma'_{P}(\mm)=b\in
  S_{Z}$ by $b_{\wdom(q)}=\sigma_{q}(\mm_{q})\in S_{\wdom(q)}$ for all $q\in P$.  For
  $a\in S_{Y}$, the update function is defined by $\mu_{P}(\mm,a)=(\mm'_{q})_{q\in P}$
  where, for all $q\in P$ we have $\mm'_{q}=\mu_{q}(\mm_{q},\proj{a'}{\rdom(q)})$ 
  with $a'=(a,\sigma'_{P}(\mm))$.
  
  We show by induction that for all $u\in S_{Y}^{*}$ we have 
  $\memory_{P}(u)=(\memory_{q}(\proj{\widehat{\sigma_{P}}(u)}{\rdom(q)}))_{q\in P}$. This 
  is clear for $u=\varepsilon$. Assume that the property holds for $u\in S_{Y}^{*}$ 
  and let $a\in S_{Y}$. Let $\memory_{P}(u)=\mm=(\mm_{q})_{q\in P}$. By induction we know 
  that $\mm_{q}=\memory_{q}(\proj{\widehat{\sigma_{P}}(u)}{\rdom(q)})$. 
  Since $\sigma_{q}$ is defined using $\memory_{q}$ we get 
  $\sigma_{q}(\proj{\widehat{\sigma_{P}}(u)}{\rdom(q)})=\sigma_{q}(\mm_{q})$. From the 
  definitions of $\widehat{\sigma_{P}}$ and $\sigma'_{P}$ we get 
  $\widehat{\sigma_{P}}(ua)=\widehat{\sigma_{P}}(u)(a,\sigma'_{P}(\mm))$.
  Finally, using the definition of $\mu_{P}$ we obtain $\memory_{P}(ua)=\mu_{P}(\mm,a)
  =(\mu_{q}(\mm_{q},\proj{(a,\sigma'_{P}(\mm))}{\rdom(q)}))_{q\in P}
  =(\memory_{q}(\proj{\widehat{\sigma_{P}}(ua)}{\rdom(q)}))_{q\in P}$.
  
  It is now easy to see that $\sigma_{P}=\sigma'_{P}$ is defined using memory structure
  $\memory_{P}$. This concludes the proof of the lemma.
\end{proof}

\section{Information preorder}
\label{10-sec:information-preorder}

We want to compare processes based on how much information they have access to.
Since the strategies of the players can be assumed a common knowledge in the setting of
synchronous distributed games, if a player $p$ knows the values of all the locations in
$\rdom(q)$ then it can also infer all the values of the locations in $\wdom(q)$.  
Clearly, if $p$ knows or can infer the values of all locations in $\rdom(q)$ then we can
say that $p$ is more informed than $q$.  We formalise this intuition below.

Let $\archi=(\loc\uplus\players,E)$ be an architecture of a synchronous distributed game.
We are interested in the information that a player may get from the environment.  The
information may be sent along an \emph{information chain}, which is a sequence
$\bar{u}=u_{0}u_{1}\cdots u_{n}$ of locations starting at an external input location
$u_{0}\in\wdom(\env)$ written by the environment and such that for all $0<i\leq n$, we
have $u_{i-1} \mathrel{E}^{2} u_{i}$, \textit{i.e.}, location $u_{i-1}$ is read by the unique
player $p_{i}$ writing to location $u_{i}$.

\begin{figure}[tbp]
  \centering
  \includegraphics[page=4]{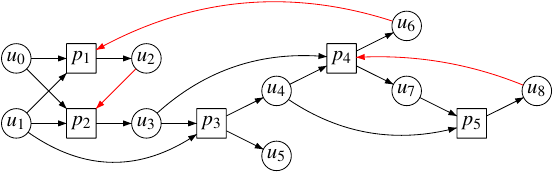}
  \caption{An architecture with a \emph{total} information preorder.}
  \label{10-fig:order1}
\commentAlt{Figure~\ref{10-fig:order1}: A directed graph with various nodes (circles and squares) and edges, some of which are highlighted in a distinct color, illustrating a network of connections. See long description.}
\commentLongAlt{Figure~\ref{10-fig:order1}: The image shows a directed graph with a mix of circular and square nodes. Circular nodes are labeled 'u0' through 'u8', and square nodes are labeled 'p1' through 'p5'.

The graph starts with 'u0' and 'u1'.

From 'u0', arrows point to 'p1' and 'p2'.
From 'u1', an arrow points to 'p2'.
From 'p1':

A red arrow points to 'u6'.
A black arrow points to 'u2'.
From 'p2':

A red arrow points to 'u2'.
A black arrow points to 'u3'.
Further connections:

From 'u3', an arrow points to 'p3'.
From 'u4', an arrow points to 'p4'.
From 'u7', an arrow points to 'p5'.
From 'u8', an arrow points to 'p5'.
Square nodes 'p3', 'p4', and 'p5' are central to the right side of the graph:

From 'p3', arrows point to 'u4' and 'u5'.
From 'p4', arrows point to 'u4' and 'u7'. A red arrow points from 'p4' to 'u8'.
'p5' receives inputs from 'u7' and 'u8'.
Additionally, there are long curved black arrows indicating connections:

From 'u1' to 'p3'.
From 'u3' to 'p4'.
From 'u4' to 'p5'.
From 'p1' to 'p4'.}
\end{figure}

As an example, consider the architecture depicted in \Cref{10-fig:order1}.   
The external inputs are the locations $\wdom(\env)=\{u_{0},u_{1}\}$.
We have the following information chains: the singleton sequences $u_{0}$ and $u_{1}$ are
the shortest ones, the sequences $u_{0}u_{2}$, $u_{1}u_{3}u_{6}u_{2}$, the sequence
$u_{1}u_{4}u_{7}u_{8}u_{6}u_{2}u_{3}u_{5}$ (the longest information chain without
repetition), etc.

Consider two players $p,q\in\players$.  We say that $q$ is \emph{more informed} than $p$,
written $p\preceq q$, if no information can be transmitted from the environment to $p$
without $q$ being aware of it.  Formally, $p\preceq q$ if for all information chains
$\bar{u}=u_{0}u_{1}\cdots u_{n}$ to $p$, \textit{i.e.}, with $u_{n}\in\rdom(p)$, we have
$\set{u_{0},\ldots,u_{n}}\cap\rdom(q)\neq\emptyset$.  It is easy to check that the
relation $\preceq$ is reflexive and transitive.  It is called the \emph{information
preorder}.

In general, we may have distinct players $p,q\in\players$ which are equally informed
($p\preceq q$ and $q\preceq p$), which is written $p\approx q$.  We may also have players
with incomparable information, \textit{i.e.}, $p\not\preceq q$ and $q\not\preceq p$.
Also, directly from the definition, we see that if some player $p$ cannot be reached by an 
information chain, then $p$ is least informed: $p\preceq q$ for all $q\in\players$.

We continue the above example with the architecture depicted in \Cref{10-fig:order1}.
Since player $p_{1}$ reads all external inputs ($\wdom(\env)\subseteq\rdom(p_{1})$), it is
clear that it is more informed than all other players: $p_{i}\preceq p_{1}$ for all $i$.
Actually, player $p_{2}$ also reads both external inputs, hence it is also more informed
than all other players.  In particular, we have $p_{1}\approx p_{2}$.  Now, $p_{3}$,
$p_{4}$ and $p_{5}$ are strictly less informed than $p_{1}$ or $p_{2}$.  Indeed, the
singleton sequence $u_{0}$ is an information chain to $p_{1}$ or $p_{2}$ and $u_{0}$ is
not read by the other players.  Player $p_{3}$ is strictly more informed than $p_{4}$.
Indeed, all information chains to $p_{4}$, either starts with $u_{1}$ which is read by
$p_{3}$, or go via $u_{3}$ which is also read by $p_{3}$.  Similarly, we can check that
$p_{4}$ is strictly more informed than $p_{5}$.

In the three architectures of \Cref{10-fig:information-fork}, the players $p$ and $q$
have incomparable information: $p\not\preceq q$ and $q\not\preceq p$.  For the third
architecture, we see that the information chain $u_{0}u_{1}u_{2}u_{3}$ to $p$ does not
intersect $\rdom(q)=\{v_{2}\}$ and similarly, the information chain $u_{0}u_{1}v_{2}$ to
$q$ does not intersect $\rdom(p)=\{u_{3}\}$.


We say that the information preorder is \emph{total} if for all players $p,q\in\players$ 
we have $p\preceq q$ or $q\preceq p$ (or both). For instance, the information preorder 
for the architecture in \Cref{10-fig:order1} is total. If the information preorder is not 
total, then the architecture contains players with incomparable information.


It will be convenient to use the terminology of \emph{cuts} in the graph $(\loc,E^{2})$ to
reason about the information preorder.  Let $X,Y\subseteq\loc$ be two subsets of
locations.  We say that $X$ \emph{cuts} $Y$ from the external inputs (or simply $X$ cuts
$Y$) if all paths in $(\loc,E^{2})$ from $\wdom(\env)$ to $Y$ intersect $X$.  The
information preorder can be reformulated as follows: for players $p,q\in\players$, we
have $p\preceq q$ if $\rdom(q)$ cuts $\rdom(p)$ (from the external inputs).

\section{Solving synchronous distributed games}
\label{10-sec:decidability-overview}

The main negative result is that, if the information preorder is not total, then
\Cref{10-prob:existWS,10-prob:existWSfiniteMemory} are undecidable.  The undecidability
proofs are given in \Cref{10-sec:undecidability-synchronous}.

\begin{theorem}\label{10-thm:sync-undec-general-case}
  \Cref{10-prob:existWS,10-prob:existWSfiniteMemory} are undecidable when the architecture
  of the synchronous distributed game contains two players $p,q\in\players$ with
  incomparable information ($p\not\preceq q$ and $q\not\preceq p$),
  even when the winning condition $W$ is specified in linear temporal logic.
\end{theorem}

On the other hand, the main positive result is that a synchronous distributed game can be
solved when its information preorder is \emph{total}.

\begin{theorem}\label{10-thm:ADG-decidability}
  Let $\game=(\arena,W)$ be a synchronous distributed game with a \emph{total}
  information preorder and a \emph{regular} winning condition.  Then, we can decide
  whether the players have a winning distributed strategy.  In the positive case, we can
  compute a winning distributed strategy with finite memory.
\end{theorem}

The proof of \Cref{10-thm:ADG-decidability} relies on two reduction techniques described
below and studied in
\Cref{10-sec:reduction-equally-informed,10-sec:reduction-most-informed}.  The algorithm
for solving a synchronous distributed game $\game$ with a total information preorder is
given in \Cref{10-sec:algo} (\Cref{10-algo:solve-SDG}).  
The complexity is studied in \Cref{10-sec:complexity}.

\medskip

There are several techniques which can be used to solve synchronous distributed games.  In
this chapter, we will mainly use reductions to simpler games having fewer players, the
goal being to eventually reduce the original distributed game to a classic two-player
game.  When this is possible, it remains to solve the two-player classic game.  All
reductions will be effective.  Moreover, from a winning (distributed)
strategy (with finite memory) of the reduced game we will be able to reconstruct a winning
distributed strategy (with finite memory) of the initial game.

We will describe two completely different reduction techniques which are based on the 
information preorder defined in \Cref{10-sec:information-preorder}.
The first technique, described in \Cref{10-sec:reduction-equally-informed}, will merge
equally informed players into a single player.  Starting from a synchronous distributed
game $\game$, we will define a \emph{quotient} game $\game_{\approx}$ having the same
locations as the original game and having a single player for each equivalence class of
players which are equally informed in $\game$.  The winning condition is unchanged.  If
$\game$ contains some equally informed players, the reduction is proper resulting in fewer
players.  From each distributed strategy $\sigma$ in $\game$ we show how to construct an
equivalent distributed strategy $\sigma'$ in the quotient $\game_{\approx}$ and
vice-versa, preserving finiteness of memory.  Here equivalent means that a sequence $\pi$
of global states is consistent with $\sigma$ if and only if it is consistent with
$\sigma'$. Hence, the games $\game$ and $\game_{\approx}$ are equi-solvable.

The second reduction, described in \Cref{10-sec:reduction-most-informed}, may be applied
when the game $\game$ contains a player $p\in\players$ which is more informed than all
other players in $\players'=\players\setminus\{p\}$.  In this case, we can construct a
\emph{simpler} game $\residue{\game}{p}$ with players $\players'$ and locations
$\loc'\subseteq\loc$.  The games $\game$ and $\residue{\game}{p}$ are equi-solvable: If a
distributed strategy $\sigma=(\sigma_{q})_{q\in\players}$ is winning in $\game$ then the
distributed strategy $\sigma'=(\sigma_{q})_{q\in\players'}$ is winning in
$\residue{\game}{p}$; and conversely, if a distributed strategy
$\sigma'=(\sigma_{q})_{q\in\players'}$ is winning in $\residue{\game}{p}$ then we can
construct a strategy $\sigma_{p}$ for player $p$ in $\game$ so that the distributed
strategy $\sigma=(\sigma_{q})_{q\in\players}$ is winning in $\game$.  Again the
constructions preserve finiteness of memory.

As a consequence of the two reductions described above, a synchronous distributed game
$\game$ is solvable when the information preorder is \emph{total}.  Indeed, in this case,
the first reduction, merging equally informed players, results in a game $\game_{\approx}$
where the information preorder is a linear order.  Let $p_{1}\succ\cdots\succ p_{k}$
be the players of $\game_{\approx}$ in decreasing information order.  Then, we apply the
second reduction repeatedly, removing at each step the most informed player and
constructing the winning condition of the simpler game.  At the end, we have a two-player
game ($p_{k}$ against $\env$), that we solve using classic methods.  In the positive case,
we get a strategy $\sigma_{k}$ with \emph{finite memory} for player $p_{k}$.  Then, we
iteratively reconstruct strategies with finite memories $\sigma_{k-1},\ldots,\sigma_{1}$
for players $p_{k-1},\ldots,p_{1}$.  Finally, $(\sigma_{i})_{1\leq i\leq k}$ is a winning
distributed strategy (with finite memory) for the game $\game_{\approx}$.  From the
equivalence of $\game$ and $\game_{\approx}$ we construct a winning distributed strategy
(with finite memory) $\sigma=(\sigma_{q})_{q\in\players}$ for the game $\game$.

\section{Merging equally informed players (Reduction 1)}
\label{10-sec:reduction-equally-informed}

Let $\archi=(\loc\uplus\players,E)$ be an architecture.  The goal is to transform the
information preorder into a partial order by merging players that are equally informed.

We define the information equivalence class of a player $p\in\players$ by
$[p]=\{q\in\players\mid q\approx p\}$, \textit{i.e.}, the set of players that are equally informed.
We also define $\da p=\{q\in\players\mid q\preceq p\}$ the set of players that are less or
equally informed than $p$.  Notice that $p\approx q$ iff $\da p = \da q$.

An architecture may contain redundant edges carrying information from some location $u$ to
some player $p$.  The edge $(u,p)\in E$ is redundant when $p$ is already aware of the
information at location $u$.  Formally, we say that $(u,p)\in E$ is a \emph{feedback} edge
from location $u$ to player $p$ if $u$ is written by a player $q$ and $p$ is more
informed than $q$, \textit{i.e.}, $u\in\wdom(\da p)=\bigcup_{q\preceq p}\wdom(q)$.  In
\Cref{10-fig:order1}, the feedback edges (in red) are going from right to left.

\begin{figure}[tbp]
  \centering
  \includegraphics[page=5]{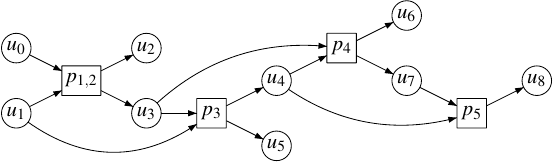}
  \caption{An architecture with total information order.}
  \label{10-fig:order2}
\commentAlt{Figure~\ref{10-fig:order2}: A directed graph with various nodes (circles and squares) and edges, illustrating a network of connections. See long description.}
\commentLongAlt{Figure~\ref{10-fig:order2}: The image shows a directed graph with a mix of circular and square nodes. Circular nodes are labeled 'u0' through 'u8', and square nodes are labeled 'p1,2', 'p3', 'p4', and 'p5'.

The graph starts with 'u0' and 'u1'.

From 'u0', an arrow points to 'p1,2'.
From 'u1', an arrow points to 'p1,2'.
From 'p1,2':

An arrow points to 'u2'.
An arrow points to 'u3'.
Further connections:

From 'u3', an arrow points to 'p3'.
From 'u4', an arrow points to 'p4'.
From 'u7', an arrow points to 'p5'.
From 'u8', an arrow points to 'p5'.
Square nodes 'p3', 'p4', and 'p5' are central to the right side of the graph:

From 'p3', arrows point to 'u4' and 'u5'.
From 'p4', arrows point to 'u6' and 'u7'.
'p5' receives inputs from 'u7' and 'u8'.
Additionally, there are long curved arrows indicating connections:

From 'u1' to 'p3'.
From 'u3' to 'p4'.
From 'u4' to 'p5'.
From 'p1,2' to 'p4'.}
\end{figure}

From $\archi$, we define the quotient architecture
$\archi_{\approx}=(\loc\uplus\players',E')$ by grouping together players that are equally
informed and removing all feedback edges.  Formally, a player in $\archi_{\approx}$
corresponds to a class of equally informed players of $\archi$:
$\players'=\players/{\approx}$.  The read and write domains in $\archi_{\approx}$ are
inherited from $\archi$ as follows.  For $p\in\players$, we let
\begin{align*}
  \wdom'([p]) &= \wdom([p])=\bigcup_{q\approx p}\wdom(q) \\
  \rdom'([p]) &= \rdom([p])\setminus\wdom(\da p)
  = \Big(\bigcup_{q\approx p}\rdom(q)\Big)\setminus\Big(\bigcup_{q\preceq p}\wdom(q)\Big) 
  \,.
\end{align*}
We write $\preceq'$ for the `more informed' relation in $\archi_{\approx}$: $[p]\preceq'[q]$ if 
$\rdom'([q])$ cuts $\rdom'([p])$.
\Cref{10-fig:order2} shows the quotient $\archi_{\approx}$ of the architecture $\archi$ of 
\Cref{10-fig:order1}.

The goal of this section is to prove the following theorem.

\begin{theorem}\label{10-thm:quotient}
  The architecture $\archi_{\approx}$ has no feedback edges and distinct players in
  $\archi_{\approx}$ are not equally informed.  Moreover, the architectures $\archi$ and
  $\archi_{\approx}$ are equivalent in the following sense: for all tuple
  $(S_{\ell})_{\ell\in\loc}$ of sets of local states,
  \begin{enumerate}
    \item Given a distributed strategy $\sigma=(\sigma_{p})_{p\in\players}$ over
    $\arena=(\archi,(S_{\ell})_{\ell\in\loc})$, we can construct a distributed strategy
    $\sigma'=(\sigma'_{c})_{c\in\players'}$ over
    $\arena_{\approx}=(\archi_{\approx},(S_{\ell})_{\ell\in\loc})$ such that for all sequences
    $\pi=s^{0}s^{1}s^{2}\cdots \in S^\infty$, we have $\pi$ consistent with $\sigma$ if
    and only if $\pi$ consistent with $\sigma'$.  
    Moreover, $\sigma'$ has finite memory when $\sigma$ has finite memory. 
  
    \item Given a distributed strategy $\sigma'=(\sigma'_{c})_{c\in\players'}$
    over $\arena_{\approx}$, we can construct a distributed strategy
    $\sigma=(\sigma_{p})_{p\in\players}$ over $\arena$ such that for all sequences
    $\pi=s^{0}s^{1}s^{2}\cdots \in S^\infty$, we have $\pi$ consistent with
    $\sigma$ if and only if $\pi$ consistent with $\sigma'$.
    
    Moreover, $\sigma$ has finite memory when $\sigma'$ has finite memory.
    More precisely, if $[p]=c$ and $\sigma'_{c}$ is defined using a finite memory
    structure $\Mem_{c}$ then $\sigma_{p}$ is defined using a memory structure $\Mem_{p}$
    having the same set of memory states as $\Mem_{c}$.
  \end{enumerate}
\end{theorem}

Before proving this theorem, we state its consequence for solving 
distributed games by merging equally informed players.

Let $\game=(\arena,W)$ be a synchronous distributed game over
$\arena=(\archi,(S_{\ell})_{\ell\in\loc})$.
Let $\archi_{\approx}=(\loc\uplus\players',E')$ be the architecture obtained from $\archi$
by merging equally informed players and removing feedback edges.  Let
$\arena_{\approx}=(\archi_{\approx},(S_{\ell})_{\ell\in\loc})$ be the corresponding arena.
Define $\game_{\approx}=(\arena_{\approx},W)$ to be the corresponding merged game.

\begin{corollary}\label{10-cor:quotient}
  The games $\game$ and $\game_{\approx}$ are equi-solvable. 
  
  More precisely, if $(\sigma_{q})_{q\in\players}$ is a winning distributed strategy in
  $\game$ then we can construct a winning distributed strategy
  $(\sigma_{c})_{c\in\players'}$ in $\game_{\approx}$.  Conversely, if
  $(\sigma_{c})_{c\in\players'}$ is a winning distributed strategy in $\game_{\approx}$
  then we can construct a winning distributed strategy
  $(\sigma_{q})_{q\in\players}$ in $\game$. Both constructions preserve finiteness of 
  memories.
\end{corollary}

\begin{remark}\label{10-rem:players-unreachable}
  Assume that architecture $\archi$ contains some player $p$ which cannot be reached by an
  information chain.  Then, the equivalence class $[p]$ is precisely the set of players
  that cannot be reached by an information chain.  We have $\da p=[p]$.  Moreover, each
  edge $(u,p')\in E$ with $p'\approx p$ is actually a feedback edge.  Therefore, in the
  quotient architecture $\archi_{\approx}$, player $[p]$ is \emph{the} least informed
  player and we have $\rdom'([p])=\emptyset$ and $\wdom'([p])=\wdom([p])$.
\end{remark}

The rest of this section is devoted to proving \Cref{10-thm:quotient}.
We start with three lemmas relating cuts and sets of players.
We say that $X\subseteq\loc$ cuts a location $u\in\loc$ if $X$ cuts $\{u\}$.  
We say that $X$ cuts a player $q\in\players$ if $X$ cuts $\rdom(q)$, and $X$ cuts a subset
$P\subseteq\players$ if $X$ cuts each player $q\in P$, \textit{i.e.}, $X$ cuts
$\rdom(P)=\bigcup_{q\in P}\rdom(q)$.

\begin{lemma}\label{10-lem:cut1}
  Let $P\subseteq\players$ be a set of players. Then, $\rdom(P)\setminus\wdom(P)$ cuts $P$.
\end{lemma}

\begin{proof}
  We show that if $X\subseteq\loc$ cuts $P\subseteq\players$, then $X\setminus\wdom(P)$ cuts $P$.
  The lemma follows since $X=\rdom(P)$ cuts $P$.
  
  Assume that $Y=X\setminus\wdom(P)$ does not cut $P$.  We find an information chain
  $\bar{u}=u_{0}u_{1}\cdots u_{n}$ with $u_{n}\in\rdom(P)$ which does not intersect $Y$.
  Since $X$ cuts $P$, the chain intersects $X$.  Take $i$ minimal with $u_{i}\in X$.
  Since $u_{i}\notin Y$ we get $u_{i}\in\wdom(P)$.  Since $u_{0}\in\wdom(\env)$ is an
  external input, we deduce $i>0$.  Therefore, $u_{i-1}\in\rdom(P)$.  Since $X$ cuts $P$,
  this is a contradiction with the minimality of $i$.
\end{proof}

\begin{lemma}\label{10-lem:cut2}
  Let $X\subseteq\loc$ and $P\subseteq\players$ be \emph{the} set of players cut by $X$.
  Then, $\rdom(P)\subseteq X\cup\wdom(P)$.
\end{lemma}

\begin{proof}
  Let $u\in\rdom(P)\setminus X$.  We have to show $u\in\wdom(P)$.  Since $X$ cuts
  $P$ and $u\in\rdom(P)\setminus X$, we deduce that $u$ is not an information chain, \textit{i.e.},
  $u\notin\wdom(\env)=\loc\setminus\wdom(\players)$. Let $q\in\players$ with 
  $u\in\wdom(q)$. Since $u\in\rdom(P)$ and $X$ cuts $P$, we deduce that $X$ cuts $u$. 
  Now, using $u\in\wdom(q)\setminus X$, we deduce that $X$ cuts $\rdom(q)$.
  Therefore, $q\in P$ and $u\in\wdom(P)$.
\end{proof}

\begin{lemma}\label{10-lem:intersect-cut}
  Let $P\subseteq\players$ be a nonempty set of equally informed players.
  Let $X=\bigcap_{q\in P}\rdom(q)$. Then, $X$ cuts $P$.
\end{lemma}

\begin{proof}
  If $X$ does not cut $P$ then there are information chains to $\rdom(P)=\bigcup_{q\in
  P}\rdom(q)$ which are not cut by $X$.  Take a minimal information chain
  $\bar{u}=u_{0}u_{1}\cdots u_{n}$ which is not cut by $X$ and with $u_{n}\in \rdom(P)$.
  Let $q\in P$ with $u_{n}\in\rdom(q)$.  Since $\bar{u}$ is not cut by $X=\bigcap_{p\in
  P}\rdom(p)$, we find $p\in P$ such that $u_{n}\notin\rdom(p)$.
  Since $q\approx p$, we know that $\rdom(p)$ cuts the chain $\bar{u}$. We get 
  $u_{i}\in\rdom(p)\subseteq \rdom(P)$ for some $i<n$. This is a contradiction with the 
  minimality of the chain $\bar{u}$.
\end{proof}

We prove now the first part of \Cref{10-thm:quotient}. We also establish important 
properties of the read domains and the information order in the quotient architecture 
$\archi_{\approx}$.

\begin{lemma}\label{10-lem:quotient}
  The architecture $\archi_{\approx}$ has no feedback edges and distinct players in $\archi_{\approx}$ are
  not equally informed.  Moreover, for all players $p,q\in\players$, we have
  \begin{enumerate}
    \item  $\rdom'([p]) = \rdom([p])\setminus\wdom(\da p)
    =\rdom(\da p)\setminus\wdom(\da p)
    =\big(\bigcap_{q\approx p}\rdom(q)\big)\setminus\wdom(\da p)$.
  
    \item $p\preceq q$ if and only if $[p]\preceq'[q]$.
  \end{enumerate}
\end{lemma}

\begin{proof}
  1.  Let $p\in\players$ be a player of $\archi$.  By \Cref{10-lem:intersect-cut} we
  know that $X=\bigcap_{q\approx p}\rdom(q)$ cuts $[p]$.  
  We show that $P=\da p$ is the set of players cut by $X$.  Indeed, if $q\preceq p$ then
  $\rdom(q)$ is cut by $\rdom(p)$, which is cut by $X$.  Hence, $X$ cuts $P=\da p$.
  Conversely, if $X$ cuts some player $q$, using $X\subseteq\rdom(p)$ we deduce that
  $\rdom(p)$ cuts $q$ and $q\preceq p$, \textit{i.e.}, $q\in\da p=P$.
  By \Cref{10-lem:cut2} we deduce that $\rdom(P)\subseteq X\cup\wdom(P)$,
  \textit{i.e.}, $\rdom(\da p)\subseteq (\bigcap_{q\approx p}\rdom(q))\cup\wdom(\da p)$. 
  Therefore, $\rdom'([p])=\rdom([p])\setminus\wdom(\da p)
  \subseteq \rdom(\da p)\setminus\wdom(\da p)
  \subseteq (\bigcap_{q\approx p}\rdom(q))\setminus\wdom(\da p)
  \subseteq \rdom([p])\setminus\wdom(\da p) = \rdom'([p])$.

  2.  Let $p,q\in\players$.  Assume that $p\preceq q$.  Then, $[p]\subseteq\da q$.
  Applying \Cref{10-lem:cut1} with $P=\da q$, we know that $\rdom'([q])=\rdom(\da
  q)\setminus\wdom(\da q)$ cuts $\da q$.  Using $[p]\subseteq\da q$, we deduce that
  $\rdom'([q])$ also cuts $\rdom([p])$.  
  We deduce that $\rdom'([q])$ cuts $\rdom'([p])\subseteq\rdom([p])$, \textit{i.e.},
  $[p]\preceq'[q]$.
  
  Conversely, assume that $[p]\preceq'[q]$.  Then, $\rdom'([q])$ cuts $\rdom'([p])$.
  Applying \Cref{10-lem:cut1} with $P=\da p$, we know that $\rdom'([p])=\rdom(\da
  p)\setminus\wdom(\da p)$ cuts $\da p$.  Therefore, $\rdom'([q])$ cuts $\da p$.  Since
  $\rdom'([q])\subseteq\rdom(q)$ (by 1) and $p\in\da p$, we deduce that $\rdom(q)$ cuts $p$,
  \textit{i.e.}, $p\preceq q$.
  
  \medskip From 2.  it follows that $[p]\approx'[q]$ iff $p\approx q$ iff $[p]=[q]$, \textit{i.e.},
  distinct players in $\archi_{\approx}$ are not equally informed. Using 2. again, we get 
  $$
  \wdom'(\da'[p])=\bigcup_{[q]\preceq'[p]}\wdom'([q])
  =\bigcup_{q\preceq p}\wdom'([q])  =\bigcup_{q\preceq p}\wdom(q) = \wdom(\da p) \,.
  $$
  Therefore, $\rdom'([p])=\rdom([p])\setminus\wdom(\da p)$ is disjoint from 
  $\wdom'(\da'[p])$ and $\archi_{\approx}$ does not have feedback edges.
\end{proof}

\begin{proof}[Proof of \Cref{10-thm:quotient}]
  The first part was already proved in \Cref{10-lem:quotient}. We prove now that the 
  architecture $\archi$ and its quotient $\archi_{\approx}$ are equivalent.
  This relies on \Cref{10-lem:combined-strategies}.
  
  1.  Let $\sigma=(\sigma_{p})_{p\in\players}$ be a distributed strategy of $\archi$.
  Consider a player $c=[p]$ of $\archi_{\approx}$.  Let $P=\da p$, $Z=\wdom(P)$ and
  $Y=\rdom(P)\setminus\wdom(P)=\rdom'(c)$.  Let $\sigma_{P}\colon S_{Y}^{*}\to S_{Z}$ be
  the map constructed in \Cref{10-lem:combined-strategies} from $(\sigma_{q})_{q\in P}$.
  We define $\sigma'_{c}$ as the projection of $\sigma_{P}$ to
  $\wdom'(c)=\wdom([p])\subseteq Z$.  Formally, for each $u\in
  S_{Y}^{*}=S_{\rdom'(c)}^{*}$ we set $\sigma'_{c}(u)=\proj{\sigma_{P}(u)}{\wdom'(c)}$.
  Notice that if the distributed strategy $\sigma$ has finite memory, then by 
  \Cref{10-lem:combined-strategies} the strategy $\sigma_{P}$ has finite memory, hence 
  the strategy $\sigma'_{c}$ also has finite memory.
  
  Let $\pi\in S^\infty$ be a sequence of global states consistent with $\sigma$.  By
  \Cref{10-lem:combined-strategies}, $\pi$ is consistent with each $\sigma_{\da p}$ for
  $p\in\players$.  Hence, $\pi$ is also consistent with each $\sigma'_{c}$ for
  $c\in\players$.

  Conversely, let $\pi'\in S^\infty$ be a sequence of global states consistent with $\sigma'$.  
  Consider the \emph{unique} sequence $\pi$ of global states consistent with $\sigma$ and
  which coincide with $\pi'$ on the external input locations, \textit{i.e.}, such that
  $\proj{\pi}{\wdom(\env)}=\proj{\pi'}{\wdom(\env)}$.  We have seen above that $\pi$ is
  consistent with $\sigma'$.  We deduce that $\pi=\pi'$ (unicity of a sequence of global
  states consistent with a distributed strategy when the external input sequence is
  fixed).  Hence, $\pi'$ is consistent with $\sigma$.
  
  2.  Let $\sigma'=(\sigma'_{c})_{c\in\players'}$ be a distributed strategy of
  $\archi_{\approx}$.  Consider a player $p\in\players$ of $\archi$ and let $c=[p]$.
  By definition, we have $\wdom(p)\subseteq\wdom'(c)$ and by \Cref{10-lem:quotient} we 
  know that  $\rdom'(c)\subseteq\rdom(p)$.
  We define $\sigma_{p}$ from $\sigma'_{c}$ by widening the input from $\rdom'(c)$ to
  $\rdom(p)$ and projecting the output from $\wdom'(c)$ to $\wdom(p)$.  More precisely,
  for each $u\in S_{\rdom(p)}^{*}$ we define
  $\sigma_{p}(u)=\proj{\sigma'_{c}(\proj{u}{\rdom'(c)})}{\wdom(p)}$.
  
  From the definition of the distributed strategy $\sigma=(\sigma_{p})_{p\in\players}$, 
  it is clear that any sequence $\pi\in S^\infty$ consistent with $\sigma'$ is also 
  consistent with $\sigma$.
  The converse is shown as above. Starting from a sequence $\pi\in S^\infty$ consistent 
  with $\sigma$, we consider the unique sequence $\pi'$ consistent with $\sigma'$ and 
  which coincides with $\pi$ on external inputs. Since $\pi'$ is consistent with $\sigma$ 
  we conclude that $\pi=\pi'$.

  Assume that the local strategy $\sigma'_{c}$ is defined using a finite memory
  structure $\Mem_{c}=(\MM_{c},\mu_{c},\initmm_{c})$.  We have $\mu_{c}\colon\MM\times
  S_{\rdom'(c)}\to\MM$ and $\sigma'_{c}\colon\MM_{c}\to S_{\wdom'(c)}$.  We show that
  $\sigma_{p}$ is defined using the memory structure
  $\Mem_{p}=(\MM_{c},\mu_{p},\initmm_{c})$ using the same memory states as $\Mem_{c}$.
  The update function $\mu_{p}$ implements the widening of the input from $\rdom'(c)$ to
  $\rdom(p)$: $\mu_{p}(\mm,a)=\mu_{c}(\mm,\proj{a}{\rdom'(c)})$ for all $\mm\in\MM$ and
  $a\in S_{\rdom(p)}$.  Then, we define
  $\sigma_{p}(\mm)=\proj{\sigma'_{c}(\mm)}{\wdom(p)}$.
\end{proof}

\section{Removing the most informed player (Reduction 2)}
\label{10-sec:reduction-most-informed}

\begin{figure}
	\centering
	\includegraphics[page=10]{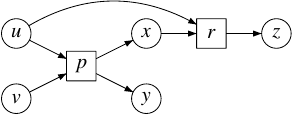}
	\hfil
	\includegraphics[page=11]{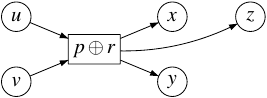}
	\caption{(left) The architecture $\archipr$ with two players $p$ and $r$, where $p$ is
	more informed than $r$.  (right) Architecture $\fusedarchipr$ where $p$ and $r$ are
	fused into a single player $\merge{p}{r}$.}
  \label{10-fig:archi-distr-strat}
\commentAlt{Figure~\ref{10-fig:archi-distr-strat}: Two diagrams illustrating a transformation from a detailed network of nodes and transitions to a more abstract, combined representation. See long description.}
\commentLongAlt{Figure~\ref{10-fig:archi-distr-strat}: The image displays two distinct network diagrams, implying a transformation or simplification from the left to the right.

Left Diagram:

Two circular nodes, 'u' and 'v', point to a square node labeled 'p'.
From 'p', two arrows emanate: one to a circular node 'x' and another to a circular node 'y'.
From 'x', an arrow points to a square node labeled 'r'.
From 'r', an arrow points to a circular node 'z'.
A curved arrow connects 'u' directly to 'r'.
Right Diagram:

Two circular nodes, 'u' and 'v', point to a combined square node labeled 'p (+) r'.
From this combined node, two arrows emanate: one to a circular node 'x' and another to a circular node 'y'.
A curved arrow connects the 'p (+) r' node directly to the circular node 'z'.
The right diagram appears to be a consolidated version of the left, where 'p' and 'r' are combined into a single node, and the path through 'x' to 'r' then 'z' is simplified to a direct connection from the combined node to 'z'.}
\end{figure}

As running example throughout this section, we consider the architecture $\archipr$ with
two players $p$ and $r$ depicted in Figure~\ref{10-fig:archi-distr-strat} (left) and the
fused architecture $\fusedarchipr$ on the right.  We wish to solve a \emph{distributed}
game $\game=(\arena,W)$ with arena $\arena=(\archipr,(S_{\ell})_{\ell\in\loc})$.  The induced
\emph{global} game is $\game^{f}=(\arena^{f},W)$ with $\arena^{f}=(\archiprf,(S_{\ell})_{\ell\in\loc})$.
By \Cref{10-lem:combined-strategies}, we know that if there is a winning
\emph{distributed} strategy $(\sigma_{p},\sigma_{r})$ for players $p$ and $r$ in $\game$,
then there is a \emph{global} strategy $\sigma_{p}\oplus\sigma_{r}$ for player
$\merge{p}{r}$ in $\game^{f}$ which is also winning.  However the converse is not true
always: we may have (winning) global strategies $\tau$ that are not \emph{distributable}, 
\textit{i.e.}, which are not of the form $\sigma_{p}\oplus\sigma_{r}$. In this
section, we provide a constructive procedure to determine whether there is a winning
\emph{distributed} strategy for $p$ and $r$ in $\game$, given the winning strategies for
$\merge{p}{r}$ in $\game^{f}$.

\medskip

Consider an arena $\arena=(\archi,(S_{\ell})_{\ell\in\loc})$ over an architecture
$\archi=(\loc\uplus\players,E)$ with a player $p\in\players$ strictly more 
informed than all other
players in $\players'=\players\setminus\{p\}$: $q\prec p$ for all $q\in\players'$.
Let $\game=(\arena,W)$ with a regular winning condition $W\subseteq S_{\loc}^{\omega}$.  
We consider the simpler architecture $\archi'=(\loc'\uplus\players',E')$ obtained from
$\archi$ by removing player $p$ and the locations that are \emph{only} connected to $p$ in
$\archi$.  The goal in this section is to construct an equi-solvable game $\game'$ over the
simpler arena $\arena'=(\archi',(S_{\ell})_{\ell\in\loc'})$.
The difficulty in this reduction is to construct a suitable winning condition for the game
$\game'$.  This requires a \emph{generalisation} of the winning condition from 
a regular \emph{word} language $W\subseteq S_{\loc}^{\omega}$ to a regular 
\emph{tree} language $T$ describing directly the winning \emph{global} strategies of the 
players.

We view strategies as trees: a strategy $\sigma_{q}\colon S_{\rdom(q)}^{*}\to
S_{\wdom(q)}$ for player $q$ is a full $(S_{\rdom(q)},S_{\wdom(q)})$-tree with directions
from $S_{\rdom(q)}$ and labels from $S_{\wdom(q)}$.  Similarly, a \emph{global} strategy
$\tau\colon S_{\wdom(\env)}^{*}\to S_{\wdom(\players)}$ for all the players 
fused together is a $ S_{\wdom(\env)}$-directed $S_{\wdom(\players)}$-labelled tree.
A branch of $\tau$ is an infinite word
$B=d_{0}d_{1}d_{2}\cdots\in S_{\wdom(\env)}^{\omega}$. 
Taking delay 1 into account, it induces an infinite play
$\overline{B}=(d_{0},\ell_{0})(d_{1},\ell_{1})(d_{2},\ell_{2})\cdots
\in(S_{\wdom(\env)}\times S_{\wdom(\players)})^{\omega}=S_{\loc}^{\omega}$ of $\arena$
with $\ell_{i}=\tau(d_{0}\cdots d_{i-1})$ for all $i\geq0$.
Notice that $\overline{B}$ is the unique infinite play of $\arena$ which is consistent
with $\tau$ and such that $\proj{\overline{B}}{\wdom(\env)}=B$.  
Therefore, a \emph{global} strategy $\tau$ is winning for a regular word
language $W\subseteq S^{\omega}$ if and only if $\tau$ is in the regular tree
language $T=\Branch(W)$, which is the set of trees $t$ such that $\overline{B}\in W$ for 
all branches $B$ of $\tau$ (\Cref{10-thm:words2trees}).

Given a distributed strategy $\sigma=(\sigma_{q})_{q\in\players}$ over $\arena$,
\Cref{10-lem:combined-strategies} allows to construct a global strategy
$\sigma_{\players}=\bigoplus_{q\in\players}\sigma_{q}$ such that an infinite play $\pi$ of
$\arena$ is consistent with $\sigma$ if and only if it is consistent with
$\sigma_{\players}$.  Therefore, the distributed strategy $\sigma$ is \emph{winning} for
$W$ if and only if $\sigma_{\players}$ is winning for $W$ if and only if
$\sigma_{\players}\in T=\Branch(W)$.

Hence, we assume in this section that the winning condition of $\game$ is given by a
regular tree language $T$, defining the winning \emph{global} strategies.  This means that
a distributed strategy $\sigma=(\sigma_{q})_{q\in\players}$ is winning if the induced
global strategy $\sigma_{\players}=\bigoplus_{q\in\players}\sigma_{q}\in T$.
We will construct a regular tree language $T'$
so that players in
$\players$ have a winning distributed strategy in $\game=(\arena,T)$ if and only if
players in $\players'$ have a winning distributed strategy in the simpler game
$\game'=(\arena',T')$.
This reduction incurs an exponential blow-up.  If the regular tree language $T$ is given
by a \emph{non-deterministic} parity tree automaton (NTA) $\nta$, we will first construct
an \emph{alternating} parity tree automaton (ATA) $\ata$ accepting the tree language 
$T'$, and then we obtain a language-equivalent NTA $\nta'$ of exponential size
by \Cref{10-thm:ata2nta}.

\medskip

\begin{figure}[t!]
  \centering
  \includegraphics[page=17]{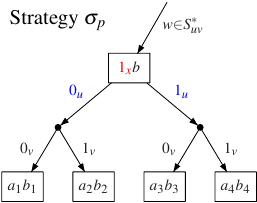}
  \hfil
  \includegraphics[page=18]{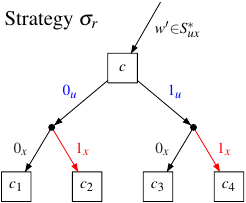}
  \\[5mm]
  \includegraphics[page=19]{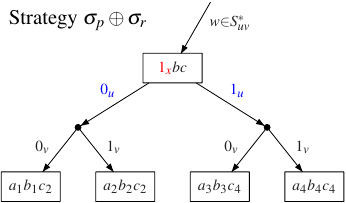}
  \caption{Composition of strategies on the architectures of \Cref{10-fig:archi-distr-strat}. 
  \\
  We assume that $S_{u}=\{0_{u},1_{u}\}$, 
  $S_{v}=\{0_{v},1_{v}\}$, $S_{x}=\{0_{x},1_{x}\}$, etc. 
  \\
  (top-left) Extract from an $S_{uv}$-directed $S_{xy}$-labelled tree (strategy)
  $\sigma_{p}$.  The top node is $w\in S_{uv}^{*}$, \textit{i.e.}, it is reached from the root
  following the sequence of directions (branch) $w$.  Since directions are pairs in
  $S_{uv}=S_{u}\times S_{v}$, we depict them in two steps: first branching with respect to
  $u$ and then to $v$.  Hence, the $0_{u}1_{v}$-child is the node labelled
  $a_{2}b_{2}$.  \\
  (top-right) Extract from an $S_{ux}$-directed $S_{z}$-labelled tree (strategy) $\sigma_{r}$.  
  \\
  (bottom) Extract from the $S_{uv}$-directed $S_{xyz}$-labelled tree (strategy)
  $\sigma_{p}\oplus\sigma_{r}$.  
  \\
  The top nodes $w$ in $\sigma_{p}$ and $w'$ in $\sigma_{r}$ correspond in the composition
  $\sigma_{p}\oplus\sigma_{r}$ when $w'=\proj{\widehat{\sigma}_{p}(w)}{u,x}$ (recall from
  the proof of \Cref{10-lem:combined-strategies} that $\widehat{\sigma}_{p}$ is defined
  inductively by $\widehat{\sigma}_{p}(\varepsilon)=\varepsilon$ and
  $\widehat{\sigma}_{p}(w\cdot
  d_{uv})=\widehat{\sigma}_{p}(w)\cdot(d_{uv},\sigma_{p}(w))$).  Now, since the $x$-value
  of $\sigma_{p}(w)$ is $\textcolor{red}{1_{x}}$, only the children
  $\textcolor{blue}{0_{u}}\textcolor{red}{1_{x}}$ and
  $\textcolor{blue}{1_{u}}\textcolor{red}{1_{x}}$ of $\sigma_{r}$ are used in the
  composition $\sigma_{p}\oplus\sigma_{r}$.  This is why the $z$-value is $c_{2}$ in both
  children $\textcolor{blue}{0_{u}}0_{v}$ and
  $\textcolor{blue}{0_{u}}1_{v}$ in $\sigma_{p}\oplus\sigma_{r}$.
  \\
  Finally, notice that $\sigma_{p}$ is the projection on $S_{xy}$ of 
  $\sigma_{p}\oplus\sigma_{r}$, but $\sigma_{r}$ cannot be recovered from the composition 
  since some of its branches are insignificant and not represented.
  }
  \label{10-fig:strategies}
\end{figure}

Coming back to our running example, given a non-deterministic tree automaton (NTA) $\nta$
recognising the winning global strategies $\tau$ of the fused player $\merge{p}{r}$, we
construct an alternating tree automaton (ATA) $\ata$ describing the strategies
$\sigma_{r}$ of player $r$ (less informed) which are \emph{factors} of the
\emph{distributable} winning strategies $\tau$ of $\merge{p}{r}$, \textit{i.e.}, such that
$\tau=\sigma_{p}\oplus\sigma_{r}$ for some strategy $\sigma_{p}$ of player $p$.
Before we go into the construction of the ATA $\ata$, it is very important to 
fully understand the composition of strategies (\Cref{10-lem:combined-strategies}).
This is explained and illustrated in \Cref{10-fig:strategies}.

\paragraph*{Distributing a strategy} 

Consider an architecture $\archi=(\loc\uplus\players,E)$ with no feedback edges and with a
player $p\in\players$ strictly more informed than all other players in
$R=\players\setminus\{p\}$.  Since $p$ is the most informed player and $\archi$ has no
feedback edges, we have $\rdom(p)=\wdom(\env)=\loc\setminus\wdom(\players)$.
Considering players in $R$ to be fused together, we let 
$\rdom(R)=\bigcup_{q\in R}\rdom(q)$ and $\wdom(R)=\bigcup_{q\in R}\wdom(q)$.
Let the locations that are read by both $p$ and $R$ be $U=\rdom(p)\cap\rdom(R)$, and the
locations read by $p$ only be $V=\rdom(p)\setminus\rdom(R)$.  Player $p$ may write to some
locations $X=\wdom(p)\cap\rdom(R)$ that are read by $R$ and to
other locations $Y=\wdom(p)\setminus\rdom(R)$ that are not read by $R$.  Finally player
$R$ writes to locations $Z=\wdom(R)$.  
We have $\rdom(p)=U\uplus V=\wdom(\env)$, $\wdom(p)=X\uplus Y$ and 
$\wdom(\players)=X\uplus Y\uplus Z$. 
Moreover, $\rdom(R)\setminus\wdom(R)=U\uplus X$.
Such an architecture ($\archipr$) is illustrated in Figure~\ref{10-fig:archi-distr-strat}
with a single player $r$ in $R$ and only one location in each set.  

Let $(S_{\ell})_{\ell\in\loc}$ be sets of local states for locations in 
$\loc=U\uplus V\uplus X\uplus Y\uplus Z$. The architecture $\archi$ induces the arena
$\arena=(\archi,(S_{\ell})_{\ell\in\loc})$.
To simplify notation, we often write $S_{UV}$ for $S_{U\cup V}=S_{U}\times S_{V}
=S_{\wdom(\env)}$, and $S_{XYZ}$ for 
$S_{X\cup Y\cup Z}=S_{X}\times S_{Y}\times S_{Z}=S_{\wdom(\players)}$, etc.
We may also omit parentheses and commas in tuples: we may write $ab$ instead of $(a,b)$, 
and $abc$ instead of $(a,b,c)$, etc. 

Let $\tau\colon S_{UV}^\ast \to S_{XYZ}$ be a \emph{global} strategy in $\arena$.
We say that $\tau$ is $p$-\emph{distributable} if $\tau=\sigma_p\oplus\sigma_R$ 
for some strategies $\sigma_p\colon S_{UV}^\ast\to S_{XY}$ and 
$\sigma_R\colon S_{UX}^\ast\to S_{Z}$ of players $p$ and $R$, respectively.
In this case, we say that $\sigma_R$ is a $p$-residue of $\tau$.
Notice that if $\tau=\sigma_p\oplus\sigma_R$ then $\sigma_p=\proj{\tau}{S_{XY}}$ is the 
projection of $\tau$ on $S_{XY}$.

Let $T$ be a set of full $(S_{UV},S_{XYZ})$-trees, \textit{i.e.}, global strategies in $\arena$.
All strategies in $T$ need not be $p$-distributable.  We denote by $\residue{T}{p}$ the
set of strategies of player $R$ that are $p$-residues of some strategies from $T$.  That
is,
$$
\residue{T}{p}= \{(S_{UX},S_{Z})\text{-trees } \sigma_R \mid
\sigma_p\oplus\sigma_R\in T \text{ for some } (S_{UV},S_{XY})\text{-tree } \sigma_{p} \} \,.
$$

\begin{example}\label{10-ex:phi3}
  We continue with the arena defined in \Cref{10-ex:distributable} on the architecture
  depicted in \Cref{10-fig:order3}.  Several global strategies are winning for the
  specification:
  $$
  \varphi_3 = \G (1_{0}\lequiv \X\X 1_{4}) \wedge \G(1_{2}\lequiv \X\X1_{5}) \wedge
  \G((1_{2} \S \Y 1_{0}) \lequiv 1_{3}) \wedge \G(1_{0}\limplies \F 1_{2}) 
  $$
  but not all winning global strategies are $p$-distributable.
  Consider for instance the global strategy $\tau$ that always writes 1 to locations
  $u_{2}$ and $u_{5}$ to satisfy the second and last constraints, copies with delay 2 the
  value of $u_{0}$ to $u_{4}$ for the first constraint, and writes $1$ to location $u_{3}$
  if and only if there was a $1$ in the past on location $u_{0}$ so that the third
  constraint is also satisfied.  This global strategy is winning but not distributable: it
  does not allow $u_3$ to transmit the sequence of values from $u_0$ to $u_4$.
  Formally, a $p$-residue $\sigma_R\colon S_{u_1u_3}^{*}\to S_{u_4u_5}$ of $\tau$ 
  should satisfy $\tau=\sigma_p\oplus\sigma_R$ where $\sigma_p$ is the projection of
  $\tau$ on the locations written by player $p$.
  This means that for all $w\in S_{\rdom(p)}^*$, we have $\tau(w)_{u_4}
  =(\sigma_p\oplus\sigma_R)(w)_{u_4}
  =\sigma_R(\proj{\widehat{\sigma_p}(w)}{u_1u_3})_{u_4}$.
  Consider $w,w'\in S^*_{\rdom(p)}$ with $\proj{w}{u_0}=100$ and $\proj{w'}{u_0} = 110$, and
  $\proj{w}{u_1}=\proj{w'}{u_1}$.  Since $\sigma_{p}$ is the projection of $\tau$ on 
  $S_{\wdom(p)}$ we get $\widehat{\sigma_p}(w)_{u_3}=011=\widehat{\sigma_p}(w')_{u_3}$ and 
  we deduce that $(\sigma_p\oplus\sigma_R)(w)_{u_4}=(\sigma_p\oplus\sigma_R)(w')_{u_4}$.
  This is a contradiction since $\tau(w)_{u_{4}}=0$ whereas $\tau(w')_{u_{4}}=1$.
\end{example}

The main result of this section is that if $T$ is a \emph{regular} tree language then 
$\residue{T}{p}$ is also \emph{regular}. 

\begin{theorem}\label{10-thm:residue}
	Let $T$ be a \emph{regular} language of $(S_{UV},S_{XYZ})$-trees. Then, the language
	$\residue{T}{p}$ of $(S_{UX},S_{Z})$-trees is also \emph{regular}.
	
  Moreover, if $T$ is recognized by an NTA $\nta=(\QA,\TransA,\IA,\clrA)$, then
  $\residue{T}{p}$ is recognized by an ATA $\ata$ with set of states
  $\QB=S_{V}\times\QA\times S_{XY}$.
  
  The ATA $\ata$ can be constructed from the NTA $\nta$ in 
  time $\mathcal{O}(M^{LM^{L}}|\nta|)$ where $L=|\loc|$ and 
  $M=\max\{|S_{\ell}| \mid \ell\in\loc\}$.
  Also, $|\QB|\leq M^{L}|\QA|$.
\end{theorem}

Before tackling the proof of this theorem, we state its consequence for solving 
distributed games by removing the most informed player.

Let $\game=(\arena,T)$ be a synchronous distributed game over
$\arena=(\archi,(S_{\ell})_{\ell\in\loc})$ where the winning condition is a regular tree
language $T$.
Let $\archi'=(\loc'\uplus\players',E')$ be the architecture obtained from $\archi$ by
removing the most informed player $p$ and the locations that are only connected to $p$.
Let $\arena'=(\archi',(S_{\ell})_{\ell\in\loc'})$ be the corresponding arena. 
Define $\residue{\game}{p}=(\arena',T')$ to be the synchronous distributed game over
$\arena'$ with winning condition the regular tree language $T'=\residue{T}{p}$.

\begin{corollary}\label{10-cor:residue}
  The games $\game$ and $\residue{\game}{p}$ are equi-solvable. 
  
  More precisely, if $(\sigma_{q})_{q\in\players}$ is a winning distributed strategy in
  $\game$ then $(\sigma_{q})_{q\in\players'}$ is a winning distributed strategy in
  $\residue{\game}{p}$.  Conversely, if $(\sigma_{q})_{q\in\players'}$ is a winning
  distributed strategy in $\residue{\game}{p}$ then there is a local strategy $\sigma_{p}$
  for player $p$ in $\arena$ such that $(\sigma_{q})_{q\in\players}$ is a winning
  distributed strategy in $\game$.
\end{corollary}

\begin{proof}
  Assume that $(\sigma_{q})_{q\in\players}$ is a winning distributed strategy in $\game$.
  By definition, we have $\sigma_{\players}\in T$.  Since
  $\sigma_{\players}=\sigma_{p}\oplus\sigma_{\players'}$ we get $\sigma_{\players'}\in
  T'=\residue{T}{p}$.  We deduce that $(\sigma_{q})_{q\in\players'}$ is a winning
  distributed strategy in $\residue{\game}{p}$.
  
  Conversely, let $(\sigma_{q})_{q\in\players'}$ be a winning distributed strategy in
  $\residue{\game}{p}$.  We have $\sigma_{\players'}\in T'=\residue{T}{p}$.  By
  definition, this means that $\sigma_{p}\oplus\sigma_{\players'}\in T$ for some local
  strategy $\sigma_{p}$ of player $p$.  Therefore, $(\sigma_{q})_{q\in\players}$ is a
  winning distributed strategy in $\game$.
\end{proof}

The rest of this section is devoted to proving \Cref{10-thm:residue}.
Let $T$ be a language of $(S_{UV},S_{XYZ})$-trees recognized by an NTA $\nta$.  Our aim is to
construct the ATA $\ata$ that recognizes $\residue{T}{p}$ which is a set of
$(S_{UX},S_{Z})$-trees.  First we discuss some challenges, then provide the formal
construction along with inline explanation of how it tackles the said challenges, then we
give a proof of correctness of the construction.  

\paragraph*{Challenges}
	
The ATA $\ata$ runs on $S_{UX}$-directed $S_Z$-labelled trees.  Suppose $\sigmaR$
is such a tree.  The run-trees of $\ata$ on $\sigmaR$ need to ensure the existence
of a $(S_{UV},S_{XY})$-tree $\sigma_p$ such that the composition
$\sigma=\sigma_{p}\oplus\sigmaR$ is a $S_{UV}$-directed $S_{XYZ}$-labelled tree 
accepted by the NTA $\nta$.
The nodes of $\sigmaR$ are labelled only by $S_Z$, whereas that of $\sigma$ are
labelled by $S_{XYZ}$.  The ATA $\ata$ on a tree $\sigmaR$ guesses the missing
components of the labels in its states.
Furthermore, $\sigma$ is $S_{UV}$-directed whereas $\sigmaR$ has only
information about the $S_{U}$ part of it.  
The ATA $\ata$ on a tree $\sigmaR$ uses its power of alternation to simulate the
run of the NTA $\nta$ on some $\sigma$ along \emph{all} the $S_{UV}$-directions.  
 
The construction of the ATA $\ata$ is driven by the following main idea.
From a run $\rhoB$ of $\ata$ over some strategy $\sigmaR$, one 
should be able to extract both a strategy $\sigma_{p}$ and a run $\rhoA$ of $\nta$ over 
the composition $\sigma_{p}\oplus\sigmaR$. Actually, $\rhoB$ will be a 
full $S_{UV}$-directed tree, and suitable projections of $\rhoB$ will give 
$\sigma_{p}$ and $\rhoA$.
Conversely, given a strategy $\sigma_{p}$ and a run $\rhoA$ of $\nta$ over 
$\sigma_{p}\oplus\sigmaR$ there should be a corresponding run $\rhoB$ of 
$\ata$ over $\sigmaR$. 
Thus a run of $\ata$ guesses a strategy $\sigma_p$ and at the same time also
checks if the composition $\sigma=\sigma_p\oplus\sigmaR$ belongs to $T$.

\begin{figure}[tbp]
  \centering
  \includegraphics[page=20]{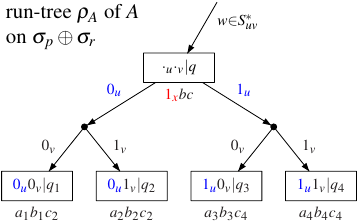}
  \caption{Extract of a run-tree $\rhoA$ of an NTA $\nta$ on the tree 
  $\sigma_{p}\oplus\sigma_{r}$ from \Cref{10-fig:strategies}.
  For clarity, below each node of $\rhoA$ is added the label of the corresponding 
  node of $\sigma_{p}\oplus\sigma_{r}$.
  We assume that the disjunct
  $(0_{u}0_{v},q_{1})\wedge(0_{u}1_{v},q_{2})\wedge(1_{u}0_{v},q_{3})\wedge(1_{u}1_{v},q_{4})$ 
  occurs in $\TransA(q,1_{x}bc)$.
  }
  \label{10-fig:run-tree-NTA}
\commentAlt{Figure~\ref{10-fig:run-tree-NTA}: A run-tree diagram showing a hierarchical branching structure with labeled nodes and edges, representing a complex process or strategy.}
\commentLongAlt{Figure~\ref{10-fig:run-tree-NTA}: The image displays a run-tree diagram, labeled "run-tree rho_A of A on sigma_p (+) sigma_r", starting from a root node.

The root node is a horizontal rectangle labeled '-_x v | q'. An incoming arrow from the top right is labeled 'w in S_AV*'.
From the root node, two main branches descend:

Left branch: Labeled '0_u' (blue), leading to an unlabeled circular node. This node then branches into two leaf nodes (rectangular):
Left leaf: Labeled '0_u 0_r | q1' (with '0_u' in blue), and 'a1b1c2' below it.
Right leaf: Labeled '0_u 1_r | q2' (with '0_u' in blue), and 'a2b2c2' below it.
Right branch: Labeled '1_x bc' (with '1_x' in red), also leading to an unlabeled circular node. This node then branches into two leaf nodes (rectangular):
Left leaf: Labeled '1_u 0_v | q3' (with '1_u' in blue), and 'a3b3c4' below it.
Right leaf: Labeled '1_u 1_v | q4' (with '1_u' in blue), and 'a4b4c4' below it.
The overall diagram illustrates a decision-making process or a sequence of operations based on various inputs and conditions, leading to different final states or outcomes.}
\end{figure}

\begin{figure}[tbp]
  \centering
  \includegraphics[page=21]{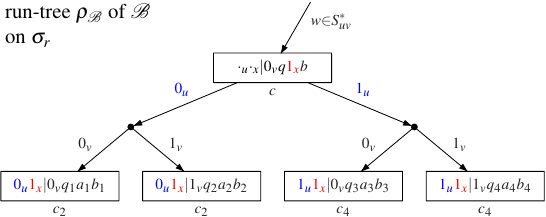}
  \caption{Extract of a run-tree $\rhoB$ of an ATA $\ata$ on the tree 
  $\sigma_{r}$ from \Cref{10-fig:strategies}.
  \\
  The ATA $\ata$ is the one constructed in the proof of \Cref{10-thm:residue} from 
  an NTA $\nta$. The set of states of $\ata$ is $\QB=S_{v}\times\QA\times S_{xy}$.
  \\
  The disjunct of $\TransA(q,1_{x}bc)$ used in \Cref{10-fig:run-tree-NTA} gives rise
  in $\Trans_{\ata}(0_{v}q1_{x}b,c)$ to the disjunct
  $(0_{u}1_{x},0_{v}q_{1}a_{1}b_{1})\wedge(0_{u}1_{x},1_{v}q_{2}a_{2}b_{2})
  \wedge(1_{u}1_{x},0_{v}q_{3}a_{3}b_{3})\wedge(1_{u}1_{x},1_{v}q_{4}a_{4}b_{4})$.
  \\
  For clarity, below each node of $\rhoB$ is added the label of the
  corresponding node of $\sigma_{r}$.  
  \\
  Note that, the projection of $\rhoB$ on $S_{xy}$ is the strategy
  $\sigma_{p}$ of \Cref{10-fig:strategies} and the projection of $\rhoB$ on
  $S_{u}S_{v}\QA$ is the run-tree $\rhoA$ of \Cref{10-fig:run-tree-NTA}.
  }
  \label{10-fig:run-tree-ATA}
\commentAlt{Figure~\ref{10-fig:run-tree-ATA}: A run-tree diagram showing a hierarchical branching structure with labeled nodes and edges, representing a complex process or strategy.}
\commentLongAlt{Figure~\ref{10-fig:run-tree-ATA}: The image displays a run-tree diagram, labeled "run-tree rho_B of B on sigma_r", starting from a root node.

The root node is a horizontal rectangle labeled '-_x | 0_v q 1_x b' (with '0_v' in red). An incoming arrow from the top right is labeled 'w in S_AV*'.
From the root node, two main branches descend:

Left branch: Labeled '0_u' (blue), leading to an unlabeled circular node. This node then branches into two leaf nodes (rectangular):
Left leaf: Labeled '0_u 1_x | 0_v q1 a1b1' (with '0_u' and '1_x' in blue), and 'c2' below it.
Right leaf: Labeled '0_u 1_x | 1_v q2 a2b2' (with '0_u' and '1_x' in blue), and 'c2' below it.
Right branch: Labeled '1_u' (blue), leading to an unlabeled circular node. This node then branches into two leaf nodes (rectangular):
Left leaf: Labeled '1_u 0_v | 0_v q3 a3b3' (with '1_u' in blue), and 'c4' below it.
Right leaf: Labeled '1_u 1_x | 1_v q4 a4b4' (with '1_u' and '1_x' in blue), and 'c4' below it.
The overall diagram illustrates a decision-making process or a sequence of operations based on various inputs and conditions, leading to different final states or outcomes.}
\end{figure}

An extract of a run-tree of an NTA $\nta$ over $(S_{uv},S_{xyz})$-trees is given in
\Cref{10-fig:run-tree-NTA}.  An extract of a run-tree of the corresponding ATA
$\ata$ is given in \Cref{10-fig:run-tree-ATA}.

\paragraph*{The ATA $\ata$}

Suppose the NTA $\nta$ is given by $\nta=(\QA,\TransA,\IA,\clrA)$.
We obtain the ATA $\ata = (\QB, \TransB, \IB, \clrB)$ as follows.  
The set of states of the ATA $\ata$ is $\QB=S_{V}\times\QA\times S_{XY}$.  
That is, along with the states of $\nta$, it guesses the $S_{XY}$-values of the labels of
$\sigma$ which are missing in $\sigmaR$. The $S_{V}$ component is not important, it is 
added to make sure that the run-trees of $\ata$ are full $S_{UV}$-directed trees.
The set of initial states is $\IB=S_{V}\times\IA\times S_{XY}$.
The acceptance condition is lifted from that of $\nta$.  That is, 
$\clrB(e,q,ab)=\clrA(q)$ for all $(e,q,ab)\in S_{V}\times\QA\times S_{XY}$.

Next we define transitions.  We give the intuitive explanation first, with the help of
\Cref{10-fig:run-tree-NTA,10-fig:run-tree-ATA}, and then give the formal definition.
The run $\rhoA$ on \Cref{10-fig:run-tree-NTA} illustrates a transition
taken by the NTA $\nta$ when it is in state $q$ at some node $w\in S_{uv}^{*}$ labelled
$\textcolor{red}{1_{x}}bc$.  The transition $\TransA(q,\textcolor{red}{1_{x}}bc)$
contains the disjunct
$(0_{u}0_{v},q_{1})\wedge(0_{u}1_{v},q_{2})\wedge(1_{u}0_{v},q_{3})\wedge(1_{u}1_{v},q_{4})$
and the choice picked by the run $\rhoA$ (and illustrated in the picture) is this
disjunct.
The simulation of this transition by the ATA $\ata$ on a corresponding node $w'\in
S_{ux}^{*}$ of $\sigma_r$ is depicted in \Cref{10-fig:run-tree-ATA}.  The node $w'$ in
$\sigma_r$ has only the $z$-value $c$ in its label.  The run of $\ata$ guesses the
$xy$-values $\textcolor{red}{1_{x}}b$ as well. The corresponding disjunct in 
$\TransB(0_{v}q\textcolor{red}{1_{x}}b,c)$ is
$(0_{u}\textcolor{red}{1_{x}},0_{v}q_{1}a_{1}b_{1})\wedge(0_{u}\textcolor{red}{1_{x}},1_{v}q_{2}a_{2}b_{2})
\wedge(1_{u}\textcolor{red}{1_{x}},0_{v}q_{3}a_{3}b_{3})\wedge(1_{u}\textcolor{red}{1_{x}},1_{v}q_{4}a_{4}b_{4})$.
Now, since the guessed $x$-value for node $w$ is $\textcolor{red}{1_{x}}$, the branches of
$\sigma_{r}$ with $0_{x}$ are inconsequential.  On the other hand, two states are sent by
the transition of $\ata$ in the direction $0_{u}\textcolor{red}{1_{x}}$ of
$\sigma_{r}$, and similarly, two states are sent in the direction
$1_{u}\textcolor{red}{1_{x}}$.
	
The transitions of $\ata$ are defined by lifting the idea described above.  Since
$\nta$ is an NTA running on $S_{UV}$-directed $S_{XYZ}$-labelled trees, for each $q\in\QA$
and $\ell=\ell_X\ell_Y\ell_Z\in S_{XYZ}$, each disjunct in a transition
$\TransA(q,\ell)$ of $\nta$ is of the form $\bigwedge_{d\in S_{UV}}(d,q_{d})$.
Now, for each $e\in S_{V}$, the transition $\TransB(eq\ell_X\ell_Y,\ell_Z)$ is
obtained from $\TransA(q,\ell)$ by replacing each disjunct 
$\bigwedge_{d\in S_{UV}}(d,q_{d})$ with 
$$
\bigvee_{(x_{d}y_{d})_{d\in S_{UV}}} ~~ 
\bigwedge_{d\in S_{UV}} (d_{U}\ell_{X}, d_{V}q_{d}x_{d}y_{d})
$$
where the first disjunction ranges over $(S_{XY})^{S_{UV}}$, \textit{i.e.}, over all tuples
$(x_{d}y_{d})_{d\in S_{UV}}$ with $x_{d}\in S_{X}$ and $y_{d}\in S_{Y}$.

\begin{enumerate}
	\item Suppose the current node of $\sigmaR$ is labelled $\ell_Z$ and the current
	state of $\ata$ is $eq\ell_X\ell_Y$.  $\ata$ aims at simulating a run of
	$\nta$ from state $q$ on a node labelled $\ell=\ell_X\ell_Y\ell_Z$.  It guesses a choice of
	the transition $\TransA(q,\ell)$ of the NTA $\nta$ which is compatible with the current
	state of $\ata$.  This amounts to guessing 
  the tuple $(q_{d})_{d\in S_{UV}}$ of states that the automaton $\nta$ will send
  along each direction $d\in S_{UV}$.
	
  \item Since the labels of the nodes in $\sigmaR$ only contain values of locations in
  $Z$, and not in $X$ or $Y$, the ATA $\ata$ also guesses the values $x_{d}\in X$
  and $y_{d}\in Y$ along each of the directions $d\in S_{UV}$, to be transmitted as part of
  the states guessed above.

  \item Since the directions $d$ in $S_{UV}$ are not present as such in the
  $S_{UX}$-directed tree $\sigmaR$, $\ata$ uses alternation to send the multitude.
  In particular, along a direction $d_U\ell_X$ of $\sigmaR$, it will send $S_V$ many
  copies of pairs in $Q\times S_{XY}$ so as to simulate the run of $\nta$ faithfully.
  Direction $d_{V}$ is added to each $q_{d}x_{d}y_{d}\in Q\times S_{XY}$ sent in 
  direction $d_{U}\ell_{X}$ to ensure that they are all different (we may have 
  $q_{d}x_{d}y_{d}=q_{d'}x_{d'}y_{d'}$ with $d_{U}=d'_{U}$ and $d_{V}\neq d'_{V}$).
  This explains the $S_{V}$ component of states in $\QB$.
\end{enumerate}

The proof of \Cref{10-thm:residue} is split in 
\Cref{10-lem:time-complexity-ata,10-lem:residue1,10-lem:residue2} below.

\begin{lemma}\label{10-lem:time-complexity-ata}
  The ATA $\ata$ can be constructed from the NTA $\nta$ in time
  $\mathcal{O}(M^{LM^{L}}|\nta|)$ 
  where $L=|\loc|$ and $M=\max\{|S_{\ell}| \mid \ell\in\loc\}$.
  Also, $|\QB|\leq M^{L}|\QA|$.
\end{lemma}

\begin{proof}
  Since $\QB=S_{V}\times\QA\times S_{XY}$ we immediately get $|\QB|\leq M^{L}|\QA|$.

  The time complexity is dominated by the time needed to construct $\TransB$ from 
  $\TransA$, which is proportional to the size of $\TransB$.
  From the definition of $\TransB$, we see that 
  \begin{align*}
    |\TransB|
    &\leq |S_{V}|\times|S_{XY}|^{|S_{UV}|}\times|\TransA| \\
    &\leq M^{|V|}\times M^{|XY|M^{|UV|}}\times|\TransA| 
     \leq M^{LM^{L}}\times|\TransA| \,.
  \end{align*}
  This concludes the proof.
\end{proof}

\begin{lemma}\label{10-lem:residue1}
  $\residue{T}{p}\subseteq\Lang(\ata)$.
\end{lemma}

\begin{proof}
  Consider $\sigmaR\in\residue{T}{p}$.  By definition, there exists $\sigma_p\colon
  S_{UV}^\ast \to S_{XY}$ such that $\sigma=\sigma_p\oplus\sigmaR\in T$.  Since
  $\sigma\in T$, we consider an accepting run $\rho_A$ of $\nta$ on $\sigma$.  Since
  $\nta$ is an NTA, we may choose $S_{UV}$ as direction set for $\rhoA$, \textit{i.e.},
  $\rhoA\colon S_{UV}^{*}\to(S_{UV}\times\QA)$ with $\rhoA(\varepsilon)=(\din,\qin)\in
  S_{UV}\times\IA$, and for all $w\in S_{UV}^{*}$ and $d\in S_{UV}$, we have
  $\rhoA(wd)=(d,q)$ for some $q\in\QA$.
  
  We construct from $\sigma_{p}$ and $\rhoA$ an accepting run $\rhoB$ of $\ata$ on
  $\sigmaR$.  See \Cref{10-fig:strategies,10-fig:run-tree-NTA,10-fig:run-tree-ATA} for an
  example.  We use the same direction set for the run $\rhoB$, so we define $\rhoB\colon
  S_{UV}^{*}\to(S_{UX}\times\QB)$.  Recall that $\QB=S_{V}\times\QA\times S_{XY}$.
  \begin{align*}
    \rhoB(\varepsilon) &= (\din_{U}a,\din_{V}\qin\sigma_{p}(\varepsilon)) 
    && \text{for some } a\in S_{X}, \\
    \rhoB(wd) &= (d_{U}\proj{\sigma_{p}(w)}{X},d_{V}\proj{\rhoA(wd)}{\QA}\sigma_{p}(wd)) 
    && \text{if $w\in S_{UV}^{*}$ and $d\in S_{UV}$.}
  \end{align*}
  
  We first show that for all nodes $w\in S_{UV}^{*}$ of $\rhoB$, the corresponding node 
  $f_{\rhoB}(w)\in S_{UX}^{*}$ of $\sigmaR$ is
  $f_{\rhoB}(w)=\proj{\widehat{\sigma}_{p}(w)}{UX}$ (recall the definition of $f_{\rhoB}$
  in \Cref{10-sec:tree-automata} and of $\widehat{\sigma}_{p}$ in
  \Cref{10-lem:combined-strategies}).  The proof is by induction on $w$. 
  The property is clear for $w=\varepsilon$ since $f_{\rhoB}(\varepsilon)=\varepsilon$.  
  Let $w\in S_{UV}^{*}$ and $d\in S_{UV}$.
  By definition of $\rhoB(wd)$, we have $f_{\rhoB}(wd)=f_{\rhoB}(w)(d_{U}\proj{\sigma_{p}(w)}{X})$.
  Now, since $\widehat{\sigma}_{p}(wd)=\widehat{\sigma}_{p}(w)(d\sigma_{p}(w))$, we 
  deduce
  $$
  \proj{\widehat{\sigma}_{p}(wd)}{UX}=
  \proj{\widehat{\sigma}_{p}(w)}{UX}(d_{U}\proj{\sigma_{p}(w)}{X})=
  f_{\rhoB}(w)(d_{U}\proj{\sigma_{p}(w)}{X})=f_{\rhoB}(wd) \,.
  $$
  
  Now, we show that $\rhoB$ is a run-tree of $\ata$ over $\sigmaR$.
  First, $\rhoB(\varepsilon)$ satisfies the initial condition of run-trees.
  Next, let $w\in S_{UV}^{*}$ be a node of $\rhoB$. Let $\rhoA(w)=(d',q)$ and 
  $\rhoA(wd)=(d,q_{d})$ for all $d\in S_{UV}$. Since $\rhoA$ is a run-tree of $\nta$ over 
  $\sigma$, we know that $\TransA(q,\sigma(w))$ contains the disjunct 
  $\bigwedge_{d\in S_{UV}}(d,q_{d})$. Let $\sigma(w)=\ell_{X}\ell_{Y}\ell_{Z}$. Since 
  $\sigma=\sigma_{p}\oplus\sigmaR$, we have $\ell_{X}\ell_{Y}=\sigma_{p}(w)$ and 
  $\ell_{Z}=\sigmaR(w')$ where $w'=f_{\rhoB}(w)=\proj{\widehat{\sigma}_{p}(w)}{UX}$.
  By definition of $\rhoB$, we have $\proj{\rhoB(w)}{\QB}=eq\ell_{X}\ell_{Y}$ for some 
  $e\in S_{V}$.  By definition of $\TransB$, we deduce that 
  $\bigwedge_{d\in S_{UV}}\rhoB(wd)$ 
  is a disjunct of $\TransB(eq\ell_{X}\ell_{Y},\ell_{Z})$.
  We deduce that $\rhoB$ satisfies the transition condition of run-trees on $\sigmaR$.
  
  Finally, both $\rhoA$ and $\rhoB$ have the same projection on $\QA$: 
  $\proj{\rhoA(w)}{\QA}=\proj{\rhoB(w)}{\QA}$ for all $w\in S_{UV}^{*}$.
  Therefore, using the definition of $\clrB$, we deduce that $\rhoB$ is an accepting
  run-tree of $\ata$ on $\sigmaR$ and $\sigmaR\in\Lang(\ata)$.
\end{proof}

\begin{lemma}\label{10-lem:residue2}
  $\Lang(\ata)\subseteq\residue{T}{p}$.
  
  Moreover, if $\sigmaR\in\Lang(\ata)$ is a regular tree using a finite memory
  structure $\Mem_{R}=(\MM_{R},\mu_{R},\initmm_{R})$, then we can construct a finite
  memory structure $\Mem_{p}$ with set of memory states $\MM_{R}\times\QB$ and a strategy
  $\sigma_{p}$ using memory $\mem_{p}$ such that $\sigma_{p}\oplus\sigmaR\in T$ and
  $\sigma_{p}\oplus\sigmaR$ is also defined using memory $\Mem_{p}$.
\end{lemma}
  
\begin{proof}
  Let $\sigmaR\in\Lang(\ata)$ and let $\rhoB$ be an accepting run of 
  $\ata$ over $\sigmaR$. By definition of $\TransB$, each disjunct of a 
  transition is a conjunction over $d\in S_{UV}$ of the following form
  $\bigwedge_{d\in S_{UV}} (d_{U}\ell_{X}, d_{V}q_{d}x_{d}y_{d})$.
  Therefore, we may use the direction set $S_{UV}$ for $\rhoB$. 
  We have $\rhoB\colon S_{UV}^{*}\to(S_{UX}\times\QB)$
  and we may assume that $\rhoB(w\textcolor{blue}{d})=(\textcolor{blue}{d_{U}}\ell_{X}, 
  \textcolor{blue}{d_{V}}q_{d}x_{d}y_{d})$ for all
  $w\in S_{UV}^{*}$ and $\textcolor{blue}{d\in S_{UV}}$.
  
  Let $\sigma_{p}\colon S_{UV}^{*}\to S_{XY}$ be the projection of $\rhoB$ on $S_{XY}$
  and $\rhoA\colon S_{UV}^{*}\to(S_{UV}\times\QA)$ be the projection of $\rhoB$ on 
  $S_{UV}\times\QA$: if $\rhoB(w)=(\textcolor{blue}{d_{U}}d_{X}, 
  \textcolor{blue}{d_{V}q}\textcolor{red}{xy})$ then $\sigma_{p}(w)=\textcolor{red}{xy}$ 
  and $\rhoA(w)=(\textcolor{blue}{d_{U}d_{V},q})$.
  We prove below that $\rhoA$ is an accepting run-tree of $\nta$ over
  $\sigma=\sigma_{p}\oplus\sigmaR$.  We deduce that $\sigma\in T=\Lang(\nta)$ and
  $\sigmaR\in\residue{T}{p}$ which proves that $\Lang(\ata)\subseteq\residue{T}{p}$.
  
  We first show by induction that for all nodes $w\in S_{UV}^{*}$ of $\rhoB$, the
  corresponding node $f_{\rhoB}(w)\in S_{UX}^{*}$ of $\sigmaR$ is
  $f_{\rhoB}(w)=\proj{\widehat{\sigma}_{p}(w)}{UX}$. 
  The property is clear for $w=\varepsilon$ since
  $f_{\rhoB}(\varepsilon)=\varepsilon$. Let $w\in S_{UV}^{*}$ and $d\in S_{UV}$. Let 
  $\rhoB(w)=(ux,eq\ell_{X}\ell_{Y})$. By definition of $\TransB$ we get
  $\rhoB(wd)=(d_{U}\ell_{X},d_{V}q_{d}x'y')$. Hence, $f_{\rhoB}(wd)=f_{\rhoB}(w)(d_{U}\ell_{X})$.
  Since $\sigma_{p}$ is the projection of $\rhoB$ on $S_{XY}$ we get 
  $\sigma_{p}(w)=\ell_{X}\ell_{Y}$. Therefore, 
  $\widehat{\sigma}_{p}(wd)=\widehat{\sigma}_{p}(w)(d\ell_{X}\ell_{Y})$ and 
  $\proj{\widehat{\sigma}_{p}(wd)}{UX}=
  \proj{\widehat{\sigma}_{p}(w)}{UX}(d_{U}\ell_{X})=f_{\rhoB}(w)(d_{U}\ell_{X})=f_{\rhoB}(wd)$.
  
  We show that $\rhoA$ is a run of $\nta$ over $\sigma=\sigma_{p}\oplus\sigmaR$.  We
  have $\rhoB(\varepsilon)=(\din_{U}\din_{X},\din_{V}\qin xy)$ for some
  $\din_{U}\din_{X}\din_{V}\in S_{UXV}$, $\qin\in\IA$ and $xy\in S_{XY}$ since $\rhoB$
  satisfies the initial condition of run-trees.  Hence,
  $\rhoA(\varepsilon)=(\din_{U}\din_{V},\qin)$ also satisfies the initial condition of
  run-trees.
  
  Now, let $w\in S_{UV}^{*}$ be a node of $\rhoB$ and let
  $\rhoB(w)=(ux,eq\ell_{X}\ell_{Y})$. We have $\sigma_{p}(w)=\ell_{X}\ell_{Y}$.
  Let $w'\in S_{UX}^{*}$ be the corresponding node in $\sigmaR$. 
  We have $w'=f_{\rhoB}(w)=\proj{\widehat{\sigma}_{p}(w)}{UX}$.  Hence, 
  $\ell_{Z}=\sigmaR(w')=\sigmaR(\proj{\widehat{\sigma}_{p}(w)}{UX})=\proj{\sigma(w)}{Z}$. 
  Therefore, $\sigma(w)=\ell=\ell_{X}\ell_{Y}\ell_{Z}$.
  
  Since $\rhoB$ satisfies the transition condition of run-trees over $\sigmaR$, the
  children of $w$ form a disjunct $\bigwedge_{d\in S_{UV}}\rhoB(wd)$ of
  $\TransB(eq\ell_{X}\ell_{Y},\ell_{Z})$.  Using the definition of $\TransB$, we deduce
  that for each $d\in S_{UV}$, we have $\rhoB(wd)=(d_{U}\ell_{X},d_{V}q_{d}x_{d}y_{d})$.
  Moreover, $\bigwedge_{d\in S_{UV}}(d,q_{d})$ is a disjunct of $\TransA(q,\ell)$.  Since
  $\rhoA(w)=(d',q)$, and $\rhoA(wd)=(d,q_{d})$ for all $d\in S_{UV}$, and
  $\sigma(w)=\ell$, we deduce that $\rhoA$ satisfies the transition condition of run-trees
  on $\sigma=\sigma_{p}\oplus\sigmaR$.
  
  Finally, since $\rhoB$ and $\rhoA$ have the same projection on $\QA$, using the 
  definition of $\clrB$ we deduce that $\rhoA$ is an accepting run-tree of $\nta$ over 
  $\sigmaR$.
  
  \medskip
  
  We know from \Cref{10-thm:ata-membership} that if a regular tree is accepted by
  $\ata$ then we can construct an accepting run-tree from a positional winning
  strategy of \Ev in the membership game.  The second part of the lemma is a consequence
  of this result.  Hence, we first define the membership game.
  
  Let $\sigmaR\in\Lang(\ata)$ be a regular $(S_{UX},S_{Z})$-tree defined using a
  finite memory structure $\Mem_{R}=(\MM_{R},\mu_{R},\initmm_{R})$, \textit{i.e.},
  $\mu_{R}\colon\MM_{R}\times S_{UX}\to\MM_{R}$ and $\sigmaR\colon\MM_{R}\to S_{Z}$.
  
  In the membership game, \Ev's positions are pairs
  $(\mm,eqxy)\in\MM_{R}\times\QB=V_{\Ev}$.  From such a position, \Ev may choose a
  disjunct in $\TransB(eqxy,\sigmaR(\mm))$ and move to a pair 
  $\big( \mm,\bigwedge_{d\in S_{UV}}(d_{U}x,d_{V}q_{d}x_{d}y_{d}) \big) \in V_{\Ad}$
  where the second component is the disjunct chosen by \Ev.  This is now a position of
  \Ad, from which he chooses a direction $d\in S_{UV}$ and moves to
  $(\mu_{R}(\mm,d_{U}x),d_{V}q_{d}x_{d}y_{d})\in V_{\Ev}$. The regular tree $\sigmaR$ 
  is accepted by $\ata$ if and only if \Ev wins the membership game from some initial 
  position $(\initmm_{R},eqxy)$ with $eqxy\in\IB$. Since parity games are uniformly 
  bi-positionally determined, we consider a positional optimal strategy $\sigma_{\Ev}$ 
  which is winning on her winning region $W_{\Ev}$.
  
  We define a finite memory structure $\Mem_{p}=(\MM_{p},\mu_{p},\initmm_{p})$.
  For the memory states, we let $\MM_{p}=\MM_{R}\times\QB=V_{\Ev}$. 
  Since $\sigmaR\in\Lang(\ata)$ we may choose some initial winning position
  $\initmm_{p}=(\initmm_{R},e^{0}q^{0}x^{0}y^{0})\in W_{\Ev}$ with $e^{0}q^{0}x^{0}y^{0}\in\IB$.
  Let $(\mm,eqxy)\in\MM_{p}=V_{\Ev}$ and let $\sigma_{\Ev}(\mm,eqxy)=
  \big( \mm,\bigwedge_{d\in S_{UV}}(d_{U}x,d_{V}q_{d}x_{d}y_{d}) \big) \in V_{\Ad}$.
  The update function is defined by $\mu_{p}((\mm,eqxy),d)
  =(\mu_{R}(\mm,d_{U}x),d_{V}q_{d}x_{d}y_{d})$ for all $d\in S_{UV}$.
  
  We define a $(S_{UV},S_{UX}\times\QB)$-tree $\rhoB$ by
  $\rhoB(\varepsilon)=(\din_{U}\din_{X},e^{0}q^{0}x^{0}y^{0})$ for some arbitrary
  $\din_{U}\din_{X}\in S_{UX}$ and $\rhoB(wd)=(d_{U}x,d_{V}q_{d}x_{d}y_{d})$
  for $w\in S_{UV}^{*}$ and $d\in S_{UV}$ if
  $\mu_{p}(\initmm_{p},w)=(\mm,eqxy)$ and $\sigma_{\Ev}(\mm,eqxy)
  =\big( \mm,\bigwedge_{d\in S_{UV}}(d_{U}x,d_{V}q_{d}x_{d}y_{d}) \big)$.
  From the definition of the membership game, it is easy to see that $\rhoB$ is a run-tree
  of $\ata$ over $\sigmaR$. Moreover, since $\initmm_{p}\in\Win_{\Ev}$, the run-tree 
  $\rhoB$ is accepting.
  
  From the definition of $\rhoB$, we see that its projection $\sigma_{p}$ on
  $S_{XY}$ can be defined using memory $\Mem_{p}$ by $\sigma_{p}(\mm,eqxy)=xy\in S_{XY}$
  for all $(\mm,eqxy)\in\MM_{p}$.  From the first part of the proof of the lemma, we know
  that $\sigma=\sigma_{p}\oplus\sigmaR\in T$.  Finally, we see that
  $\sigma=\sigma_{p}\oplus\sigmaR$ can also be defined using memory $\Mem_{p}$ by
  $\sigma(\mm,eqxy)=xyz$ with $z=\sigmaR(\mm)$.
\end{proof}

\section{Algorithm for games with total information order}
\label{10-sec:algo}

\Cref{10-thm:ADG-decidability} follows from \Cref{10-cor:quotient,10-cor:residue}: we can
decide the existence of a winning distributed strategy for a synchronous distributed game
with a total information preorder.

The algorithm to solve a synchronous distributed game $\game$ with a total information
preorder is given in \Cref{10-algo:solve-SDG}.  When players win the game $\game$, the
algorithm returns a winning distributed strategy $(\sigma_{q})_{q\in\players}$ using
finite memory structures $(\mem_{q})_{q\in\players}$.

The algorithm first computes a tree automaton $\ata_{1}$ accepting the winning
\emph{global} strategies $\sigma_{\players}$ for $\game$
(\Cref{10-thm:NBWtoDPW,10-thm:words2trees}).  If $\Lang(\ata_{1})=\emptyset$ then there
are no winning strategies (global or distributed).  Otherwise, there are global winning
strategies and we should check whether some of these could be distributed.

We compute the quotient game $\game_{\approx}$ by merging equally informed players (and 
removing feedback edges) as explained in \Cref{10-sec:reduction-equally-informed}.
The game $\game_{\approx}$ has a totally ordered set $\players'$ of players:
$p_{1}\succ\cdots\succ p_{k}$ in decreasing information order.

Since $T_{1}=\Lang(\ata_{1})\neq\emptyset$, there are winning \emph{global} strategies
when all players in $\players'$ are fused together. 
To check whether some of these global strategies can be distributed, we proceed
step-by-step, computing for
$1\leq i<k$ tree automata $\ata_{i+1}$ accepting the languages
$T_{i+1}=\residue{T_{i}}{p_{i}}$ as long as these languages are nonempty.

\begin{example}\label{10-ex:phi3-b}
  We continue \Cref{10-ex:phi3} with the specification $\varphi_{3}$.  Consider the global
  strategy $\tau$ which copies with delay 1 the values of $u_{0}$ to both $u_{2}$ and
  $u_{3}$, copies with delay 2 the values of $u_{0}$ to $u_{4}$, and copies with delay 3 
  the values of $u_{0}$ to $u_{5}$.
  Abusing notation, $\tau$ implements $u_{2}=\Y u_{0}$, $u_{3}=\Y u_{0}$,
  $u_{4}=\Y\Y u_{0}$ and $u_{5}=\Y\Y\Y u_{0}$.  Clearly, the global strategy $\tau$
  satisfies the first, second and last conjunct of $\varphi_{3}$.  Since $\Y u_{0} \S \Y
  u_{0}$ is equivalent to $\Y u_{0}$, the strategy $\tau$ also satisfies the third
  conjunct of $\varphi_{3}$.  Hence, $\tau\in T_{1}$ is a winning global strategy.
  
  Let $\sigma_{p}$ be the projection of $\tau$ to the locations $u_{2}$ and $u_{3}$:
  $\sigma_{p}$ copies with delay 1 the values of $u_{0}$ to both $u_{2}$ and $u_{3}$.  Let
  $\sigma_{R}$ be the strategy for the fused players $R=\{q,r\}$ which copies with delay 1
  the values of $u_{3}$ to $u_{4}$, and copies with delay 2 the values of $u_{3}$ to
  $u_{5}$.  We can check that $\tau=\sigma_{p}\oplus\sigma_{R}$.  Hence, the strategy
  $\tau$ is $p$-distributable and $\sigma_{R}\in T_{2}=\residue{T_{1}}{p}\neq\emptyset$.
  
  Finally, let $\sigma_{q}$ be the strategy of player $q$ which copies with delay 1 the
  values of $u_{3}$ to $u_{4}$ and let $\sigma_{r}$ be the strategy of player $r$ which
  copies with delay 1 the values of $u_{4}$ to $u_{5}$.  We can check that
  $\sigma_{R}=\sigma_{q}\oplus\sigma_{r}$.  Hence, the strategy $\sigma_{R}$ is
  $q$-distributable and $\sigma_{r}\in T_{3}=\residue{T_{2}}{q}\neq\emptyset$.
  
  Since $T_{3}\neq\emptyset$ we conclude that the distributed game 
  $\game=(\arena,\varphi_{3})$ admits winning distributed strategies. One such strategy 
  is $(\sigma_{p},\sigma_{q},\sigma_{r})$ since 
  $\tau=\sigma_{p}\oplus\sigma_{q}\oplus\sigma_{r}\in T_{1}$ is a winning global 
  strategy.
\end{example}

Coming back to the algorithm for solving distributed games, the first step is to see if
the winning global strategies in $T_{1}$ can be distributed between the most informed
player $p_{1}$ and the remaining players $R_{1}=\{p_{2},\ldots,p_{k}\}$ fused together.
To do so, as explained in \Cref{10-sec:reduction-most-informed}, we compute an alternating
tree automaton $\ata'_{1}$ for the $p_{1}$-residue of $T_{1}=\Lang(\ata_{1})$.  Then,
$\ata'_{1}$ accepts the strategies $\sigma_{R_{1}}$ for players in $R_{1}$ fused together
such that for some strategy $\sigma_{p_{1}}$ of player $p_{1}$ we have
$\sigma_{p_{1}}\oplus\sigma_{R_{1}}\in\Lang(\ata_{1})$.  We compute an NTA $\ata_{2}$
which is language equivalent to $\ata'_{1}$.  If $T_{2}=\Lang(\ata_{2})\neq\emptyset$ we
repeat the procedure to see if some of the strategies accepted by $\ata_{2}$ can be
distributed between the next most informed player $p_{2}$ and the remaining players
$R_{2}=\{p_{3},\ldots,p_{k}\}$ fused together.  We repeat this construction until, we either
obtain an NTA $\ata_{i}$ whose language is empty, in which case we conclude that there are
no winning \emph{distributed} strategies for $\game_{\approx}$, hence also for $\game$, or
$\Lang(\ata_{k})\neq\emptyset$, in which case there are winning distributed strategies for
$\game_{\approx}$, hence also for $\game$.

\SetKw{Downto}{downto}
\begin{algorithm}[ptbh]
  \caption{Solving a synchronous distributed game with a total information preorder.}
  \label{10-algo:solve-SDG}

  \KwIn{A synchronous distributed game $\Game=(\arena,\ata)$ on arena
  $\arena=(\archi,(S_{\ell})_{\ell\in\loc})$ where architecture $\archi$ has total
  information preorder, and the non-determinitic Büchi word automaton $\ata$ defines the 
  winning condition $W=\Lang(\ata)\subseteq S_{\loc}^{\omega}$.}

  \KwOut{No if the players do not have a winning distributed strategy in $\game$. 
  Otherwise, the algorithm returns a winning distributed strategy
  $(\sigma_{q})_{q\in\players}$ with the corresponding finite memory structures
  $(\Mem_{q})_{q\in\players}$.}
  
  Compute from $\ata$ the \emph{deterministic} parity tree automaton $\ata_{1}$ with set
  of states $Q_{1}$ which accepts $\Branch(W)$ (\Cref{10-thm:NBWtoDPW,10-thm:words2trees}).
  
  \If{$\Lang(\ata_{1})=\emptyset$}{\Return{No winning global (or distributed) strategies}}
  
  Compute the quotient game $\game_{\approx}$ by merging equally informed players and removing 
  feedback edges (\Cref{10-sec:reduction-equally-informed}).
  
  Let $p_{1}\succ\cdots\succ p_{k}$ be the players of $\game_{\approx}$ in decreasing
  information order.
  
  \For{$i\gets 1$ \KwTo $k-1$}{
  
    Let $R=\{p_{i+1},\ldots,p_{k}\}$ be the set of players less informed than
    $p_{i}$ and fused together.  Let $\rdom(R)=\bigcup_{j>i}\rdom(p_{j})$.  Let
    $U=\rdom(p_{i})\cap\rdom(R)$, $V=\rdom(p_{i})\setminus\rdom(R)$,
    $X=\wdom(p_{i})\cap\rdom(R)$, $Y=\wdom(p_{i})\setminus\rdom(R)$ and
    $Z=\wdom(R)=\bigcup_{j>i}\wdom(p_{j})$.  This is the situation of 
    \Cref{10-sec:reduction-most-informed} with $p=p_{i}$.
    
    Compute from $\ata_{i}$ an alternating parity tree automaton $\ata'_{i}$ with
    set of states $Q'_{i}$ which accepts the tree language
    $\residue{\Lang(\ata_{i})}{p_{i}}$ (\Cref{10-sec:reduction-most-informed}).
  
    Compute from $\ata'_{i}$ a language-equivalent non-deterministic parity tree
    automaton $\ata_{i+1}$ with set of states $Q_{i+1}$ (\Cref{10-thm:ata2nta}).

    \If{$\Lang(\ata_{i+1})=\emptyset$}{\Return{No winning distributed strategies}}
 
  }
  
  {\small\tcc{Players win the game. We compute a winning distributed strategy with 
  finite memory structures.}}
  
  Compute a finite memory structure $\Mem_{k}=(\MM_{k},\mu_{k},\initmm_{k})$ where
  $\MM_{k}=Q_{k}$ is the set of states of $\ata_{k}$ and a regular strategy $\sigma_{k}$
  for player $p_{k}$ using memory $\Mem_{k}$ such that $\sigma_{k}\in\Lang(\ata_{k})$
  (\Cref{10-thm:nta-emptiness}).
  
  \For{$i\gets k-1$ \Downto $1$}{ 
    
    {\small\tcc{Invariant: $\sigma_{>i}=\sigma_{i+1}\oplus\cdots\oplus\sigma_{k}\in
    \Lang(\ata_{i+1})=\Lang(\ata'_{i})$ and both $\sigma_{i+1}$ and $\sigma_{>i}$ are
    defined using the finite memory structure
    $\Mem_{i+1}=(\MM_{i+1},\mu_{i+1},\initmm_{i+1})$.}}
  
    Using \Cref{10-lem:residue2}, compute a finite memory structure
    $\Mem_{i}=(\MM_{i},\mu_{i},\initmm_{i})$ where $\MM_{i}=\MM_{i+1}\times Q'_{i}$ and a
    regular strategy $\sigma_{i}$ for player $p_{i}$ using memory $\Mem_{i}$ such that 
    $\sigma_{\geq i}=\sigma_{i}\oplus\sigma_{>i}\in\Lang(\ata_{i})$ and $\sigma_{\geq i}$ 
    is also defined using memory $\Mem_{i}$.
    
  }
  
  {\small\tcc{$\sigma_{1}\oplus\cdots\oplus\sigma_{k}\in\Lang(\ata_{1})$, hence
  $(\sigma_{i})_{1\leq i\leq k}$ is a winning distributed strategy for $\game_{\approx}$
  using memory structures $(\Mem_{i})_{1\leq i\leq k}$.}}
  
  Compute from $(\Mem_{i})_{1\leq i\leq k}$ and $(\sigma_{i})_{1\leq i\leq k}$, the memory
  structures $(\Mem_{q})_{q\in\players}$ and a winning distributed strategy
  $(\sigma_{q})_{q\in\players}$ for $\game$ using the memory structures
  $(\Mem_{q})_{q\in\players}$ (\Cref{10-thm:quotient}).
  
  {\small\tcc{If $[q]_{\approx}=p_{i}$ then $\Mem_{q}=(\MM_{i},\mu_{q},\initmm_{i})$.}}
  
  \Return{The winning distributed strategy $(\sigma_{q})_{q\in\players}$ with the 
  corresponding finite memory structures $(\Mem_{q})_{q\in\players}$.}
\end{algorithm}

\begin{example}
  We have seen in \Cref{10-ex:phi3-b} an example of a winning global strategy that can be 
  fully distributed. This is not always the case as we show here. Consider a global 
  strategy $\tau$ which implements: $u_{3}=\Y u_{0}$, $u_{4}=\Y\Y u_{0}$,
  $u_{2}=\Y u_{0} \vee \Y(u_{1}\wedge\neg\Y u_{0})$ and 
  $u_{5}=\Y\Y(\Y u_{0} \vee \Y(u_{1}\wedge\neg\Y u_{0}))$.
  Clearly, the global strategy $\tau$ satisfies the first, second and last conjunct of
  $\varphi_{3}$. We can check that $(a\vee\Y(b\wedge\neg a))\S a$ is equivalent to $a$. 
  Applying it with $a=\Y u_{0}$ and $b=u_{1}$ we deduce that $\tau$ satisfies the third 
  conjunct of $\varphi_{3}$.  Hence, $\tau\in T_{1}$ is a winning global strategy.

  Let $\sigma_{p}$ be the projection of $\tau$ to the locations $u_{2}$ and $u_{3}$:
  $\sigma_{p}$ implements $u_{3}=\Y u_{0}$ and $u_{2}=\Y u_{0} \vee\Y(u_{1}\wedge\neg\Y u_{0})$.  
  Let $\sigma_{R}$ be the strategy for the fused players $R=\{q,r\}$ which implements
  $u_{4}=\Y u_{3}$ and $u_{5}=\Y\Y(u_{3} \vee \Y(u_{1}\wedge\neg u_{3}))$.
  We can check that $\tau=\sigma_{p}\oplus\sigma_{R}$.  Hence, the strategy $\tau$ is
  $p$-distributable and $\sigma_{R}\in T_{2}=\residue{T_{1}}{p}\neq\emptyset$.
  
  However, the strategy $\sigma_{R}$ is not $q$-distributable, \textit{i.e.},
  $\residue{\{\sigma_{R}\}}{q}=\emptyset$.  Indeed, towards a contradiction, assume that
  $\sigma_{R}=\sigma_q\oplus\sigma_r$.  Then, $\sigma_{q}$ is the projection of
  $\sigma_{R}$ on $u_{4}$ and it implements $u_{4}=\Y u_{3}$.
  Consider two plays $\pi$ and $\pi'$ consistent with $\sigma_q\oplus\sigma_r$ such that
  $\proj{\pi}{u_1}=0^\omega$, $\proj{\pi'}{u_1}=1^\omega$, and
  $\proj{\pi}{u_3}=\proj{\pi'}{u_3}=0^\omega$.  We have
  $\proj{\pi}{u_4}=\proj{\pi'}{u_4}=0^\omega$ since $\sigma_{q}$ implements $u_{4}=\Y
  u_{3}$.  Then $\proj{\pi}{u_5}=\proj{\pi'}{u_5}$ because $\sigma_r$ only depends on
  the values of $u_4$.  This is a contradiction with
  $\sigma_R=\sigma_p\oplus\sigma_r$: if $\pi$ and $\pi'$ are consistent with $\sigma_R$,
  we should have $\proj{\pi}{u_5}=0^\omega$ and $\proj{\pi'}{u_5}=0001^\omega$.
\end{example}

Coming back to the algorithm for solving distributed games, when
$\Lang(\ata_{k})\neq\emptyset$, the second part of the algorithm computes finite memory
structures $\Mem_{k},\ldots,\Mem_{1}$
and a winning distributed strategy $\sigma_{k},\ldots,\sigma_{1}$
using these finite memory structures.  This is achieved
using \Cref{10-thm:nta-emptiness} for $\Mem_{k}$ and $\sigma_{k}$ since $\Lang(\ata_{k})$
contains a regular tree, then iteratively using \Cref{10-lem:residue2} for players
$p_{k-1},\ldots,p_{1}$.  Finally, \Cref{10-thm:quotient} allows to construct a winning
distributed strategy using finite memory structures for the initial game $\game$.

\begin{remark}
  We continue \Cref{10-rem:players-unreachable}.  Assume that the initial game $\game$ has
  a player which cannot be reached by an information chain.  Then, in the quotient game
  $\game_{\approx}$, the least informed player $p_{k}$ has an empty read domain (feedback
  edges have been removed).  Player $p_{k}$ is writing to locations in
  $Z=\wdom_{\approx}(p_{k})$ and since the game is \emph{synchronous}, $p_{k}$ is writing
  values in each round. 
  By convention, we defined $S_{\rdom(p_{k})}=S_{\emptyset}=\{\mathsf{tick}\}$ and a
  strategy for player $p_{k}$ is an infinite word $\sigma_{k}\in(S_{Z})^{\omega}$ and
  $\ata_{k}$ is a word automaton.
\end{remark}

\section{Complexity}
\label{10-sec:complexity}

In this section, we discuss the complexity of solving synchronous distributed games with
total information preorder.
Since this section focuses on complexity, we explain more precisely how a distributed game
is presented.  An architecture $\archi=(\loc\uplus\players,E)$ is simply given by the list
of locations, the list of players and the edge relation $E$.  To describe the arena, we
add the list of local states for each location $\ell\in\loc$.  Since we aim at a
nonelementary upper-bound for the complexity, the formalism used to describe the winning
condition does not really matter.  It could be any kind of word automaton on plays
(alternating, non-deterministic, deterministic) or logic on plays (temporal logic or even
monadic second-order logic).

\begin{theorem}\label{10-thm:complexity}
  Solving a synchronous distributed game with a total information preorder and a regular
  winning condition is nonelementary-complete.
\end{theorem}

The upper-bound follows from \Cref{10-algo:solve-SDG}.
Starting from a non-deterministic Büchi word automaton $\ata$ with $n$ states over
the alphabet $\Sigma=S_{\loc}$ defining the winning plays, we apply \Cref{10-thm:NBWtoDPW}
to obtain a language equivalent deterministic parity word automaton and then
\Cref{10-thm:words2trees} to compute the first tree automaton $\ata_{1}$ accepting the 
winning global strategies. The NTA $\ata_{1}$ has $n_{1}=2n^{n}n!$ states and 
$c_{1}=2n$ priorities. It is constructed in time 
$\mathcal{O}(|\Sigma|n^{2n})=\mathcal{O}(M^{L}n^{2n})$ where $L=|\loc|$ is the number of 
locations and $M=\max\{|S_{\ell}| \mid \ell\in\loc\}$.
The information preorder and the quotient architecture $\archi_{\approx}$ (hence also the
quotient game $\game_{\approx}$) is computed in polynomial time with respect to $\archi$.
Emptiness of a non-deterministic parity tree automaton $\ata_{i}$ with $n_{i}$ states and
$c_{i}$ priorities takes time $\mathcal{O}(|\ata_{i}|^{c_{i}})$ or
$|\ata_{i}|^{\mathcal{O}(\log(c_{i}))}$ (\Cref{10-thm:nta-emptiness}).
The next non-deterministic parity tree automaton $\ata_{i+1}$ is computed in exponential 
time\footnote{A careful analysis shows that $\ata_{i+1}$ is constructed from $\ata_{i}$ 
  in time $2^{\mathcal{O}(M^{3L}n_{i}^{2}+M^{L}n_{i}c_{i}\log(M^{L}n_{i}c_{i}))}$.
  The number of priorities of $\ata_{i+1}$ is $c_{i+1}=2M^{L}n_{i}c_{i}$ and its number 
  of states $n_{i+1}$ is $2^{\mathcal{O}(M^{L}n_{i}c_{i}\log(M^{L}n_{i}c_{i}))}$.}
from the NTA $\ata_{i}$: we first construct the alternating parity tree
automaton $\ata'_{i}$ (\Cref{10-lem:time-complexity-ata}) and then use
\Cref{10-thm:ata2nta}.
Notice that during the iteration, only the sizes of the tree automata grow. The number of 
locations decreases at each iteration and the set of local states in the arenas are 
preserved.
Overall, the first part of \Cref{10-algo:solve-SDG} solves the synchronous distributed
game in $k$-\textsc{ExpTime}, where $k$ is the number of players of the quotient 
game~$\game_{\approx}$.

\medskip

Conversely, we prove below a $k$-\textsc{ExpTime} lower bound for the decidability of a
synchronous distributed game with $k$ players, a linear information order, and a winning
condition described by a formula $\Phi$ in the linear temporal logic $\LTL$.
The set of atomic propositions is defined to be $\ap=\biguplus_{\ell\in\loc}\ap_{\ell}$.
For each atomic proposition $a\in\ap_{\ell}$ and each local state $r\in\S_{\ell}$ we will
define whether $r\models a$.  This is extended to global states $s=(s_{i})_{i\in\loc}\in
S_{\loc}$ by saying that $s$ satisfies $a\in\ap_{\ell}$ if $s_{\ell}\models a$.

\begin{theorem}\label{10-thm:lower-bound}
  Solving a synchronous distributed game with $k$ players and a total information preorder
  is $k$-\textsc{ExpTime}-hard when the winning condition is given in linear temporal
  logic.
\end{theorem}

The rest of this section is devoted to the proof of this theorem, which is by reduction of
the membership problem of a $k$-\textsc{ExpTime} Turing machine $M$.

\begin{figure}[tbp]
  \centering
  \includegraphics[page=6]{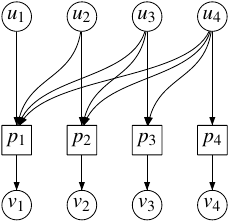}
  \caption{The architecture $\archi_{4}$ with total information order.}
  \label{10-fig:linear-order1}
\commentAlt{Figure~\ref{10-fig:linear-order1}: A directed graph with four layers of nodes (circular, square, circular), showing a feed-forward structure with multiple connections from upper circular nodes to lower square nodes.}
\commentLongAlt{Figure~\ref{10-fig:linear-order1}: The image displays a directed graph with three horizontal layers of nodes.

The top layer consists of four circular nodes labeled 'u1', 'u2', 'u3', and 'u4'.
The middle layer consists of four square nodes labeled 'p1', 'p2', 'p3', and 'p4'.
The bottom layer consists of four circular nodes labeled 'v1', 'v2', 'v3', and 'v4'.

Connections:

'u1' points to 'p1'.
'u2' points to 'p1' and 'p2'.
'u3' points to 'p1', 'p2', and 'p3'.
'u4' points to 'p1', 'p2', 'p3', and 'p4'. (Note: the lines from 'u4' to 'p1', 'p2', 'p3' appear as curved lines originating from 'u4' and passing over the other 'u' nodes before reaching their destinations).
'p1' points to 'v1'.
'p2' points to 'v2'.
'p3' points to 'v3'.
'p4' points to 'v4'.
The diagram illustrates a fan-out connection pattern from the upper 'u' nodes to the middle 'p' nodes, and a one-to-one connection from 'p' nodes to the lower 'v' nodes.}
\end{figure}

We consider a typical architecture $\archi_{k}$ with $k$ players and a linear information
order.  The architecture $\archi_{4}$ is depicted in \Cref{10-fig:linear-order1}.
The architecture $\archi_{k}=(\loc\uplus\players,E)$ consists of $k$ players
$\players=\{p_{1},\ldots,p_{k}\}$ and $2k$ locations
$\loc=\{u_{1},v_{1},\ldots,u_{k},v_{k}\}$.  The external input locations are
$\wdom(\env)=\{u_{1},u_{2},\ldots,u_{k}\}$ and the external output locations are
$\rdom(\env)=\{v_{1},v_{2},\ldots,v_{k}\}$.  Each player $p_{i}$ reads from locations
$\rdom(p_{i})=\{u_{i},\ldots,u_{k}\}$ and writes to location $\wdom(p_{i})=\{v_{i}\}$.
Hence, the information preorder is a linear order with $p_{1}\succ\cdots\succ p_{k}$.

Let $\ell,n\in\mathbb{N}$ be natural numbers.  We define the function $\tower(\ell,n)$
inductively on $\ell$.  We let $\tower(0,n)=n$ and $\tower(\ell,n)=2^{\tower(\ell-1,n)}$
for $\ell>0$.  We have
$\tower(\ell,n)=2^{\,^{\cdot^{\cdot^{\cdot^{2^{n}}}}}}\!\!\!\!\!^{\big|\ell}$.

We fix some natural number $n\geq1$ (which later will be a polynomial function of the
length of some input word $w$ of a $k$-\textsc{ExpTime} Turing machine $M$).  In the first
part of the proof of~\Cref{10-thm:lower-bound} we will show how to write winning
conditions which enforce player $\ell<k$ to count modulo $\tower(\ell,n)$.  An $\LTL$
formula $\Phi_{1}$ of size $\mathcal{O}(n)$ will directly force a winning strategy of the
most informed player $p_{1}$ to count modulo $2^{n}$.  Then, for $1<\ell<k$, we construct
a constant size formula $\Phi_{\ell}$ which forces a winning strategy of player $p_{\ell}$
to count modulo $\tower(\ell,n)$, assuming that player $p_{\ell-1}$ counts modulo
$\tower(\ell-1,n)$ under its winning stratgegy.  Finally, we construct an $\LTL$ formula
$\Phi_{k}$ of size polynomial in $|M|+|w|$ which allows a winning strategy for the least
informed player $p_{k}$ if and only if $M$ accepts $w$.

We first define domains for locations $u_{1},v_{1},\ldots,u_{k-1},v_{k-1}$.
The domain of each output variable $v_{i}$ with $i<k$ is $S_{v_{i}}=\{\skpout,0,1,\#\}$
and the domain of each input variable $u_{i}$ with $i<k$ is
$S_{u_{i}}=\{\skp,\nxt\}$ where $\skp$ means \emph{skip} and $\nxt$ means \emph{next}.
Let $\mathbb{B}=\{0,1\}$ and $\mathbb{B}_{\#}=\{0,1,\#\}$.  The environment uses $\nxt$ or
$\skp$ respectively to control when the player should output a useful value from
$\mathbb{B}_{\#}$ or stay idle with output $\skpout$.  For $i<k$, we also define the
atomic propositions $\ap_{u_{i}}=\{\skp_{i},\nxt_{i}\}$ and
$\ap_{v_{i}}=\{\skpout_{i},0_{i},1_{i},\#_{i}\}$. 
A local state $r\in S_{u_{i}}\uplus S_{v_{i}}$ satisfies an atomic proposition
$a\in\ap_{u_{i}}\uplus\ap_{v_{i}}$ if $r_{i}=a$. For instance, the local state $\nxt\in 
S_{u_{i}}$ satisfies $\nxt_{i}\in\ap_{u_{i}}$ and the local state $1\in S_{v_{i}}$ satisfies 
$1_{i}\in\ap_{v_{i}}$.

Coming back to skip and next, the winning condition will enforce that for each $i$, taking
the delay 1 into account, a winning sequence of global states should satisfy the $\LTL$
formula $\Phi_{0}$ below.
\begin{align*}
  \Phi_{0} & = \bigwedge_{1\leq i\leq k} \skpout_{i} \wedge 
  \always ( \nxt_{i} \lequiv \X \neg\skpout_{i} ) \,.
\end{align*}
For $m\geq1$, we say that player $p_{\ell}$ counts modulo $2^{m}$ if the projection on
$\mathbb{B}_{\#}$ of the output produced by $p_{\ell}$ on location $v_{\ell}$ is the
$\omega$-word $(\#0^{m}\#0^{m-1}1\#0^{m-2}10\#\cdots\#1^{m})^{\omega}$.
We will also use values $f(\ell)$ with $1\leq\ell\leq k$ defined inductively by:
$f(1)=0$ and $f(\ell+1)=f(\ell)+(f(\ell)+1)\tower(\ell-1,n)$ for $\ell\geq1$.  In
particular, $f(2)=n$ and $f(3)=n+(n+1)2^{n}$.
Now, the goal is to define $\LTL$ formulas $\Phi_{1},\ldots,\Phi_{k-1}$ such that the
following lemma holds.

\begin{lemma}\label{10-lem:counting}
  Let $\sigma$ be a distributed strategy and assume that \emph{all} plays consistent with
  $\sigma$ satisfy the condition $\bigwedge_{i=0}^{k-1}\Phi_{i}$.
  
  Let $1\leq\ell\leq k-1$.  Then, player $p_{\ell}$ counts modulo $\tower(\ell,n)$ on all
  plays consistent with $\sigma$ such that the triggering sequence of the environment on
  location $u_{\ell}$ contains infinitely many $\nxt$ symbols and sufficiently many
  (at least $f(\ell)$) $\skp$ symbols between any pair of consecutive $\nxt$
  symbols.
\end{lemma}

We start with the formula $\Phi_{1}$ which forces player $p_{1}$ to count modulo $2^{n}$
if the input sequence on $u_{1}$ contains infinitely many triggers for a useful symbol.
That is, we want the projection of the output of player $p_{1}$ on
$\mathbb{B}_{\#}$ to be the $\omega$-word
$(\#0^{n}\#0^{n-1}1\#0^{n-2}10\#\cdots\#1^{n})^{\omega}$ for all plays satisfying 
$\Phi_{0}\wedge\Phi_{1}\wedge\always\F\nxt_{1}$.  This can be
enforced by an $\LTL$ formula $\Phi_{1}$ of size $\mathcal{O}(n)$.  Let us first define 
some macros. For each $1\leq i\leq k$, we use $\X_{i}$ to move to the next useful output 
from $\mathbb{B}_{\#}$ produced by player $p_{i}$: 
$\X_{i}\varphi=\skpout_{i}\SU(\neg\skpout_{i}\wedge\varphi)$. We also define by induction
a formula $\mathsf{Init}_{i}^{\#0^{m}}$ of size $\mathcal{O}(m)$ stating that $\#0^{m}$ is a
prefix of the useful outputs of player $p_{i}$:
$\mathsf{Init}_{i}^{\#0^{m}}=\X_{i}(\#_{i}\wedge\mathsf{Init}_{i}^{0^{m}})$,
$\mathsf{Init}_{i}^{0^{1}}=\X_{i}0_{i}$, and for $m>1$,
$\mathsf{Init}_{i}^{0^{m}}=\X_{i}(0_{i}\wedge\mathsf{Init}_{i}^{0^{m-1}})$.
We recall that by $\Phi_{0}$ the output of player $p_{i}$ starts with $\skpout$.
We let $\mathbb{B}_{i}$ stand for $0_{i}\vee 1_{i}$. 
The formula enforcing player $p_{1}$ to count modulo $2^{n}$ is
\begin{align*}
  \Phi_{1} &= \big( \always\F\nxt_{1} \limplies ( \mathsf{Init}_{1}^{\#0^{n}} 
  \wedge \mathsf{Length}_{1} \wedge \mathsf{Inc}_{1} ) \big) \\
  \intertext{where}
  \mathsf{Length}_{1} &= 
  \always \big( \neg\skpout_{1} \limplies (\#_{1} \lequiv \X_{1}^{n+1}\#_{1}) \big)
  \\
  \mathsf{Inc}_{1} &= \always \big( \mathbb{B}_{1} \limplies 
  \big( \X_{1}^{n+1} 1_{1} \lequiv
  ( 0_{1} \lequiv (1_{1}\vee\skpout_{1})\SU\#_{1} ) \big) \big) \,.
\end{align*}
It is easy to see that, assuming that the sequence of visible outputs produced by $p_{1}$
is infinite, the conjunction $\mathsf{Init}_{1}^{\#0^{n}} \wedge \mathsf{Length}_{1}$
forces this sequence to be in $\#0^{n}(\#\mathbb{B}^{n})^\omega$.
Then $\mathsf{Inc}_{1}$ ensures correct incrementation (modulo $2^{n}$).  More precisely,
in two consecutive counter values on $v_{1}$, each bit has been correctly changed to reflect
incrementation: it is $1$ in the second counter value if and only if in the first counter
value it was either $0$ followed by $1$'s only or it was $1$ not followed by $1$'s only.

Therefore, $p_{1}$ counts modulo $2^{n}$ on all plays satisfying
$\Phi_{0}\wedge\Phi_{1}\wedge\always\F\nxt_{1}$.
This proves \Cref{10-lem:counting} for $\ell=1$.

Let $\ell>1$.  We will now explain how to write a formula
\begin{align*}
  \Phi_{\ell} &= \big( \always\F\nxt_{\ell} \limplies
  ( \mathsf{Init}_{\ell}\wedge\mathsf{Length}_{\ell}\wedge\mathsf{Inc}_{\ell} ) \big)
\end{align*}
which will
force player $p_{\ell}$ to count modulo $\tower(\ell,n)$.  This relies on the fact that
player $p_{\ell-1}$ is forced to count modulo $\tower(\ell-1,n)$
and also on the fact that player $p_{\ell}$ has no
information about the triggering sequence in $\{\skp,\nxt\}^\omega$ that $p_{\ell-1}$
receives via the input variable $u_{\ell-1}$.

First, the output sequence of $p_{\ell}$ should start with
$\skpout^{+}\#(\skpout^{*}0)^{*}\skpout^{*}\#$ which is enforced by the following formula of
constant size:
\begin{align*}
  \mathsf{Init}_{\ell} &= 
  \skpout_{\ell}\wedge\X_{\ell}(\#_{\ell}\wedge(\skpout_{\ell}\vee 
  0_{\ell})\SU\#_{\ell})) \,.
\end{align*}
Now, we want to enforce that, in the output sequence of player $p_{\ell}$, the number of
bits from $\mathbb{B}$ between consecutive $\#$'s is precisely $\tower(\ell-1,n)$.  This
could be enforced by the formula $\mathsf{Init}_{\ell}^{\#0^{\tower(\ell-1,n)}} \wedge
\always\big(\neg\skpout_{\ell}\limplies 
(\#_{\ell}\lequiv\X_{\ell}^{\tower(\ell-1,n)}\X_{\ell}\#_{\ell})\big)$,
similarly to what we did in $\Phi_{1}$.  The problem is that the formula above is not of
polynomial size with respect to $n$.  

We will show how to simulate $\X_{\ell}^{\tower(\ell-1,n)}$ with a formula of constant
size.  To do this, the environment will trigger a counting sequence of player
$p_{\ell-1}$, synchronizing the $\#$ outputs of player $p_{\ell-1}$ with
$\mathbb{B}_{\#}$ outputs of player $p_{\ell}$ (by keeping player $p_{\ell}$ idle between
consecutive $\#$'s written by player $p_{\ell-1}$). See \Cref{10-fig:check-ell} for an 
example with $\ell=2$. 
Assuming that player $p_{\ell-1}$ indeed counts modulo $\tower(\ell-1,n)$, this will
simulate $\X_{\ell}^{\tower(\ell-1,n)}$.
\begin{figure}[htbp]
  \centering
  $\begin{array}{cccccccccccl}
    v_{1} & \skpout \cdots \skpout & \# & 0\cdots 0 & \# & 0\cdots 01 & \# & \cdots & \# 
    & 1\cdots 1 & \# & 0\cdots 0\#\cdots  \\
    v_{2} & \cdots\cdots & b_{0} & \skpout\cdots\skpout & b_{1} & \skpout\cdots\skpout & 
    b_{2} & \cdots & b_{2^{n}-1} & \skpout\cdots\skpout & 
    b_{2^{n}} & \skpout^{*} b_{2^{n}+1} \cdots  
  \end{array}$
  \caption{Synchronization of the $\#$ outputs of player $p_{1}$ with the useful outputs 
  $b_{i}\in\mathbb{B}_{\#}$ of player $p_{2}$. This is enforced with $\mathsf{Check}_{2}$.}
  \label{10-fig:check-ell}
\commentAlt{Figure~\ref{10-fig:check-ell}: Two rows of symbolic notation, labeled 'v1' and 'v2', showing sequences of characters, numbers, and variables, implying a relationship or transformation between them. See long description.}
\commentLongAlt{Figure~\ref{10-fig:check-ell}: The image displays two rows of alphanumeric and symbolic characters, labeled 'v1' and 'v2'.

Row 'v1': "$...$" # 0...0 # 0...01 # ... # 1...1 # 0...0#...
This row shows a sequence of repeating patterns. It starts with '$...$', followed by a '#', then '0...0', another '#', '0...01', '#', and then an ellipsis '...' followed by another '#'. The pattern then continues with '1...1', '#', and '0...0#...', concluding with another ellipsis.

Row 'v2': ....... b0 \$...\$ b1 \$...\$ b2 ... b_{2N-1} \$...\$ b_{2N} \$' b_{2N+1}...
This row also shows a sequence, starting with '.......', then 'b0', '\$...\$', 'b1', '\$...\$', 'b2', and then an ellipsis '...'. The pattern continues with 'b_{2N-1}', '\$...\$', 'b_{2N}', then a prime symbol followed by 'b_{2N+1}', and finally another ellipsis.

The arrangement suggests a correspondence or mapping between the elements in row 'v1' and row 'v2'.}
\end{figure}

We define $\mathsf{Zero}_{i}=\#_{i}\wedge(\skpout_{i}\vee 0_{i})\SU\#_{i}$
which states that the next value of the counter produced by player $p_{i}$ is $0$.
 We
also define the macro $\XZero_{i}\varphi= \neg\mathsf{Zero}_{i} \SU
(\mathsf{Zero}_{i}\wedge\varphi)$ which states that $\varphi$ holds at the next position
where player $p_{i}$ produces a counter with value $0$.  Now, we define
\begin{align*}
  \mathsf{Check}_{\ell} &= (\mathsf{Zero}_{\ell-1} \wedge \neg\skpout_{\ell}) \wedge 
  (\#_{\ell-1} \lequiv \neg\skpout_{\ell}) \SU 
  (\mathsf{Zero}_{\ell-1}\wedge\neg\skpout_{\ell}) 
\end{align*}
which states that the $\mathbb{B}_{\#}$ outputs of player $p_{\ell}$ are synchronized with
the $\#$ outputs of player $p_{\ell-1}$ along a counting sequence (modulo
$\tower(\ell-1,n)$ for winning strategies) of player $p_{\ell-1}$ from $0$ back to $0$.
The environment will enforce this synchronization whenever it wants to check some property
of player $p_{\ell}$.  Next, we define
\begin{align*}
  \mathsf{Length}_{\ell} &= \always \Big( ( \mathsf{Check}_{\ell} \wedge \#_{\ell} )
  \limplies 
  \big( ( \neg\#_{\ell} \SU \mathsf{Zero}_{\ell-1} ) \wedge
  \XZero_{\ell-1} ( \mathbb{B}_{\ell} \wedge \X_{\ell} \#_{\ell} ) \big)
  \Big) \,.
\end{align*}
The formula $\mathsf{Length}_{\ell}$ is used to force player $p_{\ell}$ to output a correct
number of bits from $\mathbb{B}$ between consecutive $\#$ symbols.  Intuitively, whenever
player $p_{\ell}$ outputs $\#$, the environment may decide to check the number of bits
written by $p_{\ell}$ after this $\#$.  To do so, the environment will synchronize this
$\#$ output of player $p_{\ell}$ with $\mathsf{Check}_{\ell}$.  To satisfy the conclusion
of the implication in $\mathsf{Length}_{\ell}$, player $p_{\ell}$ must output bits during the
full counting sequence of player $p_{\ell-1}$, and then it must output a $\#$.

Consider again the example of \Cref{10-fig:check-ell} and assume that $b_{0}=\#$.  Then,
$\mathsf{Check}_{\ell} \wedge \#_{\ell}$ is satisfied at this position.  Now,
$\neg\#_{\ell} \SU \mathsf{Zero}_{\ell-1}$ implies that $b_{1},\ldots,b_{2^{n}-1}\neq\#$.
Since each $b_{i}\neq\skpout$ by $\mathsf{Check}_{\ell}$, this implies
$b_{1},\ldots,b_{2^{n}-1}\in\mathbb{B}$.  Then, $\XZero_{\ell-1} ( \mathbb{B}_{\ell}
\wedge \X_{\ell} \#_{\ell} )$ implies that $b_{2^{n}}\in\mathbb{B}$ and $b_{2^{n}+1}=\#$.

Finally, we have to check that player $p_{\ell}$ keeps incrementing the counter values on
its output location.  Again we could write a formula similar to $\mathsf{Inc}_{1}$ in
$\Phi_{1}$ using $\X_{\ell}^{\tower(\ell-1,n)+1}$ instead of $\X_{1}^{n+1}$, but this
formula would not be of polynomial size.  Instead, we define a formula
$\mathsf{Inc}_{\ell}$ of constant size, using the same idea as in $\mathsf{Length}_{\ell}$
to simulate $\X_{\ell}^{\tower(\ell-1,n)}$.
\begin{align*}
  \mathsf{IncBit}_{\ell} &= 
  \big( \XZero_{\ell-1} \X_{\ell} 1_{\ell} \big)
  \lequiv 
  \big( 0_{\ell} \lequiv (1_{\ell}\vee\skpout_{\ell})\SU\#_{\ell} \big)
  \\
  \mathsf{Inc}_{\ell} &= \always \Big( ( \mathsf{Check}_{\ell} \wedge \mathbb{B}_{\ell} )
  \limplies \mathsf{IncBit}_{\ell} \Big) \,.
\end{align*}
The formula $\mathsf{IncBit}_{\ell}$ states that, in two consecutive counter values of
player $p_{\ell}$, a bit has been correctly changed to reflect incrementation: it is $1$
in the second counter value if and only if it was $0$ followed by $1$'s only or it was $1$
not followed by $1$'s only in the first counter value.
Whenever the environment wants to check that a bit written by $p_{\ell}$ has been changed
correctly with respect to incrementation of the counter, it will synchronize this bit with
$\mathsf{Check}_{\ell}$ and formula $\mathsf{Inc}_{\ell}$ forces $p_{\ell}$ to change the bit
correctly.

\medskip

Notice that, when the triggering sequence of the environment has few $\skp$ symbols
between consecutive $\nxt$ symbols, then player $p_{\ell}$ knows that it is not being
checked (formula $\mathsf{Check}_{\ell}$ does not hold) and it may cheat when producing
the next useful output from $\mathbb{B}_{\#}$.  But if the triggering sequence of the
environment always contains sufficiently many
$\skp$ symbols between consecutive $\nxt$ symbols, then player
$p_{\ell}$ does not know whether and where it is being checked and in this case, its only
winning strategy is to count modulo $\tower(\ell,n)$.

\begin{proof}[Proof of \Cref{10-lem:counting}]
  Consider a distributed strategy $\sigma$
  such that \emph{all} plays consistent with $\sigma$ satisfy the condition
  $\bigwedge_{i=0}^{k-1}\Phi_{i}$.  The proof is by induction on $\ell$.  We have already
  seen that the property holds for $\ell=1$.  Let now $1<\ell<k$.  Fix a play $\pi$
  consistent with $\sigma$ and such that the triggering sequence of the environment on
  location $u_{\ell}$ contains infinitely many $\nxt$ symbols and at least $f(\ell)$ many
  $\skp$ symbols between any pair of consecutive $\nxt$ symbols.  Let
  $w\in\mathbb{B}_{\#}^{\infty}$ be the sequence obtained from $\proj{\pi}{v_{\ell}}$ by
  removing $\skpout$ symbols.  We have to show that $w$ is a counting sequence modulo
  $\tower(\ell,n)$, \textit{i.e.}, $w=(\#0^{m}\#0^{m-1}1\#0^{m-2}10\#\cdots\#1^{m})^{\omega}$ where
  $m=\tower(\ell-1,n)$.  Since $\pi\models\always\F\nxt_{\ell}$, from $\Phi_{0}$ and
  $\mathsf{Init}_{\ell}$, we already know that $w\in\#0^{*}\#\mathbb{B}_{\#}^{\omega}$.
  
  Note that, due to the architecture $\archi_{k}$, for all plays $\pi'$ 
  consistent with $\sigma$, if 
  $\proj{\pi'}{u_{\ell},\ldots,u_{k}}=\proj{\pi}{u_{\ell},\ldots,u_{k}}$ then 
  $\proj{\pi'}{v_{\ell}}=\proj{\pi}{v_{\ell}}$. So we can freely change the input 
  sequence on $u_{\ell-1}$ of $\pi$ without changing the output sequence on $v_{\ell}$.
  
  We show now that each occurrence of $\#$ in $w$ is followed by a word in 
  $\mathbb{B}^{\tower(\ell-1,n)}\#$.
  Consider an occurrence of $\#$ which is at position $r+1$ in $\proj{\pi}{v_{\ell}}$.
  The input sequence $\proj{\pi}{u_{\ell}}$ is of the form
  $z\nxt\skp^{j_{0}}\nxt\skp^{j_{1}}\nxt\skp^{j_{2}}\cdots$ with $|z|=r$ and $j_{i}\geq
  f(\ell)$ for all $i\geq0$.  We construct an input sequence on $u_{\ell-1}$ such that
  $\mathsf{Check}_{\ell}$ is satisfied at the considered occurrence of $\#$.
  
  Let $y=(\skp^{f(\ell-1)}\nxt)^{\tower(\ell-2,n)}$ which is of length
  $(f(\ell-1)+1)\tower(\ell-2,n)=f(\ell)-f(\ell-1)$.  For all $i\geq0$, we let
  $h_{i}=j_{i}-|y|$.  Since $j_{i}\geq f(\ell)$ we get $h_{i}\geq f(\ell-1)$.  Let $\pi'$
  be a play consistent with $\sigma$ which coincides with $\pi$ on all input locations
  except $u_{\ell-1}$ where it is $\skp^{r}\nxt y \skp^{h_{0}} \nxt y \skp^{h_{1}} \nxt y
  \skp^{h_{2}} \cdots$.  This input sequence on $u_{\ell-1}$ has infinitely many $\nxt$
  symbols, and at least $f(\ell-1)$ many $\skp$ symbols between consecutive $\nxt$
  symbols.  By induction hypothesis, the output sequence $\proj{\pi'}{v_{\ell-1}}$ counts
  modulo $\tower(\ell-1,n)$.  We deduce that $\pi'$ satisfies $\mathsf{Check}_{\ell}$ at
  the considered occurrence of $\#$ in $\proj{\pi}{v_{\ell}}=\proj{\pi'}{v_{\ell}}$.  The
  conclusion of $\mathsf{Length}_{\ell}$ at this occurrence of $\#$ implies that, ignoring
  the $\skpout$ symbols, it is followed by exactly $\tower(\ell-1,n)$ bits and then
  another $\#$.  Therefore, we have proved that $w\in\#0^{m}(\#\mathbb{B}^{m})^{\omega}$
  where $m=\tower(\ell-1,n)$.
  
  We can prove similarly, using $\mathsf{Inc}_{\ell}$ instead of $\mathsf{Length}_{\ell}$, 
  that each bit in $w$ is changed correctly to reflect incrementation in the next counter 
  value. Therefore, $\proj{\pi}{v_{\ell}}$ counts modulo $\tower(\ell,n)$.
\end{proof}

Notice that, by changing the winning condition, we may also enforce a counter value to be
decremented, or to be kept unchanged, or even to be incremented twice.  This will be used
below for counter values of player $p_{k}$.

\medskip

To complete the lower bound proof, we show how to reduce the membership problem of a 
$k$--\textsc{ExpTime} Turing machine to the existence of a winning distributed strategy 
in a distributed game with $k$ players. Now that we have the counting mechanism in place, 
this reduction is rather standard.

Let $M$ be a deterministic Turing machine which accepts words of length $m$ in time
$\tower(k,g(m))$, where $g$ is some polynomial taking non-negative values.  Given
$M$ and some input word $w$, we construct in polynomial time a distributed game
$\game(M,w)=(\arena,\Phi)$ with distributed arena
$\arena=(\archi_{k},(S_{\ell})_{\ell\in\loc})$ on architecture
$\archi_{k}=(\loc\uplus\players,E)$ as defined above.  
For locations $\{u_{1},v_{1},\ldots,u_{k-1},v_{k-1}\}$ we use the local states and atomic
propositions defined above.
For location $u_{k}$, we define $S_{u_{k}}=\{\skp,\nxt,-1,+0,+1,+2\}$ where
$\skp$ and $\nxt$ are used as before to trigger outputs of player $p_{k}$ and
$-1,+0,+1,+2$ are used to control the behaviour of player $p_{k}$.  We let
$\ap_{u_{k}}=\{\skp_{k},\nxt_{k},-1,+0,+1,+2\}$ with the obvious semantics: a local state 
$r\in S_{u_{k}}$ satisfies $a\in\ap_{u_{k}}$ if either $r=a\in\{-1,+0,+1,+2\}$ or if
$r\in\{\skp,\nxt\}$ and $r_{k}=a$.
The local states of location $v_{k}$ depend on the Turing machine $M$.

Let $Q$ be the set of states of $M$ with initial state $q_{0}$ and accepting state 
$q_a\neq q_{0}$.
Let $\Sigma$ be the input alphabet of $M$ and $\Gamma$ its tape alphabet.  We assume that
$M$ has a single tape, which is infinite on the right and with $\leftend\in\Gamma$ as left
endmarker and $\blank\in\Gamma$ as blank symbol.  A configuration of $M$ is an infinite
word of the form $\leftend u q v \blank^{\omega}$ where
$uv\in(\Gamma\setminus\{\leftend\})^{*}$ and $q\in Q$ is the current state.  Let
$\Delta=\Gamma\cup Q$.  

The set of local states of location $v_{k}$ is $S_{v_{k}}=\{\skpout,0,1,\#\}\uplus\Delta$.
We also add $\Delta$ to the atomic propositions $\{\skpout_{k},0_{k},1_{k},\#_{k}\}$ and a
local state $r\in S_{v_{k}}$ satisfies $a\in\ap_{v_{k}}$ if either $r=a\in\Delta$ or
$r\in\{\skpout,0,1,\#\}$ and $r_{k}=a$.  Recall that by $\Phi_{0}$, the environment
triggers the useful outputs in $S_{v_{k}}\setminus\{\skpout\}$ of player $p_{k}$ by
writing $\nxt$ on location $u_{k}$.

We consider the time-space diagram of the computation of $M$ on
$w\in\Sigma^{*}$.  For $t,j\geq0$, let $\gamma_{t,j}\in\Delta$ be the $j$-th symbol of the
$t$-th configuration of the computation of $M$ on $w$.  We have $\gamma_{t,0}=\leftend$
for all $t\geq0$ and when $t=0$ the sequence $(\gamma_{0,j})_{j\geq0}$ is the initial
configuration $\leftend q_{0} w \blank^{\omega}$.
For $t,j>0$, whether or not a symbol $\gamma_{t,j}$ is correct only depends on the four
symbols $\gamma_{t-1,j-1}\gamma_{t-1,j+0}\gamma_{t-1,j+1}\gamma_{t-1,j+2}$.  Formally,
since $M$ is deterministic, there is a partial function $\delta\colon\Delta^{4}\to\Delta$
such that 
$\gamma_{t,j}=\delta(\gamma_{t-1,j-1}\gamma_{t-1,j+0}\gamma_{t-1,j+1}\gamma_{t-1,j+2})$ 
for all $t,j>0$.
Note that $\delta$ depends on $M$ but not on the input word $w$.  Also, $M$ accepts $w$ if
and only if the accepting state $q_{a}$ occurs in the time-space diagram at some
$\gamma_{t,j}$ with $0\leq t\leq \tower(k,g(|w|))$, the
time bound of $M$, and $0<j\leq 1+ \tower(k,g(|w|))$.

\medskip

We explain first how the membership problem of a Turing machine can be reduced to
the existence of a winning strategy in a two-player game.  This should help
understanding our construction below which is based on the same idea, but necessitates a
much more involved reduction.

Let $h\colon\mathbb{N}\to\mathbb{N}$ be a time bound function.
Consider a \emph{deterministic} Turing machine $M$ which accepts words of length $m$ in 
time $h(m)$. Fix an input word $w$.
We construct a reachability game where \Ev has a winning strategy if and only if $M$
accepts $w$ in time at most $T=h(|w|)$.  

A position of \Ev is a tuple $(t,j,\beta)$ with $0\leq t\leq T$, $j\geq 0$ and
$\beta\in\Delta$.  The position is immediately winning for \Ev if $j=0$ and
$\beta=\leftend$, or if $j>T+1$ and $\beta=\blank$, or if $t=0$ and $\beta$ is the $j$-th
symbol of the initial configuration $\leftend q_{0} w \blank^{\omega}$. The position is 
immediately losing for \Ev if we are not in the case above and $t=0$ or $j=0$ or 
$j>T+1$.
Otherwise, $0<t\leq T$ and $0<j\leq T+1$, then from $(t,j,\beta)$, \Ev has to move to a
position of $\Adam$ of the form $(t,j,\beta\alpha_{-1}\alpha_{+0}\alpha_{+1}\alpha_{+2})$
with $\beta=\delta(\alpha_{-1}\alpha_{+0}\alpha_{+1}\alpha_{+2})$.  Then, $\Adam$ answers
by moving to a position $(t-1,j+d,\alpha_{d})$ for some $d\in\{-1,+0,+1,+2\}$.

\begin{claim}\label{10-claim:game-M-w}
  \Ev has a winning strategy from some $(t,j,q_{a})$ with $0\leq t\leq T$ and $0<j\leq
  T+1$ if and only if $M$ accepts $w$ in time at most $T$.
\end{claim}

For the proof, we can show that \Ev has a winning strategy from position $(t,j,\beta)$
with $0\leq t\leq T$ iff $\beta=\gamma_{t,j}$ in the time-space diagram of the run of $M$
on $w$.  This is clear for immediate winning positions of \Ev.  It follows directly by
induction when $0<t\leq T$ and $0<j\leq T+1$.

\medskip

Let $T=\tower(k,g(|w|))$. We fix $n=\max(g(|w|)+3,|w|+2)$.
We will only use the facts that $n$ can be computed in polynomial time and satisfies the two 
conditions: $|w|+1<\frac{1}{2}\tower(k,n)$ and $T+1<\frac{1}{2}\tower(k,n)$ (when $k\geq1$).
Notice that the second condition implies that $\gamma_{t,j}=\blank$ when $t\leq T$ and 
$j\geq\frac{1}{2}\tower(k,n)$.
For $0\leq m<\tower(k,n)$ we write $\mathsf{bin}_{k}(m)$ for the 
binary encoding of $m$ of fixed length $\tower(k-1,n)$.

To win the distributed game, player $p_{k}$ should prove that the input word $w$
is accepted by $M$.  To do so, ignoring the $\skpout$ symbols, player $p_{k}$ starts by
writing $\#\mathsf{bin}_{k}(t)\#\mathsf{bin}_{k}(j)\#q_{a}$ on location $v_{k}$ for some
$0\leq t\leq T$ and $0<j\leq T+1$.  The intended meaning is that $\gamma_{t,j}=q_{a}$ in
the time-space diagram of the computation of $M$ on $w$.

Assume that the output sequence (with $\skpout$ removed) on location $v_{k}$ ends with
some $\#\mathsf{bin}_{k}(t)\#\mathsf{bin}_{k}(j)\#\beta$ with time-space values $t,j$ and
$\beta\in\Delta$.
Player $p_{k}$ wins if $j=0$ and $\beta=\leftend$, or if $j\geq\frac{1}{2}\tower(k,n)$ and
$\beta=\blank$, or if $t=0$ and $\beta=\gamma_{0,j}$.
Player $p_{k}$ loses if $j=0$ and $\beta\neq\leftend$, or if $j\geq\frac{1}{2}\tower(k,n)$
and $\beta\neq\blank$, or if $t=0$ and $\beta\neq\gamma_{0,j}$.
If this is not the case, then player $p_{k}$ should write on $v_{k}$ four symbols
$\alpha_{-1}\alpha_{+0}\alpha_{+1}\alpha_{+2}$ with
$\beta=\delta(\alpha_{-1}\alpha_{+0}\alpha_{+1}\alpha_{+2})$.  Then, the environment will
challenge any of the $\alpha$ symbols by writing $d\in\{-1,+0,+1,+2\}$ on the input
location $u_{k}$ of player $p_{k}$.
Player $p_{k}$ has to answer the challenge by writing the sequence
$\#\mathsf{bin}_{k}(t-1)\#\mathsf{bin}_{k}(j+d)\#\alpha_{d}$ on location $v_{k}$.
Again, the intended meaning is that $\gamma_{t-1,j+d}=\alpha_{d}$.

From the general reduction given above of the membership problem of a Turing machine
to a two-player game (\Cref{10-claim:game-M-w}), it is easy to see that $p_{k}$ wins iff
the input word $w$ is accepted by $M$.
It remains to explain how to write an $\LTL$ formula $\Phi_{k}$ of size polynomial in
$|M|+|w|$ so that player $p_{k}$ is forced to follow the rules described above in order to
win the game $\game(M,w)$.  We assume that player $p_{k-1}$ counts modulo $\tower(k-1,n)$,
which, by Lemma~\ref{10-lem:counting}, can be enforced by the conjunction
$\bigwedge_{i=0}^{k-1}\Phi_{i}$.  The formula 
$\Phi_{k}=\big(\G\F\nxt_{k}\limplies\bigwedge_{j=1}^{8}\Phi_{k}^{j}\big)$
states the following:

  \smallskip\noindent
  $\Phi_{k}^{1}$: Projecting away the $\skpout$'s, the output sequence written by player
  $p_{k}$ on location $v_{k}$ should start with $\#0\mathbb{B}^{*}\#0\mathbb{B}^{*}\#q_{a}$.
  Formula $\Phi_{k}^{1}$ is of constant size.

  \smallskip\noindent
  $\Phi_{k}^{2}$: Projecting away the $\skpout$'s, the output sequence produced by
  player $p_{k}$ on location $v_{k}$ belongs to
  $(\#\mathbb{B}^{+}\#\mathbb{B}^{+}\#\Delta^{5})^{\omega}$.
  We can state that the current symbol on $v_{k}$ is in $\Delta$ with the macro
  $\Delta_{k}=\neg(\skpout_{k}\vee 0_{k}\vee 1_{k}\vee \#_{k})$.
  Hence, $\Phi_{k}^{2}$ is of constant size.
  
  \smallskip\noindent
  $\Phi_{k}^{3}$: Every sequence of bits between consecutive $\#$ symbols written by
  $p_{k}$ is of length $\tower(k-1,n)$.  This is a formula of constant size, similar to
  the formula $\mathsf{Length}_{\ell}$ described above.  The only difference is that the
  environment triggers this check with the premise
  $\mathsf{Check}_{k}\wedge\#_{k}\wedge\X_{k}\mathbb{B}_{k}$ to make sure that the $\#$
  written by $p_{k}$ is followed by a bit and not by a letter from $\Delta$.
  
  Together with the above conditions $\Phi_{k}^{1}\wedge\Phi_{k}^{2}$, this forces player
  $p_{k}$ to start by writing $\#\mathsf{bin}_{k}(t)\#\mathsf{bin}_{k}(j)\#q_{a}$ for some
  $0\leq t,j<\frac{1}{2}\tower(k,n)$.
  By doing so, player $p_{k}$ claims that $\gamma_{t,j}=q_{a}$ in the time-space diagram
  of the computation of $M$ on $w$.
  
  \smallskip\noindent
  $\Phi_{k}^{4}=\neg\mathsf{Lose}\SU\mathsf{Win}$ says that player $p_{k}$ should
  eventually write a \emph{winning} sequence in
  $\#\mathbb{B}^{+}\#\mathbb{B}^{+}\#\Delta$, without writing a losing sequence before.
  The formula $\mathsf{Win}$ holds when (and only when) it is evaluated at the first $\#$
  of a sequence of the form $\#\mathsf{bin}_{k}(t)\#\mathsf{bin}_{k}(j)\#\beta$ such that
  $j=0$ and $\beta=\leftend$, or $j\geq\frac{1}{2}\tower(k,n)$ and $\beta=\blank$, or
  $t=0$ and $\beta=\gamma_{0,j}$.
  
  We can write formulas of constant sizes for checking each of the conditions $t=0$, or
  $j=0$, or $j\geq\frac{1}{2}\tower(k,n)$ (most significant bit of $\mathsf{bin}_{k}(j)$
  is 1), or any condition $\beta=\gamma$ for some $\gamma\in\Delta$, in particular, 
  $\beta=\leftend$ and $\beta=\blank$.
  
  We explain now how to write a formula checking if $\beta=\gamma_{0,j}$. 
  Recall that the initial configuration is $\leftend q_{0} w \blank^{\omega}$.
  Let $h>0$ be minimal with $|w|+1<2^{h}$. 
  Notice that if $j\geq2^{h}$ then $\gamma_{0,j}=\blank$. Hence,
  $\beta=\gamma_{0,j}$ is equivalent to 
  $$
  \big( (j\geq2^{h})\wedge(\beta=\blank) \big) \vee
  \bigvee_{0\leq m<2^{h}} \big( (j=m)\wedge(\beta=\gamma_{0,m}) \big) \,.
  $$
  We can write formulas of size $\mathcal{O}(h)$ for checking $j\geq2^{h}$ or any of the 
  $j=m$ for $0\leq m<2^{h}$.
  Therefore, we can write a formula $\mathsf{Win}$ of size polynomial in $|w|$ with the
  expected semantics.  
  
  Similarly, we can write a formula $\mathsf{Lose}$ of size
  polynomial in $|w|$ corresponding to $j=0$ and $\beta\neq\leftend$, or
  $j\geq\frac{1}{2}\tower(k,n)$ and $\beta\neq\blank$, or $t=0$ and
  $\beta\neq\gamma_{0,j}$ (when it is evaluated at the first $\#$
  of a sequence of the form $\#\mathsf{bin}_{k}(t)\#\mathsf{bin}_{k}(j)\#\beta$).
  
  The remaining conditions enforcing player $p_{k}$ to follow the rules are required as
  long as player $p_{k}$ has not won (which implies that player $p_{k}$ has not lost in
  order to satisfy $\Phi_{k}^{4}$).  Hence, we use the \emph{release} modality of $\LTL$
  to formulate these conditions.  Recall that \emph{release} is the dual of \emph{until}:
  $\varphi\Release\psi=\neg(\neg\varphi\U\neg\psi)=(\always\psi) \vee
  (\psi\U(\varphi\wedge\psi))$.
  
  \smallskip\noindent
  $\Phi_{k}^{5}=\mathsf{Win}\Release\mathsf{Tile}_{k}$.  Here,
  $\mathsf{Tile}_{k}$ says that if we see some $\#\beta$ it should be followed by some
  $\alpha_{-1}\alpha_{+0}\alpha_{+1}\alpha_{+2}$ such that
  $\beta=\delta(\alpha_{-1}\alpha_{+0}\alpha_{+1}\alpha_{+2})$.  
  The formula $\mathsf{Tile}_{k}$ is of size polynomial in $|M|$.

  \smallskip\noindent
  $\Phi_{k}^{6}=\mathsf{Win}\Release\mathsf{DecTime}_{k}$: The
  sequence of time values is decremented.  More precisely, if we see some 
  $\#\mathsf{bin}_{k}(t)\#\mathbb{B}^{+}\#\Delta^{5}\#\mathsf{bin}_{k}(t')\#$ then $t'=t-1$.
  The formula $\mathsf{DecTime}_{k}$ of constant size checks that the bits in
  $\mathsf{bin}_{k}(t)$ are correctly modified in $\mathsf{bin}_{k}(t')$ to reflect
  decrement.  It is similar to the formula $\mathsf{Inc}_{\ell}$ described before.  To
  trigger this check, the environment uses a modified version $\mathsf{Check}'_{k}$
  allowing to synchronize \emph{the relevant bits} written by $p_{k}$ in two consecutive
  \emph{time counters} with the counting sequence of $p_{k-1}$.  In particular, the bits
  of the second (space) counter written by $p_{k}$ should be skipped, as well as the
  tile in $\Delta^{5}$.  See \Cref{10-fig:check'-k}.
  \begin{figure}[tbp]
    \centering
    $\begin{array}{cclclccccl}
      v_{k-1} & \skpout^{+} & \#\;\;z_{0}\;\,\#\;\;z_{1}\;\cdots\;\#\;\;z'_{i} 
      & \skpout \cdots\cdots \skpout \cdots\cdots \skpout 
      & z''_{i}\;\;\;\;\#\;\;z_{i+1} \,\cdots\;\#\;\;z_{m} \;\cdots  \\
      v_{k} & \cdots & b_{0}\skpout^{+}b_{1}\skpout^{+}\cdots\;b_{i}\,\skpout^{+} 
      & \#(\neg\#)^{+}\#(\neg\#)^{+}\# 
      & \skpout^{+}b_{i+1}\;\skpout^{+} \;\cdots\, b_{m}\,\skpout^{+} \cdots
    \end{array}$
    \caption{Synchronization corresponding to $\mathsf{Check}'_{k}$ of the $\#$ outputs of
    player $p_{k-1}$ with the useful outputs $b_{i}\in\mathbb{B}$ of player $p_{k}$ in 
    two consecutive time (resp.\ space) counters. In this picture, $m=\tower(k-1,n)$ and, 
    with $z_{i}=z'_{i}z''_{i}$, for each $0\leq j\leq m$, the word $z_{j}$ is
    $\mathsf{bin}_{k-1}(j\mod m)$ shuffled with $\skpout$ symbols.  Therefore,
    $b_{0},b_{m}$ are corresponding bits in two consecutive time (resp.\ space) counters.}
    \label{10-fig:check'-k}
\commentAlt{Figure~\ref{10-fig:check'-k}: Two rows of symbolic notation, labeled 'v_k-1' and 'v_k', showing complex sequences of characters, variables, and regular expression-like patterns.}
\commentLongAlt{Figure~\ref{10-fig:check'-k}: The image displays two rows of symbolic expressions, labeled 'v_k-1' and 'v_k', suggesting a relationship or transformation between them.

Row 'v_k-1': This row shows a sequence of symbols and variables. It starts with '\$S^{+\prime}\$', followed by '#', then 'z_0 # z_1 ... # z_j'. This is followed by an ellipsis '...' and then '\$...S\$'. After a series of dots '.......' and a '\$' sign, it continues with 'z_i^{prime\prime}', '# z_{i+1} ... # z_M', concluding with an ellipsis.

Row 'v_k': This row also shows a sequence, starting with an ellipsis '...'. It then shows 'b_0 S^+ b_1 S^+ ... b_j S^+'. This is followed by '#(~#)^+ #(-#)^+ #'. The sequence then continues with '\$S^+ b_{i+1} S^+ ... b_M S^+', concluding with an ellipsis.

The notation includes superscripts, primes, and mathematical symbols, indicating complex patterns or regular expressions.}
  \end{figure}
 
  The purpose of $\mathsf{Skip}_{k}$ below is to verify that
  player $p_{k-1}$ is paused until the next $\#$ written by $p_{k}$.  This is used twice
  in $\mathsf{Check}'_{k}$ to skip the space counter and then the tile.
  \begin{align*}
    \mathsf{Skip}_{k}\varphi &= (\neg\#_{k} \wedge \skpout_{k-1}) \SU 
    ( \#_{k} \wedge \skpout_{k-1} \wedge \varphi )
    \\
    \X_{\#_{k}}\varphi &= \neg\#_{k}\SU(\#_{k}\wedge\varphi)
    \\
    \mathsf{Check}'_{k} &= (\mathsf{Zero}_{k-1} \wedge \mathbb{B}_{k}) \wedge 
    (\neg\#_{k} \wedge (\#_{k-1} \lequiv \mathbb{B}_{k})) \SU {}
    \\
    &\qquad \Big( \#_{k} \wedge \skpout_{k-1} \wedge \mathsf{Skip}_{k} \mathsf{Skip}_{k}
    \big( (\#_{k-1} \lequiv \mathbb{B}_{k}) \SU 
    (\mathsf{Zero}_{k-1}\wedge\mathbb{B}_{k}) \big) \Big) 
    \\
    \mathsf{DecTime}_{k} &= 
    \big( \mathsf{Check}'_{k} \wedge
    \X_{\#_{k}}\X_{k}\mathbb{B}_{k} \big)
    \limplies \mathsf{DecBit}_{k} 
    \\
    \mathsf{DecBit}_{k} &= 
    \big( \XZero_{k-1} 1_{k} \big)
    \lequiv 
    \big( 0_{k} \lequiv (0_{k}\vee\skpout_{k})\SU\#_{k} \big)
    \,.
  \end{align*}
  The subformula $\X_{\#_{k}}\X_{k}\mathbb{B}_{k}$ of $\mathsf{DecTime}_{k}$ makes sure
  that the environment is checking the first (time) counter and not the second one
  (space).
  
  \smallskip\noindent
  $\Phi_{k}^{7}=\mathsf{Win}\Release\mathsf{UpdateSpace}_{k}$:
  The sequence of space values should agree with the challenges placed by
  the environment:  if we see a sequence of the form
  $\#\mathsf{bin}_{k}(j)\#\Delta^{5}\#\mathbb{B}^{+}\#\mathsf{bin}_{k}(j')\#$ and the
  environment writes $d\in\{-1,+0,+1,+2\}$ on location $u_{k}$ simultaneously
  with the third $\#$ of the sequence above\footnote{If 
    the environment does not challenge player $p_{k}$, \textit{i.e.}, writes $\skp$ or $\nxt$ on
    $u_{k}$ instead of $d$, then player $p_{k}$ can cheat and immediately win by writing 
    $\mathsf{bin}_{k}(t-1)\#\mathsf{bin}_{k}(0)\#\leftend$.},
  then $j'=j+d$.
  This is enforced by the constant size formula $\mathsf{UpdateSpace}_{k}$ below. The
  subformula $\mathsf{Challenge}_{d}$ makes sure that the environment is checking the
  second (space) counter with respect to the challenge $d\in\{-1,+0,+1,+2\}$.
  \begin{align*}
    \mathsf{Challenge}_{d} &= \X_{\#_{k}}
    \big( \X_{k}\neg\mathbb{B}_{k} \wedge \X_{\#_{k}} d \big)
    \\
    \mathsf{UpdateSpace}_{k} &= 
    \bigwedge_{d\in\{-1,+0,+1,+2\}} \hspace{-2em}
    \big( \mathsf{Check}'_{k} \wedge \mathsf{Challenge}_{d} \big)
    \limplies \mathsf{UpdateBit}_{k,d}  
    \\
    \mathsf{UpdateBit}_{k,-1} &= \mathsf{DecBit}_{k}
    \\
    \mathsf{UpdateBit}_{k,+0} &=
    \big( \XZero_{k-1} 1_{k} \big) \lequiv \big( 1_{k} \big)
    \\
    \mathsf{UpdateBit}_{k,+1} &=
    \big( \XZero_{k-1} 1_{k} \big)
    \lequiv 
    \big( 0_{k} \lequiv 
    (1_{k}\vee\skpout_{k})\SU\#_{k} \big)
    \\
    \mathsf{UpdateBit}_{k,+2} &=
    \big( \XZero_{k-1} 1_{k} \big)
    \lequiv 
    \big( 0_{k} \lequiv 
    (1_{k}\vee\skpout_{k})\SU(\mathbb{B}_{k}\wedge\X_{k}\#_{k}) \big)
    \,.
  \end{align*}

  \smallskip\noindent
  $\Phi_{k}^{8}=\mathsf{Win}\Release\mathsf{UpdateSymbol}_{k}$:
  When a challenge $d\in\{-1,+0,+1,+2\}$ is placed by the environment right after a tile
  $\beta\alpha_{-1}\alpha_{+0}\alpha_{+1}\alpha_{+2}$, the
  next symbol written by $p_{k}$ should be $\alpha_{d}$.  The formula
  $\mathsf{UpdateSymbol}_{k}$ is of size $\mathcal{O}(|M|)$.
  \begin{align*}
    \mathsf{UpdateSymbol}_{k} &= 
    \bigwedge_{\substack{\alpha\in\Delta\\ d\in\{-1,+0,+1,+2\}}} \hspace{-2em}
    \big( \alpha \wedge \X_{k}^{3-d}(\#_{k}\wedge d) \big)
    \limplies \X_{\#_{k}} \X_{\#_{k}} \X_{\#_{k}} \X_{k} \alpha
    \,.
  \end{align*}

To summarize, we have explained how to reduce the membership problem for an input word 
$w$ and the Turing machine $M$ to the existence of a winning distributed strategy in the 
game $\game(M,w)=(\arena,\Phi)$ with distributed arena
$\arena=(\archi_{k},(S_{\ell})_{\ell\in\loc})$ on architecture
$\archi_{k}=(\loc\uplus\players,E)$. We can see that this reduction can be computed in 
polynomial time. Note that the set of local states $S_{v_{k}}$ is of size 
$\mathcal{O}(|M|)$, the other sets of local states are independent from $M$ and $w$. The 
winning condition $\Phi=\bigwedge_{i=0}^{k}\Phi_{i}$ is of size polynomial in $|M|+|w|$.
This concludes the proof of \Cref{10-thm:lower-bound}.

\section{Undecidability results}
\label{10-sec:undecidability-synchronous}

\begin{figure}[tbp]
  \centering
  \includegraphics[page=1]{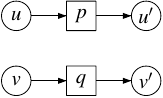}
  \hfil
  \includegraphics[page=2]{10_gastex-pictures-pics.pdf}

  \bigskip
  \includegraphics[page=3]{10_gastex-pictures-pics.pdf}
  \caption{Architectures $\archi_{1}$ (top left), $\archi_{2}$ (top right) and
  $\archi_{3}$ (bottom).  In all 3 architectures, players $p$ and $q$ have incomparable
  information.  In $\archi_{3}$, the information chains $u_{0}u_{1}u_{2}u_{3}$ to player
  $p$ and $u_{0}u_{1}v_{2}$ to player $q$ witness that $q\not\preceq p$ and $p\not\preceq
  q$ respectively.}
  \label{10-fig:information-fork}
\commentAlt{Figure~\ref{10-fig:information-fork}: A set of three distinct directed graphs illustrating different network structures and transformations, using circular and square nodes. See long description.}
\commentLongAlt{Figure~\ref{10-fig:information-fork}: The image displays three separate directed graph diagrams.

Top Left Diagram:

A linear sequence: circular node 'u' points to square node 'p', which then points to circular node 'u''.
Middle Left Diagram:

A linear sequence: circular node 'v' points to square node 'q', which then points to circular node 'v''.
Top Right Diagram:

This diagram shows an interaction between the two simpler sequences from the left.
A circular node 'u' points to square node 'p', which then points to circular node 'u''.
A circular node 'v' points to square node 'q', which then points to circular node 'v''.
Additionally, a curved arrow connects 'p' to 'v'' and another curved arrow connects 'q' to 'u''.
Bottom Diagram:

This is a more complex, larger graph.
It starts with 'u0' (circle) pointing to 'p1' (square), which points to 'u1' (circle).
'u1' points to 'p2' (square).
From 'p2', two paths diverge:
One path goes to 'u2' (circle), which then points to 'p3' (square).
The other path goes to 'v2' (circle).
From 'p3', an arrow points to 'u3' (circle).
From 'u3', an arrow points to 'p' (square), which then points to 'u'' (circle).
From 'v2', an arrow points to 'q' (square), which then points to 'v'' (circle).
A curved arrow connects 'p' to 'v''.
A curved arrow connects 'q' to 'u''.}
\end{figure}

We have seen in the above sections that synchronous distributed games are decidable when
the information preorder is total, \textit{i.e.}, when for all pairs of players $p,q$ we have
$p\preceq q$ or $q\preceq p$ (or both).
We will show in this section that solving a synchronous distributed game is an 
undecidable problem in general: it is neither recursively enumerable nor co-recursively 
enumerable.
We also show that, when we restrict to the existence of a winning distributed strategy
\emph{with finite memory}, the problem becomes recursively enumerable but is still
undecidable in general.
We prove that, as soon as the architecture contains two players $p,q$ with
incomparable information ($q\not\preceq p$ and $p\not\preceq q$), the problem of solving 
a synchronous distributed game is undecidable. 
Hence, a criterion for decidability of solving synchronous distributed games is the fact 
that the information preorder is \emph{total}.  
In the three architectures of \Cref{10-fig:information-fork}, the players $p,q$ have 
incomparable information.

We first show that the distributed architecture $\archi_1$ of
\Cref{10-fig:information-fork} yields an undecidable synchronous distributed game problem.
More precisely, we consider the distributed arena
$\arena_1=(\archi_{1},(S_{\ell})_{\ell\in\loc})$ on
architecture $\archi_1=(\loc\uplus\players,E)$, where $S_{\ell}=\mathbb{B}=\set{0,1}$ for all
locations $\ell\in\loc$.

\begin{theorem}\label{10-thm:undecidable-A1}
  \Cref{10-prob:existWS} is neither recursively enumerable nor co-recursively enumerable,
  even when restricted to synchronous distributed games over the distributed arena
  $\arena_{1}$ and winning conditions given in linear temporal logic.
\end{theorem}

\begin{proof}
  We reduce the existential non-halting problem to the problem of finding a winning
  distributed strategy, \textit{i.e.}, given a Turing machine $M$, we will define a winning
  condition such that there is a winning distributed strategy if and only if there exists
  a word $w$ over which $M$ does not halt.  As the latter problem is neither recursively
  enumerable nor co-recursively enumerable, this gives the expected result.

  Let $M=(Q,\Sigma, \Gamma, q_0, \delta)$ be a Turing machine with set of states $Q$, with 
  initial state $q_{0}$ and halting state $q_{h}$,
  input alphabet $\Sigma$, tape alphabet $\Gamma$ with blank symbol $\blank$ and left end
  marker $\leftend$, and transition function $\delta$.
  Let $\Delta=Q\uplus\Gamma$.

  We first reduce the existential non-halting problem of $M$ to the existence of a winning
  distributed strategy for a distributed game $\game_{M}=(\arena_M,W)$ over the
  distributed arena
  $\arena_M=(\archi_{1},(S'_{\ell})_{\ell\in\loc})$ where
  the output locations have a larger set of states.  More precisely,
  $S'_{u}=S'_{v}=\mathbb{B}$, $S'_{u'}=S'_{v'}=\Delta\uplus\{\#\}$ where $\#$ is a new
  symbol.  We will describle the winning condition $W\subseteq S'^\omega_\loc$ by an
  $\LTL$ formula $\Phi$.

  A word $\gamma_1q\gamma_2\in \Gamma^*Q\Gamma^{+}\blank^{\omega}$ represents a
  configuration of the Turing machine $M$ when it is in the state $q\in Q$ with
  $\gamma_1\gamma_2\in\leftend\Gamma^{\omega}$ as tape content, and its head is reading the
  first symbol of $\gamma_2$.  For a given word $w\in\Sigma^*$, we write $C_1[w]=\leftend
  q_0w\blank^{\omega}$ the initial configuration of $M$ on $w$, and, for all $k> 1$, we
  write $C_k[w]$ the configuration reached by $M$ on input $w$ after $k-1$ steps of
  computation, \textit{i.e.}, $C_1[w]\succTM C_2[w]\succTM\dots\succTM C_k[w]$.

  An input sequence written by the environment on $u$ (resp.\ $v$) encodes a natural
  number: an input sequence in $0^*1^k0\mathbb{B}^\omega$ encodes the number $k\geq1$.
  In order to win, the players $p,q$ should guess an input word $w$ on which the Turing
  machine $M$ does not halt.  Then, when the input sequence on $u$ (resp.\ $v$) encodes
  $k$, player $p$ (resp.\ $q$) should write a word which represents $C_{k}[w]$ on its
  output location $u'$ (resp.\ $v'$).  To express these rules in $\LTL$, we align input
  and output sequences in the following way: if the input sequence belongs to
  $0^{\ell}1^{k}0\mathbb{B}^{\omega}$ then the output sequence should be in
  $\#^{\ell+k+1}\Delta^{\omega}$.  The leading $0$'s in the input sequences are used to
  synchronize the output sequences written by $p$ and $q$.  This allows to write
  specifications in $\LTL$ saying that, ($=$) if the input sequences on $u,v$ both start
  with $0^{\ell}1^{k}0$ then the output sequences on $u',v'$ are equal, and ($\succTM$) if
  the input sequences on $u,v$ start with $0^{\ell+1}1^{k}0$ and $0^{\ell}1^{k+1}0$
  respectively, then the output sequences on $u',v'$ are of the form $\#^{\ell+k+2}C$ and
  $\#^{\ell+k+2}C'$ where $C,C'$ are configurations of $M$ such that $C\succTM C'$.

  Let $\pi\in S_\loc^\omega$ be an infinite sequence of global states.  We define the
  winning condition $W$ by an $\LTL$ formula $\Phi$ which is the conjunction of the
  following requirements.
  \begin{enumerate}
    \item\label{10-it:initial-config} $\Phi_{1}$: When $\proj{\pi}{u}$ encodes $1$, the
    sequence $\proj{\pi}{u'}$ represents the initial configuration of $M$ on some word
    $w\in\Sigma^{*}$.  More precisely, if $\proj{\pi}{u}\in 0^\ell 10\mathbb{B}^\omega$ with
    $\ell\geq0$, then $\proj{\pi}{u'}\in \#^{\ell+2}\leftend q_0\Sigma^*\blank^\omega$.
    
    \item\label{10-it:equal} $\Phi_{2}$: If $\proj{\pi}{u},\proj{\pi}{v}\in 0^\ell 1^k
    0\mathbb{B}^\omega$ with $\ell\geq0$ and $k>0$, then $\proj{\pi}{u'}=\proj{\pi}{v'}$.
    
    \item\label{10-it:any-initial-config} $\Phi_{3}$: If $\proj{\pi}{u}\in 0^{\ell}1^{k}
    0\mathbb{B}^\omega$ with $\ell\geq0$, $k>0$ and $\proj{\pi}{v}=0\proj{\pi}{u}$,
    then $\proj{\pi}{v'}=\#\proj{\pi}{u'}$.
    
    \item\label{10-it:successor} $\Phi_{4}$: If $\proj{\pi}{u}\in
    0^{\ell+1}1^{k}0\mathbb{B}^\omega$ and $\proj{\pi}{v}\in 0^{\ell}1^{k+1}0\mathbb{B}^\omega$
    with $\ell\geq0$ and $k>0$,
    then $\proj{\pi}{u'}=\#^{\ell+k+2}C$, $\proj{\pi}{v'}=\#^{\ell+k+2}C'$ where
    $C,C'\in \Gamma^*Q\Gamma^+\blank^{\omega}$ are configurations of the Turing machine
    $M$ and $C\succTM C'$ (successor).
    
    \item\label{10-it:halting} $\Phi_{5}$: If $\proj{\pi}{u}\in 0^{*}1^{+}0\mathbb{B}^{\omega}$ 
    then $\proj{\pi}{u'}\notin\#^*\Gamma^*q_h\Gamma^{\omega}$ ($M$ does not halt).
  \end{enumerate}
  
\begin{lemma}\label{10-lem:undecidability-1}
  If there is a word $w\in\Sigma^*$ over which $M$ does not halt, there is a
  winning distributed strategy for $\game_{M}$.
\end{lemma}

\begin{proof}
  The strategy for player $p$ (resp.  $q$) is to obey the request placed by the
  environment on its input location, \textit{i.e.}, to output $\#^{\ell+k+1}C_{k}[w]$ when the
  input is in $0^{\ell}1^{k}0\mathbb{B}^{\omega}$.  More precisely, we let $\sigma_{p}(x)=\#$
  for all words $x\in 0^{*}1^{*}$.  Now, for $x\in 0^{\ell}1^{k}0\mathbb{B}^{j-1}$ with $j>0$
  we let $\sigma_{p}(x)$ be the $j$-th symbol of $C_{k}[w]$.
  
  It is easy to check that all plays $\pi$ consistent with the strategy 
  $\sigma=(\sigma_{p},\sigma_{q})$ satisfy $\Phi=\bigwedge_{i=1}^{5}\Phi_{i}$.
  Hence, $\sigma$ is a winning strategy.  
\end{proof}

\begin{lemma}\label{10-lem:undecidability-2}
  If there is a winning distributed strategy for $\game_{M}$, there is a word
  $w\in\Sigma^{*}$ over which $M$ does not halt.
\end{lemma}

\begin{proof} 
  Let $\sigma=(\sigma_{p},\sigma_{q})$ be a winning distributed strategy.  The proof
  relies on the two following claims.  The first one says that, when the input encodes 1,
  the initial configuration written by a player does not depend on the number of leading
  $0$'s.

  \begin{claim}\label{10-cl:initial-configuration}
    There exists $w\in \Sigma^*$ such that for all $\ell\geq 0$, for all play $\pi\in
    S_\loc^\omega$ consistent with $\sigma$, if $\proj{\pi}{u}\in 0^\ell
    10\mathbb{B}^\omega$ then $\proj{\pi}{u'}=\#^{\ell+2}\leftend q_0w\blank^\omega =
    \#^{\ell+2} C_{1}[w]$.
  \end{claim}
    
  \begin{proof}
    Notice first that, since the strategy is distributed, for every play $\pi$ consistent
    with $\sigma$, the projection $\proj{\pi}{u'}$ only depends on $\proj{\pi}{u}$, and
    similarly for $v',v$.  Now, by $\Phi_{1}$, for each $\ell\geq0$, there is a word
    $w_{\ell}\in\Sigma^{*}$ such that for every play $\pi$ consistent with $\sigma$ and such
    that $\proj{\pi}{u}=0^{\ell}10^{\omega}$ we have $\proj{\pi}{u'}=\#^{\ell+2}\leftend
    q_{0}w_{\ell}\blank^\omega$.  Next, using $\Phi_{2}$ with $k=1$, we deduce that, for
    every play $\pi$ consistent with $\sigma$, 
    if $\proj{\pi}{v}\in 0^{\ell}10\mathbb{B}^{\omega}$ then
    $\proj{\pi}{v'}=\#^{\ell+2}\leftend q_{0}w_{\ell}\blank^\omega$, and
    if $\proj{\pi}{u}\in 0^{\ell}10\mathbb{B}^{\omega}$ then
    $\proj{\pi}{u'}=\#^{\ell+2}\leftend q_{0}w_{\ell}\blank^\omega$.
    Finally, using $\Phi_{3}$ with $k=1$, we get $w_{\ell}=w_{\ell+1}$ for all $\ell\geq0$,
    which completes the proof of \Cref{10-cl:initial-configuration}.    
  \end{proof}

  \begin{claim}\label{10-cl:p-configuration}
    There exists $w\in\Sigma^*$ such that for all $k>0$ and $\ell\geq 0$, for all play
    $\pi\in S_\loc^\omega$ consistent with $\sigma$, if $\proj{\pi}{u}\in 0^\ell
    1^k0\mathbb{B}^\omega$, then $\proj{\pi}{u'}=\#^{\ell+k+1} C_k[w]$.
  \end{claim}
  
  \begin{proof}
    Let $w\in\Sigma^*$ be the word given by \Cref{10-cl:initial-configuration}.  We prove
    \Cref{10-cl:p-configuration} by induction on $k$.  The case $k=1$ follows
    from~\Cref{10-cl:initial-configuration}.  Let $k\geq 1$ and $\ell\geq 0$, and let
    $\pi\in S_\loc^\omega$ be a play consistent with $\sigma$ such that 
    $\proj{\pi}{u}\in 0^\ell 1^{k+1}0\mathbb{B}^{\omega}$.
    Consider another play $\pi'$ consistent with $\sigma$, such that
    $\proj{\pi'}{u}\in 0^{\ell+1}1^{k}0\mathbb{B}^{\omega}$, and
    $\proj{\pi'}{v}\in 0^{\ell}1^{k+1}0\mathbb{B}^{\omega}$.  
    By induction hypothesis, we get $\proj{\pi'}{u'}=\#^{\ell+k+2} C_k[w]$
    and using $\Phi_{4}$ we deduce that $\proj{\pi'}{v'}=\#^{\ell+k+2} C_{k+1}[w]$.
    
    Now, consider a third play $\pi''$ consistent with $\sigma$ and such that
    $\proj{\pi''}{u}=\proj{\pi}{u}$ and $\proj{\pi''}{v}=\proj{\pi'}{v}$.
    Since the strategy $\sigma$ is distributed, we have
    $\proj{\pi''}{u'}=\proj{\pi}{u'}$ and $\proj{\pi''}{v'}=\proj{\pi'}{v'}$.
    Since $\sigma$ is a winning strategy we get $\pi''\models\Phi_{2}$ and we obtain
    $\proj{\pi''}{u'}=\proj{\pi''}{v'}$. Finally,
    $\proj{\pi}{u'}=\proj{\pi''}{u'}=\proj{\pi''}{v'}=\proj{\pi'}{v'}=\#^{\ell+k+2} 
    C_{k+1}[w]$. This concludes the proof of \Cref{10-cl:p-configuration}.
  \end{proof}

  Let $w\in\Sigma^*$ be the word given by \Cref{10-cl:p-configuration}.
  For each $k\geq 1$, let $\pi^{k}$ be a play consistent with $\sigma$ such that 
  $\proj{\pi^{k}}{u}=1^{k}0^{\omega}$. By \Cref{10-cl:p-configuration} we get 
  $\proj{\pi^{k}}{u'}=\#^{k+1}C_{k}[w]$. Since $\sigma$ is winning, we have 
  $\pi^{k}\models\Phi_{5}$ and $C_{k}[w]$ does not contain the halting state $q_{h}$.
  Therefore, $M$ does not halt on $w$. This concludes the proof of 
  \Cref{10-lem:undecidability-2}.
\end{proof}

  To summarize, we have constructed a distributed synchronous game $\game_{M}$ over the
  arena $\arena_{M}$ such that the players have a winning distributed strategy if and only
  if there is an input word $w$ on which the computation of $M$ does not halt.

\begin{corollary}\label{10-cor:WSwithInfiniteMem}
  There are synchronous distributed games having a winning distributed strategy but no
  winning distributed strategies with finite memory.
\end{corollary}

\begin{proof}
  Consider the game $\game_{M}$ with a specific Turing machine $M$ which ignores its input
  word and keeps writing some $\$\notin\Sigma\cup\{\blank\}$ on its tape and moving right.
  The machine $M$ does not halt on the empty input word $\varepsilon$, hence by
  \Cref{10-lem:undecidability-1} there is a winning distributed strategy for $\game_{M}$.
  We show now that there are no winning distributed strategy \emph{with finite memory}.
  
  Let $\sigma=(\sigma_{p},\sigma_{q})$ be a winning distributed strategy.  Assume that
  $\sigma_{p}$ is with finite memory.  Then we find $k,k'>0$ and $\ell,\ell'\geq0$ such
  that $k+\ell=k'+\ell'$, $k\neq k'$ and the memory structure reaches the same memory
  state after reading $0^{\ell}1^{k}$ and $0^{\ell'}1^{k'}$.  Consider two plays
  $\pi,\pi'$ consistent with $\sigma$ such that $\proj{\pi}{u}=0^{\ell}1^{k}0^{\omega}$
  and $\proj{\pi'}{u}=0^{\ell'}1^{k'}0^{\omega}$.
  By \Cref{10-cl:p-configuration}, there is an input word $w\in\Sigma^{*}$ such that
  $\proj{\pi}{u'}=\#^{\ell+k+1} C_k[w]$ and $\proj{\pi'}{u'}=\#^{\ell'+k'+1} C_{k'}[w]$.
  Since for each $n\geq0$, the memory structure reaches the same memory state after
  reading $0^{\ell}1^{k}0^{n}$ and $0^{\ell'}1^{k'}0^{n}$ and $\pi,\pi'$ are consistent
  with $\sigma_{p}$, we deduce that $C_{k}[w]=C_{k'}[w]$, a contradiction for the 
  specific Turing machine considered above.  
\end{proof}

  To complete the proof of \Cref{10-thm:undecidable-A1}, we explain now how to modify the
  reduction above so that it applies to the distributed arena $\arena_{1}$, \textit{i.e.}, when the
  ouput locations only use binary values.  We use a fixed-length unary encoding for the
  letters in $\Delta$.  Assume that $\Delta=\{a_{1},\ldots,a_{n}\}$ has $n>1$ elements.
  The letter $a_{i}$ is encoded as $f(a_{i})=01^{i}0^{n-i}$.  Let
  $\Theta=f(\Delta)=01^{+}0^{*}\cap\mathbb{B}^{n+1}$.  The function
  $f\colon\Delta\to\Theta\subseteq\mathbb{B}^{n+1}$ is a bijection.  We extend $f$ to 
  $\Delta\cup\{\#\}$ by setting $f(\#)=0$. Then, $f$ is extended to infinite words in 
  $\#^{*}\Delta^{\omega}$. We get a bijection
  $f\colon\#^{*}\Delta^\omega\to 0^{*}\Theta^{\omega}\subseteq\mathbb{B}^{\omega}$.
  
  We modify the reduction above by applying $f$ to the output sequences written by the
  players.  More precisely, we have to modify the formula $\Phi$ defining the winning
  condition in such a way that, if the input sequence written by the environment is in
  $0^{\ell}1^{k}0\mathbb{B}^\omega$ then the player must write
  $f(\#^{\ell+k+1}C_{k}[w])=0^{\ell+k+1}f(C_{k}[w])$ on its output location.  Each
  formulas $\Phi_{i}$ with $1\leq i\leq 5$ is modified in order to take into account the
  encoding by $f$ of the ouput sequence.  This is not difficult thanks to the fixed length
  encoding of the letters in $\Delta$ and the fact that the beginning of the encoding of a
  letter in $\Delta$ is determined by the factor $01$.
  
  As an example, we write below some useful formulas.  For $\ell\in\loc$, we use the
  atomic propositions $\ap_{\ell}=\{0_{\ell},1_{\ell}\}$ to state that the $\ell$-th
  component of a global state is $0$ or $1$.  Now, for $z\in\mathbb{B}^{*}$ we define by
  induction the formula $\varphi^{z}_{\ell}$ which states that the projection
  $\proj{\pi}{\ell}$ on location $\ell$ starts with $z$:
  $\varphi^{\varepsilon}_{\ell}=\True$ and
  $\varphi^{bz}_{\ell}=b_{\ell}\wedge\X\varphi^{z}_{\ell}$.  The following formula states
  that $\proj{\pi}{\ell}$ is in $\Theta^{\omega}$:
  $$
  \varphi^{\Theta^{\omega}}_{\ell}= \varphi^{01}_{\ell} \wedge
  \always \Big( \varphi^{01}_{\ell} \limplies 
  \big( (\X^{n+1} \varphi^{01}_{\ell}) \wedge \bigvee_{z\in\Theta} \varphi^{z}_{\ell} \big) \Big) \,.
  $$
  Now, assuming that $\proj{\pi}{\ell}$ is in $\Theta^{\omega}$, we can state that
  $\proj{\pi}{\ell}$ belongs to $L=f(\leftend q_{0} \Sigma^{*} \blank^{\omega})$ with the
  formula 
  $$
  \varphi^{L}_{\ell} =
  \varphi^{f(\leftend)f(q_{0})}_{\ell} 
  \wedge \X^{2n+2}
  \Big( \varphi^{01}_{\ell} \limplies \bigvee_{a\in\Sigma} \varphi^{f(a)}_{\ell} \Big) 
  \U \Big( \varphi^{01}_{\ell} \wedge 
  \always \big( \varphi^{01}_{\ell} \limplies \varphi^{f(\blank)}_{\ell} \big)
  \Big) \,.
  $$
  It should be clear now that we can write a formula $\Phi'$ corresponding to the 
  formula $\Phi$ up to the encoding of the output sequences defined by $f$.
  For instance, the first condition can be written
  $$
  \Phi'_{1}=\Big( 0_{u} \U (1_{u}\wedge\X 0_{u}) \Big) \limplies
  \Big( (0_{u}\wedge 0_{u'}) \U 
  \big( 1_{u}\wedge 0_{u'}\wedge \X(0_{u'} \wedge \X\varphi_{u'}^{L}) \big) \Big) \,.
  $$
  This concludes the proof of \Cref{10-thm:undecidable-A1}.
\end{proof}

Next, we prove that solving a synchronous distributed game remains an undecidable 
problem, even when we restrict to distributed strategies \emph{with finite memory}.

\begin{theorem}\label{10-thm:undecidable-P2bis}~
  \begin{enumerate}
    \item  \Cref{10-prob:existWSfiniteMemory} is recursively enumerable.
  
    \item \Cref{10-prob:existWSfiniteMemory} is not co-recursively enumerable, even when
    restricted to synchronous distributed games over the 
    architecture $\archi_{1}$ of \Cref{10-fig:information-fork}
    and winning conditions given in linear temporal logic.
  \end{enumerate}
\end{theorem}

\begin{proof}
  1.  We may enumerate the distributed strategies with finite memory and, for each of
  them, check if it is winning.  As explained in \Cref{10-sec:synchronous}, given a
  distributed strategy $\sigma=(\sigma_{p})_{p\in\players}$ for the players,
  \Cref{10-lem:combined-strategies} allows to construct a global strategy
  $\sigma_{\players}\colon S_{\wdom(\env)}^{*}\to S_{\wdom(\players)}$ for an imaginary
  single player corresponding to all players fused together, such that the distributed
  strategy $\sigma$ is winning if and only if the global strategy $\sigma_{\players}$ is
  winning.  Moreover, if the distributed strategy is with finite memory then
  \Cref{10-lem:combined-strategies} shows how to construct the corresponding global
  strategy also with finite memory.  When the winning condition is regular, it is easy to
  check whether a global strategy with finite memory is winning.  For instance, we may
  construct a tree automaton accepting the winning global strategies
  (\Cref{10-thm:words2trees}) and check whether a given global strategy with finite memory
  is accepted by this tree automaton (\Cref{10-thm:ata-membership}).
  
  2.  We reduce the halting problem of a Turing machine $M$ on the empty input word to the
  existence of a winning distributed strategy in a distributed game $\game'_{M}$ over the
  arena $\arena_{M}$ defined in the proof of \Cref{10-thm:undecidable-A1}.  We only sketch
  this proof which is similar to the one of \Cref{10-thm:undecidable-A1}.
  
  Let $C_{k}\in\Gamma^{*}Q\Gamma^{+}$ be the prefix of length $k+2$ of the 
  $k$-th configuration of $M$ on the empty word. We have $C_{1}=\leftend q_{0}\blank$ and 
  $C_{1}\succTM C_{2}\succTM C_{3}\succTM \cdots$. In order to win, the players will be 
  forced to write a sequence of the form $\#^{+}C_{1}\#^{+}C_{2}\#^{+}\cdots$ and 
  eventually write an halting configuration. The environment will control when a new 
  configuration should be started by writing $1$ on the input location while the player 
  is writing $\#$ on its output location. 
  
  Let $\pi\in S_\loc^\omega$ be an infinite sequence of global states.  We define the
  winning condition $W$ by an $\LTL$ formula $\Phi$ which is the conjunction of the
  following requirements.
  \begin{enumerate}
    \item $\Phi_{1}$: Same strategy for both players.
    If $\proj{\pi}{v}=\proj{\pi}{u}$ then $\proj{\pi}{v'}=\proj{\pi}{u'}$.
  
    \item $\Phi_{2}$: Shift.
    If $\proj{\pi}{u}=xx'$, $\proj{\pi}{v}=x\Red{0}x'$ and
    $\proj{\pi}{u'}=y\#y'$ with $|x|=|y|$ then $\proj{\pi}{v'}=y\Red{\#}\#y'$.
    
    \item $\Phi_{3}$: Initial configuration.
    If $\proj{\pi}{u}$ starts with $1$ then $\proj{\pi}{u'}$ starts with $\#C_{1}\#$.

    \item $\Phi_{4}$: Next configuration.
    If (a) $\proj{\pi}{u}$, $\proj{\pi}{v}$ start with some $10^{3}x1$
    and $00^{3}x1$ respectively, and if (b) $\proj{\pi}{u'}$, $\proj{\pi}{v'}$ start with
    some $\#C_{1}y\#$ and $\#\#^{3}y'\#$ respectively with $|x|=|y|=|y'|$ 
    (recall that $|C_1|=3$), and if (c)
    whenever the $\ell$-th letter of $x$ is $1$ then the $\ell$-th letter of both $y$ and
    $y'$ is $\#$, and if (d) $y$ does not contain the halting state $q_{h}$, then
    $\proj{\pi}{u'}$, $\proj{\pi}{v'}$ start with some $\#C_{1}y\#\Red{C'}\#$ and 
    $\#\#^{3}y'\#\Red{C}\#$ respectively with $C,C'\in\Gamma^{*}Q\Gamma^{+}$ and
    $\Red{C\succTM C'}$.
    
    \item $\Phi_{5}$: Halting.
    If $\proj{\pi}{u}\in(0^{*}1)^{\omega}$ contains infinitely many
    triggers, then $\proj{\pi}{u'}$ contains the halting state $q_{h}$.
  \end{enumerate}
  The Turing machine $M$ halts on the empty input word if and only if
  the players have a winning distributed strategy \emph{with finite memory} in the game
  $\game'_{M}=(\arena_{M},\Phi)$.  
  \begin{enumerate}
    \item When $M$ halts on the empty input word, we get a winning strategy which obeys
    the triggering sequence given by the environment by writing the next configuration of
    the computation of $M$ on the empty input word until it eventually writes the halting
    configuration, then it may always write the same configuration.  Such a strategy only
    requires finite memory.
  
    \item Assume that $M$ does not halt on the empty input word and let 
    $\sigma=(\sigma_{p},\sigma_{q})$ be a distributed strategy satisfying 
    $\Phi_{1}\wedge\Phi_{2}\wedge\Phi_{3}\wedge\Phi_{4}$.
    We claim that any play $\pi$ consistent with $\sigma$ and such that
    $\proj{\pi}{u}=10^{3}10^{4}10^{5}10^{6}\cdots$ must satisfy 
    $\proj{\pi}{u'}=\#C_{1}\#C_{2}\#C_{3}\#C_{4}\#\cdots$.
    We prove that $\proj{\pi}{u'}$ starts with $\#C_{1}\#C_{2}\#\cdots\#C_{k}\#$ by
    induction on $k\geq1$.  The base case $k=1$ follows from $\Phi_{3}$.  Assume that the
    property holds for $k\geq1$.  Consider the play $\pi$ consistent with $\sigma$ and
    such that
    \begin{align*}
      \proj{\pi}{u} &= \phantom{0^{4}}10^{3}\phantom{0}10^{4}\phantom{0}\cdots 
      10^{k+1}\phantom{0}10^{k+2}\phantom{0}1\cdots \\
      \proj{\pi}{v} &= \Red{0^{4}}10^{3}\Red{0}10^{4}\Red{0}\cdots 
      10^{k+1}\Red{0}10^{k+2}\Red{0}1\cdots \\
      \intertext{From the induction hypothesis, we know that $\proj{\pi}{u'}$ starts with}
      \proj{\pi}{u'} &= \phantom{0^{4}}\#C_{1}\phantom{0}\#C_{2}\phantom{0}\cdots 
      \#C_{k-1}\phantom{0}\#C_{k}\phantom{0}\#\cdots \\
      \intertext{Applying $\Phi_{2}$ on 
      the leading occurrences of $\Red{0}$  
      and on each $\Red{0}$ occurring before a $1$
      (last to first using also $\Phi_{1}$), we get}
      \proj{\pi}{v'} &= \Red{\#^{4}}\#C_{1}\Red{\#}\#C_{2}\Red{\#}\cdots 
      \#C_{k-1}\Red{\#}\#C_{k}\Red{\#}\#\cdots
    \end{align*}
    Now, since $M$ does not halt on the empty input word, 
    we may apply $\Phi_{4}$ with $x=10^{4}\cdots10^{k+1}10^{k+2}$, 
    $y=\#C_{2}\cdots\#C_{k-1}\#C_{k}$ and 
    $y'=\#C_{1}\Red{\#}\cdots\#C_{k-2}\Red{\#}\#C_{k-1}\Red{\#}$, and we obtain
    $\proj{\pi}{u'}=\#C_{1}y\#\Red{C'}\#\cdots$, 
    $\proj{\pi}{v'}=\#^{4}y'\#\Red{C}\#\cdots$ with 
    $\Red{C\succTM C'}$. Since $C=C_{k}$, we get $C'=C_{k+1}$ which proves the claim.
    
    Finally, since $M$ does not halt on the empty word, we deduce that
    $\pi\not\models\Phi_{5}$ and $\sigma$ is not winning.
    \qedhere
  \end{enumerate}
\end{proof}

The rest of this section is devoted to the decidability criterion based on the 
information preorder. 
The undecidability proof for the architecture $\archi_{1}$ relies on the fact that player
$p$ does not know what player $q$ receives or writes on its input/output locations.  We
have to lift this undecidability proof to any architecture $\archi$ having two players
$p$ and $q$ with incomparable information.  There are two difficulties to overcome.

First, player $p$ may be aware of what player $q$ writes to its output location, and vice
versa.  This is the case for instance in the architecture $\archi_{2}$ of
\Cref{10-fig:information-fork}.
Nevertheless, we prove below that architecture $\archi_{2}$ is also undecidable.  Let
$\arena_2=(\archi_{2},(S_{\ell})_{\ell\in\loc})$ be the
distributed arena on architecture $\archi_2=(\loc\uplus\players,E)$, where
$S_{\ell}=\mathbb{B}=\set{0,1}$ for all locations $\ell\in\loc$.

\begin{theorem}\label{10-thm:undecidable-A2}
  \Cref{10-prob:existWS} is undecidable,
  even when restricted to synchronous distributed games over the distributed arena
  $\arena_{2}$ and winning conditions given in linear temporal logic.
\end{theorem}

\begin{proof}
  We reduce the non-halting problem of a Turing machine $M$ on the empty input word.  The
  proof is a modification of the proof of \Cref{10-thm:undecidable-A1}.  We use the
  notations introduced in this proof.  Recall that in this proof, a winning distributed
  strategy guesses an input word $w\in\Sigma^{*}$ on which the Turing machine $M$ does not
  halt and, when the environment provides an input sequence $0^{\ell}1^{k}0x$ with
  $\ell\geq0$, $k>0$ and $x\in\mathbb{B}^{\omega}$ 
  the player answers by writing $0^{\ell+k+1}f(C_{k}[w])\in\mathbb{B}^{\omega}$ on its
  output location ($f$ is the encoding from $\Delta=Q\cup\Gamma$ to $\mathbb{B}^{n+1}\cap 
  01^{+}0^{*}$ with $n=|\Delta|$).
  Notice that, after the prefix $0^{\ell}1^{k}0$ of the input sequence, the remaining
  sequence of bits $x\in\mathbb{B}^{\omega}$ has no influence on the output produced by
  the player.  In our modified reduction, we force the input word of the Turing machine to
  be $w=\varepsilon$ and we use the trailing sequence of bits $x$ to obfuscate the output
  produced by player $p$ (resp.\ $q$), so that player $q$ (resp.\ $p$) cannot take
  advantage of reading location $u'$ (resp.\ $v'$).
  
  Let $\xor\colon\mathbb{B}^{2}\to\mathbb{B}$ be the XOR operation on bits. This is 
  extended to pairs of infinite sequences of bits: if 
  $x=b_{1}b_{2}\cdots\in\mathbb{B}^{\omega}$ and 
  $y=b'_{1}b'_{2}\cdots\in\mathbb{B}^{\omega}$ then $x\xor y=b''_{1}b''_{2}\cdots$ where 
  $b''_{i}=b_{i}\xor b'_{i}$ for all $i>0$.
  
  Let $C_{k}$ be the $k$-th configuration of $M$ starting from the empty word.  When
  $k=1$, this is the initial configuration $C_{1}=\leftend q_{0} \blank^{\omega}$.  Now,
  we define a winning condition such that, when the environment provides an input sequence
  $0^{\ell}1^{k}0x$ with $\ell\geq0$, $k>0$ and $x\in\mathbb{B}^{\omega}$, the player is
  forced to write on its output location $0^{\ell+k+2}x'$ where $x'=x\xor
  f(C_{k})\in\mathbb{B}^{\omega}$.  As in the proof of \Cref{10-thm:undecidable-A1}, the
  winning condition $\Psi$ is a conjunction of several conditions.
  \begin{enumerate}
    \item $\Psi_{1}$: If $\proj{\pi}{u}=0^\ell 10x$ with $\ell\geq0$ and
    $x\in\mathbb{B}^\omega$, then $\proj{\pi}{u'}=0^{\ell+3}x'$ with $x'=x\xor f(C_{1})$.
    
    \item $\Psi_{2}$: If $\proj{\pi}{u}=0^{\ell}1^{k}0x$ and
    $\proj{\pi}{v}=0^{\ell}1^{k}0y$ with $\ell\geq0$, $k>0$ and $x,y\in\mathbb{B}^\omega$,
    then $\proj{\pi}{u'}=0^{\ell+k+2}x'$ and $\proj{\pi}{v'}=0^{\ell+k+2}y'$ with $x\xor
    x'=y\xor y'$.
        
    \item $\Psi_{3}$: If $\proj{\pi}{u}=0^{\ell+1}1^{k}0x$ and
    $\proj{\pi}{v}=0^{\ell}1^{k+1}0y$ with $\ell\geq0$, $k>0$ and
    $x,y\in\mathbb{B}^\omega$, then $\proj{\pi}{u'}=0^{\ell+k+3}x'$ and
    $\proj{\pi}{v'}=0^{\ell+k+3}y'$ with $x'=x\xor f(C)$ and $y'=y\xor f(C')$ where
    $C,C'$ are configurations of the Turing machine $M$ and $C\succTM C'$ (successor).
        
    \item $\Psi_{4}$: If $\proj{\pi}{u}=0^{\ell}1^{k}0x$ with $\ell\geq0$, $k>0$ and
    $x\in\mathbb{B}^\omega$, then $\proj{\pi}{u'}=0^{\ell+k+2}x'$ and $x\xor x'\notin
    f(\Gamma^*q_h\Gamma^{\omega})$ ($M$ does not halt).
  \end{enumerate}
  It is not difficult to see that the conditions above can all be expressed in $\LTL$.  As
  an example, we give below the $\LTL$ formula $\Psi_{1}$ which express the first
  condition.  We use again the notations of the proof of \Cref{10-thm:undecidable-A1}.
  For all $z\in\mathbb{B}^{*}$, we define by induction a formula $\psi_{p}^{z}$ which states
  that if $p$ receives $x\in\mathbb{B}^{\omega}$ on its input location $u$ then, after
  delay $1$, $p$ writes some $x'\in\mathbb{B}^{\omega}$ on its output location $u'$ such
  that $z$ is a prefix of $x\xor x'$: $\psi_{p}^{\epsilon}=\True$ and $\psi_{p}^{bz}=\big(
  (0_{u} \wedge \X b_{u'}) \vee (1_{u} \wedge \X \neg b_{u'}) \big) \wedge
  \X\psi_{p}^{z}$.  Now we define
  \begin{align*}
    \Psi_{1} &= \Big( 0_{u} \U (1_{u}\wedge\X 0_{u}) \Big) \to 
    \Big( (0_{u} \wedge 0_{u'}) \U 
    \big( 1_{u} \wedge 0_{u'} \wedge \X 0_{u'} \wedge \X\X 0_{u'} 
    \wedge \X\X\Psi'_{1} \big) \Big)
    \\
    \intertext{where}
    \Psi'_{1} &= 
    \psi^{f(\leftend)f(q_0)}_{p} 
    \wedge \X^{2n+2} \Big( \psi_{p}^{01} \wedge
    \always \big( \psi^{01}_{p} \to (\psi^{f(\blank)}_{p} \wedge \X^{n+1}\psi^{01}_{p}) \big)
    \Big) \,.
  \end{align*}
  
  If the Turing machine does not halt on the empty word, we obtain a winning distributed
  strategy by obeying the requests of the environment, including the XOR part, ignoring
  the output produced by the other player.  To define the strategy, we abuse notation by
  identifying $(\mathbb{B}\times\mathbb{B})^{m}$ and $\mathbb{B}^{m}\times\mathbb{B}^{m}$.
  We let $\sigma_{p}(x,y)=0$ if $x\in 0^{*} \cup 0^{*}1^{+} \cup 0^{*}1^{+}0$ and
  $|y|=|x|$.  
  When $x\in\mathbb{B}^{*}$, $b\in\mathbb{B}$ and $|y|=|0^{\ell}1^{k}0xb|$, we let
  $\sigma_{p}(0^{\ell}1^{k}0xb,y)=b\xor b'$ where $b'$ is the $|xb|$-th symbol of
  $f(C_{k})$.  The strategy $\sigma_{q}$ is defined similarly.  It is easy to see that all
  plays consistent with $\sigma=(\sigma_{p},\sigma_{q})$ satisfy
  $\Psi=\bigwedge_{i=1}^{4}\Psi_{i}$.  Hence, the distributed strategy $\sigma$ is
  winning.
  
  Conversely, assume that there is a winning distributed strategy
  $\sigma=(\sigma_{p},\sigma_{q})$.  The following crucial lemma states that,
  \emph{essentially}, $\sigma_{p}$ does not depend on what it is reading from location
  $v'$.
  
  \begin{lemma}\label{10-lem:obfuscate}
    For all $k>0$, $\ell\geq0$ and $\pi$ consistent with $\sigma$, if 
    $\proj{\pi}{u}=0^{\ell}1^{k}0x$ with $x\in\mathbb{B}^{\omega}$ and 
    $\proj{\pi}{v'}\in 0^{\ell+k+2}\mathbb{B}^{\omega}$ then
    $\proj{\pi}{u'}=0^{\ell+k+2}(x\xor f(C_{k}))$.
  \end{lemma}
  
  \begin{proof}
    The proof is by induction on $k$. The base case $k=1$ directly follows from 
    $\Psi_{1}$, even without any assumption on $\proj{\pi}{v'}$. Assume now that the 
    property holds for some $k>0$. We prove it for $k+1$. First, we establish a symmetric 
    property for $v$ and $k+1$ using the induction hypothesis and $\Psi_{3}$. Then, we 
    transfer it to $u$ and $k+1$ using $\Psi_{2}$.
    
    \begin{claim}
      Let $\ell\geq0$ and $\pi$ be consistent with $\sigma$ such that
      $\proj{\pi}{v}=0^{\ell}1^{k+1}0y$ with $y\in\mathbb{B}^{\omega}$ and
      $\proj{\pi}{u'}=0^{\ell+k+3}\mathbb{B}^{\omega}$.  Then,
      $\proj{\pi}{v'}=0^{\ell+k+3}(y\xor f(C_{k+1}))$.
    \end{claim}

    \begin{proof}
      Let $x'\in\mathbb{B}^{\omega}$ be such that $\proj{\pi}{u'}=0^{\ell+k+3}x'$.
      Let $x=x'\xor f(C_{k})$. Consider the unique play 
      $\pi'$ consistent with $\sigma$ such that $\proj{\pi'}{u}=0^{\ell+1}1^{k}0x$ and 
      $\proj{\pi'}{v}=\proj{\pi}{v}$.  By $\Psi_{3}$ we find configurations $C,C'$ of $M$
      such that $C\succTM C'$, $\proj{\pi'}{u'}=0^{\ell+k+3}(x\xor f(C))$ and
      $\proj{\pi'}{v'}=0^{\ell+k+3}(y\xor f(C'))$.
      Using the induction hypothesis, we get $\proj{\pi'}{u'}=0^{\ell+k+3}(x\xor
      f(C_{k}))$. Therefore, $C=C_{k}$ and $C'=C_{k+1}$.
      Using the definition of $x$ we deduce that $x\xor f(C_{k})=x'$.  Therefore,
      $\proj{\pi'}{u',v}=\proj{\pi}{u',v}$ and since the strategy $\sigma_{q}$ is
      functional, we deduce that $\proj{\pi}{v'}=\proj{\pi'}{v'}=0^{\ell+k+3}(y\xor 
      f(C_{k+1}))$. The claim is proved.
    \end{proof}
    
    Now, let $\ell\geq0$ and $\pi$ be consistent with $\sigma$ such that
    $\proj{\pi}{u}=0^{\ell}1^{k+1}0x$ and $\proj{\pi}{v'}\in 0^{\ell+k+3}y'$ with
    $x,y'\in\mathbb{B}^{\omega}$. Let $y=y'\xor f(C_{k+1})$.
    Consider the unique play $\pi'$ consistent with $\sigma$ such that
    $\proj{\pi'}{u}=\proj{\pi}{u}$ and $\proj{\pi'}{v}=0^{\ell}1^{k+1}0y$.  
    By $\Psi_{2}$ we get $\proj{\pi'}{u'}=0^{\ell+k+3}x'$ and
    $\proj{\pi'}{v'}=0^{\ell+k+3}y''$ with $x\xor x'=y\xor y''$.
    Applying the claim to $\pi'$ we get $y''=y\xor f(C_{k+1})$. 
    Therefore, $x\xor x'=f(C_{k+1})$ and also $y''=y'$.
    We deduce that $\proj{\pi'}{u,v'}=\proj{\pi}{u,v'}$ and finally
    $\proj{\pi}{u'}=\proj{\pi'}{u'}=0^{\ell+k+3}x'=0^{\ell+k+3}(x\xor f(C_{k+1}))$.
    This ends the proof of \Cref{10-lem:obfuscate}. 
  \end{proof}
  
  Finally, assume that the Turing machine halts on the empty word.
  There is $k>0$ such that $C_{k}\in \Gamma^{*}q_{h}\Gamma^\omega$.
  Let $\pi$ be the unique play consistent with $\sigma$ such that 
  $\proj{\pi}{u}=1^{k}0^{\omega}$ and $\proj{\pi}{v}=0^{\omega}$.
  By \Cref{10-lem:obfuscate}, we have $\proj{\pi}{u'}=0^{k+2}f(C_{k})$.
  Therefore, $\pi\not\models\Psi_{4}$, a contradiction since $\sigma$ is a winning 
  distributed strategy. This concludes the proof of \Cref{10-thm:undecidable-A2}.
\end{proof}

In the proof above, we reduced the non-halting problem of a Turing machine on the empty
input word to \Cref{10-prob:existWS}.  
It is also possible to modify the proof of \Cref{10-thm:undecidable-P2bis} so that 
it applies to architecture $\archi_{2}$.

Finally, we prove that \Cref{10-prob:existWS} is undecidable as soon as the architecture
contains two players $p$, $q$ with incomparable information.  The second difficulty to
overcome is when the variables $u,v$ read by players $p,q$ respectively are not directly
written by the environment.  In that case, we have two information chains from
$\wdom(\env)$ to $u$ and $v$ respectively  There is an additional difficulty when these
two chains have a comon prefix as in architecture $\archi_{3}$ of
\Cref{10-fig:information-fork}.

\begin{theorem}\label{10-thm:undecidable-P1-incomparable-info}
  Let $\archi$ be an architecture with two players $p,q$ having incomparable information.
  \Cref{10-prob:existWS} is undecidable,
  even when restricted to synchronous distributed games over architecture 
  $\archi$ and winning conditions given in linear temporal logic.
\end{theorem}

\begin{proof}
  Let $\overline{u}=u_0u_1\dots u_n$ be an information chain to $p$ ($u_n\in\rdom(p)$) 
  which is not read by $q$ ($\{u_0,\dots,u_n\}\cap\rdom(q)=\emptyset$), and of minimal 
  length. By minimality of the chain, we get 
  $\{u_0,\dots,u_{n-1}\}\cap\rdom(p)=\emptyset$ (hence also 
  $\{u_0,\dots,u_{n}\}\cap\wdom(p)=\emptyset$).
  Similarly, let $\overline{v}=v_0v_1\dots v_m$ be an information chain to $q$
  ($v_m\in\rdom(q)$) which is not read by $p$ ($\{v_0,\dots,v_m\}\cap\rdom(p)=\emptyset$),
  and of minimal length.  As above, we get $\{v_0,\dots,v_{n-1}\}\cap\rdom(q)=\emptyset$
  and $\{v_0,\dots,v_m\}\cap\wdom(q)=\emptyset$.
  Obviously, from the definitions, $u_n\neq v_m$.  
  Still by minimality of the chains, if $u_{i}=v_{j}$ for some $i,j$ then $i=j<\min(n,m)$.
  Moreover, if $u_{j}=v_{j}$ we may use in $\overline{v}$ the prefix $u_{0}\cdots u_{j}$ 
  instead of $v_{0}\cdots v_{j}$. Therefore, we may assume without loss of generality 
  that the two chains share a common prefix (possibly empty) after which they are disjoint.
  Formally, there is $0\leq i\leq\min(n,m)$ such that, $u_j=v_j$ for all $0\leq j<i$, and
  $\{u_i, \dots, u_n\}\cap\{v_i,\dots, v_m\}=\emptyset$.
  If $i=0$, then there is no common prefix to the two chains.
  These two information chains will be used to convey the information from the environment
  to the players $p$ and $q$ in such a way that none of them know what the other is
  receiving.  Consider two other locations, $u'\in\wdom(p)$ and $v'\in\wdom(q)$.
  By~\Cref{10-thm:quotient}, we may assume that $\archi$ does not have feedback edges, so 
  $u'\notin\rdom(p)$ and $v'\notin\rdom(q)$.
  Let $C=\{u_0,\dots, u_{i-1},u_i, \dots, u_n, v_i,\dots, v_m,u', v'\}$ be the set locations
  involved in the two information chains.  

  We again reduce the non-halting problem of a Turing machine M on the empty input word
  to the existence of a winning distributed strategy.
  
  At first, in order to handle easily the common prefix in the information chains, we let
  $S_w=\mathbb{B}\times\mathbb{B}=\{0,1\}\times\{0,1\}$ for all
  $w\in\{u_{0},\ldots,u_{i-1}\}$.  We let $S_{w}=\mathbb{B}$ for all other locations
  $w\in\loc\setminus\{u_{0},\ldots,u_{i-1}\}$.  We will explain later how to adapt the
  undecidability proof when $i>0$ and $S_w=\mathbb{B}$ for all $w\in\loc$.
  We identify $(\mathbb{B}\times\mathbb{B})^\omega$ and
  $\mathbb{B}^\omega\times\mathbb{B}^\omega$. 
  If $i>0$, for an infinite sequence of global states
  $\pi\in S_\loc^\omega$ and a location $w\in\{u_{0},\ldots,u_{i-1}\}$, we write
  $\proj{\pi}{w}=(\proj{\pi}{w}^0,\proj{\pi}{w}^1)\in\mathbb{B}^\omega\times\mathbb{B}^\omega$.
  The specification is split into three parts.
  We assume without loss of generality that $m\leq n$.

  \begin{enumerate}

    \item $\varphi_1$: 
    The first part forbids to use a location outside of the information chains to
    transmit some information, \textit{i.e.}, the players will have to play only value 0 on
    locations outside $C$.  
    More precisely, $\varphi_{1}$ states that $\proj{\pi}{w}=0^\omega$ for
    all $w\in\wdom(\players)\setminus C$.

    \item $\varphi_2$: The second part of the specification imposes 
    to transmit the information from $u_{0}$ if $i>0$ (resp.\ from
    $(u_{0},v_{0})$ if $i=0$) to $(u_n,v_m)$, with an appropriate delay.
    It also forces players $p,q$ to start by writing $0^{n}$ on $u',v'$.
    This allows us to simulate the distributed game on arena $\arena_{2}$ 
    of \Cref{10-thm:undecidable-A2}.
    Formally, $\varphi_{2}$ states that 
    $\proj{\pi}{u_{n},v_{m}}=(0^{n}x,0^{m}y)$ with $x,y\in\mathbb{B}^\omega$ with
    $$
    (x,y)=(\proj{\pi}{u_0},\proj{\pi}{v_0}) \text{ if } i=0
    \qquad\text{and}\qquad
    (x,y)=\proj{\pi}{u_0}=(\proj{\pi}{u_0}^0,\proj{\pi}{u_0}^1) \text{ if } i>0 \,.
    $$

    \item $\varphi_{3}$: Finally, the last part enforces the requirements of
    \Cref{10-thm:undecidable-A2} over the locations $u_n,v_m,u',v'$ of a play 
    $\pi'\in S_{\loc}^\omega$ and after delay $n$.
    Formally, let $\pi\in S_{\loc}^\omega$ and let $\pi'=(x,y,x',y')$ be the suffix of
    $\proj{\pi}{u_{n},v_{m},u',v'}$ obtained by removing the first $n$ states:
    $\proj{\pi}{u_{n},v_{m},u',v'}\in(\mathbb{B}^{4})^{n}\pi'$.
    Then, $\varphi_{3}$ states that 
    $\proj{\pi}{u'},\proj{\pi}{v'}\in0^{n}\mathbb{B}^{\omega}$, and    
    $\pi'$ satisfies the specification $\Psi$ of \Cref{10-thm:undecidable-A2} in which all
    constraints on the projections on $u$, $v$, $u'$ and $v'$ respectively, are replaced
    by constraints on $x$, $y$, $x'$ and $y'$ respectively.
  \end{enumerate}

  We can prove now that $M$ does not halt on the empty word if and only if there is a
  winning distributed strategy for $\varphi=\varphi_1\wedge\varphi_2\wedge\varphi_3$.

  Assume that $M$ does not halt on the empty word.  Let $\sigma'=(\sigma'_p, \sigma'_q)$
  be the winning strategy for $(\arena_{2},\Psi)$ defined in the proof of
  \Cref{10-thm:undecidable-A2}.  Then, a winning strategy
  $\sigma=(\sigma_{p'})_{p'\in\players}$ on $\arena$ can be obtained by obeying
  specification $\varphi_{1}\wedge\varphi_{2}$ and mimicking $(\sigma'_{p},\sigma'_{q})$
  on the locations $\{u_{n},v_{m},u',v'\}$.
  More precisely, for all $p'\in\players$ and $w\in\wdom(p')\setminus C$ we define
  $\proj{\sigma_{p'}(z)}{w}=0$ for all $z\in S_{\rdom(p')}^{*}$ so that $\varphi_{1}$ is
  satisfied.  Next, the player between $u_{j-1}$ and $u_{j}$ with $0<j<i$ copies with
  delay 1 the pair of values of $u_{j-1}$ to $u_{j}$.  The players reading $u_{i-1}$ and
  writing to $u_{i},v_{i}$ (if $i>0$) split the pair $(b^{0},b^{1})$ of values of
  $u_{i-1}$ and write $b^{0}$ to $u_{i}$ and $b^{1}$ to $v_{i}$ with delay $1$.  The
  player between $u_{j-1}$ and $u_{j}$ with $i<j\leq n$ copies with delay 1 the value of
  $u_{j-1}$ to $u_{j}$, and similarly for the player between $v_{j-1}$ and $v_{j}$ for
  $i<j\leq m$.  We also define the strategy of the player (if any) writing to
  location $u_{n}$ (resp.\ $v_{m}$) to start by writing $0^{n}$ (resp.\ $0^{m}$).
  It is easy to see that plays consistent with the strategies as defined so far
  satisfy $\varphi_{1}\wedge\varphi_{2}$.
  Finally, we define
  \begin{align*}
    \text{for all $z\in S_{\rdom(p)}^{*}$,} &&
    \proj{\sigma_{p}(z)}{u'} &= 
    \begin{cases}
      0 & \text{if } |z|<n \\
      \sigma'_{p}(x,0^{|x|}) & \text{if } \proj{z}{u_{n}}\in\mathbb{B}^{n}x
    \end{cases}
    \\
    \text{for all $z\in S_{\rdom(q)}^{*}$,} &&
    \proj{\sigma_{q}(z)}{v'} &= 
    \begin{cases}
      0 & \text{if } |z|<n \\
      \sigma'_{q}(y,0^{|y|}) & \text{if } \proj{z}{v_{m}}\in\mathbb{B}^{n}y \,.
    \end{cases}
  \end{align*}
  Let $\pi\in S_{\loc}^\omega$ be a play of $\arena$ consistent with $\sigma$ 
  and let $\pi'=(x,y,x',y')$ be such that 
  $\proj{\pi}{u_{n},v_{m},u',v'}\in(\mathbb{B}^{4})^{n}\pi'$.
  Since the strategies $\sigma'_p,\sigma'_q$ defined in the proof of
  \Cref{10-thm:undecidable-A2} do not depend on their second argument, it is not difficult
  to check that $\pi'=(x,y,x',y')$ is a play of $\arena_{2}$ consistent with $\sigma'$,
  hence it satisfies $\Psi$.  We deduce that $\pi$ satisfies $\varphi_{3}$.  Therefore,
  the strategy $\sigma$ is winning the game $\game=(\arena,\varphi)$.

  \medskip
  
  Conversely, assume that we have a winning strategy
  $\sigma=(\sigma_{p'})_{p'\in\players}$ for the game $\game=(\arena,\varphi)$.  We 
  define below a strategy $\sigma'=(\sigma'_{p},\sigma'_{q})$ and prove that it is 
  winning the game $(\arena_{2},\Psi)$. We deduce from the proof of 
  \Cref{10-thm:undecidable-A2} that the Turing machine $M$ does not halt on the empty 
  word.
  
  To define $\sigma'_{p}$, we map an input in $(S_{u}\times S_{v'})^{*}$ of $p$ in
  $\arena_{2}$ to an input in $S_{\rdom(p)}^{*}$ of $p$ in $\arena$ and then apply
  $\sigma_{p}$.  We introduce some notation.  Again, we identify
  $(\mathbb{B}\times\mathbb{B})^{k}$ with $\mathbb{B}^{k}\times\mathbb{B}^{k}$ for all
  $k\geq0$.  An input of $p$ in $\arena_{2}$ is a pair 
  $(x,y')\in\mathbb{B}^{k}\times\mathbb{B}^{k}$ for some $k\geq0$.
  Below we use $x,y,x',y'\in\mathbb{B}^{*}$ for sequences of states of locations
  $u,v,u',v'$ respectively.  We define
  $\alpha_{p}\colon(S_{u}\times S_{v'})^{*}\to S_{\rdom(p)}^{*}$ and 
  $\sigma'_{p}\colon(S_{u}\times S_{v'})^{*}\to S_{u'}$ by
  \begin{align*}
    \proj{\alpha_{p}(x,y')}{w} &= 
    \begin{cases}
      0^{n}x & \text{if } w=u_{n} \\
      0^{n}0^{|x|} & \text{if } w\in\rdom(p)\setminus\{u_{n},v'\} \\
      0^{n}y' & \text{if } v'\in\rdom(p) \text{ and } w=v' 
    \end{cases}
    &
    \sigma'_{p}(x,y') &= \sigma_{p}(\alpha_{p}(x,y'))
    \\
    \intertext{and define $\alpha_{q}\colon(S_{v}\times S_{u'})^{*}\to S_{\rdom(q)}^{*}$ and 
    $\sigma'_{q}\colon(S_{v}\times S_{u'})^{*}\to S_{v'}$ by}
    \proj{\alpha_{q}(y,x')}{w} &= 
    \begin{cases}
      0^{n}y & \text{if } w=v_{m} \\
      0^{n}0^{|y|} & \text{if } w\in\rdom(q)\setminus\{v_{m},u'\} \\
      0^{n}x' & \text{if } u'\in\rdom(q) \text{ and } w=u' 
    \end{cases}
    &
    \sigma'_{q}(y,x') &= \sigma_{q}(\alpha_{q}(y,x')) \,.
  \end{align*}
  
  We show that $\sigma'$ is winning the game $(\arena_{2},\Psi)$.
  Let $\pi'=(x,y,x',y')\in S_{u,v,u',v'}^{\omega}$ be a play of $\arena_{2}$ consistent 
  with $\sigma'$. We have to show that $\pi'\models\Psi$.

  Let $\pi\in S_{\loc}^{\omega}$ be the unique play of $\arena$ consistent with $\sigma$
  such that $\proj{\pi}{u_{0}}=(x,0^{n-m}y)$ if $i>0$,
  $\proj{\pi}{u_{0},v_{0}}=(x,0^{n-m}y)$ if $i=0$, and $\proj{\pi}{w}=0^{\omega}$ for all
  other external inputs $w\in\wdom(\env)\setminus\{u_{0},v_{0}\}$.
  Since $\sigma$ is winning the game $(\arena,\varphi)$ we have $\pi\models\varphi$.  
  By definition of $\pi$ and since $\pi\models\varphi_{1}$ we get
  $\proj{\pi}{w}=0^{\omega}$ for all
  $w\in(\rdom(p)\cup\rdom(q))\setminus\{u_{n},v_{m},u',v'\}$.
  By definition of $\pi$ and since $\pi\models\varphi$ we get
  $\proj{\pi}{u_{n},v_{m},u',v'}=(0^{n}x,0^{n}y,0^{n}x'',0^{n}y'')$ for some 
  $x'',y''\in\mathbb{B}^\omega$.
  Now, since $\pi'$ is consistent with $\sigma'$ and $\pi$ is 
  consistent with $\sigma$, by definition of $\sigma'$ and induction on the length we 
  can easily verify that $x''=x'$ and $y''=y'$. 
  Finally, since $\pi\models\varphi_{3}$, we deduce that $\pi'\models\Psi$.

  \medskip
  
  We now explain how to adapt the proof when $S_w=\mathbb{B}$ for all $w\in\loc$.  This is
  needed only if $i>0$, \textit{i.e.}, if the information chains have a nonempty common prefix
  $u_{0},\ldots,u_{i-1}$.  The environment will write on $u_0$ the pair of values that
  will be transmitted to $u_i$ and $v_i$, and then to $u_n$ and $v_m$.  To do this, it
  will encode the sequence of pairs of bits into a sequence of bits.  As in the proof of
  \Cref{10-thm:undecidable-A1}, we encode a pair $(x,y)\in\mathbb{B}\times\mathbb{B}$ by a
  word $f(x,y)\in\mathbb{B}^5$.  
  We let $f(0,0)=\textcolor{red}{01}000$, $f(0,1)=\textcolor{red}{01}100$,
  $f(1,0)=\textcolor{red}{01}110$ and $f(1,1)=\textcolor{red}{01}111$.
  The prefix $\textcolor{red}{01}$ allows for an easy decoding: whenever we see a factor 
  $\textcolor{red}{01}$ we know that it starts an encoded value $f(x,y)$ and the next 
  three bits allow to recover the pair $(x,y)$.
  The encoding $f$ will be used for the locations on the nonempty common prefix 
  $u_{0},\dots,u_{i-1}$.
  
  Since the system is \emph{synchronous}, we also use an encoding $g(x)\in\mathbb{B}^{5}$ 
  of fixed length $5$ even for single bits $x\in\mathbb{B}$. 
  We let $g(0)=\textcolor{red}{01}000$ and $g(1)=\textcolor{red}{01}100$.
  The encoding $g$ will be used for the other locations in $C$, \textit{i.e.}, 
  $u_{i},\dots,u_{n},v_{i},\ldots,v_{m},u',v'$.
  
  The specification is modified in order to force the environment to write on location
  $u_{0}$ a sequence of encodings in $\Theta^{\omega}$ where
  $\Theta=f(\mathbb{B}\times\mathbb{B})=\textcolor{red}{01}1^{*}0^{*}\cap\mathbb{B}^{5}$.
  Thanks to the prefix $\textcolor{red}{01}$ of the encoding, it is easy to write an $\LTL$
  formula $\varphi_{0}$ which states that $\proj{\pi}{u_{0}}\in\Theta^{\omega}$.
  The specification will be 
  $\widetilde{\varphi}=\varphi_{0}\limplies
  (\widetilde{\varphi}_{1}\wedge\widetilde{\varphi}_{2}\wedge\widetilde{\varphi}_{3})$.
  The first condition is unchanged: $\widetilde{\varphi}_{1}=\varphi_{1}$.
  The second and third conditions are modified to take the encodings into account.
  
  Thanks to $\varphi_{0}$ we may assume that the input provided by the environment on
  location $u_{0}$ is a sequence
  $z=f(x_{1},y_{1})f(x_{2},y_{2})f(x_{3},y_{3})\cdots\in\mathbb{B}^\omega$ of encodings.
  The player between $u_{j-1}$ and $u_{j}$ for $0<j<i$ is expected to copy with delay 1
  the sequence of bits from $u_{j-1}$ to $u_{j}$.  Hence, the sequence of bits on location
  $u_{i-1}$ is expected to be $\textcolor{blue}{0^{i-1}}z$.  
  
  Now the player $p_{i}$ reading $u_{i-1}$ and writing to $u_{i}$ should decode the
  sequence $z$ into $x=g(x_{1})g(x_{2})g(x_{3})\cdots$.  To do so, delay 3 is required, so
  the sequence of bits written by $p_{i}$ on $u_{i}$ should be
  $\textcolor{blue}{0^{i-1+3}}x$.  Similarly, the sequence of bits written on $v_{i}$ should
  be $\textcolor{blue}{0^{i-1+n-m+3}}y$ where $y=g(y_{1})g(y_{2})g(y_{3})\cdots$.

  For instance, if the sequence of bits written by the environment on location $u_{0}$ is
  $z=f(1,0)f(1,1)f(0,1)f(1,1)f(0,0)\cdots$ then the sequences of bits on $u_{i-1}$, $u_{i}$ 
  and $v_{i}$ are given below:
  \begin{align*}
    u_{i-1} &\colon\quad \textcolor{blue}{\overbrace{0\cdots 0}^{i-1}} \textcolor{red}{01} 110 
    \textcolor{red}{01} 111 \textcolor{red}{01} 100 
    \textcolor{red}{01} 111 \textcolor{red}{01} 000 ~ \cdots
    \\
    u_{i} &\colon\quad \textcolor{blue}{0\cdots 0 000} 
    \textcolor{red}{01} 1 \textcolor{blue}{00} 
    \textcolor{red}{01} 1 \textcolor{blue}{00} 
    \textcolor{red}{01} 0 \textcolor{blue}{00} 
    \textcolor{red}{01} 1 \textcolor{blue}{00} 
    \textcolor{red}{01} 0 \textcolor{blue}{00} ~ \cdots
    \\
    v_{i} &\colon\quad \textcolor{blue}{0\cdots 0 \underbrace{0\cdots 0}_{n-m} 000} 
    \textcolor{red}{01} 0 \textcolor{blue}{00} 
    \textcolor{red}{01} 1 \textcolor{blue}{00} 
    \textcolor{red}{01} 1 \textcolor{blue}{00} 
    \textcolor{red}{01} 1 \textcolor{blue}{00} 
    \textcolor{red}{01} 0 \textcolor{blue}{00} ~ \cdots
  \end{align*}
  The players should then transmit the values from $u_{i}$ to $u_{n}$ with delay $n-i$ 
  and from $v_{i}$ to $v_{m}$ with delay $m-i$. Therefore, $\widetilde{\varphi}_{2}$ 
  states that, if $\proj{\pi}{u_{0}}=f(x_{1},y_{1})f(x_{2},y_{2})f(x_{3},y_{3})\cdots$ 
  then $\proj{\pi}{u_{n}}=\textcolor{blue}{0^{n+2}}g(x_{1})g(x_{2})g(x_{3})\cdots$ and
  $\proj{\pi}{v_{m}}=\textcolor{blue}{0^{n+2}}g(y_{1})g(y_{2})g(y_{3})\cdots$. Thanks to 
  our encodings, this can be written in $\LTL$. 
  
  Similarly, the third condition is modified to take the encodings into account.
  The formula $\widetilde{\varphi}_{3}$ requires that 
  $\proj{\pi}{u'},\proj{\pi}{v'}\in\textcolor{blue}{0^{n+2}}(g(\mathbb{B}))^{\omega}$
  and states that if 
  \begin{align*}
    \proj{\pi}{u_{n}} &= 0^{n+2}g(x_{1})g(x_{2})g(x_{3})\cdots &
    \proj{\pi}{u'} &= 0^{n+2}g(x'_{1})g(x'_{2})g(x'_{3})\cdots \\
    \proj{\pi}{v_{m}} &= 0^{n+2}g(y_{1})g(y_{2})g(y_{3})\cdots &
    \proj{\pi}{v'} &= 0^{n+2}g(y'_{1})g(y'_{2})g(y'_{3})\cdots
  \end{align*}
  then the specification $\Psi$ of \Cref{10-thm:undecidable-A2} is satisfied by
  $\pi'=(x,y,x',y')$ with $x=x_{1}x_{2}x_{3}\cdots$, $y=y_{1}y_{2}y_{3}\cdots$,
  $x'=x'_{1}x'_{2}x'_{3}\cdots$ and $y'=y'_{1}y'_{2}y'_{3}\cdots$.  Again, thanks to our
  encoding $g$, the formula $\widetilde{\varphi}_{3}$ can be written in $\LTL$.
  This concludes the proof of \Cref{10-thm:undecidable-P1-incomparable-info}.
\end{proof}

In the proof above, we reduced the non-halting problem of a Turing machine on the empty
input word to \Cref{10-prob:existWS}.  
It is also possible to modify the proof of \Cref{10-thm:undecidable-P2bis} and reduce the
halting problem of a Turing machine on the empty input word to
\Cref{10-prob:existWSfiniteMemory} (resp.\ \Cref{10-prob:existWS}) for synchronous
distributed games on any architecture with two players having incomparable information.
Hence, \Cref{10-prob:existWSfiniteMemory} is also undecidable when restricted to 
synchronous distributed games on a fixed architecture $\archi$ with two players having 
incomparable information.



\section*{Bibliographic references}
\label{10-sec:references}

Distributed games have been studied by Peterson and Reif in
1979~\cite{Peterson.Reif:1979}, when they introduced a new computation model:
multiple-person alternation machines.  In 1990, Pnueli and
Rosner~\cite{Pnueli.Rosner:1990} studied \emph{synchronous} distributed games in the
framework of Church's synthesis problem.  Their view of the problem inspired the
definitions of games we have adopted in this chapter.  In this seminal paper, Pnueli and
Rosner give two important results, one positive, one negative.  First, they prove that a
synchronous distributed game over architecture $\archi_1$ of
\Cref{10-fig:information-fork} is undecidable.
Their proof (similar to our proof of \Cref{10-thm:undecidable-P2bis}) is inspired from an
earlier proof from Peterson and Reif~\cite{Peterson.Reif:1979}, and shows that the
problem is not co-recursively enumerable.  Actually, they show that solving the game with
\emph{finite-memory} (\Cref{10-prob:existWSfiniteMemory}) is recursively
enumerable-complete.  The proof of Pnueli and Rosner also shows that solving the
game (\Cref{10-prob:existWS}) is undecidable (not co-recursively enumerable).
We extend their idea in \Cref{10-thm:undecidable-A1} and show that \Cref{10-prob:existWS}
is actually neither recursively enumerable, nor co-recursively enumerable.  Moreover, our
proof gives an example of a distributed game for which there exists a winning strategy,
but no winning strategy with finite memory (\Cref{10-cor:WSwithInfiniteMem}).

\begin{figure}[tbp]
  \centering
  \includegraphics[page=29]{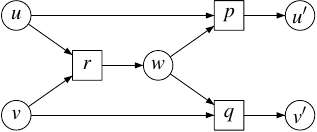}
  \hfil
  \includegraphics[page=30]{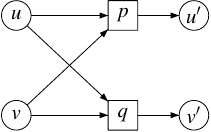}
  \caption{Left: architecture $\archi$ is undecidable for winning conditions on \emph{all}
  locations since it has an information fork.
  \\
  Right: architecture $\archi'$ is decidable since both players are equally informed.
  \\
  But, assuming delay 0 and that all locations have boolean values, the two architectures
  are equivalent (and decidable) for winning conditions restricted to \emph{external}
  locations.}
  \label{10-fig:adequate}
\commentAlt{Figure~\ref{10-fig:adequate}: Two diagrams illustrating a transformation from a detailed network of nodes and transitions to a more abstract, combined representation. See long description.}
\commentLongAlt{Figure~\ref{10-fig:adequate}: The image displays two distinct network diagrams, implying a transformation or simplification from the left to the right.

Left Diagram:

Two circular nodes, 'u' and 'v', serve as inputs.
From 'u', a direct arrow points to a square node 'p'.
From 'v', a direct arrow points to a square node 'q'.
Both 'u' and 'v' also point to a square node 'r'.
From 'r', an arrow points to a circular node 'w'.
From 'w', two arrows emanate: one to 'p' and another to 'q'.
From 'p', an arrow points to a circular node 'u''.
From 'q', an arrow points to a circular node 'v''.
Right Diagram:

Two circular nodes, 'u' and 'v', serve as inputs.
From 'u', an arrow points to a square node 'q'.
From 'v', an arrow points to a square node 'p'.
From 'p', an arrow points to a circular node 'u''.
From 'q', an arrow points to a circular node 'v''.
The right diagram appears to be a simplified version of the left, where the intermediate nodes 'r' and 'w' are removed, and the connections from 'u' and 'v' are re-routed to 'p' and 'q' in a cross-over pattern.}
\end{figure}

At the other end of the spectrum, Pnueli and Rosner~\cite{Pnueli.Rosner:1990} show that
solving synchronous distributed games over \emph{hierarchical architectures} is a
decidable problem, when the specification is given in $\LTL$.  Hierarchical architectures
are a bit different from the games with total information preorder defined in this
chapter.  They are formed by merging players, as we did in
\Cref{10-sec:reduction-equally-informed}, except that players are not merged when they are
equally informed, but when the group of players is \emph{adequately connected}.
A group of players is adequately connected,
if there exists a way to transmit the values of the locations from the read domain of the
group to the locations in the write domain of the group, simply by boolean combinations of
values. For instance, assuming delay 0 and that all locations have boolean values, the 
architecture on the left of \Cref{10-fig:adequate} is adequately connected (player $r$ 
writes to $w$ the xor of $u$ and $v$).
An architecture is hierarchical if the reduced architecture in which the
processes are merged according to adequate connectivity is a pipeline (\textit{i.e.}, the players
are organized in a chain $p_1,\ldots,p_k$ such that the read domain of player $p_{i}$ with
$1<i\leq k$ is contained in the write domain of the previous player $p_{i-1}$).
Note that a pipeline is a special case of an architecture with total information preorder.

Several modelling choices have been made by Pnueli and Rosner.  Among them, three aspects
are different from what we have presented.
\begin{itemize}
  \item They have assumed that the read domains of the players are pairwise disjoint,
  \textit{i.e.}, each location is single-read and single-write.
  
  \item They have considered that the players reacted without any delay (in this chapter
  we used delay 1), and hence have restricted their work to acyclic architectures.

  \item The winning condition is not expressed over \emph{all} locations but only over
  \emph{external} locations.
\end{itemize}
Only the third restriction has an impact on the decidability results.  

In a following paper on the subject, Kupferman and Vardi~\cite{Kupferman.Vardi:2001}
dropped the first constraint: some locations may be read by several players, but they stay
single-write.  They also introduced delays and they considered architectures that may
contain cycles.
In general, delays in the architecture do not change the decidability status of the
corresponding distributed game (Gastin, Sznajder and Zeitoun have shown
in~\cite{Gastin.Sznajder.ea:2009} how to adapt the tree automata constructions to
decide architectures with arbitrary delays).

The major contribution of Kupferman and Vardi~\cite{Kupferman.Vardi:2001} is a new proof
technique for decidability using constructions on tree automata, which inspired the
construction presented in \Cref{10-sec:reduction-most-informed} for removing the most
informed player.  Using their new technique, Kupferman and Vardi proved that solving
synchronous distributed games is decidable for pipeline, two-way pipeline and one-way
ring architectures; and winning conditions defined in the \emph{branching-time} logic
$\mathsf{CTL}^{*}$ over \emph{all} locations.

As explained in \Cref{10-sec:algo}, the decision procedure based on tree automata starts
by turning the winning condition into a tree automaton that accepts the set of winning
\emph{global} strategies.  Hence, the extension from linear-time to branching-time winning
conditions comes at almost no cost.  However, in the setting of games, the winning
condition usually defines the subset of winning plays.  This is why in this chapter we
considered (linear-time) $\omega$-regular conditions over sequences of global states, and
not branching-time specifications.

Finkbeiner and Schewe~\cite{Finkbeiner.Schewe:2005} have generalized these results and
established a decidability criterion for winning conditions\footnote{branching time
specifications defined by $\mu$-calculus formulas} on \emph{all} locations: solving a
distributed game is decidable if and only if the information preorder is total.

A decision procedure which applies to winning conditions which may depend on \emph{all}
locations as in \cite{Kupferman.Vardi:2001,Finkbeiner.Schewe:2005} is indeed more
powerful than if the winning condition is restricted to \emph{external} locations as
in~\cite{Pnueli.Rosner:1990}.  On the other hand, an undecidability result is stronger
when the winning condition is restricted to \emph{external} locations.  Moreover, there
are architectures such as the hierarchical architectures of \cite{Pnueli.Rosner:1990},
that are decidable for \emph{external} winning conditions but undecidable for full winning
conditions on \emph{all} locations.
For instance, assuming\footnote{We assume delay 0 to simplify the figure, but we get a
similar example with delay 1 by adding a player and a location between $u$ and $p$, and
similarly between $v$ and $q$.} delay 0 and that all locations have boolean values, the
architecture on the left of \Cref{10-fig:adequate} is decidable for winning conditions
restricted to \emph{external} locations (it is hierarchical).  But it is undecidable for
winning conditions on all locations (it suffices to force location $w$ to always have
value $0$ to be in the situation of the undecidable architecture $\archi_{1}$ of
\Cref{10-fig:information-fork}).

Observe that the decidability criterion of~\cite{Finkbeiner.Schewe:2005} is only valid
for full winning conditions.  As seen above, if we restrict to external winning
conditions, hierarchical architectures are decidable even when the information preorder is
not total.  Gastin, Sznajder and Zeitoun (\cite{Gastin.Sznajder.ea:2009}) have
generalized the notion of hierarchical architectures to obtain a class of distributed
games that are decidable for external winning conditions.  Establishing a criterion for
decidable architectures with external winning conditions is still an open problem.

Madhusudan and Thiagarajan~\cite{Parthasarathy.Thiagarajan:2001} studied another restriction:
\emph{local} winning conditions in the context of distributed controller
synthesis.
In that case, the winning condition is a conjunction of \emph{local}
conditions for each player.  A local condition is expressed only over the locations this
player can read or write.  Such a restriction prevents to express the winning condition of
\Cref{10-thm:undecidable-A1}.  However, they show that, for local winning conditions, the
architectures that are decidable are exactly the ones where each connected component is
a \emph{clean pipeline} (\textit{i.e.}, a pipeline in which the read domain of the last player of
the chain may include locations written by the environment, in addition to the locations
written by its predecessor).

Mohalik and Walukiewicz~\cite{Mohalik.Walukiewicz:2003} have presented a different
model of distributed games, that is able to encode both \emph{synchronous} games and
\emph{asynchronous} games (the latter are out of the scope of this chapter).  To prove
decidability, they use other techniques than the ones given in this chapter.
When two players are totally informed, they are merged into a single player
(\textsc{Division} operation).  Another operation that is necessary to solve the games is
the \textsc{Gluing} operation, that allows a player to become totally informed, by
modifying the arena through a subset construction.  In the specific case of synchronous
games, the winning condition is encoded in player $0$, which is totally informed.  When the
information preorder of the players is total, alternating between the division operation
and the gluing operation allows to obtain a classical two-player game.  The most informed
player and player $0$ (monitoring the winning condition) are merged together, giving a new
player $0$ (hence a new winning condition).  Then, the next most informed player is not
necessarily totally informed about the different states of the new player $0$.  The
\textsc{Gluing} operation will modify the arena in a way that will allow the most informed
player to become totally informed, allowing the \textsc{Division} operation to be applied
again.  The players will be successively merged with player $0$ until we reach a two-player game:
player $0$ against an environment.

Results presented in this chapter come from different papers.
\Cref{10-lem:combined-strategies} has already been stated, without proof,
in~\cite{Pnueli.Rosner:1990}, while the definition of the strategy $\sigma_P$ for a team
$P$ of players has been given in~\cite{Kupferman.Vardi:2001}.  The notion of information
preorder defined in \Cref{10-sec:information-preorder} comes
from~\cite{Finkbeiner.Schewe:2005}.  The merging of equally informed players in
\Cref{10-sec:reduction-equally-informed} comes from the general decidability result
described in~\cite{Finkbeiner.Schewe:2005}, while the second reduction described in
\Cref{10-sec:reduction-most-informed} comes mostly from~\cite{Kupferman.Vardi:2001}, and
was reused for the general result of~\cite{Finkbeiner.Schewe:2005}.

The complexity lower bound explained in \Cref{10-sec:complexity} was sketched
in~\cite{Peterson.Reif:1979} for the model of multiple-person alternation machines
(private or blindfold).  The fact that these proof ideas can be adapted to the framework
of synchronous distributed games was announced in~\cite{Pnueli.Rosner:1990} (and cited in
following papers), but we are not aware of a place where the full proof was described
until now.  For the negative results, \Cref{10-thm:undecidable-A1} is our own.  As we have
said, a variation of this proof was already published in~\cite{Pnueli.Rosner:1990},
allowing to conclude that solving a distributed game, and solving a game with
finite-memory was not co-recursively enumerable.  In \cite{Finkbeiner.Schewe:2005},
another proof was given to obtain undecidability when the winning condition is given in
$\mathsf{CTL}$.  \Cref{10-thm:undecidable-A2} comes from~\cite{Finkbeiner.Schewe:2005},
with a slightly different proof, along with the general result given in
\Cref{10-thm:sync-undec-general-case}.

\section*{Acknowledgments}
This work was partially supported by CNRS IRL 2000 ReLaX.


\part{Infinite}
\label{part:infinite}

\ifpictures
\includepdf{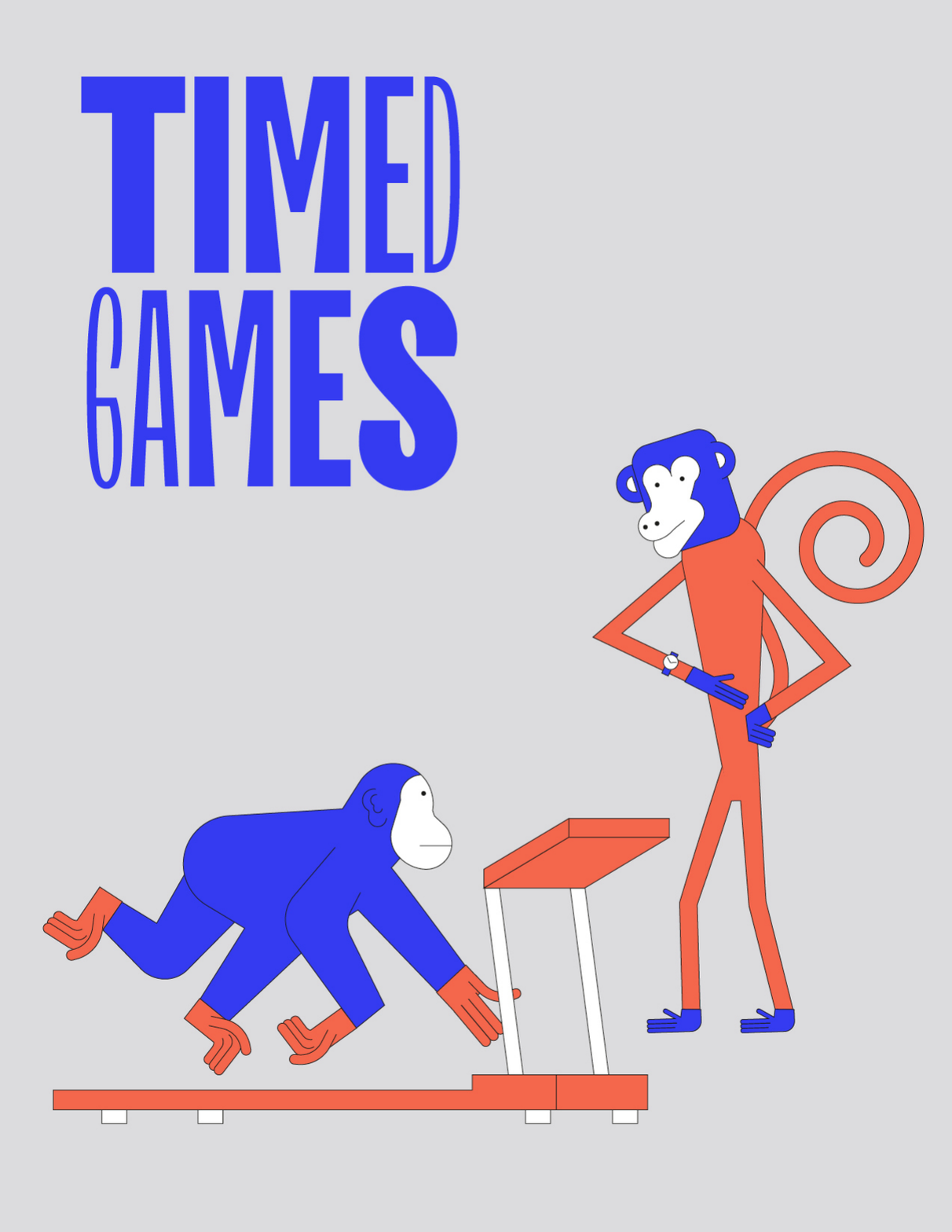}
\fi
\author[Nicolas Markey, Ocan Sankur]{Nicolas Markey, Ocan Sankur}
\copyrightline{Copyright by Nicolas Markey and Ocan Sankur 2025, to be published by Cambridge University Press in the volume \textit{Games on Graphs} edited by Nathana\"el Fijalkow}

\chapter{Timed Games}
\chapterauthor{Nicolas Markey, Ocan Sankur}
\label{11-chap:timed}

\newcommand{\Realnn}{\mathbb{R}_{\geq 0}}
\newcommand{\Clocks}{\mathcal{C}}
\newcommand{\TA}{\ensuremath{\mathcal{A}}}
\newcommand{\Locs}{\mathcal{L}}
\newcommand{\Clocksz}{\mathcal{C}_0}
\newcommand{\calQ}{\mathcal{Q}}
\newcommand{\state}{\mathsf{state}}
\newcommand{\trans}{\mathsf{trans}}
\newcommand{\post}{\mathsf{post}}
\newcommand{\step}{\mathsf{step}}

\newcommand{\postta}{\ensuremath{\mathsf{Post}}}
\newcommand{\preta}{\ensuremath{\mathsf{Pre}}}
\newcommand{\unreset}{\ensuremath{\mathsf{Unreset}}}
\newcommand{\posttime}{\ensuremath{\mathsf{Post}}_{\geq 0}}
\newcommand{\pretime}{\ensuremath{\mathsf{Pre}}_{\geq 0}} 
\newcommand{\reset}{\mathsf{Reset}}
\providecommand{\sem}[1]{\ensuremath{#1}}
\providecommand{\size}[1]{\ensuremath{|#1|}}

\newcommand{\predc}{\mathsf{Pred}_c}
\newcommand{\predt}{\mathsf{Pred}_{\geq 0}} 
\newcommand{\predu}{\mathsf{Pred}_u}

\newcommand{\calP}{\mathcal{P}}
\newcommand{\calC}{\mathcal{C}}
\newcommand{\calT}{\mathcal{T}}
\newcommand{\Dep}{\textsf{Dep}}
\newcommand{\Wait}{\textsf{Wait}}
\newcommand{\Passed}{\textsf{Passed}}
\providecommand{\Act}{\textsf{Act}}
\renewcommand{\Act}{\textsf{Act}}

\newcommand{\EA}{E_{\Adam}}
\renewcommand{\EE}{E_{\Eve}}

\newcommand{\zone}[1]{\ensuremath{\left\llbracket#1\right\rrbracket}}



The ability to model real-time constraints is crucial when automata
and games are used for verification and synthesis. Timed
automata~\cite{Alur.Dill:1994} are a model of choice for reasoning about
real-time systems: they~extend finite-state automata with a
finite number of \emph{clocks}, which are real-valued variables all
growing at the same rate, used to measure and constrain the elapse of
time between various transitions in the automaton. Because these
clocks can take arbitrary non-negative values, the~set of
configurations of a timed automaton is infinite. Still, reachability
(and~many other problems) remain decidable in timed
automata. The~interested reader can find more background
in~\cite{Alur.Dill:1994}, but we will try to keep our presentation self-contained.

In this chapter, we consider game models based on timed automata;
we~call them \emph{timed games} throughout this chapter. In~timed
games, besides choosing which transitions should be played,
the~players also decide how much time will elapse before each
transition. The elapsed time is determined using clocks, and the edges have
guards which determine clock values for which the edge can be taken.



\begin{example}\label{11-ex:intro}
\Cref{11-fig:ta1} contains a timed game with clock~$x$, where Eve's objective is to reach the vertex~$G$.
We will define these arenas as concurrent arenas: dashed edges belong to Adam,
and plain edges to Eve. Both players can take any edge at any time as long as
the guard is satisfied. For instance, Eve's edge from $\ell_1$ to~$\ell_2$ is only available if clock~$x$ has value at most~$1$, while Adam's edge from~$\ell_1$ to~$\ell_3$ is available if~$x$ is less than~$1$. 

\begin{figure}[ht]
  \centering
  \begin{tikzpicture}[node distance=2.5cm,auto]
    \node[state] at (1,0) (l1) {$\ell_1$};
    \node[state, below of=l1, node distance=1.5cm] (l5) {$\ell_5$};

    \node[state,right of=l1] (l2) { $\ell_2$};
    \node[state, accepting, below of=l2, node distance=1.5cm] (g) {G};
    \node[state,right of=l2] (l3) {$\ell_3$};
    \node[state,right of=l3] (l4) {$\ell_4$};
    \path[initial]
    (l1.-135) edge +(-135:3mm);
    \path[arrow]
    (l1) edge node[above,pos=.65] {$x\leq 1$} (l2)
    (l2) edge[dashed] node[above,pos=.35] {$x< 1$} (l3)
    (l3) edge node[above] {} (l4)
    (l4) edge[bend left] node[below] {$x\leq 1$} (l2)
    (l1) edge[dashed, bend left] node[above] {$x< 1,x:=0$} (l3)
    (l1) edge[dashed] node[left] {$x>1$} (l5)
    (l2) edge node[right] {$x\geq 2$} (g);
  \end{tikzpicture}
  \caption{Timed game~$\TA_1$.}
  \label{11-fig:ta1}
\commentAlt{Figure~\ref{11-fig:ta1}: A directed graph with five circular nodes (l1 to l5) and one double-ringed node (G), showing transitions based on conditions related to variable 'x'. See long description.}
\commentLongAlt{Figure~\ref{11-fig:ta1}: The image displays a directed graph representing a state machine or automaton.

The starting point is indicated by an incoming arrow to node 'l1' (circle).
From 'l1', there are three outgoing edges:
A solid arrow to 'l2' (circle), labeled 'x <= 1'.
A dashed arrow to 'l5' (circle), labeled 'x > 1'.
A dashed arrow curving to 'l3' (circle), labeled 'x < 1, x := 0'.
From 'l2' (circle):
A solid arrow points downwards to a double-ringed node 'G', labeled 'x >= 2'.
A dashed arrow points to 'l3' (circle), labeled 'x < 1'.
A solid arrow points to 'l4' (circle).
From 'l3' (circle), a solid arrow points to 'l4' (circle).
From 'l4' (circle), a solid curved arrow points back to 'l2' (circle), labeled 'x <= 1'.}
\end{figure}
\end{example}

In the next section, we define timed games and their semantics formally.
Then we introduce some classical tools needed to reason
about the space of clock valuations, and finally present an efficient
algorithm for deciding the winner in timed games with reachability
objectives.

\section{Notations}
\label{11-sec:notations}
We~fix a finite set~$\Clocks$ of clock variables to be used
in our timed games. Elements of~$\Realnn^\Clocks$, which assign a
value to each clock, are called \emph{valuations}.

Clocks will be used to define \emph{clock constraints}, which in turn
are used in timed automata to restrict the set of allowed behaviours:
edges are decorated with clock constraints defining conditions for
their availability.
An~\emph{atomic clock constraint} is a formula of the form $k \preceq
x \preceq' l$ or $k \preceq x - y \preceq' l$ where $x,y \in \Clocks$,
$k,l \in \mathbb{Z}\cup\{-\infty,\infty\}$ and
${\mathord\preceq,\mathord\preceq' \in
  \{\mathord<,\mathord\leq\}}$. A~\emph{clock constraint} is a
conjunction of atomic clock constraints.  A~valuation~$\nu\colon
\Clocks\to\Realnn$ satisfies a clock constraint~$g$, denoted $\nu \models g$,
if~all atomic clock constraints are satisfied when each $x\in \Clocks$
is replaced with its value~$\nu(x)$.  We~write $\Phi_\Clocks$ for the
set of clock constraints built on the clock set~$\Clocks$.

For a subset $R\subseteq \Clocks$ and a valuation~$\nu$, we~denote
with ${\nu[R \leftarrow 0]}$ the valuation defined by ${\nu[R
    \leftarrow 0](x) = 0}$ if $x \in R$ and ${\nu[R\leftarrow 0](x) =
  \nu(x)}$ otherwise. Given $d \in \mathbb{R}_{\geq 0}$ and a
valuation~$\nu$, the~valuation~$\nu+d$ is defined by $(\nu+d)(x) =
\nu(x)+d$ for all $x\in \Clocks$. We~extend these operations to sets
of valuations in the obvious~way.


%


We now formally define \emph{timed games}, which are finite
representations that define infinite-state, non-stochastic concurrent
games. 

\begin{definition}
  A~\emph{timed arena} $\calT$ is a tuple $(\Locs,
  \Clocks,\EE,\EA, c)$, where $\Locs$ is a finite set of locations,
  $\Clocks$~is a finite set of clocks, $\EE,\EA \subseteq \Locs \times
  \Phi_\Clocks \times 2^\Clocks \times \mathcal{L}$ are the sets of
  edges respectively controlled by Eve and~\Adam,
  and $c\colon \EE\cup\EA \to C$ is the colouring function.
  A~\emph{timed game} is a pair~$(\calT,\Omega)$ where
  $\Omega \subseteq C^\omega$ a qualitative objective.
\end{definition}


A~configuration of such a timed automaton is a pair~$(\ell,\nu)$ where
$\ell\in\Locs$ and $\nu\colon \Clocks\to\Realnn$.  The~set of
configurations is the set of vertices of the infinite-state game
defined by~$\calT$.

The~actions of both players are pairs~$(d,e)$ where $d\in\Realnn$ is a
delay they want to wait before playing their controlled
action~$e$. Action~$(d,e)$ is available for Eve (resp. Adam) in
configuration~$(\ell,\nu)$ if $e\in\EE(\ell)$ (resp.~$e\in\EA(\ell)$)
and, writing $e=(\ell,g,R,\ell')$, if additionally $\nu+d\models g$;
in~other terms, the~clock constraint (called~\emph{guard}~hereafter)
on~$e$ must hold true under the clock valuation reached after elapsing
$d$ time units. We~add an extra dummy action~$\bot$ for the case where
some player has no available action (this~action is only available if
no other actions~are).

Once both players have selected an action available from
configuration~$(\ell,\nu)$, the action~$(d,e)$ with smallest delay is
performed (by breaking ties in favor of Adam), leading to configuration $(\ell',(\nu+d)[R\leftarrow 0])$:
this corresponds to letting $d$ time units elapse, thereby reaching
configuration~$(\ell,\nu+d)$, and to following edge~$e$ (which by
construction is available from~$(\ell,\nu+d)$).  We~define
$\step((\ell,\nu),(d,e))$ for the configuration
$(\ell',(\nu+d)[R\leftarrow 0])$ reached from~$(\ell,\nu)$ by applying
action~$(d,e)$.



This definition captures the concurrent nature of the interaction
between a controller (Eve) and its environment (Adam) in a real-time
system, since none of the players knows in advance how long the
opponent will want to wait before performing her transition.
The~semantics of a timed arena $(\Locs, \Clocks,\EE,\EA)$ can then
formally be defined in terms of a concurrent arena (following the
definition of \Cref{9-sec:notations}).  The~underlying
graph~$(V,E)$ is such that $V=\Locs\times \Realnn^{\Clocks}$, and $E=
V\times C\times V$; the~set of actions of Eve is $\Realnn\times\EE$, and that
of Adam is $\Realnn\times\EA$; finally, the transition function,
which is not stochastic in our case, maps any
configuration~$(\ell,\nu)$ and pair of actions~$(d_\Eve,e_\Eve)$ and
$(d_\Adam,e_\Adam)$ to the edge $((\ell,\nu),\gamma,
\step((\ell,\nu),(d,e)))$, where $(d,e)=(d_\Eve,e_\Eve)$ and $\gamma=c(e_\mEve)$
if
$d_\Eve<d_\Adam$,  and $(d,a)=(d_\Adam,e_\Adam)$ and $\gamma=c(e_\mAdam)$ otherwise.

A~path in a timed arena~$\calT$ then is a path in its
associated infinite-state concurrent arena. The~qualitative
objective~$\Omega$ can then be evaluated along runs of a timed arena in
the natural way.





Contrary to~\Cref{8-chap:concurrent}, in this chapter we only consider deterministic strategies\footnote{Adding randomization over the infinite sets of actions is beyond the scope of this chapter.}.
  As~a result, timed games are not
determined, as illustrated in the following example.

\begin{figure}[ht]
  \centering
  \begin{tikzpicture}[node distance=2cm,auto]
    \tikzstyle{every state}=[minimum size=17pt,inner sep=0pt]
    \node[state] at (1,0) (A) {$\ell_0$};
    \node[state,right of=A] (B) { $\ell_1$};
    \node[state,left of=A] (C) {$\ell_2$};
    \path[initial] (A.-135) edge +(-135:3mm);
    \path[arrow] 
    (A) edge node[above] {$0<x<1$} (B)
    (A) edge[dashed] node[above] {$0<x < 1$} (C);
  \end{tikzpicture}
  \caption{Timed arena~$\TA_2$. Solid arrow is Eve's, dashed one is Adam's.}
  \label{11-fig:ta2}
\commentAlt{Figure~\ref{11-fig:ta2}: A linear sequence of three circular nodes, l2, l0, and l1, with arrows showing transitions based on a numerical condition for 'x'.}
\commentLongAlt{Figure~\ref{11-fig:ta2}: The image displays three circular nodes arranged horizontally: 'l2' on the left, 'l0' in the middle, and 'l1' on the right. An incoming arrow points to 'l0', indicating it as a starting point. From 'l0', a solid arrow points to 'l1', labeled '0 < x < 1'. Also from 'l0', a dashed arrow points to 'l2', also labeled '0 < x < 1'.}
\end{figure}

\begin{example}[Timed Games are not determined]
  In the timed arena~$\TA_2$ defined in~\Cref{11-fig:ta2}, from configuration~$(\ell_0,\vec{0})$, Eve does not have a winning strategy to reach location~$\ell_1$, but Adam does not have a winning strategy either to avoid~$\ell_1$.
  In fact, available moves for both players consist in~$(d,(\ell_0,\ell_1))$ with~$0< d<1$ for Eve, and~$(d',(\ell_0,\ell_2))$ with~$0< d' < 1$ for~\Adam.
  Thus, for any particular delay~$0<d<1$ chosen by~\Eve, Adam has a possible delay $d<d'<1$ which leads to~$\ell_2$, which is losing for~\Eve.
  This shows that Eve does not have a winning strategy.
  The argument is however symmetric, and Adam also does not have a winning strategy to avoid~$\ell_1$.
  Timed games are thus non determined.
\end{example}






\begin{example}[Winning strategy on running example]
Let us consider again the example of \Cref{11-fig:ta1} and see whether Eve
has a winning strategy from the initial configuration.
At the initial configuration~$\ell_1,x=0$, Eve needs to make a move towards~$\ell_2$
while~$x\leq 1$ since whenever~$x>1$, Adam can move to~$\ell_5$ which guarantees Eve's lost.
Assume the game proceeds to~$\ell_2$ with any value~$x\leq 1$. Here, Eve can try to wait
until~$x\geq 2$ and go to~$G$. However, if~$x<1$, then Adam can move to~$\ell_3$.
From this location, Eve can guarantee to come back to~$\ell_2$ with~$x=1$, and then move to~$G$ and win the game.
Assume now that from~$\ell_1$, the game proceeds to~$\ell_3$ with~$x=0$ due to Adam's move.
Again, Eve can wait for 1 time unit, and go back to~$\ell_2$ with~$x=1$, and win the game.
Hence, Eve has a winning strategy from $\ell_1$ for all~$x\leq 1$, and from~$\ell_2$
for all values of~$x$.
\end{example}

In this chapter, the main problem we are interested in is determining
whether Eve has a strategy for her reachability objective.
Let~$\vec{0}$ denote the clock valuation assigning~$0$ to all clocks.

\decpb[Solving a timed reachability game]{A timed arena~$\calT$, initial location~$\ell_0$, and a reachability objective~$\Reach(\Win)$}{Does Eve has a winning strategy in~$(\calT,\Reach(\Win))$ from configuration~$(\ell_0,\vec{0})$.}

The difficulty of this problem is that the concurrent
game~$((V,E),\Delta,\Omega)$ has an infinite state-space, and players
have infinitely many actions.  We~thus start by studying a data
structure to represent sets of states and operations to compute
successors and predecessors on these sets.  We then give two
algorithms to solve the above problem.  We~also show how such a
strategy for Eve can be computed and finitely represented.

\section{State-space representation}
\label{11-sec:state_space_representation}
We introduce a data structure to represent sets of clock
valuations and manipulate them efficiently in order to compute
successors and predecessors in a given timed game. This will allow us
to use a fixed-point characterisation of the winning states analogous to
that in finite games as in \Cref{3-chap:regular}.

A~\emph{zone} is any subset of~$\Realnn^\Clocks$ that can be defined
using a clock constraint (hence a~zone is convex).  We will see that
sets of states that appear when exploring the state space of a timed
game can be represented using zones.  We~use the
\emph{difference-bound matrices} to represent zones:
this is one of the main data structures used in timed-automata
verification~\cite{Dill:1990}. The~idea is to store, in a matrix,
upper bounds on clocks and on differences of pairs of clocks.
Formally, given a clock
set~$\Clocks=\{x_1,\ldots,x_m\}$, we define $\Clocksz = \Clocks \cup \{x_0\}$
where~$x_0$ is seen as a
special clock which is always~$0$.
A~difference-bound matrix~(DBM) is a $|\Clocksz|\times |\Clocksz|$
matrix with coefficients in $\{\mathord\leq,\mathord<\} \times
\mathbb{Z}$.  For any DBM~$M$, the $(i,j)$-component of the matrix~$M$
will be written $(\prec^M_{i,j}, M_{i,j})$ where $\prec^M_{i,j}$ is
the inequality in~$\{\mathord\leq,\mathord<\}$, and $M_{i,j}$ the
integer coefficient. A~DBM~$M$ defines the zone
\[
  [M] = \Bigl\{v\in \Realnn^{\Clocks}\Bigm|
  \bigwedge_{0\leq i,j \leq |\Clocksz|} v(x_i)-v(x_j) \prec^M_{i,j} M_{i,j}\Bigr\},
\]
where $v(x_0) = 0$.

\begin{example}[An example of a DBM]]
\label{11-ex:DBM}
  Consider the clock set $\Clocks=\{x_1,x_2\}$
  and the zone $Z$ defined by 
  $x_1\leq 1 \land x_1-x_2 \leq 0 \land x_2\leq 3\land x_2-x_1 \leq 2$, which can be
  written as the following DBM:
  \begin{figure}
  \[
    M=\begin{pmatrix}
      (\leq,0) &(\leq,0) &(\leq,0)\\
      (\leq,1) &(\leq,0) &(\leq,0)\\
      (\leq,3) &(\leq,2) &(\leq,0)
    \end{pmatrix}
    \qquad
  \tikz[scale=.45]{ 
    \path[use as bounding box] (-1,1.2) -- (2,4);
    \begin{scope}
      \draw[latex'-latex'] (3.5,0) node[right] {$x_1$}
        -| (0,3.8) node[left] {$x_2$};
      \begin{scope}
        \path[clip] (0,0) -- (3.5,0) [rounded corners=12mm]
          -- (3.5,3.9) [rounded corners=0mm] -| (0,0);
        \draw[dgrey,fillarea] (0,0) -- (1,1) |- (0,3) -- cycle;
        \begin{scope}[opacity=.3]
          \foreach \x in {-4,...,4}
                 {\draw (\x,0) -- +(9,9);
                   \draw (0,\x) -- +(9,0);
                   \draw (\x,0) -- +(0,9);}
        \end{scope}
      \end{scope}
    \end{scope}
    }
  \]
  \caption{Example of a DBM}
  \label{11-fig:example_dbm}
\commentAlt{Figure~\ref{11-fig:example_dbm}: A matrix labeled 'M' on the left, containing pairs of symbols and numbers, and a 2D coordinate plane with a shaded region on the right.}
\commentLongAlt{Figure~\ref{11-fig:example_dbm}: The image displays two distinct elements. On the left is a 3x3 matrix labeled 'M'. Each element of the matrix is a tuple (<=, number):

Row 1: (<=,0), (<=,0), (<=,0)
Row 2: (<=,1), (<=,0), (<=,0)
Row 3: (<=,3), (<=,2), (<=,0)
On the right is a 2D Cartesian coordinate plane with x1 and x2 axes. A grid of diagonal lines fills the positive quadrant. A shaded polygon is present in the top-left portion of the grid, near the x2 axis.}
  \end{figure}
  For instance, $M[2,0]=(\leq, 3)$ represents the
  constraint~$x_2-x_0\leq 3$, \textit{i.e.}, $x_2\leq 3$.
  The~diagram to the right of the figure represents the set~$[M]$.
\end{example}

We now define elementary operations on DBMs which are used to explore
the state space of timed games. We start by giving set-theoretic
definitions and then comment on their computation with~DBMs.

Let $\posttime(Z)$ denote the zone describing the
\emph{time-successors} of~$Z$, and~$\pretime(Z)$ the
\emph{time-predecessors} of~$Z$. Formally,
\begin{xalignat*}1
    \posttime(Z)&= \{v \in \Realnn^{\Clocks}
    \mid \exists t\geq 0.\  v-t \in Z\}
    \\
    \pretime(Z) &=
    \{v \in \Realnn^{\Clocks} \mid \exists t \geq 0.\ v + t \in Z\}.
\end{xalignat*}

Given~$R\subseteq\Clocks$, we also define
\begin{xalignat*}1
  \reset_R(Z) &= \{v \in \Realnn^{\Clocks} \mid \exists v' \in Z.\
  v=v'[R\leftarrow 0]\} \\
  \unreset_R(Z) &= \{ v \in \Realnn^{\Clocks} \mid \exists v' \in Z.\
  v' = v[R \leftarrow 0]\}.
\end{xalignat*}

These operations, together with intersection, suffice to describe
one-step successors and predecessors by an edge of a timed automaton.
For instance, given edge~$e=(\ell,g,R,\ell')$ and
set~$S \subseteq \Realnn^\Clocks$, the~set of states that are reached
after letting time elapse and taking edge~$e$ can be obtained as
\[
  \postta_e(S) = \reset_R(\posttime(S)\cap G),
\]
where~$G$ denotes the zone corresponding to the guard~$g$.
Similarly, we can compute the predecessors of~$S$ by edge~$e$ as
\[
\preta_e(S) = \pretime(G \cap \unreset_R(S)).
\]
We~illustrate these constructions on \Cref{11-fig:opzones}.
\begin{figure}[ht]
  \centering
  \begin{tikzpicture}[scale=.45]
    \begin{scope}
      \draw[latex'-latex'] (4.5,0) node[right] {$x_1$}
        -| (0,4.5) node[left] {$x_2$};
      \begin{scope}
        \path[clip] (0,0) -- (4.5,0) [rounded corners=12mm]
          -- (4.5,4.5) [rounded corners=0mm] -| (0,0);
        \fill[dgrey,fillarea] (1,1) -- (2,2) |- (1,3) -- cycle;
        \fill[opacity=.6,dgrey,hatcharea] (1,1) -- (5,5) -| cycle;
        \begin{scope}[opacity=.3]
          \foreach \x in {-4,...,4}
                 {\draw (\x,0) -- +(9,9);
                   \draw (0,\x) -- +(9,0);
                   \draw (\x,0) -- +(0,9);}
        \end{scope}
      \end{scope}
      \path (3,0.5) node (Z) {$Z$} edge[-latex',bend left=20] (1.6,1.4);
      \path (4,1.5) node (Y) {$\posttime(Z)$} edge[-latex',bend right=20] (2.9,2.7);
    \end{scope}
    \begin{scope}[xshift=6.5cm]
      \draw[latex'-latex'] (4.5,0) node[right] {$x_1$}
        -| (0,4.5) node[left] {$x_2$};
      \begin{scope}
        \path[clip] (0,0) -- (4.5,0) [rounded corners=12mm]
          -- (4.5,4.5) [rounded corners=0mm] -| (0,0);
        \fill[dgrey,fillarea] (1,1) -- (2,2) |- (1,3) -- cycle;
        \fill[opacity=.6,dgrey,hatcharea] (0,0) -- (2,2) |- (1,3) -- (0,2) -- cycle;
          \begin{scope}[opacity=.3]
            \foreach \x in {-4,...,4}
                 {\draw (\x,0) -- +(9,9);
                   \draw (0,\x) -- +(9,0);
                   \draw (\x,0) -- +(0,9);}
          \end{scope}
      \end{scope}
      \path (3,1.5) node (Z) {$Z$} edge[-latex',bend right] (2.1,2.6);
      \path (3.5,0.5) node (Y) {$\pretime(Z)$} edge[-latex',bend left=10] (.6,0.5);
    \end{scope}
    \begin{scope}[xshift=13cm]
      \draw[latex'-latex'] (4.5,0) node[right] {$x_1$}
        -| (0,4.5) node[left] {$x_2$};
      \begin{scope}
        \path[clip] (0,0) -- (4.5,0) [rounded corners=12mm]
          -- (4.5,4.5) [rounded corners=0mm] -| (0,0);
        \fill[dgrey,fillarea] (1,1) -- (2,2) |- (1,3) -- cycle;
        \draw[line width=3.2pt,dgrey] (1,0) -- (2,0);
        \begin{scope}[opacity=.3]
          \foreach \x in {-4,...,4}
                 {\draw (\x,0) -- +(9,9);
                   \draw (0,\x) -- +(9,0);
                   \draw (\x,0) -- +(0,9);}
          \end{scope}
      \end{scope}
      \path (3.2,1.8) node (Z) {$Z$} edge[-latex',bend right] (2.1,2.6);
      \path (3.5,0.8) node (Y) {$\reset_{x_2}(Z)$} edge[-latex',bend right=20] (1.4,0.1);
    \end{scope}
    \begin{scope}[xshift=19.5cm]
      \draw[latex'-latex'] (4.5,0) node[right] {$x_1$}
        -| (0,4.5) node[left] {$x_2$};
      \begin{scope}
        \path[clip] (0,0) -- (4.5,0) [rounded corners=12mm]
          -- (4.5,4.5) [rounded corners=0mm] -| (0,0);
        \fill[dgrey,fillarea] (1,0) -- (2,0) -- (3,1) |- (2,3) -- (1,2) -- cycle;
        \fill[opacity=.6,dgrey,hatcharea] (1,5) -| (2,0) -| cycle;
        \begin{scope}[opacity=.3]
          \foreach \x in {-4,...,4}
                 {\draw (\x,0) -- +(9,9);
                   \draw (0,\x) -- +(9,0);
                   \draw (\x,0) -- +(0,9);}
        \end{scope}
      \end{scope}
      \path (5,1.5) node (Z) {$Z$} edge[-latex',bend right] (3.1,2.6);
      \path (3.9,4) node (Y) {$\unreset_{x_2}(Z)$} edge[-latex',bend left=12] (2.1,3.4);
    \end{scope}
        
  \end{tikzpicture}
  \caption{Operations on zones}
  \label{11-fig:opzones}
\commentAlt{Figure~\ref{11-fig:opzones}: Four 2D coordinate planes arranged horizontally, each showing a shaded region (Z) and additional patterned or transformed regions, illustrating different operations on Z.}
\commentLongAlt{Figure~\ref{11-fig:opzones}: The image displays four distinct 2D Cartesian coordinate planes arranged side-by-side, each with x1 and x2 axes and a grid of diagonal lines.

Leftmost Plane: Contains a shaded polygonal region labeled 'Z'. An area above and to the right of 'Z' is filled with diagonal hatching and labeled 'Post>=0(Z)'. Curved arrows indicate a transformation from 'Z' to 'Post>=0(Z)'.

Second from Left Plane: Contains a shaded polygonal region labeled 'Z'. An area below and to the left of 'Z' is filled with diagonal hatching and labeled 'Pre>=0(Z)'. Curved arrows indicate a transformation from 'Pre>=0(Z)' to 'Z'.

Third from Left Plane: Contains a shaded polygonal region labeled 'Z'. A thin, horizontally shaded rectangle is below 'Z' and labeled 'Reset_v1(Z)'. Curved arrows indicate a transformation from 'Z' to 'Reset_v1(Z)'.

Rightmost Plane: Contains a shaded polygonal region labeled 'Z'. An area to the left and above 'Z' is filled with diagonal hatching and labeled 'Unreset_v2(Z)'. Curved arrows indicate a transformation from 'Unreset_v2(Z)' to 'Z'.}
\end{figure}

It is not hard to prove that the above operations preserve zones: if~$S$ is a
zone, then so is the result of any of these operations. Moreover, each
single operation can be computed in time~$O(|\Clocks|^3)$ using the
DBM representation. The~underlying algorithms often modify some
elements of the matrix and run an all-pairs shortest path algorithm,
namely the Floyd-Roy-Warshall algorithm, on a graph whose adjacency
matrix is the given~DBM.  Computing the shortest paths renders the DBM
\emph{canonical}; in~fact, this allows one to compute the tightest
constraints on all differences of clock pairs, and this can be shown
to yield a unique representation of a given zone.

Let us call the above operations \emph{basic operations} on DBMs~\cite{Bengtsson.Yi:2004}.
\begin{theorem}[Complexity of basic operations on DBMs]
\label{11-thm:complexity_basic_operations_DBMs}
  Given a DBM of size~$n\times n$, any basic operation yields a DBM
  and can be computed in time~$O(n^3)$.
\end{theorem}



Observe that a DBM always describes a convex subset
of~$\Realnn^\Clocks$ since it is a conjunction of convex clock
constraints. However, the set of winning states is in general
non-convex in timed games. The~simple arena
of~\Cref{11-fig:non-convex} provides an example: if Eve's objective
is to reach~$\ell_1$, then it should just avoid the configurations
satisfying~$1\leq x_1,x_2\leq 2$. But this set of predecessors is then
non-convex as shown in~\Cref{11-fig:non-convex}.
We~thus have to work with unions of zones, also called
\emph{federations} of zones, or \emph{federations} for~short.

\begin{figure}[ht]
  \centering
  \begin{tikzpicture}[node distance=2.5cm,auto]
    \node[state] at (0,0) (A) {$\ell$};
    \draw[initial] (A.-135) -- +(-135:3mm);
    \node[state,double] at (2.5,0) (B) {$\ell_1$};
    \node[state] at (-2.5,0) (C) {$\ell_2$};
    \path[arrow] 
    (A) edge node[above] {$\genfrac{}{}{0pt}{0}{-2 \leq x_1-x_2 \leq 1}{{}\wedge x_2 \leq 3}$} (B)
    (A) edge[dashed] node[above] {$1\leq x_1,x_2 \leq 2$} (C);
    \begin{scope}[xshift=4cm,yshift=-1cm,scale=.45]
      \draw[latex'-latex'] (4.5,0) node[right] {$x_1$}
        -| (0,4.5) node[left] {$x_2$};
      \begin{scope}
        \path[clip] (0,0) -- (4.5,0) [rounded corners=12mm]
          -- (4.5,4.5) [rounded corners=0mm] -| (0,0);
        \fill[dgrey,fillarea] (1,0) -- (2,1) -| (1,2) -| (2,1) -- (4,3) -- (1,3) -- (0,2) |- cycle;
          \begin{scope}[opacity=.3]
            \foreach \x in {-4,...,4}
                 {\draw (\x,0) -- +(9,9);
                   \draw (0,\x) -- +(9,0);
                   \draw (\x,0) -- +(0,9);}
          \end{scope}
      \end{scope}
    \end{scope}

  \end{tikzpicture}
  \caption{Winning configurations (in~$\ell$)
    for Eve to ensure reaching~$\ell_1$.
  }
  \label{11-fig:non-convex}
  \commentAlt{Figure~\ref{11-fig:non-convex}: A diagram combining a finite automaton-like structure with a 2D coordinate plane showing a shaded region with a cutout. See long description.}
\commentLongAlt{Figure~\ref{11-fig:non-convex}: The image consists of two distinct parts.

Left Part: An automaton-like diagram with three circular nodes.

A central circular node labeled 'l' has an incoming arrow, indicating it as a starting state.
From 'l', a dashed arrow points to the left to a circular node labeled 'l2'. This arrow is labeled '1 <= x1, x2 <= 2'.
From 'l', a solid arrow points to the right to a double-ringed circular node labeled 'l1'. This arrow has two labels: '-2 <= x1 - x2 <= 1' above and 'AND x2 <= 3' below.
Right Part: A 2D Cartesian coordinate plane with x1 and x2 axes. A grid of diagonal lines fills the positive quadrant. A shaded, irregular quadrilateral shape is present, with an unshaded square cutout in its center.}
\end{figure}

One particular operation that we need is complementation.
The~complement of a convex set is of course not convex in general, but
we can still compute, in polynomial time, the complement of a DBM~$M$,
written~$M^c$, as a federation of~DBMs.
\begin{theorem}[Complement of DBMs]
\label{11-thm:complement_DBM}
  The complement of a DBM of size~$n\times n$ can be computed as a
  federation of at most~$n(n-1)$ DBMs.
\end{theorem}
The above theorem is seen easily as follows. Since a DBM represents a
conjunction of constraints, the complement is computed easily as the
disjunction of the complement of each elementary constraint appearing
in~it.  For~instance, the~complement of $x_1\leq 1 \land x_2 \geq 2$ is
$x_1>1 \lor x_2<2$, which can be represented as the federation of two
zones.


In the rest of this chapter, we describe two algorithms to solve timed
games using the DBM data structure and the operations introduced
above. Our~algorithms are extensions of those used for finite games,
but we explore the set of zones instead of the set of vertices, and
predecessor and successor operations are replaced by their zone-based
counterparts.

As for finite games, we are interested in computing a fixed point to
determine whether a given configuration is winning for Eve. We~start
by introducing the zone-based counterparts of the controllable predecessors operator
which
is the main tool in the algorithms.

\section{Controllable-predecessor operator}
\label{11-sec:controllable_predecessor_operator}
We present the controllable-predecessor operator which, given sets of
states $X$ and~$Y$, returns the set of states from which Eve can
reach~$X$ in one step, while avoiding states of~$Y$ during Eve's
delay. Intuitively, the states in~$Y$ are the states from which Adam
may force an action leading outside of~$X$, which Eve would better avoid.

Recall that $\vertices = \Locs \times \Realnn^\Clocks$.
The set of \emph{safe time-predecessors} to reach~$X \subseteq \vertices$ while
avoiding~$Y \subseteq \vertices$
is defined as follows:
\[
\predt(X,Y) = \{(\ell,\nu) \in \vertices \mid \exists d\geq
    0.\ (\ell,\nu+d) \in X \wedge{}
    \forall d' \in[0,d).\ (\ell,\nu+d') \not  \in Y\}.
\]
In words, from any configuration of $\predt(X,Y)$,
the set~$X$ can be reached by
delaying while never crossing any configuration of~$Y$ on the way. 
For any $X \subseteq \vertices$, let us denote
\[
  \predc(X) = \{ (\ell,\nu) \in \vertices \mid \exists e \in \EE(\ell).\ 
\exists (\ell',\nu') \in X.\ (\ell,\nu) \xrightarrow{e} (\ell',\nu')\}.
\]
This is the set of immediate predecessors of~$X$ by edges in~$\EE$.
Symmetrically, we~define $\predu(X)$ using~$\EA$ instead of~$\EE$.
Our controllable-predecessor operator is then defined as 
%
\[
  \pi(X) = \predt(X\cup \predc(X), \predu(\vertices \setminus X)).
\]

Intuitively, the states in~$\pi(X)$ are those from which Eve~can wait
until she
can take a controllable transition to~reach~$X$, and so that
no transitions that Adam could take while Eve is waiting may lead
outside of~$X$.

\begin{lemma}\label{11-lem:pimonotonic}
  The operator $\pi$ is 
  order-preserving: if~$X\subseteq Y$, then
  $\pi(X) \subseteq \pi(Y)$.  
\end{lemma}

\begin{example}[Example for the controllable predecessor operator]
\label{11-ex:controllable_predecessor_operator}
Consider the timed game to the left of
\Cref{11-fig:contpred}. In~that game, Eve can reach her target when
clock~$x_2$ reaches value~$3$ while~$x_1$ is in~$[1;4]$. However,
Adam can take the game to a bad state when simultaneously
$x_1\in[1;2]$ and $x_2\leq 2$. The~diagram to the right of
\Cref{11-fig:contpred} shows the set of winning valuations for Eve:
it~is computed as $\predt(X,Y)$ where $X=\predc(\{W\}\times\Realnn)$
and $Y=\predu(\{\ell_0,\ell_1\}\times\Realnn)$.

\begin{figure}[h]
  \centering
  \begin{tikzpicture}
    \draw (0,1) node[state] (a) {$\ell_0$};
    \draw (0,-1) node[state] (b) {$\ell_1$};
    \draw (3,0) node[state,double] (W) {$W$};
    \draw[arrow] (a) -- (W) node[midway,above right] {$\genfrac{}{}{0pt}{0}{1\leq x_1\leq 4}{{}\wedge x_2=3}$};
    \draw (a) edge[arrow,dashed] (b) node[midway,left] {$\genfrac{}{}{0pt}{0}{1\leq x_1\leq 2}{{}\wedge x_2\leq 2}$};
    \draw[initial] (a.135) -- +(135:3mm);
    \begin{scope}[xshift=5cm,yshift=-1cm,scale=.55]
      \draw[latex'-latex'] (4.5,0) node[right] {$x_1$}
      -| (0,4.5) node[left] {$x_2$};
      \begin{scope}
        \path[clip] (0,0) -- (4.5,0) [rounded corners=10mm]
          -- (4.5,4.5) [rounded corners=0mm] -| (0,0);
        \fill[opacity=.6,dgrey,hatcharea] (1,0) -| (2,2) -| cycle;
        \fill[dgrey,fillarea] (0,1) -- (1,2) -| (2,1) -- (4,3) -- (1,3) -- (0,2) -- cycle;
        \draw[line width=3.2pt,dgrey] (1,3) -- (4,3);
        \begin{scope}[opacity=.3]
        \foreach \x in {-4,...,4}
                 {\draw (\x,0) -- +(9,9);
                   \draw (0,\x) -- +(9,0);
                   \draw (\x,0) -- +(0,9);}
        \end{scope}
      \end{scope}
      \path (3.5,4) node {$X$} edge[-latex',bend right=12] (2.8,3.2);
      \path (3.5,0.5) node {$Y$} edge[-latex',bend left=12] (2.2,0.6);
      \path (4.5,1.5) node {$\pi(W)$} edge[-latex',bend left=12] (2.6,1.8);
    \end{scope}
  \end{tikzpicture}
  \caption{Controllable predecessors}
  \label{11-fig:contpred}
\commentAlt{Figure~\ref{11-fig:contpred}: A diagram combining a finite automaton-like structure with a 2D coordinate plane showing shaded and patterned regions.}
\commentLongAlt{Figure~\ref{11-fig:contpred}: The image consists of two distinct parts.

Left Part: An automaton-like diagram with two circular nodes, 'l0' and 'l1', and one double-ringed circular node 'W'.

An incoming arrow points to 'l0', indicating it as a starting state.
From 'l0', a dashed arrow points downwards to 'l1'. This arrow has two labels: '1 <= x1 <= 2' above and 'AND x2 <= 2' below.
From 'l0', a solid arrow points to the right to 'W'. This arrow also has two labels: '1 <= x1 <= 4' above and 'AND x2 = 3' below.
Right Part: A 2D Cartesian coordinate plane with x1 and x2 axes. A grid of diagonal lines fills the positive quadrant.

A shaded polygonal region is present in the upper part, labeled 'X'. A thick horizontal line segment is part of this shaded region, labeled 'l'.
Below and to the right of 'X', another region is shaded with horizontal lines, labeled 'Y'.
A curved arrow originates from the shaded region 'X' and points to the right, labeled 'pi(W)'. A horizontal arrow originates from the region 'Y' and points to the right.}
\end{figure}
\end{example}



\begin{remark}[Reduction for zero-sum objectives]
\label{11-rmk:reduction_zero_sum}
When considering zero-sum objectives (which we do in this chapter),
a~turn-based arena can be built to decide whether Eve has a winning
strategy (and~possibly construct~one).  Intuitively, the turn-based
arena is obtained by letting Eve first pick a pair~$(d,e)$ with~$e
\in \EE$, and then letting Adam either respect this choice (\textit{i.e.},
play a larger delay) or preempt Eve's action by choosing $(d',e')$
with~$d' \leq d$ and~$e' \in \EA$.
Eve has a winning strategy in this turn-based arena if, and only if she has one in the original game.
However, this does not apply to Adam; if he has a winning strategy, this only means that Eve
does not have a winning strategy in the original game.

Given~$\calT = (\Locs, \Clocks,\EE,\EA, c)$, we define the arena~$\TA
= (\vertices,\VE,\VA,E, c'')$ 
with objective~$\Omega$, where~$\VE = \Locs\times \Realnn^\Clocks$ and
$\VA=\Locs\times\Realnn^\Clocks\times(\Realnn \times \EE
\cup\{\bot\})$.  The~set of successors of configuration~$(\ell,\nu)
\in \VE$ is the set $\{(\ell,\nu,a)\mid a \text{ available at
}(\ell,\nu)\}$.
From a configuration~$((\ell,\nu),(d,e))\in\VA$, several cases may
occur:
\begin{itemize}
\item if Adam has no available action from~$(\ell,\nu)$, or if he can
  only play actions~$(d',e')$ with $d'>d$, then the only transition
  from~$((\ell,\nu),(d,e))$ goes to $\step((\ell,\nu),(d,e))$;
\item Otherwise, for all actions~$(d',e')$ for Adam satisfying $d'\leq d$,
  there is a transition from $((\ell,\nu),(d,e))$
  to $\step((\ell,\nu),(d',e'))$.
  Moreover, if Adam has an available action~$(d',e')$ with~$d'\geq d$,
  then there is also a transition
  from~$((\ell,\nu),(d,e))$ to $\step((\ell,\nu),(d,e))$.
\end{itemize}
From a configuration~$((\ell,\nu),\bot)$, there is a transition
to~$\step((\ell,\nu),(d',e'))$ for each available action~$(d',e')$
of Adam.
The colouring function is defined as~$c''((\ell,\nu)) = c(\ell)$
and $c''((\ell,\nu),(d,e)) = c(\ell)$.

\end{remark}

\section{Backward algorithm}
\label{11-sec:backward_algorithm}
Given the DBM data structure we presented,
a backward fixed-point algorithm follows for computing the winning configurations in a timed game:

\begin{theorem}[Correctness of the backward algorithm]
   \label{11-thm:timed-control}
   For any timed game~$\TA$ and target set~$G \subseteq \VE$, 
   the~limit $\lim_{k \rightarrow \infty} \pi^k(G)$ exists, and is reached 
   in a finite number of steps. This limit is precisely the set of states from
   which the controller has a winning strategy for reaching~$G$.
\end{theorem}

\begin{proof}[Sketch]
  We~briefly explain why the fixed-point computation terminates and refer
  to~\cite{Asarin.Maler.ea:1998} for more details.  The~proof relies on the notion of
  \emph{regions}. Intuitively, regions are minimal zones, not
  containing any other zone. More precisely, writing~$K$ for the maximal
  (absolute) integer constant appearing in the timed game, the~set of regions is
  the set of zones associated with
  DBMs~$(\mathord\prec_{i,j},M_{i,j})$ such that
  \begin{itemize}
  \item $M_{i,j}\in [-K; K]\cup\{-\infty;+\infty\}$ for all~$i$ and~$j$;
  \item for all~$i\not=j$,
     \begin{itemize}
    \item either $M_{i,j}=-M_{j,i}$ and
      $\mathord\prec_{i,j}=\mathord\prec_{j,i}=\mathord\leq$,
     \item or $|M_{i,j}+M_{j,i}|=1$ and 
      $\mathord\prec_{i,j}=\mathord\prec_{j,i}=\mathord<$,
    \item or $(\prec_{i,j},M_{i,j})=(<,+\infty)$ and $(\prec_{j,i},M_{j,i})=(<,-K)$.
     \end{itemize}
  \end{itemize}
  The set of regions form a finite partition of~$\Realnn^{\Clocks}$.
  The~main argument for the proof is that the
  image by~$\pi$ of any finite union of regions is a finite union of
  regions. Since~$\pi$ is non-decreasing, the~sequence~$\pi^k(G)$
  converges after finitely many steps.

  The~fact that $\pi^k(G)$ may
  only contain winning configurations follows from our construction,
  and can be proved easily by induction on~$k$.
  One can also show that for all configurations outside of~$\pi^k(G)$,
  there is no strategy that is winning within~$k$ steps. This follows since
  the definition of~$\pi(\cdot)$ gives the actions to be played by Adam to
  prevent Eve from winning.
  Note that this does not necessarily yield a winning strategy for Adam,
  since the game is not determined.
\end{proof}


\section{Forward algorithm}
\label{11-sec:forward_algorithm}
The backward algorithm we just presented is conceptually simple, but
it~is often not very efficient in practice,
as federations tend to grow too much in size in each iteration of the
computation.
The forward algorithm we will now present is more efficient in practice.
It performs a forward exploration and only applies the controllable predecessor along branches
that actually reach the target state from the initial state. If the witness
trace is not excessively long, which is often the case in practice,
this limits the size of the federations.


We present below the algorithm proposed in~\cite{Cassez.David.ea:2005},
and as a first step, we explain the untimed version of that algorithm,
based on an algorithm of Liu and Smolka for computing
fixed points~\cite{Liu.Smolka:1998}.

\subsection*{A Forward Algorithm for Finite-State Games}

The original algorithm of Liu and Smolka is expressed in terms of
\emph{pre-fixed points} in \emph{dependency graphs}: a~dependency graph
is a pair~$G=(V,E)$ in which $E \subseteq V \times 2^V$ relates
states with sets of states.
For any order-preserving 
function~$f\colon 2^V\to2^V$
(\emph{order-preserving} meaning non-decreasing for the $\subseteq$-relation),
a~\emph{pre-fixed point} is a set~$X\subseteq V$ for which $f(X)\subseteq X$; 
it~is a \emph{fixed point} if $f(X)=X$. By~Knaster-Tarski theorem, such functions always admit a least pre-fixed point which is also the least fixed point.

Fix a dependency graph~$G=(V,E)$. 
For~$W\subseteq V$,
we~define a mapping $f_W\colon 2^V\to 2^V$: for
each~$X\subseteq V$, we~let $f_W(X)=W\cup \{v\in V\mid \exists (v,Y)\in
E.\ Y\subseteq X\}$.
%
Clearly, $X\subseteq X'$ implies $f_W(X)\subseteq f_W(X')$, so that $f_W$ admits
a least (pre-)fixed point.
The~Liu-Smolka
algorithm aims at deciding whether a given vertex~$v_0\in V$ belongs
to the least fixed point of~$f_W$.  Classical algorithms for computing
least fixed points consist in iteratively
computing~$(f_W^i(\emptyset))_i$ until convergence (assuming~$V$~is
finite). Observe that this corresponds to the algorithm of \Cref{11-thm:timed-control}
for timed games.
The~Liu-Smolka algorithm proceeds from~$v_0$, and explores
the dependency graph until it can claim that~$v_0$~is, or that it
is~not, in the least fixed point of~$f_W$.

Before tackling the algorithm, let us link least fixed points in
dependency graphs and winning sets in concurrent games (with
reachability objectives): with a concurrent arena
$\calC=(V,\Act,\delta,c')$ and a target set~$\Win$, we~associate
the dependency graph $G=(V,E)$, where $(v,T)\in E$ whenever $v\in V$
and $T\subseteq V$ is such that there exists an action~$a$ for which
$T=\{v' \mid \exists a'\in\Act.\ v'=\delta(v,a,a')\}$.  Then for any
set~$X\subseteq V$, the~set~$f_{\Win}(X)$ contains $\Win$ and all the
states from which Eve can force a visit to~$X$ in one step.
We~then~have:
\begin{proposition}\label{11-prop:fixp-game}
The least fixed point of~$f_{\Win}$ in~$G$ corresponds to the set~$W$ of
winning states for Eve in~$\calC$.
\end{proposition}

\begin{proof}
  The winning states of Eve form a pre-fixed point of~$f_{\Win}$
  containing~$\Win$: indeed, for any~$v\in f_{\Win}(W)$, either $v\in
  \Win$, or Eve has an action to move from~$v$ to some state
  in~$W$. Hence $v$ is winning, \textit{i.e.}, $v\in W$.
  

  Conversely, from any state~$v$ that is not in the least
  pre-fixed point~$X$, for any~edge~$(v,T)$, there is a state~$v'\in T$
  that again is not in~$X$. This defines a strategy for Adam to avoid
  reaching~$\Win$, so that Eve does not have a winning strategy from~$v$.
\end{proof}

\begin{algorithm}
  \SetKwFunction{Pop}{pop}\SetAlgoNoEnd
  \KwData{A dependency graph~$G=(V,E)$, a~set~$W\subseteq V$, a~node $v_0\in V$}
  \KwResult{Is $v_0$ in the least fixed point of~$f_W$?}

  \For{$v\in V$}{\lIf{$v\in W$}{$F(v):=1$}
    \lElseIf{$v==v_0$}{$F(v):=0$}
            \lElse{$F(v):=\bot$}}
  
  $\Dep(v_0):=\emptyset$;

  $\Wait:=\{(v,T)\in E \mid v=v_0\}$;

  \While{($\Wait\not=\emptyset$ and $F(v_0)==0$)}
        {$(v,T):=\Pop{\Wait}$;\\
          \If(\hfill{// case 1}){$F(v)==0$ and $\forall v'\in T.\ F(v')==1$}
             {$F(v):=1$;\\
              $\Wait:=\Wait\cup\Dep(v)$;}
          \ElseIf(\hfill{// case 2}){$\exists v'\in T.\ F(v')==0$}
                {$\Dep(v'):=\Dep(v') \cup \{(v,T)\}$;}
                \ElseIf(\hfill{// case 3}){$\exists v'\in T.\ F(v')==\bot$}
                  {$F(v'):=0$; \\
                  $\Dep(v'):=\{(v,T)\}$;\\
                  $\Wait:=\Wait\cup\{(w,U)\in E \mid w=v'\}$;
                  }
                  }
                  \Return{$F(v_0)$}
  \caption{Liu-Smolka algorithm for least fixed point of~$f_W$}
  \label{11-algo:liu.smolka:1998}
\end{algorithm}

\footnote{Our version of the algorithm slightly differs from the
  original one~\cite{Liu.Smolka:1998}: we~let $\Dep(v'):=\{(v,T)\}$ at the penultimate line
  of the \textbf{while} loop (which otherwise results in a wrong
  result, as~already noticed in~\cite{Jensen.Larsen.ea:2013}), reinforce the
  condition for case~$1$ (which otherwise would not guarantee
  termination), and reinforce the condition of the \textbf{while} loop
  to get earlier termination.}

\Cref{11-algo:liu.smolka:1998} can be seen as an alternation of forward
exploration and backward propagation. Intuitively, the algorithm first
explores the graph in a forward manner, remembering for each node~$v$
the set~$\Dep(v)$ of nodes that \emph{depend} on~$v$, and have to be
reexplored if the status of~$v$ is updated.
For~each~$v$, the algorithm maintains a
value~$F(v)$, which is~$\bot$ if~$v$ has not been explored~yet,
$0$~if~$v$~has been explored but not yet been shown to be winning, 
and~$1$ if~$v$ is known to be winning. 
Whenever a vertex~$v$ whose all successors are winning is found, the value of~$v$ is set to~$1$, and its parents (in~$\Dep(v)$)
are scheduled to be visited again to check whether their statuses have to be changed.
This is how the backward propagation is triggered; in~fact, the~search will climb in the tree as long as the values of
vertices can be updated to~$1$.



The correctness of this algorithm relies on the following lemma~\cite{Liu.Smolka:1998}:
\begin{lemma}[Invariant]\label{11-lemma:liu.smolka:1998}
  The following properties hold at the end of each run in the
  \textbf{while} loop of \Cref{11-algo:liu.smolka:1998}:
  \begin{itemize}
  \item for any $v\in V$, if $F(v)=1$, then $v$ belongs to the least
    fixed point containing~$W$;
  \item for any $v\in V$ with $F(v)=0$, and any $(v,T)\in E$, either
    $(v,T)\in \Wait$ or $(v,T)\in\Dep(v')$ for some $v'\in T$ with
    $F(v')=0$.
  \end{itemize}
\end{lemma}

\begin{proof}
  We~fix an execution of \Cref{11-algo:liu.smolka:1998}, and prove that both
  claims are true at the beginning and end of each iteration through
  the \textbf{while} loop. To~clarify the presentation, we~use
  superscript~$i$ to indicate the value of variables at the end of the
  $i$-th run through the loop, so that $\Wait^3$ is the value of
  variable~$\Wait$ after three iterations (and $\Wait^0$ is its value
  after initialization). In~particular, $(v^i,T^i)$ is the
  symbolic transition popped from the $\Wait$ list during the $i$-th
  iteration (and is not defined for~$i=0$).

  Let us first prove the first property: the initialization phase
  clearly enforces that if $F^0(x)=1$, then $x\in W$, which is
  included in any fixed point of~$f_W$. Now, assume that the property
  holds true at the beginning of the $i$-th run through the loop
  (\textit{i.e.},~if~$F^{i-1}(x)=1$, then $x$ is in the least fixed point), and
  pick some~$x\in V$ such that $F^i(x)=1$. If~$F^{i-1}(x)=1$, our
  result follows; if~not, then the $i$-th iteration of the loop must
  have run via case~$1$, hence $F^{i-1}(x')=1$ for all~$x'\in
  T^i$. By~our induction hypothesis, this indicates that $T^i$ is part
  of the least fixed point, and by definition of~$f_W$, $x$~must also
  belong to the least fixed point.


  \smallskip
  The second statement also clearly holds after initialization:
  initially, $F^0(x)=0$ only for~$x=v_0$, and all transitions
  from~$v_0$ have been stored in~$\Wait^0$. We~now assume that the
  property holds when entering the \textbf{while} loop for the $i$-th
  time, and consider $x$ such that $F^i(x)=0$ at the end of that
  loop. We~pick $(x,Y)\in E$.
  \begin{itemize}
  \item if the $i$-th run through the loop visits case~$1$, then
    already $F^{i-1}(x)=0$ (hence $x\not=v^i$). Then either
    $(x,Y)\in\Wait^{i-1}$, or $(x,Y)\in\Dep^{i-1}(x')$ for some~$x'\in Y$
    with $F(x')=0$. In~the former case: since $x\not=v^i$,
    if~$(x,Y)\in\Wait^{i-1}$ then also $(x,Y)\in \Wait^i$;
    in the latter case: if $(x,Y)\in\Dep^{i-1}(x')$ for some~$x'\in Y$
    with~$F(x')=0$, then \emph{(i)}~either $x'=v^i$, and
    $(x,Y)\in\Wait^i$ because $\Dep^{i-1}(v^i)\subseteq \Wait^i$
    (last~line of case~$1$), \emph{(ii)}~or~${x'\not=v^i}$, and
    $(x,Y)\in \Dep^{i-1}(x')=\Dep^{i}(x')$ and $F^i(x')=F^{i-1}(x')=0$.

  \item if the $i$-th run goes to case~$2$, then again $F^{i-1}(x)=0$,
    and the induction hypothesis applies: either $(x,Y)$ is
    in~$\Wait^{i-1}$, or~it is in~$\Dep^{i-1}(x')$ for some~$x'\in Y$
    with $F^{i-1}(x')=0$. For the latter case, observe that
    $\Dep^{i-1}(x')\subseteq \Dep^i(x')$, and $F^{i}(x')=F^{i-1}(x')$
    when running case~$2$; for the former case, we~have
    $\Wait^i=\Wait^{i-1}\setminus\{(v^i,T^i)\}$, so that $(x,Y)$
    remain in~$\Wait^i$ if~$(x,Y)\not=(v^i,T^i)$; finally,
    if~$(x,Y)=(v^i,T^i)$, then case~$2$ precisely adds~$(v^i,T^i)$
    to~$\Dep^i(v')$ for some~$v'\in T^i$ with~$F^i(v')=0$, which concludes
    the proof for this case.

  \item for case~$3$, we~first consider the case where $x$ is the
    state~$v'$ selected at the beginning of case~$3$: in~that case,
    all transitions from~$x$ are added to~$\Wait$, so that
    $(x,Y)\in\Wait^i$. Now, if~$x$ is not the selected state~$v'$,
    then $F^{i-1}(x)=0$, and again either $x\in\Wait^{i-1}$ or
    $x\in\Dep^{i-1}(x')$ for some~$x'\in Y$ with
    $F^{i-1}(x')=0$. The~latter case is preserved when running
    case~$3$; if~$x\in\Wait^{i-1}$: if~$(x,Y)\not=(v^i,T^i)$, then
    $x\in\Wait^i$, while if~$(x,Y)=(v^i,T^i)$, then
    $(x,Y)\in\Dep^i(v')$ for the state~$v'\in T^i$ selected at the
    beginning of case~$3$ (and~for which $F^i(v')=0$).
  \end{itemize}
\end{proof}

As a corollary, if the algorithm terminates after~$n$ rounds of the
\textbf{while} loop, then either~$F^n(v_0)=1$, or
$F^n(v_0)=0$ and $\Wait^n=\emptyset$.
\begin{itemize}
\item From the first claim of the lemma above, the former case entails
  that $v_0$ belongs to the least fixed point.
\item Now consider the second case, and let $B=\left\{v\in V\mid
  F^n(v)\in \{\bot,1\}\right\}$. We prove that $f_W(B)\subseteq
  B$. For this, we~pick $v\in f_W(B)$:
  \begin{itemize}
  \item if~$v\in W$, then $F(v)$~is set to~$1$ initially, and may
    never be changed, so that $v\in B$;
  \item otherwise, there is a transition $(v,T)$ such that $T\subseteq
    B$. If~$v\notin B$, then $(v,T)\in\Dep^n(v')$ for some~$v'\in T$
    with~$F^n(v')=0$, which contradicts the fact that $T\subseteq B$.
  \end{itemize}
  This proves that $f_W(B)\subseteq B$, so that $B$ is a pre-fixed point
  of~$f_W$. It~thus contains the least (pre-)fixed point of~$f_W$, so
  that any state~$v$ not in~$B$ (\textit{i.e.}, with $F^n(v)=0$) for sure does
  not belong to that least fixed point. In~particular, $v_0$ is not in
  the fixed point.
\end{itemize}



\medskip
It~remains to prove termination. For this, we~first notice that, for
any hyper-edge~$(v,S)$, if $(v,S)\in\Dep^i(v')$ then $(v,S)\in\Wait^j$
for some~$j < i$, and if $(v,S)\in\Wait^j$ or $(v,S)\in\Dep^j(v')$ for
some~$v'$, then $F^j(v)\not=\bot$.

We~then set
\[
M= 
2\size{\Wait} + 
2\hskip-4pt\sum_{v\text{ s.t. }F(v)=\bot}\hskip-4pt \size{\{(v,S)\in E\}} 
+ \hskip-4pt\sum_{v\text{ s.t. }F(v)=0}\hskip-4pt \size{\Dep(v)}
- \hskip-4pt\sum_{v\text{ s.t. }F(v)=1}\hskip-4pt \size{\Dep(v)}
\]
again writing~$M^i$ for the value of~$M$ after the $i$-th run through
the \textbf{while} loop.  This value is at most~$2\size{E}$ when
entering the \textbf{while} loop for the first time; clearly, it~can
never go below~$-\size{E}$. We~now prove that $M^{i}<M^{i-1}$,
which implies termination of the algorithm.

Consider the $i$-th run through the loop, popping~$(v^i,T^i)$
from~$\Wait^{i-1}$ (which decreases~$M$ by~$2$). We~consider all three cases:
\begin{itemize}
\item if~case~$1$ applies, $F^i(v)$ is set to~$1$.
  The~set~$\Dep^{i-1}(v)$ is added to~$\Wait^i$. This globally
  leaves~$M$ unchanged, so that globally $M^i=M^{i-1}-2$;
\item if~case~$2$ applies, $\Dep^{i-1}(v')$~is augmented by (at~most) one edge,
  and since $F^i(v')=0$, this increases~$M$ by at most~$1$. Hence globally
  $M^i\leq M^{i-1}-1$;
\item finally, case~$3$ increases~$M$ by~$1$: the edges added
  to~$\Wait$ are compensated by the fact that~$F(v')$ no longer
  is~$\bot$, and only one extra edge has to be considered in the
  second sum. Hence again $M^i\leq M^{i-1}-1$.
\end{itemize}
This concludes the correctness proof of~\Cref{11-algo:liu.smolka:1998}.

\begin{theorem}
\Cref{11-algo:liu.smolka:1998} terminates, and returns~$1$ if, and only~if,
$v_0$~belongs to the least fixed point of~$f_W$.
\end{theorem}

Using~\Cref{11-prop:fixp-game}, we~get a \emph{forward} algorithm for
deciding if a given state of a concurrent game is winning
for~Eve. This corresponds to the OTFUR algorithm
of~\cite{Cassez.David.ea:2005}.

\subsection*{Extension to Timed Games}

We now explain how to adapt the algorithm above to (infinite-state)
timed games. 
For~efficiency, the~algorithm relies on zones (and~DBMs); instead of
computing whether a given zone~$(\ell,Z)$ is winning, the~algorithm
maintains, for each~zone~$S=(\ell,Z)$ it~explores, a~subzone~$(\ell,Z')$ of
configurations that it knows are winning; this subzone is stored as
$F(S)$, and is updated during the execution.
As~in~\Cref{11-algo:liu.smolka:1998}, a~waiting list keeps track of the
zones to be explored, and a dependency list stores the list of nodes
to be revisited upon update of the winning subzone of a zone.
The algorithm is given in~\Cref{11-algo:sotftr}.


\begin{algorithm}
  \SetKwFunction{Pop}{pop}\SetAlgoNoEnd
  \KwData{A reachability timed game~$\game=(\arena, \Reach(\Win))$,
    a~location $\ell_0\in \Locs$}
  \KwResult{Is $(\ell_0,\mathbf{0})$ winning for Eve?}

  $S_0:=\posttime(\ell_0,\mathbf 0)$;


  \leIf{$c(S_0)==\Win$}{$F(S_0):=S_0$}{$F(S_0):=\emptyset$}

  $\Passed:=\{S_0\}$;  \hfill
  // $\Passed$ stores all configurations for which $F$ is defined

  $\Dep(S_0):=\emptyset$;

  $\Wait:=\{(S_0,\alpha,T) \mid 
    T=\posttime(\postta_{\alpha}(S_0))\not=\emptyset, \alpha\text{ transition of }\game\}$;

  \While{($\Wait\not=\emptyset$ and $(v_0,\mathbf 0)\notin F(S_0)$)}
        {$(S,\alpha,T)):=\Pop{\Wait}$;\\
          \If(\hfill{// case $A$}){$T\in\Passed$}
          {$\Dep(T):=\Dep(T) \cup \{(S,\alpha,T)\}$;\\
          $W:=\predt\left(F(S)\cup \bigcup_{\substack{S\xrightarrow{c} V\\ \!\!V\in\Passed\!\!}} \predc(F(V)),\ \bigcup_{\substack{S\xrightarrow{u} V\\ \!\!V\in\Passed\!\!}}\predu(V\setminus F(V))\right) \cap S$;\\
          \If{$F(S) \subsetneq W$}{$F(S):=W$;\\
              $\Wait:=\Wait\cup\Dep(S)$;}}
          \Else(\hfill{// case $B$})
                  {$\Passed:=\Passed \cup\{T\}$; \\
                    \If{$c(T)==\Win$}{$F(T):=T$\\ $\Wait:=\Wait\cup\{(S,\alpha,T)\}$}
                       \Else{$F(T):=\emptyset$}
                  $\Dep(T):=\{(S,\alpha,T)\}$;\\
                  $\Wait:=\Wait\cup\{(T,\alpha,U) \mid U=\posttime(\postta_{\alpha}(T))\not=\emptyset, \alpha\text{ transition of }\game\}$;
        }
        }
  \leIf{$(v_0,\mathbf 0)\in F(S_0)$}{\Return 1}{\Return 0}

        \caption{Symbolic on-the-fly algorithm for timed reachability}
  \label{11-algo:sotftr}
\end{algorithm}

The correctness of the algorithm can be proven using the following
lemma. We~omit the proof of this lemma, as it is tedious and and does
not contain any difficult argument.

\begin{lemma}\label{11-lem:sotftr}
  The following properties hold at the end of each run through
  the \textbf{while} loop of~\Cref{11-algo:sotftr}:
  \begin{itemize}
  \item for any $S\in\Passed$ and any transition~$\alpha$, if
    $T=\posttime(\postta_\alpha(S))\not=\emptyset$,
    then either $(S,\alpha,T)\in\Wait$, or
    $T\in\Passed$ and $(S,\alpha,T)\in\Dep(T)$;
  \item for any~$S\in\Passed$ and $q\in F(S)$,  $q$ is winning for~Eve;
  \item for any $S\in\Passed$ and $q\in S\setminus F(S)$,
    either
    $\Wait$ contains a symbolic transition~$(S,\alpha,S')$ from~$S$
    with~$S'\in\Passed$,
    or
    \[
      q\notin \predt\Bigl(F(S)\cup \bigcup_{\substack{S\xrightarrow{c} V\\ V\in\Passed}} \predc(F(V)),\ \bigcup_{\substack{S\xrightarrow{u} V\\ V\in\Passed}} \predu(V\setminus F(V))\Bigr) \cap S.
    \]
  \end{itemize}
\end{lemma}
The proof is omitted but can be found in~\cite{Cassez.David.ea:2005}.

Using these invariants, we get the following result:
\begin{lemma}
  If \Cref{11-algo:sotftr} terminates, all configurations in~$F(S)$ for
  any~$S\in\Passed$ are winning for~Eve;  if~additionally
  $\Wait=\emptyset$, then all configurations in~$S\setminus F(S)$, for
  any~$S\in\Passed$, are losing for~Eve.
\end{lemma}

\begin{proof}
The~first statement corresponds to the second statement of
\Cref{11-lem:sotftr}. Now, assume $\Wait^n=\emptyset$ after
termination at the $n$-th step,
and
let $L=\{q\in \Locs\times\mathbb{R}_{\geq 0}^{\Clocks} \mid \exists
S\in\Passed^n.\ q\in S\setminus F^n(S)\}$, and $M$ be the complement
of~$L$. For~any~$S\in\Passed^n$, we~then have $M\cap S\subseteq
F^n(S)$.

Pick $q\in \pi(M)\cup \Win$, where we abusively write $\Win$ to denote
all configurations with location coloured~$\Win$. We~prove that $q\in M$:
\begin{itemize}
\item if $q\in\Win$: assume $q\in L$, and pick $S\in\Passed^n$ such
  that $q\in S\setminus F^n(S)$. Since $q\in S$, it~holds $c(S)=\Win$;
  but then $F^n(S)$ is defined (since $S\in\Passed^n$), and it
  equals~$S$ by initialization of~$F$. This contradicts the fact that
  $q\in S\setminus F^n(S)$, hence $q\in M$.
\item if $q\in\pi(M)$: we again assume $q\in L$. Then for some
  $S\in\Passed^n$, $q\in S\setminus F^n(S)$. By~the third property of
  \Cref{11-lem:sotftr}, $q\notin W^n$
  (because $\Wait^n$ is
  empty).  Since $M\cap U \subseteq F^n(U)$ for all~$U\in\Passed^n$
  and $F^n(U)=\emptyset$ for all~$U\notin\Passed^n$,
  and by monotonicity, we~get
  \[
  q\notin\predt\Biggl((M\cap S)\cup \bigcup_{\substack{S\xrightarrow{c}
    V}} \predc(M\cap V),\ \bigcup_{\substack{S\xrightarrow{u}
    V}} \predu(V\setminus (M\cap V))\Biggr) \cap S.
  \]
%
  Now, notice that any $S\in\Passed^n$ is closed under~$\posttime$. Then
  \begin{xalignat*}1
    \pi(M)\cap S &= \predt(M\cup\predc(M), \predu(\overline M)) \\
     &= \predt((M\cap S)\cup (\predc(M)\cap S), \predu(\overline M)).
  \end{xalignat*}
  Now, it~can be checked that $\predc(M)\cap S \subseteq
  \bigcup_{\substack{S\xrightarrow{c} V}} \predc(M\cap V)$ and
  $\predu(\overline M) \supseteq \bigcup_{\substack{S\xrightarrow{u}
      V}} \predu(V\setminus (M\cap V))$.  So the fact that $q\in
  \pi(M)$ and $q\in S$ leads to a contradiction. This entails that
  $q\notin L$, hence $q\in M$.
\end{itemize}
In the end, we~have proven that $\pi(M)\cup\Win \subseteq M$, so that
$M$ is a pre-fixed point of $X\mapsto \pi(X)\cup\Win$, hence it~contains
all winning configurations of Eve and $L$ only contains losing
configurations.
\end{proof}


The procedure given in \Cref{11-algo:sotftr} will in general not
terminate: as in the case of timed automata, the number of zones
generated by the algorithm may be infinite. This is classically
avoided using \emph{extrapolation}: this consists in abstracting the
zones being considered by larger zones, defined by only using integer
constants less than the maximal constant appearing in the timed arena
(as we did for regions in the proof
of~\Cref{11-thm:timed-control}). This can be proven to preserve
correctness, and makes the~number of zones finite. Termination follows
by noticing that any triple~$(S,\alpha,T)$ may be added to~$\Wait$ only
a finite number of times (bounded by the number of regions
in~$T$). 
We~can then conclude:
\begin{theorem}
\Cref{11-algo:sotftr} terminates when extrapolation is used, and returns~$1$ if, and only~if,
$(\ell_0,\mathbf{0})$~is winning for Eve.
\end{theorem}

\section*{Bibliographic references}
\label{11-sec:references}
Timed games were first introduced by Asarin, Maler, Pnueli and Sifakis~\cite{Maler.Pnueli.ea:1995}, in a
slightly different form.  Our presentation is based on the algorithms
proposed by Liu and Smolka~\cite{Liu.Smolka:1998}, and extended to timed games by
Cassez, David, Fleury, Larsen and Lime~\cite{Cassez.David.ea:2005};
its~main advantage is that it runs \emph{on-the-fly}, building (part
of) the arena while exploring it, and terminating as soon as a winning
strategy is found.  A~first on-the-fly algorithm was proposed by Tripakis and Altisen
in~\cite{Tripakis.Altisen:1999}, but it was not fully on-the-fly as it would run on a
quotient graph of the timed arena, which involves an expensive
preprocessing step.

Timed games---and already timed automata---may exhibit unrealistic
behaviours, such as finite-duration executions containing infinitely
many transitions (often referred to as \emph{Zeno} behaviours).
In~our semantics, Adam may always prevent Eve from playing her move
by choosing a shorter delay than hers, even if it means selecting a convergent
sequence of delays. \Cref{11-fig:exzeno} displays a simple example of
such a situation: in~that game, Adam~may prevent Eve from reaching her
winning state~$q_1$ by always selecting a delay shorter than the delay
proposed by Eve.

\begin{figure}[ht]
  \centering
  \begin{tikzpicture}
    \draw (0,0) node[state] (a) {$q_0$};
    \draw[initial] (a.-135) -- +(-135:3mm);
    \draw (2,0) node[state,accepting] (b) {$q_1$};
    \draw[arrow] (a) -- (b) node[midway,above] {$x>0$};
    \draw (a) edge[arrow,selfloop=90,dashed]
      node[above] {$x>0;x:=0$} (a);
  \end{tikzpicture}
  \caption{A timed game where Adam may prevent Eve from reaching~$q_1$}
  \label{11-fig:exzeno}
\commentAlt{Figure~\ref{11-fig:exzeno}: A finite automaton-like diagram with two circular nodes, q0 and q1, showing transitions and a self-loop based on a condition for variable 'x'.}
\commentLongAlt{Figure~\ref{11-fig:exzeno}: The image displays a directed graph with two circular nodes, 'q0' and 'q1'. An incoming arrow points to 'q0', indicating it as a starting state. Node 'q0' has a dashed self-loop labeled 'x > 0; x := 0'. A solid arrow connects 'q0' to 'q1', labeled 'x > 0'. Node 'q1' is represented with a double ring, typically indicating an accepting or final state.}
\end{figure}

An alternative semantics of timed games was proposed by de~Alfaro,
Faella, Henzinger, Majumdar, and Stoelinga~\cite{Alfaro.Faella.ea:2003} to
circumvent this problem: it~consists in \emph{blaming} at each round
the player playing the shortest delay, and declaring any player losing
(even~if she reaches her goal) in case the infinite sequence of delays
converges and that player received infinitely many blames.  For~such a
semantics, Eve~has a winning strategy in the game
of~\Cref{11-fig:exzeno}, which consists in proposing a converging
sequence of delays, until her action is applied; such a strategy
requires infinite memory, but strategies with randomized delays can
circumvent this~\cite{Chatterjee.Henzinger.ea:2008}. Other~approaches to avoid
arbitrary-precision strategies have been explored
in~\cite{Bouyer.Markey.ea:2015}.

Timed games have been extended with weights, in order to model other
quantities besides time (\textit{e.g.} energy consumption). This is somewhat
similar to the finite-state games extended with payoffs
of~\Cref{5-chap:payoffs}; in~the timed setting however, besides evolving
along transitions in the game, the payoff is also modified when timed
elapses, and the change is proportional to the time spent. As~proven
in~\cite{Brihaye.Bruyere.ea:2005}, optimal reachability (a.k.a. the shortest-path
problem) is undecidable in that setting, but arbitrary-precise
approximation of the optimal cost can be computed~\cite{Bouyer.Jaziri.ea:2015}.

Timed games with partial observability have been investigated
in~\cite{Bouyer.DSouza.ea:2003}: in~that setting, Eve only has partial observation
of the state and clock valuation of the arena; she~also owns a finite
set of clocks she can use to measure other delays. Whether Eve has a
winning strategy in such a setting has been proved decidable if the
set of clocks and their precision are fixed; it~is undecidable if they
are not fixed.
A~different setting was developed in~\cite{Cassez.David.ea:2007}, with
\emph{stuttering-invariant strategies}, which are triggered by
observation changes. The~on-the-fly algorithm presented in this
chapter can be adapted to that setting~\cite{Cassez.David.ea:2007}. 

The~algorithm we presented in this chapter is implemented in the tool
Uppaal TiGa~\cite{Behrmann.Cougnard.ea:2007}. Uppaal~TiGa has been applied on various
cases~\cite{Cassez.Jensen.ea:2009}, and combined with
reinforcement-learning techniques to efficiently synthesise, optimise
and evaluate strategies for stochastic timed games~\cite{David.Jensen.ea:2015}.
Another approach, relying on
abstraction-refinement techniques, was proposed in~\cite{Ehlers.Mattmuller.ea:2010};
it~merges locations so as to obtain simpler games, while weakening one
player and strengthening her opponent. Solving the smaller games
provides approximations on the winning set, which are used to refine
the abstractions and accelerate their analyses. This~approach is
implemented in the tool Synthia~\cite{Peter.Ehlers.ea:2011}.

\ifpictures
\includepdf{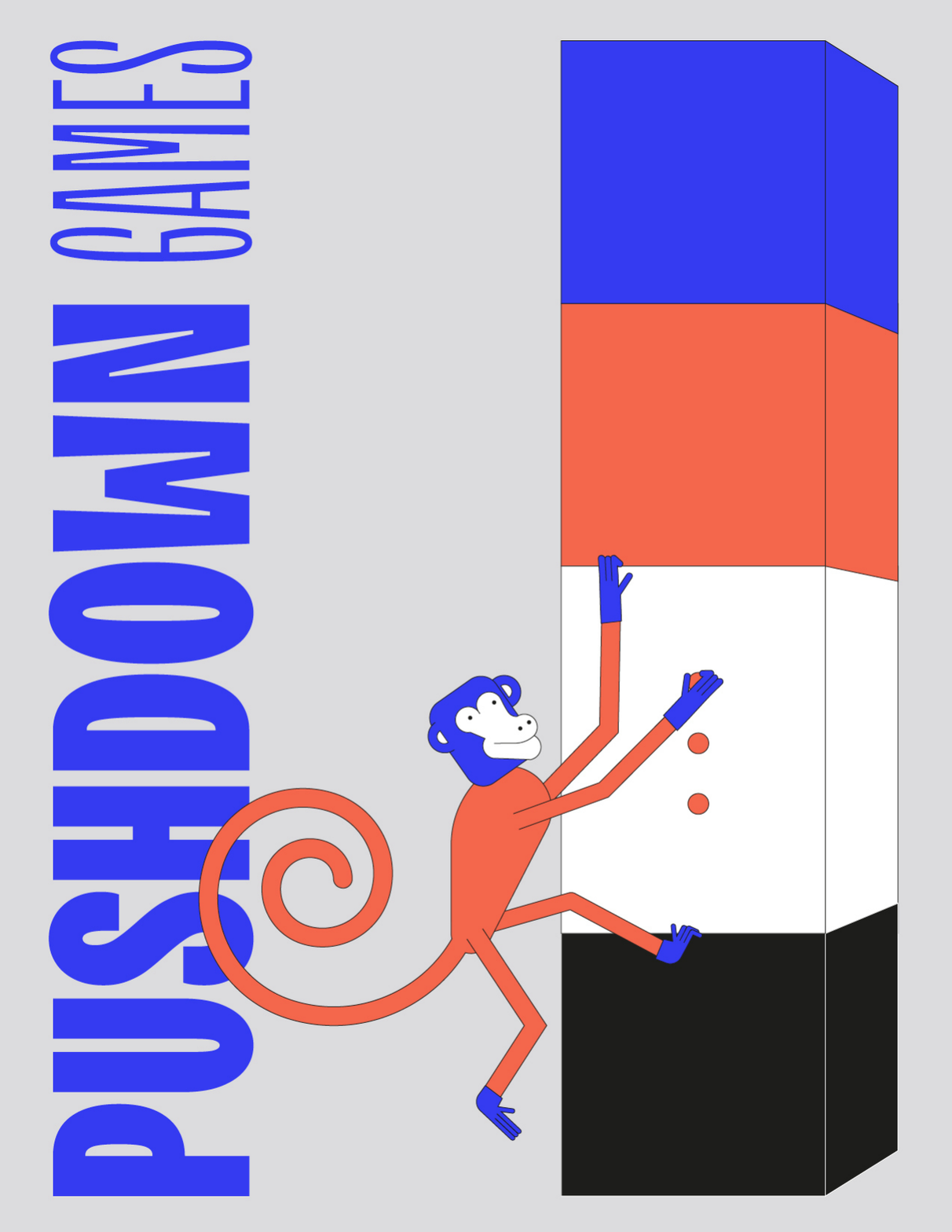}
\fi
\author[Arnaud Carayol, Olivier Serre]{Arnaud Carayol, Olivier Serre}
\copyrightline{Copyright by Arnaud Carayol and Olivier Serre 2025, to be published by Cambridge University Press in the volume \textit{Games on Graphs} edited by Nathana\"el Fijalkow}

\chapter{Pushdown Games}
\chapterauthor{Arnaud Carayol, Olivier Serre}
\label{12-chap:pushdown}

\def\AC#1{\textcolor{green!50!black}{\checkmark}\marginpar{\color{green!50!black}AC: #1}} 
\def\acchanged#1{{#1}}
\def\OS#1{\textcolor{red}{\checkmark}\marginpar{\color{red}OS: #1}} 

\renewcommand{\qed}{\hfill$\square$}
\newcommand{\pop}{\mathrm{pop}}
\newcommand{\push}[1]{\mathrm{push}(#1)}

\newcommand{\PDS}{\mathcal{P}}
\newcommand{\pdscol}[1]{\col(#1)}

\newcommand{\vect}[1]{\overrightarrow{#1}}
\newcommand{\ttrue}{t\! t}
\newcommand{\ffalse}{f\!\! f}

\newcommand {\Stepsg}[1]{\ensuremath{\mathit{Steps}_{#1}}}
\newcommand{\sh}{\mathrm{sh}}

\newcommand{\fgraph}{\widetilde{G}}
\newcommand{\farena}{\widetilde{\arena}}
\newcommand{\fgame}{\widetilde{\game}}
\newcommand{\fplay}{\widetilde{\play}}
\newcommand{\fsigma}{\widetilde{\sigma}}

\newcommand {\Rounds}[1]{\ensuremath{\mathit{Rounds}_{#1}}}

\newcommand{\kind}[2]{kind(#1,#2)}
\newcommand{\factcol}[2]{\mathrm{mcol}(#1,#2)}
\newcommand{\fac}[1]{kind^{#1}}
\newcommand{\pcol}[1]{\mathrm{mcol}^{#1}}

\newcommand{\hgraph}{\widehat{G}}
\newcommand{\harena}{\widehat{\arena}}
\newcommand{\hgame}{\widehat{\game}}
\newcommand{\hplay}{\widehat{\play}}
\newcommand{\hsigma}{\widehat{\sigma}}

This chapter studies two-player games whose arena is defined by a pushdown system\footnote{We use here the term pushdown system rather than pushdown automaton to stress the fact that we are not considering these devices as language acceptors but rather focus on the transitions systems they define.}. The vertices of the arena are the configurations of the pushdown system (\textit{i.e.}, pairs composed of a control state and a word representing the content of the stack) and the edges of the arena are defined by the pushdown system's transitions. For simplicity, both the ownership of a configuration and the objective will only depend on the control state of the configuration. Hence the partition of all the configurations between the two players will simply be given by a partition the control states. We will mainly consider the parity objective. Via a standard reduction (using a deterministic parity automaton, see~\Cref{1-sec:automata}), pushdown parity game can be used to solve any pushdown game with an $\omega$-regular objectives (which in our setting is simply an $\omega$-regular set of infinite words over  the alphabet of control states). 

The main conceptual novelty of this chapter is that the arena is no longer finite. However as these games are described by a finite amount of information: the pushdown system, the ownership partition and the $\omega$-regular objective, they are amenable to algorithmic treatment. The first natural problem in this line is to decide the winner of the game from a given configuration. We will also consider the computation of \emph{finite representations} for the winning regions and the winning strategies.

\section{Notations}
\label{12-sec:notations}
A \emph{pushdown system} is a tuple $\PDS = (Q,Q_{\mEve}, Q_{\mAdam}, \Gamma,\Delta,C)$ where:
\begin{itemize} 
	\item $Q$ is a finite set of control states with $Q = Q_{\mEve} \uplus Q_{\mAdam}$ which is partitioned between the two players \Eve\ and \Adam. Moreover with each state $q$ is associated a colour $\pdscol{q} \in C$;	\item $\Gamma$ is the stack alphabet. There is a special bottom-of-stack symbol, denoted $\bot$, which does not belong to $\Gamma$; we let $\Gamma_\bot$ denote the alphabet $\Gamma \cup \set{\bot}$;
	\item $\Delta:Q\times \Gamma_\bot\rightarrow 2^{Q\times\{\pop,\push{\gamma}\mid \gamma\in\Gamma\}}$ is the transition relation. We additionally require that for all states $p,q\in Q$, $(q,\pop)\notin\Delta(p,\bot)$, \textit{i.e.}, the bottom of stack symbol is never popped.
\end{itemize}

We call a pair $(p,s)\in Q \times \bot\Gamma^*$ a \emph{configuration} of $\PDS$: $p$ is the control state of the configuration while $s$ is its stack content. We let $\sh((p,s))=|s|-1$ denote the \emph{stack height} of the  configuration $(p,s)$. Intuitively, if $(q,\push{\gamma'})$ belongs to $\Delta(p,\gamma)$, the pushdown system
in any configuration of the form $(p,s\gamma)$ can go to state $q$ after pushing the symbol $\gamma$ on top of the stack, leading to the configuration $(q,s\gamma\gamma')$. Similarly, if $(q,\pop)$ belongs to $\Delta(p,\gamma)$, the pushdown system
in any configuration of the form $(p,s\gamma)$ can go to the configuration $(q,s)$ after \emph{poping} the top symbol from the stack.


A pushdown system induces an arena $\arena = (G,\VE,\VA)$ called a \emph{pushdown arena} where
\begin{itemize}
	\item the set of vertices is the set $V = Q \times \bot\Gamma^* $ of configurations of $\PDS$ with $\VE = Q_{\mEve} \times \bot\Gamma^*$ 
	and $\VA = Q_{\mAdam} \times \bot\Gamma^* $;
	\item the set $E$ of edges induced by $\Delta$ is
\begin{equation*}
\begin{split}
	E  = & \{((p,s\gamma),\pdscol{p},(q,s)) \mid (q,\pop)\in\Delta(p,\gamma)\}\quad \cup \\ 
	&  \{((p,s\gamma),\pdscol{p},(q,s\gamma\gamma')) \mid (q,\push{\gamma'})\in\Delta(p,\gamma)\}.
\end{split}	
\end{equation*}
\end{itemize}


\begin{remark}
In this chapter, we deviate slightly from the general setting used in the book as we colour vertices and not edges. 
Because we only consider qualitative objectives, it is more convenient to consider the equivalent setting where we label vertices by colours rather than edges which is the usual convention in pushdown games.  
In the definition of a pushdown arena, the colour of an edge is uniquely determined by the colour of the control state of the source vertex. Therefore,  we also view in this chapter plays as sequences of vertices rather than sequences of edges.
\end{remark}

Finally, a \emph{pushdown game} is a game played on a pushdown arena. In this chapter we only consider qualitative objectives of the form $\Omega\subseteq C^\omega$ where $C$ is the set of colours. As in our definition, 
colours are associated to control states, the objective we consider only depend on the sequence of control states of the configurations visited along a play. By a slight abuse of notation, for a play $\pi \in V^*$, we let $\pi=(p_1,s_1)(p_2,s_2)\cdots \in \Omega$ denotes the fact that the sequences of colours $(\pdscol{p_i})_{i\geq 1}$ belongs to $\Omega$.

\begin{example}\label{12-ex:pushdown-game-1}
Consider the pushdown system $\PDS = (Q,Q_{\mEve}, Q_{\mAdam}, \Gamma,\Delta)$ where:
\begin{itemize}
	\item $Q_{\mEve}=\{q,r\}$ and $Q_{\mAdam}=\{p\}$; $\pdscol{p} = 1$, $\pdscol{q} = 2$ and $\pdscol{r} = 0$.
	\item $\Gamma=\{\gamma\}$ is a singleton.
	\item $\Delta(p,\bot) = \{(p,\push{\gamma})\}$, $\Delta(p,\gamma) = \{(p,\push{\gamma}),(q,\pop)\}$, $\Delta(q,\bot) = \{(r,\push{\gamma})\}$, $\Delta(q,\gamma) = \{(q,\pop)\}$ and $\Delta(r,\gamma) = \{(q,\pop)\}$.
\end{itemize}

\Cref{12-fig:example-pushdown-game-1} depicts the part of the pushdown arena $\arena$ induced by $\PDS$ when restricted to the vertices reachable from $(p,\bot)$.

Consider the reachability game $(\arena, \Reach(\{0\}))$. Then, every vertex of the form $(q,\bot v)$ is winning for \Eve\ and every vertex of the form $(p,\bot v)$ is winning for \Adam\ as \Adam\ can always choose to push a $\gamma$-symbol while remaining in state $p$ hence always avoiding the state $r$ (which is the only state with colour $0$). If instead we consider, the B{\"u}chi game $\Buchi(\{0,2\})$, then \Eve\ is winning form all vertices. The strategy for \Adam\ consisting in always pushing a $\gamma$-symbol while remaining in state $p$ results in a play that infinitely often sees the colour $2$.

\begin{figure}[htb]
\begin{center}
\begin{tikzpicture}[>=stealth',thick,scale=1,transform shape]
\tikzstyle{Adam}=[draw,inner sep=4]
\tikzstyle{Eve}=[draw,rounded rectangle,inner sep=4]
\tikzstyle{Nature}=[draw,diamond,scale = .65,font=\Large]
\tikzstyle{AdamH}=[]
\tikzstyle{EveH}=[circle]
\tikzset{every loop/.style={min distance=10mm,looseness=10}}
\tikzstyle{loopleft}=[in=150,out=210]
\tikzstyle{loopright}=[in=-30,out=30]
\node[Adam] (A1) at (1,0) {$(p,\bot)$};
\node[Adam] (A2) at (3.2,0) {$(p,\bot\gamma)$};
\node[Adam] (A3) at (5.5,0) {$(p,\bot\gamma\gamma)$};
\node[Adam] (A4) at (7.7,0) {$(p,\bot\gamma\gamma\gamma)$};
\node[AdamH] (A5) at (9.6,0) {};
\node[Eve] (E0) at (-1,-1.5) {$(r,\bot\gamma)$};
\node[Eve] (E1) at (1,-1.5) {$(q,\bot)$};
\node[Eve] (E2) at (3.2,-1.5) {$(q,\bot\gamma)$};
\node[Eve] (E3) at (5.5,-1.5) {$(q,\bot\gamma\gamma)$};
\node[Eve] (E4) at (7.7,-1.5) {$(q,\bot\gamma\gamma\gamma)$};
\node[EveH] (E5) at (9.6,-1.5) {};

%
%

\path[->] (A1) edge    (A2);
\path[->] (A2) edge  (A3);
\path[->] (A2) edge (A3);
\path[->] (A3) edge (A4);
\path[->,dotted] (A4) edge (A5);

\path[->] (A2) edge (E1);
\path[->] (A3) edge (E2);
\path[->] (A4) edge (E3);
\path[->,dotted] (A5) edge (E4);

\path[->] (E0) edge[bend right]  (E1);
\path[->] (E1) edge (E0);
\path[->] (E2) edge (E1);
\path[->] (E3) edge (E2);
\path[->] (E4) edge (E3);
\path[->,dotted] (E5) edge (E4);

\end{tikzpicture}
\end{center}
\caption{Pushdown arena from \Cref{12-ex:pushdown-game-1}}
\label{12-fig:example-pushdown-game-1}
\end{figure}

\end{example}

As a pushdown arena is in general infinite, the \emph{winning region} for \Eve, \textit{i.e.}, the set of winning vertices for \Eve\ may not admit a finite presentation. Similarly, for objectives for which finite-memory strategies exists, the question of whether such a strategy can be finitely presented (and computed) is raised. Hence, we will in general distinguish the following three algorithmic problems.

\decpb[Solving a pushdown game]{A pushdown game $\game$ and an initial vertex $v_0$}{Does Eve win $\game$ from $v_0$?}

\decpb[Computing the winning region]{A pushdown game $\game$}{Output a finite presentation of the set $v$ of vertices from which  Eve wins $\game$}

In \Cref{12-thm:regularity-wr}, we will show that the winning region can be described by a finite-state automaton  for a large class of qualitative winning conditions.

\decpb[Computing a winning strategy]{A pushdown game $\game$}{Output a finite presentation of a strategy for \Eve\ that is winning from any vertex in the winning region for Eve in $\game$}


We will show, for pushdown parity games, that the winning strategy can be described using either a finite-state automaton  or a pushdown automaton (see \Cref{12-subsec:regular-strat}).

\section{Profiles and regularity of the winning regions}
\label{12-sec:profiles}
In this section, we consider a large class of objectives called \emph{prefix-independent}. For these objectives, a pushdown game can be meaningfully decomposed by considering the part of the game between the moment a symbol is pushed onto the stack and stopping as soon as it is popped.  As a consequence, we will see that for prefix-independent
objectives, the winning region can be described  using finite state automata.

To this extent, we introduce \emph{reduced games} which start with a stack containing only one symbol $\gamma$ and stop as soon as this symbol is popped from the stack. If the symbol is never popped, the objective is unchanged and otherwise the winner of the game is determined by the state reached when popping the symbol.

Let $\game=(\arena,\Omega)$ be a pushdown game played on an arena $\arena = (G,\VE,\VA)$ generated by a pushdown system $\PDS = (Q,Q_{\mEve}, Q_{\mAdam}, \Gamma,\Delta)$. For any subset $R\subseteq Q$ of control states of $\PDS$, we define a new objective $\Omega(R)$ such that a play $\play$ belongs to $\Omega(R)$ if one of the following happens:
\begin{itemize}
\item the play $\play$ belongs to $\Omega$ and does not contain
any configuration with an empty stack ( {\textit{i.e.}}, of the form $(q,\bot)$ for some state $q \in Q$),
\item the play $\play$ contains a configuration with the empty stack and the first such configuration has a state in R.
\end{itemize}

More formally, letting $V=V_\mEve\cup V_\mAdam$, $V_R = R \times \bot \Gamma^*$ and $V_\bot = Q \times \{\bot\}$ 
$$ \Omega(R) = (\Omega \setminus V^*V_\bot V^\omega)\cup (V \setminus V_\bot)^* V_R V^\omega$$
Finally, we let $\game(R)$ denote the game $(\arena,\Omega(R))$.

Remark that contrarily to the rest of the objectives considered in this chapter, $\Omega(R)$ does depend on the sequences of vertices visited by the play and not only on their colour. It would have been possible by a slight modification of the arena to only express the objective on their colours. However, our choice simplifies the presentation of the reduced games.

For any state $q\in Q$ and any stack letter $\gamma\in\Gamma$, we denote by $\mathcal{R}(q,\gamma)$ the set of subsets $R\subseteq Q$ for which \Eve\ wins in $\game(R)$ from $(q,\bot\gamma)$:
$$\mathcal{R}(q,\gamma)=\{
R\subseteq Q\mid (q,\bot\gamma) \text{ is winning for \Eve\ in } \game(R)
\}$$ and we refer to $\mathcal{R}(q,\gamma)$ as the \emph{$(q,\gamma)$-profile} of $\game$.

An objective $\Omega\subseteq C^\omega$ is \emph{prefix-independent} if the following holds: for every $u\in C^\omega$ and for every $v\in C^*$,  $u\in \Omega$ if and only if $vu\in \Omega$. The Büchi, co-Büchi and parity objectives are examples of prefix-independent objectives and the reachability objective is not.

\begin{remark} 
\label{12-rmk:stay-staying-alive}
For a prefix-independent objective, a play respecting a winning strategy for \Eve\ that starts in \Eve's winning region always stays in this region. Obviously this is no longer true if the objective is not prefix-independent. For instance in a reachability game, a play following a winning strategy for \Eve\ can leave the winning region of \Eve\ once the target has been reached. 	
\end{remark}

This simple property allows to use profiles to give an inductive characterisation of the winning region for \Eve\ when the objective is prefix-independent.

\begin{proposition}\label{12-prop:returning} Assume that $\Omega\subseteq C^\omega$ is prefix-independent. 
Let $s\in \Gamma^*$, $q\in Q$ and $\gamma\in\Gamma$. Then \Eve\ has a winning strategy in $\game$ from $(q,\bot s\gamma)$ if and only if there exists some $R\in\mathcal{R}(q,\gamma)$ such that $(r,\bot s)$ is winning for \Eve\ in $\game$ for every $r\in R$.
\end{proposition}

\begin{proof}
Assume \Eve\ has a winning strategy $\sigma$ from $(q,\bot s\gamma)$ in $\game$. Consider the set $\Pi_\sigma$ of all plays in $\game$ that starts from $(q,\bot s\gamma)$ and where \Eve\ respects $\sigma$. Define $R$ to be the (possibly empty) set that consists of all $r\in Q$ such that there is a play in $\Pi_\sigma$ of the form $v_0\cdots v_k (r,\bot s) v_{k+1}\cdots$ where each $v_i$ for $0\leq i\leq k$ is of the form $(p_i,\bot s t_i)$ for some non-empty $t_i$. In other words, $R$ consists of all states that can be reached on popping $\gamma$ for the first time in a play where \Eve\ respects $\sigma$. As seen in \Cref{12-rmk:stay-staying-alive}, \Eve\ is winning from $(r,\bot s)$ for all $r \in R$. It remains to show that $R\in\mathcal{R}(q,\gamma)$.

To this end, define a (partial) function $\xi$ as $\xi((p,\bot s t))=(p,\bot t)$ for every $p\in Q$ and set $\xi^{-1}((p,\bot t))=(p,\bot s t)$. Then $\xi^{-1}$ is extended as a morphism over $V^*$.  Now a winning strategy for \Eve\ in $\game(R)$ is defined as follows:
\begin{itemize}
\item if some empty stack configuration has already been visited play any valid move, 
\item otherwise go to $\xi(\sigma(\xi^{-1}(\pi))$, where $\pi$ is the current play.
\end{itemize}
By definition of $\Pi_\sigma$ and $R$, it easily follows that the previous strategy is winning for \Eve\ in $\game(R)$, and therefore $R\in\mathcal{R}(p,\gamma)$. 

Conversely, let us assume that there is some $R\in\mathcal{R}(q,\gamma)$ such that $(r,\bot s)$ is winning for \Eve\ in $\game$ for every $r\in R$. For every $r\in R$, let us denote by $\sigma_r$ a winning strategy for \Eve\ from $(r,\bot s)$ in $\game$. Let $\sigma_R$ be a winning strategy for \Eve\ in $\game(R)$ from $(q,\bot\gamma)$. Let us define $\xi$ and $\xi^{-1}$ as in the direct implication and extend them as (partial) morphism over $V^*$. Define the following strategy $\sigma$ for \Eve\ in $\game$ for plays starting from $(q,\bot s\gamma)$. For any such play $\pi$, 
\begin{itemize}
\item if $\pi$ does not contain a configuration of the form $(p,\bot s)$ then we take $\sigma(\pi)=\xi^{-1}(\sigma_R(\xi(\pi)))$;
\item otherwise let $\pi = \pi'\cdot(r,\bot s)\cdot \pi''$ where $\pi'$ does not contain any configuration of the form $(p,\bot s)$. If $r$ does not belong to $R$, $\sigma$ is undefined. Note that this situation will never be encountered in a play respecting $\sigma$ as $\sigma_R$ ensures that $r \in R$. If $r \in R$, one finally sets $\sigma(\pi)=\sigma_r((r,\bot s)\pi'')$.
\end{itemize} 
The strategy $\sigma$ is a winning strategy for \Eve\ in $\game$ from $(q,\bot s\gamma)$. To see this, consider a play $\pi$ starting from $(p,\bot s \gamma)$ and respecting $\sigma$.

 If the play $\pi$ does not contain configurations of the form $(r,\bot s)$ for some $r \in Q$, then the play $\xi(\pi)$ starting in $(p,\bot \gamma)$ respects $\sigma_R$ and is won by \Eve.
As $\xi(\pi)$ does not contain configurations with an empty stack, it must be the case that $\xi(\pi) \in \Omega$. As $\Omega$ only depends on the colours of the states, it is also the case that $\pi \in \Omega$ and hence, $\pi$ is winning for $\Eve$. 

If the play $\pi$ can be decomposed as $\pi' (r,\bot s) \pi''$ where $\pi'$ does not contain any configuration with the stack $\bot s$, the play $\xi(\pi' (r,\bot s))$ respects $\sigma_R$ in $\game(R)$. As $\sigma_R$ is winning, it follows that $r \in R$. By definition of $\sigma$, $(r,\bot s)\pi''$ respects $\sigma_r$ which being winning for \Eve\ implies that $(r,\bot s) \pi'' \in \Omega$. As $\Omega$ is prefix-independent, it follows that $\pi \in \Omega$.
\end{proof}


\Cref{12-prop:returning} implies that the winning region of any pushdown game equipped with a prefix-independent objectives can be described by regular languages.

\begin{theorem}\label{12-thm:regularity-wr}
Let $\game=(\arena,\Omega)$ be a pushdown game played on an arena $\arena = (G,\VE,\VA)$ generated by a pushdown system $\PDS = (Q,Q_{\mEve}, Q_{\mAdam}, \Gamma,\Delta)$, and such that $\Omega\subseteq C^\omega$ is prefix-independent. Then for any state $q\in Q$, the set \[
L_q=\{u\in \Gamma^*\mid (q,\bot u)\in W_\mEve \}
\] is a regular language over the alphabet $\Gamma$.
\end{theorem}

\begin{proof}
Fix a control state $q \in Q$, we consider a deterministic finite state automaton defined as follows. Its set of control states consists of the subsets of $Q$ and the initial state is $S_{in}=\{p\mid (p,\bot)\in W_\mEve\}$. From the state $S$ upon reading the letter $\gamma$ the automaton goes to the state $\{p\mid S\in\mathcal{R}(p,\gamma)\}$. Finally a state $S$ is final if and only if $q\in S$. It is then an immediate consequence of \Cref{12-prop:returning} that this automaton accepts the language $L_q$.
\end{proof}

\begin{remark}\label{12-rmk:automata-winning-region}
By a slight abuse, we can think of the $|Q|$ automata in \Cref{12-thm:regularity-wr}	as a single automaton that is design to first read the stack content and finally reads the control state $q$ (in this latter step, from state $S$ it either go to a final state if $q\in S$ or to a rejecting one otherwise).
\end{remark}

\begin{remark}
Note that the characterisation in \Cref{12-thm:regularity-wr} is \emph{a priori} not effective. Indeed, to construct automata for the languages $L_q$ one needs to be able to compute all the $(q,\gamma)$-profiles $\mathcal{R}(q,\gamma)$ of $\game$ and compute
the winner from configurations of the form $(q,\bot)$. 

\newcommand{\Omegapar}{\Omega_{\textrm{pal}}} 
Consider, for instance, the objective over the set of the colours $C=\{0,1,\#,\$\}$
\[
\begin{array}{l}
\Omegapar = \{ w \in C^\omega \mid w \;\textrm{contains infinitely many factors of the form}\\
\quad\quad\quad\quad\quad\quad\quad\;\;\textrm{$\#u\$\tilde{u}\#$ with $u \in \{0,1\}^*$} \} \\
\end{array}
\]
where $\tilde{u}$ denotes the mirror of the word $u$.

As a language of $\omega$-words, this objective is accepted by a deterministic $\omega$-pushdown automaton with a Büchi acceptance condition. Between two consecutive occurrences of the $\#$-symbol, the automaton checks that the word $w$ appearing in between these two occurrences is of the form
$u\$\tilde{u}$ for some word $u \in \{0,1\}^*$. This can be done in a deterministic manner as follows. First the automaton pushes onto the stack a symbol $\bot'$ (which will play the same role as the bottom of stack symbol) then it pushes onto the stack all symbols in $\{0,1\}$ that are read. Then when the first $\$$-symbol is read, it only allows to read the symbol that is on the top of the stack before popping it. Finally when a $\#$-symbol is read, if the top-most symbol of the stack is $\bot'$  the unique final state is visited. Hence ensuring that a final state is visited if the word $w$ is of the required form. If $w$ contains several $\$$ symbols or if the symbol read does not correspond to the top of the stack, the automaton enters a non-final state is in which it waits for the next $\#$-symbol.

For games with finite arenas and the $\Omegapar$ objective,  deciding the winner reduces to deciding the winner in a  pushdown game with the Büchi objective which is decidable as we will prove later in this chapter. The pushdown game is essentially a synchronized product between the finite arena and the $\omega$-pushdown automaton described previously.

However the problem of deciding the winner in a pushdown game with the $\Omegapar$ objective is undecidable even if all vertices belong to \Eve. The undecidability is proved by a reduction from  Post correspondance problem (PCP) which is a well-known to be undecidable. Recall that an instance of PCP is a finite sequence $(r_1,\ell_1),\ldots,(r_n,\ell_n)$ of pairs of words over $\{0,1\}$. Such an instance is said to admit a solution if there exists a sequence of indices $i_1\cdots i_k \in [1,n]^*$ such that:
\[
 r_{i_1} r_{i_2} \cdots r_{i_k} = \ell_{i_1} \ell_{i_2} \cdots \ell_{i_k}.
\]
The PCP problem is, given an instance, to decide if it admits a solution.
 
For an instance $I=(\ell_1,r_1),\ldots,(\ell_n,r_n)$ of PCP, we construct a pushdown game $G_I$  with the objective $\Omegapar$ such that \Eve\ wins $G_I$ from $(p_\star,\bot)$ if and only if $I$ admits a solution. In this game, \Eve\ plays alone and the play is decomposed in two phases that will repeat:
\begin{itemize}
\item in phase 1, \Eve\ can push any index $i$ in $[1,n]$	while producing the sequence of colours $r_i$. As soon as at least one index has been pushed, she can also choose to move to the sequence phase while producing the colour $\$$.
\item in phase 2, \Eve\ must (until the bottom of stack symbol is reached) pop the top most element of the stack $i$ while producing the sequence of colours $\tilde{\ell_i}$. When the bottom of stack symbol is encountered, \Eve\ goes back to the first phase while producing the colour $\#$.
\end{itemize}

If $I$ has a solution $i_1 \cdots i_k$ then the strategy in which \Eve\ always pushes this sequence in phase 1 is winning for her. As $i_1 \cdots i_k$ is a solution of $I$,
we have $u=r_{i_1}\cdot r_{i_k}=\ell_{i_1}\cdots\ell_{i_k} $ and by construction of the game, the sequence of colours associated with the play is $(u\$\tilde{u})^\omega \in \Omegapar$. Conversely if \Eve\ has a winning strategy from $(q_\star,\bot)$ then the sequence of colours associated with the winning play belongs to $\Omegapar$. In particular, it must contain a factor of the form $\#u\$\tilde{u}\#$. By construction of the game the sequence of indices $i_1 \cdots i_k$ pushed while producing $u$ is a solution of $I$.
\end{remark}

\subsection{Pushdown reachability games}
\label{ssec:reachability-pushdown-games}

We will see in the following section that for the parity condition and more generally for any $\omega$-regular winning conditions the $(q,\gamma)$-profiles can be computed for any $q \in Q$ and $\gamma \in \Gamma$. We start by the simpler case of the reachability objective.

At first sight, the reachability objective is not captured by \Cref{12-thm:regularity-wr} as it is not prefix-independent. However with a slight adaptation of the reduced game $\game(R)$ \Cref{12-thm:regularity-wr}  for the reachability condition. Intuitively we ask that the play stops as soon as a target vertex is reached.

More formally, for a reachability objective $\Reach(F)$ and letting $V=V_\mEve\cup V_\mAdam$ and $V_F = \{ v \in V \mid \pdscol{v} \in F\}$:
\[ \overline{\Omega}(R) = [(V \setminus Q \times \{\bot\})^* \cdot V_F \cdot V^\omega]\cup [(V \setminus (Q\times\{\bot\} \cup V_F))^* \cdot (R\times \{\bot\}) \cdot V^\omega]
\]

It is easily shown that with this modification to the definition of profiles both \Cref{12-prop:returning} and \Cref{12-thm:regularity-wr} are true for the reachability objective.
\newcommand{\Profs}{\mathrm{Profs}}
In the special case of the reachability objective, the set $\Profs$ of triples $(p,\gamma,R)$ such that $R \in \mathcal{R}(p,\gamma)$ can be expressed as a least fixed point.

More precisely, $\Profs$ is the smallest subset of $Q\times \Gamma \times \mathcal{P}(Q)$ such that for $p \in Q$, $\gamma \in \Gamma$ and $R \subseteq Q$, $(p,\gamma,R)$ belongs $\Profs$ if either: 
\begin{enumerate}
\item $p \in Q_F=\{ q \in Q \mid \pdscol{q} \in F\}$,
\item or $p \in Q_{\mEve}$ and for some $q \in Q$,
\begin{itemize}
\item either $(q,\pop) \in \Delta(p,\gamma)$ and $q \in R$,
\item or $(q,\push{\gamma'}) \in \Delta(p,\gamma)$  and $(q,\gamma',R') \in \Profs$ for some $R' \subseteq Q$ such that for all $p' \in R'$, $(p',\gamma,R) \in \Profs$.
\end{itemize}
\item or $p \in Q_A$ and for all $q \in Q$ the following hold: 
\begin{itemize}
\item $(q,\pop) \in \Delta(p,\gamma)$ implies $q \in R$,
\item  $(q,\push{\gamma'}) \in \Delta(p,\gamma)$ implies that there exists $R' \subseteq Q$ such that $(q,\gamma',R') \in \Profs$ and for all $p' \in R'$, $(p',\gamma,R) \in \Profs$. 
\end{itemize}
\end{enumerate}


Using this characterisation, the set $\Profs$ can be computed using the standard method for computing the least fixed point of a monotonic function by computing the sequence of approximants $\Profs_0 = \emptyset \subseteq \Profs_1 \subseteq \Profs_2 \cdots$ until it stabilizes. More precisely, for all $i \geq 0$, $\Profs_{i+1}$ is obtained by adding to $\Profs_{i}$ all the tuples that can be inferred using the properties $(1)$, $(2)$ and $(3)$ above applied to $\Profs_i$. As at most $|Q|\cdot |\Gamma|\cdot 2^{|Q|}$ tuples can be added, the sequence must stabilize in at most $|Q| \cdot |\Gamma| \cdot 2^{|Q|}$ steps. As the computation of $\Profs_{i+1}$ from $\Profs_i$ can be performed in polynomial time, the profiles in a pushdown reachability game can be computed in time  $p(|Q| \cdot |\Gamma| \cdot 2^{|Q|})$ for some polynomial $p$.

\begin{remark}
In the case where \Eve\ plays alone (\textit{i.e.}, $Q=Q_\mEve$), there is only on play respecting a fixed strategy for \Eve\ and as a result, one only need to compute profiles of the form $(p,\gamma,R)$ with $|R|\leq 1$. In this setting, the fixed-point characterisation yields a polynomial-time algorithm to compute the set of profiles.
\end{remark}




\section{Pushdown parity games}
\label{12-sec:parity}
We now focus on the central case of parity objectives. 
In \Cref{12-subsec:computing-profiles}, we show how to compute the set of profiles using a reduction to finite parity game.

For the rest of this section, we fix a pushdown parity game $\game$ played on an arena $\arena = (G,\VE,\VA)$ generated by a pushdown system $\PDS = (Q,Q_{\mEve}, Q_{\mAdam}, \Gamma,\Delta)$. We also let $V=\VE\cup\VA$ and we let the colours used in the game be $\{0,\dots,d\}$.

\subsection{Computing the profiles}
\label{12-subsec:computing-profiles}

In this section we show how to build a parity game played on a \emph{finite} arena that permits to compute the profiles in $\game$. 

We first start with some terminology and a basic result. For an infinite play $\play=v_0v_1\cdots$, let
$\Stepsg{\play}$ be the set of indices of positions where no
configuration of strictly smaller stack height is visited later in the
play. More formally, $$\Stepsg{\play}=\{i\in\mathbb{N}\mid \forall
j\geq i, \sh(v_j)\geq \sh(v_i)\}.$$ Note that $\Stepsg{\play}$ is always
infinite and hence induces a factorisation of the play $\play$ into
finite pieces.

In the factorisation induced by $\Stepsg{\play}$, a factor $v_i\cdots v_j$ is called a
\emph{bump} if $\sh(v_j)=\sh(v_i)$, called a \emph{Stair} otherwise (that is, if $\sh(v_j)=\sh(v_i)+1$).

For any play $\play$ with $\Stepsg{\play}=\{n_0<n_1<\cdots\}$, we can define the sequence $(\pdscol{\play}_i)_{i\geq
0}\in\{0,\dots,d\}^{\mathbb{N}}$ by setting $\pdscol{\play}_i=\max\{\pdscol{v_k}\mid n_i\leq k\leq n_{i+1}\}$.
This sequence fully characterises the parity objective.

\begin{proposition}
\label{12-prop:trans_cond}
Let $\play$ be a play. Then $\play$ satisfies the parity condition  if and only if $\limsup((\pdscol{\play}_i)_{i\geq 0})$ is even.
\end{proposition}

\subsubsection{Simulation Game}

In the sequel, we build a new parity game $\fgame$ over a \emph{finite} arena $\farena$.
This new game
\emph{simulates} the original pushdown game, in the sense that the
sequence of visited colours during a correct simulation of some play $\play$ in $\game$ is
exactly the sequence $(\pdscol{\play}_i)_{i\geq 0}$. Moreover, a play in which
a player does not correctly simulate the pushdown game is losing
for that player. We shall see that the winning region in $\fgame$ allows us to compute the set of profiles $\{\mathcal{R}(q,\gamma) \mid q\in Q \text{ and } \gamma\in\Gamma\}$. Hence, by \Cref{12-thm:regularity-wr}, it will imply that one can solve a pushdown game as well as compute its winning region.

\begin{figure}[htb]
\begin{center}
\begin{tikzpicture}[>=stealth',thick,scale=1,transform shape]
\tikzstyle{Adam}=[draw,inner sep=4]
\tikzstyle{Eve}=[draw,rounded rectangle,inner sep=4]
\tikzstyle{AnyPlayer}=[inner sep=4]
\node[AnyPlayer] (current) at (0,0) {$(p,\alpha,\vect{R},c)$};
\node[Eve] (nextEve) at (0,-1.5) {$(p,\alpha,\vect{R},c,q,\beta)$};
\node[Eve] (nextEveG) at (-3.5,-1.5) {\phantom{$(p,\alpha,\vect{R},c,q,\beta)$}};
\node[Eve] (nextEveD) at (3.5,-1.5) {\phantom{$(p,\alpha,\vect{R},c,q,\beta)$}};

\node at (3.4,-0.4) {$\forall (q,\push{\beta})\in\Delta(p,\alpha)$};

\node[Eve] (ntrue) at (-3,1) {$\ttrue$};
\node at (-3.4,2) {If $\exists (r,\pop)\in\Delta(p,\alpha)$};
\node at (-4.1,1.6) {s.t. $r\in R_c$};

\node[Eve] (nfalse) at (3,1) {$\ffalse$};
\node at (3.4,2) {If $\exists (r,\pop)\in\Delta(p,\alpha)$};
\node at (4.1,1.6) {s.t. $r\notin R_c$};

\node[Adam] (nextAdam) at (0,-3.5) {$(p,\alpha,\vect{R},c,q,\beta,\vect{S})$};
\node[Adam] (nextAdamG) at (-3.5,-3.5) {\phantom{$(p,\alpha,\vect{R},c,q,\beta,\vect{S})$}};
\node[Adam] (nextAdamD) at (3.5,-3.5) {\phantom{$(p,\alpha,\vect{R},c,q,\beta,\vect{S})$}};

\node at (3.4,-2.6) {$\forall \vect{S}\in (2^{Q})^{d+1}$};

\node[AnyPlayer] (currentJ) at (-3,-6) {$(q,\beta,\vect{S},\pdscol{q})$};
\node[AnyPlayer] (currentB) at (3,-6) {$(r,\alpha,\vect{R},\max(c,i,\pdscol{r})$};

\node at (3,-6.6) {$\forall r \in S_i$};

\path[->] (current) edge (nextEve);\path[->] (current) edge (nextEveG);\path[->] (current) edge (nextEveD);
\path[->] (current) edge (ntrue);\path[->] (current) edge (nfalse);
\path[->] (nextEve) edge (nextAdam);\path[->] (nextEve) edge (nextAdamG);\path[->] (nextEve) edge (nextAdamD);

\path[->] (nextAdam) edge node[above right] {$i$}  (currentB);
\path[->] (nextAdam) edge node[above left] {$\pdscol{q}$} (currentJ);

\path[->] (ntrue) edge  [loop left, loop] node[left] {$0$} (ntrue);
\path[->] (nfalse) edge  [loop right, loop] node[right] {$1$} (nfalse);

\end{tikzpicture}
\end{center}
\caption{Local structure of the arena $\farena$}
\label{12-fig:reduced-arena}
\end{figure}

Before providing a precise description of the arena
$\farena$, let us consider the following informal description of
this simulation game. We aim at simulating a play in the pushdown game from some initial vertex $(p_{in},\bot)$. In $\farena$ we
keep track of only the control state and the top stack symbol of
the currently simulated configuration.

The interesting case is when it is in a control state $p$ with top
stack symbol $\alpha$, and the player owning $p$ wants to push a
symbol $\beta$ onto the stack and change the control state to $q$. For
every strategy of \Eve, there is a certain set of possible
(finite) continuations of the play that will end with popping
$\beta$ from the stack. We require \Eve\ to declare a vector
$\vect{S}=(S_0,\dots,S_d)$ of $(d+1)$ subsets of
$Q$, where $S_i$ is the set of all
states the game can be in after popping $\beta$ along those
plays where in addition the largest visited colour while $\beta$ was on
the stack is $i$.

\Adam\ has then two choices. He can continue the game by pushing
$\beta$ onto the stack and updating the state (we call this a
\emph{pursue} move). Otherwise, he can pick a set $S_i$
and a state $r\in S_i$, and
continue the simulation from that state $r$ (we call this a
\emph{jump} move). If he does a pursue move, then he remembers the
vector $\vect{S}$ claimed by \Eve; if later on, a pop transition is simulated, the play
goes in a sink vertex and \Eve wins if and only if the resulting state is in
$S_c$ where $c$ is the largest colour seen in
the current stack level (this information is encoded in the vertex, reset after each pursue move and updated after each jump
move). If \Adam
does a jump move to a state $r$ in $S_{i}$, the currently
stored value for $c$ is updated to $\max(c,i,\pdscol{r})$,
which is the largest colour seen since the current stack level was
reached.


Therefore the main vertices of this new arena are of the form $(p,\alpha,\vect{R},c)$, which are controlled by the player who controls $p$.
Intermediate vertices are used to handle the previously
described intermediate steps. The local structure is given in
\Cref{12-fig:reduced-arena}. Two special sink vertices $\ttrue$ and $\ffalse$ are
used to simulate pop moves. This arena is equipped with a
colouring function on the edges: an edge from a vertex
$(p,\alpha,\vect{R},c,q,\beta,\vect{S})$ to a vertex $(r,\alpha,\vect{R},\max(c,i,\pdscol{r})$ has colour $i$ where $i$ is the colour of the simulated bump, an edge from a vertex
$(p,\alpha,\vect{R},c,q,\beta,\vect{S})$ to a vertex $(q,\beta,\vect{S}\pdscol{q})$ simulating a jump move has colour $\pdscol{q}$, the loop on $\ttrue$ has colour $0$ while the loop on $\ffalse$ has colour $1$; all other edges get the irrelevant colour $0$.

We now formally describe arena
$\farena$ (we refer to \Cref{12-fig:reduced-arena}) and provide some extra insight.

\begin{itemize}
\item The main vertices of $\farena$ are those of the form
$(p,\alpha,\vect{R},c)$, where $p\in Q$, $\alpha\in
\Gamma$, $\vect{R}=(R_0,\dots,R_d)\in
(2^Q)^{d+1}$ and $c\in\{0,\dots,d\}$. A vertex $(p,\alpha,\vect{R},c)$ is reached when
simulating a finite play $\play$ in $\game$ such that:
\begin{itemize}

\item The last vertex in $\play$ is $(p,\bot s\alpha)$ for some $s\in \Gamma^*$.

\item \Eve claims that she has a strategy to continue $\play$ in such
  a way that if $\alpha$ is eventually popped, the control state
  reached after popping belongs to $R_m$, where $m$ is
  the largest colour visited since the stack height was at least $|s\alpha|$.

\item The colour $c$ is the largest one since the current stack level was reached from a lower stack level.
\end{itemize}

A vertex $(p,\alpha,\vect{R},c)$ is controlled by \Eve\ if
and only if $p\in Q_\mEve$.

\item The vertices $\ttrue$ and $\ffalse$ are here to ensure
  that the vectors $\vect{R}$ encoded in the main vertices are correct. They are both controlled by \Eve\ and are sink vertices with a self loop with colour $0$ for $\ttrue$ and $1$ for $\ffalse$.

There is a transition from some vertex
$(p,\alpha,\vect{R},c)$ to $\ttrue$, if and
only if there exists a transition rule $(r,pop)\in\Delta(p,\alpha)$,
such that $r\in R_{c}$ (this means that $\vect{R}$ is correct with respect
to this transition rule).
Dually, there is a transition from a vertex
$(p,\alpha,\vect{R},c)$ to $\ffalse$
if and only if there exists a transition rule
$(r,pop)\in\Delta(p,\alpha)$ such that $r\notin R_{c}$ (this means that
$\vect{R}$ is not correct with respect to this transition rule).


\item To simulate a transition rule
  $(q,push(\beta))\in\Delta(p,\alpha)$, the player that controls
  $(p,\alpha,\vect{R},c)$ moves to
  $(p,\alpha,\vect{R},c,q,\beta)$. This vertex is
  controlled by \Eve\ who has to give a vector
$\vect{S}=(S_0,\dots,S_d)\in
(2^Q)^{d+1}$ that describes the control states that can be
  reached if $\beta$ is eventually popped. To describe this vector,
  she goes to the corresponding vertex $(p,\alpha,\vect{R},c,q,\beta,\vect{S})$.

Any vertex $(p,\alpha,\vect{R},c,q,\beta,\vect{S})$ is
controlled by \Adam\ who chooses either to simulate a bump or a
stair. In the first case, he additionally has to pick the maximal colour of the
bump. To simulate a bump with maximal
colour $i$, he goes, through an edge coloured by $i$, to a vertex
$(r,\alpha,\vect{R},\max(c,i,\pdscol{r}))$, for some $r\in
S_i$.

To simulate a stair, \Adam\ goes, through an edge coloured by $\pdscol{q}$, to the vertex
$(q,\beta,\vect{S},\pdscol{q})$.

The last component of the vertex (that stores the
largest colour seen since the currently simulated stack level was
reached) has to be updated in all those cases. After simulating a bump
of maximal colour $i$, the maximal colour is
$\max(c,i,\pdscol{r})$. After simulating a stair, this colour has to
be initialized (since a new stack level is simulated). Its value, is
therefore $\pdscol{q}$, which is the unique colour since the (new) stack
level was reached.

\end{itemize}

The edges for which we did not precise the colour are assigned colour $0$.

The following theorem relates this new game $\fgame$ and the profiles in the pushdown game $\game$.

\begin{theorem}\label{12-thm:games}
The following holds.
\begin{enumerate}
\item[(i)] A configuration $(p_{in},\bot)$ is winning for \Eve\ in $\game$
if and only if
$(p_{in},\bot,(\emptyset,\dots,\emptyset),\pdscol{p_{in}})$
is winning for \Eve\ in
$\fgame$.

\item[(ii)] For every $q\in Q$, $\gamma\in\Gamma$ and $R\subseteq Q$, $R\in\mathcal{R}(q,\gamma)$ if and only if
$(q,\gamma,(R,\dots,R),\pdscol{q})$
is winning for \Eve\ in
$\fgame$.
\end{enumerate}
\end{theorem}

The rest of the section is devoted to the proof of \Cref{12-thm:games}. We only prove point $(i)$ as the proof of point $(ii)$ is a subpart of the proof of $(i)$.

\subsubsection{Factorisation of a play in $\game$.}

Recall that, for an infinite play $\play=v_0v_1\cdots$ in $\game$, 
$\Stepsg{\play}$ denotes the set of indices of positions where no
configuration of strictly smaller stack height is visited later in the
play. 
Note that $\Stepsg{\play}$ is always
infinite and hence induces a factorisation of the play $\play$ into
finite pieces. 

Indeed, for any play $\play$ with $\Stepsg{\play}=\{n_0<n_1<\cdots\}$, one can define the sequence $(\play_i)_{i\geq
0}$ by setting ${\play}_i=v_{n_i}\cdots v_{n_{i+1}}$. Note that each of the $\Lambda_i$ is either a bump or a stair.
We designate $(\play_i)_{i\geq 0}$ as the \emph{rounds factorisation} of $\play$ and we let $\pdscol{\play_i}$ denotes the largest colour in $\play_i$.

\subsubsection{Factorisation of a play in $\fgame$.}

Recall that in $\farena$ only some edges have a relevant colour while all others get colour $0$. Hence, to represent a play, we
only keep the relevant colours of edges. More precisely, we only need to encode the
colours in $\{0,\dots,d\}$ that appears when simulating a bump: a play will be
represented as a sequence of vertices together with colours in
$\{0,\dots,d\}$ that correspond to (relevant) colours appearing on edges.

For any play in $\fgame$, a \emph{round} is a factor between two
visits through vertices of the form
$(p,\alpha,\vect{R},c)$. We have the following possible forms for a round:

\begin{itemize}

\item The round is of the form
$$(p,\alpha,\vect{R},c)(p,\alpha,\vect{R},c,q,\beta)(p,\alpha,\vect{R},c,q,\beta,\vect{S})i (r,\alpha,\vect{R},\max(c,i,\pdscol{s}))$$ and
corresponds therefore to the simulation of a rule pushing $\beta$
followed by a sequence of moves that ends by popping $\beta$. Moreover $i$ is the largest colour encountered while $\beta$ was on the stack.

\item The round is of the form
$$(p,\alpha,\vect{R},c)(p,\alpha,\vect{R},c,q,\beta)(p,\alpha,\vect{R},c,q,\beta,\vect{S})\pdscol{q}(q,\beta,\vect{S},\pdscol{q})$$ and
corresponds therefore to the simulation of a rule pushing a symbol $\beta$
leading to a new stack level below which the play will never go. We designate it as a \emph{stair}.

\end{itemize}

For any play $\fplay=v_0v_1v_2\cdots$ in $\fgame$, we consider the
subset of indices corresponding to vertices of the form
$(p,\alpha,\vect{R},c)$. More precisely:
\begin{equation*}
\Rounds{\fplay}=\{n\mid v_n=(p,\alpha,\vect{R},c),\ p\in Q,\
\alpha\in\Gamma,\\ \vect{R}\in(2^Q)^{d+1},\
0\leq c\leq d\}
\end{equation*}

Therefore, the set $\Rounds{\fplay}$ induces a natural factorisation
of $\fplay$ into rounds.

\begin{definition}[Rounds factorisation]
For a (possibly finite) play $\fplay=v_0v_1v_2\cdots$, we call
\emph{rounds factorisation} of $\fplay$, the (possibly finite) sequence
$(\fplay_i)_{i\geq 0}$ of rounds defined as follows. Let
$\Rounds{\fplay}=\{n_0<n_1<n_2<\cdots\}$, then for all $0\leq i<|\Rounds{\fplay}|$,
define $\fplay_i=v_{n_i}\cdots v_{n_{i+1}}$.

Therefore, for every $i\geq 0$, the first vertex in $\fplay_{i+1}$ equals
the last one in $\fplay_i$. Moreover,
$\fplay=\fplay_1\odot\fplay_2\odot\fplay_3\odot\cdots$, where
$\fplay_i\odot\fplay_{i+1}$ denotes the concatenation of $\fplay_i$
with $\fplay_{i+1}$ without its first vertex.
Finally, the \emph{colour} of a round is the unique colour in
$\{0,\dots,d\}$ appearing in the round.
\end{definition}

In order to prove both implications of \Cref{12-thm:games} , we
build from a winning strategy for \Eve\ in one game a winning strategy
for her in the other game. The main argument to prove that the new
strategy is winning is to prove a correspondence between the
factorisations of plays in both games.

\subsubsection{Proof of the Direct Implication of \Cref{12-thm:games}}

Assume that the configuration $(p_{in},\bot)$ is winning for \Eve\ in $\game$,
and let $\sigma$ be a corresponding winning strategy for her.

Using $\sigma$, we define a strategy $\fsigma$ for \Eve\ in
$\fgame$ from $(p_{in},\bot,(\emptyset,\dots,\emptyset),\pdscol{p_{in}})$.
This strategy stores a finite play in $\game$, that is an element in
$V^*$. This memory will
be denoted $\play$. At the beginning $\play$ is initialized to
the vertex $(p_{in},\bot)$. We first describe $\fsigma$, and then we
explain how $\play$ is updated. Both the strategy $\fsigma$ and the
update of $\play$, are described for a round.

\vspace{0.1cm}
\textbf{Choice of the move. } Assume that the play is in some
vertex $(p,\alpha,\vect{R},c)$ for $p\in Q_\mEve$. The
move given by $\fsigma$ depends on $\sigma(\play)$:
\begin{itemize}
\item If $\sigma(\play)=(r,pop)$, then \Eve\ goes to
$\ttrue$ (Proposition \Cref{12-prop:par_dir_dep_paritexp} will
  prove that this move is always possible).


\item If $\sigma(\play)=(q,push(\beta))$, then \Eve\ goes to
$(p,\alpha,\vect{R},c,q,\beta)$.
\end{itemize}

In this last case, or in the case where $p\in Q_\mAdam$ and \Adam\ goes
to $(p,\alpha,\vect{R},c,q,\beta)$, we also have to explain
how \Eve\ behaves from
$(p,\alpha,\vect{R},c,q,\beta)$. She has to provide a
vector $\vect{S}\in (2^Q)^{d+1}$ that describes which states
can be reached if $\beta$ is eventually popped, depending on the largest colour visited in the
meantime. In order to define $\vect{S}$, \Eve\ considers the set of
all possible continuations of $\play\cdot(q,s\alpha\beta)$ (where
$(p,s\alpha)$ denotes the last vertex of $\play$) where she
respects her strategy $\sigma$. For each such play, she checks whether some
configuration of the form $(r,s\alpha)$ is visited after $\play\cdot
(q,s\alpha\beta)$, that is if the stack level of $\beta$ is eventually left. If it
is the case, she considers the first configuration $(r,s\alpha)$
appearing after $\play\cdot (q,\sigma\alpha\beta)$ and the largest
colour $i$ since
$\beta$ was on the stack.
For every $i\in\{0,\dots d\}$, $S_i$, is exactly the set of states
$r\in Q$ such that the preceding case happens. 
More formally, 
\begin{equation*}
\begin{split}
S_i=&\{r\mid \exists\ \pi\cdot(q,s\alpha\beta)v_0\cdots v_k(r,s\alpha)\cdots\text{ play in } \game  \text{ where \Eve\ respects } \sigma \text{ and s.t. }\\ & \sh(v_j)>|\sigma\alpha|,\ \forall j=0,\dots,k  \text{, and }\max(\{\pdscol{v_j}\mid j=0,\dots,k\}\cup\{\pdscol{q}\})=i\}
\end{split}
\end{equation*}
Finally, we set $\vect{S}=(S_0,\dots,S_d)$ and \Eve\ moves to
$(p,\alpha,\vect{R},c,q,\beta,\vect{S})$.

\vspace{0.1cm}
\textbf{Update of $\play$. } The memory $\play$ is updated after
each visit to a vertex of the form $(p,\alpha,\vect{R},c)$.
We have two cases depending on the kind of the last round:

\begin{itemize}
\item The round is a bump, and therefore a bump
  of colour $i$ (where $i$ is the colour of the round) starting with some
  transition $(q,push(\beta))$ and ending in a state $r\in S_i$ was simulated. Let $(p,s\alpha)$ be the last vertex in
  $\play$. Then the memory becomes $\play$ extended by
  $(q,s\alpha\beta)$ followed by a sequence of moves, where \Eve\
  respects $\sigma$, that ends by popping $\beta$ and reach
  $(r,s\alpha)$ while having $i$ as largest colour. By definition of $S_i$ such a sequence of moves always exists.

\item The round is a stair and therefore we have simulated a transition 
  $(q,push(\beta))$. If $(p,s\alpha)$ denotes the last vertex in $\play$, then the updated memory is $\play\cdot (q,s\alpha\beta)$.
\end{itemize}

Therefore, with any finite play $\fplay$ in $\fgame$ in which \Eve\
respects her strategy $\fsigma$, is associated a finite play $\play$ in
$\game$. An immediate induction shows that \Eve\ respects $\sigma$ in
$\play$. The same arguments works for an infinite play $\fplay$, and the
corresponding play $\play$ is therefore infinite, starts from
$(p_{in},\bot)$ and \Eve\ respects $\sigma$ in that play. Therefore it is
a winning play.

The following proposition is a direct consequence of how $\fsigma$ was defined.

\begin{proposition}
\label{12-prop:par_dir_dep_paritexp}
Let $\fplay$ be a finite play in $\fgame$ that starts from
$(p_{in},\bot,(\emptyset,\dots,\emptyset),\pdscol{p_{in}})$,
ends in a vertex of the form $(p,\alpha,\vect{R},c)$,
and where \Eve\ respects $\fsigma$. Let $\play$ be the play associated with $\fplay$
built by the strategy $\fsigma$. Then the following holds:
\begin{enumerate}
\item $\play$ ends in a vertex of the form $(p,s\alpha)$ for some $s\in\Gamma^*$.

\item $c$ is the largest colour visited in $\play$ since $\alpha$ was pushed.

\item Assume that $\play$ is extended, that \Eve\ keeps respecting
  $\sigma$ and that the next move after $(p,\sigma\alpha)$ is to some
  vertex $(r,\sigma)$. Then $r\in R_c$.
\end{enumerate}
\end{proposition}

Proposition \Cref{12-prop:par_dir_dep_paritexp} implies that the
strategy $\fsigma$ is well defined when it provides a move to
$\ttrue$. Moreover, one can deduce that, if \Eve\ respects $\fsigma$, $\ffalse$ is never reached.

For plays that do not visits $\ttrue$ nor $\ffalse$, using the definitions of $\farena$ and $\fsigma$, we easily deduce the
following proposition.

\begin{proposition}
\label{12-prop:toto}
Let $\fplay$ be an infinite play in $\fgame$ that starts from
$(p_{in},\bot,(\emptyset,\dots,\emptyset),$ $\pdscol{p_{in}})$,
and where \Eve\ respects $\fsigma$. Let $\play$ be the associated
play built by the strategy $\fsigma$, and let $(\play_i)_{i\geq 0}$ be its rounds factorisation. Let $(\fplay_i)_{i\geq 0}$ be
the rounds factorisation of $\fplay$. Then, for every $i\geq 1$ the
following hold:
\begin{enumerate}
\item $\fplay_i$ is a bump if and only if $\play_i$ is a bump

\item $\fplay_i$ has colour $\pdscol{\play_i}$.
\end{enumerate}
\end{proposition}

\Cref{12-prop:toto} implies that for any infinite play
$\fplay$ in $\fgame$ starting from
$(p_{in},\bot,(\emptyset,\dots,\emptyset),$ $\pdscol{p_{in}})$
where \Eve\ respects $\fsigma$, the sequence of visited colours in $\fplay$ is
$(\pdscol{\play}_i)_{i\geq 0}$ for the corresponding play $\play$
in $\game$.
Hence, using \Cref{12-prop:trans_cond} we conclude that
$\fplay$ is winning if
and only if $\play$ is winning. As $\play$
is winning for \Eve, it follows that $\fplay$ is also winning for
her.

\subsubsection{Proof of the Converse Implication of \Cref{12-thm:games}}
\label{12-subsec:strategy-pushdown}

Note that in order to prove the converse implication of \Cref{12-thm:games} one could follow the direct implication and consider the point of view of \Adam. Nevertheless the proof we give here starts from a winning strategy for \Eve\ in $\fgame$ and deduces a strategy for her in $\game$: this induces a more involved proof but has the advantage to lead to an effective construction of a winning strategy for \Eve\ in $\game$ if one has an effective strategy for her in $\fgame$

Assume now that \Eve\ has a winning strategy $\fsigma$ in $\fgame$
from $(p_{\mathit in},\bot,(\emptyset,\dots,\emptyset),\pdscol{p_{in}})$.
Using $\fsigma$, we build a strategy $\sigma$ for \Eve\ in
$\game$ for plays starting from $(p_{\mathit in},\bot)$.

The strategy $\sigma$ uses, as memory, a stack $\Pi$, to store the complete
description of a play in $\fgame$. Recall here that a play in
$\fgame$ is represented as a sequence of vertices together with
colours in $\{0,\dots d\}$. Up to coding we can assume that we distinguish for free between stairs and bumps for transitions from vertices of the form $(p,\alpha,\vect{R},c,q,\beta,\vect{S})$. 

The stack alphabet of $\Pi$ is the set of vertices of 
$\farena$ together with the colours $\{0,\dots,d\}$. In the following, $top(\Pi)$ will denote
the top stack symbol of $\Pi$ while $StCont(\Pi)$ will be
the word obtained by reading $\Pi$ from bottom to top (without
considering the bottom-of-stack symbol of $\Pi$). In any play
where \Eve\ respects $\sigma$, $StCont(\Pi)$ will be a play
in $\fgame$ that starts from $(p_{\mathit
in},\bot,(\emptyset,\dots,\emptyset),\pdscol{p_{in}})$ and where
\Eve\ respects her winning strategy $\fsigma$. Moreover, for any
play $\play$ where \Eve\ respects $\sigma$, we will always have that
$top(\Pi)=(p,\alpha,\vect{R},c)$ if and only if
the current configuration in $\play$ is of the form
$(p,s\alpha)$. Finally, if \Eve\ keeps respecting $\sigma$, and
if $\alpha$  is eventually popped the configuration reached
will be of the form $(r,s)$ for some $r\in R_i$, where
$i$ is the largest visited colour since $\alpha$  was on the stack.
Initially, $\Pi$ only contains $(p_{in},\bot,(\emptyset,\dots,\emptyset),\pdscol{p_{in}})$.

In order to describe $\sigma$, we assume that we are in some
configuration $(p,s\alpha)$ and that
$top(\Pi)=(p,\alpha,\vect{R},c)$. We first
describe how \Eve\ plays if $p\in Q_\mEve$, and then we explain how
the stack is updated.

\begin{itemize}
\item \textbf{Choice of the move.} Assume that $p\in Q_\mEve$ and
that \Eve\ has to play from some vertex $(p,s\alpha)$. For
this, she considers the value of $\fsigma$ on $StCont(\Pi)$.

If it is a move to $\ttrue$, \Eve\ plays a transition
$(r,pop)$ for some state $r\in R_c$. 
\Cref{12-lem:games_ReturningSets_paritexp} will prove that such
an $r$ always exists.


If the move given by $\fsigma$ is to go to some vertex
$(p,\alpha,\vect{R},c,q,\beta)$, then \Eve\ applies
the transition $(q,push(\beta))$.

\item \textbf{Update of $\Pi$.} Assume that the last move,
played by \Eve\ or \Adam, was to go from $(p,s\alpha)$ to some
configuration $(r,s)$. The update of $\Pi$ is illustrated
by \Cref{12-fig:mise_a_jour_pile_strategie} and
explained in what follows. \Eve\ pops in $\Pi$ until she finds
some configuration of the form
$(p',\alpha',\vect{R'},c',p'',\alpha,\vect{R})$
that is part of a stair. This
configuration is therefore in the stair that simulates the pushing
of $\alpha$ onto the stack. \Eve\ updates $\Pi$ by pushing
$c$ in $\Pi$ followed by
$(r,\alpha',\vect{R'},\max(c',c,\pdscol{r}))$.


Assume that the last move, played by \Eve\ or \Adam, was to go from
$(p,s\alpha)$ to some configuration $(q,s\alpha\beta)$, and let
$(p,\alpha,\vect{R},c,q,\beta,\vect{S})=\fsigma(StCont(\Pi)\cdot(p,\alpha,\vect{R},c,q,\beta))$.
Intuitively, $\vect{S}$ describes which states \Eve\ can force a play to
reach if $\beta$ is eventually popped. \Eve\ updates $\Pi$ by
successively pushing
$(p,\alpha,\vect{R},c,q,\beta)$,
$(p,\alpha,\vect{R},c,q,\beta,\vect{S})$, and
$(q,\beta,\vect{S},\pdscol{q})$.
\end{itemize}

\begin{figure}
\begin{center}
\begin{tikzpicture}
\draw (0,0)--(0,3.6);
\draw (0,0)--(7.3,0);
\draw (0,3.6) -- (7.3,3.6);

\draw (1.3,0) -- (1.3,3.6);
\draw (1.9,0) -- (1.9,3.6);
\draw (2.5,0) -- (2.5,3.6);
\draw (3.1,0) -- (3.1,3.6);
\draw (3.7,0) -- (3.7,3.6);
\draw (4.3,0) -- (4.3,3.6);
\draw (5.6,0) -- (5.6,3.6);
\draw (6.2,0) -- (6.2,3.6);

\node at (0.65,1.8) {$\cdots$};
\node [rotate=270] at (2.2,1.8) {$(p',\alpha',\overrightarrow{R'},c')$};
\node [rotate=270] at (2.8,1.8) {$(p',\alpha',\overrightarrow{R'},c',p'',\alpha)$};
\node [rotate=270] at (3.4,1.8) {$(p',\alpha',\overrightarrow{R'},c',p'',\alpha,\overrightarrow{R})$};
\node [rotate=270] at (4.0,1.8) {$(p'',\alpha,\overrightarrow{R},col(p''))$};
\node [rotate=270] at (5.9,1.8) {$(p,\alpha,\overrightarrow{R},c)$};
\node at (4.95,1.8) {$\cdots$};

\draw [-latex](0,3.9) -- (7.3,3.9);
\draw [-latex](0,3.9) -- (0,8.1);
\draw (0.0,4.6) .. controls (0.2,5.1) and (0.5,5.1) .. (0.7,4.6) -- (1.4,6.0) .. controls (1.6,6.5) and (1.9,6.5) .. (2.1,6.0) -- (2.8,6.7) .. controls (3.1,7.4) and (3.5,7.4) .. (3.8,6.7) .. controls (4.1,8.3) and (4.6,8.3) .. (4.9,6.7) .. controls (5.1,7.2) and (5.4,7.2) .. (5.6,6.7);

\filldraw[black] (2.1,6.0) circle (2pt);
\node [align=left] at (2.8,5.5) {$(p',s)$\\$s=s'\alpha'$};
\filldraw[black] (2.8,6.7) circle (2pt); 
\node at (2,6.8) {$(p'',s\alpha)$};
\filldraw[black] (5.6,6.7) circle (2pt);
\node at (6.3,6.8){$(p,s\alpha)$};
\node at (8.0,6.8){$\text{max. col.} = c$};
\node at (7.4,6.1){$\text{max. col.} = c'$};

\draw[densely dotted] (1.9,3.6) -- (1.9,3.9) -- (2.1,6.0) -- (2.5,3.9) -- (2.5,3.6);
\draw[densely dotted] (3.7,3.6) -- (3.7,3.9) -- (2.8,6.7) -- (4.3,3.9) -- (4.3,3.6);
\draw[densely dotted] (5.6,3.6) -- (5.6,3.9) -- (5.6,6.7) -- (6.2,3.9) -- (6.2,3.6);

\draw[densely dotted] (2.8,6.7) -- (5.6,6.7);
\draw[densely dotted] (1.4,6.0) -- (5.6,6.0);
\end{tikzpicture}

\begin{tikzpicture}
\draw (0,-1)--(0,3.6);
\draw (0,-1)--(7.3,-1);
\draw (0,3.6) -- (7.3,3.6);

\draw (1.3,-1) -- (1.3,3.6);
\draw (1.9,-1) -- (1.9,3.6);
\draw (2.5,-1) -- (2.5,3.6);
\draw (3.5,-1) -- (3.5,3.6);
\draw (4.9,-1) -- (4.9,3.6);
\draw (6.3,-1) -- (6.3,3.6);
\draw (6.9,-1) -- (6.9,3.6);

\node at (0.65,1.3) {$\cdots$};
\node [rotate=270] at (2.2,1.3) {$(p',\alpha',\overrightarrow{R'},c')$};
\node [rotate=270] at (3,1.3) {$(p',\alpha',\overrightarrow{R'},c',p'',\alpha)$};
\node [rotate=270] at (4.2,1.3) {$(p',\alpha',\overrightarrow{R'},c',p'',\alpha,\overrightarrow{R})$};
\node [rotate=270] at (5.6,1.3) {$c$};
\node [rotate=270] at (6.55,1.3) {$(r,\alpha',\overrightarrow{R'},\text{max}(c',c,col(r)))$};

\draw [-latex](0,3.9) -- (7.3,3.9);
\draw [-latex](0,3.9) -- (0,8.1);
\draw (0.0,4.6) .. controls (0.2,5.1) and (0.5,5.1) .. (0.7,4.6) -- (1.4,6.0) .. controls (1.6,6.5) and (1.9,6.5) .. (2.1,6.0) -- (2.8,6.7) .. controls (3.1,7.4) and (3.5,7.4) .. (3.8,6.7) .. controls (4.1,8.3) and (4.6,8.3) .. (4.9,6.7) .. controls (5.1,7.2) and (5.4,7.2) .. (5.6,6.7) -- (6.3,6.0);

\filldraw[black] (2.1,6.0) circle (2pt);
\node [align=left] at (2.8,5.5) {$(p',s)$\\$s=s'\alpha'$};
\filldraw[black] (6.3,6.0) circle (2pt);
\node at (6.9,6.1){$(r,s)$};
\node [align=left] at (7.3,5.3){max. col.=\\$\text{max}(c',c,col(r))$};

\draw[densely dotted] (1.9,3.6) -- (1.9,3.9) -- (2.1,6.0) -- (2.5,3.9) -- (2.5,3.6);
\draw[densely dotted] (6.3,3.6) -- (6.3,3.9) -- (6.3,6.0) -- (6.9,3.9) -- (6.9,3.6);

\draw[densely dotted] (2.8,6.7) -- (5.6,6.7);
\draw[densely dotted] (1.4,6.0) -- (6.3,6.0);

\end{tikzpicture}
\caption{Updating the strategy's stack $\Pi$}
\label{12-fig:mise_a_jour_pile_strategie}
\end{center}
\end{figure}


The following lemma gives the meaning of the information stored
in $\Pi$.

\begin{lemma}
\label{12-lem:games_ReturningSets_paritexp}
Let $\play$ be a finite play in $\game$, where \Eve\ respects
$\sigma$, that starts from $(p_{\mathit in},\bot)$ and
that ends in a configuration $(p,s\alpha)$. We have the
following facts:

\begin{enumerate}

\item $top(\Pi)=(p,\alpha,\vect{R},c)$ with
$\vect{R}\in(2^Q)^{d+1}$ and $0\leq c\leq d$.

\item $StCont(\Pi)$ is a finite play in $\fgame$ that starts
from $(p_{\mathit in},\bot,(\emptyset,\dots,\emptyset),\pdscol{p_{in}})$,
that ends with $(p,\alpha,\vect{R},c)$ and where
\Eve\ respects $\fsigma$.

\item $c$ is the largest colour visited since $\alpha$ was pushed.

\item If $\play$ is extended by some move
that pops $\alpha$, the configuration $(r,s)$ that is reached
is such that $r\in R_c$.
\end{enumerate}
\end{lemma}

\begin{proof}
The proof goes by induction on $\play$. We first show that the
last point is a consequence of the second and third points. To aid readability, one can refer to 
\Cref{12-fig:mise_a_jour_pile_strategie}. Assume that
the next move after $(p,s\alpha)$ is to apply a transition
$(r,pop)\in\Delta(p,\alpha)$. The second point implies that
$(p,\alpha,\vect{R},c)$ is winning for \Eve\ in
$\fgame$. If $p\in Q_\mEve$, by definition of $\sigma$, there is
some edge from that vertex to $\ttrue$, which means that
$r\in R_c$ and allows us to conclude. If $p\in Q_\mAdam$, note that there is no
edge from $(p,\alpha,\vect{R},c)$ (winning position
for \Eve) to the (losing) vertex $\ffalse$. Hence we
conclude in the same way.

Let us now prove the other points. For this, assume that the
result is proved for some play $\play$, and let $\play'$ be an
extension of $\play$. We have two cases, depending on how $\play'$
extends $\play$:

\begin{itemize}
\item $\play'$ is obtained by applying a push transition. The result is trivial in that case.

\item $\play'$ is obtained by applying a pop transition. Let
$(p,s\alpha)$ be the last configuration in $\play$, and let
$\vect{R}$ be the last vector component in $top(\Pi)$ when
in configuration $(p,s\alpha)$. By the induction
hypothesis, it follows that $\play'=\play\cdot(r,s)$ with
$r\in R_c$. Considering how $\Pi$ is updated, and
using the fourth point, we easily deduce that the new strategy
stack $\Pi$ is as desired (one can have a look at 
\Cref{12-fig:mise_a_jour_pile_strategie} for more
intuition).
\end{itemize}
\end{proof}

Actually, we easily deduce a more precise result.

\begin{lemma}
\label{12-lem:toto_paritexp}
Let $\play$ be a finite play in $\game$ starting from
$(p_{in},\bot)$ and where \Eve\ respects $\sigma$. Let $(\play_i)_{i\geq 0}$ be its rounds factorisation. Let
$\play=StCont(\Pi)$, where $\Pi$ denotes the strategy's
stack in the last vertex of $\play$. Let
$(\play_i)_{i=0,\dots,k}$ be the rounds factorisation of $\play$.
Then the following holds:
\begin{itemize}
\item $\play_i$ is a bump if and only if ${\play}_i$ is a bump.

\item $\play_i$ has colour $\pdscol{\play}_i$.
\end{itemize}
\end{lemma}

Both \Cref{12-lem:games_ReturningSets_paritexp} and
\Cref{12-lem:toto_paritexp} are for finite plays. A version for
infinite plays would allow us to conclude. Let $\play$
be an infinite play in $\game$. We define an
infinite version of $\play$ by considering the limit of the stack
contents $(StCont(\Pi_i))_{i\geq 0}$ where $\Pi_i$ is the
strategy's stack after the first $i$ moves in $\play$. 
It is easily seen that such a limit
always exists, is infinite and corresponds to a play won by \Eve\ in $\fgame$.
Moreover the results of \Cref{12-lem:toto_paritexp} apply.

Let $\play$ be a play in $\game$ with initial
vertex $(p_{in},\bot)$, and where \Eve\ respects $\sigma$,
and let $\play$ be the associated infinite play in $\fgame$.
Therefore $\play$ is won by \Eve. Using \Cref{12-lem:toto_paritexp} and \Cref{12-prop:trans_cond},
we conclude, as in the direct implication that $\play$ is
winning.

\subsection{Solving the games and computing the winning regions and strategies}
\label{12-subsec:computing-all}

Combining \Cref{12-thm:games} and \Cref{12-thm:regularity-wr}, we obtain the following upper bounds regarding the problem of deciding the winner in a pushdown game and on constructing a finite state automaton recognising the winning region (in the sense of \Cref{12-rmk:automata-winning-region}).

\begin{theorem}\label{12-thm:solving-upper-bound}
	Let $\game$ be a pushdown parity game using colours $\{0,\dots,d-1\}$ and played on an arena generated by a pushdown system with $n$ control states and with a stack alphabet of size $m$. Then the following holds
	\begin{enumerate}
		\item[(i)] One can construct in time $\mathcal{O}(m^d 2^{nd^2})$ a deterministic finite state automaton with $2^n$ states recognising the winning region of \Eve\ in $\game$.
		\item[(ii)] One can decide in time $\mathcal{O}(m^d 2^{nd^2}+|s|)$, for any configuration $(p,\bot s)$, whether it is winning for \Eve\ in $\game$. 
	\end{enumerate}
\end{theorem}

\begin{proof}
	Consider the parity game $\fgame$ from \Cref{12-subsec:computing-profiles}. Let $n=|Q|$ and $m=|\Gamma|$. Then $\fgame$ is played on an arena with $\mathcal{O}(n m 2^{2nd})$ vertices and it uses $d$ colours. Hence, computing the winning region of this later game can be achieved in time $\mathcal{O}(m^d 2^{nd^2})$, see~\Cref{3-chap:regular}.
		
	Using, \Cref{12-thm:games}, it follows that the set of profiles can be computed in time $\mathcal{O}(m^d 2^{nd^2})$, and by \Cref{12-thm:regularity-wr} (and its proof) we know that we can construct (in the same time complexity) a deterministic finite state automaton with $2^n$ states recognising the winning region $W_\mEve$.
	
	The upper bound for the second item simply follows from the fact that running a deterministic automaton on a word is performed in linear time in the length of the word.
\end{proof}

Regarding lower bound, the following result shows that the previous upper bound is optimal. Note that it is enough to consider a reachability objective.

\begin{theorem}\label{12-thm:solving-lower-bound}
	Let $\game$ be a pushdown reachability game. Then the following problem is hard for \EXP: decide whether $(p,\bot)$ is winning for \Eve\ in $\game$.
\end{theorem}

\begin{proof}
The lower bound is established by reducing the halting problem for alternating linear space bounded Turing machine. 

Consider an alternating linear space bounded Turing machine $\mathcal{M}$. We can safely assume that $\mathcal{M}$ has a unique tape and on an input of size $n$ it uses at most $n$ tape squares. Let $Q=Q_\exists\cup Q_\forall$ be the states of the Turing machine where $Q_\exists$ are the existential states and $Q_\forall$ are the universal ones; we let $q_a$ be the (unique) accepting state of the machine. Call $A$ the tape alphabet and let $T\subseteq Q\times A\times Q\times A\times \{\leftarrow,\rightarrow\}$ the transition table of the machine. A configuration of $\mathcal{M}$ is a word $C$ of the form $uqv\in A^*QA^*$ of length $n+1$ (the meaning being that $\mathcal{M}$ is in state $q$ and that the tape contains $uv$).


We now informally describe a two-player game simulating a computation of $\mathcal{M}$ and argue that it can be encoded as a pushdown reachability game. Think first of $\mathcal{M}$ as being non-deterministic, \textit{i.e.} $Q_\forall=\emptyset$ and call $C_0$ the initial configuration. A run of $\mathcal{M}$ can be encoded as a word $r=C_0\sharp t_0\sharp C_1\sharp t_1\sharp C_2\sharp t_2\sharp C_3\sharp\cdots$ where for every $i\geq 0$, $t_i\in T$ is a transition of $\mathcal{M}$ that can be applied in configuration $C_i$ and $C_{i+1}$ is the configuration reached from $C_i$ by applying $t_i$; it is accepting if some $t_i$ is a transition to the accepting state of $\mathcal{M}$. A way to encode such $r$ with a pushdown game is that \Eve\ pushes symbols in the stack to describe $r$ and to use the control states to impose some structural constraints on the sequence of pushed symbols:
\begin{itemize}
	\item The first pushed configuration is $C_0$.
	\item Every configuration pushed has the right form, \textit{i.e.} it is a word in $A^*QA^*$ of length $n+1$. 
	\item Every configuration $C$ is followed by some pattern $\sharp t\sharp$ and the transition $t$ in this pattern can be applied from the configuration. This is ensured by storing in the control state of the pushdown process the state of $\mathcal{M}$ and the content of the currently read cell in $C$. 
\end{itemize}
To ensure these properties a linear number of control states suffices. 

Of course, this is not enough because \Eve\ could cheat and push a configuration $C_{i+1}=x_0x_1\cdots x_n$ which is not the successor of $C_i=y_0y_1\cdots y_n$ by $t_i$. To avoid this, after she described a configuration (say $C_{i+1}$) and pushed the $\sharp$ symbol, \Adam\ can stop the simulation and claim a mistake by indicating the index $k$ of a wrong update in $C_{i+1}$. If so, the game goes to a special mode where the following is performed:
\begin{itemize}
	\item the $\sharp$ symbol is popped as well as the next $n-i$ symbols;
	\item the current top symbol is $x_i$ and it is stored in the control state of the pushdown process
	\item the players keep popping until a $\sharp$ symbol is seen and the next symbol $t_i$ is also stored in the control state
	\item then the players pop $n-i-1$ symbols and then considering the next three symbols they can check whether the update was correct or not (there are several cases depending whether the reading tape was at distance at most $1$ of the position of index $i$).
\end{itemize}
Again, this can be implemented thanks to a linear number of control states in the pushdown process.

In case \Eve\ cheats the play loops in a non-final sink configuration. Otherwise it loops in a final sink configuration.

Now, if the Turing Machine $\mathcal{M}$ is alternating, the only difference is that the choice of the transition $t_i$ is made by \Eve\ if the control state in $C_i$ is existential and by \Adam\ if it is universal. The rest of the game is unchanged (in particular \Eve\ is still in charge of describing all configurations, regardless of whom picks the transition).

It is then immediate to check that \Eve\ has a winning strategy in this game if and only if the Turing machine accepts from its initial configuration.	
\end{proof}



\subsection{Pushdown and regular winning strategies}
\label{12-subsec:regular-strat}

In \Cref{12-subsec:strategy-pushdown}, we have seen that the proof of \Cref{12-thm:games} shows that a winning strategy for \Eve\ (when it exists) can be implemented by pushdown automaton that reads the pushdown system transitions chosen by the players and indicates \Eve\ moves by a function depending only of the current control state and the top-most stack symbol of the strategy automaton. 

In this section, we present a different reduction whose aim is to be able to compute a positional winning strategy for \Eve\ which furthermore can be implemented by a finite state automata. As an added benefit, we will see that this strategy is uniform in the sense that it is winning from every vertex of the winning region of \Eve.

For the rest of this section, we fix a pushdown parity game $\game$ played on an arena $\arena = (G,\VE,\VA)$ generated by a pushdown system $\PDS = (Q,Q_{\mEve}, Q_{\mAdam}, \Gamma,\Delta)$. We also let $V=\VE\cup\VA$ and we let the colours used in the game be $\{0,\dots,d\}$.

A \emph{summary} is a triple $(p,c,q)\in Q\times \{0,\dots,d\} \times Q$. A set $S$ of summaries is \emph{complete} if $(p_1,c_1,q),(q,c_2,p_2)\in S$ implies that $(p_1,\max(c_1,c_2),p_2)\in S$; it is winning if $(p,c,p)\in S$ implies that $c$ is even. For $R\subseteq Q$, a set of $R$-summaries is a set of summaries $S\subseteq R\times\{0,\dots,d\}\times R$. Associated with some stack content $s$, a summary $(p,c,q)$ aims to encode the existence of a sequence of moves from $(p,s)$ to $(q,s)$ where the top symbol of $s$ is never removed and where $c$ is the largest colour visited in the sequence.

Let $P\subseteq Q_\mEve$ and $\gamma\in\Gamma_\bot$. A \emph{$(P,\gamma)$-local strategy} for \Eve\ is a partial function $\sigma_\gamma:P\rightarrow {Q\times\{\pop,\push{\gamma}\mid \gamma\in\Gamma\}}$ such that $\sigma_\gamma(p)\in \Delta(p,\gamma)$ for all $p\in P$. Equivalently it is a selection for every state in $P$ of a consistent transition of $\PDS$ when the top symbol is $\gamma$. For a subset $R\subseteq Q$, we say that $\sigma_\gamma$ \emph{pops in $R$} if $\sigma_\gamma(q)=(r,pop)$ implies $r\in R$.
We say that a $(P,\gamma)$-local strategy is \emph{safe} if $\sigma_\gamma(p)=(q,push(\alpha))$ implies that $P\in\mathcal{R}(q,\alpha)$. From now on, we only allowed safe local strategies. 

Let $R\subseteq Q$ and $\gamma\in\Gamma$. We associate with $(R,\gamma)$ the subset $$W(R,\gamma)=\{q\mid R\in\mathcal{R}(q,\gamma)\}$$ By a small abuse of notation we let $W(\emptyset,\bot) = \{q\mid (q,\bot)\in W_\mEve\}$. Remark that for every $q\in W(R,\gamma)\cap Q_\mAdam$ and $(r,pop)\in \Delta(q,\gamma)$ one has $r\in R$.

We now define a new game $\hgame$ played on a finite arena and equipped with an $\omega$-regular objective. We start by an informal description of plays in $\hgame$ and later formally describe the arena and the objective. 

A play in $\hgame$ begins by an initialisation phase:
\begin{itemize}
	\item The play starts in $(\bot, R)$ where $R=W(\emptyset,\bot)=\{q\mid (q,\bot)\in W_\mEve\}$.
	\item From there, \Eve\ chooses $\sigma_\bot$ an $(R,\bot)$-local strategy and a set of $R$-summaries that is both complete and winning. Then, the play goes to $(\bot,R,\sigma_\bot,S)$.
\end{itemize}

Then, the plays goes for rounds of the following form:
\begin{itemize}
	\item From a vertex $(\gamma,R,\sigma_\gamma,S)$, where $\gamma\in \Gamma$, $R\subseteq Q$, $\sigma_\gamma$ is an $(R,\gamma)$-local strategy and $S$ is a set of $R$-summaries, \Eve\ chooses for every $\alpha\in\Gamma$ a $(W(R,\alpha),\alpha))$-local strategy $\sigma_\alpha$ that pops in $R$ and a set of $ W(R,\alpha)$-summaries $S_\alpha$ that is both complete and winning. The play then goes in $(\gamma,R,\sigma_\gamma,S,(\sigma_\alpha,S_\alpha)_{\alpha\in\Gamma})$.
	\item Then, \Adam\ chooses some $\alpha$ in $\Gamma$ and the play goes in $(\alpha,W(R,\alpha),\sigma_\alpha,S_\alpha)$.
\end{itemize}

Consider a tuple $(\gamma,R,\sigma_\gamma,S,(\sigma_\alpha,S_\alpha)_{\alpha\in\Gamma})$ where $\gamma\in\Gamma$, $R\subseteq Q$, $\sigma_\gamma$ is an $(R,\gamma)$-local strategy, $S$ is a set of $R$-summaries, and, for every $\alpha\in\Gamma$, $\sigma_\alpha$ is a $(W(R,\alpha),\alpha))$-local strategy $\sigma_\alpha$ that pops in $R$ and $S_\alpha$ is a set of $W(R,\alpha)$-summaries that is both complete and winning. The tuple $(\gamma,R,\sigma_\gamma,(\sigma_\alpha,S_\alpha)_{\alpha\in\Gamma})$ is \emph{consistent} if, for every $(p,r)\in R^2$, one has $(p,\max(\pdscol{p},c,\pdscol{r}),r)\in S$ as soon as we are in one of the following  two situations (the second one being the degenerated version of the first one).
\begin{itemize}
\item There exists $\alpha\in\Gamma$, $(q,c,q')\in S_\alpha$ such that
\begin{enumerate}
	\item[(i)] either $p\in Q_\mEve$ and $(q,\push{\alpha})=\sigma_\gamma(p)$, or $p\in Q_\mAdam$ and $(q,\push{\alpha})\in \Delta(p,\gamma)$, and 
	\item[(ii)] either $q'\in Q_\mEve$ and $(r,\pop)=\sigma_\alpha(q')$, or $q'\in Q_\mAdam$ and $(r,\pop)\in \Delta(q',\alpha)$.
\end{enumerate}
Intuitively, if with state $p$ and top symbol $\gamma$ one can push $\alpha$ and go to state $q$ from which we know that we can later go back to the same stack content with state $q'$ and maximal colour $c$, and finally pop $\gamma$ and end in state $r$, then we conclude that we can go from $p$ to $r$ while seeing $\max(\pdscol{p},c,\pdscol{r})$ as the maximal colour.
\item There exists $\alpha\in\Gamma$ and $q\in W(R,\alpha)$ such that
\begin{enumerate}
	\item[(i)] either $p\in Q_\mEve$ and $(q,\push{\alpha})=\sigma_\gamma(p)$, or $p\in Q_\mAdam$ and $(q,\push{\alpha})\in \Delta(p,\gamma)$, and 
	\item[(ii)] either $q\in Q_\mEve$ and $(r,\pop)=\sigma_\alpha(q)$, or $q\in Q_\mAdam$ and $(r,\pop)\in \Delta(q,\alpha)$,
	\item[(iii)] $c=\pdscol{q}$.
\end{enumerate}
Intuitively, if with state $p$ and top symbol $\gamma$ one can push $\alpha$ and go to state $q$ and directly pops $\gamma$ and end in state $r$, then we conclude that we can go from $p$ to $r$ while seeing $\max(\pdscol{p},\pdscol{q},\pdscol{r})$ as the maximal colour.
\end{itemize}

In the previous informal description, the only allowed choices for $(\sigma_\alpha,S_\alpha)_{\alpha\in\Gamma}$ are those that leads to consistent tuples.
Formally, we define the arena $\harena$ as follows:
\begin{itemize}
	\item There is a special initial vertex $(\bot,W(\emptyset,\bot))$ controlled by \Eve.
	\item For every $\gamma\in \Gamma_\bot$, every $R\subseteq Q$, every $(R,\gamma)$-local strategy $\sigma_\gamma$ and every set of $R$-summaries that is both complete and winning there is a vertex $(\gamma,R,\sigma_\gamma,S)$ controlled by \Eve.
	\item There is a vertex $(\gamma,R,\sigma_\gamma,S,(\sigma_\alpha,S_\alpha)_{\alpha\in\Gamma})$ controlled by \Adam\  for every consistent such tuple.
	\item From every vertex $(\gamma,R,\sigma_\gamma,S)$ there is an edge to every vertex of the form $(\gamma,R,\sigma_\gamma,S,(\sigma_\alpha,S_\alpha)_{\alpha\in\Gamma})$.
	\item From every vertex $(\gamma,R,\sigma_\gamma,S,(\sigma_\alpha,S_\alpha)_{\alpha\in\Gamma})$ there is an edge to $(\alpha,W(R,\alpha),\sigma_\alpha,S_\alpha)$ for every $\alpha\in\Gamma$.
\end{itemize}

Hence, a play in $\hgame$ from the initial vertex $(\gamma_0,R_0)=(\bot,W(\emptyset,\bot))$ is a sequence of vertices 
\begin{equation*}
\begin{split}
\hplay = & (\bot,R_0)(\gamma_0,R_0,\sigma_0,S_0)(\gamma_0,R_0,\sigma_0,S_0,(\sigma^1_\alpha,S^1_\alpha)_{\alpha\in\Gamma}))(\gamma_1,R_1,\sigma_1,S_1)\\ &\quad (\gamma_1,R_1,\sigma_1,S_1,(\sigma^2_\alpha,S^2_\alpha)_{\alpha\in\Gamma}))(\gamma_2,R_2,\sigma_2,S_2)\cdots	
\end{split}
\end{equation*}
with $\sigma_i=\sigma_{\gamma_i}^i$ and $S_i=S_{\gamma_i}^i$ for every $i\geq 1$.
	
It is losing for \Eve\ if there exists $(q_i)_{i\geq 0},(p_i)_{i\geq 0}\in Q^{\mathbb{N}}$ and $(c_i)_{i\geq 0}\in \{0,\dots,d\}^{\mathbb{N}}$ such that $\limsup(c_i)_{i\geq 0}$ is odd and for every $i\geq 0$ one has
\begin{itemize}
	\item $(q_i,c_i,p_i)\in S_i$; and 
	\item either $p_i\in Q_\mEve$ and $(q_{i+1},\push{\gamma_{i+1}})=\sigma_i(p_i,\gamma_i)$ or $p_i\in Q_\mAdam$ and $(q_{i+1},\push{\gamma_{i+1}})\in \Delta((p_i,\gamma_i)$.
\end{itemize}

Note that it is easily seen that the previous objective is an $\omega$-regular one.

We denote by $\hgame$ the previous game. The following result relies on the connection between $\hgame$ and the original pushdown game $\game$.

\begin{theorem}\label{12-thm:hgame}
	\Eve\ has a finite memory winning strategy in $\hgame$ from $(\bot,W(\emptyset,\bot))$.
\end{theorem}

\begin{proof}
	As the game $\hgame$ is played on a finite arena and equipped with an $\omega$-regular objective, it suffices to prove that \Eve\ has a winning strategy from $(\bot,W(\emptyset,\bot))$.

	Consider a \emph{positional} strategy $\sigma$ for \Eve\ in $\game$ that is winning on the whole winning region $W_\mEve$. Note that existence of positional winning strategies is ensure because $\game$ is a parity game.
	
	Let $s\in\bot\Gamma^*$ be some stack content. We define a set of summaries $S_s^\sigma$ associated with $s$ (and $\sigma$) by letting 
	\begin{equation*}
	\begin{split}
	S^s_\sigma=& \{(p,c,q) \mid \exists \play=v_0\cdots v_k \text{ with $k>0$, $v_0=(p,s)$, $v_k=(q,s)$, $\sh(v_i)\geq \sh(v_0)$ }\\  & \text{for all $0\leq i\leq k$, and such that \Eve\ respects $\sigma$ in $\play$ and $c$ is the largest}\\ & \text{colour visited in $\play$}\}
	\end{split}
	\end{equation*}	
	and, for every $(p,c,q)\in S^s_\sigma$, we select a play $\play_{(p,c,q)}^s$ that witnesses $(p,c,q)\in S_\sigma^s$. 
	
	Using $\sigma$ we define a strategy $\hsigma$ for \Eve\ in $\hgame$ and we later argue that it is winning for her from $(\bot,W(\emptyset,\bot))$. 
	\begin{itemize}
		\item At the beginning of the play in $(\bot,R)$ with $R=W(\emptyset,\bot)$, \Eve\ moves to $(\bot,R,\sigma_\bot,S)$ where $\sigma_\bot(r) = \sigma((r,\bot))$ for every $r\in R$, and $S=S_\sigma^\bot$
		\item Assume the current play is $$\hplay=(\bot,R_0)(\gamma_0,R_0,\sigma_0,S_0)(\gamma_0,R_0,\sigma_0,S_0,(\sigma^1_\alpha,S^1_\alpha)_{\alpha\in\Gamma}))(\gamma_1,R_1,\sigma_1,S_1)\cdots (\gamma_k,R_k,\sigma_k,S_k)$$ and let $s_{\hplay} = \gamma_0\cdots\gamma_k$. Then, \Eve\ goes to $(\gamma_k,R_k,\sigma_k,S_k,(\sigma^{k+1}_\alpha,S^{k+1}_\alpha)_{\alpha\in\Gamma}))$ with $\sigma^{k+1}_{\alpha}(r) = \sigma((r,s_{\hplay}\alpha))$ for every $r$ such that $(r,s_{\hplay}\alpha)\in W_\mEve$ and $S^{k+1}_\alpha = S_\sigma^{\hplay\alpha}$. 
	\end{itemize}
	Assume now by contradiction that $\hsigma$ is not winning and consider a losing play \begin{equation*}
\begin{split}
\hplay = & (\bot,R_0)(\gamma_0,R_0,\sigma_0,S_0)(\gamma_0,R_0,\sigma_0,S_0,(\sigma^1_\alpha,S^1_\alpha)_{\alpha\in\Gamma}))(\gamma_1,R_1,\sigma_1,S_1)\\ &\quad (\gamma_1,R_1,\sigma_1,S_1,(\sigma^2_\alpha,S^2_\alpha)_{\alpha\in\Gamma}))(\gamma_2,R_2,\sigma_2,S_2)\cdots	
\end{split}
\end{equation*}
	
Hence, there exists $(q_i)_{i\geq 0},(p_i)_{i\geq 0}\in Q^{\mathbb{N}}$ and $(c_i)_{i\geq 0}\in \{0,\dots,d\}^{\mathbb{N}}$ such that $\limsup(c_i)_{i\geq 0}$ is odd and for every $i\geq 0$ one has
\begin{itemize}
	\item $(q_i,c_i,p_i)\in S_i$; and 
	\item either $p_i\in Q_\mEve$ and $(q_{i+1},\push{\gamma_{i+1}})=\sigma_i(p_i,\gamma_i)$ or $p_i\in Q_\mAdam$ and $(q_{i+1},\push{\gamma_{i+1}})\in \Delta((p_i,\gamma_i)$.
\end{itemize}
Now, consider the play $$\play=\play^{\gamma_0}_{(q_0,c_0,p_0)}\play^{\gamma_0\gamma_1}_{(q_1,c_1,p_1)}\play^{\gamma_0\gamma_1\gamma_2}_{(q_2,c_2,p_2)}\play^{\gamma_0\gamma_1\gamma_2\gamma_3}_{(q_3,c_3,p_3)}\cdots$$
Then it is easily seen by definition that $\play$ is losing (because $\hplay$ is) while \Eve\ respects her winning strategy $\sigma$, which leads a contradiction and concludes the proof.
\end{proof}

Following \Cref{12-thm:hgame}, fix a finite memory winning strategy $\hsigma$ for \Eve\ in $\hgame$ from $(\bot,W(\emptyset,\bot))$. Using $\hsigma$ we define a positional strategy $\sigma$ for \Eve\ in $\game$. 

First, we inductively associate, with any word $s\in \bot \Gamma^*$, a finite play $\hplay_s$ in $\hgame$ where \Eve\ respects her strategy $\hsigma$: 
\begin{itemize}
	\item If $s=\bot$, we let $\hplay_s = (\bot,W(\emptyset,\bot))(\bot,W(\emptyset,\bot),\sigma_\bot,S)$ where $(\bot,W(\emptyset,\bot),\sigma_\bot,S) = \hsigma((\bot,W(\emptyset,\bot)))$.
	\item If $s=s'\beta$ for some $\beta\in\Gamma$, let $\hsigma(\hplay_{s'}) = (\gamma,R,\sigma_\gamma,S,(\sigma_\alpha,S_\alpha)_{\alpha\in\Gamma}))$ and define $$\hplay_{s}= \hplay_{s'}(\gamma,R,\sigma_\gamma,S,(\sigma_\alpha,S_\alpha)_{\alpha\in\Gamma}))(\beta,W(R,\beta),\sigma_\beta,S_\beta).$$
\end{itemize}

Now, for every configuration $(q,s)$ with $q\in Q_\mEve$, we let $(\gamma,R,\sigma_\gamma,S_\gamma)$ be the last vertex in $\hplay_s$ and if $q\in R$ we let $\sigma((q,s)) = \sigma_\gamma(q)$ and otherwise we pick an arbitrary transition for $\sigma((q,s))$ as $(q,s)$ will be a losing position for \Eve\ (see \Cref{12-prop:R-winning-states} below).

The following is a direct rephrasing of the proof of \Cref{12-thm:regularity-wr}.

\begin{proposition}\label{12-prop:R-winning-states}
Let $s\in\Gamma^*\bot$ and let $(\gamma,R,\sigma_\gamma,S)$ be the last vertex in $\hplay_s$. Then $R=\{p\mid (p,s)\in W_\mEve\}$.
\end{proposition}

The following is a consequence of \Cref{12-prop:R-winning-states} and of the requirement that in a vertex $(\gamma,R,\sigma_\gamma,S)$ \Eve\ should only propose $(W(R,\alpha),\alpha)$-local strategies that pops in~$R$.

\begin{proposition}\label{12-prop:stays-in-winning-region}
Let $\play$ be an infinite play in $\game$ starting from some winning position for \Eve\ and where \Eve\ respects strategy $\sigma$. Then any vertex visited in $\play$ is a winning position for \Eve.
\end{proposition}

\begin{proof}
	It suffices to prove that the property is true for the second vertex in $\play$ (and then conclude by induction as the strategy $\play$ is positional). If the initial vertex belongs to \Adam, then by definition all possible successors are winning for \Eve\ (otherwise the initial one would be winning for \Adam\ as well by prefix-independence of the parity objective). If the initial vertex is controlled by \Eve\ there are two cases depending whether her move is to push or pop a symbol. If the move is to pop a symbol then, by construction, the state reached belong to $R$ where the last vertex in $\hplay_s$ is $(\gamma,R,\sigma_\gamma,S)$, if $s$ denote the stack content after popping: hence, by \Cref{12-prop:R-winning-states} we conclude. If the move is to push a symbol then the result follows directly from the fact that we only consider safe local strategies and by \Cref{12-prop:R-winning-states}.
\end{proof}

The following is an easy consequence of the notion of consistent tuples.

\begin{proposition}\label{12-prop:bumps}
Let $(p,s)\in V$ and let $(\gamma,R,\sigma_\gamma,S)$ be the last vertex in $\hplay_s$. Assume that $p\in R$. Let $\play=v_0\cdots v_k$ be a finite play in $\game$ such that $k> 0$, $v_0=(p,s)$, $\sh(v_i)\geq\sh(v_0)$ for every $0< i< k$ and $v_k=(q,s)$ for some $q\in Q$. Then $(p,c,q)\in S$ where $c$ is the largest colour visited in $\play$. 
\end{proposition}

\begin{proof}
	We do the proof only when we assume that the inequality $\sh(v_i)\geq\sh(v_0)$ is strict. The case where it is large is then a consequence of the fact that $S$ is complete with successive application of the strict case.
	The proof is by induction on $k$. The base case is when $k=2$, and it corresponds to the degenerated case in the definition of consistent tuple. Now for the general case, when $k>2$, one simply considers the play $v_1\cdots v_{k-1}$, applies the induction hypothesis and conclude with the definition of consistent tuple again.
\end{proof}

We are now ready to conclude and prove that $\sigma$ is a winning strategy for \Eve.

\begin{theorem}\label{12-thm:positional-strategy}
	The positional strategy $\sigma$ is winning for \Eve\ on the whole winning region in $\game$.
\end{theorem}

\begin{proof}
	Consider an infinite play $\play=v_0v_1\cdots$ starting from some winning position for \Eve. Then by \Cref{12-prop:stays-in-winning-region} we know that the play stays in the winning region. 
	
	By contradiction assume that $\play$ is losing. We distinguish between two cases depending whether there is some vertex that is infinitely visited or not in $\play$. 
	\begin{itemize}
		\item Assume that there is a vertex $v=(q,s)$ that appears infinitely often in $\play$ and choose one of minimal stack height. Let $k_0$ be such that $v_{k_0}=v$ and such that $\sh(v_j)\geq \sh(v)$ for every $j\geq k_0$. Let $(k_i)_{i\geq 0}$ be the increasing sequence of integers $k_i\geq k_0$ such that $v_{k_i}=v$. We claim that the largest colour visited in the segment $v_{k_i}\cdots v_{k_{i+1}}$ is even: indeed, it is a direct consequence of \Cref{12-prop:bumps} and of the fact that the set of summaries we consider are winning. We then conclude that the largest colour infinitely visited in $\play$ is even hence, leading a contradiction.
		\item Assume that no vertex is infinitely often visited in $\play$. As the parity objective is prefix-independent we can assume without loss of generality that there is no visited vertex with stack-height strictly smaller than $h=\sh(v_0)$. 
			Factorise $\play$ as $v_{i_0}\cdots v_{i_1-1}v_{i_1}\cdots v_{i_2-1}v_{i_2}\cdots v_{i_3-1}\cdots$ where $\sh(v_{i_j})=\sh(v_{i_{j+1}-1})=h+j$ and $\sh(v_k)>h+j$ for all $k\geq j+1$ (equivalently stack height $h+j$ is left forever in $v_{j+1}$). Call $s_j$ the stack content in $v_{i_j}$ and consider the infinite play $\hplay$ defined as the limit of the increasing (for prefix ordering) sequence of finite plays $(\hplay_{s_j})_{j\geq 0}$: it is a play in $\hgame$ where \Eve\ respects $\hsigma$ hence, it is winning for her. 
			Now let $v_{i_j}=(q_j,s_j)$, $v_{i_{j+1}-1}=(p_j,s_j)$ and let $c_j$ be the largest colour visited in $v_{i_j}\cdots v_{i_{j+1}-1}$. Then, as we assume that $\play$ is losing one has $\limsup(c_i)_{i\geq 0}$ is odd. Moreover, if one lets \begin{equation*}
\begin{split}
\hplay = & (\bot,R_0)(\gamma_0,R_0,\sigma_0,S_0)(\gamma_0,R_0,\sigma_0,S_0,(\sigma^1_\alpha,S^1_\alpha)_{\alpha\in\Gamma}))(\gamma_1,R_1,\sigma_1,S_1)\\ &\quad (\gamma_1,R_1,\sigma_1,S_1,(\sigma^2_\alpha,S^2_\alpha)_{\alpha\in\Gamma}))(\gamma_2,R_2,\sigma_2,S_2)\cdots	
\end{split}
\end{equation*}
one has that 
\begin{itemize}
	\item $(q_i,c_i,p_i)\in S_i$ (by \Cref{12-prop:bumps}); and 
	\item either $p_i\in Q_\mEve$ and $(q_{i+1},\push{\gamma_{i+1}})=\sigma_i(p_i,\gamma_i)$ (by definition of $\sigma$) or $p_i\in Q_\mAdam$ and $(q_{i+1},\push{\gamma_{i+1}})\in \Delta((p_i,\gamma_i)$  (by definition of $\sigma$).
\end{itemize}
This means that $\hplay$ is losing, leading a contradiction.
	\end{itemize}
Hence, we conclude that $\play$ is winning which concludes the proof.
\end{proof}

\begin{remark}
\label{12-rmk:finite_state_automaton}
Note that the previous strategy $\sigma$ can be computed by a finite state automaton.
\end{remark}

\section*{Bibliographic references}
\label{12-sec:references}
The decidability of pushdown parity games is a consequence of the decidability of monadic second-order logic (MSO) on the infinite complete binary tree \cite{Rabin:1969}. The key observation is that any pushdown arena can be defined using MSO in the infinite complete binary tree $\Delta_2$. Indeed every node in $\Delta_2$ can be identified with the binary word encoding the path from the root to this node. Now, if one chooses a fixed-length encoding for the stack alphabet $\Gamma$ and for the set of states, a configuration $(p,s)$ can be associated to the node of $\Delta_2$ encoded by ${sp}$. Using this representation, the effect of a transition of the pushdown system can easily be captured by an MSO-formula. As a consequence, any property expressible in MSO on the arena can be effectively translated into equivalent MSO-property on the full binary tree. 

This observation can be used to decide the winner of the game from any given vertex (\textit{i.e.}, Problem~1). It is indeed possible to write an MSO formula $\varphi_\mEve(x)$ which expresses that \Eve\ wins the games from $x$, see~\cite{Walukiewicz:2002} for parity games with possibly uncountable arenas. In the simpler case of pushdown arenas in which all vertices have a bounded out-degree writing such a formula is straightforward (see for instance \cite[Section~2.3.4]{Cachat:2003}). 

Rabin's lemma can be used to show that the winning regions are regular sets of configurations, as defined in \Cref{12-thm:regularity-wr}. Recall that Rabin's lemma states that if the infinite tree $\Delta_2$ satisfies an existential MSO-formula $\exists X,\, \varphi(X)$ then there exists a regular set of nodes $R$ such that $\Delta_2$ satisfies $\varphi[R]$. Furthermore, a finite automaton accepting $R$ can be effectively constructed from $\varphi$. If we consider the formula $\varphi(X) = \forall x,\, x \in X \Leftrightarrow \varphi_{\Eve}(x)$, we immediately obtain a finite automaton accepting the binary encodings of the configurations in the winning region for \Eve. This automaton can easily be transformed to accept the configurations as in \Cref{12-thm:regularity-wr}.
 
Using similar arguments, one can derive the existence of a regular positional winning strategy for both players in the sense of \Cref{12-subsec:regular-strat}. 

Although effective, these results do not provide tight upper-bounds as they rely on the translation of MSO properties into equivalent parity tree automata which is non-elementary in the alternation rank of the formula. In the sequel of this bibliographic note, we will focus on the history of decision procedures for pushdown games which provide explicit complexity bounds. 

The first result on pushdown games can be traced back to the work of B{\"u}chi, which shows that the set of configurations reachable from the initial configuration of a pushdown system forms a regular language and hence can be represented by a finite state automaton \cite{Buchi:1964}. While B{\"u}chi's procedure is exponential, Caucal showed that this problem can be solved in polynomial time \cite{Caucal:1988}. The improved algorithm is a saturation process where transitions are incrementally added to a finite automaton. This technique was simplified and adapted to pushdown reachability games where the target set is a regular set of configurations by Bouajjani et al. in \cite{Bouajjani.Esparza.ea:1997} and independently in \cite{Finkel.Willems.ea:1997}. This saturation approach corresponds to the fixpoint characterisation of strategy profiles in \Cref{ssec:reachability-pushdown-games}. This method was latter extended to B{\"u}chi pushdown games \cite{Cachat:2002} and finally to pushdown parity games \cite{Hague.Ong:2009}. We refer the reader to \cite{Carayol.Hague:2014} for a survey of this method.

The decidability of pushdown parity games was established in~\cite{Walukiewicz:2001} where Walukiewicz gives algorithms to compute winning strategies for both players from a given configuration based on a deterministic pushdown automaton reading the moves of the play (as in \Cref{12-subsec:regular-strat}). In this article, it is also shown that deciding the winner in a two-player pushdown reachability game is hard for \EXP. In \cite{Serre:2003}, the winning region for both players was shown to be regular based on similar arguments as those presented in \Cref{12-sec:profiles}. The proof presented in this chapter follows \cite{Serre:2004}, except for the construction of a regular winning strategy given in Section~\ref{12-subsec:regular-strat} which is novel.

Kupferman and Vardi in \cite{Kupferman.Vardi:2000} reduced deciding the winner in a pushdown parity game to the emptiness problem for two-way alternating parity tree automata. In \cite{Vardi:1998}, Vardi had previously shown that this emptiness problem for this model of automata is in \EXP. In addition, this approach permits to compute a finite automaton accepting the winning region for both players, as well as regular winning strategies for both players. The reduction is very close to the reduction to the decidability of MSO on the full binary tree sketched in the beginning of this section, but by replacing MSO-formulas with two-way alternating parity tree automata, one can obtain optimal complexity (see \cite{Cachat:2002} for a survey of this approach). A similar approach was used by Serre in \cite{Serre:2006} to study parity one-counter games (\emph{\textit{i.e.}} pushdown parity games with a one-letter stack alphabet): by a reduction to the emptiness problem for two-way alternating parity \emph{word} automaton, one obtains a \PSPACE~upper bound (which is also a matching lower bound).

They were many subsequent works on extensions of pushdown parity games. One line of research considered non-regular winning conditions, based on properties of the stack behaviour along the play \cite{Cachat.Duparc.ea:2002}. Another line of research focused on considering games played on infinite arenas generated by extensions of pushdown automata, mainly higher-order pushdown automata and collapsible  pushdown automata, mostly motivated by connections with higher-order recursion schemes (for further references on the topic we refer to~\cite{Broadbent.Carayol.ea:2021}).

%
%
%
%
%
%
%
%
%
%
%
%
%
%
%
%
%
%

\ifpictures
\includepdf{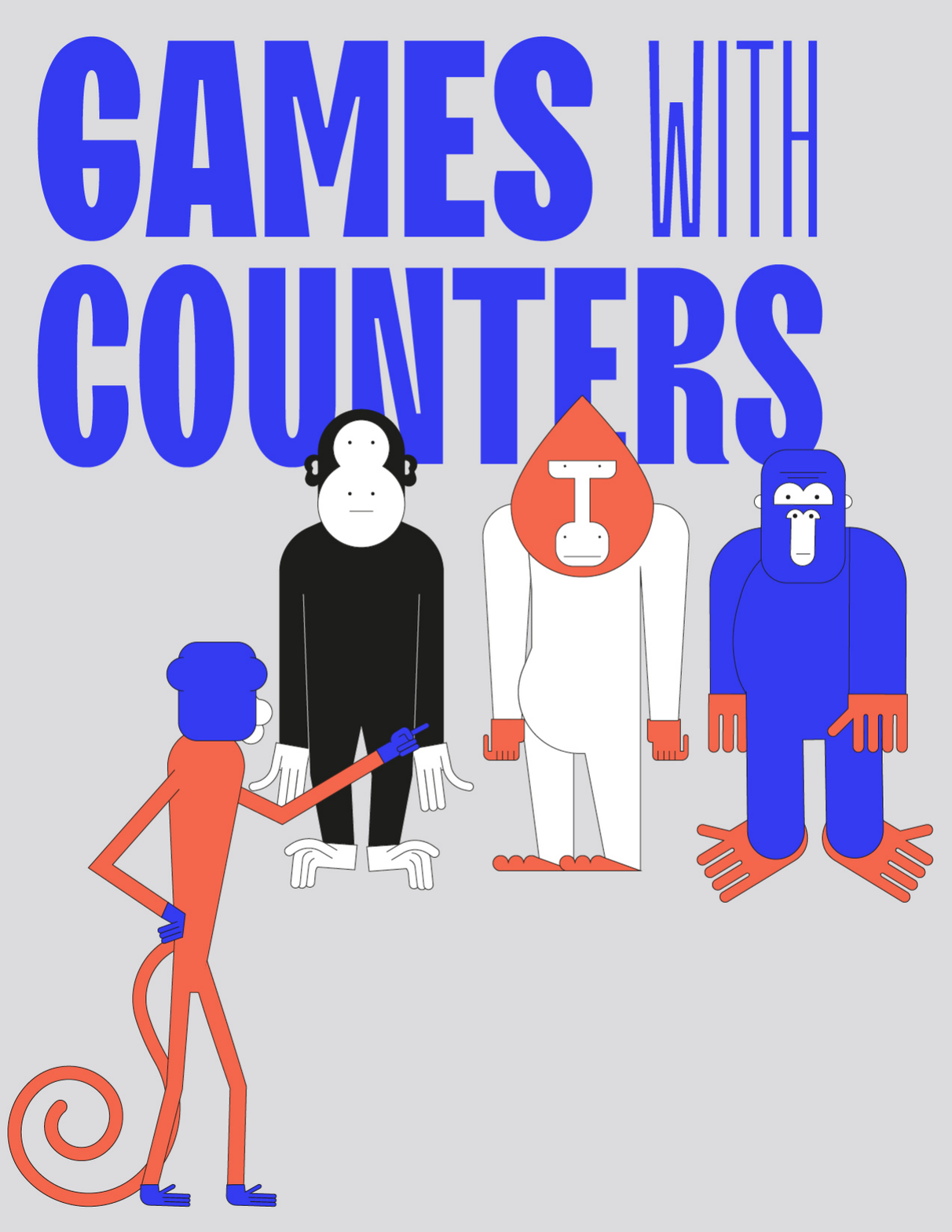}
\fi
\author[Sylvain Schmitz]{Sylvain Schmitz}
\copyrightline{Copyright by Sylvain Schmitz 2025, to be published by Cambridge University Press in the volume \textit{Games on Graphs} edited by Nathana\"el Fijalkow}

\chapter{Games with Counters}
\chapterauthor{Sylvain Schmitz}
\label{13-chap:counters}

\providecommand{\AP}{}

\newcommand{\tup}[1]{\langle #1\rangle}
\newcommand{\eqby}[1]{\stackrel{\!\,\!\,\raisebox{-.15ex}{\scalebox{.5}{\textrm{#1}}}}{=}}
\newcommand{\eqdef}{\eqby{def}}
\newcommand{\Loc}{\?L}
\providecommand{\Act}{A}
\renewcommand{\Act}{A}
\providecommand{\dom}{\mathrm{dom}\,}
\newcommand{\pto}{\mathrel{\ooalign{\hfil$\mapstochar\mkern5mu$\hfil\cr$\to$\cr}}}
\providecommand{\weight}{w}
\renewcommand{\weight}{w}
\providecommand{\loc}{\ell}
\newcommand{\sink}{\bot}
\newcommand{\dd}{k}
\newcommand{\CounterReach}{\textsf{CounterReach}\xspace}
\newcommand{\Cover}{\textsf{Cover}\xspace}
\newcommand{\NonTerm}{\textsf{NonTerm}\xspace}
\providecommand{\step}[1]{\xrightarrow{\,\raisebox{-1pt}[0pt][0pt]{\ensuremath{#1}}\,}}
\renewcommand{\step}[1]{\xrightarrow{\,\raisebox{-1pt}[0pt][0pt]{\ensuremath{#1}}\,}}
\newcommand{\mstep}[1]{\xrightarrow{\,\raisebox{-1pt}[6pt][0pt]{\ensuremath{#1}}\,}}
\newcommand{\inst}[1]{\mathrel{\mathtt{#1}}}
\providecommand{\pop}{\mathrm{pop}}
\renewcommand{\pop}{\mathrm{pop}}
\providecommand{\push}[1]{\mathrm{push}(#1)}
\renewcommand{\push}[1]{\mathrm{push}(#1)}
\providecommand{\blank}{\Box}
\newcommand{\emkl}{\triangleright}
\newcommand{\emkr}{\triangleleft}
\renewcommand{\natural}{\arena_\+N}
\newcommand{\energy}{\arena_\+E}
\newcommand{\bounded}{\arena_B}
\newcommand{\capped}{\arena_C}
\newcommand{\capp}[2][C]{\overline{\vec #2}^{#1}}
\newcommand{\lcol}{\mathrm{lcol}}
\newcommand{\vcol}{\mathrm{vcol}}
\newcommand{\litt}{\loc}
\newcommand{\Effect}{\Delta}

\newcommand{\?}{\mathcal}
\newcommand{\+}{\mathbb}

\providecommand{\qedhere}{\hfill\ensuremath\Box}

\let\oldcite\cite
\renewcommand{\cite}{\citep}
\providecommand{\citep}{\oldcite}
\providecommand{\citet}{\cite}
\providecommand{\citem}[2][1]{#1~\cite{#2}}

\providecommand{\mymoot}[1]{}

Just like timed games arise from timed systems and pushdown games
from pushdown systems, counter games arise from (multi-)counter
systems.  Those are finite-state systems further endowed with a
finite number of counters whose values range over the natural numbers,
and are widely used to model and reason about systems handling
discrete resources.  Such resources include for instance money on a
bank account, items on a factory line, molecules in chemical
reactions, organisms in biological ones, replicated processes in
distributed computing, etc.  As with timed or pushdown systems,
counter systems give rise to infinite graphs that can be turned into
infinite game arenas.

\AP One could populate a zoo with the many variants of counter systems,
depending on the available counter operations.  One of the best known
specimens in this zoo are Minsky machines~\cite{Minsky:1967},
where the operations are incrementing a counter, decrementing it, or
testing whether its value is zero.  "Minsky machines" are a universal
model of computation: their reachability problem is undecidable,
already with only two counters.  From the algorithmic perspective we
promote in this book, this means that the counter games arising from
"Minsky machines" are not going to be very interesting, unless perhaps
if we restrict ourselves to a single counter.  A more promising
species in our zoo are \emph{"vector addition systems with
  states"}~\cite{Greibach:1978}---or,
equivalently, Petri nets~\cite{Petri:1962}---, where the only
available operations are increments and decrements.  "Vector addition
systems with states" enjoy a decidable reachability
problem~\cite{Mayr:1981}, which
makes them a much better candidate for studying the associated games.

In this chapter, we focus on "vector games", that is, on games defined
on arenas defined by "vector addition systems with states" with a
partition of states controlled by~Eve and Adam.  As we are going to
see in \Cref{13-sec:counters}, those games turn out to be undecidable
already for quite restricted objectives and just two counters.  We
then investigate two restricted classes of "vector games".
\begin{enumerate}
\item In \Cref{13-sec:dim1}, we consider \emph{"one-counter games"}.  These can
  be reduced to the pushdown games of \Cref{12-chap:pushdown} and are
  therefore decidable.  Most of the section is thus devoted to proving
  sharp complexity lower bounds, already in the case of so-called
  \emph{"countdown games"}.
\item In \Cref{13-sec:avag}, we turn our attention to the main results of
  this chapter.  By suitably restricting both  the systems, with an
  \emph{"asymmetry"} condition that forbids Adam to manipulate the
  counters, and  the "objective", with a \emph{"monotonicity@monotonic objective"}
    condition that ensures that Eve's winning region is "upwards
    closed"---meaning that larger counter values make it easier for
    her to win---, one obtains a class of decidable "vector games" where "finite
  memory" strategies are sufficient.
  \begin{itemize}
  \item   This class is still rich enough to find many applications, and we
  zoom in on the connections with resource-conscious games like
  "\emph{energy} games" and "\emph{bounding} games" in
  \Cref{13-sec:resource}---a subject that will be taken further in
  \Cref{14-chap:multiobjective}.
  
  \item The computational complexity of "asymmetric" "monotonic@monotonic
  objective" "vector games" is now well-understood, and we devote
  \Cref{13-sec:complexity} to the topic; \Cref{13-tbl:cmplx} at the end of
  the chapter summarises these results.
  \end{itemize}
\end{enumerate}


\section{Vector games}
\label{13-sec:counters}
\AP A vector system is a finite directed graph with a partition of
the vertices and weighted edges.  Formally, it is a tuple
$\?V=(\Loc,\Act,\Loc_\mEve,\Loc_\mAdam,\dd)$ where $\dd\in\+N$ is a
dimension, $\Loc$ is a finite set of locations partitioned into the
locations controlled by Eve and Adam, \textit{i.e.},
$\Loc=\Loc_\mEve\uplus \Loc_\mAdam$, and
$\Act\subseteq \Loc\times\+Z^\dd\times\Loc$ is a finite set of
weighted actions.  We write $\loc\step{\vec u}\loc'$
rather than $(\loc,\vec u,\loc')$ for actions in~$\Act$.  A
vector addition system with states is a "vector system" where
$\Loc_\mAdam=\emptyset$, \textit{i.e.}, it corresponds to the one-player case.

\begin{example}[vector system]
\label{13-ex:mwg}
  \Cref{13-fig:mwg} presents a "vector system" of
  dimension two with locations $\{\loc,\loc'\}$ where~$\loc$ is
  controlled by Eve and $\loc'$ by Adam.
\end{example}
\begin{figure}[htbp]
  \centering
  \begin{tikzpicture}[auto,on grid,node distance=2.5cm]
    \node[s-eve](0){$\loc$};
    \node[s-adam,right=of 0](1){$\loc'$};
    \path[arrow,every node/.style={font=\footnotesize,inner sep=1}]
    (0) edge[loop left] node {$-1,-1$} ()
    (0) edge[bend right=10] node {$-1,0$} (1)
    (1) edge[bend left=30] node {$-1,0$} (0)
    (1) edge[bend right=30,swap] node {$2,1$} (0);
  \end{tikzpicture}
  \caption{A "vector system".}
  \label{13-fig:mwg}
\commentAlt{Figure~\ref{13-fig:mwg}: A directed graph with two nodes, a circular node labeled 'L' and a square node labeled 'L'', showing bidirectional transitions with numerical labels, and a self-loop.}
\commentLongAlt{Figure~\ref{13-fig:mwg}: The image displays a directed graph with two nodes. The node on the left is a circle labeled 'L', and the node on the right is a square labeled 'L''.
- Node 'L' has a self-loop labeled '-1, -1'.
- There are two directed arrows from 'L' to 'L'': one labeled '-1, 0' and another that is part of a bidirectional pair.
- There are two directed arrows from 'L'' to 'L': one labeled '2, 1' and another labeled '-1, 0'. The two pairs of arrows suggest bidirectional connections between the nodes, each with distinct labels for the forward and reverse directions, and separate labels for each arrow.}
\end{figure}

The intuition behind a "vector system" is that it
maintains~$\dd$ counters $\mathtt{c}_1,\dots,\mathtt{c}_\dd$ assigned
to integer values.  An action $\loc\step{\vec u}\loc'\in\Act$ then
updates each counter by adding the corresponding entry of~$\vec u$,
that is for all $1\leq j\leq\dd$, the action performs the update
$\mathtt{c}_j := \mathtt{c}_j+\vec u(j)$.

\medskip \AP Before we proceed any further, let us fix some notations
for vectors in $\+Z^\dd$.  We write `$\vec 0$' for the zero vector
with $\vec 0(j)\eqdef 0$ for all $1\leq j\leq\dd$.  For all
$1\leq j\leq\dd$, we write `$\vec e_j$' for the unit vector with
$\vec e_j(j)\eqdef 1$ and $\vec e_{j}(j')\eqdef 0$ for all $j'\neq j$.
Addition and comparison are defined componentwise, so that for
instance $\vec u\leq\vec u'$ if and only if for all $1\leq j\leq\dd$,
$\vec u(j)\leq\vec u'(j)$.  We write
$\weight(\loc\step{\vec u}\loc')\eqdef\vec u$ for the weight of an
action and $\weight(\pi)\eqdef\sum_{1\leq j\leq |\pi|}\weight(\pi_j)$
for the cumulative weight of a finite sequence of actions
$\pi\in\Act^\ast$.  For a vector $\vec u\in\+Z^\dd$, we use its
infinity norm $\|\vec u\|\eqdef\max_{1\leq j\leq\dd}|\vec u(j)|$,
hence $\|\vec 0\|=0$ and $\|\vec e_j\|=\|-\vec e_j\|=1$, and we let
$\|\loc\step{\vec u}\loc'\|\eqdef\|\weight(\loc\step{\vec
  u}\loc')\|=\|\vec u\|$
and $\|\Act\|\eqdef\max_{a\in\Act}\|\weight(a)\|$.  Unless stated
otherwise, we assume that all our vectors are represented in binary,
hence $\|\Act\|$ may be exponential in the size of~$\?V$.

\subsection{Arenas and games}
\AP A "vector system" gives rise to an infinite graph
$G_\+N\eqdef(V,E)$ over the set of vertices
$V\eqdef(\Loc\times\+N^\dd)\uplus\{\sink\}$.  The vertices of the
graph are either \emph{configurations} $\loc(\vec v)$ consisting of a
location $\loc\in \Loc$ and a vector of non-negative integers
$\vec v\in\+N^\dd$---such a vector represents a valuation of the
counters $\mathtt{c}_1,\dots,\mathtt c_\dd$---, or the
sink~$\sink$.

\AP Consider an action in~$a=(\loc\step{\vec u}\loc')$ in~$\Act$: we
see it as a partial function
$a{:}\,\Loc\times\+N^\dd\,\pto \Loc\times\+N^\dd$ with domain
$\dom a\eqdef\{\loc(\vec v)\mid \vec v+\vec u\geq\vec 0\}$ and image
$a(\loc(\vec v))\eqdef \loc'(\vec v+\vec u)$; let also
$\dom\Act\eqdef\bigcup_{a\in\Act}\dom a$.  This allows us to define
the set~$E$ of edges as a set of pairs
\begin{align*}
  E&\eqdef\{(\loc(\vec v),a(\loc(\vec v)))\mid a\in\Act\text{ and
     }\loc(\vec v)\in\dom a\}\\
  &\:\cup\:\{(\loc(\vec v),\sink)\mid\loc(\vec v)\not\in\dom\Act\}\cup\{(\sink,\sink)\}\;,
\end{align*}
where $\ing((v,v'))\eqdef v$ and $\out((v,v'))\eqdef v'$ for all
edges~$(v,v')\in E$.  There is therefore an edge $(v,v')$ between two
configurations $v=\loc(\vec v)$ and $v'=\loc'(\vec v')$ if there
exists an action $\loc\step{\vec u}\loc'\in\Act$ such that
$\vec v'=\vec v+\vec u$.  Note that, quite importantly,
$\vec v+\vec u$ must be non-negative on every coordinate for this edge
to exist.  If no action can be applied, there is an edge to the
"sink"~$\sink$, which ensures that $E$ is total: for all $v\in V$,
there exists an edge $(v,v')\in E$ for some $v'$, and thus there are
no `deadlocks' in the graph.

The configurations are naturally partitioned into those in
$\VE\eqdef\Loc_\mEve\times\+N^\dd$ controlled by~Eve and those in
$\VA\eqdef\Loc_\mAdam\times\+N^\dd$ controlled by Adam.  Regarding
the "sink", the only edge starting from~$\sink$ loops back
to it, and it does not matter who of Eve or Adam controls it.  This
gives rise to an infinite arena $\arena_\+N\eqdef(G_\+N,\VE,\VA)$ called
the natural semantics of~$\?V$.

\medskip Although we work in a turn-based setting with perfect
information, it is sometimes enlightening to consider the partial map
$\dest{:}\,V\times A\pto E$ defined by
$\dest(\loc(\vec v),a)\eqdef(\loc(\vec v),a(\loc(\vec v)))$ if
$\loc(\vec v)\in\dom a$ and
$\dest(\loc(\vec v),a)\eqdef(\loc(\vec v),\sink)$ if
$\loc(\vec v)\not\in\dom\Act$.  Note that a sequence~$\pi$ over $E$
that avoids the "sink" can also be described by an initial
configuration $\loc_0(\vec v_0)$ paired with a sequence
over~$\Act$.

\begin{example}[natural semantics]
\label{13-ex:sem}
  \Cref{13-fig:sem} illustrates the "natural semantics" of the system
  of~\Cref{13-fig:mwg}; observe that all the configurations $\loc(0,n)$
  for $n\in\+N$ lead to the "sink".
\end{example}

\begin{figure}[htbp]
  \centering\scalebox{.77}{
  \begin{tikzpicture}[auto,on grid,node distance=2.5cm]
    \draw[step=1,lightgray!50,dotted] (-5.7,0) grid (5.7,3.8);
    \node at (0,3.9) (sink) {\boldmath$\sink$};
    \draw[step=1,lightgray!50] (1,0) grid (5.5,3.5);
    \draw[step=1,lightgray!50] (-1,0) grid (-5.5,3.5);
    \draw[color=white](0,-.3) -- (0,3.8);
    \node at (0,0)[lightgray,font=\scriptsize,fill=white] {0};
    \node at (0,1)[lightgray,font=\scriptsize,fill=white] {1};
    \node at (0,2)[lightgray,font=\scriptsize,fill=white] {2};
    \node at (0,3)[lightgray,font=\scriptsize,fill=white] {3};
    \node at (1,3.9)[lightgray,font=\scriptsize,fill=white] {0};
    \node at (2,3.9)[lightgray,font=\scriptsize,fill=white] {1};
    \node at (3,3.9)[lightgray,font=\scriptsize,fill=white] {2};
    \node at (4,3.9)[lightgray,font=\scriptsize,fill=white] {3};
    \node at (5,3.9)[lightgray,font=\scriptsize,fill=white] {4};
    \node at (-1,3.9)[lightgray,font=\scriptsize,fill=white] {0};
    \node at (-2,3.9)[lightgray,font=\scriptsize,fill=white] {1};
    \node at (-3,3.9)[lightgray,font=\scriptsize,fill=white] {2};
    \node at (-4,3.9)[lightgray,font=\scriptsize,fill=white] {3};
    \node at (-5,3.9)[lightgray,font=\scriptsize,fill=white] {4};
    \node at (1,0)[s-eve-small] (e00) {};
    \node at (1,1)[s-adam-small](a01){};
    \node at (1,2)[s-eve-small] (e02){};
    \node at (1,3)[s-adam-small](a03){};
    \node at (2,0)[s-adam-small](a10){};
    \node at (2,1)[s-eve-small] (e11){};
    \node at (2,2)[s-adam-small](a12){};
    \node at (2,3)[s-eve-small] (e13){};
    \node at (3,0)[s-eve-small] (e20){};
    \node at (3,1)[s-adam-small](a21){};
    \node at (3,2)[s-eve-small] (e22){};
    \node at (3,3)[s-adam-small](a23){};
    \node at (4,0)[s-adam-small](a30){};
    \node at (4,1)[s-eve-small] (e31){};
    \node at (4,2)[s-adam-small](a32){};
    \node at (4,3)[s-eve-small] (e33){};
    \node at (5,0)[s-eve-small] (e40){};
    \node at (5,1)[s-adam-small](a41){};
    \node at (5,2)[s-eve-small] (e42){};
    \node at (5,3)[s-adam-small](a43){};
    \node at (-1,0)[s-adam-small](a00){};
    \node at (-1,1)[s-eve-small] (e01){};
    \node at (-1,2)[s-adam-small](a02){};
    \node at (-1,3)[s-eve-small] (e03){};
    \node at (-2,0)[s-eve-small] (e10){};
    \node at (-2,1)[s-adam-small](a11){};
    \node at (-2,2)[s-eve-small] (e12){};
    \node at (-2,3)[s-adam-small](a13){};
    \node at (-3,0)[s-adam-small](a20){};
    \node at (-3,1)[s-eve-small] (e21){};
    \node at (-3,2)[s-adam-small](a22){};
    \node at (-3,3)[s-eve-small] (e23){};
    \node at (-4,0)[s-eve-small] (e30){};
    \node at (-4,1)[s-adam-small](a31){};
    \node at (-4,2)[s-eve-small] (e32){};
    \node at (-4,3)[s-adam-small](a33){};
    \node at (-5,0)[s-adam-small](a40){};
    \node at (-5,1)[s-eve-small] (e41){};
    \node at (-5,2)[s-adam-small](a42){};
    \node at (-5,3)[s-eve-small] (e43){};
    \path[arrow] 
    (e11) edge (e00)
    (e22) edge (e11)
    (e31) edge (e20)
    (e32) edge (e21)
    (e21) edge (e10)
    (e12) edge (e01)
    (e23) edge (e12)
    (e33) edge (e22)
    (e13) edge (e02)
    (e43) edge (e32)
    (e42) edge (e31)
    (e41) edge (e30);
    \path[arrow] 
    (e11) edge (a01)
    (e20) edge (a10)
    (e22) edge (a12)
    (e31) edge (a21)
    (e32) edge (a22)
    (e21) edge (a11)
    (e12) edge (a02)
    (e30) edge (a20)
    (e10) edge (a00)
    (e13) edge (a03)
    (e23) edge (a13)
    (e33) edge (a23)
    (e43) edge (a33)
    (e42) edge (a32)
    (e41) edge (a31)
    (e40) edge (a30);
    \path[arrow] 
    (a11) edge (e01)
    (a20) edge (e10)
    (a22) edge (e12)
    (a31) edge (e21)
    (a32) edge (e22)
    (a21) edge (e11)
    (a12) edge (e02)
    (a30) edge (e20)
    (a10) edge (e00)
    (a33) edge (e23)
    (a23) edge (e13)
    (a13) edge (e03)
    (a43) edge (e33)
    (a42) edge (e32)
    (a41) edge (e31)
    (a40) edge (e30);
    \path[arrow] 
    (a01) edge (e22)
    (a10) edge (e31)
    (a11) edge (e32)
    (a00) edge (e21)
    (a02) edge (e23)
    (a12) edge (e33)
    (a22) edge (e43)
    (a21) edge (e42)
    (a20) edge (e41);
    \path[arrow] 
    (-5.5,3.5) edge (e43)
    (5.5,2.5) edge (e42)
    (2.5,3.5) edge (e13)
    (5.5,0.5) edge (e40)
    (-5.5,1.5) edge (e41)
    (-3.5,3.5) edge (e23)
    (-1.5,3.5) edge (e03)
    (4.5,3.5) edge (e33)
    (5.5,0) edge (e40)
    (5.5,2) edge (e42)
    (-5.5,1) edge (e41)
    (-5.5,3) edge (e43);
    \path[dotted]
    (-5.7,3.7) edge (-5.5,3.5)
    (5.7,2.7) edge (5.5,2.5)
    (2.7,3.7) edge (2.5,3.5)
    (5.7,0.7) edge (5.5,0.5)
    (-3.7,3.7) edge (-3.5,3.5)
    (-1.7,3.7) edge (-1.5,3.5)
    (4.7,3.7) edge (4.5,3.5)
    (-5.7,1.7) edge (-5.5,1.5)
    (5.75,0) edge (5.5,0)
    (5.75,2) edge (5.5,2)
    (-5.75,1) edge (-5.5,1)
    (-5.75,3) edge (-5.5,3);
    \path[arrow]
    (5.5,1) edge (a41)
    (-5.5,2) edge (a42)
    (-5.5,0) edge (a40)
    (5.5,3) edge (a43);
    \path[dotted]
    (5.75,1) edge (5.5,1)
    (-5.75,2) edge (-5.5,2)
    (-5.75,0) edge (-5.5,0)
    (5.75,3) edge (5.5,3);
    \path[-]
    (a30) edge (5.5,.75)
    (a32) edge (5.5,2.75)
    (a31) edge (-5.5,1.75)
    (a23) edge (4,3.5)
    (a03) edge (2,3.5)
    (a13) edge (-3,3.5)
    (a33) edge (-5,3.5)
    (a43) edge (5.5,3.25)
    (a41) edge (5.5,1.25)
    (a40) edge (-5.5,0.25)
    (a42) edge (-5.5,2.25);
    \path[dotted]
    (5.5,.75) edge (5.8,.9)
    (5.5,2.75) edge (5.8,2.9)
    (-5.5,1.75) edge (-5.8,1.9)
    (4,3.5) edge (4.4,3.7)
    (2,3.5) edge (2.4,3.7)
    (-3,3.5) edge (-3.4,3.7)
    (-5,3.5) edge (-5.4,3.7)
    (5.5,3.25) edge (5.8,3.4)
    (5.5,1.25) edge (5.8,1.4)
    (-5.5,.25) edge (-5.8,0.4)
    (-5.5,2.25) edge (-5.8,2.4);
    \path[arrow]
    (sink) edge[loop left] ()
    (e00) edge[bend left=8] (sink)
    (e01) edge[bend right=8] (sink)
    (e02) edge[bend left=8] (sink)
    (e03) edge[bend right=8] (sink);
  \end{tikzpicture}}
  \caption{The "natural semantics" of the
    "vector system" of \Cref{13-fig:mwg}: a circle (resp.\
    a square) at position $(i,j)$ of the grid denotes a configuration
    $\loc(i,j)$ (resp.\ $\loc'(i,j)$) controlled by~Eve (resp. Adam).}
   \label{13-fig:sem}
\end{figure}


\AP A "vector system" $\?V=(\Loc,\Act,\Loc_\mEve,\Loc_\mAdam,\dd)$, a
colouring~$\col{:}\,E\to C$, and an
objective~$\Omega\subseteq C^\omega$ together define a vector game
$\game=(\natural(\?V),\col,\Omega)$.  Because $\natural(\?V)$ is an
infinite arena, we need to impose restrictions on our "colourings"
$\col{:}\,E\to C$ and the "qualitative
objectives"~$\Omega\subseteq C^\omega$; at the very least, they should
be recursive.

There are then two variants of the associated decision problem:
\begin{itemize}
\item\AP the given initial credit variant, where we are given $\?V$,
  $\col$, $\Omega$, a location $\loc_0\in\Loc$ and an initial credit
  $\vec v_0\in\+N^\dd$, and ask whether Eve wins~$\game$ from the
  initial configuration~$\loc_0(\vec v_0)$;
\item\AP the existential initial credit variant, where we are given
  $\?V$, $\col$, $\Omega$, and a location $\loc_0\in\Loc$, and ask
  whether there exists an initial credit $\vec v_0\in\+N^\dd$ such
  that Eve wins~$\game$ from the initial
  configuration~$\loc_0(\vec v_0)$.
\end{itemize}

Let us instantiate the previous abstract definition of "vector games".
We first consider two `"reachability"-like'
\index{reachability!\emph{see also} vector game\protect\mymoot|mymoot}
objectives, where $C\eqdef\{\varepsilon,\Win\}$ and
$\Omega\eqdef\Reach$, namely "configuration reachability" and
"coverability".  The difference between the two is that, in the
"configuration reachability" problem, a specific configuration
$\loc_f(\vec v_f)$ should be visited, whereas in the "coverability"
problem, Eve attempts to visit $\loc_f(\vec v')$ for some
vector~$\vec v'$ componentwise larger or equal to
$\vec v_f$.\footnote{The name `"coverability"' comes from the the
  literature on "Petri nets" and "vector addition systems with
  states", because Eve is attempting to \emph{cover}
  $\loc_f(\vec v_f)$, \textit{i.e.}, to reach a configuration $\loc_f(\vec v')$
  with $\vec v'\geq\vec v_f$.}

\decpb["configuration reachability vector game" with "given initial credit"]
{\label{13-pb:reach} A "vector system"
  $\?V=(\Loc,\Act,\Loc_\mEve,\Loc_\mAdam,\dd)$, an initial location
  $\loc_0\in\Loc$, an initial credit $\vec v_0\in\+N^\dd$, and a
  target configuration $\loc_f(\vec v_f)\in\Loc\times\+N^\dd$.}
{Does Eve have a strategy to reach $\loc(\vec v)$ from
  $\loc_0(\vec v_0)$?
  That is, does she win the configuration
  reachability game $(\natural(\?V),\col,\Reach)$ from
  $\loc_0(\vec v_0)$, where $\col(e)= \Win$ if and only if
  $\ing(e)=\loc_f(\vec v_f)$?}

\decpb["coverability vector game" with "given initial credit"]%
{\label{13-pb:cov} A "vector system"
  $\?V=(\Loc,\Act,\Loc_\mEve,\Loc_\mAdam,\dd)$, an initial location
  $\loc_0\in\Loc$, an initial credit $\vec v_0\in\+N^\dd$, and a
  target configuration $\loc_f(\vec v_f)\in\Loc\times\+N^\dd$.}%
{Does Eve have a strategy to reach $\loc(\vec v')$ for some
  $\vec v'\geq\vec v_f$ from $\loc_0(\vec v_0)$?
  That is, does she win
  the coverability game $(\natural(\?V),\col,\Reach)$ from
  $\loc_0(\vec v_0)$, where $\col(e)= \Win$ if and only if
  $\ing(e)=\loc_f(\vec v')$ for some $\vec v'\geq\vec v_f$?}

\begin{example}[Objectives]
\label{13-ex:cov}
  Consider the target configuration $\loc(2,2)$ in
  \Cref{13-fig:mwg,13-fig:sem}.  Eve's "winning region" in the
  "configuration reachability" "vector game" is
  $\WE=\{\loc(n+1,n+1)\mid n\in\+N\}\cup\{\loc'(0,1)\}$, displayed on the left
  in \Cref{13-fig:cov}.  Eve has indeed an obvious winning strategy
  from any configuration $\loc(n,n)$ with $n\geq 2$, which is to use
  the action $\loc\step{-1,-1}\loc$ until she reaches~$\loc(2,2)$.
  Furthermore, in $\loc'(0,1)$---due to the "natural semantics"---,
  Adam has no choice but to use the action $\loc'\step{2,1}\loc$:
  therefore $\loc'(0,1)$ and $\loc(1,1)$ are also winning for Eve.
\begin{figure}[htbp]
  \centering\scalebox{.48}{
  \begin{tikzpicture}[auto,on grid,node distance=2.5cm]
    \draw[step=1,lightgray!50,dotted] (-5.7,0) grid (5.7,3.8);
    \draw[color=white](0,-.3) -- (0,3.8);
    \node at (0,3.9) (sink) {\color{red!70!black}\boldmath$\sink$};
    \draw[step=1,lightgray!50] (1,0) grid (5.5,3.5);
    \draw[step=1,lightgray!50] (-1,0) grid (-5.5,3.5);
    \node at (0,0)[lightgray,font=\scriptsize,fill=white] {0};
    \node at (0,1)[lightgray,font=\scriptsize,fill=white] {1};
    \node at (0,2)[lightgray,font=\scriptsize,fill=white] {2};
    \node at (0,3)[lightgray,font=\scriptsize,fill=white] {3};
    \node at (1,3.9)[lightgray,font=\scriptsize,fill=white] {0};
    \node at (2,3.9)[lightgray,font=\scriptsize,fill=white] {1};
    \node at (3,3.9)[lightgray,font=\scriptsize,fill=white] {2};
    \node at (4,3.9)[lightgray,font=\scriptsize,fill=white] {3};
    \node at (5,3.9)[lightgray,font=\scriptsize,fill=white] {4};
    \node at (-1,3.9)[lightgray,font=\scriptsize,fill=white] {0};
    \node at (-2,3.9)[lightgray,font=\scriptsize,fill=white] {1};
    \node at (-3,3.9)[lightgray,font=\scriptsize,fill=white] {2};
    \node at (-4,3.9)[lightgray,font=\scriptsize,fill=white] {3};
    \node at (-5,3.9)[lightgray,font=\scriptsize,fill=white] {4};
    \node at (1,0)[s-eve-small,lose] (e00) {};
    \node at (1,1)[s-adam-small,win](a01){};
    \node at (1,2)[s-eve-small,lose] (e02){};
    \node at (1,3)[s-adam-small,lose](a03){};
    \node at (2,0)[s-adam-small,lose](a10){};
    \node at (2,1)[s-eve-small,win] (e11){};
    \node at (2,2)[s-adam-small,lose](a12){};
    \node at (2,3)[s-eve-small,lose] (e13){};
    \node at (3,0)[s-eve-small,lose] (e20){};
    \node at (3,1)[s-adam-small,lose](a21){};
    \node at (3,2)[s-eve-small,win] (e22){};
    \node at (3,3)[s-adam-small,lose](a23){};
    \node at (4,0)[s-adam-small,lose](a30){};
    \node at (4,1)[s-eve-small,lose] (e31){};
    \node at (4,2)[s-adam-small,lose](a32){};
    \node at (4,3)[s-eve-small,win] (e33){};
    \node at (5,0)[s-eve-small,lose] (e40){};
    \node at (5,1)[s-adam-small,lose](a41){};
    \node at (5,2)[s-eve-small,lose] (e42){};
    \node at (5,3)[s-adam-small,lose](a43){};
    \node at (-1,0)[s-adam-small,lose](a00){};
    \node at (-1,1)[s-eve-small,lose] (e01){};
    \node at (-1,2)[s-adam-small,lose](a02){};
    \node at (-1,3)[s-eve-small,lose] (e03){};
    \node at (-2,0)[s-eve-small,lose] (e10){};
    \node at (-2,1)[s-adam-small,lose](a11){};
    \node at (-2,2)[s-eve-small,lose] (e12){};
    \node at (-2,3)[s-adam-small,lose](a13){};
    \node at (-3,0)[s-adam-small,lose](a20){};
    \node at (-3,1)[s-eve-small,lose] (e21){};
    \node at (-3,2)[s-adam-small,lose](a22){};
    \node at (-3,3)[s-eve-small,lose] (e23){};
    \node at (-4,0)[s-eve-small,lose] (e30){};
    \node at (-4,1)[s-adam-small,lose](a31){};
    \node at (-4,2)[s-eve-small,lose] (e32){};
    \node at (-4,3)[s-adam-small,lose](a33){};
    \node at (-5,0)[s-adam-small,lose](a40){};
    \node at (-5,1)[s-eve-small,lose] (e41){};
    \node at (-5,2)[s-adam-small,lose](a42){};
    \node at (-5,3)[s-eve-small,lose] (e43){};
    \path[arrow] 
    (e11) edge (e00)
    (e22) edge (e11)
    (e31) edge (e20)
    (e32) edge (e21)
    (e21) edge (e10)
    (e12) edge (e01)
    (e23) edge (e12)
    (e33) edge (e22)
    (e13) edge (e02)
    (e43) edge (e32)
    (e42) edge (e31)
    (e41) edge (e30);
    \path[arrow] 
    (e11) edge (a01)
    (e20) edge (a10)
    (e22) edge (a12)
    (e31) edge (a21)
    (e32) edge (a22)
    (e21) edge (a11)
    (e12) edge (a02)
    (e30) edge (a20)
    (e10) edge (a00)
    (e13) edge (a03)
    (e23) edge (a13)
    (e33) edge (a23)
    (e43) edge (a33)
    (e42) edge (a32)
    (e41) edge (a31)
    (e40) edge (a30);
    \path[arrow] 
    (a11) edge (e01)
    (a20) edge (e10)
    (a22) edge (e12)
    (a31) edge (e21)
    (a32) edge (e22)
    (a21) edge (e11)
    (a12) edge (e02)
    (a30) edge (e20)
    (a10) edge (e00)
    (a33) edge (e23)
    (a23) edge (e13)
    (a13) edge (e03)
    (a43) edge (e33)
    (a42) edge (e32)
    (a41) edge (e31)
    (a40) edge (e30);
    \path[arrow] 
    (a01) edge (e22)
    (a10) edge (e31)
    (a11) edge (e32)
    (a00) edge (e21)
    (a02) edge (e23)
    (a12) edge (e33)
    (a22) edge (e43)
    (a21) edge (e42)
    (a20) edge (e41);
    \path[arrow] 
    (-5.5,3.5) edge (e43)
    (5.5,2.5) edge (e42)
    (2.5,3.5) edge (e13)
    (5.5,0.5) edge (e40)
    (-5.5,1.5) edge (e41)
    (-3.5,3.5) edge (e23)
    (-1.5,3.5) edge (e03)
    (4.5,3.5) edge (e33)
    (5.5,0) edge (e40)
    (5.5,2) edge (e42)
    (-5.5,1) edge (e41)
    (-5.5,3) edge (e43);
    \path[dotted]
    (-5.7,3.7) edge (-5.5,3.5)
    (5.7,2.7) edge (5.5,2.5)
    (2.7,3.7) edge (2.5,3.5)
    (5.7,0.7) edge (5.5,0.5)
    (-3.7,3.7) edge (-3.5,3.5)
    (-1.7,3.7) edge (-1.5,3.5)
    (4.7,3.7) edge (4.5,3.5)
    (-5.7,1.7) edge (-5.5,1.5)
    (5.75,0) edge (5.5,0)
    (5.75,2) edge (5.5,2)
    (-5.75,1) edge (-5.5,1)
    (-5.75,3) edge (-5.5,3);
    \path[arrow]
    (5.5,1) edge (a41)
    (-5.5,2) edge (a42)
    (-5.5,0) edge (a40)
    (5.5,3) edge (a43);
    \path[dotted]
    (5.75,1) edge (5.5,1)
    (-5.75,2) edge (-5.5,2)
    (-5.75,0) edge (-5.5,0)
    (5.75,3) edge (5.5,3);
    \path[-]
    (a30) edge (5.5,.75)
    (a32) edge (5.5,2.75)
    (a31) edge (-5.5,1.75)
    (a23) edge (4,3.5)
    (a03) edge (2,3.5)
    (a13) edge (-3,3.5)
    (a33) edge (-5,3.5)
    (a43) edge (5.5,3.25)
    (a41) edge (5.5,1.25)
    (a40) edge (-5.5,0.25)
    (a42) edge (-5.5,2.25);
    \path[dotted]
    (5.5,.75) edge (5.8,.9)
    (5.5,2.75) edge (5.8,2.9)
    (-5.5,1.75) edge (-5.8,1.9)
    (4,3.5) edge (4.4,3.7)
    (2,3.5) edge (2.4,3.7)
    (-3,3.5) edge (-3.4,3.7)
    (-5,3.5) edge (-5.4,3.7)
    (5.5,3.25) edge (5.8,3.4)
    (5.5,1.25) edge (5.8,1.4)
    (-5.5,.25) edge (-5.8,0.4)
    (-5.5,2.25) edge (-5.8,2.4);
    \path[arrow]
    (sink) edge[loop left] ()
    (e00) edge[bend left=8] (sink)
    (e01) edge[bend right=8] (sink)
    (e02) edge[bend left=8] (sink)
    (e03) edge[bend right=8] (sink);
  \end{tikzpicture}}\quad~~\scalebox{.48}{
  \begin{tikzpicture}[auto,on grid,node distance=2.5cm]
    \draw[step=1,lightgray!50,dotted] (-5.7,0) grid (5.7,3.8);
    \draw[color=white](0,-.3) -- (0,3.8);
    \node at (0,3.9) (sink) {\color{red!70!black}\boldmath$\sink$};
    \draw[step=1,lightgray!50] (1,0) grid (5.5,3.5);
    \draw[step=1,lightgray!50] (-1,0) grid (-5.5,3.5);
    \node at (0,0)[lightgray,font=\scriptsize,fill=white] {0};
    \node at (0,1)[lightgray,font=\scriptsize,fill=white] {1};
    \node at (0,2)[lightgray,font=\scriptsize,fill=white] {2};
    \node at (0,3)[lightgray,font=\scriptsize,fill=white] {3};
    \node at (1,3.9)[lightgray,font=\scriptsize,fill=white] {0};
    \node at (2,3.9)[lightgray,font=\scriptsize,fill=white] {1};
    \node at (3,3.9)[lightgray,font=\scriptsize,fill=white] {2};
    \node at (4,3.9)[lightgray,font=\scriptsize,fill=white] {3};
    \node at (5,3.9)[lightgray,font=\scriptsize,fill=white] {4};
    \node at (-1,3.9)[lightgray,font=\scriptsize,fill=white] {0};
    \node at (-2,3.9)[lightgray,font=\scriptsize,fill=white] {1};
    \node at (-3,3.9)[lightgray,font=\scriptsize,fill=white] {2};
    \node at (-4,3.9)[lightgray,font=\scriptsize,fill=white] {3};
    \node at (-5,3.9)[lightgray,font=\scriptsize,fill=white] {4};
    \node at (1,0)[s-eve-small,lose] (e00) {};
    \node at (1,1)[s-adam-small,win](a01){};
    \node at (1,2)[s-eve-small,lose] (e02){};
    \node at (1,3)[s-adam-small,win](a03){};
    \node at (2,0)[s-adam-small,lose](a10){};
    \node at (2,1)[s-eve-small,win] (e11){};
    \node at (2,2)[s-adam-small,lose](a12){};
    \node at (2,3)[s-eve-small,win] (e13){};
    \node at (3,0)[s-eve-small,lose] (e20){};
    \node at (3,1)[s-adam-small,win](a21){};
    \node at (3,2)[s-eve-small,win] (e22){};
    \node at (3,3)[s-adam-small,win](a23){};
    \node at (4,0)[s-adam-small,lose](a30){};
    \node at (4,1)[s-eve-small,win] (e31){};
    \node at (4,2)[s-adam-small,win](a32){};
    \node at (4,3)[s-eve-small,win] (e33){};
    \node at (5,0)[s-eve-small,lose] (e40){};
    \node at (5,1)[s-adam-small,win](a41){};
    \node at (5,2)[s-eve-small,win] (e42){};
    \node at (5,3)[s-adam-small,win](a43){};
    \node at (-1,0)[s-adam-small,lose](a00){};
    \node at (-1,1)[s-eve-small,lose] (e01){};
    \node at (-1,2)[s-adam-small,win](a02){};
    \node at (-1,3)[s-eve-small,lose] (e03){};
    \node at (-2,0)[s-eve-small,lose] (e10){};
    \node at (-2,1)[s-adam-small,lose](a11){};
    \node at (-2,2)[s-eve-small,win] (e12){};
    \node at (-2,3)[s-adam-small,lose](a13){};
    \node at (-3,0)[s-adam-small,lose](a20){};
    \node at (-3,1)[s-eve-small,lose] (e21){};
    \node at (-3,2)[s-adam-small,win](a22){};
    \node at (-3,3)[s-eve-small,win] (e23){};
    \node at (-4,0)[s-eve-small,lose] (e30){};
    \node at (-4,1)[s-adam-small,lose](a31){};
    \node at (-4,2)[s-eve-small,win] (e32){};
    \node at (-4,3)[s-adam-small,win](a33){};
    \node at (-5,0)[s-adam-small,lose](a40){};
    \node at (-5,1)[s-eve-small,lose] (e41){};
    \node at (-5,2)[s-adam-small,win](a42){};
    \node at (-5,3)[s-eve-small,win] (e43){};
    \path[arrow] 
    (e11) edge (e00)
    (e22) edge (e11)
    (e31) edge (e20)
    (e32) edge (e21)
    (e21) edge (e10)
    (e12) edge (e01)
    (e23) edge (e12)
    (e33) edge (e22)
    (e13) edge (e02)
    (e43) edge (e32)
    (e42) edge (e31)
    (e41) edge (e30);
    \path[arrow] 
    (e11) edge (a01)
    (e20) edge (a10)
    (e22) edge (a12)
    (e31) edge (a21)
    (e32) edge (a22)
    (e21) edge (a11)
    (e12) edge (a02)
    (e30) edge (a20)
    (e10) edge (a00)
    (e13) edge (a03)
    (e23) edge (a13)
    (e33) edge (a23)
    (e43) edge (a33)
    (e42) edge (a32)
    (e41) edge (a31)
    (e40) edge (a30);
    \path[arrow] 
    (a11) edge (e01)
    (a20) edge (e10)
    (a22) edge (e12)
    (a31) edge (e21)
    (a32) edge (e22)
    (a21) edge (e11)
    (a12) edge (e02)
    (a30) edge (e20)
    (a10) edge (e00)
    (a33) edge (e23)
    (a23) edge (e13)
    (a13) edge (e03)
    (a43) edge (e33)
    (a42) edge (e32)
    (a41) edge (e31)
    (a40) edge (e30);
    \path[arrow] 
    (a01) edge (e22)
    (a10) edge (e31)
    (a11) edge (e32)
    (a00) edge (e21)
    (a02) edge (e23)
    (a12) edge (e33)
    (a22) edge (e43)
    (a21) edge (e42)
    (a20) edge (e41);
    \path[arrow] 
    (-5.5,3.5) edge (e43)
    (5.5,2.5) edge (e42)
    (2.5,3.5) edge (e13)
    (5.5,0.5) edge (e40)
    (-5.5,1.5) edge (e41)
    (-3.5,3.5) edge (e23)
    (-1.5,3.5) edge (e03)
    (4.5,3.5) edge (e33)
    (5.5,0) edge (e40)
    (5.5,2) edge (e42)
    (-5.5,1) edge (e41)
    (-5.5,3) edge (e43);
    \path[dotted]
    (-5.7,3.7) edge (-5.5,3.5)
    (5.7,2.7) edge (5.5,2.5)
    (2.7,3.7) edge (2.5,3.5)
    (5.7,0.7) edge (5.5,0.5)
    (-3.7,3.7) edge (-3.5,3.5)
    (-1.7,3.7) edge (-1.5,3.5)
    (4.7,3.7) edge (4.5,3.5)
    (-5.7,1.7) edge (-5.5,1.5)
    (5.75,0) edge (5.5,0)
    (5.75,2) edge (5.5,2)
    (-5.75,1) edge (-5.5,1)
    (-5.75,3) edge (-5.5,3);
    \path[arrow]
    (5.5,1) edge (a41)
    (-5.5,2) edge (a42)
    (-5.5,0) edge (a40)
    (5.5,3) edge (a43);
    \path[dotted]
    (5.75,1) edge (5.5,1)
    (-5.75,2) edge (-5.5,2)
    (-5.75,0) edge (-5.5,0)
    (5.75,3) edge (5.5,3);
    \path[-]
    (a30) edge (5.5,.75)
    (a32) edge (5.5,2.75)
    (a31) edge (-5.5,1.75)
    (a23) edge (4,3.5)
    (a03) edge (2,3.5)
    (a13) edge (-3,3.5)
    (a33) edge (-5,3.5)
    (a43) edge (5.5,3.25)
    (a41) edge (5.5,1.25)
    (a40) edge (-5.5,0.25)
    (a42) edge (-5.5,2.25);
    \path[dotted]
    (5.5,.75) edge (5.8,.9)
    (5.5,2.75) edge (5.8,2.9)
    (-5.5,1.75) edge (-5.8,1.9)
    (4,3.5) edge (4.4,3.7)
    (2,3.5) edge (2.4,3.7)
    (-3,3.5) edge (-3.4,3.7)
    (-5,3.5) edge (-5.4,3.7)
    (5.5,3.25) edge (5.8,3.4)
    (5.5,1.25) edge (5.8,1.4)
    (-5.5,.25) edge (-5.8,0.4)
    (-5.5,2.25) edge (-5.8,2.4);
    \path[arrow]
    (sink) edge[loop left] ()
    (e00) edge[bend left=8] (sink)
    (e01) edge[bend right=8] (sink)
    (e02) edge[bend left=8] (sink)
    (e03) edge[bend right=8] (sink);
  \end{tikzpicture}}
  \caption{The "winning regions" of Eve in the
    "configuration reachability" game (left) and the "coverability" game
    (right) on the graphs of \Cref{13-fig:mwg,13-fig:sem} with target
    configuration~$\ell(2,2)$.  The winning vertices are in filled in
    green, while the losing ones are filled with white with a red
    border; the "sink" is always losing.}
    \label{13-fig:cov}
\commentAlt{Figure~\ref{13-fig:cov}: Two similar network diagrams, each with multiple layers of alternating circular and square nodes, showing dense interconnections, and two central vertical lines converging at a top point labeled with a symbol, with one diagram showing some nodes highlighted.}
\commentLongAlt{Figure~\ref{13-fig:cov}: The image displays two side-by-side diagrams, both depicting complex, symmetrical network structures. Each network is composed of multiple layers of alternating circular and square nodes, arranged in a grid-like fashion. Vertical and horizontal lines form the grid, with arrows indicating directed connections between nodes.

In both diagrams:
- Nodes in one layer are connected to nodes in adjacent layers, forming a dense, crisscrossing pattern.
- There's a central vertical axis where lines from the innermost nodes converge upwards to a single point labeled with an upward arrow and a perpendicular symbol.

The left diagram features nodes outlined in red, distinguishing them from the filled grey nodes. The right diagram shows the same network structure, but with a different set of nodes highlighted in red, suggesting a change in state or path.}
\end{figure}

In the "coverability" "vector game", Eve's "winning region" is
$\WE=\{\loc(m+2,n+2),\loc'(m+2,n+2),\loc'(0,n+1),\loc(1,n+2),\loc'(2m+2,1),\loc(2m+3,1)\mid
m,n\in\+N\}$
displayed on the right in \Cref{13-fig:cov}.  Observe in particular
that Adam is forced to use the action $\ell'\step{2,1}\ell$ from
the configurations of the form $\loc'(0,n+1)$, which brings him to a
configuration $\ell(2,n+2)$ coloured~$\Win$ in the game, and thus the
configurations of the form $\loc(1,n+1)$ are also winning for Eve 
since she can play $\loc\step{-1,0}\loc'$.  Thus the configurations of
the form $\loc(2m+3,n+1)$ are also winning for Eve, as she can play
the action $\loc\step{-1,0}\loc'$ to a winning configuration
$\loc'(2m+2,n+1)$ where all the actions available to Adam go into
her winning region.
\end{example}

\begin{remark}[Location reachability]
\label{13-rmk:cov2cov}
  One can notice that "coverability" is equivalent to \emph{location
  reachability}, where we are given a target location~$\loc_f$ but no
  target vector, and want to know whether Eve have a strategy to
  reach $\loc_f(\vec v)$ for some~$\vec v$.

  Indeed, in both "configuration reachability" and "coverability", we
  can assume without loss of generality that $\loc_f\in\Loc_\mEve$ is
  controlled by Eve and that $\vec v_f=\vec 0$ is the "zero
  vector". Here is a $\LOGSPACE$ reduction to that case.  If
  $\loc_0(\vec v_0)=\loc_f(\vec v_f)$ in the case of "configuration
  reachability", or $\loc_0=\loc_f$ and $\vec v_0\geq\vec v_f$ in the
  case of "coverability", the problem is trivial.
  Otherwise, any winning play must use at least one action.  For
  each incoming action $a=(\loc\step{\vec u}\loc_f)$ of~$\loc_f$,
  create a new location~$\loc_a$ controlled by Eve and replace~$a$ by
  $\loc\step{\vec u}\loc_a\step{\vec 0}\loc_f$, so that Eve gains the
  control right before any play reaches~$\loc_f$.  Also add a new
  location~$\smiley$ controlled by Eve with actions
  $\loc_a\step{-\vec v_f}\smiley$, and use $\smiley(\vec 0)$ as target
  configuration.
\end{remark}

\begin{remark}[Coverability to reachability]
\label{13-rmk:cov2reach}
  There is a $\LOGSPACE$ reduction from "coverability" to
  "configuration reachability".  By \Cref{13-rmk:cov2cov}, we can assume
  without loss of generality that $\loc_f\in\Loc_\mEve$ is controlled
  by Eve and that $\vec v_f=\vec 0$ is the "zero vector". It suffices
  therefore to add an action $\loc_f\step{-\vec e_j}\loc_f$ for
  all $1\leq j\leq\dd$.
\end{remark}

Departing from "reachability" games, the
following is a very simple kind of "safety"
games where $C\eqdef\{\varepsilon,\Lose\}$ and $\Omega\eqdef\Safe$;
\Cref{13-fig:nonterm} shows Eve's "winning region" in the case of the
graphs of \Cref{13-fig:mwg,13-fig:sem}.

\decpb["non-termination vector game" with "given initial credit"]%
{\label{13-pb:nonterm} A "vector system"
  $\?V=(\Loc,\Act,\Loc_\mEve,\Loc_\mAdam,\dd)$, an initial location
  $\loc_0\in\Loc$, and an initial credit $\vec v_0\in\+N^\dd$.}%
{Does Eve have a strategy to avoid the "sink"~$\sink$ from
  $\loc_0(\vec v_0)$?
  That is, does she win the non-termination
  game $(\natural(\?V),\col,\Safe)$ from $\loc_0(\vec v_0)$, where
  $\col(e)=\Lose$ if and only if $\ing(e)=\sink$?} 

  \begin{figure}[bhtp]
  \centering\scalebox{.48}{
  \begin{tikzpicture}[auto,on grid,node distance=2.5cm]
    \draw[step=1,lightgray!50,dotted] (-5.7,0) grid (5.7,3.8);
    \draw[color=white](0,-.3) -- (0,3.8);
    \node at (0,3.9) (sink) {\color{red!70!black}\boldmath$\sink$};
    \draw[step=1,lightgray!50] (1,0) grid (5.5,3.5);
    \draw[step=1,lightgray!50] (-1,0) grid (-5.5,3.5);
    \node at (0,0)[lightgray,font=\scriptsize,fill=white] {0};
    \node at (0,1)[lightgray,font=\scriptsize,fill=white] {1};
    \node at (0,2)[lightgray,font=\scriptsize,fill=white] {2};
    \node at (0,3)[lightgray,font=\scriptsize,fill=white] {3};
    \node at (1,3.9)[lightgray,font=\scriptsize,fill=white] {0};
    \node at (2,3.9)[lightgray,font=\scriptsize,fill=white] {1};
    \node at (3,3.9)[lightgray,font=\scriptsize,fill=white] {2};
    \node at (4,3.9)[lightgray,font=\scriptsize,fill=white] {3};
    \node at (5,3.9)[lightgray,font=\scriptsize,fill=white] {4};
    \node at (-1,3.9)[lightgray,font=\scriptsize,fill=white] {0};
    \node at (-2,3.9)[lightgray,font=\scriptsize,fill=white] {1};
    \node at (-3,3.9)[lightgray,font=\scriptsize,fill=white] {2};
    \node at (-4,3.9)[lightgray,font=\scriptsize,fill=white] {3};
    \node at (-5,3.9)[lightgray,font=\scriptsize,fill=white] {4};
    \node at (1,0)[s-eve-small,lose] (e00) {};
    \node at (1,1)[s-adam-small,win](a01){};
    \node at (1,2)[s-eve-small,lose] (e02){};
    \node at (1,3)[s-adam-small,win](a03){};
    \node at (2,0)[s-adam-small,lose](a10){};
    \node at (2,1)[s-eve-small,win] (e11){};
    \node at (2,2)[s-adam-small,lose](a12){};
    \node at (2,3)[s-eve-small,win] (e13){};
    \node at (3,0)[s-eve-small,lose] (e20){};
    \node at (3,1)[s-adam-small,win](a21){};
    \node at (3,2)[s-eve-small,win] (e22){};
    \node at (3,3)[s-adam-small,win](a23){};
    \node at (4,0)[s-adam-small,lose](a30){};
    \node at (4,1)[s-eve-small,win] (e31){};
    \node at (4,2)[s-adam-small,win](a32){};
    \node at (4,3)[s-eve-small,win] (e33){};
    \node at (5,0)[s-eve-small,lose] (e40){};
    \node at (5,1)[s-adam-small,win](a41){};
    \node at (5,2)[s-eve-small,win] (e42){};
    \node at (5,3)[s-adam-small,win](a43){};
    \node at (-1,0)[s-adam-small,win](a00){};
    \node at (-1,1)[s-eve-small,lose] (e01){};
    \node at (-1,2)[s-adam-small,win](a02){};
    \node at (-1,3)[s-eve-small,lose] (e03){};
    \node at (-2,0)[s-eve-small,win] (e10){};
    \node at (-2,1)[s-adam-small,lose](a11){};
    \node at (-2,2)[s-eve-small,win] (e12){};
    \node at (-2,3)[s-adam-small,lose](a13){};
    \node at (-3,0)[s-adam-small,win](a20){};
    \node at (-3,1)[s-eve-small,win] (e21){};
    \node at (-3,2)[s-adam-small,win](a22){};
    \node at (-3,3)[s-eve-small,win] (e23){};
    \node at (-4,0)[s-eve-small,win] (e30){};
    \node at (-4,1)[s-adam-small,win](a31){};
    \node at (-4,2)[s-eve-small,win] (e32){};
    \node at (-4,3)[s-adam-small,win](a33){};
    \node at (-5,0)[s-adam-small,win](a40){};
    \node at (-5,1)[s-eve-small,win] (e41){};
    \node at (-5,2)[s-adam-small,win](a42){};
    \node at (-5,3)[s-eve-small,win] (e43){};
    \path[arrow] 
    (e11) edge (e00)
    (e22) edge (e11)
    (e31) edge (e20)
    (e32) edge (e21)
    (e21) edge (e10)
    (e12) edge (e01)
    (e23) edge (e12)
    (e33) edge (e22)
    (e13) edge (e02)
    (e43) edge (e32)
    (e42) edge (e31)
    (e41) edge (e30);
    \path[arrow] 
    (e11) edge (a01)
    (e20) edge (a10)
    (e22) edge (a12)
    (e31) edge (a21)
    (e32) edge (a22)
    (e21) edge (a11)
    (e12) edge (a02)
    (e30) edge (a20)
    (e10) edge (a00)
    (e13) edge (a03)
    (e23) edge (a13)
    (e33) edge (a23)
    (e43) edge (a33)
    (e42) edge (a32)
    (e41) edge (a31)
    (e40) edge (a30);
    \path[arrow] 
    (a11) edge (e01)
    (a20) edge (e10)
    (a22) edge (e12)
    (a31) edge (e21)
    (a32) edge (e22)
    (a21) edge (e11)
    (a12) edge (e02)
    (a30) edge (e20)
    (a10) edge (e00)
    (a33) edge (e23)
    (a23) edge (e13)
    (a13) edge (e03)
    (a43) edge (e33)
    (a42) edge (e32)
    (a41) edge (e31)
    (a40) edge (e30);
    \path[arrow] 
    (a01) edge (e22)
    (a10) edge (e31)
    (a11) edge (e32)
    (a00) edge (e21)
    (a02) edge (e23)
    (a12) edge (e33)
    (a22) edge (e43)
    (a21) edge (e42)
    (a20) edge (e41);
    \path[arrow] 
    (-5.5,3.5) edge (e43)
    (5.5,2.5) edge (e42)
    (2.5,3.5) edge (e13)
    (5.5,0.5) edge (e40)
    (-5.5,1.5) edge (e41)
    (-3.5,3.5) edge (e23)
    (-1.5,3.5) edge (e03)
    (4.5,3.5) edge (e33)
    (5.5,0) edge (e40)
    (5.5,2) edge (e42)
    (-5.5,1) edge (e41)
    (-5.5,3) edge (e43);
    \path[dotted]
    (-5.7,3.7) edge (-5.5,3.5)
    (5.7,2.7) edge (5.5,2.5)
    (2.7,3.7) edge (2.5,3.5)
    (5.7,0.7) edge (5.5,0.5)
    (-3.7,3.7) edge (-3.5,3.5)
    (-1.7,3.7) edge (-1.5,3.5)
    (4.7,3.7) edge (4.5,3.5)
    (-5.7,1.7) edge (-5.5,1.5)
    (5.75,0) edge (5.5,0)
    (5.75,2) edge (5.5,2)
    (-5.75,1) edge (-5.5,1)
    (-5.75,3) edge (-5.5,3);
    \path[arrow]
    (5.5,1) edge (a41)
    (-5.5,2) edge (a42)
    (-5.5,0) edge (a40)
    (5.5,3) edge (a43);
    \path[dotted]
    (5.75,1) edge (5.5,1)
    (-5.75,2) edge (-5.5,2)
    (-5.75,0) edge (-5.5,0)
    (5.75,3) edge (5.5,3);
    \path[-]
    (a30) edge (5.5,.75)
    (a32) edge (5.5,2.75)
    (a31) edge (-5.5,1.75)
    (a23) edge (4,3.5)
    (a03) edge (2,3.5)
    (a13) edge (-3,3.5)
    (a33) edge (-5,3.5)
    (a43) edge (5.5,3.25)
    (a41) edge (5.5,1.25)
    (a40) edge (-5.5,0.25)
    (a42) edge (-5.5,2.25);
    \path[dotted]
    (5.5,.75) edge (5.8,.9)
    (5.5,2.75) edge (5.8,2.9)
    (-5.5,1.75) edge (-5.8,1.9)
    (4,3.5) edge (4.4,3.7)
    (2,3.5) edge (2.4,3.7)
    (-3,3.5) edge (-3.4,3.7)
    (-5,3.5) edge (-5.4,3.7)
    (5.5,3.25) edge (5.8,3.4)
    (5.5,1.25) edge (5.8,1.4)
    (-5.5,.25) edge (-5.8,0.4)
    (-5.5,2.25) edge (-5.8,2.4);
    \path[arrow]
    (sink) edge[loop left] ()
    (e00) edge[bend left=8] (sink)
    (e01) edge[bend right=8] (sink)
    (e02) edge[bend left=8] (sink)
    (e03) edge[bend right=8] (sink);
  \end{tikzpicture}}
  \caption{The "winning region" of Eve in the
    "non-termination" game on the graphs of \Cref{13-fig:mwg,13-fig:sem}.}
    \label{13-fig:nonterm} 
\commentAlt{Figure~\ref{13-fig:nonterm}: A complex network diagram with multiple layers of alternating circular and square nodes, showing dense interconnections, and two central vertical lines converging at a top point labeled with a symbol, with some nodes highlighted in red.}
\commentLongAlt{Figure~\ref{13-fig:nonterm}: The image displays a complex, symmetrical network structure composed of alternating rows of circular (some outlined in red) and square (some outlined in red) nodes, arranged in layers. Vertical and horizontal lines form a grid, with dashed lines indicating horizontal positions labeled '0' through '4' on both sides, and vertical layers. Arrows indicate directed connections between nodes.

The network is symmetrical around a central vertical axis, which connects upwards to a single point labeled with an upward arrow and a perpendicular symbol.

- Most nodes are filled grey. However, certain circular and square nodes are outlined in red, indicating a distinct path or set of activated nodes within the network. These red-outlined nodes are primarily visible on the right side of the diagram, forming a distinct sequence in the bottom layer, and isolated instances in higher layers near the center.
- Nodes in one layer are connected to nodes in adjacent layers by multiple directed arrows, forming a dense, crisscrossing pattern. Many arrows point from nodes in an upper layer to nodes in a lower layer, and from nodes to their immediate right or left.
- Dotted lines extend outwards from the outermost nodes, indicating that the network continues.}
\end{figure}


Finally, one of the most general "vector games" are "parity@parity
vector game" games, where $C\eqdef\{1,\dots,d\}$ and
$\Omega\eqdef\Parity$.  In order to define a colouring of the "natural
semantics", we assume that we are provided with a \emph{location
  colouring} $\lcol{:}\,\Loc\to\{1,\dots,d\}$.
\decpb["parity vector game" with "given initial credit"]%
{\label{13-pb:parity}A "vector system"
  $\?V=(\Loc,\Act,\Loc_\mEve,\Loc_\mAdam,\dd)$, an initial location
  $\loc_0\in\Loc$, an initial credit $\vec v_0\in\+N^\dd$, and a
  location colouring $\lcol{:}\,\Loc\to\{1,\dots,d\}$ for some $d>0$.}%
  {Does Eve have a strategy to simultaneously avoid the
  "sink"~$\sink$ and fulfil the \index{parity!\emph{see also} vector
  game\protect\mymoot|mymoot} parity objective from $\loc_0(\vec
  v_0)$? That is, does she win the parity@parity vector game game
  $(\natural(\?V),\col,\Parity)$ from $\loc_0(\vec v_0)$, where
  $\col(e)\eqdef\lcol(\loc)$ if $\ing(e)=\loc(\vec v)$ for
  some~$\vec v\in\+N^\dd$, and $\col(e)\eqdef 1$ if $\ing(e)=\sink$?}

\begin{remark}[Non termination to parity]
\label{13-rmk:nonterm2parity}
  There is a $\LOGSPACE$ reduction from "non-termination" to
  "parity@parity vector game".
  Indeed, the two games coincide if we pick the constant location
  "colouring" defined by $\lcol(\loc)\eqdef 2$ for all $\loc\in\Loc$ in
  the parity game.
\end{remark}
\begin{remark}[Coverability to parity]
\label{13-rmk:cov2parity}
  There is a $\LOGSPACE$ reduction from "coverability" to
  "parity@parity vector game".  Indeed, by \Cref{13-rmk:cov2cov}, we can assume
  that $\loc_f\in\Loc_\mEve$ is controlled by Eve and that the target
  credit is $\vec v_f=\vec 0$ the "zero vector".  It suffices
  therefore to add an action $\loc_f\step{\vec 0}\loc_f$ and to colour
  every location $\loc\neq\loc_f$ with $\lcol(\loc)\eqdef 1$ and
  to set $\lcol(\loc_f)\eqdef 2$.
\end{remark}

The "existential initial credit" variants of
\Crefrange{13-pb:reach}{13-pb:parity} are defined similarly, where
$\vec v_0$ is not given as part of the input, but existentially
quantified in the question.

\subsection{Undecidability}
\label{13-sec:undec}
The bad news is that, although \Crefrange{13-pb:reach}{13-pb:parity}
are all decidable in the one-player case---see
the bibliographic notes~\Cref{13-sec:references} at the end of the
chapter---, they become undecidable in the two-player setting.

\begin{theorem}[Undecidability of vector games]
\label{13-thm:undec}
  "Configuration reachability", "coverability", "non-termination", and
  "parity@parity vector game" "vector games", both with "given" and
  with "existential initial credit", are undecidable in any dimension
  $\dd\geq 2$.
\end{theorem}
\begin{proof}
  By \Cref{13-rmk:cov2reach,13-rmk:nonterm2parity}, it suffices to prove the
  undecidability of "coverability" and "non-termination".  For this,
  we exhibit reductions from the "halting problem" of "deterministic
  Minsky machines" with at least two counters.

  \AP Formally, a deterministic Minsky machine with $\dd$~counters
  $\?M=(\Loc,\Act,\dd)$ is defined similarly to a "vector addition
  system with states" with additional zero test actions
  $a=(\loc\step{i\eqby{?0}}\loc')$.  The set of locations contains a
  distinguished `halt' location~$\loc_\mathtt{halt}$, and for every
  $\loc\in\Loc$, exactly one of the following holds: either (i)
  $(\loc\step{\vec e_i}\loc')\in\Act$ for some $0<i\leq\dd$ and
  $\loc'\in\Loc$, or (ii) $(\loc\step{i\eqby{?0}}\loc')\in\Act$ and
  $(\loc\step{-\vec e_i}\loc'')\in\Act$ for some $0<i\leq\dd$ and
  $\loc',\loc''\in\Loc$, or (iii) $\loc=\loc_\mathtt{halt}$.  The
  semantics of~$\?M$ extends the "natural semantics" by
  handling "zero tests" actions $a=(\loc\step{i\eqby{?0}}\loc')$: we
  define the domain as $\dom a\eqdef\{\loc(\vec v)\mid \vec v(i)=0\}$
  and the image by $a(\loc(\vec v))\eqdef \loc(\vec v)$.  This
  semantics is deterministic, and from any starting vertex of $\natural(\?M)$,
  there is a unique "play", which either eventually visits
  $\loc_\mathtt{halt}$ and then the "sink" in the next step, or keeps
  avoiding both $\loc_\mathtt{halt}$ and the "sink"
  indefinitely. 

  \AP The halting problem asks, given a "deterministic Minsky machine"
  and an initial location $\loc_0$, whether it halts, that is, whether
  $\loc_\mathtt{halt}(\vec v)$ is reachable for
  some~$\vec v\in\+N^\dd$ starting from $\loc_0(\vec 0)$.  The
  "halting problem" is undecidable in any dimension
  $\dd\geq 2$~\cite{Minsky:1967}.
  Thus the halting problem is akin to the "coverability" of
  $\loc_\mathtt{halt}(\vec 0)$ with "given initial credit"~$\vec 0$,
  but on the one hand there is only one player and on the other hand
  the machine can perform "zero tests".

  \begin{figure}[htbp]
    \centering
    \begin{tikzpicture}[auto,on grid,node distance=1.5cm]
      \node(to){$\mapsto$};
      \node[anchor=east,left=2.5cm of to](mm){deterministic Minsky machine};
      \node[anchor=west,right=2.5cm of to](mwg){vector system};
      \node[below=.7cm of to](map){$\rightsquigarrow$};
      \node[left=2.75cm of map](0){$\loc$};
      \node[right=of 0](1){$\loc'$};
      \node[right=1.25cm of map,s-eve](2){$\loc$};
      \node[right=of 2,s-eve](3){$\loc'$};
      \node[below=1.5cm of map](map2){$\rightsquigarrow$};
      \node[left=2.75cm of map2](4){$\loc$};
      \node[below right=.5cm and 1.5cm of 4](5){$\loc''$};
      \node[above right=.5cm and 1.5cm of 4](6){$\loc'$};
      \node[right=1.25cm of map2,s-eve](7){$\loc$};
      \node[below right=.5cm and 1.5cm of 7,s-eve](8){$\loc''$};
      \node[above right=.5cm and 1.5cm of 7,s-adam,inner sep=-1.5pt](9){$\loc'_{i\eqby{?0}}$};
      \node[below right=.5cm and 1.5cm of 9,s-eve](10){$\loc'$};
      \node[above right=.5cm and 1.5cm of 9,s-adam](11){$\frownie$};
      \path[arrow,every node/.style={font=\scriptsize}]
      (0) edge node{$\vec e_i$} (1)
      (2) edge node{$\vec e_i$} (3)
      (4) edge[swap] node{$-\vec e_i$} (5)
      (4) edge node{$i\eqby{?0}$} (6)
      (7) edge[swap] node{$-\vec e_i$} (8)
      (7) edge node{$\vec 0$} (9)
      (9) edge[swap] node{$\vec 0$} (10)
      (9) edge node{$-\vec e_i$} (11);
    \end{tikzpicture}
    \caption{Schema of the reduction in the proof of \Cref{13-thm:undec}.}
    \label{13-fig:undec}
\commentAlt{Figure~\ref{13-fig:undec}: A diagram showing the transformation from two deterministic Minsky machine transitions to their equivalent vector system representations. See long description.}
\commentLongAlt{Figure~\ref{13-fig:undec}: The image illustrates the translation of two types of deterministic Minsky machine transitions into equivalent vector system representations, separated by a double-headed arrow symbol resembling a map.

**Top Row:**
- **Deterministic Minsky machine (left):** Shows a state 'L' transitioning to 'L'' with an arrow labeled 'e_i_vector'.
- **Vector system (right):** Shows a circular node 'L' transitioning to a circular node 'L'' with an arrow labeled 'e_i_vector'.

**Bottom Row:**
- **Deterministic Minsky machine (left):** Shows a state 'L' with two diverging transitions:
    - One arrow labeled 'e_i_tilde' pointing to 'L''.
    - Another arrow labeled '-e_i_vector' pointing to 'L'''.
- **Vector system (right):** Shows a more complex structure:
    - A circular node 'L'.
    - From 'L', an arrow labeled '0_vector' points to a square node labeled 'L''_i_equals_e_i_tilde'. From this square node, two arrows diverge: one to a happy face in a square (top right) and another to a circular node 'L''' (bottom right), both labeled '0_vector'.
    - From 'L', another arrow labeled '-e_i_vector' points to a circular node 'L''' (bottom left).}
      \end{figure}
  Here is now a reduction to
  \Cref{13-pb:cov}.  Given an instance of the "halting problem", \textit{i.e.},
  given a "deterministic Minsky machine" $\?M=(\Loc,\Act,\dd)$ and an
  initial location $\loc_0$, we construct a "vector system"
  $\?V\eqdef(\Loc\uplus\Loc_{\eqby{?0}}\uplus\{\frownie\},\Act',\Loc,\Loc_{\eqby{?0}}\uplus\{\frownie\},\dd)$
  where all the original locations are controlled by~Eve and
  $\Loc_{\eqby{?0}}\uplus\{\frownie\}$ is a set of new locations
  controlled by Adam.  We use $\Loc_{\eqby{?0}}$ as a set of
  locations defined by
  \begin{align*}
    \Loc_{\eqby{?0}}&\eqdef\{\loc'_{i\eqby{?0}}\mid\exists\loc\in\Loc\mathbin.(\loc\step{i\eqby{?0}}\loc')\in\Act\}\intertext{and
                   define the set of actions by (see \Cref{13-fig:undec})}
    \Act'&\eqdef\{\loc\step{\vec
          e_i}\loc'\mid(\loc\step{\vec e_i}\loc')\in\Act\}\cup\{\loc\step{-\vec e_i}\loc''\mid(\loc\step{-\vec e_i}\loc'')\in\Act\}\\
    &\:\cup\:\{\loc\step{\vec
      0}\loc'_{i\eqby{?0}},\;\;\:\loc'_{i\eqby{?0}}\!\!\step{\vec 0}\loc',\;\;\:\loc'_{i\eqby{?0}}\!\!\step{-\vec e_i}\frownie\mid(\loc\step{i\eqby{?0}}\loc')\in\Act\}\;.
  \end{align*}
  We use $\loc_0(\vec 0)$ as initial configuration and
  $\loc_\mathtt{halt}(\vec 0)$ as target configuration for the
  constructed "coverability" instance.  Here is the crux of the
  argument why Eve has a winning strategy to cover
  $\loc_\mathtt{halt}(\vec 0)$ from $\loc_0(\vec 0)$ if and only if
  the "Minsky machine@deterministic Minsky machine" halts.
  Consider any configuration $\loc(\vec v)$.  If
  $(\loc\step{\vec e_i}\loc')\in\Act$, Eve has no choice but to apply
  $\loc\step{\vec e_i}\loc'$ and go to the configuration
  $\loc'(\vec v+\vec e_i)$ also reached in one step in~$\?M$.  If
  $\{\loc\step{i\eqby{?0}}\loc',\loc\step{-\vec e_i}\loc''\}\in\Act$ and
  $\vec v(i)=0$, due to the "natural semantics", Eve cannot use the
  action $\loc\step{-\vec e_i}\loc''$, thus she must use
  $\loc\step{\vec 0}\loc'_{i\eqby{?0}}$.  Still due to the "natural
  semantics", Adam cannot use
  $\loc'_{i\eqby{?0}}\!\!\step{-\vec e_i}\frownie$, thus he must use
  $\loc'_{i\eqby{?0}}\!\!\step{\vec 0}\loc'$.  Hence Eve regains the
  control in $\loc'(\vec v)$, which was also the configuration reached
  in one step in~$\?M$.  Finally, if
  $\{\loc\step{i\eqby{?0}}\loc',\loc\step{-\vec e_i}\loc''\}\in\Act$ and
  $\vec v(i)>0$, Eve can choose: if she uses
  $\loc\step{-\vec e_i}\loc''$, she ends in the configuration
  $\loc''(\vec v-\vec e_i)$ also reached in one step in~$\?M$.  In
  fact, she should not use $\loc\step{\vec 0}\loc'_{i\eqby{?0}}$,
  because Adam would then have the opportunity to apply
  $\loc'_{i\eqby{?0}}\!\!\step{-\vec e_i}\frownie$ and to win, as
  $\frownie$ is a deadlock location and all the subsequent moves end
  in the "sink".  Thus, if $\?M$ halts, then Eve has a winning
  strategy that simply follows the unique "play" of~$\?M$, and
  conversely, if Eve wins, then necessarily she had to follow the
  "play" of~$\?M$ and thus the machine halts.
    
  \medskip Further note that, in a "deterministic Minsky machine" the
  "halting problem" is similarly akin to the \emph{complement} of
  "non-termination" with "given initial credit"~$\vec 0$.  This means
  that, in the "vector system"
  $\?V=(\Loc\uplus\Loc_{\eqby{?0}}\uplus\{\frownie\},\Act',\Loc,\Loc_{\eqby{?0}}\uplus\{\frownie\},\dd)$
  defined earlier, Eve has a winning strategy to avoid the "sink"
  from~$\loc_0(\vec 0)$ if and only if the given "Minsky
  machine@deterministic Minsky machine" does
  not halt from~$\loc_0(\vec 0)$, which shows the undecidability of
  \Cref{13-pb:nonterm}.

  \medskip Finally, let us observe that both the "existential" and the
  universal initial credit variants of the "halting problem" are also
  undecidable.  Indeed, given an instance of the "halting problem",
  \textit{i.e.}, given a "deterministic Minsky machine" $\?M=(\Loc,\Act,\dd)$
  and an initial location $\loc_0$, we add $\dd$~new locations
  $\loc_\dd,\loc_{\dd-1},\dots,\loc_1$ with respective actions
  $\loc_j\step{-\vec e_j}\loc_j$ and $\loc_j\step{j\eqby{?0}}\loc_{j-1}$
  for all $\dd\geq j>0$.  This modified machine first resets all its
  counters to zero before reaching $\loc_0(\vec 0)$ and then performs
  the same execution as the original machine.  Thus there exists an
  initial credit~$\vec v$ such that the modified machine
  reaches~$\loc_\mathtt{halt}$ from $\loc_\dd(\vec v)$ if and only if
  for all initial credits~$\vec v$ the modified machine
  reaches~$\loc_\mathtt{halt}$ from $\loc_\dd(\vec v)$, if and only if
  $\loc_\mathtt{halt}$ was reachable from~$\loc_0(\vec 0)$ in the
  original machine.  The previous construction of a "vector system"
  applied to the modified machine then shows the undecidability of the
  "existential initial credit" variants of
  \Cref{13-pb:cov,13-pb:nonterm}
  .
\end{proof}


\section{Games in dimension one}
\label{13-sec:dim1}
\AP \Cref{13-thm:undec} leaves open whether "vector games" might be
decidable in dimension one.  They are indeed decidable, and more
generally we learned in \Cref{12-chap:pushdown} that "one-counter
games"---with the additional ability to test the counter for
zero---were decidable and in fact $\PSPACE$-complete.  This might seem
to settle the case of "vector games" in dimension one, except that the
"one-counter games" of \Cref{12-chap:pushdown} only allow integer
weights in $\{-1,1\}$, whereas we allow arbitrary updates in~$\+Z$
with a binary encoding.  Hence the $\PSPACE$  upper bound of
\Cref{12-chap:pushdown} becomes an~$\EXPSPACE$ one for succinct
one-counter games.

\begin{corollary}[One-dimensional vector games are in $\EXPSPACE$]
\label{13-cor:dim1}
  "Configuration reachability", "coverability", "non-termination", and
  "parity@parity vector game" "vector games", both with "given" and with "existential
  initial credit", are in $\EXPSPACE$  in dimension one.
\end{corollary}

The goal of this section is therefore to establish that this
$\EXPSPACE$ upper bound is tight (in most cases), by proving a matching
lower bound in \Cref{13-sec:one-counter}.  But first, we will study a
class of one-dimensional "vector games" of independent interest in
\Cref{13-sec:countdown}: "countdown games".

\subsection{Countdown games}
\label{13-sec:countdown}

\AP A one-dimensional "vector system"
$\?V=(\Loc,\Act,\Loc_\mEve,\Loc_\mAdam,1)$ is called a countdown
system if $\Act\subseteq\Loc\times\+Z_{<0}\times\Loc$, that is, if
for all $(\loc\step{z}\loc')\in\Act$, $z<0$.  We consider the games
defined by "countdown systems", both with "given" and with
"existential initial credit", and call the resulting games countdown
games.

\begin{theorem}[Countdown games are $\EXP$-complete]
\label{13-thm:countdown-given}
  "Configuration reachability" and "coverability" "countdown games"
  with "given initial credit" are $\EXP$-complete.
\end{theorem}
\begin{proof}
  For the upper bound, consider an instance, \textit{i.e.}, a "countdown
  system" \[
  \?V=(\Loc,\Act,\Loc_\mEve,\Loc_\mAdam,1),
  \]
  an initial location $\loc_0\in\Loc$, an initial credit $n_0\in\+N$, and a
  target configuration $\loc_f(n_f)\in\Loc\times\+N$.  Because every
  action decreases strictly the counter value, the reachable part
  of the "natural semantics" of $\?V$ starting from $\loc_0(n_0)$ is
  finite and of size at most $1+|\Loc|\cdot (n_0+1)$, and because~$n_0$
  is encoded in binary, this is at most exponential in the size of the
  instance.  As seen in \Cref{3-chap:regular}, such a "reachability"
  game can be solved in time polynomial in the size of the finite
  graph, thus in $\EXP$\ overall.

  \medskip For the lower bound, we start by considering a game played
  over an exponential-time Turing machine, before showing how to
  implement this game as a "countdown game".  Let us consider for this
  an arbitrary Turing machine~$\?M$ working in deterministic
  exponential time~$2^{p(n)}$ for some fixed polynomial~$p$ and an
  input word~$w=a_1\cdots a_n$ of length~$n$, which we assume to be
  positive.  Let $m\eqdef 2^{p(n)}\geq n$.  The computation of~$\?M$
  on~$w$ is a sequence of configurations $C_1,C_2,\dots,C_t$ of
  length~$t\leq m$.  Each configuration $C_i$ is of the form
  $\emkl \gamma_{i,1}\cdots\gamma_{i,m}\emkr$ where $\emkl$ and
  $\emkr$ are endmarkers and the symbols $\gamma_{i,j}$ are either
  taken from the finite tape alphabet~$\Gamma$ (which includes a blank
  symbol~$\blank$) or a pair $(q,a)$ of a state from~$Q$ and a tape
  symbol~$a$.  We assume that the set of states~$Q$ contains a single
  accepting state~$q_\mathrm{final}$.  The entire computation can be
  arranged over a $t\times m$ grid where each line corresponds to a
  configuration~$C_i$, as shown in \Cref{13-fig:exp}.

  \begin{figure}[htbp]
    \centering
    \hspace*{-.5ex}\begin{tikzpicture}[on grid,every node/.style={anchor=base}]
      \draw[step=1,lightgray!50,dotted] (-.5,-0.8) grid (10.5,-5.2);
      \node[anchor=east] at (-.5,-5) {$C_1$};
      \node[anchor=east] at (-.5,-4) {$C_2$};
      \node[anchor=east] at (-.5,-3.4) {$\vdots~$};
      \node[anchor=east] at (-.5,-3) {$C_{i-1}$};
      \node[anchor=east] at (-.5,-2) {$C_i$};
      \node[anchor=east] at (-.5,-1.4) {$\vdots~$};
      \node[anchor=east] at (-.5,-1) {$C_t$};
      \draw[color=white](4,-.5) -- (4,-5.2) (8,-.5) -- (8,-5.2);
      \node[lightgray] at (0,-.5) {$0$};
      \node[lightgray] at (1,-.5) {$1$};
      \node[lightgray] at (2,-.5) {$2$};
      \node[lightgray] at (3,-.5) {$3$};
      \node[lightgray] at (4,-.5) {$\cdots$};
      \node[lightgray] at (5,-.5) {$j-1$};
      \node[lightgray] at (6,-.5) {$j$};
      \node[lightgray] at (7,-.5) {$j+1$};
      \node[lightgray] at (8,-.5) {$\cdots$};
      \node[lightgray] at (9,-.5) {$m$};
      \node[lightgray] at (10,-.5) {$m+1$};
      \node at (0,-1.1) {$\emkl$};
      \node at (0,-2.1) {$\emkl$};
      \node at (0,-3.1) {$\emkl$};
      \node at (0,-4.1) {$\emkl$};
      \node at (0,-5.1) {$\emkl$};
      \node at (10,-1.1) {$\emkr$};
      \node at (10,-2.1) {$\emkr$};
      \node at (10,-3.1) {$\emkr$};
      \node at (10,-4.1) {$\emkr$};
      \node at (10,-5.1) {$\emkr$};
      \node at (1,-5.1) {$q_0,a_1$};
      \node at (2,-5.1) {$a_2$};
      \node at (3,-5.1) {$a_3$};
      \node at (4,-5.1) {$\cdots$};
      \node at (5,-5.1) {$\blank$};
      \node at (6,-5.1) {$\blank$};
      \node at (7,-5.1) {$\blank$};
      \node at (8,-5.1) {$\cdots$};
      \node at (9,-5.1) {$\blank$};
      \node at (1,-4.1) {$a'_1$};
      \node at (2,-4.1) {$q_1,a_2$};
      \node at (3,-4.1) {$a_3$};
      \node at (4,-4.1) {$\cdots$};
      \node at (5,-4.1) {$\blank$};
      \node at (6,-4.1) {$\blank$};
      \node at (7,-4.1) {$\blank$};
      \node at (8,-4.1) {$\cdots$};
      \node at (9,-4.1) {$\blank$};
      \node at (5,-3.7) {$\vdots$};
      \node at (6,-3.7) {$\vdots$};
      \node at (7,-3.7) {$\vdots$};
      \node at (4,-3.1) {$\cdots$};
      \node at (5,-3.1) {$\gamma_{i-1,j-1}$};
      \node at (6,-3.1) {$\gamma_{i-1,j}$};
      \node at (7,-3.1) {$\gamma_{i-1,j+1}$};
      \node at (8,-3.1) {$\cdots$};
      \node at (5,-2.1) {$\cdots$};
      \node at (6,-2.1) {$\gamma_{i,j}$};
      \node at (7,-2.1) {$\cdots$};
      \node at (6,-1.7) {$\vdots$};
      \node at (1,-1.1) {$q_\mathrm{final},\blank$};
      \node at (2,-1.1) {$\blank$};
      \node at (3,-1.1) {$\blank$};
      \node at (4,-1.1) {$\cdots$};
      \node at (5,-1.1) {$\blank$};
      \node at (6,-1.1) {$\blank$};
      \node at (7,-1.1) {$\blank$};
      \node at (8,-1.1) {$\cdots$};
      \node at (9,-1.1) {$\blank$};      
      \end{tikzpicture}
    \caption{The computation of~$\?M$ on
  input~$w=a_1\cdots a_n$.  This particular picture assumes~$\?M$
  starts by rewriting~$a_1$ into $a'_1$ and moving to the right in a
  state~$q_1$, and empties its tape before accepting its input by going
  to state~$q_\mathrm{final}$.}
  \label{13-fig:exp}
\commentAlt{Figure~\ref{13-fig:exp}: A grid-like diagram with labeled rows (C1 to C_f) and columns (0 to m+1), displaying various symbols and variables, including arrows and ellipses.}
\commentLongAlt{Figure~\ref{13-fig:exp}: The image displays a grid structure with rows labeled C_f, ..., C_i, C_i-1, ..., C2, C1 on the left side, and columns labeled 0, 1, 2, 3, ..., j-1, j, j+1, ..., m, m+1 at the top. Most of the labels at the top and bottom rows/columns appear in a lighter shade, suggesting they might be implicit or contextual rather than primary data points.

The grid contains various symbols, numbers, and variables:
- In row C_f, at column 1, is 'q_final,b'. Other cells in this row and column j, j+1 contain 'b' or are empty. The last column (m+1) for C_f shows a right-pointing triangle.
- In row C_i, at column j, is 'gamma_i,j'.
- In row C_i-1, at columns j-1, j, j+1, are 'gamma_i-1,j-1', 'gamma_i-1,j', 'gamma_i-1,j+1' respectively.
- In row C2, at column 1, is 'a_1'', at column 2 is 'a_1, a_2', and at column 3 is 'a_3'.
- In row C1, at column 1, is 'q_0, a_1', at column 2 is 'a_2', and at column 3 is 'a_3'.

Many cells contain ellipses '...' or triangular symbols pointing right, indicating continuation of patterns or sequences. Vertical dotted lines connect related cells in different rows, particularly around column j.}
  \end{figure}

  We now set up a two-player game where Eve wants to prove that the
  input~$w$ is accepted.  Let
  $\Gamma'\eqdef \{\emkl,\emkr\}\cup\Gamma\cup(Q\times\Gamma)$. Rather
  than exhibiting the full computation from \Cref{13-fig:exp}, the
  game will be played over positions $(i,j,\gamma_{i,j})$ where
  $0<i\leq m$, $0\leq j\leq m+1$, and $\gamma_{i,j}\in\Gamma'$.  Eve 
  wants to show that, in the computation of~$\?M$ over~$w$ as depicted
  in \Cref{13-fig:exp}, the $j$th cell of the $i$th
  configuration~$C_i$ contains~$\gamma_{i,j}$.  In order to
  substantiate this claim, observe that the content of any cell
  $\gamma_{i,j}$ in the grid is determined by the actions of~$\?M$
  and the contents of (up to) three cells in the previous
  configuration.  Thus, if $i>1$ and $0<j<m+1$, Eve provides a triple
  $(\gamma_{i-1,j-1},\gamma_{i-1,j},\gamma_{i-1,j+1})$ of symbols
  in~$\Gamma'$ that yield $\gamma_{i,j}$ according to the actions
  of~$\?M$, which we denote by
  $\gamma_{i-1,j-1},\gamma_{i-1,j},\gamma_{i-1,j+1}\vdash\gamma_{i,j}$,
  and Adam chooses $j'\in\{j-1,j,j+1\}$ and returns the control
  to Eve in position~$(i-1,j',\gamma_{i-1,j'})$.  Regarding the
  boundary cases where $i=0$ or $j=0$ or $j=m+1$, Eve wins
  immediately if $j=0$ and $\gamma={\emkl}$, or if $j=m+1$ and
  $\gamma={\emkr}$, or if $i=0$ and $0<j\leq n$ and $\gamma=a_j$, or if
  $i=0$ and $n<j\leq m$ and $\gamma={\blank}$, and otherwise Adam wins
  immediately.  The game starts in a position
  $(t,j,(q_\mathrm{final},a))$ for some $0<t\leq m$, $0< j\leq m$,
  and~$a\in\Gamma$ of Eve's choosing.  It should be clear that Eve 
  has a winning strategy in this game if and only if~$w$ is accepted
  by~$\?M$.

  We now implement the previous game as a "coverability" game over a
  "countdown system" $\?V\eqdef(\Loc,\Act,\Loc_\mEve,\Loc_\mAdam,1)$.
  The idea is that the pair $(i,j)$ will be encoded as
  $(i-1)\cdot(m+2)+j+2$ in the counter value, while the
  symbol~$\gamma_{i,j}$ will be encoded in the location.  For
  instance, the endmarker $\emkl$ at position $(1,0)$ will be
  represented by configuration $\loc_{\emkl}(2)$, the first input
  $(q_0,a_1)$ at position~$(1,1)$ by $\loc_{(q_0,a_1)}(3)$, and the
  endmarker $\emkr$ at position $(m,m+1)$ by
  $\loc_{\emkr}(m\cdot(m+2)+1)$. The game starts from the initial
  configuration $\loc_0(n_0)$ where $n_0\eqdef m\cdot(m+2)+1$ and the
  target location is~$\smiley$.

  We define for this the sets of locations
  \begin{align*}
    \Loc_\mEve&\eqdef\{\loc_0,\smiley,\frownie\}
               \cup\{\loc_\gamma\mid\gamma\in\Gamma'\}\;,\\
    \Loc_\mAdam&\eqdef\{\loc_{(\gamma_1,\gamma_2,\gamma_3)}\mid\gamma_1,\gamma_2,\gamma_3\in\Gamma'\}
               \cup\{\loc_{=j}\mid 0<j\leq n\}
               \cup\{\loc_{1\leq?\leq m-n+1}\}\;.
  \end{align*}
  The intention behind the locations $\loc_{=j}\in\Loc_\mAdam$ is
  that Eve can reach~$\smiley$ from a configuration $\loc_{=j}(c)$ if
  and only if $c=j$; we accordingly define~$\Act$ with the following
  actions, where~$\frownie$ is a deadlock location:
  \begin{align*}
    \loc_{=j}&\step{-j-1}\frownie\;,&\loc_{=j}&\step{-j}\smiley\;.
  \intertext{Similarly, Eve should be able to reach~$\smiley$ from
  $\loc_{1\leq?\leq m-n+1}(c)$ if and only if $1\leq c\leq m-n+1$,
  which is implemented by the actions}
    \loc_{1\leq?\leq m-n+1}&\step{-m+n-2}\frownie\;,&
    \loc_{1\leq?\leq m-n+1}&\step{-1}\smiley\;,&
    \smiley&\step{-1}\smiley\;.
  \end{align*}
  Note this last action also ensures that Eve can reach the
  location~$\smiley$ if and only if she can reach the configuration
  $\smiley(0)$, thus the game can equivalently be seen as a
  "configuration reachability" game.

  Regarding initialisation, Eve can choose her initial position,
  which we implement by the actions
  \begin{align*}
    \loc_0 &\step{-1} \loc_0 & \loc_0 &\step{-1}\loc_{(q_\mathrm{final},a)}&&\text{for $a\in\Gamma$}\;.
  \end{align*}
  Outside the boundary cases, the game is implemented by the following actions:
  \begin{align*}
    \loc_\gamma&\step{-m}\loc_{(\gamma_1,\gamma_2,\gamma_3)}&&&&\text{for $\gamma_1,\gamma_2,\gamma_3\vdash\gamma$}\;,\\ \loc_{(\gamma_1,\gamma_2,\gamma_3)}&\step{-k}\loc_{\gamma_k}&&&&\text{for $k\in\{1,2,3\}$}\;.
  \end{align*}
  We handle the endmarker positions via the following actions, where Eve proceeds along the left edge of \Cref{13-fig:exp} until she reaches the initial left endmarker:
  \begin{align*}  
   \loc_\emkl&\step{-m-2}\loc_\emkl\;,& \loc_\emkl&\step{-1}\loc_{=1}\;,& \loc_\emkr&\step{-m-1}\loc_\emkl\;.
  \end{align*}
  For the positions inside the input word $w=a_1\cdots a_n$, we use the actions
  \begin{align*}  
  \loc_{(q_0,a_1)}&\step{-2}\loc_{=1}\;,&\loc_{a_j}&\step{-2}\loc_{=j}&&\text{for $1<j\leq n$}\;.
  \end{align*}
  Finally, for the blank symbols of~$C_1$, which should be associated with a counter value~$c$ such that $n+3\leq c\leq m+3$, we use the
  action
  \begin{align*}  
  \loc_\blank&\step{-n-2}\loc_{1\leq?\leq m-n+1}\;.&&&&&
  \end{align*}
\qedhere
\end{proof}

\begin{theorem}[Existential countdown games are $\EXPSPACE$-complete]
\label{13-thm:countdown-exist}
  "Configuration reachability" and "coverability" "countdown games"
  with "existential initial credit" are $\EXPSPACE$-complete.
\end{theorem}
\begin{proof}
   For the upper bound, consider an instance, \textit{i.e.}, a "countdown
   system" \[
   \?V=(\Loc,\Act,\Loc_\mEve,\Loc_\mAdam,1),\] an initial
   location~$\loc_0$, and a target configuration $\loc_f\in\Loc$.  We
   reduce this to an instance of "configuration reachability" with
   "given initial credit" in a one-dimensional "vector system" by
   adding a new location $\loc'_0$ controlled by~Eve with actions
   $\loc'_0\step{1}\loc'_0$ and $\loc'_0\step{0}\loc_0$, and asking
   whether Eve has a winning strategy starting from $\loc'_0(0)$ in
   the new system.  By \Cref{13-cor:dim1}, this "configuration
   reachability" game can be solved in $\EXPSPACE$.

   \medskip For the lower bound, we reduce from the acceptance problem
   of a deterministic Turing machine working in exponential space.
   The reduction is the same as in the proof
   of \Cref{13-thm:countdown-given}, except that now the length~$t$ of the
   computation is not bounded a priori, but this is compensated by the
   fact that we are playing the "existential initial credit" version
   of the "countdown game".  \qedhere
\end{proof}

\medskip
Originally, "countdown games" were introduced with a slightly
different objective, which corresponds to the following decision
problem.
\AP\decpb["zero reachability" with "given initial credit"]
  {A "countdown system" $\?V=(\Loc,T,\Loc_\mEve,\Loc_\mAdam,1)$, an
  initial location $\loc_0\in\Loc$, and an initial credit
  $n_0\in\+N$.}
  {Does Eve have a strategy to reach a configuration $\loc(0)$ for
  some $\loc\in\Loc$?
  That is, does she win the zero reachability\index{zero reachability|see{countdown game}}
  game $(\?A_\+N(\?V),\col,\Reach)$ from $\loc_0(n_0)$, where
  $\col(e)=\Win$ if and only if $\ing(e)=\loc(0)$ for some $\loc\in\Loc$?}
\begin{theorem}[Countdown to zero games are $\EXP$-complete]
\label{13-thm:countdown-zero}
  "Zero reachability" "countdown games" with "given initial credit"
  are $\EXP$-complete.
\end{theorem}
\begin{proof}
  The upper bound of \Cref{13-thm:countdown-given} applies in the same
  way.  Regarding the lower bound, we modify the lower bound
  construction of \Cref{13-thm:countdown-given} in the following way: we
  use $\loc_0(2\cdot n_0+1)$ as initial configuration, multiply all
  the action weights in~$\Act$ by two, and add a new
  location~$\loc_\mathrm{zero}$ with an action
  $\smiley\step{-1}\loc_\mathrm{zero}$.  Because all the counter
  values in the new game are odd unless we reach $\loc_\mathrm{zero}$,
  the only way for Eve to bring the counter to zero in this new game
  is to first reach $\smiley(1)$, which occurs if and only if she
  could reach $\smiley(0)$ in the original game.
\end{proof}

\subsection{Vector games in dimension one}
\label{13-sec:one-counter}

"Countdown games" are frequently employed to prove complexity lower
bounds.  Here, we use them to show that the $\EXPSPACE$  upper bounds
from \Cref{13-cor:dim1} are tight in most cases.
\begin{theorem}[The complexity of vector games in dimension one]
\label{13-thm:dim1}
  "Configuration reachability", "coverability", and "parity@parity
  vector game" "vector games", both with "given" and with "existential
  initial credit", are $\EXPSPACE$-complete in dimension one;
  "non-termination" "vector games" in dimension one are $\EXP$-hard with
  "given initial credit" and $\EXPSPACE$-complete with "existential
  initial credit".
\end{theorem}
\begin{proof}
  By \Cref{13-thm:countdown-exist}, "configuration reachability" and
  "coverability" "vector games" with existential initial credit
  are $\EXPSPACE$-hard in dimension one.
  Furthermore, \Cref{13-rmk:cov2parity} allows to deduce that
  "parity@parity vector game" is also $\EXPSPACE$-hard.  Finally, we can
  argue as in the upper bound proof of \Cref{13-thm:countdown-exist} that
  all these games are also hard with "given initial credit": we add a
  new initial location $\loc'_0$ controlled by Eve with actions
  $\loc'_0\step{1}\loc'_0$ and $\loc'_0\step{0}\loc_0$ and play the game
  starting from $\loc'_0(0)$.

  Regarding "non-termination", we can add a self loop $\smiley\step{0}\smiley$ to the construction
  of \Cref{13-thm:countdown-given,13-thm:countdown-exist}: then the only way
  to build an infinite play that avoids the "sink" is to reach the
  target location $\smiley$.  This shows that the games are $\EXP$-hard
  with "given initial credit" and $\EXPSPACE$-hard with "existential
  initial credit".  Note that the trick of reducing "existential" to
  "given initial credit" with an initial incrementing loop $\loc'_0\step{1}\loc'_0$ does not work, because Eve would have a trivial winning
  strategy that consists in just playing this loop forever.
\end{proof}



\section{Asymmetric games}
\label{13-sec:avag}
\Cref{13-thm:undec} shows that "vector games" are too powerful to be
algorithmically relevant, except in dimension one where
\Cref{13-thm:dim1} applies.  This prompts the study of restricted kinds
of "vector games", which might be decidable in arbitrary dimension.
This section introduces one such restriction, called
\emph{"asymmetry"}, which turns out to be very fruitful: it yields
decidable games (see \Cref{13-sec:complexity}), and is
related to another class of games on counter systems called "energy
games" (see \Cref{13-sec:resource}).

\paragraph{Asymmetric Games} A "vector system"
$\?V=(\Loc,\Act,\Loc_\mEve,\Loc_\mAdam,\dd)$ is
asymmetric\index{asymmetry|see{vector system}} if, for all
locations $\loc\in\Loc_\mAdam$ controlled by Adam and all actions
$(\loc\step{\vec u}\loc')\in\Act$ originating from those,
$\vec u=\vec 0$ the "zero vector".  In other words, Adam may only
change the current location, and cannot interact directly with the
counters.

\begin{example}[Asymmetric vector system]
\label{13-ex:avg}
  \Cref{13-fig:avg} presents an "asymmetric vector system" of
  dimension two with locations partitioned as $\Loc_\mEve=\{\loc,\loc_{2,1},\loc_{\text-1,0}\}$ and $\Loc_\mAdam=\{\loc'\}$.  
  We omit the labels on the actions originating from Adam\'s
  locations, since those are necessarily the "zero vector".  It is
  worth observing that this "vector system" behaves quite differently
  from the one of \Cref{13-ex:mwg} on \cpageref{13-ex:mwg}: for
  instance, in $\loc'(0,1)$, Adam can now ensure that the "sink" will
  be reached by playing the action $\loc'\step{0,0}\loc_{\text-1,0}$,
  whereas in \Cref{13-ex:mwg}, the action $\loc'\step{-1,0}\loc$
  was just inhibited by the "natural semantics".
\end{example}
\begin{figure}[htbp]
  \centering
  \begin{tikzpicture}[auto,on grid,node distance=2.5cm]
    \node[s-eve,inner sep=3pt](0){$\loc$};
    \node[s-adam,right=of 0,inner sep=2pt](1){$\loc'$};
    \node[s-eve,above left=1cm and 1.2cm of 1](2){$\loc_{2,1}$};
    \node[s-eve,below left=1cm and 1.2cm of 1](3){$\loc_{\text-1,0}$};
    \path[arrow,every node/.style={font=\footnotesize,inner sep=1}]
    (0) edge[loop left] node {$-1,-1$} ()
    (0) edge[bend right=10] node {$-1,0$} (1)
    (1) edge[bend right=10]  (2)
    (1) edge[bend left=10] (3)
    (2) edge[swap,bend right=10] node{$2,1$} (0)
    (3) edge[bend left=10] node{$-1,0$} (0);
  \end{tikzpicture}
  \caption{An "asymmetric vector system".}
  \label{13-fig:avg}
\commentAlt{Figure~\ref{13-fig:avg}: A directed graph with four nodes (L, L', L_negative-2,1, L_negative-1,0), showing various labeled transitions, including a self-loop.}
\commentLongAlt{Figure~\ref{13-fig:avg}: The image displays a directed graph with four nodes. Node 'L' is a circle, 'L'' is a square, and 'L_negative-2,1' and 'L_negative-1,0' are circles.
- Node 'L' has a self-loop labeled '-1, -1'.
- An arrow points from 'L' to 'L'' labeled '-1, 0'.
- An arrow points from 'L'' to 'L_negative-2,1' labeled '2, 1'.
- An arrow points from 'L_negative-2,1' to 'L'.
- An arrow points from 'L'' to 'L_negative-1,0'.
- An arrow points from 'L_negative-1,0' to 'L' labeled '-1, 0'.}
\end{figure}

\subsection{The case of configuration reachability}
\label{13-sec:reach}

In spite of the restriction to "asymmetric" "vector systems",
"configuration reachability" remains undecidable.
\begin{theorem}[Reachability in asymmetric vector games is undecidable]
\label{13-thm:asym-undec}
  "Configuration reachability" "asymmetric vector games", both with
  "given" and with "existential initial credit", are undecidable in
  any dimension $\dd\geq 2$.
\end{theorem}
\begin{proof}
  We first reduce from the "halting problem" of "deterministic Minsky
  machines" to "configuration reachability" with "given initial
  credit".  Given an instance of the "halting problem", \textit{i.e.}, given
  $\?M=(\Loc,\Act,\dd)$ and an initial location $\loc_0$ where we
  assume without loss of generality that $\?M$ checks that all its
  counters are zero before going to $\loc_\mathtt{halt}$, we construct
  an "asymmetric vector system"
  $\?V\eqdef(\Loc\uplus\Loc_{\eqby{?0}}\uplus\Loc_{\dd},\Act',\Loc\uplus\Loc'_{\eqby{?0}},\Loc_{\eqby{?0}},\dd)$
  where all the original locations and $\Loc_{\dd}$ are
  controlled by~Eve and $\Loc_{\eqby{?0}}$ is controlled by Adam.

  \begin{figure}[htbp]
    \centering
    \begin{tikzpicture}[auto,on grid,node distance=1.5cm]
      \node(to){$\mapsto$};
      \node[anchor=east,left=2.5cm of to](mm){deterministic Minsky machine};
      \node[anchor=west,right=2.5cm of to](mwg){asymmetric vector system};
      \node[below=.7cm of to](map){$\rightsquigarrow$};
      \node[left=2.75cm of map](0){$\loc$};
      \node[right=of 0](1){$\loc'$};
      \node[right=1.25cm of map,s-eve](2){$\loc$};
      \node[right=of 2,s-eve](3){$\loc'$};
      \node[below=2.5cm of map](map2){$\rightsquigarrow$};
      \node[left=2.75cm of map2](4){$\loc$};
      \node[below right=.5cm and 1.5cm of 4](5){$\loc''$};
      \node[above right=.5cm and 1.5cm of 4](6){$\loc'$};
      \node[right=1.25cm of map2,s-eve](7){$\loc$};
      \node[below right=.5cm and 1.5cm of 7,s-eve](8){$\loc''$};
      \node[above right=.5cm and 1.5cm of 7,s-adam,inner sep=-1.5pt](9){$\loc'_{i\eqby{?0}}$};
      \node[below right=.5cm and 1.5cm of 9,s-eve](10){$\loc'$};
      \node[above right=.5cm and 1.5cm of 9,s-eve](11){$\loc_{i}$};
      \node[right=of 11,s-eve,inner sep=0pt](12){$\loc_{\mathtt{halt}}$};
      \path[arrow,every node/.style={font=\scriptsize}]
      (0) edge node{$\vec e_i$} (1)
      (2) edge node{$\vec e_i$} (3)
      (4) edge[swap] node{$-\vec e_i$} (5)
      (4) edge node{$i\eqby{?0}$} (6)
      (7) edge[swap] node{$-\vec e_i$} (8)
      (7) edge node{$\vec 0$} (9)
      (9) edge[swap] node{$\vec 0$} (10)
      (9) edge node{$\vec 0$} (11)
      (11) edge node{$\vec 0$} (12)
      (11) edge[loop above] node{$\forall j\neq i\mathbin.-\vec e_j$}();
    \end{tikzpicture}
    \caption{Schema of the reduction in the proof of \Cref{13-thm:asym-undec}.}
    \label{13-fig:asym-undec}
      \end{figure}

  We
  use $\Loc_{\eqby{?0}}$ and $\Loc_{\dd}$ as two sets of locations disjoint from~$\Loc$ defined by
  \begin{align*}
    \Loc_{\eqby{?0}}&\eqdef\{\loc'_{i\eqby{?0}}\in\Loc\times\{1,\dots,\dd\}\mid\exists\loc\in\Loc\mathbin.(\loc\step{i\eqby{?0}}\loc')\in\Act\}\\
    \Loc_{\dd}&\eqdef\{\loc_{i}\mid 1\leq i\leq \dd\}
    \intertext{and define the set of actions by (see \Cref{13-fig:asym-undec})}
    \Act'&\eqdef\{\loc\step{\vec
          e_i}\loc'\mid(\loc\step{\vec e_i}\loc')\in\Act\}\cup\{\loc\step{-\vec e_i}\loc''\mid(\loc\step{-\vec e_i}\loc'')\in\Act\}\\
    &\:\cup\:\{\loc\step{\vec
      0}\loc'_{i\eqby{?0}},\;\;\:\loc'_{i\eqby{?0}}\!\!\step{\vec
      0}\loc',\;\;\:\loc'_{i\eqby{?0}}\!\!\step{\vec 0}\loc_{i}\mid
      (\loc\step{i\eqby{?0}}\loc')\in\Act\}\\
    &\:\cup\:\{\loc_i\!\step{-\vec e_j}\loc_{i},\;\;\:\loc_{i}\!\step{\vec
      0}\loc_\mathtt{halt}\mid 1\leq i,j\leq\dd, j\neq i\}\;.
  \end{align*}
  We use $\loc_0(\vec 0)$ as initial configuration and
  $\loc_\mathtt{halt}(\vec 0)$ as target configuration for the
  constructed "configuration reachability" instance.  Here is the crux
  of the argument why Eve has a winning strategy to reach
  $\loc_\mathtt{halt}(\vec 0)$ from $\loc_0(\vec 0)$ if and only if
  the "Minsky machine@deterministic Minsky machine" halts, \textit{i.e.}, if
  and only if the "Minsky machine@deterministic Minsky machine"
  reaches $\loc_\mathtt{halt}(\vec 0)$.
  Consider any configuration $\loc(\vec v)$.  If
  $(\loc\step{\vec e_i}\loc')\in\Act$, Eve has no choice but to apply
  $\loc\step{\vec e_i}\loc'$ and go to the configuration
  $\loc'(\vec v+\vec e_i)$ also reached in one step in~$\?M$.  If
  $\{\loc\step{i\eqby{?0}}\loc',\loc\step{-\vec e_i}\loc''\}\in\Act$ and
  $\vec v(i)=0$, due to the "natural semantics", Eve cannot use the
  action $\loc\step{-\vec e_i}\loc''$, thus she must use
  $\loc\step{\vec 0}\loc'_{i\eqby{?0}}$.  Then, either Adam plays
  $\loc'_{i\eqby{?0}}\!\!\step{\vec 0}\loc'$ and Eve regains the
  control in $\loc'(\vec v)$, which was also the configuration reached
  in one step in~$\?M$, or Adam plays
  $\loc'_{i\eqby{?0}}\!\!\step{\vec 0}\loc_{i}$ and Eve 
  regains the control in $\loc_{i}(\vec v)$ with
  $\vec v(i)=0$.  Using the actions
  $\loc_{i}\!\step{-\vec e_j}\loc_{i}$ for
  $j\neq i$, Eve can then reach $\loc_{i}(\vec 0)$ and move
  to $\loc_\mathtt{halt}(\vec 0)$.  Finally, if
  $\{\loc\step{i\eqby{?0}}\loc',\loc\step{-\vec e_i}\loc''\}\in\Act$ and
  $\vec v(i)>0$, Eve can choose: if she uses
  $\loc\step{-\vec e_i}\loc''$, she ends in the configuration
  $\loc''(\vec v-\vec e_i)$ also reached in one step in~$\?M$.  In
  fact, she should not use $\loc\step{\vec 0}\loc'_{i\eqby{?0}}$,
  because Adam would then have the opportunity to apply
  $\loc'_{i\eqby{?0}}\!\!\step{\vec 0}\loc_{i}$, and in
  $\loc_{i}(\vec v)$ with $\vec v(i)>0$, there is no way to
  reach a configuration with an empty $i$th component, let alone to
  reach $\loc_\mathtt{halt}(\vec 0)$.  Thus, if $\?M$ halts, then Eve 
  has a winning strategy that mimics the unique "play"
  of~$\?M$, and conversely, if Eve wins, then necessarily she had to
  follow the "play" of~$\?M$ and thus the machine halts.

  \medskip Finally, regarding the "existential initial credit"
  variant, the arguments used in the proof of \Cref{13-thm:undec} apply
  similarly to show that it is also undecidable.
\end{proof}

In dimension~one, \Cref{13-thm:dim1} applies, thus "configuration
reachability" is decidable in $\EXPSPACE$.  This bound is actually
tight.
\begin{theorem}[Asymmetric vector games are $\EXPSPACE$-complete in dimension~one]
\label{13-thm:asym-dim1}
  "Configuration reachability" "asymmetric vector games", both with
  "given" and with "existential initial credit",
  are $\EXPSPACE$-complete in dimension~one.
\end{theorem}
\begin{proof}
  Let us first consider the "existential initial credit" variant.  We
  proceed as in \Cref{13-thm:countdown-given,13-thm:countdown-exist} and
  reduce from the acceptance problem for a deterministic Turing
  machine working in exponential space $m=2^{p(n)}$.  The reduction is
  mostly the same as in \Cref{13-thm:countdown-given}, with a few changes.
  Consider the integer $m-n$ from that reduction.  While this is an
  exponential value, it can be written as $m-n=\sum_{0\leq e\leq
  p(n)}2^{e}\cdot b_e$ for a polynomial number of bits $b_0,\dots,b_{p(n)}$.
  For all $0\leq d\leq p(n)$, we define $m_d\eqdef \sum_{0\leq e\leq
  d}2^{e}\cdot b_e$; thus $m-n+1=m_{p(n)}+1$.

  We define now the sets of
  locations
  \begin{align*}
    \Loc_\mEve&\eqdef\{\loc_0,\smiley\}
      \cup\{\loc_\gamma\mid\gamma\in\Gamma'\}
      \cup\{\loc_\gamma^k\mid 1\leq k\leq 3\}
      \cup\{\loc_{=j}\mid 0<j\leq n\}\\
      &\:\cup\:\{\loc_{1\leq?\leq m_d+1}\mid 0\leq d\leq
  p(n)\}\cup\{\loc_{1\leq?\leq 2^d}\mid 1\leq d\leq p(n)\}\;,\\
    \Loc_\mAdam&\eqdef\{\loc_{(\gamma_1,\gamma_2,\gamma_3)}\mid\gamma_1,\gamma_2,\gamma_3\in\Gamma'\}\;.
  \end{align*}
  The intention behind the locations $\loc_{=j}\in\Loc_\mEve$ is
  that Eve can reach~$\smiley(0)$ from a configuration $\loc_{=j}(c)$ if
  and only if $c=j$; we define accordingly~$\Act$ with the
  action $\loc_{=j}\step{-j}\smiley$.
  Similarly, Eve should be able to reach~$\smiley(0)$ from
  $\loc_{1\leq?\leq m_d+1}(c)$ for $0\leq d\leq p(n)$ if and only if
  $1\leq c\leq m_d+1$,
  which is implemented by the following actions: if $b_{d+1}=1$, then
  \begin{align*}
    \loc_{1\leq?\leq m_{d+1}+1}&\step{0}\loc_{1\leq?\leq 2^{d+1}}\;,&
    \loc_{1\leq?\leq m_{d+1}+1}&\step{-2^{d+1}}\loc_{1\leq ?\leq m_{d}+1}\;,
    \intertext{and if $b_{d+1}=0$,}
    \loc_{1\leq?\leq m_{d+1}+1}&\step{0}\loc_{1\leq ?\leq m_{d}+1}\;,
    \intertext{and finally}
    \loc_{1\leq?\leq m_0+1}&\step{-b_0}\loc_{=1}\;,&\loc_{1\leq?\leq m_0+1}&\step{0}\loc_{=1}\;,
    \intertext{where for all $1\leq d\leq p(n)$, $\loc_{1\leq?\leq 2^d}(c)$ allows to
    reach $\smiley(0)$ if and only if $1\leq c\leq 2^d$:}
    \loc_{1\leq?\leq 2^{d+1}}&\step{-2^{d}}\loc_{1\leq?\leq
                               2^d}\;,&\loc_{1\leq?\leq
                                        2^{d+1}}&\step{0}\loc_{1\leq?\leq
                                                  2^d}\;,\\\loc_{1\leq?\leq
    2^1}&\step{-1}\loc_{=1}\;,&\loc_{1\leq?\leq 2^1}&\step{0}\loc_{=1}\;.
  \end{align*}

  The remainder of the reduction is now very similar to the reduction shown
  in \Cref{13-thm:countdown-given}.
  Regarding initialisation, Eve can choose her initial position,
  which we implement by the actions
  \begin{align*}
    \loc_0 &\step{-1} \loc_0 & \loc_0 &\step{-1}\loc_{(q_\mathrm{final},a)}&&\text{for $a\in\Gamma$}\;.
    \intertext{Outside the boundary cases, the game is implemented by
    the following actions:}
    \loc_\gamma&\step{-m}\loc_{(\gamma_1,\gamma_2,\gamma_3)}&&&&\text{for
  $\gamma_1,\gamma_2,\gamma_3\vdash\gamma$}\;,\\ \loc_{(\gamma_1,\gamma_2,\gamma_3)}&\step{0}\loc^k_{\gamma_k}&\loc^k_{\gamma_k}&\step{-k}\loc_{\gamma_k}&&\text{for
  $k\in\{1,2,3\}$}\;.
  \intertext{We handle the endmarker positions via the following
  actions, where Eve proceeds along the left edge
  of \Cref{13-fig:exp} until she reaches the initial left endmarker:}
   \loc_\emkl&\step{-m-2}\loc_\emkl\;,& \loc_\emkl&\step{-1}\loc_{=1}\;,& \loc_\emkr&\step{-m-1}\loc_\emkl\;.
  \intertext{For the positions inside the input word $w=a_1\cdots
  a_n$, we use the actions}
  \loc_{(q_0,a_1)}&\step{-2}\loc_{=1}\;,&\loc_{a_j}&\step{-2}\loc_{=j}&&\text{for
  $1<j\leq n$}\;.
  \intertext{Finally, for the blank symbols of~$C_1$, which should be
  associated with a counter value~$c$ such that $n+3\leq c\leq m+3$,
  \textit{i.e.}, such that $1\leq c-n-2\leq m-n+1=m_{p(n)}+1$, we use the
  action}
  \loc_\blank&\step{-n-2}\loc_{1\leq?\leq m_{p(n)}+1}\;.
  \end{align*}

  Regarding the "given initial credit" variant, we add a new location
  $\loc'_0$ controlled by Eve and let her choose her initial credit
  when starting from $\loc'_0(0)$ by using the new actions
  $\loc'_0\step{1}\loc'_0$ and $\loc'_0\step{0}\loc_0$.
\end{proof}

\subsection{Asymmetric monotone games}
\label{13-sec:mono}

The results on "configuration reachability" might give the impression
that "asymmetry" does not help much for solving "vector games": we
obtained in \Cref{13-sec:reach} exactly the same results as in the
general case.  Thankfully, the situation changes drastically if we
consider the other types of "vector games": "coverability",
"non-termination", and "parity@parity vector games" become decidable
in "asymmetric vector games".  The main rationale for this comes from
order theory, which prompts the following definitions.

\paragraph{Quasi-orders}\AP A quasi-order $(X,{\leq})$ is a
set~$X$ together with a reflexive and transitive
relation~${\leq}\subseteq X\times X$.  Two elements $x,y\in X$ are
incomparable if $x\not\leq y$ and $y\not\leq x$, and they are
equivalent if $x\leq y$ and $y\leq x$.  The associated strict relation
$x<y$ holds if $x\leq y$ and $y\not\leq x$.

The upward closure of a subset $S\subseteq X$ is the set of
elements greater or equal to the elements of S:
${\uparrow}S\eqdef\{x\in X\mid\exists y\in S\mathbin.y\leq x\}$.  A
subset $U\subseteq X$ is upwards closed if ${\uparrow}U=U$.  When
$S=\{x\}$ is a singleton, we write more simply ${\uparrow}x$ for its
upward closure and call the resulting "upwards closed" set a
principal filter.  Dually, the downward closure of~$S$ is
${\downarrow}S\eqdef\{x\in X\mid\exists y\in S\mathbin.x\leq y\}$, a
downwards closed set is a subset $D\subseteq X$ such that
$D={\downarrow}D$, and ${\downarrow}x$ is called a principal
ideal.  Note that the complement $X\setminus U$ of an upwards closed
set~$U$ is downwards closed and vice versa.

\paragraph{Monotone Games}\AP
Let us consider again the "natural semantics" $\natural(\?V)$ of a
"vector system".  The set of vertices
$V=\Loc\times\+N^\dd\cup\{\sink\}$ is naturally equipped with a
partial ordering: $v\leq v'$ if either $v=v'=\sink$, or $v=\loc(\vec
v)$ and $v'=\loc(\vec v')$ are two configurations that share the same
location and satisfy $\vec v(i)\leq\vec v'(i)$ for all $1\leq
i\leq\dd$, \textit{i.e.}, if $\vec v\leq\vec v'$ for the componentwise
ordering.

Consider a set of colours $C$ and a vertex colouring $\vcol{:}\,V\to C$
of the "natural semantics" $\natural(\?V)$ of a "vector system", which
defines a colouring $\col{:}\,E\to C$ where
$\col(e)\eqdef\vcol(\ing(e))$.  We
say that the "colouring"~$\vcol$ is monotonic if $C$ is finite and,
for every colour $p\in C$, the set $\vcol^{-1}(p)$ of vertices coloured
by~$p$ is "upwards closed" with respect to ${\leq}$.  Clearly, the
"colourings" of "coverability", "non-termination", and "parity@parity
vector games" "vector games" are "monotonic", whereas those of
"configuration reachability" "vector games" are not.  By extension, we
call a "vector game" \emph{"monotonic"} if its underlying "colouring"
is "monotonic".

\begin{lemma}[Simulation]
\label{13-lem:mono}
  In a "monotonic" "asymmetric vector game", if Eve wins from a
  vertex~$v_0$, then she also wins from~$v'_0$ for all $v'_0\geq
  v_0$.
\end{lemma}
\begin{proof}
  It suffices for this to check that, for all $v_1\leq v_2$ in $V$,
  \begin{description}
  \item[(colours)] $\vcol(v_1)=\vcol(v_2)$ since $\vcol$ is "monotonic";
  \item[(zig Eve)] if $v_1,v_2\in V_\mEve$, $a\in\Act$, and
    $\dest(v_1,a)=v'_1\neq\sink$ is defined, then
    $v'_2\eqdef\dest(v_2,a)$ is such that $v'_2\geq v'_1$: indeed,
    $v'_1\neq\sink$ entails that $v_1$ is a configuration
    $\loc(\vec v_1)$ and $v'_1=\loc'(\vec v_1+\vec u)$ for the action
    $a=(\loc\step{\vec u}\loc')\in\Act$, but then $v_2=\loc(\vec v_2)$
    for some $\vec v_2\geq\vec v_1$ and
    $v'_2=\loc'(\vec v_2+\vec u)\geq v'_1$;
  \item[(zig Adam)] if $v_1,v_2\in V_\mAdam$, $a\in\Act$, and
    $\dest(v_2,a)=v'_2$ is defined, then
    $v'_1\eqdef\dest(v_1,a)\leq v'_2$: indeed, either $v'_2=\sink$ and
    then $v'_1=\sink$, or $v'_2\neq\sink$, thus
    $v_2=\loc(\vec v_2)$, $v'_2=\loc'(\vec v_2)$, and
    $a=(\loc\step{\vec 0}\loc')\in\Act$ (recall that the game is
    "asymmetric"), but then $v_1=\loc(\vec v_1)$ for some
    $\vec v_1\leq\vec v_2$ and thus $v'_1=\loc'(\vec v_1)\leq v'_2$.
  \end{description}
  The above conditions show that, if $\sigma{:}\,E^\ast\to\Act$ is a
  strategy of Eve that wins from~$v_0$, then by
  simulating~$\sigma$ starting from~$v'_0$---\textit{i.e.}, by applying the
  same actions when given a pointwise larger or equal history---she
  will also win.
\end{proof}

Note that \Cref{13-lem:mono} implies that $\WE$ is "upwards closed":
$v_0\in\WE$ and $v_0\leq v'_0$ imply $v_0'\in\WE$.  \Cref{13-lem:mono}
does not necessarily hold in "vector games" without the "asymmetry"
condition.  For instance, in both \Cref{13-fig:cov,13-fig:nonterm} on
\cpageref{13-fig:cov}, $\loc'(0,1)\in\WE$ but $\loc'(1,2)\in\WA$ for
the "coverability" and "non-termination" objectives.  This is due to
the fact that the action $\loc'\step{-1,0}\loc$ is available
in~$\loc'(1,2)$ but not in~$\loc'(0,1)$.

\paragraph{Well-quasi-orders}\AP What makes "monotonic" "vector games" so
interesting is that the partial order $(V,{\leq})$ associated with the
"natural semantics" of a "vector system" is a well-quasi-order.  A
"quasi-order" $(X,{\leq})$ is "well@well-quasi-order" (a \emph{"wqo"})
if any of the following equivalent characterisations
hold~\cite{Kruskal:1972}:
\begin{itemize}
  \item\AP in any infinite sequence $x_0,x_1,\cdots$ of elements
    of~$X$, there exists an infinite sequence of indices
    $n_0<n_1<\cdots$ such that $x_{n_0}\leq
    x_{n_1}\leq\cdots$---infinite sequences in $X$ are good---,
  \item\AP any strictly ascending sequence $U_0\subsetneq
    U_1\subsetneq\cdots$ of "upwards closed" sets $U_i\subseteq X$ is
    finite---$X$ has the ascending chain condition---,
  \item\AP any non-empty "upwards closed" $U\subseteq X$ has at least
    one, and at most finitely many minimal elements up to equivalence;
    therefore any "upwards closed" $U\subseteq X$ is a finite union
    $U=\bigcup_{1\leq j\leq n}{\uparrow}x_j$ of finitely many
    "principal filters"~${\uparrow}x_j$---$X$ has the finite basis
    property.
\end{itemize}

The fact that $(V,{\leq})$ satisfies all of the above is an easy
consequence of \emph{Dickson's Lemma}~\cite{Dickson:1913}.


\paragraph{Pareto Limits}\AP By the "finite basis property" of
$(V,{\leq})$ and \Cref{13-lem:mono}, in a "monotonic" "asymmetric
vector game", $\WE=\bigcup_{1\leq j\leq n}{\uparrow}\loc_j(\vec v_j)$
is a finite union of "principal filters".  The set
$\mathsf{Pareto}\eqdef\{\loc_1(\vec v_1),\dots,\loc_n(\vec v_n)\}$ is
called the Pareto limit or \emph{Pareto frontier} of the game.
Both the "existential" and the "given initial credit" variants of the
game can be reduced to computing this "Pareto limit": with
"existential initial credit" and an initial location $\loc_0$, check
whether $\loc_0(\vec v)\in\mathsf{Pareto}$ for some $\vec v$, and with
"given initial credit" and an initial configuration $\loc_0(\vec v_0)$, check
whether $\loc_0(\vec v)\in\mathsf{Pareto}$ for some $\vec v\leq\vec
v_0$.
\begin{example}[Pareto limit]
  Consider the "asymmetric vector system" from \Cref{13-fig:avg} on
  \cpageref{13-fig:avg}.  For the "coverability game" with target
  configuration $\loc(2,2)$, the "Pareto limit" is
  $\mathsf{Pareto}=\{\loc(2,2),\loc'(3,2),\loc_{2,1}(0,1),\loc_{\text-1,0}(3,2)\}$,
  while for the "non-termination game", $\mathsf{Pareto}=\emptyset$:
  Eve loses from all the vertices.  Observe that this is consistent
  with Eve's "winning region" in the "coverability" "energy game"
  shown in \Cref{13-fig:cov-nrg}.
\end{example}


\begin{example}[Doubly exponential Pareto limit]
\label{13-ex:pareto}
  Consider the one-player "vector system" of \Cref{13-fig:pareto},
  where the "meta-decrement" from~$\loc_0$ to~$\loc_1$ can be
  implemented using $O(n)$ additional counters and a set~$\Loc'$ of
  $O(n)$ additional locations by the arguments of the
  forthcoming \Cref{13-thm:avag-hard}.
  
  \begin{figure}[htbp]
    \centering
  \begin{tikzpicture}[auto,on grid,node distance=2.5cm]
    \node[s-eve](0){$\loc_0$};
    \node[s-eve,right=of 0](1){$\loc_1$};
    \node[s-eve,below right=1.5 and 1.25 of 0](2){$\loc_f$};
    \path[arrow,every node/.style={font=\footnotesize,inner sep=1}]
    (0) edge node {$-2^{2^n}\cdot\vec e_1$} (1)
    (0) edge[bend right=10,swap] node {$-\vec e_2$} (2)
    (1) edge[bend left=10] node {$\vec 0$} (2);
  \end{tikzpicture}
  \caption{A one-player "vector system"
  with a large "Pareto limit".}
  \label{13-fig:pareto}
\commentAlt{Figure~\ref{13-fig:pareto}: A directed graph with three circular nodes, L0, L1, and Lf, forming a triangular arrangement with labeled edges.}
\commentLongAlt{Figure~\ref{13-fig:pareto}: The image displays a directed graph with three circular nodes arranged in a triangle. The top-left node is labeled 'L0', the top-right node is labeled 'L1', and the bottom node is labeled 'Lf'.
- An arrow points from 'L0' to 'L1', labeled '-2^e * e_1_vector'.
- An arrow points from 'L0' to 'Lf', labeled '-e_2_vector'.
- An arrow points from 'L1' to 'Lf', labeled '0_vector'.}
  \end{figure}
  For the "coverability game" with target
  configuration~$\loc_f(\vec 0)$, if $\loc_0$ is the initial location
  and we are "given initial credit" $m\cdot\vec e_1$, Eve wins if and
  only if $m\geq 2^{2^n}$, but with "existential initial credit" she
  can start from $\loc_0(\vec e_2)$ instead.  We have indeed
  $\mathsf{Pareto}\cap(\{\loc_0,\loc_1,\loc_f\}\times\+N^\dd)=\{\loc_0(\vec
  e_2),\loc_0(2^{2^n}\cdot\vec e_1),\loc_1(\vec 0),\loc_f(\vec 0)\}$.
  Looking more in-depth into the construction of \Cref{13-thm:avag-hard},
  there is also an at least double exponential number of distinct
  minimal configurations in~$\mathsf{Pareto}$.
\end{example}

\paragraph{Finite Memory} 
Besides having a finitely represented "winning region", Eve also has
finite memory strategies in "asymmetric vector games" with "parity"
objectives; the following argument is straightforward to adapt to
the other regular objectives from \Cref{3-chap:regular}.
\begin{lemma}[Finite memory suffices in parity asymmetric vector games]
\label{13-lem:finmem}
  If Eve has a "strategy" winning from some vertex~$v_0$ in a
  "parity@parity vector game" "asymmetric vector game", then she has a
  "finite-memory" one.
\end{lemma}
\begin{proof}
  Assume~$\sigma$ is a winning strategy from~$v_0$.  Consider the tree
  of vertices visited by plays consistent with~$\sigma$: each branch
  is an infinite sequence $v_0,v_1,\dots$ of elements of~$V$ where the
  maximal priority occuring infinitely often is some even number~$p$.
  Since $(V,{\leq})$ is a "wqo", this is a "good sequence": there
  exists infinitely many indices $n_0<n_1<\cdots$ such that
  $v_{n_0}\leq v_{n_1}\leq\cdots$.  There exists $i<j$ such
  that~$p=\max_{n_i\leq n<n_j}\vcol(v_n)$ is the maximal priority
  occurring in some interval $v_{n_i},v_{n_{i+1}},\dots,v_{n_{j-1}}$.
  Then Eve can play in~$v_{n_j}$ as if she were in~$v_{n_i}$, in
  $v_{n_j+1}$ as if she were in $v_{n_i+1}$ and so on, and we prune
  the tree at index~$n_j$ along this branch so that $v_{n_j}$ is a
  leaf, and we call~$v_{n_i}$ the return node of that leaf.  We
  therefore obtain a finitely branching tree with finite branches,
  which by K{\"{o}}nig's Lemma is finite.

  The finite tree we obtain this way is sometimes called a
  self-covering tree.  
  It is relatively straightforward to construct a finite "memory
  structure"~$(M,m_0,\delta)$ (as defined in \Cref{1-sec:memory}) from a
  "self-covering tree", using its internal nodes as memory states plus
  an additional sink memory state~$m_\bot$; the initial memory
  state~$m_0$ is the root of the tree.  In a node~$m$ labelled by
  $\loc(\vec v)$, given an edge
  $e=(\loc(\vec v'),\loc'(\vec v'+\vec u))$ arising from an
  action~$\loc\step{\vec u}\loc'\in\Act$, if $\vec v'\geq\vec v$ and
  $m$~has a child~$m'$ labelled by $\loc'(\vec v+\vec u)$ in the
  "self-covering tree", then either~$m'$ is a leaf with "return
  node"~$m''$ and we set $\delta(m,e)\eqdef m''$, or $m'$~is an
  internal node and we set $\delta(m,e)\eqdef m'$; in all the other
  cases, $\delta(m,e)\eqdef m_\bot$. 
\end{proof}


\begin{example}[doubly exponential memory]
  Consider the one-player "vector system" of \Cref{13-fig:finitemem},
  where the "meta-decrement" from~$\loc_1$ to~$\loc_0$ can be
  implemented using $O(n)$ additional counters and $O(n)$ additional
  locations by the arguments of the forthcoming \Cref{13-thm:avag-hard}
  on \cpageref{13-thm:avag-hard}.
  
  \begin{figure}[htbp]
    \centering
  \begin{tikzpicture}[auto,on grid,node distance=2.5cm]
    \node[s-eve](0){$\loc_0$};
    \node[s-eve,right=of 0](1){$\loc_1$};
    \node[black!50,above=.5 of 0,font=\scriptsize]{$2$};
    \node[black!50,above=.5 of 1,font=\scriptsize]{$1$};
    \path[arrow,every node/.style={font=\footnotesize,inner sep=2}]
    (1) edge[bend left=15] node {$-2^{2^n}\cdot\vec e_1$} (0)
    (0) edge[bend left=15] node {$\vec 0$} (1)
    (1) edge[loop right] node{$\vec e_1$} ();
  \end{tikzpicture}
  \caption{A one-player "vector system"
  witnessing the need for double exponential memory.}
  \label{13-fig:finitemem}
\commentAlt{Figure~\ref{13-fig:finitemem}: A directed graph with two circular nodes, L0 and L1, showing bidirectional transitions with labels and a self-loop.}
\commentLongAlt{Figure~\ref{13-fig:finitemem}: The image displays a directed graph with two circular nodes, 'L0' on the left and 'L1' on the right.
- Node 'L0' has a label '2' above it.
- Node 'L1' has a label '1' above it.
- A bidirectional arrow connects 'L0' and 'L1'. The arrow from 'L0' to 'L1' is labeled '0_vector'. The arrow from 'L1' to 'L0' is labeled '-2^x * e_1_vector'.
- Node 'L1' also has a self-loop labeled 'e_1_vector'.}
  \end{figure}

  For the "parity@parity vector game" game with location colouring
  $\lcol(\loc_0)\eqdef 2$ and $\lcol(\loc_1)\eqdef 1$, note that Eve 
  must visit $\loc_0$ infinitely often in order to fulfil the parity
  requirements.  Starting from the initial
  configuration~$\loc_0(\vec 0)$, any winning play of Eve begins
  by 
  \begin{equation*} \loc_0(\vec 0)\step{0}\loc_1(\vec 0)\step{\vec
      e_1}\loc_1(\vec e_1)\step{\vec e_1}\cdots\step{\vec
      e_1}\loc_1(m\cdot\vec
    e_1)\mstep{-2^{2^n}}\loc_0((m-2^{2^n})\cdot\vec
    e_1) 
  \end{equation*} 
  for some~$m\geq 2^{2^n}$ before she visits again a
  configuration---namely~$\loc_0((m-2^{2^n})\cdot\vec e_1)$---greater
  or equal than a previous configuration---namely
  $\loc_0(\vec 0)$---\emph{and} witnesses a maximal even parity in the
  meantime.  She then has a winning strategy that simply repeats this
  sequence of actions, allowing her to visit successively
  $\loc_0(2(m-2^{2^n})\cdot\vec e_1)$,
  $\loc_0(3(m-2^{2^n})\cdot\vec e_1)$, etc.  In this example, she
  needs at least $2^{2^n}$ memory to remember how many times the
  $\loc_1\step{\vec e_1}\loc_1$ loop should be taken.
\end{example}

\subsubsection{Attractor Computation for Coverability}
\label{13-sec:attr}
So far, we have not seen how to compute the "Pareto limit" derived
from \Cref{13-lem:mono} nor the finite "memory structure" derived
from \Cref{13-lem:finmem}.  These objects are not merely finite but
also computable.  The simplest case is the one of "coverability"
"asymmetric" "monotonic vector games": the fixed-point computation of
\Cref{1-sec:attractors} for "reachability" objectives can be turned into
an algorithm computing the "Pareto limit" of the game.

\begin{fact}[Computable Pareto limit]
\label{13-fact:pareto-cov}
  The "Pareto limit" of a "coverability" "asymmetric vector game" is
  computable.
\end{fact}
\begin{proof}
Let $\loc_f(\vec v_f)$ be the target configuration.  We define a
chain $U_0\subseteq U_1\subseteq\cdots$ of sets $U_i\subseteq V$ by
\begin{align*}
  U_0&\eqdef{\uparrow}\loc_f(\vec v_f)\;,&
  U_{i+1}&\eqdef U_i\cup\mathrm{Pre}(U_i)\;.
\end{align*}
Observe that for all~$i$, $U_i$ is "upwards closed".  This can be
checked by induction over~$i$: it holds initially in~$U_0$, and for
the induction step, if $v\in U_{i+1}$ and $v'\geq v$, then either
\begin{itemize}
\item $v=\loc(\vec v)\in\mathrm{Pre}(U_i)\cap\VE$ thanks to some
  $\loc\step{\vec u}\loc'\in\Act$ such that
  $\loc'(\vec v+\vec u)\in U_i$; therefore $v'=\loc(\vec v')$ for some
  $\vec v'\geq \vec v$ is such that $\loc'(\vec v'+\vec u)\in U_i$ as
  well, thus $v'\in \mathrm{Pre}(U_i)\subseteq U_{i+1}$, or
\item $v=\loc(\vec v)\in\mathrm{Pre}(U_i)\cap\VA$ because for all
  $\loc\step{\vec 0}\loc'\in\Act$, $\loc'(\vec v)\in U_i$; therefore
  $v'=\loc(\vec v')$ for some $\vec v'\geq \vec v$ is such that
  $\loc'(\vec v')\in U_i$ as well, thus
  $v'\in \mathrm{Pre}(U_i)\subseteq U_{i+1}$, or
\item $v\in U_i$ and therefore $v'\in U_i\subseteq U_{i+1}$.
\end{itemize}

By the "ascending chain condition", there is a finite rank~$i$ such
that $U_{i+1}\subseteq U_i$ and then $\WE=U_i$.  Thus the
"Pareto limit" is obtained after finitely many steps.
In order to turn this idea into an algorithm, we need a way of
representing those infinite "upwards closed" sets $U_i$.  Thankfully,
by the "finite basis property", each $U_i$ has a finite basis $B_i$
such that ${\uparrow}B_i=U_i$.  We therefore compute the following
sequence of sets
\begin{align*}
  B_0&\eqdef\{\loc_f(\vec v_f)\}&B_{i+1}&\eqdef
                                       B_i\cup\min\mathrm{Pre}({\uparrow}B_i)\;.
\end{align*}
Indeed, given a finite basis~$B_i$ for~$U_i$, it is straightforward to
compute a finite basis for the "upwards closed" $\mathrm{Pre}(U_i)$.
This results in \Cref{13-algo:cov} below.
\end{proof}

\begin{algorithm}
 \KwData{A "vector system" and a target configuration $\loc_f(\vec v_f)$}

$B_0 \leftarrow \{\loc_f(\vec v_f)\}$ ;

$i \leftarrow 0$ ;
     
\Repeat{${\uparrow}B_i \supseteq B_{i+1}$}{

$B_{i+1} \leftarrow B_i \cup \min\mathrm{Pre}({\uparrow}B_i)$ ;

$i \leftarrow i + 1$ ;}

\Return{$\min B_i = \mathsf{Pareto}(\game)$}
\caption{Fixed-point algorithm for "coverability" in "asymmetric" "vector
  games".}
\label{13-algo:cov}
\end{algorithm}

While this algorithm terminates thanks to the "ascending chain
condition", it may take quite long.  For instance, in
\Cref{13-ex:pareto}, it requires at least~$2^{2^n}$ steps before it
reaches its fixed point.  This is a worst-case instance, as it turns
out that this algorithm works in \kEXP[2]; see the bibliographic notes
at the end of the chapter.  Note that such a
fixed-point computation does not work directly for "non-termination"
or "parity vector games", due to the need for greatest fixed points.


\section{Resource-conscious games}
\label{13-sec:resource}
"Vector games" are very well suited for reasoning about systems
manipulating discrete resources, modelled as counters.  However, in
the "natural semantics", actions that would deplete some resource,
\textit{i.e.}, that would make some counter go negative, are simply inhibited.
In models of real-world systems monitoring resources like a gas
tank or a battery, a depleted resource would be considered as a system
failure.  In the "energy games" of \Cref{13-sec:energy}, those situations
are accordingly considered as winning for Adam.  Moreover, if we are
modelling systems with a bounded capacity for storing resources, a
counter exceeding some bound might also be considered as a failure,
which will be considered with "bounding games" in \Cref{13-sec:bounding}.

These resource-conscious games can be seen as providing alternative
semantics for "vector systems".  They will also be instrumental in
establishing complexity upper bounds for "monotonic" "asymmetric vector
games" later in \Cref{13-sec:complexity}, and are strongly related to
"multidimensional" "mean-payoff" games, as will be explained in
\Cref{14-sec:mean_payoff_energy} of \Cref{14-chap:multiobjective}.

\subsection{Energy semantics}
\label{13-sec:energy}

"Energy games" model systems where the depletion of a resource
allows Adam to win.  This is captured by an energy semantics
$\energy(\?V)\eqdef(V,E_\+E,\VE,\VA)$ associated with a "vector
system" $\?V$: we let as before
$V\eqdef(\Loc\times\+N^\dd)\uplus\{\sink\}$, but define instead
\begin{align*}
  E_\+E&\eqdef \{(\loc(\vec v), \loc'(\vec v+\vec u)\mid
         \loc\step{\vec u}\loc'\in\Act\text{
      and }\vec v+\vec u\geq\vec 0\}\\
    &\:\cup\:\{(\loc(\vec v),\sink)\mid\forall\loc\step{\vec
      u}\loc'\in\Act\mathbin.\vec v+\vec u\not\geq\vec 0\}
    \cup\{(\sink,\sink)\}\;.
\end{align*}
In the "energy semantics", moves that would result in a negative
component lead to the "sink" instead of being inhibited.

\begin{example}[Energy semantics]
\label{13-ex:nrg}
  \Cref{13-fig:nrg} illustrates the "energy semantics" of the vector
  system depicted in~\Cref{13-fig:mwg} on \cpageref{13-fig:mwg}.  Observe that,
  by contrast with the "natural semantics" of the same system depicted
  in \Cref{13-fig:sem}, all the configurations $\loc'(0,n)$ controlled
  by Adam can now move to the "sink".
\end{example}
\begin{figure}[thbp]
  \centering\scalebox{.77}{
  \begin{tikzpicture}[auto,on grid,node distance=2.5cm]
    \draw[step=1,lightgray!50,dotted] (-5.7,0) grid (5.7,3.8);
    \draw[color=white](0,-.3) -- (0,3.8);
    \node at (0,3.9) (sink) {\boldmath$\sink$};
    \draw[step=1,lightgray!50] (1,0) grid (5.5,3.5);
    \draw[step=1,lightgray!50] (-1,0) grid (-5.5,3.5);
    \node at (0,0)[lightgray,font=\scriptsize,fill=white] {0};
    \node at (0,1)[lightgray,font=\scriptsize,fill=white] {1};
    \node at (0,2)[lightgray,font=\scriptsize,fill=white] {2};
    \node at (0,3)[lightgray,font=\scriptsize,fill=white] {3};
    \node at (1,3.9)[lightgray,font=\scriptsize,fill=white] {0};
    \node at (2,3.9)[lightgray,font=\scriptsize,fill=white] {1};
    \node at (3,3.9)[lightgray,font=\scriptsize,fill=white] {2};
    \node at (4,3.9)[lightgray,font=\scriptsize,fill=white] {3};
    \node at (5,3.9)[lightgray,font=\scriptsize,fill=white] {4};
    \node at (-1,3.9)[lightgray,font=\scriptsize,fill=white] {0};
    \node at (-2,3.9)[lightgray,font=\scriptsize,fill=white] {1};
    \node at (-3,3.9)[lightgray,font=\scriptsize,fill=white] {2};
    \node at (-4,3.9)[lightgray,font=\scriptsize,fill=white] {3};
    \node at (-5,3.9)[lightgray,font=\scriptsize,fill=white] {4};
    \node at (1,0)[s-eve-small] (e00) {};
    \node at (1,1)[s-adam-small](a01){};
    \node at (1,2)[s-eve-small] (e02){};
    \node at (1,3)[s-adam-small](a03){};
    \node at (2,0)[s-adam-small](a10){};
    \node at (2,1)[s-eve-small] (e11){};
    \node at (2,2)[s-adam-small](a12){};
    \node at (2,3)[s-eve-small] (e13){};
    \node at (3,0)[s-eve-small] (e20){};
    \node at (3,1)[s-adam-small](a21){};
    \node at (3,2)[s-eve-small] (e22){};
    \node at (3,3)[s-adam-small](a23){};
    \node at (4,0)[s-adam-small](a30){};
    \node at (4,1)[s-eve-small] (e31){};
    \node at (4,2)[s-adam-small](a32){};
    \node at (4,3)[s-eve-small] (e33){};
    \node at (5,0)[s-eve-small] (e40){};
    \node at (5,1)[s-adam-small](a41){};
    \node at (5,2)[s-eve-small] (e42){};
    \node at (5,3)[s-adam-small](a43){};
    \node at (-1,0)[s-adam-small](a00){};
    \node at (-1,1)[s-eve-small] (e01){};
    \node at (-1,2)[s-adam-small](a02){};
    \node at (-1,3)[s-eve-small] (e03){};
    \node at (-2,0)[s-eve-small] (e10){};
    \node at (-2,1)[s-adam-small](a11){};
    \node at (-2,2)[s-eve-small] (e12){};
    \node at (-2,3)[s-adam-small](a13){};
    \node at (-3,0)[s-adam-small](a20){};
    \node at (-3,1)[s-eve-small] (e21){};
    \node at (-3,2)[s-adam-small](a22){};
    \node at (-3,3)[s-eve-small] (e23){};
    \node at (-4,0)[s-eve-small] (e30){};
    \node at (-4,1)[s-adam-small](a31){};
    \node at (-4,2)[s-eve-small] (e32){};
    \node at (-4,3)[s-adam-small](a33){};
    \node at (-5,0)[s-adam-small](a40){};
    \node at (-5,1)[s-eve-small] (e41){};
    \node at (-5,2)[s-adam-small](a42){};
    \node at (-5,3)[s-eve-small] (e43){};
    \path[arrow] 
    (e11) edge (e00)
    (e22) edge (e11)
    (e31) edge (e20)
    (e32) edge (e21)
    (e21) edge (e10)
    (e12) edge (e01)
    (e23) edge (e12)
    (e33) edge (e22)
    (e13) edge (e02)
    (e43) edge (e32)
    (e42) edge (e31)
    (e41) edge (e30);
    \path[arrow] 
    (e11) edge (a01)
    (e20) edge (a10)
    (e22) edge (a12)
    (e31) edge (a21)
    (e32) edge (a22)
    (e21) edge (a11)
    (e12) edge (a02)
    (e30) edge (a20)
    (e10) edge (a00)
    (e13) edge (a03)
    (e23) edge (a13)
    (e33) edge (a23)
    (e43) edge (a33)
    (e42) edge (a32)
    (e41) edge (a31)
    (e40) edge (a30);
    \path[arrow] 
    (a11) edge (e01)
    (a20) edge (e10)
    (a22) edge (e12)
    (a31) edge (e21)
    (a32) edge (e22)
    (a21) edge (e11)
    (a12) edge (e02)
    (a30) edge (e20)
    (a10) edge (e00)
    (a33) edge (e23)
    (a23) edge (e13)
    (a13) edge (e03)
    (a43) edge (e33)
    (a42) edge (e32)
    (a41) edge (e31)
    (a40) edge (e30);
    \path[arrow] 
    (a01) edge (e22)
    (a10) edge (e31)
    (a11) edge (e32)
    (a00) edge (e21)
    (a02) edge (e23)
    (a12) edge (e33)
    (a22) edge (e43)
    (a21) edge (e42)
    (a20) edge (e41);
    \path[arrow] 
    (-5.5,3.5) edge (e43)
    (5.5,2.5) edge (e42)
    (2.5,3.5) edge (e13)
    (5.5,0.5) edge (e40)
    (-5.5,1.5) edge (e41)
    (-3.5,3.5) edge (e23)
    (-1.5,3.5) edge (e03)
    (4.5,3.5) edge (e33)
    (5.5,0) edge (e40)
    (5.5,2) edge (e42)
    (-5.5,1) edge (e41)
    (-5.5,3) edge (e43);
    \path[dotted]
    (-5.7,3.7) edge (-5.5,3.5)
    (5.7,2.7) edge (5.5,2.5)
    (2.7,3.7) edge (2.5,3.5)
    (5.7,0.7) edge (5.5,0.5)
    (-3.7,3.7) edge (-3.5,3.5)
    (-1.7,3.7) edge (-1.5,3.5)
    (4.7,3.7) edge (4.5,3.5)
    (-5.7,1.7) edge (-5.5,1.5)
    (5.75,0) edge (5.5,0)
    (5.75,2) edge (5.5,2)
    (-5.75,1) edge (-5.5,1)
    (-5.75,3) edge (-5.5,3);
    \path[arrow]
    (5.5,1) edge (a41)
    (-5.5,2) edge (a42)
    (-5.5,0) edge (a40)
    (5.5,3) edge (a43);
    \path[dotted]
    (5.75,1) edge (5.5,1)
    (-5.75,2) edge (-5.5,2)
    (-5.75,0) edge (-5.5,0)
    (5.75,3) edge (5.5,3);
    \path[-]
    (a30) edge (5.5,.75)
    (a32) edge (5.5,2.75)
    (a31) edge (-5.5,1.75)
    (a23) edge (4,3.5)
    (a03) edge (2,3.5)
    (a13) edge (-3,3.5)
    (a33) edge (-5,3.5)
    (a43) edge (5.5,3.25)
    (a41) edge (5.5,1.25)
    (a40) edge (-5.5,0.25)
    (a42) edge (-5.5,2.25);
    \path[dotted]
    (5.5,.75) edge (5.8,.9)
    (5.5,2.75) edge (5.8,2.9)
    (-5.5,1.75) edge (-5.8,1.9)
    (4,3.5) edge (4.4,3.7)
    (2,3.5) edge (2.4,3.7)
    (-3,3.5) edge (-3.4,3.7)
    (-5,3.5) edge (-5.4,3.7)
    (5.5,3.25) edge (5.8,3.4)
    (5.5,1.25) edge (5.8,1.4)
    (-5.5,.25) edge (-5.8,0.4)
    (-5.5,2.25) edge (-5.8,2.4);
    \path[arrow]
    (sink) edge[loop left] ()
    (e00) edge[bend left=8] (sink)
    (e01) edge[bend right=8] (sink)
    (e02) edge[bend left=8] (sink)
    (e03) edge[bend right=8] (sink)
    (a00) edge[bend right=8] (sink)
    (a01) edge[bend left=8] (sink)
    (a02) edge[bend right=8] (sink)
    (a03) edge[bend left=8] (sink);
  \end{tikzpicture}}
  \caption{The "energy semantics" of the
    "vector system" of \Cref{13-fig:mwg}: a circle (resp.\
    a square) at position $(i,j)$ of the grid denotes a configuration
    $\loc(i,j)$ (resp.\ $\loc'(i,j)$) controlled by~Eve (resp. Adam).}
    \label{13-fig:nrg}
\end{figure}

Given a "colouring" $\col{:}\,E\to C$ and an objective~$\Omega$, we
call the resulting game $(\energy(\?V),\col,\Omega)$ an energy
game.  In particular, we shall speak of "configuration
reachability", "coverability", "non-termination", and "parity@parity
vector game" "energy games" when replacing $\natural(\?V)$ by
$\energy(\?V)$ in \Crefrange{13-pb:reach}{13-pb:parity}; the
"existential initial credit" variants are defined similarly.

\begin{example}[Energy games]
\label{13-ex:cov-nrg}
  Consider the target configuration $\loc(2,2)$ in
  \Cref{13-fig:mwg,13-fig:nrg}.  Eve's "winning region" in the
  "configuration reachability" "energy game" is
  $\WE=\{\loc(n+2,n+2)\mid n\in\+N\}$, displayed on the left in
  \Cref{13-fig:cov-nrg}.  In the "coverability" "energy game", Eve's
  "winning region" is
  $\WE=\{\loc(m+2,n+2),\loc'(m+3,n+2)\mid m,n\in\+N\}$ displayed on
  the right in \Cref{13-fig:cov-nrg}.
\end{example}
\begin{figure}[htbp]
  \centering\scalebox{.48}{
  \begin{tikzpicture}[auto,on grid,node distance=2.5cm]
    \draw[step=1,lightgray!50,dotted] (-5.7,0) grid (5.7,3.8);
    \draw[color=white](0,-.3) -- (0,3.8);
    \node at (0,3.9) (sink) {\color{red!70!black}\boldmath$\sink$};
    \draw[step=1,lightgray!50] (1,0) grid (5.5,3.5);
    \draw[step=1,lightgray!50] (-1,0) grid (-5.5,3.5);
    \node at (0,0)[lightgray,font=\scriptsize,fill=white] {0};
    \node at (0,1)[lightgray,font=\scriptsize,fill=white] {1};
    \node at (0,2)[lightgray,font=\scriptsize,fill=white] {2};
    \node at (0,3)[lightgray,font=\scriptsize,fill=white] {3};
    \node at (1,3.9)[lightgray,font=\scriptsize,fill=white] {0};
    \node at (2,3.9)[lightgray,font=\scriptsize,fill=white] {1};
    \node at (3,3.9)[lightgray,font=\scriptsize,fill=white] {2};
    \node at (4,3.9)[lightgray,font=\scriptsize,fill=white] {3};
    \node at (5,3.9)[lightgray,font=\scriptsize,fill=white] {4};
    \node at (-1,3.9)[lightgray,font=\scriptsize,fill=white] {0};
    \node at (-2,3.9)[lightgray,font=\scriptsize,fill=white] {1};
    \node at (-3,3.9)[lightgray,font=\scriptsize,fill=white] {2};
    \node at (-4,3.9)[lightgray,font=\scriptsize,fill=white] {3};
    \node at (-5,3.9)[lightgray,font=\scriptsize,fill=white] {4};
    \node at (1,0)[s-eve-small,lose] (e00) {};
    \node at (1,1)[s-adam-small,lose](a01){};
    \node at (1,2)[s-eve-small,lose] (e02){};
    \node at (1,3)[s-adam-small,lose](a03){};
    \node at (2,0)[s-adam-small,lose](a10){};
    \node at (2,1)[s-eve-small,lose] (e11){};
    \node at (2,2)[s-adam-small,lose](a12){};
    \node at (2,3)[s-eve-small,lose] (e13){};
    \node at (3,0)[s-eve-small,lose] (e20){};
    \node at (3,1)[s-adam-small,lose](a21){};
    \node at (3,2)[s-eve-small,win] (e22){};
    \node at (3,3)[s-adam-small,lose](a23){};
    \node at (4,0)[s-adam-small,lose](a30){};
    \node at (4,1)[s-eve-small,lose] (e31){};
    \node at (4,2)[s-adam-small,lose](a32){};
    \node at (4,3)[s-eve-small,win] (e33){};
    \node at (5,0)[s-eve-small,lose] (e40){};
    \node at (5,1)[s-adam-small,lose](a41){};
    \node at (5,2)[s-eve-small,lose] (e42){};
    \node at (5,3)[s-adam-small,lose](a43){};
    \node at (-1,0)[s-adam-small,lose](a00){};
    \node at (-1,1)[s-eve-small,lose] (e01){};
    \node at (-1,2)[s-adam-small,lose](a02){};
    \node at (-1,3)[s-eve-small,lose] (e03){};
    \node at (-2,0)[s-eve-small,lose] (e10){};
    \node at (-2,1)[s-adam-small,lose](a11){};
    \node at (-2,2)[s-eve-small,lose] (e12){};
    \node at (-2,3)[s-adam-small,lose](a13){};
    \node at (-3,0)[s-adam-small,lose](a20){};
    \node at (-3,1)[s-eve-small,lose] (e21){};
    \node at (-3,2)[s-adam-small,lose](a22){};
    \node at (-3,3)[s-eve-small,lose] (e23){};
    \node at (-4,0)[s-eve-small,lose] (e30){};
    \node at (-4,1)[s-adam-small,lose](a31){};
    \node at (-4,2)[s-eve-small,lose] (e32){};
    \node at (-4,3)[s-adam-small,lose](a33){};
    \node at (-5,0)[s-adam-small,lose](a40){};
    \node at (-5,1)[s-eve-small,lose] (e41){};
    \node at (-5,2)[s-adam-small,lose](a42){};
    \node at (-5,3)[s-eve-small,lose] (e43){};
    \path[arrow] 
    (e11) edge (e00)
    (e22) edge (e11)
    (e31) edge (e20)
    (e32) edge (e21)
    (e21) edge (e10)
    (e12) edge (e01)
    (e23) edge (e12)
    (e33) edge (e22)
    (e13) edge (e02)
    (e43) edge (e32)
    (e42) edge (e31)
    (e41) edge (e30);
    \path[arrow] 
    (e11) edge (a01)
    (e20) edge (a10)
    (e22) edge (a12)
    (e31) edge (a21)
    (e32) edge (a22)
    (e21) edge (a11)
    (e12) edge (a02)
    (e30) edge (a20)
    (e10) edge (a00)
    (e13) edge (a03)
    (e23) edge (a13)
    (e33) edge (a23)
    (e43) edge (a33)
    (e42) edge (a32)
    (e41) edge (a31)
    (e40) edge (a30);
    \path[arrow] 
    (a11) edge (e01)
    (a20) edge (e10)
    (a22) edge (e12)
    (a31) edge (e21)
    (a32) edge (e22)
    (a21) edge (e11)
    (a12) edge (e02)
    (a30) edge (e20)
    (a10) edge (e00)
    (a33) edge (e23)
    (a23) edge (e13)
    (a13) edge (e03)
    (a43) edge (e33)
    (a42) edge (e32)
    (a41) edge (e31)
    (a40) edge (e30);
    \path[arrow] 
    (a01) edge (e22)
    (a10) edge (e31)
    (a11) edge (e32)
    (a00) edge (e21)
    (a02) edge (e23)
    (a12) edge (e33)
    (a22) edge (e43)
    (a21) edge (e42)
    (a20) edge (e41);
    \path[arrow] 
    (-5.5,3.5) edge (e43)
    (5.5,2.5) edge (e42)
    (2.5,3.5) edge (e13)
    (5.5,0.5) edge (e40)
    (-5.5,1.5) edge (e41)
    (-3.5,3.5) edge (e23)
    (-1.5,3.5) edge (e03)
    (4.5,3.5) edge (e33)
    (5.5,0) edge (e40)
    (5.5,2) edge (e42)
    (-5.5,1) edge (e41)
    (-5.5,3) edge (e43);
    \path[dotted]
    (-5.7,3.7) edge (-5.5,3.5)
    (5.7,2.7) edge (5.5,2.5)
    (2.7,3.7) edge (2.5,3.5)
    (5.7,0.7) edge (5.5,0.5)
    (-3.7,3.7) edge (-3.5,3.5)
    (-1.7,3.7) edge (-1.5,3.5)
    (4.7,3.7) edge (4.5,3.5)
    (-5.7,1.7) edge (-5.5,1.5)
    (5.75,0) edge (5.5,0)
    (5.75,2) edge (5.5,2)
    (-5.75,1) edge (-5.5,1)
    (-5.75,3) edge (-5.5,3);
    \path[arrow]
    (5.5,1) edge (a41)
    (-5.5,2) edge (a42)
    (-5.5,0) edge (a40)
    (5.5,3) edge (a43);
    \path[dotted]
    (5.75,1) edge (5.5,1)
    (-5.75,2) edge (-5.5,2)
    (-5.75,0) edge (-5.5,0)
    (5.75,3) edge (5.5,3);
    \path[-]
    (a30) edge (5.5,.75)
    (a32) edge (5.5,2.75)
    (a31) edge (-5.5,1.75)
    (a23) edge (4,3.5)
    (a03) edge (2,3.5)
    (a13) edge (-3,3.5)
    (a33) edge (-5,3.5)
    (a43) edge (5.5,3.25)
    (a41) edge (5.5,1.25)
    (a40) edge (-5.5,0.25)
    (a42) edge (-5.5,2.25);
    \path[dotted]
    (5.5,.75) edge (5.8,.9)
    (5.5,2.75) edge (5.8,2.9)
    (-5.5,1.75) edge (-5.8,1.9)
    (4,3.5) edge (4.4,3.7)
    (2,3.5) edge (2.4,3.7)
    (-3,3.5) edge (-3.4,3.7)
    (-5,3.5) edge (-5.4,3.7)
    (5.5,3.25) edge (5.8,3.4)
    (5.5,1.25) edge (5.8,1.4)
    (-5.5,.25) edge (-5.8,0.4)
    (-5.5,2.25) edge (-5.8,2.4);
    \path[arrow]
    (sink) edge[loop left] ()
    (e00) edge[bend left=8] (sink)
    (e01) edge[bend right=8] (sink)
    (e02) edge[bend left=8] (sink)
    (e03) edge[bend right=8] (sink)
    (a00) edge[bend right=8] (sink)
    (a01) edge[bend left=8] (sink)
    (a02) edge[bend right=8] (sink)
    (a03) edge[bend left=8] (sink);
  \end{tikzpicture}}\quad~~\scalebox{.48}{
  \begin{tikzpicture}[auto,on grid,node distance=2.5cm]
    \draw[step=1,lightgray!50,dotted] (-5.7,0) grid (5.7,3.8);
    \draw[color=white](0,-.3) -- (0,3.8);
    \node at (0,3.9) (sink) {\color{red!70!black}\boldmath$\sink$};
    \draw[step=1,lightgray!50] (1,0) grid (5.5,3.5);
    \draw[step=1,lightgray!50] (-1,0) grid (-5.5,3.5);
    \node at (0,0)[lightgray,font=\scriptsize,fill=white] {0};
    \node at (0,1)[lightgray,font=\scriptsize,fill=white] {1};
    \node at (0,2)[lightgray,font=\scriptsize,fill=white] {2};
    \node at (0,3)[lightgray,font=\scriptsize,fill=white] {3};
    \node at (1,3.9)[lightgray,font=\scriptsize,fill=white] {0};
    \node at (2,3.9)[lightgray,font=\scriptsize,fill=white] {1};
    \node at (3,3.9)[lightgray,font=\scriptsize,fill=white] {2};
    \node at (4,3.9)[lightgray,font=\scriptsize,fill=white] {3};
    \node at (5,3.9)[lightgray,font=\scriptsize,fill=white] {4};
    \node at (-1,3.9)[lightgray,font=\scriptsize,fill=white] {0};
    \node at (-2,3.9)[lightgray,font=\scriptsize,fill=white] {1};
    \node at (-3,3.9)[lightgray,font=\scriptsize,fill=white] {2};
    \node at (-4,3.9)[lightgray,font=\scriptsize,fill=white] {3};
    \node at (-5,3.9)[lightgray,font=\scriptsize,fill=white] {4};
    \node at (1,0)[s-eve-small,lose] (e00) {};
    \node at (1,1)[s-adam-small,lose](a01){};
    \node at (1,2)[s-eve-small,lose] (e02){};
    \node at (1,3)[s-adam-small,lose](a03){};
    \node at (2,0)[s-adam-small,lose](a10){};
    \node at (2,1)[s-eve-small,lose] (e11){};
    \node at (2,2)[s-adam-small,lose](a12){};
    \node at (2,3)[s-eve-small,lose] (e13){};
    \node at (3,0)[s-eve-small,lose] (e20){};
    \node at (3,1)[s-adam-small,lose](a21){};
    \node at (3,2)[s-eve-small,win] (e22){};
    \node at (3,3)[s-adam-small,lose](a23){};
    \node at (4,0)[s-adam-small,lose](a30){};
    \node at (4,1)[s-eve-small,lose] (e31){};
    \node at (4,2)[s-adam-small,win](a32){};
    \node at (4,3)[s-eve-small,win] (e33){};
    \node at (5,0)[s-eve-small,lose] (e40){};
    \node at (5,1)[s-adam-small,lose](a41){};
    \node at (5,2)[s-eve-small,win] (e42){};
    \node at (5,3)[s-adam-small,win](a43){};
    \node at (-1,0)[s-adam-small,lose](a00){};
    \node at (-1,1)[s-eve-small,lose] (e01){};
    \node at (-1,2)[s-adam-small,lose](a02){};
    \node at (-1,3)[s-eve-small,lose] (e03){};
    \node at (-2,0)[s-eve-small,lose] (e10){};
    \node at (-2,1)[s-adam-small,lose](a11){};
    \node at (-2,2)[s-eve-small,lose] (e12){};
    \node at (-2,3)[s-adam-small,lose](a13){};
    \node at (-3,0)[s-adam-small,lose](a20){};
    \node at (-3,1)[s-eve-small,lose] (e21){};
    \node at (-3,2)[s-adam-small,lose](a22){};
    \node at (-3,3)[s-eve-small,win] (e23){};
    \node at (-4,0)[s-eve-small,lose] (e30){};
    \node at (-4,1)[s-adam-small,lose](a31){};
    \node at (-4,2)[s-eve-small,win] (e32){};
    \node at (-4,3)[s-adam-small,win](a33){};
    \node at (-5,0)[s-adam-small,lose](a40){};
    \node at (-5,1)[s-eve-small,lose] (e41){};
    \node at (-5,2)[s-adam-small,win](a42){};
    \node at (-5,3)[s-eve-small,win] (e43){};
    \path[arrow] 
    (e11) edge (e00)
    (e22) edge (e11)
    (e31) edge (e20)
    (e32) edge (e21)
    (e21) edge (e10)
    (e12) edge (e01)
    (e23) edge (e12)
    (e33) edge (e22)
    (e13) edge (e02)
    (e43) edge (e32)
    (e42) edge (e31)
    (e41) edge (e30);
    \path[arrow] 
    (e11) edge (a01)
    (e20) edge (a10)
    (e22) edge (a12)
    (e31) edge (a21)
    (e32) edge (a22)
    (e21) edge (a11)
    (e12) edge (a02)
    (e30) edge (a20)
    (e10) edge (a00)
    (e13) edge (a03)
    (e23) edge (a13)
    (e33) edge (a23)
    (e43) edge (a33)
    (e42) edge (a32)
    (e41) edge (a31)
    (e40) edge (a30);
    \path[arrow] 
    (a11) edge (e01)
    (a20) edge (e10)
    (a22) edge (e12)
    (a31) edge (e21)
    (a32) edge (e22)
    (a21) edge (e11)
    (a12) edge (e02)
    (a30) edge (e20)
    (a10) edge (e00)
    (a33) edge (e23)
    (a23) edge (e13)
    (a13) edge (e03)
    (a43) edge (e33)
    (a42) edge (e32)
    (a41) edge (e31)
    (a40) edge (e30);
    \path[arrow] 
    (a01) edge (e22)
    (a10) edge (e31)
    (a11) edge (e32)
    (a00) edge (e21)
    (a02) edge (e23)
    (a12) edge (e33)
    (a22) edge (e43)
    (a21) edge (e42)
    (a20) edge (e41);
    \path[arrow] 
    (-5.5,3.5) edge (e43)
    (5.5,2.5) edge (e42)
    (2.5,3.5) edge (e13)
    (5.5,0.5) edge (e40)
    (-5.5,1.5) edge (e41)
    (-3.5,3.5) edge (e23)
    (-1.5,3.5) edge (e03)
    (4.5,3.5) edge (e33)
    (5.5,0) edge (e40)
    (5.5,2) edge (e42)
    (-5.5,1) edge (e41)
    (-5.5,3) edge (e43);
    \path[dotted]
    (-5.7,3.7) edge (-5.5,3.5)
    (5.7,2.7) edge (5.5,2.5)
    (2.7,3.7) edge (2.5,3.5)
    (5.7,0.7) edge (5.5,0.5)
    (-3.7,3.7) edge (-3.5,3.5)
    (-1.7,3.7) edge (-1.5,3.5)
    (4.7,3.7) edge (4.5,3.5)
    (-5.7,1.7) edge (-5.5,1.5)
    (5.75,0) edge (5.5,0)
    (5.75,2) edge (5.5,2)
    (-5.75,1) edge (-5.5,1)
    (-5.75,3) edge (-5.5,3);
    \path[arrow]
    (5.5,1) edge (a41)
    (-5.5,2) edge (a42)
    (-5.5,0) edge (a40)
    (5.5,3) edge (a43);
    \path[dotted]
    (5.75,1) edge (5.5,1)
    (-5.75,2) edge (-5.5,2)
    (-5.75,0) edge (-5.5,0)
    (5.75,3) edge (5.5,3);
    \path[-]
    (a30) edge (5.5,.75)
    (a32) edge (5.5,2.75)
    (a31) edge (-5.5,1.75)
    (a23) edge (4,3.5)
    (a03) edge (2,3.5)
    (a13) edge (-3,3.5)
    (a33) edge (-5,3.5)
    (a43) edge (5.5,3.25)
    (a41) edge (5.5,1.25)
    (a40) edge (-5.5,0.25)
    (a42) edge (-5.5,2.25);
    \path[dotted]
    (5.5,.75) edge (5.8,.9)
    (5.5,2.75) edge (5.8,2.9)
    (-5.5,1.75) edge (-5.8,1.9)
    (4,3.5) edge (4.4,3.7)
    (2,3.5) edge (2.4,3.7)
    (-3,3.5) edge (-3.4,3.7)
    (-5,3.5) edge (-5.4,3.7)
    (5.5,3.25) edge (5.8,3.4)
    (5.5,1.25) edge (5.8,1.4)
    (-5.5,.25) edge (-5.8,0.4)
    (-5.5,2.25) edge (-5.8,2.4);
    \path[arrow]
    (sink) edge[loop left] ()
    (e00) edge[bend left=8] (sink)
    (e01) edge[bend right=8] (sink)
    (e02) edge[bend left=8] (sink)
    (e03) edge[bend right=8] (sink)
    (a00) edge[bend right=8] (sink)
    (a01) edge[bend left=8] (sink)
    (a02) edge[bend right=8] (sink)
    (a03) edge[bend left=8] (sink);
  \end{tikzpicture}}
  \caption{The "winning regions" of Eve in the
    "configuration reachability" "energy game" (left) and the
    "coverability" "energy game"
    (right) on the graphs of \Cref{13-fig:mwg,13-fig:nrg} with target
    configuration~$\ell(2,2)$.  The winning vertices are in filled in
    green, while the losing ones are filled with white with a red
    border; the "sink" is always losing.}
    \label{13-fig:cov-nrg}
\commentAlt{Figure~\ref{13-fig:cov-nrg}: Two similar network diagrams, each with multiple layers of alternating circular and square nodes, showing dense interconnections, and two central vertical lines converging at a top point labeled with a symbol, with some nodes highlighted.}
\commentLongAlt{Figure~\ref{13-fig:cov-nrg}: The image displays two side-by-side diagrams, both depicting complex, symmetrical network structures. Each network is composed of multiple layers of alternating circular and square nodes, arranged in a grid-like fashion. Vertical and horizontal lines form the grid, with arrows indicating directed connections between nodes.

In both diagrams:
- Nodes in one layer are connected to nodes in adjacent layers, forming a dense, crisscrossing pattern.
- There's a central vertical axis where lines from the innermost nodes converge upwards to a single point labeled with an upward arrow and a perpendicular symbol.

The left diagram features nodes outlined in red, distinguishing them from the filled grey nodes. The right diagram shows the same network structure, but with a different set of nodes highlighted in red, suggesting a change in state or path compared to the left, and visually distinct from the previous similar images. In the right diagram, the top-right quarter contains filled grey nodes, while the bottom-left quarter has red-outlined nodes.}
\end{figure}

The reader might have noticed that the "natural semantics" of the
"asymmetric" system of \Cref{13-fig:avg} and the "energy semantics" of
the system of \Cref{13-fig:mwg} are essentially the same.  This
correspondence is quite general.
\begin{lemma}[Energy vs.\ asymmetric vector games]
\label{13-lem:nrg}
  "Energy games" and "asymmetric" "vector games" are
  \logspace-equivalent for "configuration reachability",
  "coverability", "non-termination", and "parity@parity vector games",
  both with "given" and with "existential initial credit".
\end{lemma}
\begin{proof}
  Let us first reduce "asymmetric vector games" to "energy games".
  Given $\?V$, $\col$, and $\Omega$ where $\?V$ is "asymmetric" and
  $\Eve$ loses if the play ever visits the "sink"~$\sink$, we see that
  $\Eve$ wins $(\natural(\?V),\col,\Omega)$ from some $v\in V$ if and
  only if she wins $(\energy(\?V),\col,\Omega)$ from $v$.  Of course,
  this might not be true if~$\?V$ is not "asymmetric", as seen for
  instance in \Cref{13-ex:cov,13-ex:cov-nrg}.

  \medskip Conversely, let us reduce "energy games" to "asymmetric
  vector games".  Consider
  $\?V=(\Loc,\Act,\Loc_\mEve,\Loc_\mAdam,\dd)$, a colouring $\col$
  defined from a vertex colouring $\vcol$ by
  $\col(e)\eqdef\vcol(\ing(e))$, and an objective $\Omega$, where
  $\vcol$ and $\Omega$ are such that $\Eve$ loses if the play ever
  visits the "sink"~$\sink$ and such that, for all $\pi\in C^\ast$,
  $p\in C$, and $\pi'\in C^\omega$, $\pi p\pi'\in\Omega$ if and only
  if $\pi pp\pi'\in\Omega$ (we shall call $\Omega$
  \emph{stutter-invariant}, and the objectives in the statement are
  indeed stutter-invariant).  We construct an "asymmetric vector
  system"
  $\?V'\eqdef(\Loc\uplus\Loc_\Act,\Act',\Loc_\mEve\uplus\Loc_\Act,\Loc_\mAdam,\dd)$
  where we add the following locations controlled by Eve:
    \begin{align*}
      \Loc_\Act&\eqdef\{\loc_a\mid a=(\loc\step{\vec
                 u}\loc')\in\Act\text{ and }\loc\in\Loc_\mAdam\}\;.
      \intertext{We also modify the set of actions:}
      \Act'&\eqdef\{\loc\step{\vec u}\loc'\mid \loc\step{\vec
             u}\loc'\in\Act\text{ and }\loc\in\Loc_\mEve\}\\
      &\:\cup\:\{\loc\step{\vec 0}\loc_a,\;\loc_a\step{\vec u}\loc'\mid a=(\loc\step{\vec u}\loc')\in\Act\text{ and }\loc\in\Loc_\mAdam\}\;.
    \end{align*}
    \Cref{13-fig:avg} presents the result of this reduction on the
    system of \Cref{13-fig:mwg}.  We define a vertex colouring
    $\vcol'$ of $\arena_\+N(\?V')$ with $\vcol'(v)\eqdef\vcol(v)$ for
    all $v\in \Loc\times\+N^\dd\uplus\{\sink\}$ and
    $\vcol'(\loc_a(\vec v))\eqdef\vcol(\loc(\vec v))$ if
    $a=(\loc\step{\vec u}\loc')\in\Act$.  Then, for all vertices
    $v\in V$, Eve wins from~$v$ in the "energy game"
    $(\energy(\?V),\col,\Omega)$ if and only if she wins from~$v$ in
    the "vector game" $(\natural(\?V'),\col',\Omega)$.  The crux of
    the argument is that, in a configuration $\loc(\vec v)$ where
    $\loc\in\Loc_\mAdam$, if $a=(\loc\step{\vec u}\loc')\in\Act$ is an
    action with $\vec v+\vec u\not\geq\vec 0$, in the "energy
    semantics", Adam can force the play into the "sink" by
    playing~$a$; the same occurs in $\?V'$ with the "natural
    semantics", as Adam can now choose to play
    $\loc\step{\vec 0}\loc_a$ where Eve has only
    $\loc_a\step{\vec u}\loc'$ at her disposal, which leads to the sink.
\end{proof}

In turn, "energy games" with "existential initial credit" are related
to the "multi-dimensional mean-payoff games" of
\Cref{14-chap:multiobjective}.


\subsection{Bounded semantics}
\label{13-sec:bounding}

While Adam wins immediately in an "energy game" if a resource gets
depleted, he also wins in a "bounding game" if a resource reaches a
certain bound~$B$.  
This is
a \emph{hard upper bound}, allowing to model systems where exceeding a
capacity results in failure, like a dam that overflows and floods the
area.  We define for a bound~$B\in\+N$ the bounded semantics
$\bounded(\?V)=(V^B,E^B,\VE^B,\VA^B)$ of a "vector system"~$\?V$ by
\begin{align*}
  V^B&\eqdef\{\loc(\vec v)\mid\loc\in\Loc\text{ and }\|\vec v\|<B\}\;,\\
  E^B&\eqdef \{(\loc(\vec v),\loc'(\vec v+\vec u))\mid\loc\step{\vec
       u}\loc'\in\Act,\vec v+\vec u\geq\vec 0,\text{ and }\|\vec
       v+\vec u\|<B\}\\
     &\:\cup\:\{(\loc(\vec v),\sink)\mid\forall\loc\step{\vec
               u}\loc'\in\Act\mathbin.\vec v+\vec u\not\geq\vec
               0\text{ or }\|\vec v+\vec u\|\geq B\}
     \cup\{(\sink,\sink)\}\;.
\end{align*}
As usual, $\VE^B\eqdef V^B\cap\Loc_\mEve\times\+N^\dd$ and
$\VA^B\eqdef V^B\cap\Loc_\mAdam\times\+N^\dd$.  Any edge from the
"energy semantics" that would bring to a configuration $\loc(\vec v)$
with $\vec v(i)\geq B$ for some $1\leq i\leq\dd$ leads instead to the
sink.  All the configurations in this arena have "norm" less than~$B$,
thus $|V^B|=|\Loc| B^\dd+1$, and the qualitative games of
\Cref{3-chap:regular} are decidable over this "arena".

Our focus here is on "non-termination" games played on the "bounded
semantics" where~$B$ is not given as part of the input, but quantified
existentially.  As usual, the "existential initial credit" variant
of \Cref{13-pb:bounding} is obtained by quantifying~$\vec v_0$
existentially in the question.
\decpb["bounding game" with "given initial credit"]%
{\label{13-pb:bounding} A "vector system"
  $\?V=(\Loc,\Act,\Loc_\mEve,\Loc_\mAdam,\dd)$, an initial location
  $\loc_0\in\Loc$, and an initial credit $\vec v_0\in\+N^\dd$.}%
  {Does there exist $B\in\+N$ such that Eve has a strategy to avoid the
  "sink"~$\sink$ from $\loc_0(\vec v_0)$ in the "bounded
  semantics"?  That is, does there exist $B\in\+N$ such that she wins
  the bounding game $(\bounded(\?V),\col,\Safe)$ from
  $\loc_0(\vec v_0)$, where $\col(e)\eqdef\Lose$ if and only if $\ing(e)=\sink$?}

\begin{lemma}\label{13-lem:parity2bounding}
  There is a \logspace\ reduction from "parity@parity vector games"
  "asymmetric" "vector games" to "bounding games", both with "given"
  and with "existential initial credit".
\end{lemma}
\begin{proof}
  Given an "asymmetric vector system"
  $\?V=(\Loc,\Act,\Loc_\mEve,\Loc_\mAdam,\dd)$, a location colouring
  $\lcol{:}\,\Loc\to\{1,\dots,2d\}$, and an initial location
  $\loc_0\in\Loc$, we construct a "vector system" $\?V'$ of dimension
  $\dd'\eqdef\dd+d$ as described in \Cref{13-fig:bounding}, where the
  priorities in~$\?V$ for $p\in\{1,\dots,d\}$ are indicated above the
  corresponding locations.
  
  \begin{figure}[htbp]
    \centering
    \begin{tikzpicture}[auto,on grid,node distance=1.5cm]
      \node(to){$\mapsto$};
      \node[anchor=east,left=2.5cm of to](mm){"asymmetric vector system"~$\?V$};
      \node[anchor=west,right=2.5cm of to](mwg){"vector system"~$\?V'$};
      \node[below=1.3cm of to](imap){$\rightsquigarrow$};
      \node[s-eve,left=2.75cm of imap](i0){$\loc$};
      \node[black!50,above=.4 of i0,font=\scriptsize]{$2p$};
      \node[right=of i0](i1){$\loc'$};
      \node[right=1.25cm of imap,s-eve](i2){$\loc$};
      \node[right=1.8 of i2,s-eve-small](i3){};      
      \node[right=1.8 of i3](i4){$\loc'$};      
      \path[arrow,every node/.style={font=\footnotesize}]
      (i0) edge node{$\vec u$} (i1)
      (i2) edge[loop above] node{$\forall 1\leq i\leq\dd\mathbin.-\vec e_i$} ()
      (i2) edge node{$\vec u$} (i3)
      (i3) edge[loop below] node{$\forall 1\leq j\leq p\mathbin.\vec e_{\dd+j}$} ()
      (i3) edge node{$\vec 0$} (i4);
      \node[below=2cm of imap](dmap){$\rightsquigarrow$};
      \node[s-eve,left=2.75cm of dmap](d0){$\loc$};
      \node[black!50,above=.4 of d0,font=\scriptsize]{$2p-1$};
      \node[right=of d0](d1){$\loc'$};
      \node[right=1.25cm of dmap,s-eve](d2){$\loc$};
      \node[right=2 of d2](d3){$\loc'$};
      \path[arrow,every node/.style={font=\footnotesize}]
      (d0) edge node{$\vec u$} (d1)
      (d2) edge[loop above] node{$\forall 1\leq i\leq\dd\mathbin.-\vec e_i$} ()
      (d2) edge node{$\vec u-\vec e_{\dd+p}$} (d3);
      \node[below=1.1cm of dmap](zmap){$\rightsquigarrow$};
      \node[s-adam,left=2.75cm of zmap](z0){$\loc$};
      \node[black!50,above=.4 of z0,font=\scriptsize]{$2p$};
      \node[right=of z0](z1){$\loc'$};
      \node[right=1.25cm of zmap,s-adam](z2){$\loc$};
      \node[right=of z2,s-eve-small](z3){};
      \node[right=of z3](z4){$\loc'$};
      \path[arrow,every node/.style={font=\footnotesize}]
      (z0) edge node{$\vec 0$} (z1)
      (z2) edge node{$\vec 0$} (z3)
      (z3) edge node{$\vec 0$} (z4)
      (z3) edge[loop below] node{$\forall 1\leq j\leq p\mathbin.\vec e_{\dd+j}$} ();
      \node[below=1.6cm of zmap](amap){$\rightsquigarrow$};
      \node[s-adam,left=2.75cm of amap](a0){$\loc$};
      \node[black!50,above=.4 of a0,font=\scriptsize]{$2p-1$};
      \node[right=of a0](a1){$\loc'$};
      \node[right=1.25cm of amap,s-adam](a2){$\loc$};
      \node[right=2 of a2](a3){$\loc'$};
      \path[arrow,every node/.style={font=\footnotesize}]
      (a0) edge node{$\vec 0$} (a1)
      (a2) edge node{$-\vec e_{\dd+p}$} (a3);
    \end{tikzpicture}
    \caption{Schema of the reduction to
      "bounding games" in the proof of \Cref{13-lem:parity2bounding}.}
      \label{13-fig:bounding}
  \end{figure}
  
  If Eve wins the "bounding game" played over $\?V'$ from some
  configuration $\loc_0(\vec v_0)$, then she also wins the "parity
  vector game" played over~$\?V$ from the configuration $\loc_0(\vec
  v'_0)$ where $\vec v'_0$ is the projection of $\vec v_0$
  to~$\+N^\dd$.  Indeed, she can play essentially the same strategy:
  by \Cref{13-lem:mono} she can simply ignore the new decrement
  self loops, while the actions on the components in
  $\{\dd+1,\dots,\dd+d\}$ ensure that the maximal priority visited
  infinitely often is even---otherwise some decrement $-\vec
  e_{\dd+p}$ would be played infinitely often but the increment $\vec
  e_{\dd+p}$ only finitely often.

  \medskip
  Conversely, consider the "parity@parity vector game" game~$\game$ played over
  $\natural(\?V)$ with the colouring defined by~$\lcol$.  Then the
  "Pareto limit" of the game is finite, thus there exists a natural
  number
  \begin{equation}\label{13-eq:b0}
    B_0\eqdef 1+\max_{\loc_0(\vec v_0)\in\mathsf{Pareto}(\?G)}\|\vec
  v_0\|
  \end{equation} bounding the "norms" of the minimal winning configurations.
  For a vector~$\vec v$ in~$\+N^\dd$, let us write $\capp[B_0]v$ for
  the vector `capped' at~$B$: for all~$1\leq i\leq\dd$,
  $\capp[B_0]v(i)\eqdef\vec v(i)$ if $\vec v(i)<B_0$ and
  $\capp[B_0]v\eqdef B_0$ if $\vec v(i)\geq B_0$.


  Consider now some configuration $\loc_0(\vec
  v_0)\in\mathsf{Pareto}(\game)$.  As seen in \Cref{13-lem:finmem},
  since $\loc_0(\vec v_0)\in\WE(\game)$, there is a finite
  "self-covering tree" witnessing the fact, and an associated winning
  strategy.  Let $H(\loc_0(\vec v_0))$ denote the height of this
  "self-covering tree" and observe that all the configurations in this
  tree have norm bounded by $\|\vec v_0\|+\|\Act\|\cdot H(\loc_0(\vec
  v_0))$.
  Let us define
  \begin{equation}\label{13-eq:b}
   B\eqdef B_0+(\|\Act\|+1)\cdot \max_{\loc_0(\vec
  v_0)\in\mathsf{Pareto}(\?G)}H(\loc_0(\vec v_0))\;.
  \end{equation}
  This is a bound on the norm of the configurations appearing on the
  (finitely many) self-covering trees spawned by the elements
  of~$\mathsf{Pareto}(\game)$.  Note that $B\geq B_0+(\|\Act\|+1)$ since
  a self-covering tree has height at least~one.

  Consider the "non-termination" game
  $\game_B\eqdef(\bounded(\?V'),\col',\Safe)$ played over the
  "bounded semantics" defined by~$B$, where $\col'(e)=\Lose$ if and
  only if $\ing(e)=\sink$.  Let $\vec b\eqdef\sum_{1\leq p\leq
  d}(B-1)\cdot\vec e_{\dd+p}$.
  {\renewcommand{\qedsymbol}{}
  \begin{claim}\label{13-cl:parity2bounding} If $\loc_0(\vec
    v)\in\WE(\game)$, then
    $\loc_0(\capp[B_0]{v}+\vec b)\in\WE(\game_B)$.
  \end{claim}}
  Indeed, by definition of the "Pareto
  limit"~$\mathsf{Pareto}(\game)$, if $\loc_0(\vec v)\in\WE(\game)$,
  then there exists~$\vec v_0\leq\vec v$ such that $\loc_0(\vec
  v_0)\in\mathsf{Pareto}(\game)$.  By definition of the bound~$B_0$,
  $\|\vec v_0\|<B_0$, thus $\vec v_0\leq\capp[B_0]v$.  Consider the
  "self-covering tree" of height~$H(\loc_0(\vec v_0))$ associated
  to~$\loc_0(\vec v_0)$, and the strategy~$\sigma'$ defined by the
  memory structure from the
  proof of \Cref{13-lem:finmem}.  This is a winning strategy for Eve 
  in $\game$ starting from $\loc_0(\vec v_0)$, and
  by \Cref{13-lem:mono}, it is also winning
  from~$\loc_0(\capp[B_0]v)$.
    
  Here is how Eve wins $\game_B$ from~$\loc_0(\capp[B_0]v+\vec b)$.
  She essentially follows the strategy~$\sigma'$, with two
  modifications.  First, whenever $\sigma'$ goes to a "return node"
  $\loc(\vec v)$ instead of a leaf $\loc(\vec v')$---thus $\vec
  v\leq\vec v'$---, the next time Eve has the control, she uses the
  self loops to decrement the current configuration by $\vec v'-\vec
  v$.  This ensures that any play consistent with the modified
  strategy remains between zero and $B-1$ on the components
  in~$\{1,\dots,\dd\}$.  (Note that if she never regains the control,
  the current vector never changes any more since~$\?V$ is
  "asymmetric".)

  Second, whenever a play in~$\game$ visits a location with even
  parity~$2p$ for some~$p$ in~$\{1,\dots,d\}$, Eve has the opportunity
  to increase the coordinates in~$\{\dd+1,\dots,\dd+p\}$ in~$\game_B$.
  She does so and increments until all these components reach~$B-1$.
  This ensures that any play consistent with the modified strategy
  remains between zero and $B-1$ on the components
  in~$\{\dd+1,\dots,\dd+p\}$.  Indeed, $\sigma'$ guarantees that the
  longest sequence of moves before a play visits a location with
  maximal even priority is bounded by $H(\loc_0(\vec v_0))$, thus the
  decrements $-\vec e_{\dd+p}$ introduced in~$\game_B$ by the
  locations from~$\game$ with odd parity~$2p-1$ will never force the
  play to go negative.
\end{proof}

The bound~$B$ defined in~\Cref{13-eq:b} in the previous proof is not
constructive, and possibly much larger than really required.
Nevertheless, one can sometimes show that an explicit~$B$ suffices in
a "bounding game".
A simple example is provided by the "coverability" "asymmetric"
"vector games" with "existential initial credit" arising from
\Cref{13-rmk:cov2parity}, \textit{i.e.}, where the objective is to reach some
location~$\loc_f$.  Indeed, it is rather straightforward that there
exists a suitable initial credit such that Eve wins the game if and
only if she wins the finite reachability game played over the
underlying directed graph over~$\Loc$ where we ignore the counters.
Thus, for an initial location~$\loc_0$, $B_0=|\Loc|\cdot\|\Act\|+1$
bounds the norm of the necessary initial credit, while a simple path
may visit at most~$|\Loc|$ locations, thus
$B=B_0+|\Loc|\cdot\|\Act\|$ suffices for Eve to win the constructed
"bounding game".

In the general case of "bounding games" with "existential initial
credit", an explicit bound can be established.  The proof goes
along very different lines and is too involved to fit in this chapter,
but we refer the reader
to \cite{Jurdzinski.Lazic.ea:2015}
for details.
\begin{theorem}[Bounds on bounding]
\label{13-thm:bounding}
  If Eve wins a "bounding game" with "existential initial credit"
  defined by a "vector
  system"~$\?V=(\Loc,\Act,\Loc_\mEve,\Loc_\mAdam,\dd)$, then an
  initial credit $\vec v_0$ with $\|\vec
  v_0\|=(4|\Loc|\cdot\|\Act\|)^{2(\dd+2)^3}$ and a bound
  $B=2(4|\Loc|\cdot\|\Act\|)^{2(\dd+2)^3}+1$ suffice for this.
\end{theorem}

\Cref{13-thm:bounding} also yields a way of handling "bounding games"
with "given initial credit".  
  

\section{The complexity of asymmetric monotone games}
\label{13-sec:complexity}
Unlike general "vector games" and "configuration reachability"
"asymmetric" ones, "coverability", "non-termination", and
"parity@parity vector game" "asymmetric vector games" are decidable.
We survey in this section the best known complexity bounds for every
case; see \Cref{13-tbl:cmplx} at the end of the chapter for a summary.

\subsection{Upper bounds}
\label{13-sec:up}
We begin with complexity upper bounds.  The main results are that
"parity@parity vector game" games with "existential initial credit"
can be solved in \coNP, but are in \kEXP[2] with "given initial
credit".  In both cases however, the complexity is pseudo-polynomial
if both the dimension~$\dd$ and the number of priorities~$d$ are
fixed, which is rather good news: one can hope that, in practice, both
the number of different resources (encoded by the counters) and the
complexity of the functional specification (encoded by the parity
condition) are tiny compared to the size of the system.

\subsubsection{Existential initial credit}
\label{13-sec:up-exist}

\paragraph{Counterless Strategies}
Consider a "strategy"~$\tau$ of Adam in a "vector game".  In all the
games we consider, "uniform" "positional" strategies suffice over the
infinite "arena" \[\natural(\?V)=(V,E,\VE,\VA):\] $\tau$ maps vertices
in~$V$ to edges in~$E$.  We call~$\tau$ counterless if, for all
locations $\loc\in\Loc_\mAdam$ and all vectors
$\vec v,\vec v'\in\+N^\dd$, $\tau(\loc(\vec v))=\tau(\loc(\vec v'))$.
A "counterless" strategy thus only considers the current location of
the play.
\begin{lemma}[Counterless strategies]
\label{13-lem:counterless}
  Let $\?V=(\Loc,\Act,\Loc_\mEve,\Loc_\mAdam,\dd)$ be an "asymmetric
  vector system", $\loc_0\in\Loc$ be a location, and
  $\lcol{:}\,\Loc\to\{1,\dots,d\}$ be a location colouring.  If Adam 
  wins from $\loc_0(\vec v)$ for every initial credit~$\vec v$ in the
  "parity@parity vector game" game played over $\?V$ with~$\lcol$, then
  he has a single "counterless strategy" such that he wins from
  $\loc_0(\vec v)$ for every initial credit~$\vec v$.
\end{lemma}
\begin{proof}
  Let $\Act_\mAdam\eqdef\{(\loc\step{\vec
    u}\loc')\in\Act\mid\loc\in\Loc_\mAdam\}$ be the set of actions
  controlled by Adam.  We assume without loss of generality that
  every location $\loc\in\Loc_\mAdam$ has either one or two outgoing
  actions, thus $|\Loc_\mAdam|\leq|\Act_\mAdam|\leq
  2|\Loc_\mAdam|$.  We proceed by induction over $|\Act_\mAdam|$.  For
  the base case, if $|\Act_\mAdam|=|\Loc_\mAdam|$ then every location
  controlled by Adam has a single outgoing action, thus any
  strategy for Adam is trivially "counterless".

  For the induction step, consider some location
  $\hat\loc\in\Loc_\mAdam$ with two outgoing actions
  $a_l\eqdef\hat\loc\step{\vec 0}\loc_l$ and
  $a_r\eqdef\hat\loc\step{\vec 0}\loc_r$.  Let $\?V_l$ and $\?V_r$ be
  the "vector systems" obtained from~$\?V$ by removing
  respectively~$a_r$ and~$a_l$ from~$\Act$, \textit{i.e.}, by using
  $\Act_l\eqdef\Act\setminus\{a_r\}$ and
  $\Act_r\eqdef\Act\setminus\{a_l\}$.  If $\Adam$ wins the
  "parity@parity vector game" game from $\loc(\vec v)$ for every
  initial credit~$\vec v$ in either $\?V_l$ or $\?V_r$, then by
  induction hypothesis he has a "counterless" winning strategy winning
  from $\loc(\vec v)$ for every initial credit~$\vec v$, and the same
  strategy is winning in~$\?V$ from $\loc(\vec v)$ for every initial
  credit~$\vec v$.

  In order to conclude the proof, we show that, if Adam loses in
  $\?V_l$ from $\loc_0(\vec v_l)$ for some $\vec v_l\in\+N^\dd$ and in
  $\?V_r$ from $\loc_0(\vec v_r)$ for some $\vec v_r\in\+N^\dd$, then
  there exists $\vec v_0\in\+N^\dd$ such that Eve wins from
  $\loc_0(\vec v_0)$ in~$\?V$.  Let $\sigma_l$ and $\sigma_r$ denote
  Eve's winning strategies in the two games.  By a slight abuse of
  notations (justified by the fact that we are only interested in a
  few initial vertices), we see plays as sequences of actions and
  strategies as maps $\Act^\ast\to\Act$.
  Consider the set of
  plays consistent with~$\sigma_r$ starting from $\loc_0(\vec v_r)$.
  If none of those plays visits $\hat\loc$, then $\Eve$ wins in $\?V$
  from $\loc_0(\vec v_r)$ and we conclude.  Otherwise, there is some
  finite prefix~$\hat\pi$ of a play that
  visits~$\hat\loc(\hat{\vec v})$ for some vector
  $\hat{\vec v}=\vec v_r+\weight(\hat\pi)$.  We let
  $\vec v_0\eqdef\vec v_l+\hat{\vec v}$ and show that Eve wins from
  $\loc_0(\vec v_0)$.

    We define now a strategy $\sigma$ for $\Eve$ over~$\?V$ that
    switches between applying~$\sigma_l$ and~$\sigma_r$ each time
    $a_r$ is used and switches back each time~$a_l$ is used.  More
    precisely, given a finite or infinite sequence~$\pi$ of actions,
    we decompose $\pi$ as $\pi_1 a_1 \pi_2 a_2 \pi_3\cdots$ where each
    segment $\pi_j\in(\Act\setminus\{a_l,a_r\})^\ast$ does not use
    either~$a_l$ nor~$a_r$ and each $a_j\in\{a_l,a_r\}$.  The
    associated mode $m(j)\in\{l,r\}$ of a segment~$\pi_j$
    is~$m(1)\eqdef l$ for the initial segment and otherwise
    $m(j)\eqdef l$ if $e_{j-1}=a_l$ and $m(j)\eqdef r$ otherwise.  The
    $l$-subsequence associated with $\pi$ is the sequence of segments
    $\pi(l)\eqdef\pi_{l_1}a_{l_2-1}\pi_{l_2}a_{l_3-1}\pi_{l_3}\cdots$
    with "mode"~$m(l_i)=l$, while the $r$-subsequence is the sequence
    $\pi(r)\eqdef\hat\pi a_{r_1-1}\pi_{r_1}a_{r_2-1}\pi_{r_2}\cdots$
    with "mode"~$m(r_i)=r$ prefixed by~$\hat\pi$.  Then we let
    $\sigma(\pi)\eqdef\sigma_{m}(\pi(m))$ where $m\in\{l,r\}$ is the
    "mode" of the last segment of~$\pi$.

    Consider an infinite play $\pi$ consistent with~$\sigma$ starting
    from~$\loc_0(\vec v_0)$.  Since $\vec v_0\geq\vec v_l$ and
    $\vec v_0\geq \vec v_r+\weight(\hat\pi)$, $\pi(l)$ and $\pi(r)$
    starting from~$\loc_0(\vec v_0)$ are consistent with
    "simulating"---in the sense of \Cref{13-lem:mono}---$\sigma_l$
    from $\loc_0(\vec v_l)$ and $\sigma_r$ from $\loc_0(\vec v_r)$.
    Let $\pi'$ be a finite prefix of~$\pi$.  Then
    $\weight(\pi')=\weight(\pi'(l))+\weight(\pi'(r))$ where $\pi'(l)$
    is a prefix of~$\pi(l)$ and $\pi'(r)$ of~$\pi(r)$, thus
    $\weight(\pi'(l))\leq\vec v_l$ and
    $\weight(\pi'(r))\leq\vec v_r+\weight(\hat\pi)$, thus
    $\weight(\pi')\leq\vec v_0$: the play~$\pi$ avoids the "sink".
    Furthermore, the maximal priority seen infinitely often along
    $\pi(l)$ and $\pi(r)$ is even (note that one of~$\pi(l)$
    and~$\pi(r)$ might not be infinite), thus the maximal priority
    seen infinitely often along~$\pi$ is also even.  This shows
    that~$\sigma$ is winning for Eve from $\loc_0(\vec v_0)$.
\end{proof}

We are going to exploit \Cref{13-lem:counterless}
in \Cref{13-thm:exist-easy} in order to prove a~\coNP\ upper bound for
"asymmetric games" with "existential initial credit": it suffices in
order to decide those games to guess a "counterless" winning
strategy~$\tau$ for Adam and check that it is indeed winning by
checking that Eve loses the one-player game arising from~$\tau$.
This last step requires an algorithmic result of independent interest.

\paragraph{One-player Case}
Let $\?V=(\Loc,\Act,\dd)$ be a "vector addition system with states",
$\lcol{:}\,\Loc\to\{1,\dots,d\}$ a location colouring, and
$\loc_0\in\Loc$ an initial location.  Then Eve wins the
"parity@parity vector game" one-player game from~$\loc_0(\vec v_0)$
for some initial credit~$\vec v_0$ if and only if there exists some
location such that
\begin{itemize}
\item $\loc$ is reachable from~$\loc_0$ in the directed graph
  underlying~$\?V$ and
\item there is a cycle~$\pi\in\Act^\ast$ from $\loc$ to itself such
  that $\weight(\pi)\geq 0$ and the maximal priority occurring
  along~$\pi$ is even.
\end{itemize}
Indeed, assume we can find such a location~$\loc$.  Let
$\hat\pi\in\Act^\ast$ be a path from~$\loc_0$ to~$\loc$ and $\vec
v_0(i)\eqdef\max\{\|\weight(\pi')\|\mid\pi'\text{ is a prefix of
}\hat\pi\pi\}$ for all $1\leq i\leq\dd$.  Then $\loc_0(\vec v_0)$ can
reach $\loc(\vec v_0+\weight(\hat\pi))$ in the "natural semantics"
of~$\?V$ by following~$\hat\pi$, and then $\loc(\vec v_0+\vec
W(\hat\pi)+n\weight(\pi))\geq \loc(\vec v_0+\weight(\hat\pi))$ after
$n$~repetitions of the cycle~$\pi$.  The infinite play arising from
this strategy has an even maximal priority.

Conversely, if Eve wins, then there is a winning play
$\pi\in\Act^\omega$ from $\loc_0(\vec v_0)$ for some $\vec v_0$.
Recall that $(V,{\leq})$ is a "wqo", and we argue as in
\Cref{13-lem:finmem} that there is indeed such a location~$\loc$.

\medskip
Therefore, solving one-player "parity vector games" boils down to
determining the existence of a cycle with non-negative effect and even
maximal priority.  We shall use linear programming techniques in order
to check the existence of such a cycle in polynomial
time~\cite{Kosaraju.Sullivan:1988}.

\medskip
Let us start with a relaxed problem: we call a
multi-cycle a non-empty finite set of cycles~$\Pi$ and let
$\weight(\Pi)\eqdef\sum_{\pi\in\Pi}\weight(\pi)$ be its weight; we write
$t\in\Pi$ if~$t\in\pi$ for some $\pi\in\Pi$.
Let $M\in 2^{\Act}$ be a set of `mandatory' subsets of actions and
$F\subseteq\Act$ a set of `forbidden' actions.  Then we say that
$\Pi$ is non-negative if $\weight(\Pi)\geq\vec 0$, and that it is
suitable for~$(M,F)$ if for all $\Act'\in M$ there exists
$t\in\Act'$ such that $t\in\Pi$, and if for all $t\in F$,
$t\not\in\Pi$.  We use the same terminology for a single cycle~$\pi$.

\begin{lemma}[Linear programs for suitable non-negative multi-cycles]
\label{13-lem:zmulticycle}
  Let $\?V$ be a "vector addition system with states",
  $M\in 2^{\Act}$, and $F\subseteq\Act$.  We can check in polynomial
  time whether~$\?V$ contains a "non-negative" "multi-cycle"~$\Pi$
  "suitable" for~$(M,F)$.
\end{lemma}
\begin{proof}
  We reduce the problem to solving a linear program.  For a
  location~$\loc$, let
  $\mathrm{in}(\loc)\eqdef\{(\loc'\step{\vec u}\loc)\in\Act\mid
  \loc'\in\Loc\}$
  and
  $\mathrm{out}(\loc)\eqdef\{(\loc\step{\vec u}\loc')\in\Act\mid
  \loc'\in\Loc\}$ be its sets of incoming and outgoing actions.  The
  linear program has a variable $x_a$ for each action $a\in\Act$,
  which represents the number of times the action~$a$ occurs in
  the "multi-cycle".  It consists of the following constraints:
  \begin{align*}
    \forall\loc&\in\Loc,&\sum_{a\in\mathrm{in}(\loc)}x_a&=\sum_{a\in\mathrm{out}(\loc)}x_a\;,\tag{"multi-cycle"}\\
    \forall a&\in\Act,&x_a&\geq 0\;,\tag{non-negative uses}\\
    \forall i&\in\{1,\dots,\dd\},&\sum_{a\in\Act} x_a\cdot\weight(t)(i)&\geq
                                            0\;,\tag{"non-negative" weight}\\
    &&\sum_{a\in\Act}x_a&\geq 0\tag{non-empty}\\
    \forall \Act'&\in M,&\sum_{a\in\Act'}x_a&\geq 0\;,\tag{every subset
                                               in~$M$ is used}\\
    \forall a&\in F,&x_a&= 0\;.\tag{no forbidden actions}
  \end{align*}
  As solving a linear program is in polynomial time~\Cref{1-thm:linear_programming}, the result follows.
\end{proof}

Of course, what we are aiming for is finding a "non-negative"
\emph{cycle} "suitable" for $(M,F)$ rather than a "multi-cycle".
Let us define for this the relation $\loc\sim\loc'$ over~$\Loc$ if
$\loc=\loc'$ or if there exists a "non-negative" "multi-cycle"~$\Pi$
"suitable" for~$(M,F)$ such that~$\loc$ and~$\loc'$ belong to some
cycle~$\pi\in\Pi$.
\begin{fact}
\label{13-fact:sim} 
The relation~$\sim$ is an equivalence relation.
\end{fact}
\begin{proof}
  Symmetry and reflexivity are trivial, and if $\loc\sim\loc'$ and
  $\loc'\sim\loc''$ because~$\loc$ and~$\loc'$ appear in some cycle
  $\pi\in\Pi$ and $\loc'$ and~$\loc''$ in some cycle $\pi'\in\Pi'$ for
  two "non-negative" "multi-cycles"~$\Pi$ and~$\Pi'$ "suitable"
  for~$(M,F)$, then up to a circular shift $\pi$ and~$\pi'$ can be
  assumed to start and end with $\loc'$, and then
  $(\Pi\setminus\{\pi\})\cup(\Pi'\setminus\{\pi'\})\cup\{\pi\pi'\}$ is
  also a "non-negative" "multi-cycle" "suitable" for~$(M,F)$.
\end{proof}

Thus~$\sim$ defines a partition~$\Loc/{\sim}$ of~$\Loc$.
In order to find a "non-negative" cycle~$\pi$ "suitable" for~$(M,F)$,
we are going to compute the partition~$\Loc/{\sim}$ of~$\Loc$
according to~$\sim$.  If we obtain a partition with a single
equivalence class, we are done: there exists such a cycle.  Otherwise,
such a cycle if it exists must be included in one of the subsystems
$(P,\Act\cap(P\times\+Z^\dd\times P),\dd)$ induced by the equivalence
classes $P\in\Loc/{\sim}$.  This yields \Cref{13-algo:zcycle}, which
assumes that we know how to compute the partition~$\Loc/{\sim}$.  Note
that the depth of the recursion in \Cref{13-algo:zcycle} is bounded
by~$|\Loc|$ and that recursive calls operate over disjoint subsets
of~$\Loc$, thus assuming that we can compute the partition in
polynomial time, then \Cref{13-algo:zcycle} also works in polynomial
time.

\begin{algorithm}
 \KwData{A "vector addition system with states"
   $\?V=(\Loc,\Act,\dd)$, $M\in 2^\Act$, $F\subseteq\Act$}

\If{$|\Loc|=1$}
  {\If{$\?V$ has a "non-negative" "multi-cycle" "suitable" for~$(M,F)$}
    {\Return{true}}}

$\Loc/{\sim} \leftarrow \mathrm{partition}(\?V,M,F)$ ;

\If{$|\Loc/{\sim}|=1$}{\Return{true}}

\ForEach{$P\in\Loc/{\sim}$}{\If{$\mathrm{cycle}((P,\Act\cap(P\times\+Z^\dd\times
    P),\dd),M,F)$}{\Return{true}}}

\Return{false}
\caption{$\text{cycle}(\?V,M,F)$}
\label{13-algo:zcycle}
\end{algorithm}

It remains to see how to compute the partition $\Loc/{\sim}$. Consider
for this the set of actions
$\Act'\eqdef\{a\mid\exists\Pi\text{ a "non-negative" "multi-cycle"
  "suitable" for $(M,F)$ with $a\in\Pi$}\}$ and
$\?V'=(\Loc',\Act',\dd)$ the subsystem induced by $\Act'$.
\begin{claim}
\label{13-fact:part}
  There exists a path from~$\loc$ to~$\loc'$ in $\?V'$
  if and only if $\loc\sim\loc'$.
\end{claim}
\begin{proof}
  If $\loc\sim\loc'$, then either $\loc=\loc'$ and there is an empty
  path, or there exist~$\Pi$ and~$\pi\in\Pi$ such that $\loc$
  and~$\loc'$ belong to~$\pi$ and $\Pi$ is a "non-negative"
  "multi-cycle" "suitable" for $(M,F)$, thus every action of~$\pi$ is
  in~$\Act'$ and there is a path in~$\?V'$.  

  Conversely, if there is a path $\pi\in{\Act'}^\ast$ from~$\loc$
  to~$\loc'$, then $\loc\sim\loc'$ by induction on~$\pi$.  Indeed, if
  $|\pi|=0$ then $\loc=\loc'$.  For the induction step, $\pi=\pi' a$
  with $\pi'\in{\Act'}^\ast$ a path from $\loc$ to $\loc''$ and
  $a=(\loc''\step{\vec u}\loc')\in\Act'$ for some~$\vec u$.  By
  induction hypothesis, $\loc\sim\loc''$ and since $a\in\Act'$,
  $\loc''\sim\loc'$, thus $\loc\sim\loc'$ by transitivity shown
  in~\Cref{13-fact:sim}. 
\end{proof}

By \Cref{13-fact:part}, the equivalence classes of~$\sim$ are the
strongly connected components of~$\?V'$.  This yields the following
polynomial-time algorithm for computing~$\Loc/{\sim}$.

\begin{algorithm}
 \KwData{A "vector addition system with states"
   $\?V=(\Loc,\Act,\dd)$, $M\in 2^\Act$, $F\subseteq\Act$}

$\Act'\leftarrow\emptyset$;

\ForEach{$a\in\Act$}{\If{$\?V$ has a "non-negative" "multi-cycle"
    "suitable"
    for~$(M\cup\{\{a\}\},F)$}{$\Act'\leftarrow\Act'\cup\{a\}$}}

$\?V'\leftarrow \text{subsystem induced by~$\Act'$}$ ;

\Return{$\mathrm{SCC}(\?V')$}
\caption{$\text{partition}(\?V,M,F)$}
\label{13-algo:part}
\end{algorithm}

Together, \Cref{13-lem:zmulticycle} and \Cref{13-algo:part,13-algo:zcycle} yield the following.

\begin{lemma}[Polynomial-time detection of suitable non-negative cycles]
\label{13-lem:zcycle}
  Let $\?V$ be a "vector addition system with states",
  $M\in 2^{\Act}$, and $F\subseteq\Act$.  We can check in polynomial
  time whether~$\?V$ contains a "non-negative" cycle~$\pi$
  "suitable" for~$(M,F)$.
\end{lemma}

Finally, we obtain the desired polynomial time upper bound for
"parity@parity vector games" in "vector addition systems with states".
\begin{theorem}[Existential one-player parity vector games are in~\P]
\label{13-thm:zcycle}
  Whether Eve wins a one-player "parity vector game" with
  "existential initial credit" is in~\P.
\end{theorem}
\begin{proof}
  Let $\?V=(\Loc,\Act,\dd)$ be a "vector addition system with states",
  $\lcol{:}\,\Loc\to\{1,\dots,$ $d\}$ a location colouring, and
  $\loc_0\in\Loc$ an initial location.  We start by trimming~$\?V$ to
  only keep the locations reachable from~$\loc_0$ in the underlying
  directed graph.  Then, for every even priority $p\in\{1,\dots,d\}$,
  we use \Cref{13-lem:zcycle} to check for the existence of a
  "non-negative" cycle with maximal priority~$p$: it suffices for this
  to set $M\eqdef\{\lcol^{-1}(p)\}$ and
  $F\eqdef\lcol^{-1}(\{p+1,\dots,d\})$.
\end{proof}

\paragraph{Upper Bounds}
We are now equipped to prove our upper bounds.  We begin with a nearly
trivial case.  In a "coverability" "asymmetric vector game" with
"existential initial credit", the counters play no role at all: Eve 
has a winning strategy for some initial credit in the "vector game" if
and only if she has one to reach the target location~$\loc_f$ in the
finite game played over~$\Loc$ and edges~$(\loc,\loc')$ whenever
$\loc\step{\vec u}\loc'\in\Act$ for some~$\vec u$.  This entails that
"coverability" "asymmetric vector games" are quite easy to solve.

\begin{theorem}[Existential coverability asymmetric vector games are in~\P]
\label{13-thm:cov-exist-P}
  "Coverability" "asymmetric" "vector games" with "existential initial
  credit" are \P-complete.
\end{theorem}

Regarding "non-termination" and "parity@parity vector game", we
exploit \Cref{13-lem:counterless,13-thm:zcycle}.

\begin{theorem}[Existential parity asymmetric vector games are in~\coNP]
\label{13-thm:exist-easy}
  "Non-termination" and "parity@parity vector game" "asymmetric"
  "vector games" with "existential initial credit" are in~\coNP.
\end{theorem}
\begin{proof}
  By \Cref{13-rmk:nonterm2parity}, it suffices to prove the statement for
  "parity@parity vector games" games.  By \Cref{13-lem:counterless},
  if Adam wins the game, we can guess a "counterless" winning
  strategy~$\tau$ telling which action to choose for every location.
  This strategy yields a one-player game, and by \Cref{13-thm:zcycle}
  we can check in polynomial time that~$\tau$ was indeed winning
  for Adam.
\end{proof}

Finally, in fixed dimension and with a fixed number of priorities, we
can simply apply the results of \Cref{13-sec:bounding}.
\begin{corollary}[Existential fixed-dimensional parity asymmetric vector games are pseudo-polynomial]
\label{13-cor:exist-pseudop}
  "Parity@parity vector game" "asymmetric" "vector games" with
  "existential initial credit" are in pseudo-polynomial time if the
  dimension and the number of priorities are fixed.
\end{corollary}
\begin{proof}
  Consider an "asymmetric vector system"
  $\?V=(\Loc,\Act,\Loc_\mEve,\Loc_\mAdam,\dd)$ and a location
  colouring $\lcol{:}\,\Loc\to\{1,\dots,2d\}$.
  By \Cref{13-lem:parity2bounding}, the "parity vector game" with
  "existential initial credit" over~$\?V$ problem reduces to a
  "bounding game" with "existential initial credit" over a "vector
  system"~$\?V'=(\Loc',\Act',\Loc'_\mEve,\Loc'_\mAdam,\dd+d)$ where
  $\Loc'\in O(|\Loc|)$ and $\|\Act'\|=\|\Act\|$.
  By \Cref{13-thm:bounding}, it suffices to consider the case of a
  "non-termination" game with "existential initial credit" played over
  the "bounded semantics" $\bounded(\?V')$ where $B$ is in
  $(|\Loc'|\cdot\|\Act'\|)^{O(\dd+d)^3}$.  Such a game can be solved in
  linear time in the size of the bounded arena using attractor
  techniques, thus in $O(|\Loc|\cdot B)^{\dd+d}$, which is in
  $(|\Loc|\cdot\|\Act\|)^{O(\dd+d)^4}$ in terms of the original instance.
\end{proof}

\subsubsection{Given initial credit}
\label{13-sec:up-given}

\begin{theorem}[Upper bounds for asymmetric vector games]
\label{13-thm:avag-easy}
  "Coverability", "non-termination", and "parity@parity vector game"
  "asymmetric" "vector games" with "given initial credit" are in
  \kEXP[2].  If the dimension is fixed, they are in \EXP, and if the
  number of priorities is also fixed, they are in pseudo-polynomial
  time.
\end{theorem}


\subsection{Lower bounds}
\label{13-sec:low}
Let us turn our attention to complexity lower bounds for "monotonic"
"asymmetric vector games".  It turns out that most of the upper bounds
shown in \Cref{13-sec:up} are tight.
\subsubsection{Existential initial credit}
In the "existential initial credit" variant of our games, we have the
following lower bound matching \Cref{13-thm:exist-easy}, already with a
unary encoding.

\begin{theorem}[Existential non-termination asymmetric vector games are \coNP-hard]
\label{13-thm:exist-hard}
  "Non-termination", and "parity@parity vector game"
  "asymmetric" "vector games" with "existential initial credit" are
  \coNP-hard.
\end{theorem}
\begin{proof}
  By \Cref{13-rmk:nonterm2parity}, it suffices to show hardness for
  "non-termination games".  We reduce from the \lang{3SAT} problem:
  given a formula $\varphi=\bigwedge_{1\leq i\leq m}C_i$ where each
  clause $C_i$ is a disjunction of the form
  $\litt_{i,1}\vee\litt_{i,2}\vee\litt_{i,3}$ of literals taken from
  $X=\{x_1,\neg x_1,x_2,$ $\neg x_2,\dots,x_k,\neg x_k\}$, we construct
  an "asymmetric" "vector system" $\?V$ where Eve wins the
  "non-termination game" with "existential initial credit" if and only
  if~$\varphi$ is not satisfiable; since the game is determined, we
  actually show that Adam wins the game if and only if~$\varphi$ is
  satisfiable.

  Our "vector system" has dimension~$2k$, and for a literal
  $\litt\in X$, we define the vector
  \begin{equation*}
    \vec u_\litt\eqdef\begin{cases}
      \vec e_{2n-1}-\vec e_{2n}&\text{if }\litt=x_n\;,\\
      \vec e_{2n}-\vec e_{2n-1}&\text{if }\litt=\neg x_n\;.
    \end{cases}
  \end{equation*}
  We define $\?V\eqdef(\Loc,\Act,\Loc_\mEve,\Loc_\mAdam,2k)$ where
  \begin{align*}
    \Loc_\mEve&\eqdef\{\varphi\}\cup\{\litt_{i,j}\mid 1\leq i\leq m,1\leq j\leq
                3\}\;,\\
    \Loc_\mAdam&\eqdef\{C_i\mid 1\leq i\leq m\}\;,\\
    \Act&\eqdef\{\varphi\step{\vec 0}C_i\mid 1\leq i\leq m\}\cup\{C_i\step{\vec 0}\litt_{i,j},\;\;\litt_{i,j}\xrightarrow{\vec u_{\litt_{i,j}}}\varphi\mid 1\leq i\leq m,1\leq j\leq 3\}\;.
  \end{align*}
    We use~$\varphi$ as our initial location.
    Let us call a map $v{:}\,X\to\{0,1\}$ a literal assignment; we
    call it conflicting if there exists $1\leq n\leq k$ such that
    $v(x_n)=v(\neg x_n)$.

    Assume that~$\varphi$ is satisfiable.  Then there exists a
    non-"conflicting" "literal assignment"~$v$ that satisfies all the
    clauses: for each $1\leq i\leq m$, there exists $1\leq j\leq 3$
    such that $v(\litt_{i,j})=1$; this yields a "counterless" strategy
    for Adam, which selects $(C_i,\litt_{i,j})$ for each
    $1\leq i\leq m$.  Consider any infinite "play" consistent with
    this strategy.  This "play" only visits literals $\litt$ where
    $v(\litt)=1$.  There exists a literal $\litt\in X$ that is visited
    infinitely often along the "play", say $\litt=x_n$.  Because~$v$ is
    non-"conflicting", $v(\neg x_n)=0$, thus the location $\neg x_n$
    is never visited.  Thus the play uses the action
    $\litt\step{\vec e_{2n-1}-\vec e_{2n}}\varphi$ infinitely often,
    and never uses any action with a positive effect on
    component~$2n$.  Hence the play is losing from any initial credit.

    Conversely, assume that~$\varphi$ is not satisfiable.  By
    contradiction, assume that Adam wins the game for all initial
    credits.  By \Cref{13-lem:counterless}, he has a "counterless" winning
    strategy~$\tau$ that selects a literal in every clause.  Consider
    a "literal assignment" that maps each one of the selected literals
    to~$1$ and the remaining ones in a non-conflicting manner.  By
    definition, this "literal assignment" satisfies all the clauses,
    but because~$\varphi$ is not satisfiable, it is "conflicting":
    necessarily, there exist $1\leq n\leq k$ and $1\leq i,i'\leq m$,
    such that $\tau$ selects $x_n$ in $C_i$ and $\neg x_n$ in
    $C_{i'}$.  But this yields a winning strategy for Eve, which
    alternates in the initial location $\varphi$ between $C_{i}$
    and $C_{i'}$, and for which an initial credit
    $\vec e_{2n-1}+\vec e_{2n}$ suffices: a contradiction.
\end{proof}

Note that \Cref{13-thm:exist-hard} does not apply to fixed dimensions
$\dd\geq 2$.  We know by \Cref{13-cor:exist-pseudop} that those games can
be solved in pseudo-polynomial time if the number of priorities is
fixed, and by \Cref{13-thm:exist-easy} that they are in \coNP.

\subsubsection{Given initial credit}
With "given initial credit", we have a lower bound matching the
\kEXP[2] upper bound of \Cref{13-thm:avag-easy}, already with a unary
encoding.  The proof itself is an adaptation of the proof by
\citem[Lipton]{Lipton:1976} of $\EXPSPACE$-hardness of "coverability" in
the one-player case.

\begin{theorem}[Coverability and non-termination asymmetric vector games are {\kEXP[2]-hard}]
\label{13-thm:avag-hard}
  "Coverability", "non-termination", and "parity@parity vector game"
  "asymmetric" "vector games" with "given initial credit" are
  \kEXP[2]-hard.
\end{theorem}
\begin{proof}
  We reduce from the "halting problem" of an alternating Minsky
  machine $\?M=(\Loc,\Act,\Loc_\mEve,\Loc_\mAdam,\dd)$ with counters
  bounded by $B\eqdef 2^{2^n}$ for $n\eqdef|\?M|$.  Such a machine is
  similar to an "asymmetric" "vector system" with increments
  $\loc\step{\vec e_i}\loc'$, decrements $\loc\step{-\vec e_i}\loc'$,
  and "zero test" actions $\loc\step{i\eqby{?0}}\loc'$, all
  restricted to locations $\loc\in\Loc_\mEve$; the only actions
  available to Adam are actions $\loc\step{\vec 0}\loc'$.  The
  set of locations contains a distinguished `halt' location
  $\loc_\mathtt{halt}\in\Loc$ with no outgoing action.  The
  machine comes with the promise that, along any "play", the norm of
  all the visited configurations $\loc(\vec v)$ satisfies
  $\|\vec v\|<B$.  The "halting problem" asks, given an initial
  location $\loc_0\in\Loc$, whether Eve has a winning strategy to
  visit $\loc_\mathtt{halt}(\vec v)$ for some $\vec v\in\+N^\dd$ from
  the initial configuration $\loc_0(\vec 0)$.  This problem is
  \kEXP[2]-complete if $\dd\geq 3$ by standard
  arguments~\cite{Fischer.Meyer.ea:1968}.

    Let us start by a quick refresher on Lipton's construction~\cite{Lipton:1976};
    see also~\cite{Esparza:1998} for a nice exposition.  At the heart
    of the construction lies a collection of one-player gadgets
    implementing \emph{level~$j$} meta-increments
    $\loc\mstep{2^{2^j}\cdot\vec c}\loc'$ and \emph{level~$j$}
    meta-decrements $\loc\mstep{-2^{2^j}\cdot\vec c}\loc'$ for
    some "unit vector"~$\vec c$ using $O(j)$ auxiliary counters and
    $\poly(j)$ actions, with precondition that the auxiliary counters
    are initially empty in~$\loc$ and post relation that they are empty
    again in~$\loc'$.  The construction is by induction over~$j$; let
    us first see a naive implementation for "meta-increments".  For
    the base case~$j=0$, this is just a standard action
    $\loc\step{2\vec c}\loc'$.  For the induction step $j+1$, we use
    the gadget of \Cref{13-fig:meta-incr} below, where
    $\vec x_{j},\bar{\vec x}_{j},\vec z_{j},\bar{\vec z}_{j}$ are
    distinct fresh "unit vectors": the gadget performs two nested
    loops, each of $2^{2^j}$ iterations, thus iterates the unit
    increment of~$\vec c$ a total of $\big(2^{2^j}\big)^2=2^{2^{j+1}}$
    times.  A "meta-decrement" is obtained similarly.

    \begin{figure}[htbp]
      \centering
      \begin{tikzpicture}[auto,on grid,node distance=1.55cm]
      \node[s-eve](0){$\loc$};
      \node[s-eve-small,right=of 0](1){};
      \node[s-eve-small,right=of 1](2){};
      \node[s-eve-small,right=of 2](3){};
      \node[s-eve-small,right=of 3](4){};
      \node[s-eve-small,right=of 4](5){};
      \node[s-eve-small,right=of 5](6){};
      \node[s-eve,right=of 6](7){$\loc'$};
      \path[arrow,every node/.style={font=\footnotesize,inner sep=2pt}]
      (0) edge node{$2^{2^j}\cdot\vec x_{j}$} (1)
      (1) edge node{$2^{2^j}\cdot\vec z_{j}$} (2)
      (2) edge node{$\bar{\vec x}_{j}-\vec x_{j}$} (3)
      (3) edge node{$\bar{\vec z}_{j}-\vec z_{j}$} (4)
      (4) edge node{$\vec c$} (5)
      (5) edge node{$-2^{2^j}\cdot\bar{\vec z}_{j}$} (6)
      (6) edge node{$-2^{2^j}\cdot\bar{\vec x}_{j}$} (7);
      \draw[->,rounded corners=10pt,>=stealth'] (5) --
      (7.4,.65) -- (5,.65) -- (3);
      \node[font=\footnotesize,inner sep=2pt] at (6.2,.75) {$\vec 0$};
      \draw[->,rounded corners=10pt,>=stealth'] (6) --
      (8.95,1.25) -- (1.9,1.25) -- (1);
      \node[font=\footnotesize,inner sep=2pt] at (5.43,1.35) {$\vec 0$};
    \end{tikzpicture}
    \caption{A naive implementation of the
      "meta-increment" $\loc\mstep{2^{2^{j+1}}\cdot\vec c}\loc'$.}
      \label{13-fig:meta-incr}
\commentAlt{Figure~\ref{13-fig:meta-incr}: A linear sequence of seven circular nodes, L to L', with various labeled and curved arrows indicating transitions.}
\commentLongAlt{Figure~\ref{13-fig:meta-incr}: The image displays a horizontal sequence of seven circular nodes, starting with 'L' on the left and ending with 'L'' on the right. Directed arrows connect consecutive nodes, each with a label:
- From L to the next unlabeled node: '2^(2^i) * x_j'
- To the next unlabeled node: '2^(2^j) * x_i'
- To the next unlabeled node: 'x_i - x_j'
- To the next unlabeled node: 'x_j - x_i'
- To the next unlabeled node: 'e_vector'
- To the next unlabeled node: '-2^(2^j) * x_i'
- To L': '-2^(2^i) * x_j'

Additionally, there are two long curved arrows above the linear sequence:
- One from the second node from the left (receiving '2^(2^i) * x_j') to the second node from the right (receiving '-2^(2^j) * x_i'), labeled '0_vector'.
- Another from the fourth node from the left (receiving 'x_j - x_i') to the third node from the right (receiving 'e_vector'), also labeled '0_vector'.}
  \end{figure}
  
  Note that this level~$(j+1)$ gadget contains two copies of the
  level~$j$ "meta-increment" and two of the level~$j$
  "meta-decrement", hence this naive implementation has
  size~$\mathsf{exp}(j)$.  In order to obtain a polynomial size, we would like
  to use a single \emph{shared} level~$j$ gadget for each~$j$, instead
  of hard-wiring multiple copies.  The idea is to use a `dispatch
  mechanism,' using extra counters, to encode the choice of "unit
  vector"~$\vec c$ and of return location~$\loc'$.  Let us see how to
  do this in the case of the return location~$\loc'$; the mechanism
  for the vector~$\vec c$ is similar.  We enumerate the (finitely many)
  possible return locations~$\loc_0,\dots,\loc_{m-1}$ of the gadget
  implementing $\loc\mstep{2^{2^{j+1}}\cdot\vec c}\loc'$.  We use two
  auxiliary counters with "unit vectors" $\vec r_j$
  and~$\bar{\vec r}_j$ to encode the return location.  Assume $\loc'$
  is the $i$th possible return location, \textit{i.e.}, $\loc'=\loc_i$ in our
  enumeration: before entering the shared gadget implementation, we
  initialise~$\vec r_j$ and~$\bar{\vec r}_j$ by performing the action
  $\loc\step{i\cdot\vec r_j+(m-i)\cdot\bar{\vec r}_j}\cdots$.  Then,
  where we would simply go to~$\loc'$ in \Cref{13-fig:meta-incr} at
  the end of the gadget, the shared gadget has a final action
  $\cdots\step{\vec 0}\loc_{\mathrm{return}_j}$ leading to a dispatch
  location for returns: for all $0\leq i<m$, we have an action
  $\loc_{\mathrm{return}_j}\step{-i\cdot\vec r_j-(m-i)\cdot\bar{\vec
      r}_j}\loc_i$
  that leads to the desired return location.

  \bigskip Let us return to the proof.  Consider an instance of the
  "halting problem".  We first exhibit a reduction to "coverability";
  by \Cref{13-rmk:cov2parity}, this will also entail the \kEXP[2]-hardness
  of "parity@parity vector game" "asymmetric" "vector games".  We
  build an "asymmetric vector system"
  \[
  \?V=(\Loc',\Act',\Loc'_\mEve,\Loc_\mAdam,\dd')
  \] with
  $\dd'=2\dd+O(n)$.  Each of the counters~$\mathtt{c}_i$ of $\?M$ is
  paired with a \emph{complementary} counter~$\bar{\mathtt{c}_i}$ such
  that their sum is~$B$ throughout the simulation of~$\?M$.  We
  denote by $\vec c_i$ and $\bar{\vec c}_i$ the corresponding "unit
  vectors" for $1\leq i\leq\dd$.  The "vector system"~$\?V$ starts by
  initialising the counters $\bar{\mathtt{c}}_i$ to~$B$ by a sequence
  of "meta-increments"
  $\loc'_{i-1}\mstep{2^{2^n}\cdot\bar{\vec c}_i}\loc'_i$ for
  $1\leq i\leq\dd$, before starting the simulation by an action
  $\loc'_\dd\step{\vec 0}\loc_0$.  The simulation of~$\?M$ uses the
  actions depicted in \Cref{13-fig:lipton}.  Those maintain the
  invariant on the complement counters.  Regarding "zero tests", Eve 
  yields the control to Adam, who has a choice between performing a
  "meta-decrement" that will fail if $\bar{\mathtt c}_i< 2^{2^n}$,
  which by the invariant is if and only if $\mathtt{c}_i>0$, or going
  to~$\loc'$.

  \begin{figure}[htbp]
    \centering
    \begin{tikzpicture}[auto,on grid,node distance=1.5cm]
      \node(to){$\mapsto$};
      \node[anchor=east,left=2.5cm of to](mm){"alternating Minsky machine"};
      \node[anchor=west,right=2.5cm of to](mwg){"asymmetric vector system"};
      \node[below=.7cm of to](imap){$\rightsquigarrow$};
      \node[s-eve,left=2.75cm of imap](i0){$\loc$};
      \node[right=of i0](i1){$\loc'$};
      \node[right=1.25cm of imap,s-eve](i2){$\loc$};
      \node[right=1.8 of i2](i3){$\loc'$};      
      \path[arrow,every node/.style={font=\footnotesize}]
      (i0) edge node{$\vec e_i$} (i1)
      (i2) edge node{$\vec c_i-\bar{\vec c}_i$} (i3);
      \node[below=1cm of imap](dmap){$\rightsquigarrow$};
      \node[s-eve,left=2.75cm of dmap](d0){$\loc$};
      \node[right=of d0](d1){$\loc'$};
      \node[right=1.25cm of dmap,s-eve](d2){$\loc$};
      \node[right=1.8 of d2](d3){$\loc'$};
      \path[arrow,every node/.style={font=\footnotesize}]
      (d0) edge node{$-\vec e_i$} (d1)
      (d2) edge node{$-\vec c_i+\bar{\vec c}_i$} (d3);
      \node[below=1.5cm of dmap](zmap){$\rightsquigarrow$};
      \node[s-eve,left=2.75cm of zmap](z0){$\loc$};
      \node[right=of z0](z1){$\loc'$};
      \node[right=1.25cm of zmap,s-eve](z2){$\loc$};
      \node[right=of z2,s-adam-small](z3){};
      \node[above right=.8 and 1.1 of z3,s-eve-small](z4){};
      \node[below right=.8 and 1.1 of z3,inner sep=0pt](z5){$\loc'$};
      \node[right=1.8 of z4](z6){$\loc_\mathtt{halt}$};
      \path[arrow,every node/.style={font=\footnotesize}]
      (z0) edge node{$i\eqby{?0}$} (z1)
      (z2) edge node{$\vec 0$} (z3)
      (z3) edge node{$\vec 0$} (z4)
      (z3) edge[swap] node{$\vec 0$} (z5)
      (z4) edge node{$-2^{2^n}\cdot\bar{\vec c}_i$} (z6);
      \node[below=1.5cm of zmap](amap){$\rightsquigarrow$};
      \node[s-adam,left=2.75cm of amap](a0){$\loc$};
      \node[right=of a0](a1){$\loc'$};
      \node[right=1.25cm of amap,s-adam](a2){$\loc$};
      \node[right=of a2](a3){$\loc'$};
      \path[arrow,every node/.style={font=\footnotesize}]
      (a0) edge node{$\vec 0$} (a1)
      (a2) edge node{$\vec 0$} (a3);
    \end{tikzpicture}
    \caption{Schema of the reduction to
      "coverability" in the proof of \Cref{13-thm:avag-hard}.}
      \label{13-fig:lipton}
  \end{figure}
  
  It is hopefully clear that Eve wins the "coverability game" played
  on~$\?V$ starting from $\loc'_0(\vec 0)$ and with target
  configuration $\loc_\mathtt{halt}(\vec 0)$ if and only if the
  "alternating Minsky machine" halts.

  \medskip Regarding "non-termination" games, we use essentially the
  same reduction.  First observe that, if Eve can ensure reaching
  $\loc_\mathtt{halt}$ in the "alternating Minsky machine", then she
  can do so after at most $|\Loc|B^\dd$ steps.  We therefore use a
  `time budget': this is an additional component in $\?V$ with
  associated "unit vector"~$\vec t$.  This component is initialised to
  $|\Loc|B^\dd=|\Loc|2^{\dd 2^n}$ before the simulation, and decreases
  by~one at every step; see \Cref{13-fig:lipton-nonterm}.  We also add
  a self loop $\loc_\mathtt{halt}\step{\vec 0}\loc_\mathtt{halt}$.
  Then the only way to avoid the "sink" and thus to win the
  "non-termination" game is to reach $\loc_\mathtt{halt}$.
  \begin{figure}[htbp]
    \centering
    \begin{tikzpicture}[auto,on grid,node distance=1.5cm]
      \node(to){$\mapsto$};
      \node[anchor=east,left=2.5cm of to](mm){"alternating Minsky machine"};
      \node[anchor=west,right=2.5cm of to](mwg){"asymmetric vector system"};
      \node[below=.7cm of to](imap){$\rightsquigarrow$};
      \node[s-eve,left=2.75cm of imap](i0){$\loc$};
      \node[right=of i0](i1){$\loc'$};
      \node[right=1.25cm of imap,s-eve](i2){$\loc$};
      \node[right=1.8 of i2](i3){$\loc'$};      
      \path[arrow,every node/.style={font=\footnotesize}]
      (i0) edge node{$\vec e_i$} (i1)
      (i2) edge node{$\vec c_i-\bar{\vec c}_i-\vec t$} (i3);
      \node[below=1cm of imap](dmap){$\rightsquigarrow$};
      \node[s-eve,left=2.75cm of dmap](d0){$\loc$};
      \node[right=of d0](d1){$\loc'$};
      \node[right=1.25cm of dmap,s-eve](d2){$\loc$};
      \node[right=1.8 of d2](d3){$\loc'$};
      \path[arrow,every node/.style={font=\footnotesize}]
      (d0) edge node{$-\vec e_i$} (d1)
      (d2) edge node{$-\vec c_i+\bar{\vec c}_i-\vec t$} (d3);
      \node[below=1.5cm of dmap](zmap){$\rightsquigarrow$};
      \node[s-eve,left=2.75cm of zmap](z0){$\loc$};
      \node[right=of z0](z1){$\loc'$};
      \node[right=1.25cm of zmap,s-eve](z2){$\loc$};
      \node[right=of z2,s-adam-small](z3){};
      \node[above right=.8 and 1.1 of z3,s-eve-small](z4){};
      \node[below right=.8 and 1.1 of z3,inner sep=0pt](z5){$\loc'$};
      \node[right=1.8 of z4](z6){$\loc_\mathtt{halt}$};
      \path[arrow,every node/.style={font=\footnotesize}]
      (z0) edge node{$i\eqby{?0}$} (z1)
      (z2) edge node{$-\vec t$} (z3)
      (z3) edge node{$\vec 0$} (z4)
      (z3) edge[swap] node{$\vec 0$} (z5)
      (z4) edge node{$-2^{2^n}\cdot\bar{\vec c}_i$} (z6);
      \node[below=1.5cm of zmap](amap){$\rightsquigarrow$};
      \node[s-adam,left=2.75cm of amap](a0){$\loc$};
      \node[right=of a0](a1){$\loc'$};
      \node[right=1.25cm of amap,s-adam](a2){$\loc$};
      \node[right=of a2,s-eve-small](a3){};
      \node[right=of a3](a4){$\loc'$};
      \path[arrow,every node/.style={font=\footnotesize}]
      (a0) edge node{$\vec 0$} (a1)
      (a2) edge node{$\vec 0$} (a3)
      (a3) edge node{$-\vec t$} (a4);
    \end{tikzpicture}
    \caption{Schema of the reduction to
      "non-termination" in the proof of \Cref{13-thm:avag-hard}.}
      \label{13-fig:lipton-nonterm}
  \end{figure}

  We still need to extend our initialisation phase.  It suffices for
  this to implement a gadget for $\dd$-"meta-increments"
  $\loc\mstep{2^{\dd 2^j}\cdot\vec c}\loc'$ and $\dd$-"meta-decrements"
  $\loc\mstep{-2^{\dd 2^j}\cdot\vec c}\loc'$; this is the same argument as
  in Lipton's construction, with a base case $\loc\mstep{2^\dd}\loc'$
  for $j=0$.  Then we initialise our time budget through $|\Loc|$
  successive $\dd$-"meta-increments"
  $\loc\mstep{2^{\dd 2^n}\cdot\vec t}\loc'$.
\end{proof}

The proof of \Cref{13-thm:avag-hard} relies crucially on the fact that the
dimension is not fixed: although $\dd\geq 3$ suffices in the
"alternating Minsky machine", we need $O(|\?M|)$ additional counters
to carry out the reduction.  A separate argument is thus needed in
order to match the \EXP\ upper bound of \Cref{13-thm:avag-easy} in fixed
dimension.

\begin{theorem}[Fixed-dimensional coverability and non-termination asymmetric vector games are \EXP-hard]
\label{13-thm:avag-two}
  "Coverability", "non-termination", and "parity@parity vector game"
  "asymmetric" "vector games" with "given initial credit" are
  \EXP-hard in dimension $\dd\geq 2$.
\end{theorem}
\begin{proof}
  We exhibit a reduction from "countdown games" with "given initial
  credit", which are \EXP-complete by \Cref{13-thm:countdown-given}.
  Consider an instance of a "configuration reachability" countdown
  game: a "countdown system"
  $\?V=(\Loc,\Act,\Loc_\mEve,\Loc_\mAdam,1)$ with initial
  configuration $\loc_0(n_0)$ and target
  configuration~$\smiley(0)$---as seen in the proof
  of \Cref{13-thm:countdown-given}, we can indeed assume that the target
  credit is zero; we will also assume that Eve controls~$\smiley$ and
  that the only action available in~$\smiley$ is
  $\smiley\step{-1}\smiley$.  We construct an "asymmetric" "vector
  system" $\?V'$ of dimension~2 such that Eve can ensure
  reaching~$\smiley(0,n_0)$ from $\loc_0(n_0,0)$ in~$\?V'$ if and only
  if she could ensure reaching $\smiley(0)$ from $\loc_0(n_0)$
  in~$\?V$.  The translation is depicted in \Cref{13-fig:dim2}.
  
  \begin{figure}[htbp]
    \centering
    \begin{tikzpicture}[auto,on grid,node distance=1.5cm]
      \node(to){$\mapsto$};
      \node[anchor=east,left=2.5cm of to](mm){"countdown system"};
      \node[anchor=west,right=2.5cm of to](mwg){"asymmetric vector system"};
      \node[below=.7cm of to](imap){$\rightsquigarrow$};
      \node[s-eve,left=2.75cm of imap](i0){$\loc$};
      \node[right=of i0](i1){$\loc'$};
      \node[right=1.25cm of imap,s-eve](i2){$\loc$};
      \node[right=1.8 of i2](i3){$\loc'$};      
      \path[arrow,every node/.style={font=\footnotesize,inner sep=1pt}]
      (i0) edge node{$-n$} (i1)
      (i2) edge node{$-n,n$} (i3);
      \node[below=1cm of imap](dmap){$\rightsquigarrow$};
      \node[s-adam,left=2.75cm of dmap](d0){$\loc$};
      \node[right=of d0](d1){$\loc'$};
      \node[below=.5 of d0]{$n=\min\{n'\mid\exists\loc''\in\Loc\mathbin.\loc\step{-n'}\loc''\in\Act\}$};
      \node[right=1.25cm of dmap,s-adam](d2){$\loc$};
      \node[right=1.8 of d2,s-eve-small](d3){};
      \node[right=1.8 of d3](d4){$\loc'$};
      \path[arrow,every node/.style={font=\footnotesize,inner sep=1pt}]
      (d0) edge node{$-n$} (d1)
      (d2) edge node{$0,0$} (d3)
      (d3) edge node{$-n,n$} (d4);
      \node[below=1.5cm of dmap](zmap){$\rightsquigarrow$};
      \node[s-adam,left=2.75cm of zmap](z0){$\loc$};
      \node[right=of z0](z1){$\loc'$};
      \node[below=.5 of z0]{$n\neq\min\{n'\mid\exists\loc''\in\Loc\mathbin.\loc\step{-n'}\loc''\in\Act\}$};
      \node[right=1.25cm of zmap,s-adam](z2){$\loc$};
      \node[right=of z2,s-eve-small](z3){};
      \node[above right=.8 and 2.1 of z3](z4){$\loc'$};
      \node[below right=.8 and 2.1 of z3,s-eve](z5){$\smiley$};
      \path[arrow,every node/.style={font=\footnotesize,inner sep=1pt}]
      (z0) edge node{$-n$} (z1)
      (z2) edge node{$0,0$} (z3)
      (z3) edge[bend left=8] node{$-n,n$} (z4)
      (z3) edge[swap,bend right=8] node{$n_0-n+1,-n_0+n-1$} (z5)
      (z5) edge[loop above] node{$-1,1$} ()
      (z5) edge[loop right] node{$\,0,0$} ();
    \end{tikzpicture}
    \caption{Schema of the reduction in the proof
    of \Cref{13-thm:avag-two}.}
    \label{13-fig:dim2}
      \end{figure}
    
  The idea behind this translation is that a configuration $\loc(c)$
  of~$\?V$ is simulated by a configuration $\loc(c,n_0-c)$ in~$\?V'$.
  The crucial point is how to handle Adam\'s moves.  In a
  configuration $\loc(c,n_0-c)$ with $\loc\in\Loc_\mAdam$, according
  to the "natural semantics" of $\?V$, Adam should be able to
  simulate an action $\loc\step{-n}\loc'$ if and only if $c\geq n$.
  Observe that otherwise if $c<n$ and thus $n_0-c>n_0-n$, Eve can
  play to reach~$\smiley$ and win immediately.  An exception to the
  above is if $n$ is minimal among the decrements in~$\loc$, because
  according to the "natural semantics" of~$\?V$, if $c<n$ there should
  be an edge to the "sink", and this is handled in the second line
  of \Cref{13-fig:dim2}.

  Then Eve can reach $\smiley(0,n_0)$ if and only if she can cover
  $\smiley(0,n_0)$, if and only if she can avoid the "sink" thanks to
  the self loop $\smiley\step{0,0}\smiley$.  This
  shows the \EXP-hardness of "coverability" and "non-termination"
  "asymmetric" "vector games" in dimension~two; the hardness of
  "parity@parity vector game" follows
  from \Cref{13-rmk:cov2parity,13-rmk:nonterm2parity}.
\end{proof}
  

\section*{Bibliographic references}
\label{13-sec:references}
\paragraph{Vector Addition Systems with States}
In their one-player version, \textit{i.e.}\ in "vector addition systems with
states", all the games presented in \Cref{13-sec:counters} are decidable.
With "given initial credit", "configuration reachability" is simply
called `reachability' and was first shown decidable by
\citem[Mayr]{Mayr:1981} (with simpler proofs
in~\cite{Kosaraju:1982}) and recently shown
to be of non-elementary complexity~\cite{Czerwinski.Lasota.ea:2019}.  "Coverability" and
"non-termination" are considerably easier, as they are
$\EXPSPACE$-complete~\cite{Lipton:1976} and so is
"parity@parity vector game"~\cite{Habermehl:1997}.  With "existential
initial credit", the problems are markedly simpler: "configuration
reachability" becomes $\EXPSPACE$-complete, while "coverability" is in
\NL\ and "non-termination" and "parity" can be solved in polynomial
time by~\Cref{13-thm:zcycle} using linear programming
techniques~\cite{Kosaraju.Sullivan:1988}.

\paragraph{Undecidability of Vector Games}
The undecidability results of \Cref{13-sec:undec} are folklore.  One
can find undecidability proofs
in~\cite{Abdulla.Bouajjani.ea:2003};
"non-termination" was called `deadlock-freedom' by \citem[Raskin et
al.]{Raskin.Samuelides.ea:2005}. "Configuration reachability" is
undecidable even in very restricted cases, like the robot games
of~\citet{Niskanen.Potapov.ea:2016}.

\paragraph{Succinct One-Counter Games}
One-dimensional "vector systems" are often called \emph{one-counter
nets} in the literature, by contrast with \emph{one-counter automata}
where zero tests are allowed.  The $\EXPSPACE$-completeness of "succinct
one-counter games" was shown by~\citem[Hunter]{Hunter:2015}.  "Countdown games"
were originally defined with "given initial credit" and a "zero
reachability" objective, and shown
\EXP-complete in \cite{Jurdzinski.Laroussinie.ea:2008}; see also
\citet{Kiefer:2013} for a variant called hit-or-run games.  The
hardness proofs for \Cref{13-thm:countdown-given,13-thm:countdown-exist} are
adapted from~\citet{Jancar.Osicka.ea:2018}, where "countdown games"
with "existential initial credit" were first introduced.

\paragraph{Asymmetric Vector Games}
The "asymmetric" "vector games" of \Cref{13-sec:avag} appear
under many guises in the literature: as `and-branching' "vector
addition systems with states"
in~\cite{Lincoln.Mitchell.ea:1992}, as `vector games'
in~\cite{Kanovich:1995}, as `B-games'
in~\cite{Raskin.Samuelides.ea:2005}, as `single sided' vector
addition games in~\cite{Abdulla.Mayr.ea:2013}, and as `alternating'
"vector addition systems with states" in~\cite{Courtois.Schmitz:2014}.

The undecidability of "configuration reachability" shown
in~\Cref{13-sec:reach} was already proven by \citem[Lincoln et
al.]{Lincoln.Mitchell.ea:1992} and used to show the
undecidability of propositional linear logic;
\citem[Kanovich]{Kanovich:1995} refines this result to
show the undecidability of the $(!,\oplus)$-Horn fragment of linear
logic.  Similar proof ideas are used for Boolean BI and separation
logic in~\cite{Larchey-Wendling.Galmiche:2013}.

\paragraph{Asymmetric Monotone Vector Games}
The notion of "asymmetric" infinite games over a "well-quasi-ordered"
"arena" constitutes a natural extension of the notion of
well-structured systems of
\citet{Abdulla.Cerans.ea:2000} and
\citet{Finkel.Schnoebelen:2001}, and was undertaken
in~\cite{Abdulla.Bouajjani.ea:2003}.
The decidability of "coverability" and "non-termination" through "wqo"
arguments like those of \Cref{13-fact:pareto-cov} was shown
by~\citem[Raskin et al.]{Raskin.Samuelides.ea:2005}.  More
advanced "wqo" techniques were needed for the first decidability proof
of "parity@parity vector game" in~\cite{Abdulla.Mayr.ea:2013}.  See
also \cite{Schmitz.Schnoebelen:2012} for more on the algorithmic uses of
"wqos".

By analysing the attractor computation of \Cref{13-sec:attr}, one can
show that \Cref{13-algo:cov} works in \kEXP[2], thus matching the
optimal upper bound from~\Cref{13-thm:avag-easy}: this can be done using
the Rackoff-style argument of \citet{Courtois.Schmitz:2014} and the
analysis of \citet{Bozzelli.Ganty:2011}, or by a direct analysis of the
attractor computation algorithm~\cite{Lazic.Schmitz:2019}.

\paragraph{Energy Games}
\AP An alternative take on "energy games" is to see a "vector system"
\[
\?V=(\Loc,\Act,\Loc_\mEve,\Loc_\mAdam,\dd)
\] as a finite "arena" with
edges $\loc\step{\vec u}\loc'$ coloured by $\vec u$, thus with set of
colours $C\eqdef\+Z^\dd$.  For an initial credit $\vec v_0\in\+N^\dd$
and $1\leq i\leq\dd$, the associated energy objective is then
defined as 
\begin{equation*}
  \mathsf{Energy}_{\vec v_0}(i)\eqdef\left\{\pi\in E^\omega\;\middle|\;\forall
  n\in\+N\mathbin.\left(\vec v_0(i)+\sum_{0\leq j\leq n}c(\pi)(i)\right)\geq 0\right\}\;,
\end{equation*}
that is, $\pi$ is winning if the successive sums of weights on
coordinate~$i$ are always non-negative.
\AP The multi-energy objective then asks for the "play"~$\pi$ to
belong simultaneously to $\mathsf{Energy}_{\vec v_0}(i)$ for all
$1\leq i\leq\dd$.  This is a multiobjective in the sense of the
forthcoming \Cref{14-chap:multiobjective}.  "Multi-energy games" are
equivalent to "non-termination" games played on the arena
$\energy(\?V)$ defined by the "energy semantics".


The relationship with "energy games" was first observed in~\cite{Abdulla.Mayr.ea:2013}.  
The equivalence with "mean-payoff games" in dimension~one was first noticed by~\citem[Bouyer et al.]{Bouyer.Fahrenberg.ea:2008}.  
A similar connection in the multi-dimensional case was established in~\cite{Chatterjee.Doyen.ea:2010} and will be discussed in~\Cref{14-chap:multiobjective}.

\paragraph{Complexity} \Cref{13-tbl:cmplx} summarises the complexity
results for "asymmetric vector games".  For the upper bounds with
"existential initial credit" of \Cref{13-sec:up-exist}, the existence
of "counterless" winning strategies for Adam was originally shown by
\citem[Br\'azdil et al.]{Brazdil.Jancar.ea:2010} in the case of
"non-termination games"; the proof of \Cref{13-lem:counterless} is a
straightforward adaptation using
ideas from \cite{Chatterjee.Doyen:2012} to handle "parities@parity
vector game".  An alternative proof through "bounding games" is
presented in~\cite{Colcombet.Jurdzinski.ea:2017}.

The \coNP\ upper of \Cref{13-thm:exist-easy} was shown soon after
Br\'azdil et al.'s work by
\citem[Chatterjee et al.]{Chatterjee.Doyen.ea:2010} in
the case of "non-termination games".  The extension of
\Cref{13-thm:exist-easy} to "parity@parity vector game" was shown
by~\cite{Chatterjee.Randour.ea:2014} by a reduction from
"parity@parity vector games" to "non-termination games" somewhat
reminiscent of ?.  The proof of
\Cref{13-thm:exist-easy} takes a slightly different approach using
\Cref{13-lem:zcycle} for finding non-negative cycles, which is a
trivial adaptation of a result by \citem[Kosaraju and
Sullivan]{Kosaraju.Sullivan:1988}.  The pseudo-polynomial bound
of~\Cref{13-cor:exist-pseudop} is taken
from~\cite{Colcombet.Jurdzinski.ea:2017}.

For the upper bounds with "given initial credit" of
\Cref{13-sec:up-given}, regarding "coverability", the \kEXP[2] upper
bound of \Cref{13-thm:avag-easy} was first shown by~\citem[Courtois and
Schmitz]{Courtois.Schmitz:2014} by adapting Rackoff's technique for
"vector addition systems with states"~\cite{Rackoff:1978}.  Regarding
"non-termination", the first complexity upper bounds were shown
by~\citem[Br\'azdil et al.]{Brazdil.Jancar.ea:2010} and were
in \kEXP, thus non-elementary in the size of the input.  Very
roughly, 
their argument went as follows: one can extract a pseudo-polynomial
"existential Pareto bound"~$B$ in the one-player case from the proof
of \Cref{13-thm:zcycle}, from which the proof of \Cref{13-lem:counterless}
yields a $2^{|\Act|}(B+|\Loc|)$ "existential Pareto bound" in the
two-player case, and finally by arguments similar to ? a tower
of~$\dd$ exponentials on the "given initial credit" problem.  The
two-dimensional case with a unary encoding was shown a bit later to be
in~\P\ by~\citem[Chaloupka]{Chaloupka:2013}.  Finally, a
matching \kEXP[2] upper bound (and pseudo-polynomial in any fixed
dimension) was obtained by~\citem[Jurdzi\'nski et
al.]{Jurdzinski.Lazic.ea:2015}.  Regarding "parity@parity vector
game", \citem[Jan\v{c}ar]{jancar:2015} showed how to obtain
non-elementary upper bounds by reducing to the case
of~\citet{Brazdil.Jancar.ea:2010}, before a tight \kEXP[2] upper
bound (and pseudo-polynomial in fixed dimension with a fixed number of
priorities) was shown
in~\cite{Colcombet.Jurdzinski.ea:2017}.

The \coNP\ hardness with "existential initial credit" in
\Cref{13-thm:exist-hard} originates from
\citet{Chatterjee.Doyen.ea:2010}.  The \kEXP[2]-hardness
of both "coverability" and "non-termination" games with "given initial
credit" from \Cref{13-thm:avag-hard} was shown
in~\cite{Courtois.Schmitz:2014} by adapting Lipton's construction for
"vector addition systems with states"~\cite{Lipton:1976}; similar
proofs can be found for instance
in~\cite{Demri.Jurdzinski.ea:2012}.
The hardness for \EXP-hardness in dimension two was first shown by
\cite{Fahrenberg.Juhl.ea:2011}. 

In dimension~one, \citem[Bouyer et al.]{Bouyer.Fahrenberg.ea:2008} proved $\NP\cap\coNP$ upper bounds for "given initial credit" and \citem[Chatterjee and Doyen]{Chatterjee.Doyen:2012} for "existential initial credit".

\paragraph{Some Applications}
Besides their many algorithmic applications for solving various types
of games, "vector games" have been employed in several fields to prove
decidability and complexity results, for instance for linear,
relevance, or separation
logics~\cite{Lincoln.Mitchell.ea:1992},
simulation and bisimulation problems~\cite{Kiefer:2013},
resource-bounded logics~\cite{Alechina.Bulling.ea:2018}, 
orchestration synthesis~\cite{Giacomo.Vardi.ea:2018},
as well as model-checking probabilistic timed automata~\cite{Jurdzinski.Laroussinie.ea:2008}.

    \begin{landscape}
      \centering
      \captionof{table}{The complexity of
        "asymmetric vector games".}\label{13-tbl:cmplx}
  \bigskip

  \begingroup
  \catcode`\&=12
  \catcode`!=4
  \ifstandalone
  \setlength{\tabcolsep}{3pt}
  \begin{tabular}{p{11em}cccc}
  \toprule 
  !!\multicolumn{3}{c}{Dimension}\\
  \cmidrule(l){3-5}
  Game ! Initial credit ! Fixed $\dd=1$ ! Fixed $\dd\geq 2$ ! Arbitrary\\
  \midrule
    configuration reachability%
    ! both
    ! $\EXPSPACE$-complete            
    !\multicolumn{2}{c}{undecidable}
  \\[-.5em]
    ! 
    ! {\tiny\Cref{13-thm:asym-dim1}} 
    !\multicolumn{2}{c}{\tiny\Cref{13-thm:asym-undec}~\cite{Lincoln.Mitchell.ea:1992}}%
  \\
  \addlinespace
  \multirow{3}{*}{"coverability"}               
    ! "existential"
    ! \multicolumn{3}{c}{\P-complete}
  \\[-.5em]
    ! 
    ! \multicolumn{3}{c}{\tiny\Cref{13-thm:cov-exist-P}}
  \\
    ! "given"      
    ! in $\NP\cap\coNP$             
    ! \EXP-complete
    ! \kEXP[2]-complete
  \\[-.5em]
    ! 
    ! {} 
    ! {\tiny\Cref{13-thm:avag-two,13-thm:avag-easy}}
    ! {\tiny\Cref{13-thm:avag-hard,13-thm:avag-easy}}
  \\[-.7em]
    ! 
    !
    !{\tiny\cite{Fahrenberg.Juhl.ea:2011}}
    !{\tiny\cite{Courtois.Schmitz:2014}}
  \\
  \addlinespace
  \multirow{3}{*}{"non-termination"}
    ! "existential"
    ! in $\NP\cap\coNP$             
    ! in \coNP
    ! \coNP-complete
  \\[-.5em]
    ! 
    ! {\tiny\cite{Chatterjee.Doyen:2012}} 
    !
    ! {\tiny\Cref{13-thm:exist-hard,13-thm:exist-easy}~\cite{Chatterjee.Doyen.ea:2010}}
  \\
    ! "given"
    ! in $\NP\cap\coNP$
    ! \EXP-complete
    ! \kEXP[2]-complete
  \\[-.5em]
    ! 
    ! {\tiny\cite{Bouyer.Fahrenberg.ea:2008}}
    ! {\tiny\Cref{13-thm:avag-two,13-thm:avag-easy}}
    ! {\tiny\Cref{13-thm:avag-hard,13-thm:avag-easy}}
  \\[-.7em]
    ! 
    !
    !{\tiny\cite{Fahrenberg.Juhl.ea:2011}}
    !{\tiny\cite{Courtois.Schmitz:2014}}
  \\
  \addlinespace
  \multirow{3}{*}{"parity@parity vector game"}
    ! "existential"
    ! in $\NP\cap\coNP$
    ! in \coNP
    ! \coNP-complete
  \\[-.5em]
    ! 
    ! {\tiny\cite{Chatterjee.Doyen:2012}}
    ! 
    !{\tiny\Cref{13-thm:exist-hard,13-thm:exist-easy}~\cite{Chatterjee.Doyen.ea:2010}}
  \\
    ! "given"
    ! 
    ! \EXP-complete
    ! \kEXP[2]-complete
  \\[-.5em]
    ! 
    ! 
    ! {\tiny\Cref{13-thm:avag-two,13-thm:avag-easy}}
    ! {\tiny\Cref{13-thm:avag-hard,13-thm:avag-easy}}
  \\[-.7em]
    ! 
    !
    !{\tiny\cite{Fahrenberg.Juhl.ea:2011}}
    !{\tiny\cite{Courtois.Schmitz:2014}}
  \\
  \bottomrule  
  \end{tabular}
  \else
  \setlength{\tabcolsep}{7pt}
  \begin{tabular}{p{12em}cccc}
  \toprule 
  !!\multicolumn{3}{c}{Dimension}\\
  \cmidrule(l){3-5}
  Game ! Initial credit ! Fixed $\dd=1$ ! Fixed $\dd\geq 2$ ! Arbitrary\\
  \midrule
  configuration reachability 
  ! -
  ! $\EXPSPACE$-complete
  !\multicolumn{2}{c}{undecidable}
  \\[-.5em]
  ! 
  ! {\tiny\Cref{13-thm:asym-dim1}} 
  !\multicolumn{2}{c}{{\tiny\Cref{13-thm:asym-undec}~\cite{Lincoln.Mitchell.ea:1992}}}\\
  \addlinespace
  \multirow{3}{*}{"coverability"}
  ! "existential"
  ! \multicolumn{3}{c}{\P-complete}
  \\[-.5em]
  ! 
  ! \multicolumn{3}{c}{\tiny\Cref{13-thm:cov-exist-P}}
  \\
  ! "given"
  ! in $\NP\cap\coNP$ 
  ! \EXP-complete
  ! \kEXP[2]-complete
  \\[-.5em]
  ! 
  ! 
  !
    {\tiny\Cref{13-thm:avag-two,13-thm:avag-easy}~\cite{Fahrenberg.Juhl.ea:2011}}
  ! {\tiny\Cref{13-thm:avag-hard,13-thm:avag-easy} \cite{Courtois.Schmitz:2014}}\\
  \addlinespace
  \multirow{3}{*}{"non-termination"}
  ! "existential"
  ! in $\NP\cap\coNP$
  ! in \coNP
  ! \coNP-complete
  \\[-.5em]
  ! 
  ! {\tiny\cite{Chatterjee.Doyen:2012}} 
  !
  !{\tiny\Cref{13-thm:exist-hard,13-thm:exist-easy}~\cite{Chatterjee.Doyen.ea:2010}}
  \\
  ! "given"
  ! in $\NP\cap\coNP$ 
  ! \EXP-complete
  ! \kEXP[2]-complete
  \\[-.5em]
  ! 
  ! {\tiny\cite{Bouyer.Fahrenberg.ea:2008}}
  ! {\tiny\Cref{13-thm:avag-two,13-thm:avag-easy}~\cite{Fahrenberg.Juhl.ea:2011}}
  !{\tiny\Cref{13-thm:avag-hard,13-thm:avag-easy}~\cite{Courtois.Schmitz:2014}}
  \\
  \addlinespace
  \multirow{3}{*}{"parity@parity vector game"}
  ! "existential"
  ! in $\NP\cap\coNP$
  ! in \coNP 
  ! \coNP-complete
  \\[-.5em]
  ! 
  ! {\tiny\cite{Chatterjee.Doyen:2012}} 
  ! 
  ! {\tiny\Cref{13-thm:exist-hard,13-thm:exist-easy}~\cite{Chatterjee.Doyen.ea:2010}}
  \\
  ! "given"
  ! 
  ! \EXP-complete
  ! \kEXP[2]-complete
  \\[-.5em]
  ! 
  ! {\tiny}
  ! {\tiny\Cref{13-thm:avag-two,13-thm:avag-easy}~\cite{Fahrenberg.Juhl.ea:2011}} 
  ! {\tiny\Cref{13-thm:avag-hard,13-thm:avag-easy}~\cite{Courtois.Schmitz:2014}}\\
  \bottomrule  
  \end{tabular}
  \fi
  \endgroup
    \end{landscape}


\section*{Acknowledgments}
This work was partially funded by ANR-17-CE40-0028 \textsc{BraVAS}.




\part{Multi}
\label{part:multi}

\ifpictures
\includepdf{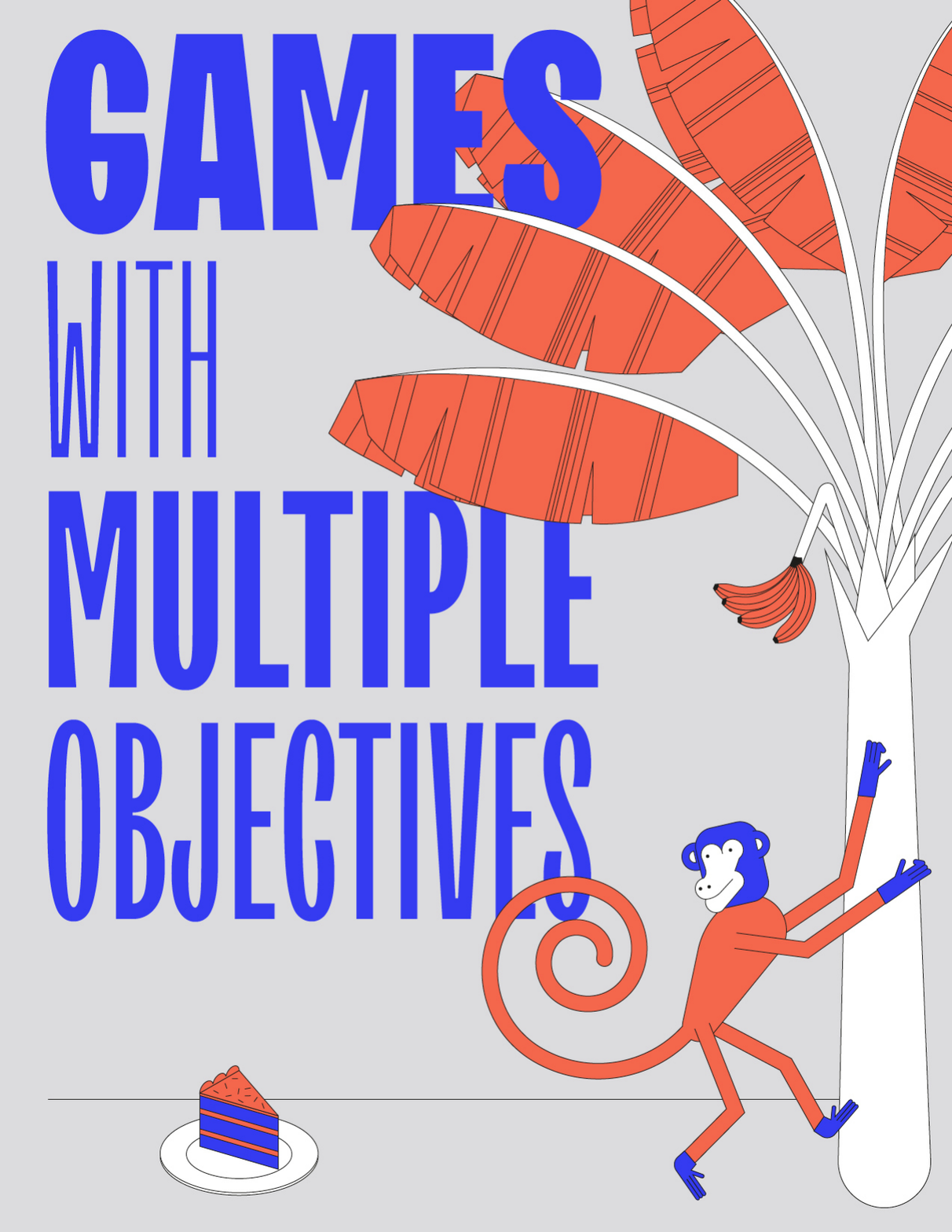}
\fi
\author[Mickael Randour]{Mickael Randour}
\copyrightline{Copyright by Mickael Randour 2025, to be published by Cambridge University Press in the volume \textit{Games on Graphs} edited by Nathana\"el Fijalkow}

\chapter{Games with Multiple Objectives}
\chapterauthor{Mickael Randour}
\label{14-chap:multiobjective}



\providecommand{\expv}{\mathbb{E}} 
\renewcommand{\expv}{\mathbb{E}} 
\newcommand{\markovProcess}{\ensuremath{{\mathcal{P}}}}
\newcommand{\stratStoch}{\ensuremath{\tau^{\mathsf{st}}}}
\newcommand{\BWC}{\text{BWC}}
\newcommand{\ecsSet}{\ensuremath{\mathcal{E}}}
\newcommand{\edgesNonZero}{\ensuremath{E_{\delta}}}
\newcommand{\ec}{\ensuremath{U}}
\newcommand{\playerOne}{\text{Eve}}
\newcommand{\playerTwo}{\text{Adam}}
\newcommand{\winningECs}{\ensuremath{\mathcal{W}}}
\newcommand{\losingECs}{\ensuremath{\mathcal{L}}}
\providecommand{\edges}{\ensuremath{E}}
\renewcommand{\edges}{\ensuremath{E}}
\newcommand{\maxWinningECs}{\ensuremath{\mathcal{U}_{\mathsf{w}}}}
\newcommand{\infVisited}[1]{\ensuremath{{\mathtt{Inf}}(#1)}}
\newcommand{\negligibleStates}{\ensuremath{V_{{\sf neg}}}}
\newcommand{\stratWC}{\ensuremath{\sigma^{\mathsf{wc}}}}
\newcommand{\stratExp}{\ensuremath{\sigma^{\mathsf{e}}}}
\newcommand{\stratComb}{\ensuremath{\sigma^{\mathsf{cmb}}}}
\newcommand{\stratSecure}{\ensuremath{\sigma^{\mathsf{sec}}}}
\newcommand{\stratWNS}{\ensuremath{\sigma^\mathsf{wns}}}
\newcommand{\stratGlobal}{\ensuremath{\sigma^{\mathsf{glb}}}}
\newcommand{\stepsWC}{\ensuremath{L}}
\newcommand{\stepsExp}{\ensuremath{K}}
\newcommand{\stepsGlobal}{\ensuremath{N}}
\newcommand{\cmbSum}{\ensuremath{\mathtt{Sum}}}
\newcommand{\typeA}{\ensuremath{\mathit{(a)}}}
\newcommand{\typeB}{\ensuremath{\mathit{(b)}}}
\newcommand{\thresholdWC}{\ensuremath{\alpha}}
\newcommand{\thresholdExp}{\ensuremath{\beta}}
\providecommand{\state}{\ensuremath{v}}
\renewcommand{\state}{\ensuremath{v}}
\newcommand{\gameNonZero}{\ensuremath{\arena_{\delta}}}
\newcommand{\reduc}{\ensuremath{\downharpoonright}}
\providecommand{\probm}{\mathbb{P}} 
\renewcommand{\probm}{\mathbb{P}} 

Up to this chapter, we have mostly been interested in finding strategies that achieve a \emph{single} objective or optimise a \emph{single} payoff function. Our goal here is to discuss what happens when one goes further and wants to build strategies that (i) ensure \emph{several objectives}, or~(ii)~provide \emph{richer guarantees} than the simple worst-case or expectation ones used respectively in zero-sum games and Markov decision processes (MDPs). 

Consider case (i). Such requirements arise naturally in applications: for instance, one may want to define a trade-off between the performance of a system and its energy consumption. A model of choice for this is the natural \textit{multidimensional} extension of the games of~\Cref{5-chap:payoffs}, where we consider weight vectors on edges and combinations of objectives.

In case (ii), we base our study on stochastic models such as MDPs (\Cref{6-chap:mdp}). We will notably present how to devise controllers that provide strong guarantees in a worst-case scenario while behaving efficiently on average (based on a stochastic model of its environment built through statistical observations); effectively reconciling the rational antagonistic behaviour of Adam, used in games, with the stochastic interpretation of uncontrollable interaction at the core of MDPs.

Stepping into the multi-objective world is like entering a jungle: the sights are amazing but the wildlife is overwhelming. Providing an exhaustive account of existing multi-objective models and the latest developments in their study is a task doomed to fail: simply consider the combinatorial explosion of all the possible combinations based on the already non-exhaustive set of games studied in the previous chapters. Hence, our goal here is to guide the reader through their first steps in the jungle, highlighting the specific dangers and challenges of the multi-objective landscape, and displaying some techniques to deal with them. To that end, we focus on models studied in~\Cref{3-chap:regular},~\Cref{5-chap:payoffs},~\Cref{6-chap:mdp} and~\Cref{13-chap:counters}, and multi-objective settings that extend them. We favour simple, natural classes of problems, that already suffice to grasp the cornerstones of multi-objective reasoning. 

\paragraph{Chapter outline.} In~\Cref{14-sec:multiple_dimensions}, we illustrate the additional complexity of multi-objective games and how relations between different classes of games that hold in the single-objec\-tive case often break as soon as we consider combinations of objectives. 

The next two sections are devoted to the \emph{simplest form} of multi-objective games: games with \emph{conjunctions} of classical objectives. In~\Cref{14-sec:mean_payoff_energy}, we present the classical case of multidimensional "mean-payoff" and "energy" games, which preserve relatively nice properties with regard to their single-objective counterparts. In~\Cref{14-sec:total_payoff_shortest_path}, we discuss the opposite situation of "total-payoff" and "shortest-path" games: their nice single-objective behaviour vanishes here.

In the last two sections, we explore a different meaning of \emph{multi-objective} through so-called rich behavioural models. Our quest here is to find strategies that provide several types of guarantees, of different nature, for the same quantitative objective. In~\Cref{14-sec:beyond_worst_case}, we address the problem of \emph{"beyond worst-case synthesis"}, which combines the rational antagonistic interpretation of two-player zero-sum games with the stochastic nature of MDPs. We will study the mean-payoff setting and see how to construct strategies that ensure a strict worst-case constraint while providing the highest expected value possible.
In~\Cref{14-sec:percentile}, we briefly present \emph{"percentile queries"}, which extend \textit{probability threshold problems} in MDPs to their multidimensional counterparts. Interestingly, \emph{"randomised strategies"} become needed in this context, whereas up to~\Cref{14-sec:percentile}, we only consider "deterministic" strategies as they suffice.

We close the chapter with the usual bibliographic discussion and pointers towards some of the many recent advances in multi-objective reasoning.

\section{From one to multiple dimensions}
\label{14-sec:multiple_dimensions}
For the first part of this chapter, we consider multidimensional quantitative games. With regard to the formalism of~\Cref{5-chap:payoffs}, the only change to the arena is the set of colours associated with edges: we now have vectors in $\R^k$ where $k \in \N_{>0}$ is the dimension of the game. As before, for computational purposes, it makes sense to restrict our colouring to rational numbers, and for the sake of simplicity, we even consider \emph{integers only} without loss of generality. Hence, $\col\colon E \rightarrow \Z^k$.

For the weighted games of~\Cref{5-chap:payoffs}, where a single quantitative objective $f$ is considered, we know that the "value" of the game exists. In most cases, optimal strategies do too, which makes the problems of computing the value and solving the game for a given threshold morally equivalent. In our simple multidimensional setting, we focus on \emph{conjunctions} of objectives. Similarly to what we did in the one-dimension case, we will write $f_{\geq \vec{x}}$ with $\vec{x} \in \Q^k$ to define the (qualitative) winning condition
\[
f_{\geq \vec{x}} = \bigcap_{i = 1}^{k} \left\lbrace \play \in \Paths_\omega(G) \mid f_i(\play) \geq \vec{x}_i\right\rbrace 
\]
where $f_i(\play)$ represents the evaluation of $f$ on the sequence of colours in the $i$-th dimension and $\vec{x}_i$ represents the $i$-th component of vector $\vec{x}$. Hence we consider the natural semantics where we want to satisfy the original objective $f$ component-wise.

\begin{example}
\label{14-ex:MMP}
Consider the simple one-player game in~\Cref{14-fig:MultiMP} fitted with the "mean-payoff" objective $\MeanPayoff^-$  (recall that two variants exist depending on the use of lim-sup or lim-inf). Let us first recall that in the single-objective case, positional strategies suffice to play optimally (\Cref{4-thm:lift_applications}). In this game, such strategies permit to achieve payoffs $(1,-1)$, $(-1,-1)$ and $(-1,1)$. Intuitively, $(-1,-1)$ is not interesting since we can do better with $(1,-1)$ or $(-1,1)$. On the other hand, these two other payoffs are incomparable and thus should not be discriminated a priori. In the multi-objective world, there is usually no total order between the outcomes of a game --- fixing a total order would actually boil down to transforming the game into a one-dimension game --- which is why there is in general no optimal strategy but rather \emph{Pareto-optimal} ones. In a nutshell, a strategy is Pareto-optimal if there exists no other strategy yielding a payoff which is as good in all dimensions and strictly better in at least one dimension.
\end{example}

\begin{definition}[Pareto-optimal strategy]
\label{14-def:ParetoStrat}
Given a $k$-dimension game $\game$ based on the conjunction of $k$ maximising (w.l.o.g.) quantitative objectives $(f_i)_{i=1}^{k}$, a strategy $\sigma$ for Eve is said to be \emph{Pareto-optimal} if it guarantees a payoff $\vec{x} \in \R^k$ such that for all other strategy $\sigma'$ of Eve ensuring payoff $\vec{x}' \neq \vec{x}$, it holds that $\vec{x}_i > \vec{x}'_i$ for some dimension $i \in \{1, \ldots, k\}$.
\end{definition}

\begin{figure}[tbp]
  \centering
  \begin{tikzpicture}[node distance=3cm,>=latex]
    \node[draw,circle](1) {$v_0$};%
    \node[draw,circle,right of=1](2) {$v_1$};%

    \path[->] (1) edge[bend left=20] node[above] {$(-1,-1)$} (2)%
    (2) edge[bend left=20] node[below] {$(-1,-1)$} (1)%
    (1) edge[loop left] node[left] {$(1,-1)$} (1)%
    (2) edge[loop right] node[right] {$(-1,1)$} (2);%

  \end{tikzpicture}
  \caption{A simple multidimensional mean-payoff game where Eve needs infinite memory to play (Pareto-)optimally.}
  \label{14-fig:MultiMP}
\commentAlt{Figure~\ref{14-fig:MultiMP}: A directed graph with two circular nodes, v0 and v1, showing bidirectional transitions with labeled pairs of numbers, and self-loops on each node.}
\commentLongAlt{Figure~\ref{14-fig:MultiMP}: The image displays a directed graph with two circular nodes, 'v0' on the left and 'v1' on the right.
- Node 'v0' has a self-loop labeled '(1, -1)'.
- A bidirectional arrow connects 'v0' and 'v1'. The arrow from 'v0' to 'v1' is labeled '(-1, -1)'. The arrow from 'v1' to 'v0' is also labeled '(-1, -1)'.
- Node 'v1' has a self-loop labeled '(-1, 1)'.}
\end{figure}

The concept of Pareto-optimality has an important consequence on multi-objective problems: the correspondence between solving a value problem and computing an optimal strategy that holds in the single-objective case does not carry over. Indeed, one may now be interested in computing the Pareto frontier consisting of all Pareto  vectors achievable by Eve. This comes at great cost complexity-wise as this frontier may include many points, and in some settings, even an \emph{infinite number of Pareto vectors} (see~\Cref{14-sec:percentile} for examples), sometimes forcing us to resort to approximation. This requires specific techniques that go beyond the focus of this chapter, hence in the following we mostly discuss the \emph{"value problem"}, also referred to as `solving the game' for a given threshold vector.

\begin{example}
\label{14-ex:MMP2}
Let us go back to~\Cref{14-ex:MMP} and fix objective $\MeanPayoff^{-}_{\geq \vec{x}}$ where $\vec{x} = (0, 0)$. As discussed before, this threshold cannot be achieved by a positional strategy. Actually, this is also the case for any \emph{finite-memory} strategy. Indeed, any finite-memory strategy induces an ultimately periodic play, where either (a) the periodic part only visits $v_0$ (resp.~$v_1$), yielding payoff $(1,-1)$ (resp.~$(-1,1)$) thanks to "prefix-independence" of the mean payoff (\Cref{5-chap:payoffs}), or (b) it visits both, in which case the mean payoff is of the form
\[
\vec{y} = \MeanPayoff^{-}(\play) = \dfrac{a \cdot (1, -1) + 2 \cdot b \cdot (-1, -1) + c \cdot (-1, 1)}{a + 2 \cdot b + c}
\]
where $a, c \in \N$ and $b \in \N_{>0}$. Observe that $\vec{y}_1 + \vec{y}_2 = -4\cdot b / (a + 2 \cdot b + c)$, which is strictly less than zero for any value of the parameters. Hence $\vec{x} = (0, 0)$ is not achievable. Now consider what happens with infinite memory: let $\sigma$ be the strategy of Eve that visits $\ell$ times $v_0$, then $\ell$ times  $v_1$, and then repeats forever with increasing values of $\ell$. The mean payoff of the resulting play is the limit of the previous equation when $a = c = \ell$ tends to infinity, with $b = 1$: intuitively, the switch between $v_0$ and $v_1$ becomes negligible in the long run and the mean payoff is $\frac{1}{2} \cdot (1,-1) + \frac{1}{2}\cdot(-1,1) = (0, 0)$.
\end{example}

\begin{remark}
While Eve cannot achieve $(0, 0)$ with finite memory, she can achieve (\textit{i.e.}, ensure at least) any payoff $(-\varepsilon, -\varepsilon)$ for $0 < \varepsilon < 1$, using sufficient memory: for instance, by taking $b = 1$ and $a = c = \lceil \frac{1}{\varepsilon} - 1\rceil$. In that sense, the payoff $\vec{x} = (0, 0)$ achievable by an infinite-memory strategy can be seen as the supremum of payoffs achievable by finite-memory strategies. Actually, this is exactly how we defined strategy $\sigma$: Eve plays according to an infinite sequence of finite-memory strategies parametrised by $\ell$, such that each strategy of the sequence ensures mean payoff $(-\varepsilon, -\varepsilon)$, with $\varepsilon \to 0$ when $\ell \to \infty$.
\end{remark}

\begin{example}
\label{14-ex:MMP3}
The reasoning above holds similarly for $\MeanPayoff^{+}$. With finite memory, the lim-sup variant coincides with the lim-inf one: because the play is \textit{ultimately periodic}, the limit exists. With infinite memory, Eve can actually achieve the payoff $\vec{x}' = (1, 1)$, witnessing a gap with the lim-inf variant. To do so, she has to play a strategy that alternates between $v_0$ and $v_1$ while staying in each vertex for a sufficiently long period such that the current mean over the corresponding dimension gets close to~$1$. Getting these means closer and closer to $1$ and using the lim-sup component-wise then suffices to achieve payoff $\vec{x}'$. This is in stark contrast to the lim-inf variant, which cannot achieve any payoff $(\varepsilon, \varepsilon)$ for $\varepsilon > 0$ (the Pareto vectors correspond to linear combinations of simple cycles, as hinted before).
\end{example}

\begin{theorem}
\label{14-thm:MMP-Eve}
Multidimensional mean-payoff games require infinite-memory strategies for Eve. Furthermore, the lim-inf and lim-sup variants are not equivalent, \textit{i.e.}, their winning regions are in general not identical.
\end{theorem}

This theorem already shows the first signs of our single-objective assumptions crumbling in the multi-objective world: we jump from positional determinacy to needing infinite memory, and objectives that were equivalent both in games and MDPs turn out to be different here. Buckle up, as this was only our first step.

\section{Mean payoff and energy}
\label{14-sec:mean_payoff_energy}
Another well-known equivalence in one-dimension is the one between "mean-payoff" and "energy" games (in the classical "existential initial credit" form, as in~\Cref{13-chap:counters}\footnote{If one adopts the quantitative view of energy games --- as in~\Cref{5-chap:payoffs}, the existence of an initial credit such that the (qualitative) energy objective is satisfied (\textit{i.e.}, the running sum of weights never drops below zero) is equivalent to having $\Energy < \infty$.}), mentioned in~\Cref{5-chap:payoffs}. The reduction is trivial: Eve has a winning strategy (and an initial credit) in the energy game if and only if she has a strategy to ensure mean payoff at least equal to zero in the mean-payoff game played over the same arena. Intuitively, the mean-payoff strategy of Eve has to reach a subgame where she can ensure that all cycles formed are non-negative. The initial credit (which can be as high as Eve wants) offsets the cost of reaching such a subgame as well as the low point of cycles in it (which can be negative but is bounded).

How does it fare in multiple dimensions? The study of "vector games" with "energy semantics" in~\Cref{13-chap:counters} gives the following result.

\begin{theorem}
\label{14-thm:MEG}
Solving multidimensional energy games is \coNP-complete. Exp\-on\-en\-tial-memory strategies suffice and are required for Eve, and positional ones suffice for Adam.
\end{theorem} 

Based on~\Cref{14-thm:MMP-Eve} and~\Cref{14-ex:MMP2}, it is clear that the aforementioned equivalence holds no more, as mean-payoff games benefit from infinite memory while energy games do not. In~\Cref{14-ex:MMP2}, the strategy that achieves $\vec{x} = (0, 0)$ for the mean payoff does so by switching infinitely often but with decreasing frequency between $v_0$ and $v_1$: the switch becomes negligible in the limit which is fine for the mean payoff. Still, this would lead the energy to drop below zero eventually, whatever the initial credit chosen by Eve, hence showing why the reduction does not carry over.

\subsection{Finite memory}

Game-theoretic models are generally used in applications, such as controller synthesis, where one actually wants to \textit{implement} a winning strategy when it exists. This is why "finite-memory" strategies have a particular appeal. Hence it is interesting to study what happens when we restrict Eve to finite-memory strategies in multidimensional mean-payoff games. 

We first observe that when both players use finite-memory strategies, the resulting play is ultimately periodic, hence the lim-inf and lim-sup variants coincide (the limit exists) and take the value of the mean over the periodic part.

\begin{proposition}
\label{14-prop:MPSI}
The lim-sup and lim-inf variants of multidimensional mean-payoff ga\-mes coincide under finite memory, \textit{i.e.}, their winning regions are identical in all games.
\end{proposition}

We now go back to the relationship with energy games. In the following, we write  $\vec{0}$ for the $k$-dimension vector $(0,\ldots{},0)$. When restricting both players to finite memory, we regain the equivalence between mean-payoff and energy games by a natural extension of the argument sketched above for one-dimension games.

\begin{theorem}
\label{14-thm:MPEG-equivalence}
For all arena and initial vertex, Eve has a winning strategy for the existential initial credit multidimensional energy game if and only if she has a finite-memory winning strategy for the multidimensional (lim-inf or lim-sup) mean-payoff game with threshold~$\vec{0}$.
\end{theorem}

\begin{proof}
Let $\arena$ be an arena coloured by integer vectors of dimension $k$ and $v_0$ be the initial vertex. We first consider the left-to-right implication. Assume that Eve has a strategy $\sigma$ and some initial credit $\vec{c}_0 \in \N^k$ such that she wins the energy objective over $\arena$. By~\Cref{14-thm:MEG}, we may assume $\sigma$ to be finite-memory and $\mem = (M, m_0, \delta)$ to be its "memory structure". Let $\arena_\sigma$ be the classical product of the arena with this memory structure ($\arena \times \mem$) restricted to the choices made by $\sigma$. We claim that any cycle in $\arena_\sigma$ is non-negative in all dimensions (we simply project paths of $\arena_\sigma$ to $C^\omega$ to interpret them as we do for paths in $\arena$). By contradiction, assume that there exists a cycle whose sum of weights is strictly negative in some dimension. Then the play reaching this cycle and looping in it forever is a play consistent with $\sigma$ that is losing for the energy objective, contradicting the hypothesis. Hence, it is indeed the case that all reachable cycles in $\arena_\sigma$ are non-negative in all dimensions. Thus, $\sigma$ ensures mean payoff at least equal to zero in all dimensions (for lim-inf and lim-sup variants).

In the opposite direction, let us assume that $\sigma$ is a finite-memory winning strategy for $\MeanPayoff^{-}_{\geq \vec{0}}$ (or equivalently $\MeanPayoff^{+}_{\geq \vec{0}}$). Using the same argument as before, we have that all cycles in $\arena_\sigma$ are non-negative. Therefore there exists some initial credit $\vec{c}_0 \in \N^k$ such that $\sigma$ satisfies the energy objective. As a trivial bound, one may take initial credit $\vert V\vert \cdot \vert M \vert \cdot W$ in all dimensions, where $\vert V\vert$ is the number of vertices of $\arena$, $\vert M \vert$ the number of memory states of $\mem$, and $W$ is the largest absolute weight appearing in the arena: this quantity bounds the lowest sum of weights achievable under an acyclic path.
\end{proof}

Observe that the finite-memory assumption is crucial to lift mean-payoff winning strategies to the energy game. Intuitively, the reasoning would break for a strategy like the one used in~\Cref{14-ex:MMP2} because the memory structure would need to be infinite and $\arena_\sigma$ would actually not contain any cycle but an infinite path of ever-decreasing energy such that no bound on the initial credit could be established.

Also, note that~\Cref{14-thm:MPEG-equivalence} makes no mention of the specific variant of mean payoff used. This is because both players play using finite-memory: Eve by hypothesis and Adam thanks to the equivalence and~\Cref{14-thm:MEG}. Hence,~\Cref{14-prop:MPSI} applies. To sum up, we obtain the following.

\begin{corollary}
Solving multidimensional mean-payoff games under finite-memory is \coNP-complete. Exponential-memory strategies suffice and are required for Eve, and positional ones suffice for Adam.
\end{corollary}

\subsection{Infinite memory}

We now turn to the general case, where Eve is allowed to use infinite memory. By~\Cref{14-ex:MMP3}, we already know that lim-sup and lim-inf variants are not equivalent. We will cover the lim-sup case in details and end with a brief overview of the lim-inf one.

\subsection*{Lim-sup variant}

Without loss of generality, we fix the objective $\MeanPayoff^{+}_{\geq \vec{0}}$ (one can always modify the weights in the arena and consider the shifted-game with threshold zero). We have seen in \Cref{14-ex:MMP3} that Eve could focus on each dimension independently and alternatively in such a way that in the limit, she obtains the supremum in each dimension. This is the core idea that we will exploit.

\begin{lemma}
\label{14-lem:MMP-Eve}
Let $\arena$ be an arena such that from all vertex $v \in \vertices$ and for all dimension~$i$, $1 \leq i \leq k$, Eve has a winning strategy for $\{\play \in \mathrm{Paths}_\omega(G) \mid \MeanPayoff^{+}_{i}(\play) \geq 0\}$. Then, from all vertex $v \in \vertices$, she has a winning strategy for $\MeanPayoff^{+}_{\geq \vec{0}}$.
\end{lemma}

Hence, being able to win in each dimension \textit{separately} suffices to guarantee winning in all dimensions \textit{simultaneously}. Note that the converse is obvious.

\begin{proof}
For each vertex $v \in \vertices$ and dimension $i$, $1 \leq i \leq k$, let $\sigma_i^v$ be a winning strategy for Eve from $v$ for $\{\play \in \Paths_\omega(G) \mid \MeanPayoff^{+}_{i}(\play) \geq 0\}$.

Let $T_{\sigma_i^v}$ be the infinite tree obtained by \textit{unfolding}  $\sigma_i^v$: it represents all plays consistent with this strategy. Formally, such a tree is obtained inductively as follows:
\begin{itemize}
\item The root of the tree represents $v$.
\item Given a node\footnote{Nodes refer to the tree, vertices to the arena.} $\eta$ representing the branch (\textit{i.e.}, prefix of play) $\rho$ starting in vertex $v$ and ending in vertex $v_\eta$, we add children as follows:
\begin{itemize}
\item if $v_\eta \in V_{\text{Eve}}$, $\eta$ has a unique child representing the vertex $\out(e)$ reached through edge $e = \sigma_i^v(\rho)$;
\item otherwise $\eta$ has one child for each possible successor of $v_\eta$, \textit{i.e.}, for each $\out(e)$ such that $e \in E$ and $\ing(e) = v_\eta$.
\end{itemize} 
\end{itemize} 
For $\varepsilon > 0$, we declare a node $\eta$ of $T_{\sigma_i^v}$ to be \textit{$\varepsilon$-good} if the mean over dimension $i$ on the path from the root to $\eta$ is at least $-\varepsilon$ (as usual, we project this path to $C^\omega$ to evaluate it). For $\ell \in \N$, let $\widehat{T}^{i, \ell}_{v, \varepsilon}$ be the tree obtained from $T_{\sigma_i^v}$ by removing all descendants of $\varepsilon$-good nodes that are at depth at least $\ell$: hence, all branches of $\widehat{T}^{i, \ell}_{v, \varepsilon}$ have length at least $\ell$ and their leaves are $\varepsilon$-good.

We first show that $\widehat{T}^{i, \ell}_{v, \varepsilon}$ is a finite tree. By K\"onig's Lemma~\cite{Konig:1936}, we only need to show that every branch is finite. By contradiction, assume it is not the case and there exists some infinite branch. By construction, it implies that this branch contains no $\varepsilon$-good node after depth $\ell$. Thus, the corresponding play $\pi$, which is consistent with $\sigma_i^v$, necessarily has $\MeanPayoff^{+}_{i}(\play) \leq -\varepsilon$. This contradicts the hypothesis that $\sigma_i^v$ is winning for dimension~~$i$. Hence the tree is indeed finite.

Based on these finite trees, we now build an infinite-memory strategy for Eve that will be winning for the conjunct objective $\MeanPayoff^{+}_{\geq \vec{0}}$: it is presented as~\Cref{14-algo:WinningStrategyForMultiMPSup}.

\SetKwBlock{Loop}{loop}{EndLoop}
\begin{algorithm}[thb]
$\varepsilon \leftarrow 1$

\Loop{
	\For{$i = 1$ to $k$}{
		Let $v$ be the current vertex, $L$ the length of the play so far
		
		$\ell \leftarrow \left\lceil\frac{L\cdot W}{\varepsilon}\right\rceil$
		
		Play according to $\sigma_i^v$ until a leaf of $\widehat{T}^{i, \ell}_{v, \varepsilon}$ is reached
	}
	$\varepsilon \leftarrow \frac{\varepsilon}{2}$
}
\caption{Winning strategy $\sigma$ for $\MeanPayoff^{+}_{\geq \vec{0}}$}
\label{14-algo:WinningStrategyForMultiMPSup}
\end{algorithm}

Recall that $W$ is the largest absolute weight in the game. Consider the situation whenever an iteration of the for-loop ends. Let $M$ be the number of steps the play followed $\sigma^v_i$ during this loop execution. Then, the mean payoff in dimension $i$ is at least $\frac{-L\cdot W - M\cdot \varepsilon}{L + M} \geq \frac{-L\cdot W - M\cdot \varepsilon}{M}$. Since $M \geq \frac{L\cdot W}{\varepsilon}$ by definition, we obtain that the mean payoff in dimension $i$ is at least $-2\cdot \varepsilon$.

Observe that since all trees are finite, we always exit the for-loop eventually, hence $\varepsilon$ tends to zero. Therefore, the supremum mean payoff is at least zero in all dimensions, which makes this strategy winning for $\MeanPayoff^{+}_{\geq \vec{0}}$.
\end{proof}

This construction is tight in the sense that infinite memory is needed for Eve, as previously proved. For Adam, we show a better situation. The proof scheme will also be the base of the upcoming algorithm.

\begin{lemma}
\label{14-lem:MMP-Adam}
Positional strategies suffice for Adam in multidimensional lim-sup mean-payoff games.
\end{lemma}

\begin{proof}
The proof works by induction on the number of vertices of the arena. The base case $\vert V\vert = 1$ is trivial. Assume the only vertex belongs to Adam. If there exists a self-loop (recall we allow several edges per pair of vertices) which has a negative weight on some dimension, Adam wins by looping on it forever. In the opposite case, he cannot win.

Now assume $\vert V\vert \geq 2$. For $i \in \{1,\ldots,k\}$, let $W^i_{\text{Adam}}$ be the winning region of Adam for the complement of $\{\play \in \Paths_\omega(G) \mid \MeanPayoff^{+}_{i}(\play) \geq 0\}$, \textit{i.e.}, the region where Adam has a strategy to force a strictly negative mean payoff in dimension $i$ (as studied in~\Cref{5-chap:payoffs}). Let $W^{\text{disj}}_{\text{Adam}} = \bigcup_{i = 1}^{k} W^i_{\text{Adam}}$. We have two cases.

First, $W^{\text{disj}}_{\text{Adam}} = \emptyset$. Then, Eve can win all one-dimension games from everywhere and by~\Cref{14-lem:MMP-Eve}, she can also win for $\MeanPayoff^{+}_{\geq \vec{0}}$. Thus, Adam has no winning strategy.

Second, $W^{\text{disj}}_{\text{Adam}} \neq \emptyset$. Then, there exists $i \in \{1,\ldots,k\}$ such that $W^i_{\text{Adam}} \neq \emptyset$. In this set, Adam has a positional strategy $\tau_i$ to avoid $\{\play \in \Paths_\omega(G) \mid \MeanPayoff^{+}_{i}(\play) \geq 0\}$ (because one-dimension mean-payoff games are positionally determined, as proved in~\Cref{4-thm:lift_applications}). This strategy also falsifies $\MeanPayoff^{+}_{\geq \vec{0}}$, hence $W^i_{\text{Adam}}$ is part of the winning region for Adam --- we denote it $W_{\text{Adam}}$, as usual.
By prefix-independence of the mean payoff, the "attractor" $W^{i, \Pre}_{\text{Adam}}= \AttrA(W^i_{\text{Adam}})$ is also part of $W_{\text{Adam}}$. We denote by $\tau_\Pre$ the corresponding attractor strategy of Adam. Moreover, the graph restricted to $\vertices \setminus W^{i, \Pre}_{\text{Adam}}$ constitutes a proper arena $\arena'$.

Let $W'_{\text{Adam}}$ be the winning region of Adam in $\arena'$ for the original winning condition $\MeanPayoff^{+}_{\geq \vec{0}}$. The arena $\arena'$ has strictly less vertices than $\arena$ since we removed the non-empty region $W^i_{\text{Adam}}$. Hence we can apply the induction hypothesis: Adam has a positional winning strategy $\tau'$ in $W'_{\text{Adam}}$. The region $V \setminus (W^{i, \Pre}_{\text{Adam}} \cup W'_{\text{Adam}})$ is winning for Eve in $\arena'$ by determinacy. But it is also winning in $\arena$ --- the original game --- since Adam cannot force the play to go in $W^{i, \Pre}_{\text{Adam}}$ from there (otherwise that region would be part of the attractor too).

We define the following positional strategy for Adam, which we claim is winning from $W_{\text{Adam}} = W^{i, \Pre}_{\text{Adam}} \cup W'_{\text{Adam}}$:
\[
\tau(v) =
\begin{cases}
\tau_\Pre(v) &\text{if } v \in W^{i, \Pre}_{\text{Adam}} \setminus W^{i}_{\text{Adam}},\\
\tau_i(v) &\text{if } v \in W^{i}_{\text{Adam}},\\
\tau'(v) &\text{if } v \in W'_{\text{Adam}}.
\end{cases}
\]
Since we already know that Eve wins from $V \setminus W_{\text{Adam}}$, it remains to prove that $\tau$ is winning from $W_{\text{Adam}}$ to conclude. Consider any play $\pi$ consistent with $\tau$ and starting in $W_{\text{Adam}}$. Two cases are possible. First, the play eventually reaches $W^{i}_{\text{Adam}}$ and Adam switches to $\tau_i$: then prefix-independence of the mean payoff guarantees that Adam wins. Second, the play never reaches $W^{i}_{\text{Adam}}$: then $\pi$ necessarily stays in $\arena'$, and $\tau'$ is winning from $W'_{\text{Adam}}$ in $\arena'$. Therefore, $\tau$ does win from everywhere in $W_{\text{Adam}}$, while being positional, which ends the proof.
\end{proof}

We use the core reasoning of this proof to build an algorithm solving multidimensional lim-sup mean-payoff games (\Cref{14-algo:MMP}). It uses as a black box a routine that computes (in "pseudo-polynomial" time) the winning vertices for Eve in one-dimension mean-payoff games. This subalgorithm, originally presented in~\Cref{5-sec:mean_payoff}, is here dubbed $\mathtt{SolveOneDimMeanPayoff}$, and takes as parameters the arena and the considered dimension.

\begin{algorithm}[tbh]
 \DontPrintSemicolon
 \KwData{Arena $\arena$ with vertices $\vertices$}
 \KwResult{$W_{\text{Eve}}$, the winning region of Eve for $\MeanPayoff^{+}_{\geq \vec{0}}$}
 $\arena' \leftarrow \arena$; $V' \leftarrow V$\\
 \Repeat{$\mathtt{LosingVertices} = \mathtt{false}$}{
  $\mathtt{LosingVertices} \leftarrow \mathtt{false}$\\
  \For{$i = 1$ to $k$}{
    $W^i_{\mathrm{Adam}} \leftarrow V' \setminus \mathtt{SolveOneDimMeanPayoff}(\arena', i)$\\
    \If{$W^i_{\mathrm{Adam}} \neq \emptyset$}{
    $V' \leftarrow V' \setminus W^i_{\text{Adam}}$\\
    $\arena' \leftarrow \arena'[V']$\tcc*{Restriction of $\arena'$ to $V'$}
    $\mathtt{LosingVertices} \leftarrow \mathtt{true}$\\
    }
  }
 }
 \Return $V'$
 \caption{Solver for multidimensional lim-sup mean-payoff games}
 \label{14-algo:MMP}
\end{algorithm}

Intuitively, we iteratively remove vertices that are declared losing for Eve because Adam can win on some dimension from them. Since removing vertices based on some dimension $i$ may decrease the power of Eve and her ability to win for another dimension~$i'$, we need the outer loop: in the end, we ensure that $V'$ contains exactly all the vertices from which Eve has a winning strategy for each dimension. By~\Cref{14-lem:MMP-Eve} and the proof of~\Cref{14-lem:MMP-Adam}, we know that this is equal to $W_{\text{Eve}}$.

We recall (\Cref{1-sec:subgames}) that, given an arena $\arena$ and a set of vertices $X$, $\arena[X]$ denotes the subarena induced by $X$.

\begin{remark}
\label{14-rmk:properArena}
The restriction $\arena'[V']$ induces a proper subarena. Indeed, we have that $W^i_{\mathrm{Adam}} = \AttrA(W^i_{\mathrm{Adam}})$ since any vertex $v$ from which Adam can force to reach $W^i_{\mathrm{Adam}}$ also belongs to $W^i_{\mathrm{Adam}}$ by prefix-independence of the mean payoff.
\end{remark}

We wrap up with the following theorem.
\begin{theorem}
\label{14-thm:MMPsup}
Solving multidimensional lim-sup mean-payoff games is in $\NP \cap \coNP$. Infinite-memory strategies are required for Eve and positional ones suffice for Adam. Furthermore, the winning regions can be computed in pseudo-polynomial time, through at most $\vert V \vert \cdot k$ calls to an algorithm solving one-dimension mean-payoff games.
\end{theorem}

\begin{proof}
The correctness of~\Cref{14-algo:MMP} follows from~\Cref{14-lem:MMP-Eve} and~\Cref{14-lem:MMP-Adam}, and its complexity is obtained using $\mathtt{SolveOneDimMeanPayoff}$ as a pseudo-polyn\-omial black-box. The memory bounds follow from~\Cref{14-lem:MMP-Eve},~\Cref{14-lem:MMP-Adam} and~\Cref{14-thm:MMP-Eve}. Hence, only the $\NP \cap \coNP$ membership remains. Recall that the decision problem under study is: given an arena $\arena$ and an initial vertex $v_0$, does $v_0$ belong to $W_{\text{Eve}}$ or not?

We first prove that the problem is in $\NP$. A non-deterministic algorithm guesses the winning region $W_{\text{Eve}}$ containing $v_0$ and witness positional strategies $\sigma_i$ for all dimensions (we know that positional strategies suffice by~\Cref{4-thm:lift_applications}). Then, it checks for every dimension $i$, for every vertex $v \in W_{\text{Eve}}$, that $\sigma_i$ is winning. This boils down to solving a polynomial number of one-player one-dimension mean-payoff games for Adam over the arenas $\arena_{\sigma_i}$ obtained by fixing $\sigma_i$. As noted in~\Cref{5-sec:mean_payoff}, it can be done in polynomial time using Karp's algorithm for finding the minimum cycle mean in a weighted digraph~\cite{Karp:1978}. By~\Cref{14-lem:MMP-Eve}, we know that if the verification checks out, Eve has a winning strategy in $W_{\text{Eve}}$ for objective $\MeanPayoff^{+}_{\geq \vec{0}}$.

Finally, we prove $\coNP$ membership. The algorithm guesses a positional winning strategy $\tau$ for Adam (from $v_0$). The verification then consists in checking that Eve has no winning strategy in the arena $\arena_{\tau}$. This can be done using~\Cref{14-algo:MMP}, through $\vert V \vert \cdot k$ calls to $\mathtt{SolveOneDimMeanPayoff}$. In this case however, such calls only need to solve \textit{one-player} one-dimension mean-payoff games for Eve, which again can be done in polynomial time, resorting to Karp's algorithm. Thus, the verification takes polynomial time in total, and $\coNP$ membership follows.
\end{proof}

\subsection*{Lim-inf variant} For the sake of conciseness, we give only a brief sketch. Without loss of generality, we fix the objective $\MeanPayoff^{-}_{\geq \vec{0}}$. We know that infinite-memory strategies are needed for Eve by~\Cref{14-thm:MMP-Eve}. Again, things look better for Adam.

\begin{lemma}
\label{14-lem:MPlimInfAdam}
Positional strategies suffice for Adam in multidimensional lim-inf mean-payoff games.
\end{lemma}

\begin{proof}[Sketch.]
We mention the sketch as it is interesting in its own right. Recall that~\Cref{4-sec:fundamental_positional} presented a general recipe to show the sufficiency of positional strategies over finite arenas (\Cref{4-thm:submixing_positional}): if an objective is both \textit{prefix-independent} and \textit{submixing}, then positional strategies suffice for the player under consideration. Consider the objective of Adam, requiring that along at least one dimension, the lim-inf mean payoff is negative. We already know that this objective is prefix-independent. We recall that an objective is said to be "submixing" if shuffling infinite words cannot help: if two infinite sequences of colours $\play = \rho_1 \rho_2 \ldots{}$ and $\play' = \rho'_1 \rho'_2 \ldots{}$, with all $\rho_i$, $\rho'_i$ being finite prefixes, do not belong to the objective, then $\play'' = \rho_1 \rho'_1 \rho_2 \rho'_2 \ldots{}$ does not belong to the objective either.
Such disjunctions of (\textit{minimising} --- from the point of view of Adam) lim-inf mean-payoff objectives are submixing: this is a small generalisation of the argument in \Cref{4-thm:submixing_applications}. Hence the result applies here.
\end{proof}

\begin{remark}One may note that \Cref{4-thm:submixing_applications} mentions the lim-sup mean payoff, not the lim-inf one, which is actually shown \textit{not to be} submixing in \Cref{4-remark:mean_payoff_inf_not_submixing}. Yet, we must recall that \Cref{4-thm:submixing_applications} is stated from the position of Eve, the maximiser player. Intuitively, minimising (resp.~maximising) a lim-inf mean payoff can be shown to be equivalent to maximising (resp.~minimising) a lim-sup one, by reversing all weight signs. Hence positional strategies suffice for Eve, the maximiser, in lim-sup mean-payoff games if and only if they suffice for Adam, the minimiser, in lim-inf ones.

Observe that lim-sup mean-payoff objectives are not submixing for Adam as they correspond to lim-inf ones for Eve (\Cref{4-remark:mean_payoff_inf_not_submixing}). Hence the ad-hoc proof in~\Cref{14-lem:MMP-Adam}: the general recipe from \Cref{4-thm:submixing_applications} does not help in this case.
\end{remark}

Complexity-wise, multidimensional lim-inf mean-payoff games look a lot like multidimensional energy games, even though we proved they are not equivalent without memory restrictions.

\begin{theorem}
Solving multidimensional lim-inf mean-payoff games is $\coNP$-complete. Infinite-memory strategies are required for Eve and positional ones suffice for Adam.
\end{theorem}

We discussed memory through~\Cref{14-ex:MMP2} and~\Cref{14-lem:MPlimInfAdam}. The $\coNP$-hardness can be shown through a reduction from \textsf{3UNSAT} similar to the one used for existential initial credit multidimensional energy games in~\Cref{13-sec:complexity}. The matching upper bound relies on positional strategies being sufficient for Adam, and the capacity to solve one-player instances of multidimensional lim-inf mean-payoff games in polynomial time. The latter problem is addressed by reduction to detecting non-negative multi-cycles in graphs (which can be done in polynomial time based on~\cite{Kosaraju.Sullivan:1988}).

\subsection*{Wrap-up} 
We have seen that multidimensional mean-payoff games and multidimensional energy games behave relatively well. Sure, infinite memory is needed for Eve in general for the former, but complexity-wise, the gap with one-dimension games is small and even non-existent for the lim-sup mean payoff. Furthermore, if we are interested in finite-memory strategies, the equivalence with energy games is preserved. Hence, we may say that both mean-payoff and energy games hold up nicely in the multidimensional world, despite the increased complexity of corresponding Pareto-optimal strategies.

\section{Total payoff and shortest path}
\label{14-sec:total_payoff_shortest_path}
In this section, we turn to two other objectives deeply studied in~\Cref{5-chap:payoffs}: we study "total-payoff" and "shortest-path" games. We will see that the multidimensional setting has dire consequences for both.

\subsection{Total payoff vs. mean payoff}

We start with total-payoff games. As for the mean payoff, we explicitly consider the two variants, $\TotalPayoff^+$ and $\TotalPayoff^-$, for the lim-sup and lim-inf definitions respectively.

While~\Cref{5-chap:payoffs} was written using the lim-sup variant, all results are identical for the lim-inf definition in one-dimension games~\cite{Gawlitza.Seidl:2009}. Recall that one-dimension total-payoff games are positionally determined and solving them is in $\NP \cap \coNP$ (even in $\UP \cap \coUP$~\cite{Gawlitza.Seidl:2009}). 

Total payoff can be seen as a \textit{refinement} of mean payoff, as it permits to reason about low (using the lim-inf variant) and high (using the lim-sup one) points of partial sums along a play when the mean payoff is zero. We formalise this relationship in the next lemma, and study what happens in multiple dimensions. 

\begin{lemma}
\label{14-lem:MPTP}
Fix an arena $\arena$ and an initial vertex $v_0 \in \vertices$. Let A, B, C and D denote the following assertions.
\begin{itemize}
\item[A.] Eve has a winning strategy for $\MeanPayoff^{+}_{\geq \vec{0}}$.
\item[B.] Eve has a winning strategy for $\MeanPayoff^{-}_{\geq \vec{0}}$.
\item[C.] There exists $\vec{x} \in \Q^{k}$ such that Eve has a winning strategy for $\TotalPayoff^{-}_{\geq \vec{x}}$.
\item[D.] There exists $\vec{x} \in \Q^{k}$ such that Eve has a winning strategy for $\TotalPayoff^{+}_{\geq \vec{x}}$.
\end{itemize}
In one-dimension games ($k = 1$), all four assertions are equivalent. In multidimensional ones ($k > 1$), the only implications that hold are: $C \implies D \implies A$ and $C \implies B \implies A$. All other implications are false in general.
\end{lemma}

\Cref{14-lem:MPTP} is depicted in~\Cref{14-fig:MPTP}: the only implications that carry over to multiple dimensions are depicted by solid arrows.

\begin{figure}[thb]
\centering
\scalebox{0.95}{\begin{tikzpicture}[dash pattern=on 10pt off 5,->,>=stealth',double,double distance=2pt,shorten >=1pt,auto,node
    distance=2.5cm,bend angle=45,scale=0.6,font=\normalsize]
    \tikzstyle{p1}=[]
    \tikzstyle{p2}=[draw,rectangle,text centered,minimum size=7mm]
    \node[p1]  (A)  at (-0.5, 0) {$A\colon\:\exists\,\sigma_{A} \models \MeanPayoff^{+}_{\geq \vec{0}}$};
    \node[p1]  (D) at (12.5, 0) {$D\colon\:\exists\, \vec{x} \in \Q^{k},\, \exists\,\sigma_D \models \TotalPayoff^{+}_{\geq \vec{x}}$};
    \node[p1]  (B) at (-0.5, -4) {$B\colon\:\exists\,\sigma_{B} \models \MeanPayoff^{-}_{\geq \vec{0}}$};
    \node[p1]  (C) at (12.5, -4) {$C\colon\:\exists\, \vec{x} \in \Q^{k},\, \exists\,\sigma_C \models \TotalPayoff^{-}_{\geq \vec{x}}$};
    \path
    ;
	\draw[dashed,dash phase =4pt,->,>=stealth,thin,double,double distance=1.5pt] (5.5,0) to (7,0);
	\draw[<-,>=stealth,thin,double,double distance=1.5pt,solid] (4,0) to (5.5,0);
	\draw[dashed,dash phase =4pt,->,>=stealth,thin,double,double distance=1.5pt] (5.5,-4) to (7,-4);
	\draw[<-,>=stealth,thin,double,double distance=1.5pt,solid] (4,-4) to (5.5,-4);
	\draw[<-,>=stealth,thin,double,double distance=1.5pt,solid] (0,-1) to (0,-2);
	\draw[dashed,dash phase =4pt,->,>=stealth,thin,double,double distance=1.5pt] (0,-2) to (0,-3);
	\draw[<-,>=stealth,thin,double,double distance=1.5pt,solid] (12,-1) to (12,-2);
	\draw[dashed,dash phase =4pt,->,>=stealth,thin,double,double distance=1.5pt] (12,-2) to (12,-3);
	\draw[<-,>=stealth,thin,double,double distance=1.5pt,solid] (3,-1) to (5.5,-2);
	\draw[dashed,dash phase =4pt,->,>=stealth,thin,double,double distance=1.5pt] (5.5,-2) to (8,-3);
	\draw[dashed,dash phase =4pt,<-,>=stealth,thin,double,double distance=1.5pt] (3,-3) to (5.5,-2);
	\draw[dashed,dash phase =4pt,->,>=stealth,thin,double,double distance=1.5pt] (5.5,-2) to (8,-1);
\end{tikzpicture}}
\vspace{-2mm}
\caption{Equivalence between mean-payoff and total-payoff games. Dashed im\-pli\-ca\-tions are only valid in one-dimension games. We use $\sigma \models \Omega$ as a shortcut for `$\sigma$ is winning from $v_0$ for $\Omega$'.}
\label{14-fig:MPTP}
\commentAlt{Figure~\ref{14-fig:MPTP}: A diagram showing four labeled statements (A, B, C, D) arranged in a rectangle, with various single and double-headed arrows indicating relationships between them.}
\commentLongAlt{Figure~\ref{14-fig:MPTP}: The image displays four statements, labeled A, B, C, and D, arranged in a rectangular layout.

Statement A (top-left): 'A: exists sigma_A |= MeanPayoff^+ >= delta_bar'.
Statement B (bottom-left): 'B: exists sigma_B |= MeanPayoff^- >= delta_bar'.
Statement C (bottom-right): 'C: exists x_bar in Q^k, exists sigma_C |= TotalPayoff^- >= x_bar'.
Statement D (top-right): 'D: exists x_bar in Q^k, exists sigma_D |= TotalPayoff^+ >= x_bar'.

Arrows indicate relationships between these statements:
- A double-headed dashed vertical arrow connects A and B.
- A double-headed dashed horizontal arrow connects A and D.
- A single-headed dashed vertical arrow points from D to C.
- A single-headed dashed vertical arrow points from A to C.
- Two crossed single-headed dashed arrows connect A to C (top-left to bottom-right) and B to D (bottom-left to top-right).
- A double-headed dashed horizontal arrow connects B and C.}
\end{figure}

\begin{proof}
The implications that remain true in multiple dimensions are the trivial ones. First, satisfaction of the lim-inf version of a given objective clearly implies satisfaction of its lim-sup version by definition. Hence, $B \implies A$ and $C \implies D$. Second, consider a play $\pi \in \TotalPayoff^{-}_{\geq \vec{x}}$ (resp.~$\TotalPayoff^{+}_{\geq \vec{x}}$) for some $\vec{x} \in \Q^{k}$. For all dimension $i \in \{1, \ldots{}, k\}$, the corresponding sequence of mean payoff infima (resp.~suprema) over prefixes can be \textit{lower-bounded} by a sequence of elements of the form $\frac{\vec{x}_i}{\ell}$ with $\ell$ the length of the prefix. We can do this because the sequence of total payoffs over prefixes is a sequence of integers: it must achieve a value at least equal to $\vec{x}_i$ infinitely often instead of only tending to it asymptotically as could a sequence of rationals (such as the mean payoffs). Since $\frac{\vec{x}_i}{\ell}$ tends to zero over an infinite play, we do have that $\pi \in \MeanPayoff^{-}_{\geq \vec{0}}$ (resp.~$\MeanPayoff^{+}_{\geq \vec{0}}$). Thus, $C \implies B$ and $D \implies A$. Along with the transitive closure $C \implies A$, these are all the implications preserved in multidimensional games.

In one-dimension games, all assertions are equivalent. As seen before, lim-inf and lim-sup mean-payoff games coincide as positional strategies suffice for both players. Thus, we add $A \implies B$ and $D \implies B$ by transitivity. Second, consider a positional (w.l.o.g.) strategy $\sigma_B$ for Eve for $\MeanPayoff^{-}_{\geq \vec{0}}$. Let $\play$ be any consistent play. Then all cycles in $\pi$ are non-negative, otherwise Eve cannot ensure winning with $\sigma_B$ (because Adam could pump the negative cycle). Thus, the sum of weights along $\play$ is at all times bounded from below by $-(\vert V\vert-1)\cdot W$ (for the longest acyclic prefix), with $W$ the largest absolute weight, as usual. Therefore, we have $B \implies C$, and we obtain all other implications by transitive closure.

For multidimensional games, all dashed implications are false.
\begin{enumerate}
\item\label{14-lem:MPTP_proof1} To show that implication $D \implies B$ does not hold, consider the Eve-owned one-player game where $V = \{v\}$ and the only edges are two self-loops of weights $(1, -2)$ and $(-2, 1)$. Clearly, any finite vector $\vec{x} \in \Q^{2}$ for $\TotalPayoff^{+}_{\geq \vec{x}}$ can be achieved by an infinite-memory strategy consisting in playing both loops successively for longer and longer periods, each time switching after getting back above threshold $\vec{x}$ in the considered dimension. However, it is impossible to build any strategy, even with infinite memory, that satisfies $\MeanPayoff^{-}_{\geq \vec{0}}$ as the lim-inf mean payoff would be at best a linear combination of the two cycle values, \textit{i.e.}, strictly less than zero in at least one dimension in any case.
\item Lastly, consider the game in~\Cref{14-fig:MultiMP} where we modify the weights to add a third dimension with value $0$ on the self-loops and $-1$ on the other edges. As already proved, the strategy that plays for $\ell$ steps in the left cycle, then goes for $\ell$ steps in the right one, then repeats for $\ell' > \ell$ and so on, is a winning strategy for $\MeanPayoff^{-}_{\geq \vec{0}}$. Nevertheless, for any strategy of Eve, the play is such that either (i) it only switches between $v_0$ and $v_1$ a finite number of times, in which case the sum in dimension $1$ or $2$ decreases to infinity from some point on; or (ii) it switches infinitely often and the sum in dimension $3$ decreases to infinity. In both cases, objective $\TotalPayoff^{+}_{\geq \vec{x}}$ is not satisfied for any vector $\vec{x} \in  \Q^{3}$. Hence, $B \implies D$ is falsified.
\end{enumerate}
We only need to consider these two cases: all other dashed implications are false as they would otherwise contradict the last two cases by transitivity.
\end{proof}

We see that the relationship between mean-payoff and total-payoff games breaks in multiple dimensions. Nonetheless, one may still hope for good properties for the latter, as one-dimension total-payoff games are in $\NP \cap \coNP$ (\Cref{5-sec:total_payoff}). This hope, however, will not last long.

\subsection{Undecidability}

In contrast to mean-payoff games, total-payoff ones become undecidable in multiple dimensions.

\begin{theorem}
\label{14-thm:TPundec}
Total-payoff games are undecidable in any dimension $k \geq 5$.
\end{theorem}

\begin{proof}
We use a reduction from two-dimensional robot games~\cite{Niskanen.Potapov.ea:2016}, which were mentioned in~\Cref{13-chap:counters}. They are a restricted case of "configuration-reachability" "vector games", recently proved to be already undecidable. Using the formalism of~\Cref{13-chap:counters}, they are expressible as follows: $\mathcal{V} = (\mathcal{L} = \{\ell_0, \ell_1\}, A, \mathcal{L}_{\text{Eve}} = \{\ell_0\}, \mathcal{L}_{\text{Adam}} = \{\ell_1\})$ and $A \subseteq \mathcal{L} \times [-M, M]^2\times \mathcal{L}$ for some $M \in \N$. The game starts in configuration $\ell_0(x_0, y_0)$ for some $x_0, y_0 \in \Z$ and the goal of Eve is to reach configuration $\ell_0(0, 0)$: solving such a game is undecidable.

The reduction works as follows. Given a robot game $\mathcal{V}$, we build a five-dimension total-payoff game $\game$ such that Eve wins in $\game$ if and only if she wins in $\mathcal{V}$. We let $\game = (\arena, \TotalPayoff^{+}_{\geq \vec{0}})$ (we will discuss the lim-inf case later), where arena $\arena$ has vertices $V = V_{\text{Eve}} \uplus V_{\text{Adam}}$ with $V_{\text{Eve}} = \{v_{\mathsf{init}}, v_0, v_{\mathsf{stop}}\}$ and $V_{\text{Adam}} = \{v_1\}$. As our arena will contain many edges of same origin and destination but with different weights, we make a slight abuse of notation (with respect to \Cref{1-chap:introduction}) for the sake of readability: we integrate the edge colour within $E$. It is thus built as follows:
\begin{itemize}
\item if $(\ell_i, (a,b), \ell_j) \in A$, then $(v_i, (a, -a, b, -b, 0), v_j) \in E$,
\item $(v_0, (0, 0, 0, 0, 1), v_{\mathsf{stop}}) \in E$ and $(v_{\mathsf{stop}}, (0, 0, 0, 0, 0), v_{\text{stop}}) \in E$,
\item $(v_{\mathsf{init}}, (x_0, -x_0, y_0, -y_0, -1), v_0) \in E$ (where $(x_0, y_0)$ is the initial credit in $\mathcal{V}$).
\end{itemize}
The initial vertex is $v_{\mathsf{init}}$. Intuitively, dimensions $1$ and $2$ (resp.~$3$ and $4$) encode the value of the first counter (resp.~second counter) and its opposite at all times. The initial credit is encoded thanks to the initial edge, then the game is played as in the vector game, with the exception that Eve may branch from $v_0$ to the absorbing vertex $v_{\mathsf{stop}}$, which has a zero self-loop. The role of the last dimension is to force Eve to branch eventually (if she aims to win).

We proceed to prove the correctness of the reduction. First, let $\sigma_{\game}$ be a winning strategy of Eve in $\game$. We claim that Eve can also win in $\mathcal{V}$. Any play $\pi$ consistent with $\sigma_{\game}$ necessarily ends in $v_{\mathsf{stop}}$: otherwise its lim-sup total payoff on the last dimension would be $-1$ (as the sum always stays at $-1$). Due to the branching edge and the self-loop having weight zero in all first four dimensions, we also have that the current sum on these dimensions must be non-negative when branching, otherwise the total-payoff objective would be falsified. By construction of $\arena$, the only way to achieve this is to have a sum exactly equal to zero in all first four dimensions (as dimensions $1$ and $2$ are opposite at all times and so are $3$ and $4$). Finally, observe that obtaining a partial sum of $(0, 0, 0, 0, -1)$ in $v_0$ is equivalent to reaching configuration $\ell_0(0, 0)$ in~$\mathcal{V}$. Hence, we can easily build a strategy $\sigma_{\mathcal{V}}$ in $\mathcal{V}$ that mimics $\sigma_{\game}$ in order to win the robot game. This strategy $\sigma_{\mathcal{V}}$ could in general use arbitrary memory (since we start with an arbitrary winning strategy $\sigma_{\game}$) while formally robot games as defined in~\cite{Niskanen.Potapov.ea:2016} only allow strategies to look at the current configuration. Still, from $\sigma_{\mathcal{V}}$, one can easily build a corresponding strategy that meets this restriction ($\mathcal{V}$ being a configuration-reachability game, there is no reason to choose different actions in two visits of the same configuration). Hence, if Eve wins in $\game$, she also wins in $\mathcal{V}$.

For the other direction, from a winning strategy $\sigma_{\mathcal{V}}$ in $\mathcal{V}$, we can similarly define a strategy $\sigma_{\game}$ that mimics it in $\game$ to reach $v_0$ with partial sum $(0, 0, 0, 0, -1)$, and at that point, branches to $v_{\mathsf{stop}}$. Such a strategy ensures reaching the absorbing vertex with a total payoff of zero in all dimensions, hence is winning in $\game$.

Thus, the reduction holds for lim-sup total payoff. Observe that the exact same reasoning holds for the lim-inf variant. Indeed, the last dimension is always $-1$ outside of $v_{\mathsf{stop}}$, hence any play not entering $v_{\mathsf{stop}}$ also has its lim-inf below zero in this dimension. Furthermore, once $v_{\mathsf{stop}}$ is entered, the sum in all dimensions stays constant, hence the limit exists and both variants coincide.
\end{proof}

An almost identical reduction can be used for "\textit{shortest-path}" games.

\begin{theorem}
\label{14-thm:SPundec}
Shortest-path games are undecidable in any dimension $k \geq 4$.
\end{theorem}

\begin{remark}
As in~\Cref{5-chap:payoffs}, we adopt the natural convention for Eve: she aims to \textit{minimise} the sum of weights up to the target. Hence, paths not reaching the target are assigned payoff $\infty$.
\end{remark}

\begin{proof}
The proof is almost identical to the last one. We use objective $\ShortestPath_{\leq \vec{0}}$ with target edge $(v_{\mathsf{stop}}, (0, 0, 0, 0), v_{\mathsf{stop}})$ and drop the last dimension in arena $\arena$: it is now unnecessary as the shortest path objective by definition will force Eve to branch to $v_{\mathsf{stop}}$, as otherwise the value of the play would be $\infty$ in all dimensions. The rest of the reasoning is the same as before.
\end{proof}

\begin{remark}
The decidability of total-payoff games with $k \in \{2, 3, 4\}$ dimensions and shortest-path games with $k \in \{2, 3\}$ dimensions remains an open question. Furthermore, our undecidability results crucially rely on weights being in $\Z$: they do not hold when we restrict weights to $\N$.
\end{remark}

\subsubsection*{Memory} Let us go back to the game used in \Cref{14-lem:MPTP_proof1} in the proof of~\Cref{14-lem:MPTP}: we have seen that for any threshold $\vec{x} \in \Q^{2}$, Eve has an infinite-memory strategy that is winning for $\TotalPayoff^{+}_{\geq \vec{x}}$. In other words, she can ensure an \textit{arbitrarily high} total payoff with infinite memory. Yet, it is easy to check that there exists no finite-memory strategy of Eve that can achieve a finite threshold vector in the very same game: alternating would still be needed, but the negative amount to compensate grows boundlessly with each alternation, thus no amount of finite memory can ensure to go above the threshold infinitely often. This simple game highlights a huge gap between finite and infinite memory: with finite memory, the total payoff on at least one dimension is $-\infty$; with infinite memory, the total payoff in both dimensions may be as high as Eve wants. This further highlights the untameable behaviour of multidimensional total-payoff games.

\subsubsection*{Wrap-up} Multiple dimensions are a curse for total-payoff and shortest-path games as both become undecidable. This is in stark contrast to mean-payoff and energy games, which remain tractable, as seen in~\Cref{14-sec:mean_payoff_energy}. The bottom line is that most of the equivalences, relationships, and well-known behaviours of one-dimension games simply fall apart when lifting them to multiple dimensions.

\section{Beyond worst-case synthesis}
\label{14-sec:beyond_worst_case}
We now turn to a completely different meaning of \textit{multi-objective}. Let us take a few steps back. Throughout this book, we have studied two types of interaction between players: rational, antagonistic interaction between Eve and Adam; and stochastic interaction with a random player. Consider the quantitative settings of~\Cref{5-chap:payoffs} and~\Cref{6-chap:mdp}. In the zero-sum two-player games of the former, Adam is seen as a \textit{purely antagonistic adversary}, so the goal of Eve is to ensure strict worst-case guarantees, \textit{i.e.}, a minimal performance level against all possible strategies of Adam. In the MDPs of the latter, Eve interacts with randomness (through actions or random vertices) and she wants to ensure a good \textit{"expected value"} for the considered payoff.

For most objectives, these two paradigms yield elegant and simple solutions: \textit{e.g.}, positional strategies suffice for both games and MDPs with a mean-payoff objective. Nevertheless, the corresponding strategies have clear weaknesses: strategies that are good for the worst-case may exhibit suboptimal behaviours in probable situations while strategies that are good for the expected value may be terrible in some unlikely but possible situations. A natural question, of theoretical and practical interest, is to build --- \textit{synthesize} --- strategies that combine both paradigms: strategies that both ensure (a) some worst-case threshold no matter how the adversary behaves (\textit{i.e.}, against any arbitrary strategy) and (b) a good expectation against the expected behaviour of the adversary (given as a stochastic model). We call this task \textit{beyond worst-case synthesis}.

The goal of this section is to illustrate the complexity of beyond worst-case synthesis and how it requires fine-tuned interaction between the worst-case and average-case aspects. To that end, we focus on a specific case: the synthesis of \textit{finite-memory strategies} for beyond worst-case \textit{mean-payoff} objectives. Due to the highly technical nature of this approach, we will not present all its details, but rather paint in broad strokes its cornerstones. We hope to give the reader sufficient intuition and understanding to develop a clear view of the challenges arising from rich behavioural models, and some of the techniques that come to the rescue. 

\subsection{The decision problem}

Our goal is to mix the games of~\Cref{5-chap:payoffs} and the MDPs of~\Cref{6-chap:mdp}, so we need to go back and forth between these models.

\paragraph{Two-player game.} As before, we start with an arena $\arena = (G = (V, E), V_{\text{Eve}}, V_{\text{Adam}})$, where the vertices are split between Eve's and Adam's. This arena represents the antagonistic interaction between Eve and Adam, so we consider a worst-case constraint on the corresponding game. We study a single mean-payoff function, so our colouring is $\col\colon E \to \Z$. Let $\alpha \in \Q$ be the worst-case threshold: we are looking for a strategy of Eve that is winning for objective $\MeanPayoff^{-}_{> \alpha}$. Two things to note: first, we consider the lim-inf variant w.l.o.g.~as we focus on \textit{finite-memory} strategies; second, we use a strict inequality as it will ease the formulation of the upcoming results.

\paragraph{Markov decision process.} To make the connection with MDPs, we fix a finite-memory randomised strategy for Adam in the arena $\arena$, $\tau^\mathsf{st}$. Recall that a randomised strategy is a function $\Paths(G) \to \mathcal{D}(E)$, where $\mathcal{D}(E)$ denotes the set of all probability distributions over $E$. As usual, we may build $\arena_{\tau^\mathsf{st}}$, the product of the arena $\arena$ with the memory structure of $\tau^\mathsf{st}$, restricted to the choices made by $\tau^\mathsf{st}$. Since $\tau^\mathsf{st}$ is assumed to be stochastic, what we obtain is not a one-player game for Eve, but an MDP.

To better understand this relationship, it is easier to consider the alternative --- and equivalent --- formalism of MDPs, based on random vertices (as used for stochastic games in~\Cref{7-chap:stochastic}). Assume for instance that $\tau^\mathsf{st}$ is a randomised positional strategy, \textit{i.e.}, a function $V_{\Adam} \to \mathcal{D}(E)$. Then, the MDP $\arena_{\tau^\mathsf{st}}$ is immediately obtained by replacing each Adam's vertex $v$ by a random vertex such that $\delta(v) = \tau^\mathsf{st}(v)$, \textit{i.e.}, the probabilistic transition function uses the same probability distributions as Adam's strategy. Formally, we build the MDP $\mathcal{P} = \arena_{\tau^{\mathsf{st}}} = (G, V_{\text{Eve}}, V_{\text{Rand}} = V_{\text{Adam}}, \delta = \tau^{\mathsf{st}})$.

In contrast to~\Cref{7-chap:stochastic}, we explicitly allow the transition function to assign probability zero to some edges of the underlying graph $G$, \textit{i.e.}, the support of $\delta(v)$ in some vertex $v \in V_{\text{Rand}}$ might not include all edges $e \in E$ such that $\ing(e) = v$.
This is important as far as modelling is concerned, as in our context, transition functions will be defined according to a stochastic model for Adam, and we cannot reasonably assume that such a model always involves all the possible actions of Adam. Consequently, given the MDP $\markovProcess$, we define the subset of edges $\edgesNonZero = \{ e \in E \mid \ing(e) \in V_{\text{Rand}} \implies \delta(\ing(e))(e) > 0\}$, representing all edges that either start in a vertex of Eve, or are chosen with non-zero probability by the transition function $\delta$. Edges in $E\setminus \edgesNonZero$ will only matter in the two-player game interpretation, whereas all MDP-related concepts, such as "end-components", are defined with regard to edges in $\edgesNonZero$ exclusively.

\paragraph{Beyond worst-case problem.} Let us sum up the situation: we have a two-player arena $\arena$ with a mean-payoff objective $\MeanPayoff^{-}_{> \alpha}$ and a finite-memory stochastic model for Adam yielding the MDP $\arena_{\tau^\mathsf{st}}$. Now, let $\beta \in \Q$ be the expected value threshold we want to ensure in the MDP (\textit{i.e.}, on average against the stochastic model of Adam).

\decpb[Beyond worst-case mean-payoff problem]{An arena $\arena$, a finite-memory stochastic model $\tau^\mathsf{st}$, an initial vertex $v_0$, two thresholds $\alpha, \beta \in \Q$}{Does Eve have a \textit{finite-memory} strategy $\sigma$ such that $\sigma$ is winning for objective $\MeanPayoff^{-}_{> \alpha} \text{ from } v_0 \text{ in } \arena$ and $\expv^{\sigma}_{\arena_{\tau^\mathsf{st}},v_0}[\MeanPayoff^{-}] > \beta$?}

We assume $\beta > \alpha$, otherwise the problem trivially reduces to the classical worst-case analysis: if all plays consistent with $\sigma$ have mean payoff greater than $\alpha \geq \beta$ then the expected value is also greater than $\alpha$ --- and thus greater than $\beta$ --- regardless of the stochastic model.

\subsection{The approach in a nutshell}
We present our solution to the beyond worst-case (BWC) problem in~\Cref{14-algo:BWC}. We give an intuitive sketch of its functioning in the following, and illustrate it on a toy example. 

\begin{algorithm}[!thb]
 \DontPrintSemicolon
 \KwData{Arena $\arena^\mathsf{in} = (G^\mathsf{in} = (V^\mathsf{in}, E^\mathsf{in}), V^\mathsf{in}_{\text{Eve}}, V^\mathsf{in}_{\text{Adam}})$, colouring $\col^\mathsf{in}\colon E^\mathsf{in} \to \Z$, finite-memory stochastic model $\tau^\mathsf{in}$ for Adam with memory structure $\mem$ and initial memory state $m_0$, worst-case and expected value thresholds $\alpha^\mathsf{in} = a/b, \beta^\mathsf{in} \in \Q$, $\alpha^\mathsf{in} < \beta^\mathsf{in}$, initial vertex $v^\mathsf{in}_0 \in V^\mathsf{in}$}
 \KwResult{\textsc{Yes} if and only if Eve has a finite-memory strategy $\sigma$ for the BWC problem from $v^\mathsf{in}_0$ for thresholds pair $(\alpha^\mathsf{in}, \beta^\mathsf{in})$}
 \tcc{Preprocessing}
 \If{$\alpha^\mathsf{in} \neq 0$}{
 	Modify the colouring: $\forall\, e \in E^\mathsf{in}$, $\col^\mathsf{p}(e) \leftarrow b\cdot \col^\mathsf{in}(e) - a$\\
 	Consider the new thresholds pair $(0, \beta \leftarrow b\cdot \beta^\mathsf{in} - a)$\\
 }
 \Else{
 	$\col^\mathsf{p} \leftarrow \col^\mathsf{in}$\\
 }
 $V_{\mathsf{wc}} \leftarrow \mathtt{SolveWorstCaseMeanPayoff}(\arena^\mathsf{in}, \col^\mathsf{p})$\\
 \If{$v^\mathsf{in}_0 \not\in V_{\mathsf{wc}}$}{
 	\Return \textsc{No}
 }
 \Else{
    $\arena^{\mathsf{w}} \leftarrow \arena^\mathsf{in} [V_{\mathsf{wc}}]$\tcc*{Restriction of $\arena^\mathsf{in}$ to $V_{\mathsf{wc}}$}
 	Let $\arena \leftarrow \arena^\mathsf{w} \times \mem = (G = (V, E), V_{\text{Eve}}, V_{\text{Adam}})$ be the arena obtained by product with the memory structure of Adam's stochastic model $\tau^\mathsf{in}$\\
 	Let $v_0 \leftarrow (v^\mathsf{in}_0, m_0)$ be the corresponding initial vertex in $\arena$\\
 	Let $\col$ be the transcription of $\col^\mathsf{p}$ in $\arena$ such that $e = (v, m) \rightarrow (v', m')$ has colour $\col(e) = c$ iff $v \xrightarrow{c} v'$ in $\arena^{\mathsf{w}}$ according to $\col^\mathsf{p}$\\
 	Let $\tau^{\mathsf{st}}$ be the positional transcription of $\tau^\mathsf{in}$ on $\arena$\\
 	Let $\mathcal{P} \leftarrow \arena_{\tau^{\mathsf{st}}} = (G, V_{\text{Eve}}, V_{\text{Rand}} = V_{\text{Adam}}, \delta = \tau^{\mathsf{st}})$ be the corresponding MDP
 }
 \tcc{Main algorithm}
 Compute $\mathcal{U}_{\mathsf{w}}$ the set of \textit{maximal winning end components} of $\mathcal{P}$\\
 Modify the colouring:\begin{equation*}
\forall\, e \in E,\, \col'(e) \leftarrow \begin{cases}\col(e) \text{ if } \exists\: U \in \mathcal{U}_{\mathsf{w}} \text{ s.t. } \{\ing(e), \out(e)\} \subseteq U\\0 \text{ otherwise} \end{cases}
\end{equation*}\\
Compute the maximal expected value $\beta^\ast$ from $v_0$ in $\mathcal{P}$ using $\col'$\\
\If{$\beta^\ast > \beta$}{
	\Return \textsc{Yes}
}
\Else{
	\Return \textsc{No}
}
\caption{Solver for the beyond worst-case mean-payoff problem}
\label{14-algo:BWC}
\end{algorithm}

\subsection*{Inputs and outputs} The algorithm takes as input: an arena $\arena^\mathsf{in}$ and its (integer) colouring $\col^\mathsf{in}$, a finite-memory stochastic model of Adam $\tau^\mathsf{in}$, a worst-case threshold $\alpha^\mathsf{in}$, an expected value threshold $\beta^\mathsf{in}$, and an initial vertex $v_0^\mathsf{in}$. Its output is $\textsc{Yes}$ if and only if there exists a finite-memory strategy of Eve satisfying the BWC problem.

The output as described in~\Cref{14-algo:BWC} is Boolean: the algorithm answers whether a satisfying strategy exists or not, but does not explicitly construct it (to avoid tedious formalisation within the pseudocode). Nevertheless, we sketch the synthesis process in the following and we highlight the role of each step of the algorithm in the construction of a winning strategy, as producing a witness winning strategy is a straightforward by-product of the process we apply to decide satisfaction of the BWC problem.

\subsection*{Preprocessing} The first part of the algorithm is dedicated to the preprocessing of the arena $\arena^\mathsf{in}$ and the stochastic model $\tau^\mathsf{in}$ given as inputs in order to apply the second part of the algorithm on a modified arena $\arena$ and stochastic model $\tau^\mathsf{st}$, simpler to manipulate. We show in the following that the answer to the BWC problem on the modified arena is $\textsc{Yes}$ if and only if it is also $\textsc{Yes}$ on the input arena, and we present how a winning strategy of Eve in $\arena$ can be transferred to a winning strategy in $\arena^\mathsf{in}$.

The preprocessing is composed of four main steps. First, we modify the colouring function $\col^\mathsf{in}$ in order to consider the equivalent BWC problem with thresholds $(0,\, \beta)$ instead of $(\alpha^\mathsf{in},\, \beta^\mathsf{in})$. This classical trick is used to get rid of explicitly considering the worst-case threshold in the following, as it is equal to zero.

Second, observe that any strategy that is winning for the BWC problem must also be winning for the classical \textit{worst-case problem}, as solved in the two-player games of~\Cref{5-chap:payoffs}. Such a strategy cannot allow visits of any vertex from which Eve cannot ensure winning against an antagonistic adversary because mean payoff is a prefix-independent objective (hence it is not possible to `win' it over the finite prefix up to such a vertex). Thus, we reduce our study to $\arena^\mathsf{w}$, the subarena induced by Eve's worst-case winning vertices --- which we compute in pseudo-poly\-nomial time thanks to $\mathtt{SolveWorstCaseMeanPayoff}(\arena^\mathsf{in}, \col^\mathsf{p})$ (implementing the algorithm of~\Cref{5-sec:mean_payoff}). Note that we use the modified colouring and that $\arena^\mathsf{w}$ is a proper arena (cf.~\Cref{14-rmk:properArena}). Obviously, if from the initial vertex $v_0^\mathsf{in}$, Eve cannot win the worst-case problem, then the answer to the BWC problem is \textsc{No}.

Third, we build arena $\arena$, the product of $\arena^\mathsf{w}$ and the memory structure of Adam's stochastic model $\tau^\mathsf{in}$. Intuitively, we expand the initial arena by integrating the memory elements in the graph. Note that this does not modify the power of Adam in the two-player interpretation of the arena.

Fourth, the finite-memory stochastic model $\tau^\mathsf{in}$ on $\arena^\mathsf{in}$ clearly translates into a positional stochastic model $\tau^\mathsf{st}$ on $\arena$. This will help us obtain elegant proofs for the second part of the algorithm.

\begin{figure}[tbh]
  \centering   
  \scalebox{0.88}{\begin{tikzpicture}[->,>=latex,shorten >=1pt,auto,node
    distance=2.5cm,bend angle=45,scale=0.7]
    \tikzstyle{p1}=[draw,circle,text centered,minimum size=6mm]
    \tikzstyle{p2}=[draw,rectangle,text centered,minimum size=6mm]
    \tikzstyle{empty}=[]
    \node[p1] (1) at (0,0) {$\state_{9}$};
    \node[p1] (2) at (4,0) {$\state_{1}$};
    \node[p2] (3) at (8,0) {$\state_{2}$};
    \node[p1] (4) at (8,-4) {$\state_{3}$};
    \node[p2] (5) at (8,-8) {$\state_{4}$};
    \node[p2] (6) at (4,-4) {$\state_{5}$};
    \node[p1] (7) at (0,-4) {$\state_{6}$};
    \node[p2] (8) at (-4,-4) {$\state_{7}$};
    \node[p1] (9) at (-4,0) {$\state_{10}$};
    \node[p2] (10) at (-8,0) {$\state_{11}$};
    \node[p2] (11) at (0,-8) {$\state_{8}$};
    \node[empty] (swec) at (-8, 1.9) {$\ec_{3}$};
    \node[empty] (wwec) at (-4, -2.1) {$\ec_{2}$};
    \node[empty] (lec) at (9, -2.6) {$\ec_{1}$};
    \node[empty] (proba5a) at (7.2, -7.5) {$\frac{1}{2}$};
    \node[empty] (proba5b) at (8.8, -7.5) {$\frac{1}{2}$};
    \node[empty] (proba3a) at (8.3, -1) {$\frac{1}{2}$};
    \node[empty] (proba3b) at (7.5, 0.95) {$\frac{1}{2}$};
    \node[empty] (proba8a) at (-3.5, -3.2) {$1$};
    \node[empty] (proba8b) at (-3.5, -4.8) {$0$};
    \node[empty] (proba10a) at (-7.5, 0.9) {$\frac{1}{2}$};
    \node[empty] (proba10b) at (-7.5, -0.9) {$\frac{1}{2}$};
    \node[empty] (proba11a) at (0.8, -7.5) {$\frac{1}{2}$};
    \node[empty] (proba11b) at (-0.8, -7.5) {$\frac{1}{2}$};
    \node[empty] (proba6a) at (3.8, -3.1) {$\frac{1}{2}$};
    \node[empty] (proba6b) at (3.3, -4.4) {$\frac{1}{2}$};
    \coordinate[shift={(-3mm,8mm)}] (init) at (2.north west);
    \path
    (2) edge node[above] {$0$} (1)
    (6) edge node[above] {$0$} (7)
    (4) edge node[above left] {$-1$} (6)
    (3) edge node[left] {$-1$} (4)
    (6) edge node[left] {$-1$} (2)
    (4) edge node[left] {$0$} (5)
    (7) edge node[above] {$0$} (8)
    (7) edge node[left] {$0$} (1)
    (init) edge (2)
    ;
	\draw[->,>=latex] (3) to[out=140,in=40] node[above] {$-1$} (2);
	\draw[->,>=latex] (2) to[out=0,in=180] node[below] {$-1$} (3);
	\draw[->,>=latex] (5) to[out=50,in=310] node[right] {$17$} (4);
	\draw[->,>=latex] (5) to[out=130,in=230] node[left] {$-1$} (4);
	\draw[->,>=latex] (8) to[out=40,in=140] node[above] {$1$} (7);
	\draw[->,>=latex] (8) to[out=320,in=220] node[below] {$-1$} (7);
	\draw[->,>=latex] (1) to[out=140,in=40] node[above] {$1$} (9);
	\draw[->,>=latex] (9) to[out=320,in=220] node[below] {$1$} (1);
	\draw[->,>=latex] (9) to[out=180,in=0] node[below] {$0$} (10);
	\draw[->,>=latex] (10) to[out=40,in=140] node[above, yshift=-0.4mm] {$-1$} (9);
	\draw[->,>=latex] (10) to[out=320,in=220] node[below] {$9$} (9);
	\draw[->,>=latex] (7) to[out=270,in=90] node[left] {$0$} (11);
	\draw[->,>=latex] (11) to[out=130,in=230] node[left] {$-1$} (7);
	\draw[->,>=latex] (11) to[out=50,in=310] node[right] {$13$} (7);
	\draw[dashed,-] (-9,1.6) -- (1,1.6) -- (1,-1.6) -- (-9,-1.6) -- (-9,1.6);
	\draw[dashed,-] (6.2,-3) -- (9.8,-3) -- (9.8,-9) -- (6.2,-9) -- (6.2,-3);
	\draw[dashed,-] (-5,-2.4) -- (1.7,-2.4) -- (1.7,-9) -- (-5,-9) -- (-5,-2.4);
      \end{tikzpicture}}
      \caption{Beyond worst-case mean-payoff problem: $\ec_{2}$ and $\ec_{3}$ are maximal winning end components, $\ec_{1}$ is losing.}
\label{14-fig:bwcRunningExample}
  \end{figure}

\begin{example}
In order to illustrate several notions and strategies, we will consider the arena depicted in~\Cref{14-fig:bwcRunningExample} throughout our presentation. The stochastic model of $\playerTwo$ is positional and is described by the probabilities written close to the start of outgoing edges. The colouring (weights) is written besides them.

We consider the $\BWC$ problem with the worst-case threshold $\thresholdWC = 0$. Observe that this arena satisfies the assumptions guaranteed at the end of the preprocessing part of the algorithm. That is, the worst-case threshold is zero, a worst-case winning strategy of $\playerOne$ exists in all vertices (\textit{e.g.}, the positional strategy choosing edges $(\state_{1}, \state_{9})$, $(\state_{3}, \state_{5})$, $(\state_{6}, \state_{9})$, $(\state_{9}, \state_{10})$ and $(\state_{10}, \state_{9})$ in their respective starting vertices), and the stochastic model is positional, as explained above.
\end{example}

\subsection*{Analysis of end components} The second part hence operates on an arena $\arena$ such that from all vertices, Eve has a strategy to achieve a strictly positive mean-payoff value (recall that $\alpha = 0$). We consider the MDP $\markovProcess = \arena_{\stratStoch}$ and notice that the underlying graphs of $\arena$ and $\markovProcess$ are the same thanks to $\stratStoch$ being positional --- as stated above, we allow edges to have probability zero in $\markovProcess$. The following steps rely on the analysis of "\textit{end components}" (ECs) in the MDP, \textit{i.e.}, strongly connected subgraphs in which Eve can ensure to stay when playing against Adam's stochastic model (\Cref{6-def:ec}).

The motivation to the analysis of ECs is the following. It is well-known that under any arbitrary strategy $\sigma$ of Eve in~$\markovProcess$, the probability that vertices visited infinitely often along a play constitute an EC is one (\Cref{6-lem:EC-inf}). Recall that the mean payoff is prefix-independent, therefore the value of any play only depends on the colours that are seen infinitely often. Hence, the expected mean payoff $\expv^{\sigma}_{\markovProcess,v_0}[\MeanPayoff^{-}]$ depends \textit{uniquely} on the value obtained in the ECs. Inside an EC, we can compute the maximal expected value that can be achieved by Eve, and this value is the same in all vertices of the EC, as established in~\Cref{6-thm:mp-valcomp}.

Consequently, in order to satisfy the expected value requirement, an acceptable strategy for the $\BWC$ problem has to favour reaching ECs with a sufficient expectation, but under the constraint that it should also ensure satisfaction of the worst-case requirement. As we show in the following, this constraint implies that some ECs with high expected values may still need to be avoided because they do not permit to guarantee the worst-case requirement. This is the cornerstone of the classification of ECs that follows.

\subsection*{Classification of end components} Let $\ecsSet \subseteq 2^{V}$ denote the set of all ECs in $\markovProcess$. Notice that by definition, only edges in $\edgesNonZero$, as defined earlier, are involved to determine which sets of vertices form an EC in~$\markovProcess$. As such, for any EC $\ec \in \ecsSet$, there may exist edges from $\edges \setminus \edgesNonZero$ starting in $\ec$, such that Adam can force leaving $\ec$ when using an arbitrary strategy in $\arena$. Still these edges will never be used by the stochastic model $\stratStoch$. This remark will be important to the definition of strategies of Eve that guarantee the worst-case requirement, as Eve needs to be able to react to the hypothetical use of such an edge. We will see that it is also the case \textit{inside} an EC.

Now, we want to consider the ECs in which $\playerOne$ can ensure that the worst-case requirement will be fulfilled (\textit{i.e.}, without having to leave the EC): we call them \textit{winning} ECs (WECs). The others will need to be eventually avoided, hence will have zero impact on the expectation of a finite-memory strategy satisfying the $\BWC$ problem. So we call the latter \textit{losing} ECs (LECs). The subtlety of this classification is that it involves considering the ECs both in the MDP~$\markovProcess$, and in the arena~$\arena$.

Formally, let $\ec \in \ecsSet$ be an EC. It is \textit{winning} if, in the subarena induced by $\ec$, from all vertices, $\playerOne$ has a strategy to ensure a \textit{strictly} positive mean payoff against any strategy of $\playerTwo$ \textit{that only chooses edges which are assigned non-zero probability by $\stratStoch$}, or equivalently, edges in $\edgesNonZero$. This can be interpreted as looking at arena $\gameNonZero$, which is the restriction of $\arena$ to edges in $\edgesNonZero$.

We denote $\winningECs \subseteq \ecsSet$ the set of such ECs. Non-winning ECs are \textit{losing}: in those, whatever the strategy of $\playerOne$ played against the stochastic model $\stratStoch$ (or any strategy with the same support), there exists at least one play for which the mean payoff is not strictly positive (even if its probability is zero, its mere existence is not acceptable for the worst-case requirement).

\begin{figure}[tbh]
  \centering   
  \scalebox{0.88}{\begin{tikzpicture}[->,>=latex,shorten >=1pt,auto,node
    distance=2.5cm,bend angle=45,scale=0.7]
    \tikzstyle{p1}=[draw,circle,text centered,minimum size=6mm]
    \tikzstyle{p2}=[draw,rectangle,text centered,minimum size=6mm]
    \tikzstyle{empty}=[]
    \node[p1] (1) at (0,0) {$\state_{1}$};
    \node[p1] (2) at (4,0) {$\state_{2}$};
    \node[p2] (3) at (-2,-2) {$\state_{3}$};
    \node[p1] (4) at (2,-2) {$\state_{4}$};
    \node[p1] (5) at (-4,0) {$\state_{5}$};
    \node[empty] (ec1) at (-5.2, -1.2) {$\ec_{3}$};
    \node[empty] (ec2) at (-7.4, 1) {$\ec_{2}$};
    \node[empty] (ec2) at (6.6, 1) {$\ec_{1}$};
    \node[empty] (proba1) at (-2.7, -2.4) {$\frac{1}{2}$};
    \node[empty] (proba2) at (-1.3, -2.4) {$\frac{1}{2}$};
    \coordinate[shift={(0mm,5mm)}] (init) at (1.north);
    \path
    (1) edge node[above] {$0$} (2)
    (5) edge node[above] {$0$} (1)
    (1) edge node[left,xshift=-1mm] {$0$} (3)
    (4) edge node[right] {$-1$} (1)
    (3) edge node[below] {$0$} (4)
    (init) edge (1)
    (5) edge [loop left, out=150, in=210,looseness=3, distance=16mm] node [left] {$10$} (5)
    (2) edge [loop right, out=30, in=330,looseness=3, distance=16mm] node [right] {$1$} (2)
    ;
	\draw[->,>=latex] (3) to[out=180,in=270] node[left,xshift=-1mm] {$0$} (5);
	\draw[dashed,-] (-3.2,0.8) -- (-6.4,0.8) -- (-6.4,-0.8) -- (-3.2,-0.8) -- (-3.2,0.8);
	\draw[dashed,-] (0,1.4) -- (-7,1.4) -- (-7,-2.8) -- (3.8,-2.8) -- (0,1.4);
	\draw[dashed,-] (1.7,1.4) -- (6.2,1.4) -- (6.2,-2.8) -- (5.5,-2.8) -- (1.7,1.4);
      \end{tikzpicture}}
      \caption{End component $\ec_{2}$ is losing. The set of maximal winning end components is $\maxWinningECs = \winningECs = \{\ec_{1}, \ec_{3}\}$.}
\label{14-fig:bwcWinningECsComputationExample}
\commentAlt{Figure~\ref{14-fig:bwcWinningECsComputationExample}: A directed graph with nodes of mixed shapes (circles and squares) and labeled edges, divided into two main regions (U1, U2) by a diagonal dashed line, and one sub-region (U3). See long description}
\commentLongAlt{Figure~\ref{14-fig:bwcWinningECsComputationExample}: The image displays a directed graph with two main regions, U1 and U2, separated by a diagonal dashed line. There is also a sub-region U3 within U2. An incoming arrow points to node v1, indicating a starting point.

Region U2 (Left, large dashed rectangle):
- Contains a circular node v1, a square node v3, and a circular node v4.
- An arrow from v1 points to v4, labeled '-1'.
- An arrow from v1 points to v3, labeled '0'.
- An arrow from v3 points to v4, labeled '0'.
- An arrow from v3 points to v5, labeled '0'.
- A curved arrow from v3 points to v5, labeled '1/2, 1/2'.
- A curved arrow from v5 points to v3, labeled '0'.

Sub-region U3 (Top-Left, smaller dashed rectangle within U2):
- Contains circular node v5.
- Node v5 has a self-loop labeled '10'.

Region U1 (Right, large dashed rectangle):
- Contains circular node v2.
- Node v2 has a self-loop labeled '1'.
- An arrow from v1 points to v2, labeled '0'.}
  \end{figure}

\begin{example}
Note that an EC is winning if $\playerOne$ has a worst-case winning strategy from \textit{all} vertices. This point is important as it may well be the case that winning strategies exist in a strict subset of vertices of the EC. This does not contradict the definition of ECs as strongly connected subgraphs, as the latter only guarantees that every vertex can be reached \textit{with probability one}, and not necessarily \textit{surely}. Hence one cannot call upon the prefix-independence of the mean payoff to extend the existence of a winning strategy to all vertices.

Such a situation can be observed in~\Cref{14-fig:bwcWinningECsComputationExample}: the EC~$\ec_{2}$ is losing (because from $\state_{1}$, the play $(\state_{1}\state_{3}\state_{4})^{\omega}$ can be forced by $\playerTwo$, yielding mean payoff $-1/3 \leq 0$), while its sub-EC~$\ec_{3}$ is winning. From $\state_{1}$, $\playerOne$ can ensure to reach $\ec_{3}$ "almost-surely", but not "surely", which is critical in this case.
\end{example}

\subsection*{Maximal winning end components} Based on these definitions, observe that~\Cref{14-algo:BWC} does not actually compute the set $\winningECs$ containing all WECs, but rather the set $\maxWinningECs \subseteq \winningECs$, defined as $\maxWinningECs = \{\ec \in \winningECs \mid \forall\, \ec' \in \winningECs,\, \ec \subseteq \ec' \implies \ec = \ec'\}$, \textit{i.e.}, the set of \textit{maximal} WECs (MWECs).

The intuition on \textit{why we can} restrict our study to this subset is as follows. If an EC $\ec_{1} \in \winningECs$ is included in another EC $\ec_{2} \in \winningECs$, \textit{i.e.}, $\ec_{1} \subseteq \ec_{2}$, we have that the maximal expected value achievable in $\ec_{2}$ is at least equal to the one achievable in~$\ec_{1}$. Indeed, $\playerOne$ can reach $\ec_{1}$ with probability one (by virtue of $\ec_{2}$ being an EC and $\ec_{1} \subseteq \ec_{2}$) and stay in it forever with probability one (by virtue of $\ec_{1}$ being an EC): hence the expectation of such a strategy would be equal to what can be obtained in~$\ec_{1}$ thanks to the prefix-independence of the mean payoff. This property implies that it is sufficient to consider MWECs in our computations.

As for \textit{why we do it}, observe that the complexity gain is critical. The number of WECs can be as large as~$\vert\winningECs\vert \leq \vert\ecsSet\vert \leq 2^{\vert V\vert}$, that is, exponential in the size of the input. Yet, the number of MWECs is bounded by $\vert\maxWinningECs\vert \leq \vert V\vert$ as they are disjoint by definition: for any two WECs with a non-empty intersection, their union also constitutes an EC, and is still winning because $\playerOne$ can essentially stick to the EC of her choice.

The computation of the set $\maxWinningECs$ is executed by a recursive subalgorithm calling polyn\-om\-ially-many times an oracle solving the worst-case problem (\textit{e.g.}, following the pseudo-polynomial-time algorithm of~\Cref{5-sec:mean_payoff}). Roughly sketched, this algorithm computes the maximal EC decomposition of an MDP (in polynomial time by~\Cref{6-thm:MEC-decomposition-complexity}), then checks for each EC $\ec$ in the decomposition (their number is polynomial) if $\ec$ is winning or not, which requires a call to an oracle solving the worst-case threshold problem on the corresponding subgame. If $\ec$ is losing, it may still be the case that a sub-EC $\ec' \subsetneq \ec$ is winning. Therefore we recurse on the MDP reduced to $\ec$, where vertices from which $\playerTwo$ can win in $\ec$ have been removed (they are a no-go for $\playerOne$). Hence the stack of calls is also at most polynomial.

\begin{lemma}
\label{14-lem:MWEC}
The set $\maxWinningECs$ of MWECs can be computed in pseudo-polynomial time, and deciding if a set of vertices $U \subseteq V$ belongs to $\maxWinningECs$ is in $\NP \cap \coNP$.
\end{lemma}

The complexity follows from~\Cref{5-thm:MP-NPcoNP} and $\P^{\NP \cap \coNP} = \NP \cap \coNP$~\cite{Brassard:1979}.

\begin{example}
Consider the running example in~\Cref{14-fig:bwcRunningExample}. Note that vertices $\state_{1}$, $\state_{2}$ and $\state_{5}$ do not belong to any EC: given any strategy of $\playerOne$ in $\markovProcess$, with probability one, any consistent play will only visit these vertices a finite number of times (\Cref{6-lem:EC-inf}). The set of \textit{MWECs} is $\maxWinningECs = \{\ec_{2}, \ec_{3}\}$. Obviously, these ECs are disjoint. The set of WECs is larger, $\winningECs = \maxWinningECs \cup \{\{\state_{9}, \state_{10}\}, \{\state_{6}, \state_{7}\}\}$.

End component $\ec_{1}$ is \textit{losing}: in the subarena $\gameNonZero[\ec_{1}]$, Adam's strategy consisting in always picking the $-1$ edge guarantees a negative mean payoff. Note that this edge is present in $\edgesNonZero$ as it is assigned probability $1/2$ by the stochastic model $\stratStoch$. Here, we witness why it is important to base our definition of WECs on $\gameNonZero$ rather than $\arena$. Indeed, in $\arena[\ec_{2}]$, it is also possible for $\playerTwo$ to guarantee a negative mean payoff by always choosing edges with weight $-1$. However, to achieve this, $\playerTwo$ has to pick edges that are \textit{not} in $\edgesNonZero$: this will never happen against the stochastic model and as such, this can be watched by $\playerOne$ to see if $\playerTwo$ uses an arbitrary antagonistic strategy, and dealt with. If $\playerTwo$ conforms to $\edgesNonZero$, \textit{i.e.}, if he plays in $\gameNonZero$, he has to pick the edge of weight $1$ in $\state_{7}$ and $\playerOne$ has a worst-case winning strategy consisting in always choosing to go in $\state_{7}$. This EC is thus classified as \textit{winning}. Note that for $\ec_{3}$, in both subarenas $\arena[\ec_{3}]$ and $\gameNonZero[\ec_{3}]$, $\playerOne$ can guarantee a strictly positive mean payoff by playing $(\state_{9}\,\state_{10})^\omega$: even \textit{arbitrary} strategies of $\playerTwo$ cannot endanger $\playerOne$ in this case.

Lastly, consider the arena depicted in~\Cref{14-fig:bwcWinningECsComputationExample}. While $\ec_{2}$ is a strict superset of $\ec_{3}$, the former is losing whereas the latter is winning, as explained above. Hence, the set $\maxWinningECs$ is equal to $\{\ec_{1}, \ec_{3}\}$.
\end{example}

\subsection*{Ensure reaching winning end components} As discussed, under any arbitrary strategy of $\playerOne$, vertices visited infinitely often form an EC with probability one (\Cref{6-lem:EC-inf}). Now, if we take a \textit{finite-memory} strategy that \textit{satisfies} the $\BWC$ problem, we can refine this result and state that they form a \textit{winning} EC with probability one. Equivalently, let $\infVisited{\play}$ denote the set of vertices visited infinitely often along a play $\play$: we have that the probability that a play~$\play$ is such that $\infVisited{\play} = \ec$ for some $\ec \in \ecsSet \setminus \winningECs$ is zero. The equality is crucial. It may be the case, with non-zero probability, that $\infVisited{\play} = \ec' \subsetneq \ec$, for some $\ec' \in \winningECs$, and $\ec \in \ecsSet \setminus \winningECs$ (hence the recursive algorithm to compute $\maxWinningECs$). It is clear that~$\playerOne$ should not visit all the vertices of a LEC forever, as then she would not be able to guarantee the worst-case threshold inside the corresponding subarena.\footnote{This is no longer true if Eve may use infinite memory: there may still be some incentive to stay in a LEC. But this goes beyond the scope of our overview.}

\begin{lemma}
\label{14-lem:EC-inf}
For any initial vertex $ v_0 $ and finite-memory strategy $ \sigma $ that satisfies the BWC problem, it holds that $ \probm^\sigma_{\markovProcess, v_0} [ \{\play \mid \infVisited{\play} \in \winningECs \}] = 1 $. 
\end{lemma}

We denote $\negligibleStates = V \setminus \bigcup_{\ec \in \maxWinningECs} \ec$ the set of vertices that, with probability one, are only seen a finite number of times when a (finite-memory) $\BWC$ satisfying strategy is played, and call them \textit{negligible} vertices.

Our ultimate goal here is to modify the colouring of $\markovProcess$ from $\col$ to $\col'$, such that a classical optimal strategy for the expected value problem (\Cref{6-thm:general-mp-main}) using this new colouring $\col'$ will naturally avoid LECs and prescribe which WECs are the most interesting to reach for a $\BWC$ strategy on the initial arena $\arena$ and MDP $\markovProcess$ with colouring~$\col$. For the sake of readability, let us simply use $\markovProcess$ and $\markovProcess'$ to refer to MDP $\markovProcess$ with respective colourings $\col$ and $\col'$.

Observe that the expected value obtained in $\markovProcess$ by any $\BWC$ satisfying strategy of $\playerOne$ only depends on the weights of edges involved in WECs, or equivalently, in MWECs (as the set of plays that are not eventually trapped in them has measure zero). Consequently, we define colouring $\col'$ as follows: we keep the weights unchanged in edges that belong to some $\ec \in \maxWinningECs$, and we put them to zero everywhere else, \textit{i.e.}, on any edge involving a negligible vertex. Weight zero is taken because it is lower than the expectation granted by WECs, which is \textit{strictly} greater than zero by definition (as~$\alpha = 0$).

\begin{example}
Consider $\ec_{1}$ in~\Cref{14-fig:bwcRunningExample}. This EC is losing as argued before. The optimal expectation achievable in $\markovProcess[\ec_{1}]$ by $\playerOne$ is $4$: this is higher than what is achievable in both $\ec_{2}$ and $\ec_{3}$. Note that there exists no WEC included in $\ec_{1}$. By~\Cref{6-chap:mdp}, we know that, from $v_1$, any strategy of $\playerOne$ will see its expectation bounded by the maximum between the optimal expectations of the ECs $\ec_{1}$, $\ec_{2}$ and $\ec_{3}$. Our previous arguments further refine this bound by restricting it to the maximum between the expectations of $\ec_{2}$ and $\ec_{3}$. Indeed, $\playerOne$ cannot benefit from the expected value of $\ec_{1}$ while using finite memory, as being trapped in~$\ec_{1}$ induces the existence of plays losing for the worst-case constraint. Hence there is no point in playing inside $\ec_{1}$ and $\playerOne$ may as well cross it directly and try to maximise its expectation using the WECs, $\ec_{2}$ and $\ec_{3}$. The set of negligible vertices in $\markovProcess$ is $\negligibleStates = V \setminus (\ec_{2} \cup \ec_{3}) = \{\state_{1}, \state_{2}, \state_{3}, \state_{4}, \state_{5}\}$. We depict $\markovProcess'$ in~\Cref{14-fig:bwc_mp_modifiedMDP}.

In the arena depicted in~\Cref{14-fig:bwcWinningECsComputationExample}, we already observed that $\ecsSet = \{\ec_{1}, \ec_{2}, \ec_{3}\}$ and $\winningECs = \maxWinningECs = \{\ec_{1}, \ec_{3}\}$. Consider the negligible vertex $\state_{1} \in \negligibleStates = \ec_{2} \setminus \ec_{3}$. A finite-memory strategy of $\playerOne$ may only take the edge $(\state_{1}, \state_{3})$ finitely often in order to ensure the worst-case requirement. If $\playerOne$ were to play this edge repeatedly, the losing play $(\state_{1}\state_{3}\state_{4})^{\omega}$ would exist (while of probability zero). Therefore, $\playerOne$ can only ensure that $\ec_{3}$ is reached with a probability arbitrarily close to one, and not equal to one, because at some point, she has to switch to edge $(\state_{1}, \state_{2})$ (after a bounded time since $\playerOne$ uses a finite-memory strategy).
\end{example}

\begin{figure}[htb]
  \centering   
  \scalebox{0.88}{\begin{tikzpicture}[->,>=latex,shorten >=1pt,auto,node
    distance=2.5cm,bend angle=45,scale=0.7]
    \tikzstyle{p1}=[draw,circle,text centered,minimum size=6mm]
    \tikzstyle{p2}=[draw,rectangle,text centered,minimum size=6mm]
    \tikzstyle{empty}=[]
    \node[p1] (1) at (0,0) {$\state_{9}$};
    \node[p1] (2) at (4,0) {$\state_{1}$};
    \node[p2] (3) at (8,0) {$\state_{2}$};
    \node[p1] (4) at (8,-4) {$\state_{3}$};
    \node[p2] (5) at (8,-8) {$\state_{4}$};
    \node[p2] (6) at (4,-4) {$\state_{5}$};
    \node[p1] (7) at (0,-4) {$\state_{6}$};
    \node[p2] (8) at (-4,-4) {$\state_{7}$};
    \node[p1] (9) at (-4,0) {$\state_{10}$};
    \node[p2] (10) at (-8,0) {$\state_{11}$};
    \node[p2] (11) at (0,-8) {$\state_{8}$};
    \node[empty] (swec) at (-6, 1.9) {WEC $\ec_{3}$ - $\expv_{\markovProcess} = \expv_{\markovProcess'} = 2$};
    \node[empty,align=center] (wwec) at (-6.6, -5.5) {WEC $\ec_{2}$\\$\expv_{\markovProcess} = \expv_{\markovProcess'} = 3$};
    \node[empty,align=center] (lec) at (5.2, -6) {LEC $\ec_{1}$\\$\expv_{\markovProcess} = 4$\\$\expv_{\markovProcess'}  = 0$};
    \node[empty] (proba5a) at (7.2, -7.5) {$\frac{1}{2}$};
    \node[empty] (proba5b) at (8.8, -7.5) {$\frac{1}{2}$};
    \node[empty] (proba3a) at (8.3, -1) {$\frac{1}{2}$};
    \node[empty] (proba3b) at (7.5, 0.95) {$\frac{1}{2}$};
    \node[empty] (proba8a) at (-3.5, -3.2) {{\large $1$}};
    \node[empty] (proba8b) at (-3.5, -4.8) {{\large $0$}};
    \node[empty] (proba10a) at (-7.5, 0.9) {$\frac{1}{2}$};
    \node[empty] (proba10b) at (-7.5, -0.9) {$\frac{1}{2}$};
    \node[empty] (proba11a) at (0.8, -7.5) {$\frac{1}{2}$};
    \node[empty] (proba11b) at (-0.8, -7.5) {$\frac{1}{2}$};
    \node[empty] (proba6a) at (3.8, -3.1) {$\frac{1}{2}$};
    \node[empty] (proba6b) at (3.3, -4.4) {$\frac{1}{2}$};
    \coordinate[shift={(-3mm,8mm)}] (init) at (2.north west);
    \path
    (2) edge node[above] {$0$} (1)
    (6) edge node[above] {$0$} (7)
    (4) edge[ultra thick] node[above left] {$0$} (6)
    (3) edge node[left] {$0$} (4)
    (6) edge node[left] {$0$} (2)
    (4) edge node[left] {$0$} (5)
    (7) edge node[above] {$0$} (8)
    (7) edge node[left] {$0$} (1)
    (init) edge (2)
    ;
	\draw[->,>=latex] (3) to[out=140,in=40] node[above] {$0$} (2);
	\draw[->,>=latex,ultra thick] (2) to[out=0,in=180] node[below] {$0$} (3);
	\draw[->,>=latex] (5) to[out=50,in=310] node[right] {$0$} (4);
	\draw[->,>=latex] (5) to[out=130,in=230] node[left] {$0$} (4);
	\draw[->,>=latex] (8) to[out=40,in=140] node[above] {$1$} (7);
	\draw[->,>=latex] (8) to[out=320,in=220] node[below] {$-1$} (7);
	\draw[->,>=latex,ultra thick] (1) to[out=140,in=40] node[above] {$1$} (9);
	\draw[->,>=latex] (9) to[out=320,in=220] node[below] {$1$} (1);
	\draw[->,>=latex,ultra thick] (9) to[out=180,in=0] node[below] {$0$} (10);
	\draw[->,>=latex] (10) to[out=40,in=140] node[above] {$-1$} (9);
	\draw[->,>=latex] (10) to[out=320,in=220] node[below] {$9$} (9);
	\draw[->,>=latex,ultra thick] (7) to[out=270,in=90] node[left] {$0$} (11);
	\draw[->,>=latex] (11) to[out=130,in=230] node[left] {$-1$} (7);
	\draw[->,>=latex] (11) to[out=50,in=310] node[right] {$13$} (7);
	\draw[dashed,-] (-9,1.6) -- (1,1.6) -- (1,-1.6) -- (-9,-1.6) -- (-9,1.6);
	\draw[dashed,-] (6.2,-3) -- (9.8,-3) -- (9.8,-9) -- (6.2,-9) -- (6.2,-3);
	\draw[dashed,-] (-5,-2.4) -- (1.7,-2.4) -- (1.7,-9) -- (-5,-9) -- (-5,-2.4);
      \end{tikzpicture}}
      \caption{Putting all weights outside MWECs to zero naturally drives the optimal expectation strategy in $\markovProcess'$, depicted by the thick edges, towards the highest valued MWECs. ECs are annotated with their corresponding optimal expectations in the original MDP $\markovProcess$ and the modified MDP $\markovProcess'$.}
\label{14-fig:bwc_mp_modifiedMDP}
  \end{figure}
  
\subsection*{Reach the highest valued winning end components} We compute the maximal expected mean payoff $\beta^{\ast}$ that can be achieved by $\playerOne$ in $\markovProcess'$, from $v_0$. This computation takes polynomial time and positional strategies are sufficient to achieve the maximal value, as established in~\Cref{6-thm:general-mp-main}.
As seen before, such a strategy reaches an EC of $\markovProcess'$ with probability one. Basically, we build a strategy that favours reaching ECs with high associated expectations in~$\markovProcess'$.

We argue that the ECs reached with probability one by this strategy are necessarily WECs in $\markovProcess$. Clearly, if a WEC is reachable instead of a losing one, it will be favoured because of the weights definition in $\markovProcess'$ (expectation is strictly higher in WECs). Thus it remains to check if the set of WECs is reachable with probability one from any vertex in~$V$. That is the case because of the preprocessing: we know that all vertices are winning for the worst-case requirement. Clearly, from any vertex in $A = V \setminus \bigcup_{\ec \in \ecsSet} \ec$, $\playerOne$ cannot ensure to stay in $A$ (otherwise it would form an EC) and thus must be able to win the worst-case requirement from reached ECs. Now for any vertex in $B = \bigcup_{\ec \in \ecsSet} \ec \setminus \bigcup_{\ec \in \maxWinningECs} \ec$, \textit{i.e.}, vertices in LECs and not in any winning sub-EC, $\playerOne$ cannot win the worst-case by staying in $B$, by definition of LEC. Since we know $\playerOne$ can ensure the worst-case constraint by hypothesis, it is clear that she must be able to reach $C = \bigcup_{\ec \in \maxWinningECs} \ec$ from any vertex in $B$, as claimed.

\subsection*{Inside winning end components} Based on that, we know that WECs of $\markovProcess$ will be reached with probability one when maximising the expected value in $\markovProcess'$. Let us first consider what we can say about such ECs if we assume that $\edgesNonZero = \edges$, \textit{i.e.}, if the stochastic model $\stratStoch$ maps all possible edges to non-zero probabilities. 
We will establish a finite-memory \textit{combined strategy} $\stratComb$ of $\playerOne$ that ensures~(i) worst-case satisfaction while yielding (ii) an expected value $\varepsilon$-close to the maximal expectation inside the EC.

For two well-chosen parameters $\stepsExp, \stepsWC \in \N$, it is informally defined as follows: during phase $\typeA$, play a positional expected value optimal strategy $\stratExp$ for $\stepsExp$ steps and memorise $\cmbSum \in \Z$, the sum of weights along these steps; in phase $\typeB$, if $\cmbSum > 0$, go to~$\typeA$, otherwise play a positional worst-case optimal strategy $\stratWC$ for~$\stepsWC$ steps, then go to $\typeA$. Intuitively, in~$\typeA$, $\playerOne$ tries to increase her expectation and approach the optimal one, while in $\typeB$, she compensates, if needed, losses that occurred in $\typeA$.

The two positional strategies exist on the subarena induced by the EC: by definition of ECs, based on~$\edgesNonZero$, the stochastic model of $\playerTwo$ will never be able to force leaving the EC against the combined strategy.

A key result to our approach is the existence of values for $\stepsExp$ and~$\stepsWC$ such that~(i) and (ii) are verified. We see plays as sequences of periods, each starting with phase~$\typeA$.

First, for any $\stepsExp$, it is possible to define $\stepsWC(\stepsExp)$ such that any period composed of both phases~$\typeA+\typeB$ ensures a mean payoff at least $1/(\stepsExp+\stepsWC) > 0$. Periods containing only phase $\typeA$ trivially induce a mean payoff at least~$1/\stepsExp$ as they are not followed by phase $\typeB$. Both rely on the weights being integers. As the length of any period is bounded by $(\stepsExp+\stepsWC)$, the inequality remains strict for the mean payoff of any play, granting~(i).

Now, consider parameter $\stepsExp$. Clearly, when~$\stepsExp \rightarrow \infty$, the expectation over phase $\typeA$ tends to the optimal one. Nevertheless, phases~$\typeB$ also contribute to the overall expectation of the combined strategy, and (in general) lower it so that it is strictly less than the optimal for any $\stepsExp, \stepsWC \in \N$. Hence to prove (ii), we not only need that the probability of playing phase $\typeB$ decreases when $\stepsExp$ increases, but also that it decreases faster than the increase of $\stepsWC$, needed to ensure~(i), so that overall, the contribution of phases~$\typeB$ tends to zero when $\stepsExp \rightarrow \infty$. This is indeed the case and is proved using (rather technical) results bounding the probability of observing a mean payoff significantly (more than some $\varepsilon$) different than the optimal expectation along a phase $\typeA$ of length $\stepsExp \in \N$: this probability decreases exponentially when~$\stepsExp$ increases, while $\stepsWC$ only needs to be polynomial in $\stepsExp$.

\begin{theorem}
\label{14-thm:insideWinning}
Let $U \in \winningECs$ be a WEC, $\stratStoch$ be such that $\edgesNonZero = \edges$, $v_0 \in U$ be the initial vertex, and let $\beta^\ast \in \Q$ be the maximal expected value achievable by $\playerOne$ in EC $U$. Then, for all~$\varepsilon > 0$, there exists a finite-memory strategy of $\playerOne$ that satisfies the $\BWC$ problem for the thresholds pair $(0,\, \beta^\ast - \varepsilon)$.
\end{theorem}

\begin{example}
Consider the subarena $\gameNonZero[\ec_{3}] = \arena[\ec_{3}]$ from~\Cref{14-fig:bwcRunningExample} and the initial vertex $\state_{10}$. The worst-case requirement can be satisfied, that is why the EC is classified as winning: alternating between~$\state_{9}$ and $\state_{10}$ is an optimal positional worst-case strategy $\stratWC$ that guarantees a mean payoff $\alpha^\ast = 1$. Its expectation is $\expv^{\stratWC}_{(\arena[\ec_{3}])_{\tau^\mathsf{st}},v_{10}}[\MeanPayoff^{-}] = 1$. On the other hand, the strategy $\stratExp$ that always selects the edge going to $\state_{11}$ is optimal regarding the expected value criterion: it induces expectation $\beta^\ast = \big(0 + \big(1/2 \cdot 9 + 1/2 \cdot (-1)\big)\big)/2 = 2$ against the stochastic model $\stratStoch$. However, it can only guarantee a mean payoff of value $-1/2$ in the worst case.

By the reasoning above, we know that it is possible to find finite-memory strategies satisfying the $\BWC$ problem for any thresholds pair $(0,\, 2 - \varepsilon)$, $\varepsilon > 0$. In particular, consider the thresholds pair $(0,\, 3/2)$. We build a combined strategy~$\stratComb$ as sketched before. Let $\stepsExp = \stepsWC = 2$: the strategy plays the edge $(\state_{10}, \state_{11})$ once, then if the edge of value $9$ has been chosen by $\playerTwo$, it chooses $(\state_{10}, \state_{11})$ again; otherwise it chooses the edge $(\state_{10}, \state_{9})$ once and then resumes choosing $(\state_{10}, \state_{11})$. This strategy satisfies the $\BWC$ problem. In the worst case,~$\playerTwo$ always chooses the $-1$ edge, but each time he does so, the $-1$ is followed by two~$+1$ thanks to the cycle $\state_{10} \state_{9} \state_{10}$. Strategy $\stratComb$ hence guarantees a mean payoff equal to $(0 - 1 + 1 + 1)/4 = 1/4 > 0$ in the worst case. For the expected value requirement, we can build the induced Markov chain $(\arena [\ec_{3}])_{\stratComb, \tau^\mathsf{st}}$ (\Cref{14-fig:mp_insideSWEC_MC}) and check that its expectation is $\expv^{\stratComb}_{(\arena[\ec_{3}])_{\tau^\mathsf{st}},v_{10}}[\MeanPayoff^{-}] = 5/3 > 3/2$ (\Cref{5-chap:payoffs}).
\end{example}

\begin{figure}[thb]
  \centering 
  \scalebox{0.88}{\begin{tikzpicture}[->,>=latex,shorten >=1pt,scale=0.8]
    \tikzstyle{p1}=[draw,circle,text centered,minimum size=6mm]
    \tikzstyle{p2}=[draw,rectangle,text centered,minimum size=14mm,text width=13mm]
    \tikzstyle{p3}=[draw,diamond,text centered,minimum size=20mm,text width=13mm]
    \tikzstyle{empty}=[]
    \node[p2] (1) at (0,0) {$\state_{10}$\\$\cmbSum > 0$};
    \node[p2] (2) at (-4,0) {$\state_{11}$\\$ $};
    \node[p2] (3) at (-4,4) {$\state_{10}$\\$\cmbSum \leq 0$};
    \node[p2] (4) at (0,4) {$\state_{9}$\\$ $};
    \node[empty] (a) at (-3.75, 1.05) {$\frac{1}{2}$};
    \node[empty] (b) at (-3.0, -1.5) {$\frac{1}{2}$};
    \coordinate[shift={(8mm,0mm)}] (init) at (1.east);
    \path
    (1) edge node[above] {$0$} (2)
    (2) edge node[left] {$-1$} (3)
    (3) edge node[above] {$1$} (4)
    (4) edge node[right] {$1$} (1)
    (init) edge (1)
    ;
	\draw[->,>=latex] (2) to[out=320,in=220] node[below] {$9$} (1);
      \end{tikzpicture}}
      \caption{Markov chain induced by the combined strategy $\stratComb$ and the stochastic model $\stratStoch$ over the WEC $\ec_{3}$ of $\arena$.}
\label{14-fig:mp_insideSWEC_MC}
\commentAlt{Figure~\ref{14-fig:mp_insideSWEC_MC}: A directed graph with four square nodes arranged in a rectangle, showing various labeled transitions, including conditions for some nodes. See long description.}
\commentLongAlt{Figure~\ref{14-fig:mp_insideSWEC_MC}: The image displays a directed graph with four square nodes arranged in a rectangular pattern.
- The top-left node is labeled 'v10' with 'Sum <= 0' below it.
- The top-right node is labeled 'v9'.
- The bottom-right node is labeled 'v10' with 'Sum > 0' below it. An incoming arrow points to this node, indicating it as a starting point.
- The bottom-left node is labeled 'v11'.

Arrows and their labels define the transitions:
- An arrow from the top-left 'v10 (Sum <= 0)' points to 'v9', labeled '1'.
- An arrow from 'v9' points to the bottom-right 'v10 (Sum > 0)', labeled '1'.
- An arrow from the bottom-right 'v10 (Sum > 0)' points to 'v11', labeled '0'.
- An arrow from 'v11' points to the top-left 'v10 (Sum <= 0)', labeled '-1' above and '1/2' below.
- A curved arrow from 'v11' points to the bottom-right 'v10 (Sum > 0)', with a label '9' on the arc and '1/2' below the arc.}
\end{figure}

\begin{remark}
\label{14-rmk:bwcPositionalNotEnough}
Positional strategies do not suffice for the $\BWC$ problem, even with randomisation. Indeed, the edge $(\state_{10}, \state_{11})$ cannot be assigned a non-zero probability as it would endanger the worst-case requirement (since the play~$(\state_{10}\state_{11})^{\omega}$ cycling on the edge of weight $-1$ would exist and have a negative mean payoff). Hence, the only acceptable positional strategy is $\stratWC$, which has only an expectation of $1$.
\end{remark}

Now, consider what happens if $\edgesNonZero \subsetneq E$. If $\playerTwo$ uses an arbitrary strategy, he can take edges of probability zero, \textit{i.e.}, in $E \setminus \edgesNonZero$, either staying in the EC, or leaving it. In both cases, this must be taken into account to satisfy the worst-case constraint as it may involve dangerous weights (recall that zero-probability edges are not considered for the EC classification). Fortunately, if this were to occur, $\playerOne$ could switch to a worst-case winning positional strategy $\stratSecure$, which exists everywhere thanks to the preprocessing, to preserve the worst-case requirement. For the expected value, this has no impact as it occurs with probability zero against $\stratStoch$. The strategy to follow in WECs hence adds this reaction procedure to the combined strategy: we call it the \textit{witness-and-secure strategy} $\stratWNS$.

\begin{theorem}
\label{14-thm:wns}
Let $U \in \winningECs$ be a WEC, $v_0 \in U$ be the initial vertex, and $\beta^\ast \in \Q$ be the maximal expected value achievable by $\playerOne$ in EC $U$. Then, for all~$\varepsilon > 0$, there exists a finite-memory strategy of $\playerOne$ that satisfies the $\BWC$ problem for the thresholds pair $(0,\, \beta^\ast - \varepsilon)$.
\end{theorem}

\begin{example}
Consider the WEC $\ec_{2}$ in~\Cref{14-fig:bwcRunningExample} and the initial vertex $\state_{6} \in \ec_{2}$. $\playerOne$ can ensure a strictly positive mean payoff in the subarena $\gameNonZero[\ec_{2}]$, but not in $\arena[\ec_{2}]$. Indeed, it is easy to see that by always choosing the $-1$ edges (which requires an edge $(\state_{7}, \state_{6}) \in \edges \setminus \edgesNonZero$), $\playerTwo$ can ensure a negative mean payoff whatever the strategy of $\playerOne$. However, there exists a strategy that ensures the worst-case constraint, \textit{i.e.}, that yields a strictly positive mean payoff against any strategy of Adam, by leaving the EC. Let $\stratSecure$ be the positional strategy that takes the edge $(\state_{6}, \state_{9})$ and then cycle on $(\state_{10}\state_{9})^{\omega}$ forever: it guarantees a mean payoff of $1 > 0$.

For a moment, consider the EC $\ec_{2}$ in $\gameNonZero$. Graphically, it means that the $-1$ edge from $\state_{7}$ to $\state_{6}$ disappears. In the subarena $\gameNonZero[\ec_{2}]$, there are two particular positional strategies. The optimal worst-case strategy $\stratWC$ guarantees a mean payoff of $1/2 > 0$ by choosing to go to $\state_{7}$. The optimal expectation strategy $\stratExp$ yields an expected mean payoff of $3$ by choosing to go to $\state_{8}$ (naturally this strategy yields the same expectation whether we consider edges in $\edgesNonZero$ or in $E$). Based on them, we build the combined strategy $\stratComb$ of Eve as defined earlier and by~\Cref{14-thm:insideWinning}, for any $\varepsilon > 0$, there are values of $\stepsExp$ and $\stepsWC$ such that it satisfies the $\BWC$ problem for thresholds $(0,\, 3-\varepsilon)$ in $\gameNonZero[\ec_{2}]$. For instance, for $\stepsExp = \stepsWC = 2$, we have $\expv^{\stratComb}_{(\arena[\ec_{2}])_{\tau^\mathsf{st}},v_{6}}[\MeanPayoff^{-}] = 13/6$.

We construct the witness-and-secure strategy $\stratWNS$ based on $\stratComb$ and $\stratSecure$ as described above. In this case, that means playing as $\stratComb$ until the $-1$ edge from $\state_{7}$ to $\state_{6}$ is taken by $\playerTwo$. This strategy ensures a worst-case mean payoff equal to $1 > 0$ thanks to $\stratSecure$ and yields expectation $\expv^{\stratWNS}_{(\arena[\ec_{2}])_{\tau^\mathsf{st}},v_{6}}[\MeanPayoff^{-}] = 13/6$ for $\stepsExp = \stepsWC = 2$.

Finally, notice that securing the mean payoff by switching to  $\stratSecure$ \textit{is needed} to satisfy the worst-case requirement if $\playerTwo$ plays in $\edges \setminus \edgesNonZero$. Also, observe that it is still necessary to alternate according to $\stratComb$ in $\gameNonZero[\ec_{2}]$ and that playing $\stratExp$ is not sufficient to ensure the worst-case (because $\playerOne$ has to deal with the $-1$ edge from $\state_{8}$ to $\state_{6}$ that remains in $\edgesNonZero$).
\end{example}

\subsection*{Global strategy synthesis} In summary, (a) LECs should be avoided and will be by a strategy that optimises the expectation on the MDP~$\markovProcess'$; (b) in WECs, $\playerOne$ can obtain ($\varepsilon$-closely) the expectation of the EC \textit{and} ensure the worst-case threshold.

Hence, we finally compare the value $\thresholdExp^{\ast}$ computed by~\Cref{14-algo:BWC} with the expected value threshold $\thresholdExp$: (i) if strictly higher, we conclude that there exists a finite-memory strategy satisfying the $\BWC$ problem, and (ii) if not, there does not exist such a strategy.

To prove (i), we establish a finite-memory strategy in $\arena$, called \textit{global strategy} $\stratGlobal$, of $\playerOne$ that ensures a strictly positive mean payoff against an antagonistic adversary, and ensures an expected mean payoff $\varepsilon$-close to $\thresholdExp^{\ast}$ (hence, strictly greater than $\thresholdExp$) against the stochastic adversary modeled by $\stratStoch$ (\textit{i.e.}, in $\markovProcess$). The intuition is as follows. We play the positional optimal strategy of $\markovProcess'$ for a sufficiently long time, defined by a parameter $\stepsGlobal \in \N$, in order to be with probability close to one in a WEC (the convergence is exponential by results on absorption times in Markov chains). Then, if we are inside a WEC, we switch to the corresponding witness-and-secure strategy (there is a different one for each MWEC) which, as sketched in the previous paragraph, ensures the worst-case and the expectation thresholds. If we are not yet in a WEC, then we switch to a worst-case winning strategy, which always exists thanks to the preprocessing. Thus the mean payoff of plays that do not reach WECs is strictly positive. Since in WECs we are $\varepsilon$-close to the maximal expected value of the EC, we can conclude that it is possible to play the optimal expectation strategy of $\markovProcess'$ for sufficiently long to obtain an overall expected value which is arbitrarily close to $\thresholdExp^{\ast}$, and still guarantee the worst-case threshold in all consistent plays.

To prove (ii), it suffices to understand that only ECs have an impact on the expectation, and that LECs cannot be used forever without endangering the worst-case requirement.

Note that given a winning strategy on $\arena$, it is possible to build a corresponding winning strategy on $\arena^{\mathsf{in}}$ by reintegrating the memory states of $\tau^{\mathsf{in}}$ in the memory structure of the winning strategy of $\playerOne$. Hence~\Cref{14-algo:BWC} is correct and complete.

\begin{theorem}
\label{14-thm:bwcCorrectAndComplete}
If~\Cref{14-algo:BWC} answers \textsc{Yes}, then there exist values of the parameters such that the pure finite-memory global strategy $\stratGlobal$ satisfies the $\BWC$ mean-payoff problem. In the opposite case, there exists no finite-memory strategy that satisfies the $\BWC$ mean-payoff problem.
\end{theorem}

\begin{example}
Consider the arena in~\Cref{14-fig:bwcRunningExample} and the associated MDP $\markovProcess$. Following~\Cref{6-chap:mdp}, analysis of the maximal ECs $\ec_{1}$, $\ec_{2}$ and $\ec_{3}$ reveals that the maximal expected mean payoff achievable in $\markovProcess$ is $4$. It is for instance obtained by the positional strategy that chooses to go to $\state_{2}$ from $\state_{1}$ and to $\state_{4}$ from $\state_{3}$. Observe that playing in $\ec_{1}$ forever is needed to achieve this expectation. By~\Cref{14-lem:EC-inf}, this should not be allowed as the worst case cannot be ensured if it is. Indeed, $\playerTwo$ can produce worst-case losing plays by playing the $-1$ edge. Clearly, the maximal expected value that $\playerOne$ can ensure while guaranteeing the worst-case requirement is thus bounded by the maximal expectation in $\markovProcess'$, \textit{i.e.}, by $3$, as depicted in~\Cref{14-fig:bwc_mp_modifiedMDP}. Let $\stratExp$ denote an optimal positional expectation strategy in $\markovProcess'$ that tries to enter~$\ec_{2}$ by playing $(\state_{1}, \state_{2})$ and $(\state_{3}, \state_{5})$, and then plays edge $(\state_{6}, \state_{8})$ forever (thick edges in~\Cref{14-fig:bwc_mp_modifiedMDP}).

Observe that~\Cref{14-algo:BWC} answers \textsc{Yes} for any thresholds pair $(0,\, \thresholdExp)$ such that $\thresholdExp < 3$. For the sake of illustration, we construct the global strategy~$\stratGlobal$ as presented earlier, with $\stepsGlobal = 6$ and $\stepsExp = \stepsWC = 2$. For the first six steps, it behaves exactly as $\stratExp$. Note that after the six steps, the probability of being in $\ec_{2}$ is $1/4 + 1/8 = 3/8$. Then, $\stratGlobal$ switches to another strategy depending on the current vertex ($\stratWNS$ or $\stratWC$) and sticks to this strategy forever. In particular, if the current vertex belongs to $\ec_{2}$, it switches to $\stratWNS$ for $\stepsExp = \stepsWC = 2$, which guarantees the worst-case threshold and induces an expectation of $13/6$. By definition of $\stratGlobal$, if the current vertex after six steps is not in $\ec_{2}$, then $\stratGlobal$ switches to $\stratWC$ which guarantees a mean payoff of $1$ by reaching vertex $\state_{9}$ and then playing $(\state_{9}\state_{10})^{\omega}$. Overall, the expected mean payoff of $\stratGlobal$ against $\stratStoch$ is
\begin{equation*}
\expv^{\stratGlobal}_{\arena_{\tau^\mathsf{st}},v_{1}}[\MeanPayoff^{-}] \geq \dfrac{3}{8}\cdot\dfrac{13}{6} + \dfrac{5}{8}\cdot 1 = \dfrac{23}{16}.
\end{equation*}
By taking $\stepsGlobal$, $\stepsExp$ and $\stepsWC$ large enough, it is possible to satisfy the $\BWC$ problem for any $\thresholdExp < 3$ with the strategy $\stratGlobal$. Also, observe that the WEC~$\ec_{2}$ is crucial to achieve expectations strictly greater than $2$, which is the upper bound when limited to EC $\ec_{3}$. For instance, $\stepsGlobal = 25$ and $\stepsExp = \stepsWC = 2$ implies an expectation strictly greater than $2$ for the global strategy.

Lastly, note that in general, the maximal expectation achievable in $\markovProcess'$ (and thus in $\markovProcess$ when limited to strategies that respect the worst-case requirement) may depend on a combination of ECs instead of a unique one. This is transparent through the solving of the expected value problem in the MDP $\markovProcess'$.
\end{example}

\subsection*{Complexity bounds} The input size of the algorithm depends on the size of the arena, the size of the memory structure for the stochastic model, and the encodings of probabilities, weights and thresholds. We can prove that all computing steps require (deterministic) polynomial time except for calls to an algorithm solving the worst-case threshold problem, which is in $\NP \cap \coNP$ and not known to be in $\P$ (\Cref{5-thm:MP-NPcoNP}). Hence, the overall complexity of the $\BWC$ problem is in $\NP \cap \coNP$ (using $\P^{\NP \cap \coNP} = \NP \cap \coNP$~\cite{Brassard:1979}) and may collapse to $\P$ if the worst-case problem were to be proved in $\P$.

The $\BWC$ problem is at least as difficult as the worst-case problem thanks to a trivial polynomial-time reduction from the latter to the former. Thus, membership to $\NP \cap \coNP$ can be seen as optimal regarding our current knowledge of mean-payoff games.

\begin{theorem}
\label{14-thm:bwcDecisionProblem}
The BWC mean-payoff problem is in $\NP \cap \coNP$ and at least as hard as solving mean-payoff games. Moreover, pseudo-polynomial-memory strategies may be necessary for Eve and are always sufficient.
\end{theorem}

The memory bounds follow from the (involved) probability results used to determine the values of parameters $K$, $L$ and $N$ in the aforementioned strategies: such parameters need to be polynomial in the size of the arena but also in the probabilities, weights and thresholds.

Thanks to the pseudo-polynomial-time algorithm of~\Cref{5-sec:mean_payoff} for mean-payoff games, we obtain the following corollary.

\begin{corollary}
\label{14-cor:BWC}
\Cref{14-algo:BWC} solves the BWC mean-payoff problem in pseudo-poly\-no\-mial time.
\end{corollary}

\subsection*{Wrap-up} As witnessed by our overview, solving the beyond worst-case problem requires much more involved techniques than solving the two individual problems, worst-case and expected value, separately. Complexity-wise, it is fortunate that the problem stays in $\NP \cap \coNP$, and is no more complex that simple mean-payoff games. The multi-objective nature of the problem still incurs a cost with regard to strategies: whereas positional strategies suffice both in mean-payoff games and mean-payoff MDPs, we here need pseudo-polynomial memory. Finally, observe that Eve does not need to use randomness: pure strategies still suffice.

\section{Percentile queries}
\label{14-sec:percentile}
We close this chapter by a quick detour to multidimensional MDPs. When considering single-dimension MDPs with payoffs, as in~\Cref{6-chap:mdp}, there are two different (yet both natural) settings that arise depending on how one decides to aggregate play values through the probability measure. Let $\markovProcess$ be an MDP, $v$ an initial vertex, and $f$ the payoff function. In the first setting, Eve's goal is to optimise the "expected value" of the payoff function, that is, to find a strategy $\sigma$ that maximises $\expv^{\sigma}_{\markovProcess,v}[f]$. In the second setting, we set a performance threshold to achieve for the payoff function, say $\alpha \in \Q$, essentially creating the qualitative objective $f_{\geq \alpha}$, and Eve aims to maximise the probability to achieve this objective, \textit{i.e.}, she is looking for a strategy $\sigma$ that maximises $\probm^{\sigma}_{\markovProcess,v}[f_{\geq \alpha}]$. The concept of \textit{percentile query} extends the latter problem to multidimensional payoffs.

From now on, assume we have an MDP $\markovProcess$ with a multidimensional colouring function $\col\colon E \rightarrow \Z^k$. Whether $\markovProcess$ uses actions as in~\Cref{6-chap:mdp} or random vertices as in~\Cref{7-chap:stochastic} does not matter for our discussion --- both are equivalent modulo slight modifications of the MDP. Recall that we denote by $f_i$, $1 \leq i \leq k$, the projection of $f$ to its $i$-th component.

\decpb[Percentile query problem]{An MDP $\markovProcess$, an initial vertex $v_0$, a payoff function $f$, $q \geq 1$ the number of percentile constraints in the query, $q$ dimensions $l_i \in \{1,\ldots, k\}$, $q$ value thresholds $\alpha_i \in \Q$, $q$ probability thresholds $\mu_i \in \Q \cap [0, 1]$.}{Does Eve have a strategy $\sigma$ such that $\sigma$ is winning for the conjunction of $q$ constraints, called percentile query, \[\mathcal{Q} = \bigwedge_{i=1}^{q} \probm^{\sigma}_{\markovProcess,v_0}[f_{l_i} \geq \alpha_i] \geq \mu_i?\]}
As usual, we also want to synthesize such a strategy $\sigma$ if one exists. Note that this percentile query framework permits to express rich properties, as each of the $q$ constraint can use a different dimension, value threshold and probability threshold. It is also possible to have different constraints related to the same dimension, for example to enforce different value thresholds for different quantiles.

The percentile query problem has been studied for a variety of payoff functions. Our aim here is not to give an exhaustive account of the corresponding results and techniques, but to highlight some new phenomena that arise in this setting, in comparison to what we have seen up to now.

\subsection{An additional leap in complexity}

The expressiveness of percentile queries asks for richer classes of strategies, even in very simple MDPs.

\begin{figure}[tbp]
  \centering
  \begin{tikzpicture}[node distance=3cm,>=latex]
    \node[draw,circle](0) at (0,0) {$v_0$};
    \node[draw,circle](1) at (-2,-1) {$v_1$};
    \node[draw,circle](2) at (2,-1) {$v_2$};

    \path[->] (0) edge node[above left] {$(0,0)$}  (1)
    (0) edge node[above right] {$(0,0)$}  (2)
    (1) edge[loop left] node[left] {$(1,0)$} (1)
    (2) edge[loop right] node[right] {$(0,1)$} (2);

  \end{tikzpicture}
  \caption{Randomised strategies are needed to achieve any percentile query of the form $\probm^{\sigma}_{\markovProcess,v_0}[\MeanPayoff^{-}_1 \geq 1] \geq \mu_1 \wedge \probm^{\sigma}_{\markovProcess,v_0}[\MeanPayoff^{-}_2 \geq 1] \geq \mu_2$ with $\mu_1, \mu_2 > 0$.}
  \label{14-fig:MultiReach}
\commentAlt{Figure~\ref{14-fig:MultiReach}: A central circular node v0 branching out to two other circular nodes, v1 and v2, each with a self-loop, with all edges and loops labeled with numerical pairs. See long description.}
\commentLongAlt{Figure~\ref{14-fig:MultiReach}: The image displays a directed graph with three circular nodes. A central node labeled 'v0' has two outgoing arrows.
- An arrow from 'v0' points to 'v1' (left), labeled '(0,0)'. Node 'v1' has a self-loop labeled '(1,0)'.
- An arrow from 'v0' points to 'v2' (right), labeled '(0,0)'. Node 'v2' has a self-loop labeled '(0,1)'.}
\end{figure}

\begin{example}
\label{14-ex:randomisedStrats}
Consider the single-player game depicted in~\Cref{14-fig:MultiReach}. Note that it is an MDP (using only Dirac distributions). Consider the payoff function $\MeanPayoff^{-}$. It is clear that due to its prefix-independence, any play $\play$ ending in $v_1$ (resp.~$v_2$) will yield $\MeanPayoff^{-}(\play) = (1,0)$ (resp.~$(0,1)$). Hence to achieve a percentile query \[\mathcal{Q} = \bigwedge_{i=1}^2\probm^{\sigma}_{\markovProcess,v_0}[\MeanPayoff^{-}_i \geq 1] \geq \mu_i,\] Eve must go towards $v_i$ with probability at least $\mu_i$. If both probability thresholds are non-zero, then this is only achievable by using randomness within Eve's strategy.
\end{example}

\Cref{14-ex:randomisedStrats} uses the mean payoff for the sake of consistency with the previous sections, but observe that it can be emulated with virtually all objectives considered in this book. In particular, using reachability with two target sets (corresponding to the edges $(1,0)$ and $(0,1)$) is sufficient.

While \textit{pure} strategies were used in most chapters\footnote{Without loss of generality, as they suffice in the respective contexts of these chapters.} of this book, \textit{randomised} strategies have already been considered in specific settings, such as in~\Cref{8-chap:concurrent}, usually to break some kind of symmetry and/or make one's strategy hard to predict. In our setting of percentile queries, we are still dealing with relatively simple models of games: we consider turn-based, perfect information games. Yet, the need for randomness arises from the expressiveness of our class of objectives, which in general require careful balance between different stochastic options.

\begin{example}
\label{14-ex:continuousPareto}
Let us have another look at the MDP in~\Cref{14-fig:MultiReach}. Consider now that the probability thresholds $\mu_1$ and $\mu_2$ in query $\mathcal{Q}$ are not fixed a priori. Instead, we are interested in the set of vectors $(\mu_1, \mu_2)$ that Eve can achieve. In particular, we want to determine the \textit{Pareto frontier}\footnote{One can easily adapt~\Cref{14-def:ParetoStrat} to this context.} and the corresponding Pareto-optimal strategies. What is interesting here is that, in our simple example, there is already an infinite, non-countable, number of Pareto vectors. Indeed, Eve can ensure any vector $(\mu_1, \mu_2)$ such that $\mu_1, \mu_2 \geq 0$ and $\mu_1 + \mu_2 = 1$ by simply taking the edge leading to $v_i$ with probability~$\mu_i$.
\end{example}

Although the Pareto frontier consists of an infinite number of points in~\Cref{14-ex:continuousPareto}, it can be represented in a finite way, as it is essentially obtained through linear combinations of two extreme vectors: $(1,0)$ and $(0,1)$. Interestingly, these two vectors correspond to what can be achieved with \textit{pure} strategies, and their convex hull yields the Pareto frontier. So, the Pareto frontier can be represented as a convex polytope whose vertex representation is given by the vectors achievable by pure strategies. This is not merely an artefact resulting from the simplicity of our example; similar phenomena occur in many settings mixing MDPs and multiple objectives.

While the continuous aspect of the Pareto frontier stems from randomness in strategies, complex Pareto frontiers are also to be expected when restricting Eve to pure strategies.

\begin{figure}[tbp]
  \centering
  \begin{tikzpicture}[node distance=3cm,>=latex]
    \tikzstyle{p1}=[draw,circle,text centered,minimum size=6mm]
    \tikzstyle{p2}=[draw,rectangle,text centered,minimum size=6mm]
    \tikzstyle{empty}=[]
    \node[draw,circle](0) at (0,0) {$v_0$};
    \node[draw,p2](1) at (-2,-1) {$r_1$};
    \node[draw,p2](2) at (2,-1) {$r_2$};
    \node[draw,p1](3) at (-2,-3) {$v_1$};
    \node[draw,p1](4) at (2,-3) {$v_2$};
    \node[empty] (pr1a) at (-2.15, -1.6) {$\frac{1}{2}$};
    \node[empty] (pr1b) at (-1.6, -1.45) {$\frac{1}{2}$};
    \node[empty] (pr2a) at (1.6, -1.45) {$\frac{1}{2}$};
    \node[empty] (pr2b) at (2.15, -1.6) {$\frac{1}{2}$};

    \path[->] (0) edge node[above left] {$(0,0)$}  (1)
    (0) edge node[above right] {$(0,0)$}  (2)
    (1) edge[bend right=45] node[below] {$(0,0)$} (0)
    (2) edge[bend left=45] node[below] {$(0,0)$} (0)
    (1) edge node[left, yshift=-2mm] {$(0,0)$} (3)
    (2) edge node[right, yshift=-2mm] {$(0,0)$} (4)
    (3) edge[loop left] node[left] {$(1,0)$} (3)
    (4) edge[loop right] node[right] {$(0,1)$} (4);
  \end{tikzpicture}
  \caption{When restricted to pure strategies, there are still infinitely many Pareto vectors for the query $\probm^{\sigma}_{\markovProcess,v_0}[\MeanPayoff^{-}_1 \geq 1] \geq \mu_1 \wedge \probm^{\sigma}_{\markovProcess,v_0}[\MeanPayoff^{-}_2 \geq 1] \geq \mu_2$. For example, $(1,0)$, $(0,1)$, $(1/2, 1/2)$, $(3/4, 1/4)$, $(1/4, 3/4)\ldots{}$}
  \label{14-fig:MultiReach2}
\commentAlt{Figure~\ref{14-fig:MultiReach2}: A central circular node v0 connected to two square nodes, r1 and r2, which in turn connect to two circular nodes, v1 and v2. See long description.}
\commentLongAlt{Figure~\ref{14-fig:MultiReach2}: The image displays a directed graph with five nodes. A central circular node labeled 'v0' is at the top. Two square nodes, 'r1' (left) and 'r2' (right), are positioned below 'v0'. Two circular nodes, 'v1' (bottom-left) and 'v2' (bottom-right), are positioned below 'r1' and 'r2' respectively.
- An arrow from 'v0' points to 'r1', labeled '(0,0)'.
- An arrow from 'r1' points to 'v0', labeled '(0,0)'.
- An arrow from 'v0' points to 'r2', labeled '(0,0)'.
- An arrow from 'r2' points to 'v0', labeled '(0,0)'.
- An arrow from 'r1' points to 'v1', labeled '(0,0)' with '1/2' below it. Node 'v1' has a self-loop labeled '(1,0)'.
- An arrow from 'r2' points to 'v2', labeled '(0,0)' with '1/2' below it. Node 'v2' has a self-loop labeled '(0,1)'.}
\end{figure}

\begin{example}
\label{14-ex:exponentialPareto}
Consider the MDP in~\Cref{14-fig:MultiReach2}. It uses the random vertices formalism as in~\Cref{14-sec:beyond_worst_case}. This MDP is a slight adaptation of the one in~\Cref{14-fig:MultiReach}, the crux being that when Eve tries to go to $v_i$ now, she has to cross $r_i$, which has probability $1/2$ to send her back to $v_0$. In the long run, it does not really matter, as Eve will almost-surely end up in $v_1$ or $v_2$ (\Cref{6-chap:mdp}). And if Eve is allowed to use randomness, we obtain the same Pareto frontier as in the previous example. Yet, these random vertices serve a purpose.

When restricted to pure strategies, Eve cannot use the inherent randomness of her strategy to achieve any given vector $(\mu_1, \mu_2)$, as she could in~\Cref{14-ex:continuousPareto}. Nonetheless, by using memory, Eve is still able to achieve infinitely many Pareto vectors. For example, by first choosing to go to $r_1$, then $r_2$ (if the play comes back to $v_0$), then $r_1$ again (and then every time the play goes back to $v_0$), Eve will achieve vector $(3/4, 1/4)$.
\end{example}

It is easy to see that infinitely many Pareto vectors can be generated with memory and no randomness in~\Cref{14-ex:exponentialPareto}; for example all vectors of the form $(1-p, p)$ where $p = 1/2^n$ for $n \in \N$. Still, all such vectors could already be generated via randomised positional strategies as sketched before.

In particular, all vectors achievable by using memory and no randomness are of the form $(\mu_1, \mu_2)$, with $\mu_1, \mu_2 \geq 0$ and $\mu_1 + \mu_2 = 1$ --- but not all such vectors can be achieved that way! Hence, by restricting the use of randomness, we have effectively created `gaps' in the Pareto frontier, and rendered its description much more difficult. In full generality, it is usually necessary to use both randomness and memory to satisfy percentile queries.

\begin{proposition}
\label{14-prop:percentileMemory}
Pareto-optimal strategies for the percentile query problem may require randomness and memory (possibly infinite depending on the payoff function).
\end{proposition}

\subsection{Complexity overview}
\label{14-subsec:percentileComplexity}

We close our discussion of percentile queries with an overview of their complexity for various payoffs studied in~\Cref{3-chap:regular},~\Cref{5-chap:payoffs}, and~\Cref{6-chap:mdp}. We sum up the situation in~\Cref{14-table:percentile}. Here $\mathcal{F} = \{\Inf$, $\Sup$, $\LimInf$, $\LimSup\}$. Parameters $\markovProcess$ and $\mathcal{Q}$ respectively represent the size of the MDP, and the size of the query; P($x$), E($x$) and P$_{\mathsf{ps}}$($x$) respectively denote polynomial, expon\-en\-tial and pseudo-polynomial time in parameter $x$. For the shortest path, only non-neg\-at\-ive weights can be used, as otherwise the problem is undecidable.

\def\arraystretch{1.2}
\begin{table}[thb]
  \footnotesize
  \centering
  \begin{tabular}{|c||c|c|c|}
    \cline{2-4} \multicolumn{1}{c||}{} & \multirow{2}{*}{~Single-constraint~} & Single-dim. & ~Multidim.~ \\
    \multicolumn{1}{c||}{} & & ~Multi-constraint~ & ~Multi-constraint~\\
    \hline
    \hline
    $\Reach$ & \P & P($\markovProcess$)$\cdot$E($\mathcal{Q}$), \PSPACE-h & --- \\
    \hline
    \multirow{2}{*}{~$\scriptsize f \in \mathcal{F}$~} & \multirow{2}{*}{\P} & \multirow{2}{*}{\P} & ~P($\markovProcess$)$\cdot$E($\mathcal{Q}$)~ \\
    & & & \PSPACE-h.\\
    \hline
    ~$\MeanPayoff^{+}$ & ~\P~ & \P & \P\\
    \hline
    ~$\MeanPayoff^{-}$ & ~\P~ & ~P($\markovProcess$)$\cdot$E($\mathcal{Q}$)~ & ~P($\markovProcess$)$\cdot$E($\mathcal{Q}$)~\\
    \hline
    \multirow{2}{*}{~$\ShortestPath$~} &  ~P($\markovProcess$)$\cdot$P$_{\mathsf{ps}}$($\mathcal{Q}$)~ & ~P($\markovProcess$)$\cdot$P$_{\mathsf{ps}}$($\mathcal{Q}$) (one target)~ & ~P($\markovProcess$)$\cdot$E($\mathcal{Q}$)~\\
    & ~\PSPACE-h.~ & ~\PSPACE-h.~ & ~\PSPACE-h.~\\
    \hline
    \multirow{2}{*}{~$\varepsilon$-gap $\DiscountedPayoff$} & ~P$_{\mathsf{ps}}$($\markovProcess, \mathcal{Q}, \varepsilon$)~ & ~P$_{\mathsf{ps}}$($\markovProcess,\varepsilon$)$\cdot$E($\mathcal{Q}$)~ & ~P$_{\mathsf{ps}}$($\markovProcess,\varepsilon$)$\cdot$E($\mathcal{Q}$)~\\
    & \NP-h. & \NP-h. & \PSPACE-h.\\
    \hline
  \end{tabular}
  \caption{Complexity of percentile query problems for various payoffs.}
  \label{14-table:percentile}
\end{table}

Some of these results are quite technical to establish, so our goal here is only to highlight interesting elements with respect to everything that has been discussed in the previous chapters and in our own. First, the payoffs present in the left column have all been discussed before; the only oddity being the notion of $\varepsilon$-gap attached to the discounted payoff. Its presence is merely technical: we do not know if percentile queries using the discounted payoff can be solved exactly (and it is linked to long-standing open questions), but a slight relaxation of the problem, called 
$\varepsilon$-gap problem, can be solved. Intuitively, this relaxation consists in allowing an $\varepsilon$-wide uncertainty area around the value thresholds ($\alpha_i$) of the query. Second, some of the expressiveness of the queries is hidden in the table. For example, when using $\Reach$ or $\ShortestPath$, one may consider different target sets in each constraint. Similarly, when using $\DiscountedPayoff$, the discount factors may vary across constraints. Finally, when meaningful, the complexity is broken down into two parts, representing the relative dependency towards the size of the MDP, and the size of the query. The interest of this approach is that, in general and for practical applications, the model size is large whereas the query, encoding a specification, is comparatively much smaller. With that in mind, the polynomial dependency in the size of the MDP for most cases can be seen as good news.

Now, let us compare to what we know outside of percentile queries. Note that single-constraint queries correspond to the probability threshold problems in MDPs (\Cref{6-chap:mdp}). We see that in most cases, the jump to multiple dimensions induces an exponential blow-up (in the number of constraints). If we compare to two-player (non-stochastic) games, as studied in~\Cref{14-sec:multiple_dimensions},~\Cref{14-sec:mean_payoff_energy}, and~\Cref{14-sec:total_payoff_shortest_path}, we see that the undecidability of shortest path objectives holds up if we replace the antagonistic player Adam by stochasticity. On the complexity side, the situation varies from payoff to payoff with, again, an interesting lack of symmetry between the two variants of mean payoff, in stark contrast to the single-dimension case.

\section*{Bibliographic references}
\label{14-sec:references}
As discussed in the introduction, the literature on multi-objective models is too vast to provide a full account here. We therefore limit ourselves to some directions particularly relevant to our focus.

\paragraph{Multidimensional games.} Our presentation of mean-payoff games is inspired by Velner et al.~\cite{Velner.Chatterjee.ea:2015}. Brenguier and Raskin studied the Pareto frontiers of these games in~\cite{Brenguier.Raskin:2015}. While we considered \textit{conjunctions} of mean-payoff objectives, Velner proved that Boolean combinations lead to undecidability~\cite{Velner:2015}.

Energy games were discussed --- through the prism of vector games --- in~\Cref{13-chap:counters}. The link between energy games and mean-payoff games under finite memory was established in~\cite{Velner.Chatterjee.ea:2015}. While triple-exponential bounds on memory for Eve's strategies could be derived from~\cite{Brazdil.Jancar.ea:2010}, the first exponential upper bounds were proved by Chatterjee et al.~\cite{Chatterjee.Randour.ea:2014}, also encompassing conjunctions with \textit{parity} objectives. These bounds have since been refined~\cite{Jurdzinski.Lazic.ea:2015} but remain exponential; indeed, it is known that exponential memory is necessary for Eve~\cite{Chatterjee.Randour.ea:2014}.
 
The undecidability of total payoff games was first established by Chatterjee et al. in~\cite{Chatterjee.Doyen.ea:2015} via reduction from the halting problem for two-counter machines: we provided here a new, simpler proof based on robot games~\cite{Niskanen.Potapov.ea:2016}. This undecidability result, along with the complexity barriers of mean-payoff and total payoff games, motivated the introduction of (multidimensional) \textit{window objectives}: conservative variants of mean-payoff and total payoff objectives that benefit from increased tractability and permit to reason about time bounds~\cite{Chatterjee.Doyen.ea:2015}. They have since been studied in a variety of contexts: variants of "parity" objectives~\cite{Bruyere.Hautem.ea:2016}, Markov decision processes~\cite{Brihaye.Delgrange.ea:2020}, timed games~\cite{Main.Randour.ea:2021}, etc.

The undecidability of shortest path games is formally new, but the result was already established for MDPs by Randour et al.~in~\cite{Randour.Raskin.ea:2017}. Here, we use the aforementioned new approach based on robot games.

Consumption games were studied by Br{\'{a}}zdil et al.~\cite{Brazdil.Chatterjee.ea:2012}: they have the flavour of energy games but are actually incomparable. In such games, only negative weights are used, and gaining energy can only be done through particular `reload edges' that refill the energy up to a predefined capacity.

Parts of our presentation of multidimensional games are inspired by~\cite{Randour:2014}.

\paragraph{Combinations of different objectives.} We focused on multidimensional games obtained by conjunction of \textit{identical} objectives. Conjunctions of \textit{heterogeneous} objectives have been studied in numerous different contexts, including mean-payoff parity games~\cite{Chatterjee.Henzinger.ea:2005}, energy parity games~\cite{Chatterjee.Doyen:2012}, average-energy games with energy constraints~\cite{Bouyer.Markey.ea:2018}, energy mean-payoff games~\cite{Bruyere.Hautem.ea:2019}, or Boolean combinations of simple quantitative objectives~\cite{Bruyere.Hautem.ea:2016}. While similarities often exist across these different settings, ad hoc techniques are still needed and complexity results may vary greatly. Developing a general framework encompassing large classes of multi-objective problems is still a vastly unexplored challenge. Some progress has been achieved recently with a focus on strategy complexity; we discuss it further below.

\paragraph{Beyond worst-case synthesis.} Our presentation is mostly inspired by Bruy\`ere et al.~in~\cite{Bruyere.Filiot.ea:2017}, which introduced the BWC synthesis problem, and where all technical details can be found. As noted in~\cite{Bruyere.Filiot.ea:2017}, allowing large inequalities in the BWC problem may require infinite-memory strategies. The case of infinite-memory strategies was studied in~\cite{Clemente.Raskin:2015} along with multidimensional BWC mean-payoff problems.

BWC problems were studied for other objectives, such as shortest path~\cite{Bruyere.Filiot.ea:2017} or parity~\cite{Berthon.Randour.ea:2017}; and on other related models (\textit{e.g.},~\cite{Brazdil.Kucera.ea:2016}). BWC principles have been implemented in the tool \textsc{Uppaal}~\cite{David.Jensen.ea:2014}. Boolean combinations of objectives akin to the BWC problem have been considered in MDPs~\cite{Berthon.Guha.ea:2020}.

The BWC framework aims to provide strategies that exhibit high performance in normal operating conditions while offering a high-level of resilience in extreme conditions. A kindred --- but softer --- approach is the study of strategies in MDPs that achieve a trade-off between the expectation and the variance over the outcomes, giving a statistical measure of the stability of the performance. Br{\'{a}}zdil et al.~have considered the mean payoff with this philosophy in~\cite{Brazdil.Chatterjee.ea:2017}.

\paragraph{Percentile queries.} The framework of percentile queries was introduced by Randour et al.~in~\cite{Randour.Raskin.ea:2017}, where they studied a variety of payoffs: all results mentioned in this chapter are from~\cite{Randour.Raskin.ea:2017}. As mentioned in~\Cref{14-subsec:percentileComplexity}, the percentile query problem was established to be undecidable for the shortest-path payoff when both positive and negative weights are allowed. The theory of percentile queries can be seen as a quantitative follow-up to the work of Etessami et al.~on reachability objectives, in~\cite{Etessami.Kwiatkowska.ea:2008}.

Several other problems have been considered on multidimensional MDPs. For example, in~\cite{Brazdil.Brozek.ea:2014}, Br{\'{a}}zdil et al.~study two different problems based on the mean payoff. On the one hand, they consider the optimisation of the expected value vector. On the other hand, they show how to optimise the probability that the payoff vector is above a threshold vector. Observe that, in comparison to percentile queries, the latter problem asks for a bound on the joint probability of the thresholds, that is, the probability of satisfying all constraints simultaneously. In contrast, percentile queries bound the marginal probabilities separately, which may allow for more modeling flexibility. Another complementary approach was considered by Haase and Kiefer in~\cite{Haase.Kiefer:2015}: whereas percentile queries allow for conjunctions between probability constraints on simple value inequalities, they consider only one probability constraint but allow for conjunctions of value constraints within this single probability constraint. Hence, both frameworks are incomparable.

Various frameworks mixing concepts from beyond worst-case synthesis and percentile queries have been developed in the recent years, both in exact and formal approaches (\textit{e.g.},~\cite{Bouyer.Gonzalez.ea:2018}), and in reinforcement learning (\textit{e.g.},~\cite{Kretnsky.Perez.ea:2018}). Comparisons between percentile queries, beyond worst-case synthesis, and other rich behavioural models can be found in~\cite{Randour.Raskin.ea:2014}.

\paragraph{Pareto frontiers.} As hinted throughout this chapter, the first step in understanding multi-objective settings is often to fix acceptable performance thresholds and to focus on the corresponding decision problem, asking if there exists a strategy to ensure these thresholds. Yet, to fully embrace the complexity of multi-objective frameworks, to be able to visualise the interplay between different (qualitative and quantitative) aspects and the corresponding trade-offs, we have to look at Pareto frontiers. This endeavour is generally difficult and one requires specific techniques to provide efficient approximations of Pareto frontiers --- due to their complexity, exact computation is often out of reach.

We already mentioned~\cite{Brenguier.Raskin:2015}, which deals with Pareto frontiers in multidimensional mean-payoff games. A seminal approach for MDPs was developed by Forejt et al.~in~\cite{Forejt.Kwiatkowska.ea:2012}. In a nutshell, it consists in obtaining successive approximations of the Pareto frontier by solving many one-dimension problems. Each of these is obtained by reinterpreting the original $k$-objective problem as a weighted sum of each individual objective. Having the weight assignment vary yields different one-dimension problems, but also permits to approximate the real Pareto frontier from a different angle, as each weight assignment morally corresponds to looking in a different direction within the $k$-dimension space. The crux is then to explore this space efficiently by a clever choice of weights at each iteration.

An excellent entry point to Pareto frontiers in MDPs --- and the rich theory of multi-objective MDPs in general --- is Quatmann's PhD thesis~\cite{Quatmann:2023}. A systematic study of the structure of Pareto frontiers  (and their underlying payoff sets) in MDPs for general classes of payoffs was recently provided by Main and Randour~\cite{Main.Randour:2025}. Interestingly, they notably show that \textit{mixing} (\textit{i.e.}, through a single toss coin at the beginning of a play, as in \Cref{14-ex:continuousPareto}) a finite number of pure strategies is sufficient to approximate any point of the Pareto frontier in most cases. That is, the simplest form of randomness (see next paragraph) suffices.

\paragraph{Complex strategies.} The additional expressive power of multi-objective games and MDPs comes at a cost in terms of algorithmic complexity, but also with regard to the classes of strategies that are necessary to play (Pareto-)optimally. We have seen various examples in this chapter where Eve had to resort to strategies with (finite or infinite) memory and/or randomness. In a similar spirit to the characterisations of positionally determined objectives discussed in the early chapters of this book, recent research has striven to characterise the need for complex strategies in broad classes of multi-objective settings.

Recently, there has been a lot of progress on understanding the power of finite-memory strategies, both in games~\cite{Bouyer.Roux.ea:2022} and in MDPs~\cite{Bouyer.Oualhadj.ea:2023}. An overview by Bouyer et al.~is given in~\cite{Bouyer.Randour.ea:2022}. \Cref{4-sec:memory_generalisations} presents some results in this direction. With a particular focus on multi-objective games, Le Roux et al.~studied general conditions under which finite-memory determinacy can be maintained through combinations of finite-memory determined objectives~\cite{Roux.Pauly.ea:2018}. A survey by Brihaye et al.~\cite{Brihaye.Goeminne.ea:2023} illustrates how the game model may influence strategy complexity, focusing on reachability-flavoured objectives.

With regard to randomness --- which we have seen to be necessary in the most general multi-objective settings, it is interesting to see that not all forms of randomness were created equal: when considering finite-memory (\Cref{1-sec:memory}) randomised strategies, one can put randomness in different parts of the memory structure (initial state, update function, and/or `next-action' function that exploits the memory to take a decision), with differences in terms of resulting power~\cite{Main.Randour:2024}. This is in contrast to the celebrated Kuhn's theorem in classical game theory, which crucially relies on the availability of infinite memory.

As complex strategies are often prohibitive in practical applications, it is sometimes interesting to consider multi-objective problems where one looks for strategies of limited complexity. That is, even if the problem requires complex strategies for Pareto-optimality, one may be interested in how good simple strategies can be. For example,~\cite{Delgrange.Katoen.ea:2020} develops techniques to explore the Pareto frontier of multi-objective MDPs under the constraint that only pure (\textit{i.e.}, without randomness) and limited-memory strategies should be considered. As seen in~\Cref{14-ex:exponentialPareto}, such constraints often break the nice structure of (unrestricted) Pareto frontiers, which renders their exploration (and representation) more difficult.

\paragraph{Alternative strategy models.} Finally, let us mention that the rise in strategy complexity that we witness in multi-objective settings is a strong motivation to consider and study alternatives to the classical \textit{Mealy machines}, prevalent in the literature --- and present as memory structures throughout this book. Indeed, different strategy models may lead to more concise representations or better understandability: a simple example being the explicit use of counters as primitives rather than their `flattening' within the memory structure. There has been a surge in research in this direction in the recent years, notably through models such as decision trees~\cite{Brazdil.Chatterjee.ea:2018}, programmatic strategies~\cite{Shabadi.Fijalkow.ea:2025}, or counter-based strategies~\cite{Ajdarów.Main.ea:2025}.

\section*{Acknowledgments}
Mickael Randour is an F.R.S.-FNRS Senior Research Associate and a member of the TRAIL Institute. His work on this chapter has been supported by the Fonds de la Recherche Scientifique – FNRS under Grants n° F.4520.18 (MIS ManySynth) and n° T.0188.23 (PDR ControlleRS).

\ifpictures
\includepdf{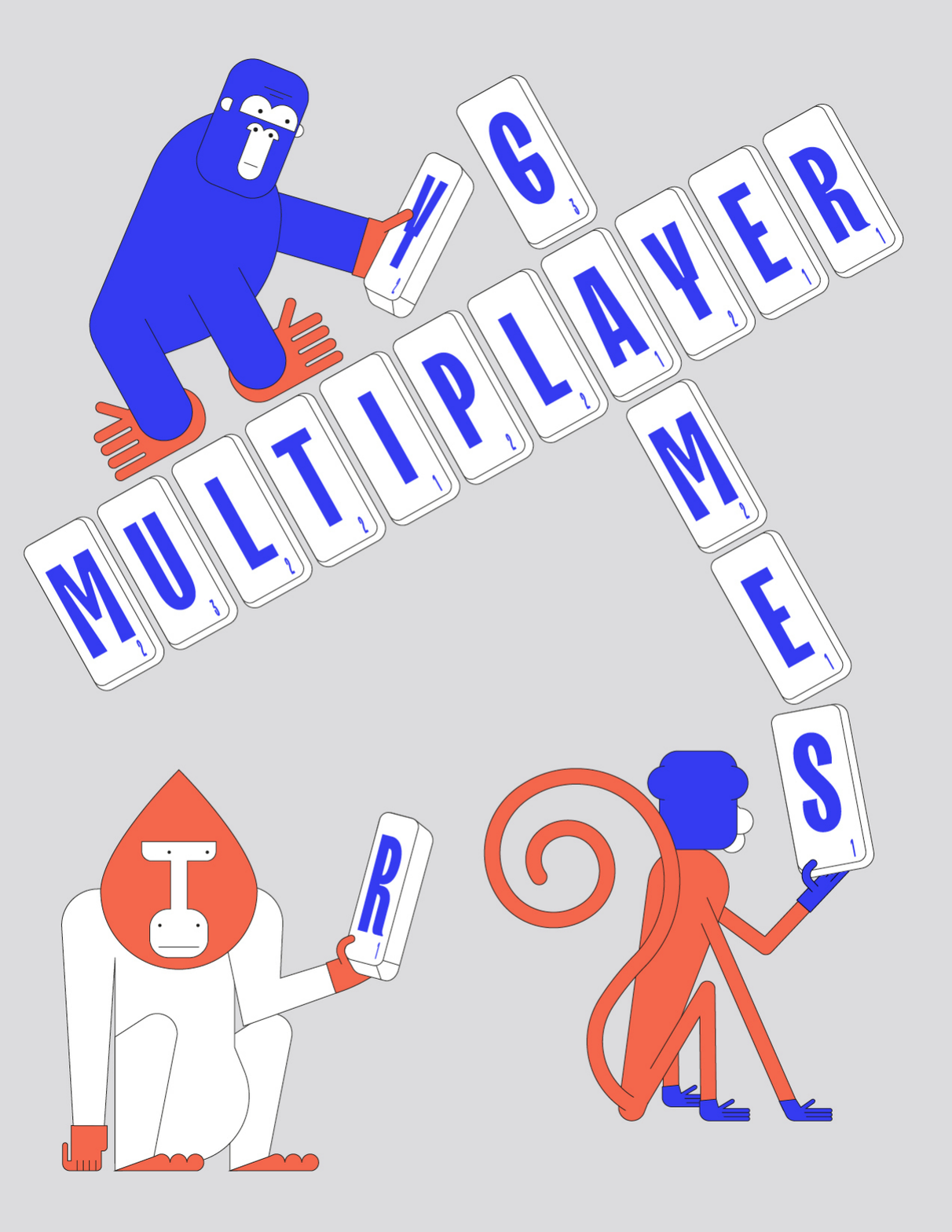}
\fi
\author[Romain Brenguier, Ocan Sankur]{Romain Brenguier, Ocan Sankur}
\copyrightline{Copyright by Romain Brenguier and Ocan Sankur 2025, to be published by Cambridge University Press in the volume \textit{Games on Graphs} edited by Nathana\"el Fijalkow}

\chapter{Multiplayer Games}
\chapterauthor{Romain Brenguier, Ocan Sankur}
\label{15-chap:multiplayer}

\def\payoff{\ensuremath{f}}
\def\Act{A}
\def\Agt{\mathcal{P}}
\def\move{\textsf{move}}
\def\Out{\textsf{Out}}
\def\Dev{\textsf{Dev}}
\def\maxinf{\text{\rm maxinf}}
\def\pes{\textsf{pes}}
\def\opt{\textsf{opt}}
\def\proj{\textsf{proj}}
\def\devg{\textsf{DevGame}}
\def\Coalition{\ensuremath{\mathcal{C}}}

In two-player games seen so far, players had objectives that are
opposite to each other's, so we were able to define them giving only
Eve's objective. Adam was seen as a purely adversarial player. Such games
are called \emph{zero-sum} games since, in a quantitative setting, the
sum of the payoffs of the two players would sum up to zero in any
outcome. However, the objectives of the players are not entirely
conflicting in all games.
In particular, in multiplayer games, that is,
games with more than two players, the binary view of zero-sum games
does not make sense;
but there are also interesting examples of non-zero-sum games with only two
players (we will see one below). In this setting, winning strategies are
no longer suitable to describe rational behaviours since the opponents
should no longer be seen as purely adversarial. In fact, when the
objectives of the players are not opposite, some cooperation becomes
possible. Then, rather than assuming that opponents are purely
adversarial, it is interesting to study the possible outcomes when they
are simply \emph{rational}, that is, follow the best strategy for their
own objectives. The notion of equilibria we will study in this chapter
aims at describing such rational behaviours.

If one is expecting for sure some specific strategies to be played by the
opponents, then the most rational response is to choose the
\emph{best response}, that is, the strategy that is optimal for the
player against the given strategies of the other players. Thus, if we
assign strategies to players, and if the players are all aware of the
strategies of the other players, then each player will be willing to
change their strategy if theirs does not turn out to be a best response.
Such a situation is seen as unstable and is undesirable in many
applications of game theory. \emph{Nash equilibrium} is defined simply
as a stable situation in such a setting: a strategy profile in which the
strategy of each player is a best response to the rest of the
strategies. Thus, no player has any incentive to change their strategy.

We will see the formal definition of a Nash equilibrium in the next
section. Let us first consider the following example.

The following Hawk-Dove game was first presented by the biologists Smith
and Price, and shown in~\Cref{15-tbl:hawk-dove}.
Here, two animals are fighting for resources and can choose to
either act as a hawk or as a dove.
If both player choose hawk they will have to fight for resources, and
thus only get payoff 0. If only one chooses hawk, they get a high payoff of
4, because they get all the valuable resources for themselves, while the dove
gets 1: they get plenty of resources but gets hunted. When they both choose
dove, they both get a payoff of 3: they have to share resources but do not get
hunted.

When a player chooses hawk then the best payoff for the opponent is
obtained by choosing dove, so as to avoid fighting for resources.
So, dove is the best response to hawk. Reciprocally, the best response to
dove is to play hawk. There are two `equilibria': (Hawk, Dove) and
(Dove, Hawk), where no player has an interest in changing their
strategy. Note that the highest payoff a player can ensure
(against all adversary strategies) is only~$1$.

Nash showed the existence of such equilibria in normal-form games\footnote{normal-form games are also called matrix games, see~\Cref{8-chap:concurrent}.}, which may require randomized strategies. This result
revolutionized the field of economics, where it is used to analyze
competitions between firms or government economic policies for example.
Game theory and the concept of Nash equilibrium are now applied to
diverse fields: in finance to analyse the evolution of market prices, in
biology to understand the evolution of some species, in political
sciences to explain public choices made by parties.

\begin{table}
  \caption{The Hawk-Dove game. Each column corresponds to a strategy of
    \(P_1\) and each line to a strategy of \(P_2\).}
  \label{15-tbl:hawk-dove}
  \begin{center}
    \begin{tabular}[c]{|@{~}l@{~}|@{~}c@{~} @{~}c@{~}|}
      \hline
      & Hawk & Dove \\
      \hline
      Hawk & 0 , 0 & 1 , 4 \\
      Dove & 4 , 1 & 3 , 3 \\
      \hline
    \end{tabular}
  \end{center}
\end{table}

\medskip
In this chapter, we will first study the computation of Nash
equilibria in multiplayer concurrent games with $\omega$-regular
objectives. The algorithms we present here differ from those that were
given for normal-form games since ours are infinite-duration with
$\omega$-regular objectives. We will then present extensions of this notion
such as secure and robust equilibria. The second result we develop is the notion of admissibility: this is a different approach to the study
of rational behaviours and consists in eliminating for each player irrational choices of strategies.

\section{Nash Equilibria for games in normal form}
\label{15-sec:nash_equilibria_normal_form}
The normal form games we consider differ from the matrix games of \Cref{8-chap:concurrent}, in that each player has their own payoff.
So for instance, when player 1 chooses column Hawk, and player 2 chooses
row Dove, the payoff for player 1 is $\payoff_{P_1}(\text{Hawk}, \text{Dove}) = 4$.
Let us call a vector of strategies specifying a strategy for each player a \emph{strategy profile}. In normal-form games, each cell of the table~$\Delta$ corresponds to a strategy profile.

\begin{definition}
  A \emph{Nash equilibrium} is a \emph{stable} strategy profile in which
  strategy is a best response against the other strategies.
\end{definition}
Thus a Nash equilibrium is a stable situation in the sense that
no player has an incentive in changing their strategy.
Nash proved that when
players are allowed to randomise among all their strategies, there always
exists a Nash equilibrium.

\begin{theorem}[Existence of Nash equilibria]
In every normal-form game with a finite number of
players, each having a finite number of pure strategies, there exists a
randomised Nash equilibrium.
\end{theorem}

Note that not all games contain pure Nash equilibria.
For example, in the rock-paper-scissors game, the best response to rock
is paper, to paper is scissors, and to scissors is rock, so none of these
pure strategies can be an equilibrium.

For finding a pure Nash equilibrium in a normal-form game, there is a simple
polynomial-time algorithm.
For each strategy profile, we look for each player whether they have a better
response than their current strategy.
If no player has a better response, the strategy profile is a Nash equilibrium,
otherwise we move to the next one, and if none satisfies the condition then there is no equilibrium.


\begin{example}[Medium Access Control]
Consider a medium access control
problem, where several users share access to a wireless channel. A
communication over the channel is successful if there are no collisions,
that is, if a single user is transmitting their message only. During each
slot, each user chooses either to transmit or to idle. Intuitively, the
number of packets transmitted without collision decreases with
the number of users emitting in the same slot. Furthermore each attempt
at transmitting has a cost. An example payoff for two players,
is represented in \Cref{15-ex:medium-access}.

\begin{table}
  \caption{A game of medium access.}
  \label{15-ex:medium-access}
  \begin{center}
    \begin{tabular}[c]{|@{\hspace{1em}}l@{\hspace{1em}}|@{\hspace{1em}}c@{\hspace{1em}}c@{\hspace{1em}}|}
      \hline
      & Emit & Wait\\
      \hline
      Emit & -1, -1 & 2, 0\\
      Wait & 0 , 2 & 0, 0\\
      \hline
    \end{tabular}
  \end{center}
\end{table}
\end{example}
We encourage the reader to find the Nash equilibria of the above game.

The game described above corresponds to a single slot of this system. In
a practical scenario, there would be a succession of slots and the payoff would be
the sum of payoffs over all slots. Normal-form games are thus not
sufficient to represent games with repetitions and to study the
evolution of the behaviours as the game evolves.

One possibility to model repetition is to use
\emph{games in extensive form} which are games played on finite trees.
However such games only model a fixed number of repetitions unlike
infinite or arbitrary duration games as studied in this book. We thus
study, in the rest of this chapter, algorithms for games played on
graphs.

\subsection{Definitions}
\label{15-subsec:definitions}

\begin{definition}
  A multiplayer \emph{arena} \(\mathcal{A}\) with~$k$ players is a tuple
  \(\langle V, \Act,\Delta,(c_P)_{P\in \Agt} \rangle \), where:

  \begin{itemize}

  \item     \(V\) is a finite set of vertices;
  \item     \(\Agt = \{1,2,\ldots,k\}\) is the set of players;
  \item     \(\Act\) is a finite set of actions, a tuple \((a_P)_{P \in \Agt}\)
    containing one action \(a_P\) for each player $A$ is called a
    \emph{move}, thus \(\Act^k\) is the set of possible moves;
  \item     \(\Delta : V \times \Act^k \to V\) is the transition function which
    associates to a pair of vertices and moves the resulting state;
  \item     \((c_P)_{P \in \Agt}\) is a tuple of colouring functions
    with~$c_P : V \rightarrow C$ for each~$P \in \Agt$.
  \end{itemize}
\end{definition}

\begin{example}
A simple three-player concurrent game is represented in \Cref{15-fig:example1}.
Vertices are $v_0$, $v_1$, $v_2$, $v_3$ and $v_4$.
Players are named $P_1$, $P_2$, $P_3$.
The set of actions is $\Act = \{ a , b\}$.
The transition relation is given by the edges in the graph, for instance
$\Delta(s_0, (a, b, a))$ is $v_1$. In our figures, $\ast$ represents
an arbitrary action.
The colouring function is represented below vertices as tuples ranging over players.
For instance, a vertex labelled by $(1,1,0)$ assigns
the first two players the colour~$1$, and the third player the colour~$0$,
In particular, $c_{P_1}(v_2) = 1$, and $c_{P_3}(v_2)= 0$.
\end{example}

\begin{figure}
  \begin{center}
    \begin{tikzpicture}
      \draw (0,0) node[draw, inner sep=7pt] (I) {$v_0$};
      \draw (I.-110) node[below] {$1, 1, 1$};
      \draw (9,1) node[draw, inner sep=7pt] (C1) {$v_1$};
      \draw (C1.-100) node[below] {$1, 1, 0$};
      \draw (9,-1) node[draw, inner sep=7pt] (C2) {$v_2$};
      \draw (C2.-90) node[below] {$1, 1, 0$};
      \draw (5, 2) node[draw, inner sep=7pt] (V3) {$v_3$};
      \draw (V3.-90) node[below] {$0, 1, 1$};
      \draw (5,-2) node[draw, inner sep=7pt] (V4) {$v_4$};
      \draw (V4.-90) node[below] {$1, 0, 1$};
      \draw[-latex'] (-1, 0) -- (I);
      \draw[-latex', rounded corners=5mm] (I) -- +(2, 0.8) -- node[pos=0.2, above]{$a, b, \ast$} node[pos=0.2,below]{$b, a, \ast$} (C1);
      \draw[-latex',rounded corners=5mm] (I) -- +(2,-0.8) -- node[pos=0.2, above]{$a, a, \ast$} node[pos=0.2,below]{$b, b, \ast$} (C2);
      \draw[-latex', rounded corners] (C1) -- +(-1,-1) -- node[left]{$\ast, \ast, a$} (C2);
      \draw[-latex'] (C1) -- node[above]{$\ast, \ast, b$} (V3);
      \draw[-latex'] (C2) -- node[below]{$\ast, b, \ast$} (V4);
      \draw[-latex', rounded corners] (C2) -- +(1,1) -- node[right]{$\ast, a, \ast$} (C1);
      \draw[-latex', rounded corners=5mm] (V3) -| node[above, pos=0.2]{$\ast, \ast, \ast$} (I.50);
      \draw[-latex', rounded corners=5mm] (V4) -| node[below, pos=0.2]{$\ast, \ast, \ast$} (I.-50);
    \end{tikzpicture}
  \end{center}
    \caption{Example of a three-player concurrent arena. The symbol $\ast$ on edges can be replaced by either $a$ or $b$.}
    \label{15-fig:example1}
\commentAlt{Figure~\ref{15-fig:example1}: A directed graph with five square nodes (v0 to v4) showing complex transitions with multiple labels on the edges and numerical labels inside the nodes. See long description.}
\commentLongAlt{Figure~\ref{15-fig:example1}: The image displays a directed graph with five square nodes, labeled v0, v1, v2, v3, and v4. An incoming arrow points to node v0, indicating it as a starting point.

Node v0 (left): Contains the label '1,1,1'.
- An arrow from v0 points to v1, labeled 'a,b,*'.
- An arrow from v0 points to v2, labeled 'a,a,*'.
- A curved arrow from v0 points to v3, labeled '*,*,*'.
- A curved arrow from v0 points to v4, labeled '*,*,*'.

Node v1 (right, top): Contains the label '1,1,0'.
- An arrow from v3 points to v1, labeled '*,*,b'.
- A bidirectional arrow connects v1 and v2. The arrow from v1 to v2 is labeled '*,*,a'.

Node v2 (right, bottom): Contains the label '1,1,0'.
- An arrow from v4 points to v2, labeled '*,b,*'.
- The bidirectional arrow from v1 points to v2, labeled '*,*,a'.

Node v3 (top-center): Contains the label '0,1,1'.
- An arrow from v3 points to v1, labeled '*,*,b'.

Node v4 (bottom-center): Contains the label '1,0,1'.
- An arrow from v4 points to v2, labeled '*,b,*'.

Additionally, there are reversed labels on some edges indicating bidirectional or alternative transitions:
- Between v0 and v1: 'b,a,*'
- Between v0 and v2: 'b,b,*'}
\end{figure}

A \emph{history} of the multiplayer arena
\({\mathcal A}\) is a finite sequence of states and moves ending with a
state, \textit{i.e.}~a word in \((V \cdot \Act^\Agt)^* \cdot V\). Note that unlike
for two-player games we include actions in the history, because knowing
the source and target vertices does not mean you know which player chose
which actions.

For a history~$\pi$, we write \(\pi_i\) the $i$-th vertex of $\pi$, starting from $0$, and
\(\move_i(\pi)\) its $i$-th move, thus
\(\pi = \pi_0 \cdot \move_0(\pi \cdot \pi_1 \cdots \move_{n-1}(\pi)\cdot \pi_n\), and
with this notation $\move_i(\pi)_P$ is the $i$-th action of player $P$ in $h$.
The length $|\pi|$ of such a history is $n + 1$. We write
$\last(\pi)$ the last vertex of h, \textit{i.e.} \(\pi_{|\pi|-1}\).
A play \(\rho\) is an
infinite sequence of vertices and moves, \textit{i.e.}~an element of
\((V \cdot \Act^{\Agt})^\omega\).

\begin{definition}[Strategy and coalition]
  A strategy is a function which associates an action to each history.
  We often write $\sigma_A$ for a strategy of player $P$.
  A coalition $\Coalition$ is a set of players in $\Agt$, and we write
  $-\Coalition$ for the remaining players, that is $-\Coalition = Agt \setminus \Coalition$.
  Let \(\Coalition\) be a coalition, a strategy \(\sigma_\Coalition\) for \(\Coalition\) is a function
  which associates a strategy \(\sigma_P\) to each player \(P\in \Coalition\).
  Given a strategy $\sigma_\Coalition$, when it is clear from the context, we simply
  write \(\sigma_P\) for \(\sigma_\Coalition(P)\).
\end{definition}

\begin{definition}[Outcomes]
  A history \(\pi\) is compatible
  with the strategy \(\sigma_\Coalition\) for coalition~$\Coalition$ if, for all \(k < |\pi| - 1\) and all
  \(P \in \Coalition\), we have 
  \((\move_k(\pi))_P = \sigma_P(\pi{\le k})\), and \(\Delta(\pi_k, \move_k(\pi)) = \pi_{k+1}\). 
  A play \(\rho\) is compatible with the strategy \(\sigma_\Coalition\) if all its prefixes are. We write
  \(\Out_{\mathcal{A}}(v_0, \sigma_\Coalition)\) for the set of plays in \(\mathcal{A}\) that
  are compatible with strategy \(\sigma_\Coalition\) and have initial vertex
  \(v_0\). Let \(\Out_{\mathcal{A}}(\sigma_\Coalition)\) denote the union
  of \(\Out_{\mathcal{A}}(v_0, \sigma_\Coalition)\) for all~$v_0$,
  and \(\Out_{\mathcal{A}}(v_0)\) the union of all \(\Out_{\mathcal{A}}(v_0, \sigma_\Coalition)\).
  The subscript~$\mathcal{A}$ can be omitted if it is clear from the context.
  These paths are called \emph{outcomes} of \(\sigma_\Coalition\) from
  \(v_0\). 
\end{definition}
Note that
when the coalition \(\Coalition\) is composed of all the players
the outcome from a given state is unique.

\begin{example}
  Consider in the example of \Cref{15-fig:example1}, the following
  strategies:
  \begin{itemize}
  \item $P_1$ always plays $a$, \textit{i.e.} $\sigma_{P_1}(\pi) = a$ for all histories $\pi$;
  \item $P_2$ plays $a$ in $v_0$ if it is the first state and then always plays $b$, \textit{i.e.} $\sigma_{P_2}(v_0) = a$ and $\sigma_{P_2}(\pi) = b$ for all $\pi \ne v_0$;
  \item $P_3$ always plays $b$, \textit{i.e.} $\sigma_{P_3}(\pi) = b$.
  \end{itemize}
  The outcome from $v_0$ in that case is
  \begin{align*}
    \Out(v_0, \sigma_{\{P_1, P_2, P_3\}}) ~ = & ~ v_0 \cdot (a, a, b) \cdot v_2 \cdot (a, b, b)
                                                \cdot v_4 \cdot (a, b, b) \cdot\\
                                              & \left(v_0 \cdot (a, b, b)
                                                \cdot v_1 \cdot (a, b, b)
                                                \cdot v_3 \cdot (a, b, b)\right)^\omega
  \end{align*}
\end{example}

\begin{definition}[Multiplayer game]
  A \emph{payoff} function associates a real number to each outcome.
  We will be mostly be interested in solving games with qualitative
  objectives, that is payoffs that take values $0$ and $1$.
  A \emph{multiplayer game} \((\mathcal{A}, (\payoff_P)_{P \in \Agt})\) is given by a multiplayer arena $\mathcal{A}$, an initial
  vertex $v_0$ and payoff function $\payoff_P$ for each player~$P$.
  When $\payoff_P$ is qualitative we simply write $\Omega_P$
  for the corresponding objective.
\end{definition}

\subsection{The Nash equilibrium problem}
\label{15-subsec:algorithm-for-finding-nash-equilibria}

In this section we will present an algorithm to compute
Nash equilibria in
multiplayer games.
The problem we are interested in is to decide the existence of a Nash
equilibrium in which the objectives of a given set of players are
satisfied.

\decpb[Existence of a constrained Nash equilibrium]%
{A multiplayer game $(\mathcal{A}, (\payoff_P)_{P \in \Agt})$, payoff bounds $(b_P)_{P\in\ \Agt}$, and an initial vertex~$v_0$}%
{Does there exist a Nash equilibrium $\sigma_{\Agt}$ such that for all
$P \in \Agt$, we have $\payoff_P(\Out({v_0,\sigma_{\Agt}})) \ge b_P$?}

The algorithm is based on a reduction to zero-sum two-players games,
which allows us to use algorithms presented in the previous chapters of
this book. More precisely, we present the \emph{deviator game}, which is
a transformation of a concurrent multiplayer game into a turn-based
zero-sum game, such that there are strong links between equilibria in
the former and winning strategies in the latter. The proofs of
this section are independent of the type of objectives we consider.

\subsection{Deviators}
\label{15-subsection:deviators}

A central notion we use is that of \emph{deviators}. These are the
players who have played different moves from those prescribed in a given
profile, thus causing a deviation from the expected outcome. Formally, a
deviator from move
\(a_{\Agt}\) to \(a'_{\Agt}\) is a player
\(D \in \Agt\) such that \(a_D \ne a'_D\) . We denote the set of
deviators by \[
        \Dev(a_{\Agt} , a'_{\Agt} ) = \{D \in \Agt \mid a_D \ne a'_D \}.
\] We extend the definition to pairs of histories and strategies by
taking the union of deviator sets of each step along the history.
Formally,
\[
\Dev(\pi, \sigma_{\Agt}) = \bigcup_{0\le i < |h|}~ \Dev(\move_i(\pi), \sigma_{\Agt}(\pi_{\le i})).
\]
For an infinite play \(\rho\), we define
\(\Dev(\rho, \sigma_{\Agt} ) = \bigcup_{i \in \mathbb{N}} \Dev(\move_i(\rho), \sigma_{\Agt}(\rho_{\le i} ))\).
Intuitively, having chosen a strategy profile \(\sigma_{\Agt}\) and
observed a play \(\rho\), deviators represent the players that must have
changed their strategies from \(\sigma_{\Agt}\) in order to generate
\(\rho\).

\begin{lemma}\label{15-lem:deviator}
Given a play \(\rho\), strategy profile~$\sigma_\Agt$, a coalition \(\Coalition\)
contains \(\Dev(\rho)\), if and only if, there exists a strategy
\(\sigma'_\Coalition\) such that \(\Out(\rho_1, \sigma_{-\Coalition}, \sigma'_\Coalition) = \rho\).
\end{lemma}
\begin{proof}
Assume that coalition \(\Coalition\) contains
\(\Dev(\rho, \sigma_{\Agt})\). We define the strategy \(\sigma_\Coalition\) to be
such that for all \(i\in \mathbb{N}\),
\(\sigma_\Coalition(\rho_{\le i} ) = (\move_i(\rho))_\Coalition\). By hypothesis, we have,
for all indices \(i\),
\(\Dev(\move_i(\rho), \sigma_{\Agt}(\rho_{\leq i})) \subseteq \Coalition\), so for
all players \(A\not\in \Coalition\),
\(\sigma_A(\rho_{\le i}) = (\move_i(\rho))_A\). Then
\(\Delta(\rho_i, \sigma'_\Coalition(\rho_{\le i}), \sigma_{-\Coalition}(\rho_{\le i})) = \rho_{i+1}\).
Hence \(\rho\) is the outcome of the profile
\((\sigma_{-\Coalition}, \sigma'_\Coalition)\).

For the other direction, let \(\sigma_{\Agt}\) be a strategy profile,
\(\sigma'_\Coalition\) a strategy for coalition \(\Coalition\), and
\(\rho \in Out_G(\rho_0 , \sigma_{-\Coalition}, \sigma'_\Coalition)\). 
For all indices \(i\), 
\(\move_i(\rho) = (\sigma_{-\Coalition}(\rho_{\le i}), \sigma'_\Coalition(\rho_{\le i}))\).
Therefore for all players \(A \not\in \Coalition\),
\((\move_i(\rho))_A = \sigma_A(\rho_{\le i})\). Consequently, this implies that
\(\Dev(\move_i(\rho), \sigma_{\Agt}(\rho_{\le i})) \subseteq \Coalition\). Hence
\(\Dev(\rho, \sigma_{\Agt}) \subseteq \Coalition\).
\end{proof}

\begin{example}
  In the example of \Cref{15-fig:example1}, we consider again the strategies,
  such that for all histories $\pi$, $\sigma_{P_1}(\pi) = a$,
  $\sigma_{P_2}(v_0) = a$ and if $\pi \ne v_0$, $\sigma_{P_2}(\pi) = b$,
  and $\sigma_{P_3}(\pi) = b$.
  Then $\Dev(v_0 \cdot (a, a, b) \cdot v_2 \cdot (a, a, b) \cdot v_1 \cdot
  (a, b, a) \cdot v_2, \sigma_{\Agt})$ is the union of:
  \begin{itemize}
  \item $\Dev(\sigma_{\Agt}(v_0), (a, a, b)) = \Dev((a, a, b), (a, a, b)) = \varnothing$
  \item $\Dev(\sigma_{\Agt}(v_0 \cdot (a, a, b) \cdot v_2), (a, a, b)) =
    \Dev( (a, b, b), (a, a, b)) = \{P_2\}$
  \item $\Dev(\sigma_{\Agt}(v_0 \cdot (a, a, b) \cdot v_2 \cdot (a, a, b) \cdot v_1), (a, b, a)) =
    \Dev( (a, b, b), (a, b, a)) = \{P_3\}$.
  \end{itemize}
  We obtain
  $\Dev(v_0 \cdot (a, a, b) \cdot v_2 \cdot (a, a, b) \cdot v_1 \cdot
  (a, b, a) \cdot v_2, \sigma_{\Agt}) = \{ P_2, P_3\}$.
  This means that both $P_2$ and $P_3$ need to change their strategies
  from $\sigma_{\Agt}$ to obtain the given history.
\end{example}

Note that Nash equilibria are defined only with respect to deviations by
single players, that is, we require all players to achieve worse or equal
payoffs than the prescribed profile when they single-handedly change
strategies. Thus, only the outcomes with singleton deviator sets
will be of interest for us in the next section where we present the algorithm.


\subsection{Deviator game}
\label{15-subsec:deviator-game}
We now present an algorithm to reduce multiplayer games to two-player games
using the notion of deviators we just defined.
Given a \emph{game}
\(\mathcal{G} = (\mathcal{A}, (\payoff_A)_{A \in \Agt})\),
we define the deviator game, denoted \(\devg(\mathcal{G})\).
Intuitively, in this game, Eve needs to play
according to an equilibrium, while Adam tries to find a profitable
deviation for any player. The vertices are \(V' = V \times 2^{\Agt}\),
where the second component, a subset of \(\Agt\), records the deviators
of the current history.

At each step, Eve chooses an action profile, and Adam chooses the move
that will apply. Adam can either respect Eve's choice, or pick a
different action profile in which case the deviators will be added to
the second component of the vertex. The game begins in
\((v_0 , \varnothing)\) and then proceeds as follows: from a vertex
\((v, D)\), Eve chooses an action profile \(a_{\Agt}\), and Adam chooses
a possibly different one \(a'_{\Agt}\). The next vertex is
\((\Delta(v, a'_{\Agt} ), D \cup \Dev(a_{\Agt} , a'_{\Agt} ))\).

\begin{example}
  An example of a partial construction of the deviator game for the
  example of \Cref{15-fig:example1}, is given in~\Cref{15-fig:ex-dev}.
  We cannot represent the full construction here, as there are 40 vertices.
\end{example}
\begin{figure}
  \begin{center}
    \begin{tikzpicture}
      \draw (0,0) node[draw, inner sep=7pt] (I) {$v_0, \varnothing$};
      \draw (8,4) node[draw, inner sep=7pt] (S10) {$v_1, \varnothing$};
      \draw (8,3) node[draw, inner sep=7pt] (S11) {$v_1, \{P_1\}$};
      \draw (8,2) node[draw, inner sep=7pt] (S12) {$v_1, \{P_2\}$};
      \draw (8,1) node[draw, inner sep=7pt] (S13) {$v_1, \{P_3\}$};
      \draw (8,0) node[draw, inner sep=7pt] (S14) {$v_1, \{P_1, P_2\}$};
      \draw (8,-1) node[draw, inner sep=7pt] (S15) {$v_1, \{P_1, P_3\}$};
      \draw (8,-2) node[draw, inner sep=7pt] (S16) {$v_1, \{P_2, p_3\}$};
      \draw (8,-3) node[draw, inner sep=7pt] (S17) {$v_1, \{P_1, P_2, P_3\}$};
      \draw (8,-4) node[draw, inner sep=7pt] (C2) {$v_2, \varnothing$};
      \draw (8,-5) node[below] {$\dots$};
      \draw[-latex'] (-1, 0) -- (I);
      \draw[rounded corners=20mm, -latex'] (I.90) |- node[pos=0.7,above]{$(a, b, a), (a, b, a)$} node[pos=0.7, below]{$(b, a, a), (b, a, a)$} (S10);
      \draw[rounded corners=16mm, -latex'] (I.75) |- node[pos=0.9,above]{$(a, a, a), (b, a, a)$} node[pos=0.9, below]{$(a, b, a), (b, a, b)$} (S11);
      \draw[rounded corners=12mm,-latex'] (I.60) |- node[pos=0.7,above]{$(a, a, a), (a, b, a)$} node [pos=0.7, below]{$(b, b, a), (b, a, a)$} (S12);
      \draw[rounded corners=6mm,-latex'] (I.45) |- node[pos=0.9,above]{$(a, b, a), (a, b, b)$} node [pos=0.9, below]{$(b, a, a), (b, a, b)$} (S13);
      \draw[-latex'] (I) -- node[pos=0.4,above]{$(a, b, a), (b, a, a)$} node [pos=0.4, below]{$(b, a, a), (a, b, a)$} (S14);
      \draw[rounded corners=6mm,-latex'] (I.-45) |- node[pos=0.9,above]{$(a, a, a), (b, a, b)$} node [pos=0.9, below]{$(b, b, a), (a, b, b)$} (S15);
      \draw[rounded corners=12mm,-latex'] (I.-60) |- node[pos=0.7,above]{$(a, a, a), (a, b, b)$} node [pos=0.7, below]{$(b, b, a), (b, a, b)$} (S16);
      \draw[rounded corners=16mm,-latex'] (I.-75) |- node[pos=0.9,above]{$(a, b, a), (b, a, b)$} node [pos=0.9, below]{$(b, a, a), (a, b, b)$} (S17);
      \draw[rounded corners=20mm,-latex'] (I.-90) |- node[pos=0.7,above]{$(a, a, a), (a, a, a)$} node [pos=0.7, below]{$(b, b, b), (b, b, b)$} (C2);
    \end{tikzpicture}
  \end{center}
    \caption{Example a deviator game construction.}
    \label{15-fig:ex-dev}
\end{figure}

We define projections \(\proj_{V}\) and \(\proj_{\Dev}\) from \(V'\) to
\(V\) and from \(V'\) to \(2^{\Agt}\) respectively, as well as
\(\proj_{\Act}\) from \(Act^{\Agt} \times Act^{\Agt}\) to \(Act^{\Agt}\) which
maps to the second component of the product, that is, Adam's action.

For a history or play \(\rho\), define \(\pi_{\Out}(\rho)\) as the play
\(\rho'\) for which, \(\rho'_i = \proj_V(\rho_i)\) and
\(\move_i(\rho') = \proj_{\Act}(\move_i(\rho))\) for all~$i$. This is thus the play
induced by Adam's actions.
Let us also denote~$\Dev(\rho) = \proj_{\Dev}(\last(\rho))$.

We can associate a strategy of Eve to each strategy profile
\(\sigma_{\Agt}\) such that she chooses the moves prescribed by
\(\sigma_{\Agt}\) at each history of \(\devg(\mathcal{G})\). Formally, we write
\(\kappa(\sigma_{\Agt})\) for the strategy defined by
\(\kappa(\sigma_{\Agt})(\pi) = \sigma_{\Agt}(\proj_{\Out}(\pi))\) for all histories~$\pi$.

The following lemma states the correctness of the construction of the
deviator game \(\devg(\mathcal{G})\), in the sense that it records the set of
deviators in the strategy profile suggested by Adam with respect to the
strategy profile suggested by Eve.

\begin{lemma}\label{15-prop:correctness-deviator-game}
  Let $\mathcal{G}$ be a multiplayer game, $v$ a vertex, \(\sigma_{\Agt}\) a
  strategy profile, and \(\sigma_{\exists} = \kappa(\sigma_{\Agt})\) the
  associated strategy in the deviator game.

  \begin{enumerate}
  \item     If \(\rho \in \Out_{\devg(\mathcal{G})}((v,\emptyset),\sigma_\exists)\), then
    \(\Dev(\proj_{\Out}(\rho), \sigma_{\Agt} ) = \Dev(\rho)\).
  \item     If \(\rho \in \Out_G(v)\) and for all index \(i\),
    \(\rho'_i = (\rho_i , \Dev(\rho_{\le i} , \sigma_{\Agt}))\) and\\
    \(\move_i(\rho') = (\sigma_{\Agt} (\rho_{\le i} ), \move_i(\rho))\), then
    \(\rho' \in \Out_{\devg(\mathcal{G})}((v,\emptyset), \sigma_\exists)\).
  \end{enumerate}
\end{lemma}
\begin{proof}
  We prove that for all $i$,
  \(\Dev(\proj_{\Out}(\rho_{\le i} , \sigma_{\Agt}) = \proj_{\Dev} (\rho_{\le i} )\),
which implies the property. The property holds for i = 0, since
initially both sets are empty. Assume now that it holds for \(i \ge 0\).
Then:
\begin{align*}
    &\ \Dev(\proj_{\Out}(\rho_{\le i+1}) , \sigma_{\Agt} ) \\
  = &\ \Dev(\proj_{\Out}(\rho_{\le i}), \sigma_{\Agt} ) \cup \Dev(\sigma_{\Agt} (\proj_{\Out}(\rho_{\le i})), \proj_{\Act} (\move_{i+1} (\rho))) \\
  & \text{(by definition of deviators)}\\
  = &\ \Dev (\rho_{\le i} ) \cup \Dev(\sigma_{\Agt} (\proj_{\Out} (\rho_{\le i}), \proj_{\Act} (\move_{i+1} (\rho))) \\
  & \text{(by induction hypothesis)} \\
  = &\ \Dev (\rho_{\le i} ) \cup \Dev(\sigma_\exists (\rho_{\le i} ), \proj_{\Act} (\move_{i+1}(\rho))) \\
  & \text{(by definition of \(\sigma_\exists\) )}\\
  = &\ \Dev (\rho_{\le i} ) \cup \Dev(\move_{i+1}(\rho)) \\
  & \text{(by assumption \(\rho \in \Out_{\devg(\mathcal{G})} ((v,\emptyset), \sigma_\exists)\))}\\
  = &\ \Dev(\rho_{\le i+1} ) \\
  & \text{(by construction of \(\devg(\mathcal{G})\))} \\
\end{align*}

Which concludes the induction.

We now prove the second part. The property is shown by induction. It
holds for \(v_0\). Assume it is true up to index \(i>0\), then
\begin{align*}
  &\ \Delta'(\rho'_i , \sigma_\exists(\rho'_{\le i}), \move_i(\rho)) \\
= &\ \Delta'((\rho_i , \Dev(\rho_{\le i} , \sigma_{\Agt})), \sigma_{\exists} (\rho'_{\le i} ), \move_i(\rho))\\
&  \text{(by definition of \(\rho'\))} \\
= &\ \Delta(\rho_i , \move_i(\rho)), \Dev(\rho_{\le i}, \sigma_{\Agt}) \cup \Dev(\sigma_{\exists}(\rho'_{\le i}), \rho_{i+1} )) \\
&  \text{(by construction of \(\Delta'\) )}\\
= &\ (\rho_{i+1} , \Dev(\rho_{\le i}, \sigma_{\Agt} ) \cup \Dev(\sigma_{\exists}(\rho'_{\le i}), \rho_{i+1} )) \\
& \text{(since \(\rho\) is an outcome of the game)} \\
= &\ (\rho_{i+1} , \Dev(\rho_{\le i} , \sigma_{\Agt} ) \cup \Dev(\sigma_{\Agt} (\rho_{\le i}), \rho_{i+1} )) \\
& \text{(by construction of \(\sigma_\exists\))} \\
= &\ (\rho_{i+1} , \Dev(\rho_{\le i+1} , \sigma_{\Agt} ))\\
& \text{(by definition of deviators)} \\
= &\ \rho'_{i+1}. \\
\end{align*}
\end{proof}


The objective of Eve in the deviator game is defined so that winning
strategies correspond to equilibria of the original game. First, as an
intermediary step, given coalition \(\Coalition\), player \(P\) and
bound \(b\), we will construct an objective stating that we can ensure
that the payoff of $P$ will not exceed $b$ even if players in $\Coalition$ change
their strategies.
Consider the following objective in \(\devg(\mathcal{G})\):
\[
  \Omega(\Coalition, P, b) = \{\rho \in \Out_{\devg(\mathcal{G})} \mid \Dev(\rho) \subseteq \Coalition \Rightarrow \payoff_P(\proj_{\Out}(\rho)) \le b\}.
\]
Intuitively, this says that if only players
in $\Coalition$ deviated from the strategy suggested by Eve, then the payoff
of $P$ is smaller than $b$.
We now show that a strategy ensuring bound \(b\) for the payoff of
$P$ against coalition \(\Coalition\) corresponds to a winning strategy for
\(\Omega(\Coalition, P, b)\) in the deviator game.

\begin{lemma} \label{15-lem:omegaCAg}
  Let \(\Coalition \subseteq \Agt\) be a coalition,
  \(\sigma_{\Agt}\) be a strategy profile, \(b \in \mathbb{R}\) a bound,
  and \(P\) a player. For all strategies \(\sigma'_\Coalition\), vertex~$v_0$,
  and coalition \Coalition, the following equivalence holds: 
  \(\payoff_P(\Out_{\devg(\mathcal{G})}(v_0, \sigma_{-\Coalition}, \sigma'_\Coalition)) \le b\) if, and
  only if, \(\kappa(\sigma_{\Agt})\) is winning in \(\devg(\mathcal{G})\) for objective
  \(\Omega(\Coalition, P, b)\).
\end{lemma}
\begin{proof} Let \(\rho\) be an outcome of
  \(\sigma_\exists=\kappa(\sigma_{\Agt}) \in \devg(\mathcal{G})\). By~\Cref{15-prop:correctness-deviator-game}, we have that
  \(\Dev(\rho) = \Dev(\proj_V(\rho),\sigma_{\Agt})\). By~\Cref{15-lem:deviator}, \(\proj_V(\rho)\) is the outcome of
  \((\sigma_{-\Dev(\rho)},\sigma'_{\Dev(\rho)})\) for some
  \(\sigma'_{\Dev(\rho)}\). If \(\Dev(\rho) \subseteq \Coalition\), then
  \[
  \payoff_P(\proj_V(\rho)) = \payoff_P(\sigma_{-\Coalition},\sigma_{\Coalition\setminus \Dev(\rho)}, \sigma'_{\Dev(\rho)}) = \payoff_P(\sigma_{-\Coalition},\sigma''_{\Coalition})\]
  where \(\sigma''_P = \sigma'_P\) if \(P \in \Dev(\rho)\) and \(\sigma_P\)
  otherwise. By hypothesis, this payoff is smaller than or equal to \(b\).
  This holds for
  all outcomes \(\rho\) of \(\sigma_\exists\), thus \(\sigma_\exists\) is
  a winning strategy for \(\Omega(\Coalition,P,b)\).

  For the other direction, assume \(\sigma_\exists = \kappa(\sigma_{\Agt})\)
  is a winning strategy in \(\devg(\mathcal{G})\) for \(\Omega(\Coalition,P,b)\). Let
  \(\sigma'_\Coalition\) be a strategy for \(\Coalition\) and \(\rho\) the outcome of
  \((\sigma'_{\Coalition},\sigma_{-{\Coalition}})\). By~\Cref{15-lem:deviator},
  \(\Dev(\rho,\sigma_{\Agt}) \subseteq \Coalition\). By~\Cref{15-prop:correctness-deviator-game},
  \(\rho'= (\rho_j, \Dev(\rho_{\le j},\sigma_{\Agt}))_{j\in \mathbb{N}}\) is
  an outcome of \(\sigma_\exists\). We have that
  \(\Dev(\rho') = \Dev(\rho,\sigma_{\Agt}) \subseteq \Coalition\). Since
  \(\sigma_\exists\) is winning, \(\rho\) is such that
  \(\payoff_P(\proj_V(\rho)) \le b\). Since
  \(\payoff_{P}(\proj_V(\rho')) = \payoff_{P}(\rho)\),
  this shows that for all strategies \(\sigma'_\Coalition\),
  \(\payoff_P(\sigma_{-\Coalition},\sigma'_\Coalition) \le b\).
\end{proof}

Now, Eve can show that there is a Nash equilibrium in a given game
by proving that whenever there is a single deviator,
the deviating player does not gain more than without the deviation,
while she does not have to prove anything on plays involving several
deviators.

\begin{theorem}\label{15-thm:dev-nash}
  Let \(\mathcal{G} = (\mathcal{A}, (\payoff_P)_{P \in \Agt})\) be a game, \(\sigma_{\Agt}\) a strategy
  profile in \(\mathcal{G}\), vertex~$v_0$, and \(F = (\payoff_P(\Out_{\mathcal{A}}(v_0,\sigma_{\Agt})))_{P\in \Agt}\)
  the payoff
  profile of \(\sigma_{\Agt}\) from~$v_0$. The strategy profile \(\sigma_{\Agt}\) is a
  Nash equilibrium if, and only if, strategy \(\kappa(\sigma_{\Agt})\) is
  winning in \(\devg(\mathcal{A})\) for the objective \(N(F)\) defined by:
  \[N(F) = \{\rho \mid |\Dev(\rho)| \ne 1\}
    \cup \bigcup_{P\in \Agt} \{\rho \mid \Dev(\rho) = \{P\}
    \land \payoff_P(\proj_{\Out}(\rho)) \le F_P\}.\]
\end{theorem}
\begin{proof} By \Cref{15-lem:omegaCAg}, \(\sigma_{\Agt}\) is a Nash
  equilibrium if, and only if, for each player \(P\),
  \(\kappa(\sigma_{\Agt})\) is winning for
  \(\Omega(\{P\}, P, F_P)\).
  So it is enough to show that for each player \(P\),
  \(\kappa(\sigma_{\Agt})\) is winning for
  \(\Omega(\{P\},P, F_P)\) if,
  and only if, \(\kappa(\sigma_{\Agt})\) is winning for \(N(F)\).

  \textbf{Implication} Let \(\rho\) be an outcome of
  \(\kappa(\sigma_{\Agt})\).

\begin{itemize}

\item   If \(|\Dev(\rho)| \ne 1\), then \(\rho\) is in \(N(F)\) by
  definition.
\item   If \(|\Dev(\rho)| = 1\), then for \(\{P\} = \Dev(\rho)\),
  \(\payoff_P(\proj_{\Out}(\rho)) \leq F_P\) because
  \(\kappa(\sigma_{\Agt})\) is winning for
  \(\Omega(\Dev(\rho), P, F_P)\). Therefore \(\rho\) is
  in \(N(F)\).
\end{itemize}

This holds for all outcomes \(\rho\) of \(\kappa(\sigma_{\Agt})\) and shows
that \(\kappa(\sigma_{\Agt})\) is winning for \(N(F)\).

\textbf{Reverse implication} 
Assume that
\(\kappa(\sigma_{\Agt})\) is winning for \(N(F)\). We now show that
\(\kappa(\sigma_{\Agt})\) is winning for
\(\Omega(\{P\},P, F_P)\) for each player \(P\). Let
\(\rho\) be an outcome of \(\kappa(\sigma_{\Agt})\), we have
\(\rho \in N(F)\). We show that \(\rho\) belongs to
\(\Omega(\{P\}, P, F_P)\):

\begin{itemize}

\item   If \(\Dev(\rho) = \varnothing\) then \(\rho = \Out(v_0, \sigma_{\Agt})\) and
  \(\payoff_P(\rho) = F_P\), so \(\rho\) is in
  \(\Omega(\{P\},P, F_P)\)
\item   If \(\Dev(\rho) \not\subseteq \{ P \}\), then
  \(\rho \in \Omega(\Coalition,P,F_P)\) by definition.
\item   Otherwise \(\Dev(\rho) = \{P\}\). Since \(\rho \in N(F)\),
  \(\payoff_P(\rho) \le F_P\). Hence
  \(\rho \in \Omega(\Coalition,P,F_P)\).
\end{itemize}

This holds for all outcomes \(\rho\) of \(\kappa(\sigma_{\Agt})\) and shows
it is winning for \(\Omega(\{P\},P,F_P)\) for each
player \(P\in \Agt\), which shows that \(\sigma_{\Agt}\) is a Nash
equilibrium.
\end{proof}


\subsubsection{Algorithm for parity games}
\label{15-subsec:algorithm-for-parity-objectives}
We now focus on the case of Parity objectives.
Recall that
Each player $P$ has
a colouring function $c_P : V \rightarrow \mathbb{N}$, inducing the parity objective $\Omega_A$.
%
Thus the payoff $\payoff_P$ assigns 1 to paths belonging to $\Omega_A$ and 0
to the others.

We now give an algorithm for the Nash equilibrium problem with parity
objectives. Given a payoff for each player \((F_P)_{P\in \Agt} \),
we can deduce from the previous theorem an algorithm that
constructs a Nash equilibrium if there exists one. We construct the
deviator game and note that we can reduce the number of vertices as
follows: since  \(\Dev(\rho_{\le k})\) is nondecreasing,
we know that Eve wins whenever this set has at least two elements.
In the construction, states with at least two deviators can be replaced by a
sink vertex that is winning for Eve. This means that the constructed
game has at most \(n \times (|\Agt| + 1) + 1\) states.

The objective can be expressed as a Parity condition in the following
way:

\begin{itemize}
\item   for each vertex \(v' = (v, \{ P \})\), \(c'(v') = c_P(v) + 1\) if
  \(F_P = 0\) and \(2 \cdot \max_v c_P(v) \) otherwise;
\item   for each vertex \(v' = (v, D)\) with \(|D| \ne 1\), \(c'(v') = 2 \cdot \max_v c_P(v)\)
  \textit{i.e.}~it is winning for Eve.
\end{itemize}

Notice that the colouring function $c'$ inverts the parity
in the case where there is a single deviator who is losing in the
prescribed strategy profile (that is, $F_P=0$). In fact,
when~$F_P=1$, the player cannot obtain more since they are already winning
so the colour is set to $2\cdot \max_v c_P(v) $ which is winning for \Eve.

\begin{lemma}
  We have \(\maxinf(c'(\rho_i)) \in 2 \mathbb{N}\) if, and
  only if, \(\rho\in N(F)\),
  where $N(F)$ is as defined in \Cref{15-thm:dev-nash}.
\end{lemma}
\begin{proof} For the implication, we will prove the contrapositive.
Let \(\rho\) be a play not in \(N(F)\), then since the deviators can only
increase along a play, we have that \(\Dev(\rho) = \{ P \}\) for some
player \(P\) and \(\payoff_P(\rho) > F_P\). This means
\(F_P = 0\) and \(\maxinf(c_P(\rho_i)) \in 2 \mathbb{N}\). By
definition of \(c'\) this implies that
\(\maxinf(c'(\rho_i)) \in 2 \mathbb{N} + 1\) which proves the
implication.

For the other implication, let \(\rho\) be such that
\(\maxinf(c'(\rho_i)) \in 2 \mathbb{N} + 1\). By definition of
\(c'\) this means \(\rho\) contains infinitely many states of the form
\((v, \{P\})\) with \(F_P = 0\). Since the deviators only increase
along the run, there is a player \(P\) such that \(\rho\) stays in the
component \(V \times \{P\}\) after some index \(k\). Then for \(i\geq k\),
\(c'(\rho_i) = c_P(\rho_i)+1\), hence
\(\maxinf(c'(\rho_i)) = \maxinf(c_P(\rho_i)) + 1\).
Therefore \(\maxinf(c_P(\rho_i)) \in 2 \mathbb{N}\), which means
\(\payoff_P(\rho) = 1 > F_P\). By definition of \(N(F)\),
\(\rho\not\in N(F)\).
\end{proof}

Given that the size of the game is polynomial and that parity games can
be decided in quasi-polynomial time (see \Cref{2-chap:parity}), the above lemma
implies the following theorem.

\begin{theorem}
For parity games, there is a quasi-polynomial algorithm to decide whether there is a Nash
equilibrium with a given payoff.
\end{theorem}

\subsection{Extensions of Nash equilibria}
\label{15-subsec:extensions-of-nash-equilibria}

\subsubsection{Subgame perfect equilibria}
\label{15-subsection:subgame-perfect-equilibria}

Nash equilibria present the disadvantage that once a player has deviated,
the others will try to punish him, forgetting everything about their own
objectives.
If we were to observe the game after this point of
deviation, it would not look like the players are playing rationally and
in fact it would not correspond to a Nash equilibrium. The concept of
\emph{subgame perfect equilibria} refines the concept of Nash
equilibrium by imposing that at each step of the history, the strategy
behaves like Nash equilibrium if we were to start the game now. Formally,
let us write \(\sigma_P \circ h\) the strategy which maps all histories
\(h'\) to \(\sigma_P(h \cdot h')\), that is the strategy that behave
like \(\sigma_P\) after \(h\). Then \((\sigma_P)_{P\in \Agt}\) is a
\emph{subgame perfect equilibrium} if for all histories \(h\),
\((\sigma_P \circ h)_{P \in \Agt}\) is a Nash equilibrium.

Imposing such a strong restriction is justified by the fact that subgame
perfect Nash equilibria exist for a large class of games. In particular
subgame perfect equilibria always exist in turn-based games with
reachability objectives.

\begin{example}
  Consider the example of \Cref{15-fig:ex-subgame}.
  There is a Nash equilibria whose outcome goes through states
  $v_0 \rightarrow v_1 \rightarrow \Omega_1$.
  In this equilibrium, Player $1$ should play $b$ in $v_2$,
  so that the best response of Player~$2$ is to play~$a$ at~$v_0$.
  Intuitively, player $1$ is threatening player $2$, to make them
  both lose from $v_2$, but this threat is not credible,
  and the profile is not a subgame perfect equilibrium.
  In fact, once $v_2$ is reached it is better for Player $1$
  to play $a$ so it is unlikely that the player will execute the said threat.
  The only subgame perfect equilibrium of this game ends in 
  the vertex satisfying both
  $\Omega_1$ and $\Omega_2$.
\end{example}
\begin{figure}
\begin{center}  
\begin{tikzpicture}
  \draw (0,0) node[draw, inner sep=7pt] (V0) {$v_0$};
  \draw (3,1) node[draw, inner sep=7pt] (V1) {$v_1$};
  \draw (3,-1) node[draw, inner sep=7pt] (V2) {$v_2$};
  \draw (6,2) node[draw, inner sep=7pt] (V3) {$\Omega_1$};
  \draw (6,0.6) node[draw, inner sep=7pt] (V4) {$\Omega_2$};
  \draw (6,-0.6) node[draw, inner sep=7pt] (V5) {$\Omega_1, \Omega_2$};
  \draw (6,-2) node[draw, inner sep=7pt] (V6) {$\varnothing$};

  \draw[-latex'] (-1, 0) -- (V0);
  \draw[-latex'] (V0) -- node[above]{$(\ast, a)$} (V1);
  \draw[-latex'] (V0) -- node[below]{$(\ast, b)$} (V2);
  \draw[-latex'] (V1) -- node[above]{$(a, \ast)$} (V3);
  \draw[-latex'] (V1) -- node[below]{$(b, \ast)$} (V4);
  \draw[-latex'] (V2) -- node[above]{$(a, \ast)$} (V5);
  \draw[-latex'] (V2) -- node[below]{$(b, \ast)$} (V6);
  \draw[-latex'] (V3) .. controls +(2,1) and +(2,-1) .. (V3);
  \draw[-latex'] (V4) .. controls +(2,1) and +(2,-1) .. (V4);
  \draw[-latex'] (V5) .. controls +(2,1) and +(2,-1) .. (V5);
  \draw[-latex'] (V6) .. controls +(2,1) and +(2,-1) .. (V6);
\end{tikzpicture}
\end{center}
\caption{Two-player game with reachability objectives. The goal of
  player 1 is to reach a state labelled with $\Omega_1$ and that of
  player 2 is to reach a state labelled with $\Omega_2$. }
\label{15-fig:ex-subgame}
\commentAlt{Figure~\ref{15-fig:ex-subgame}: A directed graph with a square root node v0 branching into two intermediate square nodes v1 and v2, which then further branch into four final square nodes, each with a self-loop. See long description.}
\commentLongAlt{Figure~\ref{15-fig:ex-subgame}: The image displays a directed graph starting from a square root node labeled 'v0', indicated by an incoming arrow from the left.
- From 'v0', two branches diverge:
    - An upper branch leads to a square node labeled 'v1', with the arrow labeled '(*,a)'.
    - A lower branch leads to a square node labeled 'v2', with the arrow labeled '(*,b)'.

- From 'v1', two branches diverge:
    - An upper branch leads to a square node labeled 'Omega_1', with the arrow labeled '(a,*)'. This node has a self-loop.
    - A lower branch leads to a square node labeled 'Omega_2', with the arrow labeled '(b,*)'. This node also has a self-loop.

- From 'v2', two branches diverge:
    - An upper branch leads to a square node labeled 'Omega_1,Omega_2', with the arrow labeled '(a,*)'. This node has a self-loop.
    - A lower branch leads to a square node labeled 'O_slash' (empty set symbol), with the arrow labeled '(b,*)'. This node also has a self-loop.

The self-loops indicate that once a state is reached, it can remain in that state indefinitely.}
\end{figure}

\subsubsection{Robust equilibria}
\label{15-subsec:robust-equilibria}
The notion of robust equilibria refines Nash equilibria in two ways:

\begin{itemize}
\item   a robust equilibrium is \emph{resilient}, \textit{i.e.}~when a small coalition
  change its strategy, none of the players of the coalition improves their
  payoff;
\item   it is \emph{immune}, \textit{i.e.}~when a small coalition changes its strategy,
  it does not decrease the payoffs of the non-deviating players.
\end{itemize}

The size of small coalitions is determined by parameter \(k\) for
resilience and \(t\) for immunity. When a strategy is both
\(k\)-resilient and \(t\)-immune, it is called a \((k,t)\)-robust
equilibrium.

The motivation behind this concept is to address these two weaknesses of
Nash equilibria:

\begin{itemize}
\item   There is no guarantee on payoffs when two (or more) players deviate together.
  Such a situation can occur in networks if the same person controls several devices
  (a laptop and a phone for instance) and can then coordinate their
  behaviours. In this case, the devices would be considered as different
  players and Nash equilibria can offer no guarantee.
\item   When a deviation occurs, the strategies of the equilibrium can punish
  the deviating user without any regard for the payoffs of the others. This
  can result in a situation where, because of a faulty device, nobody
  can use the protocol anymore.
\end{itemize}

By comparison, finding resilient equilibria with \(k>1\), 
ensures that clients have no interest in forming coalitions (up
to size \(k\)), and finding immune equilibria with \(t>0\)
ensures that other clients will not suffer from some players (up to
\(t\)) behaving differently from what was expected.

The deviator construction can be reused for finding such equilibria. We
only need to adapt the objectives. Given a game \(G=(\mathcal{A}, (\payoff_P)_{P \in \Agt})\), a
strategy profile \(\sigma_{\Agt}\), and parameters \(k\), \(t\), we have

\begin{itemize}

\item   The strategy profile \(\sigma_{\Agt}\) is \(k\)-resilient if, and only
  if, strategy \(\kappa(\sigma_{\Agt})\) is winning in \(\devg(\mathcal{A})\) for the
  \emph{resilience objective} \(\mathcal{R}(k,F)\) where
  \(F = (\payoff_P(\Out_{\mathcal{A}}(v_0, \sigma_{\Agt})))_{P \in \Agt}\) is the payoff profile of
  \(\sigma_{\Agt}\) and \(\mathcal{R}(k,F)\) is defined by:
  \[
    \begin{array}{lll}
      \mathcal{R}(k,F) & =    & \{ \rho \in \Out_{\devg(\mathcal{A})}\mid ~ |\Dev(\rho)| > k \} \\
                       & \cup & \{ \rho  \in \Out_{\devg(\mathcal{A})} \mid ~ |\Dev(\rho)| = k \land \forall P \in \Dev(\rho).\ \payoff_{P}(\proj_{\Out}(\rho)) \le F_P\} \\
                       & \cup & \{ \rho  \in \Out_{\devg(\mathcal{A})}\mid ~ |\Dev(\rho)| < k \land \forall P \in \Agt.\ \payoff_{P}(\proj_{\Out}(\rho)) \le F_P\}.
    \end{array}
  \]

\item   The strategy profile \(\sigma_{\Agt}\) is \(t\)-immune if, and only if,
  strategy \(\kappa(\sigma_{\Agt})\) is winning for the \emph{immunity
  objective} \(\mathcal{I}(t,F)\) where
  \(F = (\payoff(\Out_{\mathcal{A}}(v_0, \sigma_{\Agt})))_{P \in \Agt}\) is the payoff profile of
  \(\sigma_{\Agt}\) and \(\mathcal{I}(t,F)\) is defined by:
  \[
    \begin{array}{lll}
    \mathcal{I}(t,F) & = & \{ \rho  \in \Out_{\devg(\mathcal{A})} \mid |\Dev(\rho)| > t \}  \\
      & \cup & \{ \rho  \in \Out_{\devg(\mathcal{A})} \mid ~ \forall P \in \Agt \setminus \Dev(\rho).\  F_P \le \payoff_{P}(\proj_{\Out}(\rho)) \}.
    \end{array}
  \]
\item   The strategy profile \(\sigma_{\Agt}\) is a \((k,t)\)-robust profile in
  \(G\) if, and only if, \(\kappa(\sigma_{\Agt})\) is winning for the
  \emph{robustness objective}
  \[
        \mathcal{R}(k,t,F)= \mathcal{R}(k,F) \cap \mathcal{I}(t,F),
        \] where
        \(F = (\payoff_P(\Out_{\mathcal{A}}(v_0, \sigma_{\Agt})))_{P \in \Agt}\) is the payoff profile of
        \(\sigma_{\Agt}\).
\end{itemize}

We omit the proof and encourage the reader to do it by themselves.

\subsubsection{Extension to games with hidden actions}
\label{15-subsec:extension-to-games-with-hidden-actions}
In most practical cases, players only have a partial view of the
state of the system; so they may not be able, for instance, to
detect a deviating player immediately.
Studying equilibria in general imperfect information as in \Cref{9-chap:signal} 
would be well adapted in such situations.
Unfortunately, these games are too powerful in general since
the existence of Nash equilibria is undecidable in this case.

Nevertheless, the problem is decidable for a restricted form of imperfect
information where the players observe the visited states
but do not see the played actions; thus the actions are \emph{hidden}.

We will thus consider strategies defined as functions from \(V^*\) to \(Act\),
which represents the fact that players' decision can only depend on observed
sequence of states but not on other players' actions.

In this case, deviators cannot be defined as obviously as before, as it may
not always be possible to identify one unique deviator responsible for a
deviation. The construction will thus maintain a set of \emph{suspects}, those players that might have been
responsible for the observed deviation.
Formally, suspects for an edge \((v, v')\) with respect to a move
\((a_P)_{P\in \Agt}\) are players \(P\) such that there is \(a'_P\) and
\(\Delta(a'_P, a_{-P}) = (v, v')\). Rather than computing the
union of deviators along a history, we now consider the intersection of
suspects. That is, if at vertex~$v$, the suspect set is~$S$, and the strategy profile is~$\sigma_\Agt$,
and if the next vertex is~$v'$, then
the suspect set becomes~$S \cap \{ P \in \Agt \mid \exists a'_P, \Delta(a'_P, a_{-P}) = (v, v')\}$.


The \emph{suspect game} can be defined just like the deviator game by replacing the deviators component
by the suspects component.
The objective for Eve is that no suspect player improves their payoff. In fact, in
case of deviation, we know that the deviator belongs to the set of suspects although
we cannot know which one has deviated for sure so Eve must ensure this for all suspects.

\begin{example}
  Consider the example of~\Cref{15-fig:hidden}.
  If actions were visible there would be an equilibrium ending in the
  state labelled with $\Omega_3$: player 3 simply has to punish the player
  who would deviate from this path.
  But if we now consider hidden actions, in case of a deviation, player 3
  would observe that the play went arrives in $v_1$ instead of $v_2$
  and both Player 1 and Player 2 are suspects.
  Since Player 3 cannot punish both players at the same time, there is no
  Nash equilibrium ending satisfying $\Omega_3$.
\end{example}
\begin{figure}
\begin{center}
\begin{tikzpicture}
  \draw (0,0) node[draw, inner sep=7pt] (V0) {$v_0$};
  \draw (3,1) node[draw, inner sep=7pt] (V1) {$v_1$};
  \draw (3,-1) node[draw, inner sep=7pt] (V2) {$v_2$};
  \draw (6,2) node[draw, inner sep=7pt] (V3) {$\Omega_1$};
  \draw (6,0.6) node[draw, inner sep=7pt] (V4) {$\Omega_2$};
  \draw (6,-0.6) node[draw, inner sep=7pt] (V5) {$\Omega_3$};
  \draw (6,-2) node[draw, inner sep=7pt] (V6) {$\varnothing$};

  \draw[-latex'] (-1, 0) -- (V0);
  \draw[-latex',rounded corners=5mm] (V0) |- node[pos=0.7,above]{$(a, a, \ast)$} node[pos=0.7,below] {$(b, b, \ast)$} (V1);
  \draw[-latex',rounded corners=5mm] (V0) |- node[pos=0.7,above]{$(a, b, \ast)$} node[pos=0.7,below] {$(b, a, \ast)$} (V2);
  \draw[-latex'] (V1) -- node[above]{$(\ast, \ast, a)$} (V3);
  \draw[-latex'] (V1) -- node[below]{$(\ast, \ast, b)$} (V4);
  \draw[-latex'] (V2) -- node[above]{$(\ast, \ast, a)$} (V5);
  \draw[-latex'] (V2) -- node[below]{$(\ast, \ast, b)$} (V6);
  \draw[-latex'] (V3) .. controls +(2,1) and +(2,-1) .. (V3);
  \draw[-latex'] (V4) .. controls +(2,1) and +(2,-1) .. (V4);
  \draw[-latex'] (V5) .. controls +(2,1) and +(2,-1) .. (V5);
  \draw[-latex'] (V6) .. controls +(2,1) and +(2,-1) .. (V6);
\end{tikzpicture}
\end{center}
\caption{Three-player game with hidden actions. The goal of
  player $i$ is to reach a state labelled with $\Omega_i$.}
\label{15-fig:hidden}
\commentAlt{Figure~\ref{15-fig:hidden}: A directed graph with a square root node v0 branching into two intermediate square nodes v1 and v2, which then further branch into four final square nodes, each with a self-loop. See long description.}
\commentLongAlt{Figure~\ref{15-fig:hidden}: The image displays a directed graph starting from a square root node labeled 'v0', indicated by an incoming arrow from the left.
- From 'v0', two sets of parallel arrows diverge:
    - An upper set of two parallel arrows leads to a square node labeled 'v1'. The top arrow is labeled '(a,a,*)' and the bottom arrow is labeled '(b,b,*)'.
    - A lower set of two parallel arrows leads to a square node labeled 'v2'. The top arrow is labeled '(a,b,*)' and the bottom arrow is labeled '(b,a,*)'.

- From 'v1', two branches diverge:
    - An upper branch leads to a square node labeled 'Omega_1', with the arrow labeled '(*,*,a)'. This node has a self-loop.
    - A lower branch leads to a square node labeled 'Omega_2', with the arrow labeled '(*,*,b)'. This node also has a self-loop.

- From 'v2', two branches diverge:
    - An upper branch leads to a square node labeled 'Omega_3', with the arrow labeled '(*,*,a)'. This node has a self-loop.
    - A lower branch leads to a square node labeled 'O_slash' (empty set symbol), with the arrow labeled '(*,*,b)'. This node also has a self-loop.

The self-loops indicate that once a state is reached, it can remain in that state indefinitely. The labels on the arrows suggest input-output relationships or conditions for transitions.}
\end{figure}

\section{Admissible strategies}
\label{15-sec:admissible_strategies}
Nash equilibria and their variants seen so far describe stable situations
from which players have no incentive to deviate. This however is of limited use
in some situations. First, the stability relies on the fact that all players are informed
of the strategy profile to be played; that is, some central authority needs to publicly
announce the strategies for all players. Second, each equilibrium describes a single possible
situation. If there are several equilibria, it is not clear which one is to be chosen.

Rather than concentrating on particular equilibria, game theorists have studied
reasonings players may follow in order to exclude some strategies that are necessarily worse than others.
These worse strategies are called \emph{dominated}.
By formalizing dominated and non-dominated strategies for a given player, one can then predict the behaviour of rational players
since such a player would never use a dominated strategy but rather always pick the best one available.

In this section, we will formalize dominated strategies and show how these can be computed in games.
We then briefly show that this reasoning can be repeated, and present the iterated elimination of dominated strategies.

\subsection{Definition}\label{definition-1}
The notion of \emph{dominance} is used to compare strategies 
with respect to payoffs they yield against the rest of the players' strategies.
Consider the example of \Cref{15-tbl:normal-adm}.
Given a strategy of the second player, playing $B$ is always at least as good
as playing $A$ for the first player.
In fact, again~$C$, $B$ yields a payoff of~$4$ which is better than $3$, the payoff of~$A$;
and against $D$, both yield~$0$.
The strategy $B$ is said to dominate $A$. Intuitively, 
$B$ is either better or as good as~$A$ \emph{in all situations}, so 
playing~$B$ is the rational choice for Player~$1$.

Furthermore, by this analysis, Player~$2$ knows that Player~$1$ will player~$B$.
Given this information, the best response of Player~$1$ is to play~$C$.
By \emph{iterated elimination}, we established that
$(B, C)$ should be the only strategy profile to be played by players following this reasoning.

\begin{table}
  \caption{A normal form game solvable by iterated elimination.}
  \label{15-tbl:normal-adm}
  \begin{center}
    \begin{tabular}[c]{|@{\hspace{1em}}l@{\hspace{1em}}|@{\hspace{1em}}c@{\hspace{1em}}c@{\hspace{1em}}|}
      \hline
      & A & B \\
      \hline
      C & 3, 3 & 4, 1\\
      D & 0 , 4 & 0, 0\\
      \hline
    \end{tabular}
  \end{center}
\end{table}

Let us formalize this notion.

\begin{definition}[Dominance]
  Let \(S \subseteq \mathcal{S}^{\Agt}\) be a set of the form
  \(S = S_1 \times S_2 \times \cdots \times S_n\) which we will call a
  rectangular set. Let \(\sigma_i,\sigma_i' \in S_i\). Strategy \(\sigma_i\)
  \emph{very weakly dominates} strategy \(\sigma_i'\) with respect to \(S\),
  written \(\sigma_i \ge_S \sigma'_i\), if from all vertices \(v_0\):

  \[
    \forall \sigma_{-i} \in S_{-i}, \payoff_i(\Out(v_0,\sigma'_i,\sigma_{-i}))
    \ge
    \payoff_i(\Out(v_0, \sigma_i, \sigma_{-i})).
  \]

  Strategy \(\sigma_i\) \emph{weakly dominates} strategy \(\sigma'_i\)
  in \(S\), written \(\sigma >_S \sigma'\), if
  \(\sigma_i \ge_S \sigma_i'\) and \(\neg(\sigma_i' \le_S \sigma_i)\).
  A strategy that is
  not weakly dominated in \(S\) is \emph{admissible} in \(S\). The
  subscripts on \(\ge_S\) and \(>_S\) are omitted when the sets of
  strategies are clear from the context.
\end{definition}

Algorithms rely on the notion of \emph{optimistic} and
\emph{pessimistic value} of a history. The pessimistic value is the
maximum payoff that a player can ensure in the worst case within the strategy set~$S$.
The optimistic value is
the best the player can achieve with the help of other players, given the strategy set~$S$.

\begin{definition}[Values] The \emph{pessimistic value} of a strategy
\(\sigma_i\) for a history \(h\) with respect to a rectangular set of
strategies \(S\), is

\begin{itemize}
\item   \(\pes_i(S,h,\sigma_i) = \inf_{\sigma_{-i} \in S_{-i}} \payoff_{i}(h \cdot \Out(\last(h), \sigma_i,\sigma_{-i})).\)
\end{itemize}

The \emph{pessimistic value of a history} \(h\) for \(A_i\) with respect
to a rectangular set of strategies \(S\) is given by:

\begin{itemize}

\item   \(\pes_i(S,h) = \sup_{\sigma_i \in S_i} \pes_i(S,s,\sigma_i).\)
\end{itemize}

The \emph{optimistic value} of a strategy \(\sigma_i\) for a history
\(h\) with respect to a rectangular set of strategies \(S\) is given by:

\begin{itemize}

\item   \(\opt_i(S,h,\sigma_i) = \sup_{\sigma_{-i} \in S_{-i}} \payoff_i (h_{\le |h|-2} \cdot \Out(\last(h), \sigma_i,\sigma_{-i})).\)
\end{itemize}

The \emph{optimistic value} of a history \(h\) for \(A_{i}\) with
respect to a rectangular set of strategies \(S\) is given by:

\begin{itemize}

\item   \(\opt_i(S,h) = \sup_{\sigma_i \in S_i} \payoff_{i}(\opt_i(S,h,\sigma_i))\)
\end{itemize}
\end{definition}

We will first consider the case where~$S$ is the set of all strategies,
and omit~$S$ in the above notations.

\subsection{Simple safety games}
\label{15-subsec:simple-safety-games}

\begin{figure}
\begin{center}  
\begin{tikzpicture}
\draw (0,0) node[draw, inner sep=2pt, circle] (I) {$v_0$};
\draw (I.-90) node[below] {};
\draw (4,1) node[draw, circle, inner sep=2pt] (C1) {$v_1$};
\draw (C1.-100) node[below] {};
\draw (4,-1) node[draw, circle, inner sep=2pt, circle] (C2) {$v_2$};
\draw (8, 1) node[draw, circle, inner sep=2pt] (S1) {$v_3$};
\draw (S1.-90) node[below] {$-1, -1, 0$};
\draw (8,-1) node[draw, circle, inner sep=2pt] (S2) {$v_4$};
\draw (S2.-90) node[below] {$-1, 0, -1$};
\draw[-latex'] (-1, 0) -- (I);
\draw[-latex'] (I) -- node[above]{$(*,a,*)$} (C1);
\draw[-latex'] (I) -- node[below]{$(*,b,*)$} (C2);
\draw[-latex'] (C1) --node[left]{$(a,*,*)$} (C2);
\draw[-latex'] (C1) -- node[above]{$(b,*,*)$} (S1);
\draw[-latex'] (C2) -- node[above]{$(*,*,a)$} (S2);
\draw[-latex', rounded corners] (C2) -- +(1,1) -- node[right]{$(*,*,b)$}(C1);
\end{tikzpicture}
\end{center}
\caption{Example of a three-player turn-based simple safety game.
State $v_1$ belongs to player 1, $v_0$ to player 2, and $v_2$ to player $3$. 
Numbers below states describe the safety objective, for instance $-1, -1, 0$ is losing for player 1 and 2.
}
\label{15-fig:simple-safety}
\commentAlt{Figure~\ref{15-fig:simple-safety}: A directed graph with a circular root node v0 branching into two intermediate circular nodes v1 and v2, which then connect to each other and lead to two final circular nodes v3 and v4. See long description.}
\commentLongAlt{Figure~\ref{15-fig:simple-safety}: The image displays a directed graph starting from a circular root node labeled 'v0', indicated by an incoming arrow from the left.

From 'v0', two branches diverge:
- An upper branch leads to a circular node labeled 'v1', with the arrow labeled '(*,a,*)'.
- A lower branch leads to a circular node labeled 'v2', with the arrow labeled '(*,b,*)'.

Connections between v1 and v2:
- An arrow from 'v1' points to 'v2', labeled '(a,*,*)'.
- An arrow from 'v2' points to 'v1', labeled '(*,*,b)'.

From 'v1' and 'v2', branches lead to final nodes:
- From 'v1', an arrow points to a circular node labeled 'v3', with the arrow labeled '(b,*,*)'. Below 'v3', it is labeled '-1, -1, 0'.
- From 'v2', an arrow points to a circular node labeled 'v4', with the arrow labeled '(*,*,a)'. Below 'v4', it is labeled '-1, 0, -1'.

The labels on the arrows suggest input-output relationships or conditions for transitions, and the labels below v3 and v4 might represent output states or values.}
\end{figure}

Simple safety games, are safety games in which there are no transitions
from losing vertices to non-losing ones. Restricting to this particular
class of game makes the problem simpler because the objective becomes
prefix-independent.
Any safety game can be converted to an equivalent simple safety game
by encoding in the states which players have visited so far a losing
state. Note that this translation can be exponential in the number of
players.

For simple safety games, the pessimistic and optimistic values do not depend
on the full history but only on the last state: for all histories
\(\pes_i(h) = \pes_i(\last(h))\) and \(\opt_i(h) = \opt_i(\last(h))\).

Note that in safety games (and any qualitative games) values can be only
1 (for winning) and 0 (for losing) and since the pessimistic value is
always less than the optimistic one, the pair \((\pes_i, \opt_i)\) can
only take three values: \((0, 0)\), \((0, 1)\) and \((1, 1)\).

Intuitively, players should avoid decreasing this pair of values if they can.
In fact, the characterisation of admissible strategies we give below will be based on this
simple observation.

\begin{example}
  An example of a simple safety game is given in \Cref{15-fig:simple-safety}.
  In this game, player 1 controls $v_1$ where its optimistic
  value is $0$, as it is possible that the outcome will never reach
  $v_3$ or $v_4$.
  However $v_3$ has optimistic value $-1$ for player 1, as it is a losing
  state for player 1.
  Going from $v_1$ to $v_3$ is a bad choice for player 1, and
  it is indeed dominated by the strategy that would always choose to go
  to $v_2$.
  In state $v_3$, player 3 has pessimistic value $0$ since it can ensure not
  visiting $v_4$.
  The winning strategy for player 3 which is to always go to $v_1$ is also
  the only non-dominated strategy.
  For player 2, from state $v_0$, both choices lead to a state with values
  $(-1, 0)$ so no choice is particularly better and both strategies are
  non-dominated.
\end{example}

\begin{definition} Let \(D_i\) be the set of edges
\((v,v') \in E\) such that \(v \in V_i\) 
\(\pes_i(v) > \pes_i(v')\) or \(\opt_i(v) > \opt_i(v')\). These
are called \emph{dominated edges}.
\end{definition}

\begin{theorem}[Characterisation of admissible strategies]
  \label{15-thm:adm}
Admissible
strategies for player \(A_i\) are the strategies that never take actions
in \(D_i\).
\end{theorem}
\begin{proof} We show that if \(A_i\) plays an admissible strategy
\(\sigma_i\) then the value cannot decrease on a transition controlled by
\(A_i\). Let \(\rho \in \Out(\sigma_i,\sigma_{-i})\), and \(k\) an index
such that \(\rho_k \in V_i\). Let \((v, v') = \sigma_i(\rho_{\le k})\):

\begin{itemize}

\item   If \(\pes_i(\rho_k)=1\), then \(\sigma_i\) has to be winning against
  all strategies \(\sigma_{-i}\) of \(A_{-i}\), otherwise it would be
  weakly dominated by such a strategy. Since there is no such strategy
  from a state with value \(\pes_i \leq 0\), we must have \(\pes_i(v')=1\).
\item   If \(\opt_i(\rho_k) = 1\), then there is a profile \(\sigma'\) such that
  \(\rho = \payoff(\Out(v, \sigma'))\), which is equal to \(1\).
  Assume that $\opt_i(v')=0$. Then~$\sigma_i$ is dominated by the strategy
  $\sigma_i''$ obtained from~$\sigma_i$ by making it switch to~$\sigma'$ at~$v$.
  In fact, $\sigma_i$ is losing against all strategies of~$-i$, while~$\sigma_i''$
  is winning at least against~$\sigma_{-i}'$.
\item   If $\pes_i(v)=0$ or $\opt_i(s) = 0$, then the value cannot decrease further.
\end{itemize}

In the other direction, let \(\sigma_i,\sigma_i'\) be two strategies of
player \(A_i\) and assume \(\sigma_i' >_S \sigma_i\). We will prove
\(\sigma_i\) takes a transition in \(D_i\) at some point.

Let us fix some objects before developing the proof. There is a vertex
\(v\) and strategy profile \(\sigma_{-i} \in S_{-i}\) such that
\(\payoff(\Out(v,\sigma_i',\sigma_{-i}))=1 \wedge \payoff(\Out(v,\sigma_i,\sigma_{-i})=0\).
Let \(\rho = \Out(v,\sigma_i,\sigma_{-i})\) and
\(\rho' = \Out(v,\sigma_i',\sigma_{-i})\). Consider the first position
where these runs differ: write \(\rho = w \cdot s' \cdot s_2 \cdot w'\)
and \(\rho' = w \cdot s' \cdot s_1 \cdot w''\).

The following are simple facts that can be seen easily:

\begin{itemize}
\item   \(s' \in V_i\), because the strategy of the other players
  are identical in the two runs.
\item   \(\opt_i(s_1) = 1\) because
  \(\payoff(\Out(v,\sigma_i',\sigma_{-i}))=1\)
\item   \(\pes_i(s_2) = 0\) because
  \(\payoff(\Out(v,\sigma_i,\sigma_{-i}))=0\)
\end{itemize}

If \(\opt_i(s_2) = 0\) or \(\pes_i(s_1) = 1\) then
\(s' \rightarrow s_2 \in D_i\) so \(\sigma_i\) takes a transition of
\(D_i\). The remaining case to complete the proof is \(\opt_i(s_2) = 1\)
and \(\pes_i(s_1) = 0\).
Let us assume that~$\sigma_i$ does not take any edges from~$D_i$.
We will show that there is
a strategy for~$-i$ against which $\sigma_i$ wins and~$\sigma_i'$ loses,
which contradicts the hypothesis that~$\sigma_i'$ weakly dominates~$\sigma_i$.

We first construct a profile \(\sigma_{-i}^2 \in S_{-i}\) such that
\(\payoff(\Out(s_2,\sigma_i,\sigma_{-i}^2))=1\).
Strategy \(\sigma_{-i}^2\in S_{-i}\) never decreases
the optimistic value from \(1\) to \(0\) since the optimistic value is non-increasing.
By assumption, \(\sigma_i\)
itself does not decrease the value of \(A_i\) because it does not take
transitions of~\(D_i\). So the outcome of \((\sigma_i,\sigma_{-i}^2)\)
never reaches a state of optimistic value \(0\). Hence it never reaches
a state in \(Bad_i\) and therefore it is winning for \(A_i\).

Let us now consider a profile \(\sigma_{-i}^1 \in S_{-i}\) such that
\(\payoff(\sigma_i',\sigma_{-i}^1)=0\) from \(s_1\). Such a strategy exists
because
\(\pes_i(s_1) = 0\), so~$\sigma_i'$ is not a winning strategy.
Then there exists a
strategy profile \(\sigma_{-i}^1\) such that \(\sigma_i'\) loses from
\(s_1\).

Now consider strategy profile \(\sigma_{-i}'\) that plays like
\(\sigma_{-i}\) if the play does not start with \(w\);
and otherwise switches to
\(\sigma_{-i}^1\) at history \(ws_1\) and to \(\sigma_{-i}^2\) at history \(ws_2\).
Formally, given a history \(h\), \(\sigma_{-i}'(h) =\)

\begin{itemize}
\item   \(\sigma_{-i}^1( h')\) if \(w \cdot s_1\) is a prefix of \(h\) and
  \(w \cdot s_1 \cdot h' = h\)
\item   \(\sigma_{-i}^2( h')\) if \(w \cdot s_2\) is a prefix of \(h\) and
  \(w \cdot s_2 \cdot h' = h\)
\item   \(\sigma_{-i}(h)\) otherwise
\end{itemize}

Clearly we have
\(\payoff_i(\Out_s(\sigma_i,\sigma_{-i}'))= 1 \wedge \payoff_i(\Out_s(\sigma_i',\sigma_{-i}')) = 0\),
contradicting \(\sigma_i' \ge_S \sigma_i\).
\end{proof}

\subsection{Parity games}\label{parity-games}

The characterisation given for simple safety game is not enough for
parity objectives, as we will see in the following example.

\begin{figure}
  \begin{center}
    \begin{tikzpicture}
      \draw (0,0) node[draw, circle, inner sep=4pt] (I) {$v_0$};
      \draw (I.-90) node[below] {};
      \draw (4,0) node[draw, inner sep=4pt, circle] (C1) {$v_1$};
      \draw (C1.-100) node[below] {};
      \draw (8,0) node[draw, circle, inner sep=4pt] (C2) {$\Omega_1$};
      \draw (I.-90) node[below] {$1$};
      \draw (C1.-90) node[below] {$1$};
      \draw (C2.-90) node[below] {$2$};
      \draw[-latex'] (-1, 0) -- (I);
      \draw[-latex'] (I) -- node[above]{$(a,*)$} (C1);
      \draw[-latex'] (I) ..controls +(1,2) and +(-1,2)  .. node[above]{$(b,*)$}  (I);
      \draw[-latex'] (C1) -- node[above]{$(*,a)$}(C2);
      \draw[-latex'] (C1) .. controls +(-2, -2) ..  node[above]{$(*,b)$} (I);
      \draw[-latex', rounded corners] (C2)..controls +(1,2) and +(-1,2) .. node[above]{$(a,*)$}(C2);
    \end{tikzpicture}
  \end{center}
    \caption{Parity game where the objective for player 1 is to visit
      $\Omega_1$ infinitely often. Player 1 controls $v_0$m and player 2
      controls $v_1$.}
    \label{15-fig:adm-parity}
\commentAlt{Figure~\ref{15-fig:adm-parity}: A directed graph with three circular nodes, v0, v1, and Omega1, showing transitions with labeled pairs and self-loops.}
\commentLongAlt{Figure~\ref{15-fig:adm-parity}: The image displays a directed graph with three circular nodes arranged horizontally. From left to right, they are labeled 'v0', 'v1', and 'Omega1'. An incoming arrow points to 'v0', indicating it as a starting state. Below 'v0', it is labeled '1'. Below 'v1', it is also labeled '1'. Below 'Omega1', it is labeled '2'.
- Node 'v0' has a self-loop labeled '(b,*)'.
- A straight arrow from 'v0' points to 'v1', labeled '(a,*)'.
- A curved arrow from 'v1' points back to 'v0', labeled '(*,b)'.
- A straight arrow from 'v1' points to 'Omega1', labeled '(*,a)'.
- Node 'Omega1' has a self-loop labeled '(a,*)'.}
\end{figure}

\begin{example}
  Consider the example in~\Cref{15-fig:adm-parity}.
  In this example, although the strategy that always stays in $v_0$
  does not decrease the value of player 1, it is dominated because
  it has no chance of winning.
  By contrast the strategy that always go to $v_1$ has a chance of
  being helped by player 2 and actually reaching $\Omega_1$ it therefore
  dominates the first strategy.
\end{example}

However the fact that an admissible strategy should not decrease its own
value still holds. Assuming strategy \(\sigma_i\) of player \(P_i\) does
not decrease its own value, we can classify its outcome in three
categories according to their ultimate values.

\begin{itemize}
\item   either ultimately \(\opt_i = 0\), in which case all strategies are
  losing, and thus any strategy is admissible
\item   or ultimately \(\pes_i = 1\), in which case admissible strategies are
  exactly the winning ones
\item   or ultimately \(\pes_i = 0\) and \(\opt_i = 1\). We will focus on this
case which is more involved.
\end{itemize}

From a state of value \(0\), an admissible strategy of \(P_i\) should
allow a winning play for \(P_i\) with the help of other players.

We write \(H_i\) for set of vertices \(v\) controlled by a player
\(P_j\ne P_i\) that have at least two successors of optimistic value
\(1\). Formally, the \emph{help-states} $H_i$ of player \(P_i\) are defined
as:
\[
 \bigcup_{P_j\in \Agt \setminus\{i\}} \left\{ s \in V_j \mid \exists s',s'',\ s' \neq s'' \wedge\ s\rightarrow s' \land s \rightarrow s'' \wedge\ \opt_i(s') = 1 \wedge\ \opt_i(s'') = 1 \right\}.
\]

Intuitively, admissible strategies in the case satisfying \(\pes_i = 0\) and \(\opt_i = 1\)
are those that visit infinitely often help states. In fact, letting other players make choices
means that the player is allowing the possibility of them helping to achieve the objective.
More precisely, we have the following property whose proof is omitted.
\begin{lemma}
Let \(v\in V\), \(P_i\in \Agt\) and \(\rho\) a play be
such that \(\exists^\infty k. \opt_{i}(\rho_k) = 1\). There exists
\(\sigma_i\) admissible such that \(\rho \in \Out(v, \sigma_i)\) if, and
only if, \(\payoff_i(\rho) = 1\) or
\(\exists^\infty k. \rho_k \in H_i\).
\end{lemma}

\subsection{Iterated elimination}\label{iterated-elimination}

Once each player is restricted to use admissible strategies, they can
further refine their choices knowing that other players will not be using dominated strategies.
We already saw this in the example of \Cref{15-tbl:normal-adm}.
In fact, once player~$1$ has eliminated strategy~$A$ (which is dominated by~$B$),
player~$2$ can use this information since its best response against the remaining strategies is~$C$.
In more complex games, this reasoning can be repeated and take several steps before converging.
This repeated process is called \emph{iterated elimination of dominated strategies}.

We now define this process formally.
\begin{definition}[Iterated elimination]
The \emph{sequence of iterative elimination} is a sequence of rectangular strategy sets defined as follows.
$S^0 = (S_i^0)_{P_i \in \Agt}$ is the set of all strategies.
For $k~\geq 0$, if we write $S^k = (S_i)_{P_i \in \Agt}$,
then~$S^{k+1}_i$ is the set of strategies in~$S^{k}_i$ that are not dominated in~$S^k$.
\end{definition}

Thus, the step~$1$ of the sequence of iterative elimination corresponds
to admissible strategies defined above. Let us call them $1$-admissible.
In step 2, we again compute strategies that are dominated by only
considering $1$-admissible strategies for all players, and repeat.
Strategies that survive all step of elimination are said \emph{iteratively admissible}.

\begin{theorem}
  In parity games, the sequence of iterative elimination converges, and it reaches a non-empty fixed point.
\end{theorem}
We prove this result only for simple safety games since the case of parity conditions is too complex for the scope of this book.

Intuitively, given a game~$G$, \Cref{15-thm:adm} tells us that any strategy for player~$P_i$ that avoids using edges~$D_i$ is admissible. So if we remove all edges~$\cup_{P_i \in \Agt} D_i$ from the
game to obtain a new game called~$G_1$, then all strategies of~$G_1$ (for all players) are admissible in~$G$, and conversely.
We can then repeat this process to~$G_1$: we construct~$G_2$ by eliminating all dominated edges
in~$G_1$, and get that admissible strategies in~$G_1$ are exactly all strategies of~$G_2$,
which correspond to~$2$-admissible strategies, and so on.

Since the size of the games decrease at each step, this process necessarily stops.
It remains to show that the limit game~$G_\infty$ contains strategies. We will show that
all vertices have at least one outgoing edge in~$G_\infty$.
It suffices to show that the sets~$D_i$ never contains all edges leaving a vertex.
Let us consider any game~$G_j$.
For a vertex~$v \in V_i$ with~$\pes_i(v)=0$ and~$\opt_i(v)=0$, none of the edges are dominated.
For a vertex~$v \in V_i$ with~$\pes_i(v)=1$ (and necessarily~$\opt_i(v)=1$), there exists a winning strategy in~$G_j$ so
there must be a successor~$v'$ with~$\pes_i(v')=1$ which is not dominated.
Last, for a vertex~$v \in V_i$ with~$\pes_i(v)=0$ and~$\opt_i(v)=1$, there exists a winning play from~$v$ so for some successor~$v'$ we must have~$\opt_i(v')=1$ which is an edge that is not dominated.

\section*{Bibliographic references}
\label{15-sec:references}
Most results about equilibria fall into two-categories: they either prove
that equilibria always exist for some class of games, or they characterise
the complexity of finding a particular one.

\subsection*{Existence results}
Several authors have noticed that Nash equilibria always exist in turn-based
game for some classes of objectives, in particular this is true of
$\omega$-regular objectives.
The most general result of that kind shows that this is true for all objectives for which there exist finite memory optimal strategies~\cite{Roux.Pauly:2018}.

The notion of equilibrium we now call Nash equilibrium was
defined in the article of Nash~\cite{Nash:1950} in which he proves the
existence for a class of normal form game.
The Hawk-dove game we presented as an example in the first part of
this chapter is also called game of chicken.
The first reference to this game was by Smith and Price \cite{Smith.Price:1973}.
The example of medium of access control we presented as a motivation was
studied from a game theoric point of view in \cite{MacKenzie.Wicker:2003}.

The notion of subgame perfect equilibria is interesting because in games on
trees (or extensive games), for which they were originally introduced, they
always exist. This results can be extended to games played on graphs.
In particular subgame perfect equilibria always exist in reachability
games~\cite{Brihaye.Bruyere.ea:2012}.

\subsection*{Algorithms and complexity results}


The deviator construction and the algorithm presented in this chapter are based on \cite{Bouyer.Brenguier.ea:2011}.
Algorithms on admissible strategies on infinite games and the complexity of the related problems
were studied in
\cite{Berwanger:2007}.

Imperfect information games in the context of multiplayer games are difficult.
As soon as there are `information forks' interesting problems are undecidable.
Deciding whether two players can ensure an objective against a third player
is undecidable.
As a corollary the Nash equilibrium problem is also undecidable~\cite{Pnueli.Rosner:1990}.
The problem of Nash equilibrium is also undecidable in stochastic games even
with only three players~\cite{Bouyer.Markey.ea:2014}.

In the first section we presented a polynomial algorithm for
finding pure Nash equilibria in normal form games.
It is actually also possible to find a mixed Nash equilibria in
polynomial time using linear programming.
The same extends to finding positional mixed Nash equilibria in concurrent
games, and even resilient equilibria~\cite{Brenguier:2016}.

Action-graphs are succinct representation of matrix games. Indeed,
representing games with matrices can be costly when the number of
players increases. The size of the matrix is in fact exponential in the
number of players: when each player has two strategies there are
\(2^{\Agt}\) cells in the table.
The action-graph representation is more compact, and the representation can be
exponentially smaller.
Because of that, the algorithm is no longer polynomial.
If there are no constraints on the Nash equilibrium we are looking for, the
complexity of the problem cannot be characterised using classical classes
like \NP-completeness because equilibria always exist and thus the answer to the decision problem would always be true.
The characterisation of the complexity was done using the PPAD class~\cite{Daskalakis.Goldberg.ea:2009}.


Nash equilibria with LTL objectives is expressible in logics such as
strategy logic \cite{Chatterjee.Henzinger.ea:2010} or ATL$^\ast$~\cite{Alur.Henzinger.ea:2002}, as well as other extensions of this equilibria.
However, satisfiability in these logic is difficult: it is
2\EXP-complete for ATL$^\ast$ and undecidable for strategy logic in general.
A decidable fragment of strategy logic has been identified \cite{Mogavero.Murano.ea:2012}, but remains difficult; it is 2\EXP-complete.

\ifpictures
\includepdf{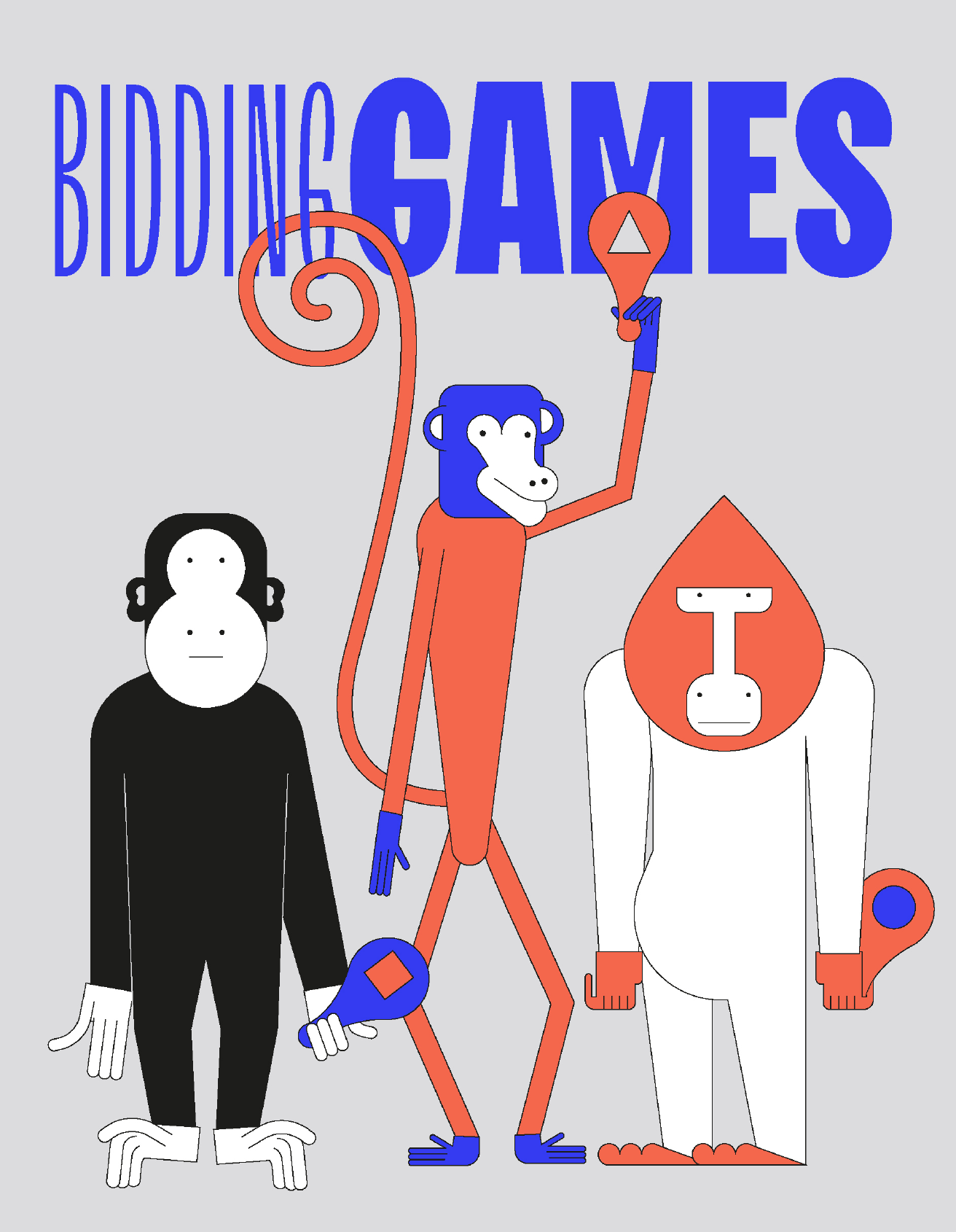}
\fi
\author[Guy Avni, Thomas A. Henzinger]{Guy Avni, Thomas A. Henzinger}
\copyrightline{Copyright by Guy Avni and Thomas A. Henzinger 2025, to be published by Cambridge University Press in the volume \textit{Games on Graphs} edited by Nathana\"el Fijalkow}

\chapter{Bidding Games}
\chapterauthor{Guy Avni, Thomas A. Henzinger}
\label{16-chap:bidding}

\newcommand{\zug}[1]{\langle #1  \rangle}
\newcommand{\stam}[1]{}
\newcommand{\floor}[1]{\lfloor #1 \rfloor}
\newcommand{\thresh}{\texttt{Th}\xspace}
\newcommand{\threshD}{\texttt{D-Th}\xspace}
\newcommand{\Trg}{\mathrm{Trg}}
\newcommand{\RT}{\mathrm{RT}}
\newcommand{\fr}{\mathrm{fr}}
\newcommand{\notF}{V \setminus F}
\renewcommand{\O}{{\cal O}}
\newcommand{\Spare}{\text{Spare}\xspace}

\providecommand{\energy}{\text{energy}}
\providecommand{\payoff}{\text{payoff}}
\newcommand{\cycles}{\text{cycles}}
\newcommand{\lolli}{\G_{\bowtie}}
\newcommand{\St}{\mbox{St}}
\newcommand{\Pot}{\mbox{Pot}}
\newcommand{\invest}{\text{Invest}}
\newcommand{\gain}{\text{Gain}}

\newcommand{\succb}[1]{#1 \oplus 0^*\xspace} 
\newcommand{\predb}[1]{#1 \ominus 0^*\xspace}

As seen in previous chapters, a graph game proceeds by placing a token on one of the vertices and allowing the players to move it throughout the graph to produce an infinite trace, which determines the winner or payoff of the game. Graph games can be categorized according to two orthogonal features: the ``mode of moving'' the token and the objective. Several modes of moving have already been introduced: the simplest and basic mode is turn-based or alternating turns (\Cref{1-chap:introduction}), stochastic moves have been considered in~\Cref{6-chap:mdp,7-chap:stochastic}, and concurrent moves in~\Cref{8-chap:concurrent}. 

In this chapter we study the {\em bidding} mode of moving: both players have budgets, and in each turn, we hold an ``auction'' (bidding) to determine which player moves the token. We will define concrete bidding mechanisms and study the properties of the resulting games when combined with various objectives. 

We adapt the central questions in graph games to bidding games:

First, the question of determinacy; namely, whether from every configuration in a qualitative game, one of the players has a pure winning strategy. 
In bidding games, we study the {\em threshold budget ratio}, which is roughly a ratio of the total budget that is both necessary and sufficient for winning the game. We stress that we focus on pure strategies. Even though the players' actions in bidding games are concurrent, a threshold ratio exists in many bidding games. For mean-payoff games, we will show existence of a value and study its dependence on the initial budget ratio. 

Second, construction of winning strategies. In turn-based games, the challenge in constructing strategies is to decide how to move the token from each vertex. 
Strategies in bidding games consist of two components. Deciding how to move the token is often relatively straightforward whereas the challenge is choosing the right bid.
On the one hand, the bid cannot be too high so that sufficient funds are available for subsequent bids, and it must be high enough so that the opponent must pay sufficiently high for winning the bidding.

Third, the complexity question of finding the winner in a graph game is adapted to the complexity question of finding the thresholds. 
 
Finally, we will show surprising equivalences between bidding games and a sub-class of stochastic games called {\em random-turn games}. 

We define concrete bidding mechanisms. In all the mechanisms that we consider, both players simultaneously submit bids that do not exceed their available budgets, the higher bidder wins the bidding, and moves the token. The mechanisms differ in three orthogonal distinctions: (1)~{\em who pays}: we consider {\em first-price} in which only the higher bidder pays their bid and {\em all-pay} bidding in which both players pay their bids. (2)~{\em who is the recipient}: in {\em Richman bidding} (named after David Richman), payments are made to the other player, and in {\em poorman bidding} the payments are made to the ``bank'' thus the money is lost. A third payment scheme called {\em taxman} spans the spectrum between Richman and poorman. (3)~{\em which bids are allowed}: in {\em continuous-bidding}, no restrictions on the granularity of the bids are imposed, whereas in {\em discrete bidding}, the budgets are given in coins and the minimal positive bid a player can make is one coin. 

This chapter focuses on first-price bidding games. We discuss results on all-pay bidding games in \Cref{16-sec:references}.
As a rule of thumb, Richman bidding is technically more approachable than the other bidding mechanisms so for ease of presentation, we will mostly focus on this bidding mechanism. 


\section{Qualitative continuous-bidding games}
\label{16-sec:qualitative_continuous}
Most of this section is devoted to the study of reachability bidding games. 
We will show that bidding games with general qualitative objectives reduce to reachability bidding games. 
We will show existence of thresholds and identify their structure. Based on the structure, we will both develop algorithms for computing thresholds and construct winning strategies based on the thresholds. We will show that reachability Richman-bidding games (and only them) are equivalence to a sub-class of stochastic games called random-turn games. 

\subsection{Definitions}
\label{16-sec:def-cont}
A bidding games is played on a graph  $\zug{\vertices, E}$, where $\vertices$ is a set of vertices and $E \subseteq \vertices \times \vertices$ are directed edges. For $v \in \vertices$, we denote by $N(v)$, the {\em neighbors} of $v$, thus $N(v) = \set{u \in \vertices: E(v, u)}$.

\subsubsection{First-price bidding} 
We make the first-price payment schemes that we consider precise. Denote Eve's budget by $B_\mEve$ and Adam's budget by $B_\mAdam$. Suppose that Eve bids $b_\mEve \in [0,B_\mEve]$ and Adam bids $b_\mAdam \in [0,\mAdam]$. 
Assume that Eve wins the bidding, and the definitions for Adam are dual. The budgets are updated as follows:
\begin{itemize}
\item {\bf Richman:} $B'_\mEve = B_\mEve - b_\mEve$ and $B'_\mAdam = B_\mAdam + b_\mEve$.
\item {\bf Poorman:} $B'_\mEve = B_\mEve - b_\mEve$ and $B'_\mAdam = B_\mAdam$.
\item {\bf Taxman:} For a fixed $\tau \in [0,1]$, we have $B'_\mEve = B_\mEve - b_\mEve$ and $B'_\mAdam = B_\mAdam+(1-\tau) \cdot b_\mEve$.
\end{itemize}

A key quantity in bidding games is the ratio between the players' budgets, or equivalently, the ratio of a player's budget from the total budget:

\begin{definition} {\bf (Budget ratio).}
When the budgets are $B_\mEve$ and $B_\mAdam$, Eve's ratio is $\frac{B_\mEve}{B_\mEve + B_\mAdam}$ and Adam's ratio is $\frac{B_\mAdam}{B_\mEve + B_\mAdam}$.
\end{definition}

We will usually normalize the sum of budgets to $1$, \textit{i.e.}, $B_\mEve+B_\mAdam=1$. In particular, in Richman bidding, the sum of budgets is constant throughout the game, so a player's budget ratio is in fact their budget. In poorman bidding, we normalize the sum to $1$ following every bidding, thus when Eve wins the bidding, the budgets are updated to $B'_\mEve = \frac{B_\mEve- b_\mEve}{B_\mEve + B_\mAdam - b_\mEve}$ and $B'_\mAdam = \frac{B_\mAdam}{B_\mEve + B_\mAdam - b_\mEve}$. 

\subsubsection{Bidding games as concurrent games}
The formal semantics of a bidding game is given by the explicit concurrent game that it represents. 
A {\em configuration} of a bidding game is $c = \zug{v, B_\mEve, B_\mAdam}$, meaning that the token is placed on $v \in \vertices$, Eve's budget is $B_\mEve$ and Adam's budget is $B_\mAdam$. An action in $c$ is a pair $\zug{u, b}$, where $u \in V$ is a neighbor of $v$ and $b \in [0,1]$ is a bid that does not exceed the available budget. 
We describe the transitions of the concurrent game. 
Suppose that the players (concurrently) choose the actions $\zug{u_\mEve, b_\mEve}$ and $\zug{u_\mAdam, b_\mAdam}$. 
Let $B'_\mEve$ and $B'_\mAdam$ be the updated budgets as defined above. Then, if Eve wins the bidding, the next configuration is $\zug{u_\mEve, B'_\mEve, B'_\mAdam}$ and otherwise it is $\zug{u_\mAdam, B'_\mEve, B'_\mAdam}$. 

Note that transitions in the concurrent game are deterministic (unlike general concurrent games that allow stochastic transitions; see~\Cref{8-chap:concurrent}). Further note that both the possible actions and the number of configurations is infinite, since in continuous bidding, we allow bids of arbitrary precision. We will revisit this construction in discrete-bidding games, where the size is finite but exponential in the representation size of the game. 

We point to the issue of bidding ties, \textit{i.e.}, when $b_\mEve = b_\mAdam$.
In continuous bidding, in the questions that we study, we will avoid the issue of bidding ties (see~\Cref{16-rem:ties}). Still, for completeness, we arbitrarily break bidding ties in favor of Eve, namely we define that she wins a bidding if $b_\mEve \geq b_\mAdam$ and otherwise Adam wins. We will handle tie breaking explicitly in discrete bidding.

\subsubsection{Threshold ratios}
Informally, the main question that we consider is:
\begin{center}
{\it What is a necessary and sufficient initial budget ratio\\ that guarantees winning the game?}
\end{center}
We illustrate a solution to this question in the following example.

\begin{figure}[ht]
\centering
\begin{tikzpicture}
\begin{scope}[every node/.style={circle,thick,draw}]
    \node (A) at (0,0) {$t_2$};
    \node (B) at (2,0) {$v_0$};
    \node (C) at (4,0) {$v_1$};
    \node (D) at (6,0) {$t_1$};
\end{scope}

\node (N) at (-3, -1) {Richman:};
\node (N) at (0,-1) {$1$};
\node (N) at (2,-1) {$\frac{2}{3}$};
\node (N) at (4,-1) {$\frac{1}{3}$};
\node (N) at (6,-1) {$0$};

\node (N) at (-3, -1.6) {Poorman:};
\node (N) at (0,-1.6) {$1$};
\node (N) at (2,-1.6) {$\frac{\sqrt{5} -1}{2}$};
\node (N) at (4,-1.6) {$\frac{3-\sqrt{5}}{2}$};
\node (N) at (6,-1.6) {$0$};

\path [->, style={very thick}] (B) edge[bend right=20] (C);
\path [->,style={very thick}] (C) edge[bend right=20] (B);
\path [->, style={very thick}] (C) edge (D);
\path [->,style={very thick}] (B) edge (A);
\end{tikzpicture}
\caption{A reachability bidding game with the threshold ratios under first-price Richman and poorman bidding.}
\label{16-fig:reach}
\commentAlt{Figure~\ref{16-fig:reach}: A linear graph with three circular nodes (t2, v0, v1, t1) and a table below showing numerical and mathematical values associated with different points in the graph for "Richman" and "Poorman" categories. See long description.}
\commentLongAlt{Figure~\ref{16-fig:reach}: The image displays a linear directed graph at the top and a table of values below it.

The graph shows four circular nodes arranged horizontally: 't2' on the far left, 'v0' next, 'v1' next, and 't1' on the far right.
- An arrow from 'v0' points to 't2'.
- A bidirectional arrow connects 'v0' and 'v1'.
- An arrow from 'v1' points to 't1'.

The table below has two rows, "Richman:" and "Poorman:", and four columns corresponding to the positions of the nodes in the graph.
- For "Richman:":
    - Under 't2': 1
    - Under 'v0': 2/3
    - Under 'v1': 1/3
    - Under 't1': 0
- For "Poorman:":
    - Under 't2': 1
    - Under 'v0': (sqrt(5)-1)/2
    - Under 'v1': (3-sqrt(5))/2
    - Under 't1': 0}    
\end{figure}

\begin{example}
\label{16-ex:reach}
Consider the reachability Richman bidding game that is depicted in~\Cref{16-fig:reach}. Eve's goal is to reach $t_\mEve$. Adam ensures winning if the game reaches $t_\mAdam$. 
Throughout the examples, for short, we describe the players' budgets as a pair $\zug{B_\mEve, B_\mAdam}$.

We start with a naive solution by showing that Eve wins when her budget exceeds $0.75$. That is, for every $\epsilon >0$, she wins from each configuration $\zug{v_0, 0.75 + \epsilon, 0.25-\epsilon}$. We describe a winning strategy. In the first turn, Eve bids $0.25$ and wins the bidding since the highest bid Adam can choose is $0.25-\epsilon$. Eve pays Adam and moves the token right, thus the next configuration is $\zug{v_1, 0.5 + \epsilon,0.5 - \epsilon}$. In the second bidding, Eve bids all her budget, which exceeds Adam's budget, thus she wins the bidding, moves the token to $t_\mEve$, and wins the game. 

A ratio of $0.75$ thus suffices for winning, but it is not necessary. The necessary and sufficient budget is in fact $\frac{2}{3}$. Formally, we show that for every $\epsilon > 0$, Eve wins from configuration $\zug{v_0, \frac{2}{3}+\epsilon, \frac{1}{3}-\epsilon}$. We describe a winning strategy. Eve's first bid is $\frac{1}{3}$. She ensures winning the bidding since she bids higher than Adam's budget, thus the second configuration is $\zug{v_1, \frac{1}{3}+ \epsilon, \frac{2}{3} - \epsilon}$. At $v_1$, it is dominant for Eve to bid her whole budget of $\frac{1}{3} + \epsilon$. If she wins the bidding, she proceeds to $t_\mEve$ and wins the game. Otherwise, Adam wins the bidding, and moves the token back to $v_0$. Note that in order to win the bidding, Adam overbids Eve's bid and pays her at least $\frac{1}{3}+\epsilon$. Thus, the next configuration is $\zug{v_0, \frac{2}{3}+2\epsilon, \frac{1}{3}-2\epsilon}$. The main idea is in the small details: we are back in the same position, but Eve's budget increases by $\epsilon$. She can repeatedly restart her strategy and force that if the game does not reach $t_\mEve$ and returns to $v_0$, her budget grows by at least $\epsilon$. Eventually her budget will exceed $0.75$ from which  she can force the game to $t_\mEve$ as described above. 

Showing that $\frac{2}{3}$ is necessary is shown in a dual manner. The same reasoning applied to Adam shows that for every $\epsilon >0$, he can force the game to $t_\mAdam$ from configurations $\zug{v_1, \frac{1}{3}-\epsilon, \frac{2}{3} + \epsilon}$ and $\zug{v_0, \frac{2}{3}-\epsilon, \frac{1}{3}+\epsilon}$. 
\hfill\qed
\end{example}

We define threshold ratios formally.

\begin{definition}
\label{def:thresholds}
{\bf (Threshold ratios).} 
Consider a bidding game over a set of vertices $V$ and let $B_\mEve \in [0,1]$ be Eve's initial budget ratio. The threshold ratio in a vertex $v \in V$, denoted $\thresh(v) \in [0,1]$ is such that
\begin{itemize}
\item when $B_\mEve > \thresh(v)$, Eve can win the game from $v$, and 
\item when $B_\mEve < \thresh(v)$, Adam can win the game.
\end{itemize}
\end{definition}

\begin{remark}
\label{16-rem:ties}
{\bf (Avoiding tie-breaking).} 
The definition of thresholds carefully avoids the case in which Eve's budget is exactly on the threshold (\textit{i.e.}, when $B_\mEve = \thresh(v)$). This is meant to circumvent the issue of ties in bidding. Indeed, the winner from a configuration $\zug{v, \thresh(v), 1-\thresh(v)}$ depends on the tie-breaking mechanism. With the current definition of thresholds, the result apply to any tie-breaking mechanism that is used. Having said that, recall that we arbitrarily break ties in favor of Eve. This implies that in fact Eve wins from configuration $\zug{v, \thresh(v), 1-\thresh(v)}$. 
\end{remark}

\begin{remark}
{\bf (Existence of thresholds).}
We stress that we restrict the players to follow pure winning strategies, namely strategies that guarantee surely winning. 
Recall that bidding games are succinctly-represented concurrent games, and the latter are not in general determined under pure strategies. For example, the game ``matching pennies'' (the players simultaneously select ``heads'' or ``tails'' and Eve wins iff they choose the same) can be formalized as a one-round concurrent reachability game in which neither player has a pure winning strategy. 
Existence of thresholds in bidding games is not trivial since it implies determinacy. Indeed, consider a configuration $c = \zug{v, B_\mEve, 1-B_\mEve}$, then Eve wins from $c$ if $B_\mEve > \thresh(v)$, Adam wins from $c$ if $B_\mEve < \thresh(v)$, and the case of $B_\mEve = \thresh(v)$ is discussed in the remark above. 
\end{remark}


\subsection{Reachability continuous-bidding games}
In this section, we study threshold budgets in reachability bidding games.
We start with games played on directed acyclic graphs (DAGs). 
Such a game is $\game = \zug{\vertices, E, \Trg}$, where the graph $\zug{\vertices, E}$ is a DAG, its set of sinks is $S \subseteq \vertices$, and $\Trg \subseteq S$ is a target set for Eve. 
Note that each play of $\game$ necessarily terminates in a sink $s \in S$. Eve wins a play iff $s \in \Trg$. 

\begin{theorem}
\label{16-thm:DAG}
Threshold budget ratios exist in every taxman-bidding game that is played on a DAG.
\end{theorem}
\begin{proof}
We describe the proof for Richman bidding. The proofs for poorman and taxman follow similar ideas, and we will point to the steps that need adaptations. 

We inductively define a function $T: \vertices \rightarrow [0,1]$, and prove that it coincides with the thresholds. 
For the base case, consider a sink $s \in S$. 
We define: 
\[T(s) = \begin{cases}0 & \text{ if } s \in \Trg\\1 & \text{ if } s \notin \Trg\end{cases}\]
Intuitively, $T(s) = \thresh(s)$ since Eve requires no budget in order to win from $s \in \Trg$ and cannot win even if she has all the budget from $s \notin \Trg$. Note that technically, in the latter case, since the sum of budgets is $1$, it is vacuously true that Eve wins from $s \notin \Trg$ with a budget of $1+\epsilon$, for $\epsilon >0$. 

For the inductive step, consider a vertex $v$ such that $T$ is defined in all of its children and for each child $u$ of $v$, we have $T(u) = \thresh(u)$. Let $v^+$ and $v^-$ be two children of $v$ that respectively achieve the maximum and minimal values of $T$, thus $T(v^+) = \argmax_{u \in N(v)} T(u)$ and $T(v^-) =  \argmin_{u \in N(v)} T(u)$. In case there are multiple such children, choose one arbitrarily. Define: \[T(v) = \frac{1}{2} \cdot \big( T(v^-) + T(v^+) \big)\]
We assume that every non-sink vertex $v \notin S$ has a path both to $\Trg$ and to $S \setminus \Trg$. If $v$ does not have such a path, then the reasoning in $v$ is as in sinks above. The assumption implies that $T(v)$ is strictly between $0$ and $1$.

We claim that $T(v) = \thresh(v)$. Suppose that Eve's budget is $T(v) + \epsilon$, and we describe a winning strategy. Her bid at v is: \[b_\mEve = \frac{1}{2} \cdot \big(T(v^+) - T(v^-)\big)\] 
Upon winning, she moves the token to $v^-$. That is, her action at $v$ is $\zug{v^-, b_\mEve}$. Let $\zug{u, b_\mAdam}$ be Adam's choice of action. We distinguish between two cases. First, Eve wins the bidding. Note that $T(v) - b_\mEve = T(v^-)$, thus the next configuration is $\zug{v^-, T(v^-) + \epsilon, 1-T(v^-) + \epsilon}$, which is winning for Eve by the induction hypothesis. Second, suppose that Adam wins the bidding. Note that $T(v) + b_\mEve = T(v^+)$ and $b_\Adam \geq b_\mEve$. Eve's budget is updated to $T(v) + b_\mAdam + \epsilon +  > T(v^+) \geq T(u)$, thus again by induction, Eve wins from the resulting configuration.

We show that Adam wins from $v$ when Eve's budget is $T(v) - \epsilon$. This is done by ``flipping'' the game and associating Adam with Eve and vice versa, and applying the solution above. Formally, define $T'(v) = 1-T(v)$. Note that in the sinks, $T'(s) = 0$ iff $s \notin \Trg$, which matches the intuition that Adam wins in $s \notin \Trg$. Further note that since $B_\mEve + B_\mAdam = 1$, we have $B_\mEve < T(v)$ implies $B_\Adam > 1-T(v) = T'(v)$. Finally, observe: \[T'(v) = 1-T(v) = 1- \frac{T(v^+) + T(v^-)}{2} = \frac{1-T(v^+) + 1-T(v^-)}{2} = \frac{T'(v^+) + T'(v^-)}{2}\]
Thus, when $B_\mAdam > T'(v)$, Adam can apply the winning strategy above to force the game to $S \setminus \Trg$. This concludes the proof for Richman bidding.

For poorman and taxman bidding, the proof follows a similar structure. We inductively define functions $P: \vertices \rightarrow [0,1]$ for poorman bidding and $X:\vertices \rightarrow [0,1]$ in taxman bidding. For $s \in S$, we define $P(s) = T(s) = 0$ if $s \in \Trg$ and $P(s)= T(s) =1$ if $s \notin \Trg$. For a non-sink vertex $v \in \vertices$, let $v^-$ and $v^+$ be two of its children that respectively obtain the minimal and maximal values according to $P$. Then, 
\[P(v) = \frac{P(v^-)}{1-P(v^-) + P(v^+)} \  \text{ and } \ b_\mEve = B_\mEve \cdot \frac{P(v) - P(v^-)}{P(v) \cdot \big(1-P(v^-)\big)} = 
B_\mEve \cdot \frac{P(v^+) - P(v)}{P(v) \cdot P(v^+)}.\]
It is useful to verify that when $B_\mEve = P(v)$, we have $P(v^-) = \frac{B_\mEve - b_\mEve}{1-b_\mEve}$, that is after winning a bidding, Eve's normalized updated is greater than $P(v^-)$, and $P(v^+) = \frac{B_{\mEve}}{1-b_\mEve}$, that is after losing a bidding, Eve's normalized updated budget is greater than $P(v^+)$.

In the case of taxman with parameter $\tau \in [0,1]$, the definition is: 
\[X(v) = \frac{X(v^-) + X(v^+) - \tau\cdot X(v^-)}{2- \tau\cdot\big(1+ X(v^-)-X(v^+) \big)} \text{ and } b_\mEve = \frac{X(v^+) - X(v^-)}{2- \tau\cdot\big(1+ X(v^-)-X(v^+)}\]
It is useful to verify that for $\tau = 0$, we have $X(v) = T(v)$, that is taxman coincides with Richman, and for $\tau=1$, we have $X(v) = P(v)$, that is taxman coincides with poorman. 
\end{proof}

We proceed to show existence of thresholds in general games.
\begin{theorem}
\label{16-thm:reach}
Threshold budgets exist in reachability taxman-bidding games.
\end{theorem}
\begin{proof}
We prove for Richman bidding and the proof for the other mechanisms follows a similar structure. 
Consider a reachability Richman-bidding game $\game = \zug{\vertices, E, \Trg_\mEve}$, where $\Trg_\mEve \subseteq \vertices$ are the target vertices for Eve. Let $\Trg_\mAdam \subseteq \vertices$ be the set of vertices from which there is no path to $\Trg_\mEve$. Clearly, Adam wins if the game reaches $\Trg_\mAdam$. In fact, we will show that Adam can only win by reaching $\Trg_\mAdam$. That is, he cannot win in an infinite play that traverse infinitely often, a vertex with a path to $\Trg_\mEve$. 

For $n \in \N$, let $\game[n]$ be a {\em truncated} game in which Eve wins iff she reaches a target vertex by turn $n$. Formally, $\game[n]$ is a game played on a DAG. Let $\thresh_n$ denote the threshold ratios in $\game[n]$ as in \Cref{16-thm:DAG}. Let $T(v) = \lim_{n \to \infty} \thresh_n(v)$. We claim that the limit exists. Indeed, the sequence is monotonically decreasing since the budget to guarantee winning in $n+1$ steps suffices to win in $n$, thus $\thresh_n(v) \geq \thresh_{n+1}(v)$. It is also bounded from below by $0$. 

We show two proofs that Eve wins from $v$ with a budget of $T(v) + \epsilon$. The first proof is simple. Since the sequence of thresholds tends to $T(v)$, there is $n \in \N$ such that $\thresh_n(v) < T(v) + \epsilon$. Eve wins in $\game$ by following a winning in $\game_n$. 

The second proof is more useful. We say that a function $f: \vertices \rightarrow [0,1]$ satisfies the {\em average property} if:
 \[f(v) = \begin{cases}
 0 & \text{ if } v \in \Trg_\mEve\\
1 & \text{ if } v \in \Trg_\mAdam\\
\frac{f(v^+) + f(v^-)}{2} & \text{ otherwise}
 \end{cases}\]
where $v^+$ and $v^-$ are neighbors of $v$ that respectively achieve the maximal and minimal values according to $f$ among the neighbors. Observe that $T$ satisfies the average property. Indeed, since each $\thresh_n$ satisfies the average property, so does the limit. 

A function that satisfies the average property gives rise to the following Eve winning strategy. Suppose that the game starts from $v$ with an Eve budget of $B_\mEve = T(v) + \epsilon$. Let $m = |\vertices|$. We call $\epsilon$, Eve's {\em spare change}. Eve's strategy guarantees that (1) when the game reaches $u \in \vertices$, her budget exceeds $T(u)$, and (2) if the game does not reach $\Trg_\mEve$, her spare change increases by a constant. Point~(1) implies that Eve does not lose. Indeed, for every $s \in \Trg_\mAdam$, we define $T(s) = 1$, thus reaching $s$ means that Eve's budget exceeds $1$, which is greater than the total budget.
Point (2) implies that her budget eventually exceeds $1- 2^{-m}$, which suffices to guarantee winning $m$ biddings in a row. 

Eve's strategy proceeds as follows. She splits $\epsilon$ into exponentially increasing parts $\epsilon_1,\ldots, \epsilon_m$, so that for every $1 \leq i \leq m$, we have $2 \sum_{1 \leq j < i} \epsilon_j \leq \epsilon_i$ and $\sum_{1 \leq j \leq m} \epsilon_j \leq \epsilon$. At turn~$i$, assuming the token is at vertex $u \in \vertices$, she bids $0.5 \cdot \big(T(u^+) - T(u^-)\big) + \epsilon_i$. Showing that the strategy ensures Point~(1) is as in the proof of \Cref{16-thm:DAG}. For Point~(2), it is not hard to show that if Eve wins $m$ consecutive biddings, she reaches $\Trg_\mEve$. Suppose that she loses the bidding at turn $k \geq 1$. In the previous turns $1,\ldots,k-1$ her spare change decreases by $\sum_{1 \leq j < k} \epsilon_j$ and in the $k$-th turn, the spare change increases by at least $\epsilon_k \geq 2 \sum_{1 \leq j < k} \epsilon_j$. Thus, her spare change increases by at least a constant $\epsilon_1 \geq \sum_{1 \leq j < k} \epsilon_j$. 

To conclude the proof of the theorem, we show that $T \equiv \thresh$ by showing that Adam wins from $v \in \vertices$ when Eve's budget is $T(v) - \epsilon$, for every $\epsilon > 0$. We flip the game. As in \Cref{16-thm:DAG}, the function $T': \vertices \rightarrow [0,1]$ defined by $T'(v) = 1-T(v)$ satisfies the average property from Adam's perspective. Applying the same construction above for Adam shows that he has a strategy that forces the game to $\Trg_\mAdam$, and we are done. 
\end{proof}

\begin{corollary}
\label{16-cor:unique}
{\bf (Uniqueness of functions that satisfy the average property).} 
Assume towards contradiction that there are two different functions $f_1$ and $f_2$ that satisfy the average property in a game $\game$. Let $v \in \vertices$ such that $f_1(v) < f_2(v)$. In the proof of \Cref{16-thm:reach}, we construct a winning strategy for Eve when her budget is $f_1(v) + \epsilon$, for every $\epsilon>0$. In addition, we show that $1-f_2$ is a function that satisfies the average property from Adam's perspective, and construct a winning strategy for him when Eve's budget is $f_2(v) - \epsilon$, for every $\epsilon>0$. Thus, there is an initial budget $B_\mEve$ with $f_1(v) < B_\mEve < f_2(v)$ for which both players have a winning strategy, which is a contradiction in a zero-sum game. 
\end{corollary}

\subsubsection{An equivalence with random-turn games}
\Cref{16-thm:reach} implies an intriguing equivalence between reachability Richman-bidding games and random processes. Before stating it in full generality, we illustrate the ``plain vanilla'' version of the equivalence, which applies to games in which all vertices have out-degree at most $2$. 

\begin{figure}[ht]
\centering

\begin{tikzpicture}

\begin{scope}[every node/.style={circle,thick,draw}]
    \node (A) at (0,0) {$t_2$};
    \node (B) at (2,0) {$v_0$};
    \node (C) at (4,0) {$v_1$};
    \node (D) at (6,0) {$t_1$};
\end{scope}

\node (N) at (-2, -1) {$\Pr[\text{Reach } t_1]$:};
\node (N) at (0,-1) {$0$};
\node (N) at (2,-1) {$\frac{1}{3}$};
\node (N) at (4,-1) {$\frac{2}{3}$};
\node (N) at (6,-1) {$1$};

\path [->, style={very thick}] (B) edge[bend right=30] node[above]{$0.5$} (C) ;
\path [->,style={very thick}] (C) edge[bend right=30] node[above]{$0.5$} (B);
\path [->, style={very thick}] (C) edge node[above]{$0.5$} (D);
\path [->,style={very thick}] (B) edge node[above]{$0.5$} (A);

\end{tikzpicture}
\caption{The random-turn game that corresponds the game in~\Cref{16-fig:reach} with the probabilities to reach $t_1$ from each vertex.}
\label{16-fig:RTreach}
\commentAlt{Figure~\ref{16-fig:RTreach}: A linear graph with four circular nodes (t2, v0, v1, t1) and a row of numbers below, indicating probabilities. See long description.}
\commentLongAlt{Figure~\ref{16-fig:RTreach}: The image displays a linear directed graph at the top and a row of values labeled "Pr[Reach t1]:" below it.

The graph shows four circular nodes arranged horizontally: 't2' on the far left, 'v0' next, 'v1' next, and 't1' on the far right.
- An arrow from 'v0' points to 't2', labeled '0.5'.
- A bidirectional arrow connects 'v0' and 'v1'. The arrow from 'v0' to 'v1' is labeled '0.5'. The arrow from 'v1' to 'v0' is also labeled '0.5'.
- An arrow from 'v1' points to 't1', labeled '0.5'.

The row of values "Pr[Reach t1]:" corresponds to the nodes above them:
- Under 't2': 0
- Under 'v0': 1/3
- Under 'v1': 2/3
- Under 't1': 1}
\end{figure}

\begin{example}
Consider the game depicted in~\Cref{16-fig:reach}. We construct a Markov chain by labeling all edges with probability $0.5$ (see \Cref{16-fig:RTreach}). Consider the probability of reaching the target $t_1$ from a vertex $u$, denoted $\Pr[\text{reach}(u, t_1)]$. Clearly, we have $\Pr[\text{reach}(t_1, t_1)]=1$ and $\Pr[\text{reach}(t_2, t_1)]=0$.  As for the other two vertices, we have $\Pr[\text{reach}(v_0, t_1)] = \frac{1}{2} \Pr[\text{reach}(v_1, t_1)] + \frac{1}{2} \Pr[\text{reach}(t_2, t_1)]$, which is technically the same as the expression in~\Cref{16-thm:reach} for threshold ratios in Richman bidding. Since the values for the targets are reversed, for every vertex $u$, we have $\thresh(u) = 1-\Pr[\text{reach}(u, t_1)]$. \hfill\qed
\end{example}

The equivalence for games with out-degree greater than $2$ relates reachability Richman-bidding games with a class of stochastic games called {\em random-turn games}. Intuitively, we require a stochastic game since unlike games with out-degree $2$, where the threshold in a vertex is simply the average of its two neighbors, when the out-degree is greater than $2$, one needs to find the neighbors that attain the extreme thresholds among the neighbors.

\begin{definition}
{\bf (Random-turn games).} Consider a bidding game $\game$ that is played on a graph over a set of vertices $\vertices$. For $p \in [0,1]$, the random-turn game that corresponds to $\game$ w.r.t. $p$, denoted $\RT(\game, p)$, is a game in which instead of bidding, in each turn we toss a (biased) coin to determine which player gets to move the token: Eve is chosen with probability $p$ and Adam with probability $1-p$. Formally, $\RT(\game, p)$ is a stochastic game (\Cref{7-chap:stochastic}). Every vertex $v\in \vertices$, is replaced by three vertices $v_N, v_\mEve$, and $v_\mAdam$. The vertex $v_N$ simulates the coin toss, thus it has two outgoing edges: one with probability $p$ to the Eve vertex $v_\mEve$ and a second with probability $1-p$ to the Adam vertex $v_\mAdam$. The outgoing edges from $v_\mEve$ and $v_\mAdam$ have the same end points: they point to $u_N$, for every neighbor $u$ of $v$. The  objective in $\RT(\game,p)$ matches that of $\game$. For example, when $\game$ is a reachability bidding game, then $\RT(\game, p)$ is a simple stochastic game. 
\end{definition}

We state the general equivalence between the two models in the following theorem. 


\begin{theorem}
\label{16-thm:RTreach}
Consider a reachability first-price Richman bidding game $\game$ over the vertices $\vertices$. For every vertex $v \in \vertices$, we have $\thresh(v) = 1-val\big(\RT(\game, 0.5), v\big)$.
\end{theorem}
\begin{proof}
The function that assigns to each vertex $v \in \vertices$, its value in $\RT(\game, 0.5)$ is a function that satisfies the average property. By \Cref{16-cor:unique}, it coincides with the thresholds. It is not hard to see that the value coincides with Adam's thresholds. 
\end{proof}

\begin{remark}
\label{16-rem:reach-poorman}
{\bf (No equivalence for poorman bidding).} 
We point out that there is no known equivalence between reachability poorman-bidding games and random-turn games. We illustrate why it is unlikely that such an equivalence exists. 
We note that values in stochastic games are rational numbers; indeed, they are a solution to a linear program. However, as shown in~\Cref{16-fig:reach}, thresholds in poorman bidding games can be irrational.
Later we will show an equivalence between mean-payoff poorman-bidding games and random-turn games. 
\end{remark}


\subsection{Parity continuous-bidding games}
\label{16-sec:continuous-parity}
In this section, we show a reduction from parity bidding games to reachability bidding games. 

We describe intuition based on the equivalence with random-turn games (\Cref{16-thm:RTreach}). We describe roughly, an algorithm to find the probability of satisfying a B\"uchi objective in a Markov chain. The algorithm relies on two properties of Markov chains: (1)~a random walk reaches a bottom strongly-connected component (BSCC) almost surely (with probability $1$), and~(2) a random walk in a BSCC visits all states infinitely almost surely. It follows from (2) that random walks that reach a BSCC that contain an accepting B\"uchi state will satisfy the B\"uchi objective almost surely. Combining with (1), we obtain a reduction to reachability Markov chains: for every vertex that is not in a BSCC, the probability of satisfying the B\"uchi objective coincides with the probability of reaching a BSCC with an accepting state. 

The counterpart of Property~(1) in bidding games has already been shown; namely, \Cref{16-thm:reach} shows that for every initial ratio (apart from the threshold) either Eve or Adam can draw the game to their target. In other words, winning Adam plays do not ``stay'' in the arena, rather they eventually reach a sink. The counterpart of Property~(2) follows from the following lemma.

\begin{lemma}
\label{16-lem:SCC}
Consider a reachability first-price taxman-bidding game $\game = \zug{\vertices, E, \Trg}$ such that every vertex in $\vertices$ has a path to $\Trg$. Then, $\thresh \equiv 0$, namely Eve wins from all vertices with a positive initial budget. 
\end{lemma}
\begin{proof}
We focus on Richman bidding and the proof for taxman is similar. As in \Cref{16-thm:reach}, with an initial budget of $\epsilon > 0$, Eve splits her budget into exponentially increasing parts $\epsilon_1,\ldots, \epsilon_m$. At turn $i \geq 1$, she bids $\epsilon_i$ and upon winning the bidding, she proceeds to a vertex that is closer to $\Trg$. Her strategy guarantees that if she loses a bidding, her budget increases. She continues until her budget exceeds $1-2^{-|\vertices|}$, which suffices to win $|\vertices|$ consecutive bids and draw the game to $\Trg$. 
\end{proof}

We proceed to solve parity games.

\begin{theorem}
\label{16-thm:parity}
Parity first-price taxman-bidding games are linearly reducible to reachability taxman-bidding games. In particular, threshold ratios exist.
\end{theorem}
\begin{proof}
Consider a bidding game $\game$. We identify the BSCCs of $\game$. We claim that in each BSCC $S$, the thresholds are all either $0$ or all $1$. We call a BSCC winning for Eve if the thresholds are $0$ and otherwise it is winning for Adam. We then construct a reachability bidding game on the rest of the graph in which each player's goal is to draw the game to a winning BSCC. 

To prove the claim, suppose the maximal parity index in a BSCC $S$ is obtained in a vertex $t$ and that it is even, and the other case is dual. We think of $S$ as a reachability game in which Eve's target is $t$ and Adam has no target. By~\Cref{16-lem:SCC}, Eve can force a visit to $t$ with any positive initial budget. To force infinite many visits, she splits her initial budget $\epsilon > 0$ into infinite many parts $\epsilon_1, \epsilon_2,\ldots$, and uses a budget of $\epsilon_i$, for $i \geq 1$, to force a visit to $t$ for the $i$-th time.
\end{proof}

\subsection{Computational complexity}
Intuitively, the computational problem that we would like to solve is finding the threshold ratio in a vertex. Formally, one way to define the corresponding decision problem is given a game $\game$ and a vertex $v$ in $\game$, decide whether $\thresh(v) > 0.5$.

\begin{theorem}
For parity Richman bidding, finding threshold ratios is in NP $\cap$ coNP, and it is in P when all vertices have out-degree at most $2$ or when the graph is undirected. For taxman and poorman bidding the problem is in PSPACE.
\end{theorem}
\begin{proof}
We start with Richman bidding. The upper bound is based on the equivalence with random-turn games (\Cref{16-thm:RTreach}): in order to find thresholds in a reachability Richman-bidding game $\game$, we construct and solve the random-turn game $\RT(\game, 0.5)$. Solving stochastic games is known to be in NP $\cap$ coNP. 
When the out-degree is $2$, the solution is obtained by solving a linear program. 
A polynomial-time algorithm for undirected graphs was shown in \cite{Lazarus.Loeb.ea:1999}. 

For taxman bidding, it is not hard to express the properties of the thresholds (see $P(\cdot)$ and $X(\cdot)$ in \Cref{16-thm:DAG}) as a non-linear program, which we solve using the {\em existential theory of the reals} \cite{Canny:1988}.
\end{proof}

\begin{remark}
{\bf (Relating strategies in bidding and random-turn games).}
Consider a bidding game $\game$. We show how to obtain winning strategies in $\game$ from a solution to $\RT(\game, 0.5)$. 
Recall that intuitively, for a vertex $v \in \vertices$, the vertex $v^-$ and $v^+$ respectively represent the optimal moves of Eve and Adam upon winning a bidding in $v$. It is not trivial to identify $v^-$ and $v^+$ when the out-degree of $v$ is greater than $2$. We obtain this information from the solution of $\RT(\game, 0.5)$. Recall that stochastic games admit optimal positional strategies (\Cref{7-chap:stochastic}). Let $f_\mEve$ and $f_\mAdam$ be two optimal strategies respectively for the two players in $\RT(\game, 0.5)$. Recall that when constructing $\RT(\game, 0.5)$, we split every vertex $v$ into three vertices $v_N$, $v_\mEve$, and $v_\mAdam$, where the outgoing edges from $v_\mEve$ and $v_\mAdam$ direct to each $u_N$, where $u$ is a neighbor of $v$ in $\game$. Let $u_N = f_\mEve(v_\mEve)$ and $u'_N = f_\mAdam(v_\mAdam)$. It can be shown that $v^- = u$ and $v^+ = u'$ achieve the extreme thresholds among the neighbors of $v$. That is, a winning Eve strategy in $\game$ proceeds from $v$ to $v^-$ upon winning the bidding. The proof of \Cref{16-thm:reach} is constructive and shows how to choose bids given the thresholds, which are again obtained from the solution to $\RT(\game, 0.5)$, namely $\thresh(v) = 1-val(v_N, \RT(\game, 0.5)$. 
\end{remark}


\section{Qualitative discrete Richman-bidding games}
\label{16-sec:qualitative_discrete}
In discrete bidding, the budgets are given in coins, and the minimal positive bid is one coin. The motivation for discrete bidding is practical. For example, in recreational games, humans cannot be required to keep track of infinite-precision budgets. Similarly, every practical application enforces some granularity on bids. 
We focus on first-price Richman discrete bidding and mention poorman discrete-bidding in \Cref{16-sec:references}. 

A second aspect in which discrete-bidding games differ from continuous-bidding is that bidding ties are resolved according to an explicit tie-breaking mechanism. We focus on the following tie-breaking mechanism: one of the players has the {\em advantage}, represented by a {\em marker}, and when a tie occurs, the player with the marker chooses between (i)~use the advantage, win the bidding, and pass the marker to the opponent, or (ii)~keep the marker and let the other player win the bidding.

\subsection{Definitions}
\subsubsection{Discrete bidding}
Fix $k \in \N$ to be the total budget in the game. 
We incorporate the marker into the players' budgets. 
We depict the marker with $*$. Define the set $\N^* = \set{0, 0^*, 1, 1^*, 2, 2^*, \ldots}$, and $[k]=\set{0,0^*,\ldots, k, k^*}$. We define an order $<$ on $\N^*$ by $0 < 0^* < 1 < 1^* < \ldots$. 
When saying that Eve has a budget of $m^* \in [k]$, we mean that Eve has the advantage, and implicitly, we mean that Adam's budget is $k-m$ and she does not have the advantage. 
We use $|m|$ to denote the integer part of $m$, \textit{i.e.}, $|m^*|=m$. 
We define operators $\oplus$ and $\ominus$ over $\N^*$.
For \(x, y \in \N\), define $x^* \oplus y = x \oplus y^* = (x + y)^*$, $x \oplus y = x+y$. For $x,y \in \N$, define \(x \ominus y = x-y\), \(x^* \ominus y = (x-y)^*\), and in particular $x^* \ominus y^* = x-y$.

\begin{definition}
{\bf (Successor and predecessor).}
For $B \in \N^*$, we denote by $\succb{B}$ and $\predb{B}$ respectively the {\em successor} and {\em predecessor} of $B$ in $\N^*$ according to $<$, defined as $\succb{B} = \min \set{x > B}$ and $\predb{B} = \max \set{x < B}$.
\end{definition}

Suppose that Eve's budget is $B_\mEve$. Implicitly, Adam's budget is $k^* \ominus B_\mEve$. Note that exactly one player has the marker. We sometimes write $B^*_\mEve$ to mean that Eve has the advantage. 
Eve can bid $b_\mEve \in [k]$ with $b_\mEve \leq B_\mEve$. In particular, her bids must be integers and in case she has the advantage, she needs to decide upfront whether she will use it. Formally, suppose that Eve's budget is $B^*_\mEve$ and a tie occurs, then she wins the bidding if her bid is $b^*_\mEve$ and if it is $b_\mEve$, she loses the bidding. 
Consider a pair of bids $b_\mEve$ and $b_\mAdam$. If Eve wins the bidding, her budget is updated to $B_\mEve \oplus b_\mAdam$ and if she loses it, her budget is updated to $B_\mEve \ominus b_\mEve$.

\subsubsection{Discrete-bidding games}
The arena of a discrete-bidding is given by $\zug{k, \vertices, E}$. 
Recall the construction of the explicit concurrent arena that corresponds to $\zug{k, \vertices, E}$ in \Cref{16-sec:def-cont}. 
Note that its size is now finite; namely, the number of vertices is $|\vertices| \cdot k$. Moreover, note that the representation size is exponentially-succinct when $k$ is given in binary.

\begin{definition}
{\bf (Discrete thresholds).} 
Consider a discrete-bidding game played on $\zug{k, \vertices, E}$ with an Eve objective ${\cal O} \subseteq \vertices^\omega$. The {\em discrete thresholds} is a function $\threshD: \vertices \rightarrow [k] \cup \set{k+1}$ such that for every configuration $c = \zug{v, B_\mEve}$:
\begin{itemize}
\item Eve wins from $c$ if $B_\mEve \geq \threshD(v)$, and 
\item Eve loses from $c$ if $B_\mEve < \threshD(v)$.
\end{itemize}
\end{definition}

\begin{remark}{\bf (Thresholds in losing vertices).}
\label{rem:losing-vertices}
Note that a game might contain {\em losing vertices} for Eve, namely vertices from which Eve loses with every initial budget. For example, in a reachability game, a sink with no path to the target is a losing vertex. We set the threshold in a losing vertex $v$ to $\thresh(v) = k+1$. Since the highest possible budget for a player in the game is $k^*$, setting $\thresh(v) = k+1$ can intuitively be understood as Eve requires more budget for winning than she can possibly obtain. Note that this intuition coincides with thresholds in continuous bidding. There, the threshold in a losing vertex $v$ is $1$ and \Cref{def:thresholds} states that Eve wins from $v$ if her budget exceeds $1$, which holds vacuously since the total budget is $1$. 
\end{remark}

We stress that we restrict the players to choose pure strategies, thus when we require a player to win it is in fact ``surely winning''. 
Note that existence of discrete thresholds immediately implies determinacy. Again, discrete-bidding games are a sub-class of concurrent games, thus existence of thresholds is not immediate. In fact, there are tie-breaking mechanisms for which discrete-bidding games are not determined, see \Cref{16-sec:qualitative_discrete}. 

\subsubsection{Frugal objectives}

A {\em frugal} objective  in bidding games generalizes a traditional objective by allowing Eve to win by reaching a collection of sink vertices with sufficient budget:


\begin{definition}
{\bf (Frugal objectives).} A frugal objective is given by a {\em target budget} $\fr: S \rightarrow [k] \cup \set{k+1}$ for Eve in a collection $S \subseteq \vertices$ of (sink) vertices. Eve wins plays that end in $s \in S$ if her budget is at least $\fr(s)$. Frugal objectives can be studied in combination with any objective, and we will study them in combination with three qualitative objectives. Let $\zug{k, \vertices, E}$ be an arena of a discrete-bidding game.
\begin{itemize}
\item A {\em frugal-reachability} bidding game is $\zug{k, \vertices, E, S, \fr}$. Consider a play $\pi$ that ends in a configuration $\zug{s, B_\mEve}$ with $s \in S$. Eve wins $\pi$ iff $B_\mEve \geq \fr(s)$. Note that a reachability bidding game is a special case of a frugal-reachability bidding game in which $\fr \equiv 0$. 
\item The {\em frugal-safety} objective is dual to frugal-reachability. A frugal-safety bidding game is $\zug{V, E, k, S, \fr}$. Eve, now the safety player, wins a play $\pi$ if: (1)~$\pi$ never reaches $S$, or (2)~$\pi$ reaches a configuration $\zug{s, B_\mEve}$ with $s \in S$ and $B_\mEve \geq \fr(s)$. Note that a safety bidding game is a special case of a frugal-safety bidding game in which $\fr \equiv k+1$. 
\item A {\em frugal-parity} bidding game is $\zug{V, E, k, p, S, \fr}$, where $p: (V \setminus S) \rightarrow \set{0,\ldots, d}$. Eve wins a play $\pi$ if (1)~$\pi$ does not reach $S$ and satisfies the parity objective, or (2)~$\pi$ satisfies a frugal-reachability objective: it ends in a configuration $\zug{s, B_\mEve}$ with $s \in S$ and $B_\mEve \geq \fr(s)$.
\end{itemize}
\end{definition}

\subsection{Frugal-reachability discrete-bidding games}
\label{16-sec:disc-reach}
We show existence of thresholds in frugal-reachability games and identify the structure of the discrete thresholds.

\begin{definition}
\label{16-def:disc-avg}
{\bf (Discrete average property).} Consider a frugal-reachability discrete-bidding game $\game = \zug{V, E, k, S, \fr}$. 
We say that a function $T: V  \rightarrow [k] \cup \set{k+1}$ has the {\em discrete average property} if for every $s \in S$, we have $T(s) = \fr(s)$, and for every $v \in V \setminus S$, 
\[
T(v) = \floor{\frac{|T(v^+)| + |T(v^-)|}{2}} +\xi \]
where
\[
\xi = 
\begin{cases}
0 &\text{if~} |T(v^+)| + |T(v^-)| \text{ is even and~} T(v^-) \in \N \\
1 &\text{if~} |T(v^+)| + |T(v^-)| \text{ is odd and~} T(v^-) \in \N^* \setminus \N \\
* &\text{otherwise}
\end{cases}
\]
and $v^+ := \arg\max_{u \in N(v)} T(u)$ and $v^- := \arg\min_{u \in N(v)} T(v)$
\end{definition}

Before relating functions that satisfy the discrete average property and discrete thresholds, we point to the following distinction between discrete- and continuous-bidding. While in continuous-bidding, functions that satisfy the average property are unique (\Cref{16-cor:unique}), we show that this is no longer the case in discrete bidding:

\begin{theorem}
\label{16-thm:non-unique}
Functions that satisfy the discrete average property are not necessarily unique. 
\end{theorem}
\begin{proof}
Consider the reachability discrete-bidding game that is depicted in \Cref{16-fig:avgbutnotthreshold}. Eve's target is $t$. Set the total budget to be $k = 5$. We represent a function $T: V \rightarrow [k]$ as a vector $\zug{T(v_0), T(v_1), T(v_2), T(t)}$. It is not hard to verify that both $\zug{4,3^*,3,2}$ and $\zug{5,4^*,3^*,2}$ both satisfy the discrete average property.
\end{proof}

\begin{figure}
	\centering
	\begin{tikzpicture}
		\colorlet{gr}{green!70!black}
		\colorlet{or}{orange!70!black}
		\draw (0,0) node[eve] (0) {$v_0$};
		\draw (2,0) node[eve] (1) {$v_1$};
		\draw (4,0) node[eve] (2) {$v_2$};
		\draw (6,0) node[adam] (t) {$t$};
		
		\draw[gr] (0,0.5) node (lab0) {\textbf{4}};
		\draw[gr] (2,0.5) node (lab1) {\textbf{3*}};
		\draw[gr] (4,0.5) node (lab2) {\textbf{3}};
		\draw (6.5,0) node (labt) {\textbf{2}};	
		
		\draw[or] (0,-0.5) node (lab0) {\textbf{5}};
		\draw[or] (2,-0.5) node (lab1) {\textbf{4*}};
		\draw[or] (4,-0.5) node (lab2) {\textbf{3*}};
		
		\draw (0) edge[->, bend left = 30] (1);
		\draw (1) edge[->, bend left = 30] (0);
		\draw (1) edge[->] (2);
		\draw (2) edge[->] (t);
		\draw (2) edge[->, bend left = 40] (0);
		\draw (0) edge[->, loop left] (0);
	\end{tikzpicture}
	\caption{A reachability discrete-bidding game with two functions that satisfy the discrete average property.}
	\label{16-fig:avgbutnotthreshold}
\commentAlt{Figure~\ref{16-fig:avgbutnotthreshold}: A directed graph with three circular nodes (v0, v1, v2) and one square node (r), showing various labeled transitions, including a self-loop. See long description.}
\commentLongAlt{Figure~\ref{16-fig:avgbutnotthreshold}: The image displays a directed graph with three circular nodes, labeled 'v0', 'v1', and 'v2', arranged horizontally, and a square node labeled 'r' to the right of 'v2'. A '2' is placed to the right of 'r'.
- Node 'v0' has a self-loop labeled '5' (orange).
- An arrow from 'v0' points to 'v1', labeled '4' (green).
- An arrow from 'v1' points to 'v0', labeled '4*' (orange).
- An arrow from 'v1' points to 'v2', labeled '3*' (green).
- An arrow from 'v2' points to 'v1', labeled '3*' (orange).
- An arrow from 'v2' points to 'r', labeled '3' (green).
The labels in green are above the arrows, and the labels in orange are below the arrows.}
\end{figure}

We show that any function $T$ that satisfies the discrete average property constitutes a lower bound on the discrete thresholds; namely, Adam wins when $B_\mEve < T(v)$. The proof combines two ingredients. We first show that when $B_\mEve \geq T(v)$, Eve can maintain an invariant on her budget: whenever the game reaches $u \in \vertices$, her budget exceeds $T(u)$. Such a strategy only guarantees ``not losing'': if the game reaches $s \in S$, Eve's budget exceeds $\fr(s) = T(s)$. It is not a winning strategy since Eve is the reachability player and must ensure that $S$ is eventually reached. Fortunately, such a construction is winning for Adam, the safety player, since not losing coincides with winning for him. In the second ingredient, we show how to apply the construction for Adam. 

Before formalizing the first ingredient, it is useful to recall continuous-bidding games. Eve maintains an invariant that her budget exceeds the threshold as follows. She bids $b = 0.5\cdot \big(\thresh(v^+) -\thresh(v^-)\big)$. If she wins, she moves to $v^-$ and the invariant is maintained since $\thresh(v) - b = \thresh(v^-)$. If she loses, the the worst that Adam can do is move to $v^+$ and the invariant is maintained since $\thresh(v) + b = \thresh(v^+)$. 

A similar idea applies to discrete bidding, though it is more intricate due to the need to round the bids. Define:
\[
\beta_T = 
	\begin{cases}
		\frac{|T(v^+)| - |T(v^-)|}{2} &\text{~When~} |T(v^+)| + |T(v^-)| \text{~is even and~} T(v^-) \in \N \\
		\floor{\frac{|T(v^+)| - |T(v^-)|}{2}}  &\text{~When~} |T(v^+)| + |T(v^-)| \text{~is odd and~} T(v^-) \in \N^*\setminus \N \\
		\predb{\frac{|T(v^+)| - |T(v^-)|}{2}} &\text{~When~} |T(v^+)| + |T(v^-)| \text{~is even and~} T(v^-) \in \N^*\setminus \N \\
		\succb{\floor{\frac{|T(v^+)| - |T(v^-)|}{2}}} &\text{~When~} |T(v^+)| + |T(v^-)| \text{~is odd and~} T(v^-) \in \N \\
	\end{cases} 
\]
Consider a configuration $c = \zug{v, B_\mEve}$ with $B_\mEve \geq T(v)$. We define Eve's bid in $c$, denoted $b_T(v,B_\mEve)$. 
Intuitively, Eve ``attempts'' to bid $\beta_T$. This is not possible when $\beta_T$ requires the advantage but Eve does not have it in $c$, \textit{i.e.}, $\beta_T \in \N^*\setminus \N$ and $B_\mEve \in \N$. In such a case, Eve bids $\succb{\beta_T} \in \N$. Formally, we define $b_T(v,B_\mEve) = \beta_T$ when both $b_T$ and $B_\mEve$ belong to either $\N$ or $\N^*\setminus \N$, and $\succb{\beta_T}$ otherwise. 

The following lemma is proven in a case by case manner. It shows that by bidding $b_T(v,B_\mEve)$ and moving to $v^-$ upon winning, Eve maintains the invariant that upon reaching $u \in \vertices$, her budget exceeds $T(u)$. Note that in the second case, in order to win the bidding, Adam must bid at least $\succb{b_T(v,B_\mEve)}$. 

\begin{lemma}
\label{16-lem:disc-inv}
For \(B_\mEve \geq T(v)\) we have $B_\mEve \ominus b_T(v,B_\mEve) \geq T(v^-)$ and $B_\mEve \oplus  (\succb{b_T(v,B_\mEve)}) \geq T(v^+)$.
\end{lemma}

We proceed to the second ingredient. We show that it is possible to ``flip the game'': we define another function $T'$ based on $T$ such that $B_\mEve < T(v)$ iff $B_\mAdam \geq T'(v)$. Moreover, we define $T'$ so that if Adam keeps his budget above $T'$, Adam wins the game. Technically, for a sink $s \in S$, we distinguish between two cases. First, $\fr(s) = k+1$ thus Eve cannot win by reaching $s$. Then, $T'(s) = 0$. Indeed, Adam wins in $s$ even when he has no budget. Second, $\fr(s) \in [k]$, thus Eve can win if the game reaches $s$ with sufficient budget. We define $T'(s) = k^* \ominus (\predb{T(v)})$. Indeed, when the game reaches $s$ with $B_\mAdam = T'(s)$, we have $B_\mEve = k^* \ominus B_\mAdam = \predb{\fr(s)} < \fr(s)$. 
We show that $T'$ satisfies the discrete average property from Adam's perspective, thus Adam can follow the strategy we construct for Eve and ensure not losing. Crucially, recall that ``not losing'' for Adam is in fact ``winning''. 

\begin{lemma}
	\label{16-lem:flippedaverage}
	Consider a discrete-bidding game $\game = \zug{V, E, k, S, \fr}$. Let \(T: V \rightarrow [k] \cup \{k+1\}\) be a function that satisfies the discrete average property. We define \(T': V  \rightarrow [k] \cup \{k+1\}\) as follows. 
		\begin{align*}
			\centering
			T'(v) = 
			\begin{cases}
				k^* \ominus (\predb{T(v)}) &\text{~If~} T(v) > 0\\
				k+1 &\text{~otherwise}
			\end{cases}
		\end{align*}
	Then, \(T'\) satisfies the average property.
\end{lemma}

We conclude this section by establishing existence of thresholds.

\begin{theorem}
\label{16-thm:disc-reach}
Consider a frugal-reachability discrete-bidding game $\game = \zug{\vertices, E, k, S, \fr}$. Discrete thresholds exist and satisfy the discrete average property.
\end{theorem}
\begin{proof}
The proof follows a similar structure as continuous-bidding games. Suppose first that $\game$ is played on a DAG with $S$ being the sinks of the graph. We find the thresholds in $\game$ in a backwards-inductive manner. For the base case, for every vertex $s \in S$, clearly $\threshD(s) = \fr(s)$. For the inductive step, consider an internal vertex $v$. Let $T(v) = \floor{\frac{|\threshD(v^+)| + |\threshD(v^-)|}{2}} +\xi$, where $v^-$ and $v^+$ are the children of $v$ that attain the extreme discrete thresholds among the children and $\xi$ is as in \Cref{16-def:disc-avg}. We claim that $T(v) = \threshD(v)$. Note that $T$ satisfies the average property. 
\Cref{16-lem:disc-inv} shows that Eve can bid and guarantee that no matter which vertex $u$ the game proceeds to, her budget exceeds $\threshD(u)$, from which she wins by induction. 

Note that in a game played on a DAG, every play necessarily ends in a leaf. Thus, a strategy that guarantees ``not losing'' is in fact a winning strategy. \Cref{16-lem:flippedaverage} shows that Adam wins from $v$ when Eve's budget is strictly below $T(v)$. 

We turn to $\G$, a game played on a general graph. For $n \in \N$, recall that we denote by $\game[n]$ the truncated game in which Eve is required to win within $n$ turns. The game $\game[n]$ is a game played on a DAG. Let $\threshD_n$ denote the thresholds in $\game[n]$. We claim that for every $v \in \vertices$, we have $\threshD(v) = \lim_{n \to \infty} \threshD_n(v)$. First, for $B_\mEve \geq \threshD(v)$, there is $n \in \N$ such that $B_\mEve \geq \threshD_n(v)$, and Eve can follow a winning strategy in $\game[n]$. Second, since each $\threshD_n$ satisfies the discrete average property, so does the limit $\threshD$. By \Cref{16-lem:flippedaverage}, Adam wins when $B_\mEve < \threshD(v)$.
\end{proof}

\begin{remark}
\Cref{16-thm:disc-reach} gives rise to a fixed-point algorithm to find discrete thresholds. Define $T_0(s)  = \fr(s)$, for $s \in S$, and $T_0(v) = k+1$, for $v \notin S$. Then, for $n \geq 1$, define $T_n$ so that it satisfies the discrete average property w.r.t. $T_{n-1}$. It is not hard to see that $T_n \equiv \threshD_n$. Computing $T_n$ from $T_{n-1}$ takes $O(|\vertices|)$ operations. Since there are $k+1$ possible values for each vertex, the sequence $T_0,T_1,\ldots$ reaches a fixed point within $(k+1)\cdot |\vertices|$ iterations. Note that when $k$ is given in binary, the algorithm is exponential.
\end{remark}

\subsection{Existence of discrete thresholds in parity discrete-bidding games}
\label{16-sec:disc-thresholds}
In this section, we prove existence of discrete threshold budgets in parity bidding games. 

We first point to a key distinction between continuous- and discrete-bidding games. Recall that the reduction from parity continuous-bidding games to reachability continuous-bidding games (\Cref{16-thm:parity}) relies on an analysis of the BSCCs of the graph. Stated for the special case of B\"uchi games, it follows from \Cref{16-lem:SCC}, that Eve wins a BSCC with any positive initial budget if it contains at least one accepting vertex. The example below that under discrete-bidding, Eve might not win a strongly-connected game with an accepting vertex even if she has all the budget.

\begin{figure}[t]
\centering

\begin{tikzpicture}
  \node[draw, circle] (v1) at (0,0) {$v_1$};
  \node[draw, circle] (v2) at (3,0) {$v_2$};
  \node[draw, circle, double] (v3) at (6,0) {$v_3$};

  \draw[->] (v1) edge[bend left] (v2);
  \draw[->] (v2) edge[bend left] (v1);
  \draw[->] (v2) -- (v3);
  \draw[->] (v1) edge[loop left] (v1);
  \draw[->] (v3) edge[bend left] (v1); 
\end{tikzpicture}
\caption{A strongly-connected B\"uchi game in which Eve wins under continuous bidding and loses under discrete bidding.}
\label{16-fig:SCC-Buchi}
\commentAlt{Figure~\ref{16-fig:SCC-Buchi}: A directed graph with three circular nodes, v1, v2, and v3, showing various transitions, including a self-loop and a double-ringed final state.}
\commentLongAlt{Figure~\ref{16-fig:SCC-Buchi}: The image displays a directed graph with three circular nodes arranged horizontally. From left to right, they are labeled 'v1', 'v2', and 'v3'. Node 'v3' is double-ringed, typically indicating an accepting or final state.
- Node 'v1' has a self-loop.
- A bidirectional arrow connects 'v1' and 'v2'.
- A straight arrow points from 'v2' to 'v3'.
- A curved arrow points from 'v3' back to 'v1'.}
\end{figure}

\begin{example}
Consider the B\"uchi game depicted in \Cref{16-fig:SCC-Buchi}. Adam plays as follows: he always bids $0$ and whenever he has the advantage in $v_1$ or $v_2$ he uses it and respectively stays in $v_1$ and moves from $v_2$ to $v_1$. Suppose Eve's initial budget is $B_\mEve$. Note that in order to visit $v_3$, Eve needs to win two biddings in a row; in $v_1$ and $v_2$. Since she can only use the advantage to pay for one of these wins, the other win must be purchased by a unit of budget, meaning that against Adam's strategy, she can visit $v_3$ at most $B_\mEve$ times, thus she loses the game. 
\end{example}

The implication of the example is that the techniques used in continuous bidding do not apply to discrete bidding, and we need to develop a new tool set.

\begin{theorem}
\label{16-thm:disc-parity-exists}
Discrete thresholds exist in parity discrete-bidding games and satisfy the discrete average property. 
\end{theorem}
\begin{proof}
We describe the proof for the special case of co-B\"uchi objectives and the generalization from co-B\"uchi to parity involves an induction on the parity indices, which we omit. Consider a co-B\"uchi discrete-bidding game $\game = \zug{k, \vertices, E, F}$; Eve wins infinite plays that visit $F \subseteq \vertices$ only finitely often. We establish existence of discrete thresholds by describing a fixed-point algorithm to compute them. 
The pseudo-code of the algorithm is described in~\Cref{16-alg:Frugal-Buchi}. The algorithm relies on two sub-routines $\textsf{Frugal-Reachability}(R)$ and $\textsf{Frugal-Safety}(S)$, which respectively return the thresholds in a frugal-reachability game $R$ and a frugal-safety game $S$. 

We refine the co-B\"uchi objective by defining an objective $\Safe_i$, for every $i \geq 0$, which intuitively requires at most $i$ visits to $F$. Clearly a play that satisfies $\Safe_i$ also satisfies the co-B\"uchi objective. More formally, a play satisfies $\Safe_i$ if it
\begin{itemize}
\item starts in $\vertices \setminus F$ and enter $F$ at most $i$ times, or
\item starts in $F$, exits $F$, and enters $F$ at most $i-1$ more times.
\end{itemize}
Note that every path that starts in $F$ violates $\Safe_0$. 

It is not hard to show that if an objective $\O_1$ is ``easier'' to satisfy that an objective $\O_2$ (formally, $\O_2 \subseteq \O_1$), then  a player requires less budget to achieve $\O_1$ than $\O_2$, which in turn means that the threshold for $\O_1$ is not higher than the threshold for $\O_2$. 
For $i \geq 0$, let $\threshD_i: V \rightarrow [k+1]$ denote the discrete thresholds for objective $\Safe_i$, and we denote by $\threshD_\game$, the discrete thresholds in the co-B\"uchi game $\game$. The following claim follows from $\Safe_i \subseteq \Safe_{i+1}$ and $\Safe_i \subseteq \CoBuchi(F)$: 

\smallskip
\noindent{\bf Claim:} For every $i \geq 0$ and $v \in \vertices$, we have $\threshD_i(v) \geq \threshD_{i+1}(v)$ and $\threshD_i(v) \geq \threshD_\game(v)$. 
\smallskip

We describe a recursive algorithm to compute $\threshD_i$. 
The algorithm constructs and solves a sequence of frugal-reachability games $R_0, R_1,\ldots$ and a sequence of frugal-safety games $\S_0, \S_1,\ldots$. 
For $i \geq 0$, recall that we denote by $\threshD_{R_i}$ and $\threshD_{S_i}$ respectively the thresholds in $R_i$ and $S_i$.  
We will use $v$ to denote a vertex in $\vertices \setminus F$ and $u$ to denote a vertex in $F$. 

\paragraph{Base case.} 
We start by characterizing $\Safe_0$. Note that for every $v \in \notF$, in order to satisfy $\Safe_0$, Eve needs to guarantee that $F$ is not visited at all. Moreover, Eve cannot win from $u \in F$. Thus, Eve's objective is a safety objective. Formally, define a safety discrete-bidding game $S_0 = \zug{V, E, k, F}$ in which Eve's goal is to avoid $F$.

\smallskip
\noindent{\bf Claim:} For $u \in F$, we have $\threshD_0(u) = k+1$ and for $v \in \notF$, we have $\threshD_0(v) = \threshD_{S_0}(v)$. 
\smallskip 

By \Cref{16-thm:disc-reach}, $\threshD_0$ satisfies the discrete average property.

\paragraph{Recusive step.}
Let $i > 0$. Assume that the discrete thresholds $\threshD_{i-1}$ are given. We characterise $\threshD_i$ as the discrete thresholds in two discrete-bidding games $S_i$ and $R_i$.

Intuitively, the frugal-reachability game $R_i$ starts from $F$ and Eve's goal is to reach $v \in \vertices \setminus F$ with a budget that exceeds $\threshD_{i-1}$.
Note that such a strategy is useful for Eve in the co-B\"uchi game $\game$; namely, when $\game$ starts from $F$, by following a winning strategy in $R_i$, Eve ensures that $F$ is exited, and upon exiting, Eve has sufficient funds to guarantee that $F$ is visited at most $i-1$ more times. In particular, $F$ is visited only finitely often in $\game$. The solution to $R_i$ characterises $\threshD_i$ in the vertices in $F$. 
Dually, the frugal-safety game $S_i$ starts from $\notF$. Now Eve is the safety player and wins if either (1)~$F$ is never reached or (2)~if $F$ is reached, Eve's budget is at least $\threshD_i$. That is, upon entering $F$, she can switch to a winning strategy in $R_i$, and continue as above. 

Formally, we construct $R_i = \zug{V, E', k, \notF, \fr_i}$ on the same arena as $\game$ only that $\notF$ are sinks, \textit{i.e.}, $E' = \set{\zug{u, u'} \in E: u \in F}$, with a frugal target budget of $\fr_i(v) = \threshD_{i-1}(v)$, for  $v \in \notF$. Let $\threshD_{R_i}: F \rightarrow [k+1]$ denote the thresholds in $R_i$. We recall the properties of $\threshD_{R_i}$. For an initial configuration $\zug{u, B}$ with $u \in F$, we have:
\begin{itemize}
\item If $B \geq \threshD_{R_i}(u)$, Eve can ensure the frugal-reachability objective; namely, a configuration $\zug{v, B'}$ is reached with $v \in \notF$ and $B' \geq \threshD_{i}(v)$.
\item If $B < \threshD_{R_i}(u)$, Adam can violate the frugal-reachability objective; namely, he can force that either (1)~$F$ is not exited or (2)~upon reaching a configuration $\zug{v, B'}$ with $v \in \notF$, then $B' < \threshD_{i}(v)$. 
\end{itemize}
Moroeover, by \Cref{16-thm:disc-reach}, $\threshD_{R_i}$ satisfies the discrete average property. We obtain the following claim by combining with the induction hypothesis. 

\smallskip
\noindent{\bf Claim:} For $u \in F$, we have $\threshD_i(u) = \threshD_{R_i}(u)$. 
\smallskip

We proceed to characterise $\threshD_i$ for vertices in $\notF$. We define a frugal-safety discrete-bidding game $S_i = \zug{V, E, k, F, \threshD_i}$. Let $\threshD_{S_i}$ denote the discrete thresholds in $S_i$. We recall the properties of $\threshD_{S_i}$.
From an initial configuration $\zug{v, B}$ with $v \in \notF$, we have:
\begin{itemize}
\item If $B \geq \threshD_{S_i}(v)$, Eve, the safety player, can force the game to either (1)~stay in $V \setminus F$, or (2)~upon reaching a configuration $\zug{u, B'}$ with $u \in F$, we have $B' \geq \threshD_{i}(u)$.
\item If $B < \threshD_{S_i}(v)$, Adam, the reachability player can guarantee the frugal-reachability objective; namely, he can force the game to reach $F$ and upon reaching $\zug{u, B'}$ with $u \in F$, we have $B' < \threshD_{i}(u)$. 
\end{itemize}
By \Cref{16-thm:disc-reach} and \Cref{16-lem:flippedaverage}, $\threshD_i$ satisfies the average property. Combining with the induction hypothesis, we obtain: 

\smallskip
\noindent{\bf Claim:} For $v \in \notF$, we have $\threshD_i(v) = \threshD_{\S_i}(v)$.
\smallskip

The first claim above implies both that the sequence $\threshD_0,\threshD_1,\ldots$ reaches a fixed point and that the fixed point is an upper bound on the co-B\"uchi discrete-thresholds. To show that the fixed-point in fact coincides with the discrete thresholds, we prove the following claim. 

\smallskip
\noindent{\bf Claim:} Let $n \in \N$ such that $\threshD_n(v) = \threshD_{n+1}(v)$. Adam, the B\"uchi player, wins from a configuration $\zug{v, B}$ with $B < \threshD_n(v)$. 
\smallskip


Suppose that $v \in \notF$ and the case of an initial vertex in $F$ is captured in the proof below. Consider a configuration $\zug{v, B}$ with $v \in \notF$ and $B < \threshD_{S_{n+1}}(v)$. Adam wins $\game$ by following a winning strategy of the reachability player in the frugal-safety game $S_{n+1}$. As mentioned above, the strategy ensures that a configuration $\zug{u, B'}$ is reached with $u \in F$ and $B' < \threshD_{R_{n}}(u)$. He then follows a winning strategy of the safety player in the frugal-reachability game $R_n$. As mentioned above, this strategy ensures that either (1)~$F$ is never exited, clearly satisfying the B\"uchi objective, or (2)~a configuration $\zug{v', B''}$ is reached with $v' \in \notF$ and $B'' < \threshD_{S_{n}}(v')$. Adam can restart his strategy from $v'$ since $\threshD_{S_{n}}(v') = \threshD_{\S_{n+1}}(v')$. By repeating this strategy Adam guarantees that every time $F$ is exited, it is entered again, and in particular, it ensures infinitely many visits to $F$.
\end{proof}

\begin{algorithm}[t]
\SetAlgoLined
\SetKwInOut{Input}{Input}
\SetKwInOut{Output}{Output}
\DontPrintSemicolon

\caption{co-B\"uchi-Thresholds$(\game)$}
\label{16-alg:Frugal-Buchi}

\Input{A game $\game$}
\Output{Thresholds $\threshD_i$} 

\BlankLine

$i := 0$\;
Define the frugal-safety game $S_0 = \zug{V, E, k, \fr_0, F}$, with $\fr_0 \equiv k+1$\;
$\threshD_{S_0} = \textsc{Frugal-Safety}(S_0)$\;
Define $\threshD_0(v) = \threshD_{S_0}(v)$, for $v \in V \setminus F$, and $\threshD_0(u) = k+1$, for $u \in F$\;

\BlankLine

\Repeat{$\threshD_{i-1} = \threshD_i$}{
    $i := i+1$\;
    Define $R_i = \zug{V, E, k, \threshD_{S_{i-1}}, V \setminus F}$\;
    $\threshD_{R_i} := \textsc{Frugal-Reachability}(R_i)$\;
    Define $S_i = \zug{V, E, k, \threshD_{R_i}, F}$\;
    $\threshD_{S_i} := \textsc{Frugal-Safety}(S_i)$\;
    Define $\threshD_i(v) = \threshD_{S_i}(v)$, for $v \in V \setminus F$, and $\threshD_i(u) = \threshD_{\game_i}(u)$, for $u \in F$\;
}
\Return{$\threshD_i$}\;

\end{algorithm}


\subsection{Finding discrete thresholds is in NP and coNP}
\label{16-sec:disc-NP-coNP}

We formalize the problem of finding threshold budgets as a decision problem: 
\begin{problem}
\label{16-prob:decision-threshold}
{\bf (Finding discrete threshold budgets).}
Given a frugal-parity discrete-bidding game $\game = \zug{\vertices, E, k, p, S, \fr}$, a vertex $v \in V$, and $\ell \in [k]$, does the following hold: $\threshD_\game(v) \geq~\ell$.
\end{problem}

We will show that \Cref{16-prob:decision-threshold} is in NP and coNP. 
We compare our situation with continuous-bidding games. 
Fortunately, unlike continuous-bidding, in discrete-bidding one can ``guess the thresholds''; namely, a function \(T : V \rightarrow [k] \cup \{k+1\}\) can be represented using $O(|V| \cdot \log(k))$ bits, which is polynomial in the input. 
A naive attempt to show membership in NP and coNP would be to guess $T$, verify that $T$ satisfies the discrete average property, and accept $\zug{\game, v, \ell}$ iff $T(v) \geq \ell$. 
Unfortunately, such an attempt fails. While \Cref{16-thm:disc-parity-exists} shows that the thresholds satisfy the discrete average property, \Cref{16-thm:non-unique} shows that there can be other functions that also satisfy it. That is, it could be the case that $T$ satisfies the average property and $T \not \equiv \thresh_\G$. 
We point out that in continuous-bidding games, if one could guess a $T$, this scheme would have succeeded since there is a unique function that satisfies the continuous average property (\Cref{16-thm:reach}). 

In the remainder of this section, we will show that the following problem is in NP and coNP by reducing it to solving a turn-based parity game of size linear in the size of the graph (and not the size of $k$). Then, an algorithm for Prob.~\ref{16-prob:decision-threshold} guesses both $T$ and winning strategies in the turn-based game. 
\begin{problem}
\label{16-prob:thresh-equiv}
{\bf (Verifying a guess of $T$).} Given a frugal-parity discrete-bidding game $\G$ with vertices $V$ and a function \(T : V \rightarrow [k] \cup \{k+1\}\) that satisfies the average property, decide whether \(T \equiv \thresh_\G\).
\end{problem}

Before describing the algorithm, we revisit the definitions in \Cref{16-sec:disc-reach}. 
Consider a function \(T : V \rightarrow [k] \cup \{k+1\}\) that satisfies the average property. 
A {\em partial strategy} is a function that given a configuration, proposes a bid and a set of {\em allowed} vertices to move to upon winning. That is, a partial strategy differs from a strategy in that the latter must choose one vertex to proceed to whereas the former chooses a set of vertices. We say that a strategy {\em agrees} with a partial strategy if they always bid the same and the strategy chooses one of the allowed vertices. We have already seen in \Cref{16-sec:disc-reach} that a  function $T$ that satisfies the discrete-average property gives rise to a partial strategy: given a configuration $\zug{v, B_\mEve}$, the partial strategy proposes the bid $b_T(v, B_\mEve)$ and the allowed vertices are $u$ such that $T(u) = \argmin_{u' \in N(v)} T(u')$. In other words, $u$ is $v^-$, and note that there can be multiple such $u$.
Recall that \Cref{16-lem:disc-inv} shows that if $B_\mEve \geq T(v)$ and Eve follows a strategy that agrees with $f_T$, then no matter how Adam responds, denoting $\zug{v', B_\mEve'}$ the next configuration, then $B'_\mEve \geq T(v')$.

The high-level idea of our algorithm to decide whether \(T \equiv \thresh_\G\) is inspired by an NP algorithm to decide whether Eve wins a turn-based parity game from an initial vertex $v_0$. We sketch that algorithm. 
First guess a positional strategy $f$, which is a function that maps each vertex $v$ that is controlled by Eve to an outgoing edge from $v$. The algorithm then verifies that $f$ is winning for Eve. 
The idea is to solve the following problem: check whether Adam has a counter strategy $g$ for $f$ such that $\play(v_0, f, g)$ is winning for Adam. Clearly, $f$ is winning iff Adam cannot counter $f$. 
Deciding whether Adam can counter $f$ is done as follows. We trim every edge in the game that does not comply with $f$. 
The resulting graph can be thought of as an automaton over a singleton alphabet in which the acceptance condition is Adam's objective. Each run of the automaton corresponds to $\play(v_0, f, g)$, for some strategy $g$.
We then check for language emptiness. The language is not empty iff Adam has a counter strategy $g$ iff $f$ is not winning. 

Our algorithm for frugal-parity discrete-bidding games follows conceptually similar steps. 
Given $T$ that satisfies the discrete average property, we construct $f_T$, and we check whether Adam can counter every Eve strategy that agrees with $f_T$. 
Unlike the algorithm in turn-based games, where a guess fixes all of Eve's choices, here we guess only a partial strategy, and Eve has the flexibility of choosing an allowed vertex in each turn. This complicates the algorithm to verify the guess $T$. Instead of checking emptiness of an automaton, we now construct and solve a turn-based game. 

We define the turn-based parity game $G_{T, \game} = \zug{\vertices_\mEve,\vertices_\mAdam, E, p}$ as follows. 
\begin{itemize}
\item $V_\mEve = \{\zug{v,T(v)}, \zug{v, \succb{T(v)}}, \zug{v, \top}: v \in (\vertices \cup S)\}$ and 
\item $V_\mAdam = \set{\zug{v, c}: v \in V}$.
\end{itemize}
We define the edges. First, we define the sinks. A vertex $\zug{v, B}$ is a sink that is winning for Eve if $v \in S$ or $B = \top$. It is a sink that is losing for Eve if $v$ is a losing vertex, i.e, $T(v) = k+1$. Second, we define edges in non-sink vertices. Consider $c = \zug{v, B} \in \vertices_\mEve$ and let $\zug{b, A} = f_T(c)$. An Eve action from $c$ is an allowed vertex $v' \in A$; formally, for every $v' \in A$, we define $\zug{v', c}$ to be a neighbor of $c$. A choice of $v'$ in $c$ intuitively models an Eve action in $\game$ of choosing $\zug{b,v'}$ at $c$; again, the bid $b$ is determined by $f_T$ and the action in $G_{T, \game}$ resolves the choice of an allowed vertex. A vertex $\zug{v', c}$ is controlled by Adam. Intuitively, Adam makes two choices: who wins the bidding and, if he decides to win the bidding, he chooses how to move the token. Formally: 
\begin{itemize}
\item Adam lets Eve win by bidding $0$, modeled by an edge from $\zug{v', c}$ to $\zug{v', B \ominus b}$.

\item Adam wins by bidding $\succb{b}$ and proceeds to $w \in N(v)$. There are restrictions. First, this is a valid choice only if Adam has sufficient budget, namely $k^* \ominus B \geq \succb{b}$. Second, let $B' = B \oplus (\succb{b})$ be Eve's updated budget. If $B' > \succb{T(w)}$, we trim the budget and set  $\zug{w, \top}$ as a neighbor of $\zug{v', c}$. Otherwise, $B' \in \set{T(w), \succb{T(w)}}$, thus $\zug{w, B'} \in \vertices_\mEve$ and we set it as a neighbor of $c$.
\end{itemize}
For ease of presentation, we define parity indices only in Eve vertices. 
A non-sink vertex in $G_{T, \G}$ ``inherits'' its parity index from the vertex in $\game$; namely, for \(c = \zug{v,B} \in \vertices_\mEve\), we define $p'(c) = p(v)$.  We set the parity index of sinks so that they are either winning (odd) or losing (even) for Eve so that they match the definition above.

\begin{lemma}
	\label{16-lem:disc-sound}
	If Eve wins from every vertex that is not a losing sink in $G_{T, \G}$, then $T \geq \threshD_\game$.
\end{lemma}
\begin{proof}
Let $f^*$ be a positional strategy for Eve that wins from all vertices in $G_{T, \game}$. We construct a winning strategy $f$ for Eve in $\game$ from a configuration $c_0 = \zug{v, T(v)}$. The strategy $f$ maintains a location in $G_{T, \game}$ such that when $\game$ is in $c = \zug{v, B_\mEve}$, then $c^* = \zug{v, B'_\mEve}$, where $B'_\mEve \in \set{T(v), \succb{T(v)}}$, $B_\mEve \geq B'_\mEve$, and $B_\mEve$ and $B'_\mEve$ agree on the status of the tie-breaking advantage. 

Initially, we set $c^*_0 = c_0$ (note that $c$ is an Eve vertex in $G_{T, \game}$). 
Let $\zug{b, A} = f_T(c)$. Recall that $f_T$ acts the same in $c$ and $c^*$, thus $\zug{b, A} = f_T(c^*)$. 
Recall that an action of $f^*$ corresponds to choosing an allowed vertex $u \in A$ (technically, $f^*$ proceeds from $c^*$ to $\zug{u, c^*}$). The action that $f$ chooses in $c$ is $\zug{u, b}$. That is, the bid agrees with $f_T$ and the choice of vertex to move to upon winning agrees with $f^*$. Adam responds to Eve's action by choosing $\zug{u', b'}$. In case Eve wins the bidding, the next configuration is $c' = \zug{u, B_\mEve \ominus b}$, which is a vertex in $G_{T, \game}$, and Eve updates the location to $c'$. Suppose that Adam wins the bidding. If $\zug{u', B_\mEve \oplus b'}$ is a vertex in $G_{T, \game}$, Eve updates as before. Otherwise, it follows from \Cref{16-lem:disc-inv} that $B_\mEve \oplus b' > \succb{T(u')}$. In such a case we say that Eve {\em restarts} the location and sets $c^* = \zug{u', B'_\mEve}$ with $B'_\mEve \in \set{T(u'), \succb{T(u')}}$ that agrees with her updated budget on the tie-breaking advantage. 

We claim that $f$ is winning. We first argue that restarting the location can occur only finitely many times. Indeed, define the \emph{spare change} in a configuration \(c = \zug{v, B}\) as \(\Spare(c)=|B| - |T(v)|\). Note that every time the location is restart, the spare change strictly increases. One can also show that in the other bidding outcomes, the spare change does not change. Moreover, the spare change is bounded by the total budget $k$. 

Consider an Adam strategy in $\game$, and let $\pi$ be the suffix of the play that includes no restarts. It is not hard to expand the sequence of locations to a play $\pi^*$ in $G_{T, \game}$ that is the outcome of $f^*$ against some Adam strategy. Since $f^*$ is winning $\pi^*$ satisfies Eve's objective. Note that $\pi$ and $\pi^*$ traverse the same sequence of vertices $\vertices$. If the two are infinite, they both satisfy Eve's parity condition. Recall that when $\game$ is in configuration $c = \zug{v, B_\mEve}$, then the location is $c^* = \zug{v, B'_\mEve}$ with $B'_\mEve \leq B_\mEve$. Suppose that the two end in a sink $s \in S$. Since $\pi^*$ satisfies the frugal objective, so does $\pi$. 
\end{proof}

The following lemma shows completeness; namely, that a correct guess of $T$ implies that Eve wins from every vertex in $G_{T, \game}$. 

\begin{lemma}
\label{16-lem:disc-complete}
If \(T \equiv \threshD_\game\), then Eve wins in \(G_{T, \game}\) from every vertex that is not a losing sink.
\end{lemma}
\begin{proof}
We sketch the proof. 
Suppose towards contradiction that $T  \equiv \threshD_{\game}$ and there is a vertex \(c^*_0 = \zug{v,B}\) in \(G_{T, \game}\) that is losing for Eve and that is not a losing sink. Let \(g^*\) be an Adam strategy that wins from \(c^*_0\) in \(G_{T, \game}\). Since $v$ is not a losing vertex, $\threshD_{\game}(v) < k+1$, and there is a winning Eve strategy $f$ from $c^*_0$ in $\game$. We reach a contradiction by constructing a counter-strategy $g$ for Adam in $\game$ based on $g^*$ that wins against $f$ from $c^*_0$. Similar to \Cref{16-lem:disc-sound}, $g$ maintains a location $c^*$ in $G_{T, \game}$ that agrees with the configuration $c$ in $\game$. Suppose that $f$ chooses action $\zug{u, b}$ in $\game$. Let $\zug{b_T(c), A} = f_T(c)$. 

We distinguish between two cases. In the first case, the action that $f$ chooses in $\game$ corresponds to an action in $G_{T, \game}$, namely $b_T(c) = b$ and $u \in A$. We say that Eve {\em agrees} with $G_{T, \game}$. We define $g$ to react as $g^*$ reacts; namely, if $g^*$ lets Eve win the bidding, so does $g$, and otherwise, $g$ bids $\succb{b}$ and moves to the neighbor chosen by $g^*$. Suppose that Eve always agrees with $G_{T, \game}$. Then, the resulting play $\pi$ in $\game$ coincides with a play $\pi^*$ in $G_{T, \game}$ that is the result of $g^*$ playing against some Eve strategy. Since $g^*$ is winning, $\pi$ is losing for Eve, which is a contraction to the assumption that $f$ is winning. Thus, at some point, $f$ must disagree with $G_{T,\game}$.

The second case handles the case in which Eve disagrees with $G_{T, \game}$. If $b < b_T(c)$, intuitively Eve is bidding too low. Adam responds by bidding $\succb{b}$ and choosing $v^+$. It can be shown that in the next configuration $\zug{v', B'_\mEve}$, we have $B'_\mEve < T(v')$, thus by the assumption that $T  \equiv \threshD_{\game}$, Adam has a winning strategy. The case of $b > b_T(c)$ is slightly more subtle. Intuitively, Eve is overbidding, and Adam lets Eve win by bidding $0$. Either the resulting configuration $\zug{v', B'_\mEve}$, has $B'_\mEve < T(v')$, and the argument above applies, or the resulting configuration is a vertex in $G_{T, \game}$ and we continue as if Eve agrees with $G_{T, \game}$. 
\end{proof}

The two lemmas above imply that \Cref{16-prob:decision-threshold} is in NP. In order to show membership in coNP, given a guess $T$, as in \Cref{16-lem:flippedaverage}, we define $T'$ by ``flipping'' $T$ and apply the reasoning above to Adam. 

\begin{theorem}
	\label{16-thm:parity-NP-coNP}
	The problem of finding discrete threshold budgets in frugal-parity discrete-bidding games is in NP and coNP. 
\end{theorem}

\begin{remark}
{\bf (Strategy complexity).}
The proofs of existence of thresholds (\Cref{16-thm:disc-reach} and \Cref{16-thm:disc-parity-exists}) construct winning strategies. The strategies constructed there are functions from configurations to actions. Thus, their implementation requires a table of size $|\vertices| \cdot k$, which is exponential when $k$ is given in binary. On the other hand, the strategy constructed in \Cref{16-lem:disc-sound} only keeps track of a location in $G_{T, \game}$, whose size is $|\vertices| \cdot 3$, which is linear in the input.
\end{remark}
\begin{remark}
We point to the intriguing state of affairs in parity discrete-bidding games. 
Deciding the winner in a turn-based parity game is a long-standing open problem, which is known to be in NP and coNP but not known to be in P. 
We describe a very easy reduction that given a turn-based parity game $\game$, constructs a parity discrete-bidding game $\game'$ with a total budget of $0$. Assume that the players alternate turns in $\game$. In order to simulate this behavior, in $\game'$, we add to every  Eve vertex $v$ an edge to a sink that is winning for Adam. Thus, Eve must use the advantage in $v$ to avoid losing and pass it to Adam. 
Similarly, for each Adam vertex, we add an edge to a sink that is winning for Eve. Clearly, an outcome of $\game$ corresponds to an outcome of $\game'$. We stress that the sum of budgets in $\game'$ is constant, in fact it is $0$. One might expect that discrete-bidding games with budgets given in binary would be (at least) exponentially harder. However, \Cref{16-thm:parity-NP-coNP} shows that all of these classes of games actually lie in NP and coNP. 
\end{remark}


\section{Mean-payoff continuous-bidding games}
\label{16-sec:mean_payoff_continuous}
In this section we study mean-payoff first-price continuous-bidding games.

\subsection{Definitions}
We recall definitions of mean-payoff games. 
A mean-payoff bidding game is played two players, called Max and Min, on a weighted graph $\zug{V, E, w}$, where $w: \vertices \rightarrow \Q$. Each infinite play has a {\em payoff}, which is Max's reward and Min's cost. We will mainly focus on games played on strongly-connected graphs. 

\begin{definition}
\label{16-def:MP}
{\bf (Payoff and energy)} Consider an infinite path $\eta = \eta_0, \eta_1, \ldots$. For $n >1$, let $\eta^n = \eta_0, \ldots, \eta_n$ be a prefix of $\eta$. The {\em energy} of $\eta^n$, denoted $\energy(\eta^n)$, is the sum of weights it traverses, thus $\energy(\eta^n) = \sum_{0 \leq i < n} w(\eta_i)$. The {\em payoff} of $\eta$, denoted $\payoff(\eta)$, is $\payoff(\eta) = \lim \inf_{n \to \infty} \frac{1}{n} \cdot \energy(\eta^n)$. Note that the use of $\lim\inf$ gives Min an advantage.
\end{definition}

We will study the following question: 
\begin{center}
{\it Given an initial budget ratio $r$, what is the optimal payoff that a player can guarantee?}
\end{center}
We formalize this question in terms of the value of a game: 

\begin{definition}
\label{16-def:MP-value}
{\bf (Mean-payoff value in bidding games).} Consider a strongly-connected mean-payoff bidding game $\game$ and a budget ratio $r \in (0,1)$. The {\em mean-payoff value} of $\game$ w.r.t. $r$, denoted $\Value(\game, r)$, is $c \in \Q$ if independent of the initial vertex:
\begin{itemize}
\item With an initial budget ratio that exceeds $r$, for every $\epsilon >0$, Max has a strategy that guarantees a payoff of at least $c-\epsilon$, and 
\item Max cannot do better: with a ratio that exceeds $1-r$, for every $\epsilon > 0$, Min has a strategy that guarantees a payoff of at most $c+\epsilon$. 
\end{itemize}
\end{definition}

\stam{
\begin{definition}
{\bf (Mean-payoff value in random-turn games).}
For $p \in [0,1]$, since $\G$ is a mean-payoff game, the random-turn game $\RT(\G, p)$ is a stochastic mean-payoff game. The expected payoff under optimal play of the two players is called the mean-payoff value, which we denote $\MP(\RT(\G, p))$. The value is known to exist \cite{Put05}, and since $\G$ is strongly-connected, the value does not depend on the initial vertex.
\end{definition}
}

\subsection{An overview of the results}
We will show equivalences between mean-payoff bidding and random-turn games in strongly-connected games. In this section we illustrate the equivalences on a specific mean-payoff game called the {\em bowtie game}, denoted $\lolli$, and depicted in \Cref{16-fig:bowtie}. 

\begin{figure}[ht]
\centering

\begin{tikzpicture}
  \node[draw, circle] (1) at (0,0) {$1$};
  \node[draw, circle] (0) at (3,0) {$0$};

  \draw[->] (1) edge[bend left] node[above] {} (0);
  \draw[->] (0) edge[bend left] node[below] {} (1);
  \draw[->] (1) edge[loop left] node[left] {} (1);
  \draw[->] (0) edge[loop right] node[right] {} (0);
  \node[below of=1, node distance=0.75cm] {$v_{\max}$};
  \node[below of=0, node distance=0.75cm] {$v_{\min}$};
\end{tikzpicture}

\caption{The bowtie mean-payoff game $\lolli$.}
\label{16-fig:bowtie}
\commentAlt{Figure~\ref{16-fig:bowtie}: A directed graph with two circular nodes, labeled '1' and '0', showing bidirectional transitions and self-loops on each node.}
\commentLongAlt{Figure~\ref{16-fig:bowtie}: The image displays a directed graph with two circular nodes arranged horizontally. The left node is labeled '1', and the right node is labeled '0'.
- Both '1' and '0' nodes have self-loops.
- A bidirectional arrow connects the '1' node and the '0' node.
- Below the '1' node, it is labeled 'v_max'. Below the '0' node, it is labeled 'v_min'.}
\end{figure}

Let $p \in [0,1]$. Recall that for a bidding game $\game$, the random-turn game $\RT(\game, p)$ is a stochastic game in which in each turn, the player who moves is determined according to a biased coin toss: Max moves with probability $p$ and Min moves with probability $1-p$. Since it is clearly optimal for Max to move left and Min to move right, we simplify $\RT(\lolli, p)$ by omitting non-optimal moves. This results in a weighted Markov chain (see \Cref{16-fig:RTlolli}). Roughly, since we expect a random walk in $\RT(\lolli, p)$ to ``stay'' in $v_\mMax$ portion $p$ of the time, the value of $\RT(\lolli, p)$ is $\Value\big(\RT(\lolli, p)\big) = p$.

\begin{figure}[ht]
\centering

\begin{tikzpicture}
  \node[draw, circle] (1) at (0,0) {$1$};
  \node[draw, circle] (0) at (3,0) {$0$};

  \draw[->] (1) edge[bend left] node[above] {$1-p$} (0);
  \draw[->] (0) edge[bend left] node[below] {$p$} (1);
  \draw[->] (1) edge[loop left] node[left] {$p$} (1);
  \draw[->] (0) edge[loop right] node[right] {$1-p$} (0);
  \node[below of=1, node distance=0.75cm] {$v_{\max}$};
  \node[below of=0, node distance=0.75cm] {$v_{\min}$};
\end{tikzpicture}

\caption{For $p \in [0,1]$, the random-turn game $\RT(\lolli, p)$.}
\label{16-fig:RTlolli}
\commentAlt{Figure~\ref{16-fig:RTlolli}: A directed graph with two circular nodes, labeled '1' and '0', showing bidirectional transitions and self-loops, with probabilistic labels involving 'p'.}
\commentLongAlt{Figure~\ref{16-fig:RTlolli}: The image displays a directed graph with two circular nodes arranged horizontally. The left node is labeled '1', and the right node is labeled '0'.
- Node '1' has a self-loop labeled 'p'. Below the node, it is labeled 'v_max'.
- Node '0' has a self-loop labeled '1 - p'. Below the node, it is labeled 'v_min'.
- A bidirectional arrow connects the '1' node and the '0' node. The arrow from '1' to '0' is labeled '1 - p'. The arrow from '0' to '1' is labeled 'p'.}
\end{figure}

Under Richman bidding, we will show that the value of a strongly-connected mean-payoff game does not depend on the initial budget ratio and only on the structure of the game. Moreover, we will show an equivalence between mean-payoff Richman-bidding games and unbiased random-turn games. Specifically, for the bowtie game, we show: 
\begin{theorem}
Under Richman bidding, for every budget ratio $r \in [0,1]$, we have $\Value(\lolli, r) = \Value\big(\RT(\lolli, 0.5)\big) = 0.5$.
\end{theorem}

The equivalence for Richman bidding has similarities with the equivalence shown in \Cref{16-thm:RTreach} for  reachability objectives. 
Recall however, that no such equivalence is known for reachability poorman-bidding games and it is unlikely to exist (\Cref{16-rem:reach-poorman}). 
We thus find it intriguing that an equivalence holds for mean-payoff poorman-bidding games, and even for the general case of taxman-bidding games. 
Moreover, the equivalence is intricate and depends on the initial budget ratio; that is, unlike Richman bidding, with a higher initial budget ratio, a player can guarantee a higher payoff.  
Specifically, in the bowtie game, we show the following:

\begin{theorem}
Under poorman bidding, for an initial budget ratio $r \in [0,1]$, we have \[\Value(\lolli, r) = \Value\big(\RT(\lolli, r)\big) = r.\]
\item Under taxman bidding with taxman parameter $\tau \in [0,1]$, we have \[\Value(\lolli, r) = \Value\big(\RT(\lolli,\frac{r+\tau\cdot (1-r)}{1+\tau})\big) = \frac{r+\tau\cdot (1-r)}{1+\tau}.\]
\end{theorem}

\stam{
Suppose first that both budgets are $1$.   for \Min that,  We assume for simplicity that \Min wins bidding ties. 

We make several observations. (1) \Min's strategy is legal: it never bids higher than the available budget; indeed, \Max bids legally, the budgets are the same, and \Min's sum of winning bids is at most \Max's sum of winning bids. (2) The size of the queue is an upper bound on the energy; indeed, every bid in the queue corresponds to a \Max winning bid that is not ``matched'' (the size is an upper bound since \Min might win biddings when the queue is empty). (3) If \Min's queue fills, it will eventually empty; indeed, if $b \in \Real$ is in the queue, in order to keep $b$ in the queue, \Max must bid at least $b$, thus eventually his budget runs out. Combining the three, since the energy is at most $0$ when the queue empties, \Min's strategy guarantees that the energy is at most $0$ infinitely often and as in Lem.~\ref{lem:FP-bowtie}, the payoff is non-positive.

Now, assume \Min's budget is $\ell \in \Nat$. \Min's strategy is the same only that whenever \Max wins with $b > 0$, \Min adds $\ell$ copies of $b$ to the queue. The strategy is legal since, intuitively, we can think of \Min's budget as if it consists of $\ell$ parts of size $1$, and each copy of the bid $b$ is paid out of a unique part. Legality then follows as in the simple case above. The other two observations above still hold, only that now, since every \Max win is matched by $\ell$ \Min wins, when the queue empties, the number of \Min wins is at least $\ell$ times as much as \Max's wins. It follows that the payoff is at most $\frac{\ell}{\ell +1}$.

}

\subsection{An optimal Min strategy under Richman bidding}
\label{16-sec:min-MP}
In this section we develop an optimal strategy for Min. 
The simplicity of the construction stems from two features. First, Richman bidding is technically simpler than poorman.  Second, the definition of mean payoff (\Cref{16-def:MP}) gives Min an advantage. We point out that due to this advantage, a Min strategy cannot be applied to Max. We will later construct $\epsilon$-optimal Max strategies that can be applied to Min.

We start by solving the bowtie game. In order to develop intuition, we will show two proofs for the following lemma.

\begin{lemma}
\label{16-lem:Min-bowtie}
For every $\epsilon >0$, Min can guarantee a payoff of $0.5$ in $\lolli$ with an initial budget of $\epsilon$. 
\end{lemma}

Throughout this section, we re-define the weights in the bowtie game to be $w(v_\mMin) = -1$ and $w(v_\mMax) = 1$.

\subsubsection{A first construction of optimal strategy in the bowtie game}
We describe a simple Min strategy in the bowtie game, called the {\em tit-for-tat} strategy.  
The tit-for-tat strategy is optimal in the bowtie game, but it is not known how to extend it to games played on general strongly-connected graphs. Still, the ideas presented here will be useful later in the constructions of more complex strategies that are extendable.

We describe the tit-for-tat strategy under Richman bidding. 
Intuitively, Min copies Max's actions. 
Formally, Min's strategy is based on a queue of numbers as follows. In each turn, if the queue is empty, Min bids $0$, and otherwise, she bids the maximal number $b$ in the queue. Let $b'$ be Max's bid. Assume WLog that Min wins bidding ties.\footnote{Min can guarantee that she wins a bidding when Max bids $0$ as follows. Assume that her initial budget is $2 \epsilon$. She uses an $\epsilon$ of her budget in the tit-for-tat strategy and the second $\epsilon$ is used as follows: if at turn $i$ the tit-for-tat strategy prescribes a bid of $0$, she bids $2^{-i} \cdot \epsilon$.} If $b' \leq b$, Min wins the bidding and decreases the energy by $1$. If $b' > b$, Min adds $b'$ to the queue and the energy by $1$. Note that Min's budget suffices to bid $b$. Indeed, $b$ corresponds to a Max winning bid in a previous round, which was paid to Min. 

We claim that the tit-for-tat strategy guarantees a non-positive payoff. Let $k$ be the energy following a finite play and $N$ be the size of the queue. Observe that $k \leq N$. Let $b_0$ be the minimal element in the queue. Observe that $N \leq \lceil \frac{1}{b_0} \rceil$. Indeed, Min's bid is at least $b_0$, any Max win must exceed $b_0$, and Max's budget is at most $1$. Note that  Min cannot impose a bound on the queue in a play. Indeed, Max can choose $b_1 > b_2 > \ldots$, bid $b_i$ when the queue is empty, fill it to $\lceil \frac{1}{b_i} \rceil$, and then let Min win to empty the queue. However, this requires Max to empty the queue infinitely often. To conclude, the tit-for-tat strategy guarantees that no matter how Max responds, in the resulting infinite play either:
\begin{itemize}
\item the energy is bounded by a constant, or
\item there are infinitely many prefixes in which the energy is $0$.
\end{itemize}
It is not hard to see that in both cases, the payoff is non-positive, we thus obtain \Cref{16-lem:Min-bowtie}.

\subsubsection{A second optimal strategy in the bowtie game}
\stam{
We start by constructing an optimal strategy in the bowtie game.

Redefine the weights to be $w(v_\mMin) = -1$ and $w(v_\mMax) = 1$. Clearly, a Min strategy that guarantees a non-negative payoff with the new rewards, guarantees payoff at least $0.5$ in the original game. 
The construction is based on the following observation:

\smallskip
\noindent{\bf Observation:} Suppose that Min's strategy guarantees that in every play $\pi$ the energy either (1)~equals $0$ infinitely often, or (2)~is bounded from above by a constant $N$. Then, $\payoff(\pi) \geq 0$.
\smallskip
}
We develop a second construction, which we will later extend to general strongly-connected games. Similar to the tit-for-tat strategy, Min's strategy guarantees that no matter how Max responds, in the resulting play $\pi$ the energy either (1)~equals $0$ infinitely often, or (2)~is bounded from above by a constant $N$.

Let $\epsilon >0$ be Min's initial budget. Assume that the initial energy is $k_0 > 0$ (that is, we consider a harder case in which the energy does not necessarily start from $0$). Min proceeds as follows. She chooses $N \in \N$ such that $\epsilon = k_0/N + \delta$, for some $\delta > 0$. This is possible since $\epsilon$ and $k_0$ are constants. 
We call $\delta$ the ``spare change'' and the rest of her budget ``main budget''. Min initially ignores the spare change. 
As long as the energy is positive, Min bids $\frac{1}{N}$. 

We claim that Min's strategy maintains the following invariant: 

\smallskip
\noindent{\bf Invariant:} Suppose that following a finite play, the energy is $k > 0$ and Min's budget is $B$. Then, $B >  \frac{k}{N}$.
\smallskip 

The invariant holds initially by the choice of $N$. For the inductive step, we distinguish between two cases. If Min wins the bidding, the energy drops to $k-1$ and since she pays $\frac{1}{N}$, her main budget drops to $\frac{k-1}{N}$. On the other hand, if she loses the bidding, the energy increases to $k+1$, and Max pays her at least $\frac{1}{N}$, thus her main budget increases to $\frac{k+1}{N}$. Note that Min ignores the spare change, hence the strict inequality in the invariant.

We claim that the invariant implies that the energy is bounded from above by $N$. Indeed, by plugging $k=N$ in the invariant, we get that an energy of $N$, implies Min's budget exceeds the total budget of $1$.

To conclude the construction, Min plays as follows with an initial budget of $\epsilon > 0$. When the energy is $0$, Min bids $0$ and ``waits'' for the energy to increase. Suppose that it increases to $k_1> 0$. Min chooses $N_1$ with $\epsilon = k_1/N_1 +\delta_1$ and $\delta_1 > 0$. She plays according to the strategy above. If the energy never reaches $0$ again, it is bounded by $N_1$. Note that since Min ignores the spare change, if the energy drops to $0$, her budget is at least $\delta_1$. Min restarts her strategy with an initial budget of $\delta_1$, waits for the energy to increase to $k_2$, selects $N_2$ accordingly, and so on. Let $\pi$ be a play that agrees with Min's strategy. Note that there are two cases: (1)~the energy hits $0$ $n-1$ times after which it stays positive, then it is bounded by $N_n$, and (2)~the energy hits $0$ infinitely often. In both cases, we have $\payoff(\pi) \leq 0$, and we obtain \Cref{16-lem:Min-bowtie}.


\subsubsection{General strongly-connected games}
The central idea in the strategy construction above is to maintain an invariant between energy and budget. 
The challenge in general games is that unlike the bowtie game, loops are not trivial, implying that the outcome of a bidding is not tightly connected with changes in energy. 

\begin{figure}[ht]
\centering
\begin{tikzpicture}
\begin{scope}[every node/.style={circle,thick,draw}]
    \node (A) at (0,0) {$-2$};
    \node (B) at (2,0) {$1$};
    \node (C) at (4,0) {$0.5$};
    \node (D) at (6,0) {$1$};
\end{scope}

\node (N) at (0,-1) {$v_1$};
\node (N) at (2,-1) {$v_2$};
\node (N) at (4,-1) {$v_3$};
\node (N) at (6,-1) {$v_4$};

\node (N) at (0,-1.5) {Po: $-4$};
\node (N) at (2,-1.5) {Po: $-1$};
\node (N) at (4,-1.5) {Po: $0$};
\node (N) at (6,-1.5) {Po: $2$};

\node (N) at (0,-2) {St: $2$};
\node (N) at (2,-2) {St: $2$};
\node (N) at (4,-2) {St: $1.5$};
\node (N) at (6,-2) {St: $1$};

\path [->, style={very thick}] (A) edge[bend right=-40] (C);
\path [->,style={very thick}] (B) edge[bend right=20] (C);
\path [->, style={draw=red,very thick}] (C) edge[bend right=20] (B);
\path [->,style={draw=red,very thick}] (B) edge (A);
\path (A) edge [->,out=200,in=160,looseness=8,style={draw=red,very thick}] node[above] {} (A);

\path [->,style={very thick}] (C) edge[bend right=20] (D);
\path [->, style={draw=red,very thick}] (D) edge[bend right=20] (C);
\path (D) edge [->,out=-20,in=20,looseness=8,style={very thick}] node[above] {} (D);
\end{tikzpicture}
\caption{A mean-payoff bidding game $\G$ with $\MP(\RT(\G)) = 0$, where we depict the weight of a vertex inside it, its potential and strength below it, and with a red edge the vertex to which Min moves the token upon winning a bidding.}
\label{16-fig:MP-example}
\commentAlt{Figure~\ref{16-fig:MP-example}: A directed graph with four circular nodes and a table below showing numerical values associated with each node for "Po" and "St" categories. See long description.}
\commentLongAlt{Figure~\ref{16-fig:MP-example}: The image displays a directed graph at the top and a table of values below it.

The graph shows four circular nodes arranged horizontally from left to right, labeled with numbers: '-2', '1', '0.5', and '1'.
- The leftmost node ('-2') has a red self-loop.
- An arrow from '-2' points to '1' (red).
- A curved black arrow from '-2' points to '0.5'.
- An arrow from '1' points to '-2' (red).
- A straight black arrow from '1' points to '0.5'.
- An arrow from '0.5' points to '1' (red).
- A straight black arrow from '0.5' points to the rightmost '1'.
- The rightmost node ('1') has a self-loop.

The table below the graph has two rows, "Po:" and "St:", and four columns corresponding to the nodes 'v1', 'v2', 'v3', and 'v4'.
- For node 'v1' (corresponding to '-2'): Po: -4, St: 2
- For node 'v2' (corresponding to '1'): Po: -1, St: 2
- For node 'v3' (corresponding to '0.5'): Po: 0, St: 1.5
- For node 'v4' (corresponding to '1'): Po: 2, St: 1}
\end{figure}

\begin{example} 
\label{16-ex:first-SCC-MP}
Consider the game depicted in \Cref{16-fig:MP-example}. We illustrate the challenge in maintaining an invariant between the budget and energy. Suppose that the token is placed on $v_3$, Min wins a bidding, and proceeds left. 
Note that unlike the bowtie game, where whenever Min wins, the energy decreases, here the energy increases following a Min win. To make things worse, Moreover, suppose that the game returns to $v_3$, then a decrease in energy occurs only if Min wins the bidding in $v_2$. 

We describe the desiderata of Min's bids. Suppose that Min's budget at $v_3$ is $B > k/N$, where $k$ is the current energy and $N$ is chosen as in the construction above and Min chooses the bids $b_{v_1}, b_{v_2}$, and $b_{v_3}$ in the respective vertices. Note that if Min loses the bidding in $v_2$ and Max moves to $v_3$, the energy when returning to $v_3$ is $k' = k-0.5 + 1 = k-0.5$ and Min's budget is updated to $B' \geq B -b_{v_3} + b_{v_2}$. Similarly, if Min wins the bidding in $v_2$, proceeds to $v_1$, and Max wins the bidding there, the updated energy is $k' = k-0.5+1-2$ and Min's budget is $B' \geq B -b_{v_3} - b_{v_2} + b_{v_1}$. We will find bids so that in both cases, we have $B' \geq k'/N$. 
\end{example}

Next, we develop a measure of how important it is to move in a vertex. 
Very roughly, suppose that the measure finds that the importance of vertex $v$ is $s \in \Q$. Then, Min (roughly) bids $s \cdot \frac{1}{N}$ in $v$. 
(The bowtie game is a special case in which both vertices are equally important). The definition of the measure is based on the concept of potential in stochastic games (see \Cref{7-chap:stochastic}).


\begin{definition}
({\bf Potentials and strengths})
Consider a strongly-connected mean-payoff game $\game$. Let $f$ and $g$ be, respectively, Min and Max optimal positional strategies in $\RT(\game, 0.5)$. Recall following a coin toss in a vertex $v$, the game $\RT(\game, 0.5)$ proceeds either to $v_{\mMin}$ and $v_{\mMax}$, which simulate the outcome of the coin toss. For $v \in V$, let $v^-,v^+ \in V$ be such that $f(v_\mMin) = v^-$ and $g(v_\mMax) = v^+$. That is, $v^-$ represents a Min optimal move upon winning the coin toss in $v$, and dually $v^+$ is a Max optimal move. The potential of $v$ is a function that satisfies the following
\[
\Pot(v) = \frac{1}{2}\big( \Pot(v^+) + \Pot(v^-)\big)  + w(v)  - val\big(\RT(\game, 0.5)\big)
\]
The {\em strength} of $v$ is the difference in potentials:
\[
\St(v) = \frac{1}{2} \big(\Pot(v^+) - \Pot(v^-)\big)
\]
\end{definition}
There are optimal strategies for which $\Pot(v^-) \leq \Pot(v') \leq \Pot(v^+)$, for every $v' \in N(v)$, which can be found for example using the strategy iteration algorithm.

\begin{example}
Consider again the game depicted in \Cref{16-fig:MP-example}. The strengths and potentials are depicted below the vertices. We revisit the outcomes described in \Cref{16-ex:first-SCC-MP}. Suppose that the game starts from $v_3$ with Min's budget exceeding $k/N$, where $k$ is the current energy and $N$ is chosen as before.
Define Min's bid in a vertex $v \in V$ to be $\St(v)/N$ and she proceeds to $v^-$ upon winning the bidding. Min's moves are depicted in red in \Cref{16-fig:MP-example}.  
Consider the outcome $v_3, v_2, v_3$, which corresponds to winning the first bidding and losing the second bidding. The updated energy is $k' = -0.5$ and the change in budget is $1/N \cdot\big(-\St(v_3) + \St(v_2)\big) = -0.5/N$. The outcome $v_3, v_2, v_1, v_3$ corresponds to two bidding wins followed by a lose. The updated energy is  $k' = 1.5$ and the change in budget is $1/N \cdot\big(-\St(v_3) - \St(v_2) + \St(v_1) \big) = 1.5/N$. In both cases, upon returning to $v_3$, we have $B' > k'/N$. We encourage the reader to verify that in every cycle in the game, the change in energy equals the (normalized) change in budget. 
\end{example}

We formalize the guarantees that lead to the invariant between energy and budget. 
Consider a strongly-connected mean-payoff bidding game $\game = \zug{V, E, w}$. Consider a finite path $\eta = v_1,\ldots, v_n$ in $\game$. It is useful to think of $\eta$ as an outcome of a bidding game in which Min bids $\St(v)$ in vertex $v$. If the successor of $v$ in $\eta$ is $v^-$, we think of Min as the ``winner'' of the bidding in $v$, and otherwise Max ``wins'' the bidding. 
We call Min wins {\em investments} and Min loses {\em gains}, where intuitively he {\em invests} in increasing the energy and {\em gains} budget whenever the energy decreases. Formally, we define the  gains and investments in $\eta$ as follows:
\[
\invest_\mMin(\eta) = \sum_{i: v_{i+1} = v_i^-} \St(v_i) \ \text{ and }\  \gain_\mMin(\eta) = \sum_{i: v_{i+1} \neq v_i^-} \St(v_i)
\]


The following lemma connects the strength, potential, and accumulated energy.

\stam{
\begin{example}
Consider the game that is depicted in \Cref{16-fig:MP-example}. 
The {\em change in budget} throughout $\eta$ is sum of gains minus We will show that in every cycle,  

For each vertex $v$, we depict $v^-$ by using a bold red edge. We illustrate Lemma~\ref{lem:magic}. Consider the path $\eta = v_3, v_2, v_1, v_1, v_4$, which intuitively corresponds to Min winning three biddings and then losing two. For example, we have $v^-_3 = v_2$ and $v^-_1 = v_1$, thus when losing the bidding in $v_1$, Max would proceed to $v_3$. The energy is $E(\eta) = -0.5+1-2-2-0.5 = -4$ (recall that the last vertex does not contribute to the accumulated energy), the gain is $G(\eta) = \St(v_1) + \St(v_4) = 3.5$, the investment is $I(\eta) = \St(v_3) + \St(v_2) + \St(v_1) = 5.5$, and the potentials of the two end points are $\Pot(v_3) = 0$ and $\Pot(v_4) = 2$. Plugging in the values, we have $\Pot(v_3) - \Pot(v_4) = 0-2 = -4 -3.5 + 5.5 = E(\eta) - G(\eta) + I(\eta)$.\hfill\qed
\end{example}
}


\begin{lemma}
\label{16-lem:magic}
Consider a strongly-connected game $\game$ with $val\big(\RT(\game, 0.5)\big)= 0$, and a finite path $\eta$ in $\game$ from $v$ to $u$. Then, $\Pot(v) - \Pot(u) \geq \energy(\eta) -  \gain_\mMin(\eta) + \invest_\mMin(\eta)$. 
\end{lemma}
\begin{proof}
We prove by induction on the length of $\eta$. For $n=1$, the claim is trivial since both sides of the equation are $0$. Suppose the claim is true for paths of length $n$ and we prove for paths of length $n+1$. We distinguish between two cases. In the first case, Min wins in $v$, thus the second vertex in $\eta$ is $v^-$. Let $\eta'$ be the suffix of $\eta$ starting from $v^-$. Note that since Min wins the first bidding, we have $\gain_\mMin(\eta) = \gain_\mMin(\eta')$ and $\invest_\mMin(\eta) = \St(v) + \invest_\mMin(\eta')$. Also, we have $\energy(\eta) = \energy(\eta') + w(v)$. Combining these, we have $\energy(\eta) - \gain_\mMin(\eta) + \invest_\mMin(\eta) = \energy(\eta') + w(v) - \gain_\mMin(\eta') + \invest_\mMin(\eta') + \St(v)$. By the induction hypothesis, we have $\Pot(v^-) - \Pot(u) \geq \energy(\eta') - \gain_\mMin(\eta') + \invest_\mMin(\eta')$. Combining these with the definition of $\St(v)$, we have the following. 
\[\energy(\eta) - \gain_\mMin(\eta) + \invest_\mMin(\eta) \leq  \St(v) + \Pot(v^-) +w(v) - \Pot(u) = \]\[=\frac{1}{2}\big(\Pot(v^+)-\Pot(v^-)\big) + \Pot(v^-)+w(v)- \Pot(u)= \Pot(v) - \Pot(u).\]

We continue to the second case in which Max wins in $v$ and let $v'$ be the second vertex in $\eta$. Recall that we have $\Pot(v^+) \geq \Pot(v')$. Dually to the first case, we have $\gain_\mMin(\eta) = \St(v) + \gain_\mMin(\eta')$ and $\invest_\mMin(\eta) = \invest_\mMin(\eta')$. 
\[ 
\energy(\eta) - \gain_\mMin(\eta) + \invest_\mMin(\eta) = \energy(\eta') - \gain_\mMin(\eta') + \invest_\mMin(\eta') - \St(v) + w(v) \leq\]
\[\leq \Pot(v') - \St(v) + w(v) - \Pot(u) \leq \Pot(v^+) - \frac{1}{2}\big(\Pot(v^+)- \Pot(v^-)\big) + w(v) - \Pot(u) =\] 
\[= \Pot(v) - \Pot(u).
\]
\end{proof}

\subsubsection{Putting it all together}
Choosing $N$ in general graphs requires some book-keeping due to paths that are not cycles. Consider a positive initial budget $B \in (0,1]$ for Min and an initial energy $k_I \in \N$. Let $\Pot_M=\max_{v \in V} |\Pot(v)|$ and $\St_M = \max_{v \in V} |\St(v)|$. We choose $N \in \N$ such that $B > \frac{k_I + \St_M + 2\Pot_M}{N}$. When the game reaches $v \in V$, Min bids $\St(v)/N$ and moves to $v^-$ upon winning. If the energy is $0$, Min bids $0$ until it rises above $0$, then Min restarts her strategy. 

The following lemma shows that Min's strategy is legal, namely Min's budget suffices to bid as the strategy requires. The proof follows from \Cref{16-lem:magic}.

\begin{lemma}
\label{16-lem:Min-invariant}
Consider a positive initial energy $k_I \in \N$. Let $\pi$ be a finite play that results from Min's strategy played against some Max strategy, where the energy throughout $\pi$ never reaches $0$, \textit{i.e.}, for every prefix $\pi^n$, for $0 \leq n \leq |\pi|$, we have $k_I + E(\pi^n) >0$. Let $k = k_I + E(\pi)$ be the energy following $\pi$. Then, Min's budget following $\pi$ is at least $\frac{k+\St_M}{N}$.
\end{lemma}
\stam{
\begin{proof}
The invariant clearly holds initially. With a slight abuse of notation, let $G(\pi)$ be the sum of ``gains'' in $\path(\pi)$, namely the sum of strengths in vertices in which \Max wins the bidding, and similarly $I(\pi)$ be the ``investments'' in $\path(\pi)$, namely the sum of strengths in vertices in which \Min wins the bidding. Let $B$ be \Min's initial budget and $B'$ his budget following $\pi$. Since \Min bids $\St(v)/N$ in a vertex $v$, we have $B' = B + \big(G(\pi) - I(\pi)\big)/N$. From Lemma~\ref{lem:magic}, we have $2\Pot_M -E(\pi) \geq I(\pi)-G(\pi)$. By combining with $k = k_I + E(\pi)$ and re-arranging, we have 
\[
2\Pot_M -E(\pi) \geq  I(\pi)-G(\pi) = N(B- B')
\]
\[
B- \frac{2\Pot_M- k+k_I}{N} \leq B'
\]
Since we define $B > \frac{k_I + \St_M + 2\Pot_M}{N}$, we obtain that $B' > \frac{k+\St_M}{N}$, and we are done.
\end{proof}
}

Since Min's budget cannot exceed $1$, \Cref{16-lem:Min-invariant} implies that if the energy does not reach $0$, then it is bounded by $N-\St_M$. The following theorem follows. 

\begin{theorem}
\label{16-thm:MP-min}
Let $\game$ be a strongly-connected mean-payoff bidding game with $\MP\big(\RT(\game)\big)=0$. Then, from every vertex in $\game$ and with any positive initial budget, Min can guarantee a non-positive payoff.
\end{theorem}

\subsection{An $\epsilon$-optimal Max strategy under Richman bidding}
We develop $\epsilon$-optimal strategies for Max in strongly-connected mean-payoff games. 

\subsubsection{Solving the bowtie game}
We first describe an $\epsilon$-optimal strategy for Max in the bowtie game. 

\begin{lemma}
\label{16-prop:Max-bowtie}
In the bowtie game (\Cref{16-fig:bowtie}), under first-price Richman bidding, for every initial ratio $r \in (0,1)$ and $\epsilon >0$, 
Max can guarantee a payoff of at least $0.5-\epsilon$.
\end{lemma}
\begin{proof}
Let $\epsilon >0$ and let $B_0>0$ be Max's initial budget. We show that Max can guarantee a payoff of at least $\frac{1}{2+\epsilon}$. We re-define the weights to be $w(v_\mMax) = c = 1+\epsilon$ and $w(v_\mMin) = -1$. It is not hard to see that if Max guarantees a non-negative payoff with the new weights, he guarantees a payoff of at least $\frac{1}{2+\epsilon}$ with the old weights. 


Let $\alpha > 0$ such that $(1+\alpha) = (1-\alpha)^{-c}$. One can show that such $\alpha$ exists since $c > 1$. Max maintains the following invariant:

\smallskip
\noindent{\bf Invariant:}  Suppose that following a finite play, the energy is $k \in \N$. Then, Max's budget exceeds $(1+\alpha)^{-k}$.
\smallskip

For the base case, we choose an initial energy of $k_0$ such that $B_0 > (1+\alpha)^{-k_0}$. Finding such $k_0$ is possible since $\lim_{k \to \infty} (1+\alpha)^{-k} = 0$. 

For the inductive step, suppose that following a finite play, the energy is $k$. Let $B = (1+\alpha)^{-k}$. The invariant states that Max's budget is $B + \delta$, for some $\delta >0$. Max bids $\alpha \cdot B$. 
We distinguish between two bidding outcomes. If Max loses the bidding, the energy decreases to $k-1$. Since Min overbids, Max's updated budget is at least 
\[B + \alpha B +\delta > \frac{1}{(1+\alpha)^{k}} + \frac{\alpha}{(1+\alpha)^{k}} = (1+\alpha)^{-(k-1)}.\] 
On the other hand, if Max wins the bidding, the energy increases to $k+c$ and his updated budget is:
\[B - \alpha B + \delta > \frac{1}{(1+\alpha)^{k}} - \frac{\alpha}{(1+\alpha)^{k}} = \frac{1 - \alpha}{(1+\alpha)^{k}}= (1+\alpha)^{-(k+c)}.\]
The last equality follows from $(1-\alpha) = (1+\alpha)^{-c}$. This concludes the proof of the invariant.

\smallskip
\noindent{\bf Claim:}  The energy in a finite play is positive. 
\smallskip

Indeed, assume towards contradiction that a play ends with $k=0$. The invariant implies that Max's budget is greater than $1$, which is a contradiction since the total budget is $1$. We thus obtain:

Recall that we choose an initial energy of $k_0$. If in fact the initial energy is $k'_0$, then the claim implies that the energy in a finite play is greater than the constant $k'_0 - k_0$. It is not hard to see that the definition of mean payoff (\Cref{16-def:MP}) implies that by keeping the energy above a constant, Max guarantees a non-negative payoff, and we are done. 
\end{proof}

\subsubsection{Generalized potentials and strengths}
Before extending the construction in \Cref{16-prop:Max-bowtie} from the bowtie game to general strongly-connected games, we generalize the technique in \Cref{16-lem:magic}. 

Let $p \in [0,1]$. Recall that $\RT(\game, p)$ is the random-turn game in which in each turn, Max is chosen to move with probability $p$. For $v \in V$, as before, we define $v^+$ and $v^-$ to be the vertex that, respectively, Max and Min proceed to upon being chosen to move in $v$. The strengths and potentials with respect to $p$ are functions that satisfy: 
\begin{equation*}
\begin{split}
&\Pot_p(v) = p \cdot \Pot_p(v^+) + (1-p) \cdot \Pot_p(v^-) + w(v) - \MP(\RT(\G,p))\\
&\St_p(v) = p\cdot (1-p) \cdot \big(\Pot_p(v^+) - \Pot_p(v^-)\big)
\end{split}
\end{equation*}
Consider a path $\eta = v_1,\ldots, v_n$ in $\game$. We define Max's gains and investments dually to the definition for Min:
\[
\invest_\mMax(\eta) = \sum_{i: v_{i+1} = v_i^+} \St(v_i) \ \text{ and }\  \gain_\mMax(\eta) = \sum_{i: v_{i+1} \neq v_i^+} \St(v_i)
\]

The following lemma generalizes \Cref{16-lem:magic} and is proven by induction, similar to that proof. We describe the intuition behind the statement. 
Suppose that Max's budget is $\nu$ and Min's budget is $\mu$, thus Max's ratio is $\nu/(\nu+\mu)$. Since the players start off with different budgets, winning and losing a bid of $b$ affects their budgets differently. One can think of the budgets as ``currencies'': when Max wins, it is as if he pays $b \cdot \mu$ in ``Min's currency'' and whenever Min pays, she pays $\nu \cdot b$ in ``Max's currency''.

\stam{
We suggest an intuitive way to read the definition of potentials. Consider a weighted Markov chain in which each vertex $v$ has two neighbors $v^+$ and $v^-$, and the probability of proceeding from $v$ to $v^+$ and $v^-$ is respectively $p$ and $1-p$. Suppose there is a target that is reached with probability $1$. Let $E(v)$ denote the expected energy in a path from $v$ to the target. We note that $E(v)$ roughly coincides with $\Pot_p(v)$. Indeed, to compute $E(v)$, we take the current energy, \textit{i.e.}, $w(v)$, and add, for each neighbor $u$, the probability of proceeding from $v$ to $u$ multiplied by the expected energy to the target from $u$, thus we obtain $E(v) = w(v) + p \cdot E(v^+) + (1-p) \cdot E(v^-)$.

The following lemma connects potential, energy, and strength and thus gives rise to the invariant between budget and energy.
We need several definitions.
Consider a finite path $\eta = v_1,\ldots, v_n$ in $\G$.
We intuitively think of $\eta$ as a play,
where for every $1 \leq i < n$,
the bid of \Max in $v_i$ is $\St_p(v_i)$
and he moves to $v_i^+$ upon winning.
Thus, when $v_{i+1} = v_i^+$, we think of \Max as {\em investing} $\St_p(v_i)$
and when $v_{i+1} \neq v_i^+$, we think of \Min winning the bid
thus \Max\ {\em gains} $\St_p(v_i)$.
We denote by $\investmax(\eta)$ and $\gainmax(\eta)$ the sum of investments and gains,
respectively. Recall that the energy of $\eta$ is $\energy(\eta) = \sum_{0 \leq i <n} w(v_i)$.
}

\begin{lemma}
\label{16-lem:magic-general}
Consider a strongly-connected game $\game$ and $p = \frac{\nu}{\nu+\mu} \in (0,1)$. For every finite path $\eta$ of length $n$ from $v$ to $u$, we have 
\[\Pot_p(v) - \Pot_p(u) + (n-1)\cdot\MP\big(\RT(\game,p)\big) \leq \energy(\eta) + \frac{\nu+\mu}{\nu\mu}\cdot\big( \gain_\mMax(\eta) \cdot \nu - \invest_\mMax(\eta)\cdot \mu \big).\]
\end{lemma}

\begin{figure}[ht]
\centering

\begin{tikzpicture}
\begin{scope}[every node/.style={circle,thick,draw}]
    \node (A) at (0,0) {$1$};
    \node (B) at (2,0) {$-1$};
    \node (C) at (4,0) {$-1$};
    \node (D) at (6,0) {$-2$};
\end{scope}

\node (N) at (0,-1) {$v_1$};
\node (N) at (2,-1) {$v_2$};
\node (N) at (4,-1) {$v_3$};
\node (N) at (6,-1) {$v_4$};

\node (N) at (0,-1.5) {Po: $6$};
\node (N) at (2,-1.5) {Po: $3$};
\node (N) at (4,-1.5) {Po: $0$};
\node (N) at (6,-1.5) {Po: $-3$};

\node (N) at (0,-2) {St: $\frac{2}{3}$};
\node (N) at (2,-2) {St: $\frac{4}{3}$};
\node (N) at (4,-2) {St: $\frac{4}{3}$};
\node (N) at (6,-2) {St: $\frac{2}{3}$};

\path [->, style={very thick}] (A) edge[bend right=20] (B);
\path [->, style={draw=red,very thick}] (B) edge[bend right=20] (A);

\path [->,style={very thick}] (B) edge[bend right=20] (C);
\path [->, style={draw=red,very thick}] (C) edge[bend right=20] (B);

\path [->,style={draw=red,very thick}] (D) edge[bend right=20] (C);
\path [->, style={very thick}] (C) edge[bend right=20] (D);

\path (A) edge [->,out=200,in=160,looseness=8,style={draw=red,very thick}] node[above] {} (A);

\path (D) edge [->,out=-20,in=20,looseness=8,style={very thick}] node[above] {} (D);
\end{tikzpicture}

\caption{A mean-payoff game $\G$ with $\MP(\RT(\G, 2/3))= 0$, Max's optimal moves are depicted in red, and $\Pot_{2/3}$ and $\St_{2/3}$ are depicted below the vertices.}
\label{16-fig:ex-2}
\commentAlt{Figure~\ref{16-fig:ex-2}: A linear sequence of four circular nodes with a mixture of black and red arrows indicating transitions and self-loops, and a table below showing numerical values associated with each node for "Po" and "St" categories. See long description.}
\commentLongAlt{Figure~\ref{16-fig:ex-2}: The image displays a linear directed graph at the top and a table of values below it.

The graph shows four circular nodes arranged horizontally from left to right, labeled with numbers: '1', '-1', '-1', and '-2'.
- The leftmost node ('1') has a red self-loop.
- A black arrow points from '1' to the second '-1' node.
- A red arrow points from the second '-1' node to '1'.
- A black arrow points from the second '-1' node to the third '-1' node.
- A red arrow points from the third '-1' node to the second '-1' node.
- A black arrow points from the third '-1' node to '-2'.
- A red arrow points from '-2' to the third '-1' node.
- The rightmost node ('-2') has a black self-loop.

The table below the graph has two rows, "Po:" and "St:", and four columns corresponding to the nodes 'v1', 'v2', 'v3', and 'v4'.
- For node 'v1' (corresponding to the leftmost '1'): Po: 6, St: 2/3
- For node 'v2' (corresponding to the second '-1'): Po: 3, St: 4/3
- For node 'v3' (corresponding to the third '-1'): Po: 0, St: 4/3
- For node 'v4' (corresponding to '-2'): Po: -3, St: 2/3}
\end{figure}

\begin{example}
Consider the game depicted in~\Cref{16-fig:ex-2}. Let $\nu = 2$ and $\mu = 1$, thus $p = \frac{2}{3}$. 
First, note that $\Value(\RT(\G, 2/3))=0$. Indeed, one can verify that the stationary distribution of $\RT(\G, 2/3)$ is $\zug{\frac{8}{15}, \frac{4}{15}, \frac{2}{15}, \frac{1}{15}}$, and the value is obtained by multiplying by the weights.
Second, one can verify the potentials and strengths, which are depicted below the respective vertex. In particular, for every vertex $v$, the vertex on its left is $v^+$, thus Max always proceeds left upon winning a bidding. For example, $\Pot_{2/3}(v_3) = 2/3 \cdot \Pot_{2/3}(v_2) + 1/3\cdot\Pot_{2/3}(v_4) - w(v_3) = 2/3 \cdot 3 + 1/3 \cdot (-3) -1 = 0$ and $\St_{2/3}(v_3) = \frac{2}{9} \cdot \big(\Pot_p(v_2) - \Pot_p(v_4)\big) = \frac{4}{3}$.


\Cref{16-lem:magic-general} is simplest to verify in cycles since the left-hand side of the inequality cancels out. Consider the cycle $\eta = v_3, v_2, v_1, v_1, v_2, v_3$ in which Max wins the first three bids and loses the last two biddings. We have $\invest_\mMax(\eta) = \frac{4}{3} + \frac{4}{3} + \frac{2}{3}$, $\gain_\mMax(\eta) = \frac{2}{3}+ \frac{4}{3}$, and $\energy(\eta) = -1$ (the last vertex does not contribute to the energy). As \Cref{16-lem:magic-general} predicts, we have $\frac{3}{2}\big(\invest_\mMax(\eta) \cdot \mu - \gain_\mMax(\eta)\cdot \nu \big)= \frac{3}{2} \cdot (\frac{10}{3} \cdot 1- \frac{6}{3} \cdot 2) = -1 = \energy(\eta)$. We encourage the reader to verify that every cycle has this property.\hfill\qed
\end{example}

We obtain the following corollary by dividing both sides by $n$, 
and letting $n$ tend towards infinity.

\begin{corollary}\label{16-cor:magic}
Consider a strongly-connected game $\game$, let $\eta$ be an infinite play and let $\mu,\nu \in \R_{> 0}$.
For every $n \in \N$, let $\eta^n$ denote the prefix of $\eta$ of size $n$.
Then
\[
\payoff(\eta)
\geq
\MP(\RT(\G, p)) +
\frac{\mu + \nu}{\mu\nu} \cdot \liminf_{n\rightarrow \infty}\frac{\mu \cdot \invest_\mMax(\eta^n) -\nu \cdot \gain_\mMax(\eta^n)}{n}.
\]
\end{corollary}

\subsubsection{Solving general strongly-connected games}

We are ready to construct $\epsilon$-optimal strategies for Max in general strongly-connected games. 

\begin{lemma}
\label{16-lem:MP-Max-Rich}
Let $\game$ be a strongly-connected mean-payoff Richman-bidding game.
With any positive initial budget, for every  $\epsilon > 0$,
Max can guarantee a payoff of at least $\MP(\RT(\G, \frac{1}{2 + \epsilon}))$.
\end{lemma}
\begin{proof}
Let $c = 1+\epsilon$, and let $p = \frac{1}{2 + \epsilon} = \frac{1}{1+c}$. Consider two optimal positional strategies in $\RT(\game, \frac{1}{2 + \epsilon})$ and use them to define vertex strengths and $v^+$ and $v^-$, for each vertex $v$.
Let $\alpha \in (0,1)$ such that 
$1-\alpha=(1+\alpha)^{-c}$. 

Max's strategy proceeds as follows. Denote by $S_{\mMax}$ the maximal strength. Suppose that Max's budget is $B$ and the token is placed on vertex $v$ with strength $s = \St_p(v)$. Then, Max bids $\alpha B\frac{s}{S_{\mMax}}$ and proceeds to $v^+$ upon winning the bidding. We show that Max's strategy maintains the following invariant:

\smallskip
\noindent{\bf Invariant:} 
Suppose that Max's budget following a finite play $\pi$ is $B$. Then, 
\begin{equation}\label{16-eq:inv_APrichman}
B \geq (1+\alpha)^{\frac{-c\cdot\invest_\mMax(\pi)+\gain_\mMax(\pi)}{2S_{\mMax}}}.
\end{equation}

We claim that the invariant implies that lemma. Indeed, since $B \leq 1$, we obtain a bound on $-c\cdot\invest_\mMax(\pi)+\gain_\mMax(\pi)$ in every finite play. Thus, by \Cref{16-cor:magic}, Max guarantees a payoff of at least $\Value\big(\RT(\game, p)\big)$.

To show the invariant, first, we assume that the initial energy $k_0$ is such that $B \geq (1+\alpha)^{-k_0}$. For the inductive step, we show the case in which Min wins the bidding following a finite play $\pi$, and the case that Max wins is dual. Let $\pi'$ be the continuation of $\pi$ after the bidding. Max's updated budget is at least:
\[
B+\alpha B\frac{s}{S_{\mMax}}= B\cdot (1+\alpha\cdot \frac{s}{S_{\mMax}}) \stackrel{\text{Bernoulli}}{\geq} B \cdot (1+\alpha)^{\frac{s}{2S_{\mMax}}} \stackrel{\text{ind. hyp.}}{\geq} \]
\[ \geq (1+\alpha)^{\frac{-c\cdot\invest_\mMax(\pi)+\gain_\mMax(\pi)}{2S_{\mMax}} + \frac{s}{2S_{\mMax}}} = (1+\alpha)^{\frac{-c\cdot\invest_\mMax(\pi')+\gain_\mMax(\pi')}{2S_{\mMax}}}
\]
Bernoulli's inequality can be used since $\alpha > -1$ and $\frac{s}{2S_{\max}}=\frac{y-x}{\alpha B}\in [0,1]$.
\end{proof}

Note that since the definition of mean-payoff (\Cref{16-def:MP}) gives Min an advantage, the construction in \Cref{16-lem:MP-Max-Rich} can be applied to construct an $\epsilon$-optimal strategy for Min, which suffices to prove the following theorem (see \Cref{16-def:MP-value}). We point out that we show a stronger claim in \Cref{16-thm:MP-min}; Min has optimal strategies rather than $\epsilon$-optimal strategies.

\begin{theorem}
\label{16-thm:MP-max}
For every strongly-connected mean-payoff Richman-bidding game $\game$, for every budget ratio $r \in [0,1]$, we have $\Value(\game, r) = \Value\big(\RT(\game, 0.5)\big)$.
\end{theorem}


\subsection{Mean-payoff poorman-bidding games}

In this section, we characterise the mean-payoff value of mean-payoff poorman-bidding games. The crux of the proof entails a construction of $\epsilon$-optimal strategies for Max. 

\subsubsection{Solutions to the bowtie game}
We describe two solutions to the bowtie game. These solutions are not complete and, unlike Richman bidding, do not directly generalize to general strongly-connected games. Still, the ideas that we develop here will be useful later. Throughout this section, we re-normalize the weights so that $w(v_\mMin) = -1$ and $w(v_\mMax) = 1$. 

First, we adapt the tit-for-tat strategy to poorman bidding, thereby constructing an optimal strategy for Min in the bowtie game. We point out that unlike Richman-bidding (\Cref{16-sec:min-MP}), we do not know of constructions of optimal strategies for Min in general strongly-connected games, and we will only construct $\epsilon$-optimal strategies based on a construction for Max.

\begin{lemma}
\label{16-lem:tit-for-tat}
Suppose that Min's initial budget is $\ell \in \N$ and Max's initial budget is $1$. Then, in $\lolli$, Min can guarantee a payoff of at most $\frac{\ell}{\ell + 1}$.
\end{lemma}
\begin{proof}
Recall the tit-for-tat strategy under Richman bidding. Min maintains a queue. When Max wins a bidding with a bid of $b$, Min enters $b$ to the queue. Min's bid $b$ is the maximal element in the queue, and she removes $b$ if she wins the bidding. Further recall that the size of the queue coincides with the energy, and that the size is bounded. 

We adapt the tit-for-tat strategy to poorman bidding. Suppose that Max's budget is $1$ and Min's budget is $\ell \in \N$. Min maintains $\ell$ queues. When Max wins with a bid of $b$, Min enters $b$ in all the queues. She bids the highest number is some queue, and removes it upon winning the bidding. It is not hard to show that Min's strategy is legal. Intuitively, we act as if Min splits her budget of $\ell$ into $\ell$ wallets each with a budget of $1$. Each time Max spends $b$, Min can spend $b$ from each of her wallets, for a total of $\ell \cdot b$. While under Richman bidding, Max can keep Min's queue non-empty, under poorman bidding, this is not the case: eventually all queues empty. This guarantees that Min wins $\ell$-times more bids that Max. 
\end{proof}

Our second construction is of an $\epsilon$-optimal Max strategy for the special case of (roughly) equal budgets. The proof structure, namely, an invariant between energy and budget and using it to show a lower bound on the energy, will continue with us to general strongly-connected games in the next section. 

\begin{lemma}
In the bowtie game, for every $\epsilon>0$, when Min's budget is $1$ and Max's budget is $1+\epsilon$, Max can guarantee payoff of at least $0$. 
\end{lemma}
\begin{proof}
We construct a Max strategy that bounds the energy from below. Similar to the constructions under Richman bidding, Max's strategy maintains an invariant between budget and energy. We describe the invariant that it maintains combinatorially. 
We need several definitions. For $k \in \N$, let $S_k$ be the square of area $k^2$ (see \Cref{16-fig:square}). We split $S_k$ into unit-area boxes. The diagonal in $S_k$ splits it into a small (black) triangle and a large (white) triangle. We respectively denote by $t_k$ and $T_k$ the areas of the two, thus $t_k = \frac{k(k-1)}{2}$ and $T_k = \frac{k(k+1)}{2}$. 

\begin{figure}[ht]
\begin{minipage}{0.4\linewidth}
\center
\begin{tikzpicture}
\draw [step=0.5cm,black,very thin] (0,0) grid (2.5,2.5);
\filldraw [fill=black] (0,0) rectangle (0.5,0.5);
\filldraw [fill=black] (0.5,0) rectangle (1,0.5);
\filldraw [fill=black] (1,0) rectangle (1.5,0.5);
\filldraw [fill=black] (1.5,0) rectangle (2,0.5);

\filldraw [fill=black] (0,0.5) rectangle (0.5,1);
\filldraw [fill=black] (0.5,0.5) rectangle (1,1);
\filldraw [fill=black] (1,0.5) rectangle (1.5,1);

\filldraw [fill=black] (0,1) rectangle (0.5,1.5);
\filldraw [fill=black] (0.5,1) rectangle (1,1.5);

\filldraw [fill=black] (0,1.5) rectangle (0.5,2);
\end{tikzpicture}

\end{minipage}
\begin{minipage}{0.4\linewidth}
\center
\begin{tabular}{|c|c|c|c|c|c|c|}
\hline
& $2$ & $3$ & $4$ & $5$ & $6$ & $\cdots$\\
\hline
 \hline
 $t_k$ & $1$ & $3$ & $6$ & $10$ & $15$ & $\cdots$\\
 \hline
 $T_k$ & $3$ & $6$ & $10$ & $15$ & $21$ & $\cdots$\\
 \hline
  $S_k$ & $4$ & $9$ & $16$ & $25$ & $36$ & $\cdots$\\
  \hline
 \end{tabular}
 \end{minipage}
\caption{Left: The square $S_5$ with area $25$, the triangle $t_5$ (black) with area $10$, and the triangle $T_5$ (white) with area $15$.}
\label{16-fig:square}
\commentAlt{Figure~\ref{16-fig:square}: A square grid on the left with some cells shaded black, and a table on the right showing sequences of numbers for different rows.}
\commentLongAlt{Figure~\ref{16-fig:square}: The image presents two distinct visual elements.

On the left, there is a square grid composed of 5 rows and 5 columns of smaller squares. The bottom-left portion of the grid is filled with black squares, forming a staircase-like pattern. Specifically, the bottom row has 5 black squares, the row above it has 4, the next has 3, then 2, and finally the top row has 1 black square on the far left. The remaining squares are white.

On the right, there is a table with multiple rows and columns. The top row contains column headers: '2', '3', '4', '5', '6', followed by an ellipsis '...'. The subsequent rows contain numerical sequences:
- Row 't_k': '1', '3', '6', '10', '15', followed by '...'.
- Row 'T_k': '3', '6', '10', '15', '21', followed by '...'.
- Row 'S_k': '4', '9', '16', '25', '36', followed by '...'.}
\end{figure}

Suppose that Max's initial budget ratio is $B_0 = \frac{1+\epsilon}{2+\epsilon}$. Choose an initial energy $k_0$ such that $B_0 > \frac{T_{k_0+1}}{S_{k_0+1}}$. Note that this is possible since $\lim_{k \to \infty} T_k/S_k = 0.5$. Max maintains the following invariant:

\smallskip
\noindent{\bf Invariant:}
When the energy is $k \geq 1$, Max's ratio of the budget is $B > \frac{T_{k+1}}{S_{k+1}}$.
\smallskip

The invariant clearly holds initially. We show how Max maintains it. Suppose that the energy is $k$ and Max's budget ratio is $B = \frac{T_{k+1}}{S_{k+1}} + \delta$. 
Intuitively, ignoring for now the ``spare change'' of $\delta$, one should think of Max's budget as if he ``owns'' $T_{k+1}$ boxes. Note that Min owns less than $t_{k+1}$ boxes. Max's bids $1/(k+1)^2$, which should be thought of as one box in the $k \times k$ square. We distinguish between two cases. Assume for simplicity that Max wins bidding ties. If Max loses the bidding, the energy decreases to $k-1$. His updated budget ratio is greater than $\frac{T_{k+1}}{S_{k+1} - 1} = \frac{T_{k}}{S_{k}}$. For example, $\frac{T_5}{S_5 -1} = \frac{15}{24} = \frac{10}{16} = \frac{T_4}{S_4}$. On the other hand, if he wins the bidding, his updated budget ratio is greater than $\frac{T_{k+1}-1}{S_{k+1} - 1} = \frac{T_{k+2}}{S_{k+2}}$. For example, $\frac{T_5-1}{S_5 -1} = \frac{14}{24} = \frac{21}{36} = \frac{T_6}{S_6}$.

We claim that the invariant implies that it is always the case that $k > 0$. To see this, assume that $k=1$. We claim that Max wins the bidding, thus in the next turn, the energy increases and will not reach $0$. By the invariant, Max's budget ratio is greater than $\frac{3}{4}$ and he bids $\frac{1}{4}$ (one box in the $2 \times 2$ square). Note that Min owns less than one box. That is, Max bids more than her entire budget, thus he necessarily wins the bidding. 
\end{proof}

\subsubsection{General strongly-connected games}

We construct $\epsilon$-optimal strategies for Max in general strongly-connected games. 

\begin{lemma}
\label{16-lem:MP-poorman}
Let $\game$ be a strongly-connected mean-payoff poorman-bidding game and $B_0$ be an initial Max budget such that $r = \frac{B_0}{B_0+1} \in (0,1)$.
For every  $\epsilon > 0$, Max can guarantee a payoff of at least $\Value\big(\RT(\game, \frac{B_0-\epsilon}{B_0+1})\big)$.
\end{lemma}
\begin{proof}
Let $B_0 \in \R$ be Max's initial budget.
Let $\epsilon> 0$, and let $W = B_0-\epsilon$. Let $B$ denote Max's budget following a finite play. 
We construct Max's strategy so that it maintains $B > W$. 
Throughout this proof, we keep Min's budget normalized to $1$.
That is, suppose that Max bids $x \in [0,1)$ and Min bids $y \in [0,1)$, then 
if Max wins the bidding ($x>y$), his budget is updated to $B'= B-b$, and if Min wins the bidding ($x \leq y$), then $B' = \frac{B}{1-y}$. 
A key insight is that 
when Max loses, we have 
\begin{equation}\label{16-equation:budget_FPP_approx}
B' = \frac{B}{1-y} > \frac{B(1-y^2)}{1-y} = B(1+y) > B + Wy.
\end{equation}
Intuitively, the property states that every cent is $W$ times more valuable to Min than it is to Max (see the tit-for-tat strategy under poorman bidding for further intuition).

\stam{
For example, if \Max's budget is $2$ and \Min's budget is $1$,
then paying $0.1$ is twice as painful for \Min as it is for \Max.
Roughly, on average, this means that \Max wins $W \sim B_0$ times more biddings than \Min,
thus he guarantees a payoff close to $\MP(\RT(\G, \frac{B_0}{B_0+1}))$.
}

Let $p= (B_0-\epsilon)/(B_0+1)$. We define a bidding strategy $f$ for Max
that guarantees a payoff of at least $\Value\big(\RT(\game, p)\big)$. We choose $\alpha \in (0,1)$ so that $(1-\alpha) = (1+\alpha)^{-(1+\epsilon)}$.
We find vertex strengths using $\RT(\game, p)$ as before. 
Let $N = \max(W,1) \cdot S_{\max}$.
The strategy $f$ is defined as follows.
\begin{itemize}
	\item When the token is placed on a vertex $v$ with strength $s = \St_p(v)$
	and Max's budget is $B$, Max bids $f(B,s) = \frac{\alpha\cdot s}{N}(B-W)$.
	\item Upon winning, Max moves the token to $v^+$.
\end{itemize}

We claim that Max's bidding strategy $f$ is legal, \textit{i.e.}, $f(B, s) \leq B$. We prove this by showing that it is always the case that $B>W$.
Indeed, observe that initially, we have $B_0 > W$,
 whenever Max loses a bidding his budget increases,
and when Max wins a bidding his updated budget is $B-f(B,s) = B - \frac{\alpha\cdot s}{N}(B-W)$,
which is still greater than $W$ since $\frac{\alpha\cdot s}{N}<1$.

For any finite play $\pi$, let
\[H(\pi) = (1+\epsilon) \cdot \invest_\mMax(\pi)- (B_0 - \epsilon) \cdot \gain_\mMax(\pi) -N \cdot \log_{1+\alpha}(\epsilon).\]
We claim that $f$ satisfies the following invariant:

\smallskip
\noindent{\bf Invariant:} Consider a finite play $\pi$ and let $B$ denote Max's budget following $\pi$. Then, 
\begin{equation}\label{16-eq:inv_FPPoorman}
    (B-W)^N \geq (1+\alpha)^{-H(\pi)}
\end{equation}
\smallskip

The proof proceeds by induction on the length of $\pi$. It is not hard to verify that the invariant holds initially by recalling that $B_0-W = \epsilon$ and both investments and gains are $0$. For the induction step, suppose that the inequality holds for $B$ and $H$, and we show that it is maintained after the bidding for the updated values $B'$ and $H'$. Suppose that Min bids $y$ and Max bids $f(B,s)$. We distinguish between two cases. 

\begin{enumerate}
    \item
    Min wins the bidding ($y \geq f(B,s)$):  
    Then, Max's gain increases by $s$, thus $H' = H - (B_0 - \epsilon) \cdot s$. 
    \[
    \begin{array}{lcl}
    (B' - W)^N & \stackrel{\eqref{16-equation:budget_FPP_approx}}{>} & (B + W\cdot x - W)^N
    \geq (B - W + W \cdot f(B,s))^N\\
    & \stackrel{{\color{white}(}}{=} & (B-W)^{N}(1 + \frac{Ws}{N}\alpha)^N
    \stackrel{\text{ind. hyp.}}{\geq} (1 + \alpha)^{-H}(1 + \frac{Ws}{N}\alpha)^N\\
    & \stackrel{Bernoulli}{\geq} & (1 + \alpha)^{-H+Ws}
    = (1 + \alpha)^{-H+(B_0-\epsilon)s}\\
    & \stackrel{{\color{white}(}}{\geq} & (1 + \alpha)^{-H'}.
    \end{array}
    \]
    Bernoulli's inequality could be used since $\alpha > -1$ and
    $\frac{Ws}{N} = \min(\frac{Ws}{S_{max}},\frac{s}{S_{max}}) \in [0,1]$.
    We also used the fact that $W = B_0 - \epsilon$.

        \item Max wins the bidding ($x < f(B,s)$): Then, Max's investment increases by $s$, thus $H' = H + (1 + \epsilon)s$. Recall that $\alpha$ is chosen so that $(1-\alpha) = (1+\alpha)^{-(1+\epsilon)}$. 
    \[
    \begin{array}{lcl}
    (B' - W)^N & \stackrel{}{=} & (B - W - f(B,s))^N\\
    & \stackrel{{\color{white}(}}{=} & (B - W)^N(1-\frac{s}{N}\alpha)^N
    \stackrel{\text{ind. hyp.}}{\geq} (1 + \alpha)^{-H}(1-\frac{s}{N}\alpha)^N\\
    & \stackrel{Bernoulli}{\geq} &  (1 + \alpha)^{-H}(1-\alpha)^s
    = (1 + \alpha)^{-H-(1+\epsilon)s}\\
    & \stackrel{{\color{white}(}}{\geq} & (1 + \alpha)^{-H'}.
    \end{array}
    \]
    Bernoulli's inequality could be used since $-\alpha > -1$ and
    $\frac{s}{N} = \min(\frac{s}{S_{max}},\frac{s}{WS_{max}}) \in [0,1]$.
\end{enumerate}

The invariant implies a lower bound on $H(\pi)$.
Intuitively, recall that Min's budget is normalized to $1$ and Max's bid is a function of $(B-W)$. Thus, as $H(\pi)$ decreases, Max's bids increase, and eventually it exceeds $1$, in which case Max wins the bidding, causing $H(\pi)$ to increase. We obtain the following claim.

\stam{
Since the left-hand side of the equation is equal to $H(\pi) + N \cdot \log_{1+\alpha}(\epsilon)$,
proving the claim is equivalent to proving a lower bound for $H(\pi)$.
To do so, we show that $H(\pi)$ cannot get too low,
as past some threshold Equation \eqref{eq:inv_FPPoorman}
guarantees that the budget $B(\pi)$ of Max is so high that
his next bid according to the strategy $f$
will be above $1$ (the whole budget of Min).
Thus Max is guaranteed to win the next bidding, which results in $H(\pi)$ going back up.
}

\smallskip
\noindent{\bf Claim:} There exists $M \in \R$ such that 
for every finite play $\pi$ coherent with $f$, we have
\begin{equation}\label{16-eq:bound_FPPoorman}
(1+\epsilon) \invest_\mMax(\pi) - (B_0-\epsilon) \gain_\mMax(\pi) \geq M.
\end{equation}
\smallskip

We combine the claim with \Cref{16-cor:magic}
(plugging $\nu = B_0-\epsilon$ and $\mu=1+\epsilon$),
to obtain that the payoff of every infinite play is at least greater than $\Value\big(\RT(\game, p)\big)$.
\stam{
Combining Claim \ref{eq:bound_FPPoorman} with Corollary~\ref{cor:magic}
(plugging $\nu = B_0-\epsilon$ and $\mu=1+\epsilon$),
we finally obtain that any infinite play coherent with the strategy $f$
has a mean-payoff greater than $\Value(\RT(\G, (B_0-\epsilon)/(B_0+1)))$.
}
\end{proof}

Since the definition of mean payoff favors Min, we can apply Max's construction in \Cref{16-lem:MP-poorman} to Min, thus we obtain the following.

\begin{theorem}
\label{16-thm:MP-poorman-max}
Consider a strongly-connected mean-payoff poorman-bidding game $\game$ and an initial budget ratio $r \in (0,1)$. Then,  $\MP(\game, r) = \Value\big(\RT(\game, r)\big)$. 
\end{theorem}

We conclude this section by stating the equivalence between strongly-connect mean-payoff taxman-bidding games and random-turn games. 

\begin{theorem}
\label{16-thm:MP-taxman}
Consider a strongly-connected mean-payoff taxman-bidding game $\game$ with taxman parameter $\tau \in [0,1]$ and an initial budget ratio $r \in (0,1)$.
Then,  \[\Value(\game, r) = \Value\big(\RT(\game,\frac{r+\tau\cdot (1-r)}{1+\tau})\big) = \frac{r+\tau\cdot (1-r)}{1+\tau}.\]
\end{theorem}

\subsection{General Mean-payoff games}

We study the following decision problem: 
\begin{problem}
MEAN-PAYOFF: Given a mean-payoff game $\game$, a vertex $v$ in $\game$, a ratio $r \in (0,1)$, and a constant $c \in \Q$, deciding whether Max can guarantee a payoff greater than $c$ from $v$ with an initial ratio that exceeds $r$.
\end{problem}

We reduce general mean-payoff games to reachability games, similar to the case of parity continuous-bidding games (\Cref{16-sec:continuous-parity}).

\begin{theorem}
\label{16-thm:complexity-MP}
The problem MEAN-PAYOFF is:
\begin{itemize}
\item In PSPACE for taxman bidding and general graphs.
\item In NP $\cap$ coNP for Richman bidding, or with taxman bidding when $\game$ is strongly-connected.
\item In P when $\game$ is strongly-connected and the out-degree of all vertices is $2$.
\end{itemize}
\end{theorem}
\begin{proof}
(Sketch) When $\game$ is strongly-connected, we construct and solve the appropriate random-turn game and obtain the value of the bidding game by \Cref{16-thm:MP-taxman}. For general Richman games, we first find the mean-payoff values in the BSCC of the game. Recall that the mean-payoff value does not depend on the initial ratio in a strongly-connected game. In the vertices that do not belong to a BSCC, we find the threshold by solving a reachability game in which Max's goal is to reach a BSCC whose mean-payoff value exceeds $c$. For poorman and taxman bidding, the ideas are similar though more complicated. For a BSCC $S$, we need to find a ratio $r(S)$ such that Max can guarantee a payoff greater than $c$ with $r(S)$. We then solve a frugal-reachability game in which Max wins if he reaches $S$ with a budget that exceeds $r(S)$. This can be done in PSPACE using by a reduction to the existential theory of the reals. 
\end{proof}


\section*{Bibliographic references}
\label{16-sec:references}

\paragraph{Reachability games.} 
Bidding games were introduced in two papers in the 90s~\cite{Lazarus.Loeb.ea:1996}. The results on reachability games in Section~\ref{16-sec:qualitative_continuous} are due to these papers, including the equivalence with random-turn games (called ``Spinner games'' there). Random-turn games have been extensively studied (see~\cite{Peres.Schramm.ea:2007}). Combinatorial bidding games have been studied in~\cite{Kant.Larsson.ea:2024}.

\paragraph{Infinite-duration bidding games.}
The study of infinite-duration bidding games was initiated in~\cite{Avni.Henzinger.ea:2019}, which establishes the equivalence between mean-payoff Richman-bidding games and (unbiased) random-turn games (\Cref{16-thm:MP-min} and~\Cref{16-thm:MP-max}). The equivalence for poorman bidding (\Cref{16-thm:MP-poorman-max}) is established in~\cite{Avni.Henzinger.ea:2018} and the generalization to taxman bidding (\Cref{16-thm:MP-taxman}) in~\cite{Avni.Jecker.ea:2021}. Recall that it suffices to construct a Max $\epsilon$-optimal strategy and that a construction consists of two ingredients: a solution to the bowtie game, which provides a {\em normalization scheme} and the technique based on potentials and strengths to extend to general strongly-connected games. The normalization scheme presented in~\cite{Avni.Henzinger.ea:2019} is explicit and in~\cite{Avni.Jecker.ea:2021} it is existential. In all schemes, the normalization is computed with respect to the current energy (see the second construction in ~\Cref{16-fig:bowtie}). On the other hand, in the normalization scheme that we present (\Cref{16-prop:Max-bowtie} and~\Cref{16-prop:Max-bowtie}), the normalization is computed based on the current budget and the credit for it is due to Isma\"el Jecker~\cite{Avni.Jecker.ea:2021}. 

\paragraph{Discrete bidding.}
The model of discrete-bidding games that we present in ~\Cref{16-sec:qualitative_discrete} is due to~\cite{Brazdil.Jancar.ea:2010}, which studies reachability games and establishes existence of threshold budgets and their characterisation (\Cref{16-thm:disc-reach}). We focus on the {\em advantage-based} tie-breaking mechanism defined there. Other tie-breaking mechanisms were defined in~\cite{Aghajohari.Avni.ea:2021}, which develops a framework for proving determinacy that does not depend on a characterisation of thresholds and that extends to Muller objectives. The characterisation of thresholds in parity discrete-bidding games (\Cref{16-sec:disc-thresholds}) and membership in NP and coNP (\Cref{16-sec:disc-NP-coNP}) is due to~\cite{Avni.Sadhukhan:2025}. 
Poorman discrete-bidding games are studied in~\cite{Avni.Meggendorfer.ea:2023}. 

\paragraph{All-pay bidding.}
All mechanisms so far are {\em first-price}: only the higher bidder pays their bid. 
In {\em all-pay bidding} both players pay their bids. Reachability all-pay bidding games were studied in~\cite{Avni.Ibsen-Jensen.ea:2020}. They naturally model a dynamic investment of bounded resources with no inherent value. Thus, they constitute a dynamic variant of the well-known Colonel Blotto games. We state below equivalences between mean-payoff all-pay bidding games and random-turn games. We distinguish between the requirement to follow {\em pure} strategies, namely strategies that always prescribe a single action as has been the case throughout this chapter, and {\em mixed} strategies, namely strategies that prescribe a probability distribution over actions. For the latter, we use  the {\em almost sure} mean-payoff value, denoted $\textsf{asMP}$, which is roughly defined as the payoff that Max can guarantee with a mixed strategy with probability $1$.

\begin{theorem}\cite{Avni.Jecker.ea:2021}
Consider a strongly-connected mean-payoff game $\G$ and a budget ratio $r \in (0,1)$. The following equivalences apply:
\begin{itemize}
\item Under all-pay Richman bidding: For pure strategies, $\MP(\G, r) = \MP\big(\RT(\G, 0)\big)$, and for  mixed strategies, $\textsf{asMP}(\G, r) = \MP\big(\RT(\G, 0.5)\big)$.
\item Under all-pay poorman bidding: For pure strategies, if $r > 0.5$, then $\MP(\G, r) = \MP\big(\RT(\G, \frac{2r-1}{r})\big)$, and if $r \leq 0.5$, then $\MP(\G, r) = \MP\big(\RT(\G, 0)\big)$.
For mixed strategies, if $r > 0.5$, then $\textsf{asMP}(\G, r) = \MP\big(\RT(\G, \frac{3r-1}{r})\big)$, and if $r \leq 0.5$, then $\textsf{asMP}(\G, r) = \MP\big(\RT(\G, \frac{1-r}{r})\big)$.
\end{itemize}
\end{theorem}

\paragraph{Non-zero-sum games.}
In a non-zero-sum bidding game played on a tree~\cite{Meir.Kalai.ea:2018}, each leaf is labelled by a pair of payoffs for the players, and~\cite{Meir.Kalai.ea:2018} studies the existence and computation of a {\em subgame perfect equilibrium}. 
{\em Auction-based scheduling}~(ABS) is a modular planning framework based on bidding games~\cite{Avni.Mallik.ea:2024}. Consider a graph $\G$ with two non-contradicting objectives $\psi_1$ and $\psi_2$. The idea is to construct and solve two bidding games, where for $i \in \set{1,2}$, the game $\G_i$ is played on $\G$ with a Eve objective $\psi_i$. If the sum of thresholds is strictly less than $1$, two winning bidding strategies can be composed at runtime and the play that they generate necessarily satisfies $\psi_1 \wedge \psi_2$. 

\paragraph{Variants.}
A bidding game with partial observation of the budgets~\cite{Avni.Jecker.ea:2023} is given by a graph and two distributions $D_1$ and $D_2$. Before the game starts, each player's budget is drawn from the corresponding distribution. The goal is to maximize the expected payoff w.r.t. $D_1$ and $D_2$. 
In a bidding games with charging~\cite{Avni.Goharshady.ea:2024}, each vertex $v$ is labelled by a pair of weights $w_1$ and $w_2$. When $v$ is entered with player budgets $B_1$ and $B_2$, first, for $i \in \set{1,2}$, each budget is {\em charged} to $B_i + w_i$, then the budgets are normalized so their sum is $1$. 
In a bidding game on {\em Markov decision process}~\cite{Avni.Henzinger.ea:2019*1}, in each turn, the players bid for which action is chosen, which gives rise to a probability distribution over successor vertices.


\section*{Acknowledgments}
We would like to thank all our collaborators Milad Aghajohari, Ventsislav Chonev, Ehsan Kafshdar Goharshady, Rasmus Ibsen-Jensen, Ism\"ael Jecker, Kaushik Mallik, Tobias Meggendorfer, Petr Novotn\'y, Suman Sadhukhan, Josef Tkadlec, and {\DJ}or{\dj}e \v{Z}ikeli\'c; we hope the collaboration was as fun and meaningful for you as it was for us. This work is supported in part by ERC Advanced Grant VAMOS 101020093 and ISF grant no. 1679/21.


\backmatter

\addtocontents{toc}{\vspace{\baselineskip}}

\renewcommand{\refname}{} 

\copyrightline{}
\chapter*{References}
\pagestyle{empty} 
\bibliography{acronyms,bib}
\bibliographystyle{cambridgeauthordate}

\end{document}